\documentclass[11pt,hidelinks,dvipsnames]{trailthesis}

\newif\iffulldoc
\fulldoctrue

\usepackage{etex}
\usepackage{graphicx}
\usepackage{makecell}
\usepackage{layout}

\usepackage[sectionbib]{bibunits}

\usepackage{pdfpages}

\usepackage[authoryear]{natbib}

\usepackage{lscape}

\usepackage{todonotes}
\usepackage{soul}

\makeatletter
\if@todonotes@disabled%

\else

\fi
\makeatother

\usepackage{changepage}

\usepackage{caption}

\usepackage{bibentry}

\usepackage{multirow}
\usepackage{booktabs}

\usepackage{pdflscape}
\usepackage{hhline}
\usepackage[shortlabels]{enumitem}
\usepackage{float}
\usepackage{threeparttable}
\usepackage{tikz} 
\usepackage{CJKutf8} 
\usepackage[utf8]{inputenc}
\usepackage{wrapfig}

\newcommand{\lamc}{$\lambda$-calculus\xspace}

\usepackage{amsfonts}
\usepackage{amsmath}
\usepackage{xspace}
\usepackage{mathtools}
\usepackage{bussproofs}
\usepackage{stmaryrd}
\usepackage{enumitem}
\usepackage{multicol}
\usepackage{tikz}
\usetikzlibrary{matrix}
\usetikzlibrary{decorations.pathreplacing,calligraphy}
\usetikzlibrary{shapes.geometric, arrows}

\usepackage{thmtools} 
\usepackage{thm-restate}
\usepackage{hyperref}
\usepackage{cleveref}

\newtheorem{lemma}{Lemma}[section]
\newtheorem{notation}{Notation}[section]

\newtheorem{remark}{Remark}[section]
\newtheorem{proposition}{Proposition}[section]
\newtheorem{corollary}{Corollary}[section]

\usepackage{hyperref}

\newcommand{\spi}{\ensuremath{\mathsf{s}\pi}\xspace}
\newcommand{\lamr}{\ensuremath{\lambda_{\oplus}}\xspace} 
\newcommand{\lamrshar}{\ensuremath{\widehat{\lambda}_{\oplus}}\xspace} 
\newcommand{\lamrfail}{\ensuremath{\lambda^{\lightning}_{\oplus}}\xspace}
\newcommand{\lamrfailunres}{\ensuremath{\lambda^{! \lightning}_{\oplus}}\xspace}
\newcommand{\lamrsharfail}{\ensuremath{\widehat{\lambda}^{\lightning}_{\oplus}}\xspace} 
\newcommand{\lamrsharfailunres}{\ensuremath{\widehat{\lambda}^{!\lightning}_{\oplus}}\xspace}

\newcommand{\unvar}[1]{ #1 ^{!}  }

\newcommand{\sep}{\ | \ } 
\newcommand{\oneb}{\mathtt{1}}
\newcommand{\perm}[1]{\ensuremath{\mathsf{PER}(#1)}}
\newcommand{\dom}[1]{\mathtt{dom}(#1)}
\newcommand{\arrt}[2]{\ensuremath{#1 \rightarrow #2}}
\newcommand{\fail}{\mathtt{fail}}

\newcommand{\concat}{\ensuremath{\diamond}}
\newcommand{\relunbag}{\ensuremath{\sim}}
\newcommand{\contexcat}{\ensuremath{\wedge}}

\newcommand{\size}[1]{\mathsf{size}(#1)} 
\newcommand{\headf}[1]{\ensuremath{\mathsf{head}(#1)}}
\newcommand{\headfsum}[1]{\ensuremath{\mathsf{head}_{\sum}}(#1)}

\newcommand{\dash}{\text{-}}

\newcommand{\shar}[2]{[#1\leftarrow #2]}

\def\subst#1#2{\{ \raisebox{.5ex}{\small$#1$}\! / \mbox{\small$#2$}\}} 
\def\linsub#1#2{\langle \raisebox{.5ex}{\small$#1$}\! / \mbox{\small$#2$}\rangle} 
\def\esubst#1#2{\langle\!\langle \raisebox{.5ex}{\small$#1$}\! / \mbox{\small$#2$}\rangle\!\rangle} 

\newcommand{\headlin}[1]{ { \{\!|} #1 { |\!\} }} 
\newcommand{\linexsub}[1]{{\langle \! |} #1 {| \! \rangle} }

\newcommand{\expr}[1]{\ensuremath{\mathbb{#1}}}
\newcommand{\lfv}[1]{\mathsf{fv}(#1)}
\newcommand{\llfv}[1]{\mathsf{lfv}(#1)}
\newcommand{\mfv}[1]{\mathsf{mfv}(#1)}
\newcommand{\mlfv}[1]{\mathsf{mlfv}(#1)}
\newcommand{\unit}{\mathbf{unit}}
\newcommand{\head}[1]{\mathsf{head}(#1)} 

\newcommand{\redd}{\longrightarrow}
\newcommand{\tred}{\stackrel{*}{\redd}}
\newcommand{\redlab}[1]{\ensuremath{\mathtt{[#1]} }}

\newcommand{\pequiv}{\equiv_\lambda}

\newcommand{\secref}[1]{$\S$\,\ref{#1}\xspace}
\newcommand{\figref}[1]{Fig.\,\ref{#1}\xspace}
\newcommand{\defref}[1]{Def.\,\ref{#1}\xspace}
\newcommand{\appref}[1]{App.\,\ref{#1}\xspace}
\newcommand{\thmref}[1]{Theorem~\ref{#1}\xspace}

\newcommand{\wfdash}{\models}
\newcommand{\core}[1]{#1^\dagger}
\newcommand{\strcore}[1]{\widehat{#1}^\dagger}

\newcommand{\out}[1]{\langle #1\rangle} 
\newcommand{\outact}[2]{\overline{#1}(#2)}
\newcommand{\outsev}[2]{\overline{#1}?(#2)}

\newcommand{\some}{\mathtt{some}}
\newcommand{\none}{\mathtt{none}}

\newcommand{\case}[2]{{#1}.\mathtt{l}_{#2}}
\newcommand{\choice}[5]{{#1}.\mathtt{case}_{#2 \in #3} \{ \mathtt{l}_{#4} : #5 \}}

\newcommand{\close}{\mathtt{close}}

\newcommand{\fn}[1]{\mathit{fn}(#1)}

\newcommand{\ampy}{\mathbin{\bindnasrepma}}
\newcommand{\with}{{\binampersand}}
\newcommand{\onef}{\mathbf{1}}
\newcommand{\dual}[1]{\overline{#1}}

\newcommand{\colorone}[1]{\textcolor{orange}{#1}}

\newcommand{\colorthree}[1]{\textcolor{purple}{#1}}

\newcommand{\encod}[2]{\llbracket#1\rrbracket_{#2}} 
\newcommand{\recencod}[1]{\llparenthesis #1 \rrparenthesis^{+}_{\bullet} } 

\newcommand{\recencodf}[1]{\colorone{\llparenthesis}  #1 \colorone{\rrparenthesis^{\bullet}}} 

\newcommand{\recencodopenf}[1]{\colorone{\llparenthesis}  #1 \colorone{\rrparenthesis^{\circ}}}

\newcommand{\piencodf}[1]{\colorthree{\llbracket}  #1 \colorthree{\rrbracket}^{\colorthree{\lightning}}}

\newcommand{\piencod}[1]{\llbracket  #1 \rrbracket}

\newcommand{\linsetminus}{\setminus \!\! \setminus}

\newcommand{\succp}[2]{\ensuremath{#1 \Downarrow_{#2}}} 

\newcommand{\bluetext}[1]{\textcolor{RoyalBlue}{#1}}
\newcommand{\cred}[1]{\textcolor{BrickRed}{#1}}

\newcommand{\revo}[2]{#2}
\newcommand{\revd}[2]{#2}

\newcommand{\revt}[2]{#2}
\newcommand{\revdaniele}[1]{#1}
 \newcommand{\secondrev}[1]{#1}
\newcommand{\srev}[1]{\secondrev{#1}}

 \usepackage{amsfonts}
\usepackage{amsmath,xspace}
\usepackage{mathtools}
\usepackage{bussproofs}
\usepackage{stmaryrd}
\usepackage{mdframed}
\usepackage{multicol}
 \mdfdefinestyle{alttight}{innertopmargin=-2pt,innerleftmargin=2pt,innerrightmargin=2pt,innerbottommargin=2pt}

\newcommand{\headsum}[1]{\ensuremath{\mathsf{head}_{\sum}}(#1)} 

\newcommand{\sclose}{\ensuremath{[\,]}}

\newcommand{\colortwo}[1]{\textcolor{RoyalBlue}{#1}}

\newcommand{\piencodftypes}[1]{\colortwo{\llbracket}  #1 \colortwo{\rrbracket}} 

\newcommand{\banged}[1]{ #1 ^{!}  }

\newcommand{\joehide}[1]{}
\newcommand{\alerthide}[1]{}

\newcommand{\colthree}[1]{\textcolor{blue}{ #1 }}

\usepackage{tikz}
\usetikzlibrary{matrix}
\usetikzlibrary{decorations.pathreplacing,calligraphy}
\usetikzlibrary{shapes.geometric, arrows}

\usepackage{amsfonts}
\usepackage{amsmath,xspace}
\usepackage{amssymb}
\usepackage{bm}
\usepackage{mathpartir}
\usepackage{mathtools}
\usepackage{thm-restate}
\usepackage{relsize}
\usepackage{bussproofs}
\usepackage{stmaryrd}
\usepackage{tikz}
\usetikzlibrary{matrix}
\usetikzlibrary{decorations.pathreplacing,calligraphy}
\usetikzlibrary{decorations.pathmorphing}
\usepackage{centernot}
\usepackage{xcolor}
\usepackage{hyperref}
\usepackage{cleveref}
\usepackage{crossreftools}

\usepackage{thmtools}

\definecolor{cblRed}{rgb}{.60,.00,.01}
\definecolor{cblBlue}{rgb}{.16,.23,.42}
\definecolor{cblBlueLt}{rgb}{.00,.56,.61}
\definecolor{cblYellow}{rgb}{1.00,.69,.00}
\definecolor{cblPink}{rgb}{.85,.00,.42}
\definecolor{cblPurple}{rgb}{.10,.00,.54}
\definecolor{cblPurpleLt}{rgb}{.76,.39,.89}
\definecolor{cblOrange}{rgb}{.74,.29,.00}
\definecolor{cblGreen}{rgb}{.38,.61,.27}

\hypersetup{
    hypertexnames=false,
    colorlinks=true,
    final,
    citecolor=cblGreen,
    linkcolor=cblRed,
    urlcolor=cblRed,
    bookmarksnumbered=true,
    bookmarksopen=true,
    bookmarksopenlevel=1,
    pdfpagemode=UseOutlines
}

\newcommand{\clrone}[1]{\textcolor{cblRed}{#1}}
\newcommand{\clrtwo}[1]{\textcolor{cblGreen}{#1}}
\newcommand{\pctxcolor}{cblBlueLt}

\newcommand{\nconf}{non-con\-flu\-ent\xspace}

\newcommand{\conf}{con\-flu\-ent\xspace}
\newcommand{\Conf}{Con\-flu\-ent\xspace}
\newcommand{\col}{\nconf}

\newcommand{\nond}{non-de\-ter\-min\-ism\xspace}
\newcommand{\nondt}{non-de\-ter\-min\-is\-tic\xspace}

\newcommand{\lamcoldet}{\ensuremath{\lambda_{\mathtt{C}}}\xspace}
\newcommand{\lamcoldetsh}{\ensuremath{\lambda_{\mathtt{C}}}\xspace}
\newcommand{\lamcoldetshlin}{\ensuremath{\lambda_{\mathtt{C}}^{\ell}}\xspace}

\newcommand{\linvar}[1]{ #1 ^{\ell}}
\newcommand{\lunvar}[1]{#1}
\newcommand{\bagsep}{\ast}

\newcommand{\bag}[1]{\lbag #1 \rbag}

\newcommand{\sharing}[2]{[#1 \leftarrow #2]}

\def\subst#1#2{\{ \raisebox{.5ex}{\small$#1$}\! / \mbox{\small$#2$}\}} 
\def\linsub#1#2{\langle \raisebox{.5ex}{\small$#1$}\! / \mbox{\small$#2$}\rangle} 
\def\esubst#1#2{\langle\!\langle \raisebox{.5ex}{\small$#1$}\! / \mbox{\small$#2$}\rangle\!\rangle} 

\newcommand{\ltalltriangle}{{\langle\!|}}
\newcommand{\rtalltriangle}{{|\!\rangle}}
\newcommand{\unexsub}[1]{\llfloor #1 \rrceil }

\renewcommand{\red}{\mathbin{\longrightarrow}}
\newcommand{\redone}{\ensuremath{\mathbin{\clrone{\longrightarrow}}}\xspace}
\newcommand{\redtwo}{\ensuremath{\mathbin{\clrtwo{\leadsto}}}\xspace}

\newcommand{\readyPrefixBisimSub}[1]{\mkern-1mu\mathtt{#1}}
\newcommand{\readyPrefixBisim}[1]{\sim_{\readyPrefixBisimSub{#1}}}
\newcommand{\nreadyPrefixBisim}[1]{\not\sim_{\readyPrefixBisimSub{#1}}}

\newcommand{\succone}{\mathbin{\clrone{\Downarrow}}}
\newcommand{\succtwo}{\mathbin{\clrtwo{\Downarrow}}}
\newcommand{\equivlam}{\equiv_\lambda}

\newcommand{\wtdash}{\vdash}

\newcommand{\para}{\mathord{\;\mathbf{|}\;}}
\newcommand{\zero}{{\bf 0}}

\newcommand{\fln}[1]{\mathit{fln}(#1)}

\newcommand{\piencodfaplas}[1]{{\textcolor{cblPink}{\llbracket}#1\textcolor{cblPink}{\rrbracket}}}
\newcommand{\piencodfscale}[1]{\color{cblPink}\left\llbracket \normalcolor #1 \color{cblPink}\right\rrbracket \normalcolor}

\newcommand{\premat}{\succeq_{\mkern-2mu \scalebox{.8}{$\nd$}} }
\newcommand{\premattwo}{\preceq_{\nd} }

\newcommand{\relalpha}{ \bowtie }

\newcommand{\readyPrefix}[1]{\downarrow_{#1}}
\newcommand{\nreadyPrefix}[1]{\centernot{\downarrow}_{\mkern-7mu#1}}
\newcommand{\dualPrefix}{\mathbin{\dual{\relalpha}}}

\newcommand{\bignd}{\scalebox{1.5}{$\nd\mkern-3mu$}}

\newcommand{\ol}[1]{\overline{#1}}
\newcommand{\0}{\bm{0}}
\newcommand{\fwd}{\mathbin{\leftrightarrow}}
\newcommand{\sepr}{\mathbin{\scalebox{1.3}{$\mid$}}}

\newcommand{\rredone}[1]{[\clrone{\shortrightarrow}_{#1}]}
\newcommand{\rredtwo}[1]{[\clrtwo{\rightsquigarrow}_{#1}]}
\newcommand{\ttype}[2][]{\text{\normalfont#1[\scc{T#2}]}}

\newcommand{\sff}[1]{\relax\ifmmode\mathsf{#1}\else\textsf{#1}\fi}
\newcommand{\mathsc}[1]{\text{\normalfont\scshape#1}}
\newcommand{\scc}[1]{\relax\ifmmode\mathsc{#1}\else\textsc{#1}\fi}

\newcommand{\tensor}{\ensuremath{\mathbin{\otimes}}}
\newcommand{\parr}{\mathbin{\rott{\with}}}
\newcommand{\rott}[1]{\mathpalette\rot{#1}}
\newcommand{\rot}[2]{\rotatebox[origin=c]{180}{$#1{#2}$}}

\renewcommand{\gets}{\mathbin{\triangleright}}

\newcommand{\subj}{\mathit{subj}}

\newcommand{\D}[1]{\textcolor{\pctxcolor}{\llparenthesis}\mkern1mu #1 \mkern1mu\textcolor{\pctxcolor}{\rrparenthesis}}

\newcommand{\hole}{[\cdot]}

\let\obigcup\bigcup
\renewcommand{\bigcup}{{\mathchoice{\textstyle}{}{}{}\obigcup}}
\let\osum\sum
\renewcommand{\sum}{{\mathchoice{\textstyle}{}{}{}\osum}}
\let\oprod\prod
\renewcommand{\prod}{\mathchoice{\textstyle}{}{}{}{\oprod}}

\newcommand{\prlplus}{\hbox{\rlap{$\mkern-.4mu{}\mathbin{|\mkern-3mu|}{}$}$-$}}
\newcommand{\nd}{\mathbin{\prlplus}}

\newcommand{\piprecong}[1]{\mathrel{\succeq_{#1}}}
\newcommand{\sub}[1]{\ensuremath{\mathit{subj}\{#1\}}}

\newcommand{\infAss}[1]{\AxiomC{#1}}

\newcommand{\infUn}[2]{\RightLabel{#2} \UnaryInfC{#1}}

\newcommand{\mtt}[1]{\mathtt{#1}}
\let\oprl\|
\renewcommand{\|}{\mathbin{|}}
\newcommand{\pclose}[1]{\ol{#1}[]}
\newcommand{\gclose}[1]{#1()}
\newcommand{\psome}[1]{\ol{#1}.\some}
\newcommand{\pnone}[1]{\ol{#1}.\none}
\newcommand{\gsome}[2]{{#1.\some}_{#2}}
\newcommand{\pname}[2]{\ol{#1}[#2]}
\newcommand{\puname}[2]{{?}\ol{#1}[#2]}
\newcommand{\gname}[2]{#1(#2)}
\newcommand{\guname}[2]{{!}#1(#2)}
\newcommand{\psel}[2]{\ol{#1}.#2}
\newcommand{\gsel}[1]{#1.\mtt{case}}
\newcommand{\pfwd}[2]{[#1 \fwd #2]}
\newcommand{\res}[1]{(\bm{\nu} #1)}
\newcommand{\1}{\bm{1}}
\newcommand{\clpi}{\ensuremath{{{\mathsf{s}\pi\mkern-2mu}^{+}\mkern-1mu}}\xspace}
\newenvironment{tarr}[2][t]{\begin{array}[#1]{@{}#2@{}}}{\end{array}}
\newcommand{\ih}[1]{IH\textsubscript{#1}\xspace}
\newcommand{\inferruleDbl}[3][]{\inferrule*[#1,fraction={===}]{#2}{#3}}
\newcommand{\bn}[1]{\mathit{bn}(#1)}
\newcommand{\fpn}[1]{\mathit{fpn}(#1)}
\newcommand{\disj}{\mathbin{\#}}
\newcommand{\refitem}[3]{\Cref{ch4#1:#2}.\labelcref{ch4i:#2#3}}
\renewcommand{\flat}[1]{\mathit{flat}(#1)}
\newcommand{\dashes}{\vspace{-1.25em}\hbox to \textwidth{\leaders\hbox to 3pt{\hss . \hss}\hfil}\smallskip}
\newcommand{\mcl}[1]{\mathcal{#1}}
\newcommand{\lctx}[1]{\mcl{#1}}
\NewDocumentCommand \pctx {s O{} m o} {\IfBooleanT{#1}{\textcolor{\pctxcolor}{\llparenthesis}\mkern1mu}\textcolor{\pctxcolor}{\mathtt{#3}}\IfBooleanT{#1}{\mkern1mu\textcolor{\pctxcolor}{\rrparenthesis}} \IfValueT{#4}{\textcolor{\pctxcolor}{#2[}#4\textcolor{\pctxcolor}{#2]}}}
\newcommand{\sucs}[1]{{\checkmark\mkern-4mu}_{#1}}

\def\shred{\mathbin{\shortrightarrow}}
\def\ndots{\hbox to 1em{.\hss.\hss.}}

\newcommand{\myDots}{\ifmmode\mathinner{\ldotp\kern-0.2em\ldotp\kern-0.2em\ldotp}\else.\kern-0.13em.\kern-0.13em.\fi}

\newcommand{\myrev}[1]{{#1}}
\newcommand{\myrevopt}[1]{}
\newcommand{\mysmall}{\small}

\usepackage[utf8]{inputenc}
\usepackage{mathpartir}
\usepackage{bussproofs}
\usepackage{arydshln}

\newcommand{\jspace}{\vspace{-2mm}}

\usepackage{color}

\usepackage[utf8]{inputenc}
\usepackage{subfigure}
\usepackage{amsmath,amsfonts}
\usepackage{proof}
\usepackage{graphicx}
\usepackage{multicol}
\usepackage{listings}
\usepackage{url}
\usepackage{algpseudocode}
\usepackage{algorithmicx}
\usepackage{wrapfig}
\usepackage{enumitem}
\usepackage{textcomp}
\usepackage{latexsym}
\usepackage{epsfig} 
\usepackage{xcolor}
\usepackage{bbm}
\usepackage{etoolbox}

\usepackage{stmaryrd}
\usepackage{xspace}

\newcommand{\wn}{\ensuremath{?}}
\newcommand{\oc}{\ensuremath{!}}
\newcommand{\vasco}{\ensuremath{\pi_\mathsf{S}}\xspace}

\newcommand{\contcomp}{\circ} 
\newcommand{\emptycont}{\ensuremath{\emptyset}}
\newcommand{\contun}[1]{\m{un} ( #1 ) }
\newcommand{\contlin}[1]{\m{lin} ( #1 )}
\newcommand{\Rec}[2]{\mu #1. #2}
\newcommand{\recur}[1]{ \ast\,#1  }
\newcommand{\dill}{ \vdash_{\textcolor{darkgray}{\ell}}}

\newcommand{\outsubj}[1]{ \m{os} ( #1 )  }
\newcommand{\nilT}{\ensuremath{{\bf end}}}
\renewcommand{\st}{ \vdash_{\textcolor{darkgray}{\mathsf{s}}}}
\newcommand{\lt}{ \vdash_{\textcolor{darkgray}{\mathsf{w}}}}

\newcommand{\rlevel}[2]{ \ast \#^{#1} #2 }
\newcommand{\lts}[1]{ \xrightarrow[]{#1} }
\newcommand{\names}[1]{\mathsf{n}({#1})}		   
\newcommand{\myl}{\ensuremath{l}}
\newcommand{\levelof}[1]{\myl({#1})}		   
\newcommand{\lvlass}{\ensuremath{\diamond}}
\newcommand{\weight}[1]{\mathsf{wt}({#1})}		   
\newcommand{\wtless}{ \prec }            
\newcommand{\wtleq}{ \preceq }            
\newcommand{\zerovec}{ \ensuremath{\mathtt{0}} }            

\newcommand{\dillcontrel}[2]{\ensuremath{\asymp^{\,#1}_{\,#2}}}
\newcommand{\vaslinunsplit}{\ensuremath{\circledast}}
\newcommand{\pidill}{\ensuremath{\pi_{\sf DILL}}\xspace}
\newcommand{\pilvl}{\ensuremath{\pi_{\sf W}}\xspace}
\newcommand{\vasout}[3]{\ensuremath{  \ov #1 \out {#2} . #3  }}
\newcommand{\vasin}[4]{\ensuremath{   #1 \  #2\inp {#3}.#4  }}
\newcommand{\vaspara}[2]{\ensuremath{ #1 \pp #2 }}
\newcommand{\vasres}[2]{\ensuremath{  \res {#1} #2  }}

\newcommand{\vassend}[2]{\ensuremath{  \oc #1 . #2  }}
\newcommand{\vasreci}[2]{\ensuremath{  \wn #1 . #2  }}
\newcommand{\vasred}{\longrightarrow}
\newcommand{\vaslang}{\ensuremath{\mathcal{S}}\xspace}

\newcommand{\fv}[1]{\mathsf{fv}(#1)}   
\newcommand{\bv}[1]{\mathsf{bv}(#1)}   

\newcommand{\vastype}[1]{\ensuremath{\mathtt{[S{:}#1]} }}

\makeatletter
\def\@tempa#1{\@xp\@tempb\meaning#1\@nil#1}
\def\@tempb#1>#2#3 #4\@nil#5{%
  \@xp\ifx\csname#3\endcsname\mathaccent
    \@tempc#4?"7777\@nil#5%
  \else
    \PackageWarningNoLine{amsmath}{%
      Unable to redefine math accent \string#5}%
  \fi
}
\def\@tempc#1"#2#3#4#5#6\@nil#7{%
  \chardef\@tempd="#3\relax\set@mathaccent\@tempd{#7}{#2}{#4#5}}

\@tempa\widehat

\newcommand{\lvlint}[3]{\ensuremath{ \#^{#1} ( #2 , #3) }}
\newcommand{\lvloutt}[3]{\ensuremath{ \#^{#1} \langle #2 , #3 \rangle  }}
\newcommand{\lvlunst}[2]{\ensuremath{ \ast \#^{#1} (#2)  }}
\newcommand{\lvlintpoly}[3]{\ensuremath{ \#^{#1} ( \tilde{#2} ) }}

\newcommand{\lvlunct}[2]{\ensuremath{ \ast \#^{ #1  } \langle #2 \rangle  }}
\newcommand{\lvlinpoly}[4]{\ensuremath{  #1 \inp {\tilde{#2}}. #4  }}
\newcommand{\lvloutpoly}[4]{\ensuremath{  \ov #1 \out {\tilde{#2} }.#4  }}
\newcommand{\lvlservpoly}[4]{\ensuremath{  \bang #1 \inp {\tilde{#2} }. #4  }}
\newcommand{\lvlin}[4]{\ensuremath{  #1 \inp {{#2}, #3}. #4  }}
\newcommand{\lvlout}[4]{\ensuremath{  \ov #1 \out {{#2}, #3 }.#4  }}
\newcommand{\lvlserv}[4]{\ensuremath{  \bang #1 \inp {{#2}, #3 }. #4  }}
\newcommand{\lvlsplit}[2]{\ensuremath{ #1 :: #2   }}
\newcommand{\dualjoin}[2]{\ensuremath{ \langle #1 , #2 \rangle    }}
\def\lvljoinsub#1#2#3{\ensuremath{(#1)\lceil \raisebox{.5ex}{\small$#2$}\! / \mbox{\small$#3$}\rfloor }}
\newcommand{\lvlnilT}{\ensuremath{{\bf unit}  }}
\newcommand{\lvllang}{\ensuremath{\mathcal{W}}\xspace}
\newcommand{\lvltype}[1]{\ensuremath{\mathtt{[W{:}#1]} }}

\newcommand{\pclosu}[1]{{{>}_{#1}^{\! *}}}
\newcommand{\abbgrdot}{{}\gtrdot \!\!\!\!>}
\newcommand{\abbgrdotstar}[1]{{\gtrdot \!\!\!\!>^*_{#1}}}

\newcommand{\dillint}[2]{\ensuremath{ #1  \lolli #2 }}
\newcommand{\dilloutt}[2]{\ensuremath{ #1 \tensor #2  }}
\newcommand{\dillunt}[1]{\ensuremath{  \bang #1 }}
\newcommand{\dillchoicet}[2]{\ensuremath{ #1 \oplus #2  }}
\newcommand{\dillselet}[2]{\ensuremath{ #1  \&   #2  }}
\newcommand{\dillnilT}{\ensuremath{ \one }}
\newcommand{\dillred}{\longrightarrow}
\newcommand{\dillin}[3]{\ensuremath{  #1 \inp {#2 }. #3  }}
\newcommand{\dillbout}[3]{\ensuremath{  \ov #1 (#2).#3 }}
\newcommand{\dillout}[3]{\ensuremath{  #1 \out { #2 }.#3 }}
\newcommand{\dillserv}[3]{\ensuremath{  \bang #1 \inp {#2  }. #3  }}
\newcommand{\dillchoice}[3]{\ensuremath{ #1. \mathbf{case} (#2 , #3)  }}
\newcommand{\dillselel}[2]{\ensuremath{ #1. \mathbf{inl} ;#2   }}
\newcommand{\dillseler}[2]{\ensuremath{ #1. \mathbf{inr} ;#2  }}
\newcommand{\dillforward}[2]{\ensuremath{ [ #1 \leftrightarrow #2 ] }}
\newcommand{\dillfwdbang}[2]{\ensuremath{ \bang [ #1 \leftrightarrow #2 ] }}
\newcommand{\dilllang}{\ensuremath{\mathcal{L}}\xspace}
\newcommand{\dillnotun}[1]{\ensuremath{{#1}^{\textcolor{red}{\dagger}} }}
\newcommand{\dillnotunsingular}[1]{\ensuremath{\dagger(#1)}}
\newcommand{\server}[1]{\ensuremath{\mathsf{svr} ( #1 )  }}
\newcommand{\client}[1]{\ensuremath{\mathsf{cli} ( #1 )  }}
\newcommand{\notserver}[1]{\ensuremath{\neg \server{#1}  }}
\newcommand{\notclient}[1]{\ensuremath{ \neg \client{#1}  }}
\newcommand{\dilltype}[1]{\ensuremath{\mathtt{[L{:}#1]} }}
\newcommand{\ttrue}{\ensuremath{\mathtt{true}}\xspace}
\newcommand{\ffalse}{\ensuremath{\mathtt{false}}\xspace}

\newcommand{\sttoltt}[3]{ \textcolor{blue}{\llparenthesis} #1 \textcolor{blue}{\rrparenthesis}_{#2}^{#3}}
\newcommand{\sttoltp}[2]{\ensuremath{\textcolor{blue}{\langle\!|} #1 \textcolor{blue}{|\!\rangle}^{#2}}}
\newcommand{\sttoltj}[3]{\textcolor{blue}{\left\llbracket \textcolor{black}{#1} \right\rrbracket}_{#2}^{#3}}

\newcommand{\sttodillt}[1]{ \textcolor{red}{\llparenthesis} #1 \textcolor{red}{\rrparenthesis}}
\newcommand{\sttodillp}[1]{\ensuremath{\textcolor{red}{\langle\!|} #1 \textcolor{red}{|\!\rangle}}}
\newcommand{\sttodillj}[1]{\textcolor{red}{\left\llbracket \textcolor{black}{#1} \right\rrbracket}}

\newcommand{\abconditionO}[6]{\ensuremath{\dagger}}
\newcommand{\abconditionT}[7]{\ensuremath{\ddagger}}
\newcommand{\abconditionTstar}[9]{\ensuremath{\ddagger^\star}}

\newcommand{\m}[1]{\mathsf{#1}}

\newcommand{\defeq}{\triangleq}

\newcommand{\bang}{\boldsymbol{!}}

\newcommand{\nil}{{\mathbf{0}}}

\newcommand{\inp}[1]{(#1)}
\newcommand{\pp}{\ {|}\ }
\newcommand{\un}{\m{un}}
\newcommand{\lin}{\m{lin}}

\newcommand{\true}{\ensuremath{\mathtt{true}}}
\newcommand{\false}{\ensuremath{\mathtt{false}}}

\newcommand{\e}{\nilT}

\newcommand{\Didtext}[1]{{\scriptsize{\textsc{#1}}}}
\newcommand{\Did}[1]{({\Didtext{#1}})}

\newcommand{\lolli}{\mathord{\multimap}}

\newcommand{\one}{\ensuremath{{\bf{ 1}}}}

\newcommand{\ov}[1]{\overline{#1}}
\def\substj#1#2{[\raisebox{.5ex}{\small$#1$}\! / \mbox{\small$#2$}]}

\newcommand{\tra}[1]{\xrightarrow{#1}}
\newcommand{\reddpp}{\tra{~~~}}

\newcommand{\cc}[2]{\ensuremath{\mathsf{cc}( #1 , #2 ) }}

\newcommand{\orvastype}[1]{\ensuremath{\mathtt{[S_>:#1]} }}
\newcommand{\guardpref}[3]{\ensuremath{ \alpha^{#3}_{(#1,#2)}   }}
\newcommand{\stpre}[2]{\ensuremath{  \vdash_{#1}^{#2}  }}
\newcommand{\partremove}[1]{\ensuremath{ \downharpoonright_{#1} }}
\newcommand{\flaten}[2]{\ensuremath{ \mathsf{flat} ( #1 , #2)   }}
\newcommand{\flatcc}[3]{\ensuremath{ \textcolor{blue}{\mathsf{flatcc}(\!}#1 , #2 , #3\textcolor{blue}{)}}}
\newcommand{\partjoin}[3]{\ensuremath{ \mathsf{join}(#1,#2,#3)   }}
\newcommand{\preserves}[2]{\ensuremath{ \mathsf{noCycle}(#1,#2)    }}
\newcommand{\compatible}[3]{\ensuremath{ \mathsf{comp}(#1,#2,#3)    }}
\newcommand{\compatiblecc}[4]{\ensuremath{ \mathsf{compcc}(#1,#2,#3, #4)  }}
\newcommand{\acyclic}[4]{\ensuremath{ \mathsf{acyclic}(#1,#2,#3, #4)   }}
\newcommand{\lvlpartord}[1]{\ensuremath{ \ell_{#1}   }}
\newcommand{\varcompat}[1]{\ensuremath{\mathsf{compN}}(#1)}
\newcommand{\spo}{\texttt{spo}}
\newcommand{\spom}{\texttt{spomax}}
\newcommand{\pairrel}{\ensuremath{  \asymp  }}

\newcommand{\partremoveabb}[1]{\textcolor{teal}{>\!\!>_{#1}}}
\newcommand{\partremdotabb}{\textcolor{orange}{\gtrdot\!\!\!\!>_{1}}}
\newcommand{\partremovestar}[1]{\textcolor{teal}{>\!\!>^*_{#1}}}

\newcommand{\lifts}{\ensuremath{  \triangleleft  }}

\setlist[description]{leftmargin=1cm,labelindent=1cm}

\setlistdepth{10}
\newlist{myEnumerate}{enumerate}{10}
\setlist[myEnumerate,1]{label=\arabic*)}
\setlist[myEnumerate,2]{label=\Roman*)}
\setlist[myEnumerate,3]{label=\Alph*)}
\setlist[myEnumerate,4]{label=\roman*)}
\setlist[myEnumerate,5]{label=[\alph*]}
\setlist[myEnumerate,6]{label=[\arabic*]}
\setlist[myEnumerate,7]{label=[\Roman*]}
\setlist[myEnumerate,8]{label=[\Alph*]}
\setlist[myEnumerate,9]{label=[\roman*]}
\setlist[myEnumerate,10]{label=[\alph*]}

\usepackage{booktabs}
\usepackage{longtable}
\usepackage{rotating}

\usepackage{silence}
\begin{document}




\frontmatter
\graphicspath{{./0-misc/images/}}


\newcommand{\thetitle}{On the Expressivity of Typed Concurrent Calculi} 
\newcommand{\thesubtitle}{} 
\newcommand{\theauthor}{Joseph William Neal Paulus}

\selectlanguage{american}
\pagestyle{fancy} 
\thispagestyle{empty}
\vspace*{30mm}

\begin{center}
    \textbf{\huge \thetitle}\\[\baselineskip]\textbf{\LARGE \thesubtitle}
\end{center}

\vspace{40mm}

\begin{center}
    \Large\theauthor
    
    \bigskip
    \iffulldoc
    \today \\ (Full version, with appendices)
    \else
    \today 
    \fi 
\end{center}

\newpage

\thispagestyle{empty}


\vfill
\noindent \begin{minipage}[b]{1.0\columnwidth}%
    Typeset with \LaTeX \\
    Printed by Gildeprint\\
    Cover designed by Jessica Moss\\
    Copyright © Joseph Paulus, 2024\\
 \end{minipage}%

\clearemptydoublepage

\mainmatter



\selectlanguage{american}

\pagestyle{fancy}
\pagenumbering{roman}


\setcounter{page}{5}

\clearemptydoublepage
\clearemptydoublepage





\tableofcontents
\clearemptydoublepage


\pagenumbering{arabic}
\setcounter{page}{1}



\chapter*{Summary}\markboth{Summary}{Summary}
\addcontentsline{toc}{chapter}{Summary}





This thesis embarks on a comprehensive exploration of formal computational models that underlie typed programming languages. We focus on programming calculi, both functional (sequential) and concurrent, as they provide a compelling rigorous framework for evaluating program semantics and for developing analyses and program verification techniques. 

More concretely, this thesis addresses the following research question: \emph{how exactly does interactive behavior generalize sequential computation?}  We seek to gauge the expressivity of the $\pi$-calculus---the paradigmatic calculus of concurrency and interaction---with respect to   sequential computation as captured by the $\lambda$-calculus. Building upon Milner's seminal work on `functions as processes', our approach contrasts these two fundamental computational models via \emph{correct translations}, which formally explain how sequential terms in $\lambda$ can be codified into concurrent processes in $\pi$. The main novelty is the use of \emph{behavioral types}, advanced type systems for terms (in $\lambda$)  and processes (in $\pi$), to define the calculi, establish the properties of typed terms/processes, and to prove the correctness   of our translations.

%

We delve into our research question along several dimensions. First, we consider   \emph{non-deterministic} computations, whereby reductions may have branching behaviors. Non-determinism brings flexibility and generality in specifications; it may be \emph{confluent} or \emph{non-confluent}: in the former case, reductions may be independent taken within alternative branches, in a non-committal way; in the latter case committing to one branch discards other alternatives. 
As another dimension, we also consider \emph{resource-aware} computation whereby resources are \emph{linear} (usable exactly once) or \emph{unrestricted} (usable zero or many times). In turn, resource-awareness paves the way to a principled, explicit treatment of \emph{failures} in computations, which may occur when there is a lack or excess of resources or when they are misused. 

One key insight is that \emph{intersection types} in the $\lambda$-calculus can precisely specify quantitative information of a term as it evolves through   computation. By providing tight, type-preserving translations between intersection types (in $\lambda$) and session types (in $\pi$) we provide an original connection between these two important and widely-studied type disciplines along the dimensions of interest.
Finally, we contrast models of sequential and concurrent computation by considering type systems for concurrency that
guarantee \emph{termination}   (strong normalization): this is a well-studied and fundamental property for sequential computation models, which is  actively studied in the concurrent setting. 

\clearemptydoublepage

\chapter*{Samenvatting}\markboth{Samenvatting}{Samenvatting}
\addcontentsline{toc}{chapter}{Samenvatting (Summary in Dutch)}

\selectlanguage{dutch}

Deze scriptie begint aan een uitgebreide verkenning van formele computationele modellen die ten grondslag liggen aan getypeerde programmeertalen. We richten ons op programmeercalculi, zowel functioneel (sequentieel) als concurrent, omdat zij een overtuigend rigoureus kader bieden voor het evalueren van programmasemantiek en voor het ontwikkelen van analyses en programmaverificatie.

Concreter richt deze scriptie zich op de volgende onderzoeksvraag: \emph{hoe generaliseert interactief gedrag precies sequentiële berekening?} We proberen de expressiviteit van de $\pi$-calculus---de paradigmatische calculus van gelijktijdigheid en interactie---te meten met betrekking tot de sequentiële berekening zoals vastgelegd door de $\lambda$-calculus. Voortbouwend op Milner's baanbrekende werk over 'functies als processen', contrasteert onze benadering deze twee fundamentele computationele modellen via \emph{correcte vertalingen}, die formeel uitleggen hoe sequentiële termen in $\lambda$ kunnen worden gecodificeerd in gelijktijdige processen in $\pi$. Onze belangrijkste bijdrage is het gebruik van \emph{gedragstypes}, geavanceerde typesystemen voor termen (in $\lambda$) en processen (in $\pi$), om de calculi te definiëren, de eigenschappen van getypeerde termen/processen vast te stellen en om de correctheid van onze vertalingen te bewijzen.

We benaderen deze onderzoeksvraag vanuit verschillende dimensies. Ten eerste beschouwen we \emph{niet-deterministische} berekeningen, waarbij reducties vertakkingsgedrag kunnen vertonen. Niet-determinisme brengt flexibiliteit en algemeenheid in specificaties; het kan \emph{confluent} of \emph{niet-confluent} zijn: in het eerste geval kunnen reducties onafhankelijk binnen alternatieve vertakkingen worden genomen, op een niet-committerende manier; in het laatste geval sluit het kiezen voor een vertakking andere alternatieven uit.
Als een andere dimensie beschouwen we ook \emph{resource-bewuste} berekeningen waarbij bronnen \emph{lineair} (exact één keer bruikbaar) of \emph{onbeperkt} (nul of meerdere keren bruikbaar) zijn. Resource-bewustzijn maakt op zijn beurt de weg vrij voor een principiële, expliciete behandeling van \emph{fouten} in berekeningen, die kunnen optreden wanneer er een tekort of overschot aan middelen is of wanneer ze verkeerd worden gebruikt.

Een belangrijk inzicht is dat \emph{intersectietypes} in de $\lambda$-calculus kwantitatieve informatie van een term nauwkeurig kunnen specificeren naarmate deze door de berekening evolueert. Door strakke, type-behoudende vertalingen te bieden tussen intersectietypes (in $\lambda$) en sessietypes (in $\pi$) leggen we een originele verbinding tussen deze twee belangrijke en veel bestudeerde typedisciplines langs de genoemde dimensies.
Ten slotte contrasteren we modellen van sequentiële en gelijktijdige berekening door typesystemen voor gelijktijdigheid te beschouwen die \emph{terminatie} garanderen (sterke normalisatie): dit is een goed bestudeerde en fundamentele eigenschap voor sequentiële berekeningsmodellen, die actief wordt bestudeerd in de gelijktijdige setting.

\selectlanguage{american}


\clearemptydoublepage







\chapter{Introduction}\label{ch1_title}

%



\section{Context and Research Question}
\label{s:rq}

This thesis embarks on a comprehensive exploration of formal computational models that underlie typed programming languages. We focus on functional programming calculi with concurrency, as they provide a compelling rigorous framework for evaluating program semantics and for developing analyses and program verification techniques. 

Alonzo Church's $\lambda$-calculus is the most significant model for sequential computing; a long-standing and firm basis for program development and verification, particularly within the functional paradigm.
On the concurrent side, our investigation centers on the $\pi$-calculus  by Robin Milner, Joachim Parrow, and David Walker (\cite{DBLP:journals/iandc/MilnerPW92a}). The $\pi$-calculus is widely accepted to be the paradigmatic model for interactive computation. This computational model transcends the expressivity of traditional input-output behaviors by modelling interactions through \emph{message passing}; this offers a flexible yet rigorous framework for expressing and reasoning about programming constructs in higher-order, functional, and object-oriented paradigms.

Contrasting sequential computation in the $\lambda$-calculus and interactive behaviour as present in the $\pi$-calculus is a natural question of great significance. The first formal comparison is due to  Milner himself (\cite{Milner92}), who showed that the $\pi$-calculus subsumes functional behaviors in the $\lambda$-calculus. More precisely, by  giving a \emph{translation} (or \emph{encoding}) of terms in $\lambda$  into processes in $\pi$, Milner showed that $\lambda$-calculus describes a specific class of well-behaved interactions, whereas processes may exhibit more diverse and fine-grained computational phenomena. 

Milner's seminal work on \emph{functions-as-processes} was developed in the \emph{untyped} setting, adopting untyped formulations of both the $\lambda$-calculus (with different reduction strategies) and the $\pi$-calculus. Later on, as type systems for the $\pi$-calculus started to emerge, this line of work was extended to consider \emph{typed} languages, thus showing how well-typed terms in $\lambda$ can be codified by well-typed processes in $\pi$. Adding types is useful, in particular to establish the correctness properties of the translation; it is also insightful, as types provide an abstract yet  complementary perspective on the fundamental connections between sequential and concurrent computation. 

Explored by a number of authors in the last decades, the line of work on `functions-as-processes' is motivated by the question: \emph{how exactly does interactive behaviour  generalize sequential computation?} In this thesis, we continue this line of work, and extend it significantly by exploring two \emph{dimensions} not studied until now. As we explain next, the first dimension concerns \emph{phenomena and properties} relevant across sequential and concurrent programming models; the second dimension concerns different \emph{(behavioral) type systems} that statically enforce them.

%

\paragraph{Phenomena and Properties} 
We consider three relevant and intertwined aspects:
\begin{itemize}
	\item 
First, \emph{non-determinism} is an important phenomenon across computational models. Non-determinism (and non-deterministic choice) is intrinsically tied to the specification and analysis of concurrent programs; it is also relevant in the sequential setting. 
	One may distinguish between \emph{confluent} and \emph{non-confluent} forms of non-determinism. In the latter formulation, selecting one of the branches discards the rest; in the former, the different  branches may coexist independently. Both formulations have different merits;  the distinction between the two is  related to different forms of \emph{commitment} expressible in specifications.
	
   \item  We also investigate forms of \emph{resource control} between the two paradigms. In $\lambda$, the resources are the terms to which functions can be applied; in $\pi$, the resources are the channels (or \emph{names}) on which interactions may occur. In either case, following Girard's \emph{linear logic}, a resource can be either \emph{linear} (usable exactly once) or \emph{unrestricted} (usable zero or infinite times). This classification immediately gives computational  steps a quantitative dimension. Moreover, such a detailed accounting of computational entities is naturally related to \emph{failures}, which, informally speaking, arise when resources are not used as intended, as in, e.g., a process that uses a linear channel twice. 
    
      \item   Finally, we study \emph{termination} (also known as \emph{strong normalization}). Termination is a cornerstone of sequential programming models: a term is terminating if all its reduction sequences are finite. Termination is also an important property in concurrency in general, and in message-passing programs in particular. In such a setting, infinite sequences of internal steps are rather undesirable, as they could jeopardize the reliable interaction between a process and its environment.
   \end{itemize}

	\paragraph{Type Systems} As famously stated by Milner, ``well-typed programs cannot go wrong'' (\cite{DBLP:journals/jcss/Milner78}). Indeed, typing enforces computations to be ``well behaved''; the way in which typing systems induce or restrict behaviors has been extensively studied. Here we study two distinct, sophisticated typing disciplines.
	 
	\begin{itemize}
		\item 
    On the sequential side, we consider  \emph{intersection types}, which offer a 
    fruitful perspective at resource-aware\-ness (see, e.g.,~\cite{DBLP:conf/tacs/Gardner94,DBLP:journals/logcom/Kfoury00,DBLP:journals/tcs/KfouryW04,DBLP:conf/icfp/NeergaardM04,BucciarelliKV17}).
    By now, intersection types have consolidated into a well-established type discipline for functional languages, but also for concurrent models (see, e.g., \cite{DBLP:conf/lics/BonoD20} for a survey).
    In particular, we adopt \emph{non-idempotent} intersection types, which  offer a convenient tool for tracking resources, as the lack of idempotency intuitively enables us to ``count'' available resources.  
\item         On the concurrent side, we consider  \emph{session types}~(\cite{DBLP:conf/concur/Honda93,DBLP:conf/esop/HondaVK98}), which (statically) enforce communication correctness in message-passing programs by organizing their interactions into structures called \emph{sessions}.  In particular, we are interested in formulations derived the \emph{Curry-Howard correspondence} between session types and linear logic (\cite{CairesP10,DBLP:conf/icfp/Wadler12}). In a nutshell, this correspondence defines a solid bridge in three layers: \begin{center}
\begin{tabular}{ r c l }
\emph{linear logic propositions} & as &  \emph{session types} 
\\ 
	\emph{proofs}  & as & \emph{message-passing processes}
	\\
	\emph{proof normalization} & as & \emph{process communication}
\end{tabular}
\end{center}
 Due to its deep logical foundations, this correspondence ensures key properties for processes, in particular \emph{deadlock-freedom} (processes do not get stuck) and termination / strong normalization.
        	\end{itemize}

        Intersection types and session types can be considered as \emph{behavioural}, in the sense that their influence on computation surely goes beyond than that of simple types for $\lambda$ and $\pi$, respectively. As we will see, the developments on `functions-as-processes' enable us to connect intersection types and session types in a brand new light.
    


\paragraph{Research Question} 
Clearly, these two dimensions (these being type systems and phenomena/properties) are related to each other, in the sense that type systems enforce the declared properties and provide consistency for the intended phenomena. The interplay between the two highlights the breadth of our study and the need for precise comparisons between the sequential and concurrent paradigms. 
This discussion brings us to the \emph{research question} addressed in this thesis:
\begin{quote}
	Can we relate formal models of sequential computation and interactive behaviors, both governed by behavioural types, considering phenomena little considered so far, such as \emph{non-determinism} and \emph{failures}, while accounting with essential properties such as \emph{deadlock-freedom}, \emph{confluence}, and \emph{termination (strong normalization)}?
\end{quote}

\medskip
\noindent
Before elaborating further on our approach to this research questions and on the contributions of this thesis, we find it useful to introduce some technical background on `functions-as-processes' as well as essential notions on translation correctness.

\section{Background}\label{intro_tech_back}

We start by giving a high-level presentation of the functions-as-processes approach pioneered by \cite{Milner92}, who considers the $\lambda$-calculus with two different reduction strategies (the lazy $\lambda$-calculus and the call-by-value $\lambda$-calculus) and a very simple $\pi$-calculus, which is  sufficient for modelling standard sequential behaviors. 

\begin{figure}
    \[
    \begin{aligned}
            \piencod{x}_u &= \overline{x}(u).\nil\\
        \piencod{\lambda x . M}_u &= u(x).u(v).\piencod{M}_v\\
        \piencod{M\,N}_u &= (\nu v)( \piencod{M}_v  \sep \overline{v}(x) . \overline{v}(u) . \piencod{x:=N}  )\\
        \piencod{x:=M} &= ~\bang x(w) . \piencod{M}_w
    \end{aligned}    
    \]
    \caption{Translation of $\lambda$ into $\pi$ as first presented by \cite{Milner92}}
    \label{fig:intro_milner}
\end{figure}

The translation is defined inductively on the structure of terms; it is  denoted $\piencod{\cdot}_u$, where $u$ denotes a channel on which the behaviour of a translated term will be provided. 
The translation consists of four parts; we explain the $\pi$-calculus notation as we go:
\begin{itemize}
	\item A variable $x$ is translated as the process $\overline{x}(u).\nil$, i.e., an output action on $x$ in which $u$ is communicated, followed by the inactive process. 
	\item An abstraction $\lambda x . M$ is translated by the process $u(x).u(v).\piencod{M}_v$, i.e., two consecutive inputs on $u$ precede (i.e., block) the translation of $M$. The first input receives a reference to the parameter $x$, whereas the second receives $v$, a reference  to the  argument of the function; both $x$ and $v$ are bound to process $\piencod{M}_v$.
	\item The translation of  ${M\,N}$ is interesting because it shows how functional application is assimilated to process synchronisation: the resulting process is $(\nu v)( \piencod{M}_v  \sep \overline{v}(x) . \overline{v}(u) . \piencod{x:=N} )$, which represents the parallel execution of the translations of the function $M$ and its argument $N$. Actually, $N$ is not immediately enabled, but is embedded in a so-called \emph{environment} $x:=M$, which is blocked by two outputs that will synchronize with the inputs in the translation of abstraction. 
	\item The translation of $x:=M$ is the process $\bang x(w) . \piencod{M}_w$, which denotes a replicated server, able to provide copies of process $\piencod{M}_w$ upon output requests on channel $x$. Intuitively, a server is appropriate in order to provide arguments for the multiple  occurrences (possibly zero) of a variable parameter in an abstraction's body. 
\end{itemize}

\begin{figure}
    \[
    \begin{aligned}
        (\lambda x.x) N & \redd_\lambda N \smallskip \\ 
        \piencod{(\lambda x.x) N}_u &= (\nu v)( v(x).v(w).\overline{x}(w).\nil \sep \overline{v}(x).\overline{v}(u).\piencod{x:=N} ) \\
        & \redd_\pi (\nu v)(\nu x)( v(w).\overline{x}(w).\nil \sep \overline{v}(u).\piencod{x:=N} ) \\
        & \redd_\pi (\nu v)(\nu x)( \overline{x}(u).\nil \sep \piencod{x:=N} ) \\
        & = (\nu v)(\nu x)( \overline{x}(u).\nil \sep \bang x(w).\piencod{N}_w ) \\
        & \redd_\pi (\nu v)(\nu x)(\piencod{N}_u    \sep     
        \bang x(w).\piencod{N}_w ) 
        \\
        & \sim  \piencod{N}_u  && (*)\\
    \end{aligned}  
    \]
    \caption{Example reduction from \cite{Milner92}}
    \label{fig:intro_milner_red}
\end{figure}

To further illustrate how the translation uses fine-grained synchronization in names to represent $\beta$-reduction,   
Milner gives the example shown in \Cref{fig:intro_milner_red}, where we use $\redd_\lambda$ and $\redd_\pi$ to denote reduction in each language.  

As the figure shows, input and output actions synchronise across the same channel, as in a handshake, sending and receiving references between each other.
The relation $\sim$ denotes a binary relation on processes (a combination of structural congruence and behavioural equivalence), which acts as a `garbage-collector': it allows us to abstract away from the replicated server (which can no longer be invoked). Overall, this example provides evidence to the fact that the translation induces a strong form of \emph{operational correspondence} between terms and processes. 

Having translations such as Milner's is significant for several reasons. Besides their conceptual merit, they allow us to transfer results across different paradigms. A salient example is the work of  \cite{DBLP:phd/ethos/Sangiorgi93}, who showed how  behavioral equivalences for $\lambda$-terms can be studied in terms of behavioral equivalences defined for processes, leveraging the translation $\piencod{\cdot}_u$ given above. 


The results in \cite{Milner92} led to a fruitful avenue of research in the formal comparison of sequential and concurrent calculi. As mentioned above, Milner's translation was in the untyped setting, and developed well before the appearance of several key ingredients in this thesis, namely: (i)~the emergence of rich type disciplines for the $\pi$-calculus; (ii)~the discovery of a \emph{Curry-Howard correspondence} for concurrency; and (iii)~the development of criteria and techniques for the study of \emph{correct translations}. 

It is insightful to briefly elaborate on (ii) and (iii) above. 
Given the Curry-Howard correspondence for concurrency, the next meaningful step in extending the relationship between concurrent and sequential  models comes from the viewpoint of logic. This begs the question: can one encode the Curry-Howard interpretation of the \lamc into the Curry-Howard interpretation of the $\pi$-calculus? This question is addressed by the work of \cite{DBLP:conf/fossacs/ToninhoCP12}, who show that proof-theoretical principles induce logically motivated translations of well-typed terms into well-typed processes. Not only are terms translated but also types themselves are transformed. Their methodology involves first giving a translation from the \lamc into the linear \lamc, a stepping stone towards a translation into the session-typed $\pi$-calculus. This second translation turns out to be very much related to that in \cite{Milner92}.


A natural question at this point is how to assess the correctness of translations such as Milner's. What methods do we use to ensure that translations between languages are non-trivial or meaningful? In fact, the possibility of transferring techniques between languages hinges very much on such a form of correctness. This question underpins the notion of \emph{relative expressiveness}, a much-studied topic in Concurrency Theory. In this respect, a widely adopted proposal is the set of  correctness criteria identified by \cite{DBLP:journals/iandc/Gorla10}. In a nutshell, this proposal allows us to argue for the correctness of translations and to reason about relations of expressive  power between a \emph{source language} (such as $\lambda$) and a \emph{target language} (such as $\pi$) connected by a given translation.

Because translation and their correctness constitute a recurring topic in this thesis, we dwell upon the definitions of different \emph{correctness criteria}. 
To describe them, we must define languages and translations:
\begin{definition}{Language}
    A language $\mathcal{L}_1$ is a triple  $ (\mathcal{P}_1, \shred_1, \asymp_1)$, where:
    \begin{itemize}
    	\item 
   $\mathcal{P}_1$ represents the set of terms (or processes) of the language (its syntax); 
   \item $\shred_1 \subset \mathcal{P}_1^2$ represents the operational semantics of the language (typically, a reduction relation)s;
   \item $\asymp_1\subset \mathcal{P}_1^2$ is an equivalence relation.
   \end{itemize}
Moreover,   $\shred^*$ represents the reflexive, transitive closure of $\shred_1$; also, $P \shred_1^\omega   $ means that there exists an infinite number of transitions emanating from $P$. 
\end{definition}

\begin{definition}{Translation}
Given a source language $\mathcal{L}_S$ and a target language $\mathcal{L}_T$, a translation from $\mathcal{L}_S$ to $\mathcal{L}_T$ is a function of the form $\piencod{\cdot}: \mathcal{P}_S \redd \mathcal{P}_T$.
\end{definition}

We shall be interested in \emph{correct} translations, i.e., translations $\piencod{\cdot}$ that satisfy certain  criteria that attest to their quality. Following Gorla, we consider all of the following criteria:
\begin{itemize}
    \item {\bf{Compositionality}}: We say $\piencod{\cdot}$ is compositional when the translation of a composite term is defined in by composing the translation of its sub-terms. In process calculi, a classic instance of this criterion arises in the homomorphic treatment of parallel composition, i.e., $\piencod{P \sep Q} = \piencod{P} \sep \piencod{Q}$. As we have seen, translations from sequential to concurrent programming models usually translate an application $M\, N$ as a process involving the parallel composition of $\piencod{M}$ and $\piencod{N}$. 
    Intuitively, compositionality ensures that actors interacting with the term do not influence the translation of the term itself.

 \item {\bf{Operational Correspondence}}: This criterion ensures that the translation preserves and reflect the behavior of source terms. It is divided into \emph{completeness} and \emph{soundness} requirements. 
 \begin{itemize}
 \item  Completeness says that all reductions in the source language are respected by the translated term. That is, $ \forall S \in \mathcal{P}_S  \text{ such that } S \shred^*_S S'  \text{ then } \exists T \in \mathcal{P}_T  \text{ with } \piencod{S} \shred^*_T T \land \piencod{S'} \asymp_T T  $.

\item Soundness says that the translation does not introduce reductions not already present in the source language. That is, $\forall S \in \mathcal{P}_S \land T \in \mathcal{P}_T   \text{ such that } 
    \piencod{S} \shred^*_T T
    \text{ then } \exists  S' \in \mathcal{P}_S \text{ with } {S} \shred^*_S S' \land \piencod{S'} \asymp_T T $. 
    
    Often, operational soundness is too strict as the target language may not match source reductions with a one-to-one correspondence and rather needs multiple reduction steps to simulate the given source behavior.
 \end{itemize}

    \item {\bf{Name Invariance}}: This criterion ensures that the translation is not dependent of  a specific choice of free names/variables. This allows for operations such as $\alpha$-conversion to not interfere with the intended behaviour of the translation.

    \item {\bf{Divergence Reflection}}: This criterion ensures that the translation does not introduce divergence behaviour. That is, divergence behavior emanating from translated terms correspond to divergent behavior already present in the source term. More precisely: $\forall P \in \mathcal{P}_S. \piencod{P} \shred_T^\omega \implies P\shred_S^\omega$.

    \item {\bf{Success Sensitiveness}}: This criterion presupposes that  $\mathcal{L}_S$ and $\mathcal{L}_T$ are both equipped with an abstract notion of success, denoted $\checkmark$. The criterion ensures that a source terms reduces in multiple steps to $\checkmark$ if and only if its corresponding translation does the same. This condition is closed under reductions.

\end{itemize}


The literature on relative expressiveness has sometimes considered \emph{full abstraction} as another correctness criterion for translations. This requirement ensures that
$P \asymp_S Q$ if and only if $\piencod{P} \asymp_T \piencod{Q}$.
Full abstraction is significant, as it allows to transfer reasoning techniques between source and target languages. However,  as \cite{DBLP:journals/mscs/GorlaN16} convincingly explain, it does not represent a good criterion for assessing the quality of translations. For this reason, we do not consider it in our developments.

To close this section, we mention that the state of the art on the relation between sequential computation in $\lambda$ and concurrent programming in $\pi$ is arguably given by the work of \cite{DBLP:conf/esop/ToninhoY18}. They present a type-preserving translation from a linear variant of System F (Linear-F) into a polymorphic session-typed $\pi$-calculus first studied by \cite{Berger2005GenericityAT}. Surprisingly, a translation in the reverse direction is also given. Their translation enjoys  {full abstraction} for relevant typed congruences in $\lambda$ and $\pi$, respectively. Their work does not consider non-determinism, confluence, and failures in the sense described above.

\section{Approach}

Having covered  essential background material, here we discuss selected aspects of the approach that we adopt to address the research question stated at the end of \Cref{s:rq}. We divide our presentation along sequential and concurrent programming models, highlighting novelties and paving the way to the outline of contributions to be given in \Cref{s:cont}.

\subsection{Sequential Models: Resource $\lambda$-Calculi}
As already said, we consider \emph{resource $\lambda$-calculi}.
A key idea in these calculi is that applications are generalized to terms of the form $M\, B$, where the argument $B$ is a \emph{resource} of possibly limited availability.
Resources can be linear or unrestricted; non-de\-ter\-mi\-nism (confluent and non-confluent) and failures arise from the (mis)use of these resources during computations. 
These notions of resource control and non-determinism come from a number of works in the literature, including \cite{DBLP:conf/concur/Boudol93,DBLP:conf/birthday/BoudolL00,PaganiR10,DBLP:journals/corr/abs-1211-4097}.

\paragraph{Non-Determinism and Explicit Failure}
In resource $\lambda$-calculi there are three syntactic categories: terms, bags, and expressions. 
We have applications of the form $M\, B$,  where $B$ denotes a finite \emph{bag} of resources (terms); non-determinism arises from the \emph{fetching} of a term from $B$.
That is, in reducing  the redex $(\lambda x. M)\, B$ a resource $N_i$ is ``fetched'' from $B$ and substituted into an occurrence of $x$ in $M$. Crucially, this fetch operation  is non-deterministic: the selection of the resource from $B$ to be used induces non-deterministic behavior. 

In this setting, reductions may fail, depending on the nature of the resource.
A linear failure arises when there is a mismatch between required and available (linear) resources; that is, because linear resources must be completely consumed during computation,  an excess or lack of these resources is deemed undesirable. 
An unrestricted failure arises when a specific (unrestricted) resource is not available.
In either case, failure is \emph{explicit}, as is expressed by a term $\fail^{\widetilde{x}}$, where $\widetilde{x}$ denotes a sequence of variables---intuitively, these are the variables involved in the failed computation.

\paragraph*{Resources.}
As we have seen, resources play a key role: they are the source of non-deterministic behavior but can also trigger failures. Computations can only be considered to be successful when the available linear resources match exactly the occurrences of variables consuming them. In the case of purely linear resources, our treatment of bags (and the associated notion of substitution) follows closely previous work by \cite{DBLP:conf/concur/Boudol93,DBLP:conf/birthday/BoudolL00,PaganiR10}.
More precisely, we adopt a weak/lazy reduction semantics, which only substitutes resources when the head variable of a term may be substituted. 
This reduction strategy closely matches the corresponding translation in the $\pi$-calculus, which  allows us to obtain direct operational correspondence results.

To illustrate our semantics, consider the identity function $\lambda x. x$ applied to a bag containing only a term $N$; this bag is denoted $\bag{N}$.
In our calculi, rather than evaluating the redex via the (usual) $\beta$-reduction $(\lambda x. x)\bag{N} \redd N$ (as in \cite{PaganiR10,DBLP:journals/corr/abs-1211-4097}), we first perform a weak $\beta$-reduction, which leads to a term with an \emph{explicit substitution}, namely $x\esubst{N}{x}$. Clearly, here there is no mismatch in resources: the abstraction contains a single occurrence of $x$ and the bag contains only the linear resource $N$. Therefore, the reduction is successful.

In the case of unrestricted resources, we adopt a strategy that allows us to express more fine-grained resource control than prior work. We start with a simple example. Consider the following terms expressible in previous literature:
\[
 (\lambda x . M \bag{x}^!)  \bag{N}  
\]
Here we have an abstraction on $x$ with an unrestricted occurrence of $x$ within a bag. This term is being applied to a bag containing the linear resource $N$. As such, $N$ must be used exactly once; however the above term says that $N$ may be substituted into an unrestricted bag and be arbitrarily used. This not only breaks the intended linearity of  $N$ but also makes failure difficult to track,  as we can no longer predict resource behaviour. One solution presented by \cite{PaganiR10} and \cite{DBLP:journals/corr/abs-1211-4097} is the following: when $N$ is substituted into the unrestricted bag, a new linear resource is spawned so that $N$ is substituted into the freshly spawned (linear) copy. This changes the behaviour of the bag $\bag{x}^!$ within the term $M \bag{x}^!$ from being used arbitrarily many times to being used \emph{at least once}. 

We choose to distinguish between \emph{linear} and \emph{unrestricted} resources, at both the bag and the variable level. Bags are composed of two parts, one linear and the other unrestricted. This design choice allows us to easily split the bag syntactically. Variables denote the type of resource they expect to consume, which is beneficial for two reasons. First, we may count linear occurrences; second, we can enforce linearity by typing by disallowing linear occurrences of variables in unrestricted behaviours.
As mentioned above, considering unrestricted resources entails an additional  source of failure, i.e., when a variable must be substituted with an unrestricted resource and no such resource is available.

In line with this design, we also formulate a separation at the level of explicit substitutions. We consider both linear and unrestricted explicit substitutions in the syntax of terms; this is different from the formulation of \cite{DBLP:conf/concur/Boudol93,DBLP:conf/birthday/BoudolL00}, where a single construct for explicit substitution handles both linear and unrestricted resources.


\paragraph*{Two Flavors of Non-determinism.}
Non-determinism is prevalent in many models of computation, allowing for richer and more expressive forms of computation. 
In the resource $\lambda$-calculus, non-determinism arises when a resource is  {fetched} from the bag. As we reduce lazily, this non-deterministic choice is made when we have an explicit substitution that acts on a variable and such a variable is at the head of the term. 

We consider both confluent and non-confluent formulations of non-determinism. To elaborate on this distinction, let us consider purely linear resources, so that non-deterministic behaviour is only due to the fetching of linear resources. Consider a term $ M \esubst{\bag{M_1,M_2}}{x}$, where $M$ has two linear occurrences of $x$, one of these being the head variable $x$. A linear \emph{explicit substitution} on $x$ acts on $M$, involving the linear bag containing $M_1$ and $M_2$. (We write $M\headlin{M_1/x}$ to denote the usual (head) substitution of $M_1$ for $x$ in $M$.)
\begin{itemize}
	\item 
The \emph{confluent} formulation of non-deterministic fetching enables the following reduction:
\[
 M \esubst{\bag{M_1,M_2}}{x} \redd M\headlin{M_1/x}\esubst{\bag{M_2}}{x} + M\headlin{M_2/x}\esubst{\bag{M_1}}{x} 	
\]
That is, the term reduces to an \emph{expression}, consisting of a nondeterministic sum of the two possibilities, i.e., to substitute $M_1$ and leave $M_2$ in the bag (left branch) and  to substitute $M_2$ and leave $M_1$ in the bag (right branch). This reduction dictates a choice of resources being substituted, similarly to the reductions given in prior works (\cite{PaganiR10,DBLP:journals/corr/abs-1211-4097}).

\item The \emph{non-confluent} formulation of fetching is as follows:
\begin{align*}
    & \mathbin{\rotatebox[origin=r]{30}{$\redd$}} ~ M\headlin{M_1/x}\esubst{\bag{M_2}}{x}
    \\[-5pt]
    M\esubst{\bag{M_1,M_2}}{x}~
    & 
    \\[-3pt]
    & \mathbin{\rotatebox[origin=r]{-30}{$\redd$}} ~ M\headlin{M_2/x}\esubst{\bag{M_1}}{x} 	
\end{align*}
That is, the reduction of the term expresses commitment: choosing one between $M_1$ and $M_2$ entails discarding the other branches, similarly to the reductions given in \cite{DBLP:conf/concur/Boudol93,DBLP:conf/birthday/BoudolL00}.
\end{itemize}

\noindent
What we have discussed thus far can be considered as an \emph{implicit} non-deterministic choice, as it is induced by the fetching of resources. There is also an \emph{explicit} non-deterministic operator, denoted `$+$', present at the level of expressions, denoting an explicit choice between terms---such an operator is also present in the languages in \cite{PaganiR10,DBLP:journals/corr/abs-1211-4097}. 

It follows naturally from the previous discussion that 
reductions may be confluent or non-confluent. In the confluent setting, we obtain the expected \emph{diamond property}, illustrated below (left-hand side). It allows terms to reduce independently within non-deterministic branches, ensuring that we are not discarding not manipulating choices within other branches. This property is strong and also convenient when considering operational correspondence properties. In contrast, in the non-confluent setting we obtain the situation illustrated in the right-hand side, which is the typical form of non-determinism in process calculi:
\[
    \begin{tikzpicture}
        \node (startmn) {$M + N$};
        \node (dummy) [right of= startmn]  {};
        \node (nextM) [above of= dummy]  {$M' + N$};
        \node (nextN) [below of= dummy] {$M + N'$};
        \node (nextfin) [right of= dummy] {$M' + N'$};
        \draw[->] (startmn) -- (nextM);
        \draw[->] (startmn) -- (nextN);
        \draw[->] (nextM) -- (nextfin);
        \draw[->] (nextN) -- (nextfin);
    \end{tikzpicture}
\qquad
    \begin{tikzpicture}
        \node (startmn) {$M + N$};
        \node (dummy) [right of= startmn]  {};
        \node (nextM) [above of= dummy]  {$M'$};
        \node (nextN) [below of= dummy] {$N'$};
        \draw[->] (startmn) -- (nextM);
        \draw[->] (startmn) -- (nextN);
    \end{tikzpicture}
\]


\paragraph*{Failure and Typing}
As we have discussed, a central question is how failure arises and how it relates to linearity. In some prior works (\cite{DBLP:conf/concur/Boudol93,DBLP:conf/birthday/BoudolL00}),  failure only arises when there is no resource to be fetched, a situation  referred to as a ``deadlock''. Still, the failure is not explictly expressed syntactically. Also, because in those works linear resources are to be used \emph{at most once}, failure cannot arise if there is an excess of linear resources.

In some other works, in particular the framework in  \cite{PaganiR10}, 
failure is   present via a term `$0$', which denotes a neutral element.
As hinted at before, here we explicitly incorporate within the syntax of terms. 
We follow the approach to failure in~\cite{PaganiR10} by decreeing that linear resources must be  used exactly once. Similarly, both ours and their approach to failure is consuming, in the sense that all computation in a non-deterministic branch is disappears when failure arises. 

There is a difference, however: the failure term in~\cite{PaganiR10} does not preserve linearity, as variables are discarded. In our failure term $\fail^{\widetilde{x}}$, the multiset $\widetilde{x}$ retains the {linear} variables that are captured by failure. For example 
$
  (\lambda x . x \bag{x} ) \oneb \redd \fail^{x,x} 
$
denotes a redex in our calculus, in which two linear occurrences of $x$ cannot be substituted as the (linear) bag is empty. As there is a mismatch in  resources the term fails, and the two linear occurrences of $x$ are preserved.

Using intersection types for typing of terms, bags and expressions is a common approach the literature; see, e.g., \cite{DBLP:conf/concur/Boudol93,BoudolL96,PaganiR10} and the related differential $\lambda$-cal\-culus of~\cite{DBLP:journals/tcs/EhrhardR03}. We use non-idempotent intersection types as they match structurally to resources; also, intersection types are \emph{quantitative} in the sense we may infer the number of occurrences of a variable (or the size of a linear bag) via its type. 

The status of the failure term $\fail^{\widetilde{x}}$ with respect to typing deserves explanation.
As we express failure explicitly at the syntax level, it is reasonable to assume that failure should somehow be typable.
To account for this, our design consists of two systems: we use intersection types to define both \emph{well-typed} terms and \emph{well-formed} terms. 
Our definition of well-typedness disallows failure, i.e., if $M$ is well-typed then $M \neq \fail^{\widetilde{x}}$. 
In contrast, well-formedness admits the failure term and assigns it an arbitrary type.
In other words, $\fail^{\widetilde{x}}$ is not well-typed, but it is well-formed.

\subsection{Concurrent Models: Session-Typed $\pi$-Calculi}

Interactions in concurrency are intrinsically non-deterministic. Accordingly, non-determinism (and non-deterministic choice) plays an important role in formulating the formal semantics for calculi for concurrency---this is the case even in formalisms such as  {CCS} (\cite{DBLP:books/daglib/0067019}). The non-deterministic choice operator `$+$' is typically \nconf, as it expresses commitment to one branch while discarding behaviours in other branches.

\paragraph{Confluent Non-Determinism}
Our focus will be on typed process calculi with non-determinism that can provide an appropriate framework, in the sense of `functions-as-processes', for the resource $\lambda$-calculi just motivated.
Non-determinism, in particular in its non-confluent variant, is often at odds with resource-control---discarding not chosen branches breaks linearity. 
Considering this, we shall adopt the work of \cite{CairesP17} as reference in our developments: it introduces a \emph{session-typed} $\pi$-calculus, which supports both (confluent) \emph{non-determinism} and \emph{failure} in a linearly-typed setting. In this framework, types arise from a Curry-Howard correspondence between session types and an extension of classical linear logic. 
Important properties for processes, such as session fidelity and deadlock-freedom, follow from the fundamental connection between cut-elimination and process synchronizaton.
It may then be easy to see that confluence in \cite{CairesP17} is a natural consequence of the underlying Curry-Howard foundations .

A salient feature of the framework in \cite{CairesP17} is the possibility of typing behaviors that are \emph{non-deterministically available}, i.e.,  session protocols that can perform as stipulated, but may also fail. There is a non-trivial balancing act involved, in order to accommodate failures while adhering to resource control via linearity.

We illustrate this point by discussing processes and their operational semantics. 
In session-typed languages, the communication actions performed on channels can be structured into \emph{sessions}, which streamline analysis. In the framework of \cite{CairesP17}, there are two special actions along a channel $x$: process `$\psome{x}; P$' confirms that the behavior along $x$ can be made available as intended and then continues as $P$, whereas the process `$\pnone{x}$' expresses a failure to provide a  behavior on $x$. These two processes are meant to interact with a process of the form `$\gsome{x}{\widetilde{w}};P$', which denotes a process $P$ that declares its \emph{dependency} on behaviors that are non-deterministically available on $x$, with  $\widetilde{w}$ exposing the  sequence of channels that depend on the availability of $x$. This sequence is crucial to hereditarily propagate a potential failure of $x$.

We postpone a formal presentation of the reduction rules for these constructs, and illustrate it instead by means of the following example:
\begin{align*}
    \res{x}(\gsome{x}{(y)}; \psome{y}; P \| \psome{x}; Q) &\redd \res{x}(\psome{y}; P \| Q)
    \\
    \res{x}(\gsome{x}{(y)}; \psome{y}; P \| \pnone{x}) &\redd \pnone{y}
\end{align*}
In the first case, the process on the left of the parallel expects to synchronize with a non-deterministic session $x$, with another session $y$ depending on it. The process on the right of the parallel confirms the availability of $x$, and so the two processes may synchronize along $x$ and proceed further. 
Differently, in the second case, the process on the left is the same as before, but the process on the right signals the failure to provide $x$. Because $y$ is dependent on the availability of $x$, the process $P$ is cancelled. As we will see, typing ensures that non-deterministic availability is appropriately handled across different branching behaviours.

In this typed $\pi$-calculus, one may think of resources as the protocol that is provided along a channel. Channels may provide \emph{linear} or \emph{unrestricted} behaviour, with the latter being assimilated to the operation of \emph{clients} that invoke \emph{servers} (persistently available processes). These server-client interactions are typed with the exponentials $!\,A$ and $? A$ in linear logic. 

\paragraph{Non-Confluent Non-Determinism}
The typed calculus in \cite{CairesP17} provides a convenient target language for a concurrent translation of resource $\lambda$-calculi under the confluent regime, with a compositional treatment of non-determinism in $\lambda$ as non-determinism in $\pi$. 
However, in the \nconf case  the situation is different: there is no satisfying literature explicitly outlining the role of \nconf \nond, in a typed setting, despite it being commonplace. 

Considering this gap, 
we introduce a 
variant of the typed $\pi$-calculus in \cite{CairesP17} with a 
\nconf \nond operator, denoted $\nd$.
We propose two different operational semantics for $\nd$, which express the level of commitment involved in selecting a branch. In the \emph{eager} semantics, the commitment of choice is determined by the branches that are performing synchronization. This is shown in the following three  reductions possible from $P_0$:
\begin{align*}
 P_0 =   \res{x}\big(\gsome{x}{(y)}; P \| (\psome{x}; Q_1 \nd \psome{x}; Q_2 \nd \pnone{x})\big) &\redd \res{x}(P \| Q_1)
    \\[-5pt]
    &\mathbin{\rotatebox[origin=r]{-15}{$\redd$}} \res{x}(P \| Q_2)
    \\[-10pt]
    &\mathbin{\rotatebox[origin=r]{-35}{$\redd$}} \, \0
\end{align*}
Intuitively, these process matches our expectation about usual, non-confluent determinism: a synchronization along channel $x$ selects one branch and discards the other two. Alternatively, a \emph{lazy} semantics realizes a more gradual approach: in such a semantics, branches of choice are grouped based on their top-level prefixes; the `collapsing' of choices is based on branches that may  synchronise. In our previous for  $P_0$, we have:
\begin{align*}
    \res{x}\big(\gsome{x}{}; P \| (\psome{x}; Q_1 \nd \psome{x}; Q_2 \nd \pnone{x})\big) &\redd \res{x}\big(P \| (Q_1 \nd Q_2)\big)
    \\[-7pt]
    &\mathbin{\rotatebox[origin=r]{-30}{$\redd$}} \0
\end{align*}
As the first two branches both confirm their behaviour along $x$ the choice between them is postponed. Notice that the placing of the choice is also preserved, i.e., the choice is not distributed, as in, e.g., $ \res{x} (P \| Q_1 ) \nd \res{x} (P \|  Q_2 )$.

\paragraph*{Termination}

Termination is a property relevant across sequential and concurrent programming models. It has been widely studied for sequential programming models, where type systems for (variants of) the $\lambda$-calculus can guarantee different \emph{normalization} properties. Termination is also relevant in concurrency, as normalization properties can play a crucial role in the verification of interacting systems. Also here we find type systems that ensure termination of (variants of) the $\pi$-calculus by typing.

As already discussed, concurrency offers a broader range of behaviours than purely sequential models; as a result, we find a wider range of typing systems governing interaction and concurrency. Specifically, different type systems for termination in the $\pi$-calculus have been developed, following vastly diverse  underlying approaches. 

We give the first comparative study of {type systems that enforce termination} for message-passing processes in the $\pi$-calculus. In  this context, interesting forms of non-terminating behavior arise when a \emph{server} process, represented by a replicated input process of the form $!x(y).P$, is invoked an infinite number of times. This sequence of infinite reductions can be caused by, e.g.,  a server that calls itself or a cycle of server calls to each other. Type systems that ensure termination use different methods to statically rule out such infinite reduction sequences. 

{Our work} concerns three type systems. The first one concerns the class of processes typable with Vasconcelos's session type system~(\citeauthor{V12}), which we denote $\vaslang$. We use this calculus as a tool for comparison due to its liberal type discipline. 
In fact, this type system offers no  termination guarantees; it only ensures session fidelity (i.e., channels always respect their protocols), and hence it may describe a vast array or ``well-behaved'' session respecting processes. 

When it comes to type systems that do ensure termination by typing, we focus on comparing the following two systems:
\begin{enumerate}
    \item We consider \lvllang, the class of processes induced by the type system in~\cite{DBLP:conf/ifipTCS/DengS04}. In this discipline, termination is enforced by \emph{weights} (or \emph{levels}) associated to channel types; roughly speaking, servers should not contain outputs that invoke a server of a ``greater'' level.

    \item We consider \dilllang, the class of processes induced by Caires and Pfenning's Curry-Howard correspondence between linear logic and session types (\cite{CairesP10}). Here termination is induced via proof-theoretical principles, similar to the way the simply-typed $\lambda$-calculus rules out infinite reduction sequences.

\end{enumerate}

Because both the types and syntax of these languages are formulated differently, in order to
enable comparisons between them (correct) translations are needed. Using these translations we may show that both $\lvllang \subset \vaslang$ and $\dilllang \subset \vaslang$ hold as expected, due to \vaslang not ensuring termination. More interestingly, we prove $\dilllang \subset \lvllang$ and $\lvllang \not \subset \dilllang$, thus explicitly determining the  exact relationship between these two classes of terminating languages. Formally, we show that there are terminating processes that belong to the class $\lvllang$ but are outside the class $\dilllang$.


\section{Outline of Contributions}
\label{s:cont}

The contributions in this thesis address our research question by showing how typed sequential computation (in resource $\lambda$-calculi with intersection types) is subsumed by typed concurrent behaviour (in session-typed $\pi$-calculi), in the style of ``functions as processes''. In the context of these languages, we provide correct translations accounting for an array of different features. The technical results are presented in four chapters (\Cref{ch2} -- \Cref{ch5}), which we briefly describe:


\begin{description}
    \item[\Cref{ch2}] We give a correct translation from a resource $\lambda$-calculus to the session-typed $\pi$-calculus in \cite{CairesP17}. The focus here is on confluent non-determinism and purely linear resources. The translation is type preserving; as mentioned earlier, we encode both well-typed (fail-free) terms and well-formed (fail-prone) terms into well-typed processes. 
    
    Confluence plays a key role in proving operational soundness, one of the two parts of the operational correspondence criterion. 
    To illustrate this point, consider the $\lambda$-term 
    $ M \esubst{\bag{M_1,M_2}}{x}$, i.e., a term $M$ subject to an explicit substitution involving a bag with two resources, $M_1$ and $M_2$. Let us assume that $M$ has a {two occurrences} of variable $x$. 
     \Cref{fig:intro_soundch2} illustrates the relationship between the source terms and their corresponding translations; for pedagogical purposes, we provide a simplified description of the translation, using \emph{contexts} $C[-]$ in $\pi$ to abstract away from details of the translation unimportant at this point. 
    
    As discussed earlier, in the confluent regime $M$ reduces into a sum involving the different possibilities for choosing a term from the bag. In our translation, we use two names, denoted $x_1,x_2$, to represent the two occurrences of $x$ in $M$. Process $P'$ represents the (simplified) translation of $M\headlin{M_1/x}\esubst{\bag{M_2}}{x} $;  similarly, process $Q'$ represents the translation of $ M\headlin{M_2/x}\esubst{\bag{M_1}}{x}$. In this example, we may reduce the left branch ($P$ to $P'$) or the right branch ($Q$ to $Q'$) independently. We use dotted arrows to represent possible other reductions available; this way, e.g., after reducing to $P' + Q$ we may reduce $Q$ or we may reduce further  $P'$. The key point in the analysis of soundness is that, due to confluence, we may freely decide on the order in which reductions are performed without interfering with the overall behaviour of the processes.

    \begin{figure}[!t]
        \[
            \begin{tikzpicture}
                \node (startmn2) {$ M \esubst{\bag{M_1,M_2}}{x} $};
                \node (nextfin) [below=0.5cm of startmn2] {$
                    \begin{aligned}
                        M\headlin{M_1/x}\esubst{\bag{M_2}}{x} +\\
                        M\headlin{M_2/x}\esubst{\bag{M_1}}{x}
                    \end{aligned}    
                    $};
                \node (dummy6) [below=0.5cm of nextfin] {$ $};
                \node (dummy7) [left of= dummy6] {$ $};
                \node (dummy8) [right of= dummy6] {$ $};
                \draw[->] (startmn2) -- (nextfin);
                \draw[dotted, ->] (nextfin) -- (dummy7);
                \draw[dotted, ->] (nextfin) -- (dummy8);
                \node (startmn) [right=3cm of startmn2] {$ 
                    P \oplus Q
                    $};
                \node (dummy) [below of= startmn] {$ $};
                \node (nextl) [left of= dummy] {$
                    P' \oplus Q
                    $};
                \node (nextr) [right of= dummy] {$
                    P \oplus Q'
                    $};
                \node (nextbelow) [below of= dummy] {$
                    P' \oplus Q'
                    $};
                \node (dummy2) [left of= nextbelow] {$ $};
                \node (dummy3) [left of= dummy2] {$ $};
                \node (dummy4) [right of= nextbelow] {$ $};
                \node (dummy5) [right of= dummy4] {$ $};
                \draw[->] (startmn) -- (nextl);
                \draw[->] (startmn) -- (nextr);
                \draw[->] (nextl) -- (nextbelow);
                \draw[->] (nextr) -- (nextbelow);
                \draw[dotted, ->] (nextl) -- (dummy3);
                \draw[dotted, ->] (nextr) -- (dummy5);
                \draw[<-,thick,red] (startmn) edge [bend right = 10] node[above,black] {$\piencod{-}_u$} (startmn2);
                \draw[<-,thick,red] (nextbelow) edge [bend right = -30] node[below,black] {$\piencod{-}_u$} (nextfin);
            \end{tikzpicture}
        \]
        \[
        \begin{aligned}
            P & = (\nu x_1,x_2) ( C[ \piencod{ M }_u ,   \piencod{ M_1 }_{x_1} , \piencod{ M_2 }_{x_2}  ]) 
           & P' &= (\nu x_2) ( C[\piencod{ M \headlin{M_1/x_1} }_u , \piencod{ M_2 }_{x_2} ] ) \\
            Q & = 
            (\nu x_1,x_2) ( C[ \piencod{ M }_u ,   \piencod{ M_2 }_{x_1} , \piencod{ M_1 }_{x_2}  ])  
          &  Q' &= (\nu x_2) ( C [\piencod{ M \headlin{M_2/x_1} }_u , \piencod{ M_1 }_{x_2} ] ) 
        \end{aligned}    
        \]
\smallskip
        \caption{Soundness under a confluent regime (\Cref{ch2})}
        \label{fig:intro_soundch2}        
    \end{figure}

    \item[\Cref{ch3}] We extend the framework of \Cref{ch2} by considering a resource $\lambda$-calculus with both linear and unrestricted resources as source language. This entails extending the definitions of the source language as well as generalizing the translation into typed processes and its corresponding proofs of correctness. Extending the language with unrestricted resources is non-trivial due to the challenge of correctly catching sources of failure. Also, a naive extension may lead to ambiguities in the consumption of resources, which in turn leads to the unpredictability of failure. We identify a syntax and semantics for unrestricted resources that builds upon the developments in \Cref{ch2} by exploiting client and server behaviors in typed processes.
    

    \item[\Cref{ch4}] Having covered the case of confluent non-determinism, here we shift our attention of the case of non-confluent nondeterminism, with source and target languages including both linear and unrestricted resources. This requires  innovations on the sequential side, but also on the concurrent side, as we now describe.
    
    On the concurrent side, we abandon the setting of \cite{CairesP17}, and introduce a session-typed $\pi$-calculus with a non-confluent non-deterministic choice operator, denoted `$P \nd Q$'. For this calculus, we give two new operational semantics, dubbed \emph{lazy} and \emph{eager}. Intuitively, they differ on how gradually they ``collapse'' branches of a non-deterministic choice in order to express commitment. It turns out that the same translation can be used in either case, and we give two proofs of correctness, one for each semantics.

    The lazy semantics allows us to prove what we call \emph{tight} correctness for our translation, whereas the eager semantics induces a \emph{loose} form of correctness. 
    Hence, we establish a tight soundness result, in which operational correspondence closely matches the computations in the sequential side, and a loose soundness result, in which non-deterministic choice is too eagerly pruned with respect to the given sequential computation. 
  
      \begin{figure}[!t]
        \[
            \begin{tikzpicture}
                \node (startmn2) {$ M \esubst{\bag{M_1,M_2}}{x} $};
                \node (dummy9) [below=0.5cm of startmn2] {$  $};
                \node (nextfin1) [left=0.5cm of dummy9] {$
                    \begin{aligned}
                        M\headlin{M_1/x}\esubst{\bag{M_2}}{x}
                    \end{aligned}    
                    $};
                \node (nextfin2) [right=0.5cm of dummy9] {$
                    \begin{aligned}
                        M\headlin{M_2/x}\esubst{\bag{M_1}}{x}
                    \end{aligned}    
                    $};
                \node (dummy7) [below of= nextfin1] {$ $};
                \node (dummy8) [below of= nextfin2] {$ $};
                \draw[->] (startmn2) -- (nextfin1);
                \draw[->] (startmn2) -- (nextfin2);
                \draw[dotted, ->] (nextfin1) -- (dummy7);
                \draw[dotted, ->] (nextfin2) -- (dummy8);
                \node (startmn) [right=5cm of startmn2] {$  P_1 $};
                \node (next1) [below of= startmn] {$ P_2 $};
                \node (dummy) [below of= next1] {$  $};
                \node (next2) [left of= dummy] {$ P_3 $};
                \node (next3) [right of= dummy] {$ P_4 $};
                \node (dummy2) [below of= next2] {$ $};
                \node (dummy3) [below of= next3] {$ $};
                \node (dummy4) [left of= next1] {$ $};
                \node (dummy5) [right of= next1] {$ $};
                \draw[->] (startmn) -- (next1);
                \draw[->] (next1) -- (next2);
                \draw[->] (next1) -- (next3);
                \draw[dotted, ->] (next2) -- (dummy2);
                \draw[dotted, ->] (next3) -- (dummy3);
                \draw[dotted, ->] (next1) -- (dummy4);
                \draw[dotted, ->] (next1) -- (dummy5);
                \draw[<-,thick,red] (startmn) edge [bend right = 10] node[above,black] {$\piencod{-}_u$} (startmn2);
                \draw[<-,thick,red] (next2) edge [bend right = -20] node[below,black] {$\piencod{-}_u$} (nextfin1);
                \draw[<-,thick,red] (next3) edge [bend right = -35] node[below,black] {$\piencod{-}_u$} (nextfin2);
            \end{tikzpicture}
            \]
            \[
        \begin{aligned}
            P_1 & = (\nu x_1,x_2) ( C[ \nd_{i \in I} \piencod{ M_i }_u    ,  \piencod{ M_1 }_{x_1} , \piencod{ M_2 }_{x_2} ] ) 
            & P_3 &= (\nu x_2) ( C[ \piencod{ M \headlin{M_1/x_1} }_u , \piencod{ M_2 }_{x_2} ] ) 
            \\
                         P_2 & = 
            (\nu x_1,x_2) ( C[ \nd_{i \in I} \piencod{ M'_i }_u ,   \piencod{ M_1 }_{x_1} , \piencod{ M_2 }_{x_2} ] )  
            & P_4 &= (\nu x_1) (  C[ \piencod{ M \headlin{M_2/x_2} }_u ,\piencod{ M_1 }_{x_1} ] ) 
        \end{aligned}    
        \]
        \smallskip
        \caption{Soundness under a non-confluent regime: the tight case  (\Cref{ch4})}
        \label{fig:intro_soundch4_tight} 
  \end{figure}
    
    \Cref{fig:intro_soundch4_tight} illustrates the situation for the case of tight soundness, with the same initial source term as in  \Cref{fig:intro_soundch2}. 
    Process $P_1$ is the translation of the source term:
    it includes the process $\nd_{i \in I} \piencod{ M_i }_u $, which expresses a non-confluent non-deterministic choice of all the permutations of $M$ with each occurrence of $x$ replaced with $x_1,x_2$. Process $P_1$ may reduce to $P_2$, which is an intermediate step needed to mimic the corresponding reduction in $\lambda$. From $P_2$, we may now reduce to $P_3$ or $P_4$ (representing the commitment to $M_1$ and $M_2$, respectively), or perform further  unrelated reduction steps. Because the reduction that leads to $P_3,P_4$ can be performed independently from any other available synchronisations, we choose to prioritise these reductions first. In this setting we cannot use the path to soundness that we exploit in \Cref{ch2,ch3}; our lazy semantics makes use of the delayed commitment (process $P_2$) to match the non-confluent fetching of resources from bags---this is important in establishing our soundness result.
    
   \Cref{fig:intro_soundch4_loose} illustrates  the case of loose soundness, associated to the eager semantics, which is less obvious and requires reconsidering the proof method. 
    We look at sets of reductions available and show that we may group them and relate them to choices in the sequential setting, almost inducing a confluence-like behaviour within the proof. This is essential, as the proof proceeds by induction on the number of reduction steps taken by the translated process; this grouping is key to apply the induction hypotheses. In the figure, $P_2$ reduces to one of $P_3, \cdots , P_6$;  we use 
 $M' , \cdots, M''''$ to denote $M$ with a specific permutation of each occurrence of $x$ replaced with $x_1,x_2$. 
    We can group the  alternatives $P_3, \cdots , P_6$ into sets of ``similar'' substitutions; when $P_2$ realizes commitment, the corresponding choice discards branches  eagerly---this corresponds to steps in the sequential side in which not only a resource is fetched but actually the entire order for fetching resources is decided. 
    In the example, $P_3, P_4, \ldots$ represent processes that synchronise across $x_1$;
    processes in this group are related to the translation of $M\headlin{M_1/x}\esubst{\bag{M_2}}{x}$ (i.e., where $M_1$ is substituted for $x_1$).
Similarly, $P_5, P_6, \ldots$ represent processes that synchronise across $x_2$ and are related to the translation of $M\headlin{M_2/x}\esubst{\bag{M_1}}{x}$.

    \begin{figure}[!t]
        \[
            \begin{tikzpicture}
                \node (startmn2) {$ M \esubst{\bag{M_1,M_2}}{x} $};
                \node (dummy9) [below=0.5cm of startmn2] {$  $};
                \node (nextfin1) [left=0.5cm of dummy9] {$
                    \begin{aligned}
                        M\headlin{M_1/x}\esubst{\bag{M_2}}{x}
                    \end{aligned}    
                    $};
                \node (nextfin2) [right=0.5cm of dummy9] {$
                    \begin{aligned}
                        M\headlin{M_2/x}\esubst{\bag{M_1}}{x}
                    \end{aligned}    
                    $};
                \node (dummy7) [below of= nextfin1] {$ $};
                \node (dummy8) [below of= nextfin2] {$ $};
                \draw[->] (startmn2) -- (nextfin1);
                \draw[->] (startmn2) -- (nextfin2);
                \draw[dotted, ->] (nextfin1) -- (dummy7);
                \draw[dotted, ->] (nextfin2) -- (dummy8);
                \node (startmn) [right=5cm of startmn2]{$  P_1 $};
                \node (next1) [below of= startmn] {$ P_2 $};
                \node (dummy) [below of= next1] {$  $};
                \node (dots1) [left of= dummy] {$ , \dots , \textcolor{red}{ \} }$};
                \node (next2) [left=12mm of  dummy] {$ P_4  $};
                \node (next21) [left=17mm of dummy ] {$\textcolor{red}{ \{ } P_3,  $};
                \node (next3) [right of= dummy] {$ \textcolor{red}{ \{ } P_5,   $};
                \node (next31) [right=12mm of dummy] {$ P_6  $};
                \node (dots2) [right=16mm of dummy] {$ , \dots , \textcolor{red}{ \} }$};
                \node (dummy2) [below of= next2] {$ $};
                \node (dummy3) [below of= next3] {$ $};
                \node (dummy21) [below of= next21] {$ $};
                \node (dummy31) [below of= next31] {$ $};
                \node (dummy4) [left of= next1] {$ $};
                \node (dummy5) [right of= next1] {$ $};
                \draw[->] (startmn) -- (next1);
                \draw[->] (next1) -- (next2);
                \draw[->] (next1) -- (next21);
                \draw[->] (next1) -- (next3);
                \draw[->] (next1) -- (next31);
                \draw[dotted, ->] (next2) -- (dummy2);
                \draw[dotted, ->] (next3) -- (dummy3);
                \draw[dotted, ->] (next21) -- (dummy21);
                \draw[dotted, ->] (next31) -- (dummy31);
                \draw[dotted, ->] (next1) -- (dummy4);
                \draw[dotted, ->] (next1) -- (dummy5);
                \draw[<-,thick,red] (startmn) edge [bend right = 10] node[above,black] {$\piencod{-}_u$} (startmn2);
                \draw[<-,thick,red] (dots1) edge [bend right = -30] node[below,black] {$\piencod{-}_u$} (nextfin1);
                \draw[<-,thick,red] (dots2) edge [bend right = -50] node[below,black] {$\piencod{-}_u$} (nextfin2);
            \end{tikzpicture}
        \]
        \[
        \begin{aligned}
            P_1 & = (\nu x_1,x_2) ( C[ \nd_{i \in I} \piencod{ M_i }_u    ,  \piencod{ M_1 }_{x_1} , \piencod{ M_2 }_{x_2} ]) 
            & P_3 &= (\nu x_2) ( C[ \piencod{ M' \headlin{M_1/x_1} }_u ,\piencod{ M_2 }_{x_2} ] ) 
            \\
                        P_2 & = 
            (\nu x_1,x_2) ( C[ \nd_{i \in I} \piencod{ M'_i }_u , \piencod{ M_1 }_{x_1} , \piencod{ M_2 }_{x_2} ])  
                        & P_4 &= (\nu x_2) ( C[ \piencod{ M'' \headlin{M_1/x_1} }_u , \piencod{ M_2 }_{x_2}  ]) 
          \\
                    &   & P_5 &= (\nu x_1) ( C[ \piencod{ M''' \headlin{M_2/x_2} }_u , \piencod{ M_1 }_{x_1} ])
                    \\
                      &        &  P_6 &= (\nu x_1) ( C[\piencod{ M'''' \headlin{M_2/x_2} }_u , \piencod{ M_1 }_{x_1}] ) 
        \end{aligned}    
        \]
\smallskip
        \caption{Soundness under a non-confluent regime: the loose case  (\Cref{ch4})}
        \label{fig:intro_soundch4_loose}     
    \end{figure}

    \item[\Cref{ch5}] In the final chapter, we turn our attention to the termination property. 
    We consider two different type systems that enforce termination of $\pi$-calculus processes by typing, and compare them using (correct) translations: one is the logically motivated system of \cite{CairesP10}, the other is the type system by \cite{DBLP:conf/ifipTCS/DengS04}, which exploits approaches from rewriting systems to exclude processes with infinite reduction sequences. 
    
    Rather than translating one system into the other directly, we find it convenient to fix a reference framework for comparisons. 
  For this purpose, we use the session-typed framework by  \cite{V12}, which is a liberal discipline that provides a convenient, broad framework for rigorous comparisons. A key technical aspect in this chapter is proving that the system of Caires and Pfenning's  induces a class of typable processes that is a strict subset of the class induced by Deng and Sangiorgi's. 
  To this end, we show that typability under the Curry-Howard correspondences induces a strict partial order on processes, a sort of hierarchy on names that excludes infinite reductions by construction. Because Deng and Sangiorgi's strictly subsumes these  strict partial orders, the strict inclusion between the classes follows. 
  \end{description}

\iffulldoc 
\Cref{ch2appendices_ch1} to \Cref{ch5appendices_ch5} contain omitted proofs to \Cref{ch2} to \Cref{ch5} respectively. \Cref{ch3appendix_types} also contains examples in well-formed derivations and auxiliary definitions. \Cref{ch4appendix_aplas} gives the full $\pi$-calculus utilising client and server behaviour along with an alternative eager semantics and the application of unrestricted bags in the resource $\lambda$-calculus. \Cref{ch5appendices_ch5} contains the auxiliary definitions needed to construct strict partial orders on processes that are essential to the proofs within \Cref{ch5ssecapp:levels}.
\else
We often provide proof sketches for main results; 
omitted technical details and proofs can be found in the full version of this document.
\fi

\section{Associated Publications}

Most of the material of this thesis has been previously reported in the following peer-reviewed publications.

The content reported in \Cref{ch2} is derived from the following papers:
\begin{itemize}

    \item \citet*{DBLP:conf/fscd/PaulusN021} \emph{Non-deterministic functions
    as non-deterministic processes}, in: Kobayashi, N., ed., 6th International Conference on Formal Structures for Computation and Deduction, \textbf{FSCD 2021}, July 17-24, 2021, Buenos Aires, Argentina (Virtual Conference), vol. 195 of LIPIcs, Schloss
    Dagstuhl - Leibniz-Zentrum f\"{u}r Informatik, pp. 21:1–21:22. \\
    \url{https://doi.org/10.4230/LIPIcs.FSCD.2021.21}

    \item \citet*{DBLP:journals/lmcs/PaulusNP23} \emph{Non-deterministic functions as non-deterministic processes (Extended Version)}, \textbf{Logical Methods in Computer Science},    19(4).\\
    \url{https://doi.org/10.46298/lmcs-19(4:1)2023}
	
\end{itemize}

The content reported in \Cref{ch3,ch4,ch5} is derived from the following papers:

\begin{itemize}
    \item \citet*{DBLP:conf/types/PaulusN021} \emph{Types and terms translated:
    Unrestricted resources in encoding functions as processes}, in: Basold, H., J. Cockx,
    S. Ghilezan, eds., 27th International Conference on Types for Proofs and Programs,
    \textbf{TYPES 2021}, June 14-18, 2021, Leiden, The Netherlands (Virtual Conference), vol.
    239 of LIPIcs, Schloss Dagstuhl - Leibniz-Zentrum f\"{u}r Informatik, pp. 11:1–11:24. \\
    \url{https://doi.org/10.4230/LIPIcs.TYPES.2021.11}

    \item \citet*{DBLP:conf/aplas/HeuvelPNP23} \emph{Typed
    non-determinism in functional and concurrent calculi}, in: Hur, C., ed., Programming Languages and Systems - 21st Asian Symposium, \textbf{APLAS 2023}, Taipei, Taiwan,
    November 26-29, 2023, Proceedings, vol. 14405 of Lecture Notes in Computer Science, Springer, pp. 112–132. \\
    \url{https://doi.org/10.1007/978-981-99-8311-7\_6}

    \item \citet*{DBLP:conf/ppdp/Paulus0N23} \emph{Termination in concurrency,
    revisited}, in: Escobar, S., V. T. Vasconcelos, eds., International Symposium on Principles and Practice of Declarative Programming, \textbf{PPDP 2023}, Lisboa, Portugal,
    October 22-23, 2023, ACM, pp. 3:1–3:14. \\
    \url{https://doi.org/10.1145/3610612.3610615}
\end{itemize}

The following paper reports complementary developments to the content in \Cref{ch4} by developing an \emph{eager} semantics:
\begin{itemize}
    \item \citet*{DBLP:conf/aplas/HeuvelPNP24} \emph{Typed non-determinism in concurrent calculi: The eager way}, in: Conference on Mathematical Foundations of Programming Semantics, \textbf{MFPS 2024}, Oxford, UK, June 19–21, 2024.
    \end{itemize}

\clearemptydoublepage 


\chapter{Non-Deterministic Functions as   Non-Deterministic Processes}\label{ch2}




\noindent We study encodings of the $\lambda$-calculus into the $\pi$-cal\-cu\-lus in the unexplored case of calculi with \emph{non-determinism}  and \emph{failures}.
On the sequential side, we consider \lamrfail, a new  non-deterministic calculus in which intersection types control resources (terms); on the concurrent side, we consider~\spi, a $\pi$-calculus in which non-determinism and failure rest upon a Curry-Howard correspondence between linear logic and session types. 
We present a typed encoding of \lamrfail into \spi and establish its correctness. 
Our encoding precisely explains the interplay of  non-deterministic and fail-prone  evaluation in \lamrfail via  typed processes in~\spi.
In particular, it shows how failures in sequential evaluation (absence/excess of resources) can be neatly codified as interaction protocols.

\section*{Introduction}\label{ch2S:one}
Milner's seminal work on encodings of the $\lambda$-calculus into the $\pi$-calculus~\cite{Milner92} explains how \emph{interaction} in $\pi$ subsumes \emph{evaluation} in~$\lambda$. 
It opened a research strand on formal connections between sequential and concurrent calculi, covering untyped and typed regimes (see, e.g.,~\cite{DBLP:journals/mscs/Sangiorgi99,DBLP:conf/birthday/BoudolL00,DBLP:conf/fossacs/BergerHY03,DBLP:conf/fossacs/ToninhoCP12,DBLP:conf/rta/HondaYB14,DBLP:conf/popl/OrchardY16,DBLP:conf/esop/ToninhoY18}). 
This chapter extends this line of work by tackling a hitherto unexplored angle, namely encodability of  calculi in which computation is \emph{non-deterministic} and may be subject to \emph{failures}---two relevant features in sequential and  concurrent programming models.

We focus on \emph{typed} calculi and study how non-determinism and failures interact with \emph{resource-aware} computation. 
In sequential calculi, \emph{non-idempotent intersection types}  offer one fruitful perspective at resource-aware\-ness \revt{}{(see, e.g.,~\cite{DBLP:conf/tacs/Gardner94,DBLP:journals/logcom/Kfoury00,DBLP:journals/tcs/KfouryW04,DBLP:conf/icfp/NeergaardM04,BucciarelliKV17})}.
Because non-idempotency amounts to distinguish between types $\sigma$ and $\sigma \land \sigma$, this class of intersection types can ``count'' different resources and enforce quantitative guarantees. 
In  concurrent calculi, resource-awareness has been much studied using \emph{linear types}. Linearity ensures that process actions occur exactly once, which is key to enforce protocol correctness. 
 \revd{B3}{In particular, \emph{session types}~\cite{DBLP:conf/concur/Honda93,DBLP:conf/esop/HondaVK98} 
specify the protocols that channels must respect; 
this typing discipline exploits linearity to  
ensure absence of communication errors and stuck processes.}
To our knowledge, connections between calculi adopting these two distinct views of resource-awareness  via types are still to be established. 
We aim to develop such connections by relating models of sequential and concurrent computation.

On the sequential side, we introduce \lamrfail: a   $\lambda$-calculus with resources, non-de\-ter\-mi\-nism, and failures, which
distills key elements from  $\lambda$-calculi studied in~\cite{DBLP:conf/concur/Boudol93,PaganiR10}.
Evaluation in  \lamrfail considers \emph{bags} of resources, and determines alternative executions governed by non-determinism.
Failure results from a lack  or excess of resources (terms), and is   captured by the term $\fail^{\widetilde{x}}$, where $\widetilde{x}$ denotes a sequence of variables.
Non-determinism  in  \lamrfail  is \emph{non-collapsing} (i.e., confluent): intuitively, given   $M$ and $N$ with reductions $M \redd M'$ and $N \redd N'$, the non-deterministic sum $M + N$ reduces to $M' + N'$.
In contrast, under a \emph{collapsing} (i.e., non-confluent) approach, as in, e.g.,~\cite{DBLP:conf/mfcs/Dezani-CiancaglinidP93},  the non-deterministic sum  $M + N$ reduces to either $M$ or $N$. 

On the concurrent side, we consider \spi: a  session-typed $\pi$-calculus with (non-collap\-sing) non-de\-ter\-mi\-nism and failure, proposed in~\cite{CairesP17}.
\spi rests upon a Curry-Howard correspondence between session types and   (classical) linear logic, extended with modalities that express   \emph{non-deterministic protocols} that may succeed or fail. Non-determinism in \spi is non-collapsing, which ensures confluent process reductions. 

\paragraph{Contributions} 
This chapter presents \secondrev{the first formal connection between a $\lambda$-calculus with non-idempotent intersection types and a $\pi$-calculus with session types. Specifically, the chapter presents} the following contributions:
\begin{enumerate}
    \item \textbf{The resource calculus \lamrfail}, a new calculus that distills the distinctive elements from previous resource calculi~\cite{DBLP:conf/birthday/BoudolL00,PaganiR10}, while offering an explicit treatment of failures in a setting with non-collapsing non-determinism. 
    
    \secondrev{We develop the syntax, semantics, and essential meta-theoretical results for \lamrfail. In particular, }
    using intersection types, we  define \emph{well-typed} (fail-free) expressions and \emph{well-formed} (fail-prone) expressions in \lamrfail \secondrev{and establish their properties}. 
    
    \item \textbf{An encoding of \lamrfail into \spi},  proven correct following established  criteria \secondrev{in the realm of relative expressiveness for concurrency}~\cite{DBLP:journals/iandc/Gorla10,DBLP:journals/iandc/KouzapasPY19}. 
     These criteria attest to an encoding's quality;  we consider
\emph{type preservation},  \emph{operational correspondence} \srev{(including \;{completeness} and \emph{soundness})},  \emph{success sensitiveness}, and \emph{compositionality}.

Thanks to these correctness properties, our encoding precisely describes how typed interaction protocols \secondrev{(given by session types)} can codify sequential evaluation in which absence and excess of resources leads to failures \secondrev{(as governed  by intersection types)}. 
\end{enumerate}

\smallskip

These contributions entail different challenges. 
The first is bridging the different mechanisms for resource-awareness involved  (i.e., intersection types in \lamrfail, session types in \spi).
A direct encoding of \lamrfail into \spi is far from obvious, as multiple occurrences of a variable in \lamrfail must  be accommodated into the linear setting of \spi. 
\srev{To overcome this challenge, 
we introduce a variant of  \lamrfail, dubbed  \lamrsharfail.
The distinctive feature of \lamrsharfail is a 
\emph{sharing} construct, which we adopt following the \emph{atomic} $\lambda$-calculus presented in~\cite{DBLP:conf/lics/GundersenHP13}.
}
Our encoding of \lamrfail  expressions into \spi processes is then in two steps.
We first define a correct encoding from \lamrfail to \lamrsharfail, which relies on the sharing construct to ``atomize''  occurrences of the same variable.
Then, we define another correct encoding, from \lamrsharfail to \spi, which extends Milner's with constructs for non-determinism. 

Another challenge is framing failures  in \lamrfail (undesirable computations)  as well-typed \spi processes. 
Using intersection types, we define \emph{well-formed} \lamrfail expressions,  which   can fail, in two stages. 
First, we consider \lamr, the sub-language of \lamrfail without   $\fail^{\widetilde{x}}$. 
We give an intersection type system for \lamr to regulate fail-free  evaluation. 
Well-formed expressions are then defined on top of well-typed \lamr expressions.
We show that \spi can correctly encode the fail-free \lamr but, more interestingly, also well-formed \lamrfail expressions, which are fail-prone. 

\begin{figure}[!t]
    \centering
\includegraphics[scale=0.9]{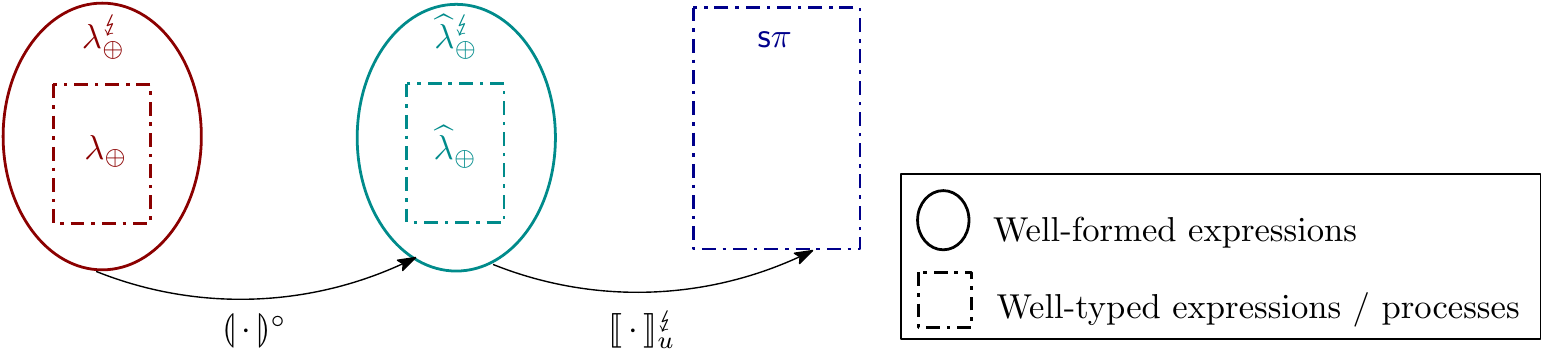}
\vspace{0.5cm}
    \caption{Overview of our approach.}
    \label{ch2fig:summary}
\end{figure}

\figref{ch2fig:summary} summarizes our approach: the encoding 
from \lamrfail to \lamrsharfail is denoted 
 $\recencodopenf{\cdot}$, whereas the encoding from \lamrsharfail to \spi is denoted $\piencodf{\cdot}_u$.

\paragraph{Organization}

 Next, \secref{ch2s:key} informally discusses  key ideas in our work. 
 \secref{ch2s:lambda} introduces the syntax and semantics of \lamrfail, and defines its intersection type system.
\secref{ch2sec:lamsharfail} introduces \lamrsharfail, the variant of \lamrfail with sharing. It also presents its associated intersection type system, and defines an encoding from $\lamrfail $ to $\lamrsharfail$.
In \secref{ch2s:pi} we summarize the syntax, semantics, and session type system of \spi, following~\cite{CairesP17}. 
\secref{ch2s:encoding} establishes the correctness of the encoding of $\lamrfail $ into $\lamrsharfail$ and presents and proves correct the encoding of \lamrshar into \spi. 
\secref{ch2s:rw} presents comparisons with related works.
\secref{ch2s:disc} closes with a discussion about our approach and results.
  
\section{Overview of Key Ideas}
\label{ch2s:key}

\revd{}{Before embarking into our technical developments, we discuss some key ideas in the definition of \lamrfail and its correct encodability into \spi.} 

\paragraph{Non-determinism.} Our source language \lamrfail has three syntactic categories: terms ($M,M'$), bags ($B,B'$) and expressions ($\expr{M}, \expr{L}$). Terms can be variables, abstractions $\lambda x. M$, 
applications $(M\ B)$, explicit substitutions $M\esubst{B}{x}$, or the explicit failure term $\fail$ (see below). 
Bags are multisets of terms (the resources); this way, e.g., $B= \bag{M_1,M_1,M_2}$ is a bag with 
three resources ($M_1$, $M_1$, and $M_2$). Expressions are sums of terms, written $M_1+M_2$; they denote a non-deterministic choice between different ways of \emph{fetching} resources from the bag.

In  $\lamrfail$, reduction is lazy: first, a $\beta$-reduction evolves to an explicit substitution, which will then fetch the elements in the bag to be 
substituted for the corresponding variable, when some conditions are satisfied: we interpret this  as ``consuming a resource". 
For instance, 
given a $\lamrfail$-term $M$
with head variable $x$ and two occurrences of $x$, we have the  reduction:
\begin{eqnarray}
\lambda x. M \bag{M_1,M_2}  & \redd &
 M \esubst{\bag{M_1,M_2}}{x} \nonumber
 \\
  & \redd & M\headlin{M_1/x}\esubst{\bag{M_2}}{x} + M\headlin{M_2/x}\esubst{\bag{M_1}}{x} = M'
	\label{ch2key:ex1} 
\end{eqnarray}
The resulting expression $M'$ is a sum that gathers two alternative computations: it may reduce by either (i)~first fetching $M_1$ from the bag and \secondrev{linearly} substituting it for $x$ in \secondrev{the head position of} $M$  (this is denoted with $M\headlin{M_1/x}$) and then continue with the rest of the bag ($M_2$, wrapped in an explicit substitution), or (ii)~fetching  and 
\secondrev{linearly} substituting $M_2$ \secondrev{in head position}, leaving $M_1$ in an explicit substitution.

\paragraph{Successful Reductions} We consider a computation as successful only  when the number of elements in the bag
matches the number of occurrences of the variable to be substituted; otherwise the computation fails.
 As an example, consider the previous example,  now with  $ M= x\bag{x\bag{I}}$ 
where $I=\lambda z.z$ is the identity. The reduction in (\ref{ch2key:ex1}) is then 
$$
 ( \lambda x. x\bag{x\bag{I}}) \bag{M_1,M_2}  \redd^* M_1\bag{x\bag{I}}\esubst{\bag{M_2}}{x} + M_2\bag{x\bag{I}}\esubst{\bag{M_1}}{x}
 $$

Hence,    when $\lambda x. M$ is applied to a bag with two resources, it evolves successfully.  
However, if $\lambda x. M$ is applied to a bag with less (or more) than two resources, the computation  evolves to the \emph{explicit failure} term $\fail^{\widetilde{z}}$,  where $\widetilde{z}$ is a multiset of variables, as we explain next.

\paragraph{Explicit Failure.}
A construct for failure is present in the resource $\lambda$-calculus in~\cite{PaganiR10}. 
In this formulation, the failure term `$0$' is consumed by sums and disappears at the end of the computation; as such, it gives no information about the failed computation and its origins.

Following~\cite{PaganiR10}, a design decision in $\lamrfail$ is to have $\fail^{\widetilde{x}}$ in the syntax of terms. 
The sequence $\widetilde{x}$ denotes the variables captured by failure; this provides useful information on the origins of a failure. 
As an example, consider a term $M$ with free variables $\widetilde{y}$ and in which the number of occurrences of $x$ is  different from 2. 
Given a bag $B=\bag{M_1,M_2}$, reduction leads to a failure, as follows:
$$
    (\lambda x. M) B \redd M\ \esubst{\bag{M_1,M_2}}{x}\redd {}  \displaystyle\sum_{\perm{\bag{M_1,M_2}}} \fail^{\widetilde{y}} = M'
    $$
In this case, $M'$ is the sum $\fail^{\widetilde{y}}+\fail^{\widetilde{y}}$, which has as many summands as the permutations of the elements of $B$. Intuitively, it means that it does not matter if one replaces the occurrence(s) of $x$ first with $M_1$ (or $M_2$), then the other occurrence (if any), with $M_2$ (or $M_1$), the result will be the same, i.e., $\fail^{\widetilde{y}}$. Here again both possibilities are expressed in a sum. The precise semantics of failure will be presented in \secref{ch2ssec:red_sem}.

\paragraph{Typability and Well-formedness}

We define an intersection type system for $\lamrfail$. This choice follows a well-established tradition of coupling resource $\lambda$-calculi with intersection types~\cite{DBLP:conf/concur/Boudol93,BoudolL96,PaganiR10}. Intersection types are also adopted in related calculi~\cite{DBLP:journals/tcs/EhrhardR03}. Intersection types are a natural typing structure for resources: they have similar mathematical properties of non-idempotency and commutativity, and can help to ``count'' the number of occurrences of a variable in a term, as well as the number of components in a bag. 

In our type systems, each element of a bag must have the same type. 
This way, e.g., a well-typed bag $B=\bag{M_1,M_2,M_3}$ has type $ \sigma\wedge \sigma \wedge \sigma$, where $\sigma$ is a strict type (cf. \defref{ch2d:typeslamrfail}). Then,  an application $M\ B$ is well-typed, say, with  type $\tau$, only if $M:\sigma\wedge  \sigma \wedge \sigma\to \tau$. 
We shall write $\sigma^k$ to denote the intersection type $\sigma\wedge \ldots \wedge \sigma$, with $k\geq 0$ copies of $\sigma$.
Notice that $\sigma^0$ denotes the empty type $\omega$.
The typing rule for application is then as expected:

\begin{prooftree}
  \AxiomC{\(  \Gamma \vdash M :  \sigma^k \to \tau \)}
    \AxiomC{\( \Gamma\vdash B : \sigma^k \)}
        \LeftLabel{\redlab{T:app}}
    \BinaryInfC{\(  \Gamma\vdash M\ B : \tau\)}
\end{prooftree}
where $\Gamma$ is a type context assigning types to variables.  

We chose to express explicit failing terms and computation.
To properly account for these computations, we define  a separate type system with  so-called \emph{well-formedness} rules, with notation `$\wfdash$'. 
Unlike rules for typability, rules for well-formedness capture computations that fail due to a mismatch  of resources (lack or excess). 
This entails some increased flexibility in selected rules. 
This way, e.g, the following is the well-formedness rule for application:
\begin{prooftree}
  \AxiomC{\( \Gamma \wfdash M :  \sigma^j \to \tau \)}
    \AxiomC{\( \Gamma \wfdash B : \sigma^k\)}
        \LeftLabel{\redlab{F:app}}
    \BinaryInfC{\( \Gamma \wfdash M\ B : \tau\)}
\end{prooftree}
Here the added flexibility is that we do not require $k=j$; hence, the rule can capture successful \emph{and} failing computations, depending on whether $k=j$ or not. As expected, the term $\fail^{\widetilde{z}}$ is not well-typed, but it is well-formed: the judgement $\Gamma\wfdash \fail^{\widetilde{z}}:\tau$ holds for an arbitrary type $\tau$ and a $\Gamma$ consisting of variable assignments 
for the variables in $\widetilde{z}$. 

Therefore, we consider two intersection type systems: one captures exclusively successful computations (see   \figref{ch2fig:app_typingrepeat}); the other, which we call the well-formedness system (see  \figref{ch2fig:app_wf_rules}), subsumes the first one by admitting both successful and failing computations. 
The weakening rule is admissible in both systems (see below).
Both systems enjoy subject reduction, whereas only well-typed terms satisfy subject expansion.

\paragraph{Controlling resources via sharing}
In order to better control the use of resources, i.e., substituting variables for terms with a careful form of duplication, we borrow ideas from the {\em sharing graphs} by \cite{GUERRINI199999,GUERRINI2003379} and define the calculus $\lamrsharfail$.
The key idea is as follows: whenever a bound variable $x$ occurs multiple times within a term, these occurrences, say \(x_1,\ldots, x_n\), are temporarily assigned new names (think aliases). This assignment is indicated with the {\em sharing construct}   $\shar{x_1,\ldots, x_n}{x}$, which we adopt following~\cite{DBLP:conf/lics/GundersenHP13}. 
This way, for instance, the $\lamrfail$-term $\lambda x. x \bag{x}$ would correspond to $\lambda x. x_1\bag{x_2}\shar{x_1,x_2}{x}$ in  $\lamrsharfail$.

 We also carefully treat the ``erasing'' of resources: if a term has vacuous abstractions, this is also indicated with the sharing construct, where the bound variable maps to ``empty''. Hence, the $\lamrfail$-term $\lambda x. y \bag{z}$ is expressed as $\lambda x. y\bag{z}\shar{}{x}$ in  $\lamrsharfail$.
The tight control of resources in $\lamrsharfail$ turns out to be very convenient to encode $\lamrfail$ into \spi, as we discuss next.

\paragraph{Encoding $\lamrfail$ into \spi.}
The central result of our work is a correct translation of $\lamrfail$ into \spi. 
In defining our translation we use $\lamrsharfail$ as a stepping stone. 
This is advantageous, because (i)~the relation between $\lamrshar$ and $\lamrsharfail$ is fairly direct and (ii)~the sharing construct in $\lamrsharfail$ makes it explicit the variable occurrences that should be treated as linear names in \spi. 

The encoding of $\lamrfail$ into $\lamrsharfail$ is denoted $\recencodf{\cdot }$ and given in \secref{ch2ss:auxtrans}.
The encoding of $\lamrsharfail$ into \spi, denoted $\piencodf{\cdot}_u$ and presented in \secref{ch2ss:second_trans}, is arguably more interesting---we discuss it below. 


The definition of $\piencodf{\cdot}_u$ considers well-formed source terms in $\lamrsharfail$ which are translated into well-typed \spi processes.
As usual, the translation is parametric on a channel name $u$, which is used to provide the behavior of the source term. 

The calculus \spi includes a non-deterministic choice operator $P \oplus Q$ and formalizes sessions which are \emph{non-deterministically available}. Intuitively, this means that a given session protocol along a name can either be available and proceed as prescribed by the corresponding session type, or fail to be available. Clearly, such a failure may have repercussions on other sessions that depend on it. 
To this end, \spi includes prefixes $x.\overline{\some}$ and $x.\overline{\none}$, which are used to confirm the availability of $x$ and to signal its failure, respectively. 
Process $x.\some_{(w_1, \cdots, w_k)};Q$ declares the dependency of sessions $w_1, \ldots, w_k$ in $Q$ on an external session along $x$. 
The corresponding reduction rules are then:
\begin{eqnarray*}
	\label{ch2eq:motiv_some}
x.\overline{\some} \para x.\some_{(w_1, \cdots, w_k)};Q  & \redd &
Q
\\
	\label{ch2eq:motiv_none}
x.\overline{\none} \para x.\some_{(w_1, \cdots, w_k)};Q  & \redd &
w_1.\overline{\none} \para \cdots  \para w_k.\overline{\none}
\end{eqnarray*}

%
\noindent
Following Milner, $\piencodf{\cdot}_u$ maps computation in $\lamrsharfail$ into session communication in \spi; non-deterministic sessions are used to codify the non-deterministic fetching of resources in $\lamrsharfail$.
This way, the translation of $(\lambda x. M\shar{x_1,x_2}{x}) B$ will enable synchronizations between the translations of $M\shar{x_1,x_2}{x}$ and $B$. 
More in details, the translation of a bag $B=\bag{M_1,M_2}$ is as follows: 
\begin{eqnarray*}
\piencodf{\bag{M_1}\cdot \bag{M_2}}_x & = &   x.\some_{\widetilde{z_1},\widetilde{z_2}} ; x(y_i). x.\some_{y_i,\widetilde{z_1},\widetilde{z_2}};x.\overline{\some} ; 
\\
& & \quad 
 \secondrev{\outact{x}{x_i}. (x_i.\some_{\widetilde{z_1}} ; \piencodf{M_1}_{x_i} \mid \piencodf{\bag{M_2}}_x \mid y_i. \overline{\none}) }
\end{eqnarray*}
where $\widetilde{z_1}$ and $\widetilde{z_2}$ denote the free variables of $M_1$ and $M_2$, respectively. 
Process $\piencodf{\bag{M_1}\cdot \bag{M_2}}_x$  first expects confirmation of session $x$; then, the translation of each resource $M_i$ is made available in a dedicated name $x_i$, which will be communicated to other processes.
Accordingly, the translation of  $\piencodf{M\shar{x_1,x_2}{x}}_u$ is expected to synchronize with the translation of the bag $B$: indeed, it confirms behavior along $x$, before receiving the names, one for each shared copy of $x$ that should be used throughout the synchronizations:
\[
\begin{aligned}
\piencodf{M\shar{x_1,x_2}{x}}_{u}&= x.\overline{\some}. \outact{x}{y_1}. \Big(y_1 . \some_{\emptyset} ;y_{1}.\close 
       \mid x.\overline{\some}; x.\some_{u, (\lfv{M} \setminus \{x_1 ,  x_2\} )};  \\
       & \hspace{2.0em }x(x_1) .. x.\overline{\some}. \outact{x}{y_2} . \big(y_2 . \some_{\emptyset} ; y_{2}.\close  \mid x.\overline{\some};\\
       & \hspace{2.0em} x.\some_{u,( \lfv{M} \setminus \{x_2\} ) };  x(x_2)
      . x.\overline{\some}; \outact{x}{y_{}}. ( y_{} . \some_{u, \lfv{M} } ; \\
       & \hspace{2.0em}y_{}.\close; \piencodf{M}_u \mid x.\overline{\none} )~ \big)  \Big) 
\end{aligned}
\]
Several confirmations take place along the channel names involved in the synchronizations; see  \secref{ch2ss:second_trans} for details. 

Non-determinism plays a key role in the translation of an application $M' B$. In this case, we consider the permutations of the elements of $B$ using non-deterministic choice in~$\spi$. 
When $B=\bag{M_1,M_2}$,  the translation is:

\[
\begin{aligned} 
\piencodf{M'\, B}_u  =~~ & (\nu v)(\piencodf{M'}_v \mid v.\some_{u , \lfv{B}} ; \outact{v}{x} . ([v \leftrightarrow u] \mid \piencodf{\bag{M_1,M_2}}_x ) ) 
\\
& \qquad  \oplus 
\\
&  (\nu v)(\piencodf{M'}_v \mid v.\some_{u , \lfv{B}} ; \outact{v}{x} . ([v \leftrightarrow u] \mid \piencodf{\bag{M_2,M_1}}_x ) )
\end{aligned}
\]
A synchronization occurs when process $\piencodf{M'}_v$ can confirm its behavior along $v$. For instance, when $M'=\lambda x. M\shar{x_1,x_2}{x}$ the translation is as
\[   \piencodf{\lambda x.M[x_1,x_2 \leftarrow x]}_v  = v.\overline{\some}; v(x).\piencodf{M[x_2,x_2 \leftarrow x]}_v
\]
and the synchronization may be possible; it depends on the translations of $M$, $M_1$, and $M_2$. 

We close this section by observing that our translations $\recencodf{\cdot }$ and $\piencodf{\cdot}_u$ satisfy well-known  correctness criteria, as formulated by \cite{DBLP:journals/iandc/Gorla10} and \cite{DBLP:journals/iandc/KouzapasPY19} (see \secref{ch2ss:criteria} for details).

\section{ \texorpdfstring{$\lamrfail$}{?}: A \texorpdfstring{$\lambda$}{}-calculus with Non-Determinism and Failure}\label{ch2s:lambda}

We define the syntax and reduction semantics of \lamrfail, our new resource calculus with non-determinism and failure. \revd{B4}{We then equip it with} non-idempotent session types, and establish the subject reduction property for well-typed and well-formed expressions (Theorems~\ref{ch2t:app_lamrsr} and \ref{ch2t:app_lamrfailsr}, respectively).
\revt{}{We also consider the subject expansion property, which holds for well-typed expressions (Theorem~\ref{ch2t:app_lamrexp}) but not for well-formed ones (Theorem~\ref{ch2t:app_lamrfailse}).}

\subsection{Syntax}\hfill
The syntax of \lamrfail combines elements from calculi introduced and studied by \cite{DBLP:conf/birthday/BoudolL00} and by \cite{PaganiR10}.
We use $x, y, \ldots$ to range over the set of \emph{variables}.
 We write $\widetilde{x}$ to denote the sequence of pairwise distinct variables $x_1,\ldots,x_k$, for some $k\geq 0$.
 {We write $|\widetilde{x}|$ to denote the length of $\widetilde{x}$}.
 
\begin{definition}{Syntax of \lamrfail }
\label{ch2def:rsyntaxfail}
The \lamrfail calculus is defined by the following grammar:
\[
\begin{array}{l@{\hspace{10mm}}lll}
\mbox{(Terms)} &M,N, L&::=& x\sep \lambda x . M \sep (M\ B) \sep  M \esubst{B}{x} \sep \fail^{\widetilde{x}}\\
\mbox{(Bags)} &A, B&::=& \oneb \sep \bag{M} \sep A \cdot B \\
\mbox{(Expressions)} & \expr{M}, \expr{N}, \expr{L}&::=&  M \sep \expr{M}+\expr{N}\\
\end{array}
\]
\end{definition}
\noindent
We have three syntactic categories: \emph{terms} (in functional position); \emph{bags} (in argument position), which denote multisets of resources; and \emph{expressions}, which are finite formal sums that represent possible results of a computation. 
Terms are unary expressions: they can be  variables, abstractions, and applications. 
Following~\cite{DBLP:conf/concur/Boudol93,DBLP:conf/birthday/BoudolL00}, the \emph{explicit substitution} of a bag $B$ for a variable $x$ \revd{B5}{in a term $M$, written $M\esubst{B}{x}$}, is also a term.
The term $\fail^{\widetilde{x}}$ results from a reduction in which there is a lack or excess of resources to be substituted, where  $\widetilde{x}$ denotes a multiset of free variables that are encapsulated within failure.

The empty bag is denoted $\oneb$. 
The bag enclosing the term $M$ is  $\bag{M}$.
The concatenation of bags $B_1$ and $B_2$ is denoted as   $B_1 \cdot B_2$; the concatenation operator `$\cdot$' is associative and  commutative, with  $\oneb$ as its identity. 
\secondrev{To ease readability, we rely on a shorthand notation for bags: we often write $\bag{N_1, N_2}$ rather than $\bag{N_1}\cdot \bag{N_2}$.}

We treat expressions as \emph{sums}, and use notations such as $\sum_{i}^{n} N_i$ for them. Sums are associative and commutative; reordering of the terms in a sum is performed silently.

\begin{example}{}\label{ch2ex:terms}
We give some examples of terms and expressions in \lamrfail:

\begin{multicols}{2}
    \begin{itemize}
        \item $M_1 = (\lambda x. x ) \bag{y}$
        \item $M_2 = (\lambda x. x ) (\bag{y,z} )$
        \item $M_3 = (\lambda x. x ) \oneb  $
        \item $M_4 = (\lambda x. y ) \oneb $
        \item $M_5 = \fail^{\emptyset} $
        \item $M_6 = (\lambda x. x ) \bag{y} + (\lambda x. x ) \bag{z} $
    \end{itemize}
\end{multicols}
Terms $M_1$, $M_2$, and $M_3$ illustrate the application of the identity function  $I=\lambda x.x$ to bags with different formats: a bag with one component, two components, and the empty bag, respectively. Special attention should be given to the fact that the $x$ has only one occurrence in $I$, whereas the bags contain zero or more components (resources). This way:
\begin{itemize}
    \item $M_1$ represents a term with a \emph{correct} number of resources;
    \item $M_2$ denotes a term with an \emph{excess} of resources; and 
    \item $M_3$ denotes a term with a \emph{lack} of resources
\end{itemize}
 This resource interpretation will become clearer once the  reduction semantics is introduced in the next subsection (cf. Example~\ref{ch2ex:reducts}).

    Term $M_4$ denotes the application of a vacuous abstraction on $x$  to the empty bag $\oneb$.
    Term $M_5$ denotes a failure term with no associated variables.
    Expression $M_6$ denotes the non-deterministic sum between two terms, each of which denotes an application of $I$ to a bag containing one element.
\end{example}

\begin{notation}[Expressions]
Notation $N \in \expr{M}$ denotes that 
$N$ is part of the sum denoted by $\expr{M}$. 
Similarly, we write $N_i \in B$ to denote that $N_i$ occurs in the bag $B$, and $B \linsetminus N_i$ to denote the   bag that is obtained by removing one occurrence of the term $N_i$ from $B$. 
\end{notation}

\subsection{Reduction Semantics}\label{ch2ssec:red_sem}\hfill

Reduction in \lamrfail is defined in terms of the relation $\redd$, defined  in \figref{ch2fig:reductions_lamrfail}; it operates lazily on expressions,  and  will be described after introducing some auxiliary notions. 

\begin{notation}
  We write  $\perm{B}$ to denote the set of all permutations of bag $B$.
Also, $B_i(n)$ denotes the $n$-th term in the (permuted)  $B_i$.
We define $\size{B}$ to denote the number of terms in bag $B$. 
That is, $\size{\oneb} = 0$
and 
$\size{\bag{M}  \cdot B} = 1 + \size{B}$.
\end{notation}

\begin{definition}{Set and Multiset of Free Variables}
\label{ch2def:fvfail}
The set of free variables of a term, bag, and expression, is defined as
    \[
    \begin{array}{l@{\hspace{1cm}}l}
\begin{array}{r@{\hspace{-.01mm}}l}
\lfv{x} \; &= \{ x \}   \\
\lfv{\lambda x . M} \; &= \lfv{M}\!\setminus\! \{x\} \\
\lfv{M\ B} \; &=  \lfv{M} \cup \lfv{B}\\
 \lfv{M \esubst{B}{x}} \; &= (\lfv{M}\setminus \{x\}) \cup \lfv{B} 
\end{array}
         & 
 \begin{array}{r@{\hspace{-.01mm}}l}
\lfv{\oneb} \; &= \emptyset \\
\lfv{\bag{M}} \; &= \lfv{M} \\
 \lfv{B_1 \cdot B_2} \; &= \lfv{B_1} \cup \lfv{B_2}\\
  \lfv{\fail^{x_1, \cdots , x_n}}\; & = \{ x_1, \cdots , x_n \}\\
  \lfv{\expr{M}+\expr{N}}\; & = \lfv{\expr{M}} \cup \lfv{\expr{N}}
 \end{array}
    \end{array}
    \]

We use $\mfv{M}$ or $\mfv{B}$ to denote a multiset of free variables, defined similarly. 
We sometimes treat the sequence $\widetilde{x}$ as a (multi)set.
We write $\widetilde{x}\uplus \widetilde{y}$ to denote the multiset union of $\widetilde{x}$ and $\widetilde{y}$ and $\widetilde{x} \setminus y$ to express that every occurrence of $y$ is removed from $\widetilde{x}$.
A term $M$ is \emph{closed} if $\lfv{M} = \emptyset$ (and similarly for expressions). \revdaniele{As usual, we shall consider $\lamrfail$-terms modulo $\alpha$-equivalence.}
\end{definition}

\begin{notation}
 $\#(x , M)$ denotes the number of (free) occurrences of $x$ in $M$. 
Similarly, we write $\#(x,\widetilde{y}) $ to denote the number of occurrences of $x$ in the multiset $\widetilde{y}$. 
\end{notation}

\begin{definition}{Head}
\label{ch2def:headfailure}
Given a term $M$, we define $\headf{M}$ inductively as:
\[
\begin{array}{l@{\hspace{0.5cm}}l}
\begin{array}{l}
 \headf{x}  = x     \\
  \headf{\lambda x.M}  = \lambda x.M \\
  \headf{M\ B}  = \headf{M}
\end{array}
     & 
\begin{array}{l}
\headf{\fail^{\widetilde{x}}}  = \fail^{\widetilde{x}}\\
\headf{M \esubst{ B }{x}} = 
\begin{cases}
    \headf{M} & \text{if $\#(x,M) = \size{B}$}\\
    \fail^{\emptyset} & \text{otherwise}
\end{cases}
\end{array}
\end{array}
\]
\end{definition}

\begin{definition}{Linear Head Substitution}
\label{ch2def:linsubfail}
Let $M$ be a term such that $\headf{M}=x$. 
The \emph{linear head substitution} of a term $N$ for $x$ in $M$, denoted  $M\headlin{ N/x }$,  is  defined  as:
\[
\begin{aligned}
     x \headlin{ N / x}   &= N \\
    (M\ B)\headlin{ N/x}  &= (M \headlin{ N/x })\ B \\
    (M\ \esubst{B}{y})\headlin{ N/x }  &= (M\headlin{ N/x })\ \esubst{B}{y} \qquad \text{where } x \not = y
\end{aligned}
\]
\end{definition}

\noindent 
Finally, we define contexts for terms and expressions and convenient notations:

\begin{definition}{Term and Expression Contexts}\label{ch2def:context_lamrfail}
Contexts for terms (CTerm) and expressions (CExpr) are defined by the following grammar:
\[
\begin{array}{c@{\hspace{0.5cm}}c}
 \text{(CTerm)}\quad  C[\cdot] ,  C'[\cdot] ::=  ([\cdot])B \mid ([\cdot])\esubst{B}{x}     \\ \text{(CExpr)}  \quad  D[\cdot] , D'[\cdot] ::= M + [\cdot] \mid   [\cdot] + M 
\end{array}
\]
\end{definition}


\begin{figure}[t]
    \centering
    
\begin{prooftree}
    \AxiomC{}
    \LeftLabel{\redlab{R:Beta}}
    \UnaryInfC{\((\lambda x. M) B \redd M\ \esubst{B}{x}\)}
    \end{prooftree}

\begin{prooftree}
    \AxiomC{$\headf{M} = x$}
    \AxiomC{$B = \bag{N_1, \dots , N_k} \ , \ k\geq 1 $}
    \AxiomC{$ \#(x,M) = k $}
    \LeftLabel{\redlab{R:Fetch}}
    \TrinaryInfC{\(
    M\ \esubst{ B}{x } \redd M \headlin{ N_{1}/x } \esubst{ (B\linsetminus N_1)}{ x }  + \cdots + M \headlin{ N_{k}/x } \esubst{ (B\linsetminus N_k)}{x}
    \)}
\end{prooftree}   

\begin{prooftree}
    \AxiomC{$\#(x,M) \neq \size{B} \qquad \widetilde{y} = (\mfv{M} \setminus x) \uplus \mfv{B} $}
    \LeftLabel{\redlab{R:Fail}}
    \UnaryInfC{\(  M\ \esubst{ B}{x } \redd {}  \displaystyle\sum_{\perm{B}} \fail^{\widetilde{y}} \)}
\end{prooftree}

\begin{prooftree}
    \AxiomC{$  \widetilde{y} = \mfv{B} $}
    \LeftLabel{$\redlab{R:Cons_1}$}
    \UnaryInfC{\(  \fail^{\widetilde{x}}\ B \redd{}  \displaystyle\sum_{\perm{B}} \fail^{\widetilde{x} \uplus \widetilde{y}} \)}
\end{prooftree}

\begin{prooftree}   
\AxiomC{$\size{B} = k  \qquad \#(z , \widetilde{x}) + k  \not= 0 \qquad  \widetilde{y} = \mfv{B}$}
    \LeftLabel{$\redlab{R:Cons_2}$}
    \UnaryInfC{\( \fail^{\widetilde{x}}\ \esubst{B}{z}  \redd {} \displaystyle \sum_{\perm{B}} \fail^{(\widetilde{x} \setminus z) \uplus \widetilde{y}}  \)}
\end{prooftree}

\begin{prooftree}
        \AxiomC{$   M \redd M'_{1} + \cdots + M'_{k} $}
        \LeftLabel{\redlab{R:TCont}}
        \UnaryInfC{$ C[M] \redd  C[M'_{1}] + \cdots +  C[M'_{k}] $}
\DisplayProof\hfill%
        \AxiomC{$ \expr{M}  \redd \expr{M}'  $}
        \LeftLabel{\redlab{R:ECont}}
        \UnaryInfC{$D[\expr{M}]  \redd D[\expr{M}']  $}
\end{prooftree}

    \caption{Reduction Rules for $\lamrfail$}
    \label{ch2fig:reductions_lamrfail}
    \hfill \break
\end{figure}

\revdaniele{The reduction relation on $\lamrfail$ is defined by the rules in 
\figref{ch2fig:reductions_lamrfail}. 
} Intuitively, reductions in $\lamrfail$ work as follows: A $\beta$-reduction induces an explicit substitution of a bag $B$ for a variable $x$ \revdaniele{in a term $M$}, denoted $\revdaniele{M}\esubst{B}{x}$. \revdaniele{In the case the head of the term $M$ is $x$ and the size of the bag $B$ coincides with the number of occurrences of $x$ in $M$, }this explicit substitution is expanded into a sum of terms, each of which features a \emph{linear head substitution} $M\headlin{ N_i/x }$, where $N_i$ is a term in $B$, \revdaniele{which will replace the variable $x$ occurring in the head of $M$}; the rest of the bag ($B\linsetminus N_i$) is kept in an explicit substitution. However, if there is a mismatch between the number of occurrences of the variable to be substituted and the number of resources available, then the reduction leads to the failure term. Formally,

\begin{itemize}
\item {\bf Rule~$\redlab{R:Beta}$} is standard and admits a  bag (possibly empty) as parameter. 

\item {\bf Rule~$\redlab{R:Fetch}$} transforms a term into an expression: it opens up an explicit substitution into a sum of terms with linear head  substitutions, each denoting the partial evaluation of an element from the bag, \revdaniele{considering all the possible choices for substituting an element $N_i$ of the bag for $x$}. 
Hence, the size of the bag will determine the number of summands in the resulting expression.
\end{itemize}

\noindent
There are three rules reduce to the failure term: their objective is to accumulate all (free) variables involved in failed reductions. 

\begin{itemize}
\item  {\bf Rule~$\redlab{R:Fail}$} formalizes failure in the evaluation of an explicit substitution $M\ \esubst{ B}{x }$, which occurs if there is a mismatch between the resources (terms) present in $B$ and the number of occurrences of $x$ to be substituted. 
The resulting failure preserves all free variables in $M$ and $B$ within its attached multiset $\widetilde{y}$, \revdaniele{and all possible computations that could have failed, via permutation of the bags, are captured  in a non-deterministic sum}.
\item {\bf Rules~$\redlab{R:Cons_1}$ and $\redlab{R:Cons_2}$} describe reductions that lazily consume the failure term, when a term has $\fail^{\widetilde{x}}$ at its head position. 
The former rule consumes bags attached to it whilst preserving all its free variables.
The latter rule is similar but for the case of explicit substitutions; its second premise ensures that either (i)~the bag in the substitution is not empty or (ii)~the number of occurrences of $z$ in the current multiset of accumulated variables is not zero. 
\end{itemize}
\revdaniele{
Notice that our Rule~\redlab{R:Fail} rule evolves to  a sum of failure terms, where each summand accounts for a permutation of the elements of the bag. As our reduction strategy fails eagerly this may  not be evident at first; however, there is still a non-deterministic choice of elements in $B$ that are waiting to be substituted at the point of failure (see Example~\ref{ch2exa:fail_sum}).
}

Finally, we describe the contextual rules:
\begin{itemize}
\item {\bf Rule~$\redlab{R:TCont}$} describes the reduction of sub-terms within an expression; in this rule, summations are expanded outside of term  contexts.

\item {\bf Rule $\redlab{R:ECont}$} says that reduction of expressions is closed by expression contexts.
\end{itemize}

\begin{notation}
As standard, 
 $\redd$ denotes one step reduction; 
 $\redd^+$ and  $\redd^*$ denote the transitive and the reflexive-transitive closure of $\redd$, respectively. 
 We write $\expr{N}\redd_{\redlab{R}} \expr{M}$ to denote that $\redlab{R}$ is the last (non-contextual) rule used in inferring the step from $\expr{N}$ to $\expr{M}$.
\end{notation}

\begin{example}{Cont. Example~\ref{ch2ex:terms}}
\label{ch2ex:reducts}
We show how the terms in Example~\ref{ch2ex:terms} can 
 reduce: 

\begin{itemize}
    \item Reduction of the term $M_1$ with an adequate number of resources:  
    \[
    \begin{aligned} 
    (\lambda x. x ) \bag{y} &\redd_\redlab{R:Beta} x \esubst{\bag{y}}{x}\\
    &\redd_\redlab{R:Fetch} y\esubst{\oneb}{x}    &\text{ since }  \#(x,x) =\size{\bag{y}} =1
    \end{aligned}
    \]
    \item  Reduction of  term $M_2$ with  excess  of resources:  
    \[
    \begin{aligned}
    (\lambda x. x ) (\bag{y,z}  ) &\redd_\redlab{R:Beta} x \esubst{(\bag{y,z}  )}{x} \\
    &\redd_{\redlab{R:Fail}} \revd{B6}{\fail^{y,z} +  \fail^{y,z}} , \ \text{ since }  \#(x,x)  =1 \neq \size{\bag{y,z}}=2
    \end{aligned}
    \]
    \item Reduction of  term $M_3$ with lack  of resources:  
    \[
    \begin{aligned} 
    (\lambda x. x ) \oneb &\redd_\redlab{R:Beta} x \esubst{\oneb}{x}\\
    &\redd_{\redlab{R:Fail}} \fail^{\emptyset}, \ \text{ since }  \#(x,x)  =1 \neq \size{\oneb}=0
    \end{aligned}
    \]
    \item  Reduction of  term $M_4$ which is a vacuous abstraction applied to an empty bag:
    \[
    \begin{aligned} 
    (\lambda x. y ) \oneb & \redd_\redlab{R:Beta} y\esubst{\oneb}{x}\\
    \end{aligned}
    \]

    
    
    \item $M_5 = \fail^{\emptyset} $ is unable to perform any reductions, i.e., it is irreducible.
    \item Reductions of the expression $M_6 = (\lambda x. x ) \bag{y} + (\lambda x. x ) \bag{z} $ :
    
        \begin{tikzpicture}
          \matrix (m) [matrix of math nodes, row sep=0.5em, column sep=0.5em,ampersand replacement=\&]
            { 
            \node(A){ }; \& 
                \node(B){ x \esubst{\bag{y}}{x} + (\lambda x. x ) \bag{z} }; \\
            \node(C){  (\lambda x. x ) \bag{y} + (\lambda x. x ) \bag{z} }; \& 
                \node(D){ }; \& 
                \node(G){x \esubst{\bag{y}}{x} +  x \esubst{\bag{z}}{x} }; \\
            \node(E){ }; \&
                \node(F){ (\lambda x. x ) \bag{y} + x \esubst{\bag{z}}{x} }; \\};
        \path (C) edge[->](B);
        \path (C) edge[->](F);
        \path (B) edge[->](G);
        \path (F) edge[->](G);
        \end{tikzpicture} 
\end{itemize}

\end{example}

The following example illustrates the use of $\perm{B}$ in Rule~\redlab{R:Fail}: independently of the order in which the resources in the bag are used, the computation fails. 

\begin{example}{}\label{ch2exa:fail_sum}
Let $M = ( \lambda x . x \bag{x \bag{y}} )\  B$, with 
$B = \bag{z_1,z_2,z_1}$.
We have: 
  \[
    \begin{aligned}
       M \redd_\redlab{R:Beta}& x \bag{x \bag{y}} \esubst{ \bag{z_1,z_2,z_1} }{ x } \\
         \redd_\redlab{R:Fail}& \sum_{\perm{B}} \fail^{y , z_1, z_2, z_1 }
    \end{aligned}
 \]
 The number of occurrences of $x$ in the term 
obtained after  $\beta$-reduction (2) does not match the size of the bag (3). Therefore, the reduction leads to failure. 
 \revo{:(A5)}{Notice that 
 $ \sum_{\perm{B}} \fail^{y , z_1, z_2, z_1 }$
 expands to a sum between six instances of 
$\fail^{y , z_1, z_2, z_1}$,
corresponding to \revdaniele{permutation of  3 elements of} the bag $B$.}
 \end{example}
 
 Notice that the left-hand sides of the reduction rules in $\lamrfail$  do not interfere with each other. 
Therefore, reduction in \lamrfail satisfies a \emph{diamond property}:

\begin{restatable}[Diamond Property for \lamrfail]{proposition}{diamondone}
\label{ch2prop:conf1_lamrfail}
     For all $\expr{N}$, $\expr{N}_1$, $\expr{N}_2$ in $\lamrfail$ s.t. $\expr{N} \redd \expr{N}_1$, $\expr{N} \redd \expr{N}_2$ { with } $\expr{N}_1 \neq \expr{N}_2$ { then } there exists $\expr{M}$ such that  $\expr{N}_1 \redd \expr{M}$, $\expr{N}_2 \redd \expr{M}$.
\end{restatable}

\begin{proof}[Proof (Sketch)]
By inspecting the rules of \figref{ch2fig:reductions_lamrfail} one can check that the left-hand sides only clash in a non-variable position with Rules~\redlab{R:Fail} and \redlab{R:Cons2}. The clash does not generate a critical pair: in fact, when applied to the $\lamrfail$-term $\fail^{z,\widetilde{x}}\esubst{\oneb}{z}$ both rules reduce to $\fail^{\widetilde{x}}$.
For all the other rules, whenever they have the same shape, the side conditions of the rules determine which rule can be applied. Therefore,
   an expression can only perform a choice of reduction steps when it is a sum of terms in which multiple summands can perform independent reductions. Without loss of generality, consider an expression $\expr{N} = N + M $ such that  $N \redd N'$ and $M \redd M'$. Then we let $\expr{N}_1 = N' + M$ and $\expr{N}_2 = N + M'$ by Rule~$\redlab{R:ECont}$.  The result follows for $\expr{M} = N' + M' $, since $\mathbb{N}_1
   \redd \mathbb{M}$ and $\mathbb{N}_2
   \redd \mathbb{M}$.
\end{proof}

\begin{remark}[A Sub-calculus without Failure (\lamr)]\label{ch2rem:lamr}
We find it convenient to define \lamr, the sub-calculus of \lamrfail without  explicit failure. 
The syntax of \lamr is obtained from Definition~\ref{ch2def:rsyntaxfail} by excluding $\fail^{\widetilde{x}}$ from the syntax of terms.
Accordingly, the  reduction relation for \lamr is given by  Rules~\redlab{R:Beta}, \redlab{R:Fetch}, \redlab{R:ECont}, and \redlab{R:TCont} in \figref{ch2fig:reductions_lamrfail}. 
Finally, Definition~\ref{ch2def:headfailure} is kept unchanged with the provision that $\headf{M \esubst{ B }{x}}$ is undefined when $\#(x,M) \not = \size{B}$.
\end{remark}


\subsection{Well-formed \texorpdfstring{$\lamrfail$}{}-Expressions}
\label{ch2sec:lamfailintertypes}\hfill 

As mentioned in \secref{ch2s:key}, we define a notion of {\em well-formed expressions}  for $\lamrfail$ by relying on a non-idempotent intersection type system, similar to the one given by \cite{PaganiR10}. Our system for well-formed expressions will be defined in two stages:
\begin{enumerate}
\item First we define a intersection type system for the sub-language $\lamr$ (cf. Rem.~\ref{ch2rem:lamr}), given in  \figref{ch2fig:app_typingrepeat}. 
\revd{:B7}{Unlike the system in \cite{PaganiR10}, our type system includes a weakening rule and a rule for typing explicit substitutions.}
\item Second, we define well-formed expressions for the full language \lamrfail, via \defref{ch2d:wellf}.
\end{enumerate}

We say that we check for ``well-formedness'' (of terms, bags, and expressions) to stress that, unlike standard type systems, our system is able to account for terms that may reduce to the failure term.

\subsubsection{Intersection Types}\label{ch2ss:lamrfail_types}\hfill

Intersection types allow us to reason about types of resources in bags but also about every occurrence of a variable. 
That is, non-idempotent intersection types enable us to distinguish expressions not only by measuring the size of a bag but also by counting the number of times a variable occurs within a term.
\begin{definition}{Types for \lamrfail}
\label{ch2d:typeslamrfail}
We define {\em strict} and {\em multiset types} by the   grammar:
\[
\begin{array}{c@{\hspace{1.2cm}}c}
  \text{(Strict)}\quad  \sigma, \tau, \delta ::= \unit \sep \arrt{\pi}{\sigma}   & \text{(Multiset)} \quad \pi ::=  \secondrev{\sigma^k} \sep \omega
\end{array}
\]
\secondrev{where $\sigma^k$ stands 
 for  $\sigma\wedge \cdots \wedge \sigma$
($k$ times, for some $k>0$).} 
\end{definition}
A strict type can be the unit type  $\unit$  or a functional type  \arrt{\pi}{\sigma}, where 
$\pi$ is a multiset type and $\sigma$ is  a strict type. 
Multiset types can be either 
an intersection  of strict types
\secondrev{ $\sigma^k$
(if $k>0$)
or the empty type $\omega$, which would correspond to $\secondrev{\sigma^k}$ with $k = 0$. Hence, $\sigma^k$ denotes an intersection;}
 the operator $\wedge $ is commutative, associative, and non-idempotent, that is, $\sigma \wedge \sigma \neq \sigma$. The empty type is the type of the empty bag; it acts as  the identity element to~$\wedge$.



\begin{definition}{}
\label{ch2d:tcontsource}
\revo{}{
{\em Type contexts}  $\Gamma , \Delta, \ldots $ are sets of type assignments  $x: \pi$, as defined by the grammar:}
\revd{B9}{
\[
    \begin{aligned}
        \Gamma, \Delta & = \dash \sep \Gamma , x:\pi 
    \end{aligned}
\]}
\revo{}{
The set of variables in $\Gamma$ is denoted as $\dom{\Gamma}$.
In writing $\Gamma, x:\pi$  we assume that $x \not \in \dom{\Gamma}$. 
}
\revo{}{
We generalize \revdaniele{the operator} $\wedge$ from types to contexts, and define  $\Gamma \contexcat \Delta$ as follows:
$$(\Gamma_1 \contexcat \Gamma_2)(x) = 
\begin{cases}
    x: \pi_1 \wedge \pi_2 &  x:\pi_i \in \Gamma_i ,\  \pi_i \not = \omega , \  i \in \{ 1 , 2 \} \\
    x: \pi_i  &  x:\pi_i \in \Gamma_i, x \not \in \dom{\Gamma_j} ,\  i \not = j,\ i,j \in \{1,2\} \\
    \text{undefined} & \text{otherwise}
\end{cases}
$$
 }

\end{definition}

{\em Type judgements} are of the form $\Gamma \vdash \expr{M}:\sigma$, where $\Gamma$ is a type context. We write $\vdash \expr{M}:\sigma$ to denote $\dash \vdash \expr{M}:\sigma$.

\begin{definition}{Well-typed Expressions}
An expression $\expr{M} \in \lamr$ is {\em well-typed} (or typable) if there exist $\Gamma$ and  $\tau$ such that $\Gamma \vdash \expr{M} : \tau$ is entailed via the rules in \figref{ch2fig:app_typingrepeat}.
\end{definition}

\begin{figure*}[!t]
    \centering
    
\begin{prooftree}
    \AxiomC{}
    \LeftLabel{\redlab{T:var}}
    \UnaryInfC{\( x: \sigma \vdash x : \sigma\)}
    \DisplayProof
    \hfill
    \AxiomC{\(  \)}
    \LeftLabel{\redlab{T:\oneb}}
    \UnaryInfC{\( \vdash \oneb : \omega \)}
    \DisplayProof
    \hfill
    \AxiomC{$ \Gamma \vdash M: \sigma$}
    \AxiomC{$ $}
    \LeftLabel{\redlab{T:weak}}
    \BinaryInfC{$ \Gamma, x:\omega \vdash M: \sigma $}
\end{prooftree}

\begin{prooftree}
    \AxiomC{\( \revo{}{\Gamma , {x}: \sigma^k \vdash M : \tau} \)}
    \LeftLabel{\redlab{T:abs}}
    \UnaryInfC{\( \Gamma \vdash \lambda x . M :  \sigma^k  \rightarrow \tau \)}
\DisplayProof
\hfill
  \AxiomC{\( \Gamma \vdash M : \sigma\)}
    \AxiomC{\( \Delta \vdash B : \sigma^k\)}
    \LeftLabel{\redlab{T:bag}}
    \BinaryInfC{\( \revo{}{\Gamma \contexcat \Delta \vdash \bag{M}\cdot B:\sigma^{k+1}} \)}
\end{prooftree}

\begin{prooftree}
  \AxiomC{\( \Gamma \vdash M : \pi \rightarrow \tau \)}
    \AxiomC{\( \Delta \vdash B : \pi \)}
        \LeftLabel{\redlab{T:app}}
    \BinaryInfC{\( \revo{}{\Gamma \contexcat \Delta \vdash M\ B : \tau}\)}
 \DisplayProof
 \hfill
   \AxiomC{\( \revo{}{\Gamma ,  {x}:\sigma^{k} \vdash M : \tau} \)}
         \AxiomC{\( \Delta \vdash B : \sigma^{k} \)}
    \LeftLabel{\redlab{T:ex \dash sub}}    
    \BinaryInfC{\( \revo{}{\Gamma \contexcat \Delta \vdash M \esubst{ B }{ x } : \tau} \)}
\end{prooftree}

\begin{prooftree}
    \AxiomC{$ \Gamma \vdash \expr{M} : \sigma$}
    \AxiomC{$ \Gamma \vdash \expr{N} : \sigma$}
    \LeftLabel{\redlab{T:sum}}
    \BinaryInfC{$ \Gamma \vdash \expr{M}+\expr{N}: \sigma$}
\end{prooftree}

    \caption{Typing Rules for \lamr }
    \label{ch2fig:app_typingrepeat}
\end{figure*}

The rules are standard. We only consider intersections of the same strict type, say $\sigma$, since the current objective is to count the number of occurrences of a variable in a term, and measure the size of a bag. We now give a brief description of the rules in \figref{ch2fig:app_typingrepeat}:
\begin{itemize}
\item {\bf Rules~\redlab{T{:}var}, \redlab{T{:}\oneb} and \redlab{T{:}weak}} are as expected: the first assigns a type to a variable, the second 
assigns the empty bag $\oneb$ the empty type $\omega$, and the third introduces a  useful weakening principle. 
\item {\bf Rule~\redlab{T:abs}} types an abstraction $\lambda x. M$ with $\sigma^k\to \tau$, as long as the variable assignment \revo{}{$x:\sigma^k$ has an intersection type with $\sigma $ occurring exactly $k$ times.}
\item {\bf Rule~\redlab{T:bag}} types a bag $B$ with a type $\sigma^{k+1}$ as long as every component of $B$ is typed with same type $\sigma$, a defined amount of times.
\item {\bf Rule~\redlab{T{:}app}} types an application $M\ B$ with $\tau$ as long as $M$ and $B$ match on the multiset type $\pi$, i.e., $M:\pi\to \tau$ and $B:\pi$. Intuitively, this means that $M$ expects a fixed amount of resources, and $B$ has exactly this number of resources.
\item {\bf Rule~\redlab{T{:}ex \dash sub}} types an explicit substitution $M\esubst{B}{x}$ with $\tau$ as long as the bag $B$ consists of elements of the same type as $x$ and the size of $B$ matches the number of times $x$ occurs in $M$, i.e., $B:\sigma^k$ and $x:\sigma^k$ types the assignment of $M:\tau$.
\item {\bf Rule~\redlab{T:sum}} types an expression (a sum) with a type $\sigma$, if each summand has type $\sigma$.
\end{itemize}

Notice that with the typing rules for $\lamr$ the failure term $\fail$ cannot be typed.  We could consider this set of rules as a type system for  \lamrfail, i.e. the extension of $\lamr$ with failure,  in which failure can be expressed but not typed.

\begin{example}{Cont. Example~\ref{ch2ex:reducts}}
\label{ch2ex:welltyped}
We explore the typability of some of the terms given in previous examples:
\begin{enumerate}
    \item Term $M_1 = (\lambda x. x ) \bag{y} $ is typable, as we have:
  \begin{prooftree}
  \AxiomC{}
  \LeftLabel{\redlab{T:var}}
  \UnaryInfC{\( x: \sigma \vdash x : \sigma\)}
  \LeftLabel{\redlab{T:abs}}
  \UnaryInfC{\(  \vdash \lambda x . x :  \sigma \rightarrow \sigma \)}
  \AxiomC{}
  \LeftLabel{\redlab{T:var}}
  \UnaryInfC{\(  y : \sigma \vdash y : \sigma\)}
  \AxiomC{\(  \)}
  \LeftLabel{\redlab{T:\oneb}}
  \UnaryInfC{\( \vdash \oneb : \omega \)}
  \LeftLabel{\redlab{T:bag}}
  \BinaryInfC{\( y : \sigma \vdash \bag{y}\cdot \oneb:\sigma \)}
  \LeftLabel{\redlab{T:app}}
  \BinaryInfC{\( y : \sigma \vdash (\lambda x. x ) \bag{y} : \sigma\)}
  \end{prooftree}
    \item Term $M_2 = (\lambda x. x ) (\bag{y,z}) $ is not typable.
    \begin{itemize}
    \item The function $\lambda x. x$ has a functional type $\sigma \to \sigma$;
        \item  The bag has an intersection type of size two: $y:\sigma, z:\sigma \vdash(\bag{y,z}):\sigma^2$;
        \item  Rule~$\redlab{T:app}$  requires a match between the type of the bag and the left of the arrow: it can only consume a bag of type $\sigma$.
    \end{itemize}
    
    \item Similarly,  $M_3 = (\lambda x. x ) \oneb $ is not typable: since $\lambda x.x$ has type $\sigma\to \sigma$, to apply the Rule~$\redlab{T:app}$  the bag must have a type $\sigma$, but the empty bag $\oneb$ can only be typed with~$\omega$.
    \item Term $M_4 = (\lambda x. y ) \oneb $ is typable, as follows:
   \begin{prooftree}
    \AxiomC{}
   \LeftLabel{\redlab{T:var}}
  \UnaryInfC{$ y: \sigma  \vdash y : \sigma$}
  \AxiomC{$ $}
  \LeftLabel{\redlab{T:weak}}
  \BinaryInfC{$y: \sigma, x:\omega  \vdash y : \sigma  $}
  \LeftLabel{\redlab{T:abs}}
  \UnaryInfC{\( y : \sigma \vdash \lambda x . y :  \omega \rightarrow \sigma \)}
  \AxiomC{\(  \)}
  \LeftLabel{\redlab{T:\oneb}}
  \UnaryInfC{\( \vdash \oneb : \omega \)}
  \LeftLabel{\redlab{T:app}}
  \BinaryInfC{\( y : \sigma \vdash (\lambda x. y ) \oneb : \sigma\)}
  \end{prooftree}
    \end{enumerate}
\end{example}

Our typing system for \lamr satisfies standard properties,  such as subject reduction, which follows from the {\em Linear} Substitution Lemma. We stress `linearity' because the lemma is stated in terms of the head linear substitution $\headlin{\cdot}$.

\begin{restatable}[Linear Substitution Lemma for \lamr]{lemma}{subtlemfailnofail}
\label{ch2lem:subt_lem}
\revo{}{
If $\Gamma , x:\sigma^k \vdash M: \tau$ (with $k \geq 1$), $\headf{M} = x$, and $\Delta \vdash N : \sigma$ 
then 
$\Gamma \contexcat \Delta, x:\sigma^{k-1} \vdash M \headlin{ N / x }: \tau $.
}
\end{restatable}

\begin{proof}
Standard, by induction on the rule applied in $\Gamma, x:\sigma \vdash M:\tau$. 
\end{proof}

\begin{restatable}[Subject Reduction for \lamr]{theorem}{subredone}
\label{ch2t:app_lamrsr}
If $\Gamma \vdash \expr{M}:\tau$ and $\expr{M} \redd \expr{M}'$ then $\Gamma \vdash \expr{M}' :\tau$.
\end{restatable}

\begin{proof} By  induction on the reduction rule (\figref{ch2fig:reductions_lamrfail}) applied in $\expr{M}$. 
\end{proof}

\revo{}{
\begin{restatable}[Linear Anti-substitution Lemma for \lamr]{lemma}{explemfailnofail}
\label{ch2lem:antisubt_lem}
\revdaniele{Let $M$ and $N$ be $\lamr$-terms such that} $\headf{M} = x$, then we have:
\begin{itemize}
    \item $\Gamma, x:\sigma^{k-1} \vdash M \headlin{ N / x }: \tau$, with $k > 1$, then  \revdaniele{there exist} $  \Gamma_1, \Gamma_2$  such that  $\Gamma_1 , x:\sigma^k \vdash M: \tau$, and $\Gamma_2 \vdash N : \sigma$, where $\Gamma = \Gamma_1 \contexcat \Gamma_2$.
    \item $\Gamma \vdash M \headlin{ N / x }: \tau$, with $x \not \in \dom{\Gamma}$, then \revdaniele{there exist} $ \Gamma_1, \Gamma_2$  such that  $\Gamma_1 , x:\sigma \vdash M: \tau$ , and $\Gamma_2 \vdash N : \sigma$, where $\Gamma = \Gamma_1 \contexcat \Gamma_2$.
\end{itemize}
%
%
\end{restatable}
}

\begin{proof}
 \revdaniele{Standard, by structural induction.}
\end{proof}

\revo{}{
\begin{restatable}[Subject Expansion for \lamr]{theorem}{subexpone}
\label{ch2t:app_lamrexp}
If $\Gamma \vdash \expr{M}':\tau$ and $\expr{M} \redd \expr{M}'$ then $\Gamma \vdash \expr{M} :\tau$.
\end{restatable}
}

\begin{proof}
 \revdaniele{Standard, by structural induction.} 
 \iffulldoc
 See \appref{ch2app:typeshar} for details.
 \else
 See the full version for details.
 \fi 
\end{proof}

\subsubsection{Well-formed Expressions (in \texorpdfstring{$\lamrfail$}{})}\label{ch2ss:lamrfail_wf}\hfill

Building upon the type system for \lamr, we now define a type system for checking {\em well-formed} \lamrfail-expressions. This approach enables us to admit expressions with a failing  computational behavior, may it be due to the mismatch in the number of resources required and available, or be due to consumption of a failing behavior by another expression.
 \secondrev{Such definition relies on the \emph{core context} which is the key to the well-formedness of failure terms:  free variables that are result of weakening will disregarded in the typing of the failure term.}
 

\secondrev{
\begin{definition}{Core Context}
    Given a context $\Gamma$, the associated  \emph{core context} is defined as 
$\core{\Gamma} = \{ x:\pi \in \Gamma \,|\, \pi \not = \omega\}$. 
\end{definition}
}


\begin{definition}{Well-formed \lamrfail expressions}
\label{ch2d:wellf}
An expression $ \expr{M}$ is \revd{B11}{\emph{well-formed}} if  there exist  $\Gamma$ and  $\tau$ such that  $ \Gamma \wfdash  \expr{M} : \tau  $ is entailed via the rules in \figref{ch2fig:app_wf_rules}.
\end{definition}

\begin{figure}
    \centering

\begin{prooftree}
\AxiomC{\( \Gamma \vdash \expr{M} : \tau \)}
\LeftLabel{\redlab{F:wf \dash expr}}
\UnaryInfC{\( \Gamma \wfdash  \expr{M} : \tau \)}
\DisplayProof
\hfill
\AxiomC{\( \Gamma \vdash B : \pi \)}
\LeftLabel{\redlab{F:wf \dash bag}}
\UnaryInfC{\( \Gamma \wfdash  B : \pi \)}
\DisplayProof
\hfill
\AxiomC{\( \Delta  \wfdash M : \tau\)}
\LeftLabel{ \redlab{F:weak}}
\UnaryInfC{\( \Delta , x: \omega \wfdash M: \tau \)}
\end{prooftree}

 \begin{prooftree}
 \AxiomC{\( \Gamma , {x}: \sigma^n \wfdash M : \tau \quad x\notin \dom{\Gamma} \)}
 \LeftLabel{\redlab{F:abs}}
 \UnaryInfC{\( \Gamma \wfdash \lambda x . M : \sigma^n  \rightarrow \tau \)}
 \DisplayProof\hfill
 \AxiomC{\( \Gamma \wfdash M : \sigma\)}
 \AxiomC{\( \Delta \wfdash B : \sigma^k\)}
 \LeftLabel{\redlab{F:bag}}
 \BinaryInfC{\( \Gamma \contexcat \Delta \wfdash \bag{M}\cdot B:\sigma^{k+1}\)}
 \end{prooftree}
        
  \begin{prooftree}
\AxiomC{$ \Gamma \wfdash \expr{M} : \sigma$}
\AxiomC{$ \Gamma \wfdash \expr{N} : \sigma$}
\LeftLabel{\redlab{F:sum}}
\BinaryInfC{$ \Gamma \wfdash \expr{M}+\expr{N}: \sigma$}
\DisplayProof
\hfill
\AxiomC{\(\secondrev{ \dom{\core{\Gamma}} = \widetilde{x} } \)}
\LeftLabel{\redlab{F:fail}}
\UnaryInfC{\( \secondrev{ {\Gamma} \wfdash  \fail^{\widetilde{x}} : \tau } \)}
\end{prooftree}

\begin{prooftree}
\AxiomC{\( \Gamma , {x}:\sigma^{k} \wfdash M : \tau \) }
\AxiomC{\( \Delta \wfdash B : \sigma^{j} \) \ \( k, j \geq 0 \)}
\LeftLabel{\redlab{F:ex \dash sub}}  
\BinaryInfC{\( \Gamma \contexcat \Delta \wfdash M \esubst{ B }{ x } : \tau \)}
\end{prooftree}
    
    \begin{prooftree}
        \AxiomC{\( \Gamma \wfdash M : \sigma^j \rightarrow \tau \)}
        \AxiomC{\( \Delta \wfdash B : \sigma^{k} \)}
        \AxiomC{\( k, j \geq 0 \)}
            \LeftLabel{\redlab{F:app}}
        \TrinaryInfC{\( \Gamma \contexcat \Delta \wfdash M\ B : \tau\)}
    \end{prooftree}
    \caption{Well-Formed Rules for \lamrfail}\label{ch2fig:app_wf_rules}
\end{figure}
    
Below we give a brief description of the rules in~\figref{ch2fig:app_wf_rules}. Essentially, they differ from the ones in \figref{ch2fig:app_typingrepeat}, by allowing mismatches between the number of copies of a variable in a functional position and the number of components in a bag. 
\begin{itemize}
\item {\bf Rules~$\redlab{F{:}wf \dash expr}$ and $\redlab{F{:}wf \dash bag}$} derive that well-typed expressions and bags in $\lamr$ are well-formed. 
\item {\bf Rules~\redlab{F{:}abs}, \redlab{F{:}bag}, and \redlab{F{:}sum}} are as in the  type system for \lamr, but extended to the system of well-formed expressions. 
\item {\bf Rules~$\redlab{F{:}ex \dash sub}$ and  $\redlab{F{:}app}$} differ from the similar typing rules as the size of the bags (as declared in their types) is no longer required to match the number of occurrences of the variable assignment in the typing context (\redlab{F:ex\dash sub}), or the type of the term in the functional position (\redlab{F:app}).
\item {\bf Rule~$\redlab{F{:}fail}$} has no analogue in the type system: we allow $\fail^{\widetilde{x}}$ to be well-formed with any strict type, provided that the core context contains the types of the variables in $\widetilde{x}$ (i.e., none of the variables in $\widetilde{x}$ is typed with $\omega$). 
\end{itemize}
Clearly, the set of {well-typed} expressions is strictly included in the set of {well-formed} expressions. 
Take, e.g., $M=x\esubst{ \bag{N_1,N_2} }{ x }$, where both $N_1$ and $N_2$ are well-typed.
    It is easy to see that  $M$  is well-formed. However, $M$ is not well-typed.

\begin{example}{Cont. Example~\ref{ch2ex:welltyped}}
\label{ch2ex:wellformed}
We explore the well-formedness of some of the terms motivated in previous examples:

\begin{enumerate}
    \item Term $M_1 = (\lambda x. x ) \bag{y} $ is well-typed and also well-formed, as we have:
    \begin{prooftree}
           \AxiomC{\( y:\sigma \vdash (\lambda x. x ) \bag{y} : \sigma \)}
           \LeftLabel{\redlab{F:wf \dash expr}}
           \UnaryInfC{\( y:\sigma \wfdash  (\lambda x. x ) \bag{y} : \sigma \)}
    \end{prooftree}

    \item We saw that term $M_2 = (\lambda x. x ) (\bag{y,z}) $ is not typable; however, it is well-formed:        
    \begin{prooftree}
            \AxiomC{\(   \vdash \lambda x. x : \sigma^1 \rightarrow \sigma \)}
            \LeftLabel{\redlab{F:wf \dash expr}}
            \UnaryInfC{\(  \wfdash \lambda x. x : \sigma^1 \rightarrow \sigma \)}
            \AxiomC{\( y:\sigma, z:\sigma  \vdash \bag{y,z} : \sigma^{2} \)}
            \LeftLabel{\redlab{F:wf \dash bag}}
            \UnaryInfC{\( y:\sigma, z:\sigma \wfdash \bag{y,z} : \sigma^{2} \)}
            \LeftLabel{\redlab{F:app}}
        \BinaryInfC{\( y:\sigma, z:\sigma \wfdash (\lambda x. x ) (\bag{y,z})  : \sigma \)}
    \end{prooftree}
  Notice that   both $\vdash \lambda x. x : \sigma^1 \rightarrow \sigma$ and $ \Gamma \vdash \bag{y,z} : \sigma^{2} $ are well-typed as $1, 2 \geq 0$.
    
    \item Similarly, the term $M_3 = (\lambda x. x ) \oneb $ is also well-formed. The corresponding derivation is as above, but uses an empty context as well as the well-formedness rule for bags: 
    \begin{prooftree}
            \AxiomC{\( \ \vdash \oneb : \sigma^{0} \)}
            \LeftLabel{\redlab{F:wf \dash bag}}
            \UnaryInfC{\(  \wfdash \oneb : \sigma^{0} \)}
    \end{prooftree}
    Notice how $\sigma^0 = \omega$ and that $  \wfdash \oneb : \omega$.
    
    \item Term $M_4 = (\lambda x. y ) \oneb $ is well-typed and also well-formed.
    
    \item Interestingly, term $M_5 = \fail^{\emptyset} $ is well-formed as:
    \begin{prooftree}
        \AxiomC{\(  \)}
        \LeftLabel{\redlab{F:fail}}
        \UnaryInfC{\(  \wfdash  \fail^{\emptyset} : \tau  \)}
    \end{prooftree}
\end{enumerate}
\end{example}

\begin{example}{}
Let us consider an expression that is not well-formed:
 $$ \lambda x . x \bag{\lambda y . y,~\lambda z . z_1 \bag{z_1\bag{z_2}}}.$$
Notice that $\lambda x.x$ is applied to bags of two different types:  
\begin{itemize}
\item The first bag containing $\lambda y. y$ is well-typed, thus well-formed. Consider the derivation $\Pi_1$:

\begin{prooftree}
    \AxiomC{}
    \LeftLabel{\redlab{T:var}}
    \UnaryInfC{\(y:\sigma\vdash y: \sigma\)}
        \LeftLabel{\redlab{T:abs}}
    \UnaryInfC{\(\vdash  \lambda y.y: \sigma\to \sigma\)}
    \AxiomC{}
    \LeftLabel{\redlab{T:\oneb}}
    \UnaryInfC{\(\vdash\oneb:\omega\)}
        \LeftLabel{\redlab{T:bag}}
    \BinaryInfC{\(\vdash \bag{\lambda y.y}\cdot \oneb:\sigma\to \sigma\)}
        \LeftLabel{\redlab{F:wf\dash bag}}
     \UnaryInfC{\(\wfdash \bag{\lambda y.y}\cdot \oneb:\sigma\to \sigma\)}
\end{prooftree}
\end{itemize}
 In the rest of the example we will omit the labels of rule applications, and concatenations with the empty bag $\oneb$ (i.e., $\bag{\lambda y. y}\cdot \oneb$ will be written simply as $\bag{\lambda y. y}$) and corresponding sub-derivations consisting of applications of Rule~\redlab{T:\oneb}.
\begin{itemize}
\item The second bag  contains $\lambda z . z_1 \bag{z_1\bag{z_2}} $  contains an abstraction that acts as a weakening as $z$ does not appear within $z_1 \bag{z_1\bag{z_2}}$. Consider the derivation $\Pi_2$:
 
\begin{prooftree}
     \AxiomC{\(z_1:\sigma\to\sigma\vdash z_1:\sigma\to \sigma\)} 
  \AxiomC{\(z_1:\sigma\to\sigma\vdash z_1:\sigma\to \sigma\)}  
     \AxiomC{\(z_2:\sigma\vdash z_2:\sigma\)} 
     \UnaryInfC{\(z_2:\sigma\vdash \bag{z_2}:\sigma\)}
     \BinaryInfC{\(z_1:\sigma\to\sigma, z_2:\sigma\vdash z_1\bag{z_2}:\sigma\)}
     \UnaryInfC{\(z_1:\sigma\to\sigma, z_2:\sigma\vdash \bag{z_1\bag{z_2}}:\sigma\)}
    \BinaryInfC{\(z_1:\sigma\to\sigma\wedge\sigma\to\sigma, z_2:\sigma\vdash z_1\bag{z_1\bag{z_2}}:\sigma\)}
\UnaryInfC{\(z_1:\sigma\to\sigma\wedge\sigma\to\sigma, z_2:\sigma, z:\omega\vdash z_1\bag{z_1\bag{z_2}}:\sigma\)}
\UnaryInfC{\(z_1:\sigma\to\sigma\wedge\sigma\to\sigma, z_2:\sigma\vdash\lambda z. z_1\bag{z_1\bag{z_2}}:\omega\to \sigma\)}
\UnaryInfC{\(z_1:\sigma\to\sigma\wedge\sigma\to\sigma, z_2:\sigma\wfdash\lambda z. z_1\bag{z_1\bag{z_2}}:\omega\to \sigma\)}
\UnaryInfC{\(\underbrace{z_1:\sigma\to\sigma \wedge \sigma\to\sigma, z_2:\sigma}_{\Gamma}\wfdash\bag{\lambda z. z_1\bag{z_1\bag{z_2}}}:\omega\to \sigma\)}
\end{prooftree}

\item The concatenation of these two bags is not well-formed since each component has a different type: $\sigma\to \sigma$ and $\omega\to \sigma$. Therefore,  $ \lambda x . x \bag{ ~ \lambda y . y ~ , ~ \lambda z . z_1 \bag{z_1\bag{z_2}} ~}$ is not well-formed.
\end{itemize}
Notice that if we change $\lambda y.y$ to $\lambda y.y_1$ in the first bag, we would have a derivation $\Pi_1'$ for $y_1:\sigma\wfdash\lambda y.y_1:\omega \to \sigma$. This would allow us to concatenate the bags with derivation $\Pi_3$:

\begin{prooftree}
     \AxiomC{$\Pi_1'$}
     \noLine
     \UnaryInfC{$y_1:\sigma\wfdash \lambda y.y_1:\omega \to \sigma$}
     \AxiomC{$\Pi_2$}
     \noLine
     \UnaryInfC{\(\Gamma\wfdash\lambda z. z_1\bag{z_1\bag{z_2}}:\omega\to \sigma\)}
      \AxiomC{}
     \UnaryInfC{\(\oneb:\omega\)}
     \BinaryInfC{\(\Gamma\wfdash\bag{\lambda z. z_1\bag{z_1\bag{z_2}}}\cdot \oneb:\omega\to \sigma\)}
     \BinaryInfC{\(\Gamma, y_1:\sigma\wfdash\bag{\lambda y.y_1}\cdot \bag{\lambda z. z_1\bag{z_1\bag{z_2}}}\cdot \oneb:(\omega\to \sigma)^2\)}
\end{prooftree}

Thus, the whole term becomes well-formed:
\begin{prooftree}
     \AxiomC{}
     \noLine
     \UnaryInfC{$x:\omega\to \sigma \vdash x:\omega\to \sigma$}
\UnaryInfC{$\vdash \lambda x. x : (\omega\to \sigma)\to \omega \to \sigma$}
\UnaryInfC{$\wfdash \lambda x. x : (\omega\to \sigma)\to \omega \to \sigma$}
\AxiomC{$\Pi_3$}
\noLine
\UnaryInfC{\(\Gamma, y_1:\sigma\wfdash\bag{ ~ \lambda y.y_1 ~,~ \lambda z. z_1\bag{z_1\bag{z_2}} ~ }:(\omega\to \sigma)^2\)}
\BinaryInfC{\(\Gamma, y_1:\sigma\wfdash\lambda x. x\bag{ ~ \lambda y.y_1 ~,~ \lambda z. z_1\bag{z_1\bag{z_2}} ~ }:\omega\to \sigma\)}
\end{prooftree}
\end{example}

Well-formedness  rules satisfy subject reduction with respect to the rules in~\figref{ch2fig:reductions_lamrfail} and relies on the linear substitution lemma for $\lamrfail$:

\begin{restatable}[Substitution Lemma for \lamrfail]{lemma}{subtlemfail}
\label{ch2lem:subt_lem_fail}

\revo{}{
If $\Gamma , x:\sigma^k \wfdash M: \tau$ (with $k \geq 1$), $\headf{M} = x$, and $\Delta \wfdash N : \sigma$ 
then 
$\Gamma \contexcat \Delta, x:\sigma^{k-1}  \wfdash M \headlin{ N / x }$.
}
\end{restatable}

We now show subject reduction on well formed expressions in \lamrfail. We use our results of subject reduction for well-typed \lamr (Theorem~\ref{ch2t:app_lamrsr}) and extend them to \lamrfail.

\begin{restatable}[Subject Reduction in \lamrfail]{theorem}{applamrfailsr}
\label{ch2t:app_lamrfailsr}
If $\Gamma \wfdash \expr{M}:\tau$ and $\expr{M} \redd \expr{M}'$ then $\Gamma \wfdash \expr{M}' :\tau$.
\end{restatable}

\begin{proof}[Proof (Sketch)]
By structural induction on the reduction rules. 
\iffulldoc	
See \appref{ch2app:lamfailintertypes} for details.
\else
See the full version for details.
\fi
\end{proof}

\revo{}{Differently from $\lamr$,  subject expansion fails for $\lamrfail$. This is due to the possibility of failure in the use of resources. In $\lamr$, if a resource is substituted within a term it is always done once, hence the term substituted must always be well-typed; however, in reductions that lead to the failure term, resources within a bag may be discarded before ever being substituted and hence, there is  no requirement to be well-formed. Formally, we have:}

\revo{}{
\begin{restatable}[Failure of Subject Expansion in \lamrfail]{theorem}{applamrfailexp}
\label{ch2t:app_lamrfailse}
If $\Gamma \wfdash \expr{M}':\tau$ and $\expr{M} \redd \expr{M}'$ then \revdaniele{it is not necessarily the case that} $ \Gamma \wfdash \expr{M} :\tau$.
\end{restatable}
}

\revo{}{
\begin{proof}
A counter-example suffices here. Consider the term $  \fail^\emptyset$, which is well-formed but not well-typed, and let $ \Omega^l$  be the term $( \lambda x. x \bag{x} ) \bag{ \lambda x. x \bag{x} } $.
 \revdaniele{Notice that $\dash \wfdash \fail^\emptyset:\tau$ and 
 $ \fail^x \esubst{\bag{\Omega^l}}{x}\redd\fail^\emptyset$, but
 $\fail^x \esubst{\bag{\Omega^l}}{x}$ is not well-formed (nor well-typed)}
 .
\end{proof}
}


\section[A Resource Calculus With Sharing]{\texorpdfstring{$ \lamrsharfail$}{}: A Resource Calculus With Sharing}\label{ch2sec:lamsharfail}

We define $\lamrsharfail$, a variant of $\lamrfail$ with a sharing construct, \srev{which we adopt following the {atomic} $\lambda$-calculus in~\cite{DBLP:conf/lics/GundersenHP13}.}
In $\lamrsharfail$, a variable is only allowed to appear once in a term:  
multiple occurrences of the same variable are  atomized, i.e., they are given new different variable names. 
\srev{The ``atomization'' of variable occurrences realized in $\lamrsharfail$ via sharing will turn out to be very convenient to define our encoding into \spi.}

 Our language $\lamrsharfail$, defined in \secref{ch2ss:syntaxshar},  includes also a form of explicit substitution, called \emph{explicit linear substitution}, which enables a refined analysis of the consumption of linear resources. Later, in \secref{ch2ssec:lamshar_semantics}, we introduce the reduction semantics that implements a lazy evaluation.
 In \secref{ch2ss:typeshar}, we present a non-idempotent intersection type system to control the use of resources.
 Finally, in \secref{ch2ss:auxtrans} we give an encoding from $\lamrfail$ into $\lamrsharfail$, denoted
  $\recencodopenf{\cdot}$, whose correctness is established in \secref{ch2s:encoding}.
 


\subsection{Syntax}
\label{ch2ss:syntaxshar}
\hfill

The syntax of \lamrsharfail only modifies the syntax of \lamrfail-terms, which is defined by the grammar below; the syntax of bags $B$ and expressions $\expr{M}$ is  as in \defref{ch2def:rsyntaxfail}.
\begin{align*}
\mbox{(Terms)} \quad  M,N, L ::= &
~~x 
\sep \lambda x . (M[ \widetilde{x} \leftarrow x ]) 
\sep (M\ B) 
\sep M \linexsub {N /x} 
\sep \fail^{\widetilde{x}}
\\
& \sep 
M [ \widetilde{x} \leftarrow x ] 
\sep (M[\widetilde{x} \leftarrow x])\esubst{ B }{ x } 
\end{align*}
Distinctive aspects are the \emph{sharing construct}  
$M [ \widetilde{x} \leftarrow x ]$ 
and the \emph{explicit linear substitution} 
$M \linexsub{ N /x}$.
The term  $M [ \widetilde{x} \leftarrow x ]$ defines the sharing of variables $\widetilde{x}$ occurring in $M$ using $x$. 
We shall refer to $x$ as \emph{sharing variable} and to $\widetilde{x}$ as \emph{shared variables}. 
Notice that $\widetilde{x}$ can be empty: $M[\leftarrow x]$ expresses that $x$ does not share any variables in $M$. \revdaniele{The sharing construct}  $M [ \widetilde{x} \leftarrow x ]$ \revdaniele{binds the variables in \(\widetilde{x}\);} the occurrence of $x_i$ can appear within the fail term $\fail^{\widetilde{y}}$, if $x_i \in \widetilde{y}$.
 In the explicit linear substitution $M \linexsub{ N /x}$ \revdaniele{binds $x$ in $M$}. 
 As in $\lamrfail$, the term $\fail^{\widetilde{x}}$ explicitly accounts for failed attempts at substituting the variables $\widetilde{x}$, due to an excess or lack of resources.  \secondrev{A variable that is not explicitly sharing/shared is called {\em independent}}.

\begin{example}{}\label{ch2ex:lambshar_terms}
The following are examples of  $\lamrsharfail$-terms.
\begin{itemize}
\item (Shared identity) $\hat{{\bf I}}=\lambda x.x_1[x_1\leftarrow x]$ 
\item (Independent  variables) \secondrev{An independent variable $x$ applied to a 1-component bag (another independent variable): $x\bag{x_1}$}
\item $\hat{{\bf I}}$ applied to a 1-component bag: $ \hat{{\bf I}} \bag{y_1}[y_1\leftarrow y]$
    \item  $\hat{{\bf I}}$ applied to a 2-component bag: \(\hat{{\bf I}}(\bag{y_1,y_2}) \shar{y_1,y_2}{y}\)
    \item Shared vacuous abstraction: $(\lambda y. x_1\bag{x_2}\shar{}{y})\shar{x_1,x_2}{x}$
    \item $\hat{{\bf I}}$ applied to a bag containing an explicit substitution of a failure term that does not share the  variable $y$: \(\hat{{\bf I}} \bag{ \fail^{\emptyset}[ \leftarrow y] \esubst{\bag{N}}{y} }  \)
    \item An abstraction on $x$ of two shared occurrences of $x$: $\hat{D}=\lambda x. x_1 \bag{x_2}\shar{x_1,x_2}{x}$
\end{itemize}
\end{example}

\srev{The syntax of terms is subject to some natural conditions on variable occurrences and on the structure of the sharing construct and the {explicit linear substitution}. We formalize these conditions as \emph{consistency}, defined as follows:}


\secondrev{
\begin{definition}{Consistent Terms, Bags, and Expressions}
\label{ch2d:consistent}
	We say that the expression $\expr{M}$ is 
\emph{consistent} 
if each subterm $M_0$ of $\expr{M}$ satisfies the following conditions:
\begin{enumerate}
	\item If $M_0 = M [ \widetilde{x} \leftarrow x ]$ then: (i) 
	$\widetilde{x}$ contains pairwise distinct variables; 
	(ii)~every $x_i \in \widetilde{x}$ must occur exactly once in $M$; (iii) $x_i$ is not a sharing variable;
	(iv)~$M$ is consistent. 
	\item If $M_0 = M \linexsub{ N /x}$ then: (i) the variable $x$ must occur exactly once in $M$;
	(ii) $x$ cannot be a sharing variable; 	(iii)~$M$ and $N$ are consistent; (iv)~$\lfv{M} \cap \lfv{N} = \emptyset$. 
    \item Otherwise, for other forms of $M_0$, variables must occur exactly once, i.e.,:
        \begin{itemize}
            \item If $M_0 = \lambda x . (M[ \widetilde{x} \leftarrow x ])$ then: $x \not \in \lfv{M} $; $\widetilde{x}$ contains pairwise distinct variables; every $x_i \in \widetilde{x}$ must occur exactly once in $M$ and is not a sharing variable; $M$ is consistent.            
            \item If $M_0 = (M\ B)$ then $\lfv{M} \cap \lfv{B} = \emptyset$ and $M$ and $B$ are consistent.
            \item If $M_0 = \fail^{\widetilde{x}}$ then $\widetilde{x}$ contains pairwise distinct variables.
            \item If $M_0 =(M[\widetilde{x} \leftarrow x])\esubst{ B }{ x } $  then: $x \not \in \lfv{M} $; $\widetilde{x}$ contains pairwise distinct variables; every $x_i \in \widetilde{x}$ must occur exactly once in $M$ and is not a sharing variable;   $\lfv{M} \cap \lfv{B} = \emptyset$; and $M$ and $B$ are consistent.
        \end{itemize}
\end{enumerate}
Consistency extends to bags as follows. 
The bag $\oneb$ is always consistent. 
The bag $\bag{M}$ is consistent if $M$ is consistent.
The bag $A \cdot B$ is consistent  if (i)~$A$ and $B$ are consistent and (ii)~$\lfv{A} \cap \lfv{B} = \emptyset$.
\end{definition}  
}

\srev{We now discuss the consistency conditions for the sharing construct $M [ \widetilde{x} \leftarrow x ]$. Condition~1(ii) enforces that variables cannot have more than one linear occurrence in the subject of a sharing construct: this condition rules out terms such as 
    $x_1 \bag{x_1\bag{y}}[x_1 \leftarrow x]$. Condition~1(iii), which rules out terms of the form $x_1 \bag{x_2\bag{x_3\bag{y}}}[{x_1 , x_2 }\\ \leftarrow x'] [x' , x_3 \leftarrow x]$,  is for convenience: by requiring that sharing occurrences appear at the top level in bindings, we can easily deduce the number of occurrences of a variable by measuring the size of $\widetilde{x}$ in $[\widetilde{x} \leftarrow x]$, rather than inductively having to measure the occurrences of each $x' \in \widetilde{x}$ in multiple sharing constructs.}
    
    \srev{Conditions on the explicit linear substitution  $M\linexsub{N/x}$ formalize our design choice:
    an explicit linear substitution is defined when the number of 
    variables to be substituted coincides with the number of available resources. 
    In particular, Condition~2(i) rules out 
    terms of the form $y\linexsub{M/x}$, where an explicit linear substitution has no variable to perform a substitution. Condition~2(ii) rules  out terms such as $M[x_1,x_2 \leftarrow x]\linexsub{M/x}$, in which a term is to be linearly substituted for a single variable $x$; however, as the variable is shared twice within $M$, there are less available terms to be substituted than it is necessary. }
 
   \secondrev{ Finally, Condition 3 enforces that each variable occurs only once  in a consistent term, and also that in 
 $\fail^{\widetilde{x}}$, the $\widetilde{x}$  denotes a  set of variables (rather than a multiset), as variables can appear at most once within consistent terms. Thus, consistent terms also excludes terms such as $\fail^{x,x}$}.
   
 \secondrev{In what follows, we shall be working with consistent terms only, which we will call simply terms in 
our definitions and results. As we will see, consistency will be preserved by 
reduction (\thmref{ch2thm:consistency_reductions})
and ensured by typing  (\thmref{ch2thm:consistency_type}) and a structural congruence on terms~(\thmref{ch2thm:term_consistency}).}

\subsection{Reduction Semantics}\label{ch2ssec:lamshar_semantics}\hfill

Similarly to $\lamrfail$, the reduction semantics of \lamrsharfail is given by a relation $\redd$, defined by the rules in \figref{ch2fig:share-reductfailure}; it consists of an extension of reductions in $\lamrfail$ that deals with the sharing construct $\shar{\cdot}{\cdot }$ and  with the explicit linear substitution $\cdot \linexsub{\cdot /\cdot }$. In order to define the reduction rules formally,  we require some auxiliary notions:
the free variables of an expression/term, 
the head of a term, linear head substitution, and contexts.

\begin{definition}{Free Variables}
\label{ch2d:fvsh}
The set of free variables of a term, bag and expressions in \lamrsharfail, is defined inductively as

       \[
       \begin{array}{l}
       \begin{array}{l@{\hspace{1.0cm}}l}
          \lfv{x} = \{ x \}  &  \lfv{ \fail^{\widetilde{x}}} =\{ \widetilde{x}\} \\
          \lfv{ \bag{M}} = \lfv{M} & \lfv{B_1 \cdot B_2} = \lfv{B_1} \cup \lfv{B_2} \\
          \lfv{M\ B} = \lfv{M}  \cup \lfv{B} & \lfv{\oneb} = \emptyset \\
          \lfv{M \linexsub {N /x}} = (\lfv{M}\setminus \{x\})\cup \lfv{N} &
          \lfv{M [ \widetilde{x} \leftarrow x ]} = (\lfv{M}  \setminus \{\widetilde{x}\}) \cup \{ x \}\\
          \lfv{\lambda x . (M[ \widetilde{x} \leftarrow x ])} = \lfv{M[ \widetilde{x} \leftarrow x ]}\setminus \{x\}&  \lfv{\expr{M}+\expr{N}} = \lfv{\expr{M}} \cup \lfv{\expr{N}}
          \end{array}\\
          \ \lfv{ (M[\widetilde{x} \leftarrow x])\esubst{ B }{ x }} = (\lfv{M[\widetilde{x} \leftarrow x]}\setminus \{x\})\cup \lfv{B}
    \end{array}
    \]
    
 As usual, a term $M$ is {\em closed} if $\lfv{M}=\emptyset$.    
\end{definition}

\begin{definition}{Head}
\label{ch2d:headshar}
The head of a term $M$, denoted $\headf{M}$, is defined inductively:
\[
\begin{array}{l}
\begin{array}{l@{\hspace{3cm}}l}
\headf{x}  = x   &  \headf{\lambda x . (M[ \widetilde{x} \leftarrow x ])}  = \lambda x . (M[ \widetilde{x} \leftarrow x ])
\\
\headf{M\ B}  = \headf{M} & \headf{M \linexsub{N /x}}  = \headf{M}
\\
\headf{\fail^{\widetilde{x}}}  = \fail^{\widetilde{x}}&
\end{array}
\\
\ \headf{M[\widetilde{x} \leftarrow x]} = 
\begin{cases}
    x & \text{If $\headf{M} = y \text{ and } y \in \widetilde{x}$}\\
    \headf{M} & \text{Otherwise}
\end{cases}
\\
\ \revd{B29}{ \headf{(M[\widetilde{x} \leftarrow x])\esubst{ B }{ x }} = 
\begin{cases}
    \fail^{\emptyset} & \text{If $ | \widetilde{x} | \not = \size{B}$}\\
    \headf{M[ \leftarrow x]} & \text{If $ \widetilde{x} = \emptyset$ and $B = \oneb$}\\
    (M[\widetilde{x} \leftarrow x])\esubst{ B }{ x } & 
    \text{Otherwise} 
\end{cases} }
\end{array}
\]
\end{definition}


\revd{B29}{The most notable difference between $\headf{\cdot}$ in $\lamrfail$ (cf. Definition \ref{ch2def:headfailure}) and in $\lamrsharfail$ concerns explicit substitution. Both definitions return $\fail^\emptyset$ in a mismatch of resources; in $\lamrsharfail$, the head term of an explicit substitution is only defined in the case of empty sharing (weakening). As we will see, this allows us to prioritize explicit substitution reductions over fetch reductions, as the head variable will block until an explicit substitution is separated into its linear component. 
}

\begin{definition}{Linear Head Substitution}\label{ch2def:headlinfail}
Given a term $M$ with $\headf{M} = x$, the linear substitution of a term $N$ for   $x$ in $M$, written $M\headlin{ N / x}$ is inductively defined as:
\begin{align*}
x \headlin{ N / x}   & = N 
\\
(M\ B)\headlin{ N/x }  & = (M \headlin{ N/x })\ B  
\\
(M \linexsub{L /y} ) \headlin{ N/x } &= (M\headlin{ N/x })\ \linexsub{L /y}  & x \not = y\\
((M[\widetilde{y} \leftarrow y])\esubst{ B }{ y })\headlin{ N/x } &= (M[\widetilde{y} \leftarrow y]\headlin{ N/x })\ \esubst{ B }{ y }  
& x \not = y \\
(M[\widetilde{y} \leftarrow y]) \headlin{ N/x } &=  (M\headlin{ N/x }) [\widetilde{y} \leftarrow y] & x \not = y
\end{align*}

\end{definition}

We now define contexts for terms and expressions in  \lamrsharfail.  Term contexts involve an explicit linear substitution, rather than an explicit substitution: this is due to the reduction strategy we have chosen to adopt (cf. Rule~\redlab{RS:Ex\dash Sub} in \figref{ch2fig:share-reductfailure}), as we always wish to evaluate explicit substitutions first. Expression contexts can be seen as sums with holes. 

\begin{definition}{Term and Expression Contexts in \lamrsharfail}\label{ch2def:ctxt_lamsharfail}
Let $[\cdot]$ denote a hole.
Contexts for terms and expressions are defined by the following grammar:
\[
\begin{array}{l@{\hspace{1.3cm}}rl}
\text{(Term Contexts)}     &  C[\cdot] ,  C'[\cdot] &::=([\cdot])B \mid ([\cdot])\linexsub{N/x} \mid 
([\cdot])[\widetilde{x} \leftarrow x] \mid \\ && \quad ([\cdot])[ \leftarrow x]\esubst{\oneb}{ x} \\
 \text{(Expression Contexts)}& D[\cdot] , D'[\cdot] & ::= M + [\cdot] \mid [\cdot] + M
\end{array}
\]
The substitution of a hole with a 
term $M$ in a context $C[\cdot]$, denoted  $C[M]$, must be a 
\lamrsharfail-term. 
\end{definition}

We assume that the terms that fill in the holes respect 
\secondrev{consistency}
(i.e., variables appear in a term only once, shared variables must occur in the context).

\begin{example}{}
    This example illustrates that certain contexts cannot be filled with certain terms. Consider the hole in context $C[\cdot ]= ([\cdot])\linexsub{N/x}$. 
    \begin{itemize} 
     \item  The hole cannot be filled with $y$, since  $C[y]= y\linexsub{N/x}$ is not a consistent term.
    Indeed, $M\linexsub{N/x}$ requires that $x$ occurs exactly once within $M$.
    \item Similarly, the hole cannot be filled with $\fail^{z}$ with $z\neq x$, since $C[\fail^{z}]= (\fail^{z})\linexsub{N/x}$ and $x$ does not occur in the $\fail^z$, thus, the result is  not a consistent term.
    \end{itemize}
\end{example}

\begin{figure}[!t]
\centering
  \begin{prooftree}
    \AxiomC{$\raisebox{17.0pt}{}$}
    \LeftLabel{\redlab{RS{:}Beta}}
    \UnaryInfC{\(  (\lambda x. M[\widetilde{x} \leftarrow x]) B  \redd M[\widetilde{x} \leftarrow x]\esubst{ B }{ x }  \)}
 \end{prooftree}

 \begin{prooftree}
    \AxiomC{$B = \bag{M_1}
    \cdots  \bag{M_k} \qquad k \geq  1 $}
    \AxiomC{$ M \not= \fail^{\widetilde{y}} $}
    \LeftLabel{\redlab{RS{:}Ex \dash Sub}}
    \BinaryInfC{\( \!M[x_1,\ldots, x_k \leftarrow x]\esubst{ B }{ x } \redd \sum_{B_i \in \perm{B}}M\linexsub{B_i(1)/x_1} \cdots \linexsub{B_i(k)/x_k}    \)}
 \end{prooftree}

 \begin{prooftree}
    \AxiomC{$ \headf{M} = x$}
     \LeftLabel{\redlab{RS{:}Lin\dash Fetch}}
     \UnaryInfC{\(  M \linexsub{N/x} \redd  M \headlin{ N/x } \)}
\end{prooftree}
     
\begin{prooftree}
       \AxiomC{$ k \neq \size{B} \quad   \widetilde{y} = (\lfv{M} \setminus \{ x_1,\ldots, x_k \} ) \cup \lfv{B}$}
   \LeftLabel{\redlab{RS{:}Fail}}
   \UnaryInfC{\( M[x_1,\ldots, x_k\leftarrow x]\esubst{ B }{ x } \redd \sum_{\perm{B}}  \fail^{\widetilde{y}} \)}
 \end{prooftree}

\begin{prooftree}
    \AxiomC{\( \widetilde{y} = \lfv{B} \)}
    \LeftLabel{$\redlab{RS{:}Cons_1}$}
    \UnaryInfC{\(  \fail^{\widetilde{x}} B  \redd {} \hspace{-4mm} \displaystyle\sum_{\perm{B}} \hspace{-3mm}\fail^{\widetilde{x} \cup \widetilde{y}}  \)}
        \DisplayProof\hspace{-3mm}
      \AxiomC{\(  \size{B} = k \quad k  +  | \widetilde{x} | \not= 0  \quad  \widetilde{z} = \lfv{B}\)}
    \LeftLabel{$\redlab{RS{:}Cons_2}$}
    \UnaryInfC{\(  (\fail^{\widetilde{x}\cup \widetilde{y}} [ \widetilde{x} \leftarrow x])\esubst{ B }{ x }  {} \redd\hspace{-3mm} \displaystyle \sum_{\perm{B}}\hspace{-3mm}\fail^{\widetilde{y}\cup \widetilde{z}} \)}
\end{prooftree}

\begin{prooftree}
    \AxiomC{\( \widetilde{z} = \lfv{N} \)}
    \LeftLabel{$\redlab{RS{:}Cons_3}$}
    \UnaryInfC{\( \fail^{\widetilde{y}\cup x} \linexsub{N/x}  {} \redd  \fail^{\widetilde{y} \cup \widetilde{z}}  \)}
\end{prooftree}

\begin{prooftree}
        \AxiomC{$   M \redd M'_{1} + \cdots + M'_{k} $}
        \LeftLabel{$\redlab{RS{:}TCont}$}
        \UnaryInfC{$ C[M] \redd  C[M'_{1}] + \cdots +  C[M'_{k}] $}
\DisplayProof\hfill%
        \AxiomC{$ \expr{M}  \redd \expr{M}'  $}
        \LeftLabel{$\redlab{RS{:}ECont}$}
        \UnaryInfC{$D[\expr{M}]  \redd D[\expr{M}']  $}
\end{prooftree}

    \caption{Reduction Rules for \lamrsharfail.}
\label{ch2fig:share-reductfailure}
\end{figure}

Now we are ready to describe the rules in ~\figref{ch2fig:share-reductfailure}. Intuitively, the lazy reduction relation $\redd$ on expressions works as follows: 
a $\beta$-reduction in \lamrsharfail results into an explicit substitution $M\shar{\widetilde{x}}{x}\esubst{ B }{ x }$, which then evolves, as an in intermediate step, to an expression consisting of explicit linear substitutions,  which are the ones reducing to a linear head substitution $\headlin{ N / x}$ (with $N \in B$) when the size of $B$ coincides with the number of occurrences of  $x$ in $M$. 
The term reduces to failure when there is a mismatch between the size of $B$ and the number of shared variables to be substituted.
More in details, we have:
\begin{itemize}
    \item {\bf Rule~\redlab{RS{:}Beta}} is standard and reduces to an explicit substitution.
    \item {\bf Rule~\redlab{RS{:}Ex \dash Sub}} applies when the size $k$ of the bag coincides with the length of the list  $\widetilde{x}=x_1,\ldots,x_k$. Intuitively, this rule ``distributes'' an  explicit substitution into a sum of terms involving explicit linear substitutions; it considers all possible permutations of the elements in the bag among all shared variables.
   \item  {\bf Rule~\redlab{RS{:}Lin \dash Fetch}} specifies the evaluation of a term with an explicit linear substitution into a linear head substitution.
\end{itemize}

We have three rules that reduce to the failure term---their objective is to accumulate all (free) variables involved in failed reductions. 
Accordingly:
\begin{itemize} 
\item {\bf Rule~$\redlab{RS{:}Fail}$} formalizes failure in the evaluation of an explicit substitution $M[\widetilde{x}\leftarrow x]\esubst{ B}{x }$, which occurs if there is a mismatch between the resources (terms) present in $B$ and the number of occurrences of $x$ to be substituted. 
The resulting failure term preserves all free variables in $M$ and $B$ within its attached set $\widetilde{y}$.
\item {\bf Rules~$\redlab{RS{:}Cons_1}$ and~$\redlab{RS{:}Cons_2}$}  describe reductions that lazily consume the failure term, when a term has $\fail^{\widetilde{x}}$ at its head position. 
The former rule consumes bags attached to it whilst preserving all its free variables. 
\item  {\bf  Rule~\redlab{RS{:}Cons_3}} accumulates into the failure term the free variables involved in an explicit linear substitution. 
\end{itemize}
The contextual Rules~$\redlab{RS{:}TCont}$ and $\redlab{RS{:}Econt}$ are standard. 

\begin{example}{}
We show how a term $M = (\lambda x . x_1 [x_1 \leftarrow x]) \bag{ \fail^{\emptyset}[ \leftarrow y] \esubst{\bag{N}}{y} }$ can reduce using Rule~$\redlab{RS{:}Cons_2}$.

\[
      \begin{aligned}
            M  &\redd_{\redlab{RS{:}Beta}} x_1 [x_1 \leftarrow x]  \esubst{\bag{ \fail^{\emptyset}[ \leftarrow y] \esubst{\bag{N}}{y} }}{x} \\
            & \redd_{\redlab{RS{:}Ex \dash Sub}} x_1  \linexsub{ \fail^{\emptyset}[ \leftarrow y] \esubst{\bag{N}}{y}  / x_1} \\
             &\redd_{\redlab{RS{:}Lin \dash Fetch}} \fail^{\emptyset}[ \leftarrow y] \esubst{\bag{N}}{y}  \\ &\redd_{\redlab{RS{:}Cons_2}} \fail^{\lfv{N}}
        \end{aligned}
 \]

 \end{example}

   \begin{example}{}
We illustrate how Rule~$\redlab{RS{:}Fail} $ can introduce $\fail^{\widetilde{x}}$ into a term. It also shows how Rule~$\redlab{RS{:}Cons_3} $ consumes an explicit linear substitution:
 \[
      \begin{aligned}
            x_1 [\leftarrow y] \esubst{\bag{N}}{y}[x_1 \leftarrow x] \esubst{\bag{M}}{x}  &\redd_{\redlab{RS{:}Ex \dash Sub}} x_1 [\leftarrow y] \esubst{\bag{N}}{y}\linexsub{M/x_1}  \\
            &\redd_{\redlab{RS{:}Fail}} \fail^{ \{ x_1 \} \cup \lfv{N}  }\linexsub{M/x_1} \\ &\redd_{\redlab{RS{:}Cons_3}} \fail^{\lfv{M} \cup \lfv{N} }
        \end{aligned}
  \]
 \end{example}

Similarly to \lamrfail, reduction in \lamrsharfail satisfies a \emph{diamond property}. Therefore, we have the analogue of Proposition~\ref{ch2prop:conf1_lamrfail}: 


\begin{proposition}[Diamond Property for \lamrsharfail]
\label{ch2prop:conf1_lamrsharfail}
     For all $\expr{N}$, $\expr{N}_1$, $\expr{N}_2$ in $\lamrsharfail$ s.t. $\expr{N} \redd \expr{N}_1$, $\expr{N} \redd \expr{N}_2$ { with } $\expr{N}_1 \neq \expr{N}_2$  then there exists   $\expr{M}$ such that $\expr{N}_1 \redd \expr{M}$ and $\expr{N}_2 \redd \expr{M}$.
\end{proposition}

\begin{proof}
The thesis follows as in $\lamrfail$ since  the left-hand sides of the reduction rules in $\lamrsharfail$ do not interfere with each other. 
\end{proof}


\begin{remark}[A Calculus with Sharing but Without Failure (\lamrshar)]
\label{ch2r:lamrshar}
As we did in Remark~\ref{ch2rem:lamr}, we define a sub-calculus of \lamrsharfail  in which failure is not explicit. The calculus   \lamrshar is obtained from the syntax of \lamrsharfail by disallowing the term $\fail^{\widetilde{x}}$. The relevant reduction rules from \figref{ch2fig:share-reductfailure} are \redlab{RS:Beta}, \redlab{RS{:}Ex \dash Sub}, \redlab{RS{:}Lin\dash Fetch}, and the two contextual rules. We keep \defref{ch2d:headshar} unchanged with the provision that $\headf{M \esubst{ B }{x}}$ is undefined when $| \widetilde{x} | \not = \size{B}$. 
\end{remark}

\subsection{Non-Idempotent Intersection Types}
\label{ch2ss:typeshar}\hfill

Similarly to $\lamrfail$, we now define \emph{well-formed} \lamrsharfail expressions and a system of rules for checking {well-formedness} by modifying the rules in~\figref{ch2fig:app_wf_rules}. The grammar of strict and multiset types,  the notions of typing assignments, 
and judgements are  the same as in Section~\ref{ch2sec:lamfailintertypes}.
\secondrev{We need an extension to the notion of typing context: whereas in \lamrfail variables were only assigned to multiset types, now sharing variables are assigned to multiset types, shared and independent variables  are assigned to strict types.}
\begin{definition}{}
\label{ch2d:tcont}
\secondrev{
We extend the definition of typing contexts (\defref{ch2d:tcontsource}) as follows:
\[
    \begin{aligned}
        \Gamma, \Delta & = \dash \sep \Gamma, x:\pi \sep \Gamma , x:\sigma
    \end{aligned}
\]
 The definition of core contexts is extended accordingly, and also denoted as $\core{\Gamma}$.}
\end{definition}

The presentation is in two phases:
\begin{enumerate}
\item We consider the intersection type system given in \figref{ch2fig:typing_sharing} for which we consider the sub-calculus \lamrshar, the sharing calculus excluding failure (cf. Rem.~\ref{ch2r:lamrshar}).
\item We define well-formed expressions for the full language \lamrsharfail, via \defref{ch2def:wf_sharlam} (see below).
\end{enumerate}

\secondrev{To avoid ambiguities, we write $x:\sigma^1$  to make it explicit that the type assignment involves an intersection type (and a sharing variable), rather than a strict type.}

    

\subsubsection{Well-typed Expressions (in \texorpdfstring{$\lamrshar$}{})}\hfill 

 The typing rules in \figref{ch2fig:typing_sharing} are essentially the same as the ones in \figref{ch2fig:app_typingrepeat},  but now taking into account the sharing construct $M [\widetilde{x}\leftarrow x ]$ and the explicit linear substitution. We discuss selected rules:

\begin{itemize}
\item {\bf Rules~\redlab{TS{:}var}, \redlab{TS{:}\oneb}, \redlab{TS{:}bag}, \redlab{TS{:}app}, and \redlab{TS{:}sum}} are the same as in \figref{ch2fig:app_typingrepeat}, considering sharing within the terms and bags.

\item {\bf Rule~\redlab{TS{:}weak}} deals with $k=0$, typing the term $M[\leftarrow x] $, when there are no occurrences of $x$ in $M$, as long as $M$ is typable.
\item {\bf Rule~\redlab{TS{:}abs\dash sh}} is as expected: it requires that the sharing variable is assigned the $k$-fold intersection type $\sigma^k$. 
\item {\bf Rule~\redlab{TS{:}ex \dash lin \dash sub}} supports explicit linear substitutions and consumes one occurrence of $x:\sigma$ from the context.
\item {\bf Rule~\redlab{TS{:}ex \dash sub}} types explicit substitutions where a bag must consist of both the same type and length of the shared variable it is being substituted for.
\item {\bf Rule~\redlab{TS{:}share}} requires that the shared variables $x_1,\ldots, x_k$ have the same type as the sharing variable $x$, for $k\neq 0$. \secondrev{This rule justifies the need for the extension of contexts with assignments of the form $x:\sigma$. This way, e.g.,   Example~\ref{ch2ex:termwelltyped} below gives  an application of Rule~\redlab{TS:share} with $k=1$).}
\end{itemize}

\begin{definition}{Well-typed Expressions}
An expression $\expr{M} \in \lamrsharfail$ is {\em well-typed} (or typable) if there exist $\Gamma$ and  $\tau$ such that $\Gamma \vdash \expr{M} : \tau$ is entailed via the rules in \figref{ch2fig:typing_sharing}.
\end{definition}

Again, the failure term $\fail$  in $\lamrsharfail$ is not typable via this typing system. The following examples illustrate the typing rules.

\begin{example}{}
\label{ch2ex:termwelltyped}
The term $ ((\lambda x. x_1 [x_1 \leftarrow x] ) \bag{y_1} ) [ y_1 \leftarrow y] $ is well-typed, as follows:

            \begin{prooftree}
                    \AxiomC{}
                    \LeftLabel{\redlab{TS{:}var}}
                    \UnaryInfC{\( x_1: \sigma \vdash  x_1 : \sigma\)}
                    \LeftLabel{ \redlab{TS{:}share}}
                    \UnaryInfC{\( x: \sigma^1 \vdash  x_1 [x_1 \leftarrow x] : \sigma\)}
                    \LeftLabel{\redlab{TS{:}abs\dash sh}}
                    \UnaryInfC{\(  \vdash \lambda x. x_1 [x_1 \leftarrow x] :  \sigma^1 \rightarrow \sigma \)}
                                            \AxiomC{}
                        \LeftLabel{\redlab{TS{:}var}}
                        \UnaryInfC{\(  y_1 : \sigma \vdash y_1 : \sigma\)}
                            \AxiomC{\(  \)}
                        \LeftLabel{\redlab{TS{:}\oneb}}
                        \UnaryInfC{\( \vdash \oneb : \omega \)}
                    \LeftLabel{\redlab{TS{:}bag}}
                    \BinaryInfC{\( y_1 : \sigma \vdash \bag{y_1}\cdot \oneb:\sigma^{1} \)}
                    \LeftLabel{\redlab{TS{:}app\dash sh}}
                \BinaryInfC{\( y_1 : \sigma \vdash ((\lambda x. x_1 [x_1 \leftarrow x] ) \bag{y_1} ) : \sigma\)}
                \LeftLabel{ \redlab{TS{:}share}}
                \UnaryInfC{\(  y: \sigma^1 \vdash ((\lambda x. x_1 [x_1 \leftarrow x] ) \bag{y_1} ) [ y_1 \leftarrow y] : \sigma \)}
            \end{prooftree}
       \end{example}

\begin{figure}[t]
    \centering

\begin{prooftree}
    \AxiomC{}
    \LeftLabel{\redlab{TS{:}var}}
    \UnaryInfC{\( x: \sigma \vdash x : \sigma\)}
    \DisplayProof
    \hfill
    \AxiomC{\(  \)}
    \RightLabel{\(\)}
    \LeftLabel{\redlab{TS{:}\oneb}}
    \UnaryInfC{\( \vdash \oneb : \omega \)}
\DisplayProof
    \hfill
      \AxiomC{\( \Delta  \vdash M : \tau\)}
    \LeftLabel{ \redlab{TS{:}weak}}
    \UnaryInfC{\( \Delta , x: \omega \vdash M[\leftarrow x]: \tau \)}
\end{prooftree}

\begin{prooftree}
    \AxiomC{\( \Delta , x: \sigma^k \vdash M[\widetilde{x} \leftarrow x] : \tau \)}
    \LeftLabel{ \redlab{TS{:}abs \dash sh}}
    \UnaryInfC{\( \Delta \vdash \lambda x . (M[\widetilde{x} \leftarrow x]) : \sigma^k \rightarrow \tau \)}
    \DisplayProof
    \hfill
    \AxiomC{\( \Gamma \vdash M : \pi \rightarrow \tau \)}
    \AxiomC{\( \Delta \vdash B : \pi \)}
        \LeftLabel{\redlab{TS{:}app}}
    \BinaryInfC{\( \Gamma , \Delta \vdash M\ B : \tau\)}
    \end{prooftree}

\begin{prooftree}
     \AxiomC{\( \Gamma \vdash M : \sigma\)}
    \AxiomC{\( \Delta \vdash B : \sigma^{k}\)}
    \LeftLabel{\redlab{TS{:}bag}}
    \BinaryInfC{\( \Gamma , \Delta \vdash \bag{M}\cdot B:\sigma^{k+1} \)}
\DisplayProof\hfill
    \AxiomC{\( \Delta \vdash N : \sigma \qquad \Gamma  , x:\sigma \vdash M : \tau \)}
    \LeftLabel{\redlab{TS\!:\!ex\dash lin\dash sub}}
    \UnaryInfC{\( \Gamma , \Delta \vdash M \linexsub{N / x} : \tau \)}
\end{prooftree}

\begin{prooftree}
    \AxiomC{\( \Delta \vdash B : \sigma^k \qquad \Gamma , x:\sigma^k \vdash M [\widetilde{x} \leftarrow x]: \tau \)}
    \LeftLabel{\redlab{TS\!:ex \dash sub}}    
    \UnaryInfC{\( \Gamma , \Delta \vdash M[\widetilde{x} \leftarrow x] \esubst{ B }{ x } : \tau \)}
    \DisplayProof\hfill
     \AxiomC{$ \Gamma \vdash \expr{M} : \sigma$}
    \AxiomC{$ \Gamma \vdash \expr{N} : \sigma$}
    \LeftLabel{\redlab{TS{:}sum}}
    \BinaryInfC{$ \Gamma \vdash \expr{M}+\expr{N}: \sigma$}
\end{prooftree}

\begin{prooftree}
    \AxiomC{\( \Delta , x_1: \sigma, \cdots, x_k: \sigma \vdash M : \tau \quad x\notin \dom{\Delta} \quad k \not = 0\)}
    \LeftLabel{ \redlab{TS{:}share}}
    \UnaryInfC{\( \Delta , x: \sigma^k \vdash M[x_1 , \cdots , x_k \leftarrow x] : \tau \)}
    \end{prooftree}

    \caption{Typing Rules for \lamrshar.}
    \label{ch2fig:typing_sharing}
\end{figure}


\secondrev{
\begin{restatable}[Consistency Stability Under $\redd$]{theorem}{Consistencyreductions}
\label{ch2thm:consistency_reductions}
    If $\expr{M}$ is a consistent $\lamrsharfail$-expression and $\expr{M} \redd \expr{M}'$ then $\expr{M}'$ is consistent.
\end{restatable}
}

\begin{proof}
   \secondrev{By structural induction, and analyzing  the reduction rules applied in $\expr{M}$.
   \iffulldoc
   See \Cref{ch2app:typeshar} for details.}
   \else
   See the full version for details.
   \fi
\end{proof}

As expected, the typing system satisfies the subject reduction property w.r.t. the reduction relation given in \figref{ch2fig:share-reductfailure}, excluding rules for failure.
\begin{theorem}[Subject Reduction in \lamrshar]
\label{ch2t:app_lamrsrshar}
If $\Gamma \vdash \expr{M}:\tau$ and $\expr{M} \redd \expr{M}'$ then $\Gamma \vdash \expr{M}' :\tau$.
\end{theorem}

\begin{proof}
    Standard by induction on the rule applied in $\expr{M}$.
\end{proof}

\revo{}{
\begin{restatable}[Linear Anti-substitution Lemma for \lamrshar]{lemma}{explemfailnofailshar}
\label{ch2lem:antisubt_lem_shar}
Let $M$ and $N$ be $\lamrshar$-terms such that $\headf{M} = x$. The following hold:
\begin{itemize}
    \item If $\Gamma, x:\sigma^{k-1} \vdash M \headlin{ N / x }: \tau$, with $k > 1$, then  there exist $  \Gamma_1, \Gamma_2$  such that  $\Gamma_1 , x:\sigma^k \vdash M: \tau$, and $\Gamma_2 \vdash N : \sigma$, where $\Gamma = \Gamma_1 \contexcat \Gamma_2$.
    \item If $\Gamma \vdash M \headlin{ N / x }: \tau$, with $x \not \in \dom{\Gamma}$, then there exist $ \Gamma_1, \Gamma_2$  such that  $\Gamma_1 , x:\sigma \vdash M: \tau$ , and $\Gamma_2 \vdash N : \sigma$, where $\Gamma = \Gamma_1 \contexcat \Gamma_2$.
\end{itemize}
\end{restatable}
}

\begin{proof}
By structural induction on the reduction rule from Fig.~\ref{ch2fig:typing_sharing}.
\iffulldoc
See \appref{ch2app:typeshar} for details.
\else
See the full version for details.
\fi
\end{proof}

\revo{}{
\begin{restatable}[Subject Expansion for \lamrshar]{theorem}{subexponeshar}
\label{ch2t:app_lamrexpshar}
If $\Gamma \vdash \expr{M}':\tau$ and $\expr{M} \redd \expr{M}'$ then $\Gamma \vdash \expr{M} :\tau$.
\end{restatable}
}

\begin{proof}
Standard, by induction on the reduction rule applied. 
\iffulldoc
See \appref{ch2app:typeshar} for details.
\else
See the full version for details.
\fi
\end{proof}

\subsubsection{Well-formed Expressions (in \texorpdfstring{$\lamrsharfail$}{})}\hfill

On top of the intersection type system for \lamrshar, we  define well-formed expressions: {$\lamrsharfail$-terms whose computation may lead to failure}. 

\begin{definition}{Well-formedness in $\lamrsharfail$}\label{ch2def:wf_sharlam}
An expression $ \expr{M}$ is well formed if there exist  $\Gamma$ and  $\tau$ such that $\Gamma \wfdash  \expr{M} : \tau $ is entailed via the rules in \figref{ch2fig:wfsh_rules}.
\end{definition}

\begin{figure}[!t]
    \centering

    \begin{prooftree}
            \AxiomC{\( \Gamma \vdash \expr{M} : \tau \)}
            \LeftLabel{\redlab{FS\!:\!wf \dash expr}}
            \UnaryInfC{\( \Gamma \wfdash  \expr{M} : \tau \)}
            \DisplayProof\hfill
            \AxiomC{\( \Gamma \vdash B : \pi \)}
            \LeftLabel{\redlab{FS\!:\!wf \dash bag}}
            \UnaryInfC{\( \Gamma \wfdash  B : \pi \)}
            \DisplayProof\hfill
             \AxiomC{\( \Gamma  \wfdash M : \tau\)}
        \LeftLabel{ \redlab{FS\!:\!weak}}
        \UnaryInfC{\( \Gamma , x: \omega \wfdash M[\leftarrow x]: \tau \)}
    \end{prooftree}

\begin{prooftree}
        \AxiomC{\( \Gamma , x: \sigma^k \wfdash M[\widetilde{x} \leftarrow x] : \tau \quad x\notin \dom{\Gamma} \)}
        \LeftLabel{\redlab{FS{:}abs\dash sh}}
        \UnaryInfC{\( \Gamma \wfdash \lambda x . (M[\widetilde{x} \leftarrow x]) : \sigma^k  \rightarrow \tau \)}
         \DisplayProof
    \hfill
    \AxiomC{\( \secondrev{ \dom{\core{\Gamma}} = \widetilde{x} }\)}
        \LeftLabel{\redlab{FS{:}fail}}
        \UnaryInfC{\( \secondrev{ \Gamma \wfdash  \fail^{\widetilde{x}} : \tau } \)}
   \end{prooftree}

   \begin{prooftree}
        \AxiomC{\( \Gamma \wfdash M : \sigma^{j} \rightarrow \tau \quad \Delta \wfdash B : \sigma^{k} \)}
            \LeftLabel{\redlab{FS{:}app}}
        \UnaryInfC{\( \Gamma, \Delta \wfdash M\ B : \tau\)}
        \DisplayProof
\hfill
           \AxiomC{\( \Gamma \wfdash M : \sigma\quad \Delta \wfdash B : \sigma^{k} \)}
        \LeftLabel{\redlab{FS{:}bag}}
        \UnaryInfC{\( \Gamma, \Delta \wfdash \bag{M}\cdot B:\sigma^{k+1} \)}
    \end{prooftree}

\begin{prooftree}
\AxiomC{\( \Gamma  , x:\sigma \wfdash M : \tau \quad \Delta \wfdash N : \sigma \)}
\LeftLabel{\redlab{FS{:}ex \dash lin \dash sub}}
\UnaryInfC{\( \Gamma, \Delta \wfdash M \linexsub{N / x} : \tau \)}
 \DisplayProof
\hfill
\AxiomC{$ \Gamma \wfdash \expr{M} : \sigma \quad  \Gamma \wfdash \expr{N} : \sigma$}
\LeftLabel{\redlab{FS{:}sum}}
\UnaryInfC{$ \Gamma \wfdash \expr{M}+\expr{N}: \sigma$}
 \end{prooftree}

    \begin{prooftree}
            \AxiomC{\( \Gamma , x:\sigma^{k} \wfdash M[\widetilde{x} \leftarrow x] : \tau \)}
             \AxiomC{\( \Delta \wfdash B : \sigma^{j} \)}
        \LeftLabel{\redlab{FS{:}ex \dash sub}}    
        \BinaryInfC{\( \Gamma, \Delta \wfdash M [\widetilde{x} \leftarrow x]\esubst{ B }{ x } : \tau \)}
    \end{prooftree}

    \begin{prooftree}
        \AxiomC{\( \Gamma , x_1: \sigma, \cdots, x_k: \sigma \wfdash M : \tau \quad x\notin \dom{\Gamma} \quad k \not = 0\)}
        \LeftLabel{ \redlab{FS{:}share}}
        \UnaryInfC{\( \Gamma , x: \sigma^{k} \wfdash M[x_1 , \cdots , x_k \leftarrow x] : \tau \)}  
    \end{prooftree}

    \caption{Well-formedness Rules for \lamrsharfail.}\label{ch2fig:wfsh_rules}
\end{figure}

Rules~$\redlab{FS{:}wf\dash expr}$ and $\redlab{FS{:}wf\dash bag}$ guarantee that every well-typed expression and bag, respectively, is well-formed. {Since our language is expressive enough to account for failing computations, we include rules for checking the structure of these ill-behaved terms---terms that can be well-formed, but not typable. For instance,} 

\begin{itemize} 
\item {\bf Rules~$\redlab{FS{:}ex \dash sub}$ and  $\redlab{FS{:}app}$} differ from similar typing rules in \figref{ch2fig:typing_sharing}: the size of the bags (as declared in their types) is no longer required to match. 
\item {\bf Rule~$\redlab{FS{:}fail}$} has no analogue in the type system: we allow the failure term $\fail^{\widetilde{x}}$ to be well-formed with any type, provided that the core context contains types for the variables in $\widetilde{x}$.
\end{itemize}
The other rules are similar to their corresponding ones in~\figref{ch2fig:app_wf_rules} and \figref{ch2fig:typing_sharing}.

The following example illustrates a $\lamrsharfail$ expression that is well-formed but not well-typed.
\begin{example}{Cont. Example~\ref{ch2ex:shar-wf}}\label{ch2ex:shar-wf}
The $\lamrsharfail$ expression consisting of an application of $\hat{I}$ to a bag containing a failure term 
 \(\lambda x . x_1 [x_1 \leftarrow x]) \bag{ \fail^{\emptyset}[ \leftarrow y] \esubst{ \oneb }{y} }\) is well-formed with type $\sigma$. The derivation, with omitted rule labels, is the following:
 
 \begin{prooftree}
 \AxiomC{\(x_1:\sigma \vdash x_1:\sigma\)}
 \UnaryInfC{\(x_1:\sigma \wfdash x_1:\sigma\)}
 \UnaryInfC{\(x:\sigma^1 \wfdash x_1\shar{x_1}{x}:\sigma\)}
  \UnaryInfC{\( \wfdash \lambda x. x_1\shar{x_1}{x}:\sigma\to \sigma\)}
    \AxiomC{}
    \UnaryInfC{\(\wfdash \fail^{\emptyset}:\sigma\)}
    \UnaryInfC{\(y:\omega\wfdash
     \fail^{\emptyset}\shar{}{y}:\sigma\)} 
    \AxiomC{}
    \UnaryInfC{\(\vdash \oneb:\omega\)}
    \UnaryInfC{\(\wfdash \oneb:\omega\)}
    \BinaryInfC{\(\wfdash
     \fail^{\emptyset}\shar{}{y}\esubst{1}{y}:\sigma\)}
     \AxiomC{}
    \UnaryInfC{\(\vdash \oneb:\omega\)}
     \UnaryInfC{\(\wfdash \oneb:\omega\)}
     \BinaryInfC{\(\wfdash
     \bag{\fail^{\emptyset}\shar{}{y}\esubst{1}{y}}:\sigma^1\)}
     \BinaryInfC{\( \wfdash \lambda x. x_1\shar{x_1}{x}\bag{\fail^{\emptyset}\shar{}{y}\esubst{1}{y}}:\sigma\)}
     \end{prooftree}
     
Besides, we have 
$\lambda x . x_1 [x_1 \leftarrow x]) \bag{ \fail^{\emptyset}[ \leftarrow y] \esubst{ \oneb }{y} }\redd^* \fail^{\emptyset}[ \leftarrow y] \esubst{ \oneb }{y}$.
\end{example}

Well-formed $\lamrsharfail$ expressions satisfy the subject reduction property; as usual, the  proof relies on a linear substitution lemma for $\lamrsharfail$.

\begin{restatable}[Substitution Lemma for \lamrsharfail]{lemma}{lamrsharfailsubs}
\label{ch2l:lamrsharfailsubs}
If $\Gamma , x:\sigma \wfdash M: \tau$, $\headf{M} = x$, and $\Delta \wfdash N : \sigma$ 
then 
$\Gamma , \Delta \wfdash M \headlin{ N / x }:\tau$.
\end{restatable}
\begin{proof}
	By structural induction on $M$. 
    \iffulldoc
    See \appref{ch2app:typeshar} for details.
    \else
    See the full version for details.
    \fi
\end{proof}

\begin{restatable}[Subject Reduction in \lamrsharfail]{theorem}{applamrsharfailsr}
\label{ch2t:app_lamrsharfailsr}
If $\Gamma \wfdash \expr{M}:\tau$ and $\expr{M} \redd \expr{M}'$ then $\Gamma \wfdash \expr{M}' :\tau$.
\end{restatable}

\begin{proof}
By structural induction on the reduction rule from Fig.~\ref{ch2fig:share-reductfailure}.
\iffulldoc
See \appref{ch2app:typeshar} for details.
\else
See the full version for details.
\fi 
\end{proof}

We close this part by stating the failure of subject expansion for well-formed expressions.
 
\begin{restatable}[Failure of Subject Expansion in \lamrsharfail]{theorem}{applamrfailexpshar}
\label{ch2t:app_lamrfailsrshar}
If $\Gamma \wfdash \expr{M}':\tau$ and $\expr{M} \redd \expr{M}'$ then it is not necessarily the case that $\Gamma \wfdash \expr{M} :\tau$.
\end{restatable}

\begin{proof}
We adapt the counter-example from the proof of \thmref{ch2t:app_lamrfailse}.
 Consider the term $ \fail^{\emptyset}$, which is well-formed but not well-typed, and let $ \Omega^l = ( \lambda x. x_1 \bag{x_2}[x_1, x_2 \leftarrow x] ) \bag{ \lambda x. x_1 \bag{x_2}[x_1, x_2 \leftarrow x]} $.
 \revdaniele{Notice that  
 $ \fail^{x_1}[x_1 \leftarrow x] \esubst{\bag{\Omega^l}}{x}\redd\fail^\emptyset$
 and $\dash \wfdash \fail^\emptyset:\tau$, 
  but
 $\fail^{x_1}[x_1 \leftarrow x] \esubst{\bag{\Omega^l}}{x}$ is not well-formed (nor well-typed).}
\end{proof}

\secondrev{
\begin{restatable}[Consistency enforced by typing]{theorem}{consistencytype}
\label{ch2thm:consistency_type}
    Let $\expr{M}$ be a $\lamrsharfail$-expression. If $\Gamma \wfdash \expr{M}$ then ${\expr{M}}$ is consistent.
\end{restatable}
\begin{proof}
    By induction on the type derivation. 
    \iffulldoc
    See \Cref{ch2app:typeshar} for details.
    \else
    See the full version for details.
    \fi 
\end{proof}
}

\subsubsection*{Taking Stock}
Up to here, we have presented our source language \lamrfail---a new resource $\lambda$-calculus with failure---and its fail-free sub-calculus \lamr. Based on them we defined well-typed and well-formed expressions. Similarly, we defined the intermediate calculus \lamrsharfail and its sub-calculus \lamrshar. We now move on to define 
a translation of $\lamrfail$ into $\lamrsharfail$. 

\subsection{From \texorpdfstring{\lamrfail}{} into \texorpdfstring{\lamrsharfail}{}}
\label{ch2ss:auxtrans}
\secondrev{Borrowing inspiration from translations given in~\cite{DBLP:conf/lics/GundersenHP13} for the atomic $\lambda$-calculus, we now define a translation  $\recencodopenf{\cdot}$ from well-formed expressions in $\lamrfail$ into $\lamrsharfail$.  It relies on an auxiliary translation $\recencodf{\cdot}$ on 
$\lamrfail$-terms, which depends on 
 the notion of (simultaneous) linear substitution (\defref{ch2def:lin_subst}) which, intuitively,  forces all bound variables in \lamrfail to become shared variables in \lamrsharfail.
}
The correctness of $\recencodopenf{\cdot}$ will be addressed in \secref{ch2ss:firststep}.



\begin{definition}{Linear substitution}\label{ch2def:lin_subst}
    \secondrev{Suppose given a \lamrfail-term  $M$, a variable $x$, and a sequence of variables $\widetilde{w} = y,\widetilde{z}$.
    When \(\#(x , M) = |\widetilde{w} |\)  and  \(\{y\} \cap \widetilde{z}=\emptyset \), 
    the {\em linear substitution} $M\linsub{y,\widetilde{z}}{x}$ of  variable $x$ for variables $\widetilde{w}$ in $M$ is defined inductively as follows:}

    \secondrev{
    \[
        \begin{aligned}
            x\linsub{y}{x} & = 
                    y\\
            (\lambda z . M )\linsub{y}{x} & =  
                    \lambda z . (M\linsub{y}{x}) \quad \text{if } x \in \lfv{M}\\
            (M\ B)\linsub{y}{x} & = 
                \begin{cases}
                    ((M\linsub{y}{x})\ B) &\text{if } x \in \lfv{M} \\ 
                    (M\ (B\linsub{y}{x})) &\text{if } x \not \in \lfv{M}, x \in \lfv{B}\\
                \end{cases}
                \\
            \fail^{\widetilde{z}}\linsub{y}{x} & = 
                    \fail^{\widetilde{z}',y} \quad \text{if } x \in \widetilde{z} \text{ and } \widetilde{z} = \widetilde{z}' , x\\
            (M\esubst{ B }{ z })\linsub{y}{x} & = 
                \begin{cases}
                    (M\linsub{y}{x})\esubst{ B }{ z } &\text{if } x \in \lfv{M} \\ 
                    M\esubst{ B\linsub{y}{x} }{ z } &\text{if }  x \not \in \lfv{M} ,x \in \lfv{B}\\
                \end{cases}
                \\
            \oneb\linsub{y}{x} & = \text{undefined }\\
            \bag{M}\linsub{y}{x} & = 
                    \bag{M\linsub{y}{x}} \quad \text{if } x \in \lfv{M} \\
            (A \cdot B)\linsub{y}{x} & = 
                \begin{cases}
                    ((A\linsub{y}{x}) \cdot B) &\text{if } x \in \lfv{A} \\ 
                    (A \cdot (B\linsub{y}{x})) &\text{if } x \not \in \lfv{A} , x \in \lfv{B}
                    \\
                \end{cases}\\
             M\linsub{y,\widetilde{z}}{x} &= (M\linsub{y}{x})\linsub{\widetilde{z}}{x} 
        \end{aligned}
    \]
    }
    
    Otherwise, in all other cases, the substitution is undefined. 
    We write $M\linsub{z_1,z_2, \cdots, z_k}{x}$ to stand for 
    $( \cdots ((M\linsub{z_1}{x}) \linsub{z_2}{x})\cdots \linsub{z_k}{x})$.
    %
    %
    %
    %
\end{definition}

\secondrev{Notice that for a \lamrfail-term with multiple occurrences of the variable to be substituted for, this linear substitution fixes an ordering of instantiation. For example,  $\lambda x. y\bag{y,x}\linsub{z_1,z_2}{y}$  results in $\lambda x. z_1\bag{z_2,x}$, and a permutation of variables as in  $\lambda x. z_2\bag{z_1,x}$ is not accounted for. This is not restrictive; actually it is enough for our purposes since this substitution will only be used in \defref{ch2def:enctolamrsharfail} and the variables being substituted will be bound by sharing, and therefore could be $\alpha$-renamed.}

\begin{definition}{From $\lamrfail$ to $\lamrsharfail$}
\label{ch2def:enctolamrsharfail}
Let $M \in \lamrfail$.
Suppose $\Gamma \wfdash {M} : \tau$, with
$\dom{\Gamma} = \lfv{M}=\{x_1,\cdots,x_k\}$ and  $\#(x_i,M)=j_i$.  
We define $\recencodopenf{M}$ as
\[
 \recencodopenf{M} = 
\recencodf{M\linsub{\widetilde{y_{1}}}{x_1}\cdots \linsub{\widetilde{y_k}}{x_k}}[\widetilde{y_1}\leftarrow x_1]\cdots [\widetilde{y_k}\leftarrow x_k] 
\]
   where  $\widetilde{y_i}=y_{i_1},\cdots, y_{i_{j_i}}$
   and the translation $\recencodf{\cdot}: \lamrfail \to \lamrsharfail$  is defined in~\figref{ch2fig:auxencfail}.
   The translation $\recencodopenf{\cdot}$ extends homomorphically to expressions.
   \end{definition}

\srev{As already mentioned, } 
the translation $\recencodopenf{\cdot}$ ``atomizes'' occurrences of variables, \srev{in the spirit of~\cite{DBLP:conf/lics/GundersenHP13}}: it converts $n$ occurrences of a variable $x$ in a term into $n$ distinct variables $y_1, \ldots, y_n$.
The sharing construct coordinates the occurrences of these variables by constraining each to occur exactly once within a term. 
We proceed in two stages:
\begin{enumerate} 
\item First, we use $\recencodopenf{\cdot}$ to ensure that each free variable (say, $y$) is replaced by a shared variable (say, $y_i \in \widetilde{y}$),  \revo{}{which is externally bound by the $y$ in $[\widetilde{y}\leftarrow y]$}. 
\item Second, we apply the auxiliary translation $\recencodf{\cdot}$ on the corresponding  \revo{A20}{to the sharing of bound variables.}
\end{enumerate}
We now describe the two cases of \figref{ch2fig:auxencfail} that are noteworthy. 
\begin{itemize}
\item In $\recencodf{  \lambda x . M  }$, the occurrences of $x$ are replaced with fresh shared variables that only occur once  in $M$.
\item The definition of $\recencodf{  M \esubst{ B }{ x }  }$ considers two possibilities.
If the bag being translated is non-empty and the explicit substitution would not lead to failure (the number of occurrences of $x$ and the size of the bag coincide) then we translate the explicit substitution as a sum of explicit linear substitutions. 
Otherwise, the explicit substitution will lead to a failure, and the translation proceeds inductively. 
As we will see, doing this will enable a tight operational correspondence result with $\spi$.
\end{itemize}

\begin{figure*}[t]
\[
\begin{aligned}
    \recencodf{ x  } & =  x  \hspace{3.2cm} \recencodf{  \oneb  } =  \oneb  \hspace{3cm}
    \recencodf{  \fail^{\widetilde{x}} }  =  \fail^{\widetilde{x}}
    \\
    \recencodf{  M\ B }  &=  \recencodf{M}\ \recencodf{B} \hspace{5cm}  \recencodf{\bag{M}\cdot B} =  \bag{\recencodf{M}}\cdot \recencodf{B} 
    \\
    \recencodf{  \lambda x . M  }  &=   \lambda x . (\recencodf{M\langle \widetilde{y} / x  \rangle} [\widetilde{y} \leftarrow x]) 
    \quad  \text{ $\#(x,M) = n$, each $y_i\in\widetilde{y}$ is fresh} 
    \\
    \recencodf{  M \esubst{ B }{ x }  } &= 
    \begin{cases}
        \displaystyle\sum_{B_i \in \perm{\recencodf{ B }}}\hspace{-.5cm}\recencodf{ M \langle \widetilde{y}/ x  \rangle } \linexsub{B_i(1)/x_1} \cdots \linexsub{B_i(k)/x_k} &
        (*) 
        \\
        \recencodf{M\langle y_1. \cdots , y_k / x  \rangle} [\widetilde{y} \leftarrow x] \esubst{ \recencodf{B} }{ x } 
        &
        (**)
    \end{cases}
    \\
    (*) & \quad  \#(x,M) = \size{B} = k \geq 1 
    \\
    (**) & \quad \text{otherwise, } \#(x,M) = k\geq 0
\end{aligned}
\]
    \caption{Auxiliary Translation: \lamrfail into \lamrsharfail.}
    \label{ch2fig:auxencfail}
\end{figure*}

\begin{example}{Cont. Example~\ref{ch2ex:terms}}
\label{ch2ex:termencoded}
We illustrate the translation $\recencodopenf{\cdot}$ on previously discussed examples. In all cases, we start by ensuring that the free variables are shared. This explains the occurrence of $[ y_1 \leftarrow y]$ in the translation of $M_1$ as well as $ [ y_1 \leftarrow y]$ and $ [ z_1 \leftarrow z] $ in the translation of $M_2$. Then, the auxiliary translation $\recencodf{\cdot}$ ensures that bound variables that are guarded by an abstraction are shared. This explains, e.g., the occurrence of  $[x_1 \leftarrow x] $ in the translation of $M_1$.

    \begin{itemize}
        \item The translation of a $\lamrfail$-term with one occurrence of a bound variable and one occurrence of a free variable: $M_1=(\lambda x. x ) \bag{y}$.
        
        \[
        \begin{aligned} 
        \recencodopenf{M_1} &=  \recencodopenf{ (\lambda x. x ) \bag{y} } \\
        &= \recencodf{ (\lambda x. x ) \bag{y_1}   } [ y_1 \leftarrow y]\\
        &= ((\lambda x. x_1 [x_1 \leftarrow x] ) \bag{y_1} ) [ y_1 \leftarrow y] 
        \end{aligned}
        \]
         \item The translation of a $\lamrfail$-term with one bound and two  different free variables: $M_2= (\lambda x. x ) (\bag{y,z} )$.
        \[
        \begin{aligned}
            \recencodopenf{M_2} &=\recencodopenf{ (\lambda x. x ) (\bag{y,z} ) }\\
            &= \recencodf{ (\lambda x. x ) (\bag{y_1,z_1} )  } [ y_1 \leftarrow y] [ z_1 \leftarrow z]\\
            & = ( (\lambda x. x_1 [x_1 \leftarrow x] ) (\bag{y_1,z_1} )  )[ y_1 \leftarrow y] [ z_1 \leftarrow z]
        \end{aligned}
        \]
        \item The translation of a $\lamrfail$-term with a vacuous abstraction: $M_4=(\lambda x. y ) \oneb$.
        \[
        \begin{aligned}
        \recencodopenf{M_4} &= \recencodopenf{(\lambda x. y ) \oneb } \\
        &= \recencodf{(\lambda x. y_1 ) \oneb  } [y_1 \leftarrow y]\\ &= ((\lambda x. y_1 [ \leftarrow x] ) \oneb )[y_1 \leftarrow y]
        \end{aligned}
        \]
        \item The translation of a $\lamrfail$-expression: $M_6=(\lambda x. x ) \bag{y} + (\lambda x. x ) \bag{z}$.
        \[
        \begin{aligned}
            \recencodopenf{M_6} &= \recencodopenf{(\lambda x. x ) \bag{y} + (\lambda x. x ) \bag{z} } \\
            &= \recencodopenf{ (\lambda x. x ) \bag{y}} + \recencodopenf{ (\lambda x. x ) \bag{z}}\\
            & = ((\lambda x. x_1 [x_1 \leftarrow x] ) \bag{y_1} ) [ y_1 \leftarrow y] 
              + ((\lambda x. x_1 [x_1 \leftarrow x] ) \bag{z_1} ) [ z_1 \leftarrow y]
        \end{aligned}
        \]
    \end{itemize}

\end{example}

\begin{example}{}
The translation of a $\lamrfail$-term with two occurrences of a bound variable and two occurrences of a free variable: $M= (\lambda x. x \bag{x}) (\bag{y,y} )$.
        \[
        \begin{aligned}
            \recencodopenf{M} &=\recencodopenf{ (\lambda x. x\bag{x} ) (\bag{y,y} ) } \\
            &= \recencodf{ (\lambda x. x\bag{x} ) (\bag{y_1,y_2} )  } [ y_1,y_2 \leftarrow y] \\
            & = ( (\lambda x. x_1 \bag{x_2}\shar{x_1,x_2}{x} ) (\bag{y_1,y_2} )  )\shar{ y_1,y_2}{y}
        \end{aligned}
        \]
\end{example}

\begin{example}{}
Now consider the translation of $y \esubst{B}{x}$, with $\lfv{B}=\emptyset$ and $y \neq x$:
 \[
 \begin{aligned} 
 \recencodopenf{y\esubst{B}{x}}&=\recencodf{y_0 \esubst{B}{x}}[y_0\leftarrow y]\\
 &=y_0[\leftarrow x]\esubst{\recencodf{B}}{x}[y_0\leftarrow y].
 \end{aligned}
 \]
Hence, the translation induces (empty) sharing on $x$, even if $x$ does not occur in the term $y$.
\end{example}

\secondrev{
\begin{restatable}[$\recencodopenf{\cdot }$ Preserves  Consistency]{proposition}{consistencyencode}
\label{ch2thm:consistency_encod}
Let $\expr{M}$ be a $\lamrfail$-expression. Then
    $\recencodopenf{\expr{M}}$ is a consistent 
    $\lamrsharfail$-expression.
\end{restatable}
}
\begin{proof}
\secondrev{
  By induction on the structure of $\expr{M}$. 
  \iffulldoc
  See \appref{ch2app:typeshar} for details.
  \else
 See the full version for details.
 \fi 
}
\end{proof}

\section[A Session-Typed Calculus]{\texorpdfstring{\spi}{}: A Session-Typed \texorpdfstring{$\pi$}{}-Calculus with \\ Non-Determinism}
\label{ch2s:pi}

The $\pi$-calculus~\cite{DBLP:journals/iandc/MilnerPW92a} is a model of concurrency in which \emph{processes} interact via \emph{names} (or \emph{channels}) to exchange values, which  can be  themselves names.
Here we overview \spi, introduced by \cite{CairesP17}, in which \emph{session types}~\cite{DBLP:conf/concur/Honda93,DBLP:conf/esop/HondaVK98}  ensure that the two endpoints of a channel perform matching actions:
when one endpoint sends, the other receives; when an endpoint closes, the other closes too.
Following~\cite{CairesP10,DBLP:conf/icfp/Wadler12},
\spi defines a Curry-Howard correspondence between session types and a  linear logic with two dual modalities  ($\with A$ and $\oplus A$),  which define \emph{non-deterministic} sessions.
In \spi, cut elimination corresponds to process communication, proofs correspond to processes, and propositions correspond to session types. 

\subsection{Syntax and Semantics}
We use $x, y,z, w \ldots$ to denote {names}   implementing the \emph{(session) endpoints} of protocols specified by session types. 
We consider the sub-language of~\cite{CairesP17}
without labeled choices and replication, which is actually sufficient to encode $\lamrfail$.

\begin{definition}{Processes}\label{ch2d:spi}
The syntax of \spi processes is given by the grammar in Fig.~\ref{ch2f:spi}. 
\end{definition}

\begin{figure}[!t]
\[
\begin{array}{rcl@{\hspace{1.5cm}}l}
    P,Q &::=  & \zero & \text{(inaction)} \\
      &\sep &\overline{x}(y).P& \text{(output)}\\
      &\sep &  x(y).P & \text{(input)}\\
      &\sep &  (P \para Q) & \text{(parallel)}\\
      &\sep &  (\nu x)P & \text{(restriction)}\\
      &\sep & [x \leftrightarrow y] & \text{(forwarder)}\\
      &\sep &x.\overline{\close} & \text{(session close)}\\
      &\sep &x.\close;P & \text{(complementary close)}\\
      &\sep & x.\overline{\some};P &\text{(session confirmation)}\\
      &\sep & x.\overline{\none} & \text{(session failure)}\\
      &\sep & x.\some_{(w_1, \cdots, w_n)};P & \text{(session dependency)}\\
      &\sep & P \oplus Q & \text{(non-deterministic choice)}
\end{array}
\]
\caption{Syntax of \spi. \label{ch2f:spi}}
\end{figure}


As standard, $\zero$ is the inactive process. Session communication is performed using the pair of primitives output and input: the output process $\overline{x}(y).P$ sends a fresh name $y$ along session $x$ and then continues as $P$; the input process $x(y).P$ receives a name $z$ along $x$ and then continues as  $P\subst{z}{y}$, which denotes the capture-avoiding substitution of $z$ for $y$ in $P$.
Process $P \para Q$ denotes the parallel execution of $P$ and $Q$. 
Process $(\nu x)P$ denotes the process $P$ in which name $x$ has been restricted, i.e., $x$ is kept private to $P$. The forwarder process $[x \leftrightarrow y]$ denotes a bi-directional link between sessions $x$ and $y$.
 Processes $x.\overline{\close}$ and $x.\close;P$ denote complementary actions for   closing session $x$.

The following constructs introduce non-determi\-nis\-tic sessions which, intuitively, \emph{may} provide a session protocol  \emph{or} fail. 
    \begin{itemize}
    \item Process $x.\overline{\some};P$ confirms that the session  on $x$ will execute and  continues as $P$.
    \item  Process $x.\overline{\none}$ signals the failure of implementing the session on $x$.
    
    \item Process $x.\some_{(w_1, \cdots,w_n)};P$ specifies a dependency on a non-deterministic session $x$. 
    This process can  either (i)~synchronize with an action $x.\overline{\some}$ and continue as $P$, or (ii)~synchronize with an action $x.\overline{\none}$, discard $P$, and propagate the failure on $x$ to $(w_1, \cdots, w_n)$, which are sessions implemented in $P$.
    When $x$ is the only session implemented in $P$, the tuple of dependencies is empty and so we write simply $x.\some;P$.
    
    \item $P \oplus Q$ denotes a \emph{non-deterministic choice} between $P$ and $Q$. 
            We shall often write $\bigoplus_{i \in I} P_i$ to stand for $P_1 \oplus \cdots \oplus P_n$.
            
\end{itemize}
\noindent
In  $(\nu y)P$ and $x(y).P$ the distinguished occurrence of name $y$ is binding, with scope $P$.
The set of free names of $P$ is denoted by $\fn{P}$. 
We identify process up to consistent renaming of bound names, writing $\equiv_\alpha$ for this congruence. 
We omit trailing occurrences of $\zero$; this way, e.g., we write $x.\close$ instead of $x.\close;\zero$.

\emph{Structural congruence}, denoted $\equiv$, expresses basic identities on the structure of processes and the non-collapsing nature of non-determinism.   

\begin{definition}{Structural Congruence}
\label{ch2def:spistructcong}
 Structural congruence 
is defined as the least congruence relation on processes such that:
\[
\begin{array}{l@{\hspace{0.0cm}}l}
\begin{array}{rcl}
P \para \zero \! &\equiv& \! \zero\\
P \para Q &\equiv& Q \para P \\
(P \para Q) \para R  &\equiv& P \para (Q \para R)\\
\left[x\leftrightarrow y\right]&\equiv&\left[y \leftrightarrow x\right]\\
  ((\nu x )P) \para Q  &\equiv& (\nu x)(P \para Q), x \not \in \fn{P}\\
(\nu x)(P \para (Q \oplus R))  &\equiv& (\nu x)(P \para Q) \oplus (\nu x)(P \para R)\\
\end{array}
&
\begin{array}{rcl}
\zero \oplus \zero  &\equiv&  \zero\\
P \oplus Q &\equiv & Q \oplus P
\\
(P \oplus Q) \oplus R  &\equiv& P \oplus (Q \oplus R)\\
(\nu x)\zero  &\equiv& \zero\\
(\nu x)(\nu y)P  &\equiv& (\nu y)(\nu x)P\\
   P \equiv_\alpha Q &\Longrightarrow& P  \equiv Q
\end{array}
\end{array}
\]

\end{definition}

\subsection{Operational Semantics}\hfill

The {operational semantics} of \spi  is given by a reduction relation, denoted $P\redd Q$, which is the smallest relation on processes generated by the rules in~\figref{ch2fig:redspi}. These rules specify the computations that a process performs on its own. We now explain each rule.

\begin{figure}[!t]

{\
\[
  \begin{array}{l@{\hspace{1.5cm}}rcl}
  \redlab{Comm}&\overline{x}{(y)}.Q \para x(y).P    & \redd & 
  (\nu y) (Q \para P)
  \\
  \redlab{Forw}& (\nu x)( [x \leftrightarrow y] \para P) & \redd &  P \subst{y}{x}  \quad (x \neq y)
 \\
 \redlab{Close}&x.\overline{\close} \para x.\close;P   & \redd &  P 
 \\
 \redlab{Some}&x.\overline{\some};P \para x.\some_{(w_1, \cdots, w_n)};Q  & \redd & P \para Q
 \\
\redlab{None}&x.\overline{\none} \para x.\some_{(w_1, \cdots, w_n)};Q  & \redd & 
w_1.\overline{\none} \para \cdots \para w_n.\overline{\none}\\
\redlab{Cong}& P\equiv P'\wedge P' \redd Q' \wedge Q'\equiv Q &\Longrightarrow& P  \redd Q\\
\redlab{Par}& Q \redd Q' &\Longrightarrow& P \para Q  \redd P \para Q'\\
  \redlab{Res}&P \redd Q  &\Longrightarrow& (\nu y)P \redd (\nu y)Q \\
  \redlab{NChoice}&Q \redd Q' &\Longrightarrow& P \oplus Q  \redd P \oplus Q' 
\\[2mm]
\end{array}
\]}
\caption{Reduction for \spi.}
    \label{ch2fig:redspi}
\end{figure}

\begin{itemize}
\item {\bf Rule~\redlab{Comm}}  formalizes communication, which concerns bound names only (internal mobility): name $y$ is bound in both $\overline{x}{(y)}.Q$ and $x(y).P$.
\item  {\bf Rule~\redlab{Forw}}  implements the  forwarder process  that leads to a name  substitution.
\item {\bf Rule~\redlab{Close}} formalizes session closure and is self-explanatory. 
\item {\bf Rule~\redlab{Some}} describes the synchronization of a process, that is dependent on a non-deterministic session $x$, with the complementary  process  $x.\overline{\some}$ that confirms the availability of such non-deterministic session. 
\item {\bf Rule~\redlab{None}} applies when the non-deterministic session is not available, prefix $x.\overline{\none}$ triggers this failure to all dependent sessions $w_1, \ldots, w_n$; this may in turn trigger further failures (i.e., on sessions that depend on $w_1, \ldots, w_n$). 
\item {\bf Rule~\redlab{NChoice}} defines the closure of reduction with respect to non-collapsing non-deterministic choice.
\item {\bf Rules~\redlab{Cong}, \redlab{Par} and
\redlab{Res}} are standard and formalize that reduction is closed under structural congruence, and also contextual closure of  parallel and restriction constructs.
\end{itemize}

\begin{example}{}
We illustrate confluent reductions starting in a non-deterministic process $R$ which will fail during communication due to unavailability of a session: 
\begin{align*}
R=& (\nu x) ( x.\some_{(y_1,y_2)};y_1(z).y_2(w).\zero \para  ( x.\overline{\some};P \oplus x.\overline{\none}  ) )
\\	
  \equiv  & (\nu x) ( x.\some_{(y_1,y_2)};y_1(z).y_2(w).\zero  \para x.\overline{\some};P ) \\
  &   \oplus  (\nu x) ( x.\some_{(y_1,y_2)};y_1(z).y_2(w).\zero  \para x.\overline{\none}  ) 
\end{align*}
Letting $Q = y_1(z).y_2(w).\zero$, we have:

    \begin{center}
        {\small 
        \begin{tikzpicture}
          \matrix (m) [matrix of math nodes, row sep=2em, column sep=-10em,ampersand replacement=\&]
            { 
                \node(A){ }; \& 
                \node(B){ (\nu x) ( x.\some_{(y_1,y_2)};Q \para 
                         x.\overline{\some};P)  \oplus (y_1.\overline{\none} \para y_2.\overline{\none} )
                     }; \\
                \node(C){R= \begin{aligned} 
                        &(\nu x) ( x.\some_{(y_1,y_2)};Q \para \\
                        & ( x.\overline{\some};P \oplus x.\overline{\none}  ) ) \end{aligned} }; \& 
                \node(D){ }; \& 
                \node(G){ \begin{aligned} 
                    &(\nu x)(Q \para P) \oplus
                    \\
                    & (y_1.\overline{\none} \para y_2.\overline{\none} ) 
                    \end{aligned}
                    }; \\
                \node(E){ }; \&
                \node(F){
                            (\nu x)(Q \para P) \oplus (\nu x) ( x.\some_{(y_1,y_2)};Q \para x.\overline{\none} )
                     }; \\};
                
            \path (C) edge[->](B);
            \path (C) edge[->](F);
            \path (B) edge[->](G);
            \path (F) edge[->](G);
        \end{tikzpicture}
    }
      
    \end{center}

Observe that  reduction is confluent. 
The resulting term 
$(\nu x)(Q \para P) \oplus (y_1.\overline{\none} \para \\ y_2.\overline{\none} )$
includes both alternatives for the interaction on $x$, namely the successful one (i.e., $(\nu x)(Q \para P)$) but also the failure of $x$, which is then propagated to $y_1$ and $y_2$, i.e., $y_1.\overline{\none} \para y_2.\overline{\none}$. 
\end{example}

\subsection{Type System}\hfill
The type discipline for  $\spi$  is based on the type system given in~\cite{CairesP17}, which contains modalities $\with A$ and $\oplus A$, as dual types for non-deterministic sessions.

\begin{definition}{Session Types}
\label{ch2d:sts}
Session types are given by 
\begin{align*}
A,B & ::=  \bot \sep   \onef \sep 
A \otimes B  \sep A \ampy B  
 \sep  \with A \sep \oplus A  
\end{align*} 
\end{definition}

\noindent
Types are assigned to names: an \emph{assignment} $x:A$ enforces the use of name $x$ according to the   protocol specified by $A$.
The multiplicative units  $\bot$ and  $\onef$ are used to type terminated (closed) endpoints.
 $A \otimes B$ types a name that first outputs a name of type $A$ before proceeding as specified by $B$.
Similarly, $A \ampy B $ types a name that first inputs a name of type $A$ before proceeding as specified by $B$.
Then we have the two modalities introduced in~\cite{CairesP17}.
We use $\with A$ as the type of a (non-deterministic) session that \emph{may  produce} a behavior of type $A$.
Dually, $\oplus A$ denotes the type of a session that \emph{may consume} a behavior of type $A$.

The two endpoints of a  session must be \emph{dual} to ensure  absence of communication errors. 
The dual of a type $A$ is denoted $\dual{A}$. 
Duality corresponds to negation $(\cdot)^\bot$ in linear logic:

\begin{definition}{Duality}
\label{ch2def:duality}
The duality relation on types is given by:
\begin{align*}
\dual{\onef} & =  \bot 
&
\dual{\bot}   & =  \onef
&
\dual{A\otimes B}  & = \dual{A} \ampy \dual{B}
&
\dual{A \ampy B}  & = \dual{A} \otimes \dual{B} 
&
\dual{\oplus A}   & =    \with \dual{A}
&
\dual{\with A}  & =   \oplus \overline {A}  
\end{align*}
\end{definition}

Typing judgments are of the form $P \vdash \Delta$, where $P$ is a process and $\Delta$ is a context of 
\srev{the form $x_1:A_1, \ldots, x_n:A_n$, which defines the assignment of type $A_i$ to name $x_i$ (with $1 \leq i \leq n$); all names $x_i$ must be distinct. The context $\Delta$ is \emph{linear} in that it is subject to exchange (the ordering of assignments does not matter), but not to weakening and contraction. In writing `$\Delta, x:A$', we assume that $x$ does not occur in~$\Delta$; also, in writing `$\Delta_1, \Delta_2$', we assume that the names in $\Delta_1$ are distinct from those in $\Delta_2$.}
The empty context is denoted `$\cdot$'. 
We write $\with \Delta$ to denote that all assignments in $\Delta$ have a non-deterministic type, i.e., $\with \Delta = w_1:\with A_1, \ldots, w_n:\with A_n$, for some $A_1, \ldots, A_n$. 
The typing judgment $P \vdash \Delta$ corresponds to the logical sequent $ \vdash \Delta$ for classical linear logic, which can be recovered by erasing processes and name assignments.

\begin{figure}[!t]
\begin{prooftree}
\AxiomC{}
\LeftLabel{\redlab{T\cdot}}
\UnaryInfC{$\zero \vdash $}
\DisplayProof
\hfill
\AxiomC{}
\LeftLabel{\redlab{Tid}}
\UnaryInfC{$[x \leftrightarrow y] \vdash x{:}A, y{:}\dual{A}$}
\end{prooftree}

\begin{prooftree}
\AxiomC{$P \vdash  \Delta, y:{A} \quad Q \vdash \Delta', x:B $}
\LeftLabel{\redlab{T\otimes}}
\UnaryInfC{$\dual{x}(y). (P \mid Q) \vdash  \Delta, \Delta', x: A\otimes B$}
\DisplayProof
\hfill
\AxiomC{$P \vdash \Gamma, y:C, x:D$}
\LeftLabel{\redlab{T\ampy}}
\UnaryInfC{$x(y).P \vdash \Gamma, x: C\ampy D $}
\end{prooftree}

\begin{prooftree}
\AxiomC{\mbox{\ }}
\LeftLabel{\redlab{T\onef}}
\UnaryInfC{$x.\dual{\close} \vdash x: \onef$}
\DisplayProof
\hfill
\AxiomC{$P\vdash \Delta$}
\LeftLabel{\redlab{T\bot}}
\UnaryInfC{$x.\close;P \vdash x{:}\bot, \Delta$}
\end{prooftree}

\begin{prooftree}
\AxiomC{$P \vdash  \Delta \quad Q \vdash  \Delta'$}
\LeftLabel{\redlab{T\mid}}
\UnaryInfC{$P\mid Q \vdash \Delta, \Delta'$}
\DisplayProof
\hfill
\AxiomC{$P \vdash \Delta, x:\dual{A} \quad  Q \vdash  \Delta', x:A$}
\LeftLabel{\redlab{Tcut}}
\UnaryInfC{$(\nu x)(P \mid Q) \vdash\Delta, \Delta'$}
\end{prooftree}

\begin{prooftree}
\AxiomC{$P \vdash \Delta, x:A $}
\LeftLabel{\redlab{T\with_d^x}}
\UnaryInfC{$x.\dual{\some};P \vdash \Delta, x :\with A$}
\DisplayProof
\hfill
\AxiomC{$P \;{ \vdash} \widetilde{w}:\with\Delta, x:A$}
\LeftLabel{\redlab{T\oplus^x_{\widetilde{w}}}}
\UnaryInfC{$x.\some_{\widetilde{w}};P \vdash \widetilde{w}{:}\with\Delta, x{:}\oplus A$}
\end{prooftree}

\begin{prooftree}
\AxiomC{}
\LeftLabel{\redlab{T\with^x}}
\UnaryInfC{$x.\dual{\none} \vdash x :\with A$}
\DisplayProof
\hfill
\AxiomC{$P \vdash \with\Delta \qquad Q  \;{\vdash} \with\Delta$}
\LeftLabel{\redlab{T\with}}
\UnaryInfC{$P\oplus Q \vdash \with\Delta$}
\end{prooftree}

\caption{Typing rules for \spi.}
\label{ch2fig:trulespifull}
\end{figure}

Typing rules for processes correspond to proof rules in the logic; see \figref{ch2fig:trulespifull}. 
This way, Rule~$\redlab{T\cdot}$ allows us to introduce the inactive process $\zero$. Rule~$\redlab{Tid}$ interprets the identity axiom using the forwarder process. Rules~$\redlab{T\otimes}$ and $\redlab{T \ampy }$ type output and input of a name along a session, respectively. 
Rules~$\redlab{T \onef}$ and $\redlab{T \bot}$ type the process constructs for session termination.
  Rules~$\redlab{T cut}$ and  $\redlab{T \mid }$  define cut and mix principles in the logic, which induce typing rules for independent and dependent parallel composition, respectively.

The last four rules in \figref{ch2fig:trulespifull} are used to type process constructs related to non-de\-ter\-mi\-nism and failure. 
 Rules~$ \redlab{T \with_d^x}$ and $ \redlab{T \with^x}$ introduce a session of type $\with A$, which may produce a behavior of type $A$: while the former rule covers the case in which $x:A$ is indeed available, the latter rule formalizes the case in which $x:A$ is not available (i.e., a failure).
 Rule~$\redlab{T \oplus^x_{\widetilde{w}}}$, accounts for the possibility of not being able 
to consume the session $x:A$  by considering sessions, the sequence of names $\widetilde{w} = w_1, \ldots, w_n$,  different from $x$ as potentially not available. 
 Rule~$\redlab{T \with }$ expresses non-deterministic choice of processes $P$ and $Q$ that implement non-deterministic behaviors only.

The type system enjoys type preservation, a result that
follows directly from the cut elimination property in the underlying logic; it ensures that the observable interface of a system is invariant under reduction.
The type system also ensures other properties for well-typed processes (e.g. global progress  and confluence); see~\cite{CairesP17} for details.

\begin{theorem}[Type Preservation~\cite{CairesP17}]
If $P \vdash \Delta$ and $P \redd Q$ then $Q \vdash \Delta$.
\end{theorem}

Having defined  \spi, we now move on to define a correct translation from \lamrfail to \spi.


\section{A Correct Encoding}
\label{ch2s:encoding}
\srev{Having introduced the typed sequential calculi \lamrfail and \lamrsharfail (as well as the translation~$\recencodopenf{~\cdot~} : \lamrfail \to \lamrsharfail $) and the typed concurrent calculus \spi, in this section we show how to correctly translate \lamrfail into \spi, using \lamrsharfail as a stepping stone.} 

\srev{Before delving into technical details, we briefly discuss the significance of our encoding. As in Milner's seminal work, our translation explains how interaction in $\pi$ provides a principled interpretation of evaluation in $\lambda$. We tackle the challenging case in which evaluation and interaction are fail-prone and non-deterministic, effectively generalizing previous translations. Because our encoding preserves types, our developments also delineate a new connection between non-idempotent intersection types and logically motivated session types---indeed, our translation of functions as processes goes hand-in-hand with a translation on types (\figref{ch2fig:enc_sestypfail}), which reveals a new protocol-oriented interpretation of the non-idempotent intersections that govern functional resources. }

\srev{As already mentioned, we shall proceed in two steps. We rely on the translation $\recencodopenf{\cdot}$ from well-formed expressions in \lamrfail to well-formed expressions in \lamrsharfail given in \secref{ch2ss:auxtrans}.
As \lamrfail and \lamrsharfail share the same syntax of types, in this case the translation of types  is the identity. 
    Then, the translation $\piencodf{\cdot}_u$ (for some name $u$) transforms well-formed expressions in \lamrsharfail to well-typed processes in \spi (cf. Fig.~\ref{ch2f:sum}).
We first define \emph{encodability criteria} for translations, which include type preservation; these criteria lead to the notion of \emph{correct encoding} (\secref{ch2ss:criteria}). Then, in \secref{ch2ss:firststep} we establish the correctness of the translation  $\recencodopenf{\cdot}$ (Corollary~\ref{ch2cor:one}); 
finally, in \secref{ch2ss:secondstep}, we present the translation $\piencodf{\cdot}_u$ and establish its correctness (Corollary~\ref{ch2cor:two}).
    }
    
\tikzstyle{mynode1} = [rectangle, rounded corners, minimum width=2cm, minimum height=1cm,text centered, draw=black, fill=brown!80!purple!40]
\tikzstyle{arrow} = [thick,->,>=stealth]

\tikzstyle{mynode2} = [rectangle, rounded corners, minimum width=2cm, minimum height=1cm,text centered, draw=black, fill=teal!20]

\tikzstyle{mynode3} = [rectangle, rounded corners, minimum width=2cm, minimum height=1cm,text centered, draw=black, fill=violet!20]
\tikzstyle{arrow} = [thick,->,>=stealth]

\begin{figure}[!t]
\begin{center}
\begin{tikzpicture}[node distance=2cm]
\node (source) [mynode1] {$\lamrfail$};
\node (interm) [mynode2, right of=source,  xshift=3cm] {$\lamrsharfail$};
\node (target) [mynode3, right of=interm,  xshift=3cm] {$\spi$};
\draw[arrow] (source) --  node[anchor=south] {$ \recencodopenf{\cdot}$} (interm);
\node (enc1) [right of= source, xshift=.4cm, yshift=-.5cm] {\secref{ch2ss:auxtrans}};
\node (enc2) [right of= interm, xshift=.4cm, yshift=-.5cm] {\secref{ch2ss:secondstep}};
\draw[arrow] (interm) -- node[anchor=south] {$ \piencodf{\cdot }_u $ } (target);
\end{tikzpicture}
\end{center}
\caption{Summary of our approach.\label{ch2f:sum}}
\end{figure}

\subsection{Encodability Criteria}
\label{ch2ss:criteria}\hfill

We follow most of the criteria defined by \cite{DBLP:journals/iandc/Gorla10}, a widely studied abstract framework for establishing the \emph{quality} of translations.
A \emph{language} $\mathcal{L}$ is defined as a pair containing a set of terms $\mathcal{M}$ and a reduction semantics $\redd$ on terms (with reflexive, transitive closure denoted $\tred$). \srev{A behavioral equivalence on terms, denoted $\approx$, is also assumed.} 
Then, a \emph{correct encoding}, defined next, concerns a translation of terms of a source language $\mathcal{L}_1$ into terms of a target language  $\mathcal{L}_2$ that respects  certain criteria. 
The criteria in~\cite{DBLP:journals/iandc/Gorla10} concern \emph{untyped} languages; because we consider \emph{typed} languages,  we follow \cite{DBLP:journals/iandc/KouzapasPY19} in requiring also that translations preserve typability. 

\begin{definition}{\secondrev{Correct Encoding}}
\label{ch2d:encoding}
Let $\mathcal{L}_1 = (\mathcal{M}, \redd_1)$
and 
$\mathcal{L}_2 = (\mathcal{P}, \redd_2)$
be two languages and let $\approx_1$ be a behavioral equivalence on terms in $\mathcal{M}$.
We use $M, M', \ldots$ and $P, P', \ldots$ to range over elements in  $\mathcal{M}$ and $\mathcal{P}$.
We say that a translation  $\encod{\cdot}{}: \mathcal{M} \to \mathcal{P}$ is a \emph{correct encoding} if it satisfies the following criteria:
\begin{enumerate}
\item {\it Type preservation:} For every well-typed $M$, it holds that $\encod{M}{}$ is well-typed.

    \item {\it Operational Completeness:} For every ${M}, {M}'$, and ${M}''$ such that ${M} \tred_1 {M}' \approx_1 {M}'' $, it holds that $\encod{{M}}{} \tred_2 \encod{{M}''}{}$.
    
    \item {\it Operational Soundness:} For every $M$ and $P$ such that $\encod{M}{} \tred_2 P$, there exist $M'$ and $M''$ such that $M \redd^*_1 M' \approx_1 M''$ and $P \tred_2 \encod{M''}{}  $.
    
    \item {\it Success Sensitiveness:} Let $\checkmark_1$ and $\checkmark_2$ denote a success predicate in $\mathcal{M}$ and $\mathcal{P}$, respectively. 
For every ${M}$, it holds that $M \checkmark_1$ if and only if $\encod{M}{} \checkmark_2$.     
    
\end{enumerate}
\end{definition}

We briefly describe  the criteria. First, type preservation is a natural requirement and a distinguishing aspect of our work, given that we always consider source and target calculi with types.
Operational completeness formalizes how reduction steps of a source term are mimicked by its corresponding translation in the target language; $\approx_1$ conveniently abstracts away from source terms useful in the translation but which are not meaningful in comparisons. 
Operational soundness concerns the opposite direction: it formalizes the correspondence between (i)~the reductions of a target term obtained via the translation and (ii)~the reductions of the corresponding source term. The role of $\approx_1$ can be explained as in completeness.
\srev{Our use of the equivalence $\approx_1$ for $\mathcal{M}_1$, rather than of an equivalence on $\mathcal{M}_2$, is a minor difference with respect to~\cite{DBLP:journals/iandc/Gorla10}.}
Finally, success sensitiveness complements completeness and soundness, which concern reductions and therefore do not contain information about observable behaviors. 
The so-called success predicates $\checkmark_1$ and $\checkmark_2$ serve as a minimal notion of \emph{observables}; the criterion then says that observability of success of a source term implies observability of success in the corresponding target term, and vice versa.


Besides these semantic criteria, we also consider \emph{compositionality}, a syntactic criterion that requires that a composite source term is translated as the combination of the translations of its sub-terms. 

 \subsection{Correctness of \texorpdfstring{$\recencodopenf{\cdot}$}{}}
 \label{ch2ss:firststep}\hfill

We prove that the translation  $\recencodopenf{\cdot}$ from $\lamrfail$ into $\lamrsharfail$ in \secref{ch2ss:auxtrans} is a correct encoding, in the sense of \defref{ch2d:encoding}.
Because our translation $\recencodopenf{\cdot}$ is defined in terms of $\recencodf{\cdot}$, it satisfies \emph{weak compositionality}, in the sense of \cite{DBLP:journals/entcs/Parrow08}.

\subsubsection{Type Preservation}\hfill

We now prove that 
$\recencodopenf{\cdot}$
translates well-formed $\lamrfail$-expressions
into
well-formed expressions \lamrsharfail-expressions
(Theorem~\ref{ch2thm:preservencintolamrfail2}). 
Notice that because \lamrfail and \lamrsharfail share the same type syntax, \revo{}{there is no translation on types/contexts involved (i.e., an identity translation applies)}. 


Next we define well formed preservation in the translation $\recencodf{\cdot}$ from $\lamrfail$ to $\lamrsharfail$. We rely on the prerequisite proof of type preservation in the translation $\recencodf{\cdot}$ on the sub-calculi from $\lamr$ to $\lamrshar$, and also on syntactic properties of the translation such as: (i) the property below guarantees that the translation $\recencodf{\cdot }$ commutes with the linear head  substitution; (ii) preservation of typability/well-formedness w.r.t. linear substitutions in \lamrsharfail.
\begin{proposition}
\label{ch2prop:linhed_encfail}
Let $M, N$ be $\lamrfail$-terms. We have:
 \begin{enumerate}
 \item $ \recencodf{M\headlin{N/x}}=\recencodf{M}\headlin{\recencodf{N}/x}$.
 \item $ \recencodf{M\linsub{\widetilde{x}}{x}}=\recencodf{M}\linsub{\widetilde{x}}{x}$, where $\widetilde{x}=x_1,\ldots, x_k$ is sequence of pairwise distinct fresh variables.
 \end{enumerate}
\end{proposition}

\begin{proof}
By induction of the structure of $M$.
\end{proof}


\begin{lemma}[Preservation under Linear Substitutions in $\lamrsharfail$] Let ${M} \in \lamrsharfail$.\label{ch2lem:preser_linsub}
    \begin{enumerate}
        \item Typing: If $\Gamma, x:\sigma^{k} \vdash {M} : \tau$
    \revo{A6}{}
    then 
     $\Gamma, x_i:\sigma^{k-1} \vdash {M}\linsub{x_i}{x} : \tau$.
        \item Well-formedness: If $\Gamma, x:\sigma^{k} \wfdash {M} : \tau$
     \revo{A6}{}
     then $\Gamma, x_i:\sigma^{k-1} \wfdash {M}\linsub{x_i}{x} : \tau$.
    \end{enumerate}
\end{lemma}
\begin{proof}
Standard by induction on the rules from \figref{ch2fig:typing_sharing} for item (1), and ~\figref{ch2fig:wfsh_rules} for item~(2).
\end{proof}

 The following example illustrates that the translation of a well-formed expression in $\lamrfail$ is a well-formed $\lamrsharfail$-expression.
\begin{example}{Cont. Example~\ref{ch2ex:wellformed}}
 Term $M_2=(\lambda x. x ) (\bag{y,z})$ is well-formed with a well-formedness judgment  \( y:\sigma, z:\sigma \wfdash (\lambda x. x ) (\bag{y,z})  : \sigma \). In Example~\ref{ch2ex:termencoded} we showed that:
 \[
    \recencodopenf{M_2} = ( (\lambda x. x_1 [x_1 \leftarrow x] ) (\bag{y_1,z_1})) [ y_1 \leftarrow y] [ z_1 \leftarrow z] \]
which is well-formed with translated well-formed judgment $ y:\sigma^1, z:\sigma^1 \wfdash \recencodopenf{M_2}:\sigma$.
The derivation is given below (using rules from \figref{ch2fig:wfsh_rules});  we omit the labels of rule applications and concatenations with the empty bag, i.e., we write $\bag{y_1}$ instead of $\bag{y_1}\cdot \oneb$.
\begin{prooftree}
\AxiomC{}
\UnaryInfC{\(x_1:\sigma\vdash x_1:\sigma\)}
\UnaryInfC{\(x_1:\sigma\wfdash x_1:\sigma\)}
\UnaryInfC{\(x:\sigma^1 \wfdash x\shar{x_1}{x}:\sigma\)}
\UnaryInfC{\( \wfdash \lambda x.(x\shar{x_1}{x}):\sigma\to\sigma \)}
\AxiomC{}
\UnaryInfC{\(y_1:\sigma\vdash y_1:\sigma\)}
\UnaryInfC{\(y_1:\sigma\wfdash y_1:\sigma\)}
\UnaryInfC{\(y_1:\sigma^1\wfdash \bag{y_1}:\sigma^1\)}
\AxiomC{}
\UnaryInfC{\(z_1:\sigma\vdash z_1:\sigma\)}
\UnaryInfC{\(z_1:\sigma\wfdash z_1:\sigma\)}
\UnaryInfC{\(z_1:\sigma^1\wfdash \bag{z_1}:\sigma^1\)}
\BinaryInfC{\(y_1:\sigma^1,z_1 :\sigma^1\wfdash \bag{y_1}\cdot \bag{z_1}:\sigma^2\)}
\BinaryInfC{\(y_1:\sigma^1,z_1:\sigma^1\wfdash \lambda x. (x_1\shar{x_1}{x})\bag{y_1,z_1}:\sigma\)}
\UnaryInfC{\(y:\sigma^1,z_1:\sigma^1\wfdash \lambda x. (x_1\shar{x_1}{x})\bag{y_1,z_1}\shar{y_1}{y}:\sigma\)}
\UnaryInfC{\(y:\sigma^1,z:\sigma^1\wfdash \lambda x. (x_1\shar{x_1}{x})\bag{y_1,z_1}\shar{y_1}{y}\shar{z_1}{z}:\sigma\)}
\end{prooftree}
\end{example}

    
     


 

 
 As the translation $\recencodopenf{\cdot }$ for 
 \revo{}{} 
 $\lamrfail$-terms is defined in terms of $\recencodf{\cdot}$,
 \revo{A20}{}
 it is natural that  preservation of well-formedness under $\recencodopenf{\cdot }$ (Theorem~\ref{ch2thm:preservencintolamrfail2}) relies on the preservation of well-formedness  under $\recencodf{\cdot}$, given next.
 
 \secondrev{To state well-formedness preservation, we use $\core{\Gamma}$, the core context of $\Gamma$ (\defref{ch2d:tcont}). In the following property, we use an additional condition on $\core{\Gamma}$, which reflects the fact that intersection types get ``flattened'' by virtue of the translation. The condition, denoted~$\strcore{\Gamma}$, is defined whenever $\core{\Gamma}$ contains only unary multisets as follows: if
$x: \sigma^1 \in \core{\Gamma}$ for all $x \in \dom{\core{\Gamma}}$, then $x: \sigma \in \strcore{\Gamma}$. }

\begin{restatable}[Well-formedness preservation for $\recencodf{\cdot}$]{lemma}{preservencintolamrfail}
\label{ch2lem:wfpreserv_closedtrans}
\revo{A20,A21,A22}{
Let $B$ and  $\expr{M}$  be a  bag and an expression in $\lamrfail$, respectively. \secondrev{Also, let $\Gamma$ be a context such that $\strcore{\Gamma}$ is defined. We have:}
\begin{enumerate}
\item
    \revo{A8}{If $\Gamma \wfdash B:\pi$  
then $\strcore{\Gamma} \wfdash \recencodf{B}:\pi$.}
    \item 
    \revo{A9}{If $\Gamma \wfdash \expr{M}:\sigma$ 
then $\strcore{\Gamma} \wfdash \recencodf{\expr{M}}:\sigma$.}
\end{enumerate}
}
\end{restatable}

\begin{proof}[Proof (Sketch)]
By mutual induction on the typing derivations $\Gamma\wfdash B:\sigma$ and $\Gamma\wfdash \expr{M}:\sigma$. The proof of item (1) follows mostly by induction hypothesis, by analyzing the rule applied (\figref{ch2fig:app_wf_rules}).
The proof of item (2), also follows by analyzing the rule applied, but it is more delicate, especially when treating  cases involving Rules~\redlab{FS:app} or \redlab{FS:ex\dash sub}, for which the size of the bag does not match the number of occurrences of variables in the expression.
\iffulldoc
See  \appref{ch2app:encodingprop} for full details.
\else
 See the full version for details.
\fi 
\end{proof}

\begin{restatable}[Well-formedness Preservation for $\recencodopenf{\cdot}$]{theorem}{preservencintolamrfailtwo}
\label{ch2thm:preservencintolamrfail2}
Let $B$ and  $\expr{M}$  be a bag and an expression in $\lamrfail$, respectively. 
\begin{enumerate}

\item
    \revo{A10}{If $\Gamma \wfdash B:\pi$ 
then $\core{\Gamma} \wfdash \recencodopenf{B}:\pi$}.

    \item 
    \revo{A11}{If $\Gamma \wfdash \expr{M}:\sigma$ 
then $\core{\Gamma}\wfdash \recencodopenf{\expr{M}}:\sigma$}.

\end{enumerate}
\end{restatable}

\begin{proof}[Proof (Sketch)]
By mutual induction on the typing derivations $\Gamma\wfdash B:\sigma$ and $\Gamma\wfdash \expr{M}:\sigma$. Note that for a bag $B$, since the first part of translation consists in sharing the free variables of $B$, we will work with the translated bag:
 \[
    \recencodopenf{B}=\recencodf{B\linsub{\widetilde{x_1}}{x_1}\ldots \linsub{\widetilde{x_k}}{x_k}}\shar{\widetilde{x_1}}{x_1}\ldots \shar{\widetilde{x_k}}{x_k}
\]
and the rest of the proof depends on Proposition~\ref{ch2prop:linhed_encfail} that moves linear substitutions outside $\recencodf{\cdot}$, then Lemma~\ref{ch2lem:preser_linsub} that guarantees preservation of typability/well-formedness under linear substitutions,  and Lemma~\ref{ch2lem:wfpreserv_closedtrans} for treating the closed translation. The dependency extends to the proof of item (2), for expressions. 
\iffulldoc
The full proof can be found in \appref{ch2app:encodingprop}.
\else
See the full version for details.
\fi 
\end{proof}

\subsubsection{Operational Correspondence: Completeness and Soundness}\label{ch2app:ss:compsound}\hfill

Def.~\ref{ch2d:encoding} states operational completeness and soundness over the reflexive, transitive closure of the reduction rules. However, in the case of $\recencodopenf{\cdot}$, we prove completeness and soundness for a single reduction step (cf. Fig.~\ref{ch2f:opcf}).  This is sufficient: by the diamond property (Proposition~\ref{ch2prop:conf1_lamrfail}) a result stated for $\redd$ can be extended easily to $\tred$, by induction on the length of the reduction sequence. (The result is immediate when the length is zero.)

\revdaniele{We rely on a {\em structural equivalence} over $\lamrfail$-expressions, denoted $\pequiv$, which is the least congruence satisfying $\alpha$-conversion and satisfying the identities in ~\figref{ch2fig:rPrecongruencefail}. This congruence allows us to move explicit substitutions to the right of the term and to ignore explicit substitutions of a variable $x$ for empty bags in a term that does not contain $x$. }

\begin{example}{}
   Consider the failure term $M= \fail^{y,y,z} \esubst{\oneb}{x}$.
   Since $\size{\oneb} = 0$, the term $M$ cannot reduce using Rule~$\redlab{R:Cons_2}$, which requires that the size of the bag is greater than 0. Instead, \revdaniele{we use the structural equivalence identity in \figref{ch2fig:rPrecongruencefail}: 
   $\fail^{y,y,z}\esubst{\oneb}{x} \pequiv \fail^{y,y,z}$.}
 \end{example}

\begin{figure}[!t]
    \[
    \begin{array}{rcl@{\hspace{0.5cm}}l}
      M \esubst{\oneb}{x} &\revdaniele{\pequiv}& M & \text{(if  $x \not \in \lfv{M}$)} 
    \\
    M B_1\esubst{B_2}{x} & \pequiv & (M\esubst{B_2}{x})B_1 & \text{(if $x \not \in \lfv{B_1}$)}
    \\
        M\esubst{B_1}{y}\esubst{B_2}{x} & \pequiv& (M\esubst{B_2}{x})\esubst{B_1}{y} & \revt{C12}{\text{(if  $ x\neq y, x \not \in \lfv{B_1}$ and $y \not \in \lfv{B_2}$)}}\\
         M \revdaniele{\pequiv} M' &\Rightarrow&  C[M] \revdaniele{\pequiv} C[M']\\
    \expr{M} \revdaniele{\pequiv} \expr{M}' &\Rightarrow&      D[\expr{M}]  \revdaniele{\pequiv} D[\expr{M}'] 
        \end{array}
    \]
\caption{Congruence in \lamrfail}
    \label{ch2fig:rPrecongruencefail}
\end{figure}


\begin{figure}[!t]
\begin{tikzpicture}[scale=.9pt]
\node (opcom) at (4.3,7.0){Operational Completeness};
\draw[rounded corners, color=brown!80!purple!90!black, fill=brown!80!purple!40] (0,5.2) rectangle (15.5,6.6);
\draw[rounded corners, color=teal, fill=teal!20] (0,1.4) rectangle (15.5,2.8);
\node (lamrfail) at (1.2,6) {$\lamrfail$:};
\node (expr1) [right of=lamrfail, xshift=.3cm] {$\mathbb{N}$};
\node (expr2) [right of=expr1, xshift=2cm] {$\mathbb{M} \pequiv \mathbb{M}'$};
\node (expr3) [right of=expr2, xshift=-0.5cm, yshift = -0.1cm] {};
\draw[arrow] (expr1) -- (expr2);
 \node at (5.0,5.6) {\redlab{R}};
\node (lamrsharfail) at (1.2,2) {$\lamrsharfail$:};
\node (transl1) [right of=lamrsharfail, xshift=.3cm] {$\recencodopenf{\mathbb{N}}$};
\node (transl2) [right of=transl1, xshift=2.5cm] {$\recencodopenf{\mathbb{M'}}$};
\draw[arrow, dotted] (transl1) -- node[anchor= south] {$*$ }(transl2);
\node (enc1) at (2,4) {$\recencodopenf{\cdot }$};
\node (refcomp) [right of=enc1, xshift=1.4cm]{Thm~\ref{ch2l:app_completenessone}};
\node  at (7.0,4) {$\recencodopenf{\cdot }$};
\draw[arrow, dotted] (expr1) -- (transl1);
\draw[arrow, dotted] (expr3) -- (transl2);
\node (opcom) at (11,7.0){Operational Soundness};
\node (expr1shar) [right of=expr2, xshift=2.3cm] {$\mathbb{N}$};
\node (expr2shar) [right of=expr1shar, xshift=2.8cm] {$\mathbb{N'}\pequiv \mathbb{N}''$};
\node (expr3shar) [right of=expr2shar, xshift=-0.5cm, yshift = -0.1cm] {};
\draw[arrow, dotted] (expr1shar) -- (expr2shar);
\node at (12.5,5.6) {\redlab{R}};
\node (transl1shar) [right of=transl2, xshift=1.8cm] {$\recencodopenf{\mathbb{N}}$};
\node (transl2shar) [right of=transl1shar, xshift=1.25cm] {$\mathbb{L}$};
\node (transl3shar) [right of=transl2shar, xshift=1.05cm]{$\recencodopenf{\mathbb{N'}}$};
\draw[arrow,dotted] (expr3shar) -- (transl3shar);
\draw[arrow,dotted] (expr1shar) -- (transl1shar);
\node (enc2) at (8.5,4) {$\recencodopenf{\cdot }$};
\node (refsound) [right of=enc2, xshift=1.5cm]{Thm~\ref{ch2l:soundnessone}};
\node  at (15,4) {$\recencodopenf{\cdot }$};
\draw[arrow] (transl1shar) -- (transl2shar);
\draw[arrow,dotted] (transl2shar) -- node[anchor= south] {*}(transl3shar);
\end{tikzpicture}
\vspace{1em}
\caption{Operational correspondence for $\recencodopenf{\cdot }$ \label{ch2f:opcf}}	
\end{figure}

\begin{restatable}[Operational Completeness]{theorem}{appcompletenessone}
\label{ch2l:app_completenessone}
Let $\expr{M}, \expr{N}$ be well-formed $\lamrfail$ expressions. 
Suppose $\expr{N}\redd_{\redlab{R}} \expr{M}$.
\begin{enumerate}
\item If $\redlab{R} =  \redlab{R:Beta}$  then $ \recencodopenf{\expr{N}}  \redd^{\leq 2}\recencodopenf{\expr{M}}$;

\item If $\redlab{R} = \redlab{R:Fetch}$   then   $ \recencodopenf{\expr{N}}  \redd^+ \recencodopenf{\expr{M}'}$, for some $ \expr{M}'$ such that  $\expr{M} \pequiv \expr{M}'$. 
\item If $\redlab{R} \neq \redlab{R:Beta}$    and $\redlab{R} \neq \redlab{R:Fetch}$ then   $ \recencodopenf{\expr{N}}  \redd \recencodopenf{\expr{M}}$.
\end{enumerate}
\end{restatable}

\begin{proof}[Proof (Sketch)]
By induction on the rules from Figure~\ref{ch2fig:reductions_lamrfail} applied to infer $\expr{N}\redd_{\redlab{R}} \expr{M}$. We analyse the reduction depending on whether $\redlab{R}$ is either $\redlab{R:Beta}$, or $\redlab{R:Fetch}$,  or neither. In the case the rule applied is $\redlab{Beta}$, then $\mathbb{N}=(\lambda x. M')\bag{B}$ and $\mathbb{M}=M'\esubst{B}{x}$.  When  applying the translation $\recencodopenf{\cdot}$ to $\mathbb{N}$ and $\mathbb{M}$ we obtain:
\begin{itemize} 
\item $\recencodopenf{\mathbb{N}}=((\lambda x.\recencodf{ M^{''}\langle{\widetilde{y}/x}\rangle}[\widetilde{y}\leftarrow x])\recencodf{B'})[\widetilde{x_1}\leftarrow x_1]\ldots [\widetilde{x_k}\leftarrow x_k]$
\item $\recencodopenf{\mathbb{M}}=\recencodf{M^{''}\esubst{B'}{x}} [\widetilde{x_1}\leftarrow x_1]\ldots [\widetilde{x_k}\leftarrow x_k]$
\end{itemize}
\revd{B17}{where $B'$ and $M^{''}$ stand for the renamings of $B$ and $M'$, respectively,} after sharing the multiple occurrences of their free/bound variables (\defref{ch2def:enctolamrsharfail}). Note that 
\[\recencodopenf{\mathbb{N}}\redd{}_{\redlab{RS:Beta}}(\recencodf{ M^{''} \langle{\widetilde{y}/x} \rangle} [\widetilde{y} \leftarrow x] \esubst{\recencodf{B'}}{x}) [\widetilde{x_1}\leftarrow x_1]\ldots [\widetilde{x_k}\leftarrow x_k]:=\mathbb{L},\] and according to rules in \figref{ch2fig:share-reductfailure}, the remaining reduction depends upon the characteristics of the bag $\recencodf{B'}$:
\begin{enumerate}[(i)]
    \item $\size{\recencodf{B'}}=\#(x,M^{''})=k\geq 1$. 
    Then, 
    $\recencodopenf{\mathbb{N}}\redd{}_{\redlab{RS:Beta}}\mathbb{L}\redd_{\redlab{RS:ex\dash sub}}\recencodopenf{\mathbb{M}}$.
    \item Otherwise, $\mathbb{L}$ can be further expanded, the ``otherwise case'' of the translation of explicit substitutions,  such that 
    \[
       \begin{aligned}
        \recencodopenf{\mathbb{N}}\redd{}_{\redlab{RS:Beta}} & (\recencodf{ M^{''} \langle{\widetilde{y}/x} \rangle} [\widetilde{y} \leftarrow x] \esubst{\recencodf{B'}}{x}) [\widetilde{x_1}\leftarrow x_1]\ldots [\widetilde{x_k}\leftarrow x_k]\\
        & = \mathbb{L}  = \recencodopenf{\mathbb{M}}
       \end{aligned} 
    \]
\end{enumerate}
In the case the rule applied is $\redlab{R:Fetch}$, the proof depends on the size $n$ of the bag. The interesting case is when the bag $B$  has only one component (i.e., $n=1$):  from $\mathbb{N}\redd_{\redlab{F:Fetch}} \mathbb{N}$  we have that  $\mathbb{N}=M\esubst{\bag{N_1}}{x}$ and $\mathbb{M}=M\headlin{N_1/x}\esubst{1}{x}$. We need to use the congruence $\pequiv$ to obtain $\mathbb{M}=M\headlin{N_1/x}\esubst{1}{x}\pequiv M\headlin{N_1/x}:=\mathbb{M'}$ and then conclude that $\recencodopenf{\mathbb{N}}\redd\recencodopenf{\mathbb{M'}}$. 
The analysis for the other cases is also done by inspecting the structure of expressions and bags. 
\iffulldoc
The full proof can be found in \appref{ch2app:compandsoundone}.
\else
See the full version for details.
\fi 
\end{proof}

We establish soundness for a single reduction step. As  we discussed for completeness, the property generalizes to multiple steps.

\begin{restatable}[Operational Soundness]{theorem}{soundnessone}
\label{ch2l:soundnessone}
Let $\expr{N}$ be a well-formed $\lamrfail$ expression. 
Suppose $ \recencodopenf{\expr{N}}  \redd \expr{L}$. Then, there exists $ \expr{N}' $ such that $ \expr{N}  \redd_{\redlab{R}} \expr{N}'$ and 

\begin{enumerate}
 \item If $\redlab{R} = \redlab{R:Beta}$ then $\expr{ L } \redd^{\leq 1} \recencodopenf{\expr{N}'}$;

   \item If $\redlab{R} \neq \redlab{R:Beta}$ then $\expr{ L } \redd^*  \recencodopenf{\expr{N}''}$, for $ \expr{N}''$ such that  $\expr{N}' \pequiv \expr{N}''$.
\end{enumerate}
\end{restatable}

\begin{proof}[Proof (Sketch)]
By induction on the structure of $\mathbb{N}$ and inspecting the rules from \figref{ch2fig:share-reductfailure} that can be applied in $\recencodopenf{\mathbb{N}}$. The interesting cases happen when $\mathbb{N}$ is either an application  $\mathbb{N}=(M\ B)$ or an explicit substitution $\mathbb{N}=M\esubst{B}{x})$. The former is reducible when $\mathbb{N}$ is an instance of $\redlab{R:Beta}$ or when $M=\fail^{\widetilde{x}}$ and $\mathbb{N}$ is an instance of $\redlab{R:Cons_1}$. The latter, for $\mathbb{N}=M\esubst{B}{x})$, the proof is split in several subcases depending whether: (i) size of the bag $\size{B}=\#(x,M)\geq 1$, and three possible reductions can take place $\redlab{RS:lin\dash fetch}$, $\redlab{RS:Cons_3}$ and $\redlab{RS:Cont}$, depending if $M$ is a failing term or not; (ii)~$\size{B}\neq \#(x,M)$ or $\size{B}=0$,  and the proof follows either applying Rule~$\redlab{RS:Fail}$ or  by induction hypothesis. 
\iffulldoc
The full proof can be found in \appref{ch2app:compandsoundone}.
\else
See the full version for details.
\fi 
\end{proof}

\subsubsection{Success Sensitiveness}\hfill

We now consider success sensitiveness, a property that complements (and relies on) operational completeness and soundness. For the purposes of the proof, we consider the extension of $\lamrfail$ and $\lamrsharfail$ with dedicated constructs and predicates that specify success. 

\begin{definition}{}\label{ch2def:ext_succ}
We extend the syntax of terms for $\lamrfail$ and $\lamrsharfail$ with the same $\checkmark$ construct. 
In both cases, we assume $\checkmark$ is well formed. 
Also, we define $\headf{\checkmark} = \checkmark$ and $\recencodf{\checkmark} = \checkmark$
\end{definition}

An expression $\expr{M}$ has success, denoted \succp{\expr{M}}{\checkmark}, when there is a sequence of reductions from \expr{M} that leads to an expression that includes a summand that contains an occurrence of $\checkmark$ in head position.

\begin{definition}{Success in \lamrfail and \lamrsharfail}
\label{ch2def:app_Suc3}
In $\lamrfail$ and $\lamrsharfail $, we define 
\[\succp{\expr{M}}{\checkmark}\iff \exists M_1 , \cdots , M_k. ~\expr{M} \redd^*  M_1 + \cdots + M_k \text{ and } \headf{M_j} = \checkmark,\]
for some  $j \in \{1, \ldots, k\}$.
\end{definition}

\begin{definition}{Head of an expression}
We extend \defref{ch2d:headshar} from terms to expressions as follows:
\secondrev{
 $$
    \headfsum{\expr{M}} = 
        \begin{cases}
            \headf{M_i} & \text{if } \headf{M_i} = \headf{M_j} \text{ for all } M_i,M_j \in \expr{M}\\
            \text{undefined} & \text{otherwise}
        \end{cases}
 $$
 }
\end{definition}


\begin{restatable}[Preservation of head term]{proposition}{checkpres}
\label{ch2Prop:checkpres}
The head of a term is preserved when applying the translation $\recencodopenf{\cdot}$, i.e., 
$$\forall M \in \lamrfail. ~~ \headf{M} = \checkmark \iff \headfsum{\recencodopenf{M}} = \checkmark.$$
\end{restatable}

 \begin{proof}[Proof (Sketch)]
By induction on the structure of $M$ considering the extension of the language established in~\defref{ch2def:ext_succ}. 
\iffulldoc
See \appref{ch2app:sucessone} for details.
\else
See the full version for details.
\fi 
 \end{proof}

\begin{restatable}[Success Sensitivity]{theorem}{appsuccesssensce}
\label{ch2proof:app_successsensce}
Let  \expr{M} be a well-formed $\lamrfail$-expression.
Then,
\[\expr{M} \Downarrow_{\checkmark}\iff \recencodopenf{\expr{M}} \Downarrow_{\checkmark}.\]
\end{restatable}

 \begin{proof}[Proof (Sketch)]
By induction on the structure of $\lamrfail$ and $\lamrsharfail$ expressions. The if-case  follows from operational soundness (Thm.~\ref{ch2l:soundnessone}) by analyzing a reductions starting from $\recencodopenf{\mathbb{M}}$. Reciprocally, the only-if-case  follows by operational completeness (Thm.~\ref{ch2l:app_completenessone}),  analyzing  reductions starting from $\mathbb{M}$.
\iffulldoc
See \appref{ch2app:sucessone} for details.
\else
See the full version for details.
\fi 
\end{proof}
We have the corollary below, which follows from Theorems~\ref{ch2thm:preservencintolamrfail2}, 
\ref{ch2l:app_completenessone}, 
\ref{ch2l:soundnessone}, and 
\ref{ch2proof:app_successsensce}:

\begin{corollary}
\label{ch2cor:one}
Our translation  $ \recencodopenf{ \cdot } $ 
is a correct encoding, in the sense of \defref{ch2d:encoding}.
\end{corollary}

\subsection{From \texorpdfstring{$\lamrsharfail$}{} to \texorpdfstring{$\spi$}{}}
\label{ch2ss:secondstep}\hfill

We now define our translation of $\lamrsharfail$ into $\spi$, denoted $\piencodf{\cdot}_u$, and establish its correctness. 
As usual in translations of $\lambda$ into $\pi$, we use a name $u$ to provide the behavior of the translated expression. 
In our case, $u$ is a non-deterministic session: the translated expression can be   available or not; this is signalled by prefixes `$u.\overline{\some}$'
and 
`$u.\overline{\none}$', respectively. 
Notice that every (free) variable $x$  in a $\lamrsharfail$-term $M$ becomes a name $x$ in its corresponding process $\piencodf{M}_u$ and is assigned an appropriate session type.

\subsubsection{An Auxiliary Translation}


Before introducing $\piencodf{\cdot}_u$, we first discuss the translation $\piencod{\cdot}_u: \lamrshar \rightarrow \spi$, i.e., the translation in which the source language does not include failures. 
This auxiliary translation, shown in \figref{ch2fig:enc},  is given for pedagogical purposes: it  allows us to gradually discuss several key design decisions in $\piencodf{\cdot}_u$.

\begin{figure}[!t]
    \begin{align*}
    \piencod{ x }_u  &=   x.\overline{\some} ;[x \leftrightarrow u ]
    \\
    \piencod{ \lambda x . M[\widetilde{x} \leftarrow x]}_u    &=  u.\overline{\some};u(x).\piencod{ M[\widetilde{x} \leftarrow x] }_u 
    \\
        \piencod{ M\ B }_u   &=  \bigoplus_{B_i \in \perm{B}} (\nu v)( \piencod{ M}_v \mid v.\some_{u , \lfv{B}} ; \outact{v}{x} . ( x.\some_{\lfv{B_i}}; \piencod{ B_i}_x \mid [v \leftrightarrow u] ) ) 
    \\
    \piencod{M [ x_1, \ldots, x_k \leftarrow x ] }_u   &=   x.\overline{\some};x(x_1). \cdots. x(x_k).x.\close; \encod{M}{u} 
    \\
    \piencod{ M[ \leftarrow x] }_u   &=  x.\overline{\some};x.\close; \encod{M}{u} 
    \\
    \piencod{\bag{M}\cdot B}_x   &=   \outact{x}{x_1}. (x_1.\some_{\lfv{B}};\encod{M}{x_1} \sep \encod{B}{x} )  
    \\
    \piencod{\oneb }_x   &=   x. \overline{\close} 
    \\
    \piencod{ M[\widetilde{x} \leftarrow x] \esubst{ B }{ x} }_u   &=  \bigoplus_{B_i \in \perm{B}} (\nu x)( \piencod{ M[\widetilde{x} \leftarrow x]}_u \mid x.\some_{\lfv{B_i}}; \piencod{ B_i}_x )  
    \\
    \piencod{ M \linexsub{N / x}  }_u   &=  (\nu x) ( \piencod{ M }_u \mid   x.\some_{\lfv{N}};\piencod{ N }_x  )  
    \\
    \piencod{\expr{M}+\expr{N} }_u    &=  \piencod{ \expr{M} }_u \oplus \piencod{ \expr{N} }_u  
\end{align*}
    \caption{An auxiliary translation of \lamrshar into \spi, without failures}
    \label{ch2fig:enc}
\end{figure}
We describe each of the cases from the translation $\piencod{\cdot}_u$, focusing on the role of non-deterministic sessions (expressed using prefixes `$x.\overline{\some}$' and `$x.\some_{(w_1, \cdots, w_n)}$' in \spi):
\begin{itemize}
    \item $\piencod{ x }_u$: Because sessions are non-deterministically available, the translation first confirms that the behavior along $x$ is available; subsequently, the forwarder process induces a substitution $\subst{x}{u}$.

    \item $\piencod{ \lambda x . M[\widetilde{x} \leftarrow x]}_u$: As in the case of variables, the translation first confirms the behavior along $u$ before receiving a name, which will be used in the translation of $M[\widetilde{x} \leftarrow x]$, discussed next.

        \item $\piencod{ M\ B }_u$: This process  models the application of $M$ to bag $B$ as a non-deterministic choice in the order in which the elements of $B$ are substituted into $M$. Substituting each $B_i$ involves a protocol in which the translation of a term  $\lambda x . M'[\widetilde{x} \leftarrow x]$ within $M$ confirms its own availability, before and after the exchange of the name $x$, on which the translation of $B_i$ is spawned. This protocol uses the fact that $M\ B$ does not reduce to failure, i.e., there is no lack or excess of resources in $B$.

    \item $\piencod{M [ x_1, \ldots, x_k \leftarrow x ] }_u$: The translation first confirms the availability of the behavior along $x$. 
    Then, it receives along $x$ a name for each $x_i$: these received names will be used to synchronize with the translation of bags (see below). Subsequently, the protocol on $x$ safely terminates and the translation of $M$ is executed.
        
    \item $ \piencod{ M[ \leftarrow x] }_u$: When there are no variables to be shared with $x$, the translation simply confirms the behavior on $x$, close the protocol immediately after, and executes the translation of $M$.

    \item $\piencod{\bag{M}\cdot B}_x$: The translation of a non-empty bag essentially makes each element available in its corresponding order. 
    This way, for the first element $M$ a name $x_1$ is sent over $x$; the translation of ${M[x_1,\cdots , x_n \leftarrow x]}$, discussed above, must send a confirmation on $x_1$ before the translation of $M$ is executed. After these exchanges, the translation of the rest of the bag is spawned.

    \item $\piencod{\oneb }_x$: In line with the previous item, the translation of the empty bag simply closes the name $x$; this signals that there are no (further) elements in the bag and that all synchronizations are complete.

    \item $\piencod{ M[\widetilde{x} \leftarrow x] \esubst{ B }{ x} }_u$: In this case, the translation is a sum involving the parallel composition of (i)~the translation of each element $B_i$ in the bag and (ii)~the translation of $M$. Observe that a fresh name $x$ is created to enable synchronization between these two processes. Also, as in previous cases, notice how the translation of $B_i$ must first confirm its availability along $x$.
    
    \item $\piencod{ M \linexsub{N / x}  }_u$: This translation essentially executes the translations of $M$ and $N$ in parallel, with a caveat: the translation of $N$ depends on the availability of a behavior along $x$, to be produced within the translation of $M$. 
    
    \item $\piencod{\expr{M}+\expr{N} }_u$: This translation  homomorphically preserves the non-determinism between $M$ and $N$. 
    
\end{itemize}

\begin{example}{}
\label{ch2ex:failfreered}
Consider the \lamrshar-term
$M_0 = ( \lambda x. M [x_1, x_2 \leftarrow x]) \bag{N_1,N_2}$. 
Writing $\lfv{B}$ to denote the free variables in $N_1$ and $N_2$, the process 
$\piencod{M_0}_u$  is as follows:
\begin{align*}
  = 
    & \piencod{ ( \lambda x. M [x_1, x_2 \leftarrow x] ) \bag{N_1,N_2} }_u 
    \\
     = & (\nu v)( \piencod{ \lambda x. M [x_1, x_2 \leftarrow x] }_v \mid 
      \underbrace{v.\some_{u , \lfv{B}} ; \outact{v}{x} . ( x.\some_{\lfv{B}}; \piencod{ \bag{N_1,N_2}  }_x \mid [v \leftrightarrow u] )}_{P_1} ) 
     \\
    & \oplus
    \\ 
    & (\nu v)( \piencod{ \lambda x. M [x_1, x_2 \leftarrow x] }_v \mid
    \underbrace{v.\some_{u , \lfv{B}} ; \outact{v}{x} . ( x.\some_{\lfv{B}}; \piencod{ \bag{N_2,N_1}  }_x \mid [v \leftrightarrow u] )}_{P_2} ) 
    \\
    =~ & (\nu v)( v.\overline{\some};v(x).x.\overline{\some};x(x_1). x(x_2).x.\close; \piencod{M}{v}  \mid P_1)
    \\
    & \oplus 
    \\
    & (\nu v)( v.\overline{\some};v(x).x.\overline{\some};x(x_1). x(x_2).x.\close; \piencod{M}{v}   \mid P_2)\\
\end{align*}
The translation immediately opens up a non-deterministic choice with two alternatives, corresponding to the bag of size 2.
Because of non-collapsing non-determinism, after some reductions, this amounts to accounting for the two different orders in which $N_1$ and $N_2$ can be extracted from the bag.

\[
\begin{aligned}
 \revd{B19}{
 \piencod{M_0}_u} \redd^*~ & (\nu  x)( x(x_1). x(x_2).x.\close; \revd{B19}{\piencod{M}_{u}}  \mid   \piencod{ \bag{N_1,N_2}  }_x  ) 
    \\
    & \oplus \\
    & (\nu  x)( x(x_1). x(x_2).x.\close; \piencod{M}_{u}   \mid   \piencod{ \bag{N_2,N_1}  }_x  )
    \end{aligned}
\]
We show further reductions for one of the processes, which we will denote $R$,  for $R=(\nu  x)( x(x_1). x(x_2).x.\close; \piencod{M}_{u}  \mid   \piencod{ \bag{N_1,N_2}  }_x  )$, in the resulting sum (reductions for the other process are similar):
\begin{align*}
 R= &  (\nu  x)( x(x_1). x(x_2).x.\close; \piencod{M}_{u}  \mid   \piencod{ \bag{N_1,N_2}  }_x  )
  \\
    =~ & 
   (\nu  x)( x(x_1). x(x_2).x.\close; \piencod{M}_{u}  \mid   \outact{x}{x_1}. (x_1.\some_{\lfv{N_1}};\piencod{N_1}_{x_1} \sep 
   \\
   & \qquad \qquad \qquad \qquad \outact{x}{x_2}. (x_2.\some_{\lfv{N_2}};\piencod{N_2}_{x_2} \sep x. \overline{\close} ) )     )\\
   \redd^*~ & (\nu  x_1,x_2)( \piencod{M}_{u}  \mid x_1.\some_{\lfv{N_1}};\piencod{N_1}_{x_1} \sep 
    x_2.\some_{\lfv{N_2}};\piencod{N_2}_{x_2}  )
\end{align*}
\end{example}

\subsubsection{The Translation}\label{ch2ss:second_trans}
\hfill 

The translation $\piencod{\cdot}_{x}$ leverages non-deterministic sessions in \spi to give a concurrent interpretation of \lamrshar, the non-deterministic (but fail-free) sub-calculus of \lamrsharfail. In a nutshell, non-deterministic sessions entail the explicit confirmation of the availability of  a name's behavior, via synchronizations of a prefix `$x.\some_{(w_1, \cdots, w_n)}$'   with a corresponding prefix `$x.\overline{\some}$'. Clearly, $\piencod{\cdot}_{x}$ under-utilizes the expressivity of \spi: in processes resulting from $\piencod{\cdot}_{x}$, no prefix `$x.\some_{(w_1, \cdots, w_n)}$' will ever synchronize with a prefix `$x.\overline{\none}$'. 
Indeed, because terms in \lamrshar never reduce to failure,  $\piencod{\cdot}_{x}$ should not account for such failures.

We may now introduce $\piencodf{\cdot }_u$, our translation of the fail-prone calculus $\lamrsharfail$ into $\spi$. It builds upon the structure of $\piencod{\cdot}_{x}$ to account for failures in expressions due to the lack or excess of resources. To this end, as we will see, $\piencodf{\cdot }_u$ does exploit prefixes `$x.\overline{\none}$' to signal failures.

\paragraph{Translating Expressions}
We introduce the translation  $\piencodf{\cdot }_u$, which will be shown to be a correct encoding, according to the criteria given in \secref{ch2ss:criteria}.

\begin{definition}{From $\lamrsharfail$ into $\spi$: Expressions}
\label{ch2def:enc_lamrsharpifail}
Let $u$ be a name.
The translation $\piencodf{\cdot }_u: \lamrsharfail \rightarrow \spi$ is defined in \figref{ch2fig:encfail}.
\end{definition}


We discuss the most interesting aspects of the translation in \figref{ch2fig:encfail}, in particular how the possibility of failure (lack or excess of resources in bags) induces differences with respect to the translation in \figref{ch2fig:enc}.

\begin{figure}[t!]

{
\begin{align*}	
   \piencodf{x}_u & = x.\overline{\some} ; [x \leftrightarrow u]  
  \\[1mm] 
   \piencodf{\lambda x.M[\widetilde{x} \leftarrow x]}_u & = u.\overline{\some}; u(x).\piencodf{M[\widetilde{x} \leftarrow x]}_u
\\[1mm]
  \piencodf{M\, B}_u & = \bigoplus_{B_i \in \perm{B}} (\nu v)(\piencodf{M}_v \mid v.\some_{u , \lfv{B}} ; \outact{v}{x} . ([v \leftrightarrow u] \mid \piencodf{B_i}_x ) )  
     \\[1mm]
      \piencodf{ M[\widetilde{x} \leftarrow x] \esubst{ B }{ x} }_u & =  \bigoplus_{B_i \in \perm{B}} (\nu x)( \piencodf{ M[\widetilde{x} \leftarrow x]}_u \mid \piencodf{ B_i}_x )  
      \\[1mm]
    \piencodf{M[x_1, x_2 \leftarrow x]}_u 
    & = 
       x.\overline{\some}. \outact{x}{y_1}. \Big(y_1 . \some_{\emptyset} ;y_{1}.\close 
       \mid x.\overline{\some};x.\some_{u, (\lfv{M} \setminus \{x_1 ,  x_2\} )}; \\
       & \hspace{2.0em} x(x_1) . x.\overline{\some}. \outact{x}{y_2} . \big(y_2 . \some_{\emptyset} ; y_{2}.\close  \mid x.\overline{\some};x.\some_{u,( \lfv{M} \setminus \{x_2\} ) };
      \\
       & \hspace{2.8em} x(x_2). x.\overline{\some}; \outact{x}{y_{}}. ( y_{} . \some_{u, \lfv{M} } ;y_{}.\close; \piencodf{M}_u \mid x.\overline{\none} )~ \big)  \Big) 
    \\[1mm]
    \piencodf{M[ \leftarrow x]}_u & = x. \overline{\some}. \outact{x}{y} . ( y . \some_{u,\lfv{M}} ;y_{}.\close; \piencodf{M}_u \mid x. \overline{\none}) \\[1mm]
    \piencodf{\bag{M} \cdot B}_x & =
    \begin{array}[t]{l}
       x.\some_{\lfv{\bag{M} \cdot B} } ; x(y_i). x.\some_{y_i, \lfv{\bag{M} \cdot B}};x.\overline{\some} ; \outact{x}{x_i}
       \\[1mm]
       \qquad . (x_i.\some_{\lfv{M}} ; \piencodf{M}_{x_i} \mid \piencodf{B}_x \mid y_i. \overline{\none})
   \end{array} 
   \\[3mm]    
      \piencodf{\oneb}_x & = x.\some_{\emptyset} ; x(y). (y.\overline{\some};y. \overline{\close} \mid x.\some_{\emptyset} ; x. \overline{\none}) 
      \\[1mm]
      \piencodf{\fail^{x_1 , \cdots , x_k}}_u & = u.\overline{\none} \mid x_1.\overline{\none} \mid \cdots \mid x_k.\overline{\none} 
   \\[1mm]
           \piencodf{ M \linexsub{N / x}  }_u   & =   (\nu x) ( \piencodf{ M }_u \mid   x.\some_{\lfv{N}};\piencodf{ N }_x  )  
        \\[1mm]
     \piencodf{\expr{M}+\expr{N} }_u    & =  \piencodf{ \expr{M} }_u \oplus \piencodf{ \expr{N} }_u   
\end{align*}
}
    \caption{Translating \lamrsharfail expressions into \spi processes.}
    \label{ch2fig:encfail}
\end{figure}

Most salient differences can be explained by looking at the translation of the application $M \, B$. Indeed, the sources of failure in $\lamrsharfail$ concern a mismatch between the number of variable occurrences in $M$ and the number of resources present in $B$. Both $M$ and $B$ can fail on their own, and our translation into \spi must capture this mutual dependency. Let us recall the translation given in  \figref{ch2fig:enc}:
$$
    \piencod{ M\ B }_u   =  \bigoplus_{B_i \in \perm{B}} (\nu v)( \piencod{ M}_v \mid v.\some_{u , \lfv{B}} ; \outact{v}{x} . ( x.\some_{\lfv{B_i}}; \piencod{ B_i}_x \mid [v \leftrightarrow u] ) ) 
$$
The corresponding translation in \figref{ch2fig:encfail} is seemingly simpler:
$$
  \piencodf{M\, B}_u  = \bigoplus_{B_i \in \perm{B}} (\nu v)(\piencodf{M}_v \mid v.\some_{u , \lfv{B}} ; \outact{v}{x} . ([v \leftrightarrow u] \mid \piencodf{B_i}_x ) )  
$$
Indeed, the main difference is the prefix `$x.\some_{\lfv{B_i}}$', which is present in process $\piencod{ M\ B }_u$ but is not explicit in process $\piencodf{M\, B}_u$. Intuitively, such a prefix denotes the dependency of $B$ on $M$; because terms in $\lamrshar$ do not fail, we can be certain that a corresponding confirming prefix `$x.\overline{\some}$' will be available to spawn every $\piencod{B_i}_x$. When moving to $\lamrsharfail$, however, this is not the case:  $\piencodf{M}_v$ may fail to provide the expected number of corresponding confirmations. 
For this reason, the role of prefix `$x.\some_{\lfv{B_i}}$' in $\piencod{ M\ B }_u$ is implemented within process $\piencodf{B_i}_x$. As a consequence, the translations for sharing terms ($M[\widetilde{x} \leftarrow x]$ and $ M[ \leftarrow x]$) and for bags ($\bag{M}\cdot B$ and $\oneb$) are more involved in the case of failure.

With this motivation for $\piencodf{M\, B}_u$ in mind, we discuss the remaining entries in \figref{ch2fig:encfail}:

\begin{itemize}
\item Translations for $x$ and $\lambda x.M[\widetilde{x} \leftarrow x]$ are exactly as in \figref{ch2fig:enc}:
$$
    \piencodf{ x }_u  =   x.\overline{\some} ;[x \leftrightarrow u ]
    \qquad \qquad 
    \piencodf{ \lambda x . M[\widetilde{x} \leftarrow x]}_u    =  u.\overline{\some};u(x).\piencodf{ M[\widetilde{x} \leftarrow x] }_u 
$$

    \item Similarly as $\piencodf{M \, B}_u$, discussed above, the translation of $M[\widetilde{x} \leftarrow x] \esubst{ B }{ x}$ is more compact than the one in \figref{ch2fig:enc}, because confirmations for each of the elements of the bag are handled within their respective translations:
    $$      \piencodf{ M[\widetilde{x} \leftarrow x] \esubst{ B }{ x} }_u  =  \bigoplus_{B_i \in \perm{B}} (\nu x)( \piencodf{ M[\widetilde{x} \leftarrow x]}_u \mid \piencodf{ B_i}_x )  
$$
    
    
       \item As anticipated, the translation of $M[x_1, \ldots, x_k \leftarrow x]$ is more involved than before. For simplicity, let us discuss the representative case when $k = 2$ (two shared variables):
       \begin{align*}
       	          \piencodf{M[x_1, x_2 \leftarrow x]}_u & = 
            x.\overline{\some}. \outact{x}{y_1}. \Big(y_1 . \some_{\emptyset} ;y_{1}.\close 
       \mid x.\overline{\some};
      \\
      & \hspace{1.5em} x.\some_{u, (\lfv{M} \setminus \{x_1 ,  x_2\} )}; x(x_1) .  x.\overline{\some}. \outact{x}{y_2} . 
      \\
      & \hspace{2.em} \big(y_2 . \some_{\emptyset} ; y_{2}.\close  \mid x.\overline{\some};x.\some_{u,( \lfv{M} \setminus \{x_2\} ) };\\
      & \hspace{2.5em}  x(x_2).x.\overline{\some}; \outact{x}{y_{}}. ( y_{} . \some_{u, \lfv{M} } ;\\
      & \hspace{3.0em} y_{}.\close; \piencodf{M}_u \mid x.\overline{\none} )~ \big)  \Big)
       \end{align*}
       This process is meant to synchronize with the translation of a bag. After confirming the presence of a behavior on name $x$, an auxiliary name $y_i$ is sent to signal that there are elements to be substituted. This name implements a short protocol that allows us to check for lack of resources in the bag. These steps on $y_i$ are followed by another confirmation and also a request for confirmation of behavior along $x$; this represents that the name can fail in one of two ways, capturing the mutual dependency between $M$ and the bag mentioned above. Once these two steps on $x$ have succeeded, it is finally safe for the process to receive a name $x_i$. This process is repeated for each shared variable to ensure safe communication of the elements of the bag. The last line shows the very final step: a name $y$ is communicated to ensure that there are no further elements in the bag; in such a case, $y$ fails and the failure is propagated to  $\piencodf{M}_u$. The prefix `$x.\overline{\none}$' signals the end of the shared variables, and is meant to synchronize with the translation of $\oneb$, the last element of the bag. If the bag has elements that still need to be synchronized then the failure along $x$ is propagated to the remaining resources within the translation of the  bag.
    
    \item The translation of $M[ \leftarrow x]$  corresponds to the final step in the translation  just discussed:
    $$        \piencodf{M[ \leftarrow x]}_u  = x. \overline{\some}. \outact{x}{y} . ( y . \some_{u,\lfv{M}} ;y_{}.\close; \piencodf{M}_u \mid x. \overline{\none})
$$


    \item The translation of the non-empty bag $\bag{M} \cdot B$ is as follows: 
    \begin{align*}
    	         \piencodf{\bag{M} \cdot B}_x & =
       x.\some_{\lfv{\bag{M} \cdot B} } ; x(y_i). x.\some_{y_i, \lfv{\bag{M} \cdot B}};x.\overline{\some} ; \outact{x}{x_i}
       \\[1mm]
       & \qquad . (x_i.\some_{\lfv{M}} ; \piencodf{M}_{x_i} \mid \piencodf{B}_x \mid y_i. \overline{\none})
    \end{align*}
    Notice how this process operates hand in hand with the translation of $M[x_1, \ldots, x_k \\ \leftarrow x]$.
    The process first waits for its behavior to be confirmed; then, the auxiliary name $y_i$ is received from the translation of $M[x_1, \ldots, x_k \leftarrow x]$. The name $y_i$ fails immediately to signal that there are more resources in the bag. Name $x$ then confirms its behavior and awaits its behavior to be confirmed. Subsequently, a name $x_i$ is sent: this is the name on which the translation of $M$ will be made available to the application. After that, name $x$ is used in the translation of $B$, the rest of the bag.
    
    \item The translation of $\oneb$ operates aligned with the translations just discussed, exploiting the fact that in fail-free reductions the last element of the bag must be $\oneb$:
    $$
          \piencodf{\oneb}_x  = x.\some_{\emptyset} ; x(y). (y.\overline{\some};y. \overline{\close} \mid x.\some_{\emptyset} ; x. \overline{\none}) 
    $$
    This process relays the information that the translated empty bag is no longer able to provide resources for further substitutions. It first waits upon a correct behavior followed by the reception of a name $y$. The process then confirms its behavior along $y$: this signals that there are no further resources. Concurrently, name $x$ waits for a confirmation of a behavior and ends with `$x. \overline{\none}$', thus signaling the failure of producing further behaviors.
    
    \item The explicit failure term $\fail^{x_1 , \cdots , x_k}$ is not part of $\lamrshar$ and so it was not covered in  \figref{ch2fig:enc}. Its translation is straightforward:
    $$      \piencodf{\fail^{x_1 , \cdots , x_k}}_u  = u.\overline{\none} \mid x_1.\overline{\none} \mid \cdots \mid x_k.\overline{\none} 
$$
The failure term is translated as the non-availability of a behavior along name $u$, composed with the non-availability of sessions along the names/variables $x_1, \ldots, x_n$ encapsulated by the source failure term.


\item The translations for $M \linexsub{N / x} $ and $\expr{M}+\expr{N}$ are exactly as before:
$$
           \piencodf{ M \linexsub{N / x}  }_u    =   (\nu x) ( \piencodf{ M }_u \mid   x.\some_{\lfv{N}};\piencodf{ N }_x  )  
        \qquad 
     \piencodf{\expr{M}+\expr{N} }_u     =  \piencodf{ \expr{M} }_u \oplus \piencodf{ \expr{N} }_u   
$$
\end{itemize}

\newcommand{\cnum}[1]{[\mathsf{#1}]}

\subsubsection{Examples} Before presenting the session types associated to our translation $ \piencodf{\cdot}_u$, we present a series of examples that illustrate different possibilities in a step-by-step fashion:
\begin{itemize}
    \item No failure: an explicit substitution that is provided an adequate amount of resources;
    \item Failure due to excess of resources in the bag;
    \item Failure due to lack of resources in the bag.
\end{itemize}
We first discuss the translation of a term in which there is no failure. In that follows, we refer to a specific reduction by adding a number as in, e.g., `$\redd_{\cnum{3}}$'.

\begin{example}{No Failure}\label{ch2ex:encsucc}
Let us consider the well-formed $\lamrsharfail$-term $N [x_1 \leftarrow x] \esubst{ \bag{M} }{x}$, where, for simplicity, we assume that $\revo{A12}{\lfv{N} \setminus \{ x_1 \} = \lfv{M} = \emptyset}$. 
As we have seen, $N [x_1 \leftarrow x] \esubst{ \bag{M} }{x} \redd N \linexsub{M / x}$. 
We discuss reduction steps for $ \piencodf{ N[ x_1 \leftarrow x] \esubst{ \bag{M} }{ x} }_u $, highlighting in \bluetext{blue} relevant prefixes.
First, we have:
  \[
  \small
  \begin{aligned} 
   \piencodf{ N[ x_1 \leftarrow x] \esubst{ \bag{M} }{ x} }_u 
               =~ &  (\nu x)( \piencodf{ N[x_1 \leftarrow x]}_u \mid \piencodf{\bag{M} }_x ) \\
       =~ &  (\nu x)\big( \bluetext{x.\overline{\some}}. \outact{x}{y_1}. (y_1 . \some_{\emptyset} ;y_{1}.\close 
       \mid x.\overline{\some};x.\some_{u}; 
       \\
       & \hspace{2.8em} . x(x_1).x.\overline{\some}; \outact{x}{y_{}}. ( y_{} . \some_{u, x_1 } ;y_{}.\close; \piencodf{N}_u \mid x.\overline{\none} )~ ) \\
       & \hspace{2em} \mid \bluetext{x.\some_{\emptyset}} ; x(y_1). x.\some_{y_1};x.\overline{\some} ; \outact{x}{x_1}\\
       & \hspace{2.8em} . (x_1.\some_{\emptyset} ; \piencodf{M}_{x_1} \mid y_1. \overline{\none} \mid x.\some_{\emptyset} ; x(y).    (y.\overline{\some};y. \overline{\close} \\
       &\hspace{2.8em} \mid x.\some_{\emptyset} ; x. \overline{\none}) ) \big) 
       \end{aligned}
       \]
A detailed description of the reduction steps follows:
\begin{itemize}
    \item Reduction $\redd_{\cnum{1}}$ concerns the name $x$ confirming its behavior (see highlighted prefixes above), and reduction $\redd_{\cnum{2}}$ concerns the communication of name $y_1$:
    \[
    \small
    \begin{aligned}
     \piencodf{ N[ x_1 \leftarrow x] \esubst{ \bag{M} }{ x} }_u\redd_{\cnum{1}}~ &  (\nu x)( \bluetext{\outact{x}{y_1}}. \big(y_1 . \some_{\emptyset} ;y_{1}.\close 
       \mid x.\overline{\some};x.\some_{u};x(x_1) . 
       \\
       & \hspace{1em} . x.\overline{\some}; \outact{x}{y_{}}. ( y_{} . \some_{u,x_1} ;y_{}.\close; \piencodf{N}_u \mid x.\overline{\none} )~ \big) \\
       & \hspace{1em} \mid \bluetext{ x(y_1)}. x.\some_{y_1 };x.\overline{\some} ; \outact{x}{x_1}. (x_1.\some_{\emptyset} ; \piencodf{M}_{x_1}\\
       & \hspace{1em}  \mid y_1. \overline{\none} \mid x.\some_{\emptyset} ; x(y).  (y.\overline{\some};y. \overline{\close}\mid \\
       & \hspace{1em} x.\some_{\emptyset} ; x. \overline{\none}) ) ) 
       \\
       \redd_{\cnum{2}}~ &  (\nu x, y_1)( y_1 . \some_{\emptyset} ;y_{1}.\close 
       \mid \bluetext{x.\overline{\some}};x.\some_{u };x(x_1) . 
       \\
       & \hspace{1em} . x.\overline{\some}; \outact{x}{y_{}}. ( y_{} . \some_{u, x_1 } ;y_{}.\close; \piencodf{N}_u \mid x.\overline{\none} )~ \\
       & \hspace{1em} \mid \bluetext{ x.\some_{y_1} };x.\overline{\some} ; \outact{x}{x_1} . (x_1.\some_{\emptyset} ; \piencodf{M}_{x_1} \\
       & \hspace{1em}\mid y_1. \overline{\none} \mid x.\some_{\emptyset} ; x(y).    (y.\overline{\some};y. \overline{\close} \mid \\
       & \hspace{1em} x.\some_{\emptyset} ; x. \overline{\none}) ) ) \\
       := & P
    \end{aligned}
    \]
    \item Reduction $\redd_{\cnum{3}}$ concerns $x$ confirming its behavior, which signals that there are variables free for substitution in the translated term. In the opposite direction, 
    reduction $\redd_{\cnum{4}}$ signals that there are elements in the bag which are available for substitution in the translated term.
    \[
    \small
    \begin{aligned}
     P \redd_{\cnum{3}}~ &  (\nu x, y_1)( y_1 . \some_{\emptyset} ;y_{1}.\close 
       \bluetext{\mid x.\some_{u}};x(x_1) . 
       \\
       & \hspace{1em} . x.\overline{\some}; \outact{x}{y_{}}. ( y_{} . \some_{u, x_1 } ;y_{}.\close; \piencodf{N}_u \mid x.\overline{\none} )~ \\
       & \hspace{1em} \mid \bluetext{x.\overline{\some}} ; \outact{x}{x_1} . (x_1.\some_{\emptyset} ; \piencodf{M}_{x_1} \mid y_1. \overline{\none} \mid x.\some_{\emptyset} ; x(y). \\
       & \hspace{1em} (y.\overline{\some};y. \overline{\close} \mid x.\some_{\emptyset} ; x. \overline{\none}) ) ) 
       \\
     \redd_{\cnum{4}}~ &  (\nu x, y_1)( y_1 . \some_{\emptyset} ;y_{1}.\close 
       \para  \bluetext{x(x_1)} .  x.\overline{\some}; \outact{x}{y_{}}. ( y_{} . \some_{u, x_1 } ;y_{}.\close; \\
       & \hspace{1em} \piencodf{N}_u \mid x.\overline{\none} ) \mid  \bluetext{\outact{x}{x_1}}. (x_1.\some_{\emptyset} ; \piencodf{M}_{x_1}\mid y_1. \overline{\none} \\
       & \hspace{2em}   \mid x.\some_{\emptyset} ; x(y).   (y.\overline{\some};y. \overline{\close} \mid x.\some_{\emptyset} ; x. \overline{\none}) ) ) \hspace{1cm}  (:= Q)
    \end{aligned}
    \]
    \item Given the confirmations in the previous two steps, reduction $\redd_{\cnum{5}}$ can now safely communicate a name $x_1$. This reduction synchronizes the shared variable $x_1$ with the first element in the bag.
    \[
    \small
    \begin{aligned}
    Q   \redd_{\cnum{5}}~ &  (\nu x, y_1, x_1)( y_1 . \some_{\emptyset} ;y_{1}.\close   \mid  \bluetext{ x.\overline{\some}}; \outact{x}{y_{}}. ( y_{} . \some_{u, x_1 } ; y_{}.\close;  \piencodf{N}_u \\
       & \hspace{2em}\mid x.\overline{\none} ) \mid  x_1.\some_{\emptyset} ; \piencodf{M}_{x_1} \mid y_1. \overline{\none} \mid \bluetext{x.\some_{\emptyset}} ; x(y).   (y.\overline{\some};y. \overline{\close} \\
       & \hspace{2.5em} \mid x.\some_{\emptyset} ; x. \overline{\none}) ) \hspace{1cm}  (:= R)
    \end{aligned}
    \]
    \item Reduction $\redd_{\cnum{6}}$ concerns  $x$ confirming its behavior. At this point, we could have alternatively performed a reduction on name $y_1$. We chose to discuss all reductions on  $x$ first; thanks to confluence this choice has no effect on the overall behavior.  Reduction $\redd_{\cnum{7}}$ communicates name $y$ along $x$. 
    \[\small
    \begin{aligned}
        R  \redd_{\cnum{6}}~ &  (\nu x, y_1, x_1)( y_1 . \some_{\emptyset} ;y_{1}.\close 
       \mid \bluetext{ \outact{x}{y_{}}}. ( y_{} . \some_{u, x_1 } ;y_{}.\close; \piencodf{N}_u ~ \\
       & \hspace{2em} \mid x.\overline{\none} )\mid  x_1.\some_{\emptyset} ;\piencodf{M}_{x_1} \mid y_1. \overline{\none} \mid \bluetext{x(y)}. (y.\overline{\some};y. \overline{\close}  
       \\
       & \hspace{2em}\mid x.\some_{\emptyset} ; x. \overline{\none}) )\\
       \redd_{\cnum{7}}~ &  (\nu x, y, y_1, x_1)( y_1 . \some_{\emptyset} ;y_{1}.\close 
       \mid  y_{} . \some_{u, x_1 } ;y_{}.\close; \piencodf{N}_u \mid \bluetext{x.\overline{\none}} ~ \\
       & \hspace{2em} \mid  x_1.\some_{\emptyset} ; \piencodf{M}_{x_1} \mid y_1. \overline{\none} \mid  y.\overline{\some};y. \overline{\close} \mid \bluetext{x.\some_{\emptyset}} ; x. \overline{\none} ) \hspace{.8cm} 
       \\
       := & S
    \end{aligned}
    \]
    \item Reduction $\redd_{\cnum{8}}$ cancels the behavior along $x$, meaning that there are no more free variables to synchronize with. Subsequently, reduction $\redd_{\cnum{9}}$ cancels the behavior along $y_1$: \revd{B22}{at the beginning, when $y_1$ was received, the encoded bag had the element $M$ left to be synchronized; at this point, the failure on $y_1$ signals that the bag still has elements to be synchronized with}. 
    \[
    \small
    \begin{aligned}
    S %
       \redd_{\cnum{8}}~ &   (\nu  y, y_1, x_1)(\bluetext{ y_1 . \some_{\emptyset}} ;y_{1}.\close 
       \mid  y_{} . \some_{u, x_1 } ;y_{}.\close; \piencodf{N}_u \mid  x_1.\some_{\emptyset} ; \piencodf{M}_{x_1}  \\
       & \hspace{2em}  \mid \bluetext{y_1. \overline{\none}} \mid  y.\overline{\some};y. \overline{\close} ) 
       \\
       \redd_{\cnum{9}}~ & (\nu  y,x_1)(  \bluetext{y_{} . \some_{u, x_1 }} ;y_{}.\close; \piencodf{N}_u  \mid  x_1.\some_{\emptyset} ; \piencodf{M}_{x_1} x\mid  \bluetext{y.\overline{\some}};y. \overline{\close} )  
       \\
       :=&T
    \end{aligned}
    \]
    \item Finally, reductions $\redd_{\cnum{10}}$ and $\redd_{\cnum{11}}$ concern name $y$: the former signals that the bag has no more elements to be synchronized for substitution; the latter closes the session, as it has served its purpose of correctly synchronizing the translated term. The resulting process corresponds to the translation of $N \linexsub{M / x}$.
    \[
    \begin{aligned}
    T  \redd_{\cnum{10}}~ &  (\nu  y, x_1)(  \bluetext{y_{}.\close}; \piencodf{N}_u  \mid  x_1.\some_{\emptyset} ; \piencodf{M}_{x_1} \mid  \bluetext{y. \overline{\close}}  ) 
       \\
       \redd_{\cnum{11}}~ &  (\nu   x_1)( \piencodf{N}_u \mid  x_1.\some_{\emptyset} ; \piencodf{M}_{x_1}  ) 
       =\piencodf{ N \linexsub{M / x}  }_u
    \end{aligned}
    \]
\end{itemize}
\end{example}

We now discuss the translation of a term that fails due to an excess of resources.

\begin{example}{Excess of Resources}\label{ch2ex:encfailexcess}
Let us consider the well-formed $\lamrsharfail$-term that does not share occurrences of $x$, i.e., $N [ \leftarrow x] \esubst{ \bag{M} }{x}$, where $M, N$ are closed (i.e. $\lfv{N} = \lfv{M} = \emptyset$). This term's translation is:
	  \[\small
    \begin{aligned}
       \piencodf{ N[ \leftarrow x] \esubst{ \bag{M} }{ x} }_u & =   (\nu x)( \piencodf{ N[ \leftarrow x]}_u \mid \piencodf{\bag{M} }_x ) \\
       & = (\nu x)( x. \overline{\some}. \outact{x}{y_1} . ( y_1 . \some_{u } ;y_1.\close; \piencodf{N}_u \mid x. \overline{\none}) \mid x.\some_{\emptyset} ;
       \\
       & \qquad  x(y_1). x.\some_{y_1};x.\overline{\some} ; \outact{x}{x_i}
      . (x_i.\some_{\emptyset} ; \piencodf{M}_{x_i} \mid \piencodf{\oneb}_x \mid y_1. \overline{\none}) )
      \\
    \end{aligned}
    \]

      \begin{itemize}
        \item Reductions $\redd_{\cnum{1}}$ and $\redd_{\cnum{2}}$ follow as in  Example \ref{ch2ex:encsucc}.
    	  \[\small
    \begin{aligned}
      \piencodf{ N[ \leftarrow x] \esubst{ \bag{M} }{ x} }_u & \redd_{\cnum{1}} (\nu x)(  \outact{x}{y_1} . ( y_1 . \some_{u } ;y_1.\close; \piencodf{N}_u \mid x. \overline{\none}) \mid  x(y_1). 
       \\
       & \qquad \quad x.\some_{y_1};x.\overline{\some} ; \outact{x}{x_i}
           . (x_i.\some_{\emptyset} ; \piencodf{M}_{x_i} \mid \piencodf{\oneb}_x \mid y_1. \overline{\none}) )\\
       & \redd_{\cnum{2}} (\nu x, y_1)(   y_1 . \some_{u } ;y_1.\close; \piencodf{N}_u \mid {\cred{  x. \overline{\none}}} \mid 
       \\
       & \quad \quad  { \cred{x.\some_{y_1};}}x.\overline{\some} ; \outact{x}{x_i} . (x_i.\some_{\emptyset} ; \piencodf{M}_{x_i} \mid \piencodf{\oneb}_x \mid y_1. \overline{\none}) )\qquad 
       \\
       :=& P
    \end{aligned}
    \]
  Notice how the translation of the term first triggers the failure: prefix $x.\overline{\none}$ (highlighted in \cred{red}) signals that there are no (more) occurrences of $x$ within the process; nevertheless, the translation of the bag is still trying to communicate the translation of $M$. This failure along $x$ causes the chain reaction of the failure along $y_1$, which eventually triggers across the translation of $N$.
  
        \item Reduction $\redd_{\cnum{3}}$ differs from $\redd_{\cnum{3}}$ in Example~\ref{ch2ex:encsucc}, as the translation of the shared variable is empty, we abort along the name $x$;  as the translated bag still contains elements to synchronize, the abortion of the bag triggers that failure of the dependant name $y_1$.
        \[
        \begin{aligned}
            P& \redd_{\cnum{3}} (\nu  y_1)(   {\cred{  y_1 . \some_{u } ;}}y_1.\close; \piencodf{N}_u \mid { \redd{ y_1. \overline{\none}}}) 
       \\
       & \redd_{\cnum{4}}   u. \overline{\none}  = \piencodf{ \fail^{\emptyset} }_u
        \end{aligned}
        \]
        \item Reduction $\redd_{\cnum{4}}$ differs from that of $\redd_{\cnum{9}}$ and $\redd_{\cnum{10}}$ from Example \ref{ch2ex:encsucc}: the name $y_1$ fails signaling that there was an element in the bag that was to be sent; as the translation of the term $N$ is guarded by the confirmation along $y_1$, it aborts.
    \end{itemize}
    
  \end{example}

Finally, we illustrate how $ \piencodf{\cdot}_u$ acts on a term that fails due to lack of resources in a bag.

\begin{example}{Lack of Resources}\label{ch2ex:encfaillack}
Consider the well-formed $\lamrsharfail$-term $N [ x_1 \leftarrow x] \esubst{ \oneb }{x}$, where $N$ is a closed  term (i.e. $\lfv{N} =  \emptyset$). This term's translation is:
  \[\small
    \begin{aligned}
        \piencodf{ N [ x_1 \leftarrow x] \esubst{ \oneb }{x} }_u  
       =~ &   (\nu x)( \piencodf{ N [ x_1 \leftarrow x] }_u \mid \piencodf{ \oneb }_x ) \\
        =~ & (\nu x)( x.\overline{\some}. \outact{x}{y_1}. (y_1 . \some_{\emptyset} ;y_{1}.\close \para x.\overline{\some};x.\some_{u};
       \\
       & \quad  x(x_1) .  x.\overline{\some}; \outact{x}{y_{2}}. ( y_{2} . \some_{u, x_1 } ;y_{2}.\close; \piencodf{N}_u \mid x.\overline{\none} )~ )  \para \\
       & \quad  x.\some_{\emptyset} ; x(y_1). ( y_1.\overline{\some};y_1 . \overline{\close} \mid x.\some_{\emptyset} ; x. \overline{\none}) )\qquad (:= P) \\
       \end{aligned}
      \]
        Notice how the translation of the empty bag $\oneb$ triggers the failure: prefix `$x.\overline{\none}$' signals that there are no (more) elements in the bag; however, the translated term aims to synchronize, as it (still) requires resources.
    
    \begin{itemize}
        \item Reductions $\redd_{\cnum{1}}$ and $\redd_{\cnum{2}}$ follow from Example \ref{ch2ex:encsucc}.
      \[\small
    \begin{aligned}
       P \redd_{\cnum{1}} ~ & (\nu x)( \outact{x}{y_1}. (y_1 . \some_{\emptyset} ;y_{1}.\close \mid  x.\overline{\some};x.\some_{u};x(x_1) .
       \\
       &  \quad   x.\overline{\some}; \outact{x}{y_{2}}. ( y_{2} . \some_{u, x_1 } ;y_{2}.\close; \piencodf{N}_u \mid x.\overline{\none} )~ )  \para \\
       &  \quad  x(y_1). ( y_1.\overline{\some};y_1 . \overline{\close} \mid x.\some_{\emptyset} ; x. \overline{\none}) ) \\
       \redd_{\cnum{2}} ~ & (\nu x, y_1)(  y_1 . \some_{\emptyset} ;y_{1}.\close \mid x.\overline{\some};x.\some_{u};x(x_1) . 
       \\
       &  \quad   x.\overline{\some}; \outact{x}{y_{2}}. ( y_{2} . \some_{u, x_1 } ;y_{2}.\close; \piencodf{N}_u \mid x.\overline{\none} )  \para \\
       &  \quad  y_1.\overline{\some};y_1 . \overline{\close} \mid x.\some_{\emptyset} ; x. \overline{\none} ) 
    \end{aligned}
    \]
  
        \item Reductions $\redd_{\cnum{3}}$ and $\redd_{\cnum{4}}$ follow from that of $\redd_{\cnum{9}}$ and $\redd_{\cnum{10}}$ in Example \ref{ch2ex:encsucc}; as the term contains the element $x_1$ for synchronization, the encoding of $N$ is not guarded by $y_1$.
        \[\small
        \begin{aligned}
        Q    \redd_{\cnum{3}} ~ & (\nu x, y_1)(  y_{1}.\close \mid x.\overline{\some};x.\some_{u};x(x_1) . 
       \\
       &  \quad  x.\overline{\some}; \outact{x}{y_{2}}. ( y_{2} . \some_{u, x_1 } ;y_{2}.\close; \piencodf{N}_u \mid x.\overline{\none} )  \para \\
       &  \quad  y_1 . \overline{\close} \mid x.\some_{\emptyset} ; x. \overline{\none} ) \\
       \redd_{\cnum{4}} ~ & (\nu x )(  x.\overline{\some};x.\some_{u};x(x_1) .  x.\overline{\some}; \outact{x}{y_{2}}. ( y_{2} . \some_{u, x_1 } ;y_{2}.\close;  \\
       &  \quad  \piencodf{N}_u \mid x.\overline{\none} )  \mid x.\some_{\emptyset} ; x. \overline{\none} ) \\
       \redd_{\cnum{5}} ~ & (\nu x )(  { \cred{ x.\some_{u}}};x(x_1) .  x.\overline{\some}; \outact{x}{y_{2}}. ( y_{2} . \some_{u, x_1 } ;y_{2}.\close; \piencodf{N}_u \mid x.\overline{\none} )  \mid \\
       & \quad { \cred{  x. \overline{\none} }} ) \\
       \redd_{\cnum{6}}~ &  u. \overline{\none}
       =~   \piencodf{ \fail^{\emptyset}  }_u
        \end{aligned}
        \]
        \item Reduction $\redd_{\cnum{5}}$ follows from reduction $ \redd_{\cnum{3}} $ in Example \ref{ch2ex:encsucc}.
        \item Reduction $\redd_{\cnum{6}}$ differs from that of $\redd_{\cnum{4}}$ from Example \ref{ch2ex:encsucc}: the bag contains no elements, and signals this by aborting along the name $x$; still, the term expects to receive an element of the bag, and prematurely aborts.
    \end{itemize}
    
  \end{example}

\paragraph{Translating Types}
\srev{In describing our translation $\piencodf{\cdot}_{-}$ we have informally referred to (non-deterministic) session protocols in \spi that implement (non-deterministic) expressions in \lamrsharfail. We are actually able to make these intuitions precise and give a translation of intersection types (for \lamrsharfail, cf. \defref{ch2d:typeslamrfail}) into session types (for \spi, cf. \defref{ch2d:sts}). This provides the protocol-oriented interpretation of intersections mentioned earlier. Intuitively speaking, given an intersection type $\pi$, we will have a  corresponding session type $\piencodf{\pi}$ that determines a protocol tied to the evaluation of a (fail-prone, non-deterministic) expression with type $\pi$.}

\begin{definition}{From $\lamrsharfail$ into $\spi$: Types}
\label{ch2def:enc_sestypfail}
The translation  $\piencodf{\cdot}$ on types is defined 
in \figref{ch2fig:enc_sestypfail}.
 \revo{}{Let  
$\Gamma = x_1: \sigma_1, \cdots, x_m : \sigma_k, v_1: \pi_1 , \cdots , v_n: \pi_n$
be as in \defref{ch2d:tcont}.}
            
For some strict types $\tau_1,\cdots,\tau_n$ and $i_1,\cdots,i_n \geq 0$ we define: 
$$\piencodf{\Gamma} = x_1: \with \overline{\piencodf{\sigma_1}} , \cdots ,  x_k : \with \overline{\piencodf{\sigma_k}} ,
 v_1: \with \overline{\piencodf{\pi_1}_{(\tau_1, i_1)}}, \cdots , v_n: \with \overline{\piencodf{\pi_n}_{(\tau_n, i_n)}}$$
\end{definition}

\begin{figure}[!t]
    \centering

\begin{align*}
 \piencodf{\unit} & = \with \onef 
 \\[1mm]
 \piencodf{\pi \rightarrow \tau}  & = \with((  \overline{\piencodf{\pi }_{(\sigma, i)}} ) \ampy \piencodf{\tau})  \quad \text{(for some strict type $\sigma$, with $i \geq 0$)}  
 \\[1mm]
   \piencodf{ \sigma \wedge \pi }_{(\tau, i)} &= \overline{   \with(( \oplus \bot) \otimes ( \with  \oplus (( \with  \overline{\piencodf{ \sigma }} )  \ampy (\overline{\piencodf{\pi}_{(\tau, i)}})))) } \\
     & = \oplus(( \with \onef) \ampy ( \oplus  \with (( \oplus \piencodf{\sigma} ) \otimes (\piencodf{\pi}_{(\tau, i)})))) 
     \\[1mm]
      \piencodf{\omega}_{(\sigma, i)} &=  
 \begin{cases}
     \overline{\with(( \oplus \bot )\otimes ( \with \oplus \bot )))} & \text{if $i = 0$}
     \\
\overline{   \with(( \oplus \bot) \otimes ( \with  \oplus (( \with  \overline{\piencodf{ \sigma }} )  \ampy (\overline{\piencodf{\omega}_{(\sigma, i - 1)}})))) } & \text{if $i > 0$}
\end{cases}
\end{align*}

    \caption{Translating intersection types as session types.}
    \label{ch2fig:enc_sestypfail}
\end{figure}

As we will see, given a well-formedness judgement $\Gamma \wfdash \expr{M} : \tau$, with the translations on types and assignments defined above, we will have $\piencodf{\expr{M}}_u \vdash 
\piencodf{\Gamma}, 
u : \piencodf{\tau}$; this is the content of the \emph{type preservation} property (\thmref{ch2t:preservationtwo}).

The translation of types in \figref{ch2fig:enc_sestypfail} leverages non-deterministic session protocols (typed with `$\with$') to represent non-deterministic fetching and fail-prone evaluation in \lamrsharfail. 
Notice that the translation of the multiset type $\pi$ depends on two arguments (a strict type $\tau$ and a number $i \geq 0$) which are left unspecified above, \secondrev{but are appropriately specified in Proposition~\ref{ch2prop:app_aux}}.
This is crucial to represent mismatches in \lamrsharfail (i.e., sources of failures) as typable processes in \spi. 
For instance, in \figref{ch2fig:wfsh_rules}, Rule~\redlab{FS{:}app} admits a mismatch between 
$\sigma^{j} \rightarrow \tau$ and $ \sigma^{k}$, for it allows $j \neq k$.
In our proof of type preservation, these two arguments are instantiated appropriately, enabling typability as session-typed processes.

 We are now ready to consider correctness  for $\piencodf{\cdot}_u$, in the sense of  \defref{ch2d:encoding}.
First, the compositionality property follows directly from   \figref{ch2fig:encfail}. 
In the following sections, we state the remaining properties in \defref{ch2d:encoding}: type preservation, operational correspondence, and success sensitiveness.

 \subsubsection{Type Preservation}
 We prove that our translation from $\lamrsharfail$ to $\spi$  maps well-formed $\lamrsharfail$ expressions to session-typed processes in $\spi$. \secondrev{First, we show that translated multiset types can be ``lengthened'' by setting appropriate parameters to the encoding. } 

        
        
    

\begin{restatable}{proposition}{appaux}
\label{ch2prop:app_aux}
 \revo{A27}{Suppose $\sigma^j$ and $\sigma^k$ are arbitrary strict types (\defref{ch2d:typeslamrfail}), for some $j, k \geq 0$. 
Following \figref{ch2fig:enc_sestypfail}, consider their encoding into session types $\piencodf{\sigma^{j}}_{(\tau_1, m)} $ and $\piencodf{\sigma^{k}}_{(\tau_2, n)}$, respectively, where  $\tau_1, \tau_2$ are strict types and $n, m \geq 0$. \\  We have 
 $\piencodf{\sigma^{j}}_{(\tau_1, m)} = \piencodf{\sigma^{k}}_{(\tau_2, n)}$
  under the following conditions:}
         \begin{enumerate}
         \item If $j > k$ then we take $\tau_1 $ to be an arbitrary strict type and $m = 0$; also, we take $\tau_2 $ to be $\sigma$ and $n = j-k$.
        
         \item If $j < k$ then we take $\tau_1 $ to be $\sigma$ and $m = k-j$; also,  we take $\tau_2 $ to be an arbitrary strict type and $n = 0$. 
        
         \item Otherwise, if $j = k$ then we take $m = n = 0$. Also, $\tau_1 , \tau_2 $ are arbitrary strict types.
     \end{enumerate}
\end{restatable}

\begin{proof}
\secondrev{
Immediate by unfolding the translation. 
\iffulldoc
The full analysis can be found in \appref{ch2app:typeprestwo}.
\else
See the full version for details.
\fi 
}
\end{proof}

\secondrev{
Given Proposition \ref{ch2prop:app_aux} we now show that the translation preserves types:
}

\begin{restatable}[Type Preservation for $\piencodf{\cdot}_u$]{theorem}{preservationtwo}
\label{ch2t:preservationtwo}
Let $B$ and $\expr{M}$ be a bag and an expression in $\lamrsharfail$, respectively.
\begin{enumerate}
\item If $\secondrev{\core{\Gamma}} \wfdash B : \pi$
then 
$\piencodf{B}_u \vdash  \piencodf{\secondrev{\core{\Gamma}}}, u : \piencodf{\pi}_{(\sigma, i)}$, \revo{A13}{for some strict type $\sigma$ and index $ i \geq 0 $.}

\item If $\secondrev{\core{\Gamma}} \wfdash \expr{M} : \tau$
then 
$\piencodf{\expr{M}}_u \vdash  \piencodf{\secondrev{\core{\Gamma}}}, u :\piencodf{\tau}$.
\end{enumerate}

\end{restatable}

\begin{proof}
 By mutual induction on the typing derivation of $B$ and $\mathbb{M}$ , with an analysis of the last rule applied in $\Gamma\wfdash B:\pi$ and in $\Gamma\wfdash \mathbb{M}:\tau$. \revd{B26}{One key aspect of this proof is the application of Proposition~\ref{ch2prop:app_aux} to ensure duality of types. Intuitively, the conditions given by Proposition~\ref{ch2prop:app_aux} are used to instantiate the parameters in the encoding of intersection types, so as to ensure that when intersection types have different types the smaller type can be correctly ``padded'' to match the size of the larger type---Example~\ref{ch2ex:padd}, given below, illustrates this padding. } 
 \iffulldoc
 The full proof can be found in \appref{ch2app:typeprestwo}.
 \else
 See the full version for details.
 \fi 
\end{proof}

\begin{example}{\revo{A2}{Parameters in the encoding of types}}
\label{ch2ex:padd}
We give the dual types when encoding intersection types, namely the case of $\piencodf{ \sigma \wedge \pi }_{(\sigma, i)}$, to express the encoding of intersection typed behavior into session typed behavior. 
The application of dual types is most evident in the application of a bag into an abstraction: the bag providing the intersection type and the abstraction consuming it. In session types the interaction between these is expressed by dual session types where one channel provides a behavior and and the dual channel provides the dual session type behavior via the cut rule.
Let us consider the term \( (  \lambda x . M [x_1, x_2 \leftarrow x]  ) B \) typed with the well-formedness rules by:
   \begin{prooftree}
        \AxiomC{\( \Gamma \wfdash \lambda x . M [x_1, x_2 \leftarrow x] : (\sigma \wedge \sigma) \rightarrow \tau \quad \Delta \wfdash B : \sigma^k \)}
            \LeftLabel{\redlab{FS{:}app}}
        \UnaryInfC{\( \Gamma, \Delta \wfdash (  \lambda x . M [x_1, x_2 \leftarrow x]  ) B : \tau\)}
    \end{prooftree}
When applying the translation of \figref{ch2fig:encfail} to the term we obtain:
\[
\bigoplus_{B_i \in \perm{B}} (\nu v)(\piencodf{\lambda x . M [x_1, x_2 \leftarrow x]}_v \mid v.\some_{u , \lfv{B}} ; \outact{v}{x} . ([v \leftrightarrow u] \mid \piencodf{B_i}_x ) )  
\]
By appealing to Type Preservation (\thmref{ch2t:preservationtwo}) we obtain both $ \piencodf{\lambda x . M [x_1, x_2 \leftarrow x]}_v \vdash  \piencodf{\Gamma}, v :\piencodf{(\sigma \wedge \sigma) \rightarrow \tau}  $ and $\piencodf{B}_x \vdash \piencodf{\Delta} , x: \piencodf{\sigma^k }_{(\delta_2, i_2)} $.
We give the typing for one non-deterministic branch where we take an arbitrary permutation of $B$ is as follows by applying the rules of \figref{ch2fig:trulespifull} and that $\Pi_1$ is derived to be:
\begin{prooftree}
\AxiomC{}
\LeftLabel{\redlab{Tid}}
\UnaryInfC{$[v \leftrightarrow u] \vdash   v : \overline{\piencodf{\tau}} , u:\piencodf{\tau}  $}
\AxiomC{$\piencodf{B}_x \vdash \piencodf{\Delta} , x: \piencodf{\sigma^k }_{(\delta_2, i_2)} $}
\LeftLabel{\redlab{T\otimes}}
\BinaryInfC{$\outact{v}{x} . ([v \leftrightarrow u] \mid \piencodf{B}_x ) \vdash \piencodf{\Delta}, v :   \piencodf{\sigma^k }_{(\delta_2, i_2)} \ampy \overline{\piencodf{\tau}} , u:\piencodf{\tau}  $}
\LeftLabel{\redlab{T\oplus^x_{\widetilde{w}}}}
\UnaryInfC{$ v.\some_{u , \lfv{B}} ; \outact{v}{x} . ([v \leftrightarrow u] \mid \piencodf{B}_x ) \vdash \piencodf{\Delta}, v :\dual{\piencodf{(\sigma^k) \rightarrow \tau}} , u:\piencodf{\tau} $}
\end{prooftree}
Hence we obtain the derivation applying the $\redlab{Tcut}$ rule:
\begin{prooftree}
\AxiomC{$ \piencodf{\lambda x . M [x_1, x_2 \leftarrow x]}_v \vdash  \piencodf{\Gamma}, v :\piencodf{(\sigma \wedge \sigma) \rightarrow \tau}  $}
\AxiomC{$ \Pi_1 $}
\BinaryInfC{$(\nu v)(\piencodf{\lambda x . M [x_1, x_2 \leftarrow x]}_v \mid v.\some_{u , \lfv{B}} ; \outact{v}{x} . ([v \leftrightarrow u] \mid \piencodf{B}_x )) \vdash \piencodf{\Gamma}, \piencodf{\Delta}, u:\piencodf{\tau} $}
\end{prooftree}
Now we shall focus on the typing of the channel $v$ and $x$ in this process as these channel describes the behavior of the encoded intersection type which we are trying to match via duality. By the translation on types from \figref{ch2fig:enc_sestypfail} we have that

\[
\piencodf{(\sigma \wedge \sigma) \rightarrow \tau} = \with((  \overline{\piencodf{(\sigma \wedge \sigma) }_{(\delta_1, i_1)}} ) \ampy \piencodf{\tau})
\]

\begin{itemize}
    \item When $B = \oneb$ we have derivation:
    \[
        \piencodf{ \oneb }_x \wfdash  \piencodf{\Delta}, x :\piencodf{\omega }_{(\delta_2, i_2)}
    \]
    To obtain duality from Rule~\redlab{Tcut}  we must have that $\piencodf{\sigma^2 }_{(\delta_1, i_1)} = \piencodf{\omega }_{(\delta_2, i_2)} $. By Proposition \ref{ch2prop:app_aux} we can take $\delta_1$ to be an arbitrary strict type, $i_1 = 0$, $i_2=2$ , $\delta_2 = \sigma$. We have $\piencodf{\omega }_{(\sigma, 2)}$ evaluated as:
    
    \[
    \begin{aligned}
         & = \overline{   \with(( \oplus \bot) \otimes ( \with  \oplus (( \with  \overline{\piencodf{ \sigma }} )  \ampy (\overline{\piencodf{\omega}_{(\sigma, 1)}})))) }\\
        & = \overline{   \with(( \oplus \bot) \otimes ( \with  \oplus (( \with  \overline{\piencodf{ \sigma }} )  \ampy ( \with(( \oplus \bot) \otimes ( \with  \oplus (( \with  \overline{\piencodf{ \sigma }} )  \ampy (\overline{\piencodf{\omega}_{(\sigma, 0)}})))) )))) }\\
        & = \piencodf{\sigma^2 }_{(\delta_1, i_1)}
    \end{aligned}
    \]
    
    \item When $B = \bag{N_1,N_2}$ we have derivation:
    \[
        \piencodf{ \bag{N_1,N_2} }_x \wfdash  \piencodf{\Delta}, x :\piencodf{\sigma^2 }_{(\delta_2, i_2)}
    \]
   To obtain duality from Rule~\redlab{Tcut}  we must have that $\piencodf{\sigma^2 }_{(\delta_1, i_1)} = \piencodf{\sigma^2 }_{(\delta_2, i_2)} $. By Proposition \ref{ch2prop:app_aux} we can take $\delta_1$ and $\delta_2$ to be an arbitrary strict type and  $i_1 = i_2 = 0$ . We then obtain $\piencodf{\sigma^2 }_{(\delta_1, 0)} = \piencodf{\sigma^2 }_{(\delta_2, 0)} $, as $\piencodf{\omega}_{(\delta_1, 0)} = \piencodf{\omega}_{(\delta_2, 0)} $ for any two strict types $\delta_1,\delta_2$.
    
    \item When $B = \bag{N_1,N_2,N_3}$ we have derivation:
    \[
        \piencodf{ \bag{N_1,N_2,N_3} }_x \wfdash  \piencodf{\Delta}, x :\piencodf{\sigma^3 }_{(\delta_2, i_2)}
    \]
    To obtain duality from Rule~\redlab{Tcut}  we must have that $\piencodf{\sigma^2 }_{(\delta_1, i_1)} = \piencodf{\sigma^3 }_{(\delta_2, i_2)} $. By Proposition \ref{ch2prop:app_aux} we can take $\delta_2$ to be an arbitrary strict type, $i_2 = 0$, $i_1=2$ , $\delta_1 = \sigma$. Then the case proceeds similarly to when $B = \oneb$.
    \end{itemize}
\end{example}

\subsubsection{Operational Correspondence: Completeness and Soundness}\label{ch2app:s:compsound}

We now state our operational correspondence results (completeness and soundness, cf. Fig.~\ref{ch2f:opcs}).

\paragraph*{A Congruence} 

We will identify some $\lamrsharfail$-terms such as $M\shar{}{x}\esubst{\oneb}{x}$ and $M$. The identification is natural, as the former is a term $M$ with no occurrences of $x$ in which $x$ is going to be replaced with $\oneb$, which clearly describes a substitution that ``does nothing'', and would result in $M$ itself.  With this intuition, other terms are identified via a {\em congruence} (denoted $\pequiv  
$) on terms and expressions that is formally defined in Fig.~\ref{ch2fig:rsPrecongruencefailure}.

\begin{figure}[!t]
\[
\begin{array}{rll}
  M [ \leftarrow x] \esubst{\oneb}{x} \!\!\!\! &\revdaniele{\pequiv} M &
  \\
  MB \linexsub{N/x}  \!\!\!\!&\pequiv (M\linexsub{N/x})B &  \text{with } x \not \in \lfv{B} 

         \\
        MA[\widetilde{x} \leftarrow x]\esubst{B}{x} 
         \!\!\!\!&\pequiv (M[\widetilde{x} \leftarrow x]\esubst{B}{x})A
         &  \text{with } x_i \in \widetilde{x} \Rightarrow x_i \not \in \lfv{A}
         \\
M[\widetilde{y} \leftarrow y]\esubst{A}{y}[\widetilde{x} \leftarrow x]\esubst{B}{x} \!\!\!\! &\revdaniele{\pequiv} 
(M[\widetilde{x} \leftarrow x]\esubst{B}{x})[\widetilde{y} \leftarrow y]\esubst{A}{y} &   \text{with } x_i \in \widetilde{x} \Rightarrow x_i \not \in \lfv{A}
\\
  M \linexsub{N_2/y}\linexsub{N_1/x} 
    \!\!\!\!&\pequiv M\linexsub{N_1/x}\linexsub{N_2/y} &
    \text{with } x \not \in \lfv{N_2},\revdaniele{ y \notin \lfv{N_1}}
\\
     C[M] \!\!\!\! & \revdaniele{\pequiv} C[M'] 
     &
        \text{with } M \revdaniele{\pequiv} M'
    \\
        D[\expr{M}] 
        \!\!\!\! & \revdaniele{\pequiv} D[\expr{M}'] 
         & 
         \text{with } \expr{M} \revdaniele{\pequiv} \expr{M}'
\end{array}
\]

\caption{Congruence in \lamrsharfail.}
    \label{ch2fig:rsPrecongruencefailure}
\end{figure}

  \begin{example}{Cont. Example~\ref{ch2ex:shar-wf}}\label{ch2ex:precong_fail}
We illustrate the congruence in case of  failure:
\[
      \begin{aligned}
            (\lambda x . x_1 [x_1 \leftarrow x]) \bag{ \fail^{\emptyset}[ \leftarrow y] \esubst{ \oneb }{y} } & \redd_{\redlab{RS{:}Beta}} x_1 [x_1 \leftarrow x]  \esubst{\bag{ \fail^{\emptyset}[ \leftarrow y] \esubst{\oneb}{y} }}{x} \\
           & \redd_{\redlab{RS{:}Ex \dash Sub}} x_1  \linexsub{ \fail^{\emptyset}[ \leftarrow y] \esubst{\oneb}{y}  / x_1} \\
            & \redd_{\redlab{RS{:}Lin \dash Fetch}} \fail^{\emptyset}[ \leftarrow y] \esubst{\oneb}{y}   \\
            & \pequiv \fail^{\emptyset} 
        \end{aligned}
  \]
  
  In the last step, Rule~$\redlab{RS{:}Cons_2}$ cannot be  applied:
  $y$ is sharing with no shared variables and the explicit substitution involves the bag $\oneb$.
  \end{example}

\secondrev{
\begin{restatable}[Consistency Stability Under \(\equiv\)]{theorem}{consistnequiv}
\label{ch2thm:term_consistency}
    Let  ${\expr{M}}$ be a consistent $\lamrsharfail$-expression. If $\expr{M} \equiv \expr{M}'$ then ${\expr{M}'}$ is consistent.
\end{restatable}
}

\begin{proof}
\secondrev{
    By induction on the structure of $\expr{M}$; 
    \iffulldoc
    see \Cref{ch2compandsucctwo} for details.
    \else
    see the full version for details.
    \fi 
}
\end{proof}

\begin{figure}[t!]
\begin{tikzpicture}[scale=.9pt]
\node (opcom) at (4.3,7.0){Operational Completeness};
\draw[rounded corners, color=teal,fill=teal!20] (0,5.2) rectangle (14.5,6.6);
\draw[rounded corners, color=violet!80!black, fill=violet!10] (0,1.4) rectangle (14.5,2.8);
\node (lamrfail) at (1.2,6) {$\lamrsharfail$:};
\node (expr1) [right of=lamrfail, xshift=.3cm] {$\mathbb{N}$};
\node (expr2) [right of=expr1, xshift=2cm] {$\mathbb{M} \pequiv \mathbb{M}' $};
\draw[arrow] (expr1) --  (expr2);
\node (lamrsharfail) at (1.2,2) {$\spi$:};
\node (transl1) [right of=lamrsharfail, xshift=.3cm] {$\piencodf{\mathbb{N}}$};
\node (transl2) [right of=transl1, xshift=2cm] {$Q=\piencodf{\mathbb{M}'}$};
\draw[arrow, dotted] (transl1) -- node[anchor= south] {$*$ }(transl2);
\node (enc1) at (2,4) {$\piencodf{\cdot }$};
\node (refcomp) [right of=enc1, xshift=.9cm]{Thm~\ref{ch2l:app_completenesstwo}};
\node  at (6.5,4) {$\piencodf{\cdot }$};
\draw[arrow, dotted] (expr1) -- (transl1);
\draw[arrow, dotted] (expr2) -- (transl2);
\node (opcom) at (11,7.0){Operational Soundness};
\node (expr1shar) [right of=expr2, xshift=1.8cm] {$\mathbb{N}$};
\node (expr2shar) [right of=expr1shar, xshift=2.8cm] {$\mathbb{N'}$};
\node (equivilencebit) [right of=expr1shar, xshift=2.4cm , yshift = -0.3cm] {$\pequiv$};
\draw[arrow, dotted] (expr1shar) -- node[anchor= south] {*} (expr2shar);
\node (transl1shar) [right of=transl2, xshift=1.8cm] {$\piencodf{\mathbb{N}}$};
\node (transl2shar) [right of=transl1shar, xshift=1cm] {$Q$};
\node (expr3shar) [right of=transl2shar, xshift=.8cm]{$Q'=\piencodf{\mathbb{N'}}$};
\draw[arrow,dotted] (expr2shar) --  (expr3shar);
\draw[arrow,dotted] (expr1shar) -- (transl1shar);
\node (enc2) at (8.5,4) {$\piencodf{\cdot }$};
\node (refsound) [right of=enc2, xshift=1.5cm]{Thm~\ref{ch2l:app_soundnesstwo}};
\node  at (13.8,4) {$\piencodf{\cdot }$};
\draw[arrow] (transl1shar) -- node[anchor= south] {*} (transl2shar);
\draw[arrow,dotted] (transl2shar) -- node[anchor= south] {*}(expr3shar);
\end{tikzpicture}
\vspace*{0.5cm}
\caption{Operational Correspondence for $\piencodf{\cdot }$\label{ch2f:opcs}}	
\end{figure}

\begin{definition}{Partially Open Terms}
We say that a $\lamrsharfail$-term $M$ is \emph{partially open} if $\forall x \in \lfv{M}$ (cf. \defref{ch2d:fvsh}) implies that $x$ is not a sharing variable.
\end{definition}

Notice that the class of open terms (no conditions on free variables) subsumes the class of partially open terms, which in turn  subsumes the class of closed terms. Consider the following example.

\begin{example}{Partially Open Terms}
    We give three examples of well-formed $\lamrsharfail$-terms:
    \[
    \begin{aligned}
        M_1 &= \lambda x. x_1 [x_1 \leftarrow x] 
        &
        \quad M_2 &= \lambda x. (x_1 \bag{y}) [x_1 \leftarrow x] 
        &
        \quad M_3 &= (x_1 \bag{y}) [x_1 \leftarrow x]     
    \end{aligned}
   \]
    Here the only closed term is $M_1$ as $M_2$ has one free variable (i.e., $y$)  and $M_3$ has two free variables ($y$ and $x$). 
    While $M_2$ is partially open, $M_3$ is not because $x$ is a sharing variable. 
\end{example}

The following proposition  will be used in the proof of operational completeness (Theorem~\ref{ch2l:app_completenesstwo}) and operational soundness (Theorem~\ref{ch2l:app_soundnesstwo}). The proposition relies on well-formed partially open terms; however, in the proof of operational correspondence we only consider closed terms rather then partially open terms.

\secondrev{
\begin{restatable}[]{proposition}{encodingreduces}
\label{ch2prop:NEEDTONAME}
    Suppose $N$ is a well-formed, partially open $\lamrsharfail$-term with $\headf{N} = x$.
     Then, there exist an index set $I$, names $\widetilde{y}$ and $n$, and processes $P_i$ such that the following four conditions hold:
    \begin{enumerate}
        \item $$\piencodf{ N }_{u} \redd^* \bigoplus_{i \in I}(\nu \widetilde{y})(\piencodf{ x }_{n} \mid P_i) $$
    \end{enumerate}
    \begin{enumerate}[resume]
        \item There exists a \lamrsharfail-term $N'$ such that $N \pequiv N'$ and:
            $$\piencodf{ N' }_{u} = \bigoplus_{i \in I}(\nu \widetilde{y})(\piencodf{ x }_{n} \mid P_i) $$
    \end{enumerate}
    \begin{enumerate}[resume]
        \item For any well-formed and partially open \lamrsharfail-term $M$:
        $$\piencodf{ N\headlin{ M/x } }_{u} \redd^* \bigoplus_{i \in I}(\nu \widetilde{y})(\piencodf{ M }_{n} \mid P_i) $$
    \end{enumerate}
    \begin{enumerate}[resume]
        \item There exists a \lamrsharfail-term $M'$ such that $M' \pequiv N\headlin{ M/x }$ and:
            $$\piencodf{ M' }_{u} = \bigoplus_{i \in I}(\nu \widetilde{y})(\piencodf{ M }_{n} \mid P_i)  $$
    \end{enumerate}
    %
    %
    %
    %
\end{restatable}
}

\secondrev{
\begin{proof}
    By induction on the structure of $N$. We briefly sketch the strategy for proving it case below, 
    \iffulldoc
    but 
     the  complete proof can be found in \appref{ch2compandsucctwo}.
     \else
     but 
     the  complete proof can be found in the full version.
     \fi 
    \begin{enumerate}
        \item The interesting cases are for $N=M\linexsub{N'/x}$ and $N=M\shar{\widetilde{y}}{y}\esubst{B}{y}$, when   $\size{B}=\size{\widetilde{y}} = 0$ and  $\headf{M}=x$. 
            Notice that $N=M\shar{\widetilde{y}}{y}$ is not a case, because of the definition of partially open term: $y$ is a sharing variable in $N$ and $y \in \lfv{N}$.
            The other cases follow easily by the induction hypothesis.
        \item  Reductions are only introduced by explicit weakening, which can be eliminated via the precongruence.
        \item Follows from (1) and the fact that linear head substitution can be placed deeper within the term until it reaches the head variable.
        \item Follows from (2) and (3).
    \end{enumerate}
\end{proof}
}




Because of the diamond property (Proposition \ref{ch2prop:conf1_lamrsharfail}), it suffices to consider a completeness result based on a single reduction step in $\lamrsharfail$:

\begin{restatable}[Operational Completeness]{theorem}{opcomplete}
\label{ch2l:app_completenesstwo}
Let $\expr{N} $ and $ \expr{M} $ be well-formed, \srev{partially open} $\lamrsharfail $ expressions. If $ \expr{N}\redd \expr{M} $ then there exist $Q$ and $\expr{M}'$
such that $\expr{M}' \pequiv \expr{M}$, $\piencodf{\expr{N}}_u  \redd^* Q = \piencodf{\expr{M}'}_u$.
\end{restatable}
\begin{proof}
By induction on the reduction rule applied to infer $\mathbb{N}\redd \mathbb{M}$. The case in which
\(\mathbb{N}\redd_{\redlab{RS:Lin\dash~Fetch}} \mathbb{M}\) happens for $\mathbb{N}=M\linexsub{N'/x}$ with $\headf{M}=x$, and $\mathbb{M}=M\headlin{N'/x}$. The translation of $\mathbb{N}$ is of the form (omitting details):
\[
\begin{aligned}
\piencodf{\revd{B29}{\expr{N}}}_u&= (\nu x) (\piencodf{M}_u\mid x.\some_{\lfv{N'}};\piencodf{N'}_x)\\
 &\redd^* (\nu x) (\bigoplus_{i \in I}(\nu \widetilde{y})(\piencodf{ x }_{n} \mid P_i)\mid x.\some_{\lfv{N'}};\piencodf{N'}_x), ~\text{by Proposition~\ref{ch2prop:NEEDTONAME}}\\
 &\redd^* \revd{B29}{\bigoplus_{i \in I}(\nu \widetilde{y})(P_i\mid \piencodf{N'}_n )}= \piencodf{\mathbb{M}}_{u}
\end{aligned}
\]
The other cases follow by analyzing reductions from the translation of $\mathbb{N}$. 
\iffulldoc
The full proof can be found in \appref{ch2compandsucctwo}.
\else
See the full version for details.
\fi 
\end{proof}

\revd{B41}{Notice how Proposition~\ref{ch2prop:NEEDTONAME} requires a term to be partially open; however, we prove operational correspondence for closed terms. The reason for this is that we start from a source closed term in \lamrfail, which is translated by $\recencodopenf{\cdot}$ into a closed $\lamrsharfail$-term.
}

\begin{example}{Cont. Example~\ref{ch2ex:encfailexcess}}
\label{ch2ex:enc2}
\revd{B29}{Recall that $M$ and $N$ are well-formed with $\lfv{N} = \lfv{M} = \emptyset$}, we can verify that  $ N [ \leftarrow x] \esubst{ \bag{M} }{x}$ and $\fail^{\lfv{N} \cup \lfv{M}}$  are also well-formed. We have
 $$
 N [ \leftarrow x] \esubst{ \bag{M} }{x} \redd_{\redlab{RS{:}Fail}} \fail^{\lfv{N} \cup \lfv{M}}
 $$  
In \spi, this reduction is mimicked as $$\piencodf{ N[ \leftarrow x] \esubst{ \bag{M} }{ x} }_u \redd^*  \piencodf{ \fail^{\lfv{N} \cup \lfv{M}} }_u.$$
In fact, 
\[\small
     \begin{aligned}
        \piencodf{ N[ \leftarrow x] \esubst{ \bag{M} }{ x} }_u 
        =~&   (\nu x)( \piencodf{ N[ \leftarrow x]}_u \mid \piencodf{\bag{M} }_x ) \\
        =~& (\nu x)( x. \overline{\some}. \outact{x}{y_i} . ( y_i . \some_{u} ;y_{i}.\close; \piencodf{N}_u \mid x. \overline{\none}) \mid 
        \\
        =~&  x.\some_{\emptyset } ; x(y_i). x.\some_{y_i};x.\overline{\some} ; \outact{x}{x_i}. \\
         &\hspace{.8cm} (x_i.\some_{\emptyset} ; \piencodf{M}_{x_i} \mid \piencodf{\oneb}_x \mid y_i. \overline{\none}) )
       \\
         \redd~& (\nu x)(  \outact{x}{y_i} . ( y_i . \some_{u} ;y_{i}.\close; \piencodf{N}_u \mid x. \overline{\none}) \mid \\
       & \hspace{.8cm}  x(y_i). x.\some_{y_i};x.\overline{\some} ; \outact{x}{x_i}. (x_i.\some_{\emptyset} ; \piencodf{M}_{x_i} \mid \piencodf{\oneb}_x \mid y_i. \overline{\none}) )
       \\
       \redd~& (\nu x)(   y_i . \some_{u} ;y_{i}.\close; \piencodf{N}_u \mid x. \overline{\none} \mid \\
       & \hspace{.8cm}  x.\some_{y_i };x.\overline{\some} ; \outact{x}{x_i} . (x_i.\some_{\emptyset} ; \piencodf{M}_{x_i} \mid \piencodf{\oneb}_x \mid y_i. \overline{\none}) )
       \\
       \redd~& (\nu x)(   y_i . \some_{u} ;y_{i}.\close; \piencodf{N}_u \mid   y_i.\overline{\none} )
       \\
       \redd~& u. \overline{\none} \\
       =~&  \piencodf{ \fail^{\lfv{N} \cup \lfv{M}} }_u
     \end{aligned}
     \]    
    \end{example}

To state soundness we rely on the congruence relation  $\pequiv$, given in \figref{ch2fig:rsPrecongruencefailure}.

\begin{notation}
Recall the congruence $\pequiv$ for \lamrsharfail, given in Figure~\ref{ch2fig:rsPrecongruencefailure}.
We write $N \redd_{\pequiv} N'$ iff $N \pequiv N_1 \redd N_2 \pequiv N' $, for some $N_1, N_2$. Then, $\redd_{\pequiv}^*$ is the reflexive, transitive closure of $\redd_{\pequiv}$. 
We use the notation $ M \redd_{\pequiv}^i N$ to state that $M$ performs $i$ steps of $\redd_{\pequiv}$ to $N$ in $i \geq 0$ steps. When $i = 0$ it refers to no reduction taking place. 
\end{notation}


\begin{restatable}[Operational Soundness]{theorem}{opsound}
\label{ch2l:app_soundnesstwo}
Let $\expr{N}$ be a 
well-formed, \secondrev{partially open}  $ \lamrsharfail$ expression. 
If $ \piencodf{\expr{N}}_u \redd^* Q$
then there exist $Q'$  and $\expr{N}' $ such that 
$Q \redd^* Q'$, $\expr{N}  \redd^*_{\pequiv} \expr{N}'$ 
and 
$\piencodf{\expr{N}'}_u = Q'$.
\end{restatable}

\begin{proof}[Proof (Sketch)]
By induction on the structure of $\mathbb{N}$ with sub-induction on the number of reduction steps in $ \piencodf{\expr{N}}_u \redd^* Q$. The cases  in which $\mathbb{N}=x$, or $\mathbb{N}=\fail^{\widetilde{x}}$, or $\mathbb{N}=\lambda x. M\shar{\widetilde{x}}{x}$, are easy since there are no reductions starting from $\piencodf{\mathbb{N}}_u$, i.e., $ \piencodf{\expr{N}}_u \redd^0 Q$  which implies $\piencodf{\mathbb{N}}_u=\piencodf{\mathbb{N'}}_u=Q=Q'$ and the result follows trivially. The analysis for some  cases are exhaustive, for instance, when   $\mathbb{N}=(M \ B)$ or $\mathbb{N}=M\shar{\widetilde{x}}{x} \esubst{B}{x}$, there are several sub-cases  to be considered: (i) $B$ being equal to  $\oneb$ or not; (ii) $\size{B}$   matching the number of occurrences of the variable in $M$ or not; (iii) $M$ being a failure term or not.

We now discuss one of these cases to illustrate the recurring idea used in the proof: let $\mathbb{N}=(M \ B)$ and suppose that we are able to perform $k> 1$ steps  to a process $Q$, i.e., 
 
 \begin{equation}\label{ch2eq:enc_n}
\piencodf{\mathbb{N}}_{u}= \piencodf{(M \ B)}_{u}= \bigoplus_{B_i \in \perm{B}} (\nu v)(\piencodf{M}_v \mid v.\some_{u , \lfv{B}} ; \outact{v}{x} . ([v \leftrightarrow u] \mid \piencodf{B_i}_x ) )  \redd^k Q
 \end{equation}
 
 Then there exist an  $\spi$ process $R$ and integers $n,m$ such that $k=m+n$ and
   \[
            \begin{aligned}
               \piencodf{\expr{N}}_u &\redd^m  \bigoplus_{B_i \in \perm{B}} (\nu v)( R \mid v.\some_{u , \lfv{B}} ; \outact{v}{x} . ( \piencodf{ B_i}_x \mid [v \leftrightarrow u] ) ) \redd^n  Q\\
            \end{aligned}
            \]
           where the first $m \geq 0$ reduction steps are  internal to $\piencodf{ M}_v$; type preservation in \spi ensures that, if they occur,  these reductions  do not discard the possibility of synchronizing with $v.\some$. Then, the first of the $n \geq 0$ reduction steps towards $Q$ is a synchronization between $R$ and $v.\some_{u, \lfv{B}}$. 
           
           We will  consider the case when $m = 0$ and $n \geq 1$. Then  $R = \piencodf{\expr{M}}_u \redd^0 \piencodf{\expr{M}}_u$ and  there are two possibilities of having an unguarded $v.\overline{\some}$ or $v.\overline{\none}$ without internal reductions:
           \begin{enumerate}[(i)]
               \item $M = (\lambda x . M' [\widetilde{x} \leftarrow x]) \linexsub{N_1 / y_1} \cdots \linexsub{N_p / y_p} \qquad (p \geq 0)$
                        
            \item $M = \fail^{\widetilde{z}}$
           \end{enumerate}
           
    Firstly we use case (i) to express the need for the reduction $\expr{N}  \redd^*_{\pequiv} \expr{N}'$. In this case $\mathbb{N}=((\lambda x . M' [\widetilde{x} \leftarrow x]) \linexsub{N_1 / y_1} \cdots \linexsub{N_p / y_p} \ B)$ and $\piencod{\mathbb{N}}_u$ may perform synchronizations where both $\piencod{\lambda x . M'}_v$ and $\piencod{B}_x$ synchronize across their shared channel. Here we use the congruence relation as follows:
    \[
    \begin{aligned}
        \mathbb{N} & = ((\lambda x . M' [\widetilde{x} \leftarrow x]) \linexsub{N_1 / y_1} \cdots \linexsub{N_p / y_p} \ B) \\
        & \pequiv ((\lambda x . M'  [\widetilde{x} \leftarrow x])\ B) \linexsub{N_1 / y_1} \cdots \linexsub{N_p / y_p} \\
    \end{aligned}
    \]
    This enables the abstraction $\lambda x . M'$ to synchronize with the bag $B$.
    
    Now we will develop case (ii):
     \[
    \begin{aligned}
    \piencodf{M}_v &= \piencodf{\fail^{\widetilde{z}}}_v= \piencodf{\fail^{\widetilde{z}}}_v= v.\overline{\none} \mid \widetilde{z}.\overline{\none} \\
    \end{aligned}
    \]
 With this shape for $M$, the translation and reductions from (\ref{ch2eq:enc_n}) become
 \begin{equation}\label{ch2eq:red_n}
 \begin{aligned}
 \piencodf{\expr{N}}_u  = & \bigoplus_{B_i \in \perm{B}} (\nu v)( \piencodf{ M}_v \mid v.\some_{u, \lfv{B}} ; \outact{v}{x} . (  \piencodf{ B_i}_x \mid [v \leftrightarrow u] ) )\\
  = & \bigoplus_{B_i \in \perm{B}} (\nu v)(  v.\overline{\none}\mid \widetilde{z}.\overline{\none} \mid v.\some_{u, \lfv{B}} ; \outact{v}{x} . (  \piencodf{ B_i}_x \mid [v \leftrightarrow u] ) )\\
  \redd & \bigoplus_{B_i \in \perm{B}}   u.\overline{\none} \mid \widetilde{z}.\overline{\none}  \mid \lfv{B}.\overline{\none} 
 \end{aligned}
\end{equation}
   We also have that 
    $  \expr{N} = \fail^{\widetilde{z}} \ B  \redd \sum_{\perm{B}} \fail^{\widetilde{z} \cup \lfv{B} }  = \expr{M}$. 
 Furthermore, we have:
     \begin{equation}\label{ch2eq:red_m}
     \begin{aligned}
       \piencodf{\expr{M}}_u &= \piencodf{\sum_{\perm{B}} \fail^{\widetilde{z} \cup \lfv{B} }}_u \\
       &= \bigoplus_{\perm{B}}\piencodf{ \fail^{\widetilde{z} \cup \lfv{B} }}_u\\
     &  = \bigoplus_{\perm{B}}    u.\overline{\none} \mid \widetilde{z}.\overline{\none}  \mid \lfv{B}.\overline{\none}
        \end{aligned}
    \end{equation}
From reductions in (\ref{ch2eq:red_n}) and (\ref{ch2eq:red_m}) one has $\piencodf{\expr{N}}_u\redd \piencodf{\expr{M}}_u$, and the result follows with $n=1$ and $\piencodf{\expr{M}}_u=Q=Q'$.
\iffulldoc
 The full proof can be found in \appref{ch2compandsucctwo}.
 \else
 See the full version for details.
 \fi 
\end{proof}

\subsubsection{Success Sensitiveness}

Finally, we consider success sensitiveness. This requires extending \lamrsharfail and \spi with success predicates. 


\begin{definition}{}
We extend the syntax of $\spi$ processes (Definition~\ref{ch2d:spi}) 
with the $\checkmark$ construct, which we assume well typed. 
Also, we extend Definition~\ref{ch2def:enc_lamrsharpifail} by 
defining $\piencodf{\checkmark}_u = \checkmark$
\end{definition}

\begin{definition}{}
We say that a process occurs \emph{guarded} when it occurs behind a prefix (input, output, closing of channels and non-deterministic session behavior). That is, $P$ is guarded if  $\alpha.P$ or $ \alpha;P$, where $ \alpha = \overline{x}(y), x(y), x.\overline{\close}, x.\close,$ $ x.\overline{\some}, x.\some_{(w_1, \cdots, w_n)} $. 
We say it occurs \emph{unguarded} if it is not guarded for any prefix.
\end{definition}

\begin{definition}{Success in \spi}
\label{ch2def:Suc4}
We extend the syntax of $\spi$ processes  
with the $\checkmark$ construct, which we assume well-typed. 
We define 
$\succp{P}{\checkmark}$ to hold whenever there exists a $P'$
such that 
$P \redd^* P'$
and $P'$ contains an unguarded occurrence of $\checkmark$.
\end{definition}


\begin{restatable}[Preservation of Success]{proposition}{pressucctwo}
\label{ch2Prop:checkprespi}
The $\checkmark$ at the head of a \revd{B31}{partially open} term is preserved to an unguarded occurrence of $\checkmark$ when applying the translation $\piencodf{\cdot}_u$ up to reductions and vice-versa. That is to say:
\begin{enumerate}
    \item $\forall M \in \lamrsharfail: \quad \headf{M} = \checkmark \implies \piencodf{M}_u \redd^* (P \mid \checkmark) \oplus Q $
    \item $\forall  M \in \lamrsharfail: \quad \piencodf{M}_u =  (P \mid \checkmark) \oplus Q \implies \headf{M} = \checkmark$
\end{enumerate} 

\end{restatable}

\begin{proof}[Proof (Sketch)]
By induction on the structure of $M$.
For item (1),  consider the case $M=(N\ B)$ and  $\headf{N \ B} = \headf{N} = \checkmark$. This term's translation is
\[\piencodf{N \ B}_u = \bigoplus_{B_i \in \perm{B}} (\nu v)(\piencodf{N}_v \mid v.\some_{u, \lfv{B}} ; \outact{v}{x} . ([v \leftrightarrow u] \mid \piencodf{B_i}_x ) ).\]
By the induction hypothesis,  $\checkmark$ is unguarded in $\piencodf{N}_u$ after a sequence of reductions, i.e., $ \piencodf{N}_u\redd^* (\checkmark \mid P')\oplus Q'$, for some $\spi$ processes $P'$ and $Q'$. Thus, 
\[
\begin{aligned} 
\piencodf{N \ B}_u & \redd^* \bigoplus_{B_i \in \perm{B}} (\nu v)((\checkmark \mid P')\oplus Q' \mid v.\some_{u, \lfv{B}} ; \outact{v}{x} . ([v \leftrightarrow u] \mid \piencodf{B_i}_x ) )\\
& \equiv  \checkmark \mid (\nu v)(P'\oplus Q' \mid v.\some_{u, \lfv{B}} ; \outact{v}{x} . ([v \leftrightarrow u] \mid \piencodf{B_j}_x ) )\\ 
& \quad \oplus \Big( 
\bigoplus_{B_i \in (\perm{B} \linsetminus B_j )  } \checkmark \mid (\nu v)(P'\oplus Q' \mid v.\some_{u, \lfv{B}} ; \outact{v}{x} . ([v \leftrightarrow u] \mid \piencodf{B_i}_x ) ) 
\Big)  \\
& \equiv  (\checkmark \mid P) \oplus Q \\ 
\end{aligned}
\]
and the result follows by taking $P = (\nu v)(P'\oplus Q' \mid v.\some_{u, \lfv{B}} ; \outact{v}{x} . ([v \leftrightarrow u] \mid \piencodf{B_j}_x ) ) $ and 
$Q = \bigoplus_{B_i \in ({\perm{B}} \linsetminus B_j)} \checkmark \mid (\nu v)(P'\oplus Q' \mid v.\some_{u, \lfv{B}} ; \outact{v}{x} . ([v \leftrightarrow u] \mid \piencodf{B_i}_x ) )$.
The analysis for the other cases are similar; 
\iffulldoc
see \appref{ch2app:succtwo} for details.
\else
see the full version for details.
\fi 
\end{proof}

The translation $\piencodf{\cdot}_u:\lamrsharfail \rightarrow \spi$ is success sensitive on well-formed closed expressions. 
\begin{restatable}[Success Sensitivity]{theorem}{successsenscetwo}
\label{ch2proof:successsenscetwo}
Let  \expr{M} be a closed well-formed $\lamrsharfail$-expression.
Then,
\[\expr{M} \Downarrow_{\checkmark}\iff \succp{\piencodf{\expr{M}}_u}{\checkmark} .\]
\end{restatable}

\begin{proof}[Proof (Sketch)]
 Suppose  $\expr{M} \Downarrow_{\checkmark} $.
    By Definition \ref{ch2def:app_Suc3} there exists $\mathbb{M}'=M_1 + \cdots + M_k$ such that $\expr{M} \redd^* \expr{M}'$ and
    $\headf{M_j} = \checkmark$, for some  $j \in \{1, \ldots, k\}$ and  $M_j$. By operational completeness (Theorem~\ref{ch2l:app_completenesstwo}),  there exists $ Q$ such that $\piencodf{\expr{M}}_u  \redd^* Q = \piencodf{\expr{M}'}_u$.
   Due to compositionality of $\piencodf{\cdot}$ and the homomorphic preservation of non-determinism,  we have:
\begin{itemize}
\item \(Q = \piencodf{M_1}_u \oplus \cdots \oplus \piencodf{M_k}_u\)
\item \revt{B32}{\(\piencodf{M_j}_u  = C[ \piencodf{\checkmark}_v ]= C[ \checkmark]\)}
\end{itemize}
By Proposition \ref{ch2Prop:checkprespi}, item  (1), since $\headf{M_j} = \checkmark$ it follows that $ \piencodf{M_j}_u \redd^*  P \mid \checkmark \oplus Q'$. Hence $Q$ reduces to a process that has an unguarded occurrence of $\checkmark$. The proof of the converse  is similar and 
\iffulldoc
can be found in \appref{ch2app:succtwo}.
\else
can be found in the full version.
\fi 
\end{proof}
As main result of this sub-section, we have the corollary below, which follows from the previously stated Theorems~\ref{ch2t:preservationtwo}, 
\ref{ch2l:app_completenesstwo}, 
\ref{ch2l:app_soundnesstwo}, and 
\ref{ch2proof:successsenscetwo}:

\begin{corollary}
\label{ch2cor:two}
Our translation  $ \piencodf{ \cdot } $ is a correct encoding, in the sense of \defref{ch2d:encoding}.
\end{corollary}

\noindent
Together, Corollary~\ref{ch2cor:one} and Corollary~\ref{ch2cor:two} ensure that $\lamrfail$ can be correctly translated into \spi, using \lamrsharfail as a stepping stone.

\section{A Motivating Example}
\label{ss:exammotiv}
We motivate the expressivity of the calculi with a novel example, illustrating the use of linearity within resource consumption where we assume  access to a specification of a programming language that implements the resource $\lambda$-calculus $ \lamrsharfail $. The underling process model given vie the correct encoding presented we may assume access to some implimentation of the semantics. Consider the following protocol for a movie review company: The company sends early issues of three movies (`Jaws', `Dune', `Elf') to three reviewers (`$\sff{reviewer1}$', `$\sff{reviewer2}$', `$ \sff{reviewer3}$') and receive their reviews as the sum of their scores. As the movies are not publicly available, therefore each movie should be watched only once to ensure it is not copied or distributed (hence considering them as linear resources to be consumed exactly once) and wish not to overwork the reviewers by only allow them to watch and review one movie each. Assuming we have access to a function `$\sff{review}(x,y)$', which takes a reviewer $x$ and a film $y$, returning the reviewers score for the given movie:
\[
    \begin{aligned}
       M =  &((\lambda x_1, x_2, x_3. \lambda y_1,y_2,y_3.\sff{review(}x_1,y_1) + \sff{review(}x_2,y_2) + \sff{review(}x_3,y_3) \\
        & \quad (\sff{reviewer1} \oplus \sff{reviewer2} \oplus \sff{reviewer3}))  (\sff{Jaws} \oplus \sff{Dune} \oplus \sff{Elf}))
            \end{aligned}
\]

The behaviour of $M$ can be distilled into three parts. First, the term $\lambda x_1, x_2, x_3. \lambda y_1,y_2,y_3.\,\sff{review(}x_1,y_1) + \sff{review(}x_2,y_2) + \sff{review(}x_3,y_3)$, where two abstractions take place, the first being on $x_1, x_2, x_3$ and the second on $y_1,y_2,y_3 $. The notation $\lambda x_1, x_2, x_3.M$ denotes a function taking three linear arguments. Hence we interpret these two abstractions to both take three linear resources each and apply them within the predefined function $ \sff{review(}x,y)$. Second, this abstraction is applied to a nondeterministic choice  of three linear resources (denoted by $\oplus$), which are represented by the reviewers. This allows the following reduction:
\[
    \begin{aligned}
       M =~  &(\lambda x_1, x_2, x_3. \lambda y_1,y_2,y_3.\sff{review}(x_1,y_1) + \sff{review}(x_2,y_2) + \sff{review}(x_3,y_3) \\
        & \quad (\sff{reviewer1} \oplus \sff{reviewer2} \oplus \sff{reviewer3})) 
        \\
        \red~& (\lambda   x_2, x_3. \lambda y_1,y_2,y_3.\sff{review}( \sff{reviewer1},y_1) + \sff{review}(x_2,y_2) + \sff{review}(x_3,y_3) \\
        & \quad ( \sff{reviewer2} \oplus \sff{reviewer3})) \\
        &+ (\lambda   x_2, x_3. \lambda y_1,y_2,y_3.\sff{review}(\sff{reviewer2}  ,y_1) + \sff{review}(x_2,y_2) + \sff{review}(x_3,y_3) \\
        & \quad (\sff{reviewer1} \oplus  \sff{reviewer3})) \\
        & + (\lambda   x_2, x_3. \lambda y_1,y_2,y_3.\sff{review}(\sff{reviewer3} ,y_1) + \sff{review}(x_2,y_2) + \sff{review}(x_3,y_3) \\
        & \quad (\sff{reviewer1} \oplus \sff{reviewer2} )) 
            \end{aligned}
\]

Finally, the term is then also applied to the non-deterministic choice between the three movies, just as before. The term will then reduce as follows:
\[
    \begin{aligned}
M     \red^* 
        &  (\sff{review(}\sff{reviewer1},\sff{Jaws}) + \sff{review(}\sff{reviewer2} ,\sff{Dune}) + \sff{review(}\sff{reviewer3},\sff{Elf}) )\\
        & \qquad  \oplus 
        (\sff{review(}\sff{reviewer1},\sff{Jaws}) + \sff{review(}\sff{reviewer2} ,\sff{Elf}) + \sff{review(} \sff{reviewer3},\sff{Dune}) )\\
        &  \qquad \oplus 
        (\sff{review(}\sff{reviewer1},\sff{Dune}) + \sff{review(}\sff{reviewer2} ,\sff{Jaws}) + \sff{review(} \sff{reviewer3},\sff{Elf}) )\\
        & \qquad  \oplus 
        (\sff{review(}\sff{reviewer1},\sff{Dune}) + \sff{review(}\sff{reviewer2} ,\sff{Elf}) + \sff{review(} \sff{reviewer3},\sff{Jaws}) )\\
        & \qquad  \oplus 
        (\sff{review(}\sff{reviewer1},\sff{Elf}) + \sff{review(}\sff{reviewer2} ,\sff{Jaws}) + \sff{review(} \sff{reviewer3},\sff{Dune}) )\\
        & \qquad  \oplus 
        (\sff{review(}\sff{reviewer1},\sff{Elf}) + \sff{review(}\sff{reviewer2} ,\sff{Dune}) + \sff{review(} \sff{reviewer3},\sff{Jaws}) )\\
    \end{aligned}
\]

 This example shows that we can non-deterministically compute any combination of reviewers to scores. Modifications to this implementation can be incorporated easily. For instance, suppose that the protocol changes and that we now wish one reviewer to review all three movies. In that case, the application is applied to the resources containing three instances of a single reviewer.

As we have provided a correct encoding from sequential to concurrent computation with both sound and complete correctness results, we can exploit aspects of concurrent calculi to extend the programming language with sequencing, which is a primitive construct in $\spi$. 

Consider now the protocol that makes a decision on what movie to watch. Suppose we have a function called `$\sff{decide(}x)$' that receives a movie $x$ and makes a decision on whether or not to watch the movie, writing the result boolean result into a database. If a movie has already been decided to be watched then all future movies return false (the tracking of this may be decided by some global variable). Suppose that the decision is to watch either `Jaws' or `Elf'
  but not  `Dune'; then we can define the function
\[
    \begin{aligned}
         \lambda x. \sff{decide(}x) (\sff{Jaws} \oplus \sff{Dune} \oplus \sff{Elf}) \red^* 
         &
         \sff{decide(Jaws)} ; \sff{decide(Dune)} ; \sff{decide(Elf)}   \\
         & + \sff{decide(Jaws)}  ; \sff{decide(Elf)} ; \sff{decide(Dune)}  \\
         & + \sff{decide(Dune)} ;\sff{decide(Jaws)} ; \sff{decide(Elf)}   \\
         &  + \sff{decide(Dune)};  \sff{decide(Elf)} ;\sff{decide(Jaws)}    \\
         &  + \sff{decide(Elf)} ; \sff{decide(Jaws)} ; \sff{decide(Dune)}    \\
         & + \sff{decide(Elf)} ; \sff{decide(Dune)} ; \sff{decide(Jaws)}    
    \end{aligned}
\]
 
Here we interpret the abstractions $\lambda x. \sff{decide(}x)$ to consume linear resources $x$ and apply them non-deterministically in sequence, as represented by the rule:
\[
    (\lambda x. M ) (N_1 \oplus  N_2 ) \red  (\lambda x. M ) N_1 ; (\lambda x. M ) N_2  + (\lambda x. M ) N_2 ; (\lambda x. M ) N_1 
\]

The first three choices result in the decision `Jaws' and the last three results in `Elf'. This is a high-level example, but sufficient to show that we may exhibit more interesting non-deterministic behavior by extending functional calculi with sequencing constructs from concurrent calculi, motivating the need for our correct encodings.
We can mimic this type of behavior within the resource $\lambda$-calculus with:
\[(\lambda x .  \sff{sequence3}  \bag{ \sff{decide(}x) , \sff{decide(}x) , \sff{decide(}x)  } ) \bag{ \sff{Jaws} , \sff{Dune}  , \sff{Elf}   } \]
where  `$\sff{sequence3}$'  is a function that takes three elements in a bag and runs them one at a time in sequence (using mechanisms from into concurrent computation) and then terminating.

\section{Related Work}
\label{ch2s:rw}

Closely related works have been already discussed in the introduction and throughout the chapter; here we mention other related literature.

\subsubsection*{Intersection Types}
\revt{}{The first works on intersection types date back to the late 70s~(see, e.g.,~\cite{DBLP:journals/aml/CoppoD78,Pottinger80}) and consider intersections with the  \emph{idempotence} property (i.e., $\sigma \land \sigma = \sigma$). 
This formulation enables the analysis of \emph{qualitative} properties of $\lambda$-calculi, such as (strong) normalization and solvability.
By dropping idempotence, intersection types can   characterize \emph{quantitative} properties, such as, e.g., bounds on the number of steps needed to reach a normal form.
Early works on non-idempotent intersection types include~\cite{DBLP:conf/tacs/Gardner94,DBLP:journals/logcom/Kfoury00,DBLP:journals/tcs/KfouryW04}. 
The paper~\cite{DBLP:conf/lics/BonoD20}  overviews the origins, development, and applications of intersection types.}

\revd{}{Our work formally connects non-idempotent intersection types and classical linear logic extended with the modalities $\with$ and $\oplus$, interpreted in~\cite{CairesP17} as session types for non-deterministically available protocols. 
\srev{To the best of our knowledge, this is an unexplored angle. Prior connections between (non-idempotent) intersection types and linear logic arise in very different settings (see~\cite{DBLP:journals/pacmpl/MazzaPV18} and references therein). They include~\cite{DBLP:conf/icfp/NeergaardM04}, which presents a connection based on a correspondence between normalization and type inference;
the work~\cite{DBLP:journals/corr/abs-0905-4251,DBLP:journals/mscs/Carvalho18}, which shows a correspondence between the \emph{relational model} of linear logic and an non-idempotent intersection type system;} 
and~\cite{DBLP:conf/fossacs/Ehrhard20}, which concerns \emph{indexed} linear logic (cf.~\cite{DBLP:journals/apal/BucciarelliE00,DBLP:journals/apal/BucciarelliE01}).
}

\srev{The work~\cite{DBLP:journals/pacmpl/LagoVMY19} develops a type system for the $\pi$-calculus based on non-idempotent intersections. The type system ensures that processes are ``well-behaved''---they never produce run-time errors, and can always reduce to an idle process. Remarkably, they show that their type system is \emph{complete}: every well-behaved process is typable. Although their type system does not consider session types,  it is related to our work for it builds upon Mazza et al.'s~correspondence between linear logic and intersection types, given in terms of \emph{polyadic approximations}~(\cite{DBLP:journals/pacmpl/MazzaPV18}).}


\subsubsection*{Other Resource $\lambda$-calculi}
A fine-grained treatment of duplication and erasing---similar to our design for $\lamrsharfail$---is present in Kesner and Lengrand's  $\lambda${\tt lxr}-calculus~(\cite{DBLP:journals/iandc/KesnerL07}), a 
simply-typed, deterministic $\lambda$-calculus that is in correspondence with proof nets.
The $\lambda${\tt lxr}-calculus includes operators called weakening \(\mathcal{W}_{\_}(\_)\) and contraction \(\mathcal{C}_{\_}^{\_|\_}(\_)\) to deal with empty and non-empty sharing, respectively. 
In this approach, our terms 
$\lambda x. x \bag{x}$  and $\lambda x. y \bag{z}$ 
would be expressed as \(\mathcal{C}_{x}^{x_1|x_2}(\lambda x. x_1\bag{x_2})\) and \(\mathcal{W}_x(\lambda x.y\bag{z})\), respectively.

Our approach is convenient when expressing the sharing of more than two occurrences of a variable in a term; as in, e.g., the $\lamrfail$-term $\lambda x.( x\bag{x,x})$ which would correspond to $\lambda x. (x_1\bag{x_2,x_3})\shar{x_1,x_2,x_3}{x}$ in $\lamrsharfail$.
In the $\lambda${\tt lxr}-calculus, contractions are binary, and so representing $\lambda x.( x\bag{x,x})$ requires the composition of two binary contractions.

More substantial differences appear at the level of types. 
As we have seen, in \lamrsharfail we use intersection types to define  well-typed and well-formed expressions (see~\figref{ch2fig:typing_sharing} and~\figref{ch2fig:wfsh_rules}, respectively).   In particular, recall the well-formedness rule for the sharing construct:
\begin{prooftree}
\AxiomC{\(\Gamma, x_1:\sigma, \ldots, x_k:\sigma \wfdash M:\tau \quad x \notin \dom{\Gamma} \quad  k\neq 0\)}
\LeftLabel{\redlab{FS:share}}
\UnaryInfC{\(\Gamma, x: \sigma^k \wfdash M\shar{x_1,\ldots, x_k}{x}:\tau\)}
\end{prooftree}
where, as mentioned above, $\sigma^k$ denotes the intersection type $\sigma\wedge \ldots \wedge \sigma$. 
Differently, the typing rule for contraction in the $\lambda${\tt lxr}-calculus involves an arbitrary (simple) type $A$:
\begin{prooftree}
\AxiomC{\(\Gamma, y:A, z:A \vdash M:B \)}
\LeftLabel{(Cont)}
\UnaryInfC{\(\Gamma, x: A \vdash \mathcal{C}_{x}^{y|z}(M):B\)}
\end{prooftree}
Our weakening rule $\redlab{FS:weak}$ types the empty sharing term $M\shar{}{x}$ as follows:
\begin{prooftree}
\AxiomC{\(\Gamma \wfdash M:\tau\)}
\LeftLabel{\redlab{FS:weak}}
\UnaryInfC{\(\Gamma, x:\omega \wfdash M\shar{}{x}:\tau\)}
\end{prooftree}
Hence, the context $\Gamma$ is weakened with a variable assignment $x:\omega$, where $\omega$ denotes the empty type.  In contrast, weakening in the $\lambda${\tt lxr}-calculus involves a  (simple) type $A$: 
\begin{prooftree}
\AxiomC{\(\Gamma \vdash M:A \)}
\LeftLabel{(Weak)}
\UnaryInfC{\(\Gamma, x:B \vdash \mathcal{W}_x(M): A\)}
\end{prooftree}
Hence, the context can be weakened with an assignment $x:B$, where $B$ is a simple type.

\smallskip
\srev{Inspired by the multiplicative exponential fragment of linear logic,\cite{DBLP:journals/tcs/KesnerR11} define the so-called \emph{prismoid of resources}, a parametric framework of simply-typed $\lambda$-calculi in which each language incorporates different choices for contraction, weakening, and substitution operations. 
The prismoid defines a uniform and general setting for establishing key properties of typed terms, 
including simulation of $\beta$-reduction, confluence, and strong normalization.
One of the languages included in the prismoid is a minor variant of the $\lambda${\tt lxr}-calculus, which we have just mentioned.}

\smallskip

There are some similarities between $\lamrfail$ and the differential $\lambda$-calculus, introduced in~\cite{DBLP:journals/tcs/EhrhardR03}. Both express non-deterministic choice via sums and use linear head reduction for evaluation. In particular, our fetch rule, which consumes non-deterministically elements from a bag, is related to the derivation (which has similarities with substitution) of a differential term. However, the focus of~\cite{DBLP:journals/tcs/EhrhardR03} is not on typability nor encodings to process calculi; instead they relate the Taylor series of analysis to the linear head reduction of $\lambda$-calculus.

\subsubsection*{Functions as Processes}
A source of inspiration for our developments is the work by \cite{DBLP:conf/birthday/BoudolL00}. As far as we know, this 
is the only prior study that connects $\lambda$ and $\pi$ from a resource-oriented perspective, via an encoding of a $\lambda$-calculus with multiplicities into a $\pi$-calculus without sums.
The goal of~\cite{DBLP:conf/birthday/BoudolL00} is different from ours, as they study the discriminating power of semantics for $\lambda$ as induced by encodings into~$\pi$. In contrast, we study how typability delineates the encodability of resource-awareness across sequential and concurrent realms.
The source and target calculi in~\cite{DBLP:conf/birthday/BoudolL00} are untyped, whereas  we consider typed calculi and our encodings preserve typability. 
As a result, the encoding in~\cite{DBLP:conf/birthday/BoudolL00} is conceptually different from ours; remarkably, our encoding respects linearity and homomorphically translates sums. 

Prior works have studied encodings of  typed $\lambda$-calculi into typed $\pi$-calculi; see, e.g.,~~\cite{DBLP:journals/mscs/Sangiorgi99,DBLP:conf/birthday/BoudolL00,DBLP:books/daglib/0004377,DBLP:conf/fossacs/BergerHY03, DBLP:conf/fossacs/ToninhoCP12,DBLP:conf/rta/HondaYB14,DBLP:conf/esop/ToninhoY18}. None of these works consider {non-determinism} and {failures}; the one exception is the encoding in~\cite{CairesP17}, which involves a $\lambda$-calculus with exceptions and failures (but without non-determinism due to bags, as in \lamrfail) for which no reduction semantics is given. As a result, the encoding in~\cite{CairesP17} is different from ours, and is only shown to preserve typability:   properties such as operational completeness, operational soundness, and success sensitivity---important in our developments---are not considered.

\section{Concluding Remarks}
\label{ch2s:disc}

\subsubsection*{Summary}
We developed a correct encoding of \lamrfail, a new resource $\lambda$-calculus in which expressions feature non-determinism and explicit failure, into \spi, a session-typed $\pi$-calculus in which behavior is non-deterministically available: session protocols may perform as stipulated but also fail. 
Our encodability result is obtained by appealing to  \lamrsharfail, an intermediate language with a \emph{sharing construct} that simplifies the treatment of variables in expressions. 
To our knowledge, we are the first to relate typed $\lambda$-calculi and typed $\pi$-calculi encompassing non-determinism and   failures, while connecting intersection types and session types, two different mechanisms for resource-awareness in sequential and concurrent settings, respectively.  

 
\subsubsection*{Design of \lamrfail (and \lamrsharfail)}
\revd{}{The design of \lamrfail has been influenced by the logically justified treatment of non-determinism and explicit failure in \spi. Our correct encoding of \lamrfail into \spi makes this influence precise by connecting terms and processes but also their associated intersection types and linear logic propositions.}
We have also adopted features from previous resource $\lambda$-calculi, in particular those in~\cite{DBLP:conf/concur/Boudol93,DBLP:conf/birthday/BoudolL00,PaganiR10}.
Major similarities between \lamrfail and these calculi include: 
as in \cite{DBLP:conf/birthday/BoudolL00}, our semantics performs lazy evaluation and linear substitution on the head variable;  
as in~\cite{PaganiR10}, our reductions lead to non-deterministic sums.
A distinctive feature of \lamrfail    is its lazy treatment of failures via the   term $\fail^{\widetilde{x}}$. In contrast,  in~\cite{DBLP:conf/concur/Boudol93,DBLP:conf/birthday/BoudolL00} there is no dedicated term to represent failure. 
The non-collapsing semantics for non-de\-ter\-mi\-nism is another distinctive feature of \lamrfail. 

\srev{Our design for \lamrsharfail has been informed by the atomic $\lambda$-calculus introduced in~\cite{DBLP:conf/lics/GundersenHP13}. 
Also, our translation from \lamrfail into \lamrsharfail (\defref{ch2def:enctolamrsharfail}) borrows insights from translations given in~\cite{DBLP:conf/lics/GundersenHP13}.}
The calculus \lamrsharfail is also loosely related to the $\lambda$-calculus with sharing in~\cite{GhilezanILL11}, which considers (idempotent) intersection types.
Notice that the calculi in~\cite{DBLP:conf/lics/GundersenHP13,GhilezanILL11} do not consider explicit failure nor non-determinism.
We  distinguish between \emph{well-typed} and \emph{well-formed} expressions: this  allows us to make fail-prone evaluation in \lamrfail explicit. It is interesting that  explicit failures can be elegantly encoded as protocols in \spi---this way, we make the most out of \spi's expressivity.

Bags in \lamrfail have \emph{linear} resources, which are used exactly once. 
In \Cref{ch3}, we have defined an extension of \lamrfail in which bags contain both linear and  \emph{unrestricted} resources, as in~\cite{PaganiR10}, and 
established that our approach to encodability into \spi extends to such an enriched language. This development requires  the full typed process framework in~\cite{CairesP17}, with replicated processes and labeled choices (not needed to encode \lamrfail). 

\subsubsection*{Future Work}

The approach and results developed here enable us to tackle open questions that go beyond the scope of this work. We comment on some of them:
\begin{itemize}
	\item 
It would be useful to investigate the \emph{relative expressiveness} of \lamrfail with respect to other resource calculi, such as those in~\cite{DBLP:conf/birthday/BoudolL00,PaganiR10}. Derived 
encodability (and non-encodability) results could potentially unlock transfer of reasoning techniques between different calculi.
\item \revd{}{Besides transfer of techniques, one application of encodings between sequential and concurrent calculi is in the design of functional concurrent languages with advanced features. In this respect, it should be feasible to develop a variant of Wadler's GV~(\cite{DBLP:conf/icfp/Wadler12}) with non-determinism, resources, explicit failure, and session communication by exploiting our correct encodings from \lamrfail to \spi.}

 \item \revdaniele{It would be relevant to investigate \emph{decidability properties} of the intersection type systems for $\lamrfail$ and  $ \lamrsharfail $. Our translation is proven correct under the assumption that we consider only well-formed $\lamrfail$-terms.  The type assignment problem for intersection type systems is, in general, undecidable~\cite{DBLP:conf/popl/Leivant83}; it would be interesting to consider decidable fragments of intersection type systems via, for instance,  ranking restrictions~\cite{DBLP:journals/tcs/Bakel95}.}
\item It would be insightful to establish \emph{full abstraction}   for our translation of  \lamrfail into \spi. 
We choose not to consider it because,  as argued in~\cite{DBLP:journals/mscs/GorlaN16}, full abstraction is not an informative criterion when it comes to an encoding's quality. Establishing full abstraction requires developing the behavioral theory of \lamrfail and \spi, which is relevant and challenging in itself.
\end{itemize}

\clearemptydoublepage






\chapter{Unrestricted Resources in Encoding  Functions as Processes}
\label{ch3}

\chaptermark{Types and Terms Translated}

Type-preserving translations are effective rigorous tools in the study of core programming calculi. 
In this chapter, we develop a new typed translation that connects sequential and concurrent calculi; it is governed by type systems that control \emph{resource consumption}. 
Our main contribution is the source language, a  new resource $\lambda$-calculus with non-determinism and failures, dubbed  \lamrfailunres. 
In \lamrfailunres, resources are split into linear and unrestricted; failures are explicit and arise from this distinction. 
We define a type system  based on  intersection types
to control resources and fail-prone computation.
The target language is \spi, an existing session-typed $\pi$-calculus that results from a Curry-Howard correspondence between linear logic and session types. 
Our typed translation subsumes our prior work; interestingly, it treats  unrestricted resources in \lamrfailunres as client-server session behaviours in \spi.

\section{Introduction}

\subparagraph{Context}

\emph{Type-preserving translations} are effective rigorous tools in the study of core programming calculi. 
They can be seen as an abstract counterpart to the type-preserving compilers that enable key optimisations in the implementation of programming languages.
The goal of this chapter is to develop a new typed translation that connects sequential and concurrent calculi, and is governed by  type systems that control \emph{resource consumption}.

A central idea in the {resource} $\lambda$-calculus is to consider that in an application $M\, N$ the argument $N$ is a \emph{resource} of possibly limited availability. 
This generalisation of the $\lambda$-calculus triggers many fascinating questions, such as typability, solvability, expressiveness power, etc.,  which have been studied in different settings (see, e.g.,~\cite{DBLP:conf/concur/Boudol93,DBLP:conf/birthday/BoudolL00,PaganiR10,DBLP:journals/corr/abs-1211-4097}). 

In established resource $\lambda$-calculi, such as those by \cite{DBLP:conf/concur/Boudol93} and by \cite{PaganiR10}, a more general  form of application is considered:  a term can be applied to a bag of resources $B=\bag{N_1}\cdot \ldots \cdot \bag{N_k}$, where  $N_1, \ldots, N_k$ denote terms; then, an application $M\ B$ must take into account that each $N_i$ may be reusable or not.  Thus, non-determinism is natural in resource $\lambda$-calculi, because a  term has now  multiple ways of consuming resources from the bag. This bears a strong resemblance with  process calculi such as the $\pi$-calculus~\cite{DBLP:journals/iandc/MilnerPW92a}, in which concurrent interactions are intrinsically non-deterministic.

{There are different flavors of non-determinism. Over two decades ago, \cite{BoudolL96,DBLP:conf/birthday/BoudolL00} explored connections between a resource $\lambda$-calculus and the $\pi$-calculus. In their setting, an application $M\ B$ would branch, i.e., $M$ could consume a resource $N_j$ in $B$ (with $j \in \{1, \ldots k\}$) and discard the other $k-1$ resources in a non-confluent manner; this is what we call a {\em collapsing} approach to non-determinism.
On a different direction, 
\cite{PaganiR10} proposed $\lambda^r$, a resource $\lambda$-calculus that implements  \emph{non-collapsing}  non-determinism, whereby all the possible alternatives for resource consumption are retained together in a sum, ensuring confluence. They investigated typability and characterisations of solvability in $\lambda^r$, but no connection with the $\pi$-calculus was established. In an attempt to address this gap, \Cref{ch2} identified $\lamrfail$, a resource $\lambda$-calculus with non-collapsing non-determinism, explicit failure, and \emph{linear} resources  (to be used exactly once), and developed a correct typed translation into a session typed $\pi$-calculus~\cite{CairesP17}. } The calculus~$\lamrfail$, however, does not include \emph{unrestricted} resources (to be used zero or many times).

\subparagraph{Chapter Overview}
Here we introduce a new $\lambda$-calculus, dubbed $\lamrfailunres$, its intersection type system, and its translation into session-typed processes.
Our motivation is twofold: to elucidate the status of unrestricted resources in a functional setting with non-collapsing non-determinism, and to characterise unrestricted resources within a translation of functions into processes.
Unlike its predecessors, \lamrfailunres distinguishes between  {linear} and {unrestricted} resources. This distinction determines the semantics of terms and especially the deadlocks (\emph{failures}) that arise due to mismatches in resources. This way,  $\lamrfailunres$ subsumes   $\lamrfail$, which is  purely linear  and cannot express failures related to unrestricted resources.

Distinguishing  linear and unrestricted resources is not a new insight. This idea goes back to Boudol's $\lambda$-calculus with multiplicities~(\cite{DBLP:conf/concur/Boudol93}), where arguments can be tagged as unrestricted. 
What is new about $\lamrfailunres$ is that the distinction between linear and unrestricted resources leads to two main differences. 
First,  occurrences of a variable can be linear or unrestricted, depending on the kind of resources they should be substituted with. This way, e.g., a linear occurrence of variable must be substituted with a linear resource. 
In $\lamrfailunres$, a variable can have linear and unrestricted occurrences in the same term.
(Notice that we use the adjective `{linear}' in connection to resources used exactly once, and not to the number of occurrences of a variable in a term.)
Second, failures depend on the nature of the involved resource(s). In $\lamrfailunres$, a linear failure arises from a mismatch between required and available (linear) resources; an unrestricted failure arises when a specific (unrestricted) resource is not available. 

Accordingly, the syntax of $\lamrfailunres$ incorporates linear and unrestricted resources, enabling their consistent separation, within non-collapsing non-determinism. 
The calculus allows for linear and unrestricted occurrences of variables, as just discussed;  
bags comprise two separate zones, linear and unrestricted; 
and the \emph{failure term} $\fail^{x_1, \cdots, x_n}$ explicitly mentions the linear variables $x_1, \ldots, x_n$. 
The (lazy) reduction semantics of $\lamrfailunres$ includes two different rules for ``fetching'' terms from bags, and for consistently handling the failure term. 
\alerthide{reviewer here wants non-deterministic sums instead, I am unsure what you both think would be best here}

 We equip $\lamrfailunres$ with non-idempotent intersection types, extending the approach in~\Cref{ch2}:   in $\lamrfailunres$, intersection types  account for more than resource multiplicity, since the elements of the unrestricted bag can have different types. Using intersection types,  we define a class of \emph{well-formed} \lamrfailunres expressions, which includes terms that correctly consume resources but also terms that may reduce to the failure term. Well-formed expressions thus subsume the \emph{well-typed} expressions that can be defined in a sub-language of \lamrfailunres without the failure term.  

The calculus \lamrfailunres can express terms whose dynamic behaviour is not captured by prior works. This way, e.g., the  identity function ${\bf I}$ admits two formulations, depending on whether the variable occurrence  is linear or unrestricted. 
One can have $\lambda x. x$, as usual, but also the unrestricted variant $\lambda x. x[i]$, where `$[i]$' is an index annotation (similar to a qualifier or a tag), which indicates that $x$ should be replaced by the $i$-th element of the unrestricted zone of the bag. The behaviour of these functions will depend on the bags that are provided as their arguments.
Similarly,  we can express variants of $\Delta=\lambda x.xx$ and $\Omega=\Delta\,\Delta$ whose  behaviours again depend on linear or unrestricted occurrences of variables and bags. 
{Consider the term $\Delta_7=\lambda x. (x[2](\oneb \bagsep \bag{x[1]}^!\concat \bag{x[1]}^!))$, 
where we use `$\bagsep$' to separate linear and unrestricted resources in the bag, and `$\concat$' denotes concatenation of unrestricted resources. 
Term $\Delta_7$ is an abstraction on $x$ of an application of an unrestricted occurrence of $x$, which aims to consume the second component of an unrestricted bag, to a bag with an empty linear zone (denoted $\oneb$) and an unrestricted zone with resources $\bag{x[1]}^!$ and $\bag{x[1]}^!$. The self-application $\Delta_7\Delta_7$ produces a non-terminating behaviour and yet $ \Delta_7 $ itself is well-formed}  (see Example~\ref{ch3ex:delta7_wf1}).

  Both \lamrfailunres and \lamrfail are  \emph{logically motivated} resource $\lambda$-calculi, in the following sense: their design has been strongly influenced by $\spi$, a typed $\pi$-calculus resulting from the Curry-Howard correspondence between linear logic and session types in~\cite{CairesP17}, where proofs correspond to processes and cut elimination to process communication. 
As demonstrated in~\cite{CairesP17}, providing primitive support for explicit failures is key to expressing many useful programming idioms (such as exceptions); this insight is a leading motivation in our design for \lamrfailunres.


To attest to the logical underpinnings of \lamrfailunres, we develop a typed translation (or \emph{encoding}) of $\lamrfailunres$ into $\spi$ and establish its correctness with respect to well-established criteria \cite{DBLP:journals/iandc/Gorla10,DBLP:journals/iandc/KouzapasPY19}. 
As in~\\Cref{ch2}, we encode \lamrfail into \spi by relying on an intermediate language with \emph{sharing} constructs~\cite{DBLP:conf/lics/GundersenHP13,GhilezanILL11,DBLP:journals/iandc/KesnerL07}. 
A key idea in encoding \lamrfailunres is to codify the behaviour of unrestricted occurrences of a variable and their corresponding resources in the bag as \emph{client-server connections}, leveraging the copying semantics for the exponential ``$!A$'' induced by the Curry-Howard correspondence. 
This typed encoding into \spi  justifies the semantics of \lamrfailunres in terms of precise session protocols (i.e., linear logic propositions, because of the correspondence).

 In summary, the \textbf{main contributions} of this chapter are:
 (1)~The resource calculus \lamrfailunres of linear and unrestricted resources, and its associated intersection type system. 
    (2)~A typed encoding of \lamrfailunres into \spi,  which connects well-formed expressions (disciplined by intersection types) and well-typed concurrent processes (disciplined by session types,  under the Curry-Howard correspondence with linear logic), subsuming the results in~\Cref{ch2}.


\subparagraph{Additional Material}
\iffulldoc
The appendices contain omitted material. 
\appref{ch3appA} collects  technical details on \lamrfailunres. 
\appref{ch3appB} details the proof of subject reduction for well-formed \lamrfailunres expressions. \appref{ch3appC}--\appref{ch3app:encodingtwo} collect omitted definitions and proofs for our encoding of \lamrfailunres into \spi. 
\else
Technical details on \lamrfailunres, the proof of subject reduction for well-formed \lamrfailunres expressions and omitted definitions and proofs for our encoding of \lamrfailunres into \spi can be found in the full version.
\fi 

\section[Unrestricted Resources, Non-Determinism, and Failure]{\lamrfailunres: Unrestricted Resources, Non-Determinism, and Failure}
\label{ch3s:lambda}


\subparagraph{Syntax.} We shall use $x, y, \ldots$ to range over \emph{variables}, and $i,j\ldots$, as positive integers, to range over  {\em indices}.
Variable occurrences will be \emph{annotated} to distinguish the kind of resource  they should be substituted with (linear or unrestricted). 
With a slight abuse of terminology, we may write `linear variable' and `unrestricted variable' to refer to linear and unrestricted {occurrences} of a variable. 
As we will see, a variable's annotation will be inconsequential for binding purposes.
We write  $ {\widetilde{x}}$ to abbreviate  $ {x}_1, \ldots,  {x}_n$, for $n\geq 1$ and each $x_i$ distinct.

%

\begin{definition}{$\lamrfailunres$}\label{ch3def:rsyntaxfailunres}
We define \emph{terms} ($M,N$), \emph{bags} ($A,B$), and \emph{expressions} ($\expr{M}, \expr{N}$) as:
\begin{align}
&\mbox{(Annotations)}&[*]&::= [i] \sep [{\ell}] \qquad i\in \mathbb{N} \\
& \mbox{(Terms)} &M,N &::=   x[*] \sep \lambda x . M \sep (M\ B) \sep  M \esubst{B}{x}  \sep   \fail^{ {\widetilde{x}}}   \\
& \mbox{(Terms)} &M,N &::=  {x} \sep x[i] \sep \lambda x . M \sep (M\ B) \sep  M \esubst{B}{x}  \sep   \fail^{ {\widetilde{x}}}   \\
&\mbox{(Linear Bags)} &C, D &::= \oneb \sep \bag{M}  \cdot\, C 
\\
& \mbox{(Unrestricted Bags)} & U, V &::= \banged{\oneb} \sep \banged{\bag{M}} \sep U \concat V \\
& \mbox{(Bags)} &A, B&::=  C \bagsep U \\
&\mbox{(Expressions)} & \expr{M}, \expr{N} &::=  M \sep \expr{M}+\expr{N}
\end{align}
To lighten up notation, we shall omit the annotation for linear variables. This way, e.g., we write $(\lambda x. x)B$
rather than $(\lambda x. x[{\ell}])B$.
\end{definition}

\Cref{ch3def:rsyntaxfailunres} introduces three syntactic categories: \emph{terms} (in functional position); \emph{bags} (multisets of resources, in argument position), and \emph{expressions}, which are finite formal sums that denote possible results of a computation. Below we describe each category in details. 
\begin{itemize} 
\item Terms (unary expressions):
\begin{itemize}
\item Variables: We write $x[{\ell}]$ to denote a \emph{linear} occurrence of $x$, i.e,  an occurrence that can only be substituted for linear resources. Similarly, $x[i]$ denotes an \emph{unrestricted} occurrence of $x$, i.e., an occurrence that can only be substituted for a resource located at the $i$-th position of an unrestricted bag. 
\item Abstractions  $\lambda x. M$ of a variable $x$ in a term $M$, which may have contain linear or unrestricted occurrences of $x$. 
This way, e.g.,   $\lambda x.x$ and $\lambda x. x[i]$ are linear and unrestricted versions of the identity function. 
Notice that the scope of $x$ is $M$, as usual, and that $\lambda x. (\cdot)$ binds both linear and unrestricted occurrences of $x$. 
\item Applications  of a term $M$ to a bag $B$ (written $M\ B$) and the explicit substitution of a bag $B$ for a variable $x$ (written $\esubst{B}{x}$) are as expected (cf.~\cite{DBLP:conf/concur/Boudol93,DBLP:conf/birthday/BoudolL00}). 
Notice that in $M\esubst{B}{x}$ the occurrences of $x$ in $M$, linear and unrestricted, are bound.
Some conditions apply to $B$: this will be evident later on, after we define our operational semantics (cf. \figref{ch3fig:reductions_lamrfailunres}).
\item The failure term $\fail^{ {\widetilde{x}}}$ denotes a term that will  result from a reduction in which there is a lack or excess of resources, where  $ {\widetilde{x}}$ denotes a multiset of free linear variables that are encapsulated within failure. 
\end{itemize}

\item A bag $B$ is defined as $C\bagsep U$: the concatenation of a  bag of linear resources $C$ with a bag (actually, a list) of unrestricted resources $U$. We write $\bag{M}$  to denote the linear bag that encloses term $M$, and use $\bag{M}^!$ in the unrestricted case.
\begin{itemize}
\item Linear bags ($C, D, \ldots$) are multisets of terms. The empty linear bag is denoted~$\oneb$. We write $C_1 \cdot C_2$ to denote the concatenation of   $C_1$ and $C_2$; this is a commutative and associative operation, where $\oneb$ is the identity. 
\item Unrestricted bags ($U, V, \ldots$) are ordered lists of terms. 
The empty  unrestricted bag is denoted as $\banged{\oneb}$.
 The concatenation of $U_1$ and $U_2$ is denoted by $U_1\concat U_2$; this operation is  associative but not commutative. 
 Given $i \geq 1$, we write $U_i$ to denote the $i$-th element of the unrestricted (ordered) bag $U$.
 \end{itemize}
 \item  Expressions are sums of terms, denoted as $\sum_{i}^{n} N_i$, where $n > 0$. Sums are associative and commutative; reordering of the terms in a sum is performed silently. 
\end{itemize}
\begin{example}{}\label{ch3ex:var_lin_unr}
Consider the term  
 $M:=\lambda x.( x[1] \bag{x}\bagsep \bag{y[1]}^!)$,
 which has linear and unrestricted occurrences of the same variable.
  This  is an abstraction of an application that contains two bound occurrences of $x$ (one unrestricted with index $1$, and one linear) and one free unrestricted occurrence of $y[1]$, occurring in an unrestricted bag. 
 As we will see, in $M\ (C\bagsep U)$, the unrestricted occurrence `$x[1]$' should be replaced by the first element of $U$. 
 \end{example}


 The salient features of \lamrfailunres ---the explicit construct for failure, the index annotations on unrestricted variables,  the ordering of unrestricted bags---are \emph{design choices} that  will be responsible for interesting behaviours, as   the following examples  illustrate.

\begin{example}{}\label{ch3ex:id_term}
As already mentioned, 
    $\lamrfailunres$ admits different  variants of the usual $\lambda$-term ${\bf I}=\lambda x. x$.  We could have one in which $x$ is a linear  variable (i.e., $\lambda x. x$), but also several possibilities if $x$ is unrestricted (i.e., $\lambda x. x[i]$, for some positive integer $i$).
    Interestingly, because \lamrfailunres supports {failures}, {non-determinism}, and the {consumption} of arbitrary terms of the unrestricted bag, these two variants of ${\bf I}$ can have  behaviours that may differ from the usual interpretation of ${\bf I}$. In Example~\ref{ch3ex:id_sem} we will show that the six terms below give different behaviours:
    
\begin{minipage}{5cm}
  \begin{itemize}
        \item $M_1= (\lambda x. x )(\bag{N}\bagsep U)$
        \item $M_2= (\lambda x. x )(\bag{N_1}\cdot \bag{N_2}\bagsep U)$
        \item $M_3=(\lambda x. x[1] )(\bag{N}\bagsep \oneb^!)$
    \end{itemize}
\end{minipage} 
\hspace*{0.5cm}
\begin{minipage}{5cm}
  \begin{itemize}
        \item $M_4=(\lambda x. x[1] )( \oneb \bagsep \bag{N}^! \concat U)$
        \item $M_5=(\lambda x. x[1] )( \oneb \bagsep \oneb^! \concat U)$
        \item $M_6=(\lambda x. x[i] )( C\concat U)$
    \end{itemize}
\end{minipage}

\noindent We will see that $M_1$, $M_4$, $M_6$ reduce without failures, whereas $M_2$, $M_3$, $M_5$ reduce to failure. 
\end{example}


\begin{example}{}\label{ch3ex:delta}
Similarly, \lamrfailunres allows for several forms of the standard $\lambda$-terms such as $\Delta:=\lambda x. xx$ and $\Omega:=\Delta \Delta$, depending on whether the variable $x$ is linear or unrestricted:
\begin{enumerate}
 \item  $\Delta_1:= \lambda x. (x(\bag{x}\bagsep \oneb^!))$ consists of an abstraction of a linear occurrence of $x$ applied to a linear bag containing another linear occurrence of $x$. There are two forms of self-applications of $\Delta_1$, namely: $ \Delta_1(\bag{\Delta_1}\bagsep \oneb^!)$ and $ \Delta_1(1\bagsep \bag{\Delta_1}^!)$.
\item $\Delta_4:= \lambda x. (x[1](\bag{x}\bagsep \oneb^!))$ consists of an unrestricted occurrence of $x$ applied to a linear bag (containing a linear occurrence of $x$) that is composed with an empty unrestricted bag. Similarly, there are two self-applications of $\Delta_4$, namely: $ \Delta_4(\bag{\Delta_4}\bagsep \oneb^!)$ and $ \Delta_4(\oneb\bagsep \bag{\Delta_4}^!)$.

\item We show applications of  an unrestricted variable occurrence ($x[2]$ or $x[1]$) applied to an empty linear bag composed with a non-empty  unrestricted bag (of size two):

\begin{itemize}
    \item $\Delta_3= \lambda x.(x[1](\oneb\bagsep \bag{x[1]}^!\concat \bag{x[1]}^!))$
    \item $\Delta_5:= \lambda x. (x[2](\oneb\bagsep \bag{x[1]}^!\concat \bag{x[2]}^!))$ 
    \item $\Delta_6:= \lambda x.( x[1](\oneb\bagsep \bag{x[1]}^!\concat \bag{x[2]}^!))$ 
    \item  $\Delta_7:= \lambda x.( x[2](\oneb\bagsep \bag{x[1]}^!\concat \bag{x[1]}^!))$
\end{itemize}

Applications between these terms express behaviour, similar to a lazy evaluation of  $\Omega$: 

\begin{minipage}{6cm}
    \begin{itemize}
        \item $\Omega_5:=\Delta_5(\oneb\bagsep \bag{\Delta_5}^!\concat \bag{\Delta_5}^!)$
        \item $\Omega_{5,6}:=\Delta_5(\oneb\bagsep \bag{\Delta_5}^!\concat \bag{\Delta_6}^!)$
    \end{itemize}
\end{minipage} 
\hspace*{0.5cm}
\begin{minipage}{6cm}
    \begin{itemize}
        \item $\Omega_{6,5}:=\Delta_6(\oneb\bagsep \bag{\Delta_5}^!\concat \bag{\Delta_6}^!)$
        \item $\Omega_7:=\Delta_7(\oneb\bagsep \bag{\Delta_7}^!\concat \bag{\Delta_7}^!)$
    \end{itemize}
\end{minipage}

\end{enumerate}
The behaviour of these terms will be made explicit later on (see Examples~\ref{ch3ex:deltasem} and \ref{ch3ex:deltasemii}).
\end{example}
\subparagraph{Semantics.}
The semantics of \lamrfailunres captures that linear resources can be used only once, and that unrestricted resources can be used {\em ad libitum}. Thus, the evaluation of a function applied to a multiset of linear resources produces 
 different possible behaviours, depending on the way these resources are substituted for the linear variables. This induces non-determinism, which we formalise using a \emph{non-collapsing} approach, in which  expressions keep all the different possibilities open, and do not commit to one of them. This is in contrast to \emph{collapsing} non-determinism, in which selecting one alternative discards the rest.
 
We define a reduction relation $\redd$,   which operates lazily on expressions. Informally, a $\beta$-reduction induces an explicit substitution of a bag $B=C\bagsep U$ for a variable $x$, denoted $\esubst{B}{x}$, in a term $M$. 
This explicit substitution is then expanded 
depending on whether the head of $M$
 has a linear or an unrestricted variable. 
 Accordingly, in \lamrfailunres there are \emph{two sources of failure}: one concerns mismatches on linear resources (required vs available resources); the other concerns the unavailability of a required unrestricted resource (an empty bag $\banged{\oneb}$).
 
To formalise reduction, we require a few auxiliary notions.

\begin{definition}{}\label{ch3d:fvars}
The multiset of free linear variables of  $\mathbb{M}$, denoted $\mlfv{\mathbb{M}}$,  is defined below.
We denote by $[ {x}]$ the multiset containing the linear variable $x$ and $[x_1,\ldots, x_n]$ denotes the multiset containing $x_1,\ldots, x_n$. We write $\widetilde{x}\uplus \widetilde{y}$ to denote the multiset union of $\widetilde{x}$, and $\widetilde{y}$ and $\widetilde{x} \setminus y$ to express that every occurrence of $y$ is removed from $\widetilde{x}$.
{\normalsize
\begin{align*}
\mlfv{ {x}}  &= [  {x} ] &\mlfv{{x}[i]} &= \mlfv{ {\oneb}}  = \emptyset\\
  \mlfv{C \bagsep U}  &= \mlfv{C} &  \mlfv{M\ B}  &=  \mlfv{M} \uplus \mlfv{B}   \\
     \mlfv{ {\bag{M}}}  &= \mlfv{M} & \mlfv{\lambda x . M} & = \mlfv{M}\!\setminus\! \{  {x} \}\\ 
    \mlfv{M \esubst{B}{x}}  &= (\mlfv{M}\setminus \{  {x} \}) \uplus \mlfv{B} & \mlfv{ {\bag{M}}  \cdot C} &= \mlfv{ {M}} \uplus \mlfv{ C} \\
    \mlfv{\expr{M}+\expr{N}}  &= \mlfv{\expr{M}} \uplus \mlfv{\expr{N}} & \mlfv{\fail^{ {x}_1, \cdots ,  {x}_n}}  &= [  {x}_1, \ldots ,  {x}_n ]
\end{align*}
}
\normalsize
A term $M$ (resp. expression $\expr{M}$) is called  \emph{linearly closed} if $\mlfv{M} = \emptyset$ (resp. $\mlfv{\expr{M}} = \emptyset$).
\end{definition}

\begin{notation}
We shall use the following notations. 
\begin{itemize}
\item 
$N \in \expr{M}$ means that 
$N$ occurs in the sum  $\expr{M}$. 
Also, we write $N_i \in C$ to denote that $N_i$ occurs in the linear bag $C$, and $C \setminus N_i$ to denote the linear bag   obtained by removing one occurrence of  $N_i$ from $C$.
\item  $\#( {x}, M)$ denotes the number of (free) linear occurrences of $x$ in $M$. 
Also, $\#(x,\widetilde{y}) $ denotes the number of occurrences of $x$ in the multiset $\widetilde{y}$. 
\item 
   $\perm{C}$  is the set of all permutations of a linear  bag $C$ and $C_i(n)$ denotes the $n$-th term in the (permuted)  $C_i$.
\item $\size{C}$ denotes the number of terms in a linear bag $C$. 
That is, $\size{\oneb} = 0$
and 
$\size{\bag{M}  \cdot\, C} = 1 + \size{C}$. Given a bag $B = C \bagsep U$, we define  $\size{B}$ as $\size{C}$.
\end{itemize}
\end{notation}

\begin{definition}{Head}
\label{ch3def:headfailure}
Given a term $M$, we define $\headf{M}$ inductively as:
{\small \[  
    \begin{aligned}
        &
        \begin{aligned}
            \headf{x}  &= x & \headf{M\ B}  &= \headf{M}& \headf{\lambda x.M}  &= \lambda x.M \\
            \headf{x[i]}  &= x[i]  & \headf{\fail^{\widetilde{x}}}  &= \fail^{\widetilde{x}}& \\
        \end{aligned} \\
        &
             \headf{M \esubst{ B }{x}} =
        \begin{cases}
            \headf{M} & \text{if $\#(x,M) = \size{B}$}\\
            \fail^{\emptyset} & \text{otherwise}
        \end{cases}
    \end{aligned}
\]
}
\end{definition}
\begin{definition}{Head Substitution}
\label{ch3def:linsubfail}
Let $M$ be a term such that $\headf{M}=x$. 
The \emph{head substitution} of a term $N$ for $x$ in $M$, denoted  $M\headlin{ N/x }$,  is inductively  defined as follows (where $ x \not = y$):
\[
\begin{aligned}
&x \headlin{ N / x}   = N 
\hspace{4mm} (M\ B)\headlin{ N/x}  = (M \headlin{ N/x })\ B \\
&   (M\ \esubst{B}{y})\headlin{ N/x }  = (M\headlin{ N/x })\ \esubst{B}{y} 
\\[1mm]
\end{aligned}
\]
\end{definition}
\noindent When $\headf{M}=x[i]$, the head substitution   $M\headlin{N/x[i]}$ works as expected: $x[i]\headlin{N/x[i]}=N$ as the base case of the definition. 
Finally, we define contexts for terms and expressions:

\begin{definition}{Evaluation Contexts}\label{ch3def:context_lamrfail}
Contexts for terms (CTerm) and expressions (CExpr) are defined by the following grammar:
\[
\begin{array}{l@{\hspace{1cm}}l}
 \text{(CTerm)}\quad  C[\cdot] ,  C'[\cdot] ::=  ([\cdot])B \mid ([\cdot])\esubst{B}{x} &
 \text{(CExpr)}  \quad   D[\cdot] , D'[\cdot] ::= M + [\cdot] 
\end{array}
\]
\end{definition}

\begin{figure*}[!t]
	    \centering
	    \smallskip
	    \small
\begin{prooftree}
        \AxiomC{\ }
        \noLine
        \UnaryInfC{\ }
        \LeftLabel{\redlab{R:Beta}}
        \UnaryInfC{\((\lambda x. M) B \redd M\esubst{B}{x}\)}
\end{prooftree}
\begin{prooftree}
\AxiomC{$\headf{M} =  {x}$}
    \AxiomC{$C = {\bag{N_1}}\cdot \dots \cdot {\bag{N_k}} \ , \ k\geq 1 $}
    \AxiomC{$ \#( {x},M) = k $}
    \LeftLabel{\redlab{R:Fetch^{\ell}}}
    \TrinaryInfC{\(
    M\esubst{ C \bagsep U  }{x } \redd M \headlin{ N_{1}/ {x} } \esubst{ (C \setminus N_1)\bagsep U}{ x }  + \cdots + M \headlin{ N_{k}/ {x} } \esubst{ (C \setminus N_k)\bagsep U}{x}
    \)}
\end{prooftree}
\begin{prooftree}
    \AxiomC{$\headf{M} = {x}[i]
 \quad \#( {x},M) = \size{C}$}
    \AxiomC{$ U_i = \banged{\bag{N}}$}
    \LeftLabel{\redlab{R:Fetch^!}}
    \BinaryInfC{\(
    M\ \esubst{ C \bagsep U  }{x } \redd M \headlin{ N/{x}[i] } \esubst{ C \bagsep U}{ x } 
    \)}
    \end{prooftree}

    \begin{prooftree}
\AxiomC{$\#( {x},M) \neq \size{C} \quad \widetilde{y} = (\mlfv{M} \!\setminus x) \uplus \mlfv{C} $}
    \LeftLabel{\redlab{R:Fail^{ \ell }}}
    \UnaryInfC{\(  M\esubst{C \bagsep U}{x } \redd {}  \displaystyle\sum_{\perm{C}}\!\!\! \fail^{\widetilde{y}} \)}
\end{prooftree}
\vspace{-0.7cm}

\begin{prooftree}
    \AxiomC{$\#( {x},M) = \size{C}$}
    \AxiomC{$U_i = \banged{\oneb} \quad \headf{M} = {x}[i]  $}
    \LeftLabel{\redlab{R:Fail^!}}
    \BinaryInfC{\(  M \esubst{C \bagsep U}{x } \!\redd  \! M \headlin{ \fail^{\emptyset} /{x}[i] } \esubst{ C \bagsep U}{ x }\)}
\end{prooftree}

\begin{prooftree}
   \AxiomC{$\widetilde{y} = \mlfv{C} $}
    \LeftLabel{$\redlab{R:Cons_1}$}
    \UnaryInfC{\(  (\fail^{\widetilde{x}})\ C \bagsep U \redd {}  \displaystyle\sum_{\perm{C}} \fail^{\widetilde{x} \uplus \widetilde{y}} \)}
\end{prooftree}
\vspace{-0.7cm}

\begin{prooftree}
    \AxiomC{$ \#(z , \widetilde{x}) =  \size{C} \  \widetilde{y} = \mlfv{C} $}
    \LeftLabel{$\redlab{R:Cons_2}$}
    \UnaryInfC{$\fail^{\widetilde{x}}\ \esubst{C \bagsep U}{z}  \redd {}\displaystyle\sum_{\perm{C}} \fail^{(\widetilde{x} \setminus z) \uplus\widetilde{y}}$}
\end{prooftree}
\vspace{-0.7cm}
\begin{prooftree}
        \AxiomC{$ \expr{M}  \redd \expr{M}'  $}
        \LeftLabel{\redlab{R:ECont}}
        \UnaryInfC{$D[\expr{M}]  \redd D[\expr{M}']  $}
        \DisplayProof
        \qquad 
    \AxiomC{$   M \redd   \sum_{i=1}^k M'_{i}  $}
        \LeftLabel{\redlab{R:TCont}}
        \UnaryInfC{$ C[M] \redd  \sum_{i=1}^k C[M'_{i}] $}
\end{prooftree}
    \caption{Reduction rules for $\lamrfailunres$.}
    \label{ch3fig:reductions_lamrfailunres}
    \end{figure*}

\smallskip
\noindent
Reduction  is defined by the rules in \figref{ch3fig:reductions_lamrfailunres}.
 Rule~$\redlab{R:Beta}$ induces explicit substitutions.
Resource consumption is implemented by two fetch rules, which open up explicit substitutions:
\begin{itemize}
	\item 
Rule~$\redlab{R:Fetch^{\ell}}$, the \emph{linear fetch}, ensures that the number of required resources matches the size of the linear bag $C$.
It induces a sum of terms with head substitutions, each denoting the partial evaluation of an element from $C$. 
Thus, the size of $C$ determines the summands in the resulting expression.
\item 
 Rule~$\redlab{R:Fetch^!}$, the \emph{unrestricted fetch}, consumes a resource occurring in a specific position of the unrestricted bag $U$ via a linear head substitution of an unrestricted variable occurring in the head of the term. In this case, reduction results in an explicit substitution with $U$ kept unaltered. Note that we check for the size of the linear bag $C$: in the case $\#(x,M)\neq \size{C}$, the term evolves to a linear failure via Rule~$\redlab{R:\fail^\ell}$ (see Example~\ref{ch3ex:sizeC}). This is another design choice: linear failure is prioritised in $\lamrfailunres$.
 \end{itemize}

Four rules show reduction to failure terms, and accumulate free variables involved in failed reductions.
Rules~$\redlab{R:Fail^{\ell}}$ and $\redlab{R:Fail^!}$ formalise the failure to evaluate an explicit substitution $M\esubst{C\bagsep U}{x}$.
The former rule targets a linear failure, which  occurs when the size of $C$ does not match the number of occurrences of ${x}$. The multiset $\widetilde{y}$ preserves all free linear variables in $M$ and $C$. 
The latter rule targets an \emph{unrestricted failure}, which occurs when the head of the term is $x[i]$ and $U_i$ (i.e., the $i$-th element of $U$) is empty. 
In this case, failure preserves the free linear variables in $M$ and $C$ excluding the head unrestricted occurrence $x[i]$ which is replaced by $\fail^\emptyset$.

Rules~$\redlab{R:Cons_1}$ and~$\redlab{R:Cons_2}$ describe reductions that lazily consume the failure term, when a term has $\fail^{\widetilde{x}}$ at its head position. 
The former rule consumes bags attached to it whilst preserving all its free linear variables; the latter rule consumes explicit substitution attached to it whilst also preserving all its free linear variables. The side condition $\#(z,\widetilde{x})=\size{C}$ is necessary in Rule~$\redlab{R:Cons_2}$ to avoid a clash with the premise of Rule~$\redlab{R:Fail^{\ell}}$.
Finally, 
Rules $\redlab{R:ECont}$ and $\redlab{R:TCont}$ state  closure by  the $C$ and $D$ contexts (cf. \defref{ch3def:context_lamrfail}).


    Notice that the left-hand sides of the reduction rules in $\lamrfailunres$  do not interfere with each other. 
As a result, reduction in \lamrfailunres satisfies a \emph{diamond property}: for all $\expr{M}\in \lamrfailunres$, if there exist $\expr{M}_1,\expr{M}_2\in \lamrfailunres$ such that $\expr{M}\redd \expr{M}_1$ and  $\expr{M}\redd \expr{M}_2$, then there exists $\expr{N}\in \lamrfailunres$ such that $\expr{M}_1\redd \expr{N}\longleftarrow \expr{M}_2$ 
\iffulldoc
(see \appref{ch3appA}).
\else
(see the full version for further details).
\fi

\begin{notation}
As usual,  
  $\redd^*$ denotes  the reflexive-transitive closure of $\redd$. 
We write $\expr{N}\redd_{\redlab{R}} \expr{M}$ to denote that $\redlab{R}$ is the last (non-contextual) rule used in the step from $\expr{N}$ to $\expr{M}$.
\end{notation}


\begin{example}{Cont. Example~\ref{ch3ex:id_term}}
\label{ch3ex:id_sem}
We illustrate different reductions for $\lambda x. x$ and $\lambda x. x[i]$.  
\begin{enumerate}
    \item $M_1= (\lambda x. x )(\bag{N}\bagsep U)$ concerns a linear variable $x$ with an linear bag containing one element. This is similar to the usual {meaning} of applying an identity function to a term:
    
\(
\begin{aligned} 
(\lambda x. x )(\bag{N}\bagsep U)&\redd_{\redlab{R:Beta}} x\esubst{\bag{N}\bagsep U}{x}\\
& \redd_{\redlab{R:Fetch^{\ell}}} x\headlin{N/x}\esubst{\oneb\bagsep U}{x}=N\esubst{\oneb\bagsep U}{x},    
 \end{aligned}
\)

with a ``garbage collector'' that collects unused unrestricted resources.
    \item $M_2= (\lambda x. x )(\bag{N_1}\cdot \bag{N_2}\bagsep U)$ concerns the case in which a linear variable $x$ has a single occurrence but the linear bag has size two. Term $M_2$ reduces to a sum of failure terms:
     
     \(\begin{aligned} 
    (\lambda x. x )(\bag{N_1}\cdot \bag{N_2}\bagsep U)&\redd_{\redlab{R:Beta}} x\esubst{\bag{N_1}\cdot \bag{N_2}\bagsep U}{x}\\
    &\redd_{\redlab{R:Fail^{\ell}}} \sum_{\perm{C}}\fail^{\widetilde{y}}
    \end{aligned}\)
    
    for $C=\bag{N_1}\cdot \bag{N_2}$ and $\widetilde{y}=\mlfv{C}$.
    
    \item $M_3=(\lambda x. x[1] )(\bag{N}\bagsep \oneb^!)$ represents  an abstraction of an unrestricted variable, which aims to consume the first element of the unrestricted bag. Because this bag is empty, $M_3$ reduces to failure:
       
        \(\begin{aligned}
    (\lambda x. x[1])(\bag{N}\bagsep \oneb^!)&\redd_{\redlab{R:Beta}}x[1]\esubst{\bag{N}\bagsep \oneb^!}{x}\redd_{\redlab{R:\fail^{\ell}}}\fail^{\widetilde{y}},
    \end{aligned}\)
    
        for   $\widetilde{y}=\mlfv{N}$. Notice that $0= \#(x,x[1])\neq \size{\bag{N}}=1$, since there are no linear occurrences of  $x$ in $x[1]$.

\end{enumerate}

\end{example}

\begin{example}{}
\label{ch3ex:sizeC}
To illustrate the need to check `$\size{C}$' in $\redlab{R:Fail^!}$, consider the term $ x[1] \esubst{\bag{M} \bagsep \oneb^!}{x}$, which features both a mismatch of linear bags for the linear variables to be substituted and an empty unrestricted bag with the need for the first element to be substituted. 
We check the size of the linear bag because we wish to prioritise the reduction of Rule~$\redlab{R:Fail^{\ell}}$. Hence, in case of a mismatch of linear resources  we wish not to  perform a reduction via Rule~$\redlab{R:Fail^!}$.
 This is a design choice: our semantics collapses linear failure at the earliest moment it arises.
\end{example}



    
    \begin{example}{Cont. Example~\ref{ch3ex:delta}}
    \label{ch3ex:deltasem}
    Self-applications of $\Delta_1$ do not behave as an expected variation of a lazy reduction from $\Omega$. Both
    $ \Delta_1(\bag{\Delta_1}\bagsep \oneb^!)$ and  $ \Delta_1(\oneb \bagsep \bag{\Delta_1}^!)$ reduce to failure  since the number of linear occurrences of $x$ does not match the number of  resources in the linear bag:
    \(
    \begin{aligned}
     \Delta_1(\bag{\Delta_1}\bagsep \oneb^!) 
     \redd(x(\bag{x}\bagsep \oneb))\esubst{\bag{\Delta_1}\bagsep \oneb^!}{x} \redd\fail^\emptyset.
     \end{aligned}
     \)
     
     The term $\Delta_4(\oneb \bagsep \bag{\Delta_4}^!)$ also fails: the linear bag is empty and there is one linear occurrence of $x$ in $\Delta_4$.  Note that $\Delta_4(\bag{\Delta_4}\bagsep \bag{\Delta_4}^!)$ reduces to another application of $\Delta_4$ before failing:
     {
     \[
     \begin{aligned}
     \Delta_4(\bag{\Delta_4}\bagsep \bag{\Delta_4}^!) &= (\lambda x. (x[1] (\bag{x}\bagsep \oneb^!)))(\bag{\Delta_4}\bagsep \bag{\Delta_4}^!)\\
     &\redd_{\redlab{R{:}Beta}} (x[1] (\bag{x}\bagsep \oneb^!))\esubst{\bag{\Delta_4}\bagsep \bag{\Delta_4}^!}{x}\\
     &\redd_{\redlab{R{:}Fetch^!}} (\Delta_4 (\bag{x}\bagsep \oneb^!))\esubst{\bag{\Delta_4}\bagsep \bag{\Delta_4}^!}{x}\\
     &\redd^*    \fail^\emptyset \esubst{\bag{x}\bagsep \oneb^!}{y}\esubst{\bag{\Delta_4}\bagsep \bag{\Delta_4}^!}{x}
     \end{aligned}
     \]
     }
     \end{example} 
\noindent Differently from~\Cref{ch2}, there are terms in $\lamrfailunres$ that when applied to each other behave similarly to $\Omega$, namely $\Omega_{5,6}$, $\Omega_{6,5}$, and $\Omega_7$ (Example~\ref{ch3ex:delta}).

\begin{example}{Cont. Example~\ref{ch3ex:delta}}

\label{ch3ex:deltasemii}
The following reductions illustrate different behaviours  provided that subtle changes are made within $\lamrfailunres$-terms:
\begin{itemize}

\item An interesting behaviour of $\lamrfailunres$ is that variations of $\Delta$ can be applied to each other and appear alternately (highlighted in \textcolor{blue}{blue}) in the functional position throughout the computation---this behaviour is illustrated in \figref{ch3fig:diag_delta56}:

  \(
    \small \hspace{-1cm}
  \begin{aligned}
  \Omega_{5,6} &= \Delta_5 ( \oneb\bagsep \banged{\bag{\Delta_5}} \concat \banged{\bag{\Delta_6}} )\\
 & =(\lambda x. ( x[2] ( 1\bagsep \banged{\bag{x[1]}} \concat \banged{\bag{x[2]}})))
  ( 1\bagsep \banged{\bag{\Delta_5}} \concat \banged{\bag{\Delta_6}})\\ 
&  \redd_\redlab{R{:}Beta} (x[2] ( \oneb\bagsep \banged{\bag{x[1]}} \concat \banged{\bag{x[2]}})) \esubst{\oneb\bagsep \banged{\bag{\Delta_5}} \concat \banged{\bag{\Delta_6}} }{ x}  \\
    &\redd_\redlab{R:Fetch^!} (\colthree{\Delta_6} ( {\oneb\bagsep \banged{\bag{x[1]}} \concat \banged{\bag{x[2]}}})) \esubst{ \oneb\bagsep \banged{\bag{\Delta_5}} \concat \banged{\bag{\Delta_6}} }{ x} \\ 
     &\redd_\redlab{R{:}Beta} (y[1](\oneb\bagsep \bag{y[1]}^!\concat \bag{y[2]}^!) \esubst{(\oneb\bagsep \banged{\bag{x[1]}} \concat \banged{\bag{x[2]}})}{y} \esubst{\oneb\bagsep \banged{\bag{\Delta_5}} \concat \banged{\bag{\Delta_6}} }{ x }\\ 
    &\redd_\redlab{R{:}Fetch^!} (x[1](1\bagsep \bag{y[1]}^!\concat \bag{y[2]}^!) \esubst{(\oneb\bagsep \banged{\bag{x[1]}} \concat \banged{\bag{x[2]}})}{y} \esubst{\oneb\bagsep \banged{\bag{\Delta_5}} \concat \banged{\bag{\Delta_6}}}{ x}   \\
     &\redd_\redlab{R{:}Fetch^!} (\colthree{\Delta_5}(1\bagsep \bag{y[1]}^!\concat \bag{y[2]}^!) \esubst{(\oneb\bagsep \banged{\bag{x[1]}} \concat \banged{\bag{x[2]}})}{y} \esubst{\oneb\bagsep \banged{\bag{\Delta_5}} \concat \banged{\bag{\Delta_6}}}{ x}   \\
&\redd \ldots
        \end{aligned}
        \)
 \begin{figure}[!t]
\begin{mdframed}

\begin{center}
\begin{tikzpicture}[scale=0.8]
\draw[->] (2,1.3) --node[anchor= south,xshift=1.75mm,yshift=-0.5mm] {*}  (4.5,2.7); 
\draw[->] (4.75,2.7) --node[anchor= south,xshift=1.75mm,yshift=-1.75mm] {*} (7.75,1.3); 
\node[right] at (-1,1){$\Delta_{5}( \oneb\bagsep \banged{\bag{\Delta_5}} \concat \banged{\bag{\Delta_6}})$};
\node at (7.0,1){$\Delta_{5}\ldots \esubst{\oneb\bagsep \banged{\bag{\Delta_5}} \concat \banged{\bag{\Delta_6}}}{x}$};
\draw[->]  (8.4,1.3) --node[anchor= south,xshift=1.75mm,yshift=-0.5mm] {*} (11.5,2.7); 
\node at (4,3){$\Delta_{6}\ldots  \esubst{\oneb\bagsep \banged{\bag{\Delta_5}} \concat \banged{\bag{\Delta_6}}}{x}$};
\node at (11,3){$\Delta_{6}\ldots \esubst{\oneb\bagsep \banged{\bag{\Delta_5}} \concat \banged{\bag{\Delta_6}}}{x}$};
\draw[->]  (11.7,2.7)--node[anchor= south,xshift=1.75mm,yshift=-1.75mm] {*} (15,1.3); 
\node at (15,1) {\ldots };
\end{tikzpicture}
\end{center}
\end{mdframed}
\caption{An $\Omega$-like behaviour in $\lamrfailunres$ (cf. Example~\ref{ch3ex:deltasemii}).}\label{ch3fig:diag_delta56}
\end{figure}       
  \item Applications of $\Delta_7$ into two unrestricted copies of $\Delta_7$
  behave as $\Omega$ producing a non-terminating behaviour.
  Letting $B = 1\bagsep \bag{x[1]}^!\concat \bag{x[1]}^!$, we have:
  
\(
\begin{aligned} 
&\Omega_7=(\lambda x. (x[2](1\bagsep \bag{x[1]}^!\concat \bag{x[1]}^!)))(1\bagsep \bag{\Delta_7}^!\concat \bag{\Delta_7}^!)\\
&\redd_{\redlab{R{:}Beta}} (x[2](1\bagsep \bag{x[1]}^!\concat \bag{x[1]}^!))\esubst{1\bagsep \bag{\Delta_7}^!\concat \bag{\Delta_7}^!}{x}\\
&\redd_{\redlab{R{:}Fetch^!}} (\Delta_7({1\bagsep \bag{x[1]}^!\concat \bag{x[1]}^!}))\esubst{1\bagsep \bag{\Delta_7}^!\concat \bag{\Delta_7}^!}{x}\\
&\redd_{\redlab{R{:}Beta}} (y[2](1\bagsep \bag{y[1]}^!\concat \bag{y[1]}^!))\esubst{B}{y}\esubst{1\bagsep \bag{\Delta_7}^!\concat \bag{\Delta_7}^!}{x}\\
&\redd_{\redlab{R{:}Fetch^!}} (x[1](1\bagsep \bag{y[1]}^!\concat \bag{y[1]}^!))\esubst{B}{y}\esubst{1\bagsep \bag{\Delta_7}^!\concat \bag{\Delta_7}^!}{x}\\
&\redd_{\redlab{R{:}Fetch^!}} (\Delta_7(1\bagsep \bag{y[1]}^!\concat \bag{y[1]}^!))\esubst{B}{y}\esubst{1\bagsep \bag{\Delta_7}^!\concat \bag{\Delta_7}^!}{x}  \\
&\redd \ldots
\end{aligned}
\) 

Later on we will show that this term is well-formed (see Example~\ref{ch3ex:delta7_wf1}) with respect to the intersection type system introduced in \secref{ch3sec:lam_types}.
\end{itemize}
\end{example}


\section[Intersection Types]{Well-Formed Expressions via Intersection Types}\label{ch3sec:lam_types}

We define  \emph{well-formed} $\lamrfailunres$-expressions by relying on a non-idempotent intersection type system,  based on the system by \cite{BucciarelliKV17}.
Our system for well-formed expressions  subsumes the one in \Cref{ch2} it uses {\em strict} and {\em multiset} types to check  linear bags;  moreover, it uses  {\em list} and {\em tuple} types to check unrestricted bags. 
As in \Cref{ch2}, we write
``well-formedness'' (of terms, bags, and expressions) to stress that, unlike usual type systems, our system can account for terms that may reduce to the failure term (cf. Remark~\ref{ch3r:types}).

\begin{definition}{Types for \lamrfailunres}
\label{ch3d:typeslamrfail}
We define {\em strict}, {\em multiset}, \emph{list}, and \emph{tuple types}.
\[
\begin{array}{l@{\hspace{.8cm}}l}
\hspace{-0.3cm}
\begin{array}{l@{\hspace{.2cm}}rcl}
  \text{(Strict)} & \quad  \sigma, \tau, \delta &::=& \unit \sep \arrt{( \pi , \eta)}{\sigma}   
  \\
  \text{(Multiset)} & \quad \pi, \zeta &::=& \bigwedge_{i \in I} \sigma_i \sep \omega
\end{array}
&
\begin{array}{l@{\hspace{.2cm}}rcl}
  \text{(List)}& \quad \eta, \epsilon  &::=&  \sigma  \sep \epsilon \concat \eta 
  \\
  \text{(Tuple)}&\quad   ( \pi , \eta)
\end{array}
\end{array}
\]
\end{definition}

A strict type can be the   $\unit$ type  or a functional type  \arrt{( \pi , \eta)}{\sigma}, where 
$( \pi , \eta)$ is a tuple type and $\sigma$ is  a strict type. Multiset types can be 
either the empty type $\omega$ or 
an intersection  of strict types $\bigwedge_{i \in I} \sigma_i$, with $I$ non-empty. The operator $\wedge $ is commutative, associative, non-idempotent, that is, $\sigma \wedge \sigma \neq \sigma$, with identity $\omega$. The intersection type $\bigwedge_{i\in I} \sigma_i$ is the type of a linear bag; the cardinality of $I$ corresponds to its size. 

A list type can be either an strict type $\sigma$ or the composition $\epsilon\concat \eta$ of two list types $\epsilon$ and $\eta$. We use the list type  $\epsilon\concat \eta$ to type the concatenation of two unrestricted bags.
A tuple type $(\pi,\eta)$ types the concatenation of a linear bag of type $\pi$ with an unrestricted bag of type $\eta$.
Notice that a list type $\epsilon \concat \eta$ can be recursively unfolded into a finite composition of strict types $\sigma_1\concat   \ldots \concat \sigma_n$, for some $n\geq 1$. In this case the length of $\epsilon\concat \eta$ is $n$ and that $\sigma_i$ is its $i$-th strict type, for $1\leq i \leq n$.



\begin{notation}
\label{ch3not:lamrrepeat}
 Given $k \geq 0$, we  write $\sigma^k$ to stand for $\sigma \wedge \cdots \wedge \sigma$
($k$ times, if $k>0$) or for $\omega$ (if $k=0$).
Similarly,  $\hat{x}:\sigma^k$
stands for $x:\sigma , \cdots , x:\sigma$ ($k$ times, if $k>0$)
 or for $x:\omega$  (if $k=0$).  Given $k \geq 1$, we  write $\banged{x}:\eta   $ to stand for $x[1]:\eta_1 , \cdots , x[k]:\eta_k$.
\end{notation}

\begin{notation}[$\eta \relunbag \epsilon$]
\label{ch3not:ltypes}
 Let $\epsilon$ and $\eta$ be two list types, with the length of $\epsilon$ greater or equal to that of $\eta$. 
Let us write $\epsilon_i$ and $\eta_i$ to denote the $i$-th strict type in $\epsilon$ and $\eta$, respectively.
We write $\eta \relunbag \epsilon$ meaning the initial sublist, whenever there exist $  \epsilon'$ and $ \epsilon''$   such that: i) $ \epsilon = \epsilon' \concat \epsilon''$; ii)  the size of $\epsilon' $ is that of $\eta$; iii)  for all $i$, $\epsilon'_i = \eta_i $.
\end{notation}



\noindent {\em Linear contexts} range over  $\Gamma , \Delta, \ldots $ and {\em  unrestricted contexts} range over $\Theta, \Upsilon, \ldots$. They are defined by the following grammar:
$$
\begin{array}{c@{\hspace{.9cm}}c}
   \Gamma, \Delta ::= \dash ~|~  x:\sigma ~\mid ~\Gamma, x:\sigma & \Theta, \Upsilon ::= \dash ~|~  x^!:\eta ~|~ \Theta,x^!:\eta
\end{array}
$$

The empty linear/unrestricted type assignment is  denoted `$\dash$'.
Linear variables can occur more than once in a linear context; they are assigned only strict types. 
For instance,  $x:(\tau , \sigma) \rightarrow \tau , x: \tau$  is a valid context: it means that $x$ can be of both type $(\tau , \sigma) \rightarrow \tau$ and $\tau$. 
 In contrast, unrestricted variables can occur at most once  in unrestricted contexts; they are assigned only list types.
The multiset of linear variables in $\Gamma$ is denoted as $\dom{\Gamma}$; similarly, $\dom{\Theta}$ denotes the set of unrestricted variables in $\Theta$. 

Judgements are of the form $\Theta;\Gamma \wfdash \expr{M}:\sigma$, where the left-hand side contexts are separated by ``;'' and  $\expr{M}:\sigma$ means that    $\expr{M}$ has type $\sigma$. We write $\wfdash \expr{M}:\sigma$ to denote $\dash \,;\, \dash \wfdash \expr{M}:\sigma$.

As in \Cref{ch2} we rely on core contexts, here only linear free variables that are result of weakening will disregarded in the typing of the failure term as unrestricted variables are not captured by failure.
\begin{definition}{Core Context}
    Given a linear context $\Gamma$, the associated  \emph{core context} is defined as 
$\core{\Gamma} = \{ x:\pi \in \Gamma \,|\, \pi \not = \omega\}$. 
\end{definition}

\begin{figure*}[!t]
\small
	    \centering    
    \begin{prooftree}
        \AxiomC{}
        \LeftLabel{\redlab{F{:}var^{ \ell}}}
        \UnaryInfC{\( \Theta ;  {x}: \sigma \wfdash  {x} : \sigma\)}
        \DisplayProof
\hfill
        \AxiomC{\( \Theta , \banged{x}: \eta; {x}: \eta_i , \Delta \wfdash  {x} : \sigma\)}
        \LeftLabel{\redlab{F{:}var^!}}
        \UnaryInfC{\( \Theta , \banged{x}: \eta; \Delta \wfdash {x}[i] : \sigma\)}
        \DisplayProof
        \hfill
        \AxiomC{\(  \)}
        \LeftLabel{\redlab{F{:}\oneb^{\ell}}}
        \UnaryInfC{\( \Theta ; \dash \wfdash \oneb : \omega \)}
    \end{prooftree}
  \vspace{-5mm}
  \begin{prooftree}
        \AxiomC{\(  \)}
        \LeftLabel{\redlab{F{:}\oneb^!}}
        \UnaryInfC{\( \Theta ;  \dash  \wfdash \banged{\oneb} : \sigma \)}
          \DisplayProof
    \hfill
  \AxiomC{\(\Theta , \banged{z} : \eta ; \Gamma ,  {\hat{z}}: \sigma^{k} \wfdash M : \tau \quad z\notin \dom{\Gamma} \)}
        \LeftLabel{\redlab{F{:}abs}}
        \UnaryInfC{\( \Theta ;\Gamma \wfdash \lambda z . M : (\sigma^{k} , \eta )   \rightarrow \tau \)}
\end{prooftree}
  \vspace{-6mm}
\begin{prooftree}
     \AxiomC{\quad \( \quad \Theta ;\Gamma \wfdash M : (\sigma^{j} , \eta ) \rightarrow \tau \)}
\noLine
         \UnaryInfC{\(  \Theta ;\Delta \wfdash B : (\sigma^{k} , \epsilon )  \)}
      \AxiomC{\( \eta \relunbag \epsilon \)}
            \LeftLabel{\redlab{F{:}app}}
        \BinaryInfC{\( \Theta ; \Gamma, \Delta \wfdash M\ B : \tau\)}
 \DisplayProof
 \hfill
 \AxiomC{\qquad \( ~\quad \Theta ,\banged{x} : \eta ; \Gamma ,  {\hat{x}}: \sigma^{j} \wfdash M : \tau  \)}
    \noLine
   \UnaryInfC{\( \Theta ; \Delta \wfdash B : (\sigma^{k} , \epsilon ) \)}
     \AxiomC{\( \eta \relunbag \epsilon \)}
        \LeftLabel{\redlab{F{:}ex \dash sub}}  
        \BinaryInfC{\( \Theta ; \Gamma, \Delta \wfdash M \esubst{ B }{ x } : \tau \)}
   \end{prooftree}
  \vspace{-5mm}
    \begin{prooftree}
  \AxiomC{\( \Theta ; \Gamma\wfdash C : \sigma^k\)}
    \AxiomC{\(  \Theta ;\dash \wfdash  U : \eta \)}
        \LeftLabel{\redlab{F{:}bag}}
        \BinaryInfC{\( \Theta ; \Gamma \wfdash C \bagsep U : (\sigma^{k} , \eta  ) \)}
\DisplayProof
\hfill
   \AxiomC{\( \Theta ; \Gamma \wfdash M : \sigma\)}
        \AxiomC{\( \Theta ; \Delta \wfdash C : \sigma^k\)}
        \LeftLabel{\redlab{F{:}bag^{ \ell}}}
        \BinaryInfC{\( \Theta ; \Gamma, \Delta \wfdash \bag{M}\cdot C:\sigma^{k+1}\)}
    \end{prooftree}
  \vspace{-6mm}
    \begin{prooftree}
        \AxiomC{\( \Theta ; \dash \wfdash M : \sigma\)}
        \LeftLabel{\redlab{F{:}bag^!}}
        \UnaryInfC{\( \Theta ; \dash  \wfdash \banged{\bag{M}}:\sigma \)}
   \DisplayProof
   \hfill
        \AxiomC{\( \Theta ; \dash \wfdash U : \epsilon\)}
        \AxiomC{\( \Theta ; \dash \wfdash V : \eta\)}
        \LeftLabel{\redlab{F{:}\concat bag^!}}
        \BinaryInfC{\( \Theta ; \dash  \wfdash U \concat V :\epsilon \concat \eta \)}
    \DisplayProof
    \hfill
  \AxiomC{\( \dom{\core{\Gamma}} = \widetilde{x} \)}
        \LeftLabel{\redlab{F{:}fail}}
        \UnaryInfC{\( \Theta ; \Gamma \wfdash  \fail^{\widetilde{x}} : \tau  \)}
    \end{prooftree}
     \vspace{-6mm}
    \begin{prooftree}
         \AxiomC{$ \Theta ;\Gamma \wfdash \expr{M} : \sigma \qquad \Theta ;\Gamma \wfdash \expr{N} : \sigma$}
        \LeftLabel{\redlab{F{:}sum}}
        \UnaryInfC{$ \Theta ;\Gamma \wfdash \expr{M}+\expr{N}: \sigma$}
          \DisplayProof
        \qquad 
                \AxiomC{$ \Theta; \Gamma \wfdash M: \sigma \quad  x \not \in \dom{\Gamma}$}
        \LeftLabel{\redlab{F{:}weak}}
        \UnaryInfC{$ \Theta ; \Gamma, x:\omega \wfdash M: \sigma $}
    \end{prooftree}
    \caption{Well-formedness rules for \lamrfailunres (cf. Def.~\ref{ch3d:wellf_unres}). In Rules~\redlab{F{:}app} and \redlab{F{:}ex\dash sub}: $k, j \geq 0$.}\label{ch3fig:wf_rules_unres}
\end{figure*}

\begin{definition}{Well-formed \lamrfailunres expressions}
\label{ch3d:wellf_unres}
An expression $ \expr{M}$ is well-formed (wf, for short) if  there exist  $\Gamma$, $\Theta$ and  $\tau$ such that  $ \Theta ; \Gamma \wfdash  \expr{M} : \tau  $ is entailed via the rules in \figref{ch3fig:wf_rules_unres}.
\end{definition}


\noindent We describe the well-formedness rules in \figref{ch3fig:wf_rules_unres}. 
\begin{itemize}
\item Rules~$\redlab{F:var^{ \ell}}$ and $\redlab{F:var^!}$ assign types to linear and unrestricted variables, respectively. 
\item Rule~$\redlab{F:var^!}$ resembles the {\em copy} rule~\cite{CairesP10} where we use a linear copy of an unrestricted variable $x[i]$ of type $\sigma$, typed with $x^!:\eta$, and type the linear copy with the corresponding strict type $\eta_i$ which in this case the linear copy $x$ would have type equal to $\sigma$. 
\item Rules~$\redlab{F:\oneb^{ \ell}}$ and $\redlab{F:\oneb^!}$ assign types to the empty linear/unrestricted bag: $\oneb$ has type $\omega$, whereas  $\oneb^!$ has an arbitrary strict type $\sigma$. Arbitrariness is allowed since  the substitution of an unrestricted variable for $1^!$ leads to a \(\fail\) term (Rule \redlab{R:Fail^!}), which has an arbitrary strict type.

\item Rule $\redlab{F:abs}$ assigns type $(\sigma^k,\eta)\to \tau $ to an abstraction $\lambda z. M$, provided that the unrestricted occurrences of $z$ may be typed by the unrestricted context containing $z^!:\eta$, the linear  occurrences of $z$ are typed with the linear context containing $\hat{z}: \sigma^k$, for some $k\geq 0$, and there are no other linear occurrences of $z$ in the linear context $\Gamma$.
\item Rules $\redlab{F{:}app}$ and  $\redlab{F{:}ex \dash sub}$ (for application  and explicit substitution, resp.) use the condition $\eta \relunbag \epsilon $ (cf. Notation~\ref{ch3not:ltypes}), which 
captures the portion of the unrestricted bag that is effectively used in a term: it 
ensures that $\epsilon$ can be decomposed into some $\epsilon'$ and $\epsilon''$, such that each  type component $\epsilon_i'$ matches with $\eta_i$. If this requirement is satisfied,  Rule~$\redlab{F{:}app}$ types an application $M\ B$  given that $M$ has a functional type in which the left of the arrow is  a  tuple type $ (\sigma^{j} , \eta ) $ whereas the bag $ B$ is typed with tuple  $ (\sigma^{k} , \epsilon ) $. Similarly,  Rule~$\redlab{F{:}ex \dash sub}$ types the term $ M \esubst{ B }{ x }$ provided that  $ B$ has the tuple type $ (\sigma^{k} , \epsilon ) $ and $M$ is typed with the variable $x$ having linear type assignment $ \sigma^{j} $ and unrestricted type assignment $ \eta $.
 
\end{itemize}

\begin{remark}
 Differently from  intersection type systems~\cite{BucciarelliKV17, PaganiR10},  in Rules  $\redlab{F{:}app}$ and   $\redlab{F{:}ex \dash sub}$ there is no equality requirement  between $j$ and $k$,  as we would like to capture terms that  fail due to a mismatch in resources: we only require that the linear part of the tuples are composed of the same strict type, say  $\sigma$. 
As a term can take an unrestricted bag with arbitrary size we only require that the elements of the unrestricted bag that are used have a ``consistent'' type, i.e., the type of the unrestricted bag satisfies the relation $\relunbag$ with the unrestricted fragment of the corresponding tuple type.
\end{remark}

\noindent There are four rules for bags:
\begin{itemize}
\item Rule~$\redlab{F:bag^!}$ types an unrestricted bag $\bag{M}^!$ with  the  type $\sigma$ of $M$. Note that $\bag{x}^!$, an unrestricted bag containing a linear variable $x$, is not well-formed, whereas $\bag{x[i]}^!$ is well-formed. 
\item Rule $\redlab{F:bag}$ assigns the tuple type $(\sigma^k,\eta)$ to the concatenation of a linear bag of type $\sigma^k$ with an unrestricted bag of type $\eta$. \item Rules $\redlab{F:bag^{ \ell}}$
and $\redlab{F{:}\concat bag^!}$ type the concatenation of linear and unrestricted bags. 
\item Rule~$ \redlab{F{:}\oneb^!} $ allows an empty unrestricted bag to have an arbitrary $\sigma$ type since it may be referred to by a variable for substitution: we must be able to compare its type with the type of unrestricted variables that may  consume the empty bag (this reduction would inevitably lead to failure).
\end{itemize}
As in~\Cref{ch2}, Rule  $\redlab{F{:}\fail}$ handles the failure term, and is the main difference with respect to standard type systems. Rules for sums and weakening ($\redlab{F:sum}$ and $\redlab{F:weak}$) are standard.

\begin{example}{Cont. Example~\ref{ch3ex:deltasemii}}\label{ch3ex:delta7_wf1} Term $\Delta_7:=\lambda x. x[2]( \oneb \bagsep \banged{\bag{x[1]}} \concat \banged{\bag{x[1]}} ) $ is well-formed, as ensured by the judgement $ \Theta ; \dash \wfdash \Delta_7: (\omega ,  \sigma' \concat (\sigma^{j} , \sigma' \concat   \sigma' ) \rightarrow \tau )    \rightarrow \tau  $, whose derivation is given below:
        
\begin{itemize}
\item $\Pi_3$ is the derivation of 
\(  \Theta , \banged{x} : \eta ;\dash \wfdash  \banged{\bag{x[1]}} : \sigma',\)  for $ \eta = \sigma' \concat (\sigma^{j} , \sigma' \concat   \sigma' ) \rightarrow \tau $.
\item $\Pi_4$ is the derivation:
\( \Theta , \banged{x}: \eta ; \dash \wfdash {x}[2] : (\sigma^{j} , \sigma' \concat   \sigma' ) \rightarrow \tau \)
\item $\Pi_5$ is the derivation: \( \Theta , \banged{x} :  \eta   ; x : \omega \wfdash ( \oneb \bagsep \banged{\bag{x[1]}} \concat \banged{\bag{x[1]}} )  : (\omega , \sigma' \concat   \sigma' ) \)
\end{itemize}
Therefore, 
 \begin{prooftree}
    \AxiomC{\( \Pi_5\)}
    \AxiomC{\( \Pi_4 \)}
       \AxiomC{\( \sigma' \concat   \sigma' \relunbag \sigma' \concat   \sigma' \)}
    \LeftLabel{\redlab{F{:}app}}
            \TrinaryInfC{\( \Theta , \banged{x} :  \eta  ;  x : \omega \wfdash x[2]( \oneb \bagsep \banged{\bag{x[1]}} \concat \banged{\bag{x[1]}} ) : \tau \)}
           \LeftLabel{\redlab{F{:}abs}}
            \UnaryInfC{\( \Theta ; \dash \wfdash \underbrace{\lambda x. (x[2]( \oneb \bagsep \banged{\bag{x[1]}} \concat \banged{\bag{x[1]}} ))}_{\Delta_7}: ( \omega ,  \eta )    \rightarrow \tau  \)}
        \end{prooftree}
\end{example}

Well-formed expressions satisfy subject reduction (SR);
\iffulldoc
 see \appref{ch3appB} for a proof.
 \else
 a proof may be found in the full version.
 \fi

 \begin{theorem}[SR in \lamrfailunres]
\label{ch3t:lamrfailsr_unres}
If $\Theta ; \Gamma \wfdash \expr{M}:\tau$ and $\expr{M} \redd \expr{M}'$ then $\Theta ;\Gamma \wfdash \expr{M}' :\tau$.
\end{theorem}
\begin{proof}By structural induction on the reduction rules. 
We proceed by analysing the rule applied in $\expr{M}$. 
An interesting case occurs when the rule is $\redlab{F:Fetch^!}$:  Then $ \expr{M} = M\ \esubst{ C \bagsep U }{x }$, where $U = \banged{\bag{N_1}} \concat \cdots \concat \banged{\bag{N_l}} $ and $\headf{M} =  {x}[i]$. The reduction is as follows:

    \begin{prooftree}
        \AxiomC{$\headf{M} =  {x}[i]$}
        \AxiomC{$ U_i = \banged{\bag{N_i}} $}
        \LeftLabel{\redlab{R:Fetch^!}}
        \BinaryInfC{\(
        M\ \esubst{ C \bagsep U  }{x } \redd M \headlin{ N_i/ {x}[i] } \esubst{ C \bagsep U}{ x } 
        \)}
    \end{prooftree}
    
     By hypothesis, one has the derivation:
     \begin{prooftree}
            \AxiomC{\( \Theta , \banged{x} : \eta ; \Gamma' , \hat{x}: \sigma^{j} \wfdash M : \tau \)}
               \AxiomC{$\Pi$}
               \noLine
               \UnaryInfC{\(  \Theta ; \cdot\wfdash U : \epsilon\)}
                \AxiomC{\( \Theta ; \Delta \wfdash  {C} : \sigma^k\)}
            \LeftLabel{\redlab{F{:}bag}}
            \BinaryInfC{\( \Theta ; \Delta \wfdash C \bagsep U : (\sigma^{k} , \epsilon ) \)}
            \AxiomC{\(  \eta \relunbag \epsilon  \)}
        \LeftLabel{\redlab{F{:}ex \dash sub}}  
        \TrinaryInfC{\( \Theta ; \Gamma', \Delta \wfdash M \esubst{ C \bagsep U }{ x } : \tau \)}
    \end{prooftree}
    where $\Pi$ has the form
    {\small{
    \begin{prooftree}
            \AxiomC{\( \Theta ; \cdot \wfdash N_1 : \epsilon_1\)}
            \LeftLabel{\redlab{F{:}bag^!}}
            \UnaryInfC{\( \Theta ; \cdot \wfdash \banged{\bag{N_1}}  : \epsilon_1\)}
            \AxiomC{\( \cdots \)}
            \AxiomC{\( \Theta ; \cdot \wfdash N_l : \epsilon_l \)}
            \LeftLabel{\redlab{F{:}bag^!}}
            \UnaryInfC{\( \Theta ; \cdot \wfdash \banged{\bag{N_l}} : \epsilon_l \)}
        \LeftLabel{\redlab{F{:}\concat bag^!}}
        \TrinaryInfC{\( \Theta ; \cdot  \wfdash \banged{\bag{N_1}} \concat \cdots \concat \banged{\bag{N_l}}  :\epsilon \)}
    \end{prooftree}
    }}
    with $\Gamma = \Gamma' , \Delta $. Notice that if $\epsilon_i = \delta$ and $\eta \relunbag \epsilon$ then $\eta_i = \delta$.
    \iffulldoc
     By Lemma~\ref{ch3lem:subt_lem_failunres_un}, there 
exists a derivation
 $\Pi_1$ of  $
    \Theta , \banged{x} : \eta ; \Gamma' , \hat{x}: \sigma^{j} \wfdash   M \headlin{ N_{i}/  {x}[i] } : \tau $. 
    \else
    there 
    exists a derivation
     $\Pi_1$ of  $
        \Theta , \banged{x} : \eta ; \Gamma' , \hat{x}: \sigma^{j} \wfdash   M \headlin{ N_{i}/  {x}[i] } : \tau $ (shown in the full version). 
    \fi
    Therefore, we have: 
         \begin{prooftree}
            \small
            \AxiomC{\( \Theta , \banged{x} : \eta ; \Gamma' , \hat{x}: \sigma^{j} \wfdash M \headlin{ N_{1}/  {x}[i] } : \tau \)}
               \AxiomC{\(  \Theta ; \cdot\wfdash U : \epsilon\)}
                \AxiomC{\( \Theta ; \Delta \wfdash  {C} : \sigma^k\)}
            \LeftLabel{\redlab{F{:}bag}}
            \BinaryInfC{\( \Theta ; \Delta \wfdash C \bagsep U : (\sigma^{k} , \epsilon ) \)}
            \AxiomC{\(  \eta \relunbag \epsilon  \)}
        \LeftLabel{\redlab{F{:}ex \dash sub}}  
        \TrinaryInfC{\( \Theta ; \Gamma', \Delta \wfdash M \headlin{ N_{i}/ {x[i]} } \esubst{  C \bagsep U }{ x } : \tau \)}
    \end{prooftree}

\end{proof}

\begin{remark}[Well-Formed vs Well-Typed Expressions]
\label{ch3r:types}
Our type system (and Theorem \ref{ch3t:lamrfailsr_unres}) can be specialised to the case of \emph{well-typed} expressions that do not contain (and never reduce to) the failure term.
In particular, Rules  $\redlab{F{:}app}$ and   $\redlab{F{:}ex \dash sub}$ would need to check that $\sigma^k = \sigma^j$, as failure can be caused due to a mismatch of linear resources. {A difference between well typed and well formed expressions is that the former satisfy subject expansion, but the latter do not: expressions that lead to failure can be ill-typed yet failure itself is well-formed.}

\end{remark}



\section[A Translation into Processes]{A Typed Encoding of \lamrfailunres into Concurrent Processes}
\label{ch3s:encoding}
We encode \lamrfailunres into \spi,
a session $\pi$-calculus that stands on a Curry-Howard correspondence between linear logic and session types (\secref{ch3s:pi}). We 
extend the two-step approach  that we devised in~\Cref{ch2} for the sub-calculus \lamrfail  (with linear resources only) (cf.~\figref{ch3fig:two-step-enc}).
    First, in \secref{ch3ssec:first_enc}, we define an encoding $\recencodopenf{\cdot}$ from well-formed expressions in \lamrfailunres to well-formed expressions in  a variant of \lamrfailunres with \emph{sharing}, dubbed \lamrsharfailunres (\secref{ch3ssec:lamshar}).
    Then, in \secref{ch3ssec:second_enc}, we define an encoding $\piencodftypes{\cdot}_u$ (for a name $u$) from well-formed expressions in \lamrsharfailunres to well-typed processes in \spi.

We prove that  $\recencodopenf{\cdot}$ and $\piencodftypes{\cdot}_u$ satisfy well-established correctness criteria~\cite{DBLP:journals/iandc/Gorla10,DBLP:journals/iandc/KouzapasPY19}: \emph{type preservation}, \emph{operational  completeness},  \emph{operational soundness}, and \emph{success sensitiveness} 
\iffulldoc
(cf.~\appref{ch3ss:criteria}).
\else
(cf.~the full version)
\fi 
Because \lamrfailunres includes unrestricted resources, the results given here strictly  generalise those in~\Cref{ch2}.

\subsection[Session-Typed Calculus]{\spi: A Session-Typed $\pi$-Calculus}
\label{ch3s:pi}
\spi is a $\pi$-calculus with \emph{session types}~\cite{DBLP:conf/concur/Honda93,DBLP:conf/esop/HondaVK98}, which  ensure that the endpoints of a channel perform matching actions.
We consider the full process  framework in~\cite{CairesP17}, including constructs for specifying labelled choices and client/server connections;   they will be useful to codify   unrestricted resources and variables in \lamrfailunres.
Following~\cite{CairesP10,DBLP:conf/icfp/Wadler12},
\spi stands on a Curry-Howard correspondence between session types and a linear logic with dual modalities/types ($\with A$ and $\oplus A$),  which define \emph{non-deterministic} session behaviour.
As in~\cite{CairesP10,DBLP:conf/icfp/Wadler12},
in \spi, cut elimination corresponds to communication, proofs to processes, and propositions to session types.

 \tikzstyle{mynode1} = [rectangle, rounded corners, minimum width=1.75cm, minimum height=0.8cm,text centered, draw=black, fill=gray!30]
\tikzstyle{arrow} = [thick,->,>=stealth]

\tikzstyle{mynode2} = [rectangle, rounded corners, minimum width=1.75cm, minimum height=0.8cm,text centered, draw=black, fill=RedOrange!20]

\tikzstyle{mynode3} = [rectangle, rounded corners, minimum width=1.75cm, minimum height=0.8cm,text centered, draw=black, fill=RoyalBlue!20]
\tikzstyle{arrow} = [thick,->,>=stealth]

\begin{figure}[!t]
\begin{center}
\begin{tikzpicture}[node distance=2cm, scale=.7pt]
\node (source) [mynode1] {$\lamrfailunres$};
\node (interm) [mynode2, right of=source,  xshift=3cm] {$\lamrsharfailunres$};
\node (target) [mynode3, right of=interm,  xshift=3cm] {$\spi$};
\draw[arrow] (source) --  node[anchor=south] {$ \recencodopenf{\cdot}$} (interm);
\node (enc1) [right of= source, xshift=.4cm, yshift=-.35cm] {\secref{ch3ssec:first_enc}};
\node (enc2) [right of= interm, xshift=.4cm, yshift=-.35cm] {\secref{ch3ssec:second_enc}};
\draw[arrow] (interm) -- node[anchor=south] {$ \piencodftypes{\cdot }_u $ } (target);
\end{tikzpicture}
\end{center}
\vspace{-3mm}
\caption{Our two-step approach to encode  $\lamrfailunres$ into $\spi$.}\label{ch3fig:two-step-enc}
\end{figure}

\subparagraph{Syntax.}
\emph{Names}  $x, y,z, w \ldots$ denote the endpoints of protocols specified by session types. We write $P\subst{x}{y}$ for the capture-avoiding substitution of $x$ for $y$ in process $P$.

\begin{definition}{Processes}\label{ch3d:spi}
The syntax of \spi processes is given by the grammar below. 
	\begin{align*}
P,Q ::=  ~&  \zero \sep \overline{x}(y).P \sep  x(y).P \sep \case{x}{i};P \sep \choice{x}{i}{I}{i}{P_i} \sep  x.\overline{\close} \sep x.\close;P \\
 \sep &  (P \para Q)
  \sep  [x \leftrightarrow y]  \sep (\nu x)P \sep ~!x(y).P \sep  \outsev{x}{y}.P 
  \\
   \sep &  x.\overline{\some};P \sep x.\overline{\none}
\sep      x.\some_{(w_1, \cdots, w_n)};P \sep  (P \oplus Q)
\end{align*}
\end{definition}
\noindent
Process $\zero$ denotes inaction. 
Process $\overline{x}(y).P$ sends a fresh name $y$ along  $x$ and then continues as $P$. Process 
     $x(y).P$ receives a name $z$ along $x$ and then continues as  $P\subst{z}{y}$. Process $\choice{x}{i}{I}{i}{P_i}$ is a branching construct, with labelled alternatives indexed by the finite set $I$: it awaits a choice on $x$ with continuation $P_j$ for each  $j \in I$. Process $\case{x}{i};P$ selects on $x$ the alternative indexed by $i$ before continuing as $P$. Processes $x.\overline{\close}$ and $x.\close;P$ are complementary actions for  closing session $x$. 
     We sometimes use the  shorthand notations $\overline{y}\sclose$ and $y\sclose;P$ to stand for ${y}.\overline{\close}$ and $y.\close;P$, respectively.
      Process $P \para Q$ is the parallel execution of $P$ and $Q$. 
The forwarder process $[x \leftrightarrow y]$ denotes a bi-directional link between sessions $x$ and $y$. 
Process $(\nu x)P$ denotes the process $P$ in which name $x$  is kept private (local) to $P$. 
Process $!x(y).P$ defines a server that spawns copies of $P$ upon requests on $x$.
Process $\outsev{x}{y}.P$ denotes a client that connects to a server by sending the fresh name $y$ on $x$. 

The remaining constructs  come from~\cite{CairesP17} and introduce non-determi\-nis\-tic sessions which  \emph{may} provide a session protocol  \emph{or} fail.
    Process $x.\overline{\some};P$ confirms that the session  on $x$ will execute and  continues as $P$. Process $x.\overline{\none}$ signals the failure of implementing the session on $x$.
    Process $x.\some_{(w_1, \cdots,w_n)};P$ specifies a dependency on a non-deterministic session $x$. 
    This process can  either (i)~synchronise with an action $x.\overline{\some}$ and continue as $P$, or (ii)~synchronise with an action $x.\overline{\none}$, discard $P$, and propagate the failure on $x$ to $(w_1, \cdots, w_n)$, which are sessions implemented in $P$.
    When $x$ is the only session implemented in $P$, there is no tuple of dependencies $(w_1, \cdots,w_n)$ and so we write simply $x.\some;P$.
        Finally, process $P \oplus Q$ denotes a non-deterministic choice between $P$ and $Q$. We shall often write $\bigoplus_{i \in \{1 , \cdots , n \} } P_i$ to stand for $P_1 \oplus \cdots \oplus P_n$.
In  $(\nu y)P$ and $x(y).P$ the   occurrence of name $y$ is binding, with scope $P$.
The set of free names of $P$ is denoted by $\fn{P}$.

\subparagraph{Semantics.}
The   \emph{reduction relation} of \spi specifies the computations that a process performs on its own (cf.~\figref{ch3fig:redspi}). 
It is closed by  \emph{structural congruence}, denoted $\equiv$, which expresses basic identities for processes and the non-collapsing nature of non-determinism 
\iffulldoc
(cf. \appref{ch3appC}).
\else
(cf.~full version for details).
\fi


\begin{figure}[!t]
\small
\centering
\begin{align*}
 & \overline{x}{(y)}.Q \para x(y).P  \redd (\nu y) (Q \para P) &
 & x.\overline{\some};P \para x.\some_{(w_1, \cdots, w_n)};Q  \redd P \para Q\\
 & Q \redd Q' \Rightarrow P \oplus Q   \redd P \oplus Q' &
 & x.\overline{\close} \para x.\close;P  \redd P \\ 
 & \case{x}{i};Q \para \choice{x}{i}{I}{i}{P_i}   \redd  Q \para P_i &   
 & !x(y).Q \para  \outsev{x}{y}.P   \redd (\nu x)( !x (y).Q \para  (\nu y)( Q \para P ) ) \\
 & (\nu x)( [x \leftrightarrow y] \para P) \redd P \subst{y}{x}  \quad (x \neq y) &
 & P\equiv P'\wedge P' \redd Q' \wedge Q'\equiv Q \Rightarrow P   \redd Q\\
 & Q \redd Q' \Rightarrow P \para Q   \redd P \para Q'&
 & P \redd Q  \Rightarrow (\nu y)P  \redd (\nu y)Q \\
 & 
\end{align*}
\vspace{-1.2cm}
\[x.\overline{\none} \para x.\some_{(w_1, \cdots, w_n)};{Q} \redd w_1.\overline{\none} \para \cdots \para w_n.\overline{\none}\]
    \caption{Reduction for \spi}
    \label{ch3fig:redspi}
\end{figure}

The first reduction rule formalises communication, which concerns bound names only (internal mobility), as $y$ is bound in $\overline{x}{(y)}.Q$ and $x(y).P$.
 Reduction  for the forwarder process leads to a substitution.
The reduction rule for closing a session is self-explanatory, as is the rule in which prefix $x.\overline{\some}$ confirms the availability of a non-deterministic session. When a non-deterministic session is not available,   $x.\overline{\none}$ triggers this failure to all dependent sessions $w_1, \ldots, w_n$; this may in turn trigger further failures (i.e., on sessions that depend on $w_1, \ldots, w_n$).
The remaining rules define contextual reduction with respect to restriction,  composition, and non-deterministic choice.




\subparagraph{Type System}
Session types govern the behaviour of the names of a process.
An assignment $x:A$ enforces the use of name $x$ according to the  protocol specified by $A$.

\begin{definition}{Session Types}
Session types are given by 
\[
\begin{array}{rl}
  A,B ::= & \bot \sep   \onef \sep 
A \otimes B  \sep A \ampy B  
\sep  \oplus_{i \in I} \{ \mathtt{l}_i : A_i  \}  
 \sep \with_{i \in I} \{ \mathtt{l}_i : A_i \}  \sep ! A \sep   ? A 
 \sep  \with A \sep \oplus A  
\end{array}
\]
\end{definition}

\noindent
The multiplicative units  $\bot$ and  $\onef$ are used to type closed session endpoints.
We use $A \otimes B$ to type a name that first outputs a name of type $A$ before proceeding as specified by $B$.
Similarly, $A \ampy B $ types a name that first inputs a name of type $A$ before proceeding as specified by $B$.
Then, $! A$ types a name that repeatedly provides a service specified by $A$.
Dually, $ ? A $ is the type of a name that can connect to a server offering $A$.
Types 
$\oplus_{i \in I} \{ \mathtt{l}_i : A_i  \}$ and $\with_{i \in I} \{ \mathtt{l}_i : A_i \}$
are assigned to names that can select and offer a labelled choice, respectively.
Then we have the two modalities introduced in~\cite{CairesP17}.
We use $\with A$ as the type of a (non-deterministic) session that \emph{may  produce} a behaviour of type $A$.
Dually, $\oplus A$ denotes the type of a session that \emph{may consume} a behaviour of type $A$.

The two endpoints of a  session should be \emph{dual} to ensure  absence of communication errors. The dual of a type $A$ is denoted $\dual{A}$. 
Duality corresponds to negation $(\cdot)^\bot$ in  linear logic~\cite{CairesP17}. 

\begin{definition}{Duality}
\label{ch3def:duality}
Duality on types is given by:
\[
\small
\begin{array}{l}
\begin{array}{c@{\hspace{.5cm}}c@{\hspace{.5cm}}c@{\hspace{.5cm}}c@{\hspace{.5cm}}c}
\dual{\onef}  =  \bot 
&
\dual{\bot}    =  \onef
&
\dual{A\otimes B}   = \dual{A} \ampy \dual{B}
&
\dual{\oplus_{i \in I} \{ \mathtt{l}_i : A_i  \}}   = \with_{i \in I} \{ \mathtt{l}_i : \dual{A_i} \}
&
\dual{\oplus A}    =    \with \dual{A}
\\
\dual{!A}  =  ?A 
 &
 \dual{?A}    =  !A
 &
\dual{A \ampy B}   = \dual{A} \otimes \dual{B} 
&
\dual{\with_{i \in I} \{ \mathtt{l}_i : A_i \}}   = \oplus_{i \in I} \{ \mathtt{l}_i : \dual{A_i}  \}
&
 \dual{\with A}   =   \oplus \overline {A}
\end{array}
\end{array}
\]
\end{definition}

\noindent
Judgements are of the form $P \vdash \Delta; \Theta$, where $P$ is a process, $\Delta$ is the linear context, and $\Theta$ is the unrestricted context.
Both $\Delta$ and $\Theta$ contain assignments of types to names, but satisfy different substructural principles: while $\Theta$ satisfies weakening, contraction and exchange, $\Delta$ only satisfies exchange. The empty context is denoted `$\cdot$'. 
We write $\with \Delta$ to denote that all assignments in $\Delta$ have a non-deterministic type, i.e., $\Delta = w_1{:}\with A_1, \ldots, w_n{:}\with A_n$, for some $A_1, \ldots, A_n$. 
The typing judgement $P \vdash \Delta$ corresponds to the logical sequent for classical linear logic, which can be recovered by erasing processes and name assignments. 

Typing rules for processes in \figref{ch3fig:trulespi} correspond to proof rules in linear logic;
we discuss some of them.
Rule~$\redlab{Tid}$ interprets the identity axiom using the forwarder process. 
Rules~$\redlab{T \onef}$ and $\redlab{T \bot}$ type the process constructs for session termination.
 Rules~$\redlab{T\otimes}$ and $\redlab{T \ampy }$ type output and input of a name along a session, resp. 
The last four rules are used to type process constructs related to non-de\-ter\-mi\-nism and failure. 
Rules~$ \redlab{T \with_d^x}$ and $ \redlab{T \with^x}$ introduce a session of type $\with A$, which may produce a behaviour of type $A$: while the former rule covers the case in which $x:A$ is indeed available, the latter rule formalises the case in which $x:A$ is not available (i.e., a failure).
Given a sequence of names $\widetilde{w} = w_1, \ldots, w_n$,  Rule~$\redlab{T \oplus^x_{\widetilde{w}}}$ accounts for the possibility of not being able 
to consume the session $x:A$  by considering sessions different from $x$ as potentially not available. 
Rule~$\redlab{T \oplus }$ expresses non-deterministic choice of processes $P$ and $Q$ that implement non-deterministic behaviours only.
Finally, Rule~$\redlab{T\oplus_i} $ and $ \redlab{T\with} $ correspond, resp., to selection and branching: the former provides a selection of behaviours along $x$ as long as  $P$ is guarded with the $i$-{th} behaviour; the latter offers a labelled choice where each behaviour $A_i$ is matched to a corresponding $P_i$.

\begin{figure*}[!t]
\small
\centering
\begin{prooftree}
\AxiomC{\mbox{\ }}
    \LeftLabel{\redlab{Tid}}
    \UnaryInfC{$[x \leftrightarrow y] \vdash x{:}A, y{:}\dual{A}; \,\Theta$}
    \DisplayProof
\hfill
    \AxiomC{\mbox{\ }}
    \LeftLabel{\redlab{T\onef}}
    \UnaryInfC{$x.\dual{\close} \vdash x: \onef; \,\Theta$}
    \DisplayProof
\hfill
    \AxiomC{$P\vdash \Delta ; \Theta $}
    \LeftLabel{\redlab{T\bot}}
    \UnaryInfC{$x.\close;P \vdash x{:}\bot, \Delta; \,\Theta$}
\end{prooftree}
  \vspace{-5mm}
  \begin{prooftree}
    \AxiomC{$P \vdash  \Delta, y:{A}; \,\Theta \quad Q \vdash \Delta', x:B; \,\Theta $}
    \LeftLabel{\redlab{T\otimes}}
    \UnaryInfC{$\dual{x}(y). (P \para Q) \vdash  \Delta, \Delta', x: A\otimes B; \,\Theta$}
\DisplayProof
\hfill 
    \AxiomC{$P \vdash \Delta, y:C, x:D; \,\Theta$}
    \LeftLabel{\redlab{T\ampy}}
    \UnaryInfC{$x(y).P \vdash \Delta, x: C\ampy D; \,\Theta $}
    \end{prooftree}
  \vspace{-6mm}
\begin{prooftree}
       \AxiomC{$P \;{ \vdash} \widetilde{w}:\with\Delta, x:A; \,\Theta$}
    \LeftLabel{\redlab{T\oplus^x_{\widetilde{w}}}}
    \UnaryInfC{$x.\some_{\widetilde{w}};P \vdash \widetilde{w}{:}\with\Delta, x{:}\oplus A; \,\Theta$}
    \DisplayProof
    \hfill
    \AxiomC{$P \vdash \Delta, x:A; \,\Theta$}
    \LeftLabel{\redlab{T\with_d^x}}
    \UnaryInfC{$x.\dual{\some};P \vdash \Delta, x :\with A; \,\Theta$}
    \end{prooftree}
      \vspace{-6mm}
    \begin{prooftree}
    \AxiomC{}
    \LeftLabel{\redlab{T\with^x}}
    \UnaryInfC{$x.\dual{\none} \vdash x :\with A; \,\Theta$}
    \DisplayProof
    \hfill 
    \AxiomC{$P \vdash \with\Delta; \,\Theta \qquad Q  \;{\vdash} \with\Delta; \,\Theta$}
    \LeftLabel{\redlab{T\oplus}}
    \UnaryInfC{$P\oplus Q \vdash \with\Delta; \,\Theta$}
 \end{prooftree}
   \vspace{-6mm}
   \begin{prooftree}
    \AxiomC{$ P  \vdash  \Delta, x: A_i ; \Theta $}
    \LeftLabel{\redlab{T\oplus_i}}
    \UnaryInfC{$\case{x}{i};P \vdash  \Delta,  x: \oplus_{i \in I} \{ \mathtt{l}_i : A_i  \}; \Theta $}
\DisplayProof
\hfill
    \AxiomC{$ P_i \vdash \Delta , x: A_i ; \Theta \quad (\forall i \in I)$}
    \LeftLabel{\redlab{T\with}}
    \UnaryInfC{$ \choice{x}{i}{I}{i}{P_i} \vdash \Delta , x: \with_{i \in I} \{ \mathtt{l}_i : A_i \} ; \Theta $}
\end{prooftree}
  \vspace{-6mm}
\begin{prooftree}
    \AxiomC{$P \vdash \Delta ; x : A , \Theta$}
    \LeftLabel{\redlab{T?}}
    \UnaryInfC{$P \vdash \Delta ,x :? A ; \Theta$}
\DisplayProof
\hfill
\AxiomC{$ P \vdash y: A; \Theta$}
    \LeftLabel{\redlab{T!}}
    \UnaryInfC{$!x(y).P \vdash x: !A; \Theta $}
\DisplayProof
\hfill
       \AxiomC{$P \;{ \vdash} \Delta , y: A ; x: A , \Theta $}
    \LeftLabel{\redlab{Tcopy}}
    \UnaryInfC{$ \outsev{x}{y}.P \vdash \Delta ; x: A , \Theta $}
\end{prooftree}
\caption{Typing rules for \spi.}
\label{ch3fig:trulespi}
\end{figure*}

The type system enjoys type preservation, a result that
follows from the cut elimination property in linear logic; it ensures that the observable interface of a system is invariant under reduction.
The type system also ensures other properties for well-typed processes (e.g. global progress, strong normalisation,  and confluence); see~\cite{CairesP17} for details.

\begin{theorem}[Type Preservation~\cite{CairesP17}]
If $P \vdash \Delta; \,\Theta$ and $P \redd Q$ then $Q \vdash \Delta; \,\Theta$.
\end{theorem}

\subsection[An Auxiliary Calculus With Sharing]{\lamrsharfailunres: An Auxiliary Calculus With Sharing}\label{ch3ssec:lamshar}

To facilitate the encoding of \lamrfailunres into \spi, we define \lamrsharfailunres: an auxiliary calculus whose constructs are inspired by the work of \cite{DBLP:conf/lics/GundersenHP13}, \cite{GhilezanILL11}, and \cite{DBLP:journals/iandc/KesnerL07}. 
The syntax of \lamrsharfailunres only modifies the syntax of terms, which is defined by the grammar below; variables ${x}[*]$, bags $B$, and expressions $\expr{M}$ are  as in \Cref{ch3def:rsyntaxfailunres}.
\begin{align*}
\mbox{(Terms)} \quad  M,N, L ::= &
~~  {x}[*] 
\sep \lambda x . (M[ {\widetilde{x}} \leftarrow  {x}]) 
\sep (M\ B) \sep M \linexsub{N /  {x}} 
\sep M \unexsub{U / x}
\\
    \sep &~~
\fail^{\widetilde{x}}\sep 
M [  {\widetilde{x}} \leftarrow  {x} ] 
\sep (M[ {\widetilde{x}} \leftarrow  {x}])\esubst{ B }{ x } 
\end{align*}
\noindent
We consider the {\it sharing  construct} $M[\widetilde{x}\leftarrow x]$ and two   kinds of explicit substitutions: the {\it explicit linear substitution}, written $M\linexsub{N/ {x}}$, and the {\it explicit unrestricted substitution}, written $M\unexsub{U/\unvar{x}}$. 
The term $M[ {\widetilde{x}}\leftarrow  {x}]$ defines the sharing of variables $ {\widetilde{x}}$ occurring in $M$ using the linear variable $ {x}$. 
We shall refer to $ {x}$ as \emph{sharing variable} and to $ {\widetilde{x}}$ as \emph{shared variables}. A linear variable is only allowed to appear once in a term. 
Notice that $ {\widetilde{x}}$ can be empty: $M[\leftarrow  {x}]$ expresses that $ {x}$ does not share any variables in $M$. 
As in $\lamrfailunres$, the term $\fail^{ {\widetilde{x}}}$ explicitly accounts for failed attempts at substituting the variables in $ {\widetilde{x}}$.

We summarise some requirements. 
In  $M [ \widetilde{x} \leftarrow x ]$, we require: (i)~every $x_i \in \widetilde{x}$ occurs exactly once in $M$ and that (ii)~$x_i$ is not a sharing variable.
The occurrence of $x_i$ can appear within the fail term $\fail^{\widetilde{y}}$, if $x_i \in \widetilde{y}$.
In the explicit linear substitution $M \linexsub{ N /  {x}}$, we require: the variable $x$ has to occur in $M$; $x$ cannot be a sharing variable; and $x$ cannot be in an explicit linear substitution occurring in $M$;  {all free {\em linear} occurrences of $x$ in $M$ are bound}. In the explicit unrestricted substitution $M \unexsub{ U / \unvar{x}}$, we require: all free {\em unrestricted} occurrences of $x$  in $M$ are bound; 
$\unvar{x}$ cannot be in an explicit unrestricted substitution occurring in $M$. 
This way, e.g.,  $M'\linexsub{L/ {x}}\linexsub{N/ {x}}$ 
and $M'\linexsub{U'/\unvar{x}}\linexsub{U/\unvar{x}}$ 
are not valid terms in $\lamrsharfailunres$. We consider consistent terms as defined in \Cref{ch2d:consistent}. As consistency applies to linear variables and does not concern unrestricted variables the results of reduction preservation, ensured by typing and structural congruence on terms follow analogously. When we refer to terms we will be referring to consistent terms.

The following congruence will be important when proving encoding correctness.

\begin{definition}{}\label{ch3def:rsPrecongruencefailure}
The congruence $\pequiv$ for \lamrsharfailunres on terms and expressions  is given by the identities below.
{\small
\[
\begin{array}{rcl}
    M \unexsub{U / \unvar{x}}  &\pequiv& M, \ x \not \in M\\
    
    (MB) \linexsub{N/x} & \pequiv& (M\linexsub{N/x})B, \ x \not \in \lfv{B} \\
    
    (MB) \unexsub{U / \unvar{x}} & \pequiv & (M\unexsub{U / \unvar{x}})B, \ x \not \in \lfv{B} \\
    
    (MA)[\widetilde{x} \leftarrow x]\esubst{B}{x}  &\pequiv& (M[\widetilde{x} \leftarrow x]\esubst{B}{x})A, \   x_i \in \widetilde{x} \Rightarrow x_i \not \in \lfv{A}\\
    
    M[\widetilde{y} \leftarrow y]\esubst{A}{y}[\widetilde{x} \leftarrow x]\esubst{B}{x} & \pequiv & (M[\widetilde{x} \leftarrow x]\esubst{B}{x})[\widetilde{y} \leftarrow y]\esubst{A}{y},\  
    \begin{aligned}
    x_i \in \widetilde{x} \Rightarrow x_i \not \in \lfv{A}, \\
       y_i \in \widetilde{y} \Rightarrow y_i \not \in \lfv{B}
    \end{aligned}
     \\[1mm]
    
    M \linexsub{N_2/y}\linexsub{N_1/x} &\pequiv & M\linexsub{N_1/x}\linexsub{N_2/y}, x \not \in \lfv{N_2}, y\notin \lfv{N_1} \\[1mm]
    
    M \unexsub{U_2 / \unvar{y}}\unexsub{U_1 / \unvar{x}} & \pequiv & M\unexsub{U_1 / \unvar{x}}\unexsub{U_2 / \unvar{y}}, x \not \in \lfv{U_2}, y\notin \lfv{U_1} \\[1mm]
    
    C[M]   &\pequiv& C[M'], \ \text{with } M \pequiv M' \\
    
    D[\expr{M}]  &\pequiv& D[\expr{M}'], \ \text{with } \expr{M} \pequiv \expr{M}'
\end{array}
\]
}
\end{definition}
The first rule states that we may remove unneeded unrestricted substitutions when the variable in concern does not appear within the term. The next three identities enforce that bags can always be moved in and out of all forms of explicit substitution, which are  useful manipulate expressions and to form a redex for Rule \redlab{R:Beta}. The other rules deal with permutation of explicit substitutions and contextual closure.

Well-formedness for \lamrsharfailunres, based on intersection types, is defined as in \secref{ch3sec:lam_types};
\iffulldoc
see  \appref{ch3app:ssec:lamshar}.
\else
see the full version for details.
\fi 


\subsection[First Step]{Encoding \lamrfailunres into \lamrsharfailunres} \label{ch3ssec:first_enc}
We define an encoding  $\recencodopenf{\cdot}$ from well-formed terms in $\lamrfailunres$ into $\lamrsharfailunres$. This encoding relies on an intermediate encoding $\recencodf{\cdot}$ on $\lamrfailunres$-terms. 
 
 \begin{notation}
Given a term $M$ such that  $\#(x , M) = k$
and a sequence of pairwise distinct fresh variables $\widetilde{x} = x_1, \ldots, x_k$ we write $M \linsub{\widetilde{x}}{x}$ or $M\linsub{x_1,\cdots, x_k}{x}$ to stand for 
$M\linsub{x_1}{x}\cdots \linsub{x_k}{x}$, i.e.,  a simultaneous linear substitution whereby each distinct linear occurrence of $x$ in $M$ is replaced by a distinct $x_i \in \widetilde{x}$. 
Notice that each $x_i$ has the same type as $x$.
We use (simultaneous) linear substitutions to force all bound linear variables in \lamrfailunres to become shared variables in \lamrsharfailunres.
\end{notation}

\begin{definition}{From $\lamrfailunres$ to $\lamrsharfailunres$}
\label{ch3def:enctolamrsharfailunres}
    Let $M \in \lamrfailunres$.
    Suppose $\Theta ; \Gamma \wfdash {M} : \tau$, with
    $\dom{\Gamma} = \llfv{M}=\{ {x}_1,\cdots, {x}_k\}$ and  $\#( {x}_i,M)=j_i$.  
    We define $\recencodopenf{M}$ as
    \begin{equation*}
     \recencodopenf{M} = 
    \recencodf{M\linsub{ {\widetilde{x}_{1}}}{ {x}_1}\cdots \linsub{ {\widetilde{x}_k}}{ {x}_k}}[ {\widetilde{x}_1}\leftarrow  {x}_1]\cdots [ {\widetilde{x}_k}\leftarrow  {x}_k] 
     \end{equation*}
       where  $ {\widetilde{x}_i}= {x}_{i_1},\cdots,  {x}_{{j_i}}$
       and the encoding $\recencodf{\cdot}: \lamrfailunres \to \lamrsharfailunres$  is defined in~\figref{ch3fig:auxencfailunres} on $\lamrfailunres$-terms. 
       The encoding $\recencodopenf{\cdot}$ extends homomorphically to expressions.
    \end{definition}

The encoding $\recencodopenf{\cdot}$  converts $n$ occurrences of   $x$ in a term into $n$ distinct variables $x_1, \ldots, x_n$.
The sharing construct coordinates them by constraining each to occur exactly once within a term. 
We proceed in two stages. 
First, we share all linear free linear variables using $\recencodopenf{\cdot}$: this ensures that free variables are replaced by shared variables which are then bound by the sharing construct. 
Second, we apply the encoding $\recencodf{\cdot}$ on the corresponding term. 
The encoding is presented in \figref{ch3fig:auxencfailunres}:
$ \recencodf{\cdot}$ maintains $x[i]$ unaltered, and acts homomorphically over  concatenation  of bags   and explicit substitutions. {The encoding renames bound variables with bound shared variables.}
As we will see, this will enable a tight operational correspondence result with $\spi$.
\iffulldoc
In \appref{ch3app:encodingone} we establish the correctness of   $\recencodopenf{\cdot}$.
\else
We establish the correctness of $\recencodopenf{\cdot}$ in the full version.
\fi

\begin{figure*}[t]
\small
\[
\begin{array}{c@{\hspace{.5cm}}c@{\hspace{.5cm}}c@{\hspace{.5cm}}c}
   \recencodf{  {x}  }  =   {x}  &
    \recencodf{ {x}[i]  }  =  {x}[i]  & \recencodf{  \oneb  } =  \oneb   &\\
     \recencodf{\banged{\oneb}} = \banged{\oneb}&  \recencodf{  \fail^{\widetilde{x}} } = \fail^{\widetilde{x}} & \recencodf{  M\ B }  =  \recencodf{M}\ \recencodf{B} 
& \\
    \recencodf{\banged{\bag{M}}} = \banged{\bag{M}}&  \recencodf{  \bag{M}\cdot C}  =  \bag{ \recencodf{M}} \cdot \recencodf{C}& \recencodf{  C \bagsep U  }  = \recencodf{C} \bagsep \recencodf{ U }&\\
  \recencodf{U \concat V} = U \concat V &  \recencodf{  M \linexsub{N /  {x}} }  =  \recencodf{M} \linexsub{ \recencodf{N} /  {x}}
     &
     \recencodf{ M \unexsub{U / \unvar{x}} }  =  \recencodf{M} \unexsub{ \recencodf{U} / \unvar{x}}&
\end{array}
\]
\vspace{-7mm}
\small
\begin{align*}
	    \recencodf{  \lambda x . M  }  &=   \lambda x . (\recencodf{M\langle  {x}_1 , \cdots ,  {x}_n /  {x}  \rangle} [ {x}_1 , \cdots ,  {x}_n \leftarrow  {x}])
  \quad  \text{ $\#( {x},M) = n$, each $ {x}_i$ is fresh}
\\
   \recencodf{  M \esubst{ C \bagsep U }{ x }  } &= 
   \begin{cases}
   \displaystyle\sum_{C_i \in \perm{\recencodf{ C  } }}\recencodf{ M \langle  {\widetilde{x}}/  {x}  \rangle } \linexsub{C_i(1)/ {x}_1} \cdots \linexsub{C_i(k)/ {x}_k}\unexsub{ U/ \unvar{x}},  ~ (*)\\
    \recencodf{M\langle  {x}_1, \cdots ,  {x}_k /  {x}  \rangle} [  {x}_1, \cdots ,  {x}_k \leftarrow  {x}] \esubst{ \recencodf{ C \bagsep U } }{ x }, ~  (**)
     \end{cases} 
\end{align*}
\vspace{-7mm}
\begin{align*}
    \hspace{-2cm}(*) & \text{if $\#( {x},M) = \size{C} = k$} &
    (**) & \text{if }\#( {x},M) = k\geq 0
\end{align*}
    \caption{Auxiliary Encoding: \lamrfailunres into \lamrsharfailunres}
    \label{ch3fig:auxencfailunres}
\end{figure*}


\begin{example}{}\label{ch3ex:id_enc} We apply the encoding $\recencodf{\cdot}$ in some of the $\lamrfailunres$-terms  from Example~\ref{ch3ex:id_term}: for simplicity, we assume that $N$ and $U$ have no free variables.
 \[{\small 
    \begin{aligned}
    \recencodf{(\lambda x.x)\ \bag{N}\bagsep U}&=\recencodf{\lambda x. x}\recencodf{\bag{N}\bagsep U}=\lambda x. x_1[x_1\leftarrow x]\bag{\recencodf{N}}\bagsep \recencodf{U}\\
   \recencodf{(\lambda x. x[1] ) \oneb \bagsep \bag{N}^! \concat U}&=\recencodf{(\lambda x. x[1]}\recencodf{\oneb \bagsep \bag{N}^! \concat U}=(\lambda x. x[1] [ \leftarrow x] ) \oneb \bagsep \bag{\recencodf{N}}^! \concat \recencodf{U}
    \end{aligned}
    }
\]

\end{example}




\subsection[Second Step]{Encoding $\lamrsharfailunres$ into $\spi$}\label{ch3ssec:second_enc}



We now define our encoding of $\lamrsharfailunres$ into $\spi$, and establish its correctness. 

\begin{notation}
To help illustrate the behaviour of the encoding, 
 we use the names $x$, $\linvar{x}$, and $\banged{x}$ to denote three distinct channel names: while $\linvar{x}$ is the channel that performs the linear substitution behaviour of the encoded term, channel $\banged{x}$ performs the unrestricted behaviour. 
\end{notation}

\begin{definition}{From $\lamrsharfailunres$ into $\spi$: Expressions}
\label{ch3def:enc_lamrsharpifailunres}
Let $u$ be a name.
The encoding $\piencodftypes{\cdot }_u: \lamrsharfailunres \rightarrow \spi$ is defined in \figref{ch3fig:encfailunres}.
\end{definition}

Every (free) variable $x$  in an $\lamrsharfailunres$ expression becomes a name $x$ in its corresponding \spi process.
As customary in encodings of $\lambda$ into $\pi$, we use a name $u$ to provide the behaviour of the encoded expression. 
In our case, $u$ is a non-deterministic session: the encoded expression can be effectively available or not; this is signalled by prefixes $u.\overline{\some}$
and 
$u.\overline{\none}$, respectively. 




We discuss the most salient aspects of the encoding in \figref{ch3fig:encfailunres}.
   \begin{itemize}
   \item 
    While linear variables are encoded as in \Cref{ch2}, the encoding of an unrestricted variable $x[j]$, not treated in \Cref{ch2}, is much more interesting: it first connects to a server along channel $x$ via a request $ \outsev{\banged{x}}{{x_i}}$ followed by a selection on $ {x}_i.l_{j}$, which takes the $j$-{th} branch.
   
    \item The encoding of   $\lambda x. M[\widetilde{x} \leftarrow x] $ confirms its behaviour first followed by the receiving of a channel $x$. The channel $x$ provides a linear channel $\linvar{x}$ and an unrestricted channel $\banged{x}$ for dedicated substitutions of the linear and unrestricted bag components.

    \item We encode $M\, (C \bagsep U)$ as  a non-deterministic sum: an application involves a choice in the order in which the elements of $C$ are substituted. 
    \item 
    The encoding of $C \bagsep U$ synchronises with the encoding of ${\lambda x . M[\widetilde{x} \leftarrow x]}$. The channel $ \linvar{x}$ provides the linear behaviour of the bag $C$ while $\banged{x}$ provides the behaviour of   $U$; this is done by guarding the encoding of $U$ with a server connection such that every time a channel synchronises with $!\banged{x} (x_i)$ a fresh copy of $U$ is spawned.
    
    \item The encoding of ${\bag{M} \cdot\, C}$ synchronises with the encoding of ${M[\widetilde{x} \leftarrow x]}$, just discussed. The name $y_i$ is used to trigger a failure in the computation if there is a lack of elements in the encoding of the bag.
    \item
    The encoding of ${M[\widetilde{x} \leftarrow x]}$ first confirms the availability of the linear behaviour along $\linvar{x}$. Then it sends a name $y_i$,  which is used to collapse the process in the case of a failed reduction. 
    Subsequently, for each shared variable, the encoding receives a name, which will act as an occurrence of the shared variable. 
    At the end, a failure prefix on $x$ is used to signal that there is no further  information to send over.
    \item 
    The encoding of $U$ synchronises with the last half encoding of ${x[j]}$; the name $x_i$ selects the $j$-th term in the unrestricted bag.
    \item 
    The encoding of ${ M \linexsub{N / x}} $ is the composition of the encodings of $M$ and $N$, where we await a confirmation of a behaviour along the variable that is being substituted.
    \item 
     ${ M \unexsub{U / \unvar{x}}} $ is encoded as the composition of the encoding of $M$ and a server guarding the encoding of $U$: in order for $\piencodftypes{M}_u$ to gain access to $\piencodftypes{U}_{x_i}$ it must first synchronise with the server channel $\banged{x}$ to spawn a fresh copy of $U$.
    \item The encoding of ${\expr{M}+\expr{N} }$ homomorphically preserves non-determinism. 
    Finally, the encoding of $\fail^{x_1 , \cdots , x_k}$ simply triggers failure on $u$ and on each of $x_1 , \cdots , x_k$.
\end{itemize}


\begin{figure*}[t!]
\small
\[
    \hspace{-0.0cm}
	\begin{array}{rl}
	   \piencodftypes{ {x}}_u &\hspace{-2mm} =  
	   {x}.\overline{\some} ; [ {x} \leftrightarrow u]  
	   \\[1mm]
   \piencodftypes{{x}[j]}_u & \hspace{-2mm}= 
   \outsev{\banged{x}}{{x_i}}. {x}_i.l_{j}; [{x_i} \leftrightarrow u] 
   \\[1mm]
    \piencodftypes{\lambda x.M[ {\widetilde{x}} \leftarrow  {x}]}_u & \hspace{-2mm}= 
    u.\overline{\some}; u(x). x.\overline{\some}; x(\linvar{x}). x(\banged{x}). x. \close ; \piencodftypes{M[ {\widetilde{x}} \leftarrow  {x}]}_u
    \\[1mm]
          \piencodftypes{ M[ {\widetilde{x}} \leftarrow  {x}] \esubst{ C \bagsep U }{ x} }_u &\hspace{-2mm} =\!\! \displaystyle  
      \bigoplus_{C_i \in \perm{C}}\!\! (\nu x)( x.\overline{\some}; x(\linvar{x}). x(\banged{x}). x. \close ;\piencodftypes{ M[ {\widetilde{x}} \leftarrow  {x}]}_u \para \piencodftypes{ C_i \bagsep U}_x )  
      \\[1mm]
  \piencodftypes{M (C \bagsep U)}_u &\hspace{-2mm} =  \displaystyle
  \bigoplus_{C_i \in \perm{C}} (\nu v)(\piencodftypes{M}_v \para v.\some_{u , \llfv{C}} ; \outact{v}{x} . ([v \leftrightarrow u] \para \piencodftypes{C_i \bagsep U}_x ) )  
  \\[1mm]    
  \piencodftypes{ C \bagsep U }_x & \hspace{-2mm}=  
  x.\some_{\llfv{C}} ; \outact{x}{\linvar{x}}.\big( \piencodftypes{ C }_{\linvar{x}} \para \outact{x}{\banged{x}} .( !\banged{x} (x_i). \piencodftypes{ U }_{x_i} \para x.\overline{\close} ) \big)
    \\[1mm]
        \piencodftypes{{\bag{M}} \cdot {C}}_{\linvar{x}} & \hspace{-2mm}=
       \linvar{x}.\some_{\llfv{\bag{M} \cdot C} } ; \linvar{x}(y_i). \linvar{x}.\some_{y_i, \llfv{\bag{M} \cdot C}};\linvar{x}.\overline{\some} ; \outact{\linvar{x}}{x_i}
       .\\[1mm]
       & \qquad (x_i.\some_{\llfv{M}} ; \piencodftypes{M}_{x_i} \para \piencodftypes{C}_{\linvar{x}} \para y_i. \overline{\none}) 
 \\[1mm]
    \piencodftypes{{\oneb}}_{\linvar{x}} &\hspace{-2mm} = 
   \linvar{x}.\some_{\emptyset} ; \linvar{x}(y_n). ( y_n.\overline{\some};y_n . \overline{\close} \para \linvar{x}.\some_{\emptyset} ; \linvar{x}. \overline{\none}) 
   \\[1mm]
   \piencodftypes{\banged{\oneb}}_{x} &\hspace{-2mm} =  x.\overline{\none} 
   \\[1mm]
   \piencodftypes{\banged{\bag{N}}}_{x}  &\hspace{-2mm} =  \piencodftypes{N}_{x}
   \\[1mm]
       \piencodftypes{ U }_{x}  &\hspace{-2mm}=  \choice{x}{U_i}{U}{i}{\piencodftypes{U_i}_{x}} 
       \\[1mm] 
      \piencodftypes{ M \linexsub{N /  {x}}  }_u    &\hspace{-2mm} =   
      (\nu  {x}) ( \piencodftypes{ M }_u \para    {x}.\some_{\llfv{N}};\piencodftypes{ N }_{ {x}}  )
      \\[1mm] 
      \piencodftypes{ M \unexsub{U / \unvar{x}}  }_u   &\hspace{-2mm} =   
      (\nu \banged{x}) ( \piencodftypes{ M }_u \para   ~!\banged{x}(x_i).\piencodftypes{ U }_{x_i} ) 
      \\[1mm]
       \piencodftypes{M[  \leftarrow  {x}]}_u &\hspace{-2mm} =
  \linvar{x}. \overline{\some}. \outact{\linvar{x}}{y_i} . ( y_i . \some_{u,\llfv{M}} ;y_{i}.\close; \piencodftypes{M}_u \para \linvar{x}. \overline{\none})
  \\[1mm]
          \piencodftypes{M[ {x}_1, \cdots ,  {x}_n \leftarrow  {x}]}_u &\hspace{-2mm} =
      \linvar{x}.\overline{\some}. \outact{\linvar{x}}{y_1}. \big(y_1 . \some_{\emptyset} ;y_{1}.\close;\zero   \para 
      \\[1mm]
      & \qquad \linvar{x}.\overline{\some};\linvar{x}.\some_{u, (\llfv{M} \setminus  {x}_1 , \cdots ,  {x}_n )};\linvar{x}( {x}_1) . \piencodftypes{M[ {x}_2,\cdots ,  {x}_n \leftarrow  {x}]}_u \big)
   \\[1mm] 
     \piencodftypes{\expr{M}+\expr{N} }_u    &\hspace{-2mm} =  \piencodftypes{ \expr{M} }_u \oplus \piencodftypes{ \expr{N} }_u
     \\[1mm]
   \piencodftypes{\fail^{x_1 , \cdots , x_k}}_u  &\hspace{-2mm} = 
   u.\overline{\none} \para x_1.\overline{\none} \para \cdots \para x_k.\overline{\none} 
\end{array}
\]
    \caption{Encoding \lamrsharfailunres into \spi (cf. Def.~\ref{ch3def:enc_lamrsharpifailunres}).}
    \label{ch3fig:encfailunres}
\end{figure*}

\begin{example}{}\label{ch3ex:id_pienc}[Cont. \cref{ch3ex:id_term}]
We illustrate the encoding $\piencodftypes{\cdot }$ on the $\lamrsharfailunres$-terms/bags occurring in $ M_1=\lambda x.x_1[x_1\leftarrow x](\bag{\recencodf{N}}\bagsep \recencodf{U})$ as below:


\(
\begin{aligned}
\piencodftypes{\lambda x.x_1[x_1\leftarrow x]}_v&=v.\overline{\some}; v(x). x.\overline{\some}; x(\linvar{x}).x(\banged{x}). x\sclose ;\piencodftypes{x_1[x_1\leftarrow x]}_v
\end{aligned}
\)

 \(
\begin{aligned}
\piencodftypes{\bag{\recencodf{N}}\bagsep \recencodf{U}}_x& =x.\some_{\llfv{\bag{\recencodf{N}}}}; \outact{x}{\linvar{x}} .( \piencodftypes{ \recencodf{N}}_{\linvar{x}} \para\outact{x}{\banged{x}} .( !\banged{x} (x_i). \piencodftypes{ \recencodf{U} }_{x_i} \para \overline{x}\sclose ) )
\end{aligned}
\)



{\small 
\(
\begin{aligned}
  \piencodftypes{\recencodf{M_1}}_{u}&= 
\piencodftypes{\lambda x. x_1[x_1\leftarrow x]\bag{\recencodf{N}}\bagsep \recencodf{U}}_{u}\\
 &=(\nu v)(\piencodftypes{\lambda x.x_1[x_1\leftarrow x]}_v\para v.\some_{u,\llfv{\recencodf{N}}};\outact{v}{x}.([v\leftrightarrow u]\para \piencodftypes{\bag{\recencodf{N}}\bagsep \recencodf{U}}_x))            
 \\
&=(\nu v)(  v.\overline{\some}; v(x). x.\overline{\some}; x(\linvar{x}). x(\banged{x}). x\sclose ;  \linvar{x}.\overline{\some}. \outact{\linvar{x}}{y_1}. (y_1 . \some_{\emptyset} ;y_{1}\sclose;\zero \para
\\
& \quad  \linvar{x}.\overline{\some};\linvar{x}.\some_{u}; \linvar{x}( {x}_1) . \linvar{x}. \overline{\some}. \outact{\linvar{x}}{y_2} .  ( y_2 . \some_{u, x_1 } ;y_{2}\sclose; \piencodftypes{\colthree{x_1}}_v \para \linvar{x}. \overline{\none}) )
            \para 
            \\ 
& \quad  v.\some_{u , \llfv{\recencodf{N}}};  \outact{v}{x} . ([v \leftrightarrow u] \para \\
& \quad 
 x.\some_{\llfv{\recencodf{N}}} ;  \outact{x}{\linvar{x}} .( \linvar{x}.\some_{\llfv{\recencodf{N}} } ; \linvar{x}.\some_{y_1, \llfv{ \bag{ \recencodf{N}}}};
\\
            & \quad  \linvar{x}. \overline{\some} ; \outact{\linvar{x}}{x_1} 
            .  (x_1.\some_{\llfv{\recencodf{N}}} ;\piencodftypes{\recencodf{N}}_{x_1} \para y_1. \overline{\none} \para  \linvar{x}.\some_{\emptyset} ;\linvar{x}(y_2).
            \\
            & \quad   ( y_2.\overline{\some};  \overline{y_2} \sclose \para \linvar{x}.\some_{\emptyset} ; \linvar{x}. \overline{\none}) )  \para  \outact{x}{\banged{x}} .( !\banged{x} (x_i). \piencodftypes{ \recencodf{U} }_{x_i} \para \overline{x}\sclose ) ) )
            )   
            \end{aligned}
            \)
}
\end{example}

We now encode intersection types (for  \lamrsharfailunres) into session types (for \spi).

\begin{figure}[!t]
	{\small 
\begin{align*}
 \piencodftypes{\unit} &= \with \onef 
 \\ 
  \piencodftypes{ \eta } &= \&_{\eta_i \in \eta} \{ \mathtt{l}_i ; \piencodftypes{\eta_i} \}  
  \\
  \piencodftypes{(\sigma^{k} , \eta )   \rightarrow \tau} &= \with( \dual{\piencodftypes{ (\sigma^{k} , \eta  )  }_{(\sigma, i)}} \ampy \piencodftypes{\tau}) 
  \\
  \piencodftypes{ (\sigma^{k} , \eta  )  }_{(\sigma, i)} &= \oplus( (\piencodftypes{\sigma^{k} }_{(\sigma, i)}) \otimes ((!\piencodftypes{\eta}) \otimes (\onef))  )
  \\[1mm]
 \piencodftypes{ \sigma \wedge \pi }_{(\sigma, i)} &= \overline{   \with(( \oplus \bot) \otimes ( \with  \oplus (( \with  \overline{\piencodftypes{ \sigma }} )  \ampy (\overline{\piencodftypes{\pi}_{(\sigma, i)}})))) } \\
     & = \oplus(( \with \onef) \ampy ( \oplus  \with (( \oplus \piencodftypes{\sigma} ) \otimes (\piencodftypes{\pi}_{(\sigma, i)}))))
     \\
 \piencodftypes{\omega}_{(\sigma, i)} & =  \quad \begin{cases}
     \overline{\with(( \oplus \bot )\otimes ( \with \oplus \bot )))} &  \text{if $i = 0$}
     \\
\overline{   \with(( \oplus \bot) \otimes ( \with  \oplus (( \with  \overline{\piencodftypes{ \sigma }} )  \ampy (\overline{\piencodftypes{\omega}_{(\sigma, i - 1)}})))) } & \text{if $i > 0$}
\end{cases}
\end{align*}
}
    \caption{Encoding of intersection types into session types  (cf. Def.~\ref{ch3def:enc_sestypfailunres})}
    \label{ch3fig:enc_types}
\end{figure}

\begin{definition}{From $\lamrsharfailunres$ into $\spi$: Types}
\label{ch3def:enc_sestypfailunres}
The translation  $\piencodftypes{\cdot}$  in Figure~\ref{ch3fig:enc_types} extends as follows to a  context 
$\Gamma =  {x}_1{:} \sigma_1, \cdots,  {x}_m {:} \sigma_m,  {v}_1{:} \pi_1 , \cdots ,  {v}_n{:} \pi_n$
and a context 
$\Theta =\banged{x}_1 {:} \eta_1 , \cdots , \banged{x}_k {:} \eta_k$:
\[
\begin{aligned}
\piencodftypes{\Gamma} &= {x}_1 : \with \overline{\piencodftypes{\sigma_1}} , \cdots ,   {x}_m : \with \overline{\piencodftypes{\sigma_m}} ,
  {v}_1:  \overline{\piencodftypes{\pi_1}_{(\sigma, i_1)}}, \cdots ,  {v}_n: \overline{\piencodftypes{\pi_n}_{(\sigma, i_n)}}\\
  \piencodftypes{\Theta}&=\banged{x}_1 : \dual{\piencodftypes{\eta_1}} , \cdots , \banged{x}_k : \dual{\piencodftypes{\eta_k}} 
\end{aligned}
\]

\end{definition}

\noindent
This encoding formally expresses how non-de\-termi\-nis\-tic session protocols (typed with `$\with$') capture linear and unrestricted resource consumption in \lamrsharfailunres. 
Notice that the encoding of the multiset type $\pi$ depends on two arguments (a strict type $\sigma$ and a number $i \geq 0$) which are left unspecified above.
This is crucial to represent failures in \lamrsharfailunres as typable processes in \spi. 
For instance, given $(\sigma^{j} , \eta ) \rightarrow \tau$ and $ ( \sigma^{k} , \eta)$, the well-formedness rule for application admits a mismatch 
\iffulldoc
($j \neq k$, cf. Rule \redlab{FS{:}app} in \figref{ch3app_fig:wfsh_rulesunres}, \appref{ch3app:ssec:lamshar})
\else
($j \neq k$, cf. Rule \redlab{FS{:}app} in the full version for details)
\fi
. In our proof of type preservation, the two arguments of the encoding are instantiated appropriately. 
Notice also how the client-server behaviour of unrestricted resources appears as `$!\piencodftypes{\eta}$' in the encoding of the tuple type
$(\sigma^{k} , \eta)$.
With our encodings of expressions and types in place, we can now define our encoding of judgements:

\begin{definition}{}
If $\expr{M}$ is an \lamrsharfailunres expression such that 
$\Theta ; \Gamma \wfdash \expr{M} : \tau$
then we define the encoding of the judgement to be: 
$\piencodftypes{\expr{M}}_u \vdash 
\piencodftypes{\Gamma}, 
u : \piencodftypes{\tau} ; \piencodftypes{\Theta} $.

\end{definition}



 The correctness  of  our encoding $\piencodftypes{\cdot}_u:\lamrfailunres\to \spi$, stated in Theorem~\ref{ch3thm:op_correct}
 \iffulldoc 
 (and detailed in~\appref{ch3app:encodingtwo})
 \else
 (and detailed in the appendix of the full version)
 \fi 
 , relies on a notion of {\it success}  for both $\lamrfailunres $  and $\spi$, given by the $\checkmark$ construct:




\begin{definition}{}
\label{ch3def:Suc3}We extend the syntax of terms for $\lamrsharfailunres$  and processes for \spi with $\checkmark$:
\begin{itemize}
    \item {\bf (In \lamrsharfailunres)} \succp{\expr{M}}{\checkmark} iff
there exist  $M_1, \cdots, M_k$ such that 
$\expr{M} \redd^*  M_1 + \cdots + M_k$ and
$\headf{M_j'} = \checkmark$, for some  $j \in \{1, \ldots, k\}$ and term $M_j'$ such that $M_j\pequiv  M_j'$.
\item {\bf (In \spi)} $\succp{P}{\checkmark}$ holds whenever there exists a $P'$
such that 
$P \redd^* P'$
and $P'$ contains an unguarded occurrence of $\checkmark$
(i.e., an occurrence that does not occur behind a prefix).
\end{itemize}

\end{definition}


\noindent
We now state operational correctness.
 \figref{ch3fig:my_label} illustrates the relation between completeness and soundness that the encoding satisfies: solid arrows denote reductions assumed, dashed arrows denote the application of $\piencodftypes{\cdot }_u$, and dotted arrows denote the existing reductions that can be implied from the results.
 
We remark that since \lamrsharfailunres satisfies the diamond property, it suffices to consider  completeness based on a single reduction ($ \expr{N}\redd \expr{M}$). Soundness uses the congruence $\pequiv$ in \defref{ch3def:rsPrecongruencefailure}.
We write $N \redd_{\pequiv} N'$ iff $N \pequiv N_1 \redd N_2 \pequiv N' $, for some $N_1, N_2$. Then, $\redd_{\pequiv}^*$ is the reflexive, transitive closure of $\redd_{\pequiv}$. For success sensitivity,  
we decree $\piencodftypes{\checkmark}_u = \checkmark$. We  have:

\begin{figure}
\begin{center}
	\begin{tikzpicture}[scale=.9pt]
\draw[rounded corners, color=black,fill=RedOrange!20] (0,5.5) rectangle (14.5,6.5);
\node (opcom) at (4.0,6.8){Operational Completeness};
\draw[rounded corners, color=black, fill=RoyalBlue!20] (0,2.5) rectangle (14.5,3.5);
\node (lamrfail) at (0.7,6) {$\lamrsharfailunres$:};
\node (expr1) [right of=lamrfail, xshift=.3cm] {$\mathbb{N}$};
\node (expr2) [right of=expr1, xshift=2cm] {$\mathbb{M}$};
\draw[arrow] (expr1) --  (expr2);
\node (lamrsharfail) at (0.7,3.0) {$\spi$:};
\node (transl1) [right of=lamrsharfail, xshift=.3cm] {$\piencodftypes{\mathbb{N}}_u$};
\node (transl2) [right of=transl1, xshift=2cm] {$Q \equiv \piencodftypes{\mathbb{M}}_u$};
\draw[arrow, dotted] (transl1) -- node[anchor= south,yshift=-1.25mm] {*}(transl2);
\node (enc1) at (1.65,4.5) {$\piencodftypes{\cdot }_u$};
\node  at (6.0,4.5) {$\piencodftypes{\cdot }_u$};
\node (refcomp) [right of=enc1, xshift=.9cm]{Thm~\ref{ch3thm:op_correct} (b)};
\draw[arrow, dashed] (expr1) -- (transl1);
\draw[arrow, dashed] (expr2) -- (transl2);
\node (opcom) at (10.65,6.8){Operational Soundness};
\node (expr1shar) [right of=expr2, xshift=1.8cm] {$\mathbb{N}$};
\node (expr2shar) [right of=expr1shar, xshift=2.8cm] {$\mathbb{N'}$};
\draw[arrow, dotted] (expr1shar) -- node[anchor= south, yshift=-1.25mm] {*} (expr2shar);
\node (transl1shar) [right of=transl2, xshift=1.8cm] {$\piencodftypes{\mathbb{N}}_u$};
\node (transl2shar) [right of=transl1shar, xshift=1cm] {$Q$};
\node (expr3shar) [right of=transl2shar, xshift=.8cm]{$Q'\equiv \piencodftypes{\mathbb{N'}}_u$};
\draw[arrow,dashed] (expr2shar) --  (expr3shar);
\draw[arrow,dashed] (expr1shar) -- (transl1shar);
\node (enc2) at (8.1,4.5) {$\piencodftypes{\cdot }_u$};
\node  at (13.35,4.5) {$\piencodftypes{\cdot }_u$};
\node (refsound) [right of=enc2, xshift=1.5cm]{Thm~\ref{ch3thm:op_correct}(c)};
\draw[arrow] (transl1shar) -- node[anchor= south, yshift=-1.25mm] {*} (transl2shar);
\draw[arrow,dotted] (transl2shar) -- node[anchor= south, yshift=-1.25mm] {*}(expr3shar);
\node at (12.45,5.7) {\small $\pequiv$ };
\end{tikzpicture}
\end{center}
    \caption{An overview  of operational soundness and completeness for $\piencodftypes{\cdot }_u$.}
    \label{ch3fig:my_label}
\end{figure} 

 \begin{theorem}[Operational Correctness] \label{ch3thm:op_correct} Let $\expr{N} $ and $ \expr{M} $ be well-formed $\lamrsharfailunres $ closed expressions.
 \begin{enumerate}[(a)]
 \item (Type Preservation) Let $B$  be a bag. We have:
         \begin{enumerate}[(i)]
        \item If $\Theta ; \Gamma \wfdash B :(\sigma^{k} , \eta )$
        then 
        $\piencodftypes{B}_u \wfdash  \piencodftypes{\Gamma}, u : \piencodftypes{(\sigma^{k} , \eta )}_{(\sigma, i)} ; \piencodftypes{\Theta}$.
        
        \item If $\Theta ; \Gamma \wfdash \expr{M} : \tau$
        then 
        $\piencodftypes{\expr{M}}_u \wfdash  \piencodftypes{\Gamma}, u :\piencodftypes{\tau} ; \piencodftypes{\Theta}$.
        \end{enumerate}
  \item (Completeness)  If $ \expr{N}\redd \expr{M}$ then there exists $Q$ such that $\piencodftypes{\expr{N}}_u  \redd^* Q \pequiv \piencodftypes{\expr{M}}_u$.
 \item \label{ch3thm:op_correct_sound} (Soundness) If $ \piencodftypes{\expr{N}}_u \redd^* Q$
then 
$Q \redd^* Q'$, $\expr{N}  \redd^*_{\pequiv} \expr{N}'$ 
and 
$\piencodftypes{\expr{N}'}_u \equiv Q'$, 
for some $Q', \expr{N}'$.
\item (Success Sensitivity)  $\succp{\expr{M}}{\checkmark}$ if, and only if, $\succp{\piencodftypes{\expr{M}}_u}{\checkmark}$.
 \end{enumerate}
 \end{theorem}
 
  \begin{proof}
  Below we illustrate the most interesting case of the proof of soundness. Detailed proof can be found in 
  \iffulldoc
  \appref{ch3app:encodingtwo}.
  \else
  the full version.
  \fi 
  \end{proof}

\begin{proof}
 All items are proven by structural induction; a detailed proof can be found 
 \iffulldoc
 in~\appref{ch3app:encodingtwo}.
 \else
 in the full version.
 \fi 

 Below we present the most interesting case in the proof of {\em soundness}: the case when $\expr{N} =  M (C \bagsep U) $. Then, 
  $$ \piencodftypes{\expr{N}}_u=\piencodftypes{M (C \bagsep U)}_u = \bigoplus_{C_i \in \perm{C}} (\nu v)(\piencodftypes{M}_v \para v.\some_{u , \llfv{C}} ; \outact{v}{x} . ([v \leftrightarrow u] \para \piencodftypes{C_i \bagsep U}_x ) ).$$ 
  
  The proof then proceeds by induction on the number of reduction steps $k$ that can be taken from $\piencodftypes{\expr{N}}_u$, i.e, $\piencodftypes{\expr{N}}_u \redd^k Q$. We will consider the case when $k \geq 1$, where for some process $R$ and non-negative integers $n, m$ such that $k = n+m$, we have the following:
            \[
            \begin{aligned}
               \piencodftypes{\expr{N}}_u & \redd^m  \bigoplus_{C_i \in \perm{C}} (\nu v)(R \para v.\some_{u , \llfv{C}} ; \outact{v}{x} . ([v \leftrightarrow u] \para \piencodftypes{C_i \bagsep U}_x ) ) \redd^n  Q\\
            \end{aligned}
            \]
There are several cases to analyse depending on the values of $m$ and $n$, and the shape of $M$. We consider  $m = 0$, $n \geq 1$ and $M=(\lambda x . (M'[ {\widetilde{x}} \leftarrow  {x}])) \linexsub{N_1 / y_1} \cdots \linexsub{N_p / y_p} \unexsub{U_1 / \unvar{z_1}} \cdots \unexsub{U_q / \unvar{z_q}}$, where  $p, q \geq 0$.  Then,  $\piencodftypes{\expr{N}}_u$ can perform the following reduction:
\[
\small
 \begin{aligned}
 \piencodftypes{\expr{N}}_u & \redd^* \bigoplus_{C_i \in \perm{C}} (\nu  \widetilde{y},\widetilde{z}, x)( x.\overline{\some}; x(\linvar{x}). x(\banged{x}). x \sclose ; \piencodftypes{M'[ {\widetilde{x}} \leftarrow  {x}]}_u \para Q''\para  \piencodftypes{C_i \bagsep U}_x ) ~ (:= Q_3) \\
\end{aligned}
     \]
where $Q''$ defines the encoding of explicit substitutions within the encoded subterm $M$.
Notice that:
 \[
                    \begin{aligned}
                        \expr{N} &=(\lambda x . (M'[ {\widetilde{x}} \leftarrow  {x}])) \linexsub{N_1 / y_1} \cdots \linexsub{N_p / y_p} \unexsub{U_1 / \unvar{z_1}} \cdots \unexsub{U_q / \unvar{z_q}}(C\bagsep U)\\
                        &\pequiv (\lambda x . (M'[ {\widetilde{x}} \leftarrow  {x}]) (C \bagsep U)) \linexsub{N_1 / y_1} \cdots \linexsub{N_p / y_p} \unexsub{U_1 / \unvar{z_1}} \cdots \unexsub{U_q / \unvar{z_q}} \\
                      & \redd   M'[ {\widetilde{x}} \leftarrow  {x}] \esubst{(C \bagsep U)}{x} \linexsub{N_1 / y_1} \cdots \linexsub{N_p / y_p} \unexsub{U_1 / \unvar{z_1}} \cdots \unexsub{U_q / \unvar{z_q}} = \expr{M}
                    \end{aligned}
                    \]
where the  congruence holds assuming the necessary $\alpha$-renaming of variables. 
  Finally, one can verify that  $\piencodftypes{\expr{M}}_u = Q_3$, and the result follows. 
    \end{proof}


\begin{example}{}\label{ch3ex:opcorr}
Recall again term $M_1$  from Example~\ref{ch3ex:id_term}.
It can be shown that $\recencodf{M_1}\redd^* \recencodf{N}\unexsub{\recencodf{U}/x^!}$. 
To illustrate operational completeness, we can verify preservation of  reduction, via $\piencodftypes{\cdot}$: reductions below use the  rules for $\spi$ in Figure~\ref{ch3fig:redspi}---see Figure~\ref{ch3fig:opcorr}.
\end{example}

\begin{figure}[H]
{\small	
   \[
    \begin{aligned}
    & \piencodftypes{\recencodf{M_1}}  = 
    \\
    & (\nu v)(  \colthree{ v.\overline{\some}; v(x).} x.\overline{\some}; x(\linvar{x}). x(\banged{x}). x\sclose ; \linvar{x}.\overline{\some}. \outact{\linvar{x}}{y_1}.  (y_1 . \some_{\emptyset} ; y_{1}\sclose;\zero \para\\
    &\qquad  \linvar{x}.\overline{\some}; \linvar{x}.\some_{u}; \linvar{x}( {x}_1) . \linvar{x}. \overline{\some}. \outact{\linvar{x}}{y_2} . ( y_2 . \some_{u, x_1 } ;y_{2}.\sclose; \piencodftypes{x_1}_v \para \linvar{x}. \overline{\none}) )
    \para \\ 
    &\qquad\colthree{v.\some_{u , \llfv{\recencodf{N}}} ;\outact{v}{x}.} (\colthree{[v \leftrightarrow u] }\para   x.\some_{\llfv{\recencodf{N}}} ; \outact{x}{\linvar{x}} .(\linvar{x}.\some_{\llfv{\recencodf{N}}} ;\linvar{x}(y_1). \\
    & \qquad  \linvar{x}. \some_{y_1, \llfv{ \bag{ \recencodf{N}}}}; \linvar{x}. \overline{\some} ; \outact{\linvar{x}}{x_1}.  (x_1.\some_{\llfv{\recencodf{N}}}; \piencodftypes{\recencodf{N}}_{x_1} \para y_1. \overline{\none} \para  \linvar{x}.\some_{\emptyset} ;\\ 
    & \qquad     \linvar{x}(y_2). ( y_2.\overline{\some};\overline{y_2}  \sclose \para \linvar{x}.\some_{\emptyset} ; \linvar{x}. \overline{\none}) )  \para \outact{x}{\banged{x}} .( !\banged{x} (x_i). \piencodftypes{ \recencodf{U} }_{x_i} \para \overline{x}\sclose ) ) )
    )
    \\[1mm]
    & \redd^{3}
    (\nu x)( \colthree{ x.\overline{\some}; x(\linvar{x}).} x(\banged{x}). x \sclose ; \linvar{x}.\overline{\some}. \outact{\linvar{x}}{y_1}.  (y_1 . \some_{\emptyset} ; y_{1}\sclose;\zero \para \linvar{x}.\overline{\some};\linvar{x}.\some_{u}; \\ 
    &\qquad  \linvar{x}( {x}_1). \linvar{x}. \overline{\some}. \outact{\linvar{x}}{y_2} . ( y_2 . \some_{u, x_1 } ;y_{2}\sclose; \piencodftypes{x_1}_u \para \linvar{x}. \overline{\none}) )
    \para  (\colthree{x.\some_{\llfv{\recencodf{N}}} ; \outact{x}{\linvar{x}}}. \\
    & \qquad  (\linvar{x}.\some_{\llfv{\bag{\recencodf{N}}}};\linvar{x}(y_1). \linvar{x}.  \some_{y_1, \llfv{ \bag{ \recencodf{N}}}}; \linvar{x}. \overline{\some} ; \outact{\linvar{x}}{x_1}. (x_1.\some_{\llfv{\recencodf{N}}} ; \piencodftypes{\recencodf{N}}_{x_1} \para \\
    & \qquad  y_1. \overline{\none} \para  \linvar{x}.  \some_{\emptyset} ;\linvar{x}(y_2). ( y_2.\overline{\some};\overline{y_2} \sclose \para \linvar{x}.\some_{\emptyset} ; \linvar{x}. \overline{\none}) )  \para \\
    & \qquad \outact{x}{\banged{x}}.( !\banged{x} (x_i). \piencodftypes{ \recencodf{U} }_{x_i} \para \overline{x}\sclose ) ) )
    )
    \\[1mm]
            & \redd^{2}
    (\nu x, \linvar{x})( \colthree{x(\banged{x}).} x. \sclose ; \linvar{x}.\overline{\some}. \outact{\linvar{x}}{y_1}.  (y_1 . \some_{\emptyset} ; y_{1}\sclose;\zero  \para \linvar{x}.\overline{\some}; \linvar{x}.\some_{u}; \linvar{x}( {x}_1) .  \\ 
    &\qquad  \linvar{x}. \overline{\some}. \outact{\linvar{x}}{y_2} . ( y_2 . \some_{u, x_1 } ;y_{2}\sclose; \piencodftypes{x_1}_u \para \linvar{x}. \overline{\none}) )
    \para   (\linvar{x}.\some_{\llfv{\recencodf{N}}}  ;\linvar{x}(y_1).  \\ 
    & \qquad \linvar{x}.  \some_{y_1, \llfv{  \recencodf{N}}}; \linvar{x}.\overline{\some} ;\outact{\linvar{x}}{x_1}.    (x_1.\some_{\llfv{\recencodf{N}}} ;\piencodftypes{\recencodf{N}}_{x_1} \para y_1. \overline{\none} \para  \linvar{x}.\some_{\emptyset} ;  \linvar{x}(y_2).  \\
    & \qquad  ( y_2.\overline{\some};\overline{y_2}\sclose  \para \linvar{x}.\some_{\emptyset} ; \linvar{x}. \overline{\none}) )\para \colthree{\outact{x}{\banged{x}}.}( !\banged{x} (x_i). \piencodftypes{ \recencodf{U} }_{x_i} \para \overline{x}\sclose ) ) 
    )
    \\[1mm]
    & \redd
    (\nu x, \linvar{x}, x^!)( \colthree{x \sclose }; \linvar{x}.\overline{\some}. \outact{\linvar{x}}{y_1}.  (y_1 . \some_{\emptyset} ; y_{1}\sclose;\zero  \para \linvar{x}.\overline{\some}; \linvar{x}.\some_{u}; \linvar{x}( {x}_1) .  \linvar{x}. \overline{\some}. \\ 
    &\qquad \outact{\linvar{x}}{y_2} .( y_2 . \some_{u, x_1 } ;y_{2}\sclose; \piencodftypes{x_1}_u \para \linvar{x}. \overline{\none}) )
    \para   (\linvar{x}.\some_{\llfv{\recencodf{N}}}  ;\linvar{x}(y_1). \linvar{x}. \some_{y_1, \llfv{  \recencodf{N}}}; \\ 
    & \qquad \linvar{x}. \overline{\some} ;   \outact{\linvar{x}}{x_1}. (x_1.\some_{\llfv{\recencodf{N}}} ; \piencodftypes{\recencodf{N}}_{x_1} \para y_1. \overline{\none} \para  \\
    &
        \qquad  \linvar{x}.\some_{\emptyset} ; \linvar{x}(y_2). ( y_2.\overline{\some};\overline{y_2}  \sclose \para 
        \linvar{x}.\some_{\emptyset} ; \linvar{x}. \overline{\none}) )  \para !\banged{x} (x_i). \piencodftypes{ \recencodf{U} }_{x_i} \para \colthree{\overline{x}\sclose}  ) )
    \\[1mm]
        & \redd
    (\nu  \linvar{x}, x^!)( \colthree{\linvar{x}.\overline{\some}. \outact{\linvar{x}}{y_1}. } (y_1 . \some_{\emptyset} ; y_{1}.\sclose;\zero  \para \linvar{x}.\overline{\some}; \linvar{x}.\some_{u}; \linvar{x}( {x}_1) . \\ 
    &\qquad  \linvar{x}. \overline{\some}. \outact{\linvar{x}}{y_2} . ( y_2 . \some_{u, x_1 } ;y_{2}.\sclose; \piencodftypes{x_1}_u \para \linvar{x}. \overline{\none}) )
    \para   (\colthree{\linvar{x}.\some_{\llfv{\recencodf{N}}}  ;\linvar{x}(y_1).}\\
    & \qquad \linvar{x}.  \some_{y_1, \llfv{  \recencodf{N}}}; \linvar{x}. \overline{\some} ; \outact{\linvar{x}}{x_1}. (x_1.\some_{\llfv{\recencodf{N}}} ; \piencodftypes{\recencodf{N}}_{x_1} \para y_1. \overline{\none} \para \\ 
    & \qquad  \linvar{x}.\some_{\emptyset} ;  \linvar{x}(y_2). ( y_2.\overline{\some};\overline{y_2}  \sclose \para \linvar{x}.\some_{\emptyset} ; \linvar{x}. \overline{\none}) )  \para  !\banged{x} (x_i). \piencodftypes{ \recencodf{U} }_{x_i}  ) )
    ) 
    \\[1mm]
    & \redd
    (\nu  \linvar{x}, y_1,  x^!)( y_1 . \some_{\emptyset};
    y_{1}\sclose;\zero \para \linvar{x}.\overline{\some}; \linvar{x}.\some_{u}; \linvar{x}( {x}_1) .  \linvar{x}. \overline{\some}. \outact{\linvar{x}}{y_2}.   \\ 
    & \qquad ( y_2 . \some_{u, x_1 } ; y_{2}\sclose; \piencodftypes{x_1}_u \para \linvar{x}. \overline{\none})\para   (             \linvar{x}. \some_{y_1, \llfv{  \recencodf{N}}}; \linvar{x}. \overline{\some} ;  \outact{\linvar{x}}{x_1}. (x_1.\some_{\llfv{\recencodf{N}}} ; \\ 
    & \qquad  \piencodftypes{\recencodf{N}}_{x_1} \para y_1. \overline{\none} \para  \linvar{x}.\some_{\emptyset} ; \linvar{x}(y_2). ( y_2.\overline{\some};\overline{y_2}\sclose \para \linvar{x}.\some_{\emptyset} ; \linvar{x}. \overline{\none}) )  \para \\
    & \qquad  !\banged{x} (x_i). \piencodftypes{ \recencodf{U} }_{x_i}  ) )
    \\[1mm]
    & \redd^*  (\nu  x_1,  x^!)({x_1}.\overline{\some} ; [ {x_1} \leftrightarrow u]
    \para  x_1.\some_{\llfv{\recencodf{N}}} ; \piencodftypes{\recencodf{N}}_{x_1} \para    !\banged{x} (x_i). \piencodftypes{ \recencodf{U} }_{x_i})
    \\[1mm]
    & \redd^* (\nu x^!)(
        \piencodftypes{\recencodf{N}}_{u} \para    !\banged{x} (x_i). \piencodftypes{ \recencodf{U} }_{x_i})
        \\
    &              =\piencodftypes{\recencodf{N}\unexsub{\recencodf{U}/x^!}}_u
    \end{aligned}
    \]
}
\caption{Illustrating operational correspondence, following Example~\ref{ch3ex:opcorr}.\label{ch3fig:opcorr}}
\end{figure}

\subsection{A Motivating Example}

We use a cyclic task scheduler for automated maintenance as a motivating example. Consider a server that is running a maintenance routine; the user wants the server to be able to run independently in the background without having to worry about its condition. The server has a list of operations that it must perform, such as running virus checks ($\sff{VC}$), clearing memory ($\sff{CM}$), and defragmenting its storage ($\sff{DS}$). These tasks must be run periodically throughout while the server is running and hence they must be called multiple times. Finally, these tasks are run cyclically, with each task being executed in sequence, and the cycle repeats until a stop notification is received.

We sketch the design of functions of this nature, which we are able to conceive  due to the to underling encoding including client-server behaviours, given in this chapter.
Consider the reduction in the binary case:
\[
    \frac{x \in \fv{M}}{
    (\lambda x. M ) (N_1 \succ N_2 ) \red  (\lambda x. M \{N_1 / x\}) (N_2 \succ N_1 ) }
\]
Above, $M \{ N_1 / x \}$ represents a substitution of $N_1$ for a single occurrence of the left-most $x$ in the term $M$. Also, $N_1 \succ N_2$ denotes the non-deterministic choice interpreted as $N_1$ is the resource that must be used first then $N_2$; that is, the resources $N_1$ and $N_2$ can be seen as queued. Then this reduction substitutes the $N_1$ for $x$ and moves $N_1$ to the end of the queue. We can consider this behaviour a subset of the behaviour expressible in $\lamrsharfailunres$ as $\lamrsharfailunres$ allows for the substitution of resources in any order rather then cycling through each resource in turn. Our correct encoding gives us {guideline} that allow for the underling process model to give semantics to this function.

Now consider the following function:
\[
M = (\lambda x . \sff{sequence8} (\sff{execute}(x)))( \sff{VC} \succ \sff{CM} \succ \sff{DS} )
\]

Here we assume access to the function `$\sff{execute}(x)$', which takes in a task such as $\sff{VC}$ and executes it, and `$\sff{sequenceN}$', which   is a function that takes $N$ elements and runs them one at a time in sequence and then terminates (similar to $\sff{sequence3}$  in \Cref{ss:exammotiv}). In this term, `$\sff{execute}(x)$' is the only resource applied to `$\sff{sequence8}$' and hence the order does not matter in this substitution. In other words, we could consider `$\sff{sequence8} (\sff{execute}(x))$' to be equivalent to:
\[\sff{execute}(x);\sff{execute}(x);\sff{execute}(x);\sff{execute}(x);\sff{execute}(x);\sff{execute}(x);\sff{execute}(x);\sff{execute}(x)\]

In the underling process model both `$\sff{execute}$' and our tasks $\sff{VC},\sff{CM}$ and $\sff{DS}$ can be considered \emph{servers}. In fact `$\sff{execute}$' can be called an arbitrary number  of times and hence must be readily available. From our encoding we know that unrestricted resources act as servers within the concurrent paradigm; hence, we may interpret that the tasks (for example $\sff{DS}$) are invoked via a server call when the task needs to be performed. Similarly, $\sff{execute}$ behaves as a server providing the corresponding behaviour. Given this, we can have reductions of the form:
\[
\begin{aligned}
    M =& (\lambda x . \sff{sequence8} (\sff{execute}(x)))( \sff{VC} \succ \sff{CM} \succ \sff{DS} ) \\
    \red & (\lambda x .  \sff{execute}(\sff{x} 
    );\sff{execute}(\sff{x} 
    );\sff{execute}(\sff{x} 
    );\sff{execute}(x); \cdots )(  \sff{CM} \succ \sff{DS} \succ \sff{VC}) \\
    = & (\lambda x .  \sff{execute}(\sff{VC} 
    );\sff{execute}(\sff{x} 
    );\sff{execute}(\sff{x} 
    );\sff{execute}(x); \cdots )(  \sff{CM} \succ \sff{DS} \succ \sff{VC}) \\
    \red & (\lambda x .  \sff{execute}(\sff{VC} 
    );\sff{execute}(\sff{CM} 
    );\sff{execute}(\sff{x} 
    );\sff{execute}(x); \cdots)(  \sff{DS} \succ \sff{VC} \succ \sff{CM}) \\
    \red & (\lambda x . \sff{execute}(\sff{VC} 
     );\sff{execute}(\sff{CM} 
     );\sff{execute}(\sff{DS} 
     );\sff{execute}(x); \cdots )(  \sff{VC} \succ \sff{CM}\succ \sff{DS} ) \\
    \red & \cdots
\end{aligned}
\]

After eight iterations of executing tasks, the term $M$ terminates. The reduction of $M$ will run the given tasks in sequence until terminating, allowing for each task to be used an arbitrary amount of times.

The extension of the calculus $\lamrsharfailunres$ with sequencing as in \Cref{ss:exammotiv} with the following function also allows us to express this behaviour:
\[
(\lambda x . \sff{execute}(x[1]);\sff{execute}(x[2]);\sff{execute}(x[3]);\sff{execute}(x[1]);\cdots) \banged{\bag{\sff{VC} , \sff{CM} , \sff{DS}  }}
\]
where `$\sff{execute}(x[1]);\sff{execute}(x[2]);\sff{execute}(x[3]);\sff{execute}(x[1]);\ldots$' denotes the finite sequence of executions that occurs before terminating. Each execution takes the argument $x[i]$ and each subsequent call increases the index $i$, thus mimicking the behaviour of the unrestricted resources in $ \banged{\bag{\sff{VC} , \sff{CM} , \sff{DS}  }} $ being called in sequence.

 \section{Concluding Remarks}
\subparagraph{Summary} 
We have extended the line of work we developed in \Cref{ch2}, on resource $\lambda$-calculi with firm logical foundations via typed concurrent processes. We presented \lamrfailunres, a resource calculus with
 non-determinism and explicit failures, with dedicated treatment for linear and unrestricted resources. By means of examples, we illustrated the expressivity, (lazy) semantics, and design decisions underpinning   \lamrfailunres, and introduced a class of well-formed expressions based on intersection types, which includes fail-prone expressions. To bear witness to the logical foundations of \lamrfailunres, we defined and proved correct a typed encoding into the  concurrent calculus \spi, which subsumes the one in \Cref{ch2}.
 We plan to study key properties for \lamrfailunres (such as solvability and normalisation) by leveraging our  typed encoding into \spi.
 
\subparagraph{Related Work}
With respect to previous resource calculi, a distinctive feature of 
 \lamrfailunres  is its support of explicit failures, which may arise depending on the interplay between (i)~linear and unrestricted occurrences of variables in a term and (ii)~associated resources in the bag. This feature allows \lamrfailunres to express variants of usual $\lambda$-terms ($\mathbf{I}$, $\Delta$, $\Omega$) not expressible in other resource calculi.

Related to \lamrfailunres is Boudol's work on a $\lambda$-calculus  in which  multiplicities can be infinite~\cite{DBLP:conf/concur/Boudol93,DBLP:conf/birthday/BoudolL00}. 
An intersection type system is used to prove {\em adequacy} with respect to a testing semantics. However, failing behaviours as well as typability are not explored. Multiplicities can be expressed in $\lamrfailunres$:  a linear resource is available $m$ times when the linear bag contains $m$ copies of it; the term fails if the corresponding number of linear variables is different from $m$.

Also related is the resource $\lambda$-calculus by \cite{PaganiR10}, which includes linear and reusable resources; the latter are available in multisets, also called bags. In their setting, $M[N^!]$ denotes an application of a term $M$ to a resource $N$ that can be  used {\em ad libitum}.  Standard terms such as {\bf I}, $\Delta$ and $\Omega$ are expressed as $\lambda x.x$, $\Delta:=\lambda x. x[x^!]$,  and $\Omega:=\Delta[\Delta^!]$, respectively;  different variants are possible but cannot express the desired behaviour. A lazy reduction semantics is based on {\em baby} and {\em giant} steps: whereas the first consume one resource at each time, the second comprises several baby steps; combinations of the use of resources (by permuting resources in bags) are considered. A (non-idempotent) intersection type system is proposed: normalisation and a characterisation of solvability are investigated. Unlike our work, encodings into the $\pi$-calculus are not explored in~\cite{PaganiR10}.


\clearemptydoublepage

\chapter{Typed Non-determinism in Functional and Concurrent Calculi}\label{ch4}

    We study functional and concurrent calculi with \nond, along with  {type systems} to control resources based on  {linearity}.
The interplay between \nond and linearity is delicate: careless handling of branches can discard resources meant to be used exactly once.
Here we go beyond prior work by considering \nond in its standard sense: once a branch is selected, the rest are discarded.
Our technical contributions are three-fold.
    First, we introduce a $\pi$-calculus with non-de\-ter\-ministic choice, governed by session types.
    Second, we introduce a resource $\lambda$-calculus, governed by intersection types, in which \nond concerns fetching of resources from bags.
    Finally, we connect our two typed non-deterministic calculi via a correct translation.
    %

\section{Introduction}
\label{ch4s:intro}

\myrev{In this chapter, we present new formulations of typed programming calculi with \emph{non-determinism}.}
A classical ingredient of models of computation, \nond brings flexibility and generality in specifications.
In process calculi such as CCS and the $\pi$-calculus,  one source of non-determinism is {choice}, which is typically \emph{\nconf}: that is, given $P + Q$, we have either $P + Q \longrightarrow P$ or $P + Q \longrightarrow Q$.
Thus, committing to a branch entails discarding the rest.

\myrev{We study non-determinism as a way of increasing the expressivity of typed calculi in which resource control is based on \emph{linearity}.}
The interplay between non-determinism and linearity is delicate: a careless discarding of branches can jeopardize resources meant to be used exactly once.
On the concurrent side, we consider the $\pi$-calculus,
\myrev{the paradigmatic model of concurrency~\cite{DBLP:books/daglib/0004377}.}
We focus  on $\pi$-calculi with \emph{session types}~\cite{DBLP:conf/concur/Honda93,DBLP:conf/esop/HondaVK98}, in which linear logic principles ensure communication correctness:
here the resources are names that perform session protocols; they can be \emph{unrestricted} (used multiple times) and \emph{linear} (used exactly once).
To properly control resources, \nconf non-determinism is confined to unrestricted names; linear names can only perform {deterministic choices}.

In this context, considering \emph{confluent} forms of \nond can be appealing.
Intuitively, such formulations allow all branches to proceed independently:  given $P_1 \longrightarrow Q_1$ and $P_2 \longrightarrow Q_2$, then $P_1 + P_2 \longrightarrow  Q_1 + P_2$
and
$P_1 + P_2 \longrightarrow  P_1 + Q_2$.
Because confluent \nond does not discard branches, it is  compatible with a resource-conscious view of computation.

\Conf \nond has been studied mostly in the functional setting; it is present, e.g., in
Pagani and Ronchi della Rocca's {resource} $\lambda$-cal\-cu\-lus~(\cite{PaganiR10})
and in
Ehrhard and Regnier's differential $\lambda$-cal\-culus~(\cite{DBLP:journals/tcs/EhrhardR03}).
In~\cite{PaganiR10}, non-determinism resides in the application of a term $M$ to a \emph{bag} of available resources $C$; a $\beta$-reduction applies $M$ to a resource \emph{non-deterministically fetched} from $C$.
\Conf \nondt choice is also present in the session-typed $\pi$-calculus by \cite{CairesP17}, where it expresses a choice between different implementations of the same session protocols, which are all \emph{non-deterministically available}---they may be available but may also \emph{fail}.
In their work, a Curry-Howard correspondence between linear logic and session types (`\emph{propositions-as-sessions}'~\cite{CairesP10,DBLP:conf/icfp/Wadler12}) ensures confluence,  protocol fidelity, and deadlock-freedom.
\Cref{ch2,ch3} relate functional and concurrent calculi with confluent \nond: they give a translation
 of a resource $\lambda$-calculus into the session $\pi$-calculus from~\cite{CairesP17}, in the style of Milner's `\emph{functions-as-pro\-ce\-sses}'~(\cite{Milner90}).

Although results involving confluent \nond are most significant, usual (\nconf) \nond remains of undiscussed convenience in formal modeling; consider, e.g., specifications of distributed protocols~\cite{DBLP:journals/entcs/BergerH00,DBLP:conf/concur/NestmannFM03} in which commitment is essential.
Indeed,  \nconf \nondt choice is commonplace in verification frameworks such as mCRL2~\cite{DBLP:books/mit/GrooteM2014}.
It is also  relevant in functional calculi; a well-known framework is \cite{DBLP:journals/iandc/deLiguoroP95} (untyped) non-deterministic
 $\lambda$-calculus (see also~\cite{DezaniCiancaglini:TR-TU-96} and references therein).

\smallskip
\myrev{
To further illustrate the difference between confluent and non-confluent \nond, we consider an example adapted from~\cite{CairesP17}: a movie server that offers a choice between buying a movie or watching its trailer.
In \clpi, the typed $\pi$-calculus that we present in this chapter, this server can be specified as follows:
    \[
    \sff{Server}_s = \gsel{s} \left\{
        \begin{array}{@{}l@{}}
            \sff{buy} : \gname{s}{\textit{title}} ; \gname{s}{\textit{paym}} ; \pname{s}{\texttt{movie}} ; \pclose{s}
            ~,~
            \\
            \sff{peek} : \gname{s}{\textit{title}} ; \pname{s}{\texttt{trailer}} ;  \pclose{s}
        \end{array}
    \right\}
    \begin{array}{@{}l@{}}
        -\sff{Server}_s^{\sff{buy}}
        \\
        -\sff{Server}_s^{\sff{peek}}
    \end{array}
    \]
where $\gname{s}{-}$ and $\pname{s}{-}$ denote input and output prefixes on a name/channel $s$, respectively, and `\texttt{movie}' and `\texttt{trailer}' denote references to primitive data.
Also, the free names of a process are denoted with subscripts.
Process $\sff{Server}_s$ offers a choice on name~$s$ ($\gsel{s}\{-\}$) between labels \sff{buy} and \sff{peek}.
If \sff{buy} is received, process $\sff{Server}_s^{\sff{buy}}$ is launched: it receives the movie's title and a payment method, sends the movie, and closes the session on~$s$ ($\pclose{s}$).
If \sff{peek} is received, it proceeds as $\sff{Server}_s^{\sff{peek}}$: the server receives the title, sends the trailer, and closes the session.}

\myrev{Using the non-deterministic choice operator of \clpi, denoted `$\nd$', we can specify a process for  a client Alice who is interested in the movie `Jaws' but is undecided about buying the film or just watching its trailer for free:
\[
    \sff{Alice}_s := \begin{array}[t]{@{}r@{}lll@{}}
        &
        \psel{s}{\sff{buy}} ;
        &
        \pname{s}{\texttt{Jaws}} ; \pname{s}{\texttt{mcard}} ; \gname{s}{\textit{movie}} ; \gclose{s} ; \0
        &
        -\sff{Alice}_s^{\sff{buy}}
        \\
        {} \nd {} &
        \psel{s}{\sff{peek}} ;
        &
        \pname{s}{\texttt{Jaws}} ; \gname{s}{\textit{trailer}} ; \gclose{s} ; \0
        &
        -\sff{Alice}_s^{\sff{peek}}
    \end{array}
\]
If $\sff{Alice}_s$ selects the label $\sff{buy}$ ($\psel{s}{\sff{buy}}$), process $\sff{Alice}_s^{\sff{buy}}$ is launched: it sends title and payment method, receives the movie, waits for the session to close ($\gclose{s}$), and then terminates ($\0$).
If $\sff{Alice}_s$ selects \sff{peek}, process $\sff{Alice}_s^{\sff{peek}}$ is launched: it sends a title, receives the trailer, waits for the session to close, and terminates. Then, process $\sff{Sys} := \res{s} ( \sff{Server}_s \| \sff{Alice}_s )$ denotes the composition of client and server, connected  along $s$ (using $ \res{s}$).
Our semantics for  \clpi, denoted $\redtwo$, enforces non-confluent non-determinism, as $\sff{Sys}$ can reduce to separate processes, as expected:
\begin{mathpar}
    \sff{Sys} \redtwo \res{s} ( \sff{Server}_s^{\sff{buy}} \| \sff{Alice}_s^{\sff{buy}} )
    \and
    \text{and}
    \and
    \sff{Sys} \redtwo \res{s} ( \sff{Server}_s^{\sff{peek}} \| \sff{Alice}_s^{\sff{peek}} )
\end{mathpar}
}
\myrev{In contrast, the confluent non-deterministic choice  from \cite{CairesP17}, denoted~`$\oplus$', behaves differently: in their confluent semantics, $\sff{Sys}$ reduces to a  \emph{single} process including \emph{both alternatives}, i.e., $\sff{Sys} \red  \res{s} ( \sff{Server}_s^{\sff{buy}} \| \sff{Alice}_s^{\sff{buy}} ) \oplus \res{s} ( \sff{Server}_s^{\sff{peek}} \| \sff{Alice}_s^{\sff{peek}} )$.
}

\paragraph*{Contributions.}
We study new concurrent and functional  calculi with usual (\nconf) forms of \nond.
Framed in the typed (resource-conscious) setting,
we strive for definitions that do not exert a too drastic discarding of branches (as in the \nconf case) but also that do not exhibit a lack of commitment (as in the \conf case).
Concretely, we present:

\smallskip \noindent
(\secref{ch4s:pi})
\clpi, a variant of the session-typed $\pi$-calculus in~\cite{CairesP17}, now with \nconf \nondt choice.
Its semantics adapts to the typed setting the usual semantics of non-deterministic choice in the untyped $\pi$-calculus~\cite{DBLP:books/daglib/0004377}.
Well-typed processes enjoy type preservation and deadlock-freedom (\Cref{ch4t:srPi,ch4t:dfPi}).

\smallskip \noindent
(\secref{ch4s:lambda})
\lamcoldetshlin , a resource $\lambda$-calculus with   \nond, enhanced with constructs for expressing resource usage and failure.
Its  non-idempotent intersection type system provides a quantitative measure of the need/usage of resources.
Well-typed terms enjoy subject reduction and subject expansion (\Cref{ch4t:lamSRShort,ch4t:lamSEShort}).

\smallskip \noindent
(\secref{ch4s:trans})
A typed translation of~\lamcoldetshlin into \clpi, which
provides further validation for our \nondt calculi, and casts them in the context of `{functions-as-processes}'.
We prove that our translation is \emph{correct}, i.e., it preserves types and satisfies tight operational correspondences (\Cref{ch4def:encod_judge,ch4t:correncLazy}).

\smallskip \noindent
Moreover, \secref{ch4s:disc} closes by discussing related works.
Appendices contain (i)~omitted material (in particular, proofs of technical results); (ii)~an alternative \emph{eager} semantics for \clpi, which we  compare against the lazy semantics;
 and (iii)~extensions of \clpi and \lamcoldetshlin with  \emph{unrestricted} resources.

\section{A Typed \texorpdfstring{$\pi$}{Pi}-calculus with Non-deterministic Choice}
\label{ch4s:pi}

We introduce $\clpi$, a session-typed $\pi$-calculus with   \nondt choice.
Following~\cite{CairesP17}, session types express protocols to be executed along channels.
These protocols can be \emph{non-deterministic}: sessions may succeed but also fail.
The novelty in $\clpi$ is the \nondt choice operator `$\!P \nd Q$', whose \emph{lazily committing semantics} is
compatible with linearity.
We prove that well-typed processes satisfy two key properties: \emph{type preservation} and  \emph{deadlock-freedom}.

\begin{figure}[t] \mysmall
        \begin{align*}
            P,Q &::=
            \0 & \text{inaction}
            & \quad \sepr
            \pfwd{x}{y} & \text{forwarder}
            \\
            &~~~\sepr
            \res{x}(P \| Q) & \text{connect}
            & \quad \sepr
            P \nd Q & \text{non-determinism}
            \\ &~~~\sepr
            \pname{x}{y};(P \| Q) & \text{output}
            & \quad \sepr
            \gname{x}{y};P & \text{input}
            \\ &~~~\sepr
            \psel{x}{\ell};P & \text{select}
            & \quad \sepr
            \gsel{x}\{i:P\}_{i \in I} & \text{branch}
            \\ &~~~\sepr
            \pclose{x} & \text{close}
            & \quad \sepr
            \gclose{x};P & \text{wait}
            \\ &~~~\sepr
            \gsome{x}{w_1,\ldots,w_n};P & \text{expect}
            & \quad \sepr
            \psome{x};P & \text{available}
            \\ &~~~\sepr
            P \| Q & \text{parallel}
            & \quad \sepr
            \pnone{x} & \text{unavailable}
        \end{align*}

        \smallskip
        \dashes

        \vspace{-5ex}
        \begin{align*}
            P &\equiv P' ~ [P \equiv_\alpha P']
            &
            \pfwd{x}{y} &\equiv \pfwd{y}{x}
            &
            P \| \0 &\equiv P
            \\
            (P \| Q) \| R &\equiv P \| (Q \| R)
            &
            P \| Q &\equiv Q \| P
            &
            \res{x}(P \| Q) &\equiv \res{x}(Q \| P)
            \\
            P \nd P &\equiv P
            &
            P \nd Q &\equiv Q \nd P
            &
            (P \nd Q) \nd R &\equiv P \nd (Q \nd R)
        \end{align*}
        \vspace{-6ex}
        \begin{align*}
            \res{x}((P \| Q) \| R) &\equiv \res{x}(P \| R) \| Q
            &
            [ x \notin \fn{Q} ]
            \\
            \res{x}(\res{y}(P \| Q) \| R) &\equiv \res{y}(\res{x}(P \| R) \| Q)
            &
            [ x \notin \fn{Q}, y \notin \fn{R} ]
        \end{align*}
    \caption{ \clpi: syntax (top) and structural congruence (bottom).}\label{ch4f:pilang}
\end{figure}

\subsection{Syntax and Semantics}

We use $P, Q, \ldots$ to denote processes, and $x,y,z,\ldots$ to denote \emph{names} representing  channels. 
\Cref{ch4f:pilang} (top) gives the syntax of processes.
$P\{y/z\}$ denotes  the capture-avoiding substitution of $y$ for $z$ in  $P$.
Process~$\0$ denotes inaction, and
$\pfwd{x}{y}$ is a forwarder: a bidirectional link between $x$ and $y$.
Parallel composition appears in two forms:
while the process $P \| Q$ denotes communication-free concurrency,
process $\res{x}(P \| Q)$ uses restriction $\res{x}$ to express that $P$ and $Q$ implement complementary behaviors   on  $x$ and do not share any other names.

Process $P\nd Q$ denotes the non-deterministic choice between $P$ and $Q$: intuitively, if one choice can perform a synchronization, the other option may be discarded if it cannot.
Since $\nd$ is associative, we often omit parentheses. 
Also, we write $\bignd_{i \in I} P_i$ for the non-deterministic choice between each $P_i$ for $i \in I$.

Our output construct integrates parallel composition and restriction: process $\pname{x}{y};(P \| Q)$ sends a fresh name $y$ along $x$ and then continues as $P\| Q$.
The type system will ensure that behaviors on $y$ and $x$ are implemented by $P$ and $Q$, respectively, \myrev{which do not share any names---this separation defines communication-free concurrency and is key to ensuring deadlock-freedom}.
The  input process  $\gname{x}{y};P$ receives a name $z$ along $x$ and continues as $P\{z/y\}$, \myrev{which does not require the separation present in the output case. 
}
Process $\gsel{x}\{i:P_i\}_{i \in I}$  denotes a branch with labeled choices indexed by the finite set $I$: it awaits a choice on $x$ with continuation $ P_j$ for each $j \in I$.
The process $\psel{x}{\ell};P$ selects on $x$ the choice labeled $\ell$ before continuing as~$P$.
Processes $\pclose{x}$ and $\gclose{x}; P$ are dual actions for closing the session on $x$.
We omit replicated servers $\guname{x}{y};P$ and corresponding client requests $\puname{x}{y};P$, but they can be easily added 
\iffulldoc
(cf.~\secref{ch4as:fullPi}).
\else
(cf.~the full version).
\fi

The remaining constructs define non-deterministic sessions which
may provide a protocol or fail, following~\cite{CairesP17}.
    Process $\psome{x}; P$ confirms the availability of a session on $x$ and continues as $P$.
        Process $\pnone{x}$ signals the failure to provide the session on $x$.
    Process $\gsome{x}{w_1,\ldots,w_n}
    ; P$ specifies a dependency on a non-deterministic session on $x$ (names $w_1, \ldots , w_n$ implement sessions in~$P$). This
    process can either (i) synchronize with a `$\psome{x}$' and continue as $P$, or (ii) synchronize
    with a `$\pnone{x}$', discard $P$, and propagate the failure to $w_1, \ldots , w_n$.
To reduce eye strain, in writing $\gsome{x}{}$ we freely combine names and sets of names.
This way, e.g.,
we write $\gsome{x}{y,\fn{P},\fn{Q}}$ rather than $\gsome{x}{\{y\} \cup \fn{P} \cup \fn{Q}}$.

Name $y$ is bound in $\res{y}(P \| Q)$, $\pname{x}{y};(P \| Q)$, and $\gname{x}{y};P$.
We write $\fn{P}$ and $\bn{P}$ for the free and bound names of $P$, respectively. 
We adopt Barendregt's convention.


\paragraph*{Structural Congruence.}
Reduction defines the steps that a process performs on its own.
It relies on \emph{structural congruence} ($\equiv$), the least congruence relation on processes induced by the rules in \Cref{ch4f:pilang} (bottom).
Like the syntax of processes, the definition of $\equiv$ is aligned with the type system (defined next), such that $\equiv$ preserves typing (subject congruence, cf.\ \Cref{ch4t:srPi}).
Differently from~\cite{CairesP17}, we do not allow distributing non-deterministic choice over parallel and restriction.
As shown in 
\iffulldoc
\secref{ch4s:piEager}
\else
the appendix of the full version
\fi 
, the position of a non-deterministic choice in a process determines how it may commit, so changing its position affects commitment.

\paragraph*{Reduction: Intuitions and Prerequisites.}
Barring non-deterministic choice, our reduction rules    arise as directed interpretations of proof transformations in the  underlying linear logic.
We follow \cite{CairesP10} and \cite{DBLP:conf/icfp/Wadler12} in interpreting cut-elimination in linear logic as synchronization in $\clpi$.

Before delving into our reduction rules (\Cref{ch4f:redtwo}), it may be helpful  to consider the usual reduction axiom for the (untyped) $\pi$-calculus (e.g.,~\cite{DBLP:journals/iandc/MilnerPW92a,DBLP:books/daglib/0004377}):
\begin{equation}
  (\pname{x}{z};P_1 + M_1) \|
(\gname{x}{y};P_2 + M_2)
\longrightarrow
P_1 \| P_2\{ z / y \}
\label{ch4eq:usualpi}
\end{equation}
This axiom \myrev{captures the interaction of two (binary) choices: it} integrates the commitment of choice in  synchronization; after the reduction step,  \myrev{the two branches not involved in the synchronization, $M_1$ and $M_2$, are discarded}.
Our semantics of $\clpi$ is defined similarly: when a prefix within a branch of a choice synchronizes with its dual, that \myrev{branch reduces and the entire process commits to it}.

The key question at this point is: when and to which branches should we commit?
In~\eqref{ch4eq:usualpi}, a communication commits to a single branch.
For \clpi,  we   define a \emph{lazy semantics} that minimizes commitment as much as possible.

The intuitive idea is that multiple branches of a choice may contain the same prefix, and so all these branches represent possibilities for synchronization (``possible branches'').
Other branches with different prefixes denote different possibilities (``impossible branches'').
When one synchronization is chosen, the possible branches are maintained while the impossible ones are discarded.

\begin{example}{}
\label{ch4ex:possible}
    To distinguish possible and impossible branches, consider:
    \myrev{
    \begin{align*}
        P := \res{s} \mkern-1mu\big(\mkern-2mu \gsel{s}\{\sff{buy}:\myDots,\sff{peek}:\myDots\} \| ( \psel{s}{\sff{buy}}; \myDots \nd \psel{s}{\sff{buy}}; \myDots \nd \psel{s}{\sff{peek}} ; \myDots ) \mkern-3mu\big)
    \end{align*}
    The branch construct (case) provides the context for the non-deterministic choice.
    When the case synchronizes on the `\sff{buy}' label, the two branches prefixed by `$\psel{s}{\sff{buy}}$' are possible, whereas the branch prefixed by `$\psel{s}{\sff{peek}$}' becomes impossible, and can be discarded. The converse occurs when the  `\sff{peek}' label is selected.
    }
\end{example}

To formalize these intuitions, our reduction semantics (\Cref{ch4f:redtwo})  relies on some auxiliary definitions. First, we define contexts.

\begin{definition}{}\label{ch4d:ndctx}
    We define \emph{ND-contexts} ($\,\pctx{N},\pctx{M}$) as follows:
    \[
        \pctx{N},\pctx{M} ::= \hole \sepr \pctx{N} \| P \sepr \res{x}(\pctx{N} \| P) \sepr \pctx{N} \nd P
    \]
    The process obtained by replacing $\hole$ in $\pctx{N}$ with $P$ is denoted $\pctx{N}[P]$.
    We refer to ND-contexts that do not use the clause `\,$\pctx{N} \nd P$' as \emph{D-contexts}, denoted $\pctx{C},\pctx{D}$.
\end{definition}

Using D-contexts, we can express that, e.g.,  $\bignd_{i \in I} \pctx{C_i}[\pclose{x}]$ and $\bignd_{j \in J} \pctx{D_j}[\gclose{x};Q_j]$ should match.
To account for reductions with impossible branches, we define a precongruence on processes, denoted $\piprecong{S}$, where the parameter $S$ denotes the subject(s) of the prefix in the possible branches.
Our semantics is closed under $\piprecong{S}$.
Hence, e.g., anticipating a reduction on $x$, the possible branch $\pctx{C_1}[\gname{x}{y};P]$ can be extended with an  impossible branch to form $\pctx{C_1}[\gname{x}{y};P] \nd \pctx{C_2}[\gclose{z};Q]$.

Before defining $\piprecong{S}$ (\Cref{ch4d:rpreone}), we first define prefixes (and their subjects).
\myrev{Below, we write $\widetilde{x}$ to denote a finite tuple of names $x_1, \ldots, x_k$.}

\begin{definition}{}\label{ch4d:prefix}
    Prefixes are defined as follows:
    \[
        \alpha, \beta ::=
        \pname{x}{y} \sepr \gname{x}{y} \sepr \psel{x}{\ell} \sepr \gsel{x} 
        \sepr \pclose{x} \sepr \gclose{x} \sepr
        \psome{x} \sepr \pnone{x} \sepr \gsome{x}{\widetilde{w}} \sepr \pfwd{x}{y}
    \]
    The subjects of $\alpha$, denoted $\sub{\alpha}$, are $\{x,y\}$ in case of $\pfwd{x}{y}$, or $\{x\}$.
    By abuse of notation, we write $\alpha;P$ even when $\alpha$ takes no continuation (as in $\pclose{x}$, $\pnone{x}$, and $\pfwd{x}{y}$) and for $\pname{x}{y}$ which takes a parallel composition as continuation.
\end{definition}

\begin{definition}{}\label{ch4d:rpreone}
    Let $\relalpha$ denote the least relation on prefixes (\defref{ch4d:prefix}) defined by: \\
    (i)~$\pname{x}{y} \relalpha \pname{x}{z}$, (ii)~$\gname{x}{y} \relalpha \gname{x}{z}$, and (iii)~$\alpha \relalpha \alpha$ otherwise.

    \medskip
    Given a non-empty set $S\subseteq\{x,y\}$, the precongruence $P \piprecong{S} Q$ holds when both following conditions hold:
    \begin{enumerate}
        \item\label{ch4i:pipreSingleton}
            $S = \{x\}$
            implies
            \\
            $P = \Big( \bignd_{i \in I} \pctx{C_i}[\alpha_i;P_i] \Big) \nd \Big( \bignd_{j \in J} \pctx{C_j}[\beta_j;Q_j] \Big)$
            and
            $Q = \bignd_{i \in I} \pctx{C_i}[\alpha_i;P_i]$,
            where
            \\
            (i)~$\forall i,i' \in I.\, \alpha_i \relalpha \alpha_{i'}$ and $\sub{\alpha_i} = \{x\} $,
            and \\
            (ii)~$\forall i \in I.\, \forall j \in J.\, \alpha_i \not\relalpha \beta_j \land x \in \fn{\beta_j;Q_j}$;

        \item\label{ch4i:pipreForwarder}
            $S = \{x,y\}$
            implies
            \\
            $P = \Big( \bignd_{i \in I} \pctx{C_i}[\pfwd{x}{y}] \Big) \nd  \Big( \bignd_{j \in J} \pctx{C_j}[\pfwd{x}{z_j}] \Big) \nd \Big( \bignd_{k \in K} \pctx{C_k}[\alpha_k;P_k] \Big)$
            \\
            and
            $Q = \bignd_{i \in I} \pctx{C_i}[\pfwd{x}{y}]$,
            where
            \\
            (i)~$\forall j \in J.\, z_j \not = y$,
            and
            (ii)~$\forall k \in K.\ x \in \fn{\alpha_k;P_k} \land \forall z.\ \alpha_k \not\relalpha \pfwd{x}{z}$.
    \end{enumerate}
\end{definition}

\noindent
\myrev{
Intuitively, $\relalpha$ allows us to equate output/input prefixes with the same subject (but different object).
The rest of \Cref{ch4d:rpreone} accounts for two kinds of reduction, using $S$ to discard ``impossible'' branches.
In case $S$ is $\{x\}$ (\Cref{ch4i:pipreSingleton}), it concerns a synchronization on $x$; in case $S$ is $\{x,y\}$, it concerns forwarding on $x$ and $y$ (\Cref{ch4i:pipreForwarder}).
In both cases, $P$ and $Q$ contain matching prefixes on $x$, while $P$ may contain additional branches with different or blocked prefixes on $x$; $x$ must appear in the hole of the contexts in the additional branches in $P$ (enforced with $x \in \fn{\ldots}$), to ensure that no matching prefixes are discarded.
}

\begin{example}{}
\label{ch4ex:piprecongr}
    Recall process $P$ from \Cref{ch4ex:possible}.
    \myrev{%
    To derive a synchronization with the `\sff{buy}' alternative of the case, we can use $\piprecong{S}$ to discard the `\sff{peek}' alternative, as follows:
    $
        \psel{s}{\sff{buy}} ; \myDots \nd \psel{s}{\sff{buy}} ; \myDots \nd \psel{s}{\sff{peek}} ; \myDots ~\piprecong{s}~ \psel{s}{\sff{buy}} ; \myDots \nd \psel{s}{\sff{buy}} ; \myDots
    $  
    }%
\end{example}

\begin{figure}[!t] \mysmall
        \begin{align*}
            \mathsmaller{\rredtwo{\scc{Id}}}~
            &
            \res{x} \Big(\!\! \bignd_{i \in I}\pctx[\big]{C_i}[\pfwd{x}{y}] \| Q\Big)
            \redtwo_{x,y}
            \bignd_{i \in I} \pctx{C_i}[Q\{ y / x \}]
            \\
            \mathsmaller{\rredtwo{\tensor \parr}}~
            &
            \res{x} \Big(\!\! \bignd_{i \in I} \pctx{C_i}[\pname{x}{y_i};(P_i \| Q_i)] \| \bignd_{j \in J} \pctx{D_j}[x(z);R_j] \Big)
            \\ &
            \redtwo_x
            \bignd_{i \in I} \pctx[\Big]{C_i}[\! \res{x} \big(  Q_i \|  \res{w}( P_i\{w/y_i\} \| \bignd_{j \in J} \pctx{D_j}[R_j\{w/z\}] ) \big) \!]
            \\
            \mathsmaller{\rredtwo{\oplus \with}}~
            &
            \res{x} \Big(\!\! \bignd_{i \in I} \pctx{C_i}[\psel{x}{k'};P_i]  \| \bignd_{j \in J} \pctx{D_j}[\gsel{x}\{k:Q_j^k\}_{k \in K}] \Big)
            \\ &
            \redtwo_{x}
            \res{x} \Big( \bignd_{i \in I} \pctx{C_i}[P_i]  \| \bignd_{j \in J} \pctx{D_j}[ {Q_j^{k'}}  ]  \Big)
            \qquad [k' \in K]
            \\
            \mathsmaller{\rredtwo{\1 \bot}}~
            &
            \res{x} \Big(\!\! \bignd_{i \in I} \pctx{C_i}[ \pclose{x} ] \|  \bignd_{j \in J} \pctx{D_j}[ \gclose{x};Q_j ] \Big)
            \redtwo_{x}
            \bignd_{i \in I} \pctx{C_i}[ \0 ]  \| \bignd_{j \in J} \pctx{D_j}[Q_j]
            \\
            \mathsmaller{\rredtwo{\some}}~
            &
            \res{x} \Big(\!\! \bignd_{i \in I} \pctx{C_i}[ \psome{x};P_i ]  \| \bignd_{j \in J} \pctx{D_j}[\gsome{x}{w_1, \ldots, w_n};Q_j] \Big)
            \\ &
            \redtwo_{x}
            \res{x} \Big( \bignd_{i \in I} \pctx{C_i}[P_i]  \| \bignd_{j \in J} \pctx{D_j}[ {Q_j}  ]  \Big)
            \\
            \mathsmaller{\rredtwo{\none}}~
            &
            \res{x} \Big(\!\! \bignd_{i \in I} \pctx{C_i}[ \pnone{x} ] \| \bignd_{j \in J} \pctx{D_j}[\gsome{x}{w_1, \ldots, w_n};Q_j] \Big)
            \\ &
            \redtwo_{x}
            \bignd_{i \in I} \pctx{C_i}[\0]  \| \bignd_{j \in J} \pctx{D_j}[ \pnone{w_1} \| \ldots \| \pnone{w_n}  ]
        \end{align*}
        \begin{mathpar}
            \mathsmaller{\rredtwo{\piprecong{S}}}~~
            \inferrule{
                x \in S
                \\
                P \piprecong{S} P'
                \\
                Q \piprecong{S} Q'
                \\
                \res{x}(P' \| Q') \redtwo_S R
            }{
                \res{x}(P \| Q) \redtwo_S R
            }
                        \and
            \mathsmaller{\rredtwo{\nu\nd}}~~
            \inferrule{
                \res{x}(P \| \pctx[\big]{N}[\pctx{C}[Q_1] \nd \pctx{C}[Q_2]])\redtwo_S R
            }{
                \res{x}(P \| \pctx[\big]{N}[\pctx{C}[Q_1 \nd Q_2]])\redtwo_S R
            }
                        \and
            \mathsmaller{\rredtwo{\equiv}}~~
            \inferrule{
                P \equiv P'
                \\
                P' \redtwo_S Q'
                \\
                Q' \equiv Q
            }{
                P \redtwo_S Q
            }
            \and
            \mathsmaller{\rredtwo{\nu}}~~
            \inferrule{
                P \redtwo_S P'
            }{
                \res{x}(P \| Q) \redtwo_S \res{x}(P' \| Q)
            }
            \and
            \mathsmaller{\rredtwo{\|}}~~
            \inferrule{
                P \redtwo_S P'
            }{
                P \| Q \redtwo_S P' \| Q
            }
            \and
            \mathsmaller{\rredtwo{\nd}}~~
            \inferrule{
                P \redtwo_S P'
            }{
                P \nd Q \redtwo_S P' \nd Q
            }
        \end{mathpar}
    \caption{
       Reduction semantics for \clpi.
    }
    \label{ch4f:redtwo}
\end{figure}

\paragraph*{Reduction Rules.}
\Cref{ch4f:redtwo} gives the rules for the (lazy) reduction semantics, denoted $\redtwo_S$, where the set
$S$  contains the names involved in the interaction. We  omit the curly braces in this annotation; this way, e.g., we write `$\redtwo_{x,y}$' instead of `$\redtwo_{\{x,y\}}$'.
\myrev{
Also, we write $\redtwo_S^k$ to denote a sequence of $k \geq 0$ reductions.
}

The first six rules in \Cref{ch4f:redtwo} formalize forwarding and communication: they are defined on choices containing different D-contexts (cf. \Cref{ch4d:ndctx}), each with the same prefix but possibly different continuations; these rules preserve the non-deterministic choices.
Rule~$\rredtwo{\scc{Id}}$ fixes $S$ to the forwarder's two names, and the other rules fix $S$ to the one involved name.
\myrev{In particular, Rule~$\rredtwo{{\tensor}{\parr}}$ formalizes name communication: it involves multiple senders and multiple receivers (grouped in choices indexed by $I$ and $J$, respectively). Because they proceed in lock-step, reduction leads to substitutions involving the same (fresh) name $w$; also, the scopes of the choice and the contexts enclosing the senders is extended.}

Rule~$\rredtwo{\piprecong{S}}$ is useful to derive a synchronization that discards groups of  choices.
\myrev{
Rule~$\rredtwo{\nu\nd}$ allows inferring reductions when non-deterministic choices are not top-level: e.g., $\res{x} \big( \pclose{x} \| \res{y} ( ( \gclose{x} ; Q_1 \nd \gclose{x} ; Q_2 ) \| R ) \big) \redtwo_x \res{y} ( Q_1 \| R ) \nd \res{y} ( Q_2 \| R )$.
}
The last four rules formalize that reduction is closed under structural congruence, restriction, parallel composition, and non-deterministic choice.

\myrev{As mentioned earlier, a key motivation for our work is to have \nondt choices that effectively enforce commitment, without a too drastic discarding of alternatives.
Next we illustrate this intended form of \emph{gradual commitment}.}
\begin{example}{A Modified Movie Server}\label{ch4ex:eve}
    \myrev{
    Consider the following variant of the movie server from the introduction, where the handling of the payment   is now modeled as a branch:
    \[
        \sff{NewServer}_s := \gname{s}{\textit{title}} ; \gsel{s} \left\{
            \begin{array}{@{}l@{}}
                \sff{buy} : \gsel{s}\left\{
                    \begin{array}{@{}l@{}}
                        \sff{card} : \gname{s}{\textit{info}} ; \pname{s}{\texttt{movie}} ; \pclose{s} ,
                        \\
                        \sff{cash} : \pname{s}{\texttt{movie}} ; \pclose{s}
                    \end{array}
                \right\}
                ,
                \\
                \sff{peek} : \pname{s}{\texttt{trailer}} ; \pclose{s}
            \end{array}
        \right\}
    \]
    Consider a client, Eve, who cannot decide between buying `Oppenheimer' or watching its trailer.
    In the former case, she has two options for payment method:
    \[
        \sff{Eve}_s := \pname{s}{\texttt{Oppenheimer}} ; \left(
            \begin{array}{@{}rl@{}}
                &
                \psel{s}{\sff{buy}} ; \psel{s}{\sff{card}} ; \pname{s}{\texttt{visa}} ; \gname{s}{\textit{movie}} ; \gclose{s} ; \0
                \\ {\nd} &
                \psel{s}{\sff{buy}} ; \psel{s}{\sff{cash}} ; \gname{s}{\textit{movie}} ; \gclose{s} ; \0
                \\ {\nd} &
                \psel{s}{\sff{peek}} ; \gname{s}{\textit{link}} ; \gclose{s} ; \0
            \end{array}
        \right)
    \]
    }
    \myrev{
    Let $\sff{Sys}^* := \res{s} ( \sff{NewServer}_s \| \sff{Eve}_s )$.
    After sending the movie's title, Eve's choice (buying or watching the trailer) enables gradual commitment. We have:
    \begin{mathpar}
        \sff{Sys}^* \redtwo_s^2 \res{s} \big( \gsel{s}\{ \sff{card} : \ldots , \sff{cash} : \ldots \} \| ( \psel{s}{\sff{card}} ; \ldots \nd \psel{s}{\sff{cash}} ; \ldots ) \big) =: \sff{Sys}^*_1
        \\
        \text{and}
        \and
        \sff{Sys}^* \redtwo_s^2 \res{s} ( \pname{s}{\texttt{trailer}} ; \ldots \| \gname{s}{\textit{trailer}} ; \ldots ) =: \sff{Sys}^*_2
    \end{mathpar}
    Process $\sff{Sys}^*_1$ represents the situation for Eve after selecting $\sff{buy}$, in which case the third alternative ($\psel{s}{\sff{peek}}; \ldots$) can be discarded as an impossible branch.
    Process $\sff{Sys}^*_2$ represents the dual situation. 
    From~$\sff{Sys}^*_1$, the selection of payment method completes the commitment to one alternative; we have:
    $\sff{Sys}^*_1 \redtwo_s \res{s} ( \gname{s}{\textit{info}} ; \myDots \| \pname{s}{\texttt{visa}} ; \myDots )$
    and
    $\sff{Sys}^*_1 \redtwo_s \res{s} ( \pname{s}{\texttt{movie}} ; \myDots \| \gname{s}{\textit{movie}} ; \myDots )$.
    }
\end{example}


\iffulldoc
In~\Cref{ch4s:piEager}
 we discuss an alternative \emph{eager} semantics that commits to a single branch upon communication, as in~\eqref{ch4eq:usualpi}.
\else
In the full version
 we discuss an alternative \emph{eager} semantics that commits to a single branch upon communication, as in~\eqref{ch4eq:usualpi}.
\fi

\subsection{Resource Control for \texorpdfstring{\clpi}{spi+} via Session Types}
\label{ch4ss:piTypeSys}

We define a session type system for \clpi\kern-.7ex, following `propositions-as-sessions'~\cite{CairesP10,DBLP:conf/icfp/Wadler12}.
As already mentioned, in a session type system,
resources are names that perform protocols:
the \emph{type assignment} $x:A$ says that $x$ should conform to the protocol specified by the session type $A$.
We give the syntax of types:
\begin{align*}
    A,B &::= \1 \sepr \bot \sepr A \tensor B \sepr A \parr B \sepr {\oplus}\{i:A\}_{i \in I} 
    \sepr {\with}\{i:A\}_{i \in I} \sepr {\with}A \sepr {\oplus}A
\end{align*}
The units $\1$ and $\bot$ type closed sessions.
$A \tensor B$ types a name that first outputs a name of type~$A$ and then proceeds as   $B$.
Similarly, $A \parr B$ types a name that inputs a name of type $A$ and then proceeds as~$B$.
Types ${\oplus}\{i:A_i\}_{i \in I}$ and  ${\with}\{i:A_i\}_{i \in I}$ are given to names that can select and offer a labeled choice, respectively.
 Then, ${\with}A$ is the type of a name that \emph{may produce} a behavior of type $A$, or fail; dually, ${\oplus}A$ types a name that \emph{may consume} a behavior of type $A$.


For any type $A$ we denote its \emph{dual} as $\ol{A}$.
Intuitively,  dual types serve to avoid communication errors: the type at one end of a channel is the dual of the type at the opposite end.
Duality is an involution, defined as follows:
\begin{align*}
    \ol{\1} &= \bot
    & \ol{A \tensor B} &= \ol{A}\parr\ol{B}
    & \ol{{\oplus} \{i:A_i\}_{i\in I}} &= {\with}\{i:\ol{A_i}\}_{i\in I}
    & \ol{{\with}A} &= {\oplus}\ol{A}
    \\
    \ol{\bot} &= \1
    & \ol{A \parr B} &= \ol{A}\tensor \ol{B}
    & \ol{{\with}\{i:A_i\}_{i\in I}} &= {\oplus}\{i:\ol{A_i} \}_{i\in I}
        & \ol{{\oplus}A} &= {\with}\ol{A}
\end{align*}

Judgments are of the form $P\vdash \Gamma$, where $P$ is a process and $\Gamma$ is a context, a collection of type assignments.
In writing $\Gamma, x:A$, we assume $x \notin \dom{\Gamma}$.
We write $\dom{\Gamma}$ to denote the set of names appearing in $\Gamma$.
We write $\with \Gamma$ to denote that $\forall x:A \in \Gamma.~ \exists A'.~ A = \with  A'$.

\begin{figure}[!t] \mysmall
        \begin{mathpar}
            {\ttype[\scriptsize]{cut}}~
            \inferrule{
                P \vdash \Gamma, x{:}A
                \\
                Q \vdash \Delta, x{:}\ol{A}
            }{
                \res{x}(P \| Q) \vdash \Gamma, \Delta
            }
            \and
            \ttype[\scriptsize]{mix}~
            \inferrule{
                P \vdash \Gamma
                \\
                Q \vdash \Delta
            }{
                P \| Q \vdash \Gamma, \Delta
            }
            \and
            \ttype[\scriptsize]{$\nd$}~
            \inferrule{
                P \vdash \Gamma
                \\
                Q \vdash \Gamma
            }{
                P \nd Q \vdash \Gamma
            }
            \and
            \ttype[\scriptsize]{empty}~
            \inferrule{ }{
                \0 \vdash \emptyset
            }
            \and
            \ttype[\scriptsize]{id}~
            \inferrule{ }{
                \pfwd{x}{y} \vdash x{:}A, y{:}\ol{A}
            }
            \and
            \ttype[\scriptsize]{$\1$}~
            \inferrule{ }{
                \pclose{x} \vdash x{:}\1
            }
            \and
            \ttype[\scriptsize]{$\bot$}~
            \inferrule{
                P \vdash \Gamma
            }{
                \gclose{x};P \vdash \Gamma, x{:}\bot
            }
            \and
            \ttype[\scriptsize]{$\tensor$}~
            \inferrule{
                P \vdash \Gamma, y{:}A
                \\
                Q \vdash \Delta, x{:}B
            }{
                \pname{x}{y};(P \| Q) \vdash \Gamma, \Delta, x{:}A \tensor B
            }
            \and
            \ttype[\scriptsize]{$\parr$}~
            \inferrule{
                P \vdash \Gamma, y{:}A, x{:}B
            }{
                \gname{x}{y}; P \vdash \Gamma, x{:}A \parr B
            }
            \and
            \ttype[\scriptsize]{$\oplus$}~
            \inferrule{
                P \vdash \Gamma, x{:}A_j
                \\
                j \in I
            }{
                \psel{x}{j};P \vdash \Gamma, x{:}{\oplus}\{i:A_i\}_{i \in I}
            }
            \and
            \ttype[\scriptsize]{$\with$}~
            \inferrule{
                \forall i \in I.~ P_i \vdash \Gamma, x{:}A_i
            }{
                \gsel{x}\{i:P_i\}_{i \in I} \vdash \Gamma, x{:}{\with}\{i:A_i\}_{i \in I}
            }
            \and
            \ttype[\scriptsize]{${\with}\some$}~
            \inferrule{
                P \vdash \Gamma, x{:}A
            }{
                \psome{x};P \vdash \Gamma, x{:}{\with}A
            }
            \and
            \ttype[\scriptsize]{${\with}\none$}~
            \inferrule{ }{
                \pnone{x} \vdash x{:}{\with}A
            }
            \and
            \ttype[\scriptsize]{${\oplus}\some$}~
            \inferrule{
                P \vdash {\with}\Gamma,  x{:}A
            }{
                \gsome{x}{\dom{\Gamma}};P \vdash {\with}\Gamma,   x{:}{\oplus}A
            }
        \end{mathpar}
    \caption{Typing rules for \clpi.}
    \label{ch4fig:trulespi}
\end{figure}

\Cref{ch4fig:trulespi} gives the typing rules: they correspond to the rules in Curry-Howard interpretations of classical linear logic as session types (cf.\ \cite{DBLP:conf/icfp/Wadler12}), with the rules for ${\with}A$ and ${\oplus}A$ extracted from~\cite{CairesP17}, and the additional Rule~\ttype{$\nd$} for non-confluent non-deterministic choice, which modifies the confluent rule in~\cite{CairesP17}.

Most rules follow~\cite{DBLP:conf/icfp/Wadler12}, so we focus on those related to non-determinism.
Rule~\ttype{${\with}\some$} types a process with a name whose behavior can be provided, while Rule~\ttype{${\with}\none$} types a name whose behavior cannot.
Rule~\ttype{${\oplus}\some$} types a process with a name $x$ whose   behavior may not be available.
If the behavior is not available, all  the sessions in the process must be canceled; hence, the rule requires all names to be typed under the ${\with}A$ monad.

Rule~\ttype{$\nd$} types our new non-deterministic choice operator;  the branches must be typable under the same typing context.
Hence, all branches denote the same sessions, which may be implemented differently.
In context of a synchronization, branches that are kept are able to synchronize, whereas the discarded branches are not; nonetheless, the remaining branches still represent different implementations of the same sessions.
Compared to the rule for non-determinism in~\cite{CairesP17}, we do not require processes to be typable under the ${\with}A$ monad.
   \myrev{
   \begin{example}{}
   Consider again process $\sff{Eve}_s$ from \Cref{ch4ex:eve}.
    The three branches of the non-deterministic choice give \emph{different implementations of the same session}: 
    assuming primitive, self-dual data types $\mathtt{C}$, $\mathtt{M}$, and $\mathtt{L}$,
    all three branches on $s$ are typable by $\oplus \big\{ \sff{buy} : \oplus \{ \sff{card} : \mathtt{C} \tensor \mathtt{M} \parr \bot , \sff{cash} : \mathtt{M} \parr \bot \} , \sff{peek} : \mathtt{L} \parr \bot \big\}$.
       \end{example}
    }

    \myrev{
\begin{example}{Unavailable Movies}\label{ch4x:unavailableMovies}
    Consider now a modified movie server, 
   which offers movies that may not be yet available.
    We specify this server using non-deterministic choice and non-deterministically available sessions:
    \[
        \sff{BuyServ}_s := \gname{s}{\textit{title}} ; ( \pnone{s} \nd \psome{s} ; \gname{s}{\textit{paym}} ; \pname{s}{\texttt{movie}} ; \pclose{s} ) \vdash s:\mathtt{T} \parr \big( \with ( \mathtt{P} \parr \mathtt{M} \tensor \1 ) \big),
    \]
    where $\mathtt{T},\mathtt{P},\mathtt{M}$ denote primitive, self-dual data-types.
    While the branch `$\pnone{s}$' signals that the movie is not available, the branch `$\psome{s} ; ...$' performs the expected protocol.
    We now define a client Ada who buys a movie for Tim, using session $s$; Ada only forwards it to him (using session $u$) if it is actually available:
    \begin{align*}
        \sff{Ada}_{s,u} &:= \pname{s}{\texttt{Barbie}} ; \gsome{s}{u} ; \pname{s}{\texttt{visa}} ; \gname{s}{\textit{movie}} ; \gclose{s} ; \psome{u} ; \pname{u}{\textit{movie}} ; \pclose{u}
        \\
        &\hphantom{:= } \vdash s:\mathtt{T} \tensor \big( \oplus ( \mathtt{P} \tensor \mathtt{M} \parr \bot ) \big) , u:\with  ( \mathtt{M} \tensor \1 )
        \\
        \sff{Tim}_u &:= \gsome{u}{} ; \gname{u}{\textit{movie}} ; \gclose{u} ; \0 \vdash u:\oplus ( \mathtt{M} \parr \1 )
    \end{align*}
    Let $\sff{BuySys} := \res{s} \big( \sff{BuyServ}_s \| \res{u} ( \sff{Ada}_{s,u} \| \sff{Tim}_u ) \big)$.
    Depending on whether the server has the movie ``Barbie'' available, we have the following reductions:
    \[
        \sff{BuySys} \redtwo_s^2 \res{u} ( \pnone{u} \| \sff{Tim}_u \mkern-2mu )
        ~~
        \text{or}
        ~~
        \sff{BuySys} \redtwo_s^5 \res{u} ( \psome{u} {;} \mkern-2mu \myDots \| \sff{Tim}_u \mkern-2mu )
    \]
\end{example}
    }

Our type system ensures \emph{session fidelity} and \emph{communication safety}, but not confluence:
the former says that processes correctly follow their ascribed session protocols, and the latter that no communication errors/mismatches occur.
Both properties follow from the fact that typing is consistent across structural congruence and reduction.
\iffulldoc
See~\Cref{ch4ss:TPLazy} for details.
\else
See the full version for details.
\fi

\begin{theorem}[Type Preservation]
\label{ch4t:srPi}
    If $P \vdash \Gamma$, then both $P \equiv Q$ and $P \redtwo_S Q$ (for any $Q$ and $S$) imply $Q \vdash \Gamma$.
\end{theorem}


Another important, if often elusive, property in session types is \emph{dead\-lock-free\-dom}, which ensures that processes can reduce as long as they are not inactive.
Our type system satisfies deadlock-freedom for processes with fully connected names, i.e., typable under the empty context.
\iffulldoc
See \Cref{ch4ss:DFLazy} for details.
\else
See the full version for details.
\fi

\begin{theorem}[Deadlock-freedom]\label{ch4t:dfPi}
    If $P \vdash \emptyset$ and $P \not\equiv \0$, then there are $Q$ and $S$ such that $P \redtwo_S Q$.
\end{theorem}


\section{A Non-deterministic Resource \texorpdfstring{$\lambda$}{Lambda}-calculus}\label{ch4s:lambda}

We present \lamcoldetshlin, a resource $\lambda$-calculus
with   \nond  and lazy evaluation.
In \lamcoldetshlin,   \nond is  \col and \emph{implicit}, as it arises from the fetching of terms from {bags} of \emph{linear} resources.
This is different from  $\clpi$, where the choice operator `$\nd$'
specifies \nond \emph{explicitly}.
A mismatch between the number of variable occurrences and the size of the bag induces \emph{failure}.

In $\lamcoldetshlin$, the \emph{sharing} construct  $M\sharing{x_1,\ldots, x_n}{x}$,
expresses that
$x$  may be used in $M$ under ``aliases'' $x_1,\ldots, x_n$.
Hence, it atomizes $n$ occurrences of   $x$   in~$M$, via an explicit pointer to $n$ variables.
This way, e.g.,
the $\lambda$-term $\lambda x.(x\ x)$ is expressed in $\lamcoldetshlin$ as $\lambda x. (x_1 \bag{x_2}\sharing{x_1,x_2}{x})$, where $\bag{x_2}$ is a bag containing~$x_2$.

\begin{figure}[t]\mysmall
        \begin{align*}
            M,N,L ::=~
            & x & \text{variable}
            & \quad
            {}\sepr M \esubst{C}{x} & \text{intermediate subst.}
            \\
            \sepr~
            & (M\ C) & \text{application}
            & \quad
            {}\sepr M \linexsub{C/\widetilde{x}} & \text{explicit subst.}
            \\
            \sepr~
            & \lambda x.M & \text{abstraction}
            & \quad
            {}\sepr \fail^{\tilde{x}} & \text{failure}
            \\
            \sepr~
            & M\sharing{\widetilde{x}}{x} & \text{sharing}
            \\
            C,D ::=~
            & \oneb \sepr \bag{M} \cdot\, C
            \span\span
            & \text{ bag}
            \\
            \lctx{C} ::=~
            & \hole \sepr \lctx{C}\sharing{\widetilde{x}}{x} \sepr (\lctx{C}\ C) \sepr \lctx{C} \linexsub{C/\widetilde{x}}
            \span\span
            & \text{context}
        \end{align*}
    \caption{Syntax of $\lamcoldetshlin$: terms, bags, and contexts.}
    \label{ch4f:lambdalin}
\end{figure}

\subsection{Syntax and Reduction Semantics}

\paragraph*{Syntax.}
We use $x,y,z,\ldots $ for variables, and write $\widetilde{x}$ to denote a finite sequence of pairwise distinct $x_i$'s, with length $|\widetilde{x}|$.
\Cref{ch4f:lambdalin} gives the syntax of terms ($M,N,L$) and bags~($C, D$).
The empty bag is denoted $\oneb$.
We use $C_i$ to denote the $i$-th term in $C$, and $\size{C}$ denotes the number of elements in $C$.
To ease readability, we often write, e.g., $\bag{N_1, N_2}$
as a shorthand notation for
$\bag{N_1} \cdot \bag{N_2}$.

In  $M\sharing{\widetilde{x}}{x}$, we say that $\widetilde{x}$ are the \emph{shared variables}  and that $x$  is the \emph{sharing variable}.
We require for each $x_i \in \widetilde{x}$: (i)~$x_i$ occurs exactly once in $M$; (ii)~$x_i$ is not a sharing variable.
The sequence $\widetilde{x}$ can be empty:
$M\sharing{}{x}$ means that $x$ does not share any variables in $M$.
Sharing binds the shared variables in the term.

An abstraction $\lambda x. M$ binds
occurrences  of  $x$ in $M$.
Application $(M\ C)$ is as usual.
The term  $M \linexsub{C /  \widetilde{x}}$  is the \emph{explicit substitution}   of a bag $C$ for $\widetilde{x}$ in  $M$.
We require $\size{C} = |\widetilde{x}|$ and for each $x_i \in \widetilde{x}$: (i)~$x_i$ occurs in $M$; (ii)~$x_i$ is not a sharing variable; (iii)~$x_i$ cannot occur in another explicit substitution in $M$.
The term $M\esubst{ C }{ x }$ denotes an intermediate explicit substitution that does not (necessarily) satisfy the conditions for explicit substitutions.

The term $\fail^{\tilde{x}}$ denotes failure; the variables in $\widetilde{x}$ are ``dangling'' resources, which cannot be accounted for after failure.
We write $\lfv{M}$ to denote the free variables of $M$, defined as expected.
Term $M$ is \emph{closed} if $\lfv{M} = \emptyset$.

As in \Cref{ch2,ch3} we assume that all terms are consistent (\Cref{ch2d:consistent}).

\begin{figure}[t] \mysmall
        \begin{mathpar}
            \inferrule[$\redlab{RS{:}Beta}$]{ }{
                (\lambda x . M) C  \red M\esubst{ C }{ x }
            }
            \and
            \inferrule[$\redlab{RS{:}Ex \dash Sub}$]{
                \size{C} = |\widetilde{x}|
                \\
                M \not= \fail^{\tilde{y}}
            }{
                (M\sharing{\widetilde{x}}{x})\esubst{ C }{ x } \red  M\linexsub{C  /  \widetilde{x}}
            }
            \and
                        \inferrule[$\redlab{RS:TCont}$]{
                M \red    N
            }{
                \lctx{C}[M] \red   \lctx{C}[N]
            }
            \and
            \inferrule[$\redlab{RS{:}Fetch^{\ell}}$]{
                \headf{M} =  {x}_j
                \\
                0 < i \leq \size{C}
            }{
                M \linexsub{C /  \widetilde{x}, x_j} \red  M \headlin{ C_i / x_j }  \linexsub{(C \setminus C_i ) /  \widetilde{x}  }
            }
            \and
            \inferrule[$\redlab{RS{:}Fail^{\ell}}$]{
                \size{C} \neq |\widetilde{x}|
                \\
                \widetilde{y} = (\lfv{M} \setminus \{  \widetilde{x}\} ) \cup \lfv{C}
            }{
                (M\sharing{\widetilde{x}}{x}) \esubst{C }{ x }  \red  \fail^{\tilde{y}}
            }
            \and
            \inferrule[$\redlab{RS{:}Cons_1}$]{
                \widetilde{y} = \lfv{C}
            }{
                \fail^{\tilde{x}}\ C  \red  \fail^{\tilde{x} \cup \widetilde{y}}
            }
            \and
            \inferrule[$\redlab{RS{:}Cons_2}$]{
                \size{C} =   |  {\widetilde{x}} |
                \\
                \widetilde{z} = \lfv{C}
            }{
                (\fail^{ {\tilde{x}} \cup \tilde{y}} \sharing{\widetilde{x}}{x})\esubst{ C }{ x }  \red  \fail^{\tilde{y} \cup \widetilde{z}}
            }
            \and
            \inferrule[$\redlab{RS{:}Cons_3}$]{
                \widetilde{z} = \lfv{C}
            }{
                \fail^{\tilde{y}\cup \tilde{x}} \linexsub{C /  \widetilde{x}} \red  \fail^{\tilde{y} \cup \widetilde{z}}
            }
        \end{mathpar}
        where $\headf{M}$ is defined as follows:
        \begin{align*}
            \headf{ x } &= x
            &
            \headf{ \lambda x.M } &= \lambda x.M
            &
            \headf{ (M\ C) } &= \headf{M}
            \\
            \headf{ \fail^{\widetilde{x}} } &= \fail^{\widetilde{x}}
            &
            \headf{ M \esubst{ C }{ x } } &= M \esubst{ C }{ x }
            &
            \headf{ M \linexsub{C / \widetilde{x}} } &= \headf{M}
            \\
            \headf{ M\sharing{\widetilde{x}}{x} } = \begin{cases}
                x & \text{$\headf{ M } = y$ and $y \in  {\widetilde{x}}$}
                \\
                \headf{ M } & \text{otherwise}
            \end{cases}
            \span\span\span\span
        \end{align*}
    \caption{Reduction rules for $\lamcoldetshlin$.}
    \label{ch4fig:reduc_intermlin}\label{ch4f:lambda_redlin}
\end{figure}

\paragraph*{Semantics.}
\Cref{ch4fig:reduc_intermlin} gives the reduction semantics, denoted $\red$, and  the \emph{head variable} of term $M$, denoted $\headf{M}$.
 Rule~$\redlab{RS:Beta}$ induces an intermediate substitution.
Rule~$\redlab{RS:Ex{\dash}Sub}$  reduces an intermediate substitution to an explicit substitution, provided the size of the bag equals the number of shared variables.
In case of a mismatch, the term evolves into failure via Rule~$\redlab{RS:Fail^\ell}$.

An explicit substitution $M \linexsub{C/\widetilde{x}}$, where the head variable of $M$ is $x_j \in \widetilde{x}$, reduces via Rule~$\redlab{R:Fetch^\ell}$.
The rule extracts a $C_i$ from $C$ (for some $0 < i \leq \size{C}$) and substitutes it for $x_j$ in $M$; this is how fetching induces a non-deterministic choice between $\size{C}$ possible reductions.
Rules~$\redlab{RS:Cons_j}$ for $j \in \{1,2,3\}$ consume terms when they meet failure.
Finally, Rule~$\redlab{RS:TCont}$ closes reduction under contexts.
The following example illustrates reduction.

\begin{example}{}\label{ch4ex:syntax}\label{ch4ex:lambdaRed}
    Consider the term $M_0 = ( \lambda x. x_1 \bag{x_2 \bag{x_3\ \oneb} } \sharing{ \widetilde{x} }{x} )\ \bag{\fail^{\emptyset} , y ,I\,}$,
    where $I = \lambda x. (x_1 \sharing{x_1}{x})$ and $\widetilde{x} = x_1 , x_2 ,x_3$.
    First, $M_0$ evolves into an intermediate substitution~\eqref{ch4eq:lin_cons_sub1}.
    The bag can provide for all shared variables, so it then evolves into an explicit substitution~\eqref{ch4eq:lin_cons_sub2}:
    \begin{align}
        M_0
        &\red  (x_1 \!\bag{\!x_2 \!\bag{\!x_3\ \oneb\!} \!}  \sharing{\widetilde{x}}{x}) \esubst{ \bag{\fail^{\emptyset} , y , I}   }{x}
        \label{ch4eq:lin_cons_sub1}
        \\
        &\red
        (x_1 \!\bag{\!x_2 \!\bag{\!x_3\ \oneb\!}  \!}) \linexsub{\! \bag{\!\fail^{\emptyset} , y , I\!} \!/ \widetilde{x} } = M
        \label{ch4eq:lin_cons_sub2}
    \end{align}
    Since $ \headf{M} = x_1$, one of the three elements of the bag will be substituted.
    $M$ represents a non-deterministic choice between the following three reductions:
    \begin{minipage}{\textwidth}
         \begin{align*}
        \mathbin{\rotatebox[origin=r]{30}{$\red$}} &~ (\fail^{\emptyset} \bag{x_2 \bag{x_3\ \oneb}  }) \linexsub{  \bag{y, I} /  x_2,x_3  }
        = N_1
        \\[-5pt]
        M~
        \red &~ (y \bag{x_2 \bag{x_3\ \oneb}  }) \linexsub{ \bag{\fail^{\emptyset} , I} /  x_2,x_3  }
        = N_2
        \\[-3pt]
        \mathbin{\rotatebox[origin=r]{-30}{$\red$}} &~ (I \bag{x_2 \bag{x_3\ \oneb}}) \linexsub{ \bag{\fail^{\emptyset} , y}  /  x_2,x_3  }
        = N_3
    \end{align*}
    \end{minipage}
\end{example}

\subsection{Resource Control for \texorpdfstring{\lamcoldetshlin}{Lambda} via Intersection Types}\label{ch4sec:lamTypes}

Our type system for   $\lamcoldetshlin$ is based on non-idempotent intersection types.
As in prior works~\cite{PaganiR10,DBLP:conf/birthday/BoudolL00}, intersection types account for available resources in bags, which are unordered and have all the same type.
Because we admit the term $\fail^{\tilde x}$ as typable, we say that our system enforces \emph{well-formedness} rather than \emph{well-typedness}.
As we will see, well-typed terms form the sub-class of well-formed terms that does not include $\fail^{\tilde x}$ (see the text after \Cref{ch4t:lamSRShort}).

Strict types ($\sigma, \tau, \delta$) and multiset types ($\pi, \zeta$) are defined as follows:
\begin{align*}
    \sigma, \tau, \delta ::=~ &
    \unit \sepr \arrt{ \pi }{\sigma}
   \qquad  \pi, \zeta ::=
    \bigwedge_{i \in I} \sigma_i \sepr \omega
\end{align*}
\noindent
Given a non-empty $I$, multiset types $\bigwedge_{i\in I}\sigma_i$ are given to bags of size $|I|$.
This operator is associative, commutative,  and non-idempotent (i.e., $\sigma\wedge\sigma\neq \sigma$), with identity $\omega$.
Notation $\sigma^k$ stands for $\sigma \wedge \cdots \wedge \sigma$ ($k$ times, if $k>0$) or $\omega$ (if $k=0$).

Judgments have the form $\Gamma\wfdash M:\tau$, with contexts defined as follows:
\begin{align*}
    \Gamma,\Delta &::= \dash \sepr \Gamma , x:\pi  \sepr \Gamma, x:\sigma
\end{align*}
where $\dash$ denotes the empty context.
We write $\dom{\Gamma}$ for the set of variables in~$\Gamma$.
For $\Gamma, x:\pi$, we assume $x \not \in \dom{\Gamma}$.
To avoid ambiguities, we write $x:\sigma^1$  to denote that the assignment involves a multiset type, rather than a strict~type.
Given $\Gamma$, its \emph{core context} $\core{\Gamma}$ concerns variables with types different from $\omega$; it is defined as
$\core{\Gamma} = \{ x:\pi \in \Gamma \,|\, \pi \not = \omega\}$.

\begin{definition}{Well-formedness in $\lamcoldetshlin$}
    A  term $M$ is \emph{well-formed} if there exists a context $\Gamma$ and a  type  $\tau$ such that the rules in \Cref{ch4fig:wfsh_ruleslin} entail $\Gamma \wfdash  M : \tau $.
\end{definition}

\begin{figure}[t] \mysmall
        \begin{mathpar}
            \mprset{sep=1.8em}
            \inferrule[$\redlab{FS{:}var^{\ell}}$]{ }{
                 {x}: \sigma \wfdash  {x} : \sigma
            }
            \and
            \inferrule[$\redlab{FS{:}\oneb^{\ell}}$]{ }{
                \dash \wfdash \oneb : \omega
            }
            \and
            \inferrule[$\redlab{FS{:}bag^{\ell}}$]{
                 \Gamma \wfdash N : \sigma
                \\
                 \Delta \wfdash C : \sigma^k
            }{
                 \Gamma , \Delta \wfdash \bag{N}\cdot C:\sigma^{k+1}
            }
            \and
            \inferrule[$\redlab{FS{:}fail}$]{
                \dom{\core{\Gamma}} = \widetilde{x}
            }{
                \core{\Gamma} \wfdash  \fail^{\widetilde{x}} : \tau
            }
            \and
            \inferrule[$\redlab{FS{:}weak}$]{
                 \Gamma  \wfdash M : \tau
            }{
                \Gamma ,  {x}: \omega \wfdash M\sharing{}{x}: \tau
            }
            \and
            \inferrule[$\redlab{FS{:}shar}$]{
                \Gamma ,  {x}_1: \sigma, \ldots,  {x}_k: \sigma \wfdash M : \tau
                \\
                k \not = 0
            }{
                \Gamma ,  {x}: \sigma^{k} \wfdash M \sharing{{x}_1 , \ldots ,  {x}_k}{x}  : \tau
            }
            \and
            \inferrule[$\redlab{FS{:}abs\dash sh}$]{
                \Gamma ,  {x}: \sigma^k \wfdash M\sharing{\widetilde{x}}{x} : \tau
            }{
                 \Gamma \wfdash \lambda x . (M\sharing{\widetilde{x}}{x})  : \sigma^k \rightarrow \tau
            }
            \and
            \inferrule[$\redlab{FS{:}app}$]{
                \Gamma \wfdash M : \sigma^{j} \rightarrow \tau
                \\
                \Delta \wfdash C : \sigma^{k}
                \\
            }{
                 \Gamma , \Delta \wfdash M\ C : \tau
            }
            \and
            \inferrule[$\redlab{FS{:}Esub}$]{
                \Gamma ,  {x}: \sigma^{j} \wfdash M\sharing{\widetilde{x}}{x} : \tau
                \\\\
                \Delta \wfdash C : \sigma^{k}
            }{
                \Gamma , \Delta \wfdash (M\sharing{\widetilde{x}}{x})\esubst{ C }{ x }  : \tau
            }
            \and
            \inferrule[$\redlab{FS{:}Esub^{\ell}}$]{
                \Gamma  ,  x_1:\sigma, \cdots , x_k:\sigma \wfdash M : \tau
                \\\\
                \Delta \wfdash C : \sigma^k
            }{
                \Gamma , \Delta \wfdash M \linexsub{C /  x_1, \cdots , x_k} : \tau
            }
        \end{mathpar}
    \caption{Well-Formedness Rules for $\lamcoldetshlin$.}
    \label{ch4fig:wfsh_ruleslin}
\end{figure}

\noindent
In \Cref{ch4fig:wfsh_ruleslin},
Rule~$\redlab{FS:var^{\ell}}$ types variables.
Rule~$\redlab{FS:\oneb^{\ell}}$ types the empty bag with $\omega$.
Rule~$\redlab{FS:bag^{\ell}}$ types the concatenation of bags.
Rule~$\redlab{FS:\fail}$ types the term $\fail^{\widetilde{x}}$ with a strict type $\tau$, provided that the domain of the core context coincides with $\widetilde{x}$ (i.e., no  variable in $\widetilde{x}$ is typed with $\omega$).
Rule~$\redlab{FS:weak}$ types $M\sharing{}{x}$ by weakening the context with $x:\omega$.
Rule~$\redlab{FS:shar}$ types $M\sharing{\widetilde{x}}{x}$ with $\tau$, provided that there are assignments to the shared variables in~$\widetilde{x}$.

Rule~$\redlab{FS:abs{\dash}sh}$ types an abstraction $\lambda x.( M\sharing{\widetilde{x}}{x})$ with  $\sigma^k\to \tau$, provided that   $M\sharing{\widetilde{x}}{x}:\tau$ can be entailed from an assignment  $x:\sigma^k$.
Rule~\redlab{FS:app}  types $(M\ C)$, provided that $M$ has type $\sigma^j \to \tau$ and  $C$ has type $\sigma^k$. Note that, unlike usual intersection type systems, $j$ and $k$ may differ.
Rule~$\redlab{FS:Esub}$ types the intermediate substitution of a bag $C$ of type $\sigma^k$, provided that $x$ has type $\sigma^j$; again, $j$ and $k$ may differ.
Rule~$\redlab{FS:Esub^{\ell}}$  types  $M\linexsub{C/\widetilde{x}}$ as long as $C$ has type $\sigma^{|\widetilde{x}|}$,  and  each $x_i \in \widetilde{x}$ is of type $\sigma$.

Well-formed terms satisfy subject reduction (SR), whereas \emph{well-typed} terms, defined below, satisfy also subject expansion (SE).
\iffulldoc
See \secref{ch4a:lamTypes} and \secref{ch4a:subexpan} for details.
\else
See the full version for details.
\fi

\begin{theorem}[SR in \lamcoldetshlin]\label{ch4t:lamSRShort}
    If $ \Gamma \wfdash M:\tau$ and $M \red M'$, then $ \Gamma \wfdash M' :\tau$.
\end{theorem}


From our system for well-formedness we can extract a system for \emph{well-typed} terms, which do not include $\fail^{\tilde x}$.
Judgments for well-typedness are denoted $ \Gamma \wtdash M:\tau$, with rules copied from \Cref{ch4fig:wfsh_ruleslin} (the rule name prefix \texttt{FS} is replaced with \texttt{TS}), with the following modifications:
(i)~Rule~$\redlab{TS{:}fail}$ is removed; (ii)~Rules~$\redlab{TS{:}app}$ and $\redlab{TS{:}Esub}$ are modified to disallow a mismatch between variables and resources, i.e., multiset types should match in size.
Well-typed terms are also well-formed, and thus satisfy SR.
Moreover, \myrev{as a consequence of adopting (non-idempotent) intersection types,} they also satisfy~SE:

\begin{theorem}[SE in $\lamcoldetshlin$]\label{ch4t:lamSEShort}
    If $ \Gamma \wtdash M':\tau$ and $M \red M'$, then $ \Gamma \wtdash M :\tau$.
\end{theorem}


\section{A Typed Translation of \texorpdfstring{\lamcoldetshlin}{Lambda} into \texorpdfstring{\clpi}{Pi}
}
\label{ch4s:trans}

While \clpi features \nondt choice,
\lamcoldetshlin
\myrev{is a prototypical programming language in which implicit non-determinism}
implements  fetching of resources.
Resources are controlled using different type systems  (session types in \clpi, intersection types  in \lamcoldetshlin).
To reconcile these differences and
\myrev{illustrate the potential of \clpi to precisely model non-determinism as found in realistic programs/protocols},
we give a translation of  \lamcoldetshlin into \clpi. 
This translation 
preserves types (\thmref{ch4def:encod_judge}) and respects well-known  criteria for dynamic correctness~\cite{DBLP:journals/iandc/Gorla10,DBLP:phd/dnb/Peters12a,DBLP:journals/corr/abs-1908-08633} (\thmref{ch4t:correncLazy}).

\paragraph*{The Translation.}
Given a \lamcoldetshlin-term \(M\), its translation into \clpi is denoted \(\piencodf{M}_{u}\) and given in \Cref{ch4fig:encodinglin}.
As usual, every variable \(x\) in $M$ becomes a name \(x\) in process \(\piencodf{M}_{u}\), where
name \(u\) provides the behavior of \(M\).
A peculiarity is that, to handle failures in \lamcoldetshlin, \(u\) is a non-deterministically available session: the translated term can be available or not, as signaled by prefixes \(\psome{u}\) and \(\pnone{u}\), respectively.
As a result, reductions from \(\piencodf{M}_{u}\) include synchronizations that codify $M$'s behavior but also   synchronizations that confirm a session's availability.

\begin{figure*}[t] \mysmall
        \begin{align*}
            \piencodfaplas{ {x}}_u
            &= \psome{x}; \pfwd{x}{u}
            \\
            \piencodfaplas{\lambda x.M}_u
            &= \psome{u};\gname{u}{x}; \piencodfaplas{M}_u
            \\
            \piencodfaplas{(M\,C)}_u
            &=
            \res{v} (\piencodfaplas{M}_v \| \gsome{v}{u , \llfv{C}};\pname{v}{x}; ( \piencodfaplas{C}_x   \| \pfwd{v}{u}  ) )
            \\
            \piencodfaplas{ M \esubst{ C }{ x} }_u
            &=  \res{x}( \piencodfaplas{ M}_u \| \piencodfaplas{ C }_x )
            \\
            \piencodfaplas{\bag{N} \cdot~ C}_{{x}}
            &=
            \gsome{{x}}{\lfv{C}, \lfv{N} }; \gname{x}{y_i}; \gsome{{x}}{y_i, \lfv{C}, \lfv{N}}; \psome{{x}}; \pname{{x}}{z_i};
            \\ &\qquad
            ( \gsome{z_i}{\lfv{N}}; \piencodfaplas{N}_{z_i}
              \| \piencodfaplas{C}_{{x}} \| \pnone{y_i} )
            \\
            \piencodfaplas{{\oneb}}_{{x}}
            &=
            \gsome{{x}}{\emptyset};\gname{x}{y_n};  ( \psome{ y_n}; \pclose{y_n}  \| \gsome{{x}}{\emptyset}; \pnone{{x}} )
            \\
            \piencodfaplas{M \!\linexsub{ \!\bag{ \!N_1 , N_2\! } \!/ x_1,x_2 }}_u
            &=
            \res{z_1}( \gsome{z_1}{\lfv{N_{1}}};\piencodfaplas{ N_{1} }_{ {z_1}} \| \res{z_2} (
                \gsome{z_2}{\lfv{N_{2}}};\piencodfaplas{ N_{2} }_{ {z_2}}
            \\ & \qquad
                \| \bignd_{x_{i} \in \{ x_1 , x_2  \}} \bignd_{x_{j} \in \{ x_1, x_2 \setminus x_{i}  \}} \piencodfaplas{ M }_u \{ z_1 / x_{i} \} \{ z_2 / x_{j} \} ) )
            \\
            \piencodfaplas{M\sharing{}{x}}_u
            &=
            \psome{{x}}; \pname{{x}}{y_i}; ( \gsome{y_i}{ u , \lfv{M} }; \gclose{ y_{i} } ;\piencodfaplas{M}_u \| \pnone{ {x} } )
            \\
            \piencodfaplas{M\sharing{\widetilde{x}}{x}}_u
            &=
                \psome{{x}}; \pname{{x}}{y_i}; \big( \gsome{y_i}{ \emptyset }; \gclose{ y_{i} } ; \0
                \| \psome{{x}}; \gsome{{x}}{u, \lfv{M} \setminus  \widetilde{x} };
                \\ & \qquad
            \bignd_{x_i \in \widetilde{x}} \,\gname{{x}}{{x}_i};\piencodfaplas{M\sharing{(\widetilde{x} {\setminus} x_i)}{x}}_u \big)
            \\
            \piencodfaplas{\fail^{x_1, \ldots, x_k}}_u
            &= \pnone{ u}  \| \pnone{ x_1} \| \ldots \| \pnone{ x_k}
        \end{align*}
    \caption{Translation of $\lamcoldetshlin$ into $\clpi$.}\label{ch4fig:encodinglin}
\end{figure*}
At its core, our translation  follows Milner's.
This way, e.g., the process \(\piencodfaplas{(\lambda x.M)\ C}_u\)  enables synchronizations between \(\piencodfaplas{\lambda x.M}_v\) and \(\piencodfaplas{C}_{x}\) along
  name~$v$, resulting in the translation of an intermediate substitution.
The \emph{key novelty} is the role and treatment of   non-determinism. Accommodating \col \nond is non-trivial, as it entails translating explicit substitutions and sharing in $\lamcoldetshlin$ using the \nondt choice operator $\nd$ in $\clpi$.
Next we discuss these novel aspects, while highlighting differences with respect to a translation by \Cref{ch2,ch3}, which is given in the confluent setting (see \secref{ch4s:disc}).

   In \Cref{ch4fig:encodinglin}, non-deterministic choices occur in the translations of   $M\linexsub{C/\widetilde{x}}$ (explicit substitutions) and  $M\sharing{\widetilde{x}}{x}$ (non-empty sharing).
    Roughly speaking, the position  of $\nd$  in the translation of $M\linexsub{C/\widetilde{x}}$  represents the most desirable way of mimicking the fetching of terms from a bag.
This use of $\nd$ is a central idea in our translation: as we explain below,  it allows for appropriate commitment in  \nondt choices, but also for \emph{delayed} commitment when necessary.

For simplicity, we consider explicit substitutions \(M\mkern-3mu\linexsub{C/\widetilde{x}}\) where $C=\bag{\mkern-2mu N_1{,}N_2}$ and \(\widetilde{x}=x_1,x_2\).
The translation \(\piencodfaplas{M \linexsub{C/\widetilde{x}}}_{u}\) uses the processes \(\piencodfaplas{N_i}_{z_i}\), where each $z_i$ is fresh.
First, each bag item confirms its behavior.
Then, a variable~\(x_{i} \in \widetilde{x}\) is chosen non-deterministically;
we ensure that these choices consider all variables.
Note that writing \(\bignd_{x_{i} \in \{x_1,x_2 \}}\bignd_{x_{j}\in \{x_1,x_2\}\setminus x_{i} }\) is equivalent to non-de\-ter\-mi\-nis\-tic\-ally assigning \(x_{i},x_{j} \) to each permutation of \(x_1,x_2\).
The resulting choice involves \(\piencodfaplas{M}_{u}\) with $x_{i}, x_j$ substituted by $z_1, z_2$.
Commitment here is triggered only via synchronizations along $z_1$ or $z_2$; synchronizing with
$\gsome{z_i}{\lfv{N_{i}}};\piencodfaplas{N_{i} }_{ {z_i}}$ then represents fetching   $N_{i}$ from the bag.
\myrev{The size of the translated term $\piencodfaplas{M \linexsub{C/\widetilde{x}}}_{u}$ is exponential with respect to the size of $C$.}

The process
\(\piencodfaplas{M\sharing{\widetilde{x}}{x}}_{u}\)
proceeds as follows.
First, it confirms its behavior along \(x\).
Then it sends a name \(y_i\) on \(x\), on which a failed reduction may be handled.
Next,
the translation confirms again its behavior along \(x\) and non-deterministically receives a reference to an $x_i \in \widetilde{x}$.
Each branch consists of $\piencodfaplas{M\sharing{(\widetilde{x} {\setminus} x_i)}{x}}_u$.
The possible choices are permuted, represented by~\(\bignd_{x_i\in \widetilde{x}}\).
{Synchronizations with $\piencodfaplas{M\sharing{(\widetilde{x} {\setminus} x_i)}{x}}_u$ and bags delay commitment in this choice (we return to this point below).}
The process \(\piencodfaplas{M\sharing{}{x}}_{u}\) is similar but simpler: here the  name \(x\) fails, as it cannot take further elements to substitute.

In case of a failure (i.e., a mismatch between the size of the bag~\(C\) and the number of variables in $M$), our translation ensures that the confirmations of \(C\) will not succeed.
This is how failure in \lamcoldetshlin is correctly translated to failure in~\clpi.

\paragraph*{Translation Correctness.}
The translation is typed: intersection types in \lamcoldetshlin are translated into session types in \clpi (\Cref{ch4fig:enc_typeslin}).
This translation of types abstractly describes how non-deterministic fetches are codified as non-deterministic session protocols.
{
    It is worth noting that this translation of types is the same as in~\Cref{ch3}.
   This is not surprising: as we have seen, session types effectively abstract away from the behavior of processes, as all branches of a non-deterministic choice use the same typing context.
    Still, it is pleasant that the translation of types remains unchanged across different translations with
        our (non-confluent) non-determinism (in \Cref{ch4fig:encodinglin})
        and with
    confluent non-determinism (in~\Cref{ch3}).
}

To state \emph{static} correctness,
we require the following definition:

\begin{definition}{}\label{ch4def:enc_sestypfaillin}
    Let
    $
        \Gamma =  {{x}_1: \sigma_1}, \ndots,  {{x}_m : \sigma_m},  {{v}_1: \pi_1} , \ndots ,  {v}_n: \pi_n
    $
    be a context.
    The translation  $\piencodfaplas{\cdot}_{\_}$  in~\Cref{ch4fig:enc_typeslin} extends to contexts
    as follows:
    \begin{align*}
        \piencodfaplas{\Gamma} = {} & {x}_1 : \with \overline{\piencodfaplas{\sigma_1}} , \cdots ,   {x}_m : \with \overline{\piencodfaplas{\sigma_m}} ,
         {v}_1:  \overline{\piencodfaplas{\pi_1}_{(\sigma, i_1)}}, \cdots ,  {v}_n: \overline{\piencodfaplas{\pi_n}_{(\sigma, i_n)}}
    \end{align*}
\end{definition}

\begin{figure}[t] \mysmall
        \begin{align*}
            \piencodfaplas{\unit} &= \with \onef
            &
            \qquad \qquad
            \piencodfaplas{\sigma^{k}   \rightarrow \tau} &= \with( \dual{\piencodfaplas{ \sigma^{k}  }_{(\sigma, i)}} \ampy \piencodfaplas{\tau})
            \\
            \piencodfaplas{ \sigma \wedge \pi }_{(\tau, i)} = \oplus(( \with \onef) \ampy ( \oplus  \with (( \oplus \piencodfaplas{\sigma} ) \otimes (\piencodfaplas{\pi}_{(\tau, i)}))))
            \span\span\span
            \\
            \piencodfaplas{\omega}_{(\sigma, i)}
            &= \begin{cases}
                \oplus ((\with \1) \parr (\oplus \with \1))
                & \text{if $i = 0$}
                \\
                \oplus ((\with \1) \parr (\oplus\, \with ((\oplus \piencodfaplas{\sigma}) \tensor (\piencodfaplas{\omega}_{(\sigma, i-1)}))))
                & \text{if $i > 0$}
            \end{cases}
            \span\span
        \end{align*}
    \caption{Translation of intersection types into session types  (cf.\ \defref{ch4def:enc_sestypfaillin}).}
    \label{ch4fig:enc_typeslin}
\end{figure}

\noindent
Well-formed terms translate into well-typed processes:

\begin{theorem}
    \label{ch4def:encod_judge}
    If
    $ \Gamma \wfdash {M} : \tau$,
    then
    $
        \piencodfaplas{{M}}_u \vdash
        \piencodfaplas{\Gamma},
        u : \piencodfaplas{\tau}
    $.
\end{theorem}


To state \emph{dynamic} correctness, we rely on established  notions that (abstractly) characterize \emph{correct translations}.
A language \({\cal L}=(L,\to)\)  consists of a set of terms $L$ and a reduction relation $\to$ on $L$.
Each language ${\cal L}$ is assumed to contain a success constructor \(\sucs{}\).
A term \(T \in  L\) has {\em success}, denoted \(\succp{T}{\sucs{}}\), when there is a sequence of reductions (using \(\to\)) from \(T\) to a term satisfying success criteria.

Given ${\cal L}_1=(L_1,\to_1)$ and ${\cal L}_2=(L_2,\to_2)$, we seek translations \(\encod{\cdot }{}: {L}_1 \to {L}_2\) that are correct: they satisfy well-known correctness criteria~\cite{DBLP:journals/iandc/Gorla10,DBLP:phd/dnb/Peters12a,DBLP:journals/corr/abs-1908-08633}.
We state the set of correctness criteria that determine the correctness of a translation.

\begin{definition}{Correct Translation}\label{ch4d:encCriteria}
    Let $\mathcal{L}_1= (\mathcal{M}, \shred_1)$ and  $\mathcal{L}_2=(\mathcal{P}, \shred_2)$ be two languages.
    Let $ \asymp_2 $ be an equivalence 
  over $\mcl{L}_2$.
    We use $M,M' $ (resp.\ $P,P' $) to range over terms in $\mcl{M}$  (resp.\ $\mcl{P}$).
  Given a translation $\encod{\cdot }{}: {\cal M}\to {\cal P}$, we define:
    \begin{description}
        \item \textbf{Completeness:}
            For every ${M}, {M}' $ such that ${M} \shred_1^\ast {M}'$, there exists $ P $ such that $ \encod{{M}}{} \shred_2^\ast P \asymp_2 \encod{{M}'}{}$.

        %
        \item \textbf{Weak Soundness:}
            For every $M$ and $P$ such that $\encod{M}{} \shred_2^\ast P$, there exist  $M'$, $P' $ such that $M \shred_1^\ast M'$ and $P \shred_2^\ast P' \asymp_2 \encod{M'}{}  $.

        %
        \item \textbf{Success Sensitivity:}
            For every ${M}$, we have $\succp{M}{\sucs{}}$ if and only if $\succp{\encod{M}{}}{\sucs{}}$.
    \end{description}
    %
\end{definition}

Let us write $\Lambda$ to denote the set of well-formed \lamcoldetshlin terms, and $\Pi$ for the set of all well-typed \clpi processes, both including $\sucs{}$.
We have 
our final  result:

\begin{theorem}[Translation correctness under $\redtwo$]
    \label{ch4t:correncLazy}
    The translation $\piencodf{\cdot}_{\_}: (\Lambda, \red) \to (\Pi, \redtwo)$ is 
    correct  (cf.\ \Cref{ch4d:encCriteria}) using equivalence $\equiv$ (\Cref{ch4f:pilang}).
\end{theorem}

\noindent
The proof of \Cref{ch4t:correncLazy} involves instantiating/proving each of the parts of \mbox{\defref{ch4d:encCriteria}}.
%
%
%
%
%
%
%
%
%
%
Among these, \emph{weak soundness} is the most  challenging to prove.
Our prior work on translations of typed $\lambda$ into $\pi$ with confluent non-determinism~\Cref{ch2,ch3} rely critically on
confluence to match a behavior in $\pi$ with a corresponding behavior in $\lambda$.
Because in our setting  confluence is lost,  we must resort to a different proof.

As already discussed, our translation makes the implicit non-determinism in a $\lamcoldetshlin$-term $M$ explicit by adding non-deterministic choices in key points of $\piencodfaplas{M}_u$.
  Our reduction $\redtwo$ preserves those branches that simultaneously have the same prefix available (up to $\relalpha$). In proving
  weak soundness,
  we exploit the fact that reduction entails delayed commitment.
    To see this, consider the following terms:
\begin{eqnarray}
    \res{x}(  (\alpha_1 ; P_1 \nd \alpha_2 ; P_2) \| Q)  \label{ch4ex:sound1}
    \\
    \res{x}(  \alpha_1 ; P_1 \| Q) \nd \res{x}(\alpha_2 ; P_2 \| Q) \label{ch4ex:sound2}
\end{eqnarray}
In \eqref{ch4ex:sound1}, commitment to a choice
relies on whether $\alpha_1 \relalpha\alpha_2$ holds (cf. \Cref{ch4d:rpreone}).
If $\alpha_1 \not \relalpha \alpha_2$, a choice is made; otherwise,   commitment is delayed, and depends on $P_1$ and $P_2$.
Hence, in \eqref{ch4ex:sound1} the possibility of committing to either branch is kept open.
In contrast, in \eqref{ch4ex:sound2} commitment to a choice is independent of   $\alpha_1 \relalpha\alpha_2$.

Our translation exploits the delayed commitment of non-determinism illustrated by \eqref{ch4ex:sound1} to mimic commitment to non-deterministic choices in \lamcoldetshlin, which manifests in fetching resources from bags.
The fact that this delayed commitment preserves information about the different branches (e.g., $P_1$ and $P_2$ in \eqref{ch4ex:sound1}) is essential to establish
weak soundness,
i.e., to match a behavior in \clpi with a corresponding  step in \lamcoldetshlin.
In contrast, forms of non-determinism in $\piencodfaplas{N}_u$ that resemble \eqref{ch4ex:sound2} are useful to characterize behaviors  different from fetching.

\section{Summary and Related Work}
\label{ch4s:disc}

We studied the interplay between resource control and non-determinism in typed  calculi.
We introduced \clpi and \lamcoldetshlin, two calculi with \col \nond, both with type systems for resource control.
Inspired by the untyped $\pi$-calculus,  \nond in \clpi is lazy and explicit, with  session types defined following `propositions-as-sessions'~\cite{CairesP17}.
In \lamcoldetshlin,   \nond  arises in the fetching of resources, and is regulated by intersection types.
A correct translation of \lamcoldetshlin into \clpi precisely connects their different forms of  \nond. 

\paragraph{Related Work}
Integrating (\col) \nond within session types is non-trivial, as carelessly discarding  branches would break  typability.
Work by \cite{CairesP17}, already mentioned, develops a confluent semantics by requiring that non-determinism is only used inside the monad $\with A$; our non-confluent semantics drops this requirement.
This allows us to consider non-deterministic choices not possible in~\cite{CairesP17}, such as, e.g., selections of different labels.
We stress that linearity is not jeopardized: the branches of `$\nd$' do not represent \emph{different sessions}, but \emph{different implementations} of the same sessions.

\cite{DBLP:conf/birthday/AtkeyLM16} and \cite{journal/lmcs/KokkeMW20} extend `pro\-po\-si\-tions-as-sessions' with \nond. Their approaches are very different (conflation of the additives and bounded linear logic, respectively) and
  support non-determinism for {unrestricted names only}.
\myrevopt{This is a major difference:  
\Cref{ch4ex:piprecongr,ex:piprecongr_red,ex:safe_or_risk_pro}
can only be supported in \cite{DBLP:conf/birthday/AtkeyLM16,journal/lmcs/KokkeMW20} by dropping linearity, which is important there.}
Also, \cite{DBLP:conf/birthday/AtkeyLM16,journal/lmcs/KokkeMW20} do not connect with typed $\lambda$-calculi, as we do. 
\cite{conf/icfp/RochaC21} also consider non-determinism, relying on confluence and on unrestricted names.
\cite{DBLP:journals/tcs/CasalMV22,DBLP:conf/esop/VasconcelosCAM20} develop a type system for \emph{mixed sessions} (sessions with mixed choices), which can express non-determinism but does not ensure deadlock-freedom.
Ensuring deadlock-freedom by typing is a key feature of the `propositions-as-sessions' approach that we adopt for \clpi.

Our language \lamcoldetshlin is most related to  calculi by \cite{DBLP:conf/concur/Boudol93}, \cite{DBLP:conf/birthday/BoudolL00}, and by \cite{PaganiR10}.
Non-determinism  in the calculi in~\cite{DBLP:conf/concur/Boudol93,DBLP:conf/birthday/BoudolL00}   is committing and implicit; their linear resources can be consumed \emph{at most} once, rather than \emph{exactly} once.
The work~\cite{PaganiR10} considers non-committing \nond that is both implicit (as in \lamcoldetshlin) and explicit (via a sum operator on terms).
Both~\cite{DBLP:conf/concur/Boudol93,PaganiR10} develop (non-idempotent) intersection type systems to regulate resources.
In our type system,  all terms in a bag have the same type; the system in~\cite{PaganiR10} does not enforce this condition.
Unlike these type systems, our system for well-formedness can type terms with a lack or an excess of resources.

 \cite{DBLP:conf/birthday/BoudolL00} and \Cref{ch2,ch3} give translations of resource $\lambda$-calculi into $\pi$.
The translation in~\cite{DBLP:conf/birthday/BoudolL00}
is used to study the semantics induced upon $\lambda$-terms by a translation into \(\pi\); unlike ours, it does not consider types.
As already mentioned in~\secref{ch4s:trans},  \Cref{ch2,ch3} relate calculi with \emph{confluent} \nond: a resource $\lambda$-calculus with sums on terms, and the session $\pi$-calculus from~\cite{CairesP17}.
Our translation of terms in this chapter and that in~\Cref{ch2,ch3} are very different:
while here we use non-deterministic choice to mimic the sharing construct, the translation in~\Cref{ch2,ch3} uses it to translate bags.
Hence, our \Cref{ch4t:correncLazy} cannot be derived from~\Cref{ch2,ch3}.

\myrevopt{Milner's  seminal work~\cite{Milner90} connects
untyped, deterministic $\lambda$ and $\pi$.
Sangiorgi studies the behavioral equivalences that Milner's translations induce on $\lambda$-terms~\cite{DBLP:phd/ethos/Sangiorgi93,DBLP:journals/iandc/Sangiorgi94,DBLP:journals/mscs/Sangiorgi99};   in particular, the work~\cite{DBLP:journals/iandc/Sangiorgi94} considers an untyped, non-de\-ter\-minis\-tic $\lambda$-calculus.
Sangiorgi and Walker~\cite{DBLP:books/daglib/0004377} offer a unified presentation of (typed) translations of $\lambda$ into $\pi$: 
they consider simply-typed $\lambda$-calculi and $\pi$-calculi with input-output types;  non-deterministic choices are not considered.}

The last decade of work on `prop\-o\-sitions-as-sessions' has delivered   insightful connections with typed $\lambda$-calculi---see, e.g.,~\cite{DBLP:conf/icfp/Wadler12,DBLP:conf/fossacs/ToninhoCP12,DBLP:conf/esop/ToninhoY18}.
Excepting~\Cref{ch2,ch3}, already discussed,
none of these works consider non-deterministic $\lambda$-calculi.


\clearemptydoublepage

\chapter{Termination in Concurrency, Revisited}\label{ch5}

Termination is a central property in sequential programming models: a term is terminating if all its reduction sequences are finite. 
Termination is also important in concurrency in general, and for message-passing programs in particular.
A variety of type systems that enforce termination by typing have been developed. 
In this chapter, we rigorously compare several type systems for $\pi$-calculus processes from the unifying perspective of termination.
Adopting \emph{session types} as reference framework, we consider two different type systems: one follows Deng and Sangiorgi's weight-based approach; the other is Caires and Pfenning's Curry-Howard correspondence between linear logic and session types. 
Our technical results precisely connect these very different type systems, and shed light on the classes of client/server interactions they admit as correct.

\section{Introduction}
The purpose of this chapter is to present the first comparative study of type systems that enforce \emph{termination} for message-passing processes in the $\pi$-calculus, the paradigmatic model of concurrency.

Termination is a cornerstone of sequential programming models: a term is terminating if all its reduction sequences are finite. 
Termination is also an important property in concurrency in general, and in message-passing programs in particular. 
In such a setting, infinite sequences of internal steps are rather undesirable, as they could jeopardize the reliable interaction between a process and its environment. 
That is, we would like processes that exhibit \emph{infinite} sequences of observable actions, possibly intertwined with \emph{finite} sequences of internal/unobservable steps (i.e., reductions). 

In the (un)typed  $\pi$-calculus, infinite behavior can be  expressed via operators for recursion (or recursive definitions) or replication. 
We are interested in replication, and in particular in \emph{input-guarded} replication, denoted $!x(y).P$.
Input-guarded replication neatly captures the essence of \emph{servers} that are persistently available to spawn interactive behavior upon invocations by concurrent \emph{clients}.
This way, it   precisely expresses the controlled invocation of (shared) resources. 
To understand its operation, let us write $x\langle z\rangle$ to denote an output prefix, intended as an invocation to a server such as $!x(y).P$.
The corresponding reduction rule is then roughly as follows:
$$!x(y).P \pp x\langle z\rangle.Q ~\longrightarrow~ !x(y).P \pp P\substj{z}{y} \pp Q $$
Thus, after a synchronization on $x$, the server $!x(y).P$ continues to be available, and a copy of $P$ is spawned (where $\substj{z}{y}$ denotes the substitution of $y$ with $z$, as usual), enabling interaction with $Q$.

In this setting, an obvious source of non-terminating behaviors is when clients and servers invoke each other indefinitely. 
This situation arises, in particular,  when client invocations occur in the body of a server, which can easily trigger infinite ``ping-pong'' reductions, as in the following process (where $\nil$ denotes inaction):
\begin{equation}
!x(y).x\langle y\rangle.\nil \pp x\langle w\rangle.\nil ~\longrightarrow~ !x(y).x\langle y\rangle.\nil \pp x\langle w\rangle.\nil \pp \nil ~\longrightarrow~ \cdots 
\label{ch5eq:ping}	
\end{equation}
The challenge of statically ruling out processes such as \eqref{ch5eq:ping} while enabling expressive client/server interactions has been addressed by multiple authors via various type systems, see, e.g.,~\cite{DBLP:conf/lics/YoshidaBH01,DBLP:conf/ifipTCS/DengS04,DBLP:journals/mscs/Sangiorgi06,DBLP:conf/concur/DemangeonHS10,DBLP:journals/toplas/KobayashiS10,DBLP:journals/fuin/Piccolo12,DBLP:conf/tgc/ToninhoCP14,DBLP:journals/pacmpl/LagoVMY19}. 
Their underlying approaches are vastly diverse.
For instance, 
  \cite{DBLP:conf/lics/YoshidaBH01} adopt a type-theoretical approach  based on logical relations and linear action types.
 \cite{DBLP:conf/ifipTCS/DengS04} transport ideas from rewriting systems (well-founded measures) into a $\pi$-calculus with simple types.
Caires and Pfenning's Curry-Howard correspondence between linear logic and session types~(\cite{CairesP10}) represents yet another approach: their type system enforces termination based purely on proof-theoretical principles, by  interpreting the exponential `$!A$' as the type of a server and by connecting cut elimination with process synchronization.
Several natural questions arise. How do these type disciplines compare? What are their relative strengths? More concretely, are there terminating processes detected as such by one type system but not by some other? If so, where is the difference?

As inviting and intriguing these questions are, a technical approach to a formal comparison is far from obvious. 
An immediate obstacle concerns the underlying formal models: all the type systems mentioned above operate on \emph{different dialects} of the $\pi$-calculus, involving, e.g., synchronous/asynchronous communication, and monadic/polyadic message passing. These differences quickly escalate at the level of the respective type systems, with the presence/absence of \emph{linearity} unsurprisingly playing a key distinguishing role. How do we even start formulating the intended comparison?

We frame our formal comparison as follows.
As baseline for comparison we take the   $\pi$-calculus processes typable with Vasconcelos's session type system~(\cite{V12}).  This is a quite liberal type system, which induces a  broad class of session processes (including non-terminating ones), which is convenient for our purposes. 
In the following, this baseline class of processes is denoted $\vaslang$. 

We then consider two representative classes of processes, both terminating by typing. One is based on Deng and Sangiorgi's \emph{weight-based} type system; the other is Caires and Pfenning's linear-logic type system. Because these type systems are so different from Vasconcelos's, to connect them with \vaslang we require  typed translations. This leads to two classes of terminating processes:
\begin{itemize}
	\item  \lvllang contains all processes in \vaslang (i.e., typable under 
	Vasconcelos's type system) which are also typable (up to a translation) under the weight-based type system.
	\item  \dilllang contains all processes in \vaslang  which are also typable (up to another translation) by the Curry-Howard correspondence.
\end{itemize}

 This way, because Vasconcelos's  system can type non-terminating processes, both $\lvllang \subset \vaslang$ and $\dilllang \subset \vaslang$ hold by definition. 
Our technical contributions are two-fold. 
\begin{enumerate} 
	\item 
Because the type systems by Vasconcelos and by Deng and Sangiorgi are so different, to define \lvllang we develop a \emph{new weight-based type system} that combines elements from both: it ensures termination by enforcing well-founded measures (as Deng and Sangiorgi's) while accounting for linearity and sessions (as Vasconcelos's). The translation involved in bridging \vaslang and this new type system determines a technique for ensuring termination of session-typed processes, which is new and of independent interest.
\item We prove that $\dilllang \subset \lvllang$
but 
$\lvllang \not \subset \dilllang$, thus determining the exact relationship between  these classes of typed processes.
Our discovery is that there are terminating session-typed processes that are typable with the weight-based approach but not under the Curry-Howard correspondence. 
In other words, techniques based on well-founded measures turn out to be more powerful for enforcing termination than proof-theoretical foundations.
\end{enumerate}
Next, we introduce the class \vaslang.
\Cref{ch5s:weight} develops the new weight-based type system and \Cref{ch5s:lvllang} studies its corresponding class \lvllang.
The Curry-Howard correspondence for concurrency is recalled in \Cref{ch5s:pas}, and its corresponding class \dilllang is presented in \Cref{ch5s:dilllang}.
Finally, \Cref{ch5s:close} collects concluding remarks.

\section{The Class \texorpdfstring{$\vaslang$}{??} of Session Processes}

We present the process language that we shall consider as reference in our comparisons, and its corresponding session type system.
We distinguish between (i)~the processes induced by this process model and (ii)~the class of well-typed processes (\Cref{ch5d:vaslang}); in the following, these classes are denoted by \vasco and $\vaslang$, respectively.
We consider the type system by \cite{V12}, which ensures communication safety and session fidelity, but not progress/deadlock-freedom nor termination. 
Our presentation closely follows~\cite{V12}, pointing out differences where appropriate. 

\subsection{The Process Model \texorpdfstring{\vasco}{??}}
\label{ch5ss:pm}

\begin{definition}{Processes}
	\label{ch5def:vascosyntax}
	Let  $x, y, \ldots$ range over \emph{variables}, denoting \emph{channel names} (or \emph{session endpoints}), and $v, v', \ldots$ over \emph{values}; 
for simplicity, the sets of values and variables coincide.
Also, let  $P, Q, \ldots$ range over \emph{processes}, defined by the grammar of \Cref{ch5f:vascosyntax}, which induces the class \vasco.
\end{definition}

\begin{figure}[t!]
	\[
	\begin{aligned}
		P,Q ::= &								&& \mbox{(Processes)}\\
				&  \vasout{x}{v}{P}				&& \mbox{(output)} 
				&&  \vasin{q}{x}{y}{P}			&& \mbox{(input)} \\
				&  \vaspara{P}{Q}  		  		&& \mbox{(composition)} 
				&&  \vasres{xy}{P}				&& \mbox{(restriction)} \\
				&  \nil	     					&& \mbox{(inaction)} \\
		q   ::= &								&& \mbox{(Qualifiers)}\\
				& \lin 							&& \mbox{(linear)}
				&& \un 							&& \mbox{(unrestricted)}\\
		v	::= &								&& \mbox{(Values)}\\
				& x								&& \mbox{(variables)}\\
	\end{aligned}
	\]
	\caption{Syntax of the session $\pi$-calculus \vasco}
	\label{ch5f:vascosyntax}
\end{figure}

The output process $\vasout{x}{v}{P}$ sends value $v$ across channel $x$ and then continues as $P$. 
In the input process $ \vasin{q}{x}{y}{P}$, the qualifier $q$ can be either $\lin$ (denoting a linear input) or $\un$ (denoting an unrestricted input, i.e., a replicated server). 
In either case, $x$ expects to receive a value that will replace free occurrences of $y$ in $P$. 
Parallel composition $\vaspara{P}{Q}$ denotes the concurrent execution of processes $P$ and $Q$. 
The process $\vasres{xy}{P}$ denotes the restriction of the \emph{co-variables} $x$ and $y$ with scope $P$. This declares them as dual endpoints, which are expected to behave complementarily to each other. 
We write $\res {zv:S} P $ when either $z$ or $v$ have session type $S$ in $P$.
As we will see, a synchronization  always occurs across a pairs of co-variables.
Finally, the inactive process is denoted as $\nil$.

As usual, the set of free variables in a process $P$ is denoted $\fv{P}$, and similarly $\bv{P}$ for bound variables. 
The capture-free substitution of the variable $z$ by the value $v$ is denoted as $\substj{v}{z}$. We adopt Barendregt's variable convention.

With respect to \cite{V12}, the above the process syntax leaves out boolean values, conditional expressions, and labeled choices, which are all inessential for our comparative study of termination. 

\begin{definition}{Reduction Semantics}
The reduction relation $\vasred$ of \vasco is defined in \Cref{ch5f:vascosymantics}.
\end{definition}

\begin{figure}[t!]
	\[
		\begin{aligned}
			\vaspara{P}{Q} &\equiv \vaspara{Q}{P}
			&
			\vaspara{P}{\nil} & \equiv P
			\\
			 \vaspara{(\vaspara{P}{Q})}{R} &\equiv \vaspara{P}{(\vaspara{Q}{R})} 
			&
			\vasres{xy}{\nil} & \equiv \nil
			\\
			\vasres{xy}\vasres{zw}{P} & \equiv \vasres{zw}\vasres{xy}{P}
			&
			\!\!\vasres{xy}{P} & \equiv \vasres{yx}{P}~ 
			\\
			\vaspara{(\vasres{xy}P)}{Q} & \equiv \vasres{xy}(\vaspara{P}{Q}) \,\text{[$x,y \not \in \fv{Q}$]}
		\end{aligned}
	\]
	\hrule
	\smallskip
	\[
		\begin{array}{rll}  
		\Did{R-LinCom} & 
			\vasres{xy}{(  
					\vaspara{\vasout{x}{v}{P}}
						{\vaspara{\vasin{\lin}{y}{z}{Q}}
							{R}} 
				   )}  
			\\
			& \quad \vasred \vasres{xy}{(
					\vaspara{P}
						{\vaspara{Q\substj{v}{z}}
							{R}}
	
				)}
			\\[1mm] 
		\Did{R-UnCom} & 
			\vasres{xy}{(
					\vaspara{\vasout{x}{v}{P}}
						{\vaspara{\vasin{\un}{y}{z}{Q}}
							{R}}
				)} 
				\\
			& \qquad \vasred
			\vasres{xy}{(
					\vaspara{P}
						{\vaspara{Q\substj{v}{z}}
							{\vaspara{\vasin{\un}{y}{z}{Q}}
								{R}}
						}
				)}
			\\[1mm]  
			\Did{R-Par}&		{P\vasred Q}\Longrightarrow
								{\vaspara{P}{R} \vasred \vaspara{Q}{R}}
		\\[1mm]
		\Did{R-Res}&	{P\vasred Q} \Longrightarrow
								{\vasres{xy}{P} \vasred \vasres{xy}{Q}}
		   \\[1mm]
		\Did{R-Str}& 	{P\equiv P',\ P\vasred Q,\ Q'\equiv Q}
									\Longrightarrow
									{P'\vasred Q'} & 
		\end{array}
	\]
	\caption{Reduction semantics for \vasco}
	\label{ch5f:vascosymantics}
\end{figure}

The reduction semantics for \vasco follows standard lines for (session) $\pi$-calculi; it is closed under a structural congruence, denoted $\equiv$, which captures expected principles for parallel composition and restriction.
The reduction rule $\Did{R-LinCom}$ captures the linear communication across co-variables $x$ and $y$, appropriately declared by restriction, in which value $v$ is exchanged.
Similarly, rule $\Did{R-UnCom}$ denotes unrestricted communication across co-variables; in this case, the input prefix is persistent, and remains ready for further synchronizations after reduction.
The contextual rules $\Did{R-Par}$ and $\Did{R-Res}$ express that concurrent processes can reduce within the scope of parallel composition and restriction. Finally, rule $\Did{R-Str}$  denotes that reductions are closed under structural congruence.

\subsection{Session Types}\label{ch5ss:typesess}
%

We endow \vasco with the session type system by \cite{V12}, which ensures that well-typed processes respect their protocols but does not ensure deadlock-freedom nor termination guarantees.
With respect to the syntax of types in~\cite{V12},  we only consider channel endpoint types (no ground types such as \textsf{bool}).

\begin{definition}{Session Types}
The syntax of {session types} ($T, S, \ldots$) is given in \Cref{ch5f:vascosessiontypes}.
\end{definition}

\begin{figure}[t!]
	\[
	\begin{aligned}
		q   ::= &	&							
		 \mbox{(Qualifiers)}
		 & & 
		 		T,S ::= &	&							 \mbox{(Types)} &&
		\\
				& \lin 							& \mbox{(linear)}
				& & &
				\nilT 						& \mbox{(termination)}
				\\
				& \un 							& \mbox{(unrestricted)}
				& & & q \ p  & \mbox{(pretypes)}
				\\
		p   ::= &								& \mbox{(Pretypes)}
		& & & 
		a & \mbox{(type variable)}
		\\
				& \vasreci{T}{S}				& \mbox{(receive)}
				& & & 
				\Rec{a}{T}	 & \mbox{(recursive types)}
				\\
				& \vassend{T}{S}				& \mbox{(send)}
				& & & 
				 & 
		\\
			\Gamma	 ::= &  						& \mbox{(Contexts)}
				& & & &
				\\
				& \emptyset						& \mbox{(empty)}
				& & & &
				\\
				& \Gamma , x:T
				& \mbox{(assumption)} & 
	\end{aligned}
	\]
	\caption{Session Types of \vasco}
	\label{ch5f:vascosessiontypes}
\end{figure}

Session types $T, S$ describe protocols as \emph{sequences} of actions for an endpoint; they do not admit the parallel usage of an endpoint.
They have the following forms:
\begin{enumerate}
	\item Type $\nilT$ is given to an endpoint with a completed protocol.
	
\item Type $ q \ p $ denotes pre-type $p$ with qualifier $q$, which indicates either a linear or an unrestricted behavior ($\lin$ and $\un$, respectively). 
The pre-type $\wn T.S$ is given to an endpoint that first receives a value of type $T$ and then continues according to type~$S$.
	Dually, the pre-type $\oc T.S$ is intended for an endpoint that first outputs a value of type $T$ and then continues according to  $S$.
	
	\item Type $\Rec{a}{T}$ is a recursive type, with type variable $a$. A recursive type is required to be \emph{contractive}, {i.e., it contains no subexpression of type $\Rec{a_1}\ldots \Rec{a_n}a_1$}; and $a$ is bound with scope $T$. Notions of bound and free type variables, alpha-conversion and capture-avoiding substitutions (denoted $\substj{S}{a}$) is defined as usual. Type equality is based on regular infinite trees \cite{V12}.
\end{enumerate} 

Recursive types that are \emph{tail-recursive} are expressive enough to type servers and clients; we have a dedicated notation for them.

\begin{notation}[Server and Client Types]\label{ch5note:rectype}
	We shall write $\recur{\wn T}$ to denote the server type $ \Rec{a}{\un\  \wn T.a}$, where variable $a$ does not occur in $T$. 
	Similarly, we write $\recur{\oc T}$ to denote the client type $ \Rec{a}{\un\  \oc T.a}$
\end{notation}

In the following, we shall work  with tail-recursive types only.
A central notion in session-based concurrency is  \emph{duality}, 
which relates session types offering opposite (i.e., complementary) behaviors; it stands at the basis of communication safety and session fidelity. 

\begin{definition}{Duality}
	Given a (tail-recursive) session type $T$, its dual type $\dual{T}$ is defined as follows:
	\begin{displaymath}
		\begin{array}{rclrclrcl}
			\dual{\nilT} &=&  \nilT &\quad  \dual{\oc T.S} &=& \wn T .\dual{S} & \quad \dual{\recur{\wn {T}}} &=&  \recur{\oc {T}}
			\\
			\dual{q \ p} &=& q \ \dual{p }&
			\dual{\wn T.S} &=& \oc T.\dual{S} &
			\dual{ \recur{\oc {T}}} &=& \recur{\wn {T}} 
		\end{array}
		\end{displaymath}

\end{definition}

We now collect definitions and results from~\cite{V12} that will lead to state the main properties of typable processes. 

\begin{definition}{Predicates on Types/Contexts}\label{ch5def:pred_cont}
We consider two predicates on types, denoted $\contlin{T}$ and $\contun{T}$, defined as follows:
\begin{itemize}
	\item $\contun{T}$ if and only if $T = \nilT$ or $T = \un \ p$. 
	\item $\contlin{T}$ if and only if $\ttrue$.
\end{itemize}
The definition extends to contexts as follows: we write
$q(\Gamma)$ if and only if $x:T \in \Gamma$ implies $q(T)$.
\end{definition}

This way, to express that $T$ defines strictly linear behavior we write $\neg \contun{T}$ (and similarly for a context $\Gamma$).
The following notation is useful to separate the linear and unrestricted portions of a context:
\begin{notation}\label{ch5note:sep}
We write $\Gamma \vaslinunsplit \Gamma'$ if $\contun{\Gamma} \land \neg \contun{\Gamma'}$.
\end{notation}


\begin{definition}{Context Split and Update}\label{ch5d:splitsts}
The split and update operations on contexts, denoted $\contcomp$ and $+$, are defined as follows.
	\begin{mathpar}
		\inferrule{}
		{ \emptycont \contcomp \emptycont = \emptycont }
		\and
		\inferrule{ \Gamma_1 \contcomp \Gamma_2 = \Gamma \\ \contun{T}}
		{\Gamma , x:T = ( \Gamma_1 , x:T ) \contcomp ( \Gamma_2 , x: T ) }
		\and
		\inferrule{ \Gamma_1 \contcomp \Gamma_2 = \Gamma }
		{ \Gamma , x: \lin\  p = (\Gamma_1 , x: \lin\  p) \contcomp \Gamma_2 }
		\and
		\inferrule{  \Gamma_1 \contcomp \Gamma_2 = \Gamma }
		{ \Gamma , x: \lin \  p = \Gamma_1 \contcomp ( \Gamma_2 , x: \lin \ p ) }
		\and
		\inferrule{ x:U \not \in \Gamma }
		{ \Gamma + x:T = \Gamma , x:T }
		\and
		\inferrule{ \contun{T} }
		{ ( \Gamma , x:T ) + x:T = ( \Gamma , x:T  ) }
	\end{mathpar}
\end{definition}

The typing system considers two kinds of judgments, for processes and for variables, denoted  $\Gamma \st P$ and $\Gamma\st x:T$, respectively. 
We write $~ \st P$ when $\Gamma$ is empty.
The typing rules are given in \Cref{ch5f:vasorigrules}.
We will explain Rule \vastype{In}: it  is parametric on the qualifiers $q_1$ and $q_2$  and  covers three different behaviours depending on whether $q_i$ is $\lin$ or $\un$, for $i=1,2$. In the case $q_1=\lin$, to prove $\Gamma_1 \contcomp \Gamma_2 \st\lin\ x\inp y.P$, we need to prove $\Gamma_1 \st x: q_2 \wn T.S $ and $ (\Gamma_2 + x: S), y:T \st P$; note  that $\lin (\Gamma_1\contcomp \Gamma_2)$ is true, by \Cref{ch5def:pred_cont}.   In the case $q_2=\lin$, both judgments hold if $\Gamma_1=\Gamma_1',  x: \lin \wn T.S$, the assignment $x: \lin \wn T.S$ does not occur in $\Gamma_2$, by \Cref{ch5d:splitsts},  and $x:S$ is added to $\Gamma_2$ for the continuation.   
Differently, when $q_2=\un$, both judgments hold if $\Gamma_1=\Gamma_1',  x: \recur{\wn T}$, the assignment $x: \recur{\wn T}$ also occurs in $\Gamma_2$ which with the addition of $x:S$ in $\Gamma_2$ implies $S = \recur{\wn T}$. Notice that the case $q_1=\un$ and $q_2=\lin$ is not possible since $\un (\Gamma_1\contcomp \Gamma_2)$ implies that all assignments in $\Gamma_1\contcomp \Gamma_2$ have types $\nilT$ or with `$\un$'; thus, in that case we cannot prove $\Gamma_1 \st x: \lin \wn T.S $.

 Similarly, Rule \vastype{Out} is parametric on the qualifier $q$.

	\begin{figure}[!t]
	\begin{mathpar}
		\inferrule* [ left = \vastype{Var} ]{ \contun{\Gamma} }
		{ \Gamma , x:T \st x:T }
		\and
		\inferrule* [ left = \vastype{Nil} ]{ \contun{\Gamma} }
		{ \Gamma \st \nil  }
		\and
		\inferrule*[ left =\vastype{Par} ]{ \Gamma_1 \st P \\ \Gamma_2 \st Q }
		{ \Gamma_1 \contcomp \Gamma_2 \st P \pp Q  }
		\and
		\inferrule*[ left =\vastype{Res} ]{ \Gamma , x:T ,y: \dual{T} \st P }
		{ \Gamma \st  \res {xy} P }
		\and
		\inferrule[\vastype{In} ]{ q_1(\Gamma_1 \contcomp \Gamma_2 ) \\ \Gamma_1 \st x: q_2 \wn T.S  \\ (\Gamma_2 + x: S), y:T \st P }
		{ \Gamma_1 \contcomp \Gamma_2 \st q_1\ x\inp y.P }
		\and
		\inferrule* [ left = \vastype{Out} ]{ \Gamma_1 \st x: q \oc T.S 	\\ \Gamma_2 \st v:T  \\ \Gamma_3 + x:S \st P  }
		{ \Gamma_1 \contcomp \Gamma_2 \contcomp \Gamma_3 \st \ov x\out v.P }
	\end{mathpar}
			\caption{Typing rules for \vasco (cf.~\cite{V12}).\label{ch5f:vasorigrules}}
	\end{figure}
	
The main property of the type system concerns \emph{well-formed} processes, which are defined next. 
\begin{definition}{Redexes and Well-formedness}
	A \emph{redex} is a process of the form $ \vasin{q}{x}{v}{P} \pp \vasout{y}{z}{Q}$.  
	Processes of the form $ \vasin{q}{x}{v}{P}$ and $\vasout{y}{z}{Q}$ have prefix $x$ and $y$, respectively.

	A process is \emph{well-formed} if, for each of its structurally congruent processes of the form $ \res {x_1y_1} \cdots \res {x_ny_n} (P \pp Q \pp R) $, the following conditions hold. (1)~If $P$ and $Q$ are processes prefixed at the same variable, then they are of
the same nature (input, output). (2)~If $P$ is prefixed at $x_1$ and $Q$ is prefixed at $y_1$ then $P \pp Q$ is a redex.
\end{definition}

\begin{theorem}[Properties of the Type System]\label{ch5thm:propertiesvasco}
The type system satisfies the following properties (see \cite{V12} for details):
\begin{itemize}
	\item 
	If $\Gamma \st P$ and $P \equiv Q$, then $\Gamma \st Q$. 
	\item 
	If $\Gamma \st P$ and $P \vasred  Q$, then $\Gamma \st Q$.
	\item If $\Gamma, x : T\st P$ and $x \not \in \fn{P}$, then $\Gamma \st P$.
	\item 
	If $~\st P$ then $P$ is well-formed.
\end{itemize}
\end{theorem}

For technical convenience, we rely on the \emph{refined} typing rules for input and output in \Cref{ch5f:vasrules}, which are equivalent (but more fine-grained) than those in \Cref{ch5f:vasorigrules}.

	\begin{figure}[!t]
	\begin{mathpar}
		\inferrule[\vastype{Lin-In_1} ]
 		{ \Gamma_1 , x : \lin \vasreci{T}{S} \st x: \lin \vasreci{T}{S}  \\ 
		\Gamma_2 , x: S , y : T \st P}
		{ \Gamma_1 , x: \lin \vasreci{T}{S} \contcomp \Gamma_2 \st  \vasin{\lin}{x}{y}{P}   }
		\and
		\inferrule* [ left = \vastype{Lin-In_2} ]
		{  \Gamma_1 , x: \recur{\wn T} \st x: \recur{\wn T}  \\
		\Gamma_2 , x: \recur{\wn T} , y : T \st P
		}
		{  (\Gamma_1 , x: \recur{\wn T}) \contcomp (\Gamma_2 , x: \recur{\wn T}) \st  \vasin{\lin}{x}{y}{P} }
		\and
		\inferrule* [ left = \vastype{Un-In} ]
		{  \Gamma \st x: \recur{\wn T}  \\
		\Gamma , y : T \st P
		}
		{  \Gamma \st   \vasin{\un}{x}{y}{P}  }
		\and
		\inferrule* [ left = \vastype{Un-Out} ]
		{ \Gamma_1 \st x: \recur{\oc T}  \\  \Gamma_2 \st v: T  \\
		\Gamma_3 \st P
		}
		{  \Gamma_1 \contcomp	\Gamma_2 \contcomp \Gamma_3 \st   \vasout{x}{v}{P}} 
		\and
		\inferrule[\vastype{Lin-Out} ]
		{ \Gamma_1 \st x: \lin\ \vassend{T}{S}  \\  \Gamma_2 \st v: T  \\
		\Gamma_3, x:S \st P
		}
		{  \Gamma_1 \contcomp	\Gamma_2 \contcomp \Gamma_3 \st   \vasout{x}{v}{P}}  
	\end{mathpar}
		\caption{Refined typing rules for input and output.\label{ch5f:vasrules}}
	\end{figure}

We close this section by defining the class of processes \vaslang.
\begin{definition}{\vaslang}
\label{ch5d:vaslang}
	We define $ \vaslang = \{ P \in \vasco \mid  \exists \Gamma \ s.t. \ \Gamma \st P  \}$.
\end{definition}

\begin{example}{A Non-Terminating Process in \vaslang}
\label{ch5ex:infinite}
		Consider the process $P_{\ref{ch5ex:infinite}} =  \vasres{xy}{( \vaspara{\vasout{y}{w}{\nil}}{\vasin{\un}{x}{z}{\vasout{y}{w}{\nil}}} )}$,   which invokes itself ad infinitum. 
	Process $P_{\ref{ch5ex:infinite}}$ is in $\vaslang$ because $w: \nilT \st P_{\ref{ch5ex:infinite}}$ holds with the following derivation:
		\begin{mathpar}
	\inferrule*[left=\vastype{Res}]{
				\inferrule*[left=\vastype{Par}]{ 
					\Pi \\
					\inferrule*[left=\vastype{Un-In}]{ 
						\inferrule{
							\contun{\Gamma}
						}{
							\Gamma  \st x: \recur{\wn \nilT}  
						}
						\\
						\Pi
					}
					{  \Gamma \st   \vasin{\un}{x}{z}{\vasout{y}{w}{\nil}}  }
				}
				{ \Gamma \st \vasout{y}{w}{\nil} \pp \vasin{\un}{x}{z}{\vasout{y}{w}{\nil}} }
			}
			{ 
				w: \nilT \st  \vasres{xy}{( \vaspara{\vasout{y}{w}{\nil}}{\vasin{\un}{x}{z}{\vasout{y}{w}{\nil}}} )}
			}
		\end{mathpar}		
with 	 $  \Gamma = x:\recur{\wn \nilT} , y: \recur{\oc \nilT}, w: \nilT $ and $\Pi$ is the derivation		
		\begin{mathpar}
			\inferrule*[left=\vastype{Un-Out}]{ 
				\inferrule{
					\contun{\Gamma' }
				}
				{
					\Gamma' \st y: \recur{\oc \nilT }  
				}
				\\
				\inferrule{
					\contun{\Gamma' }
				}{
					\Gamma' \st w: \nilT  
				}\\
				\inferrule{ 
					\contun{\Gamma' } 
				}
				{ 
					\Gamma' \st \nil  
				}
			}
			{  \Gamma' \st \vasout{y}{w}{\nil} } 
		\end{mathpar}
	with $ \Gamma' = x:\recur{\wn \nilT} , y: \recur{\oc \nilT}, w: \nilT, z: \nilT $.
\end{example}

\section{A Weight-based Approach to Terminating Processes}
\label{ch5s:weight}
We move on to consider a type system that ensures termination for a class of $\pi$-calculus processes. 
Following \cite{DS06}, the type system uses \emph{weights} (or \emph{levels}) to avoid infinite reduction sequences. 
This type system will induce a class of terminating  $\vasco$ processes, denoted \lvllang (\Cref{ch5d:lvllang}), obtained via appropriate translations on processes and types.
To ease the definition of such translations, here we define a type system that mildly modifies the system of~\cite{DS06} to account for linearity and synchronous/polyadic (tuple-based) communication. 
Our main result is that the weight-based system ensures termination (\Cref{ch5t:lvlsn}). 

\subsection{Processes} 
We introduce a process model for the weight-based type system, denoted \pilvl, formally defined next.
In the following, we write $\tilde{y}$ to stand for the finite tuple $y_1,\cdots, y_n$. 

\begin{definition}{Processes}
The syntax of \pilvl processes is given by the grammar in \Cref{ch5f:lvlsyntandtype} (top).
\end{definition}

\pilvl is designed to stand in between \vasco and the process model in~\cite{DS06}.  
Communication in \pilvl is polyadic, i.e., exchanges involve a tuple of names, rather than a single name as in  \Cref{ch5def:vascosyntax} and~\cite{DS06}. 
We shall often consider tuples of length two (i.e., dyadic communication), as this suffices for a continuation-passing encoding of sessions~\cite{DGS12}. 
Another difference with respect to~\cite{DS06} is that inputs can be linear or  unrestricted; this will facilitate the formal connection with \vasco and its type system. 
The role of linearity is more prominent at the level of types, defined later on.

We give the operational semantics of \pilvl in terms of the (early)
labeled transition system (LTS), with the following labels for input, output, bound output, and silent transitions (synchronizations):
$$\alpha ::= x(\tilde{v}) \,|\, \ov x\out { \tilde{y} } \,|\, \res {y,\widetilde{b}}\ov x\out { \tilde{v} } \,|\, \tau $$ 

The rules, given in \Cref{ch5f:lvllts}, are standard. Rules \lvltype{Par} and \lvltype{Tau} can be applied symmetrically across parallel composition.
	
\begin{figure*}[!h]
	\begin{mathpar}
		\inferrule[\lvltype{In} ]{  }
		{ \lvlinpoly{x}{y}{z}{P} \lts{x \inp {\tilde{v}}} P\substj{v_1}{y_1}\substj{v_2}{y_2}  }
		\and
		\inferrule[\lvltype{Par} ]{ P \lts{\alpha} P' \\ \bn{\alpha} \cap \fn{Q} = \emptyset }
		{ P \pp Q \lts{\alpha} P'\pp Q }
		\and
		\inferrule[\lvltype{Res} ]{ P \lts{ \alpha } P' \\ x \not \in \names{\alpha} }
		{ \res {x}P \lts{ \alpha } \res {x}P' }
				\and
		\inferrule[\lvltype{Rep} ]{  }
		{ \bang x \inp { \tilde{y} }.P \lts{x \inp { \tilde{v} }} \bang x \inp { \tilde{y}}.P  \pp P\substj{v_1}{y_1} }

		\\
				\inferrule[\lvltype{Out} ]{   }
		{ \lvloutpoly{x}{y}{z}{P}  \lts{\ov x\out { \tilde{y} }}  P  }
		\and
		\inferrule[\lvltype{Tau} ]{ P  \lts{ \res {\widetilde{b}}\ov x\out { \tilde{v} } } P' \\ Q \lts{ x \inp {\tilde{v}} }  Q' \\ \widetilde{b} \cap \fn{Q} = \emptyset }
		{ P \pp Q \lts{ \tau } \res{\widetilde{b}} (P'\pp Q')}
		\and
		\inferrule[\lvltype{Open} ]{ P \lts{ \res {\widetilde{b}}\ov x\out { \tilde{v} } } P' \\ y \in (\fn{v_1} \cup \fn{v_2}) - \{ \widetilde{b} , x \} }
		{ \res {y}P \lts{ \res {y,\widetilde{b}}\ov x\out { \tilde{v} } } P' }
	\end{mathpar}
\caption{An LTS for \pilvl}
\label{ch5f:lvllts}
\end{figure*}

\subsection{Types}

\begin{definition}{Types for \pilvl}
The syntax of weight-based types for \pilvl is given by the grammar in \Cref{ch5f:lvlsyntandtype} (bottom).
\end{definition}

As in~\cite{DS06}, our link types  for \pilvl are \emph{simple}, i.e., they do not admit the sequencing of actions enabled by session types. 
Our syntax of types extends that in~\cite{DS06} to account for (i)~dyadic communication and (ii)~explicit types for clients and servers.
Concerning~(ii), we purposefully adopt the tail-recursive types for clients and servers defined for \vasco, rather than more general recursive types.

\begin{figure}[t!]
	\[
	\begin{aligned}
		P,Q ::= &								&& \mbox{(Processes)}\\
				&  \lvlout{x}{y_1}{y_2}{P}		&& \mbox{(output)}
				&&  \lvlin{x}{y_1}{y_2}{P}					&& \mbox{(linear input)}
				\\
								& P \pp Q  	    		  		&& \mbox{(parallel)}
				&&  \lvlserv{x}{y_1}{y_2}{P}			&& \mbox{(server)}\\
				& \res {x}P						&& \mbox{(restriction)}
				&& \nil	     					&& \mbox{(inaction)}\\
	\end{aligned}
	\]
	\hrule
	\smallskip
	\[
	\begin{aligned}
		S,T,V,L ::= &								&& \mbox{(Link Types)}\\
				& \lvlint{n}{V_1}{V_2}			&& \mbox{(linear input type)}\\
				& \lvloutt{n}{V_1}{V_2}			&& \mbox{(linear output type)}\\
				& \lvlunst{n}{V}				&& \mbox{(unrestricted server type)}\\
				& \lvlunct{n}{V}				&& \mbox{(unrestricted client type)}\\
				& \lvlnilT							&& \mbox{(termination)}
		\\
		n ::= & ~1, 2, \cdots && \mbox{(weights)}
	\end{aligned}
	\]
	\caption{Syntax of processes and types for \pilvl.}
	\label{ch5f:lvlsyntandtype}
\end{figure}

We introduce some notions borrowed from the  type system from \Cref{ch5ss:typesess}: duality, contexts, predicates on types, operations on contexts.

\begin{definition}{Duality}
	Duality on linked types is defined as:
		\[
		\begin{aligned}
			\dual{\lvlint{n}{V_1}{V_2}}	& = \lvloutt{n}{\dual{V_1}}{\dual{V_2}}		&
			\dual{\lvloutt{n}{V_1}{V_2}} & = \lvlint{n}{\dual{V_1}}{\dual{V_2}}	& \dual{\lvlnilT} & = \lvlnilT	
			\\
			\dual{\lvlunst{n}{V}} & = \lvlunct{n}{\dual{V}}				
			&
			\dual{\lvlunct{n}{V}} & = \lvlunst{n}{\dual{V}}				
		\end{aligned}	
	\]
\end{definition}

\begin{definition}{Contexts}
	Contexts are given by the grammar:
	\[
		\Gamma , \Delta :: = \cdot \sep \Gamma , x:V \sep \Gamma , x:\dualjoin{V}{\dual{V}}
	\]
	where $\Gamma , x:L$ and $\Gamma , x:\dualjoin{L}{\dual{L}}$ imply $x \not \in \dom{\Gamma}$. 
	
	Following the sorts of~\cite{HVK98}, the assignment 
	$x:\dualjoin{L}{\dual{L}}$ denotes the pairing of $x$  with two complementary protocols, where $\dualjoin{L}{\dual{L}} = \dualjoin{\dual{L}}{L}$. 
	We use \lvlsplit{x}{L} to stand for $x:\dualjoin{L}{\dual{L}}$ when $L$ is the main object of interest.
	We write $x\lvlass T $ if either $x:T$ or $x::T$ holds (i.e., $\lvlass \in \{ : , :: \}$). 
\end{definition}

\begin{definition}{Unrestricted Types}
	Predicate $\contun{T}$ holds if $T = \lvlunst{n}{V}$, $T = \lvlunct{n}{V}$, $T = \lvlnilT$, or x:\dualjoin{L}{\dual{L}} with $\contun{L}$. We write $\contun{\Gamma}$ if $\contun{T}$ holds for every $x \lvlass T \in \Gamma$.
\end{definition}


Following \Cref{ch5d:splitsts}, the following definitions 
gives a relation to split contexts into two parts.

\begin{definition}{Split Relation on Contexts}
The relation $\contcomp$ on contexts is defined in \Cref{ch5f:lvlcontrelation}. \end{definition}

\begin{figure}
	\begin{mathpar}
		\inferrule{}
		{ \emptycont \contcomp \emptycont = \emptycont }
		\and
		\inferrule{ \Gamma_1 \contcomp \Gamma_2 = \Gamma  \\ \contun{T}}
		{\Gamma , x: T = ( \Gamma_1 , x: T  ) \contcomp ( \Gamma_2 , x: T  ) }
		\and
		\inferrule{ \Gamma_1 \contcomp \Gamma_2 = \Gamma  \\ \contun{V}}
		{\Gamma , x:\dualjoin{V}{\dual{V}} = ( \Gamma_1 , x:\dualjoin{V}{\dual{V}}  ) \contcomp ( \Gamma_2 , x:\dualjoin{V}{\dual{V}}  ) }
		\and
		\inferrule{ \Gamma_1 \contcomp \Gamma_2 = \Gamma \\ \neg \contun{V}}
		{ \Gamma , x:\dualjoin{V}{\dual{V}} = (\Gamma_1 , x: V) \contcomp  (\Gamma_2 , x: \dual{V}) }
		\and
		\inferrule{ \Gamma_1 \contcomp \Gamma_2 = \Gamma \\ \neg \contun{V}}
		{ \Gamma , x:\dualjoin{V}{\dual{V}} = (\Gamma_1 , x:\dualjoin{V}{\dual{V}}) \contcomp  \Gamma_2  }
		\and
		\inferrule{ \Gamma_1 \contcomp \Gamma_2 = \Gamma \\ \neg \contun{V}}
		{ \Gamma , x:\dualjoin{V}{\dual{V}} = \Gamma_1  \contcomp  (\Gamma_2 , x:\dualjoin{V}{\dual{V}}) }
		\and
		\inferrule{ \Gamma_1 \contcomp \Gamma_2 = \Gamma \\ \neg \contun{T}}
		{ \Gamma , x: T = (\Gamma_1 , x: T) \contcomp \Gamma_2 }
		\and
		\inferrule{  \Gamma_1 \contcomp \Gamma_2 = \Gamma \\ \neg \contun{T}}
		{ \Gamma , x: T = \Gamma_1 \contcomp ( \Gamma_2 , x: T ) }
	\end{mathpar}
\caption{Splitting of Contexts for \pilvl}
\label{ch5f:lvlcontrelation}
\end{figure}

We now introduce notions on processes that are essential to Deng and Sangiorgi's approach to termination by typing.

\begin{definition}{Level Function, $\levelof{x}$}
Let $\mathcal{N}$ denote the set of all names. 
We define the function $\levelof{\cdot}: \mathcal{N} \to \mathbb{N}$ to map names of a process (free and bound) to naturals. 
We assume $\alpha$-conversion is silently used to avoid name capture and ensure uniqueness of bound names.
Given a (typed) process, we define this function as follows: 
$$
\levelof{x} = \begin{cases}
	n & \text{if $x : T$ or $x ::T$}
	\\
	& \text{with $T \in \{\lvlint{n}{V_1}{V_2},\lvloutt{n}{V_1}{V_2}, \lvlunst{n}{V}, \lvlunct{n}{V}\}$}
	\\
	m & \text{if $x: \lvlnilT$, for any $m \in \mathbb{N}$} 
\end{cases}
$$
\end{definition}

\begin{definition}{Active Outputs, $\outsubj{\cdot}$}
Given a process $P$, the set of names with active outputs $\outsubj{P}$ is  defined inductively:
\[
\begin{aligned}
	\outsubj{\lvloutpoly{x}{y}{z}{P}} & = \{x\} \cup \outsubj{P}
	&
	\outsubj{\lvlinpoly{x}{y}{z}{P}} & = \outsubj{P}\\
	\outsubj{P \pp Q } & = \outsubj{P} \cup \outsubj{Q}
	&
	\outsubj{\res {x}P	} & = \outsubj{P}\\
		\outsubj{\nil} & = \emptyset 
	&
	\outsubj{\lvlservpoly{x}{y}{z}{P}} & = \emptyset
\end{aligned}
\]
\end{definition}


\begin{figure*}[!t]
	\small
		\begin{mathpar}
			\inferrule[\lvltype{Var_1} ]{ \contun{\Gamma} }
			{ \Gamma , x:V \lt x:V }
			\and
			\inferrule[\lvltype{Var_2} ]{ \contun{\Gamma} }
			{ \Gamma , x: \dualjoin{V}{\dual{V}} \lt x:V }
			\and
						\inferrule[\lvltype{Nil} ]{ \contun{\Gamma} }
			{ \Gamma \lt \nil }
			\and
			\inferrule[\lvltype{Par} ]{\Gamma_1 \lt P \\ \Gamma_2 \lt Q }
			{ \Gamma_1 \contcomp \Gamma_2 \lt P \pp Q }
			\and
			\inferrule[\lvltype{Res} ]{ \Gamma, x: \dualjoin{V}{\dual{V}}  \lt P  }
			{ \Gamma \lt \res {x}P }
			\and 
			\inferrule* [ left = \lvltype{Lin-In_1} ]{ \Gamma_1 , x: \lvlint{n}{V_1}{V_2} \lt x: \lvlint{n}{V_1}{V_2} \\ \Gamma_2 ,  y_1:V_1, y_2:V_2 \lt P \\ \levelof{x} = \levelof{y_2} }
			{ \Gamma_1 , x: \lvlint{n}{V_1}{V_2} \contcomp \Gamma_2  \lt \lvlin{x}{y_1}{y_2}{P} }
			\and
			\inferrule* [ left = \lvltype{Lin-In_2} ]{ 
				\Gamma_1 , x: \lvlunst{n}{V} \lt x: \lvlunst{n}{V} \\
				\Gamma_2 , x: \lvlunst{n}{V} , y_1:V , y_2: \lvlnilT \lt P }
			{ ( \Gamma_1 , x: \lvlunst{n}{V}) \contcomp (\Gamma_2 , x: \lvlunst{n}{V}) \lt \lvlin{x}{y_1}{y_2}{P}  }
			\and
			\inferrule* [ left = \lvltype{Lin-In_3} ]{
				\Gamma , \lvlsplit{x}{\lvlunst{n}{V}} \lt x:\lvlunst{n}{V} \\
				\Gamma , \lvlsplit{x}{ \lvlunst{n}{V}} ,   y_1:V  , y_2: \lvlnilT \lt P }
			{ \Gamma , \lvlsplit{x}{\lvlunst{n}{V}} \lt \lvlin{x}{y_1}{y_2}{P}  }
			\and
			\mprset{ flushleft }
			\inferrule* [ left = \lvltype{Lin-Out} ]{
				\Gamma_1 ,  x: \lvloutt{n}{V_1}{V_2} \lt  x: \lvloutt{n}{V_1}{V_2} \\
				\Gamma_2 , y_1:V_1 \lt y_1:V_1 \\
				\Gamma_3 , y_2:V_2 \lt P \\ \levelof{x} = \levelof{y_2} }
			{ (\Gamma_1 , x: \lvloutt{n}{V_1}{V_2}) \contcomp (\Gamma_2 , y_1:V_1) \contcomp (\Gamma_3, \lvlsplit{y_2}{ V_2} ) \lt \lvlout{x}{y_1}{y_2}{P} }
			\and
			\inferrule* [ left = \lvltype{Un-Out_1} ]{ 
				\Gamma_1 , x:  \lvlunct{n}{V} \lt x:  \lvlunct{n}{V} \\
				\Gamma_2, x:  \lvlunct{n}{V} , y: V \lt y: V \\
				\Gamma_3, x:  \lvlunct{n}{V} , y_2: \lvlnilT   \lt P  
			}
			{( \Gamma_1 , x:  \lvlunct{n}{V} )\contcomp	(\Gamma_2, x:  \lvlunct{n}{V} , y: V) \contcomp (\Gamma_3, x:  \lvlunct{n}{V})
			 \lt 
			\lvlout{ x}{y_1}{y_2}{P} 
			}
			\and
			\inferrule* [ left = \lvltype{Un-Out_2} ]{  
				\Gamma_1, \lvlsplit{x}{ \lvlunct{n}{V}} \lt x:  \lvlunct{n}{V} \\
				\Gamma_2, \lvlsplit{x}{ \lvlunct{n}{V}} , y: V \lt y: V \\
				\Gamma_3 , \lvlsplit{x}{ \lvlunct{n}{V}} , y_2: \lvlnilT   \lt P  
			}
			{ (\Gamma_1 , \lvlsplit{x}{ \lvlunct{n}{V}}) \contcomp	
			(\Gamma_2, \lvlsplit{x}{ \lvlunct{n}{V}} , y: V) \contcomp 
			(\Gamma_3, \lvlsplit{x}{ \lvlunct{n}{V}}) \lt 
			\lvlout{x}{y_1}{y_2}{P} 
			}
			\\
			\inferrule* [ left = \lvltype{Un-In_1} ]{ 
				\Gamma, x: \lvlunst{n}{V} \lt x: \lvlunst{n}{V} \\
				\Gamma, x: \lvlunst{n}{V}, y_1:V , y_2: \lvlnilT \lt P \\  \forall b \in \outsubj{P}, \ \levelof{b} < n }
			{ \Gamma, x: \lvlunst{n}{V} \lt  \lvlserv{x}{y_1}{y_2}{P} }
			\and
			\inferrule* [ left = \lvltype{Un-In_2} ]{
				\Gamma, \lvlsplit{x} {\lvlunst{n}{V}} \lt x: \lvlunst{n}{V} \\
				\Gamma, \lvlsplit{x} {\lvlunst{n}{V}}, y_1:V , y_2: \lvlnilT \lt P \\  \forall b \in \outsubj{P}, \ \levelof{b} < n }
			{ \Gamma, \lvlsplit{x}{\lvlunst{n}{V}} \lt  \lvlserv{x}{y_1}{y_2}{P} }
		\end{mathpar}
\caption{Typing rules for \pilvl}
\label{ch5f:lvltyperules}
\end{figure*}

Typing judgments are of the form $\Gamma \lt P$, with corresponding typing given in \Cref{ch5f:lvltyperules}.  
Typability is contingent on a level function: we say a process $P$ is well-typed if there exists a level function $\levelof{\cdot}$ such that a typing derivation $\Gamma \lt P$ holds, for some $\Gamma$.




We comment on some of the rules in \Cref{ch5f:lvltyperules} for \pilvl, contrasting them with those in \Cref{ch5f:vasorigrules} for \vasco.
Rule \lvltype{Var_1} is similar to rule~\vastype{Var}.
Rule~\lvltype{Var_2} is the corresponding rule for complementary interaction:   if $x:\dualjoin{V}{\dual{V}}$, then we can assign the type $x:V$. Intuitively, name $x$ encapsulates the types of its two endpoints, denoted as $V$ and $\dual{V}$. As long as $x$ respects one of these types, the channel is considered correctly typed.

Rule \lvltype{Lin-In_1} acts as the linear counterpart to \vastype{In}. Importantly, there is no direct counterpart for $x$ as a linear complementary interaction. Instead, the context split $\Gamma , x:\dualjoin{V}{\dual{V}} = (\Gamma_1 , x:{V}) \contcomp (\Gamma_2 , x:{\dual{V}})$ allows for the application of rule~\lvltype{Lin-In_1}. This structural mechanism operates silently within the rules where $V$ is linear, achieved through context split. As a result, this disallows linear channels from consuming linear complementary interactions.

Rules \lvltype{Lin-In_2} and \lvltype{Lin-In_3}, the first with `:' and the second with `::',  are  counterparts to rule \vastype{In} for  unrestricted types with linear qualifier. 
Similarly, \lvltype{Lin-Out}, \lvltype{Un-Out_1}, and \lvltype{Un-Out_2} represent the rule \vastype{Out}. 
Furthermore, \lvltype{Un-In_1} and \lvltype{Un-In_2} are the unrestricted counterparts to rule \vastype{In} with unrestricted qualifier. 
These rules adopt the main condition from~\cite{DS06}, i.e., the weight of types of the active outputs must be strictly less than the weight of the type of the channel of the server providing them. 
Finally, rule \lvltype{Res} types a restricted channel through a complementary interaction.

We state the type preservation property:
\begin{theorem}[Type Preservation]
	\label{ch5thm:lvltypepres}
	Suppose $\Gamma \lt P$ for a level function $l$. If $ P \lts{\tau} P' $ then $\Gamma \lt P'$ for the same level function $l$.
\end{theorem}


\subsection{Termination by Typing}
A process terminates if all its reduction sequences are finite. 
We show that our formulation of the type system in \cite{DS06} also enforces termination by typing. The proof follows the same lines as in~\cite{DS06}: 
a weight is associated with a well-typed process; this weight is then shown to strictly reduce when the the process synchronizes. 
The weight is actually a \emph{vector} constructed from the observable active outputs  of a channel within a typed process. 

\begin{definition}{Vectors}
	We define vectors and their operations: 
\begin{itemize}
	\item Given $k \geq 1$, we write $\zerovec_i$ to denote the vector $\langle n_k , n_{k-1} , \cdots ,  n_1  \rangle $ where $n_i = 1$ and $ n_j = 0$ for every other~$j$. Also, $\zerovec$ denotes the zero vector where  $n_i = 0$ for every $i$. 
	\item 
	Given vectors $v_1 =\langle n_k , n_{k-1} ,\cdots ,  n_1  \rangle$ and $v_2 = \langle m_l , m_{l-1}  ,\cdots ,  m_1  \rangle $,  with $k \geq l$, the sum  $v_1+v_2$ is defined in two steps. 
	Firstly, if $k > l$ then the shorter vector $v_2$ is extended into $v'_2$ by adding  zeroes to match the size of $v_1$, i.e., $v'_2 =\langle m_k , m_{k-1} ,\cdots , m_l  ,\cdots ,  m_1  \rangle $, with $ \langle m_k , m_{k-1} ,\cdots , m_{l+1} \rangle = \zerovec$. 
	Then, addition of $ v_1 $ and $ v'_2 $ is applied pointwise.
	
	\item Given vectors $v_1 = \langle n_k , n_{k-1} ,\cdots ,  n_1 \rangle$ and $v_2 = \langle m_k , m_{k-1} ,\cdots ,  m_1 \rangle$ of equal size $k$, the ordering $v_1 \wtless v_2$ is defined iff $\exists i \leq k, \ n_i < m_i $ and $\forall j > i, \ n_j = m_j$.
	\end{itemize}
\end{definition}

Using vectors, we define the weight of a well-typed process:

\begin{definition}{Weights}
	Given a well-typed process $P$ with level function $l$, the weight of  $P$ is the vector defined inductively as: 
	\[
	\begin{aligned}
	\weight{\nil} & = 0 &
	\weight{\bang x \inp { \tilde{y}}.P	}&  = 0 
	\\
	\weight{x\inp { \tilde{y}}.P} & = \weight{P} &
	\weight{\ov x\out { \tilde{y} }.P} & = \weight{P} + \zerovec_{\levelof{x}} 
	\\
	\weight{P \pp Q} & = \weight{P} + \weight{Q} &
	\weight{\res {x}P} & = \weight{P} &
	\end{aligned}
	\]	
\end{definition}		

We have the following  results, whose proof is as in~\cite{DS06}:

\begin{proposition}
	\label{ch5prop:lvlweightreduce}
	If $ \Gamma \lt P $ and $ P \lts{\tau} P' $ then $ \weight{P'} \wtless \weight{P}$.
\end{proposition}


\begin{theorem}[Termination]
\label{ch5t:lvlsn}
	If $\Gamma \lt P$ then $P$ terminates.
\end{theorem}


\section{\texorpdfstring{\lvllang}{??}: A Class of Terminating Processes}
\label{ch5s:lvllang}

Here we define and study \lvllang, a class of terminating \vasco processes induced by the weight-based type system given in \Cref{ch5s:weight}, which leverages  
translations on processes and types/contexts, denoted $\sttoltp{\cdot}{}$ and $\sttoltt{\cdot}{l}{}$, respectively.
Concretely, \lvllang is defined as follows:

\begin{definition}{$\lvllang$}
\label{ch5d:lvllang}
We define:
 $$ \lvllang = \{ P \in \vasco \mid  \exists \Gamma , l \ s.t. \ (\Gamma \st P) \land  \ 
\sttoltt{\Gamma}{l}{} \lt \sttoltp{P}{}     \} $$
\end{definition}	

Hence, \lvllang contains those processes from $\vaslang$ (\Cref{ch5d:vaslang}) whose translation gives typable $\pilvl$ processes.
By \Cref{ch5t:lvlsn},   $\lvllang$ thus provides a characterization of terminating processes in $\vaslang$. 
In the following we formally define the translations $\sttoltp{\cdot}{}$ and $\sttoltt{\cdot}{l}{}$, and establish their main properties. 
Our main result is that $\lvllang \subset \vaslang$ (\Cref{ch5t:wsubsets}): there are typable processes in $\vaslang$ which are not terminating under the weight-based approach.

\subsection{The Typed Translation}
Our translation is \emph{typed}, i.e., the translation of a \vasco process depends on its associated (session) types. 
We first present the translation on processes and types separately; then, we   combine them to define the translation of a typing judgment.
\begin{definition}{Translating Processes}
\label{ch5d:encvs}
The translation   $\sttoltp{\cdot}{}: \vasco \to \pilvl $ is given in \Cref{ch5f:lvltypeencod} (top), where we assume $z$ is fresh.
\end{definition}
 
We discuss some interesting cases in the translation of processes:
\begin{itemize} 
\item The shape of process $\sttoltp{  \vasin{\lin}{x}{y}{P}  }{} $ depends on whether  $x$ has a linear or an unrestricted type: this is due to rule $\vastype{In}$ (\Cref{ch5f:vasorigrules}) which depends on a qualifier $q_2$ that can be linear or unrestricted. If $x: \lin \wn T.S$ then the translation is  $\lvlin{x}{y}{z}{\sttoltp{P\substj{z}{x}}{}}$, with the continuation along $z$; otherwise, in case $x: \recur{\wn T}$, the translation is $\lvlin{x}{y}{z}{\sttoltp{P}{}}$, since there is no continuation in $x$, as explained in the description of rule \vastype{In} in \Cref{ch5f:vasorigrules}.
 \item The process $\sttoltp{   \vasin{\un}{x}{y}{P}  }{} $ is simply an unrestricted input process $\lvlserv{x}{y}{z}{\sttoltp{P}{}} $.
\item The process $\sttoltp{ \vasout{x}{y}{P}  }{} $, the translation of a bound send, also depends on the type of $x$ and the justification for it is similar to the translation of linear inputs described above. 
\item The process $ \sttoltp{\res {xy}P}{}  $ is simply $ \res {z} \sttoltp{P\substj{z}{x}\substj{z}{y}}{}$: the co-variables $x,y$ are replaced by the restricted (fresh) name $z$. The duality between the types of $x$ and $y$, say $x:L$ and $y:\dual{L}$, must be preserved by the type of $z$ in $\pilvl$. This correspondence will become evident when discussing the translation of judgements (\Cref{ch5def:firsttransl_judgments}).

\end{itemize}

\begin{figure}[!t]
	\[
	\begin{aligned}
		\sttoltp{ \nil  }{} &=  \nil  \\
		\sttoltp{  P \pp Q  }{} 	&=  \sttoltp{P}{} \pp \sttoltp{Q}{}  \\
		\sttoltp{  \res {xy} P  }{} 	&=   \res {z} \sttoltp{P\substj{z}{x}\substj{z}{y}}{}  \\
		\sttoltp{  \vasin{\lin}{x}{y}{P}  }{} 	&=  
				\begin{cases}
						\lvlin{x}{y}{z}{\sttoltp{P\substj{z}{x}}{}} & \text{If } x: \lin \wn T.S \\
						\lvlin{x}{y}{z}{\sttoltp{P}{}} & \text{If } x: \recur{\wn T}\\
				\end{cases}  \\
				\sttoltp{   \vasin{\un}{x}{y}{P}  }{} 	&=  \lvlserv{x}{y}{z}{\sttoltp{P}{}}  \\
				\sttoltp{ \vasout{x}{y}{P}  }{} 	&= 
				\begin{cases}
						\res{z} \lvlout{x}{y}{z}{\sttoltp{P\substj{z}{x}}{}} & \text{If }x: \lin \oc T.S \\
						\lvlout{x}{y}{z}{\sttoltp{P}{}} & \text{If }x: \recur{\oc T} \\
				\end{cases}
	\end{aligned}
	\]
		\hrule
		\smallskip
		\[
	\begin{aligned}
		\sttoltt{ \Gamma }{l}{} 	&= \sttoltt{ x_1 : T_1}{l}{} , \cdots , \sttoltt{x_n : T_n }{l}{}\\
		\sttoltt{x : T}{l}{}		&= x : \sttoltt{ T }{l}{x}\\
		\sttoltt{\nilT}{l}{x} &=  \lvlnilT \\
		\sttoltt{ \lin\  \vasreci{T}{S} }{l}{x}	&= \lvlint{\levelof{x}}{\sttoltt{T}{l}{\alpha}}{\sttoltt{S}{l}{x}}\\
		\sttoltt{ \lin\  \vassend{T}{S} }{l}{x}	&= \lvloutt{\levelof{x}}{\sttoltt{T}{l}{\beta}}{\sttoltt{S}{l}{x}}\\
		\sttoltt{ \recur{\wn T} }{x}{l}	&= \lvlunst{\levelof{x}}{\sttoltt{T}{l}{\gamma}}\\
		\sttoltt{ \recur{\oc T} }{x}{l}	&=\lvlunct{\levelof{x}}{\sttoltt{T}{l}{\gamma}}\\
	\end{aligned}
	\]
\caption{From \vasco to \pilvl  (\Cref{ch5d:encvs,ch5d:encvst})}
\label{ch5f:lvltypeencod}
\end{figure}

\begin{definition}{Translating Types/Contexts}
\label{ch5d:encvst}
	The translation  $\sttoltt{-}{l}{}$ of session types and contexts is given in \Cref{ch5f:lvltypeencod} (bottom). 
	The translation of  contexts
is parametric on a level function~$l$.  In particular, the translation of a type assignment $\sttoltt{x:T}{l}{}$, relies on an auxiliary translation $x:\sttoltt{T}{l}{x}$, which is deemed to be assigned a level $l(x)$ in the translated type $\sttoltt{T}{l}{x}$, depending on the shape of $T$. Other names, denoted $\alpha, \beta,\gamma\ldots$, are necessary when translating within types.
\end{definition}

The translation $\sttoltt{-}{l}{}$ follows  the continuation-passing approach of~\cite{DGS12} to encode session types into link types. 
The translation of tail-recursive types is rather direct, and self-explanatory.


By combining the translations of types and processes in \Cref{ch5f:lvltypeencod} we obtain a translation of type judgements / derivations in $\vasco$ into type judgements / derivations in $\pilvl$. 
We use an auxiliary notation:
%


%

\begin{definition}{Translating  Judgements/Derivations}\label{ch5def:firsttransl_judgments}
The translation of a type judgment for $\vasco$ into a type judgment for $\pilvl$ is parametric on the level function  $l:\mathcal{N} \to \mathbb{N}$, and is defined as: 
\begin{align*}
\sttoltj{\Gamma \st P}{l}{}& =\sttoltt{\Gamma}{l}{} \lt \sttoltp{P}{}
\\
\sttoltj{\Gamma, x:T \st x:T}{l}{} &= \sttoltt{\Gamma}{l}{}, x:\sttoltt{T}{l}{x}  \lt x:\sttoltt{T}{l}{x} 
	\end{align*}
	This translation induces  an inductive construction of the translation  of {type  derivations} in $\pilvl$ from type derivations in $\vasco$, denoted as:
\begin{mathpar}
	\sttoltj{{\scriptsize \vastype{Rule}}~\inferrule{\Upsilon_i \\ \forall i \in I}{\Gamma \st P}}{l}{} ~=~
		{\scriptsize\lvltype{Rule}}~\inferrule{\sttoltj{\Upsilon_i}{l}{}  \\ \forall i \in I}{ \sttoltt{\Gamma}{l}{} \lt \sttoltp{P}{}}
\end{mathpar} 
where $\Upsilon_i$ denotes a set of derivations used to prove $\Gamma \st P$. 

The translation, of which  \Cref{ch5fig:encodWtoSi} gives an excerpt, relies on analyzing the last rule~\vastype{Rule} applied in the derivation $\Gamma \st P$ and the unfolding of the translation of judgements, mapping to a derivation $ \sttoltt{\Gamma}{l}{} \lt \sttoltp{P}{}$ in  $\pilvl$,  in which the last rule applied is~\lvltype{rule}.
\end{definition}

%

\begin{landscape}
\begin{figure*}
\smaller
$$
\begin{array}{c}
\hline
\textbf{Case: Input}\\
\hline
\begin{array}{lcl}
&&\\
\mprset{flushleft} 
\sttoltj{
\inferrule[\vastype{Lin-In_1} ]
{\Gamma_1' \st x: \lin \vasreci{T}{S}   \\
\Gamma_2 , x: S , y : T \st P } 
{ \Gamma_1'  \contcomp \Gamma_2 \st  \vasin{\lin}{x}{y}{P}  }
}{l}{} & = &
\mprset{flushleft}
\inferrule*[left= \lvltype{Lin-In_1}] {
					\sttoltt{\Gamma_1'}{l}{}  \lt x: \lvlint{l(x)}{\sttoltt{T}{l}{y}}{\sttoltt{S}{l}{z}} \\\\
					\sttoltt{\Gamma_2}{l}{} ,   y:\sttoltt{T}{l}{y} , z:\sttoltt{S}{l}{z}  \lt \sttoltp{P\substj{z}{x}}{}  \\ l(x) = l(z) }
				{
					\sttoltt{\Gamma_1'}{l}{} 	\contcomp
					\sttoltt{\Gamma_2}{l}{}
					\lt  x \inp {y , z}.\sttoltp{P\substj{z}{x}}{}
				}\\[.3cm] 
\text{where }\Gamma_1'=\Gamma_1,  x : \lin \vasreci{T}{S}&& 	\text{where } \sttoltt{\Gamma_1'}{l}{}=\sttoltt{\Gamma_1}{l}{} , x: \lvlint{l(x)}{\sttoltt{T}{l}{y}}{\sttoltt{S}{l}{z}}	\\[.2cm]
\hdashline
\mprset{flushleft}
&&\\
\sttoltj{
    \inferrule[\vastype{Lin-In_2}]
     {  \Gamma_1' \st x: \recur{\wn T}  \\
       \Gamma_2', y : T \st P
		}
	{  \Gamma_1' \contcomp \Gamma_2'  \st  \vasin{\lin}{x}{y}{P} }
				}{l}{}  &=&
				\mprset{flushleft}
\mprset{flushleft}
\inferrule*[left=\lvltype{Lin-In_2}]
{
\sttoltt{\Gamma_1'}{l}{}  \lt x: \lvlunst{l(x)}{\sttoltt{T}{l}{y}} \\
\sttoltt{\Gamma_2'}{l}{} ,  z:\lvlnilT  \lt \sttoltp{P}{}  }
{  \sttoltt{\Gamma_1'}{l}{} 
\contcomp
\sttoltt{\Gamma_2'}{l}{}  
					\lt x\inp {y , z}.\sttoltp{P}{} 
	}\\[8mm]
\text{ where }\Gamma_1'= \Gamma_1 , x: \recur{\wn T} \text{ and }	\Gamma_2'= \Gamma_2 , x: \recur{\wn T} &&  \text{where }\sttoltt{ \Gamma_1'}{l}{}= \sttoltt{\Gamma_1}{l}{} , x: \lvlunst{l(x)}{\sttoltt{T}{l}{y}} \text{ and } \sttoltt{ \Gamma_2'}{l}{}= \sttoltt{\Gamma_2}{l}{} , x: \lvlunst{l(x)}{\sttoltt{T}{l}{y}}\\[.2cm]
\hdashline
&&\\
\sttoltj{
\mprset{flushleft}
\inferrule[\vastype{Un-In}]
{ 
\Gamma' \st x: \recur{\wn T} \\
\Gamma' , y : T \st P }
{  \Gamma, x: \recur{\wn T} \st   \vasin{\un}{x}{y}{P}  }}{l}{}
			& = &
\mprset{flushleft}
\inferrule*[left= \lvltype{Un-In_1}]
{
\sttoltt{\Gamma'}{l}{} \lt x:\lvlunst{l(x)}{\sttoltt{T}{l}{y}} \\\\
\sttoltt{\Gamma'}{l}{} , y:\sttoltt{T}{l}{y},  z : \lvlnilT   \lt \sttoltp{P}{}  \\
 \forall b \in \outsubj{\sttoltp{P}{}}, \ \levelof{b} < \levelof{x}
}
{
\sttoltt{\Gamma}{l}{} ,  x:\lvlunst{l(x)}{\sttoltt{T}{l}{y}}  \lt \bang x \inp {y , z}.\sttoltp{P}{} 
}\\[5mm]
\text{ where } \Gamma'=\Gamma, x: \recur{\wn T}&&\text{ where }\sttoltt{\Gamma'}{l}{} =\sttoltt{\Gamma}{l}{} , x: \rlevel{l(x)}{\sttoltt{T}{l}{y}} \\[2mm]
\end{array}\\
\end{array}
$$
\end{figure*}
\begin{figure*}\ContinuedFloat
\smaller
$$
\hspace{-2cm}
\begin{array}{c}
\hline
\textbf{Case: Output}\\  
\hline
\\
\begin{array}{lcl}
\sttoltj{
\mprset{flushleft}
\inferrule[\vastype{Lin-Out} ]
{
\Gamma_1' \st x: \lin \vassend{T}{S} \\
\Gamma_2' \st y: T \\
\Gamma_3, x:S \st P
}
{ \Gamma_1' \contcomp	\Gamma_2'\contcomp \Gamma_3 \st   \vasout{x}{y}{P}} 
}{l}{}
&=&
\mprset{flushleft}
\inferrule*[left=\lvltype{Lin-Out}]
{\inferrule{
\sttoltt{\Gamma_1'}{l}{} \lt x: \lvloutt{l(x)}{\sttoltt{T}{l}{y}}{\sttoltt{S}{l}{z}}\\\\
\sttoltt{\Gamma_2'}{l}{} \lt  y:\sttoltt{T}{l}{y} \\ 
\sttoltt{\Gamma_3}{l}{} , z : \sttoltt{S}{l}{z} \lt \sttoltp{P\substj{z}{x}}{}    \\ \levelof{x} = \levelof{z}
}
{
\sttoltt{(\Gamma_1'}{l}{}\contcomp \sttoltt{\Gamma_2'}{l}{} \contcomp \sttoltt{\Gamma_3}{l}{}), 
\lvlsplit{z}{\sttoltt{S}{l}{z}}  \lt \ov x \out { y,z }.(\sttoltp{P\substj{z}{x}}{}) 
}
}
{\sttoltt{\Gamma_1'}{l}{}\contcomp \sttoltt{\Gamma_2'}{l}{} \contcomp \sttoltt{\Gamma_3}{l}{} 
					 \lt \res{z} \ov x \out { y,z }.(\sttoltp{P\substj{z}{x}}{})
				}\\[7mm]
\text{ where } \Gamma_1'=\Gamma_1, x: \lin \vassend{T}{S} \text{ and } \Gamma'_2=\Gamma_2, y: T&&\text{where }\sttoltt{\Gamma_1'}{l}{}=  \sttoltt{\Gamma_1}{l}{}, x: \lvloutt{l(x)}{\sttoltt{T}{l}{y}}{\sttoltt{S}{l}{z}} \text{ and } \sttoltt{\Gamma_2'}{l}{}= \sttoltt{\Gamma_2}{l}{}, y:\sttoltt{T}{l}{y}\\[2mm] 
\hdashline
&&\\
\sttoltj{
\mprset{flushleft}
\inferrule[\vastype{Un-Out}]
{
 \Gamma_1' \st x: \recur{\oc T}	\\
 \Gamma_2' \st y: T\\
\Gamma_3' \st P
}
{  \Gamma_1'\contcomp	\Gamma_2' \contcomp \Gamma_3' \st   \vasout{x}{y}{P}} 
				}{l}{} 
			&=&
\mprset{flushleft}
\inferrule*[left=\lvltype{Un-Out_1}]
{
\sttoltt{\Gamma_1'}{l}{} \lt x: \rlevel{l(x)}{\sttoltt{T}{l}{y}}\\
\sttoltt{\Gamma_2'}{l}{} \lt y:\sttoltt{T}{l}{y}\\\\ 
\sttoltt{\Gamma_3'}{l}{}, z: \lvlnilT\lt \sttoltp{P}{} }
{
 \sttoltt{\Gamma_1'}{l}{} \contcomp	\sttoltt{\Gamma_2'}{l}{} 
					\contcomp \sttoltt{\Gamma_3'}{l}{}
					\lt \ov x \out { y,z }.\sttoltp{P}{} 
				}\\[7mm]
\text{ where } \Gamma_1'= \Gamma_1 , x: \recur{\oc T}, \  \Gamma_2'= \Gamma_2 , x: \recur{\oc T}, y: T  && \text{where } \sttoltt{\Gamma_1'}{l}{} =\sttoltt{\Gamma_1}{l}{} , x: \rlevel{l(x)}{\sttoltt{T}{l}{y}}, \  \sttoltt{\Gamma_2'}{l}{}=\sttoltt{\Gamma_2}{l}{}, x: \rlevel{l(x)}{\sttoltt{T}{l}{y}} , y:\sttoltt{T}{l}{y}\\
\qquad \quad \Gamma_3'= \Gamma_3 , x: \recur{\oc T} &&  \qquad \quad \sttoltt{\Gamma_3'}{l}{} =\sttoltt{\Gamma_3}{l}{} , x: \rlevel{l(x)}{\sttoltt{T}{l}{y}}\\[2mm]
\end{array}\\
\hline
\end{array}
$$
\caption{From derivations in \vasco to derivations in \pilvl  (excerpt, cf.~\Cref{ch5def:firsttransl_judgments})}\label{ch5fig:encodWtoSi}
\jspace
\end{figure*}
\end{landscape}

%

\subsection{Results}
In general, the translation of a $P \in \vaslang$ is not necessarily typable in $\pilvl$; this occurs when, e.g., $P$ is non-terminating.  
We focus on processes in $\vaslang$ that are typable in $\pilvl$, and therefore, are terminating.

\begin{notation}
	We write $\sttoltt{\Gamma}{l}{} \lt \sttoltp{P}{}$ if $\sttoltj{\Gamma \st P}{l}{}$ holds, for some~$l$.
	
\end{notation}

Our translations are correct, in the following sense:

\begin{theorem}[Operational Completeness]\label{ch5thm:op_completeness}
	Let $ P \in \lvllang$ such that  $\sttoltt{\Gamma}{l}{} \lt \sttoltp{P}{}$,  for some level function $l$. Then there exists $R \in \lvllang$ such that $P \vasred Q \implies\sttoltp{P}{}{} \lts{ \tau } \sttoltp{R}{}{}$  and $R \equiv Q$.
\end{theorem}

\begin{theorem}[Operational Soundness]\label{ch5thm:op_soundness}
	Let $P \in \lvllang$ with  $\sttoltt{\Gamma}{l}{} \lt \sttoltp{P}{}$, for some level function $l$. If  $\sttoltp{P}{}{} \lts{ \tau } U$ Then there exists $R,Q \in \lvllang$ such that $P \vasred Q \land R \equiv Q \land U = \sttoltp{R}{}{} $. 
\end{theorem}
An immediate corollary of \Cref{ch5thm:op_completeness} is that our translation preserves (non-)terminating behaviour, i.e.,  does not map non-terminating processes in $\vaslang$ into terminating processes in $\pilvl$.

\begin{corollary}
$\sttoltp{\cdot}{}$ preserves (non-)terminating behaviour.
\end{corollary}

The following result corroborates our informal intuitions about $\vaslang$ and $\lvllang$. It also precisely characterizes a class of terminating processes based on our correct translations  $\sttoltp{\cdot}{}$ and $\sttoltt{\cdot}{l}{}$.
\begin{theorem}
\label{ch5t:wsubsets}
$\lvllang \subset \vaslang$.
\end{theorem}
\begin{proof}[Proof (Sketch)]
The inclusion $\lvllang \subseteq \vaslang$ is immediate by definition. 
To prove that the inclusion is strict, we consider a counterexample, i.e., a process $P$ typable in $\vasco$ but not typable in $\pilvl$.
Process $P_{\ref{ch5ex:infinite}}$ from \Cref{ch5ex:infinite} suffices for this purpose.
\end{proof}

\section{Propositions as Sessions}
\label{ch5s:pas}
We now introduce $\pidill$, the process model induced by the Curry-Howard correspondence between linear types and session types  (\emph{propositions-as-sessions})~\cite{CairesP10}.
\pidill is a synchronous $\pi$-calculus extended with (binary) guarded choice and forwarding.

\begin{definition}{Processes and Types}
	Processes  in $\pidill$ are given by the grammar in \Cref{ch5f:dillsyntandtype} (top). 	
	Types coincide with linear logic propositions, as given in the grammar in \Cref{ch5f:dillsyntandtype} (bottom). 
\end{definition}


\begin{definition}{Reduction in $\pidill$}
The reduction semantics of \pidill is defined in \Cref{ch5f:dillcongred} (bottom), relying on structural congruence, the least congruence relation defined in \Cref{ch5f:dillcongred} (top).  
\end{definition}

\begin{figure}[t!]
	\[
	\begin{aligned}
		P,Q ::= &								&& \mbox{(Processes)}\\
				&  \dillout{x}{y}{P}		&& \mbox{(output)}
				&&  \dillin{x}{y}{P}					&& \mbox{(linear input)}\\
				&  \dillserv{x}{y}{P}			&& \mbox{(server)}
				&& P \pp Q  	    		  		&& \mbox{(composition)}\\
				& \res {x}P						&& \mbox{(restriction)}
				&& \nil	     					&& \mbox{(inaction)}\\
				& \dillchoice{x}{P}{Q}			&& \mbox{(branching)}
				&& \dillseler{x}{P}				&& \mbox{(select right)}\\
				& \dillselel{x}{P}				&& \mbox{(select left)}
				&& \dillforward{x}{y}			&& \mbox{(forwarding)}
	\end{aligned}
	\]
	\hrule
	\smallskip
	\[
	\begin{aligned}
		A,B ::= &								&& \mbox{(Types)}
		\\
				& \dillnilT						&& \mbox{(Termination)}  
				&& \dillunt{A}					&& \mbox{(Shared)}
				\\
				& \dillint{A}{B}				&& \mbox{(Receive)} 
				&& \dilloutt{A}{B}				&& \mbox{(Send)} 
				\\
				& \dillchoicet{A}{B}			&& \mbox{(Selection)} 
				&& \dillselet{A}{B}				&& \mbox{(Branching)} 
	\end{aligned}
	\]
	\caption{Processes and types of the session $\pi$-calculus $\pidill$}
	\label{ch5f:dillsyntandtype}
\end{figure}

\begin{figure}[t!]
		\[
		\begin{aligned}
			&\begin{aligned}
				P \pp \nil &\equiv P 
				&
				P \equiv_\alpha Q &\implies P \equiv Q
				&
				P \pp Q &\equiv Q \pp P
				\\
				\res{x} \nil &\equiv \nil
				&
				 P \pp (Q \pp R) &\equiv (P \pp Q) \pp R
				 &\!\!\res{x} \res{y} P &\equiv \res{y} \res{x} P
			\end{aligned}\\
			&\begin{aligned}
				x \not \in \fn{P} & \implies P \pp \res{x}Q \equiv \res{x} (P \pp Q)
				&
			\end{aligned}
		\end{aligned}
		\]
		\hrule
		\smallskip
		\[
			\begin{array}{rll}  
			\Did{R$\leftrightarrow$} & 
				\res{x} (\dillforward{x}{y}	\pp P)
				\dillred 
				 P\substj{y}{x} \quad \text{if } x \not = y
				\\[1mm] 
			\Did{RC} & 
				\dillout{x}{y}{P} \pp \dillin{x}{z}{Q}	
				\dillred 
				P \pp Q\substj{y}{z}
				\\[1mm] 
			\Did{R!} & 
				\dillin{x}{y}{P} \pp \dillserv{x}{z}{Q}
				\dillred
				P \pp \vaspara{Q\substj{y}{z}}  \dillserv{x}{z}{Q}
				\\[1mm]  
			\Did{RL}&
				\dillselel{x}{P} \pp  \dillchoice{x}{Q}{R}
				\dillred
				P \pp Q
			\\[1mm]
			\Did{RR}&
				\dillseler{x}{P}  \pp  \dillchoice{x}{Q}{R}\
				\dillred
				P \pp R
			\\[1mm]
			\Did{R$\pp$}&
				Q \dillred R 
				\implies 
				P \pp Q \dillred P \pp R
			\\[1mm]
			\Did{R$\nu$}&
				P \dillred Q 
				\implies 
				\res{x} P \dillred \res{x} Q
			\\[1mm]
			\Did{R$\equiv$}&
			P \equiv P' \land P' \dillred Q' \land Q' \equiv Q 
			\implies
			P \dillred Q
			\\[1mm]
			\end{array}
		\]
	\caption{Structural congruence and reductions for $\pidill$}
	\label{ch5f:dillcongred}
\end{figure}

\begin{notation}[Process Abbreviations]
	We adopt the following abbreviations for bound outputs and replicated forwarders:
	\begin{align*}
	\dillbout{x}{z}{P} &=  \res{z}\dillout{x}{z}{P}
	\\
		\dillfwdbang{x}{y}  & = \dillserv{y}{z}{
		 \dillbout{x}{k}{	\dillforward{k}{z}}}
	\end{align*}

\end{notation}

As usual, a type environment is a collection of type assignments $x : A $ where $x$ is a name and $A$ a type, the names being pairwise disjoint.  
The empty environment is denoted `$\cdot$'.
We consider  {\em unrestricted} environments (denoted $\Gamma,\Gamma'$) and  {\em linear} environments  (denoted as $\Delta,\Delta'$); while the former satisfy weakening and contraction, the latter do not.  


We denote by $dom(\Gamma)$, the {\em domain of } $\Gamma$, the set of names whose type assignments are in $\Gamma$, i.e., $dom(\Gamma)=\{x \mid x:A \in \Gamma\}$.  Also,  $\Gamma(x)$ denotes the type of the name $x\in dom(\Gamma)$, i.e., $\Gamma(x)=A$, if $x:A\in \Gamma$. The domain of $\Delta$  and $\Delta(x)$ are similarly defined.

Typing judgments for $\pidill$ are of the form $\Gamma ; \Delta    \dill P \ :: x:A$.
Such a judgment is intuitively read as: ``$P$ provides protocol $A$ along $x$ by using the protocols described in the assignments in $\Gamma$ and $\Delta$''. The domains of  $\Gamma$, $\Delta$ and $x:A$ are pairwise disjoint.
	The corresponding type
	rules  are given  in \Cref{ch5f:dilltyperule}.
	Each logical operator is represented by right and left rules: the former explains how to \emph{offer} a behavior (according to the operator's interpretation, cf. \Cref{ch5f:dillsyntandtype} (bottom)); the latter explains how to \emph{make use} of a behavior typed with the operator.
In particular, the behavior of clients and servers is governed by four typing rules:
\dilltype{cut^!},
\dilltype{copy},
 \dilltype{\dillunt{} L}, and
 \dilltype{\dillunt{} R}.

The Curry-Howard correspondence connects the logical principle of cut elimination with process synchronization. As a result, we have  the fundamental property ensured by typing:

\begin{theorem}[Type Preservation]
If $\Gamma ; \Delta    \dill P \ :: x:A$
and $P \dillred Q$
then
$\Gamma ; \Delta    \dill Q \ :: x:A$.
\end{theorem}
		
The type system enforces also progress and termination. The latter property can be proven using logical relations~\cite{DBLP:journals/iandc/PerezCPT14}.

\begin{figure*}[t!]
	\begin{mathpar}
		\inferrule* [ left = \dilltype{\dillnilT L} ]{  \Gamma ; \Delta \dill P :: T }
		{ \Gamma ; \Delta , x : \dillnilT  \dill P :: T }
		\and
		\inferrule* [ left = \dilltype{\dillnilT R} ]{   }
		{ \Gamma ; \cdot  \dill \nil ::   x : \dillnilT }
		\and
		\inferrule* [ left = \dilltype{fwd} ]{   }
		{ \Gamma ; x:A  \dill \dillforward{x}{y} ::   y : A }
		\and
		\inferrule* [ left = \dilltype{\dilloutt{}{} L} ]{ \Gamma ; \Delta , y:A , x:B \dill P :: T  }
		{ \Gamma ; \Delta , x :\dilloutt{A}{B} \dill \dillin{x}{y}{P} :: T  }
		\and
		\inferrule* [ left = \dilltype{\dilloutt{}{} R} ]{ \Gamma ; \Delta_1 \dill P :: y : A \\ \Gamma ; \Delta_2 \dill Q :: x : B  }
		{ \Gamma ; \Delta_1, \Delta_2  \dill \dillbout{x}{y}{(P \pp Q)} ::   x : \dilloutt{A}{B} }
		\and
		%
		%
		\inferrule* [ left = \dilltype{cut} ]{ \Gamma ; \Delta_1 \dill P :: x: A \\ \Gamma ; \Delta_2 , x:A \dill Q ::T }
		{ \Gamma ; \Delta_1 ,  \Delta_2  \dill \res{x}(P \pp Q) ::  T }
		\and
		\inferrule* [ left = \dilltype{cut^!} ]{  \Gamma ; \cdot \dill P :: y: A \\ \Gamma , u:A ; \Delta  \dill Q ::T  }
		{ \Gamma ;   \Delta  \dill \res{u}(\dillserv{u}{y}{P} \pp Q) ::  T }
		\and
		\inferrule* [ left = \dilltype{copy} ]{ \Gamma , u:A ; \Delta , y:A \dill P  :: T  }
		{ \Gamma , u:A ; \Delta  \dill  \dillbout{u}{y}{P} ::  T }
		\and
		\inferrule* [ left = \dilltype{\dillunt{} L} ]{ \Gamma , u:A ; \Delta \dill P \substj{u}{x} :: T  }
		{ \Gamma ; \Delta , x: \dillunt{A} \dill P :: T }
		\and
		\inferrule* [ left = \dilltype{\dillunt{} R} ]{ \Gamma ; \cdot \dill Q :: y:A }
		{ \Gamma ; \cdot  \dill \dillserv{x}{y}{Q} ::   x : \dillunt{A} }
		\and
		\inferrule*[left=\dilltype{\dillselet{}{} L_1} ]{ \Gamma ; \Delta , x:A \dill 
		P :: T  }
		{ \Gamma ; \Delta , x:\dillselet{A}{B}  \dill \    \dillselel{x}{P} :: T }
		\and
		\inferrule* [ left = \dilltype{\dillchoicet{}{} L} ]{ \Gamma ; \Delta , x:A \dill P :: T \quad \Gamma ; \Delta , x:B \dill P :: T }
		{ \Gamma ; \Delta , x:\dillchoicet{A}{B} \dill   \dillchoice{x}{P}{Q} :: T }
		\and
		%
		%
		\inferrule*[left=\dilltype{\dillchoicet{}{} R_2} ]{ \Gamma ; \Delta \dill 
		P ::  x:B  }
		{ \Gamma ; \Delta    \dill \    \dillseler{x}{P} :: x:\dillchoicet{A}{B} }
	\end{mathpar}
	\caption{Type rules for $\pidill$ (selection)}
	\label{ch5f:dilltyperule}
\end{figure*}

\section{\texorpdfstring{\dilllang}{??}: A Class of Terminating Processes}\label{ch5s:dilllang}
We now study \dilllang, another class of terminating \vasco processes.
This class is induced by the Curry-Howard system given in \Cref{ch5s:pas}, which leverages  
translations on processes and types/contexts, denoted $\sttodillp{\cdot}{}$ and $\sttodillt{\cdot}$, respectively.
Roughly, \dilllang is defined as follows:
	$$ \dilllang = \{ P \in \vasco \mid  \ \Gamma \vaslinunsplit \Delta \st P ~\land~ 
	 	 {\sttodillt{\Gamma}};   \sttodillt{\Delta}  \dill \sttodillp{{P}}  :: u: \sttodillt{\dual{S}} 	\} 
	$$
\Cref{ch5d:dilllang} will give a formal definition.
In the following we define the translations $\sttodillp{\cdot}{}$ and $\sttodillt{\cdot}$, and establish their properties. 
Our main result is that 
$\dilllang \subset \lvllang$
but 
$\lvllang \not \subset \dilllang$
(\Cref{ch5thm:wnotinl,ch5thm:main_result}): there are terminating processes detected as such by the weight-based approach but not by the Curry-Howard correspondence.

We require some auxiliary definitions.
The following predicates say whether a session type contains client or server behaviors.

\begin{definition}{}\label{ch5d:predsts}
Given a session type $T$, we define predicates $\server{{T}}$ and $\client{{T}}$ as follows:
	\[
	\begin{array}{rl@{\hspace{2cm}}rl}
		\server{\recur{\wn T}} &  = \ttrue 
		&
		\client{\recur{\oc T}} &  = \ttrue 
		\\
		\server{\nilT} &  = \ffalse 
		&
		\client{\nilT} &  = \ffalse 
		\\
		\server{q \ \vassend{S}{T}} &  = \server{T}
		&
		\client{q \ \vassend{S}{T}} &  = \client{T}
		\\
		\server{q \ \vasreci{S}{T}} &  = \server{T} 
		&
		\client{q \ \vasreci{S}{T}} &  = \client{T}
		\\
		\server{\recur{\oc T}} &  =  \server{T} 
		&
		\client{\recur{\wn T}} &  = \client{T}
	\end{array}
\]
These predicates extend to contexts $\Gamma$ as expected. This way, e.g., $\server{\Gamma}$ stands for $\bigwedge_{x\in dom(\Gamma)} \Gamma(x)$.
Also, we write $\server{\Gamma;P}$ to stand for $ \bigwedge_{x \in (\fn{P} \cap dom(\Gamma) )} \server{\Gamma(x)}$, returning $\ttrue$ when $(\fn{P} \cap dom(\Gamma)) = \emptyset$.
Analogous definitions for $\client{\cdot}$ , $\notserver{\cdot} $, and $ \notclient{\cdot} $ arise similarly. 
\end{definition}


This way, intuitively:
\begin{itemize}
	\item $\notserver{T}\land \notclient{T}$ means that $T$ is an always-linear behavior, i.e., it does not contain server and client actions.
	\item $\server{T} \land \notclient{T}$ means that $T$ contains some server behavior and that it does not contain client behaviors.
	\item $\notserver{T}\land \client{T}$ means that $T$ will at some point exhibit client behaviors and that it does not contain server behaviors.
\end{itemize}
Also, $\server{T} \land \client{T}$ means that $T$ contains both server and client actions; this combination, however, is excluded by typing. 
 
\begin{example}
	We further illustrate \Cref{ch5d:predsts} by example:
	\begin{center}
		\begin{tabular}{ cc|c|c } 
		 &$T$ & $\server{T}$ & $\client{T}$  \\
		 \hline 
		 1& $\recur{ \oc ( \lin \  \vassend{(\recur{\oc S})}{(\lin \ \vasreci{(\recur{\wn R})}{(\recur{\wn T_0})})})}$ & \ttrue & \ttrue   \\
		 2& $\lin \ \vassend{(\recur{\oc S})}{(\lin \ \vasreci{(\recur{\wn R})}{(\recur{\wn T_0})})}$ & \ttrue  & \ffalse  \\
		 3& $\recur{\oc (\lin \ \vassend{(\recur{\oc S})}{(\lin \ \vasreci{(\recur{\wn R})}{\nilT})})}$& \ffalse & \ttrue    \\
		 4& $\lin \ \vassend{(\recur{\oc S})}{(\lin \ \vasreci{(\recur{\wn R})}{\nilT})}$ & \ffalse & \ffalse \\
		\end{tabular}
	\end{center} 
	Both $(1)$ and $(2)$ return \ttrue  for $\server{T}$ because of their final behavior (i.e., `$\recur{\wn T_0}$'), whereas $(3)$ and $(4)$ return \ffalse, because their final  behavior is $\nilT$. 
	Both $(1)$ and $(3)$ return \ttrue for $\client{T}$ as their initial type behavior (i.e., `$\recur{\oc T'}$') is that of a client, whereas $(2)$ and~$(4)$ return \ffalse as they do not contain any client behavior.
\end{example}

\subsection{The Typed Translation}

\begin{definition}{Translating Processes}
The translation $\sttodillp{\cdot}: \vasco \to \pidill$ is given in~\Cref{ch5f:encproc}. 
\end{definition}

\begin{figure*}[!t]
	\small
	\[
	\begin{aligned}
		&\sttodillp{\vasout{x}{y}{P}}  = 
			\begin{cases}
				 \dillbout{x}{z}{( \dillforward{y}{z}  \pp  \sttodillp{{P}}   )} 
					& \text{If $ x: \lin \oc (T).S \land \neg \contun{T}\land \notserver{T}$.}
					\\
				 \dillbout{x}{z}{( \dillfwdbang{y}{z}  \pp  \sttodillp{{P}}   )}
					& \text{If $ x: \lin \oc (T).S \land \contun{T} \land \notserver{T}$.}
					\\
				\dillbout{x}{z}{ 
					\dillbout{z}{w}{
						( \dillforward{y}{w}  \pp  \sttodillp{{P}}   )
					}
				} 
					&
				  \text{If $ x: \recur{\oc T} \land \neg \contun{T} \land  \notserver{T} \land \client{T}$}
					\\
				\dillbout{x}{z}{ 
					\dillbout{z}{w}{
						( \dillfwdbang{y}{w}  \pp  \sttodillp{{P}}   )
					}
				}
					&
			 \text{If $ x: \recur{\oc T} \land \contun{T} \land \notserver{T} \land \client{T}$}
			 \\
			 				\dillbout{x}{z}{ \dillselel{z}{
					\dillbout{z}{w}{
						( \dillfwdbang{y}{w}  \pp  \sttodillp{{P}}   )
					}
				}
				}
					&
					  \text{If $ x: \recur{\oc T} \land \contun{T}\land \notserver{T} \land \notclient{T} $}
			\\
			 				\dillbout{x}{z}{ \dillselel{z}{
					\dillbout{z}{w}{
						( \dillforward{y}{w}  \pp  \sttodillp{{P}}   )
					}
				}
				}
					&
					  \text{If $ x: \recur{\oc T} \land \neg \contun{T}\land \notserver{T} \land \notclient{T} $}
			\end{cases}\\
		&\sttodillp{\res{xy}\vasout{z}{y}{P}  }  = 
			\begin{cases}
				\dillbout{z}{x}{\dillseler{x}{\sttodillp{P}}}  & \text{If $z:\recur{\oc T} \land \notserver{T} \land \notclient{T} $.}\\
				\dillbout{z}{x}{\sttodillp{P}} & \text{If $z:\recur{\oc T} \land \server{T} \land \notclient{T} $.}
			\end{cases}
		\\
		&\sttodillp{\res{xy}\vasout{z}{x}{(P \pp Q)}}  = 
			\dillbout{z}{y}{(\sttodillp{P} \pp \sttodillp{Q})}
			\quad \text{If $z:  \lin \vassend{T}{S} \wedge z \not \in \fn{P} \land y \not \in \fn{Q}$}
		\\
		&\sttodillp{\vasin{\lin}{x}{y}{P}}  = 
				\dillin{x}{y}{\sttodillp{P}} \quad \text{If $x: \lin \vasreci{T}{S}$}
		\\
		&\sttodillp{\vasin{\un}{x}{{y}}{P}}  = 
			\begin{cases}
				\dillserv{x}{z}{\sttodillp{P\substj{z}{y}}} 
				& \text{If $ x: \recur{\wn T} \land\server{T}\land \notclient{T} $.} \\
				\dillserv{x}{z}{ \dillin{z}{y}{ \sttodillp{P} }
				} 
				& \text{If $ x: \recur{\wn T} \land\notserver{T}\land \client{T} $.} \\
				\dillserv{x}{z}{  
				\dillchoice{z}{	\dillin{z}{y}{ \sttodillp{P} }	}{	\sttodillp{P\substj{z}{y}} }
				} 
				& \text{If $ x: \recur{\wn T} \land\notserver{T}\land \notclient{T} $.}
			\end{cases}
		\\ 
		&	\sttodillp{\res{x y}(P \pp Q)}  = 
					\res{x}( \sttodillp{P} \pp \sttodillp{Q}\substj{x}{y}) \quad \text{If $ y \not \in \fn{P} \land x \not \in \fn{Q}$}
		\\
		&\sttodillp{P \pp Q}  = \res{w} ( \sttodillp{P} \pp \sttodillp{Q}   ) \quad \text{With $ w $ fresh}
		\\
		&\sttodillp{\nil}  = \nil 
	\end{aligned}
	\]
\caption{Translating processes in \vasco into $\pidill$ \label{ch5f:encproc}}
\end{figure*}

The translation of processes relies on type information; in particular, the translation of outputs and unrestricted inputs depends on whether the overall behavior of channels exhibits server or client behaviors (cf. \Cref{ch5d:predsts}). 
In translating  outputs, we check whether the output is free or bound.  
The translation of free outputs is further influenced by whether the sender is associated with a linear connection or acts as a client connected to a server. There are 5 cases to consider, and the translated processes are designed to preserve typability.
Similar conditions apply to the translation of bound outputs.

\begin{remark}
To ensure typability of the translated process, we explain some of the choices in~\Cref{ch5f:encproc}:
\begin{enumerate}
\item In a free output $\vasout{x}{z}{P}$ the value $z$   cannot have a server behavior. In~\Cref{ch5f:encproc}, this is ensured using the predicate $\notserver{T}$.
\item In an unrestricted bound output $\res{xy}\vasout{z}{y}{P} $,  the value $y$ cannot have a client behavior. In~\Cref{ch5f:encproc}, this is ensured using the predicate $\notclient{T}$. 
\end{enumerate}

We illustrate what we mean by ``client behavior'' above. Consider the process $P=\res{xy}( \res{wv}\vasout{x}{v}{\vasin{\un}{w}{a}{\nil}}\pp \vasin{\un}{y}{c}{\vasout{c}{b}{\nil}})$.
In $P$, the output action on $x$ is an unrestricted bound output, whose object $v$ has a client behavior: after one reduction, an output on $v$ will be ready to invoke the server on $w$. 
 Notice that $P \in \vaslang$, as $P$ is typable with $b:\nilT\st P$ and $x:\recur{\oc (\recur{\oc \nilT})}, y:\recur{\wn (\recur{\oc \nilT})}, w:\recur{\wn \nilT}$ and $ v:\recur{\oc \nilT}$. 


We want a typable  translation of the judgement $\sttodillj{\Gamma \st P}_u$. Consider the partial translation of $P$, i.e., $\sttodillp{P}=\res{x}(\sttodillp{P_1}\pp \sttodillp{Q_1}\substj{x}{y})$, 
 where we use the abbreviations
\begin{itemize}
\item  $P_1= \res{wv}\vasout{x}{v}{\vasin{\un}{w}{a}{\nil}}$, and 
\item  $Q_1= \vasin{\un}{y}{c}{\vasout{c}{b}{\nil}}$.
\end{itemize}
 Suppose we can apply \dilltype{cut}, then there are derivations $\Pi_1$ and $\Pi_2$ such that 
\begin{mathpar} 
\small
\inferrule
{
\inferrule{\Pi_1}{b:\dillnilT; \cdot \dill  \sttodillp{Q_1}\substj{x}{y}:: y: !( \dillint{\dillnilT}{\dillnilT})}\quad
\inferrule{\Pi_2}{b:\dillnilT; x:!( \dillint{\dillnilT}{\dillnilT})\dill \sttodillp{P_1} :: u:T}
}
{b:\dillnilT;\cdot \dill \res{x}( \sttodillp{P_1}\pp \sttodillp{Q_1}\substj{x}{y})::u:T}
\end{mathpar}

Consider the partial translation $\sttodillp{\res{wv}\vasout{x}{v}{\vasin{\un}{w}{a}{\nil}}}= \dillbout{x}{w}{(\dillserv{w}{a'}{P_1'})}$ for some $P_1'$ that we will leave opaque for now.
Notice, however, that the following derivation is not possible: to type $\dillserv{w}{a'}{P_1'}$ we would need $u=w$ to apply $\dilltype{\dillunt{}R}$ (above the application of $\dilltype{copy}$), but $w$ already occurs in the context and this contradicts the domain restriction of $\dill$ judgements.
\begin{mathpar}
\inferrule*[left=\dilltype{\dillunt{}L}]
{
\inferrule*[left=\dilltype{copy}]
{b:\dillnilT , x:( \dillint{\dillnilT}{\dillnilT});w:( \dillint{\dillnilT}{\dillnilT}) \not \dill !w(a).0::u:T}
{b:\dillnilT , x:( \dillint{\dillnilT}{\dillnilT});\cdot  \dill \dillserv{w}{a'}{P_1'}::u:T}}
{b:\dillnilT ; x:!( \dillint{\dillnilT}{\dillnilT})\dill \dillbout{x}{w}{(\dillserv{w}{a'}{P_1'})}::u:T}
\end{mathpar}
A similar argument and example can be used to justify the first item of this remark.
\end{remark}

While the translation of linear inputs is straightforward, in translating unrestricted inputs we check whether the synchronization concerns a bound or free output. 
When the unrestricted input cannot discern the client or server behavior from the type, it offers both behaviors using a branching construct; the synchronizing party (i.e. the translation of output, free or bound) then determines the desired behavior using a corresponding selection construct.

\begin{example}{} \label{ch5ex:trans_proc} Consider $P=\res{xy}(  \vasin{\un}{ x}{z}{\nil}\pp  \vasout{y}{w}{\nil})$, a $\vasco$ process that implements a simple  server-client communication.  As in \Cref{ch5ex:infinite}, one can verify that  
$x:\recur{\wn \nilT}, y:\recur{\oc \nilT}$, $w:\nilT$ and $z:\nilT$, which  entail $w:\nilT \st P$. Since $y\notin \fn{\vasin{\un}{x}{z}{\nil}}$ and $x\notin \fn{\vasout{y}{w}{\nil}}$, the translation of $P$ is as:
\[
\begin{aligned}
\sttodillp{P}= \res{x}(\sttodillp{\vasin{\un}{x}{z}{\nil}}\pp \sttodillp{\vasout{x}{w}{\nil}})
\end{aligned}
\]
Note that $\notclient{\nilT}\wedge \notserver{\nilT}\wedge \contun{\nilT}$ holds (cf.~\Cref{ch5def:pred_cont} and \Cref{ch5d:predsts}). Thus,
\[
\begin{aligned}
\sttodillp{\vasin{\un}{x}{z}{\nil}} &=\dillserv{x}{v}{  
				\dillchoice{v}{	\dillin{v}{z}{ \nil }	}{\nil }
				} \\
\sttodillp{\vasout{x}{w}{\nil}} &= \dillbout{x}{z}{ \dillselel{z}{
					\dillbout{z}{v}{
						( \dillfwdbang{w}{v}  \pp  \nil   )
					}
				}}
\end{aligned}
\]
\end{example}

\begin{definition}{}\label{ch5d:nobang}
Given a session type/linear logic proposition $A$, we write $\dillnotunsingular{{A}}$ to denote $A$ without 
 top-level occurrences of `\,$\bang$\,', i.e., $\dillnotunsingular{\dillunt{A}} =  A$ and is  the identity function otherwise.
\end{definition}

\begin{definition}{Translating Types/Context}
The translation $\sttodillt{\cdot }$ from session types in $\vasco$ to logic propositions in $\pidill$ is given in~\Cref{ch5f:enctype}. 
The translation of types extends to contexts as expected;
we shall write $\dillnotun{\sttodillt{\Gamma}}$ to stand for  $\dillnotunsingular{\sttodillt{\Gamma}}$.
\end{definition}

\begin{figure}[!t]
\[
	\begin{aligned}
		\sttodillt{\nilT} &  = \dillunt{\dillnilT} \\
		\sttodillt{\lin \vassend{S}{T}} &  =  \dillint{\sttodillt{S}}{\sttodillt{T}}\\
		\sttodillt{\lin \vasreci{S}{T}} &  = \dilloutt{\sttodillt{S}}{\sttodillt{T}} \\
		\sttodillt{\recur{\wn T}} &  = 
			\begin{cases}
				\dillunt{\sttodillt{T}} & \text{If $\server{T}\land \notclient{T} $.}\\
				\dillunt{(\dilloutt{\sttodillt{T}}{\dillnilT})} & \text{If $\notserver{T}\land \client{T} $.}\\
				\dillunt{(\dillchoicet{(\dilloutt{\sttodillt{T}}{\dillnilT})}{\sttodillt{T}})} & \text{If $\notserver{T}\land \notclient{T} $.}  
			\end{cases}\\
		\sttodillt{\recur{\oc T}} &  =  
			\begin{cases}
				\dillunt{\sttodillt{\dual{T}}} & \text{If $\server{T}\land \notclient{T} $.} \\
				\dillunt{({\dillint{\sttodillt{T}}{\dillnilT}})} & \text{If $\notserver{T}\land \client{T} $.}\\
				\dillunt{(\dillselet{(\dillint{\sttodillt{T}}{\dillnilT})}{\sttodillt{\dual{T}}})} & \text{If $\notserver{T}\land \notclient{T} $.}
			\end{cases}\\
	\end{aligned}
\]
\caption{Translating session types into logical propositions \label{ch5f:enctype}}
\end{figure}

The translation of $\nilT$ and linear input/output types is standard.
As for client and servers, the translation of types follows the translation of processes. 
When the type of the client or server exhibits a server behavior, the type is encoded into an unrestricted type.
Notice that a client type $\recur{\oc T}$ is translated into its dual behavior $\dillunt{\sttodillt{\dual{T}}} $, but a server is not. 
This has to do with the left/right interpretation of judgments in \pidill: servers always occur on the right-hand side; to provide a dual behavior, the client should itself be dual.

\begin{example}{Cont.~\Cref{ch5ex:trans_proc}}\label{ch5ex:transl_type}
Consider the type assignments 
$x:\recur{\wn \nilT}, y:\recur{\oc \nilT}$, $w:\nilT$ and $z:\nilT$. Since  $\notclient{\nilT}\wedge \notserver{\nilT}$, the translation in~\Cref{ch5f:enctype} gives: 
\begin{itemize} 
\item $x:\sttodillt{\recur{\wn\nilT}}=\dillunt{(\dillchoicet{(\dilloutt{\sttodillt{\nilT}}{\dillnilT})}{\sttodillt{
\nilT}})}=\dillunt{(\dillchoicet{(\dilloutt{\dillunt{\dillnilT}}{\dillnilT})}{\dillnilT})}$;
\item $y:\sttodillt{\recur{\oc \nilT}}=\dillunt{(\dillselet{(\dillint{\sttodillt{\nilT}}{\dillnilT})}{\sttodillt{\dual{\nilT}}})}=\dillunt{(\dillselet{(\dillint{\dillunt{\dillnilT}}{\dillnilT})}{~\dillnilT})}$
\end{itemize}
The translations to $z:\dillnilT$ and $w:\dillnilT$ are trivial.
\end{example}

Armed with the translations of processes and types given in~\Cref{ch5f:encproc} and  \Cref{ch5f:enctype}, 
we are now ready to translate a judgment $\Gamma, \Delta \st P$ into $\dillnotun{\sttodillt{\Gamma}{}{}} ;  \sttodillt{\Delta}{}{} \dill \sttodillp{P}{} :u:: A$, for some name $u$.
This translation requires that $\contun{\Gamma}$ and $\neg \contun{\Delta}$, i.e., $\Gamma$ is unrestricted and $\Delta$ is `not' unrestricted; this is the abbreviation $\Gamma \vaslinunsplit \Delta$ (\Cref{ch5note:sep}).

The following auxiliary notion relates contexts that may differ in exactly one assignment:
	\begin{definition}
		Given contexts  $\Gamma , \Gamma'$, we write  $\Gamma \dillcontrel{z}{T} \Gamma' $ if $(\Gamma = \Gamma' \land z \not \in \dom{\Gamma}) \lor (\Gamma = \Gamma' , z:T) $ for some type $T$. 
	\end{definition}

\begin{definition}{Translating Judgements}
\label{ch5d:transjudgdill}
Given a judgment $\Gamma \vaslinunsplit \Delta \st P$ 
and a name $u$, 	its translation   $\sttodillj{\Gamma \vaslinunsplit \Delta \st P}_{u}{}$ is defined as 
	$$ \dillnotun{\sttodillt{\Gamma'}{}{}} ;  \sttodillt{\Delta'}{}{} \dill \sttodillp{P}{} :: u: A$$ 
	where $\Gamma'$, $\Delta'$, and $A$ are subject to one of the following conditions:
	\begin{itemize}
		\item  $A= \sttodillt{\dual{T}}{}{}$ when $\{ u: T \} \subset \Gamma , \Delta$ with $ (\Gamma \dillcontrel{u}{T} \Gamma') \land (\Delta \dillcontrel{u}{T} \Delta')$;~or
		\item  $A= \dillnilT $ when $ u \not\in \dom{\Gamma , \Delta}$ with $ (\Gamma = \Gamma') \land (\Delta = \Delta')$.
	\end{itemize}
	
	\Cref{ch5d:secondtablep1} defines the translation by induction on $P$, assuming that the contexts satisfy the appropriate requirements, i.e., $\contun{\Gamma,\Gamma'} \land \neg \contun{\Delta , \Delta_1,\Delta' }$ and $A,\Gamma'$ and $\Delta'$ are as one of the cases above.  
\end{definition}


	\begin{table}[h!]
		\smaller
		\centering
		\begin{tabular}{!{\vline width 1.5pt}c!{\vline width 0.5pt}c!{\vline width 1.5pt}}
			\specialrule{.1em}{0em}{0em}
			\multirow{2}{*}{
			} & $\Gamma \vaslinunsplit \Delta \st P$  \\
			\cline{2-2}
			& $\dillnotun{\sttodillt{\Gamma}} ;  \sttodillt{\Delta} \dill \sttodillp{P} :: u : A$  \\
			\specialrule{.1em}{0em}{0em}
			\specialrule{.1em}{0em}{0em}
			\multirow{2}{*}{
				1
			} 
			&
				$
					{ \Gamma  \vaslinunsplit \cdot  \st \nil  }
				$
			\\
			\cline{2-2}
			& 
				$
					{ \dillnotun{\sttodillt{\Gamma}} ;  \cdot \dill \nil :: u : \dillnilT }
				$
			\\
			\specialrule{.1em}{0em}{0em}
			\multirow{2}{*}{
				2
			} 
			&
				$
					\Gamma \vaslinunsplit \Delta_1 , \Delta \st P \pp Q 
				$
			\\
			\cline{2-2}
			& 
				$
				\begin{array}{ll}
						\dillnotun{\sttodillt{\Gamma'}} ; \sttodillt{\Delta_1} , \sttodillt{\Delta'}   \dill \res{w} ( \sttodillp{P} \pp \sttodillp{Q}   )  ::  u: A
					&
					\text{if } w \not \in \dom{\Gamma, \Delta_1, \Delta}  \land u \notin \fn{P}
				\end{array}
				$
			\\
			\specialrule{.1em}{0em}{0em}
			\multirow{2}{*}{
				3
			} 
			&
				$
				\begin{array}{l}
					\Gamma \vaslinunsplit  \Delta_1 ,  \Delta \st 
					\res {zv:V} (P \pp Q)
				\end{array}
				$
			\\
			\cline{2-2}
			& 
				$
				\begin{array}{l}
					\text{If } 
					(u\notin \fn{P})
					\land
					( (\neg \contun{V}) \lor 
					( \contun{V} \land v \not \in \fn{P} \land z \not \in \fn{Q} )) 
					\\
						\dillnotun{\sttodillt{\Gamma'}} ; \sttodillt{\Delta_1} , \sttodillt{\Delta'}  \dill \res{z}( \sttodillp{P} \pp \sttodillp{Q}\substj{z}{v} ) ::  u: A 
				\end{array}
				$
			\\
			\specialrule{.1em}{0em}{0em}
			\multirow{2}{*}{
				4
			} 
			&
			$
			\begin{array}{ll}
				\Gamma , x: \recur{\wn T} \vaslinunsplit \cdot \st  
				  \vasin{\un}{x}{y}{P}
			\end{array}
			$
			\\
			\cline{2-2}
			& 
			$
			\begin{array}{l}
				\text{If } u = x \land x \not \in \fn{P} \text{ and one of the following holds:}
				\\
				\begin{tabular}{ll}
						$
							\dillnotun{\sttodillt{\Gamma}} ; \cdot  \dill   \dillserv{x}{w}{  
							\dillchoice{w}{	\dillin{w}{y}{ \sttodillp{P} }	}{	\sttodillp{P\substj{w}{y}} }} 
							 :: u: \dillunt{(\dillselet{(\dillint{\sttodillt{T}}{\dillnilT})}{\sttodillt{\dual{T}}})} $
							 &
							 $\text{if } \notserver{T}\land \notclient{T}
						$
					\smallskip
					\\
					\hdashline 
					\noalign{\smallskip}
					$\dillnotun{\sttodillt{\Gamma}} ; \cdot  \dill   \dillserv{x}{w}{  
						{ \sttodillp{P\substj{w}{y}} }} ::  u: \dillunt{\sttodillt{\dual{T}}} $ 
					& 
					\text{if } $\server{T} \land \notclient{T} $
					\smallskip
					\\
					\hdashline 
					\noalign{\smallskip}
					$\dillnotun{\sttodillt{\Gamma}} ; \cdot  \dill   \dillserv{x}{w}{  
						\dillin{w}{y}{ \sttodillp{P} }	} ::  u: \dillunt{(\dillint{\sttodillt{T}}{\dillnilT})}$  
					& 
					\text{if } $\notserver{T} \land \client{T}$
					\smallskip
				\end{tabular}
			\end{array}
			$
			\\
			\specialrule{.1em}{0em}{0em}
			\multirow{2}{*}{
				5
			} 
			&
			$
			\begin{array}{l}
				\Gamma \vaslinunsplit  x: \lin \vasreci{T}{S} , \Delta \st 
				   \vasin{\lin}{x}{y}{P}
			\end{array}	
			$
			\\
			\cline{2-2}
			& 
			\begin{tabular}{ll} 
				$\dillnotun{\sttodillt{\Gamma}} ; \sttodillt{\Delta} \dill \dillin{x}{y}{\sttodillp{P}}  :: u:\dillint{\sttodillt{T}}{\sttodillt{\dual{S}}}$ & \text{if $u = x$}
				\smallskip
				\\
				\hdashline 
				\noalign{\smallskip}
				$\dillnotun{\sttodillt{\Gamma'}} ; \sttodillt{\Delta'} , x:\dilloutt{\sttodillt{T}}{\sttodillt{S}}  \dill \dillin{x}{y}{\sttodillp{P}} :: u: A$ & 
				\text{otherwise}
				\smallskip
				\end{tabular}
			\\
			\specialrule{.1em}{0em}{0em}
			\multirow{2}{*}{
				6
			} 
			&
			$
			\begin{aligned}
				& \Gamma , z: \recur{\oc T} \vaslinunsplit \Delta  \st
				  \res{xy}\vasout{z}{y}{P}
			\end{aligned}	
			$
			\\
			\cline{2-2}
			& 
			$
			\begin{array}{l}
				\text{If } (\neg \contun{T}) \lor (y \not \in \fn{P} \land \contun{T}) \text{ and:}
				\\
				\begin{tabular}{ll}
							$\dillnotun{\sttodillt{\Gamma'}} , z:\dillselet{\dillint{\sttodillt{T}}{\sttodillt{\dillnilT}}}{\sttodillt{\dual{T}}}  ; \sttodillt{\Delta'}  \dill 
							\dillbout{z}{x}{
							\dillseler{x}{\sttodillp{P}}} ::  u: A  
						$
						&
						$
							\text{if } \notserver{T} \land \notclient{T}  
						$
					\smallskip
					\\
					\hdashline 
					\noalign{\smallskip}
					$ \dillnotun{\sttodillt{\Gamma'}} , z:\sttodillt{\dual{T}}  ; \sttodillt{\Delta'}  \dill  \dillbout{z}{x}{
						\sttodillp{P}} :: u: A $
					& 
					\text{if } $\server{T} \land \notclient{T} 
					$
					\smallskip
				\end{tabular}
			\end{array}
			$
			\\
			\specialrule{.1em}{0em}{0em}
			\multirow{2}{*}{
				7
			} 
			&
			$
				\Gamma \vaslinunsplit z: \lin \vassend{T}{S} , \Delta_1, \Delta \st \res{xy}\vasout{z}{x}{(P \pp Q)}
			$
			\\
			\cline{2-2}
			& 
			\begin{tabular}{ll} 
				\multicolumn{2}{l}{
						$\text{If } (\neg \contun{T}) \lor (x \not \in \fn{P} \cup \fn{Q} \land \contun{T}) \text{ and:}$
				}\\
				\multicolumn{2}{l}{
					$
					\begin{array}{ll}
						\dillnotun{\sttodillt{\Gamma}} ; \sttodillt{\Delta_1} , \sttodillt{\Delta} \dill
						\dillbout{z}{y}{(\sttodillp{P} \pp \sttodillp{Q})} ::   z : \dilloutt{\sttodillt{T}}{\sttodillt{\dual{S}}} & \qquad \quad  \text{if }z = u \land u,z \not \in \fn{P} \land y \not \in \fn{Q}
					\end{array}
					$
				}
				\smallskip
				\\
				\hdashline 
				\noalign{\smallskip}
				\multicolumn{2}{l}{
					$ 
					\begin{array}{ll}
						\dillnotun{\sttodillt{\Gamma'}} ; \sttodillt{\Delta_1} , \sttodillt{\Delta'}  , z : \dillint{\sttodillt{T}}{\sttodillt{S}} \dill 
						\dillbout{z}{y}{(\sttodillp{P} \pp \sttodillp{Q})} :: u: A  & \text{if }{z \not = u} \land {u,z \not \in \fn{P}}\land y \not \in \fn{Q} 
					\end{array}
					$
				}
				\smallskip
			\end{tabular}
			\\
			\specialrule{.1em}{0em}{0em}
			\multirow{2}{*}{
				8
			} 
			&
				$
				\begin{array}{l}
				\smallskip
					\Gamma  \vaslinunsplit v:T, x: \lin \oc (T).S , \Delta 
					\st \ov x\out v.P
				\end{array}
				$
			\\
			\cline{2-2}
			& 
				\begin{tabular}{ll}
					$ \dillnotun{\sttodillt{\Gamma}} ;  v: \sttodillt{T}, \sttodillt{\Delta}   \dill \dillbout{x}{y}{(\dillforward{v}{y}  \pp  \sttodillp{P})} ::   u : \dilloutt{\sttodillt{T}}{\sttodillt{\dual{S}}} $ & if $u = x \land \neg \contun{T} \land \notserver{T}$
					\smallskip
					\\
					\hdashline 
					\noalign{\smallskip}
					$ \dillnotun{\sttodillt{\Gamma'}} ;  v: \sttodillt{T},  \sttodillt{\Delta'}  , x : \dillint{\sttodillt{T}}{\sttodillt{S}}  \dill \dillbout{x}{y}{(\dillforward{v}{y} \pp P)} ::  u: \sttodillt{\dual{R}}  $ &	if  $\neg \contun{T} \land \notserver{T} $
					\smallskip
				\end{tabular}
			\\
			\specialrule{.1em}{0em}{0em}
			\multirow{2}{*}{
				9
			} 
			&
			$
			\begin{array}{l}
				\Gamma,  v:T \vaslinunsplit x: \lin \oc (T).S , \Delta 
				 \st \ov x\out v.P 
			\end{array}
			$
			\\
			\cline{2-2}
			& 
			\begin{tabular}{ll}
				$  \dillnotun{\sttodillt{\Gamma}} , v: \sttodillt{T} ;  \sttodillt{\Delta} \dill \dillbout{x}{y}{(\dillfwdbang{v}{y}  \pp  \sttodillp{P})} ::   x : \dilloutt{\sttodillt{T}}{\sttodillt{\dual{S}}}$ & if $u = x \land \contun{T} \land \notserver{T}$

				\smallskip
				\\
				\hdashline 
				\noalign{\smallskip}
				$ \dillnotun{\sttodillt{\Gamma'}},  v:T  ; \sttodillt{\Delta'}  , x : \dillint{\sttodillt{T}}{\sttodillt{S}}  \dill \dillbout{x}{y}{(\dillfwdbang{v}{y} \pp P)} ::  u: A $ & if  $\contun{T} \land \notserver{T}$
				\smallskip
			\end{tabular}
			\\
			\specialrule{.1em}{0em}{0em}
			\multirow{2}{*}{
				10
			} 
			&
			$
			\begin{aligned}
				& \Gamma , x: \recur{\oc T} \vaslinunsplit v:T , \Delta 
				 \st \ov x\out v.P
			\end{aligned}
			$
			\\
			\cline{2-2}
			& 
			\begin{tabular}{l} 
				If $\notserver{T} \land \notclient{T}  \land \neg \contun{T} \land z \not \in \fn{P}$  
				\\
				$ \dillnotun{\sttodillt{\Gamma'}} , x:\dillselet{\dillint{\sttodillt{T}}{\dillnilT}}{\sttodillt{\dual{T}}} ;  v:\sttodillt{T} , \sttodillt{\Delta'} \dill \dillbout{x}{z}{ \dillselel{z}{ \dillbout{z}{w}{( \dillforward{v}{w}  \pp  \sttodillp{{P}} )}}} :: u: A$
				\smallskip
				\\
				\hdashline 
				\noalign{\smallskip}
				If $\notserver{T} \land \client{T}  \land \neg \contun{T} \land z \not \in \fn{P}$ 
				\\
				$ \dillnotun{\sttodillt{\Gamma'}} , x:\dillint{\sttodillt{T}}{\dillnilT} ;  v:\sttodillt{T} , \sttodillt{\Delta'}   \dill \dillbout{x}{z}{ \dillbout{z}{w}{ ( \dillforward{v}{w}  \pp  \sttodillp{{P}} )} } ::  u: A $
				\smallskip
			\end{tabular}
			\\
			\specialrule{.1em}{0em}{0em}
			\multirow{2}{*}{
				11
			} 
			&
			$
			\begin{aligned}
				& \Gamma , x: \recur{\oc T}, v:T\vaslinunsplit \Delta 
				 \st \ov x\out v.P
			\end{aligned}
			$
			\\
			\cline{2-2}
			& 
			$
			\begin{array}{l}
			\text{If } \notserver{T} \land \notclient{T} \land \contun{T}\land z \notin \fn{P}\\ 
			\dillnotun{\sttodillt{\Gamma'}} , x:\dillselet{\dillint{\sttodillt{T}}{\dillnilT}}{\sttodillt{\dual{T}}} ; v:\sttodillt{T} ; \sttodillt{\Delta'} \dill \dillbout{x}{z}{ \dillselel{z}{ \dillbout{z}{w}{( \dillfwdbang{v}{w}  \pp  \sttodillp{{P}} )}}} :: u: A		\smallskip
			\\
			\hdashline
				\text{If  } \notserver{T} \land \client{T} \land \contun{T} \land z \not \in \fn{P}
				\\
				\begin{aligned}
					& \dillnotun{\sttodillt{\Gamma'}} , x:\dillint{\sttodillt{T}}{\dillnilT} , v:\sttodillt{T};   \sttodillt{\Delta'}  \dill \dillbout{x}{z}{ 
					\dillbout{z}{w}{
						( \dillfwdbang{v}{w}  \pp  \sttodillp{{P}}   )
						}
					} :: u: A
				\end{aligned}
			\end{array}
			$
			\\
			\specialrule{.1em}{0em}{0em}
		\end{tabular}
		\vspace{0.5cm}
		\caption{From judgments in \vasco to judgments in \pidill (\Cref{ch5d:transjudgdill}).  \label{ch5d:secondtablep1}}
	\end{table}

Using this translation of judgments, a translation of derivations can be defined exactly as in \Cref{ch5def:firsttransl_judgments}.

We discuss some entries in \Cref{ch5d:secondtablep1} with the following example.

\begin{example}{Cont.~\Cref{ch5ex:trans_proc}} 

The translation of judgments defined in~\Cref{ch5d:secondtablep1} relies on the typability in the source language (\vaslang) to determine the exact conditions to the typability of the translated process in~$\pidill$. First, the type derivation of $w:\nilT\st \res{xy}(  \vasin{\un}{ x}{z}{\nil}\pp  \vasout{y}{w}{\nil})$ is essential to build a type derivation for the translated judgment (if one exists):
\begin{mathpar}
\small
\inferrule
{
\inferrule{
\inferrule{\Gamma\st x:\recur{\wn \nilT} \\
 \Gamma, z:\nilT \st \nil}{\Gamma\st \vasin{\un}{x}{z}{\nil}}
 \\
\inferrule{\Gamma\st y:\recur{\oc \nilT}\\\\
\Gamma \st w:\nilT\\
\Gamma\st \nil}
{\Gamma\st \vasout{y}{w}{\nil}}
 }
 {\Gamma \st \vasin{\un}{ x}{z}{\nil}\pp  \vasout{y}{w}{\nil}}
 }
{w:\nilT \st \res{xy} (\vasin{\un}{ x}{z}{\nil}\pp  \vasout{y}{w}{\nil})}
\end{mathpar}
where $\Gamma= x:\recur{\wn \nilT}, y:\recur{\oc \nilT}, w:\nilT$

Second, the translation  $\sttodillj{w:\nilT\st \res{xy}(  \vasin{\un}{ x}{z}{\nil}\pp  \vasout{y}{w}{\nil})}_u$
 corresponds to entry 3 of \Cref{ch5d:secondtablep1}, 
\begin{mathpar}
\inferrule
{}
{w:{\dillnilT};\cdot \dill \res{x}(\sttodillp{\vasin{\un}{ x}{z}{\nil}}\pp \sttodillp{\vasout{x}{w}{\nil}})::u:A}
\end{mathpar}
and we have previously observed that the side conditions hold. Since $u$ is not in the type context, we have $A=\dillnilT$. Now we proceed to build a type derivation for the translated judgment by applying rule $\dilltype{cut}$:
\begin{mathpar}
\small
\inferrule
{
\inferrule{\Pi_1}
{w:{\dillnilT}; \cdot\dill \sttodillp{\vasin{\un}{ x}{z}{\nil}}::x:  \dillunt{B}
} \quad
\inferrule{\Pi_2
\\\\
w:{\dillnilT},y :B; \cdot \dill  \sttodillp{\vasout{y}{w}{\nil}})::u:\dillnilT}{w:{\dillnilT}; x :\dillunt{B} \dill  \sttodillp{\vasout{x}{w}{\nil}})::u:\dillnilT}
}
{w:{\dillnilT};\cdot \dill \res{x}(\sttodillp{\vasin{\un}{ x}{z}{\nil}}\pp \sttodillp{\vasout{x}{w}{\nil}})::u:\dillnilT}
\end{mathpar}
and we will show that there exist derivations $\Pi_1$ and $\Pi_2$  such that the derivation holds.  We recall the translations $\sttodillp{\vasin{\un}{x}{z}{0}}$ and $\sttodillp{\vasout{x}{w}{
\nil}}$ in \Cref{ch5ex:trans_proc}.  

Third, the left premise is the translation $\sttodillj{\Gamma\st \vasin{\un}{x}{z}{\nil}}_x$  and corresponds to entry (4) of the~\Cref{ch5d:secondtablep1}. We  recall \Cref{ch5ex:transl_type} for $\dillnotun{\sttodillt{\Gamma}}=w:\dillnilT, x:  (\dillchoicet{(\dilloutt{\dillunt{\dillnilT}}{\dillnilT})}{\dillnilT}) , y: (\dillselet{(\dillint{\dillunt{\dillnilT}}{\dillnilT})}{\dillnilT})$ and use the abbreviation $B=(\dillselet{(\dillint{\dillunt{\dillnilT}}{\dillnilT})}{\dillnilT})$ and strengthening of $\dillnotun{\sttodillt{\Gamma}}$. The derivation $\Pi_1$ is as follows:
\begin{mathpar}
\small
\inferrule*[left=\dilltype{\dillunt{R}}]{
\inferrule*[left=\dilltype{\dillselet{}{}R}]{
\inferrule{
\inferrule{
 w:{\dillnilT}; \cdot  \dill \nil :: v: \dillnilT
}
{ w:{\dillnilT}; z:\dillnilT \dill \nil :: v: \dillunt{\dillnilT}}
}
{ w:{\dillnilT}; \cdot\dill \dillin{v}{z}{\nil} ::v: \dillint{\dillunt{\dillnilT}}{\dillnilT}} \\
 w:{\dillnilT}; \cdot\dill \nil :: v: \dillnilT
}
{ w:{\dillnilT}; \cdot \dill   \dillchoice{v}{	\dillin{v}{z}{ \nil }	}{\nil }
							 :: v: \dillselet{\dillint{(\dillunt{\dillnilT}}{\dillnilT})}{\dillnilT} }
}
{ w:{\dillnilT}; \cdot\dill \dillserv{x}{v}{  \dillchoice{v}{	\dillin{v}{z}{ \nil }}{\nil }} ::x: \dillunt{(\dillselet{(\dillint{\dillunt{\dillnilT}}{\dillnilT})}{\dillnilT})}
}
\end{mathpar}

Fourth, the right premise is the translation $\sttodillj{\Gamma\st \vasout{y}{w}{0}\substj{x}{y}}_u$ and  corresponds to the entry (11a) of the~\Cref{ch5d:secondtablep1}. The derivation $\Pi_2$ is as follows:

\begin{mathpar}
\small
\inferrule*[left=\dilltype{copy}]{
\inferrule{
\inferrule{
\dillnotun{\sttodillt{\Gamma'}} ; \cdot  \dill  \dillfwdbang{w}{v} :: v :\dillunt{\dillnilT}\quad 
\dillnotun{\sttodillt{\Gamma'}} ; z:\dillnilT    \dill   \nil  :: u :\dillnilT}
{\dillnotun{\sttodillt{\Gamma'}} ; z:(\dillint{\dillunt{\dillnilT}}{\dillnilT})   \dill \dillbout{z}{v}{( \dillfwdbang{w}{v}  \pp  \nil )} :: u :\dillnilT
}}{
\dillnotun{\sttodillt{\Gamma'}} ; z:\dillselet{(\dillint{\dillunt{\dillnilT}}{\dillnilT})}{\dillnilT}   \dill \dillselel{z}{ \dillbout{z}{v}{( \dillfwdbang{w}{v}  \pp  \nil )}} :: u: \dillnilT
}}
{\dillnotun{\sttodillt{\Gamma'}}  ; \cdot  \dill \dillbout{y}{z}{ \dillselel{z}{ \dillbout{z}{v}{( \dillfwdbang{w}{v}  \pp  \nil )}}} :: u: \dillnilT}
\end{mathpar}
where we strengthen $\dillnotun{\sttodillt{\Gamma}}$ to  $\dillnotun{\sttodillt{\Gamma'}}= y: (\dillselet{(\dillint{\dillunt{\dillnilT}}{\dillnilT})}{\dillnilT}), w:{\dillnilT}$.
\end{example}

We have the following property, which holds by definition of the entries of \Cref{ch5d:secondtablep1}:

\begin{theorem}[Type preservation]\label{ch5thm:second_type_pres}
	If $\Gamma \vaslinunsplit \Delta \st P$ then $ \dillnotun{\sttodillt{\Gamma'}{}{}} ;  \sttodillt{\Delta'}{}{} \dill \sttodillp{P}{} :: u: A$ is well-typed in \pidill, with  $A, \Gamma'$ and $\Delta'$ as in \Cref{ch5d:transjudgdill}.
\end{theorem}

Notice that the translations of typable \vasco processes are not necessarily typable in \pidill.  
We shall concentrate on  processes in $\vaslang$ that are typable in $\pidill$:
\begin{notation}
	We write $\dillnotun{\sttodillt{\Gamma'}};   \sttodillt{\Delta'}  \dill \sttodillp{{P}}  :: u: \sttodillt{\dual{S}} $ whenever $\sttodillj{\Gamma ,\Delta  \st P}_u$  holds, with $\Gamma \dillcontrel{u}{S} \Gamma'$ and $\Delta \dillcontrel{u}{S} \Delta'$. 
\end{notation}

We can finally define $\dilllang$:

\begin{definition}{$\dilllang$}
\label{ch5d:dilllang}
Let $u$ be a name.	We define:
	\begin{align*}
	 \dilllang = \{ P \in \vasco \mid &  \ \Gamma \vaslinunsplit \Delta \st P \land 	\Gamma \dillcontrel{u}{S} \Gamma' 
	 \land  \Delta \dillcontrel{u}{S} \Delta'  \\	& \land \dillnotun{\sttodillt{\Gamma'}};   \sttodillt{\Delta'}  \dill \sttodillp{{P}}  :: u: \sttodillt{\dual{S}} 	\} 
	\end{align*}
where contexts and types mentioned are existentially quantified.
\end{definition}

\subsection{Results}
\begin{theorem}[$\dilllang \subset \lvllang$]\label{ch5thm:main_result}
	Let $P\in \vaslang$ such that $\Gamma \st P$, for some context $\Gamma$. If there exists $  u$  such that  $\sttodillj{\Gamma \st P}_u $ holds, then there exists $l$ such that $ \sttoltj{\Gamma \st P}{l}{}$ holds.
\end{theorem}

The proof of \Cref{ch5thm:main_result} is 
	by induction on the structure of $P$. 
	We exploit a number of invariant properties for the type systems for \pidill and \pilvl, including:
	\begin{itemize}
		\item In $\pidill$, judgments for typed processes never exhibit servers on the left-hand side.
		\item In \pilvl, levels for types that do not exhibit server behavior can be decreased at will.
		\item  The type system of \pidill ensures that the name on the right-hand side is not guarded by servers.
	\end{itemize}

\begin{theorem}[$\lvllang \not \subset \dilllang$]
\label{ch5thm:wnotinl}
	 $\exists P \in \lvllang$  with $\Gamma \st P$ and $\sttoltj{\Gamma \st P}{l}{} $ for some $l$  such that $ \nexists  \ z \ s.t. \ \sttodillj{\Gamma \st P}_z $.
\end{theorem}

To prove \Cref{ch5thm:wnotinl}, it suffices to consider the \vasco process
	\[
		P = 	
		\vasres{xy}{(
			\vaspara{ 
				\vasin{\lin}{x}{z}{\vasin{\un}{z}{w}{\nil}} 
			}
			{
				\vasres{st}{
					\vasout{y}{s}{(
						\vaspara{
							\vasres{uv}{(\vasout{t}{u}{\nil})}
						}
						{
							\nil
						}
					)}
				}
			}
		)}
	\]
Clearly, $P$ is terminating:
	\[
		\begin{aligned}
			P &\reddpp 
					\vasres{st}{(
						\vaspara{ 
							\vasin{\un}{s}{w}{\nil}
						}
						{
							\vasres{uv}{(\vasout{t}{u}{\nil})}
						}
					)}
				 \reddpp 
					\vasres{st}{
						\vasin{\un}{s}{w}{\nil}
					}
		\end{aligned}
	\]
Process $P$ can be typed so as to establish $P \in  \vaslang$. 
Also, there is a level function that makes its translation into \pilvl typable. Hence, $P \in \lvllang$. 
However, its translation into \pidill is not typable, so $P \not\in \dilllang$.

%
	
\section{Closing Remarks}
\label{ch5s:close}
We presented a comparative study of type systems for concurrent processes in the $\pi$-calculus, from the unifying perspective of termination and session types. 
To our knowledge, this is the first study of its kind.
Even by focusing on only three different type systems, we were confronted with technical challenges connected with the intrinsic differences between them.
The typed process model \vasco~\cite{V12}, focused on session-based concurrency, admits a rather broad class of processes, exploiting a clear distinction between linear and unrestricted resources, implemented via context splitting.  
The typed process model \pilvl combines features from type systems that target the termination property~\cite{DS06} and type systems for sessions. 
Finally, the typed process model \pidill~\cite{CairesP10} rests upon a firm logical foundation, and its control of clients and servers is directly inherited from the logical principles of the exponential $!A$.
Notice that \pidill is unique among type systems for the $\pi$-calculus in that it ensures  protocol fidelity, deadlock-freedom, confluence, and strong normalization/termination for typed processes.

The main take-away message is that the Curry-Howard correspondence is strictly weaker than weight-based approaches for enforcing the termination property. 
Hence, the  control of server/client interactions that is elegantly enabled by the copying semantics of  $!A$ turns out to be rather implicit when contrasted to weight-based techniques.
Interestingly, \cite{DardhaP15,DBLP:journals/jlap/DardhaP22} arrived to a similar conclusion in their comparative study of type systems focused on the deadlock-freedom property: type systems based on the Curry-Howard correspondence can detect strictly less deadlock-free processes than other, more sophisticated type systems. 
Notice that the study in~\cite{DardhaP15,DBLP:journals/jlap/DardhaP22} considers only finite processes, without input-guarded replication (so all process are terminating).

Immediate items for current and future work include incorporating other type systems into our formal comparisons. 
The type systems by \cite{DBLP:journals/mscs/Sangiorgi06} and by \cite{DBLP:conf/lics/YoshidaBH01} are very appealing candidates. 
Also, Deng and Sangiorgi proposed several type systems for termination.  
Here we considered only the simplest variant, which induces the class \lvllang and is already different from \dilllang; it would be interesting to consider the other variants.






\clearemptydoublepage

\chapter{Conclusions}
\label{ch6_conc}

We first remind the reader of the research question that we have strived to answer (\Cref{s:rq}):

\begin{quote}
	Can we relate formal models of sequential computation and interactive behaviors, both governed by behavioural types, considering phenomena little considered so far, such as \emph{non-determinism} and \emph{failures}, while accounting with essential properties such as \emph{deadlock-freedom}, \emph{confluence}, and \emph{termination (strong normalization)}?
\end{quote}

To answer this question simply: Yes, we have shown that the intended relationships between sequential and concurrent models of computation can be obtained by capitalising on the extensive literature  on \emph{relative expressiveness} that has been developed within the fields of concurrency theory and process calculi. Our technical approach has advanced the study of relative expressiveness by considering advanced \emph{type systems}.

One leading idea in our work is considering intersection types as the sequential analog of the behavioural types that discipline computation in the concurrent paradigm. On the sequential side we have considered our source calculi to be $\lambda$-calculi with \emph{resource control}, \emph{non-determinism}, and \emph{failure}. We introduce and study new resource $\lambda$-calculi that distill and articulate key features from other similar languages proposed in the literature. For these calculi, (non-idempotent) intersection types govern the control and behaviour of resources within a term and their evaluation/substitution;  due to their quantitative nature, intersection types can capture the evolution of a term.

On the concurrent side, the target calculi we consider are based on a session-typed $\pi$-calculus with client server interactions, non-determinism and failure, proposed by \cite{CairesP17}. Session types arise from a concurrent interpretation of classical linear logic, in the style of the Curry-Howard isomorphism (`propositions-as-types'). This reduces the intrinsic expressivity for processes, but ensures strong behavioural properties directly derived from the logical foundations (in particular, from the connection between cut-elimination and process synchronization).  By encoding resource $\lambda$-calculi into session-typed $\pi$-calculi, we demonstrated how not only non-determinism and explicit failures can be handled within a logically-motivated framework but also the precise connection between intersection types and session types themselves.

To establish our formal relationships we follow the well-known framework by \cite{DBLP:journals/iandc/Gorla10};  in fact, the definition of our source languages  and their type system is strongly guided by Gorla's correctness criteria. In particular, we purposefully aimed at source and target languages for which translations can enjoy strong formulations of the \emph{soundness} property, which 
captures how the behaviour of the encoded term (in $\pi$) reflects the behaviour of the source term (in $\lambda$). In the literature we can find different formulations of soundness, with varying levels of flexibility and strength. In our work we have strived to obtain a very general/flexible formulation of soundness, for which formal proofs are technically challenging; important for our proofs  are properties of \emph{determinacy} and \emph{confluence} that come directly from the logically-motivated process model. 
We have also considered the case in which source and target languages abandon confluent reductions, which further increases the complexity of this endeavour. 

Finally, our comparative analysis of type systems that enforce termination for concurrent processes illuminated the strengths and limitations of various approaches, providing valuable insights for future research. Also in this context, our approach to comparison relied heavily on techniques from relative expressiveness, once again framed in the typed setting, which requires translations of processes/terms but also of their corresponding types and type judgments.

Our research not only sheds light on these intricate relationships but also lays the groundwork for further advancements in this area. More in details, we have been led to several significant insights and contributions as follows:

\begin{itemize}

    \item \emph{Encoding Resource $\lambda$-calculi in Session-Typed $\pi$-Calculus (\Cref{ch2})}: \\  We developed a correct translation of \lamrfail, a resource $\lambda$-calculus featuring non-determinism and explicit failure, into \spi, a session-typed $\pi$-calculus, which not only illustrates the compatibility of these two formalisms but also provides a robust framework for modelling non-determinism and explicit failures. This encoding establishes a precise connection between terms, processes, intersection types, and linear logic propositions, providing a logical foundation for handling non-deterministic behaviors and failures. This work stands as the first to draw such connections, thereby opening new avenues for the application of intersection types and session types in understanding resource-aware computing in both sequential and concurrent settings.

    \item \emph{Encoding Resource $\lambda$-calculi with Linear and Unrestricted
    Resources (\Cref{ch3})}: \\ 
    We extended our previous work on resource $\lambda$-calculi by introducing \lamrfailunres, a calculus in which bags of resources consist of two sorts: a linear and an unrestricted component. Separating the bag in this way is influenced by linear logic and ensures that the behaviour that a resource allows (linear or unrestricted) does not change during computation. This is important to enable the predictability of failures within computation. The motivation for unrestricted resources stems from the desire to capitalize on the expressivity of the target model that already includes unrestricted behaviors in the form of \emph{client-server} interactions, related to the concurrent interpretation of the exponentials $?$ and $!$ in linear logic. We aim to demonstrate what the target process model allows us to express at the source level, exploring the extent to which we can expand our sequential model and the behaviors we can extract from the underlying target process calculus. Although other formulations of resource calculi with unrestricted resources exist (see, e.g., \cite{DBLP:conf/birthday/BoudolL00,PaganiR10}),  our formulation is novel in that it is directed from the viewpoint of encodability with flexible correctness results and the extraction of target language behaviours. 
    
    \item \emph{Exploration of Non-Confluent Non-Determinism in Typed Calculi (\Cref{ch4})}: \\ 
In this chapter, we shift our focus from confluent to non-confluent non-determinism. 
    Our investigation into \clpi and \lamcoldetshlin showcased calculi integrating resource control with non-confluent non-determinism yielded significant insights. We examined lazy and explicit non-determinism within session types, emphasizing the role of intersection types in regulating resource fetching we demonstrate the feasibility of integrating different forms of non-determinism.  The correct translation of \lamcoldetshlin into \clpi highlights the compatibility and interplay between different forms of non-determinism across calculi.
    
    Furthermore, this exploration opens new avenues for future research, including the refinement of type systems to better manage non-determinism and the extension of these principles to other computational models. By addressing the challenges posed by non-confluent non-determinism, our work contributes to the broader goal of developing reliable and efficient concurrent systems that can handle unpredictability in a controlled manner.

    \item \emph{Comparative Study of Type Systems for Concurrent Processes (\Cref{ch5})}: \\
    We conducted a comparative analysis of type systems for the $\pi$-calculus, focusing on termination. 
    While in the $\lambda$-calculus the study of termination properties via types is well-understood in general, the situation for the $\pi$-calculus is different. Indeed, because reductions in a concurrent setting allow for a wider spectrum of behaviors, we find  several advanced type systems enforcing termination at different levels of completeness, i.e., by analysing client-server interactions in various different ways. Still, formal comparisons between these advanced type systems were not studied before our work.

    The comparative analysis of type systems rests upon the idea of classifying different classes of terms induced by the typing disciplines as well as by relations of relative expressiveness, in which soundness and completeness properties play a crucial role. 
    Although this chapter may seem slightly unrelated to the previous chapters, we argue we that there is a significant relationship between them. The concepts in the sequential world mirror those in the concurrent world, with types on one side corresponding to types on the other, maintaining a cohesive framework across both paradigms.
    This study revealed intrinsic differences between typed process models such as \vasco, \pilvl, and \pidill. The Curry-Howard correspondence, while fundamental, demonstrates limitations in enforcing termination when compared to previously developed approaches based on weights/measures, yet also inducing a strict partial order that we can consider a subclass of the weights/measures system of \pilvl.

\end{itemize}

In conclusion, our exploration has contributed insights into formal models of computation that incorporate non-determinism, failures, and resource control. By establishing connections between different calculi and type systems, we provide a foundational framework that can be further explored. This foundational framework should ultimately enhance the further development the theoretical underpinnings and practical applications of concurrent programming paradigms.

Moving forward, we aim to deepen our understanding of key properties like solvability within the resource calculus (pertaining to the work in chapters \Cref{ch2,ch3,ch4}) and termination in the $\pi$-calculus (\Cref{ch5}). Immediate items for future work in the latter include incorporating other type systems into our formal comparisons. The type systems proposed by \cite{DBLP:journals/mscs/Sangiorgi06} and by \cite{DBLP:conf/lics/YoshidaBH01} are particularly promising candidates. This future research would further solidify the theoretical contributions we have presented.

All in all, our work serves as a stepping stone toward a more integrated understanding of how non-determinism, resource control, and type systems can jointly enhance the robustness and predictability of computational models in different paradigms.

\clearemptydoublepage



\phantomsection%
\addcontentsline{toc}{chapter}{Bibliography}

\bibliographystyle{TRAILbib}

\bibliography{./1-intro/referenceslmcs,./2-LMCS/referenceslmcs}





\clearemptydoublepage



\iffulldoc
{\footnotesize
\appendix



\newpage

\chapter{Appendix of Chapter 2}\label{ch2appendices_ch1}

\section{Appendix to \texorpdfstring{\secref{ch2sec:lamfailintertypes}}{}}
\label{ch2app:lamfailintertypes}

\explemfailnofail*

\begin{proof}
By induction on the structure of $ M $:

    \begin{myEnumerate}
        
        \item When $M = x$ then we have $x \headlin{ N / x}   = N$ and may derive the derivation of $ \Gamma \vdash N: \tau $ with $x \not \in \dom{\Gamma}$. By taking $\Gamma_1 = \emptyset$ and $\Gamma_2 = \Gamma$ as $\Gamma = \emptyset \contexcat \Gamma$ the case follows as $ \Gamma \vdash N : \tau$ and
            
                \begin{prooftree}
                    \AxiomC{}
                    \LeftLabel{\redlab{T:var}}
                    \UnaryInfC{\( x: \sigma \vdash x : \sigma\)}
                \end{prooftree}
        
        \item When $M = (M\ B)$ then we have that $(M\ B)\headlin{ N/x}  = (M \headlin{ N/x })\ B$. Let us consider two cases:
            
            \begin{myEnumerate}
                
                \item When $x \in \lfv{M \headlin{ N/x }}$
                    \begin{prooftree}
                        \AxiomC{\( \Gamma', x:\sigma^{k-1}  \vdash M \headlin{ N/x } : \pi \rightarrow \tau' \)}
                        \AxiomC{\( \Delta \vdash B : \pi \)}
                            \LeftLabel{\redlab{T:app}}
                        \BinaryInfC{\( (\Gamma', x:\sigma^{k-1} ) \contexcat \Delta \vdash (M \headlin{ N/x })\ B : \tau'\)}
                    \end{prooftree}
                    
                    By the IH we have that $\Gamma', x:\sigma^{k-1} \vdash M \headlin{ N/x } : \pi \rightarrow \tau'$ implies that $\exists \  \Gamma_1', \Gamma_2$ such that  $\Gamma_1' , x:\sigma^k \vdash M: \tau$, and $\Gamma_2 \vdash N : \sigma$ with $\Gamma' = \Gamma_1' \contexcat \Gamma_2$.
                    
                    \begin{prooftree}
                        \AxiomC{\( \Gamma_1', x:\sigma^{k}  \vdash M  : \pi \rightarrow \tau' \)}
                        \AxiomC{\( \Delta \vdash B : \pi \)}
                            \LeftLabel{\redlab{T:app}}
                        \BinaryInfC{\( (\Gamma_1', x:\sigma^{k} ) \contexcat \Delta \vdash M\ B : \tau'\)}
                    \end{prooftree}
                    
                \item When $x \not \in \lfv{M \headlin{ N/x }}$
                    \begin{prooftree}
                      \AxiomC{\( \Gamma'  \vdash M \headlin{ N/x } : \pi \rightarrow \tau' \)}
                        \AxiomC{\( \Delta \vdash B : \pi \)}
                            \LeftLabel{\redlab{T:app}}
                        \BinaryInfC{\( \Gamma' \contexcat \Delta \vdash (M \headlin{ N/x })\ B : \tau'\)}
                    \end{prooftree}
                    
                    By the IH we have that $\Gamma' \vdash M \headlin{ N/x } : \pi \rightarrow \tau'$ implies that $\exists \  \Gamma_1', \Gamma_2$ such that  $\Gamma_1' , x:\sigma \vdash M: \tau$, and $\Gamma_2 \vdash N : \sigma$ with $\Gamma' = \Gamma_1' \contexcat \Gamma_2$.
                    
                    \begin{prooftree}
                        \AxiomC{\( \Gamma_1', x:\sigma  \vdash M \headlin{ N/x } : \pi \rightarrow \tau' \)}
                        \AxiomC{\( \Delta \vdash B : \pi \)}
                            \LeftLabel{\redlab{T:app}}
                        \BinaryInfC{\( (\Gamma_1', x:\sigma ) \contexcat \Delta \vdash M\ B : \tau'\)}
                    \end{prooftree}
                
            \end{myEnumerate}

        \item When $M = M \esubst{B}{y}$ then we have that $(M\ \esubst{B}{y})\headlin{ N/x } = (M\headlin{ N/x })\ \esubst{B}{y}$ where $x \not = y$
            
            \begin{myEnumerate}
                
                \item When $x \in \lfv{M \headlin{ N/x }}$
                    
                    \begin{prooftree}
                       \AxiomC{\( \Gamma', x:\sigma^{k-1} ,  {y}:\delta^{j} \vdash (M\headlin{ N/x }) : \tau \)}
                             \AxiomC{\( \Delta \vdash B : \delta^{j} \)}
                        \LeftLabel{\redlab{T:ex \dash sub}}    
                        \BinaryInfC{\( \Gamma',  {y}:\delta^{j} \contexcat \Delta \vdash (M\headlin{ N/x }) \esubst{ B }{ y } : \tau \)}
                    \end{prooftree}
                    
                    By the IH we have that $\Gamma', x:\sigma^{k-1} ,  {y}:\delta^{j} \vdash (M\headlin{ N/x }) : \tau$ implies that $\exists \  \Gamma_1', \Gamma_2$ such that  $\Gamma_1' , x:\sigma^k, {y}:\delta^{j} \vdash M: \tau$, and $\Gamma_2 \vdash N : \sigma$ with $\Gamma',  {y}:\delta^{j} = (\Gamma_1',  {y}:\delta^{j}) \contexcat \Gamma_2$.
                    
                    \begin{prooftree}
                       \AxiomC{\( \Gamma'_1, x:\sigma^{k} ,  {y}:\delta^{j} \vdash M : \tau \)}
                             \AxiomC{\( \Delta \vdash B : \delta^{j} \)}
                        \LeftLabel{\redlab{T:ex \dash sub}}    
                        \BinaryInfC{\( \Gamma'_1 \contexcat \Delta \vdash M \esubst{ B }{ y } : \tau \)}
                    \end{prooftree}
                    
                \item When $x \not \in \lfv{M \headlin{ N/x }}$
                    
                    \begin{prooftree}
                       \AxiomC{\( \Gamma' ,  {y}:\delta^{k} \vdash (M\headlin{ N/x }) : \tau \)}
                             \AxiomC{\( \Delta \vdash B : \delta^{k} \)}
                        \LeftLabel{\redlab{T:ex \dash sub}}    
                        \BinaryInfC{\( \Gamma' \contexcat \Delta \vdash (M\headlin{ N/x }) \esubst{ B }{ y } : \tau \)}
                    \end{prooftree}
                    
                    By the IH we have that $\Gamma' ,  {y}:\delta^{k} \vdash (M\headlin{ N/x }) : \tau$ implies that $\exists \  \Gamma_1', \Gamma_2$ such that  $\Gamma_1' , x:\sigma \vdash M: \tau$, and $\Gamma_2 \vdash N : \sigma$ with $\Gamma',  {y}:\delta^{k} = (\Gamma_1',  {y}:\delta^{k}) \contexcat \Gamma_2$.
                    
                    \begin{prooftree}
                       \AxiomC{\( \Gamma'_1, x:\sigma ,  {y}:\delta^{k} \vdash M : \tau \)}
                             \AxiomC{\( \Delta \vdash B : \delta^{k} \)}
                        \LeftLabel{\redlab{T:ex \dash sub}}    
                        \BinaryInfC{\( \Gamma'_1 \contexcat \Delta \vdash M \esubst{ B }{ y } : \tau \)}
                    \end{prooftree}
                    
            \end{myEnumerate}

        \item When $M = \lambda y . M$ then linear head substitution is undefined on this term as $\headf{M} \not = x$.

        \item When $M = \fail^{\widetilde{x}}$ then $M$ is not well typed.

    \end{myEnumerate}

\end{proof}

\subexpone*

\begin{proof}
By induction on the reduction rule applied.
There are four possible cases. 
    
    \begin{myEnumerate}
    
        \item When $\expr{M}'$ is reduced to via the Rule~\redlab{R:Beta}
    
            \begin{prooftree}
                \AxiomC{}
                \LeftLabel{\redlab{R:Beta}}
                \UnaryInfC{\((\lambda x. M) B \redd M\ \esubst{B}{x}\)}
            \end{prooftree}
            
            Then $\expr{M}' = M\ \esubst{B}{x} $ can be type as follows:
            
            \begin{prooftree}
                    \AxiomC{\( {\Gamma ,  {x}:\sigma^{k} \vdash M : \tau} \)}
                     \AxiomC{\( \Delta \vdash B : \sigma^{k} \)}
                \LeftLabel{\redlab{T:ex \dash sub}}    
                \BinaryInfC{\( {\Gamma \contexcat \Delta \vdash M \esubst{ B }{ x } : \tau} \)}
            \end{prooftree}
            
            From the typing of $\expr{M}'  $ we can deduce that $\expr{M} = (\lambda x. M) B $ may be typed by:
            
            \begin{prooftree}
                \AxiomC{\( {\Gamma , {x}: \sigma^k \vdash M : \tau} \)}
                \LeftLabel{\redlab{T:abs}}
                \UnaryInfC{\( \Gamma \vdash \lambda x . M :  \sigma^k  \rightarrow \tau \)}
                \AxiomC{\( \Delta \vdash B : \pi \)}
                    \LeftLabel{\redlab{T:app}}
                \BinaryInfC{\( {\Gamma \contexcat \Delta \vdash (\lambda x. M) B : \tau}\)}
            \end{prooftree}
        
        \item When $\expr{M}'$ is reduced to via the Rule~\redlab{R:Fetch}
            
            \begin{prooftree}\hspace{-1cm}
                \AxiomC{$\headf{M} = x$}
                \AxiomC{$B = \bag{N_1, \dots ,N_k} \ , \ k\geq 1 $}
                \AxiomC{$ \#(x,M) = k $}
                \LeftLabel{\redlab{R:Fetch}}
                \TrinaryInfC{\(
                M\ \esubst{ B}{x } \redd M \headlin{ N_{1}/x } \esubst{ (B\linsetminus N_1)}{ x }  + \cdots + M \headlin{ N_{k}/x } \esubst{ (B\linsetminus N_k)}{x}
                \)}
            \end{prooftree}
        
        Let us consider two cases:
        
        \begin{myEnumerate}
            
            \item The bag $B$ has $k$ elements where $k > 1$, then we type $ M \headlin{ N_{i}/x } \esubst{ (B\linsetminus N_i)}{ x } $ with the derivation $\Pi_i$ to be:
            
                \begin{prooftree}
                        \AxiomC{\( \Gamma ,  {x}:\sigma^{k-1} \vdash M \headlin{ N_{1}/x } : \tau \)}
                         \AxiomC{\( \Delta \vdash (B\linsetminus N_1) : \sigma^{k-1} \)}
                    \LeftLabel{\redlab{T:ex \dash sub}}    
                    \BinaryInfC{\( \Gamma \contexcat \Delta \vdash M \headlin{ N_{1}/x } \esubst{ (B\linsetminus N_1)}{ x }  : \tau \)}
                \end{prooftree}
            
            We can type the sum with each derivation $ \Pi_i$ to be 
            
                \begin{prooftree}\hspace{-1cm}
                        \AxiomC{$ \Pi_1$}
                        \UnaryInfC{$ \Gamma \contexcat \Delta \vdash M \headlin{ N_{1}/x } \esubst{ (B\linsetminus N_1)}{ x } : \tau$}
                        \AxiomC{$ \Pi_k$}
                        \UnaryInfC{$ \Gamma \contexcat \Delta \vdash M \headlin{ N_{k}/x } \esubst{ (B\linsetminus N_k)}{x} : \tau$}
                        \UnaryInfC{$ \vdots $}
                    \BinaryInfC{$ \Gamma \contexcat \Delta \vdash M \headlin{ N_{1}/x } \esubst{ (B\linsetminus N_1)}{ x }  + \cdots + M \headlin{ N_{k}/x } \esubst{ (B\linsetminus N_k)}{x}: \tau$}
                \end{prooftree}
                
                By the anti-substitution lemma (Lemma~\ref{ch2lem:antisubt_lem}) we have that $\exists \ \Gamma_1, \Gamma_2$  such that  $\Gamma_1 , x:\sigma^k \vdash M: \tau$, and $\Gamma_2 \vdash N_i : \sigma$ with $\Gamma = \Gamma_1 \contexcat \Gamma_2$ and finally we have:
                
                \begin{prooftree}
                        \AxiomC{\( \Gamma_1 ,  {x}:\sigma^k \vdash M : \tau \)}
                         \AxiomC{\( \Delta \contexcat \Gamma_2 \vdash B : \sigma^{k} \)}
                    \LeftLabel{\redlab{T:ex \dash sub}}    
                    \BinaryInfC{\( \Gamma \contexcat \Delta \vdash M \esubst{ B }{ x }  : \tau \)}
                \end{prooftree}
            
            notice that we make use that $\Gamma_2 \vdash N_i : \sigma$ to ensure that the bag $B$ is well typed.
            
            \item The bag $B$ has one element, then we type $ M \headlin{ N_{i}/x } \esubst{ \oneb }{ x } $ with the derivation $\Pi$ to be:
            
                \begin{prooftree}
                        \AxiomC{\( \Gamma  \vdash M \headlin{ N_{1}/x } : \tau \)}
                         \AxiomC{\( \Delta \vdash \oneb : \omega \)}
                    \LeftLabel{\redlab{T:ex \dash sub}}    
                    \BinaryInfC{\( \Gamma \contexcat \Delta \vdash M \headlin{ N_{1}/x } \esubst{ \oneb}{ x }  : \tau \)}
                \end{prooftree}
            
        \end{myEnumerate}

        By the anti-substitution lemma (Lemma~\ref{ch2lem:antisubt_lem}) we have that $\exists \ \Gamma_1, \Gamma_2$  such that  $\Gamma_1 , x:\sigma \vdash M: \tau$ , and $\Gamma_2 \vdash N_1 : \sigma$ with $\Gamma = \Gamma_1 \contexcat \Gamma_2$ and finally we have:
                
            \begin{prooftree}
                    \AxiomC{\( \Gamma_1 ,  {x}:\sigma \vdash M : \tau \)}
                     \AxiomC{\( \Delta \contexcat \Gamma_2 \vdash \bag{N_1} : \sigma \)}
                \LeftLabel{\redlab{T:ex \dash sub}}    
                \BinaryInfC{\( \Gamma \contexcat \Delta \vdash M \headlin{ N_{1}/x } \esubst{ N }{ x }  : \tau \)}
            \end{prooftree}

        \item When $\expr{M}'$ is reduced to via the Rule~\redlab{R:TCont}

            \begin{prooftree}
                    \AxiomC{$   M \redd M'_{1} + \cdots + M'_{k} $}
                    \LeftLabel{\redlab{R:TCont}}
                    \UnaryInfC{$ C[M] \redd  C[M'_{1}] + \cdots +  C[M'_{k}] $}
            \end{prooftree}
        
            Hence the proof follows by the IH on $M$.
        
        \item When $\expr{M}'$ is reduced to via the Rule~\redlab{R:ECont}
        
            \begin{prooftree}
                    \AxiomC{$ \expr{M}  \redd \expr{M}'  $}
                    \LeftLabel{\redlab{R:ECont}}
                    \UnaryInfC{$D[\expr{M}]  \redd D[\expr{M}']  $}
            \end{prooftree} 
            
            Hence the proof follows by the IH on $M$.
        
        
        
            


    \end{myEnumerate}
    
\end{proof}

\subtlemfail*

\begin{proof}
By structural induction on $M$ with $\headf{M}=x$. There are three cases to be analyzed:

\begin{myEnumerate}
\item $M=x$.

This case follows trivially. First,  $x:\sigma \wfdash x:\sigma$ and $\Gamma=\emptyset$.  Second,  $x\headlin{N/x}=N$, by definition. Since $\Delta\wfdash N:\sigma$, by hypothesis, the result follows.



    \item $M = M'\ B$.
    
    In this case, $\headf{M'\ B} = \headf{M'} = x$, and by inversion of the typing derivation one has the following derivation:

    \begin{prooftree}
        \AxiomC{$\Gamma_1 , x:\sigma^m \wfdash M': \delta^{j}  \rightarrow \tau$}\
        \AxiomC{$\Gamma_2 \wfdash B :  \delta^{l} $}
    	\LeftLabel{\redlab{F{:}app}}
        \BinaryInfC{$ ( \Gamma_1 , x:\sigma^m) \contexcat \Gamma_2 \wfdash M'B:\tau $}    
    \end{prooftree}
 where $\Gamma,x:\sigma^k= ( \Gamma_1 , x:\sigma^m) \contexcat \Gamma_2$, $\delta$ is a strict type, and $j,l,m$ are non-negative  integers, possibly different with $m \geq 1$.
 
 By IH, we get $\Gamma_1 \contexcat \Delta , x:\sigma^{m-1} \wfdash M'\headlin{N/x}:\delta^{j} \rightarrow \tau $, which gives the following derivation: 
    \begin{prooftree}
        \AxiomC{$\Gamma_1 \contexcat \Delta , x:\sigma^{m-1} \wfdash M'\headlin{ N / x }:  \delta^{j}  \rightarrow \tau$}\
        \AxiomC{$\Gamma_2 \wfdash B :  \delta^{l} $}
    	\LeftLabel{\redlab{F{:}app}}
        \BinaryInfC{$(\Gamma_1 \contexcat \Delta , x:\sigma^{m-1}) \contexcat \Gamma_2  \wfdash ( M'\headlin{ N / x } ) B:\tau $}    
    \end{prooftree}
    Therefore, from \defref{ch2def:linsubfail}, one has $\Gamma \contexcat \Delta , x:\sigma^{k-1} \wfdash ( M' B) \headlin{ N / x }  :\tau $, and the result follows.
\item $M = M'\esubst{B }{ y}$.

In this case,  $\headf{M'\esubst{B }{ y}} = \headf{M'} = x$, with $x \not = y$, and by  inversion of the typing derivation one has the following derivation:

\begin{prooftree}
    \AxiomC{\( \Gamma_1 , {y}:\delta^{l} , x:\sigma^m \wfdash M' : \tau \)}
    \AxiomC{\( \Gamma_2 \wfdash B :  \delta^{j} \)}
	\LeftLabel{\redlab{F{:}ex \dash sub}}
    \BinaryInfC{\( ( \Gamma_1  ,  x:\sigma^m ) \contexcat \Gamma_2 \wfdash M' \esubst{B }{ y} : \tau \)}
\end{prooftree}
 where $\Gamma , x:\sigma^k = ( \Gamma_1  ,  x:\sigma^m ) \contexcat \Gamma_2 $, $\delta$ is a strict type and $j,l,m$ are positive integers with $m \geq 1$.
By IH, we get $ ( \Gamma_1 , {y}:\delta^{l} , x:\sigma^{m-1}) \contexcat \Delta \wfdash M'\headlin{N/x}:\tau$ and 
\begin{prooftree}
    \AxiomC{\(  ( \Gamma_1 , {y}:\delta^{l} , x:\sigma^{m-1}) \contexcat \Delta \wfdash  M' \headlin{ N / x } : \tau \)}
    \AxiomC{\( \Gamma_2 \wfdash B : \delta^{j} \)}
	\LeftLabel{\redlab{F{:}ex \dash sub}}
    \BinaryInfC{\(  ( \Gamma_1 , {y}:\delta^{l} , x:\sigma^{m-1}) \contexcat \Delta \contexcat \Gamma_2  \wfdash M' \headlin{ N / x } \esubst{ B }{ y} : \tau \)}
\end{prooftree}
\end{myEnumerate}

From \defref{ch2def:linsubfail}, $M' \esubst{ B }{ y} \headlin{ N / x } = M' \headlin{ N / x} \esubst{ B }{ y}$, therefore, $\Gamma \contexcat \Delta , x:\sigma^{k-1} \wfdash (M'\esubst{ B }{ y})\headlin{ N / x }:\tau$ and  the result follows.
\end{proof}

\applamrfailsr*

\begin{proof} By structural induction on the reduction rules. We proceed by analysing the rule applied in $\expr{M}$. There are seven cases:

\begin{myEnumerate}

	\item Rule~$\redlab{R:Beta}$.
	
	Then $\expr{M} = (\lambda x . M)B \redd M\ \esubst{B}{x}=\expr{M}'$.

 	Since $\Gamma\wfdash \expr{M}:\tau$, by \revo{A15}{ inversion of the typing derivation} one has the following derivation:
	\begin{prooftree}
			\AxiomC{$ \Gamma' , {x}:\sigma^{j}  \wfdash  M: \tau $}
			\LeftLabel{\redlab{F{:}abs}}
            \UnaryInfC{$ \Gamma' \wfdash \lambda x. M: \sigma^{j} \rightarrow \tau $}
            \AxiomC{$\Delta \wfdash B: \sigma^{k} $}
			\LeftLabel{\redlab{F{:}app}}
		\BinaryInfC{$ \Gamma' \contexcat \Delta \wfdash (\lambda x. M) B:\tau $}
	\end{prooftree}
	for $\Gamma = \Gamma' \contexcat \Delta $. Notice that
    \begin{prooftree}
                \AxiomC{$ \Gamma' , {x}:\sigma^{j} \wfdash  M: \tau $}
                      \AxiomC{$\Delta \wfdash B:\sigma^{k}  $}
                \LeftLabel{\redlab{F{:}ex \dash sub}}
            \BinaryInfC{$ \Gamma' \contexcat \Delta \wfdash M \esubst{ B }{ x }:\tau $}
    \end{prooftree}
    
    Therefore, $ \Gamma\wfdash \expr{M}':\tau$ and the result follows.

	\item  Rule~$\redlab{R:Fetch}$.

	Then $ \expr{M} = M\ \esubst{B}{x}$, where $B=  \bag{N_1, \dots ,N_k}$ , $k\geq 1$, $ \#(x,M) = k $, and $\headf{M} = x$. The reduction is as follows: 
	
	\begin{prooftree}\hspace{-1cm}
    \AxiomC{$\headf{M} = x$}
    \AxiomC{$B = \bag{N_1, \dots ,N_k} \ , \ k\geq 1 $}
    \AxiomC{$ \#(x,M) = k $}
    \LeftLabel{\redlab{R:Fetch}}
    \TrinaryInfC{\(
    M\ \esubst{ B}{x } \rightarrow M \headlin{ N_{1}/x } \esubst{ (B\linsetminus N_1)}{ x }  + \cdots + M \headlin{ N_{k}/x } \esubst{ (B\linsetminus N_k)}{x}
    \)}
    \end{prooftree}

To simplify the proof we take $k=2$, as the case $k>2$ is similar. Therefore,  by inversion of the typing derivation and $B=\bag{N_1,N_2}$: 
    \begin{prooftree}\hspace{-1cm}
    \small
             \AxiomC{\(\Gamma' , x:\sigma \wedge \sigma \wfdash  M: \tau\)}
    			\AxiomC{\(  \Delta_1 \wfdash N_1 : \sigma \)}
    				\AxiomC{\( \Delta_{2} \wfdash N_{2} : \sigma   \)}
    				\AxiomC{\(  \)}
                    \LeftLabel{\redlab{F{:}\oneb}}
                    \UnaryInfC{\( \wfdash \oneb : \omega \)}
    			\LeftLabel{\redlab{F{:}bag}}
    			\BinaryInfC{\( \Delta_2  \wfdash \bag{N_2}: \sigma   \)}
    			\LeftLabel{\redlab{F{:}bag}}
             \BinaryInfC{\(\Delta  \wfdash B: \sigma \wedge \sigma \) }
    	\LeftLabel{\redlab{F{:}ex \dash sub}}
    	\BinaryInfC{\(\Gamma' \contexcat \Delta  \wfdash   M \esubst{ B }{ x } : \tau\)}
    \end{prooftree}
    
where $\Delta= \Delta_1 \contexcat \Delta_2$ and $\Gamma = \Gamma' \contexcat \Delta $. By the Substitution Lemma (Lemma~\ref{ch2lem:subt_lem_fail}), there exists a derivation $\Pi_1$ of  $(\Gamma' , x:\sigma ) \contexcat \Delta_1 \wfdash   M \headlin{ N_{1}/x } : \tau $ and a derivation $\Pi_2$ of $(\Gamma' , x:\sigma ) \contexcat \Delta_2 \wfdash   M \headlin{ N_{2}/x } : \tau $. Therefore, one has the following derivation:

\begin{adjustwidth}{-1cm}{}
    \begin{prooftree}
        \small
            \AxiomC{\( \Pi_1 \) }
            \AxiomC{\( \Delta_2 \wfdash \bag{N_2} : \sigma \) }
        \LeftLabel{\redlab{F{:}ex \dash sub}}
        \BinaryInfC{\( \Gamma' \contexcat \Delta  \wfdash M \headlin{ N_{1}/x } \esubst{ \bag{N_2}}{ x }  : \tau\ \) }
            \AxiomC{\( \Pi_2 \) }
            \AxiomC{\( \Delta_1 \wfdash \bag{N_1} : \sigma \) }
        \LeftLabel{\redlab{F{:}ex \dash sub}}
        \BinaryInfC{\( \Gamma' \contexcat \Delta  \wfdash M \headlin{ N_2/x } \esubst{ \bag{N_1} }{x }  : \tau\  \) }
    	\LeftLabel{\redlab{F{:}sum}}
        \BinaryInfC{\(\Gamma' \contexcat \Delta  \wfdash     M \headlin{ N_{1}/x } \esubst{ \bag{N_2} }{ x } +  M \headlin{ N_{2}/x } \esubst{ \bag{N_1} }{x } : \tau\)}
    \end{prooftree}
\end{adjustwidth}

Assuming  $ \expr{M}'  =   \headlin{ N_{1}/x } \esubst{\bag{N_2}}{ x }  + M \headlin{ N_{2}/x } \esubst{  \bag{N_1}}{x }$, the result follows.

\item Rule~$ \redlab{R:Fail} $.

Then $\expr{M} =  M\ \esubst{ B}{x } $ where $B = \bag{N_1, \dots \cdot ,N_k} \ , \ k\geq 0 $ , $ \#(x,M) \not = k $ and we can perform the following reduction:

\begin{prooftree}
    \AxiomC{$\#(x,M) \neq \size{B}$}
    \AxiomC{\( \widetilde{y} = (\mfv{M}\setminus x )\uplus \mfv{B}\)}
    \LeftLabel{\redlab{R:Fail}}
    \BinaryInfC{\( M\ \esubst{ B}{x } \redd \sum_{\perm{B}} \fail^{\widetilde{y}}\)}
\end{prooftree}   
with $\expr{M}'=\sum_{\perm{B}} \fail^{\widetilde{y}}$. By hypothesis, one has the derivation:

\begin{prooftree}
            \AxiomC{\( \Delta \wfdash B :  \sigma^{j} \)}
            \AxiomC{\( \Gamma' , {x}:\sigma^{k} \wfdash M : \tau \)}
        \LeftLabel{\redlab{F{:}ex \dash sub}}    
        \BinaryInfC{\( \Gamma' \contexcat \Delta \wfdash M \esubst{ B }{ x } : \tau \)}
    \end{prooftree}
Notice that we also have from $\#(x,M) \neq \size{B}$ that $j\neq k$.Hence $\Gamma = \Gamma' \contexcat \Delta $ and we may type the following:
    
    \begin{prooftree}
        \AxiomC{\( \)}
        \LeftLabel{\redlab{F{:}fail}}
        \UnaryInfC{$ \Gamma \wfdash \fail^{\widetilde{y}} : \tau$}
        \AxiomC{\( \cdots \)}
        \AxiomC{\( \)}
        \LeftLabel{\redlab{F{:}fail}}
        \UnaryInfC{$\Gamma \wfdash \fail^{\widetilde{y}} : \tau$}
        \LeftLabel{\redlab{F{:}sum}}
        \TrinaryInfC{$ \Gamma \wfdash \sum_{\perm{B}} \fail^{\widetilde{y}}: \tau$}
    \end{prooftree}

\item Rule~$\redlab{R:Cons_1}$.

Then $\expr{M} =   \fail^{\widetilde{x}} \ B $ where $B = \bag{N_1, \dots ,N_k} $ , $k \geq 0$ and we can perform the following reduction:

\begin{prooftree}
    \AxiomC{$\size{B} = k$}
    \AxiomC{\( \widetilde{y} = \mfv{B} \)}
    \LeftLabel{$\redlab{R:Cons_1}$}
    \BinaryInfC{\( \fail^{\widetilde{x}}\ B  \redd \sum_{\perm{B}} \fail^{\widetilde{x} \uplus \widetilde{y}} \)}
\end{prooftree}
where $\expr{M}'=\sum_{\perm{B}} \fail^{\widetilde{x} \uplus \widetilde{y}}$. By hypothesis and inversion of the typing derivation, there exists the following derivation:

    \begin{prooftree}
        \AxiomC{\( \)}
        \LeftLabel{\redlab{F{:}fail}}
        \UnaryInfC{\( \Gamma' \wfdash \fail^{\widetilde{x}} : \pi' \rightarrow \tau \)}
        \AxiomC{\( \Delta \wfdash B : \pi \)}
            \LeftLabel{\redlab{F{:}app}}
        \BinaryInfC{\( \Gamma' \contexcat \Delta \wfdash \fail^{\widetilde{x}} \ B : \tau\)}
    \end{prooftree}

Hence $\Gamma = \Gamma' \contexcat \Delta $ and we may type the following:

    \begin{prooftree}
        \AxiomC{\( \)}
        \LeftLabel{\redlab{F{:}fail}}
        \UnaryInfC{$ \Gamma \wfdash \fail^{\widetilde{x} \uplus \widetilde{y}} : \tau$}
        \AxiomC{\( \cdots \)}
        \AxiomC{\( \)}
        \LeftLabel{\redlab{F{:}fail}}
        \UnaryInfC{$\Gamma \wfdash \fail^{\widetilde{x} \uplus \widetilde{y}} : \tau$}
        \LeftLabel{\redlab{F{:}sum}}
        \TrinaryInfC{$ \Gamma \wfdash \sum_{\perm{B}} \fail^{\widetilde{x} \uplus \widetilde{y}}: \tau$}
    \end{prooftree}

\item Rule~$\redlab{R:Cons_2}$.

Then $\expr{M} =   \fail^{\widetilde{z}}\ \esubst{B}{x} $ where $B = \bag{N_1, \dots ,N_k} $ , $k \geq 1$ and 
we can perform the following reduction:

\begin{prooftree}
    \AxiomC{$\size{B} = k$}
    \AxiomC{\(  \#(x , \widetilde{z}) + k  \not= 0 \)}
    \AxiomC{\( \widetilde{y} = \mfv{B} \)}
    \LeftLabel{$\redlab{R:Cons_2}$}
    \TrinaryInfC{\( \fail^{\widetilde{z}}\ \esubst{B}{x}  \redd \sum_{\perm{B}} \fail^{(\widetilde{z} \setminus x) \uplus\widetilde{y}} \)}
\end{prooftree}

where $\expr{M}'=\sum_{\perm{B}} \fail^{(\widetilde{z} \setminus x) \uplus \widetilde{y}}$. By hypothesis and inversion of the typing derivation, there exists a derivation:

    \begin{prooftree}
            \AxiomC{\( \dom{\core{(\Gamma' , {x}:\sigma^{k})}}=\widetilde{z} \)}
            \LeftLabel{\redlab{F{:}fail}}
            \UnaryInfC{\( \Gamma' ,{x}:\sigma^{k}\wfdash  \fail^{\widetilde{z}} : \tau  \)}
            \AxiomC{\( \Delta \wfdash B : \sigma^{j} \)}
        \LeftLabel{\redlab{F{:}ex \dash sub}}    
        \BinaryInfC{\( \Gamma' \contexcat \Delta \wfdash \fail^{\widetilde{z}} \esubst{ B }{ x } : \tau \)}
    \end{prooftree}

Hence $\Gamma = \Gamma' \contexcat \Delta $ and we may type the following:

    \begin{prooftree}
        \AxiomC{\( \)}
        \LeftLabel{\redlab{F{:}fail}}
        \UnaryInfC{$ {\Gamma} \wfdash \fail^{(\widetilde{z} \setminus x) \uplus\widetilde{y}} : \tau$}
        \AxiomC{$ \cdots $}
         \AxiomC{\( \)}
        \LeftLabel{\redlab{F{:}fail}}
        \UnaryInfC{$ {\Gamma} \wfdash \fail^{(\widetilde{z} \setminus x) \uplus\widetilde{y}} : \tau$}
        \LeftLabel{\redlab{F{:}sum}}
        \TrinaryInfC{$ {\Gamma} \wfdash \sum_{\perm{B}} \fail^{(\widetilde{z} \setminus x) \uplus\widetilde{y}} : \tau$}
    \end{prooftree}
    
    \item Rule~$\redlab{R:TCont}$.

Then $\expr{M} = C[M]$ and the reduction is as follows:

\begin{prooftree}
        \AxiomC{$   M \redd  M'_{1} + \cdots +  M'_{l} $}
        \LeftLabel{\redlab{R:TCont}}
        \UnaryInfC{$ C[M] \redd  C[M'_{1}] + \cdots +  C[M'_{l}] $}
\end{prooftree}

where $\expr{M}'= C[M'_{1}] + \cdots +  C[M'_{l}]$. 
The proof proceeds by analysing the context $C$:

    \begin{enumerate}
    \item $C=[\cdot]\ B$.
    
    In this case $\expr{M}=M \ B$, for some $B$, and the following derivation holds:
\begin{prooftree}
    \AxiomC{\(  \Gamma' \wfdash  M:  \sigma^{j} \rightarrow \tau \)}
    \AxiomC{\( \Delta \wfdash  B : \sigma^{k} \)}
        \LeftLabel{\redlab{F{:}app}}
    \BinaryInfC{\( \Gamma' \contexcat \Delta \wfdash  M\ B : \tau\)}
\end{prooftree}

where $\Gamma = \Gamma' \contexcat \Delta $.

Since $\Gamma'\wfdash M: \sigma^j \rightarrow \tau$ and $M\redd M_1'+\ldots + M_l'$, it follows by IH that $\Gamma'\wfdash M_1'+\ldots + M_l':\sigma^j \rightarrow \tau$. By applying \redlab{F{:}sum}, one has $\Gamma'\wfdash M_i' : \sigma^j \rightarrow \tau$, for $i=1,\ldots, l$.  Therefore, we may type the following:

\begin{prooftree}
\AxiomC{\(  \forall i \in {1 , \cdots , l} \)}
\AxiomC{\(  \Gamma' \wfdash  M'_{i}: \sigma^{j} \rightarrow \tau \)}
\AxiomC{\( \Delta \wfdash  B :  \sigma^{k} \)}
\LeftLabel{\redlab{F{:}app}}
\BinaryInfC{\(  \Gamma' \contexcat \Delta \wfdash (M'_{i}\ B):  \tau \)}
 \LeftLabel{\redlab{F{:}sum}}
    \BinaryInfC{\( \Gamma' \contexcat \Delta \wfdash (M'_{1}\ B) + \cdots +  (M'_{l} \ B) : \tau\)}
\end{prooftree}

Thus, $\Gamma\wfdash \expr{M'}: \tau$, and the result follows.

    \item $C=([\cdot])\esubst{B}{x}$.
    
    This case is similar to the previous one.
\end{enumerate}
	\item Rule~$ \redlab{R:ECont} $.
	
Then $\expr{M} = D[\expr{M}'']$ where $\expr{M}'' \rightarrow \expr{M}'''$ then we can perform the following reduction:
\begin{prooftree}
        \AxiomC{$ \expr{M}''  \redd \expr{M}'''  $}
        \LeftLabel{$\redlab{R:ECont}$}
        \UnaryInfC{$D[\expr{M}'']  \redd D[\expr{M}''']  $}
\end{prooftree}

Hence $\expr{M}' =  D[\expr{M}'''] $.
The proof proceeds by analysing the context $D$:

\begin{enumerate}
    \item $D= [\cdot] + \expr{N}$. In this case $\expr{M}= \expr{M}''+\expr{N}$ by inversion of the typing derivation:
\begin{prooftree}
    \AxiomC{$ \Gamma \wfdash  \expr{M}'' : \tau$}
    \AxiomC{$ \Gamma \wfdash  \expr{N} : \tau$}
    \LeftLabel{\redlab{F{:}sum}}
    \BinaryInfC{$ \Gamma \wfdash  \expr{M}''+\expr{N}: \tau$}
\end{prooftree}

Since $\Gamma\vdash \expr{M}^{''}:\tau $ and $\expr{M}^{''}\redd \expr{M}^{'''}$, by IH, it follows that $\Gamma\vdash \expr{M}^{'''}:\tau$  and we may type the following:

\begin{prooftree}
    \AxiomC{$ \Gamma \wfdash  \expr{M}''' : \tau$}
    \AxiomC{$ \Gamma \wfdash  \expr{N} : \tau$}
    \LeftLabel{\redlab{F{:}sum}}
    \BinaryInfC{$ \Gamma \wfdash  \expr{M}'''+\expr{N}: \tau$}
\end{prooftree}
Therefore, $\Gamma\vdash \expr{M'}:\tau$ and the  result follows.

    \item $D= \expr{N} + [\cdot]$.
      This case is similar to the previous one.
\end{enumerate}
\end{myEnumerate}
\end{proof}

\section{Appendix to \texorpdfstring{\secref{ch2ss:typeshar}}{}}
\label{ch2app:typeshar}

\Consistencyreductions*


\begin{proof}
    \secondrev{By structural induction on the reduction rules. We will consider two key reduction rules, the other cases follow analogously via  application of the IH.\\
\begin{enumerate}
 \item Rule \redlab{RS{:}Ex \dash Sub}. In this case, we have
    \begin{prooftree}\hspace{-1cm}
        \AxiomC{$B = \bag{M_1}
        \cdots  \bag{M_k} \qquad k \geq  1 $}
        \AxiomC{$ M \not= \fail^{\widetilde{y}} $}
        \LeftLabel{\redlab{RS{:}Ex \dash Sub}}
        \BinaryInfC{\( \!M[x_1,\ldots, x_k \leftarrow x]\esubst{ B }{ x } \redd \sum_{B_i \in \perm{B}}M\linexsub{B_i(1)/x_1} \cdots \linexsub{B_i(k)/x_k}    \)}
     \end{prooftree}
    Notice that if a bag is consistent then each element in the bag is consistent, that is, for any permutation $B_i$ of the bag $B$ then each $B_i(n)$ is consistent. Then, the assumption of consistency for $(M[\widetilde{x} \leftarrow x])\esubst{ B }{ x } $,  along with each element of the bag being consistent implies consistency of $ \sum_{B_i \in \perm{B}}M\linexsub{B_i(1)/x_1} \cdots \linexsub{B_i(k)/x_k} $ for each permutation of~$B$.\\
  \item Rule \redlab{RS{:}Lin \dash Fetch}. 
  In this case, we have
    \begin{prooftree}
     \AxiomC{$ \headf{M} = x$}
         \LeftLabel{\redlab{RS{:}Lin\dash Fetch}}
         \UnaryInfC{\(  M \linexsub{N/x} \redd  M \headlin{ N/x } \)}
    \end{prooftree}
    This case follows from the fact that  $M \headlin{ N/x }$  preserves consistency. The argument is by structural induction, with   base case of $M = x$ together with the fact that $N$ is consistent trivially implies that $x \headlin{ N/x }$ must also be consistent. As for the inductive step, notice that `adding' $N$ to the structure of $M$ does not break any of the consistency requirements: the consistency of $M \linexsub{N/x}$ implies that   the free variables of $M$ and $N$ are disjoint.
   \end{enumerate}
   }
\end{proof}


\explemfailnofailshar*

\begin{proof}
By induction on the structure of $ M $:

    \begin{myEnumerate}
        
        \item When $M = x$ then we have $x \headlin{ N / x}   = N$ and may derive the derivation of $ \Gamma \vdash N: \tau $ with $x \not \in \dom{\Gamma}$. By taking $\Gamma_1 = \emptyset$ and $\Gamma_2 = \Gamma$ as $\Gamma = \emptyset \contexcat \Gamma$ the case follows as $ \Gamma \vdash N : \tau$ and
            
                \begin{prooftree}
                    \AxiomC{}
                    \LeftLabel{\redlab{TS:var}}
                    \UnaryInfC{\( x: \sigma \vdash x : \sigma\)}
                \end{prooftree}
        
        \item When $M = (M\ B)$ then we have that $(M\ B)\headlin{ N/x}  = (M \headlin{ N/x })\ B$. Let us consider two cases:
            
            \begin{myEnumerate}
                
                \item When $x \in \lfv{M \headlin{ N/x }}$
                    \begin{prooftree}
                        \AxiomC{\( \Gamma', x:\sigma^{k-1}  \vdash M \headlin{ N/x } : \pi \rightarrow \tau' \)}
                        \AxiomC{\( \Delta \vdash B : \pi \)}
                            \LeftLabel{\redlab{TS:app}}
                        \BinaryInfC{\( (\Gamma', x:\sigma^{k-1} ) \contexcat \Delta \vdash (M \headlin{ N/x })\ B : \tau'\)}
                    \end{prooftree}
                    
                    By the IH we have that $\Gamma', x:\sigma^{k-1} \vdash M \headlin{ N/x } : \pi \rightarrow \tau'$ implies that $\exists \  \Gamma_1', \Gamma_2$ such that  $\Gamma_1' , x:\sigma^k \vdash M: \tau$, and $\Gamma_2 \vdash N : \sigma$ with $\Gamma' = \Gamma_1' \contexcat \Gamma_2$.
                    
                    \begin{prooftree}
                        \AxiomC{\( \Gamma_1', x:\sigma^{k}  \vdash M \headlin{ N/x } : \pi \rightarrow \tau' \)}
                        \AxiomC{\( \Delta \vdash B : \pi \)}
                            \LeftLabel{\redlab{TS:app}}
                        \BinaryInfC{\( (\Gamma_1', x:\sigma^{k} ) \contexcat \Delta \vdash M\ B : \tau'\)}
                    \end{prooftree}
                    
                \item When $x \not \in \lfv{M \headlin{ N/x }}$
                    \begin{prooftree}
                      \AxiomC{\( \Gamma'  \vdash M \headlin{ N/x } : \pi \rightarrow \tau' \)}
                        \AxiomC{\( \Delta \vdash B : \pi \)}
                            \LeftLabel{\redlab{TS:app}}
                        \BinaryInfC{\( \Gamma' \contexcat \Delta \vdash (M \headlin{ N/x })\ B : \tau'\)}
                    \end{prooftree}
                    
                    By the IH we have that $\Gamma' \vdash M \headlin{ N/x } [\widetilde{y} \leftarrow y]: \pi \rightarrow \tau'$ implies that $\exists \  \Gamma_1', \Gamma_2$ such that  $\Gamma_1' , x:\sigma \vdash M[\widetilde{y} \leftarrow y]: \tau$, and $\Gamma_2 \vdash N : \sigma$ with $\Gamma' = \Gamma_1' \contexcat \Gamma_2$.
                    
                    \begin{prooftree}
                        \AxiomC{\( \Gamma_1', x:\sigma  \vdash M : \pi \rightarrow \tau' \)}
                        \AxiomC{\( \Delta \vdash B : \pi \)}
                            \LeftLabel{\redlab{TS:app}}
                        \BinaryInfC{\( (\Gamma_1', x:\sigma ) \contexcat \Delta \vdash M\ B : \tau'\)}
                    \end{prooftree}
                
            \end{myEnumerate}

        \item When $M = M[\widetilde{y} \leftarrow y] \esubst{B}{y}$ then we have that $(M [\widetilde{y} \leftarrow y]\esubst{B}{y})\headlin{ N/x } = (M\headlin{ N/x })\ [\widetilde{y} \leftarrow y] \esubst{B}{y}$ where $x \not = y$.
            
            \begin{myEnumerate}
                
                \item When $x \in \lfv{M \headlin{ N/x }}$:
                    
                    \begin{prooftree}
                            \AxiomC{\( \Gamma', x:\sigma^{k-1} ,  {\widetilde{y}}:\delta^{j} \vdash (M\headlin{ N/x }) : \tau \)}
                            \LeftLabel{ \redlab{TS{:}share}}
                            \UnaryInfC{\( \Gamma', x:\sigma^{k-1} ,  {y}:\delta^{j} \vdash (M\headlin{ N/x })[\widetilde{y} \leftarrow y] : \tau \)}
                       
                             \AxiomC{\( \Delta \vdash B : \delta^{j} \)}
                        \LeftLabel{\redlab{TS:ex \dash sub}}    
                        \BinaryInfC{\( \Gamma' \contexcat \Delta \vdash (M\headlin{ N/x }) [\widetilde{y} \leftarrow y]\esubst{ B }{ y } : \tau \)}
                    \end{prooftree}
                    
                    By the IH we have that $\Gamma', x:\sigma^{k-1} ,  {\widetilde{y}}:\delta^{j} \vdash (M\headlin{ N/x }) : \tau$ implies that $\exists \  \Gamma_1', \Gamma_2$ such that  $\Gamma_1' , x:\sigma^k , {\widetilde{y}}:\delta^{j}\vdash M: \tau$, and $\Gamma_2 \vdash N : \sigma$ with $\Gamma',  {y}:\delta^{j} = (\Gamma_1',  {y}:\delta^{j}) \contexcat \Gamma_2$.
                    \begin{prooftree}
                            \AxiomC{\( \Gamma'_1, x:\sigma^{k} ,  {\widetilde{y}}:\delta^{j} \vdash M : \tau \)}
                            \LeftLabel{ \redlab{TS{:}share}}
                            \UnaryInfC{\( \Gamma'_1, x:\sigma^{k} ,  {y}:\delta^{j} \vdash M[\widetilde{y} \leftarrow y] : \tau \)}
                            
                             \AxiomC{\( \Delta \vdash B : \delta^{j} \)}
                        \LeftLabel{\redlab{TS:ex \dash sub}}    
                        \BinaryInfC{\( \Gamma'_1 \contexcat \Delta \vdash (M)[\widetilde{y} \leftarrow y] \esubst{ B }{ y } : \tau \)}
                    \end{prooftree}
                    
                \item When $x \not \in \lfv{M \headlin{ N/x }}$:
                                        \begin{prooftree}
                            \AxiomC{\( \Gamma',  {\widetilde{y}}:\delta^{k} \vdash (M\headlin{ N/x }) : \tau \)}
                            \LeftLabel{ \redlab{TS{:}share}}
                            \UnaryInfC{\( \Gamma' ,  {y}:\delta^{k} \vdash (M\headlin{ N/x })[\widetilde{y} \leftarrow y] : \tau \)}
                            
                             \AxiomC{\( \Delta \vdash B : \delta^{k} \)}
                        \LeftLabel{\redlab{TS:ex \dash sub}}    
                        \BinaryInfC{\( \Gamma' \contexcat \Delta \vdash (M\headlin{ N/x })[\widetilde{y} \leftarrow y] \esubst{ B }{ y } : \tau \)}
                    \end{prooftree}
                    
                    By the IH we have that $\Gamma' ,  {y}:\delta^{k} \vdash (M\headlin{ N/x }) : \tau$ implies that $\exists \  \Gamma_1', \Gamma_2$ such that  $\Gamma_1' , x:\sigma, {\widetilde{y}}:\delta^{k} \vdash M: \tau$, and $\Gamma_2 \vdash N : \sigma$ with $\Gamma',  {y}:\delta^{k} = (\Gamma_1',  {y}:\delta^{k}) \contexcat \Gamma_2$.
                    
                    \begin{prooftree}
                            \AxiomC{\( \Gamma', x:\sigma ,  {\widetilde{y}}:\delta^{k} \vdash M : \tau \)}
                            \LeftLabel{ \redlab{TS{:}share}}
                            \UnaryInfC{\( \Gamma', x:\sigma ,  {y}:\delta^{k} \vdash M[\widetilde{y} \leftarrow y] : \tau \)}
                            
                             \AxiomC{\( \Delta \vdash B : \delta^{k} \)}
                        \LeftLabel{\redlab{TS:ex \dash sub}}    
                        \BinaryInfC{\( \Gamma' \contexcat \Delta \vdash (M[\widetilde{y} \leftarrow y]\headlin{ N/x }) \esubst{ B }{ y } : \tau \)}
                    \end{prooftree}
                    
            \end{myEnumerate}

        \item When $M = M[\widetilde{y} \leftarrow y] $ then we have that $(M [\widetilde{y} \leftarrow y])\headlin{ N/x } = (M\headlin{ N/x })\ [\widetilde{y} \leftarrow y]$ where $x \not = y$.
            
            \begin{myEnumerate}
                
                \item When $x \in \lfv{M \headlin{ N/x }}$:
                    
                    \begin{prooftree}
                            \AxiomC{\( \Gamma', x:\sigma^{k-1} ,  {\widetilde{y}}:\delta^{j} \vdash (M\headlin{ N/x }) : \tau \)}
                            \LeftLabel{ \redlab{TS{:}share}}
                            \UnaryInfC{\( \Gamma', x:\sigma^{k-1} ,  {y}:\delta^{j} \vdash (M\headlin{ N/x })[\widetilde{y} \leftarrow y] : \tau \)}
                    \end{prooftree}
                    
                    By the IH we have that $\Gamma', x:\sigma^{k-1} ,  {\widetilde{y}}:\delta^{j} \vdash (M\headlin{ N/x }) : \tau$ implies that $\exists \  \Gamma_1', \Gamma_2$ such that  $\Gamma_1' , x:\sigma^k , {\widetilde{y}}:\delta^{j}\vdash M: \tau$, and $\Gamma_2 \vdash N : \sigma$ with $\Gamma',  {y}:\delta^{j} = (\Gamma_1',  {y}:\delta^{j}) \contexcat \Gamma_2$.
                    
                    \begin{prooftree}
                            \AxiomC{\( \Gamma'_1, x:\sigma^{k} ,  {\widetilde{y}}:\delta^{j} \vdash M : \tau \)}
                            \LeftLabel{ \redlab{TS{:}share}}
                            \UnaryInfC{\( \Gamma'_1, x:\sigma^{k} ,  {y}:\delta^{j} \vdash M[\widetilde{y} \leftarrow y] : \tau \)}
                    \end{prooftree}
                    
                \item When $x \not \in \lfv{M \headlin{ N/x }}$:
                    
                    \begin{prooftree}
                            \AxiomC{\( \Gamma',  {\widetilde{y}}:\delta^{k} \vdash (M\headlin{ N/x }) : \tau \)}
                            \LeftLabel{ \redlab{TS{:}share}}
                            \UnaryInfC{\( \Gamma' ,  {y}:\delta^{k} \vdash (M\headlin{ N/x })[\widetilde{y} \leftarrow y] : \tau \)}
                    \end{prooftree}
                    
                    By the IH we have that $\Gamma' ,  {y}:\delta^{k} \vdash (M\headlin{ N/x }) : \tau$ implies that $\exists \  \Gamma_1', \Gamma_2$ such that  $\Gamma_1' , x:\sigma, {\widetilde{y}}:\delta^{k} \vdash M: \tau$, and $\Gamma_2 \vdash N : \sigma$ with $\Gamma',  {y}:\delta^{k} = (\Gamma_1',  {y}:\delta^{k}) \contexcat \Gamma_2$.
                    
                    \begin{prooftree}
                            \AxiomC{\( \Gamma', x:\sigma ,  {\widetilde{y}}:\delta^{k} \vdash M : \tau \)}
                            \LeftLabel{ \redlab{TS{:}share}}
                            \UnaryInfC{\( \Gamma', x:\sigma ,  {y}:\delta^{k} \vdash M[\widetilde{y} \leftarrow y] : \tau \)}
                    \end{prooftree}
                    
            \end{myEnumerate}

        \item When $M = M \linexsub{N /y} $ then we have that $(M \linexsub{N /y})\headlin{ N/x } = (M\headlin{ N/x })\ \linexsub{N /y}$ where $x \not = y$.
            
            \begin{myEnumerate}
                
                \item When $x \in \lfv{M \headlin{ N/x }}$:
                    
                    \begin{prooftree}
                        \AxiomC{\( \Delta \vdash N : \delta \qquad \Gamma', x:\sigma^{k-1} ,  {y}:\delta \vdash (M\headlin{ N/x }) : \tau \)}
                        \LeftLabel{\redlab{TS\!:\!ex\dash lin\dash sub}}
                        \UnaryInfC{\( \Gamma', x:\sigma^{k-1} ,  {y}:\delta , \Delta \vdash (M\headlin{ N/x }) \linexsub{N / y} : \tau \)}
                    \end{prooftree}
                    
                    By the IH we have that $\Gamma', x:\sigma^{k-1} ,  {\widetilde{y}}:\delta^{j} \vdash (M\headlin{ N/x }) : \tau$ implies that $\exists \  \Gamma_1', \Gamma_2$ such that  $\Gamma_1' , x:\sigma^k , {\widetilde{y}}:\delta^{j}\vdash M: \tau$, and $\Gamma_2 \vdash N : \sigma$ with $\Gamma',  {y}:\delta^{j} = (\Gamma_1',  {y}:\delta^{j}) \contexcat \Gamma_2$.

                    \begin{prooftree}
                        \AxiomC{\( \Delta \vdash N : \delta \qquad \Gamma'_1, x:\sigma^{k} ,  {y}:\delta \vdash M : \tau \)}
                        \LeftLabel{\redlab{TS\!:\!ex\dash lin\dash sub}}
                        \UnaryInfC{\( \Gamma'_1, x:\sigma^{k} ,  {y}:\delta , \Delta \vdash M \linexsub{N / y} : \tau \)}
                    \end{prooftree}
                    
                \item When $x \not \in \lfv{M \headlin{ N/x }}$:
                    
                    \begin{prooftree}
                        \AxiomC{\( \Delta \vdash N : \delta \qquad \Gamma',  {y}:\delta \vdash (M\headlin{ N/x }) : \tau \)}
                        \LeftLabel{\redlab{TS\!:\!ex\dash lin\dash sub}}
                        \UnaryInfC{\( \Gamma' ,  {y}:\delta , \Delta \vdash (M\headlin{ N/x }) \linexsub{N / y} : \tau \)}
                    \end{prooftree}
                    
                    By the IH we have that $\Gamma' ,  {y}:\delta^{k} \vdash (M\headlin{ N/x }) : \tau$ implies that $\exists \  \Gamma_1', \Gamma_2$ such that  $\Gamma_1' , x:\sigma, {\widetilde{y}}:\delta^{k} \vdash M: \tau$, and $\Gamma_2 \vdash N : \sigma$ with $\Gamma',  {y}:\delta^{k} = (\Gamma_1',  {y}:\delta^{k}) \contexcat \Gamma_2$.
                    
                    \begin{prooftree}
                        \AxiomC{\( \Delta \vdash N : \delta \qquad \Gamma'_1, x:\sigma ,  {y}:\delta \vdash M : \tau \)}
                        \LeftLabel{\redlab{TS\!:\!ex\dash lin\dash sub}}
                        \UnaryInfC{\( \Gamma'_1, x:\sigma ,  {y}:\delta , \Delta \vdash M \linexsub{N / y} : \tau \)}
                    \end{prooftree}
                    
            \end{myEnumerate}

        \item When $M = \lambda y . M[\widetilde{y} \leftarrow y]$ then linear head substitution is undefined on this term as $\headf{M} \not = x$.

        \item When $M = \fail^{\widetilde{x}}$ then $M$ is not well typed.

    \end{myEnumerate}

\end{proof}

\subexponeshar*

\begin{proof}
By induction on the reduction rule applied.
There are five possible cases. 
    
    \begin{myEnumerate}
    
        \item When $\expr{M}'$ is reduced to via the Rule~\redlab{RS:Beta}:
    
            \begin{prooftree}
                \AxiomC{}
                \LeftLabel{\redlab{RS:Beta}}
                \UnaryInfC{\((\lambda x. M[\widetilde{x} \leftarrow x ]) B \redd M[\widetilde{x} \leftarrow x ]\ \esubst{B}{x}\)}
            \end{prooftree}
            
            Then $\expr{M}' = M[\widetilde{x} \leftarrow x ]\ \esubst{B}{x} $ can be type as followed:
            
            \begin{prooftree}
                    \AxiomC{\( {\Gamma ,  {x}:\sigma^{k} \vdash M[\widetilde{x} \leftarrow x ] : \tau} \)}
                     \AxiomC{\( \Delta \vdash B : \sigma^{k} \)}
                \LeftLabel{\redlab{TS:ex \dash sub}}    
                \BinaryInfC{\( {\Gamma \contexcat \Delta \vdash M[\widetilde{x} \leftarrow x ] \esubst{ B }{ x } : \tau} \)}
            \end{prooftree}
            
            From the typing of $\expr{M}'  $ we can deduce that $\expr{M} = (\lambda x. M[\widetilde{x} \leftarrow x ]) B $ may be typed by:
            
            \begin{prooftree}
                \AxiomC{\( {\Gamma , {x}: \sigma^k \vdash M[\widetilde{x} \leftarrow x ] : \tau} \)}
                \LeftLabel{\redlab{TS:abs}}
                \UnaryInfC{\( \Gamma \vdash \lambda x . M[\widetilde{x} \leftarrow x ] :  \sigma^k  \rightarrow \tau \)}
                \AxiomC{\( \Delta \vdash B :  \sigma^k \)}
                    \LeftLabel{\redlab{TS:app}}
                \BinaryInfC{\( {\Gamma \contexcat \Delta \vdash (\lambda x. M[\widetilde{x} \leftarrow x ]) B : \tau}\)}
            \end{prooftree}

        \item When $\expr{M}'$ is reduced to via the Rule~\redlab{RS{:}Ex \dash Sub}:

         \begin{prooftree}\hspace{-1cm}
            \AxiomC{$B = \bag{M_1}
            \cdots  \bag{M_k} \qquad k \geq  1 $}
            \AxiomC{$ M \not= \fail^{\widetilde{y}} $}
            \LeftLabel{\redlab{RS{:}Ex \dash Sub}}
            \BinaryInfC{\( \!M[x_1,\ldots, x_k \leftarrow x]\esubst{ B }{ x } \redd \sum_{B_i \in \perm{B}}M\linexsub{B_i(1)/x_1} \cdots \linexsub{B_i(k)/x_k}    \)}
         \end{prooftree}

        Then $\expr{M}' = \!M[x_1,\ldots, x_k \leftarrow x]\esubst{ B }{ x } $ can be type as followed:

         \begin{prooftree}\hspace{-1cm}
            \AxiomC{\( \Gamma  , x_1:\sigma, \cdots , x_k:\sigma \vdash M : \tau \)}
            \AxiomC{\( \Delta_1 \vdash B_i(1) : \sigma  \)}
            \BinaryInfC{\( \vdots \)}
            \AxiomC{\( \Delta_k \vdash B_i(k) : \sigma  \)}
            \LeftLabel{\redlab{TS\!:\!ex\dash lin\dash sub}}
            \BinaryInfC{\( \Gamma , \Delta_1 , \cdots , \Delta_k \vdash M\linexsub{B_i(1)/x_1} \cdots \linexsub{B_i(k)/x_k} : \tau \)}
            \AxiomC{$ \forall B_i \in \perm{B} $}
            \LeftLabel{\redlab{TS{:}sum}}
            \BinaryInfC{$ \Gamma , \Delta_1 , \cdots , \Delta_k  \vdash \sum_{B_i \in \perm{B}}M\linexsub{B_i(1)/x_1} \cdots \linexsub{B_i(k)/x_k}: \tau$}
        \end{prooftree}

         From the typing of $\expr{M}'  $ we can deduce that $\expr{M} = (\lambda x. M[\widetilde{x} \leftarrow x ]) B $ may be typed by:

        \begin{prooftree}
            \AxiomC{\( \Delta_1 \vdash M_1 : \sigma\)}
            \AxiomC{\( \Delta_k \vdash M_k : \sigma\)}
            \UnaryInfC{\(\ \vdots \)}
            \LeftLabel{\redlab{TS{:}bag}}
            \BinaryInfC{\( \Delta_1 , \cdots , \Delta_k \vdash B : \sigma^k \)}
            \AxiomC{\( \Gamma  , x_1:\sigma, \cdots , x_k:\sigma \vdash M : \tau \)}
            \UnaryInfC{\( \qquad \Gamma , x:\sigma^k \vdash M [\widetilde{x} \leftarrow x]: \tau \)}
            \LeftLabel{\redlab{TS\!:ex \dash sub}} 
            \BinaryInfC{\( \Gamma , \Delta_1 , \cdots , \Delta_k  \vdash M[\widetilde{x} \leftarrow x] \esubst{ B }{ x } : \tau \)}
        \end{prooftree}

         \item When $\expr{M}'$ is reduced to via the Rule~\redlab{RS{:}Lin\dash Fetch}:
        
         \begin{prooftree}
            \AxiomC{$ \headf{M} = x$}
             \LeftLabel{\redlab{RS{:}Lin\dash Fetch}}
             \UnaryInfC{\(  M \linexsub{N/x} \redd  M \headlin{ N/x } \)}
        \end{prooftree}
        
        The result follow from \Cref{ch2lem:antisubt_lem_shar}.

        \item When $\expr{M}'$ is reduced to via the Rule~\redlab{RS:TCont}:
            \begin{prooftree}
                    \AxiomC{$   M \redd M'_{1} + \cdots + M'_{k} $}
                    \LeftLabel{\redlab{RS:TCont}}
                    \UnaryInfC{$ C[M] \redd  C[M'_{1}] + \cdots +  C[M'_{k}] $}
            \end{prooftree}
        
            Hence the proof follows by the IH on $M$.
        
        \item When $\expr{M}'$ is reduced to via the Rule~\redlab{RS:ECont}:
            \begin{prooftree}
                    \AxiomC{$ \expr{M}  \redd \expr{M}'  $}
                    \LeftLabel{\redlab{RS:ECont}}
                    \UnaryInfC{$D[\expr{M}]  \redd D[\expr{M}']  $}
            \end{prooftree} 
            
            Hence the proof follows by the IH on $M$.
        
        
        
            

    \end{myEnumerate}
    
\end{proof}

\lamrsharfailsubs*

\begin{proof}
By structural induction on $M$ with $\headf{M}=x$.
There are six cases to be analyzed:
\begin{myEnumerate}
\item $M=x$.

In this case, $x:\sigma \wfdash x:\sigma$ and $\Gamma=\emptyset$.  Observe that $x\headlin{N/x}=N$, since $\Delta\wfdash N:\sigma$, by hypothesis, the result follows.

    \item $M = M'\ B$.
    
    Then $\headf{M'\ B} = \headf{M'} = x$, and the derivation is the following by inversion of the typing derivation:
    \begin{prooftree}
        \AxiomC{$\Gamma_1 , x:\sigma \wfdash M': \delta^{j}  \rightarrow \tau$}\
        \AxiomC{$\Gamma_2 \wfdash B : \delta^{k} $}
    	\LeftLabel{\redlab{FS{:}app}}
        \BinaryInfC{$\Gamma_1 , \Gamma_2 , x:\sigma \wfdash M'B:\tau $}    
    \end{prooftree}

    where $\Gamma=\Gamma_1,\Gamma_2$, and  $j,k$ are non-negative integers, possibly different.  Since $\Delta \vdash N : \sigma$, by IH, the result holds for $M'$, that is,
    $$\Gamma_1 , \Delta \wfdash M'\headlin{ N / x }: \delta^{j}  \rightarrow \tau$$
    which gives the  derivation:


    \begin{prooftree}
        \AxiomC{$\Gamma_1 , \Delta \wfdash M'\headlin{ N / x }: \delta^{j}  \rightarrow \tau$}\
        \AxiomC{$\Gamma_2 \wfdash B : \delta^{k} $}
    	\LeftLabel{\redlab{FS{:}app}}
        \BinaryInfC{$\Gamma_1 , \Gamma_2 , \Delta \wfdash ( M'\headlin{ N / x } ) B:\tau $}    
    \end{prooftree}
    
      From \defref{ch2def:headlinfail},   $(M'B) \headlin{ N / x } = ( M'\headlin{ N / x } ) B$, therefore, $\Gamma, \Delta \wfdash  (M' B) \headlin{ N / x } :\tau $ and the result follows.
    
    \item $M = M'[\widetilde{y} \leftarrow y] $.
    
    Then $ \headf{M'[\widetilde{y} \leftarrow y]} = \headf{M'}=x$, for  $y\neq x$. Therefore by inversion of the typing derivation, 
    \begin{prooftree}
        \AxiomC{\( \Gamma_1 , y_1: \delta, \cdots, y_k: \delta , x: \sigma \wfdash M' : \tau \quad y\notin \Gamma_1 \quad k \not = 0\)}
        \LeftLabel{ \redlab{FS{:}share}}
        \UnaryInfC{\( \Gamma_1 , y: \delta^k, x: \sigma \wfdash M'[y_1 , \cdots , y_k \leftarrow y] : \tau \)}
    \end{prooftree}
    where $\Gamma=\Gamma_1 , y: \delta^k$. 
    By IH, the result follows for $M'$, that is, 
    $$\Gamma_1 , y_1: \delta, \cdots, y_k: \delta ,\Delta \wfdash M'\headlin{N/x} : \tau $$
    
    and we have the derivation:
    
    \begin{prooftree}
        \AxiomC{\( \Gamma_1 , y_1: \delta, \cdots, y_k: \delta , \Delta \wfdash  M' \headlin{ N / x} : \tau \quad y\notin \Gamma_1 \quad k \not = 0\)}
        \LeftLabel{ \redlab{FS{:}share} }
        \UnaryInfC{\( \Gamma_1 , y: \delta^k, \Delta \wfdash M' \headlin{ N / x} [\widetilde{y} \leftarrow y] : \tau \)}
    \end{prooftree}
    From \defref{ch2def:headlinfail} one has  $M'[\widetilde{y} \leftarrow y] \headlin{ N / x } = M' \headlin{ N / x} [\widetilde{y} \leftarrow y]$. Therefore, $\Gamma,\Delta\wfdash M'[\widetilde{y} \leftarrow y] \headlin{ N / x }:\tau$ and the result follows.
    \item $M = M'[ \leftarrow y] $.
    
    Then $ \headf{M'[ \leftarrow y]} = \headf{M'}=x$ with  $x \not  = y $, 
    \begin{prooftree}
        \AxiomC{\( \Gamma  , x: \sigma  \wfdash M : \tau\)}
        \LeftLabel{ \redlab{FS{:}weak} }
        \UnaryInfC{\(  \Gamma  , y: \omega, x: \sigma  \wfdash M[\leftarrow y]: \tau \)}
    \end{prooftree}
     and $M'[ \leftarrow y] \headlin{ N / x } = M' \headlin{ N / x} [ \leftarrow y]$. Then by the IH:
    \begin{prooftree}
        \AxiomC{\(  \Gamma , \Delta  \wfdash M \headlin{ N / x}: \tau\)}
        \LeftLabel{ \redlab{FS{:}weak}}
        \UnaryInfC{\(  \Gamma  , y: \omega, \Delta \wfdash M\headlin{ N / x}[\leftarrow y]: \tau \)}
    \end{prooftree}

\item $M = M'[\widetilde{y} \leftarrow y]\esubst{B }{ y}$.

Then $\headf{M'[\widetilde{y} \leftarrow y]\esubst{B }{ y}} = \headf{M'[\widetilde{y} \leftarrow y]} = x \not = y$  by inversion of the typing derivation we have: 

\begin{prooftree}
    \AxiomC{\( \Gamma_1 , \hat{y}:\delta^{k} , x:\sigma \wfdash M'[\widetilde{y} \leftarrow y] : \tau \)}
    \AxiomC{\( \Gamma_2 \wfdash B : \delta^{j} \)}
	\LeftLabel{\redlab{FS{:}ex \dash sub}}
    \BinaryInfC{\( \Gamma_1 , \Gamma_2 ,  x:\sigma \wfdash M' [\widetilde{y} \leftarrow y] \esubst{B }{ y} : \tau \)}
\end{prooftree}
 and $M' [\widetilde{y} \leftarrow y] \esubst{ B}{ y} \headlin{ N / x } = M' [\widetilde{y} \leftarrow y] \headlin{ N / x} \esubst{ B}{ y}$. By IH:

\begin{prooftree}
    \AxiomC{\( \Gamma_1 , \hat{y}:\delta^{k}, \Delta \wfdash  M' [\widetilde{y} \leftarrow y] \headlin{ N / x } : \tau \)}
    \AxiomC{\( \Gamma_2 \wfdash B : \delta^{j} \)}
	\LeftLabel{\redlab{FS{:}ex \dash sub}}
    \BinaryInfC{\( \Gamma_1 , \Gamma_2 ,  \Delta \wfdash M' [\widetilde{y} \leftarrow y] \headlin{ N / x } \esubst{ B }{ y} : \tau \)}
\end{prooftree}

    \item $M =  M' \linexsub {M'' /y}$.
    
    Then $\headf{M' \linexsub {M'' /y}} = \headf{M'} = x \not = y$, by inversion of the typing derivation we have: 
    
    \begin{prooftree}
        \AxiomC{\( \Delta \wfdash M'' : \delta \)}
        \AxiomC{\( \Gamma  , y:\delta , x: \sigma \wfdash M : \tau \)}
        \LeftLabel{ \redlab{FS{:}ex \dash lin \dash sub} }
        \BinaryInfC{\( \Gamma_1, \Gamma_2 , x: \sigma \wfdash M' \linexsub {M'' /y} : \tau \)}
    \end{prooftree}
     and $M' \linexsub {M'' /y}  \headlin{ N / x } = M'  \headlin{N / x } \linexsub {M'' /y}$. Then by the IH:
    
    \begin{prooftree}
        \AxiomC{\( \Delta \wfdash M'' : \delta \)}
        \AxiomC{\( \Gamma  , y:\delta , \Delta  \wfdash M'  \headlin{N / x } : \tau \)}
        \LeftLabel{ \redlab{FS{:}ex \dash lin \dash sub} }
        \BinaryInfC{\( \Gamma_1, \Gamma_2 , \Delta  \wfdash M'  \headlin{N / x } \linexsub {M'' /y} : \tau \)}
    \end{prooftree}
    \end{myEnumerate}
\end{proof}

\applamrsharfailsr*

\begin{proof} By structural induction on the reduction rule from \figref{ch2fig:share-reductfailure} applied in $\expr{M}\redd \expr{N}$. There are nine cases to be analyzed:

\begin{myEnumerate}

	\item Rule~$\redlab{RS{:}Beta}$.
	
	Then $\expr{M} = (\lambda x. M[\widetilde{x} \leftarrow x]) B $  and the reduction is:
	  \begin{prooftree}
        \AxiomC{}
        \LeftLabel{\redlab{RS{:}Beta}}
        \UnaryInfC{\((\lambda x. M[\widetilde{x} \leftarrow x]) B \redd M[\widetilde{x} \leftarrow x]\ \esubst{ B }{ x }\)}
     \end{prooftree}

 	where $ \expr{M}'  =  M[\widetilde{x} \leftarrow x]\ \esubst{ B }{ x }$. Since $\Gamma\wfdash \expr{M}:\tau$ we get the following derivation by inversion of the typing derivation:
	\begin{prooftree}
			\AxiomC{$ \Gamma' , x_1:\sigma , \cdots , x_j:\sigma  \wfdash  M: \tau $}
			\LeftLabel{ \redlab{FS{:}share} }
			\UnaryInfC{$  \Gamma' , x:\sigma^{j}  \wfdash  M[\widetilde{x} \leftarrow x]: \tau $}
			\LeftLabel{ \redlab{FS{:}abs \dash sh} }
            \UnaryInfC{$ \Gamma' \wfdash \lambda x. M[\widetilde{x} \leftarrow x]: \sigma^{j} \rightarrow \tau $}
              \AxiomC{$\Delta \wfdash B: \sigma^{k} $}
			\LeftLabel{ \redlab{FS{:}app} }
		\BinaryInfC{$ \Gamma' , \Delta \wfdash (\lambda x. M[\widetilde{x} \leftarrow x]) B:\tau $}
	\end{prooftree}
	for $\Gamma = \Gamma' , \Delta $ and $x\notin \dom{\Gamma'}$. 
	Notice that: 

    \begin{prooftree}
                \AxiomC{$ \Gamma' , x_1:\sigma , \cdots , x_j:\sigma  \wfdash  M: \tau $}
			\LeftLabel{ \redlab{FS{:}share} }
			\UnaryInfC{$  \Gamma' , x:\sigma^{j}  \wfdash  M[\widetilde{x} \leftarrow x]: \tau $}
                \AxiomC{$\Delta \wfdash B:\sigma^{k}  $}
                \LeftLabel{ \redlab{FS{:}ex \dash sub} }
            \BinaryInfC{$ \Gamma' , \Delta \wfdash M[\widetilde{x} \leftarrow x]\ \esubst{ B }{ x }:\tau $}
    \end{prooftree}

    Therefore $ \Gamma',\Delta\wfdash\expr{M}' :\tau$ and the result follows.
    
    \item Rule~$ \redlab{RS{:}Ex \dash Sub}$.
    
    Then $ \expr{M} =  M[x_1, \cdots , x_k \leftarrow x]\ \esubst{ B }{ x }$ where $B=  \bag{N_1, \dots ,N_k} $. By inversion of the typing derivation the reduction is:
    \begin{prooftree}\hspace{-1cm}
        \AxiomC{$B = \bag{N_1,
        \cdots ,N_k} \quad k \geq  1 $}
        \AxiomC{$ M \not= \fail^{\widetilde{y}} $}
        \LeftLabel{\redlab{RS{:}Ex \dash Sub}}
        \BinaryInfC{\( M[x_1, \cdots , x_k \leftarrow x]\ \esubst{ B }{ x } \redd \sum_{B_i \in \perm{B}}M\ \linexsub{B_i(1)/x_1} \cdots \linexsub{B_i(k)/x_k}    \)}
    \end{prooftree}
    
    and $\expr{M'}= \sum_{B_i \in \perm{B}}M\ \linexsub{B_i(1)/x_1} \cdots \linexsub{B_i(k)/x_k}.$
    To simplify the proof we take $k=2$, as the case $k>2$ is similar. Therefore,
    
    \begin{itemize}
        \item $B=\bag{N_1,N_2}$; and
        \item $\perm{B}=\{\bag{N_1,N_2}, \bag{N_2,N_1}\}$
    \end{itemize} 
    
    Since $\Gamma\wfdash \expr{M}:\tau$ we get a derivation where we first type the bag $B$ with the derivation $\Pi$, given next:
    
    \begin{prooftree}
                    \AxiomC{\(  \Delta_1 \wfdash N_1 : \sigma \)}
                                    \AxiomC{\( \Delta_{2} \wfdash N_{2} : \sigma   \)}
        				\AxiomC{\(  \)}
                        \LeftLabel{ \redlab{FS{:}\oneb} }
                        \UnaryInfC{\( \wfdash \oneb : \omega \)}
        			\LeftLabel{ \redlab{FS{:}bag} }
        			\BinaryInfC{\( \Delta_2  \wfdash \bag{N_2}: \sigma   \)}
        			\LeftLabel{ \redlab{FS{:}bag} }
                 \BinaryInfC{\(\Delta  \wfdash B: \sigma \wedge \sigma \) }
    \end{prooftree}
The full derivation is as follows:
    \begin{prooftree}
                    \AxiomC{$ \Gamma' , x_1:\sigma ,  x_2:\sigma  \wfdash  M: \tau $}
			    \LeftLabel{ \redlab{FS{:}share} }
			    \UnaryInfC{$  \Gamma' , x:\sigma \wedge \sigma  \wfdash  M[\widetilde{x} \leftarrow x]: \tau $}
                                        \AxiomC{\(  \Pi \)}
                    \LeftLabel{ \redlab{FS{:}ex \dash sub} }
            \BinaryInfC{$ \Gamma' , \Delta \wfdash M[\widetilde{x} \leftarrow x]\ \esubst{ B }{ x }:\tau $}
    \end{prooftree}

    where $\Delta= \Delta_1,\Delta_2$ and $\Gamma = \Gamma' , \Delta $. We can build a derivation $\Pi_{1,2}$ of $ \Gamma' , \Delta \wfdash  M \linexsub{N_1/x_1} \linexsub{N_2/x_2}    : \tau$ as :
        \begin{prooftree}
        \AxiomC{\( \Gamma'  , x_1:\sigma, x_2:\sigma \wfdash M : \tau \)}
        \AxiomC{\( \Delta_1 \wfdash N_1 : \sigma \)}
        \LeftLabel{ \redlab{FS{:}ex \dash lin \dash sub} }
        \BinaryInfC{\( \Gamma , \Delta_1  , x_2:\sigma \wfdash M \linexsub{N_1 /x_1} : \tau \)}
        \AxiomC{\( \Delta_2 \wfdash N_2 : \sigma \)}
        \LeftLabel{ \redlab{FS{:}ex \dash lin \dash sub} }
        \BinaryInfC{$ \Gamma' , \Delta \wfdash  M \linexsub{N_1/x_1} \linexsub{N_2/x_2}    : \tau$}
    \end{prooftree}
    
    Similarly, we can obtain a derivation $\Pi_{2,1}$ of $ \Gamma' , \Delta \wfdash  M \linexsub{N_2/x_1} \linexsub{N_1/x_2}    : \tau$.   Finally, applying Rule~\redlab{FS{:}sum}:
    \begin{prooftree}
        \AxiomC{\( \Pi_{1,2} \)}
        \AxiomC{\( \Pi_{2,1} \)}
        \LeftLabel{ \redlab{FS{:}sum} }
        \BinaryInfC{$ \Gamma' , \Delta \wfdash  M \linexsub{N_1 /x_1} \linexsub{N_2/x_k} + M \linexsub{N_2/x_1} \linexsub{N_1/x_k}   : \tau $}
    \end{prooftree}
    
    and the result follows.
    
    \item Rule~$\redlab{RS{:}Lin \dash Fetch} $.
    
    Then $ \expr{M} =M\ \linexsub{N/x}  $ where  $\headf{M} = x$. The reduction is: 
    
    \begin{center}
     \AxiomC{$ \headf{M} = x$}
     \LeftLabel{\redlab{RS{:}Lin \dash Fetch}}
     \UnaryInfC{\(M\ \linexsub{N/x} \redd  M \headlin{ N/x } \)}
     \DisplayProof
    \end{center}
    and $\expr{M'}=M\headlin{N/x}$.     Since $\Gamma\wfdash \expr{M}:\tau$ we get the following derivation by inversion of the typing derivation:
    
    \begin{prooftree}
        \AxiomC{\( \Delta \wfdash N : \sigma \)}
        \AxiomC{\( \Gamma'  , x:\sigma \wfdash M : \tau \)}
        \LeftLabel{ \redlab{FS{:}ex \dash lin \dash sub} }
        \BinaryInfC{\( \Gamma', \Delta \wfdash M \linexsub{N / x} : \tau \)}
    \end{prooftree}
        
    where $\Gamma = \Gamma' , \Delta $. By the Substitution Lemma (Lemma~\ref{ch2l:lamrsharfailsubs}), we obtain a derivation $\Gamma' , \Delta \wfdash   M \headlin{ N/x } : \tau $, and the result follows.

\item Rule~$\redlab{RS{:}TCont}$.

Then $\expr{M} = C[M]$ and the reduction is as follows:
\begin{prooftree}
        \AxiomC{$   M \redd M'_{1} + \cdots +  M'_{k} $}
        \LeftLabel{\redlab{RS{:}TCont}}
        \UnaryInfC{$ C[M] \redd  C[M'_{1}] + \cdots +  C[M'_{k}] $}
\end{prooftree}
with $\expr{M'} =  C[M'_{1}] + \cdots +  C[M'_{k}] $. 
The proof proceeds by analysing the context $C$. \\
There are four cases:

\begin{enumerate}
    \item $C=[\cdot]\ B$.
    
    In this case $\expr{M}=M \ B$, for some $B$. Since $\Gamma\vdash \expr{M}:\tau$ by inversion of the typing derivation, one has the derivation:
\begin{prooftree}
    \AxiomC{\(  \Gamma' \wfdash  M: \sigma^{j} \rightarrow \tau \)}
    \AxiomC{\( \Delta \wfdash  B : \sigma^{k} \)}
        \LeftLabel{ \redlab{FS{:}app} }
    \BinaryInfC{\( \Gamma', \Delta \wfdash  M\ B : \tau\)}
\end{prooftree}

where $\Gamma = \Gamma' , \Delta $. From  $\Gamma'\wfdash M:\sigma^j\rightarrow\tau$ and the reduction $M \redd M'_{1} + \cdots +  M'_{k} $, one has by IH that  $\Gamma'\wfdash M_1'+\ldots, M_k':\sigma^j\rightarrow\tau$, which entails $\Gamma'\wfdash M_i':\sigma^j\rightarrow\tau$, for $i=1,\ldots, k$, via Rule~\redlab{FS{:}sum}. Finally, we may type the following:
\begin{prooftree}
            \AxiomC{\(  \forall i \in {1 , \cdots , l} \)}
			    \AxiomC{\(  \Gamma' \wfdash  M'_{i}: \sigma^{j} \rightarrow \tau \)}
    				\AxiomC{\( \Delta \wfdash  B : \sigma^{k} \)}
        		\LeftLabel{ \redlab{FS{:}app} }
			\BinaryInfC{\(  \Gamma', \Delta \wfdash (M'_{i}\ B):  \tau \)}
			
        \LeftLabel{ \redlab{FS{:}sum} }
    \BinaryInfC{\( \Gamma', \Delta \wfdash (M'_{1}\ B) + \cdots +  (M'_{l} \ B) : \tau\)}
\end{prooftree}

Since $ \expr{M}'  =   (C[M'_{1}]) + \cdots +  (C[M'_{l}]) = M_1'B+\ldots+M_k'B$, the result follows.
\item  Cases $C=[\cdot]\linexsub{N/x} $ and $C=[\cdot][\widetilde{x} \leftarrow x]$ are similar to the previous one.
\item $C= [\cdot][ \leftarrow x]\esubst{\oneb}{ x}$

In this case $\expr{M}=C[M]=M[\leftarrow x] \esubst{\oneb}{x}$. Since $\Gamma\wfdash \expr{M}:\tau$ by inversion of the typing derivation, one has a derivation

\begin{prooftree}
    \AxiomC{$\Gamma\wfdash M:\tau$}
    \LeftLabel{\redlab{FS \dash weak}}
    \UnaryInfC{\(  \Gamma, x:\omega \wfdash  M[\leftarrow x] : \tau \)}
    \AxiomC{}
    \LeftLabel{\redlab{TS{:} \oneb}}
    \UnaryInfC{\(  \vdash  \oneb: \omega \)}
    \LeftLabel{\redlab{FS{:} wf \dash bag}}
    \UnaryInfC{\(  \wfdash  \oneb: \omega \)}
        \LeftLabel{ \redlab{FS{:}ex\dash sub} }
    \BinaryInfC{\(  \Gamma \wfdash  M[\leftarrow x] \esubst{\oneb}{x}:\tau \)}
\end{prooftree}

From $M\redd M_1+\ldots+M_k$ and $\Gamma \wfdash M:\tau$, by the IH, it follows that $\Gamma\wfdash M_1+\ldots+ M_k:\tau$, and consequently, $\Gamma\wfdash M_i$, via application of \redlab{FS{:}sum}. Therefore, there exists a derivation

\begin{prooftree}
            \AxiomC{$\Gamma \wfdash M_i:\tau $}
            \UnaryInfC{$\Gamma, x:\omega\wfdash M_i[\leftarrow  x]:\tau $}
            \AxiomC{$\wfdash \oneb:\omega$}
            \BinaryInfC{$\Gamma \wfdash M_i[\leftarrow  x]\esubst{\oneb}{x}:\tau $}
\end{prooftree}
for each $i=1,\ldots,k$. By applying \redlab{FS{:}sum}, we obtain $\Gamma \wfdash M_1[\leftarrow  x]\esubst{\oneb}{x}+\ldots+ M_k[\leftarrow  x]\esubst{\oneb}{x}:\tau $, and the result follows.
\end{enumerate}

	\item Rule~$ \redlab{RS{:}ECont} $.
	
Then $\expr{M} = D[\expr{M}_1 ]$ where $\expr{M}_1  \redd \expr{M}_2$ then we can perform the following reduction:

\begin{prooftree}
        \AxiomC{$ \expr{M}_1  \redd \expr{M}_2 $}
        \LeftLabel{$\redlab{RS{:}ECont}$}
        \UnaryInfC{$D[\expr{M}_1]  \redd D[\expr{M}_2]  $}
\end{prooftree}

and $\expr{M'}=D[\expr{M}_2]$. \\
The proof proceeds by analysing the context $D$. There are two cases: 
$D= [\cdot] + \expr{N}$ and $D= \expr{N} + [\cdot]$. 
We analyze only the first one:

    $D= [\cdot] + \expr{N}$. In this case $\expr{M}= \expr{M}_1+\expr{N}$ and by inversion of the typing derivation:
\begin{prooftree}
    \AxiomC{$ \Gamma \wfdash  \expr{M}_1 : \tau$}
    \AxiomC{$ \Gamma \wfdash  \expr{N} : \tau$}
    \LeftLabel{ \redlab{FS{:}sum} }
    \BinaryInfC{$ \Gamma \wfdash  \expr{M}_1+\expr{N}: \tau$}
\end{prooftree}

From $ \Gamma \wfdash \expr{M}_1 : \tau$ and $ \expr{M}_1  \redd \expr{M}_2  $, by IH, one has that $\Gamma \wfdash  \expr{M}_2 : \tau$.
Hence we may type the following:
\begin{prooftree}
    \AxiomC{$ \Gamma \wfdash  \expr{M}_2 : \tau$}
    \AxiomC{$ \Gamma \wfdash  \expr{N} : \tau$}
    \LeftLabel{ \redlab{FS{:}sum} }
    \BinaryInfC{$ \Gamma \wfdash  \expr{M}_2+\expr{N}: \tau$}
\end{prooftree}
Since $\expr{M}'=D[\expr{M}_2]=\expr{M}_2+\expr{N}$,  the result follows.

\item Rule~$ \redlab{RS{:}Fail} $.

Then $\expr{M} =  M[x_1, \cdots , x_k \leftarrow x]\ \esubst{ B }{ x } $ where $B = \bag{N_1,\dots ,N_l}  $ and  the reduction is:
 \begin{prooftree}
    \AxiomC{$k \neq \size{B}$}
     \AxiomC{$ \widetilde{y} = (\lfv{M} \setminus \{  x_1, \cdots , x_k \} ) \cup \lfv{B} $}
    \LeftLabel{\redlab{RS{:}Fail}}
    \BinaryInfC{\( M[x_1, \cdots , x_k \leftarrow x]\ \esubst{ B }{ x } \redd \sum_{B_i \in \perm{B}}  \fail^{\widetilde{y}} \)}
 \end{prooftree}
where $\expr{M'}=\sum_{B_i \in \perm{B}}  \fail^{\widetilde{y}}$. Since $\Gamma\wfdash \expr{M}$ and by inversion of the typing derivation, one has a derivation:
\begin{prooftree}
            \AxiomC{\( \Gamma' , x_1:\sigma,\ldots, x_k:\sigma \wfdash M: \tau \)}
            \LeftLabel{ \redlab{FS{:}ex \dash sub} }    
            \UnaryInfC{\( \Gamma' , x:\sigma^{k} \wfdash M[x_1, \cdots , x_k \leftarrow x] : \tau \)}
            \AxiomC{\( \Delta \wfdash B : \sigma^{j} \)}
        \LeftLabel{ \redlab{FS{:}ex \dash sub} }    
        \BinaryInfC{\( \Gamma', \Delta \wfdash M[x_1, \cdots , x_k \leftarrow x]\ \esubst{ B }{ x }  : \tau \)}
    \end{prooftree}

where $\Gamma = \Gamma' , \Delta $. We may type the following:
    \begin{prooftree}
        \AxiomC{\( \)}
        \LeftLabel{ \redlab{FS{:}fail}}
        \UnaryInfC{\( \Gamma' , \Delta \wfdash  \fail^{\widetilde{y}} : \tau  \)}
    \end{prooftree}
since $\Gamma',\Delta$ contain assignments on the free variables in $M$ and $B$.
Therefore, $\Gamma\wfdash \fail^{\widetilde{y}}:\tau$, by applying \redlab{FS{:}sum}, it follows that $\Gamma\wfdash \sum_{B_i\in \perm{B}}\fail^{\widetilde{y}}:\tau$, as required. 
\item Rule~$\redlab{RS{:}Cons_1}$.

Then $\expr{M} =   \fail^{\widetilde{x}}\ B $ where $B = \bag{N_1, \dots ,N_k} $  and  the  reduction is:

\begin{prooftree}
    \AxiomC{\( B = \bag{N_1, \dots ,N_k} \)}
    \AxiomC{\( \widetilde{y} = \lfv{B} \)}
    \LeftLabel{$\redlab{RS{:}Cons_1}$}
    \BinaryInfC{\( \fail^{\widetilde{x}} \ B  \redd \sum_{\perm{B}} \fail^{\widetilde{x} \cup \widetilde{y}} \)}
\end{prooftree}
and $ \expr{M}'  =  \sum_{\perm{B}} \fail^{\widetilde{x} \cup \widetilde{y}} $.
Since $\Gamma\wfdash \expr{M}:\tau$ and by inversion  of the typing derivation, one has the derivation: 
    \begin{prooftree}
        \AxiomC{\( \)}
        \LeftLabel{\redlab{FS{:}fail}}
        \UnaryInfC{\( \Gamma' \wfdash \fail^{\widetilde{x}} : \omega \rightarrow \tau \)}
        \AxiomC{\( \Delta \wfdash B : \pi \)}
            \LeftLabel{\redlab{FS{:}app}}
        \BinaryInfC{\( \Gamma', \Delta \wfdash \fail^{\widetilde{x}}\ B : \tau\)}
    \end{prooftree}

where $\Gamma = \Gamma' , \Delta $. After $\perm{B}$ applications of \redlab{FS{:}sum}, we obtain $ \Gamma \wfdash \sum_{\perm{B}} \fail^{\widetilde{x} \cup \widetilde{y}}: \tau$, and the result follows.

\item Rule~$\redlab{RS{:}Cons_2}$.

Then $\expr{M} =   (\fail^{\widetilde{x}\cup \widetilde{y}} [ \widetilde{x} \leftarrow x])\esubst{ B }{ x } $for $B = \bag{N_1, \dots ,N_k} $  and the reduction is:

\begin{prooftree}
    \AxiomC{\(  B = \bag{N_1, \dots ,N_k} \quad k+|\widetilde{x}|\neq 0 \)}
    \AxiomC{\(  \widetilde{y} = \lfv{B} \)}
    \LeftLabel{$\redlab{RS{:}Cons_2}$}
    \BinaryInfC{\( (\fail^{\widetilde{x} \cup \widetilde{y}} [ \widetilde{x} \leftarrow x])\esubst{ B }{ x } \redd \sum_{\perm{B}} \fail^{\widetilde{y} \cup \widetilde{z}} \)}
\end{prooftree}
with $\expr{M'}=\sum_{\perm{B}} \fail^{\widetilde{y}\cup \widetilde{z}}$. Since $\Gamma\wfdash \expr{M}:\tau$ and by inversion  of the typing derivation, one has the derivation:

    \begin{prooftree}
              \AxiomC{\( \)}
            \LeftLabel{\redlab{FS{:}fail}}
            \UnaryInfC{\( \Delta , x_1: \sigma, \cdots, x_j: \sigma \wfdash \fail^{\widetilde{x}\cup \widetilde{y}} : \tau \quad x\notin \Delta \quad k \not = 0\)}
            \LeftLabel{ \redlab{FS{:}share}}
            \UnaryInfC{\( \Delta , x: \sigma^j \wfdash \fail^{\widetilde{x}\cup \widetilde{y}}[x_1 , \cdots , x_j \leftarrow x] : \tau \)}
             \AxiomC{\( \Delta \wfdash B : \sigma^{k} \)}
        \LeftLabel{\redlab{FS{:}ex \dash sub}}    
        \BinaryInfC{\( \Gamma, \Delta \wfdash \fail^{\widetilde{x}\cup \widetilde{y}} [\widetilde{x} \leftarrow x]\esubst{ B }{ x } : \tau \)}
    \end{prooftree}

Hence $\Gamma = \Gamma' , \Delta $ and $ \expr{M}'  =  \sum_{\perm{B}} \fail^{\widetilde{y}\cup \widetilde{z}}$ and we may type the following:
    \begin{prooftree}
        \AxiomC{\( \)}
        \LeftLabel{\redlab{FS{:}fail}}
        \UnaryInfC{$ \Gamma \wfdash \fail^{\widetilde{y}\cup \widetilde{z}} : \tau$}
        \AxiomC{$ \cdots $}
        \LeftLabel{\redlab{FS{:}sum}}
        \BinaryInfC{$ \Gamma \wfdash \sum_{\perm{B}} \fail^{\widetilde{y}\cup \widetilde{z}}: \tau$}
    \end{prooftree}
    \item Rule~$\redlab{RS{:}Cons_3}$.
    
    Then $\expr{M}=\fail^{\widetilde{y}\cup x}$ and the reduction is 
    
    \begin{prooftree}
    \AxiomC{$\widetilde{z}=\lfv{N}$}
    \LeftLabel{\redlab{RS{:}Cons_3}}
    \UnaryInfC{$\fail^{\widetilde{y}\cup x}\linexsub{N/x}\redd \fail^{\widetilde{y}\cup \widetilde{z}}$}
    \end{prooftree}
    with $\expr{M'}=\fail^{\widetilde{y}\cup \widetilde{z}}$. Since $\Gamma\wfdash \expr{M}$ and by inversion of the typing derivation, one has the derivation
    
    \begin{prooftree}
    \AxiomC{}
        \LeftLabel{\redlab{FS{:}fail}}
    \UnaryInfC{$\Gamma',x:\sigma \wfdash \fail^{\widetilde{y}\cup x}:\tau$}
    \AxiomC{$\Delta\wfdash N:\sigma$}
    \LeftLabel{\redlab{FS{:}ex\dash lin\dash sub}}
    \BinaryInfC{$\Gamma',\Delta \wfdash \fail^{\widetilde{y}\cup x}\linexsub{N/x}:\tau$}
    \end{prooftree}
    where $x\notin \dom{\Gamma'}$, $\dom{\Gamma'}=\widetilde{y}$ and $\dom{\Delta}=\widetilde{z}=\lfv{N}$.
    
    We can type the following:
    \begin{prooftree}
    \AxiomC{}
    \LeftLabel{\redlab{FS{:}fail}}
    \UnaryInfC{$\Gamma',\Delta\wfdash\fail^{\widetilde{y}\cup \widetilde{z}}:\tau $}
    \end{prooftree}
    and the result follows.
    \end{myEnumerate}
\end{proof}

\consistencytype*

\begin{proof}
\secondrev{
By induction on the typing derivation, with a case analysis on the last applied rule (\Cref{ch2fig:wfsh_rules}). We only consider the cases for the typing rules that relate to the sharing construct and the explicit substitution. 
    First, consider conditions~1(i)~to~1(iv), which are related to $M [ \widetilde{x} \leftarrow x ]$. The conditions are as follows (i) 
	$\widetilde{x}$ contains pairwise distinct variables; 
	(ii)~every $x_i \in \widetilde{x}$ must occur exactly once in $M$; (iii) $x_i$ is not a sharing variable;
	(iv)~$M$ is consistent. By considering rule~$\redlab{FS{:}share}$, we have:
    \begin{prooftree}
        \AxiomC{\( \Gamma , x_1: \sigma, \cdots, x_k: \sigma \wfdash M : \tau \quad x\notin \dom{\Gamma} \quad k \not = 0\)}
        \LeftLabel{ \redlab{FS{:}share}}
        \UnaryInfC{\( \Gamma , x: \sigma^{k} \wfdash M[x_1 , \cdots , x_k \leftarrow x] : \tau \)}  
    \end{prooftree}
    Condition~1(i) follows from uniquness of variables within the context. Condition~1(ii) follows from the premise, which ensures that $M$ is well-formed with a context including each $x_i$; linearity conditions imply that each $x_i$ must be consumed so it must occur in  $M$. Condition~1(iii) also follows directly from the well-formedness of $M$: each $x_i$ is typed with a strict type, and the rule ensures that the sharing variable $x$ is typed with the multiset type $ \sigma^{k}$. Finally condition~1(iv) is ensured by the IH.
    }
    
    \secondrev{
    For conditions~2(i)~to~2(iv) which are (i) the variable $x$ must occur exactly once in $M$;
	(ii) $x$ cannot be a sharing variable; 	(iii)~$M$ and $N$ are consistent; (iv)~$\lfv{M} \cap \lfv{N} = \emptyset$. Consider the case of rule $\redlab{FS{:}ex \dash lin \dash sub}$: 
    \begin{prooftree}
        \AxiomC{\( \Gamma  , x:\sigma \wfdash M : \tau \quad \Delta \wfdash N : \sigma \)}
        \LeftLabel{\redlab{FS{:}ex \dash lin \dash sub}}
        \UnaryInfC{\( \Gamma, \Delta \wfdash M \linexsub{N / x} : \tau \)}
    \end{prooftree}
    First, because $\Gamma$ and $\Delta$ are disjoint, $x$ cannot appear within $\Delta$ and $M$ must consume the type of $x:\sigma$; hence $x$ must occur in $M$, satisfying condition~2(i) and 2(iv). 
    Second, $\Gamma , x:\sigma$ ensures a strict type for $x$;       
    if $x$ were a sharing variable in $M$ then $x$ would have a multiset type $\pi$. Therefore, condition~2(ii) is satisfied. Finally, condition~(iii) is satisfied by induction on $M$ and $N$.
    }
\end{proof} 

\consistencyencode*
\begin{proof}
\secondrev{
  By induction on the structure of $\expr{M}$. 
  Notice that $\recencodf{\cdot}$ ensures consistency for bound variables: it replaces all occurrences of a bound variable (say $y$) with fresh bound variables (say, $y_1, \ldots, y_k$). Thus, the following hold for bound variables: (i) they occur once within a term and (ii) they are not shared themselves, as the sharing of variables only occurs when handling binders associated to explicit substitutions and abstractions. 
  As for free variables, the translation $\recencodopenf{\cdot}$ replaces each occurrence with a fresh variable, and does so before applying $\recencodf{\cdot}$; this ensures that free variables that are already shared are not shared again. Because of this design, the translations preserve consistency.
}
\end{proof}

\section{Appendix to  \texorpdfstring{\secref{ch2ss:firststep}}{}}
\label{ch2app:firststep}

\subsection{Encoding \texorpdfstring{$\recencodf{\cdot}$}{}}
\label{ch2apen:firstencod}

\subsubsection[Auxiliary Encoding]{ Auxiliary Encoding: From \texorpdfstring{$\lamr$}{} into \texorpdfstring{$\lamrshar$}{} }


\begin{restatable}[]{prop}{}
\label{ch2prop:linhed_enc}
The encoding commutes with linear substitution: $ \recencodf{M\headlin{N/x}}=\recencodf{M}\headlin{\recencodf{N}/x}$

\end{restatable}

\begin{proof}
By induction of the structure of $M$ in $M\headlin{\recencodf{N}/x}$. 
\end{proof}

 \begin{restatable}[Well-typedness preservation for $\recencodf{-}$]{prop}{typeencintolamrfail}
\label{ch2prop:typeencintolamrfail}
\revo{A17,A18,A19}{
Let $B$ and $\expr{M}$  be a  bag and an  expression in $\lamrfail$, respectively. 
\begin{enumerate}
\item
    If $\Gamma \vdash B:\sigma$ 
then $ \strcore{\Gamma} \vdash \recencodf{B}:\sigma$.
    \item 
    If $\Gamma \vdash \expr{M}:\sigma$  
then $ \strcore{\Gamma} \vdash \recencodf{\expr{M}}:\sigma$.
\end{enumerate}}
\end{restatable}

\begin{proof}

 By mutual induction on the typing derivations for $B$ and $\expr{M}$, with an analysis of the last rule applied. 

\noindent Part~(1) includes two cases:
    \begin{enumerate}[i)]
    \item Rule~$\redlab{T:\oneb}$: Then $B = \oneb$ and the thesis follows trivially, 
    because the encoding of terms/bags  
        (cf.  Figure~\ref{ch2fig:auxencfail})
        ensures that $\recencodf{\oneb}=\oneb$.
                
        \item Rule~$\redlab{T:bag}$. Then $B = \bag{M}\cdot A$, where $M$ is a term and $A$ is a bag,
        and 
        \begin{prooftree}
                    \AxiomC{\( \Gamma \vdash M : \sigma\)}
                \AxiomC{\( \Delta \vdash A : \pi\)}
                \LeftLabel{$\redlab{T:bag}$}
            \BinaryInfC{\( \Gamma \contexcat \Delta \vdash \bag{M}\cdot A:\sigma \wedge \pi\)}
            \end{prooftree}
By the IHs, we have both $\strcore{\Gamma} \vdash \recencodf{M} : \sigma$ and $\strcore{\Delta} \vdash \recencodf{A} : \pi$. The thesis then  follows by applying Rule~$\redlab{TS:bag}$ in \lamrsharfail:
        \begin{prooftree}
                \AxiomC{\( \strcore{\Gamma}   \vdash \recencodf{M} : \sigma\)}
                \AxiomC{\( \strcore{\Delta}  \vdash \recencodf{A} : \pi\)}
                \LeftLabel{$\redlab{TS:bag}$}
            \BinaryInfC{\( \strcore{\Gamma} , \strcore{\Delta}  \vdash \bag{\recencodf{M}}\cdot \recencodf{A}:\sigma \wedge \pi\)}
        \end{prooftree}
        \end{enumerate}
    
    \noindent Part~(2) considers six cases:
    
    \begin{enumerate}[i)]

                \item Rule~$\redlab{T:var}$: Then $\expr{M} = x$ 
        and 
                \begin{prooftree}
            \AxiomC{}
            \LeftLabel{$\redlab{T:var}$}
            \UnaryInfC{\( x: \sigma \vdash x : \sigma\)}    
        \end{prooftree}
        By the encoding of terms
        (cf. \figref{ch2fig:auxencfail}
        ), we infer $x: \sigma \vdash x : \sigma$ and so 
        the thesis holds immediately. 
        
        
        \item Rule~$\redlab{T:abs}$: Then  $\expr{M} = \lambda x . M $ 
        and 
          \begin{prooftree}
    \AxiomC{\( \Gamma , x: \sigma^n \vdash M : \tau \)}
    \LeftLabel{$\redlab{T:abs}$}
    \UnaryInfC{\( \Gamma \vdash \lambda x . M :  \sigma^n  \rightarrow \tau \)}
            \end{prooftree}
            
                 By the encoding of terms (cf. \figref{ch2fig:auxencfail}), we have 
        $\recencodf{  \expr{M}  }  =   \lambda x . \recencodf{M\linsub{x_1,\cdots, x_n}{x} } [\widetilde{x}\leftarrow x]$, where  $\#(x,M) = n$ and each $x_i$ is fresh.
        
            We work on the premise $\Gamma , {x}: \sigma^n \vdash M : \tau$ before appealing to the IH.
            
            Then, by  $n$ applications of Lemma~\ref{ch2lem:preser_linsub} to this judgment, we obtain 
            \begin{equation}\label{ch2eq:tp1}
             {\Gamma}, {x_1: \sigma, \cdots,  x_n: \sigma} \vdash {M}\linsub{x_1,\cdots, x_n}{x} : \tau
            \end{equation}
            By IH on \eqref{ch2eq:tp1} we have
              \begin{equation}\label{ch2eq:tp2}
             \strcore{\Gamma}, {x_1: \sigma, \cdots,  x_n: \sigma} \vdash \recencodf{M\linsub{x_1,\cdots, x_n}{x}} : \tau
            \end{equation}
            
              Starting from \eqref{ch2eq:tp2}, we then have the following type derivation for  $\recencodf{\expr{M}}$, which concludes the proof for this case:
        

        \begin{prooftree}
         \LeftLabel{$(*)$}
                    \AxiomC{\( \strcore{\Gamma} , x_1: \sigma, \cdots, x_n: \sigma \vdash \recencodf{M\linsub{x_1,\cdots, x_n}{x}} : \tau \)}
            \LeftLabel{$\redlab{TS:share}$}
            \UnaryInfC{\( \strcore{\Gamma} , x: \sigma^n  \vdash \recencodf{M\linsub{x_1,\cdots, x_n}{x}}[x_1 , \cdots , x_n \leftarrow x] : \tau \)}
            \LeftLabel{$\redlab{TS:abs \dash sh}$}
            \UnaryInfC{\( \strcore{\Gamma} \vdash \lambda x . (\recencodf{M\linsub{x_1,\cdots, x_n}{x}}[x_1 , \cdots , x_n \leftarrow x]) : \sigma^n \rightarrow \tau \)}
        \end{prooftree}

        
        
        \item Rule~$\redlab{T:app}$: Then $\expr{M} = M\ B$ and
         \begin{prooftree}
    \AxiomC{\( \Gamma \vdash M : \pi \rightarrow \tau \)}
    \AxiomC{\( \Delta \vdash B : \pi \)}
        \LeftLabel{$\redlab{T:app}$}
    \BinaryInfC{\( \Gamma \contexcat \Delta \vdash M\ B : \tau\)}
            \end{prooftree}
            By IH we have both 
            $ \strcore{\Gamma} \vdash \recencodf{M} : \pi \rightarrow \tau$
            and 
            $ \strcore{\Delta} \vdash \recencodf{B} : \pi$,
            and the thesis follows easily by Rule~$\redlab{TS:app}$ in \lamrsharfail:
              \begin{prooftree}
                \AxiomC{\( \strcore{\Gamma} \vdash \recencodf{M} : \pi \rightarrow \tau \)}
                \AxiomC{\( \strcore{\Delta} \vdash \recencodf{B} : \pi \)}
                \LeftLabel{$\redlab{TS:app}$}
            \BinaryInfC{\( \strcore{\Gamma}, \strcore{\Delta} \vdash \recencodf{M}\ \recencodf{B} : \tau\)}
        \end{prooftree}
        
            
        
        
        \item Rule~$\redlab{T:ex \dash sub}$: Then $\expr{M} = M \esubst{ B }{ x }$
        and the proof is split in two cases, depending on the shape of $B$:
        
        \begin{enumerate}
            \item $B=\oneb$. In this case, $\expr{M}=M\esubst{\oneb}{x}$ and we obtain the following type derivation:
            
            \begin{prooftree}
            \LeftLabel{$\oneb$}
        \AxiomC{$\vdash \oneb :\omega$}
        \AxiomC{$ \Gamma\vdash M:\tau$}
        \LeftLabel{$\redlab{T:weak}$}
        \UnaryInfC{$ \Gamma, x:\omega \vdash M:\tau $}
        \LeftLabel{$\redlab{T:ex \dash sub}$}
        \BinaryInfC{$\Gamma  \vdash M\esubst{\oneb}{x}:\tau$}
        \end{prooftree}
        
        By IH we have both $\vdash \oneb : \omega$ and $\strcore{\Gamma} \vdash \recencodf{M}:\tau$. 
        By the encoding of terms (Figure~\ref{ch2fig:auxencfail}), $\recencodf{M\esubst{\oneb}{x}}=\recencodf{M}[\leftarrow x]\esubst{\oneb}{x}$, and the result holds
               by the following type derivation:
        \begin{prooftree}
            \AxiomC{$\vdash \oneb :\omega$}
            \AxiomC{$\strcore{\Gamma} \vdash\recencodf{M}:\tau$}
            \LeftLabel{$\redlab{TS:weak}$}
            \UnaryInfC{ $ \strcore{\Gamma}, x:\omega \vdash\recencodf{M}[\leftarrow x]:\tau $}
            \LeftLabel{$\redlab{TS:ex \dash sub}$}
            \BinaryInfC{$ \strcore{\Gamma}  \vdash \recencodf{M}[\leftarrow x]\esubst{\oneb}{x}:\tau$}
        \end{prooftree}

            \item $B= \bag{N_1, \ldots ,N_n}$, $n\geq 1$.
            Suppose w.l.o.g. that $n=2$, then $B= \bag{N_1,N_2}$ and
        \begin{prooftree}
        \AxiomC{$\Delta_1\vdash N_1:\sigma$}
        \AxiomC{$\Delta_2\vdash N_2:\sigma$}
        \LeftLabel{$\redlab{T:bag}$}
        \BinaryInfC{$\Delta_1 \contexcat \Delta_2\vdash \bag{N_1}\cdot \bag{N_2}:\sigma^2$}
        \AxiomC{$\Gamma, {x}:\sigma^2 \vdash M:\tau $}
        \LeftLabel{$\redlab{T:ex \dash sub}$}
        \BinaryInfC{$\Gamma \contexcat \Delta_1 \contexcat \Delta_2\vdash M\esubst{B}{x}:\tau$}
        \end{prooftree}

        By IH we have 
        $ \strcore{\Delta_1} \vdash \recencodf{N_1}:\sigma$
        and
        $ \strcore{\Delta_2} \vdash \recencodf{N_2}:\sigma$.
        We can expand 
        $\Gamma, {x}:\sigma^2 \vdash M:\tau $
        into
       $\Gamma, x:\sigma \wedge \sigma \vdash M:\tau$.
       By Lemma \ref{ch2lem:preser_linsub} and the IH on this last sequent we obtain
       $$ \strcore{\Gamma}, y_1:\sigma, y_2:\sigma \vdash \recencodf{M\linsub{y_1,y_2}{x}}:\tau$$
       where $\#(x,M)=2$ and $y_1,y_2$ are fresh variables with the same type as $x$.
       
    Now, by the encoding of terms (Figure~\ref{ch2fig:auxencfail}), we have
    
        \[
        \begin{aligned}
             \recencodf{M\esubst{\bag{N_1,N_2}}{x}}
                = & ~\recencodf{M\linsub{y_1,y_2}{x}}\linexsub{\recencodf{N_1}/y_1
                }\linexsub{\recencodf{N_2}/y_2} + \\
                & \recencodf{M\linsub{y_1,y_2}{x}}\linexsub{\recencodf{N_1}/y_2
                }\linexsub{\recencodf{N_2}/y_1}\\
                 = & ~\expr{M}'
        \end{aligned}
        \]
        
        We give typing derivations in $\lamrsharfail$ for each summand.     First, let $\Pi_1$ be the following derivation:
       
         \begin{prooftree}
            \AxiomC{$ \strcore{\Delta_2}\vdash \recencodf{N_2}:\sigma$}
            \AxiomC{$ \strcore{\Delta_1}\vdash \recencodf{N_1}:\sigma$}
              \AxiomC{$ \strcore{\Gamma}, y_1:\sigma, y_2:\sigma \vdash \recencodf{M\linsub{y_1,y_2}{x}}:\tau$}     
              \BinaryInfC{$ \strcore{\Gamma}, y_2:\sigma, \strcore{\Delta_1}\vdash \recencodf{M\linsub{y_1,y_2}{x}}\linexsub{\recencodf{N_1}/y_1}:\tau$}
              \BinaryInfC{$ \strcore{\Gamma},  \strcore{\Delta_1},\strcore{\Delta_2}\vdash \recencodf{M\linsub{y_1,y_2}{x}}\linexsub{\recencodf{N_1}/y_1}\linexsub{\recencodf{N_2}/y_2}:\tau$}
         \end{prooftree}

          Similarly, we can obtain a derivation $\Pi_2$ for: 
          \[ \strcore{\Gamma},  \strcore{\Delta_1}, \strcore{\Delta_2} \vdash \recencodf{M\linsub{y_1,y_2}{x}}\linexsub{\recencodf{N_1}/y_2}\linexsub{\recencodf{N_2}/y_1}:\tau\]  
          
          From $\Pi_1$, $\Pi_2$, and Rule~$\redlab{TS:sum}$, the thesis follows:

          \begin{prooftree}
        \AxiomC{$\Pi_1 \qquad \Pi_2$}
         \LeftLabel{$\redlab{TS:sum}$}
          \UnaryInfC{$ \strcore{\Gamma},  \strcore{\Delta_1}, \strcore{\Delta_2} \vdash \expr{M}' :\tau$}
          \end{prooftree}
       
          \end{enumerate}
        \item Rule~$\redlab{T:weak}$: Then  $\expr{M}=M$ and
        \begin{prooftree}
        \AxiomC{$\Gamma\vdash M:\sigma \qquad x\notin \dom{\Gamma}$}
        \LeftLabel{$\redlab{T:weak}$}
        \UnaryInfC{$\Gamma, x:\omega \vdash M:\sigma $}
        \end{prooftree}
        Because $\redlab{TS:weak}$ is a silent typing rule in \lamrfail, we have that $ x \not \in \lfv{M}$ and so this case does not apply.

        

        \item Rule~$\redlab{T:sum}$:
        
        This case follows easily by IH.
    \end{enumerate}

\end{proof}

\subsubsection{Properties}
\label{ch2app:encodingprop}

We divide the proof of well-formedness preservation: we first prove it for $\recencodf{-}$, then we extend it to $\recencodopenf{-}$.

\preservencintolamrfail*

\begin{proof}
By mutual induction on the typing derivations for $B$ and $\expr{M}$ , with an analysis of the last rule (from \figref{ch2fig:app_wf_rules}) applied. We proceed with the following nine cases:
\begin{enumerate}
\item This case includes two subcases:
    \begin{enumerate}
    \item Rule~\redlab{F:wf \dash bag}. 
    
    Then by inversion of the typing derivation,
    \begin{prooftree}
    \AxiomC{$\Gamma\vdash B:\sigma$}     
    \LeftLabel{$\redlab{F{:}wf-bag}$}
    \UnaryInfC{$\Gamma\wfdash  B:\sigma$}
    \end{prooftree}
    
   By Propostion \ref{ch2prop:typeencintolamrfail} we have $\Gamma\vdash B:\sigma$ implies $ \strcore{\Gamma}\vdash \recencodf{B}:\sigma$. Notice that the encoding $\recencodf{\cdot}$  given in 
   \figref{ch2fig:auxencfail}, is a restriction of $\recencodf{\cdot}$ to $\lamr$. Therefore, $ \strcore{\Gamma} \vdash \recencodf{B}:\sigma$, and the result follows after an application of \redlab{FS{:}wf-bag}.
    
    
                
        \item Rule~\redlab{F:bag}. 
        
        In this case $B = \bag{M}\cdot A$, where $M$ is a term and $A$ is a bag, and we have the following derivation by inversion of the typing derivation:
        \begin{prooftree}
                    \AxiomC{\( \Gamma \wfdash M : \sigma\)}
                \AxiomC{\( \Delta \wfdash A : \sigma^k  \)}
                \LeftLabel{\redlab{F:bag}}
            \BinaryInfC{\( \Gamma \contexcat \Delta \wfdash \bag{M}\cdot A:\sigma^{k+1}  \)}
            \end{prooftree}
with $\dom{\Gamma}=\lfv{M}$ and $\dom{\Delta}=\lfv{A}$. By the IHs, we have both
\begin{itemize}
    \item $ \strcore{\Gamma} \vdash \recencodf{M} : \sigma$; and
    \item $ \strcore{\Delta} \vdash \recencodf{A} : \sigma^k$.
\end{itemize}

By applying Rule~$\redlab{FS{:}bag}$ from~\figref{ch2fig:wfsh_rules}, for $\lamrsharfail$, we obtain the following derivation:
\begin{prooftree}
\AxiomC{\( \strcore{\Gamma}  \wfdash \recencodf{M} : \sigma\)}
\AxiomC{\( \strcore{\Delta} \wfdash \recencodf{A} : \sigma^k \)}
\LeftLabel{\redlab{FS{:}bag}}
\BinaryInfC{\( \strcore{\Gamma}, \strcore{\Delta} \wfdash \bag{\recencodf{M}}\cdot \recencodf{A}:\sigma^{k+1}\)}
        \end{prooftree}
        
        Since $\recencodf{\bag{M}\cdot A}= \bag{\recencodf{M}}\cdot \recencodf{A}, $ one has $ \strcore{\Gamma}, \strcore{\Delta} \vdash \recencodf{\bag{M}\cdot A}:\sigma^{k+1}$, and the result follows.
        \end{enumerate}
    
\item    This case is divided in seven subcases:
    
    \begin{enumerate}

        \item Rule~\redlab{F:wf \dash expr}.
        
        
        Then the thesis follows trivially from type preservation in $\recencodopenf{-}$ of Proposition \ref{ch2prop:typeencintolamrfail}.
        
        \item Rule~\redlab{F:weak}.
        
        In this case, $\expr{M}=M$ and by inversion of the typing derivation we have the derivation
        
        \begin{prooftree}
                    \AxiomC{$\Gamma_1\wfdash M:\tau $}
                    \LeftLabel{\redlab{F:weak}}
                    \UnaryInfC{$\Gamma_1,x:\omega\wfdash M:\tau $}
        \end{prooftree}
        
        Because $\redlab{weak}$ is a silent well-formed rule in \lamrfail, we have that $ x \not \in \lfv{M}$ and so this case does not apply.
        
        \item Rule~\redlab{F:abs}.
        
        In this case $\expr{M} = \lambda x . M $ by inversion of the typing derivation
        and we have the derivation:
        \begin{prooftree}
            \AxiomC{\( \Gamma , {x}: \sigma^n \wfdash M : \tau \)}
            \LeftLabel{\redlab{F:abs}}
            \UnaryInfC{\( \Gamma \wfdash \lambda x . M :  \sigma^n  \rightarrow \tau \)}
        \end{prooftree}
            
        By the encoding given in~\figref{ch2fig:auxencfail}, we have 
        $\recencodf{  \expr{M}  }  =   \lambda x . \recencodf{M\linsub{x_1,\cdots, x_n}{x} } [x_1 , \cdots , x_n \leftarrow x]$, where  $\#(x,M) = n$ and each $x_i$ is fresh and has the same type as $x$.
            From ${\Gamma}, {x}: \sigma^{n} \wfdash {M} : \tau$, we obtain after $n$ applications of Proposition~\ref{ch2prop:linhed_encfail} and Lemma~\ref{ch2lem:preser_linsub}: 
            \begin{equation*}\label{ch2eq:tp1fail}
                {\Gamma}, {x_1: \sigma, \cdots,  x_n: \sigma} \wfdash {M}\linsub{x_1,\cdots, x_n}{x} : \tau
            \end{equation*}
            By IH  we have:
              \begin{equation*}\label{ch2eq:tp2fail}
             \strcore{\Gamma}, {x_1: \sigma, \cdots,  x_n: \sigma} \wfdash \recencodf{M\linsub{x_1,\cdots, x_n}{x}} : \tau
            \end{equation*}
            
        which gives us the following derivation:
        
        \begin{prooftree}
         \LeftLabel{$(*)$}
                    \AxiomC{\( \strcore{\Gamma} , x_1: \sigma, \cdots, x_n: \sigma \wfdash \recencodf{M\linsub{x_1,\cdots, x_n}{x}} : \tau \)}
            \LeftLabel{\redlab{FS{:}share}}
            \UnaryInfC{\( \strcore{\Gamma} , x : \sigma^n  \wfdash \recencodf{M\linsub{x_1,\cdots, x_n}{x}}[x_1 , \cdots , x_n \leftarrow x] : \tau \)}
            \LeftLabel{\redlab{FS{:}abs \dash sh}}
            \UnaryInfC{\( \strcore{\Gamma} \wfdash \lambda x . (\recencodf{M\linsub{x_1,\cdots, x_n}{x}}[x_1 , \cdots , x_n \leftarrow x]) : \sigma^n \rightarrow \tau \)}
        \end{prooftree}
        and the result follows.

        \item Rule~\redlab{F:app}.

        In this case $\expr{M} = M\ B$, and by inversion of the typing derivation we have the derivation:
         \begin{prooftree}
            \AxiomC{\( \Gamma \wfdash M : \sigma^{j} \rightarrow \tau \)}
            \AxiomC{\( \Delta \wfdash B : \sigma^{k} \)}
                \LeftLabel{\redlab{F:app}}
            \BinaryInfC{\( \Gamma \contexcat \Delta \wfdash M\ B : \tau\)}
        \end{prooftree}
            By IH we have both 
            $ \strcore{\Gamma} \wfdash \recencodf{M} : \sigma^{j} \rightarrow \tau$
            and 
            $ \strcore{\Delta} \wfdash \recencodf{B} : \sigma^{k}$,
            and the result follows easily by Rule~$\redlab{FS{:}app}$ in \lamrsharfail:
              \begin{prooftree}
                            \AxiomC{\( \strcore{\Gamma} \wfdash \recencodf{M} : \sigma^{j} \rightarrow \tau \)}
                \AxiomC{\( \strcore{\Delta} \wfdash \recencodf{B} : \sigma^{k} \)}
                \LeftLabel{\redlab{FS{:}app}}
            \BinaryInfC{\( \strcore{\Gamma}, \strcore{\Delta} \wfdash \recencodf{M}\ \recencodf{B} : \tau\)}
        \end{prooftree}

        \item Rule~\redlab{F:ex \dash sub}.


        Then $\expr{M} = M \esubst{ B }{ x }$
        and the proof is split in two cases, depending on the shape of~$B$:
        
  \begin{enumerate}
  \item When $\#(x,M) = \size{B} = k \geq 1 $.
  
Then we have $B= \bag{N_1, \ldots ,N_n}$, $n\geq 1$.
Suppose w.l.o.g. that $n=2$, then $B= \bag{N_1,N_2}$ and by inversion  of the typing derivation we have the following derivation:
            
\begin{prooftree}
 \AxiomC{$\Delta_1\wfdash N_1:\sigma$}
 \AxiomC{$\Delta_2\wfdash N_2:\sigma$}
 \LeftLabel{\redlab{F:bag}}      
 \BinaryInfC{$\Delta_1 \contexcat \Delta_2\wfdash \bag{N_1}\cdot \bag{N_2}:\sigma^2$}
 \AxiomC{$\Gamma_1, {x}:\sigma^2 \wfdash M:\tau $}
 \LeftLabel{\redlab{F:ex \dash sub}}
 \BinaryInfC{$\Gamma_1 \contexcat \Delta_1 \contexcat \Delta_2 \wfdash M\esubst{B}{x}:\tau$}
 \end{prooftree}     
 where $\Gamma=\Gamma_1,\Delta_1,\Delta_2$.
            By IH we have both
            \begin{itemize}
                \item $ \strcore{\Delta_1} \vdash \recencodf{N_1}:\sigma$; and 
            \item $ \strcore{\Delta_2} \vdash \recencodf{N_2}:\sigma$; and
                \item $ \strcore{\Gamma_1} , {x}:\sigma^2 \wfdash \recencodf{M}:\tau $
            \end{itemize}

   We can expand 
    $ \strcore{\Gamma_1}, {x}:\sigma^2 \wfdash \recencodf{M}:\tau $ into $ \strcore{\Gamma_1}, x:\sigma \wedge \sigma \wfdash \recencodf{M}:\tau$, which gives $\strcore{\Gamma_1}, y_1:\sigma, y_2:\sigma\wfdash \recencodf{M}\linsub{y_1,y_2}{x}:\tau$, after two applications of Proposition~\ref{ch2prop:linhed_encfail} along with the application of Lemma \ref{ch2lem:preser_linsub}, with  $y_1,y_2$  fresh variables of the same type as $x$. Since the encoding $\recencodf{\cdot }$ commutes with the linear substitution $\linsub{\cdot }{\cdot }$ (Proposition~\ref{ch2prop:linhed_encfail}), if follows that, $\strcore{\Gamma_1}, y_1:\sigma, y_2:\sigma\wfdash \recencodf{M\linsub{y_1,y_2}{x}}:\tau$.
    
    

           Let $\Pi_1$ be the derivation obtained after  two consecutive applications of Rule~$\redlab{FS{:} ex \dash lin \dash sub}$:          
\begin{prooftree}\hspace{-1cm}
 \AxiomC{$\strcore{\Gamma_1}, y_1:\sigma, y_2:\sigma \wfdash \recencodf{M\linsub{y_1,y_2}{x}}:\tau$}
  \AxiomC{$\strcore{\Delta_1} \wfdash \recencodf{N_1}:\sigma$}                  
  \BinaryInfC{$ \strcore{\Gamma_1}, y_2:\sigma, \strcore{\Delta_1} \wfdash \recencodf{M\linsub{y_1,y_2}{x}}\linexsub{\recencodf{N_1}/y_1}:\tau$} 
  \AxiomC{$ \strcore{\Delta_2}\wfdash \recencodf{N_2}:\sigma$}
  \BinaryInfC{$ \strcore{\Gamma_1}, \strcore{\Delta_1}, \strcore{\Delta_2} \wfdash \recencodf{M\linsub{y_1,y_2}{x}} \linexsub{\recencodf{N_1}/y_1}\linexsub{\recencodf{N_2}/y_2}:\tau$}
\end{prooftree}

 Similarly, we can obtain a derivation $\Pi_2$ for:
 $$ \strcore{\Gamma_1},  \strcore{\Delta_1}, \strcore{\Delta_2} \wfdash \recencodf{M\linsub{y_1,y_2}{x}}\linexsub{\recencodf{N_1}/y_2}\linexsub{\recencodf{N_2}/y_1}:\tau$$

 By the encoding given in Figure~\ref{ch2fig:auxencfail}, we have
            \[
            \begin{aligned}
                \recencodf{M\esubst{\bag{N_1,N_2}}{x}}
                 = & \recencodf{M\linsub{y_1,y_2}{x}}\linexsub{\recencodf{N_1}/y_1
                }\linexsub{\recencodf{N_2}/y_2} +\\
                & \recencodf{M\linsub{y_1,y_2}{x}}\linexsub{\recencodf{N_1}/y_2
                }\linexsub{\recencodf{N_2}/y_1}
            \end{aligned}
            \]

 Therefore, 
         
  \begin{prooftree}
 \AxiomC{$\Pi_1$}
  \AxiomC{$\Pi_2$}
  \LeftLabel{$\redlab{FS{:}sum}$}
  \BinaryInfC{$\strcore{\Gamma_1}, \strcore{\Delta_1}, \strcore{\Delta_2} \wfdash \recencodf{M\esubst{\bag{N_1,N_2}}{x}}:\tau$}
 \end{prooftree}
            
      and the result follows.

            \item  $ \#(x,M) = k \neq \size{B}$.
            
           In this case, $\size{B}=j$ for some $j\neq k$, and by inversion of the typing derivation we have the following derivation:
            
            \begin{prooftree}
                    \AxiomC{\( \Delta \wfdash B : \sigma^{j} \)}
                    \AxiomC{\( \Gamma_1 , {x}:\sigma^{k} \wfdash M : \tau \)}
                \LeftLabel{\redlab{F:ex \dash sub}}    
                \BinaryInfC{\( \Gamma_1 \contexcat \Delta \wfdash M \esubst{ B }{ x } : \tau \)}
            \end{prooftree}
 where $\Gamma=\Gamma_1 \contexcat \Delta$. By IH we have both
            
  \begin{itemize}
     \item $ \strcore{\Delta} \wfdash \recencodf{B} : \sigma^{j} $; and 
                \item  $ \strcore{\Gamma_1}, \hat{x}:\sigma^{k} \wfdash \recencodf{M}:\tau$. 
            \end{itemize}

We analyse two cases, depending on the number $k$ of occurrences of $x$ in $M$:
\begin{enumerate} 

\item $k=0$.
 
 From $\Gamma_1, {x:\omega} \wfdash M:\tau$, which we get $\Gamma_1 \wfdash M:\tau$, via Rule~$\redlab{F:weak}$. The IH gives $ \strcore{\Gamma_1} \wfdash \recencodf{M}:\tau$, which entails:
\begin{prooftree}
                \AxiomC{\( \strcore{\Gamma_1}  \wfdash \recencodf{M} : \tau\)}
                \LeftLabel{ \redlab{FS{:}weak}}
                \UnaryInfC{\( \strcore{\Gamma_1}, x: \omega \wfdash \recencodf{M}[\leftarrow x]: \tau \)}
                \AxiomC{$ \strcore{\Delta} \wfdash \recencodf{B} : \sigma^{j} $}
                \LeftLabel{ \redlab{FS{:}ex \dash sub}}
                \BinaryInfC{$ \strcore{\Gamma_1} ,\strcore{\Delta} \wfdash \recencodf{M}[\leftarrow x]\esubst{B}{x}: \tau$}
\end{prooftree}

By the encoding given in Figure~\ref{ch2fig:auxencfail},  $\recencodf{M\esubst{B}{x}}=\recencodf{M}[\leftarrow x]\esubst{\recencodf{B}}{x}$, and the result follows.
  \item  $k> 0$.
  
   By applying Proposition~\ref{ch2prop:linhed_encfail} in \( \strcore{\Gamma_1} , {x}:\sigma^{k} \wfdash \recencodf{ M} : \tau \), we obtain 
  \[ \strcore{\Gamma_1} , x_1:\sigma,\ldots,  x_k:\sigma \wfdash \recencodf{M}\linsub{x_1,\ldots, x_k}{x} : \tau \]
 From Proposition~\ref{ch2prop:linhed_encfail} and Lemma \ref{ch2lem:preser_linsub}, it follows that  $ \strcore{ \Gamma_1} , x_1:\sigma,\ldots,  x_k:\sigma \wfdash \recencodf{M\linsub{x_1,\ldots, x_k}{x}} : \tau $, which entails $x\notin \dom{\Gamma_1}$ since $k\neq 0$. First we give $\Pi$:
 
\begin{prooftree}
    \AxiomC{$ \strcore{\Gamma_1}\ , x_1:\sigma,\ldots,  x_k:\sigma \wfdash \recencodf{M\linsub{x_1,\ldots, x_k}{x}} : \tau $}
    \LeftLabel{\redlab{FS{:}share}}    
    \UnaryInfC{\( \strcore{\Gamma_1} , x:\sigma^{k} \wfdash \recencodf{M\linsub{x_1. \cdots , x_k}{x}} [x_1. \cdots , x_k \leftarrow x]  : \tau \)}
\end{prooftree}

finally we give the full derivation:

\begin{prooftree}\hspace{-1cm}
    \AxiomC{\( \strcore{\Delta} \wfdash \recencodf{B} : \sigma^{j} \)}
    \AxiomC{$ \Pi $}
    \LeftLabel{\redlab{FS{:}ex \dash sub}}    
    \BinaryInfC{\( \strcore{\Gamma}, \strcore{\Delta} \wfdash \recencodf{M\langle x_1. \cdots , x_k / x  \rangle} [x_1. \cdots , x_k \leftarrow x] \esubst{ \recencodf{B} }{ x } : \tau \)}
\end{prooftree}
            
%
    \end{enumerate}         
            
       
          \end{enumerate}

        \item Rule~\redlab{F:fail}.

        The result follows trivially, because the encoding of failure in~\figref{ch2fig:auxencfail} is such that $\recencodf{\fail^{\widetilde{x}}}=\fail^{\widetilde{x}} $.
                
        \item Rule~\redlab{F:sum}.
        
        This case follows easily by IH.
    \end{enumerate}
    \end{enumerate}

\end{proof}

\preservencintolamrfailtwo*

\begin{proof}
By mutual induction on the typing derivations $\Gamma\wfdash B:\sigma$ and $\Gamma\wfdash \expr{M}:\sigma$, exploiting both Proposition~\ref{ch2prop:linhed_encfail} and Lemma~\ref{ch2lem:preser_linsub}. The analysis for bags (Part 1) follows directly from the IHs and will be omitted.

As for Part 2, there are two main cases to consider:
\begin{enumerate}[i)]
    \item $\expr{M} = M$. 
    
Without loss of generality, assume $\lfv{M} = \{x,y\}$. Then, 
\begin{equation}\label{ch2eq:thmpres1fail}
{x}:\sigma_1^j, {y}:\sigma_2^k \wfdash M : \tau    
\end{equation}
where  $\#(x,M)=j$ and 
$\#(y,M)=k$, for some positive integers $j$ and $k$.

After $j+k$ applications of Lemma~\ref{ch2lem:preser_linsub}
we obtain:
\begin{equation*}\label{ch2eq:thmpres2fail}
x_1:\sigma_1, \cdots, x_j:\sigma_1, y_1:\sigma_2, \cdots, y_k:\sigma_2 \wfdash M\linsub{\widetilde{x}}{x}\linsub{\widetilde{y}}{y} : \tau    
\end{equation*}
 where  $\widetilde{x}=x_{1},\ldots, x_{j}$
   and $\widetilde{y}=y_{1},\ldots, y_{k}$. 
From Proposition~\ref{ch2prop:linhed_encfail} and Lemma~\ref{ch2lem:preser_linsub}
one has
\begin{equation*}\label{ch2eq:thmpres3fail}
{x_1:\sigma_1, \ldots, x_j:\sigma_1, y_1:\sigma_2, \ldots, y_k:\sigma_2} \wfdash \recencodf{M\linsub{\widetilde{x}}{x}\linsub{\widetilde{y}}{y}} : \tau  
\end{equation*}
Since ${x_1:\sigma_1, \ldots, x_j:\sigma_1, y_1:\sigma_2, \ldots, y_k:\sigma_2}= x_1:\sigma_1, \ldots, x_j:\sigma_1, y_1:\sigma_2, \ldots, y_k:\sigma_2$, 
we have the following derivation:
\begin{prooftree}
    \AxiomC{${x_1:\sigma_1, \cdots, x_j:\sigma_1, y_1:\sigma_2, \cdots, y_k:\sigma_2} \wfdash \recencodf{M\linsub{\widetilde{x}}{x}\linsub{\widetilde{y}}{y}} : \tau$}
    \LeftLabel{$\redlab{FS{:}share}$}
    \UnaryInfC{$x:\sigma_1^j, y_1:\sigma_2, \cdots, y_k:\sigma_2 \wfdash \recencodf{M\linsub{\widetilde{x}}{x}\linsub{\widetilde{y}}{y}}[\widetilde{x}\leftarrow x] : \tau$}
\LeftLabel{$\redlab{FS{:}share}$}
    \UnaryInfC{$x:\sigma_1^j, y:\sigma_2^k \wfdash \recencodf{M\linsub{\widetilde{x}}{x}\linsub{\widetilde{y}}{y}}[\widetilde{x}\leftarrow x][\widetilde{y}\leftarrow y] : \tau$}
    \end{prooftree}
    
    By expanding~\defref{ch2def:enctolamrsharfail}, we have 
$$
 \recencodopenf{M} = 
\recencodf{M\linsub{\widetilde{x}}{x}\linsub{\widetilde{y}}{y}}[\widetilde{x}\leftarrow x][\widetilde{y}\leftarrow y], 
$$

    which completes the proof for this case. 
    
    \item $\expr{M} = M_1 + \cdots + M_n$.
    
    This case proceeds easily by IH, using Rule~$\redlab{FS{:}sum}$.
    \end{enumerate}
\end{proof}

\subsection{Completeness and Soundness}
\label{ch2app:compandsoundone}

\appcompletenessone*

\begin{proof}
By induction on the rule from~\figref{ch2fig:reductions_lamrfail} applied to infer $\expr{N}\redd \expr{M}$, distinguishing  three cases. Below $\widetilde{[x_{1k}\leftarrow x_{1k}]}$ abbreviates $[\widetilde{x_1}\leftarrow x_1]\ldots [\widetilde{x_k}\leftarrow x_k]$:

\begin{myEnumerate}

    \item The rule applied is $\redlab{R}=\redlab{R:Beta}$. 
    
    In this case, 
        $\expr{N}= (\lambda x. M) B$, the reduction is 
    \begin{prooftree}
        \AxiomC{}
        \LeftLabel{\redlab{R:Beta}}
        \UnaryInfC{\((\lambda x. M) B \redd M\ \esubst{B}{x}\)}
    \end{prooftree}

          and $\expr{M}= M\esubst{B}{x}$. Below we assume $\lfv{\expr{N}}=\{x_1,\ldots, x_k\}$ and $\widetilde{x_i}=x_{i_1},\ldots, x_{i_{j_i}}$, where $j_i= \#(x_i, N)$, for $1\leq i\leq k$.
        On the one hand, we have:
        \begin{equation}\label{ch2eq:beta1fail}
            \begin{aligned}
            \recencodopenf{\expr{N}}&= \recencodopenf{(\lambda x. M)B}\\
            &=  \recencodf{((\lambda x. M)B) \langle{\widetilde{x_1}/x_1}\rangle\cdots \langle{\widetilde{x_k}/x_k}\rangle}\widetilde{[x_{1k}\leftarrow x_{1k}]}
            \\
            &=  \recencodf{(\lambda x. M^{'})B'}\widetilde{[x_{1k}\leftarrow x_{1k}]}\\
            &=  (\recencodf{\lambda x. M^{'}}\recencodf{B'})\widetilde{[x_{1k}\leftarrow x_{1k}]} \\
            &=  ((\lambda x.\recencodf{ M^{'}\langle{\widetilde{y}/x}\rangle}[\widetilde{y}\leftarrow x])\recencodf{B'})\widetilde{[x_{1k}\leftarrow x_{1k}]} \\
            &\redd_{\redlab{RS{:}Beta}} (\recencodf{ M^{'} \langle{\widetilde{y}/x} \rangle} [\widetilde{y} \leftarrow x] \esubst{\recencodf{B'}}{x}) \widetilde{[x_{1k}\leftarrow x_{1k}]}=\expr{L}
            \end{aligned}
        \end{equation}
      
        \revo{A24}{where we define $M'$ and $B'$ to be $M$ and $B$ after the substitutions of $\langle \widetilde{x_1}/x_1\rangle\cdots \langle \widetilde{x_k}/x_k\rangle$.} On the other hand, we have:
        \begin{equation}\label{ch2eq:beta2fail}
            \begin{aligned}
               \recencodopenf{\expr{M}}&=\recencodopenf{M\esubst{B}{x}}\\
               &=\recencodf{M\esubst{B}{x}\langle{\widetilde{x_1}/x_1}\rangle\cdots \langle{\widetilde{x_k}/x_k}\rangle} \widetilde{[x_{1k}\leftarrow x_{1k}]}\\
               &=\recencodf{M^{'}\esubst{B'}{x}} \widetilde{[x_{1k}\leftarrow x_{1k}]}
            \end{aligned}
        \end{equation}

        We need to analyse two sub-cases: either $\#(x,M') = \size{B} = k \geq 1 $ or $\#(x,M') = k$ and our first sub-case is not met.
        \begin{enumerate}[i)]
            \item If $\#(x,M') = \size{B} = k \geq 1$ then we can reduce $\expr{L}$ using Rule~\redlab{RS:Ex-sub}:
            \begin{equation*}
                \begin{aligned}
                    \expr{L}\redd & \sum_{B_i\in \perm{\recencodf{B}}}\recencodf{M^{'}\linsub{\widetilde{y}}{x}}\linexsub{B_i(1)/y_1}\cdots \linexsub{B_i(n)/y_n} \widetilde{[x_{1k}\leftarrow x_{1k}]} \\
                     = & \recencodopenf{\expr{M}}
                \end{aligned}
            \end{equation*}
            
            From \eqref{ch2eq:beta1fail} and \eqref{ch2eq:beta2fail} and $\widetilde{y}=y_1\ldots y_n$, one has the desired result.

            \item Otherwise,  $\#(x,M) = k$ (either $k=0$ or $k\neq \size{B}$).
            


            Expanding  the encoding in \eqref{ch2eq:beta2fail} :
            \begin{align*}
                \recencodopenf{M}&= \recencodf{M^{'}\esubst{B'}{x}} \widetilde{[x_{1k}\leftarrow x_{1k}]}\\
                & = (\recencodf{ M^{'} \langle{\widetilde{y}/x} \rangle} [\widetilde{y} \leftarrow x] \esubst{\recencodf{B'}}{x}) \widetilde{[x_{1k}\leftarrow x_{1k}]}
            \end{align*}
           Therefore  $\recencodopenf{M} =\expr{L}$ and $\recencodopenf{\expr{N}}\redd \recencodopenf{\expr{M}}$.
        
        \end{enumerate}
        

    \item The rule applied is $\redlab{R}=\redlab{R:Fetch}$.
    
Then $\expr{N}=M\esubst{B}{x}$ and  the reduction is 
            \begin{prooftree}
     \AxiomC{$\headf{M} = x \quad B = \bag{N_1, \dots ,N_n}, \ n\geq 1 \quad  \#(x,M) = n $}
      \LeftLabel{\redlab{R:Fetch}}                        \UnaryInfC{$M\esubst{B}{x} \redd \sum_{i=1}^n M\headlin{N_i/x}\esubst{B\linsetminus N_i}{x} $}
            \end{prooftree}
            
            with $\expr{M}= \sum_{i=1}^n M\headlin{N_i/x}\esubst{B\linsetminus N_i}{x}$.

            Below we assume $\lfv{\expr{N}}=\lfv{M\esubst{\bag{N1}}{x}}=\{x_1,\ldots, x_k\}$. We distinguish two cases:

            \begin{enumerate}
                    \item $n = 1$.
                    
                   Then $B = \bag{N_1}$ and  $\expr{N}=M\esubst{\bag{N_1}}{x}\redd  M\headlin{N_1/x} \esubst{\oneb}{x} = \expr{M} $.

                On the one hand, we have:
\begin{equation*}\label{ch2eq:fetch3fail}
  \begin{aligned}
  \recencodopenf{\expr{N}}&= \recencodopenf{M\esubst{\bag{N_1}}{x}}\\                      &=    \recencodf{(M\esubst{\bag{N_1}}{x})\langle \widetilde{x_1}/x_1\rangle\cdots \langle \widetilde{x_k}/x_k\rangle }\widetilde{[x_{1k}\leftarrow x_{1k}]}\\
  &=  \recencodf{M'\esubst{\bag{N_1'}}{x}}\widetilde{[x_{1k}\leftarrow x_{1k}]}\\
& = \recencodf{M'\langle y_1 / x \rangle}\linexsub{\recencodf{N_1'}/y_1} \widetilde{[x_{1k}\leftarrow x_{1k}]}, \text{ notice that }\headf{M'}=y_1\\
& = \recencodf{M^{''}}\linexsub{\recencodf{N_1'}/y_1} \widetilde{[x_{1k}\leftarrow x_{1k}]}\\
    &\redd_{\redlab{RS{:}Lin\dash Fetch}} \recencodf{M''}\headlin{\recencodf{N_1'}/y_1}\widetilde{[x_{1k}\leftarrow x_{1k}]}\\
  \end{aligned}
\end{equation*}
\revo{A24}{where we define $M'$ and $N_1'$ to be $M$ and $N_1$ after the substitutions of $\langle \widetilde{x_1}/x_1\rangle\cdots \langle \widetilde{x_k}/x_k\rangle$; similarly, we define $M''$ to be $M'$after the substitution of $y_1$ for $x$}. On the other hand,                    
\begin{equation*}\label{ch2eq:fetch4fail}
 \begin{aligned}
 \recencodopenf{\expr{M}}&= \recencodopenf{M\headlin{N_1/x} \esubst{\oneb}{x}}\\
  &= \recencodf{M\headlin{N_1/x} \esubst{\oneb}{x} \langle \widetilde{x_1}/x_1\rangle\cdots \langle \widetilde{x_k}/x_k\rangle }\widetilde{[x_{1k}\leftarrow x_{1k}]}\\
  &= \recencodf{M'\headlin{N_1'/x} \esubst{\oneb}{x}} \widetilde{[x_{1k}\leftarrow x_{1k}]}\\
  &= \recencodf{M'\headlin{N_1'/x}} [\leftarrow x]  \esubst{\oneb}{x}\widetilde{[x_{1k}\leftarrow x_{1k}]}\\
                     \end{aligned}
                    \end{equation*}
            
            By the congruence defined in~\figref{ch2fig:rPrecongruencefail} for $\lamrfail$, one has  $M\esubst{\oneb}{x}\pequiv M$.
            
            Therefore, $\expr{M} = M\headlin{N_1/x}  \esubst{\oneb}{x} \pequiv M\headlin{N_1/x} =  \expr{M}'$. Expanding $\recencodopenf{\expr{M}'}$ we have:
 \begin{equation*}\label{ch2eq:fetch5fail}                   
 \begin{aligned}
 \recencodopenf{\expr{M}'}&=\recencodopenf{M\headlin{N_1/x}} \\
 &=    \recencodf{M\headlin{N_1/x} \langle \widetilde{x_1}/x_1\rangle\cdots \langle \widetilde{x_j}/x_j\rangle } \widetilde{[x_{1k}\leftarrow x_{1k}]}\\
 &=  \recencodf{M' \headlin{N_1'/x}  } \widetilde{[x_{1k}\leftarrow x_{1k}]}\\
 &=  \recencodf{M' }\headlin{\recencodf{N_1'}/x}   \widetilde{[x_{1k}\leftarrow x_{1k}]}\\
 \end{aligned}
  \end{equation*}
                Hence, $\recencodopenf{\expr{N}}\redd\recencodopenf{\expr{M'}}$ and the result follows.
        
                    \item   $n> 1$ 
                    
                     To simplify the proof, we take $n=2$ (the analysis when $n>2$ is similar). Then $B=\bag{N_1,N_2}$ and the reduction is
                     
                $$\expr{N}=M\esubst{B}{x}\redd  M\headlin{N_1/x} \esubst{\bag{N_2}}{x}+M\headlin{N_2/x} \esubst{\bag{N_1}}{x}=\expr{M}$$
                    Notice that $\#(x,M)=2$, we take   $y_1,y_2$ fresh variables.
            On the one hand, we have:  
  \begin{equation}\label{ch2eq:fetch1fail}
 \begin{aligned}
\recencodopenf{\expr{N}}&=\recencodopenf{M\esubst{B}{x}}
=    \recencodf{M\esubst{B}{x}\langle \widetilde{x_1}/x_1\rangle\cdots \langle \widetilde{x_k}/x_k\rangle } \widetilde{[x_{1k}\leftarrow x_{1k}]}\\
 &=  \recencodf{M'\esubst{B'}{x}}\widetilde{[x_{1k}\leftarrow x_{1k}]}\\
 & = ( \recencodf{M'\langle y_1, y_2 / x \rangle}\linexsub{\recencodf{N_1'}/y_1} \linexsub{\recencodf{N_2'}/y_2}\\
 &  \qquad +
 \recencodf{M'\langle y_1, y_2 / x \rangle}\linexsub{\recencodf{N_2'}/y_1} \linexsub{\recencodf{N_1'}/y_2})\widetilde{[x_{1k}\leftarrow x_{1k}]}\\
& =( \recencodf{M^{''}}\linexsub{\recencodf{N_1'}/y_1} \linexsub{\recencodf{N_2'}/y_2}\\
 & \qquad + \recencodf{M^{''}}\linexsub{\recencodf{N_2'}/y_1} \linexsub{\recencodf{N_1'}/y_2})\widetilde{[x_{1k}\leftarrow x_{1k}]}\\
&\redd^2_{\redlab{RS{:}Lin\dash Fetch}}
     (\recencodf{M''}\headlin{\recencodf{N_1'}/y_1}\linexsub{\recencodf{N_2'}/y_2} \\
     & \qquad + \recencodf{M''}\headlin{\recencodf{N_2'}/y_1}\linexsub{\recencodf{N_1'}/y_2})\widetilde{[x_{1k}\leftarrow x_{1k}]}\\
& = \expr{L}.
\end{aligned}
\end{equation}
                    
\revo{A24}{where we define $M'$ and $B'$ to be $M$ and $B$ after the substitutions of $\langle \widetilde{x_1}/x_1\rangle\cdots \langle \widetilde{x_k}/x_k\rangle$ and $N_1' , N_2'$ are the the elements of the bag $B'$. 
Similarly, we define $M''$ to be $M'$ after the substitution $\langle y_1, y_2 / x \rangle$}.
On the other hand, we have:
  \begin{equation}\label{ch2eq:fetch2fail}
  \begin{aligned}
  \recencodopenf{\expr{M}}&=\recencodopenf{M\headlin{N_1/x} \esubst{\bag{N_2}}{x}+M\headlin{N_2/x} \esubst{\bag{N_1}}{x}}\\
  &=\recencodopenf{M\headlin{N_1/x} \esubst{\bag{N_2}}{x}}+\recencodopenf{M\headlin{N_2/x} \esubst{\bag{N_1}}{x}}\\
  &=\recencodf{M\headlin{N_1/x} \esubst{N_2}{x}}\langle \widetilde{x_1}/x_1\rangle\cdots \langle \widetilde{x_k}/x_k\rangle\widetilde{[x_{1k}\leftarrow x_{1k}]}\\
 & \qquad \qquad + \recencodf{M\headlin{N_2/x} \esubst{N_1}{x}\langle \widetilde{x_1}/x_1\rangle\cdots \langle \widetilde{x_k}/x_k\rangle }\widetilde{[x_{1k}\leftarrow x_{1k}]}\\
 &=\recencodf{M'\headlin{N_1'/x} \esubst{N_2'}{x}}\widetilde{[x_{1k}\leftarrow x_{1k}]}\\
 & \qquad  \qquad + \recencodf{M'\headlin{N_2'/x} \esubst{N_1'}{x} }\widetilde{[x_{1k}\leftarrow x_{1k}]}\\
   & =\recencodf{M'\headlin{N_1'/x}} \linexsub{\recencodf{N_2'}/y_2}\widetilde{[x_{1k}\leftarrow x_{1k}]} \\
   & \qquad \qquad + \recencodf{M'\headlin{N_2'/x}} \linexsub{\recencodf{N_1'}/y_2} \widetilde{[x_{1k}\leftarrow x_{1k}]}
   \end{aligned}        
 \end{equation}
                    The reductions in \eqref{ch2eq:fetch1fail} and \eqref{ch2eq:fetch2fail} lead to identical expressions, up to renaming of shared variables, which are taken to be fresh by definition. In both cases, we have taken the same  fresh variables.
                    
            \end{enumerate}
         
             \item The rule applied is $ \redlab{R}\neq \redlab{R:Beta}$ and $ \redlab{R}\neq \redlab{R:Fetch}$. There are two possible cases. Below $\widetilde{[x_{1n}\leftarrow x_{1n}]}$  abbreviates $[\widetilde{x_1}\leftarrow x_1]\cdots [\widetilde{x_n}\leftarrow x_n]$:
             \begin{myEnumerate}
                 
            \item  $\redlab{R}=\redlab{R:Fail}$
            
            Then $\expr{N}=M\esubst{B}{x}$ and the reduction is 
            
            \begin{prooftree}
            \AxiomC{$\#(x,M)\neq \size{B}  \qquad  \widetilde{y} = (\mfv{M}\setminus x)\uplus \mfv{B} $}
            \LeftLabel{\redlab{R:Fail}}
            \UnaryInfC{$M\ \esubst{ B}{x } \redd \sum_{\perm{B}} \fail^{\widetilde{y}}$}
            \end{prooftree}
            where $\expr{M}= \sum_{\perm{B}} \fail^{\widetilde{y}}$. Below assume $\lfv{\expr{N}}=\{x_1,\ldots, x_n\}$.

                On the one hand, we have:
            \begin{equation*}\label{ch2eq:fail1fail}
                \begin{aligned}
                \recencodopenf{\expr{N}} &= \recencodopenf{M\esubst{B}{x}}
                = \recencodf{M\esubst{B}{x}\langle \widetilde{x_1}/x_1\rangle\cdots \langle \widetilde{x_n}/x_n\rangle } \widetilde{[x_{1n}\leftarrow x_{1n}]} \\
                 &= \recencodf{M'\esubst{B'}{x} } \widetilde{[x_{1n}\leftarrow x_{1n}]} \\
                &=  \recencodf{M'\langle y_1, \cdots , y_k / x  \rangle} [y_1, \cdots , y_k \leftarrow x] \esubst{ \recencodf{B'} }{ x } \widetilde{[x_{1n}\leftarrow x_{1n}]}\\
                & \redd_{\redlab{RS{:}Fail}} \sum_{\perm{B}} \fail^{\widetilde{y}} \  \widetilde{[x_{1n}\leftarrow x_{1n}]}=\expr{L}\\
                \end{aligned}
            \end{equation*}
            
            \revo{A24}{where we define $M'$ and $B'$ to be $M$ and $B$ after the substitutions of $\langle \widetilde{x_1}/x_1\rangle\cdots \langle \widetilde{x_k}/x_k\rangle$}. On the other hand, we have:
            \begin{equation*}\label{ch2eq:fail2fail}
                \begin{aligned}
                \recencodopenf{\expr{M}} &= \recencodopenf{\sum_{\perm{B}} \fail^{\widetilde{y}}}= \sum_{\perm{B}} \recencodopenf{\fail^{\widetilde{y}}}\\
                &= \sum_{\perm{B}} \recencodf{\fail^{\widetilde{y}}} \widetilde{[x_{1n}\leftarrow x_{1n}]} 
                = \sum_{\perm{B}} \fail^{\widetilde{y}} \widetilde{[x_{1n}\leftarrow x_{1n}]} = \expr{L}\\
                \end{aligned}
            \end{equation*}
            
            Therefore, $\recencodopenf{\expr{N}} \redd \recencodopenf{\expr{M}} $ and the result follows.

            \item  $\redlab{R}= \redlab{R:Cons_1}$.
            
            Then $\expr{N}= \fail^{\widetilde{y}}\ B$ and the reduction is

            \begin{prooftree}
            \AxiomC{$\size{B}=k \qquad  \widetilde{z} = \mfv{B} $}        
            \LeftLabel{\redlab{R:Cons_1}}
            \UnaryInfC{$ \fail^{\widetilde{y}}\ B  \redd \sum_{\perm{B}} \fail^{\widetilde{y} \uplus \widetilde{z}} $}
            \end{prooftree}
            

           and $\expr{M}'= \sum_{\perm{B}} \fail^{\widetilde{y} \uplus \widetilde{z}}$. Below we assume $\lfv{\expr{N}}=\{x_1,\ldots, x_n\}$.
           
                On the one hand, we have:
            \begin{equation*}\label{ch2eq:consume1fail}
                \begin{aligned}
                \recencodopenf{N} &= \recencodopenf{\fail^{\widetilde{y}}\ B}
                = \recencodf{ \fail^{\widetilde{y}}\ B \langle \widetilde{x_1}/x_1\rangle\cdots \langle \widetilde{x_n}/x_n\rangle } \widetilde{[x_{1n}\leftarrow x_{1n}]}\\
                 &= \recencodf{ \fail^{\widetilde{y'}}\ B'  } \widetilde{[x_{1n}\leftarrow x_{1n}]}= \recencodf{ \fail^{\widetilde{y'}}} \ \recencodf{ B' } \widetilde{[x_{1n}\leftarrow x_{1n}]}\\
                &= \fail^{\widetilde{y'}} \ \recencodf{ B' } \widetilde{[x_{1n}\leftarrow x_{1n}]}\\
                & \redd_{\redlab{RS{:}Cons_1}} \sum_{\perm{B}} \fail^{\widetilde{y'} \cup \widetilde{z'}}  \widetilde{[x_{1n}\leftarrow x_{1n}]}=\expr{L}\\
                \end{aligned}
            \end{equation*}
            
            \revo{A24}{where we define $B'$ to be $B$ after the substitutions of $\langle \widetilde{x_1}/x_1\rangle\cdots \langle \widetilde{x_k}/x_k\rangle$. Similarly, $\widetilde{y'}$ and $\widetilde{z'}$ are $\widetilde{y}$ and $\widetilde{z}$ after the substitution $\langle \widetilde{x_1}/x_1\rangle\cdots \langle \widetilde{x_k}/x_k\rangle$}. 
            On the other hand, we have:
            \begin{equation*}\label{ch2eq:consume2fail}
                \begin{aligned}
                \recencodopenf{M} &= \recencodopenf{\sum_{\perm{B}} \fail^{\widetilde{y} \uplus \widetilde{z}}} \\
                & = \sum_{\perm{B}} \recencodopenf{\fail^{\widetilde{y} \uplus \widetilde{z}}}\\
                &= \sum_{\perm{B}} \recencodf{\fail^{\widetilde{y'} \uplus \widetilde{z'}}}\widetilde{[x_{1n}\leftarrow x_{1n}]} \\
                & = \sum_{\perm{B}} \fail^{\widetilde{y'} \cup \widetilde{z'}}\widetilde{[x_{1n}\leftarrow x_{1n}]}=\expr{L}\\
                \end{aligned}
            \end{equation*}
            
            Therefore, $\recencodopenf{\expr{N}}\redd \expr{L}= \recencodopenf{ \expr{M}}$, and the result follows.

            \item $\redlab{R}= \redlab{R:Cons_2}$
            
            Then $\expr{N}= \fail^{\widetilde{y}}\ \esubst{B}{x}$ and the reduction is 
        \begin{prooftree}
            \AxiomC{$\size{B} = k$}
            \AxiomC{\(  \#(x , \widetilde{y}) + k  \not= 0 \)}
            \AxiomC{\( \widetilde{z} = \mfv{B} \)}
            \LeftLabel{$\redlab{R:Cons_2}$}
            \TrinaryInfC{\( \fail^{\widetilde{y}}\ \esubst{B}{x}  \redd \sum_{\perm{B}} \fail^{(\widetilde{y} \setminus x) \uplus\widetilde{z}} \)}
        \end{prooftree}
            
      and $\expr{M}=\sum_{\perm{B}} \fail^{(\widetilde{y} \setminus x) \uplus\widetilde{z}}$. 
            Below we assume $\lfv{\expr{N}}=\{x_1,\ldots, x_n\}$.

            On the one hand, we have: (below $\widetilde{y}=y_1,\ldots, y_m$)
            \begin{equation}\label{ch2eq:consume3fail}
                \begin{aligned}
                \recencodopenf{\expr{N}} &= \recencodopenf{\fail^{\widetilde{y}}\ \esubst{B}{x}}\\
                &= \recencodf{ \fail^{\widetilde{y}}\ \esubst{B}{x} \langle \widetilde{x_1}/x_1\rangle\cdots \langle \widetilde{x_n}/x_n\rangle } \widetilde{[x_{1n}\leftarrow x_{1n}]}\\
                &= \recencodf{ \fail^{\widetilde{y'}}\langle y_1 / x  \rangle \cdots \langle y_m / x  \rangle}[\widetilde{y}\leftarrow x] \ \esubst{ \recencodf{ B' } }{x} \widetilde{[x_{1n}\leftarrow x_{1n}]}\\
                &= \fail^{\widetilde{y''}} [\widetilde{y}\leftarrow x] \ \esubst{ \recencodf{ B' } }{x} \widetilde{[x_{1n}\leftarrow x_{1n}]}\\
                & \redd_{\redlab{RS{:}Cons_2}} \fail^{(\widetilde{y'} \setminus x) \cup\widetilde{z'}} \widetilde{[x_{1n}\leftarrow x_{1n}]}\\
                \end{aligned}
            \end{equation}
            
            As $\widetilde{y}$ consists of free variables,  in $\fail^{\widetilde{y}}\ \esubst{B}{x} \langle \widetilde{x_1}/x_1\rangle\cdots \langle \widetilde{x_n}/x_n\rangle$ the substitutions also occur on $ \widetilde{y}$ resulting in a new $\widetilde{y'}$ where all $x_i$'s are replaced with their fresh components in $\widetilde{x_i}$. Similarly for \revo{A24}{$z'$ and $B'$ as well as } $\widetilde{y''}$ being $\widetilde{y'}$ with each $x$ replaced with a fresh $y_i$. On the other hand, we have:
            \begin{equation}\label{ch2eq:consume4fail}
                \begin{aligned}
                \recencodopenf{M} &= \recencodopenf{\sum_{\perm{B}} \fail^{(\widetilde{y} \setminus x) \uplus\widetilde{z}}}
                = \sum_{\perm{B}} \recencodopenf{\fail^{(\widetilde{y} \setminus x) \uplus\widetilde{z}}}\\
                &= \sum_{\perm{B}} \recencodf{\fail^{(\widetilde{y'} \setminus x) \uplus\widetilde{z}}} \widetilde{[x_{1n}\leftarrow x_{1n}]}
                 = \fail^{(\widetilde{y'} \setminus x) \cup \widetilde{z'}} \widetilde{[x_{1n}\leftarrow x_{1n}]}\\
                \end{aligned}
            \end{equation}
            
            The reductions in \eqref{ch2eq:consume3fail} and \eqref{ch2eq:consume4fail} lead to identical expressions.
        \end{myEnumerate}
    \end{myEnumerate}
    
     As before, the reduction via Rule~$\redlab{R} $ could occur inside a context (cf. Rules $\redlab{R:TCont}$ and $\redlab{R:ECont}$). We consider only the case when the contextual rule used is $\redlab{R:TCont}$. We have $\expr{N} = C[N]$. When we have $C[N] \redd_{\redlab{R}} C[M] $ such that $N \redd_{\redlab{R}} M$ we need to show that $\recencodopenf{ C[N]} \redd^j \recencodopenf{ C[M] }$for some $j$ dependent on ${\redlab{R}}$. 
     Firstly, let us assume ${\redlab{R}} = \redlab{R:Cons_2}$  then we take $j = 1$. Let us take $C[\cdot]$ to be $[\cdot]B$ and $\lfv{NB} = \{ x_1, \cdots , x_k  \}$ then 
     \[
        \begin{aligned}
            \recencodopenf{ N B}  & = \recencodf{NB\linsub{\widetilde{x_{1}}}{x_1}\cdots \linsub{\widetilde{x_k}}{x_k}}\widetilde{[x_{1k}\leftarrow x_{1k}]} \\
             & = \recencodf{N' B'}\widetilde{[x_{1k}\leftarrow x_{1k}]} = \recencodf{N'}\recencodf{ B'}\widetilde{[x_{1k}\leftarrow x_{1k}]} \\
        \end{aligned}
     \]
    We take $N'B'= NB\linsub{\widetilde{x_{1}}}{x_1}\cdots \linsub{\widetilde{x_k}}{x_k}$, and by the IH that $ \recencodf{N}\redd \recencodf{M}$ and hence we can deduce that $\recencodf{N'}\redd \recencodf{M'}$ where $M'B'= MB\linsub{\widetilde{x_{1}}}{x_1}\cdots \linsub{\widetilde{x_k}}{x_k}$. Finally,
    \[
        \begin{aligned}
            \recencodf{N'}\recencodf{ B'}\widetilde{[x_{1k}\leftarrow x_{1k}]}\redd \recencodf{M'}\recencodf{ B'}\widetilde{[x_{1k}\leftarrow x_{1k}]}
        \end{aligned}
    \]
     and hence $ \recencodopenf{C[N]} \redd \recencodopenf{C[M]} $.
\end{proof}

\soundnessone*

\begin{proof}
By induction on the structure of $\expr{N}$ with the following six cases given below, where $\widetilde{[x_{1k}\leftarrow x_{1k}]}$ abbreviates $[\widetilde{x_1}\leftarrow x_1]\ldots [\widetilde{x_k}\leftarrow x_k]$:

\begin{enumerate}[i)]
    \item $\expr{N} = x$:
    
    Then $\recencodopenf{x} = x_1 [x_1 \leftarrow x]$, and no reductions can be performed.
    
    \item $\expr{N} = \lambda x. N$:

    Suppose $\lfv{N} = \{ x_1, \cdots , x_k\}$. Then,
    \[
    \begin{aligned}
        \recencodopenf{\lambda x. N} &= \recencodf{\lambda x. N\langle \widetilde{x_1} / x_1 \rangle \cdots \langle \widetilde{x_k} / x_k \rangle} \widetilde{[x_{1k}\leftarrow x_{1k}]}\\
        &= \recencodf{\lambda x. N'}\widetilde{[x_{1k}\leftarrow x_{1k}]}= \lambda x. \recencodf{N'\langle \widetilde{y} / x \rangle} [\widetilde{y} \leftarrow x]  \widetilde{[x_{1k}\leftarrow x_{1k}]},
    \end{aligned}
    \]
    \revo{A25}{where $N'$ is $N$ after the substitutions $\langle \widetilde{x_1} / x_1 \rangle \cdots \langle \widetilde{x_k} / x_k \rangle$} and no reductions can be performed.

    \item $\expr{N} = N B$:
    
    Suppose  $\lfv{NB} = \{ x_1, \cdots , x_n\}$. Then
    \begin{equation}\label{ch2eq:app_npfail}
    \begin{aligned}
        \recencodopenf{\expr{N}}&=\recencodopenf{NB}  = \recencodf{NB\langle \widetilde{x_1} / x_1 \rangle \cdots \langle \widetilde{x_n} / x_n \rangle} \widetilde{[x_{1n}\leftarrow x_{1n}]}\\
        & = \recencodf{N' B'} \widetilde{[x_{1n}\leftarrow x_{1n}]}
         = \recencodf{N'} \recencodf{B'} \widetilde{[x_{1n}\leftarrow x_{1n}]}
    \end{aligned}
    \end{equation}
    where $\widetilde{x_i}=x_{i1},\ldots, x_{ij_i}$, for $1\leq i \leq n$ and \revo{A25}{ $N', B'$ are $N$ and $B$ after performing the substitutions $\langle \widetilde{x_1} / x_1 \rangle \cdots \langle \widetilde{x_k} / x_k \rangle$ }.
    By the reduction rules in~\figref{ch2fig:share-reductfailure} there are three possible reductions starting in $\expr{N}$:
  
    \begin{enumerate}
        \item $\recencodf{N'}\recencodf{B'}\widetilde{[x_{1n}\leftarrow x_{1n}]}$ reduces via rule $\redlab{RS{:}Beta}$.
        
        In this case  $N=\lambda x. N_1$, and the encoding in (\ref{ch2eq:app_npfail}) gives $N'= N\langle \widetilde{x_1} / x_1 \rangle \cdots \langle \widetilde{x_n} / x_n \rangle$, which  implies $N' =\lambda x. N_1^{'}$ and the following holds:
        \begin{equation*}
        \begin{aligned}
            \recencodf{N'}=\recencodf{(\lambda x. N'_1)} &= (\lambda x. \recencodf{N'_1 \langle \widetilde{y} / x \rangle} [\widetilde{y} \leftarrow x])
             = (\lambda x. \recencodf{N^{''}} [\widetilde{y} \leftarrow x])
        \end{aligned}
        \end{equation*}
        
        Thus, we have the following $\redlab{RS{:}Beta}$ reduction from \eqref{ch2eq:app_npfail}:
        \begin{equation}\label{ch2eq:sound.appfail}
            \begin{aligned}
                \recencodopenf{\expr{N}} &= \recencodf{N'} \recencodf{B'}\widetilde{[x_{1n}\leftarrow x_{1n}]}=(\lambda x. \recencodf{N''} [\widetilde{y} \leftarrow x] \recencodf{B'}) \widetilde{[x_{1n}\leftarrow x_{1n}]}\\
                &\redd_{\redlab{RS{:}Beta}}  \recencodf{N^{''}} [\widetilde{y} \leftarrow x] \esubst{\recencodf{B'}}{x}  \widetilde{[x_{1n}\leftarrow x_{1n}]} =\expr{L}
            \end{aligned}
        \end{equation}

        \revo{A25}{where $ N'' $ is $N'$ after the substitutions $\langle \widetilde{y} / x \rangle$}. Notice that the expression $\expr{N}$ can perform the following $\redlab{R:Beta}$-reduction:
        \[\expr{N}=(\lambda x. N_1) B\redd_{\redlab{R:Beta}} N_1 \esubst{B}{x} \]

        Assuming $\expr{N'}=N_1 \esubst{B}{x}$, there are two cases:
        
        \begin{enumerate}
            
            \item$\#(x,M) = \size{B} = k \geq 1$.

            On the one hand:
            \begin{equation*}\label{ch2eq:sound_appn1fail}
            \begin{aligned}
                \recencodopenf{\expr{N'}}&=\recencodopenf{N_1 \esubst{B}{x}}\\
                &= \recencodf{N_1 \esubst{B}{x}\langle \widetilde{x_1} / x_1 \rangle \cdots \langle \widetilde{x_n} / x_n \rangle} \widetilde{[x_{1n}\leftarrow x_{1n}]}\\
                & = \recencodf{N_1' \esubst{B'}{x}}\widetilde{[x_{1n}\leftarrow x_{1n}]}\\
                & = \sum_{B_i \in \perm{\recencodf{ B }}}\recencodf{ N_1' \langle y_1 , \cdots , y_k / x  \rangle } \linexsub{B_i(1)/y_1} \cdots\\
                & \qquad \linexsub{B_i(k)/y_k} \widetilde{[x_{1n}\leftarrow x_{1n}]}\\
                & = \sum_{B_i \in \perm{\recencodf{ B }}}\recencodf{ N_1''} \linexsub{B_i(1)/y_1} \cdots \linexsub{B_i(k)/y_k} \widetilde{[x_{1n}\leftarrow x_{1n}]}\\
            \end{aligned}
            \end{equation*}
            \revo{A25}{where $ N_1'' $ is $N_1'$ after the substitution $\langle \widetilde{y} / x \rangle$}. 
            
            On the other hand, after an application of Rule~\redlab{RS:Ex-Sub}:
            \begin{equation*}\label{ch2eq:sound_appn2fail}
            \begin{aligned}
                \expr{L} &= \recencodf{N''} [\widetilde{y} \leftarrow x] \esubst{\recencodf{B'}}{x}  \widetilde{[x_{1n}\leftarrow x_{1n}]} \\
                &\redd\sum_{B_i \in \perm{\recencodf{ B }}}\recencodf{ N_1''} \linexsub{B_i(1)/y_1} \cdots \linexsub{B_i(k)/y_k} \widetilde{[x_{1n}\leftarrow x_{1n}]}\\
                &= \recencodopenf{\expr{N}'}
            \end{aligned}
            \end{equation*}
            
         and the result follows.
            
            \item Otherwise, either $\#(x,N_1) = k=0$ or $\#(x,N_1)\neq \size{B}$.
            In this case:
            \begin{equation*}\label{ch2eq:sound_appn3fail}
            \begin{aligned}
                \recencodopenf{\expr{N'}}&=\recencodopenf{N_1 \esubst{B}{x}}\\
                & = \recencodf{N_1 \esubst{B}{x}\langle \widetilde{x_1} / x_1 \rangle \cdots \langle \widetilde{x_n} / x_n \rangle}\widetilde{[x_{1n}\leftarrow x_{1n}]}\\
                & = \recencodf{N_1' \esubst{B'}{x}}\widetilde{[x_{1n}\leftarrow x_{1n}]}\\
                & =  \recencodf{N^{''}} [\widetilde{y} \leftarrow x] \esubst{\recencodf{B'}}{x} \widetilde{[x_{1n}\leftarrow x_{1n}]}=\expr{L} \\
            \end{aligned}
            \end{equation*}
            
            From (\ref{ch2eq:sound.appfail}):  $\recencodopenf{\expr{N}}\redd \expr{L}=\recencodopenf{\expr{N'}}$ and the result follows.

        \end{enumerate}
        
        \item $\recencodf{N'}\recencodf{B'}\widetilde{[x_{1n}\leftarrow x_{1n}]}$ reduces via rule $\redlab{RS{:} Cons_1}$.

        In this case we would have  $N=\fail^{\widetilde{y}}$, and the encoding in (\ref{ch2eq:app_npfail}) gives $N'= N\linsub{\widetilde{x_1}}{x_1}\ldots \linsub{\widetilde{x_n}}{x_n}$, which implies $N'
        =\fail^{\widetilde{y'}} $, we let $\size{B} = k $ and the following:

        \begin{equation}\label{ch2eq:sound.consumfail}
            \begin{aligned}
                \recencodopenf{\expr{N}} &= \recencodf{N'} \recencodf{B'}\widetilde{[x_{1n}\leftarrow x_{1n}]}
                = \recencodf{\fail^{\widetilde{y'}}} \recencodf{B'}\widetilde{[x_{1n}\leftarrow x_{1n}]}\\
                &= \fail^{\widetilde{y'}} \recencodf{B'}\widetilde{[x_{1n}\leftarrow x_{1n}]}\\
                & \redd \sum_{\perm{B}} \fail^{\widetilde{y'} \uplus \widetilde{z}} \widetilde{[x_{1n}\leftarrow x_{1n}]}, \text{ where } \widetilde{z} = \lfv{B'}\\
            \end{aligned}
        \end{equation}
        
        The expression $\expr{N}$ can perform the following $\redlab{R}=\redlab{R:Cons_1}$-reduction:
        
        \begin{equation}\label{ch2eq:sound.2consumfail}
            \expr{N}=\fail^{\widetilde{y}} \  B\redd_{\redlab{R}} \sum_{\perm{B}} \fail^{\widetilde{y}\uplus \widetilde{z}} \text{  where } \widetilde{z} = \mfv{B}
        \end{equation}
        
        From (\ref{ch2eq:sound.consumfail}) and (\ref{ch2eq:sound.2consumfail}), we infer  that  $\expr{L}=\recencodopenf{\expr{N'}}$ and so the result follows.
        
        \item Suppose that $\recencodf{N'} \redd \recencodf{N''}$. \\
        This case follows from the IH.
    \end{enumerate}

    \item  $\expr{N} = N \esubst{B}{x}$:
    
    Suppose  $\lfv{N \esubst{B}{x}} = \{ x_1, \cdots , x_k\}$. Then,

    \begin{equation}\label{ch2eq:sound_expsubfail}
    \begin{aligned}
    \recencodopenf{\expr{N}}= \recencodopenf{N \esubst{B}{x}}
         &= \recencodf{N \esubst{B}{x}\langle \widetilde{x_1} / x_1 \rangle \cdots \langle \widetilde{x_k} / x_k \rangle} \widetilde{[x_{1k}\leftarrow x_{1k}]}\\
        &= \recencodf{N' \esubst{B'}{x}} \widetilde{[x_{1k}\leftarrow x_{1k}]}
        \end{aligned}
    \end{equation}
    
    \revo{A25}{where $N', B'$ are $N$ and $B$ after performing the substitutions $\langle \widetilde{x_1} / x_1 \rangle \cdots \langle \widetilde{x_k} / x_k \rangle$ }. Let us consider the two possibilities of the encoding:
    
    \begin{myEnumerate}
        
        \item $ \#(x,M) = \size{B} = k \geq 1 $.
        
        Then we continue equation \eqref{ch2eq:sound_expsubfail} as follows:
        \begin{equation}\label{ch2eq:sound_expsub2fail}
            \begin{aligned}
                \recencodopenf{\expr{N}} &= \recencodf{N' \esubst{B'}{x}} \widetilde{[x_{1k}\leftarrow x_{1k}]} \\
                &=  \sum_{B_i \in \perm{\recencodf{ B' }}}\recencodf{ N' \langle y_1 , \cdots , y_n / x  \rangle } \linexsub{B_i(1)/y_1} \cdots \\
                & \qquad \linexsub{B_i(n)/y_n}\widetilde{[x_{1k}\leftarrow x_{1k}]} \\
                &=  \sum_{B_i \in \perm{\recencodf{ B' }}}\recencodf{ N'' } \linexsub{B_i(1)/y_1} \cdots \linexsub{B_i(n)/y_n}\widetilde{[x_{1k}\leftarrow x_{1k}]} \\
            \end{aligned}
        \end{equation}
        
        \revo{A25}{where $N''$ is $N'$ after performing the substitutions $ \langle y_1 , \cdots , y_n / x  \rangle $ }. There are three possible reductions, these being from rules \redlab{RS{:}Lin \dash Fetch}, $\redlab{RS{:}Cons_3}$, and  \redlab{RS{:}Cont}.
        
        \begin{myEnumerate}
            
            \item Suppose that $\headf{N''} = y_1$.
            
             Then one has to consider the shape of the bag $B'$:
                \begin{myEnumerate}

                    \item When $B'$ has only one element $N_1$ then from (\ref{ch2eq:sound_expsub2fail}) and by letting $B = \bag{N_1}$ and $B' = \bag{N'_1}$ we have
                    \begin{equation}\label{ch2eq:sound_expsub3fail}
                        \begin{aligned}
                            \recencodopenf{\expr{N}} 
                            & = \recencodf{N^{''}}\linexsub{\recencodf{N_1'}/y_1} \widetilde{[x_{1k}\leftarrow x_{1k}]}, \text{since }\headf{M'}=y_1\\
                            &\redd \recencodf{N^{''}}\headlin{\recencodf{N_1'}/y_1}\widetilde{[x_{1k}\leftarrow x_{1k}]} = \expr{L}
                        \end{aligned}
                    \end{equation}
                    
                    We also have:
                    \begin{equation}\label{ch2eq:sound_expsub4fail}
                        \begin{aligned}
                            \expr{N} 
                            & = N\esubst{\bag{N_1}}{x}\\
                            &\redd N\headlin{N_1/x}\esubst{\oneb}{x} = \expr{N}' 
                           \pequiv N\headlin{N_1/x}
                             = \expr{N}''
                            \\
                        \end{aligned}
                    \end{equation}
                    
                    From (\ref{ch2eq:sound_expsub3fail}) and (\ref{ch2eq:sound_expsub4fail}), we infer  that  $\expr{L}'=\recencodopenf{\expr{N'}}$ and so the result follows.

                    \item When $B'$ has more then one element. Let us say that $B =  \bag{N_1,N_2}$ and $B' =  \bag{N'_1,N'_2}$ and cases for larger bags proceed similarly then from (\ref{ch2eq:sound_expsub2fail}). (Below we use the fact that $\headf{M'}=y_1$)
\begin{equation}\label{ch2eq:sound_expsub33fail}
\begin{aligned}
\recencodopenf{\expr{N}} 
& = \recencodf{N''}\linexsub{\recencodf{N_1'}/y_1}\linexsub{\recencodf{N_2'}/y_2} \widetilde{[x_{1k}\leftarrow x_{1k}]} \\
& + \recencodf{N''}\linexsub{\recencodf{N_2'}/y_1}\linexsub{\recencodf{N_1'}/y_2} \widetilde{[x_{1k}\leftarrow x_{1k}]}, \\
&\redd
\recencodf{N''} \headlin{\recencodf{N_1'}/y_1} \linexsub{\recencodf{N_2'}/y_2} [\widetilde{[x_{1k}\leftarrow x_{1k}]}\\
 & + \recencodf{N''} \headlin{\recencodf{N_2'}/y_1} \linexsub{\recencodf{N_1'}/y_2} \widetilde{[x_{1k}\leftarrow x_{1k}]} = \expr{L}
   \end{aligned}
\end{equation}
                    
     We also have:
\begin{equation}\label{ch2eq:sound_expsub44fail}
 \begin{aligned}
 \expr{N} 
  & = N\esubst{\bag{N_1,N_2}}{x}\\
& \redd N\headlin{N_1/x}\esubst{\bag{N_2}}{x} + N\headlin{N_2/x}\esubst{\bag{N_1}}{x} 
                             = \expr{N}' 
 \end{aligned}
\end{equation}
                    
                    From (\ref{ch2eq:sound_expsub33fail}) and (\ref{ch2eq:sound_expsub44fail}), we infer  that  $\expr{L}'=\recencodopenf{\expr{N'}}$ and so the result follows.
                    
                \end{myEnumerate}

            \item Suppose that $N'' = \fail^{\widetilde{z'}}$. Then we proceed similarly as from (\ref{ch2eq:sound_expsub2fail}):
                    \begin{equation}\label{ch2eq:sound_expsub99fail}
                        \begin{aligned}
                            \recencodopenf{\expr{N}} 
                            & = \sum_{B_i \in \perm{\recencodf{ B' }}}\fail^{\widetilde{z'}} \linexsub{B_i(1)/y_1} \cdots \linexsub{B_i(n)/y_n}\\
                            &\redd^* \sum_{B_i \in \perm{\recencodf{ B' }}}\fail^{(\widetilde{z'} \setminus y_1, \cdots , y_n) \uplus\widetilde{y}}, \text{ since }\headf{M'}=y_1\\
                           &= \expr{L}'
                        \end{aligned}
                    \end{equation}
                    
         where $\widetilde{y} = \lfv{B_i(1)} \uplus \cdots \uplus \lfv{B_i(n)}$.       We also have that
                    
                    \begin{equation}\label{ch2eq:sound_expsub11fail}
                        \begin{aligned}
                            \expr{N} 
                            & = \fail^{\widetilde{z}} \esubst{B}{x} 
                             \redd \fail^{(\widetilde{z} \setminus x) \uplus\widetilde{y}} 
                            = \expr{N}' 
                        \end{aligned}
                    \end{equation}
            
            \text{ where }  $\widetilde{y} = \mfv{B}$. 
                From (\ref{ch2eq:sound_expsub99fail}) and (\ref{ch2eq:sound_expsub11fail}), we infer  that  $\expr{L}'=\recencodopenf{\expr{N'}}$ and so the result follows.
            
            \item  Suppose that $N'' \redd N'''$
                
                This case follows by the IH.
            
        \end{myEnumerate}

        \item Otherwise we continue equation (\ref{ch2eq:sound_expsubfail}) as follows where $\#(x,M) = k$
            \begin{equation}
            \begin{aligned}
                \recencodopenf{\expr{N}} &= \recencodf{N' \esubst{B'}{x}}  \widetilde{[x_{1k}\leftarrow x_{1k}]} \\
                &=  \recencodf{N'\langle y_1. \cdots , y_k / x  \rangle} [y_1. \cdots , y_k \leftarrow x] \esubst{ \recencodf{B'} }{ x }  \widetilde{[x_{1k}\leftarrow x_{1k}]}\\
                &=  \recencodf{N''} [y_1. \cdots , y_k \leftarrow x] \esubst{ \recencodf{B'} }{ x }  \widetilde{[x_{1k}\leftarrow x_{1k}]} \\
            \end{aligned}
            \end{equation}
            
            Let us consider the two possible cases:
            
            \begin{myEnumerate}
                
                \item $ \#(x,M) = \size{B} = k = 0 $.
                
                Then we have:
                                    \begin{equation}\label{ch2eq:sound_expsubotherwise1}
                    \begin{aligned}
                        \recencodopenf{\expr{N}} &=  \recencodf{N'}  \esubst{ 1 }{ x }  \widetilde{[x_{1k}\leftarrow x_{1k}]} \\
                    \end{aligned}
                    \end{equation}
                    
                    Reductions can only appear in $\recencodf{N'}$ and the case follows by the IH.
                    
                \item Otherwise we can perform the reduction:
                    \begin{equation}\label{ch2eq:sound_expsubotherwise2}
                    \begin{aligned}
                        \recencodopenf{\expr{N}} &= \recencodf{N''} [y_1. \cdots , y_k \leftarrow x] \esubst{ \recencodf{B'} }{ x }  \widetilde{[x_{1k}\leftarrow x_{1k}]}\\
                        &\redd \sum_{B_i \in \perm{B}}  \fail^{\widetilde{z'}} \widetilde{[x_{1k}\leftarrow x_{1k}]}
                        = \expr{L}'
                    \end{aligned}
                    \end{equation}
                 \text{ where } $\widetilde{z'} = \lfv{N''} \uplus \lfv{B'}$.
                We also have that
                    
                    \begin{equation}\label{ch2eq:sound_expsubotherwise3}
                        \begin{aligned}
                            \expr{N} 
                            & = N \esubst{B}{x}  \redd \sum_{\perm{B}} \fail^{\widetilde{z}}  = \expr{N}' 
                        \end{aligned}
                    \end{equation}
                    
                  \text{ where } $\widetilde{z} = \mfv{M} \uplus \mfv{B}$.   \\ 
                From (\ref{ch2eq:sound_expsubotherwise2}) and (\ref{ch2eq:sound_expsubotherwise3}), we infer  that  $\expr{L}'=\recencodopenf{\expr{N'}}$ and so the result follows.
                
            \end{myEnumerate}
        
    \end{myEnumerate}

    \item $\expr{N} = \fail^{\widetilde{y}}$
        \\
        Then $\recencodopenf{\fail^{\widetilde{y}}} = \fail^{\widetilde{y}}$, and no reductions can be performed.

    \item $\expr{N} = \expr{N}_1 + \expr{N}_2$: \\ This case holds by the IH.
\end{enumerate}
\end{proof}

\subsection{Success Sensitiveness}
\label{ch2app:sucessone}

\checkpres*

\begin{proof}
By induction on the structure of $M$. We only need to consider terms of the following form.

\begin{myEnumerate}

    \item When $ M = \checkmark $ the case is immediate.
    
    \item When $ M = NB $ with $\lfv{NB} = \{x_1,\cdots,x_k\}$ and  $\#(x_i,M)=j_i$ we have that: 

        \[ 
            \begin{aligned}
                \headfsum{\recencodopenf{NB}} &= \headfsum{\recencodf{NB\linsub{\widetilde{x_{1}}}{x_1}\cdots \linsub{\widetilde{x_k}}{x_k}}[\widetilde{x_1}\leftarrow x_1]\cdots [\widetilde{x_k}\leftarrow x_k]}\\
                &= \headfsum{\recencodf{NB}}
                 = \headfsum{\recencodf{N}} 
            \end{aligned}
        \]
         and $\headf{NB}= \headf{N} $, by the IH we have $\headf{N} = \checkmark \iff \headfsum{\recencodf{N}} = \checkmark$.
         
    \item When $M = N \esubst{B}{x}$, we must have that $\#(x,M) = \size{B}$ for the head of this term to be $\checkmark$. Let $\lfv{N \esubst{B}{x}} = \{x_1,\cdots,x_k\}$ and  $\#(x_i,M)=j_i$. We have that: 
        \[
            \begin{aligned}
                \headfsum{\recencodopenf{M}} &= \headfsum{\recencodf{N \esubst{B}{x}\linsub{\widetilde{x_{1}}}{x_1}\cdots \linsub{\widetilde{x_k}}{x_k}}[\widetilde{x_1}\leftarrow x_1]\cdots [\widetilde{x_k}\leftarrow x_k]}\\
                &= \headfsum{\recencodf{N \esubst{B}{x}}}\\
                &= \headfsum{\sum_{B_i \in \perm{\recencodf{ B }}}\recencodf{ N \langle x_1 , \cdots , x_k / x  \rangle } \linexsub{B_i(1)/x_1} \cdots \linexsub{B_i(k)/x_k}}\\
                &= \headfsum{\recencodf{ N \langle x_1 , \cdots , x_k / x  \rangle } \linexsub{B_i(1)/x_1} \cdots \linexsub{B_i(k)/x_k}}\\
                &= \headfsum{\recencodf{ N \langle x_1 , \cdots , x_k / x  \rangle } } 
            \end{aligned}
        \]
        
        and $\headf{N \esubst{B}{x}} = \headf{N}$, by the IH we have $$\headf{N} = \checkmark \iff \headfsum{\recencodf{N}} = \checkmark$$

\end{myEnumerate}
\end{proof}

\appsuccesssensce*

\begin{proof}
By induction on the structure of expressions $\lamrfail$ and $\lamrsharfail$. We proceed with the proof in two parts.

\begin{myEnumerate}
    
    \item Suppose that  $\expr{M} \Downarrow_{\checkmark} $. We will prove that $\recencodopenf{\expr{M}} \Downarrow_{\checkmark}$.

    By operational completeness (\thmref{ch2l:app_completenessone}) we have that if $\expr{M}\redd_{\redlab{R}} \expr{M'}$ then

    \begin{enumerate}
        \item If $\redlab{R} =  \redlab{R:Beta}$  then $ \recencodopenf{\expr{M}}  \redd^{\leq 2}\recencodopenf{\expr{M}'}$;

        \item If $\redlab{R} =\redlab{R:Fetch}$   then   $ \recencodopenf{\expr{M}}  \redd^+ \recencodopenf{\expr{M}''}$, for some $ \expr{M}''$ such that  $\expr{M}' \pequiv \expr{M}''$. 
        \item If $\redlab{R} \neq  \redlab{R:Beta}$ and $\redlab{R}\neq \redlab{R:Fetch}$  then $ \recencodopenf{\expr{M}}  \redd\recencodopenf{\expr{M}'}$;
    \end{enumerate}
    
    Notice that  neither our  reduction rules  (in Figure ~\ref{ch2fig:share-reductfailure}), or our congruence $\pequiv$ (in Figure~\ref{ch2fig:rsPrecongruencefailure}),  or  our encoding ($\recencodopenf{\checkmark }=\checkmark$)  create or destroy a $\checkmark$ occurring in the head of term. By Proposition \ref{ch2Prop:checkpres} the encoding preserves the head of a term being $\checkmark$. The encoding acts homomorphically over sums, therefore, if a $\checkmark$ appears as the head of a term in a sum, it will stay in the encoded sum. We can iterate the operational completeness lemma and obtain the result.

    \item Suppose that $\recencodopenf{\expr{M}} \Downarrow_{\checkmark}$. We will prove that $ \expr{M} \Downarrow_{\checkmark}$. 
    
   From \defref{ch2def:app_Suc3} we have that  $\succp{\recencodopenf{\expr{M}}}{\checkmark}\implies \exists M_1 , \cdots , M_k. ~\expr{M} \redd^*  M_1 + \cdots + M_k \text{ and } \headf{M_j} = \checkmark,$
    for some  $j \in \{1, \ldots, k\}$.
    
   Notice that if $\recencodopenf{\expr{M}}$ is itself a term headed with $\checkmark$, say $\headf{\recencodopenf{\expr{M}}}=\checkmark$, then $\expr{M}$ is itself headed with $\checkmark$, from Proposition \ref{ch2Prop:checkpres}.
   
   Based on the shape of $\recencodopenf{\expr{M}}$, we consider two cases.
   The first case, when $\recencodopenf{\expr{M}} = M_1+\ldots+M_k$, $k\geq 2$, and $\checkmark$ occurs in the head of an $M_j$, follows a similar reasoning.  Then $\expr{M}$ has one of the forms:
   \begin{enumerate}
       \item   $\expr{M}= N_1$, then $N_1$ must contain the subterm $ M\esubst{B}{x}$ and $\size{B}=\#(x,M)$. 
       Since, 
       
        $\recencodopenf{M\esubst{B}{x}}=\displaystyle\sum_{B_i \in \perm{\recencodf{ B }}}\recencodf{M\linsub{\widetilde{x}}{x}}\linexsub{B_i(1)/x_i}\ldots \linexsub{B_i(k)/x_i}$,  we can apply Proposition \ref{ch2Prop:checkpres} as we may apply $ \headfsum{\recencodopenf{M\esubst{B}{x}}} $.
       
       \item $\expr{M}=N_1+\ldots+N_l$ for $l \geq 2$.
       
       The reasoning is similar and uses the fact  that the encoding distributes homomorphically over sums.
   \end{enumerate}

   \revo{A26}{The second case is when $\recencodopenf{\expr{M}}\redd^* M_1+\ldots+M_k$, and $\headf{M_j}=\checkmark$, for some $j$ and $M_j$. By operational soundness (\thmref{ch2l:soundnessone}) we have that if $ \recencodopenf{\expr{M}}  \redd \expr{L}$ then there exist $ \expr{M}' $ such that $ \expr{M}  \redd_{\redlab{R}} \expr{M}'$ and 
    \begin{enumerate}
        \item If $\redlab{R} = \redlab{R:Beta}$ then $\expr{ L } \redd^{\leq 1} \recencodopenf{\expr{M}'}$;
        \item If $\redlab{R} \neq \redlab{R:Beta}$ then $\expr{ L } \redd^*  \recencodopenf{\expr{M}^{''}} $, for $ \expr{M}''$ such that  $\expr{M}' \pequiv \expr{M}''$.
    \end{enumerate}
     The reasoning is similar to the previous case, since our reduction rules do not introduce/eliminate $\checkmark$ occurring in the head of terms and by taking $\expr{L}$ to be $M_1+\ldots+M_k$ with $\headf{M_j}=\checkmark$, for some $j$ and $M_j$ the result follows}. 
\end{myEnumerate}
\end{proof}

\section{Appendix to \texorpdfstring{\secref{ch2ss:secondstep}}{}}

\subsection{Type Preservation}
\label{ch2app:typeprestwo}

\appaux*

\begin{proof}

    We shall prove the case of $(1)$ and the case of $(2)$ follows immediately. The case of $(3)$ is immediate by the encoding on types defined in Definition \ref{ch2def:enc_sestypfail}. Hence we take $j > k$, $\tau_1 $ to be an arbitrary type and $m = 0$; also, we take $\tau_2 $ to be $\sigma$ and $n = j-k$. Hence we want to show that $ \piencodf{\sigma^{j}}_{(\tau_1, 0)} = \piencodf{\sigma^{k}}_{(\sigma, n)} $. We have the following
    \[
        \begin{aligned}
            \piencodf{\sigma^{k}}_{(\sigma, n)} &= \oplus(( \with \onef) \ampy ( \oplus  \with (( \oplus \piencodf{\sigma} ) \otimes (\piencodf{\sigma^{k-1}}_{(\sigma, n)}))))\\
            \piencodf{\sigma^{k-1}}_{(\sigma, n)} &= \oplus(( \with \onef) \ampy ( \oplus  \with (( \oplus \piencodf{\sigma} ) \otimes (\piencodf{\sigma^{k-2}}_{(\sigma, n)}))))\\
            \vdots\\
            \piencodf{\sigma^{1}}_{(\sigma, n)} &= \oplus(( \with \onef) \ampy ( \oplus  \with (( \oplus \piencodf{\sigma} ) \otimes (\piencodf{\omega}_{(\sigma, n)}))))
        \end{aligned}
    \]
    and
    \[
        \begin{aligned}
            \piencodf{\sigma^{j}}_{(\tau_1, 0)} &= \oplus(( \with \onef) \ampy ( \oplus  \with (( \oplus \piencodf{\sigma} ) \otimes (\piencodf{\sigma^{j-1}}_{(\tau_1, 0)}))))\\
            \piencodf{\sigma^{j-1}}_{(\tau_1, 0)} &= \oplus(( \with \onef) \ampy ( \oplus  \with (( \oplus \piencodf{\sigma} ) \otimes (\piencodf{\sigma^{j-2}}_{(\tau_1, 0)}))))\\
            \vdots\\
            \piencodf{\sigma^{j-k + 1}}_{(\tau_1, 0)} &= \oplus(( \with \onef) \ampy ( \oplus  \with (( \oplus \piencodf{\sigma} ) \otimes (\piencodf{\sigma^{j-k}}_{(\tau_1, 0)}))))
        \end{aligned}
    \]
    Notice that $n = j-k$, hence we wish to show that $ \piencodf{\sigma^{n}}_{(\tau_1, 0)} = \piencodf{\omega}_{(\sigma, n)} $.  Finally we have that:
    \[
        \begin{aligned}
            \piencodf{\omega}_{(\sigma, n)} & = \oplus(( \with \onef) \ampy ( \oplus  \with (( \oplus \piencodf{\sigma} ) \otimes (\piencodf{\omega}_{(\sigma, n-1)})))) \\
            \piencodf{\omega}_{(\sigma, n-1)} & = \oplus(( \with \onef) \ampy ( \oplus  \with (( \oplus \piencodf{\sigma} ) \otimes (\piencodf{\omega}_{(\sigma, n-2)})))) \\
            \vdots\\
            \piencodf{\omega}_{(\sigma, 1)} & = \oplus(( \with \onef) \ampy ( \oplus  \with (( \oplus \piencodf{\sigma} ) \otimes (\piencodf{\omega}_{(\sigma, 0)})))) \\
            \piencodf{\omega}_{(\sigma, 0)} &= \oplus(( \with \onef) \ampy ( \oplus  \with \onef ) \\
        \end{aligned}
    \]
    and 
    \[
        \begin{aligned}
            \piencodf{\sigma^{n}}_{(\tau_1, 0)} &= \oplus(( \with \onef) \ampy ( \oplus  \with (( \oplus \piencodf{\sigma} ) \otimes (\piencodf{\sigma^{n-1}}_{(\tau_1, 0)}))))\\
            \piencodf{\sigma^{n-1}}_{(\tau_1, 0)} &= \oplus(( \with \onef) \ampy ( \oplus  \with (( \oplus \piencodf{\sigma} ) \otimes (\piencodf{\sigma^{n-2}}_{(\tau_1, 0)}))))\\
            \vdots\\
            \piencodf{\sigma^{1}}_{(\tau_1, 0)} &= \oplus(( \with \onef) \ampy ( \oplus  \with (( \oplus \piencodf{\sigma} ) \otimes (\piencodf{\omega}_{(\tau_1, 0)})))) \\
            \piencodf{\omega}_{(\tau_1, 0)} &= \oplus(( \with \onef) \ampy ( \oplus  \with \onef ) \\
        \end{aligned}
    \]
\end{proof}

\preservationtwo*

\begin{proof}
By mutual induction on the typing derivation of $B$ and $\expr{M}$, with an analysis for the last rule applied.
Recall that the encoding of types ($\piencodf{-}$) has been given in 
Definition~\ref{ch2def:enc_sestypfail}.
    
    \begin{enumerate}
        \item We consider two cases:
        
        \begin{enumerate}
        
        \item Rule~$\redlab{FS{:}wf \dash bag}$:
        
        In this case we have the following derivation:
        
            \begin{prooftree}
                    \AxiomC{\( \core{\Gamma} \vdash B : \pi \)}
                    \LeftLabel{\redlab{FS{:}wf \dash bag}}
                    \UnaryInfC{\( \core{\Gamma} \wfdash  B : \pi \)}
              \end{prooftree}
        
        There are two cases to be analyzed:

        \begin{enumerate}[i)]
            
            \item We may type bags with the $\redlab{TS{:}bag}$ Rule. 
            
            This case is similar to that of $\redlab{FS{:}bag}$
            
            \item We may type bags with the $\redlab{TS{:}\oneb}$ Rule.

            That is, 
            
                \begin{prooftree}
                    \AxiomC{\(  \)}
                    \RightLabel{\(\)}
                    \LeftLabel{\redlab{TS{:}\oneb}}
                    \UnaryInfC{\( \vdash \oneb : \omega \)}
                \end{prooftree}
            Our encoding gives us:
            $$\piencodf{\oneb}_x = x.\some_{\emptyset} ; x(y_n). ( y_n.\overline{\some};y_n . \overline{\close} \mid x.\some_{\emptyset} ; x. \overline{\none})$$
            
            and  the encoding of $\omega$ can be either:
            \begin{enumerate}
            \item  $\piencodf{\omega}_{(\sigma,0)} =  \overline{\with(( \oplus \bot )\otimes ( \with \oplus \bot ))}$; or
            \item $\piencodf{\omega}_{(\sigma, i)} =  \overline{   \with(( \oplus \bot) \otimes ( \with  \oplus (( \with  \overline{\piencodf{ \sigma }} )  \ampy (\overline{\piencodf{\omega}_{(\sigma, i - 1)}})))) }$
            \end{enumerate}
and one can build the following type derivation (rules from Figure~\ref{ch2fig:trulespifull}):
        
    \begin{adjustwidth}{-2.5cm}{}
        \begin{prooftree}
                \AxiomC{\mbox{\ }}
                \LeftLabel{\redlab{T\onef}}
                \UnaryInfC{$y_n . \overline{\close} \vdash y_n: \onef$}
                \LeftLabel{\redlab{T\with_d^x}}
                \UnaryInfC{$y_n.\overline{\some};y_n . \overline{\close} \vdash  y_n :\with \onef$}
                \AxiomC{}
                \LeftLabel{\redlab{T\with^x}}
                \UnaryInfC{$x.\dual{\none} \vdash x :\with A$}
                \LeftLabel{\redlab{T\oplus^x_{\widetilde{w}}}}
                \UnaryInfC{$x.\some_{\emptyset} ; x. \overline{\none} \vdash  x{:}\oplus \with A$}
            \LeftLabel{\redlab{T\mid}}
            \BinaryInfC{$( y_n.\overline{\some};y_n . \overline{\close} \mid x.\some_{\emptyset} ; x. \overline{\none}) \vdash y_n :\with \onef, x{:}\oplus \with A$}
            \LeftLabel{\redlab{T\ampy}}
            \UnaryInfC{$x(y_n). ( y_n.\overline{\some};y_n . \overline{\close} \mid x.\some_{\emptyset} ; x. \overline{\none}) \vdash  x: (\with \onef) \ampy (\oplus \with A) $}
            \LeftLabel{\redlab{T\oplus^x_{\widetilde{w}}}}
            \UnaryInfC{$x.\some_{\emptyset} ; x(y_n). ( y_n.\overline{\some};y_n . \overline{\close} \mid x.\some_{\emptyset} ; x. \overline{\none}) \vdash  x{:}\oplus ((\with \onef) \ampy (\oplus \with A))$}
        \end{prooftree}
    \end{adjustwidth}
            
            Since $A$ is arbitrary,  we can take $A=\oneb$ for $\piencodf{\omega}_{(\sigma,0)} $ and  $A= \overline{(( \with  \overline{\piencodf{ \sigma }} )  \ampy (\overline{\piencodf{\omega}_{(\sigma, i - 1)}}))}$  for $\piencodf{\omega}_{(\sigma,i)} $, in both cases, the result follows.

        \end{enumerate}

        \item Rule $\redlab{FS{:}bag}$:
    
        Then $B = \bag{M}\cdot A$ and we have the following derivation:
        
        \begin{prooftree}
            \AxiomC{\( \core{\Gamma} \wfdash M : \sigma\)}
            \AxiomC{\( \core{\Delta} \wfdash A : \sigma^{k} \)}
            \LeftLabel{\redlab{FS{:}bag}}
            \BinaryInfC{\( \core{\Gamma}, \core{\Delta} \wfdash \bag{M}\cdot A:\sigma^{k+1} \)}
        \end{prooftree}

To simplify the proof, we will consider $k=2$ (the case  $k> 2$ follows analogously).

By IH we have
\begin{align*}
    \piencodf{M}_{x_i} & \vdash \piencodf{\core{\Gamma}}, x_i: \piencodf{\sigma}
    \\
    \piencodf{A}_x & \vdash \piencodf{\core{\Delta}}, x: \piencodf{\sigma\wedge \sigma}_{(\tau, j)}
\end{align*}
By Definition~\ref{ch2def:enc_lamrsharpifail},

\begin{equation}
\begin{aligned}
    \piencodf{\bag{M} \cdot A}_{x} =& x.\some_{\lfv{\bag{M} \cdot A} } ; x(y_i). x.\some_{y_i, \lfv{\bag{M} \cdot A}};x.\overline{\some} ; \outact{x}{x_i}.\\
    &(x_i.\some_{\lfv{M}} ; \piencodf{M}_{x_i} \mid \piencodf{A}_{x} \mid y_i. \overline{\none})
    \end{aligned}
\end{equation} 

Let $\Pi_1$ be the derivation:

\begin{prooftree}
            \AxiomC{$\piencodf{M}_{x_i} \;{ \vdash} \piencodf{\core{\Gamma}}, x_i: \piencodf{\sigma} $}
            \LeftLabel{\redlab{T\oplus^x_{\widetilde{w}}}}
            \UnaryInfC{$x_i.\some_{\lfv{M}} ; \piencodf{M}_{x_i} \vdash \piencodf{\core{\Gamma}} ,x_i: \oplus \piencodf{\sigma} $}
            
            \AxiomC{}
            \LeftLabel{\redlab{T\with^x}}
            \UnaryInfC{$ y_i. \overline{\none} \vdash y_i :\with \onef$}
            
        \LeftLabel{\redlab{T\mid}}
        \BinaryInfC{$\underbrace{x_i.\some_{\lfv{M}} ; \piencodf{M}_{x_i} \mid y_i. \overline{\none}}_{P_1} \vdash \piencodf{\core{\Gamma}} ,x_i: \oplus \piencodf{\sigma}, y_i :\with \onef $}
\end{prooftree}

Let $ P_1 = (x_i.\some_{\lfv{M}} ; \piencodf{M}_{x_i} \mid y_i. \overline{\none})$ in the the derivation $\Pi_2$ below:

\begin{adjustwidth}{-1.5cm}{}
\small
\begin{prooftree}
        \AxiomC{$ \Pi_1$} 
        
        \AxiomC{$ \piencodf{A}_{x}  \vdash  \piencodf{\core{\Delta}}, x: \piencodf{\sigma\wedge \sigma}_{(\tau, j)} $}
        
        \LeftLabel{\redlab{T\otimes}}
    \BinaryInfC{$ \outact{x}{x_i}. (P_1 \mid \piencodf{A}_{x}) \vdash  \piencodf{\core{\Gamma}}  ,  \piencodf{\core{\Delta}}, y_i :\with \onef, x: (\oplus \piencodf{\sigma})  \otimes (\piencodf{\sigma\wedge \sigma}_{(\tau, j)}) $}
    \LeftLabel{\redlab{T\with_d^x}}
    \UnaryInfC{$\underbrace{x.\overline{\some} ; \outact{x}{x_i}. (P_1 \mid \piencodf{A}_{x}  )}_{P_2} \vdash \piencodf{\core{\Gamma}}  ,  \piencodf{\core{\Delta}}, y_i :\with \onef, x: \with (( \oplus \piencodf{\sigma} ) \otimes (\piencodf{\sigma\wedge \sigma}_{(\tau, j)}))  $}
\end{prooftree}
\end{adjustwidth}

Let $P_2 = (x.\overline{\some} ; \outact{x}{x_i}. (P_1 \mid \piencodf{A}_{x} ))$ in the derivation below
(the last two rules that were applied  are \redlab{T\oplus^x_{\widetilde{w}}} and \redlab{T\ampy}):
\begin{adjustwidth}{-2cm}{}
\begin{prooftree}
\small
    \AxiomC{$ \Pi_2$} 
    \noLine
    \UnaryInfC{$\vdots$}
    \noLine
    \UnaryInfC{$P_2\vdash \piencodf{\core{\Gamma}}  ,  \piencodf{\core{\Delta}}, y_i :\with \onef, x: \with (( \oplus \piencodf{\sigma} ) \otimes (\piencodf{\sigma\wedge \sigma}_{(\tau, j)}))  $}
    \LeftLabel{\redlab{T\oplus^x_{\widetilde{w}}}}
    \UnaryInfC{$x.\some_{y_i, \lfv{\bag{M} \cdot A}};P_2  \vdash \piencodf{\core{\Gamma}}  ,  \piencodf{\core{\Delta}}, y_i :\with \onef, x:\oplus  \with (( \oplus \piencodf{\sigma} ) \otimes (\piencodf{\sigma\wedge \sigma}_{(\tau, j)}))$}
    \UnaryInfC{$x(y_i). x.\some_{y_i, \lfv{\bag{M} \cdot A}};P_2  \vdash \piencodf{\core{\Gamma}}  ,  \piencodf{\core{\Delta}}, x: ( \with \onef) \ampy ( \oplus  \with (( \oplus \piencodf{\sigma} ) \otimes (\piencodf{\sigma\wedge \sigma}_{(\tau, j)}))) $}
    \UnaryInfC{$\underbrace{x.\some_{\lfv{\bag{M} \cdot A} } ; x(y_i). x.\some_{y_i, \lfv{\bag{M} \cdot A}};P_2 }_{\piencodf{\bag{M}\cdot A}_x }\vdash   \piencodf{\core{\Gamma}}  ,  \piencodf{\core{\Delta}}, x: \oplus(( \with \onef) \ampy ( \oplus  \with (( \oplus \piencodf{\sigma} ) \otimes (\piencodf{\sigma\wedge \sigma}_{(\tau, j)})))) $}
\end{prooftree}
\end{adjustwidth}

   From Definitions~\ref{ch2def:duality} (duality) and \ref{ch2def:enc_sestypfail}, we infer:
\begin{equation*}
    \begin{aligned}
         \oplus(( \with \onef) \ampy ( \oplus  \with (( \oplus \piencodf{\sigma} ) \otimes (\piencodf{\sigma\wedge \sigma}_{(\tau, j)})))) &=\piencodf{\sigma\wedge \sigma \wedge \sigma}_{(\tau, j)}
    \end{aligned}
\end{equation*}
Therefore, $\piencodf{\bag{M}\cdot A}_x \vdash \piencodf{\core{\Gamma},\core{\Delta}}, x: \piencodf{\sigma\wedge \sigma \wedge \sigma}_{(\tau, j)} $ and the result follows.

        \end{enumerate}
        
    \item  The proof of type preservation for expressions, relies on the analysis of nine cases:
    \begin{enumerate}
            
        \item Rule \redlab{FS{:}wf \dash expr}:

        Then we have the following derivation:
        
            \begin{prooftree}
                    \AxiomC{\( \core{\Gamma} \vdash \expr{M} : \tau \)}
                    \LeftLabel{\redlab{FS{:}wf \dash expr}}
                    \UnaryInfC{\( \core{\Gamma} \wfdash  \expr{M} : \tau \)}
            \end{prooftree}
        
            Cases follow from their corresponding case from $\redlab{FS{:}\dash}$. In the case of $ \redlab{TS{:}var} $ we have:
            
            \begin{prooftree}
                \AxiomC{}
                \LeftLabel{$\redlab{TS{:}var}$}
                \UnaryInfC{\( x: \tau\vdash x : \tau  \)}
            \end{prooftree}
            
        By Definition~\ref{ch2def:enc_sestypfail},  $\piencodf{x:\tau}= x:\with \overline{\piencodf{\tau }}$, and by Figure~\ref{ch2fig:encfail},  $\piencodf{x}_u=x.\overline{\some};[x\leftrightarrow u]$. The thesis holds thanks to the following derivation:
        
            \begin{prooftree}
                \AxiomC{}
                \LeftLabel{$\redlab{ (Tid)}$}
                \UnaryInfC{$ [x \leftrightarrow u ] \vdash x:  \overline{\piencodf{\tau }}  , u :  \piencodf{ \tau } $}
                \LeftLabel{$\redlab{T\with^{x}_{d})}$}
                \UnaryInfC{$ x.\overline{\some} ;[x \leftrightarrow u ] \vdash x: \with  \overline{\piencodf{ \tau }} , u :  \piencodf{ \tau } $}
            \end{prooftree}

        \item Rule $\redlab{FS{:}abs \dash sh}$:
        
        Then $\expr{M} = \lambda x . (M[\widetilde{x} \leftarrow x])$, and the derivation is:
        
        \begin{prooftree}
            \AxiomC{\( \core{\Delta} , x_1: \sigma, \cdots, x_k: \sigma \wfdash M : \tau  \)}
            \LeftLabel{ \redlab{FS{:}share}}
            \UnaryInfC{\( \core{\Delta} , x: \sigma \wedge \cdots \wedge \sigma \wfdash M[x_1 , \cdots , x_k \leftarrow x] : \tau \quad x\notin \core{\Delta}\)}
            \LeftLabel{\redlab{FS{:}abs \dash sh}}
            \UnaryInfC{\( \core{\Delta} \wfdash \lambda x . (M[\widetilde{x} \leftarrow x]) : \sigma^k  \rightarrow \tau \)}
        \end{prooftree}

\noindent To simplify the proof we will consider $k=2$ ( $k>2$ follows similarly). 

        By the IH, we have 
        \[\piencodf{M}_u\vdash  \piencodf{\core{\Delta} , x_1:\sigma, x_2:\sigma }, u:\piencodf{\tau}.\]
        
        From 
        \defref{ch2def:enc_lamrsharpifail} and \defref{ch2def:enc_sestypfail}, it follows that 

        \[
            \begin{aligned}
            \piencodf{ \core{\Delta} , x_1: \sigma, x_2: \sigma } &= \piencodf{\core{\Delta}}, x_1:\with\overline{\piencodf{\sigma}},  x_2:\with\overline{\piencodf{\sigma}}\\[3mm]
                \piencodf{\lambda x.M[x_1, x_2 \leftarrow x]}_u &= u.\overline{\some}; u(x).\piencodf{M[x_1, x_2 \leftarrow x]}_u \\
                & =                     u.\overline{\some}; u(x). x.\overline{\some}. \outact{x}{y_1}. (y_1 . \some_{\emptyset} ;y_{1}.\close;\zero\\
                &\quad \mid x.\overline{\some};x.\some_{u , (\lfv{M} \setminus x_1 , x_2 )};x(x_1). \textcolor{red}{x.\overline{\some}.}\\
                & \quad \textcolor{red}{\outact{x}{y_2} . (y_2 . \some_{\emptyset} ; y_{2}.\close;\zero \mid x.\overline{\some};} \\
                &\quad\textcolor{red}{ x.\some_{u, (\lfv{M} \setminus x_2 )};x(x_2).}  \textcolor{blue}{x.\overline{\some}; \outact{x}{y_{3}}. } \\
                &\quad \textcolor{blue}{( y_{3} . \some_{u,  \lfv{M} } ;y_{3}.\close; \piencodf{M}_u\mid x.\overline{\none} ) ) )}
            \end{aligned}
            \]

We shall split the expression into three parts:
\[
\begin{aligned}
   \textcolor{blue}{ N_1} &= x.\overline{\some}; \outact{x}{y_{3}}. ( y_{3} . \some_{u,  \lfv{M} } ;y_{3}.\close; \piencodf{M}_u \mid x.\overline{\none} )\\
    \textcolor{red}{N_2} &= x.\overline{\some}. \outact{x}{y_2} . (y_2 . \some_{\emptyset} ; y_{2}.\close;\zero \mid x.\overline{\some};x.\some_{u, (\lfv{M} \setminus x_2 )};\\
    & \qquad x(x_2) . N_1)\\
    N_3 &= u.\overline{\some}; u(x). x.\overline{\some}. \outact{x}{y_1}. (y_1 . \some_{\emptyset} ;y_{1}.\close;\zero \mid x.\overline{\some};\\
    &\qquad x.\some_{u , (\lfv{M} \setminus x_1 , x_2 )};x(x_1) .N_2)
\end{aligned}
\]

and we obtain the  derivation for term $N_1$ as follows:
\begin{adjustwidth}{-2cm}{}
\begin{prooftree}
\small
        \AxiomC{$\piencodf{M}_u \vdash \piencodf{ \core{\Delta} , x_1: \sigma, x_2: \sigma }, u:\piencodf{\tau} $}
        \LeftLabel{\redlab{T\bot}}
        \UnaryInfC{$y_{3}.\close; \piencodf{M}_u \vdash \piencodf{ \core{\Delta} , x_1: \sigma, x_2: \sigma }, u:\piencodf{f\tau}, y_{3}{:}\bot$}
        \LeftLabel{\redlab{T\oplus^x_{\widetilde{w}}}}
        \UnaryInfC{$y_{3} . \some_{u,  \lfv{M} } ;y_{3}.\close; \piencodf{M}_u \vdash \piencodf{ \core{\Delta} , x_1: \sigma, x_2: \sigma }, u:\piencodf{\tau}, y_{3}{:}\oplus \bot $}
        \AxiomC{}
        \LeftLabel{\redlab{T\with^x}}
        \UnaryInfC{$x.\dual{\none} \vdash x :\with A$}
    \LeftLabel{\redlab{T\otimes}}
    \BinaryInfC{$ \outact{x}{y_{3}}. ( y_{3} . \some_{u,  \lfv{M} } ;y_{3}.\close; \piencodf{M}_u \mid x.\overline{\none} ) \vdash \piencodf{ \core{\Delta} , x_1: \sigma, x_2: \sigma }, u:\piencodf{\tau} , x: ( \oplus \bot )\otimes ( \with A ) $}
    \LeftLabel{\redlab{T\with_d^x}}
    \UnaryInfC{$\underbrace{x.\dual{\some}; \outact{x}{y_{3}}. ( y_{3} . \some_{u,  \lfv{M} } ;y_{3}.\close; \piencodf{M}_u \mid x.\overline{\none} )}_{N_1} \vdash \piencodf{ \core{\Delta} , x_1: \sigma, x_2: \sigma } , u:\piencodf{\tau} , x: \overline{\piencodf{\omega}_{(\sigma, i)}} $}
\end{prooftree}
\end{adjustwidth}
Notice that the last rule applied \redlab{T\with_d^x} assigns $x: \with ((\oplus \bot) \otimes (\with A))$. Again, since $A$ is arbitrary,  take $A= \oplus (( \with  \overline{\piencodf{ \sigma }} )  \ampy (\overline{\piencodf{\omega}_{(\sigma, i - 1)}}))$, obtaining $x:\overline{\piencodf{\omega}_{(\sigma,i)}}$.
In order to obtain a type derivation for $N_2$, consider the derivation $\Pi_1$:
\begin{adjustwidth}{-2cm}{}
\begin{prooftree}
\small
        \AxiomC{$N_1 \vdash \piencodf{\core{\Delta}}, x_1:\with\overline{\piencodf{\sigma}},  x_2:\with\overline{\piencodf{\sigma}} , u:\piencodf{\tau}, x: \overline{\piencodf{\omega}_{(\sigma, i)}} $}
        \LeftLabel{\redlab{T\ampy}}
        \UnaryInfC{$x(x_2) . N_1  \vdash \piencodf{\core{\Delta}}, x_1:\with\overline{\piencodf{\sigma}},  u:\piencodf{\tau}, x: ( \with\overline{\piencodf{\sigma}} ) \ampy (\overline{\piencodf{\omega}_{(\sigma, i)}}) $}
        \LeftLabel{\redlab{T\oplus^x_{\widetilde{w}}}}
        \UnaryInfC{$ x.\some_{u, (\lfv{M} \setminus x_2 )};x(x_2) . N_1 \vdash \piencodf{\core{\Delta}}, x_1:\with\overline{\piencodf{\sigma}},  u:\piencodf{\tau}, x{:}\oplus (( \with\overline{\piencodf{\sigma}} ) \ampy (\overline{\piencodf{\omega}_{(\sigma, i)}}))$}
        \LeftLabel{\redlab{T\with_d^x}}
        \UnaryInfC{$ x.\overline{\some};x.\some_{u, (\lfv{M} \setminus x_2 )};x(x_2) . N_1  \vdash \piencodf{\core{\Delta}}, x_1:\with\overline{\piencodf{\sigma}},  u:\piencodf{\tau} , x :\with \oplus (( \with\overline{\piencodf{\sigma}} ) \ampy ( \overline{\piencodf{\omega}_{(\sigma, i)}} ))$}
\end{prooftree}
\end{adjustwidth}
We take $ P_1 = x.\overline{\some};x.\some_{u, (\lfv{M} \setminus x_2 )};x(x_2) . N_1$ and $\core{\Gamma_1} =   \piencodf{\core{\Delta}}, x_1:\with\overline{\piencodf{\sigma}},  u:\piencodf{\tau} $ and continue the derivation of $ N_2 $

\begin{adjustwidth}{-1cm}{}
{\small 
\begin{prooftree}
        \AxiomC{}
        \LeftLabel{\redlab{T\cdot}}
        \UnaryInfC{$\zero\vdash $}
        \LeftLabel{\redlab{T\bot}}
        \UnaryInfC{$ y_{2}. \close;\zero \vdash y_{2} : \bot  $}
        \LeftLabel{\redlab{T\oplus^x_{\widetilde{w}}}}
        \UnaryInfC{$ y_2 . \some_{\emptyset} ; y_{2}.\close;\zero \vdash  y_2{:}\oplus \bot$}
        
        \AxiomC{$\Pi_1 $}
        \noLine
        \UnaryInfC{$\vdots$}
        \noLine
        \UnaryInfC{$P_1\vdash \core{\Gamma_1}, x:\with \oplus (( \with\overline{\piencodf{\sigma}} ) \ampy ( \overline{\piencodf{\omega}_{(\sigma, i)}} ))$}
    \LeftLabel{\redlab{T\otimes}}
    \BinaryInfC{$\outact{x}{y_2} . (y_2 . \some_{\emptyset} ; y_{2}.\close;\zero \mid P_1) \vdash \core{\Gamma_1} ,  x: (\oplus \bot)\otimes (\with \oplus (( \with\overline{\piencodf{\sigma}} ) \ampy ( \overline{\piencodf{\omega}_{(\sigma, i)}} )) ) $}
    \LeftLabel{\redlab{T\with_d^x}}
    \UnaryInfC{$\underbrace{x.\overline{\some}. \outact{x}{y_2} . (y_2 . \some_{\emptyset} ; y_{2}.\close;\zero \mid P_1)}_{N_2} \vdash \core{\Gamma_1} , x : \overline{ \piencodf{\sigma \wedge \omega}_{(\sigma, i)}} $}
\end{prooftree}
}
\end{adjustwidth}

Finally, we type $N_3$ by first having the derivation $\Pi_2 $:
\begin{adjustwidth}{-1cm}{}
{\small 
\begin{prooftree} 
        \AxiomC{$ N_2 \vdash \piencodf{\core{\Delta}}, x_1:\with\overline{\piencodf{\sigma}},  u:\piencodf{\tau} , x : \overline{ \piencodf{\sigma \wedge \omega}_{(\sigma, i)}} $}
        \LeftLabel{\redlab{T\ampy}}
        \UnaryInfC{$x(x_1) . N_2  \vdash \piencodf{\core{\Delta}},   u:\piencodf{\tau} , x: ( \with\overline{\piencodf{\sigma}} ) \ampy \overline{\piencodf{\sigma \wedge \omega}_{(\sigma, i)}} $}
        \LeftLabel{\redlab{T\oplus^x_{\widetilde{w}}}}
        \UnaryInfC{$ x.\some_{u , (\lfv{M} \setminus x_1 , x_2 )};x(x_1) . N_2 \vdash \piencodf{\core{\Delta}}, u:\piencodf{\tau} , x{:}\oplus ( ( \with\overline{\piencodf{\sigma}} ) \ampy \overline{\piencodf{\sigma \wedge \omega}_{(\sigma, i)}} ) $}
        \LeftLabel{\redlab{T\with_d^x}}
        \UnaryInfC{${P_2} \vdash \piencodf{\core{\Delta}}, u:\piencodf{\tau} , x :\with \oplus ( ( \with\overline{\piencodf{\sigma}} ) \ampy \overline{\piencodf{\sigma \wedge \omega}_{(\sigma, i)}} )$}
\end{prooftree}
}
\end{adjustwidth}

We let $ P_2 = x.\overline{\some};x.\some_{u , (\lfv{M} \setminus x_1 , x_2 )};x(x_1) .N_2$ and $\core{\Gamma_2} =   \piencodf{\core{\Delta}}, u:\piencodf{\tau}  $. We continue the derivation of $N_3={u.\overline{\some}; u(x). x.\overline{\some}. \outact{x}{y_1}. (y_1 . \some_{\emptyset} ;y_{1}.\close;\zero \mid P_2 )}$:

\begin{adjustwidth}{-2cm}{}
{\small 
\begin{prooftree}
        \AxiomC{}
        \LeftLabel{\redlab{T\cdot}}
        \UnaryInfC{$\zero\vdash $}
        \LeftLabel{\redlab{T\bot}}
        \UnaryInfC{$ y_{1}. \close;\zero \vdash y_{1} : \bot  $}
        \LeftLabel{\redlab{T\oplus^x_{\widetilde{w}}}}
        \UnaryInfC{$ y_1 . \some_{\emptyset} ; y_{1}.\close;\zero \vdash  y_1{:}\oplus \bot$}
        \AxiomC{$ \Pi_2 $}
    \LeftLabel{\redlab{T\otimes}}
    \BinaryInfC{$\outact{x}{y_1}. (y_1 . \some_{\emptyset} ;y_{1}.\close;\zero \mid P_2 ) \vdash  \core{\Gamma_2}, x: (\oplus \bot) \otimes ( \with \oplus ( ( \with\overline{\piencodf{\sigma}} ) \ampy \overline{\piencodf{\sigma \wedge \omega}_{(\sigma, i)}} ) ) $}
    \LeftLabel{\redlab{T\with_d^x}}
    \UnaryInfC{$x.\overline{\some}. \outact{x}{y_1}. (y_1 . \some_{\emptyset} ;y_{1}.\close;\zero \mid P_2 ) \vdash \piencodf{\core{\Delta}}, u:\piencodf{\tau} , x : \overline{\piencodf{\sigma \wedge \sigma}_{(\sigma, i)}} $}
    \LeftLabel{\redlab{T\ampy}}
    \UnaryInfC{$u(x). x.\overline{\some}. \outact{x}{y_1}. (y_1 . \some_{\emptyset} ;y_{1}.\close;\zero \mid P_2 ) \vdash \piencodf{\core{\Delta}} , u: ( \overline{\piencodf{\sigma \wedge \sigma}_{(\sigma, i)}} ) \ampy ( \piencodf{\tau} ) $}
    \LeftLabel{\redlab{T\with_d^x}}
    \UnaryInfC{${N_3} \vdash \piencodf{\core{\Delta}} , u :\with (( \overline{ \piencodf{\sigma \wedge \sigma}_{(\sigma, i)}} ) \ampy ( \piencodf{\tau} ))$}
\end{prooftree}
}
\end{adjustwidth}

        Since $\piencodf{\sigma\wedge \sigma\rightarrow \tau}= \with(  \overline{\piencodf{\sigma \wedge \sigma}_{(\sigma, i)}}  \ampy  \piencodf{ \tau }) $, we have proven that $\piencodf{\lambda x. M[\widetilde{x}\leftarrow x]}_u\vdash \piencodf{\core{\Delta}}, u:\piencodf{\sigma\wedge \sigma \rightarrow \tau}$  and the result follows.
        \item Rule $\redlab{FS{:}app}$:

        Then $\expr{M} = M\ B$, and the derivation is

        \begin{prooftree}
            \AxiomC{\( \core{\Gamma} \wfdash M : \sigma^{j} \rightarrow \tau \)}
            \AxiomC{\( \core{\Delta} \wfdash B : \sigma^{k} \)}
                \LeftLabel{\redlab{FS{:}app}}
            \BinaryInfC{\( \core{\Gamma}, \core{\Delta} \wfdash M\ B : \tau\)}
        \end{prooftree}

    By IH, we have both

    \begin{itemize}
        \item  $\piencodf{M}_u\vdash \piencodf{\core{\Gamma}}, u:\piencodf{\sigma^{j} \rightarrow \tau}$;
        \item and $\piencodf{B}_u\vdash \piencodf{\core{\Delta}}, u:\overline{\piencodf{\sigma^{k}}_{(\tau_2, n)}}$, for some $\tau_2$ and some $n$.
    \end{itemize}
    From the fact that $\expr{M}$ is well-formed and \defref{ch2def:enc_lamrsharpifail} and \defref{ch2def:enc_sestypfail}, we  have:
    
    \begin{itemize}
        \item $B=\bag{N_1 , \cdots , N_k}$;
        \item $\displaystyle{\piencodf{ M\ B }_u =  \bigoplus_{B_i \in \perm{B}} (\nu v)(\piencodf{M}_v \mid v.\some_{u , \lfv{B}} ; \outact{v}{x} . ([v \leftrightarrow u] \mid \piencodf{B_i}_x ) )} $;
        \item $\piencodf{\sigma^{j} \rightarrow \tau}=\with( \overline{\piencodf{\sigma^{j}}_{(\tau_1, m)}} \ampy \piencodf{\tau})$, for some $\tau_1$ and some $m$.
    \end{itemize}
    
    Also, since $\piencodf{B}_u\vdash \piencodf{\core{\Delta}}, u:\piencodf{\sigma^{k}}_{(\tau_2, n)}$, we have the following derivation $\Pi_i$:
\begin{adjustwidth}{-1.5cm}{}
  \begin{prooftree}
            \AxiomC{$\piencodf{ B_i}_x\vdash \piencodf{ \core{\Delta} }, x:{\piencodf{ \sigma^{k} }_{(\tau_2, n)}} $ }
                
            \AxiomC{\(\)}                                   
            \LeftLabel{$ \redlab{Tid}$}
            \UnaryInfC{$ [v \leftrightarrow u]                   \vdash v:  \overline{\piencodf{ \tau }} , u: \piencodf{ \tau }$}
        \LeftLabel{$\redlab{T \otimes}$}
        \BinaryInfC{$\outact{v}{x} . ([v \leftrightarrow u] \mid \piencodf{B_i}_x ) \vdash \piencodf{ \core{\Delta }}, v:\piencodf{ \sigma^{k} }_{(\tau_2, n)} \otimes \overline{ \piencodf{ \tau }} , u:\piencodf{ \tau } $}
        \LeftLabel{$\redlab{T\oplus^{v}_{w}}$}
        \UnaryInfC{$  v.\some_{u , \lfv{B}} ; \outact{v}{x} . ([v \leftrightarrow u] \mid \piencodf{B_i}_x )  \vdash\piencodf{ \core{\Delta} }, v:\oplus (\piencodf{ \sigma^{k} }_{(\tau_2, n)} \otimes  \overline{\piencodf{ \tau }}), u:\piencodf{ \tau }$} 
    \end{prooftree}
\end{adjustwidth}
    
Notice that 
\begin{equation*}
\begin{aligned}
    \oplus (\piencodf{ \sigma^{k} }_{(\tau_2, n)} \otimes  \overline{\piencodf{ \tau }}) &=  \overline{\piencodf{\sigma^{k} \rightarrow \tau} }
\end{aligned}
\end{equation*}

    Therefore, by one application of $\redlab{Tcut}$ we obtain the derivations $\nabla_i$, for each $B_i \in \perm{B}$:
    \begin{adjustwidth}{-1cm}{}
        \begin{prooftree}
        \AxiomC{\( \piencodf{M}_{v} \vdash  \piencodf{ \core{\Gamma} }, v: \with (\overline{\piencodf{ \sigma^{j}} }_{(\tau_1,m)} \ampy ( \piencodf{ \tau }))\)}
        \AxiomC{$\Pi_i$}
        \LeftLabel{\( \redlab{Tcut} \)}    
        \BinaryInfC{$ (\nu v)( \piencodf{ M}_v \mid v.\some_{u , \lfv{B}} ; \outact{v}{x} . ([v \leftrightarrow u] \mid \piencodf{B_i}_x ) ) \vdash \piencodf{ \core{\Gamma} } ,\piencodf{ \core{\Delta} } , u: \piencodf{ \tau }$}
        \end{prooftree}
    \end{adjustwidth}
        
       In order to apply \redlab{Tcut}, we must have that $\piencodf{\sigma^{j}}_{(\tau_1, m)} = \piencodf{\sigma^{k}}_{(\tau_2, n)}$, therefore, the choice of $\tau_1,\tau_2,n$ and $m$, will consider the different possibilities for $j$ and $k$, as in Proposition~\ref{ch2prop:app_aux}.

        We can then conclude that $\piencodf{M B}_u \vdash \piencodf{ \core{\Gamma}}, \piencodf{ \core{\Delta} }, u:\piencodf{ \tau }$:
        \begin{adjustwidth}{-1.5cm}{}
        \begin{prooftree}
        \small
        \AxiomC{For each $B_i \in \perm{B} \qquad  \nabla_i$}
        \LeftLabel{$\redlab{T\with}$}
        \UnaryInfC{$\displaystyle{\bigoplus_{B_i \in \perm{B}} (\nu v)( \piencodf{ M}_v \mid v.\some_{u , \lfv{B}} ; \outact{v}{x} . ([v \leftrightarrow u] \mid \piencodf{B_i}_x ) )  } \vdash \piencodf{ \core{\Gamma}}, \piencodf{ \core{\Delta} }, u:\piencodf{ \tau }$}
        \end{prooftree}
    \end{adjustwidth}
        and the result follows.
        
        \item Rule $\redlab{FS{:}share}$:
        
        Then $\expr{M} = M [ x_1, \dots x_k \leftarrow x ]$ and 
        
        \begin{prooftree}
            \AxiomC{\( \core{\Delta} , x_1: \sigma, \cdots, x_k: \sigma \wfdash M : \tau \quad x\notin \core{\Delta} \quad k \not = 0\)}
            \LeftLabel{ \redlab{FS{:}share}}
            \UnaryInfC{\( \core{\Delta} , x: \sigma_{k} \wfdash M[x_1 , \cdots , x_k \leftarrow x] : \tau \)}
        \end{prooftree}
            
         The proof for this case is contained within 2(b).  
        
        \item Rule $\redlab{FS{:}weak}$:
        
        Then $\expr{M} = M[ \leftarrow x]$ and
        
        \begin{prooftree}
            \AxiomC{\( \core{\Gamma}  \wfdash M : \tau\)}
            \LeftLabel{ \redlab{FS{:}weak}}
            \UnaryInfC{\( \core{\Gamma} , x: \omega \wfdash M[\leftarrow x]: \tau \)}
        \end{prooftree}

        However $\core{\Gamma} , x: \omega$ is not a core context hence we disallow the case.


                
                

       
        \item Rule $\redlab{FS{:}ex \dash sub}$:
        
        Then $\expr{M} = M[x_1, \cdots , x_k \leftarrow x]\ \esubst{ B }{ x }$ and
        
        \begin{prooftree}
                \AxiomC{\( \core{\Delta} \wfdash B : \sigma^{j} \)}
                
                \AxiomC{\( \core{\Gamma} , x:\sigma^{k} \wfdash  M[x_1, \cdots , x_k \leftarrow x] : \tau \)}
            \LeftLabel{\redlab{FS{:}ex \dash sub}}    
            \BinaryInfC{\( \core{\Gamma}, \core{\Delta} \wfdash M[x_1, \cdots , x_k \leftarrow x]\ \esubst{ B }{ x } : \tau \)}
        \end{prooftree}
        
        
    
    By Proposition~\ref{ch2prop:app_aux} and IH we have both 
    \[
    \begin{aligned} 
    \piencodf{ M[x_1, \cdots , x_k \leftarrow x]}_u&\vdash \piencodf{\core{\Gamma}}, x: \overline{ \piencodf{ \sigma_k }_{(\tau, n)}} , u:\piencodf{\tau}\\
    \piencodf{B}_x&\vdash \piencodf{\core{\Delta}}, x:\piencodf{ \sigma_j }_{(\tau, m)}
    \end{aligned}
    \]
    From \defref{ch2def:enc_lamrsharpifail}, we have 
    \begin{equation*}
        \begin{aligned}
       \piencodf{ M[\widetilde{x} \leftarrow x]\ \esubst{ B }{ x }}_u&= \bigoplus_{B_i \in \perm{B}} (\nu x)( \piencodf{ M[\widetilde{x} \leftarrow x]}_u \mid \piencodf{ B_i}_x )  \\
        \end{aligned}
    \end{equation*}

Therefore, for each $B_i \in \perm{B} $, we obtain the following derivation $\Pi_i$:
\begin{adjustwidth}{-1cm}{} 
\begin{prooftree}
            \AxiomC{$\piencodf{ M[\widetilde{x} \leftarrow x]}_u  
            \vdash \piencodf{ \core{\Gamma} }  , x:  \overline{ \piencodf{ \sigma_k }_{(\tau, n)}}, u:  \piencodf{ \tau }$}
            \AxiomC{$    \piencodf{ B_i}_x \vdash \piencodf{ \core{\Delta} }, x: \piencodf{ \sigma_j }_{(\tau, m)}$}
        \LeftLabel{$\redlab{Tcut}$}
        \BinaryInfC{$ (\nu x)( \piencodf{ M[\widetilde{x} \leftarrow x]}_u \mid \piencodf{ B_i}_x )  \vdash \piencodf{ \core{\Gamma}} , \piencodf{ \core{\Delta} } , u: \piencodf{ \tau } $}               
        \end{prooftree}
    \end{adjustwidth}
        We must have that $\piencodf{\sigma^{j}}_{(\tau, m)} = \piencodf{\sigma^{k}}_{(\tau, n)}$ which holds by the conditions in Proposition \ref{ch2prop:app_aux}.
        Therefore, from $\Pi_i$ and multiple applications of $\redlab{T\with}$ it follows that
        \begin{adjustwidth}{-1cm}{}
        \begin{prooftree}
                    \AxiomC{$\forall \bigoplus_{B_i \in \perm{B}} \hspace{1cm} \Pi_i$}
                    \LeftLabel{$\redlab{T\with}$}
        \UnaryInfC{$ \bigoplus_{B_i \in \perm{B}} (\nu x)( \piencodf{ M[\widetilde{x} \leftarrow x]}_u \mid x.\some_w; \piencodf{ B_i}_x )  \vdash\piencodf{ \core{\Gamma} } , \piencodf{ \core{\Delta} } , u: \piencodf{ \tau }$}
        \end{prooftree}
    \end{adjustwidth}
        that is, $\piencodf{M[x_1,x_2\leftarrow x]\esubst{B}{x}}\vdash \piencodf{\core{\Gamma}, \core{\Delta}}, u:\piencodf{\tau}$ and the result follows.
        
        \item Rule $\redlab{FS{:}ex \dash lin \dash sub}$:

        Then $\expr{M} = M \linexsub{N / x}$ and
        
        \begin{prooftree}
        \AxiomC{\( \core{\Delta} \wfdash N : \sigma \)}
        \AxiomC{\( \core{\Gamma}  , x:\sigma \wfdash M : \tau \)}
            \LeftLabel{\redlab{FS{:}ex \dash lin \dash sub}}
        \BinaryInfC{\( \core{\Gamma}, \core{\Delta} \wfdash M \linexsub{N / x} : \tau \)}
        \end{prooftree}
    
    By IH we have both 
    \[
    \begin{aligned}
        \piencodf{N}_x&\vdash \piencodf{\core{\Delta}}, x: \piencodf{\sigma}\\
        \piencodf{M}_x&\vdash \piencodf{\core{\Gamma}}, x: \with\overline{\piencodf{\sigma}}, u:\piencodf{\tau}.
    \end{aligned}
    \]
    
    From \defref{ch2def:enc_lamrsharpifail}, $\piencodf{M \linexsub{N / x} }_u=(\nu x) ( \piencodf{ M }_u \mid   x.\some_{\lfv{N}};\piencodf{ N }_x  )$ and 
    \begin{adjustwidth}{-1.5cm}{}
    \begin{prooftree}
        \small
        \AxiomC{\( \piencodf{ M }_u \vdash  \piencodf{ \core{\Gamma} } , u :  \piencodf{ \tau } , x : \with \overline{\piencodf{ \sigma }}\)}
        \AxiomC{\( \piencodf{ N }_x  \vdash  \piencodf{ \core{\Delta} } , x : \piencodf{ \sigma } \)}
                \LeftLabel{$\redlab{T\oplus^x}$}
        \UnaryInfC{\( x.\some_{\lfv{N}};\piencodf{ N }_x \vdash  \piencodf{ \core{\Delta} } , x : \oplus \piencodf{ \sigma }\)}
        \LeftLabel{$\redlab{TCut}$}
        \BinaryInfC{\((\nu x) ( \piencodf{ M }_u \mid   x.\some_{\lfv{N}};\piencodf{ N }_x  )  \vdash  \piencodf{ \core{\Gamma}} , \piencodf{ \core{\Delta} } , u : \piencodf{ \tau }  \)}
        \end{prooftree}
    \end{adjustwidth}

        Observe that for the application of Rule~$\redlab{TCut}$ we used the fact that $\overline{\oplus\piencodf{\sigma}}=\with \overline{\piencodf{\sigma}}$. Therefore, $\piencodf{M \linexsub{N / x} }_u\vdash \piencodf{ \core{\Gamma}} , \piencodf{ \core{\Delta} } , u : \piencodf{ \tau } $ and the result follows.

        \item Rule $\redlab{FS{:}fail}$:
        
        Then $\expr{M} = M \linexsub{N / x}$ and
        
            \begin{prooftree}
                \AxiomC{\( \core{(x_1:\sigma_1, \cdots , x_n:\sigma_n)} = x_1:\sigma_1, \cdots , x_n:\sigma_n  \)}
                \LeftLabel{\redlab{FS{:}fail}}
                \UnaryInfC{\( x_1:\sigma_1, \cdots , x_n:\sigma_n  \wfdash  \fail^{x_1, \cdots , x_n} : \tau \)}
            \end{prooftree}
        
        From Definition \ref{ch2def:enc_lamrsharpifail}, $\piencodf{\fail^{x_1, \cdots , x_n} }_u= u.\overline{\none} \mid x_1.\overline{\none} \mid \cdots \mid x_k.\overline{\none} $ and 
        
    \begin{adjustwidth}{-1cm}{}
    {\small 
        \begin{prooftree}
                \AxiomC{}
                \LeftLabel{\redlab{T\with^u}}
                \UnaryInfC{$u.\overline{\none} \vdash u : \piencodf{ \tau } $}

                    \AxiomC{}
                    \LeftLabel{\redlab{T\with^{x_1}}}
                    \UnaryInfC{$x_1.\overline{\none} \vdash_1 : \with \overline{\piencodf{\sigma_1}} $}
                    
                    \AxiomC{}
                    \LeftLabel{\redlab{T\with^{x_n}}}
                    \UnaryInfC{$x_n.\overline{\none} \vdash x_n : \with \overline{\piencodf{\sigma_n}} $}
                    \UnaryInfC{$\vdots$}
                \BinaryInfC{$x_1.\overline{\none} \mid \cdots \mid x_k.\overline{\none} \vdash  x_1 : \with \overline{\piencodf{\sigma_1}}, \cdots  ,x_n : \with \overline{\piencodf{\sigma_n}}$}
            \LeftLabel{\redlab{T\mid}}
            \BinaryInfC{$u.\overline{\none} \mid x_1.\overline{\none} \mid \cdots \mid x_k.\overline{\none} \vdash x_1 : \with \overline{\piencodf{\sigma_1}}, \cdots  ,x_n : \with \overline{\piencodf{\sigma_n}}, u : \piencodf{ \tau }$}
        \end{prooftree}
        }
    \end{adjustwidth}
        Therefore, $\piencodf{\fail^{x_1, \cdots , x_n} }_u\vdash  x_1 : \with \overline{\piencodf{\sigma_1}}, \cdots  ,x_n : \with \overline{\piencodf{\sigma_n}}, u : \piencodf{ \tau } $ and the result follows.

        \item Rule $\redlab{FS{:}sum}$: 
        
        This case follows easily by IH.
    \end{enumerate}
    \end{enumerate}
\end{proof}

\subsection{Completeness and Soundness}
\label{ch2compandsucctwo}


\consistnequiv*

\begin{proof}
By induction on the structure of $\expr{M}$. Let us consider first two conditions 1 and 2 as other conditions are analogous. The congruence  rules that concern the sharing construct of condition 1 are:
        \[
            \begin{aligned}
            &\begin{array}{rll}
                M [ \leftarrow x] \esubst{\oneb}{x} \!\!\!\! & \pequiv M &
                \\
                MA[\widetilde{x} \leftarrow x]\esubst{B}{x} 
                         \!\!\!\!&\pequiv (M[\widetilde{x} \leftarrow x]\esubst{B}{x})A
                         &  (*)
                \\
                M[\widetilde{y} \leftarrow y]\esubst{A}{y}[\widetilde{x} \leftarrow x]\esubst{B}{x} \!\!\!\! & \pequiv
                (M[\widetilde{x} \leftarrow x]\esubst{B}{x})[\widetilde{y} \leftarrow y]\esubst{A}{y} &  (**) 
                \\
            \end{array}
            \\
            &\begin{array}{ll}
                (*)\text{with } x_i \in \widetilde{x} \Rightarrow x_i \not \in \lfv{A} &
                (**) \text{with } x_i \in \widetilde{x} \Rightarrow x_i \not \in \lfv{A}
            \end{array}
            \end{aligned}
        \]
        Notice that these rules neither add or remove occurrences of shared variables neither do they allow shared variables to be extruded from their bindings by their side conditions. Also, they do not introduce new sharing on already shared variables. 
        Hence, conditions 1(i) to 1(iv) are preserved by these rules.
        
        Now consider the congruence rules concerning the explicit substitution of condition 2:
        \[
            \begin{array}{rll}
            MB \linexsub{N/x}  \!\!\!\!&\pequiv (M\linexsub{N/x})B &  \text{with } x \not \in \lfv{B} 
            \\
            M \linexsub{N_2/y}\linexsub{N_1/x} 
                     \!\!\!\!&\pequiv M\linexsub{N_1/x}\linexsub{N_2/y} &
                     \text{with } x \not \in \lfv{N_2},\revdaniele{ y \notin \lfv{N_1}}
            \\
            \end{array}
        \]
        As before, variables are not duplicated or eliminated from terms and by the side conditions of the rules they cannot extrude bound variables. Similarly, the rules do not introduce any sharing or new free variables.
        Hence conditions 2(i) to 2(iv) are satisfied.
\end{proof}

\encodingreduces*

\begin{proof}
Let us consider each part:
\begin{myEnumerate}

    \item We proceed by induction on the structure of $N$.

    \begin{myEnumerate}
    
        \item  $N = x$.
        
        Then $\piencodf{x}_u$. Hence $ I = \emptyset$ and $ \widetilde{y} = \emptyset$.
    
        \item $N = (M\ B)$. 
        
        Then $\headf{M\ B} = \headf{M} = x$ and 
        \[ \hspace{-1cm}\piencodf{N}_u = \piencodf{M\ B}_u  = \bigoplus_{B_i \in \perm{B}} (\nu v)(\piencodf{M}_v \mid v.\some_{u, \lfv{B}} ; \outact{v}{x} . ([v \leftrightarrow u] \mid \piencodf{B_i}_x ) ) \]
        and the result follows by induction on $\piencodf{M}_u$.
        
        \item $N = M[\widetilde{y} \leftarrow y]$. Not possible due to the assumption of partially open terms.
        
        \revd{B29}{ 
        \item $N = (M[\widetilde{y} \leftarrow y])\esubst{ B }{ y }$. \\
        Then $\headf{(M[\widetilde{y} \leftarrow y])\esubst{ B }{ y }} = \headf{(M[\widetilde{y} \leftarrow y])} = x$ when $\widetilde{y} = \emptyset,\ B = \oneb $ and $\headf{M} = x$ .
        \[
        \begin{aligned}
           \piencodf{N}_u &= \piencodf{(M[ \leftarrow y])\esubst{ \oneb }{ y }}_u = (\nu y)( \piencodf{ M[ \leftarrow y]}_u \mid  \piencodf{\oneb}_y) 
           \\
           &=(\nu y)( y. \overline{\some}. \outact{y}{z} . ( z . \some_{u,\lfv{M}} ;z_{}.\close; \piencodf{M}_u \mid y. \overline{\none})  \mid \\
           & \qquad y.\some_{\emptyset} ; y(z). (z.\overline{\some};z. \overline{\close} \mid y.\some_{\emptyset} ; y. \overline{\none})) \\
           & \redd^*  \piencodf{M}_u \\
        \end{aligned}
        \]
        Then the result follows by induction on $\piencodf{ M }_u $.
        }

        \item When $N = M \linexsub {N' /y}$, then $\headf{M \linexsub {N' /y}} = \headf{M } = x$ and 
        \[
        \begin{aligned}
           \piencodf{N}_u & = \piencodf{M \linexsub {N' /y}}_u 
            =  (\nu y) ( \piencodf{ M }_u \mid   x.\some_{\lfv{N'}};\piencodf{ N' }_x  ) 
        \end{aligned}
        \]
        Then true by induction on $\piencodf{ M }_u $

    \end{myEnumerate}

    \item In this case, notice how reductions are only introduced when $N$ has sub-term $(M[ \leftarrow y])\esubst{ \oneb }{ y }$ from case 1(IV), however from the congruence of \figref{ch2fig:rsPrecongruencefailure} we may rewrite this sub-term to be $M$ which eliminates the need for reductions. Inductively, performing this application of $\pequiv$ provides the result.

    \item This case is similar to the first, with the clear difference that linear head substitution must also be used. However, we can inductively push the linear head substitution inside the term to reach the head variable. Consider the base case when $N = x$ and we have some well-formed partially open term $M$. Then $\piencodf{N\headlin{M/x}}_u = \piencodf{x\headlin{M/x}}_u = \piencodf{M}_u$. Hence $ I = \emptyset$ and $ \widetilde{y} = \emptyset$ matching that of case 1(i).

    Next, let us consider the case of $N\headlin{M/x} = M' \linexsub {N' /y}\headlin{M/x} = M'\headlin{M/x} \linexsub {N' /y}$. By considering 1(V) we can see the evaluating the translation of creates the same process shape up to linear head substitution. Other cases follow analogously.

    \item This is a consequence of both (2) and (3).
    
\end{myEnumerate}

\end{proof}

\revd{B46}{
\begin{notation}
    We use the notation $\lfv{M}.\overline{\none}$ and $\widetilde{x}.\overline{\none}$ where $\lfv{M}$ or $\widetilde{x}$ are equal to $ x_1 , \cdots , x_k$ to describe a process of the form $x_1.\overline{\none} \mid \cdots \mid x_k.\overline{\none} $
\end{notation}
}

\opcomplete*

\begin{proof}
By induction on the reduction rule applied to infer $\expr{N}\redd \expr{M}$.  
We have five cases.

    \begin{enumerate}
        \item  Case $\redlab{RS{:}Beta}$: 
              
               Then  $ \expr{N}= (\lambda x. M[\widetilde{x} \leftarrow x]) B \redd M[\widetilde{x} \leftarrow x]\ \esubst{ B }{ x }=\expr{M}$.
               \\
        On the one hand, we have:
                \begin{equation}\label{ch2eq:compl_lsbeta1fail}
        \begin{aligned}
        \piencodf{\revd{B43}{\expr{N}}}_u &= \piencodf{(\lambda x. M[\widetilde{x} \leftarrow x]) B}_u\\
        &=  \bigoplus_{B_i \in \perm{B}} (\nu v)( \piencodf{ \lambda x. M[\widetilde{x} \leftarrow x]}_v \mid v.\some_{u,\lfv{B}} ; \outact{v}{x} . ( \piencodf{ B_i}_x \mid [v \leftrightarrow u] ) )\\
        &=  \bigoplus_{B_i \in \perm{B}} (\nu v)( v.\overline{\some}; v(x).\piencodf{M[\widetilde{x} \leftarrow x]}_v \mid \\
        & \qquad \qquad \qquad v.\some_{u,\lfv{B}} ; \outact{v}{x} . ( \piencodf{ B_i}_x \mid [v \leftrightarrow u] ) )\\
        & \redd  \bigoplus_{B_i \in \perm{B}} (\nu v)(  v(x).\piencodf{M[\widetilde{x} \leftarrow x]}_v \mid  \outact{v}{x} . ( \piencodf{ B_i}_x \mid [v \leftrightarrow u] ) )\\
        & \redd  \bigoplus_{B_i \in \perm{B}} (\nu v, x)(  \piencodf{M[\widetilde{x} \leftarrow x]}_v \mid   \piencodf{ B_i}_x \mid [v \leftrightarrow u] ) \\
        & \redd  \bigoplus_{B_i \in \perm{B}} (\nu x)(  \piencodf{M[\widetilde{x} \leftarrow x]}_u \mid   \piencodf{ B_i}_x  ) \\
        \end{aligned}
        \end{equation}
        
        On the other hand, we have:
        \begin{equation}\label{ch2eq:compl_lsbeta2fail}
            \begin{aligned}
               \piencodf{\expr{M}}_u &= \piencodf{M[\widetilde{x} \leftarrow x]\ \esubst{ B }{ x }}_u = \bigoplus_{B_i \in \perm{B}} (\nu x) (\piencodf{ M[\widetilde{x} \leftarrow x] }_u \mid  \piencodf{ B_i}_x ) \\
            \end{aligned}
        \end{equation}
        Therefore, by \eqref{ch2eq:compl_lsbeta1fail} and \eqref{ch2eq:compl_lsbeta2fail} the result follows.
        
        \item Case $ \redlab{RS{:}Ex \dash Sub}$: 
        
        Then $ N=M[x_1, \cdots , x_k \leftarrow x]\ \esubst{ B }{ x }$, with $B=\bag{N_1 , \ldots , N_k}$, $k\geq 1$ and $M \not= \fail^{\widetilde{y}}$. 
        The reduction is $$\expr{N} = M[x_1, \cdots , x_k \leftarrow x]\ \esubst{ B }{ x } \redd \sum_{B_i \in \perm{B}}M\ \linexsub{B_i(1)/x_1} \cdots \linexsub{B_i(k)/x_k} = \expr{M}.$$
        
        We detail the encodings of $\piencodf{\expr{N}}_u$ and $\piencodf{\expr{M}}_u$. To simplify the proof, we will consider $k=1$ (the case   $k> 1$ follows analogously). 
        \\
        On the one hand, we have:
        \begin{equation}\label{ch2eq:compl_lsbeta3fail}
        \begin{aligned}
        \piencodf{\expr{N}}_u &= \piencodf{M[x_1 \leftarrow x]\ \esubst{ B }{ x }}_u =  \bigoplus_{B_i \in \perm{B}} (\nu x)( \piencodf{ M[x_1\leftarrow x]}_u \mid  \piencodf{ B_i}_x ) 
        \\
        &=  \bigoplus_{B_i \in \perm{B}} (\nu x)( x.\overline{\some}. \outact{x}{y_1}. (y_1 . \some_{\emptyset} ;y_{1}.\close;\zero \mid x.\overline{\some}; \\
                & \hspace{1cm} x.\some_{u, (\lfv{M} \setminus x_1 )};x(x_1). x.\overline{\some}; \outact{x}{y_{2}}. ( y_{2} . \some_{u,\lfv{M}} ;y_{2}.\close;   \\
                &\hspace{1cm}\piencodf{M}_u\mid x.\overline{\none} ) ) \mid x.\some_{\lfv{B_i(1)}} ; x(y_1). x.\some_{y_1,\lfv{B_i(1)}};x.\overline{\some} ;  \\
                &\hspace{1cm}\outact{x}{x_1}.(x_1.\some_{\lfv{B_i(1)}};\piencodf{B_i(1)}_{x_1} \mid y_1. \overline{\none} \mid x.\some_{\emptyset} ; x(y_2). \\
                &\hspace{1cm} ( y_2.\overline{\some};y_2 . \overline{\close} \mid x.\some_{\emptyset} ; x. \overline{\none}) ) )
        \\
        & \redd^* \bigoplus_{B_i \in \perm{B}} (\nu x, y_1,x_1,y_2)( 
                  y_1 . \some_{\emptyset} ;y_{1}.\close;\zero \mid y_1. \overline{\none} \mid \\
                & \hspace{1cm}  y_{2} . \some_{u,\lfv{M}};y_{2}.\close; \piencodf{M}_u \mid y_2.\overline{\some};y_2 . \overline{\close} \mid \\
                & \hspace{1cm} x.\overline{\none}   \mid  x.\some_{\emptyset} ; x. \overline{\none} \mid  x_1.\some_{\lfv{B_i(1)}} ; \piencodf{B_i(1)}_{x_1} ) 
        \\
        & \redd^* \bigoplus_{B_i \in \perm{B}} (\nu x_1 )(  \piencodf{M}_u \mid  x_1.\some_{\lfv{B_i(1)}} ; \piencodf{B_i(1)}_{x_1} ) \\
        \end{aligned}
        \end{equation}
        
        On the other hand, we have:
        \begin{equation}\label{ch2eq:compl_lsbeta4fail}
        \begin{aligned}
            \piencodf{\expr{M}}_u &= \piencodf{\sum_{B_i \in \perm{B}}M\ \linexsub{B_i(1)/x_1}}_u = \bigoplus_{B_i \in \perm{B}} \piencodf{M\ \linexsub{B_i(1)/x_1} }_u\\
            &= \bigoplus_{B_i \in \perm{B}} (\nu x_1 )( \piencodf{M}_{u} \mid x_1.\some_{\lfv{B_i(1)}};\piencodf{B_i(1)}_{x_1} )   \\
        \end{aligned}
        \end{equation}
        Therefore, by \eqref{ch2eq:compl_lsbeta3fail}
        and  \eqref{ch2eq:compl_lsbeta4fail} the result follows.
        
        \item Case $ \redlab{RS{:}Lin \dash Fetch}$: 
       
        Then we have 
        $\expr{N} = M\ \linexsub{N'/x}$ with $\headf{M} = x$ and $\expr{N} \redd  M \headlin{ N' /x } = \expr{M}$.
        
       On the one hand, we have:
        \begin{equation}\label{ch2eq:compl_lsbeta5fail}
        \begin{aligned}
        \piencodf{N}_u = \piencodf{M\ \linexsub{N'/x}}_u&= (\nu x) ( \piencodf{ M }_u \mid   x.\some_{\lfv{N'}};\piencodf{ N' }_x  )\\
       & \redd^* (\nu x) ( \bigoplus_{i \in I}(\nu \widetilde{y})(\piencodf{ x }_{j} \mid P_i) \mid   x.\some_{\lfv{N'}};\piencodf{ N' }_x  ) \quad (*)   
        \\
        &= (\nu x) ( \bigoplus_{i \in I}(\nu \widetilde{y})(\piencodf{ x }_{j} \mid P_i)  \mid   x.\some;\piencodf{ N' }_x  )  \\
        & \redd (\nu x) ( \bigoplus_{i \in I}(\nu \widetilde{y})([x \leftrightarrow j ] \mid P_i) \mid   \piencodf{ N' }_x  ) \\
        & \redd  \bigoplus_{i \in I}(\nu \widetilde{y}) ( P_i \mid   \piencodf{ N' }_j  )     =Q
        \end{aligned}
        \end{equation}
        
       where the reductions denoted by $(*)$ are inferred via Proposition~\ref{ch2prop:NEEDTONAME}.
       
       \revd{B29}{On the other hand, we have by Proposition \ref{ch2prop:NEEDTONAME} :
        \begin{equation}\label{ch2eq:compl_lsbeta6fail}
        \begin{aligned}
        \piencodf{\expr{M}}_u &= \piencodf{M \headlin{ N'/x }}_u \redd^*  \bigoplus_{i \in I}(\nu \widetilde{y}) ( P_i \mid   \piencodf{ N' }_j  )
        \end{aligned}
        \end{equation}
        We also have by Proposition \ref{ch2prop:NEEDTONAME} and \eqref{ch2eq:compl_lsbeta6fail}
        that there exists $M'$ such that $M' \pequiv M \headlin{ N'/x }$ with:
        \begin{equation}\label{ch2eq:compl_lsbeta7fail}
            \piencodf{ M' }_{u} = \bigoplus_{i \in I}(\nu \widetilde{y}) ( P_i \mid   \piencodf{ N' }_j  )
        \end{equation}
        }
        
                Therefore, by \eqref{ch2eq:compl_lsbeta5fail}
        and  \eqref{ch2eq:compl_lsbeta7fail} the result follows.
        
        \item Case $\redlab{RS{:}TCont}$ and $\redlab{RS{:}ECont}$:
         These cases follow by IH.
         
        
        \item Case $\redlab{RS{:}Fail}$:
        
        Then, 
        $\expr{N} = M[x_1, \cdots , x_k \leftarrow x]\ \esubst{ B }{ x }$ with $k \neq \size{B}$ and
        
        $$ \expr{N} \redd  \sum_{B_i \in \perm{B}}  \fail^{\widetilde{y} } = \expr{M},$$ where $\widetilde{y} = (\lfv{M} \setminus \{  x_1, \cdots , x_k \} ) \cup \lfv{B}$. 
        
        Let us assume that $k > l$ and we proceed similarly for $k > l$. Hence $k = l + m$ for some $m \geq 1$. \revd{B44}{On the one hand, we have \eqref{ch2eq:compl_fail1-fail}, this can be seen in \figref{ch2fig:proofreductions1}}.
        
        \begin{figure}[!t]
        \hrule 
            {\small
            \begin{equation}\label{ch2eq:compl_fail1-fail}
            \begin{aligned}
                \piencodf{\revd{B45}{\expr{N}}}_u &= \piencodf{M[x_1, \cdots , x_k \leftarrow x]\ \esubst{ B }{ x }}_u\\
                &=  \bigoplus_{B_i \in \perm{B}} (\nu x)( \piencodf{ M[x_1, \cdots , x_k \leftarrow x]}_u \mid  \piencodf{ B_i}_x )  
                \\
                &=  \bigoplus_{B_i \in \perm{B}} (\nu x)( x.\overline{\some}. \outact{x}{y_1}. (y_1 . \some_{\emptyset} ;y_{1}.\close;\zero \mid x.\overline{\some};x.\some_{u,(\lfv{M} \setminus x_1 , \cdots , x_k )};\\
       & \hspace{1cm} x(x_1) . \cdots  x.\overline{\some}. \outact{x}{y_k} . (y_k . \some_{\emptyset} ; y_{k}.\close;\zero \mid x.\overline{\some};x.\some_{u,(\lfv{M} \setminus  x_k )}; \\
       & \hspace{1cm} x(x_k) . x.\overline{\some}; \outact{x}{y_{k+1}}. ( y_{k+1} . \some_{u,\lfv{M} } ;y_{k+1}.\close; \piencodf{M}_u \mid x.\overline{\none} )) \cdots )  \\
      &\hspace{1cm} \mid x.\some_{\lfv{B}} ; x(y_1). x.\some_{y_1,\lfv{B}};x.\overline{\some} ; \outact{x}{x_1}. (x_1.\some_{\lfv{B_i(1)}} ; \piencodf{B_i(1)}_{x_1} \\
       & \hspace{1cm} \mid y_1. \overline{\none} \mid \cdots  x.\some_{\lfv{B_i(l)}} ; x(y_l). x.\some_{y_l ,\lfv{B_i(l)}};x.\overline{\some} ; \outact{x}{x_l}. (x_l.\some_{\lfv{B_i(l)}} ;  \\
       &\hspace{1cm}  \piencodf{B_i(l)}_{x_l} \mid y_l. \overline{\none} \mid x.\some_{\emptyset} ; x(y_{l+1}). ( y_{l+1}.\overline{\some};y_{l+1} . \overline{\close} \mid x.\some_{\emptyset} ; x. \overline{\none}) ) ) )
                \\
     & \redd^*  \bigoplus_{B_i \in \perm{B}} (\nu x, y_1, x_1, \cdots  y_l, x_l)(  y_1 . \some_{\emptyset} ;y_{1}.\close;\zero \mid \cdots \mid y_l . \some_{\emptyset} ;y_{l}.\close;\zero \\
             & \hspace{1cm}x.\overline{\some}. \outact{x}{y_{l+1}} . (y_{l+1} . \some_{\emptyset} ; y_{l+1}.\close;\zero \mid x.\overline{\some};x.\some_{u,(\lfv{M} \setminus x_{l+1} , \cdots , x_k )}; \\
             & \hspace{1cm} x(x_{l+1}). \cdots x.\overline{\some}. \outact{x}{y_k} . (y_k . \some_{\emptyset} ; y_{k}.\close;\zero \mid x.\overline{\some};x.\some_{u,(\lfv{M} \setminus  x_k )};x(x_k) . \\
            &\hspace{1cm} x.\overline{\some}; \outact{x}{y_{k+1}}. ( y_{k+1} . \some_{u,\lfv{M} } ;y_{k+1}.\close; \piencodf{M}_u \mid x.\overline{\none} )) \cdots ) \mid \\
         & \hspace{1cm}  x_1.\some_{\lfv{B_i(1)}} ; \piencodf{B_i(1)}_{x_1} \mid \cdots \mid  x_l.\some_{\lfv{B_i(l)}} ; \piencodf{B_i(l)}_{x_l} \mid  y_1. \overline{\none} \mid \cdots \mid y_l. \overline{\none}\\
          & \hspace{1.5cm} x.\some_{\emptyset} ; x(y_{l+1}). ( y_{l+1}.\overline{\some};y_{l+1} . \overline{\close} \mid x.\some_{\emptyset} ; x. \overline{\none}) ) 
                \\
                & \redd^* 
                    \bigoplus_{B_i \in \perm{B}} (\nu x, x_1, \cdots  , x_l)(
           x.\some_{u,(\lfv{M} \setminus x_{l+1} , \cdots , x_k )};x(x_{l+1}) . \cdots \\
     & \hspace{1.5cm} x.\overline{\some}. \outact{x}{y_k} . (y_k . \some_{\emptyset} ; y_{k}.\close;\zero \mid x.\overline{\some};x.\some_{u,(\lfv{M} \setminus  x_k )};x(x_k) . \\
      &\hspace{1.5cm} x.\overline{\some}; \outact{x}{y_{k+1}}. ( y_{k+1} . \some_{u,\lfv{M}} ;y_{k+1}.\close; \piencodf{M}_u \mid x.\overline{\none} ) )  \mid \\
     &\hspace{1.5cm}   x_1.\some_{\lfv{B_i(1)}} ; \piencodf{B_i(1)}_{x_1} \mid \cdots \mid  x_l.\some_{\lfv{B_i(l)}} ; \piencodf{B_i(l)}_{x_l} \mid  x. \overline{\none} ) 
                \\
        & \redd \bigoplus_{B_i \in \perm{B}} (\nu  x_1, \cdots  , x_l)(  u . \overline{\none} \mid x_1 . \overline{\none} \mid  \cdots \mid x_{l} . \overline{\none} \mid (\lfv{M} \setminus x_{1} , \cdots , x_k ) . \overline{\none}  \mid \\
        & \hspace{1.5cm} x_1.\some_{\lfv{B_i(1)}} ; \piencodf{B_i(1)}_{x_1} \mid \cdots \mid  x_l.\some_{\lfv{B_i(l)}} ; \piencodf{B_i(l)}_{x_l}  ) 
                \\
                & \redd^*  \bigoplus_{B_i \in \perm{B}}  u . \overline{\none} \mid (\lfv{M} \setminus \{  x_1, \cdots , x_k \} ) \cup \lfv{B} . \overline{\none}  
            \end{aligned}
            \end{equation}
            }
            \hrule 
        \caption{Reductions of an encoded explicit substitution}
            \label{ch2fig:proofreductions1}
        \end{figure}

        On the other hand, we have:
        
        \begin{equation}\label{ch2eq:compl_fail2-fail}
        \begin{aligned}
            \piencodf{\expr{M}}_u &= \piencodf{\sum_{B_i \in \perm{B}}  \fail^{\widetilde{y}}}_u= \bigoplus_{B_i \in \perm{B}} \piencodf{\fail^{\widetilde{y}} }_u\\
            &= \bigoplus_{B_i \in \perm{B}} u . \overline{\none} \mid (\lfv{M} \setminus \{  x_1, \cdots , x_k \} ) \cup \lfv{B} . \overline{\none}  \\
        \end{aligned}
        \end{equation}
        
        Therefore, by \eqref{ch2eq:compl_fail1-fail}
        and  \eqref{ch2eq:compl_fail2-fail} the result follows.
        
        \item Case $\redlab{RS{:}Cons_1}$:
        
        Then, 
        $\expr{N} = \fail^{\widetilde{x}}\ B$ with $B =  \bag{N_1 , \dots , N_k} $ and $\expr{N} \redd  \sum_{\perm{B}} \fail^{\widetilde{x} \cup \widetilde{y}} = \expr{M}$, where $ \widetilde{y} = \lfv{B}$. 
        
        On the one hand, we have: 
        
        \begin{equation}\label{ch2eq:compl_cons1-fail}
        \begin{aligned}
            \piencodf{N}_u &= \piencodf{ \fail^{\widetilde{x}}\ B }_u\\
            &= \bigoplus_{B_i \in \perm{B}} (\nu v)(\piencodf{\fail^{\widetilde{x}}}_v \mid v.\some_{u,\lfv{B}} ; \outact{v}{x} . ([v \leftrightarrow u] \mid \piencodf{B_i}_x ) ) \\
            &= \bigoplus_{B_i \in \perm{B}} (\nu v)( v . \overline{\none} \mid \widetilde{x}. \overline{\none} \mid v.\some_{u, \lfv{B}} ; \outact{v}{x} . ([v \leftrightarrow u] \mid \piencodf{B_i}_x ) ) \\
            & \redd \bigoplus_{B_i \in \perm{B}} u . \overline{\none} \mid \widetilde{x}. \overline{\none} \mid \widetilde{y}. \overline{\none} \\
            &= \bigoplus_{\perm{B}} u . \overline{\none} \mid \widetilde{x}. \overline{\none} \mid \widetilde{y}. \overline{\none} \\
        \end{aligned}
        \end{equation}
        
        On the other hand, we have:
        \begin{equation}\label{ch2eq:compl_cons2-fail}
        \begin{aligned}
            \piencodf{\expr{M}}_u &= \piencodf{\sum_{\perm{B}} \fail^{\widetilde{x} \cup \widetilde{y}}}_u
            = \bigoplus_{\perm{B}} \piencodf{\fail^{\widetilde{x} \cup \widetilde{y}} }_u\\
            &= \bigoplus_{\perm{B}} u . \overline{\none} \mid \widetilde{x}. \overline{\none} \mid \widetilde{y}. \overline{\none} \\
        \end{aligned}
        \end{equation}
        
        Therefore, by \eqref{ch2eq:compl_cons1-fail}
        and  \eqref{ch2eq:compl_cons2-fail} the result follows.

        \item Cases $\redlab{RS{:}Cons_2}$ and $\redlab{RS{:}Cons_3}$: These cases follow by IH similarly to Case 7.

    \end{enumerate}
\end{proof}

\opsound*

\begin{proof}
By induction on the structure of $\expr{N} $ and then induction on the number of reductions of $\piencodf{\expr{N}} \redd_{\pequiv}^* Q$

\begin{myEnumerate}
    \item  $\expr{N} = x$, $\expr{N} = \fail^{\emptyset}$ and $\expr{N} = \lambda x . (M[ \widetilde{x} \leftarrow x ])$.
    
    These cases are trivial since no reduction can take place. 
    
    
    
    

    
    \item $\expr{N} =  (M\ B) $.

        Then, 
        $$ \piencodf{(M\ B)}_u = \bigoplus_{B_i \in \perm{B}} (\nu v)(\piencodf{M}_v \mid v.\some_{u,\lfv{B}} ; \outact{v}{x} . ([v \leftrightarrow u] \mid \piencodf{B_i}_x ) )  $$ and we are able to perform the  reductions from $\piencodf{(M\ B)}_u$. 

        We now proceed by induction on $k$, with  $\piencodf{\expr{N}}_u \redd^k Q$. 
        
        
    
            The interesting case is when $k \geq 1$ (the case $k=0$ is trivial).

            Then, for some process $R$ and $n, m$ such that $k = n+m$, we have the following:
            \[
            \begin{aligned}
               \piencodf{\expr{N}}_u & =  \bigoplus_{B_i \in \perm{B}} (\nu v)( \piencodf{ M}_v \mid v.\some_{u,\lfv{B}} ; \outact{v}{x} . (  \piencodf{ B_i}_x \mid [v \leftrightarrow u] ) )\\
               & \redd^m  \bigoplus_{B_i \in \perm{B}} (\nu v)( R \mid v.\some_{u , \lfv{B}} ; \outact{v}{x} . ( \piencodf{ B_i}_x \mid [v \leftrightarrow u] ) ) \\
               &\redd^n  Q\\
            \end{aligned}
            \]
            Thus, the first $m \geq 0$ reduction steps are  internal to $\piencodf{ M}_v$; type preservation in \spi ensures that, if they occur,  these reductions  do not discard the possibility of synchronizing with $v.\some$. Then, the first of the $n \geq 0$ reduction steps towards $Q$ is a synchronization between $R$ and $v.\some_{u, \lfv{B}}$.
            
            We consider two sub-cases, depending on the values of  $m$ and $n$:
            \begin{myEnumerate}
                \item When $m = 0$ and $n \geq 1$:
                    
                    Thus $R = \piencodf{\expr{M}}_v$, and there  are two possibilities of having an unguarded $v.\overline{\some}$ or $v.\overline{\none}$ without internal reductions.   By the diamond property (Proposition~\ref{ch2prop:conf1_lamrsharfail}) we will be reducing each non-deterministic choice of a process simultaneously.
                    Then we have the following for each case:

                    \begin{myEnumerate}
                        \item $M = (\lambda x . M' [\widetilde{x} \leftarrow x]) \linexsub{N_1 / y_1} \cdots \linexsub{N_p / y_p} \qquad (p \geq 0)$.
                        
                           $$
                        \begin{aligned}
                            \piencodf{M}_v &= \piencodf{(\lambda x . M' [\widetilde{x} \leftarrow x]) \linexsub{N_1 / y_1} \cdots  \linexsub{N_p / y_p}}_v \\
                            &= (\nu\widetilde{y}) ( \piencodf{(\lambda x . M' [\widetilde{x} \leftarrow x])}_v \mid y_1.\some_{\lfv{N_1}};  \piencodf{ N_1 }_{y_1} \mid \cdots \\
                            & \qquad \mid y_p.\some_{\lfv{N_p}};\piencodf{ N_p }_{y_p} )\\
                            &= (\nu \widetilde{y}) ( \piencodf{(\lambda x . M' [\widetilde{x} \leftarrow x])}_v \mid Q'' ), \text{ for } \widetilde{y} = y_1 , \cdots , y_p\\
                            &= (\nu \widetilde{y}) ( v.\overline{\some};v(x).\piencodf{ M'[\widetilde{x} \leftarrow x] }_v \mid Q'' )\\
                        \end{aligned}
                        $$
                        where 
                        $Q'' = y_1.\some_{\lfv{N_1}};\piencodf{ N_1 }_{y_1} \mid \cdots \mid y_p.\some_{\lfv{N_p}};\piencodf{ N_p }_{y_p}$.

                    With this shape for $M$, the encoding of $\mathbb{N}$ becomes:

                    \begin{adjustwidth}{-1cm}{}
                    \[
                    \small
                        \begin{aligned}
                        \piencodf{\expr{N}}_u & = \piencodf{(M\ B)}_u\\
                        &= \bigoplus_{B_i \in \perm{B}} (\nu v)( \piencodf{ M}_v \mid v.\some_{u,\lfv{B}} ; \outact{v}{x} . (  \piencodf{ B_i}_x \mid [v \leftrightarrow u] ) )\\
                        & = \bigoplus_{B_i \in \perm{B}} (\nu v)( (\nu \widetilde{y}) ( v.\overline{\some};v(x).\piencodf{ M'[\widetilde{x} \leftarrow x] }_v \mid Q'' ) \mid \\
                        & \hspace{2cm} v.\some_{u,\lfv{B}} ; \outact{v}{x} . ( \piencodf{ B_i}_x \mid [v \leftrightarrow u] ) )\\
                        & \redd \bigoplus_{B_i \in \perm{B}} (\nu v , \widetilde{y})( v(x).\piencodf{ M'[\widetilde{x} \leftarrow x] }_v \mid \outact{v}{x} . (  \piencodf{ B_i}_x \mid [v \leftrightarrow u] ) \mid Q'' )  = Q_1\\
                        & \redd \bigoplus_{B_i \in \perm{B}} (\nu v, \widetilde{y} ,x)( \piencodf{ M'[\widetilde{x} \leftarrow x] }_v \mid  \piencodf{ B_i}_x \mid [v \leftrightarrow u]  \mid Q'')  = Q_2\\
                        & \redd \bigoplus_{B_i \in \perm{B}} (\nu x, \widetilde{y})( \piencodf{ M'[\widetilde{x} \leftarrow x] }_u \mid  \piencodf{ B_i}_x   \mid Q'') = Q_3\\
                        \end{aligned}
                    \]
                    \end{adjustwidth}

                    We also have that 
                    
                    \[
                    \begin{aligned}
                        \expr{N} &=(\lambda x . M' [\widetilde{x} \leftarrow x]) \linexsub{N_1 / y_1} \cdots \linexsub{N_p / y_p}\ B \\
                        &\pequiv ((\lambda x . M' [\widetilde{x} \leftarrow x]) B) \linexsub{N_1 / y_1} \cdots \linexsub{N_p / y_p} \\
                       & \redd  M'[\widetilde{x} \leftarrow x] \esubst{B}{x} \linexsub{N_1 / y_1} \cdots \linexsub{N_p / y_p} = \expr{M}
                    \end{aligned}
                    \]
                    
                    Furthermore, we have: 
                    \[
                     \begin{aligned}
                           \piencodf{\expr{M}'}_v &= \piencodf{M'[\widetilde{x} \leftarrow x] \esubst{B}{x}\linexsub{N_1 / y_1} \cdots \linexsub{N_p / y_p}}_v \\
                           &= \bigoplus_{B_i \in \perm{B}} (\nu x)( \piencodf{ M'[\widetilde{x} \leftarrow x]}_v \mid  \piencodf{ B_i}_x \mid Q'' ) 
                    \end{aligned}
                    \]
                        
                    We consider different possibilities for $n \geq 1$; in all of the thesis holds: \begin{myEnumerate}
                        \item $n = 1$:
                        
                        Then  $Q = Q_1$  and  $\piencodf{\expr{N}}_u \redd^1 Q_1$. In addition,
                        
                        \begin{myEnumerate}
                            \item  \( Q_1 \redd^2 Q_3 = Q' \), 
                            \item $\expr{N} \redd^1 M'[\widetilde{x} \leftarrow x] \esubst{B}{x} = \expr{N}'$,
                            \item  $\piencodf{M'[\widetilde{x} \leftarrow x] \esubst{B}{x}}_u = Q_3$.
                        \end{myEnumerate}
                        and the result follows.

                        \item  $n = 2$:
                         
                       Then $Q = Q_2$ and $ \piencodf{\expr{N}}_u \redd^2 Q_2$. In addition, 
                        
                        \begin{itemize}
                            \item $Q_2 \redd^1 Q_3 = Q'$ , 
                            \item $\expr{N} \redd^1 M'[\widetilde{x} \leftarrow x] \esubst{B}{x} = \expr{N}'$
                            \item $\piencodf{M'[\widetilde{x} \leftarrow x]) \esubst{B}{x}}_u = Q_3$
                        \end{itemize}
                        and the result follows.
                         
                        \item $n \geq 3$:
                        
                       Then $ \piencodf{\expr{N}}_u \redd^3 Q_3 \redd^l Q$, for $l \geq 0$. In addition,  $\expr{N} \redd \expr{M}'$ and  $Q_3 = \piencodf{\expr{M}'}_u$. By the IH, there exist $ Q'$ and $\expr{N}'$ such that $Q \redd^i Q'$, $\expr{M}' \redd_{\pequiv}^j \expr{N}'$ and $\piencodf{\expr{N}'}_u = Q'$ . Finally,
                        $\piencodf{\expr{N}}_u \redd^3 Q_3 \redd^l Q \redd^i Q'$ and $\expr{N} \rightarrow \expr{M}'  \redd_{\pequiv}^j \expr{N}'$, and the result follows.
                        
                    \end{myEnumerate}
                       \item $M = \fail^{\widetilde{z}}$.

                    \[
                        \begin{aligned}
                            \piencodf{M}_v &= \piencodf{\fail^{\widetilde{z}}}_v= v.\overline{\none} \mid \widetilde{z}.\overline{\none} \\
                        \end{aligned}
                    \]
                    
                    With this shape for $M$, the encoding of $\expr{N}$ becomes:
                    \begin{adjustwidth}{-1cm}{}
                    \[
                    \begin{aligned}
                        \piencodf{\expr{N}}_u & = \piencodf{(M\ B)}_u\\
                        &= \bigoplus_{B_i \in \perm{B}} (\nu v)( \piencodf{ M}_v \mid v.\some_{u, \lfv{B}} ; \outact{v}{x} . (  \piencodf{ B_i}_x \mid [v \leftrightarrow u] ) )\\
                        & = \bigoplus_{B_i \in \perm{B}} (\nu v)(  v.\overline{\none}\mid \widetilde{z}.\overline{\none} \mid v.\some_{u, \lfv{B}} ;  \outact{v}{x} . (  \piencodf{ B_i}_x \mid [v \leftrightarrow u] ) )\\
                        & \redd \bigoplus_{B_i \in \perm{B}}   u.\overline{\none} \mid \widetilde{z}.\overline{\none}  \mid \lfv{B}.\overline{\none} \\
                        &= \bigoplus_{\perm{B}}   u.\overline{\none} \mid \widetilde{z}.\overline{\none}  \mid \lfv{B}.\overline{\none} \\
                        \end{aligned}
                    \]
                    \end{adjustwidth}

                    \end{myEnumerate}
                    
                  Also, 
                    \[  \expr{N} = \fail^{\widetilde{z}} \ B  \redd \sum_{\perm{B}} \fail^{\widetilde{z} \cup \lfv{B} }  = \expr{M}.  \]
                    
                    Furthermore, 
                    \[
                     \begin{aligned}
                           \piencodf{\expr{M}}_u &= \piencodf{\sum_{\perm{B}} \fail^{\widetilde{z} \cup \lfv{B} }}_u \\
                         &  = \bigoplus_{\perm{B}}\piencodf{ \fail^{\widetilde{z} \cup \lfv{B} }}_u\\
                          & = \bigoplus_{\perm{B}}    u.\overline{\none} \mid \widetilde{z}.\overline{\none}  \mid \lfv{B}.\overline{\none}
                    \end{aligned}
                    \]

                \item When $m \geq 1$ and $ n \geq 0$, the distinguish two cases:
                    
                    \begin{myEnumerate}
                        \item $n = 0$:
                            
                            Then, 
                            $$ 
                            \begin{aligned} &\bigoplus_{B_i \in \perm{B}} (\nu v)( R \mid v.\some_{u, \lfv{B}} ; \outact{v}{x} . (  \piencodf{ B_i}_x \mid [v \leftrightarrow u] ) ) = Q,\\ 
                            &~ \text{ and }  \piencodf{M}_u \redd^m R.
                            \end{aligned} 
                            $$ 
                        By the IH there exist $R'$  and $\expr{M}' $ such that $R \redd^i R'$, $M \redd_{\pequiv}^j \expr{M}'$, and $\piencodf{\expr{M}'}_u = R'$.  Hence, 
                            \[ 
                            \small
                            \begin{aligned}
                               \piencodf{\expr{N}}_u & =  \bigoplus_{B_i \in \perm{B}} (\nu v)( \piencodf{ M}_v \mid v.\some_{u, \lfv{B}} ; \outact{v}{x} . (  \piencodf{ B_i}_x \mid [v \leftrightarrow u] ) )\\
                               & \redd^m  \bigoplus_{B_i \in \perm{B}} (\nu v)( R \mid v.\some_{u , \lfv{B}} ; \outact{v}{x} . (  \piencodf{ B_i}_x \mid [v \leftrightarrow u] ) ) = Q\\
                               & \redd^i  \bigoplus_{B_i \in \perm{B}} (\nu v)( R' \mid v.\some_{u, \lfv{B}} ; \outact{v}{x} . (  \piencodf{ B_i}_x \mid [v \leftrightarrow u] ) ) = Q'\\
                            \end{aligned}
                            \]
                            and so the \lamrsharfail term can reduce as follows: $\expr{N} = (M\ B) \redd_{\pequiv}^j M'\ B = \expr{N}'$ and  $\piencodf{\expr{N}'}_u = Q'$.

                        \item $n \geq 1$:
                        
                            Then  $R$ has an occurrence of an unguarded $v.\overline{\some}$ or $v.\overline{\none}$, which implies it is of the form 
                            $ \piencodf{(\lambda x . M' [\widetilde{x} \leftarrow x]) \linexsub{N_1 / y_1} \cdots \linexsub{N_p / y_p}}_v $ or $ \piencodf{\fail}_v $, and the case follows by IH.

                    \end{myEnumerate}

            \end{myEnumerate}

        
        This concludes the analysis for the case $\expr{N} = (M \, B)$.
        
        \item $\expr{N} = M [ \widetilde{x} \leftarrow x ]$.
    
            The sharing variable $x$ is not free and the result follows by vacuity.
        
        \item $\expr{N} = (M[\widetilde{x} \leftarrow x])\esubst{ B }{ x }$. Then,
            
            \[
                \begin{aligned}
                    \piencodf{\expr{N}}_u &=\piencodf{ (M[\widetilde{x} \leftarrow x])\esubst{ B }{ x } }_u = \bigoplus_{B_i \in \perm{B}} (\nu x)( \piencodf{ M[\widetilde{x} \leftarrow x]}_u \mid \piencodf{ B_i}_x )
                \end{aligned}
            \]


            \begin{myEnumerate}
                \item  $\size{\widetilde{x}} = \size{B}$.
                
                    Then let us consider the shape of the bag $ B$.
                    
     \begin{myEnumerate}
                
            \item When $B = 1$
                         
                            We have the following
                            \begin{adjustwidth}{-1cm}{}
                            \[
                                \small
                            \begin{aligned}
\piencodf{\expr{N}}_u &= (\nu x)( \piencodf{ M[ \leftarrow x]}_u \mid \piencodf{ 1}_x ) \\
&=   (\nu x)(  x. \overline{\some}. \outact{x}{y_i} . ( y_i . \some_{u, \lfv{M}} ;y_{i}.\close; \piencodf{M}_u \mid x. \overline{\none}) \mid\\ &\qquad x.\some_{\emptyset} ; x(y_n). ( y_n.\overline{\some};y_n . \overline{\close} \mid x.\some_{\emptyset} ; x. \overline{\none}) )  \\
& \redd  (\nu x)(   \outact{x}{y_i} . ( y_i . \some_{u, \lfv{M}} ;y_{i}.\close; \piencodf{M}_u \mid x. \overline{\none}) \mid \\
& \hspace{1cm}  x(y_n). ( y_n.\overline{\some};y_n . \overline{\close} \mid x.\some_{\emptyset} ; x. \overline{\none}) )           \quad = Q_1                      \\
 & \redd  (\nu x, y_i)(     y_i . \some_{u, \lfv{M}} ;y_{i}.\close; \piencodf{M}_u \mid x. \overline{\none} \mid  y_n.\overline{\some};\\
 & \hspace{1cm}y_n . \overline{\close} \mid x.\some_{\emptyset} ; x. \overline{\none})
                 \quad                = Q_2 \\
& \redd  (\nu x, y_i)( y_{i}.\close; \piencodf{M}_u \mid x. \overline{\none} \mid   y_n . \overline{\close} \mid x.\some_{\emptyset} ;  x. \overline{\none})
                                     \quad = Q_3
                                \\
& \redd (\nu x)(  \piencodf{M}_u \mid x. \overline{\none} \mid   x.\some_{\emptyset} ; x. \overline{\none}) 
 \qquad = Q_4\\
& \redd   \piencodf{M}_u   \quad = Q_5
                            \end{aligned}
                            \]
                            \end{adjustwidth}
                            Notice how $Q_2$ has a choice however the $x$ name can be closed at any time so for simplicity we only perform communication across this name once all other names have completed their reductions.
                            
                           Now proceed by induction on the number of reductions $\piencodf{\expr{N}}_u \redd^k Q$.
                            
                            \begin{myEnumerate}
                                
                                \item  $k = 0$:
                                
                                This case is trivial.

                                \item  $k = 1$: ($2 \leq  k \leq 4$: is similar.)
                                
                                    Then,  $Q = Q_1$ and $ \piencodf{\expr{N}}_u \redd^1 Q_1$. In addition,  $Q_1 \redd^4 Q_5 = Q'$,   $\expr{N} \redd^0 M[ \leftarrow x] \esubst{ 1 }{ x } \pequiv M$ and $\piencodf{ M }_u = Q_5$, and the result follows.
                                    
                                
                                \item  $k \geq 5$:
                                
                                    Then $ \piencodf{\expr{N}}_u \redd^5 Q_5 \redd^l Q$, for $l \geq 0$. Since $Q_5 = \piencodf{ M }_u$,  by the IH it follows  that there exist $ Q' $ and $ \expr{N}' $ such that  $ Q \redd^i Q' ,  M \redd_{\pequiv}^j \expr{N}'$ and $\piencodf{\expr{N}'}_u = Q'$. 
                         
                         Then,  $ \piencodf{\expr{N}}_u \redd^5 Q_5 \redd^l Q \redd^i Q'$ and by the contextual reduction  one has $\expr{N} = (M[ \leftarrow x])\esubst{ 1 }{ x } \redd_{\pequiv}^j  \expr{N}' $ and the case holds.

                            \end{myEnumerate}               
                        
                        \item$B = \bag{N_1, \cdots ,N_l}$, for $l \geq 1$.
                        
                            Then,  consider the reductions in \figref{ch2fig:proofreductions2}.
                            
                            \begin{figure}[!t]
                            \hrule 
                            { \small
                            \[
                            \begin{aligned}
                                          \piencodf{\expr{N}}_u &=\piencodf{(M[\widetilde{x} \leftarrow x])\esubst{ B }{ x }}_u\\
                             & = \bigoplus_{B_i \in \perm{B}} (\nu x)( \piencodf{ M[\widetilde{x} \leftarrow x]}_u \mid \piencodf{ B_i}_x ) \\
                                 &=  \bigoplus_{B_i \in \perm{B}} (\nu x)( x.\overline{\some}. \outact{x}{y_1}. (y_1 . \some_{\emptyset} ;y_{1}.\close;\zero \mid x.\overline{\some};x.\some_{u, (\lfv{M} \setminus x_1 , \cdots , x_l )}; 
                                 \\
                               &\qquad x(x_1) . \cdots x.\overline{\some}. \outact{x}{y_l} . (y_l . \some_{\emptyset} ; y_{l}.\close;\zero \mid x.\overline{\some};x.\some_{u,(\lfv{M} \setminus x_l )};x(x_l) . \\
                              &\qquad  x.\overline{\some}; \outact{x}{y_{l+1}}. ( y_{l+1} . \some_{u,\lfv{M}} ;y_{l+1}.\close; \piencodf{M}_u \mid x.\overline{\none} )) \cdots ) \mid \\
                             & \qquad  x.\some_{\lfv{B}} ; x(y_1). x.\some_{y_1, \lfv{B} };x.\overline{\some} ; \outact{x}{x_1}. (x_1.\some_{\lfv{B_i(1)}} ; \piencodf{B_{i}(1)}_{x_1}  \\
                             &\qquad   \mid y_1. \overline{\none}\mid \cdots x.\some_{\lfv{B_{i}(l)}} ; x(y_l). x.\some_{y_l, \lfv{B_{i}(l)}} ;x.\overline{\some} ; \outact{x}{x_l}. (x_l.\some_{\lfv{B_{i}(l)}} ;   \\
                              & \qquad \piencodf{B_{i}(l)}_{x_l}\mid y_l. \overline{\none}\mid x.\some_{\emptyset} ; x(y_{l+1}). ( y_{l+1}.\overline{\some};y_{l+1} . \overline{\close} \mid x.\some_{\emptyset} ; x. \overline{\none})
                                  )
                                  )
                                  )
                               \\
                                & \redd ^{5l}
                                \bigoplus_{B_i \in \perm{B}} (\nu x , x_1,y_1, \cdots , x_l,y_1)( y_1 . \some_{\emptyset} ;y_{1}.\close;\zero \mid  \cdots  y_l . \some_{\emptyset} ; y_{l}.\close;\zero \mid  \\
                                  &\qquad x.\overline{\some}; \outact{x}{y_{l+1}}. ( y_{l+1} . \some_{u,\lfv{M}} ;y_{l+1}.\close; \piencodf{M}_u \mid x.\overline{\none} )
                                   \mid \\
                                  & \qquad  x_1.\some_{\lfv{B_{i}(1)}} ; \piencodf{B_{i}(1)}_{x_1}  \mid y_1. \overline{\none} \mid \cdots   x_l.\some_{\lfv{B_{i}(l)}} ; \piencodf{B_{i}(l)}_{x_l}  \mid y_l. \overline{\none}\mid\\
                                  & \qquad x.\some_{\emptyset} ; x(y_{l+1}). ( y_{l+1}.\overline{\some};y_{l+1} . \overline{\close} \mid x.\some_{\emptyset} ; x. \overline{\none})
                                  )
                               \\
                               & \redd ^{5}
                               \bigoplus_{B_i \in \perm{B}} (\nu  x_1,y_1, \cdots , x_l,y_1 )(  y_1 . \some_{\emptyset} ;y_{1}.\close;\zero \mid  \cdots  y_l . \some_{\emptyset} ; y_{l}.\close;\zero \\
                                 & \qquad \mid \piencodf{M}_u \mid   x_1.\some_{\lfv{B_{i}(1)}} ; \piencodf{B_{i}(1)}_{x_1}  \mid y_1. \overline{\none} \mid \cdots   x_l.\some_{\lfv{B_{i}(l)}} ; \piencodf{B_{i}(l)}_{x_l}  \mid y_l. \overline{\none} )
                           \\
                               & \redd ^{l}  \bigoplus_{B_i \in \perm{B}} (\nu x_1,\cdots , x_l )( \piencodf{M}_u \mid x_1.\some_{\lfv{B_{i}(1)}} ; \piencodf{B_{i}(1)}_{x_1}  \mid  \cdots \mid  x_l.\some_{\lfv{B_{i}(l)}} ; \piencodf{B_{i}(l)}_{x_l} )\\
                               &   
                        = Q_{6l + 5}\\
                            \end{aligned}
                            \]
                            }
                            \hrule 
                            \caption{Reductions of encoded explicit substitution}
                                \label{ch2fig:proofreductions2}
                            \end{figure}

                            The proof follows by induction on the number of reductions $\piencodf{\expr{N}}_u \redd^k Q$.
                            
                            \begin{myEnumerate}
                                \item $k = 0$:
                                
                                This case is trivial. Take $\piencodf{\expr{N}}_u=Q=Q'$ and $\expr{N}=\expr{N}'$.
                                
                                \item  $1 \leq k \leq 6l + 5$:
                                    
                                    Then,  $ \piencodf{\expr{N}}_u \redd^k Q_k$.
                                    Observing the reductions in \figref{ch2fig:proofreductions2}, one has 
                                    $Q_k \redd^{6l + 5 - k} Q_{6l + 5} = Q'$ ,

                                    $\expr{N} \redd^1 \sum_{B_i \in \perm{B}}M\ \linexsub{B_i(1)/x_1} \cdots \linexsub{B_i(l)/x_l} = \expr{N}'$ and 
                                    
                                    $\piencodf{\sum_{B_i \in \perm{B}}M\ \linexsub{B_i(1)/x_1} \cdots \linexsub{B_i(l)/x_l}}_u = Q_{6l + 5}$, and the result follows.
                                
                                \item $k > 6l + 5$:
                                
                            Then, $ \piencodf{\expr{N}}_u \redd^{6l + 5} Q_{6l + 5} \redd^n Q$, for $n \geq 1$. In addition,
                                    
                                    $\expr{N} \redd^1 \sum_{B_i \in \perm{B}}M\ \linexsub{B_i(1)/x_1} \cdots \linexsub{B_i(l)/x_l}$ and 
                                    
                                    $Q_{6l + 5} = \piencodf{\sum_{B_i \in \perm{B}}M\ \linexsub{B_i(1)/x_1} \cdots \linexsub{B_i(l)/x_l}}_u$. By the IH there exist  $ Q'$ and $\expr{N}'$ such that $  Q \redd^i Q'$, $$\sum_{B_i \in \perm{B}}M\ \linexsub{B_i(1)/x_1} \cdots \linexsub{B_i(l)/x_l} \redd_{\pequiv}^j \expr{N}'$$ and $\piencodf{\expr{N}'}_u = Q'$. Finally,
                                    
                                $\piencodf{\expr{N}}_u \redd^{6l + 5} Q_{6l + 5} \redd^n Q \redd^i Q'$ and
                                
                                $\expr{N} \rightarrow \sum_{B_i \in \perm{B}}M\ \linexsub{B_i(1)/x_1} \cdots \linexsub{B_i(l)/x_l}  \redd_{\pequiv}^j \expr{N}'$.

                            \end{myEnumerate}

                    \end{myEnumerate}

                \item  $\size{\widetilde{x}} > \size{B}$.
                
                    Then,
                    $\expr{N} = M[x_1, \cdots , x_k \leftarrow x]\ \esubst{ B }{ x }$ with $B = \bag{N_1 ,  \cdots , N_l},$ for $ k > l$. Also,
                    $$ \expr{N} \redd  \sum_{B_i \in \perm{B}}  \fail^{\widetilde{z}} = \expr{M}\  \text{  and } \ \widetilde{z} = (\lfv{M} \setminus \{  x_1, \cdots , x_k \} ) \cup \lfv{B}. $$
                    On the one hand, we have \figref{ch2fig:proofreductions3}.
                    Hence $k = l + m$ for some $m \geq 1$

                    \begin{figure}[!t]
                    \hrule 
                    { \small
                    
                    \[
                    \begin{aligned}
                        \piencodf{N}_u &= \piencodf{M[x_1, \cdots , x_k \leftarrow x]\ \esubst{ B }{ x }}_u \\
                        &=  \bigoplus_{B_i \in \perm{B}} (\nu x)( \piencodf{ M[x_1, \cdots , x_k \leftarrow x]}_u \mid  \piencodf{ B_i}_x ) 
                        \\
                        &=  \bigoplus_{B_i \in \perm{B}} (\nu x)(                            x.\overline{\some}. \outact{x}{y_1}. (y_1 . \some_{\emptyset} ;y_{1}.\close;\zero \mid x.\overline{\some};x.\some_{u, (\lfv{M} \setminus x_1 , \cdots , x_k )}; \\
                                &\qquad x(x_1) . \cdots x.\overline{\some}. \outact{x}{y_k} . (y_k . \some_{\emptyset} ; y_{k}.\close;\zero \mid x.\overline{\some};x.\some_{u,(\lfv{M} \setminus x_k )};x(x_k) . \\
                                &\qquad x.\overline{\some}; \outact{x}{y_{k+1}}. ( y_{k+1} . \some_{u,\lfv{M}} ;y_{k+1}.\close; \piencodf{M}_u \mid x.\overline{\none} )) \cdots ) \mid \\
                                &\qquad x.\some_{\lfv{B}} ; x(y_1). x.\some_{y_1, \lfv{B}} ;x.\overline{\some} ; \outact{x}{x_1}. (x_1.\some_{\lfv{B_i(1)}} ; \piencodf{B_i(1)}_{x_1} \\
                                &\qquad  \mid y_1. \overline{\none} \mid \cdots x.\some_{\lfv{B_i(l)}} ; x(y_l). x.\some_{y_l, \lfv{B_i(l)} };x.\overline{\some} ; \outact{x}{x_l}. (x_l.\some_{\lfv{B_i(l)}} ; \\
                                &\qquad   \piencodf{B_i(l)}_{x_l} \mid y_l. \overline{\none} \mid x.\some_{\emptyset} ; x(y_{l+1}). ( y_{l+1}.\overline{\some};y_{l+1} . \overline{\close} \mid x.\some_{\emptyset} ; x. \overline{\none}) ) 
                               ) )
                        \\
                        & \redd^{5l} \bigoplus_{B_i \in \perm{B}} (\nu x, y_1, x_1, \cdots  y_l, x_l)( 
                                  y_1 . \some_{\emptyset} ;y_{1}.\close;\zero \mid \cdots \mid y_l . \some_{\emptyset} ;y_{l}.\close;\zero \\
                                &\qquad x.\overline{\some}. \outact{x}{y_{l+1}} . (y_{l+1} . \some_{\emptyset} ; y_{l+1}.\close;\zero \mid x.\overline{\some};x.\some_{u,(\lfv{M} \setminus x_{l+1} , \cdots , x_k )}; \\
                                &\qquad x(x_{l+1}) . \cdots x.\overline{\some}. \outact{x}{y_k} . (y_k . \some_{\emptyset} ; y_{k}.\close;\zero \mid x.\overline{\some};x.\some_{u,(\lfv{M} \setminus x_k )};x(x_k) . \\
                                &\qquad x.\overline{\some}; \outact{x}{y_{k+1}}. ( y_{k+1} . \some_{u,\lfv{M}} ;y_{k+1}.\close; \piencodf{M}_u \mid x.\overline{\none} )) \cdots ) \mid \\
                                &\qquad   x_1.\some_{\lfv{B_i(1)}} ; \piencodf{B_i(1)}_{x_1} \mid \cdots \mid  x_l.\some_{\lfv{B_i(l)}} ; \piencodf{B_i(l)}_{x_l} \mid \\
                                &\qquad y_1. \overline{\none} \mid \cdots \mid y_l. \overline{\none}\\
                                &\qquad x.\some_{\emptyset} ; x(y_{l+1}). ( y_{l+1}.\overline{\some};y_{l+1} . \overline{\close} \mid x.\some_{\emptyset} ; x. \overline{\none}) ) \\
                        & \redd^{l+ 5} 
                           \bigoplus_{B_i \in \perm{B}} (\nu x, x_1, \cdots , x_l)( x.\some_{u,(\lfv{M} \setminus x_{l+1} , \cdots , x_k )};x(x_{l+1}) . \cdots \\
                                &\qquad x.\overline{\some}. \outact{x}{y_k} . (y_k . \some_{\emptyset} ; y_{k}.\close;\zero \mid x.\overline{\some};x.\some_{u,(\lfv{M} \setminus x_k )};x(x_k) . \\
                                &\qquad x.\overline{\some}; \outact{x}{y_{k+1}}. ( y_{k+1} . \some_{u,\lfv{M}} ;y_{k+1}.\close; \piencodf{M}_u \mid x.\overline{\none} ) )  \mid \\
                                &\qquad   x_1.\some_{\lfv{B_i(1)}} ; \piencodf{B_i(1)}_{x_1} \mid \cdots \mid  x_l.\some_{\lfv{B_i(l)}} ; \piencodf{B_i(l)}_{x_l} \mid   x. \overline{\none} ) \\
                        & \redd
                          \bigoplus_{B_i \in \perm{B}} (\nu  x_1, \cdots  , x_l)(  u . \overline{\none} \mid x_1 . \overline{\none} \mid  \cdots \mid x_{l} . \overline{\none} \mid (\lfv{M} \setminus \{  x_1, \cdots , x_k \} ). \overline{\none}  
                          \\
                               & \qquad \qquad \mid  x_1.\some_{\lfv{B_i(1)}} ; \piencodf{B_i(1)}_{x_1} \mid \cdots \mid  x_l.\some_{\lfv{B_i(l)}} ; \piencodf{B_i(l)}_{x_l}  ) 
                               \\ 
                        & \redd^{l}  \bigoplus_{B_i \in \perm{B}}  u . \overline{\none} \mid (\lfv{M} \setminus \{  x_1, \cdots , x_k \} ). \overline{\none} \mid \lfv{B}. \overline{\none} 
                        \\
                        & = Q_{7l + 6}  
                    \end{aligned}
                    \]
                    
                    }
                    \hrule 
                    \caption{Reductions of an encoded explicit substitution that leads to failure}
                        \label{ch2fig:proofreductions3}
                    \end{figure}

                    Now we proceed by induction on the number of reductions $\piencodf{\expr{N}}_u \redd^j Q$.
                            
                            \begin{myEnumerate}
                                \item $j = 0$:
                               
                               This case is trivial. 
                                
                                \item $1 \leq j \leq 7l + 6$:
                                    
                                 Then,
                                 
                                 $ \piencodf{\expr{N}}_u \redd^j Q_j \redd^{7l + 6 - j} Q_{7l + 6} = Q'$ , $\expr{N} \redd^1 \sum_{B_i \in \perm{B}}  \fail^{\widetilde{z}} = \expr{N}'$ and $\piencodf{\sum_{B_i \in \perm{B}}  \fail^{\widetilde{z}} }_u = Q_{7l + 6}$, and the result follows.
                                
                                \item  $j > 7l + 6$:
                                
                            Then, $ \piencodf{\expr{N}}_u \redd^{7l + 6} Q_{7l + 6} \redd^n Q$, for $n \geq 1$. Also,  $\expr{N} \redd^1 \sum_{B_i \in \perm{B}}  \fail^{\widetilde{z}}$. However no further reductions can be performed.

                            \end{myEnumerate}

                \item $\size{\widetilde{x}} < \size{B}$.    
                   
                   Proceeds similarly to the previous case.
                    
            \end{myEnumerate}

        \item  $\expr{N} = M \linexsub {N' /x}$. 
    
            Then,
            \[
            \begin{aligned}
                \piencodf{M \linexsub {N' /x}}_u &= (\nu x) ( \piencodf{ M }_u \mid   x.\some_{\lfv{N'}};\piencodf{ N' }_x  )  \\
            \end{aligned}
            \]
            
            Then we have
            \[
            \begin{aligned}
               \piencodf{\expr{N}}_u & =  (\nu x) ( \piencodf{ M }_u \mid   x.\some_{\lfv{N'}};\piencodf{ N' }_x  ) \\
               & \redd^m  (\nu x) ( R \mid   x.\some_{\lfv{N'}};\piencodf{ N' }_x  )\\
               & \redd^n  Q\\
            \end{aligned}
            \]
            
            for some process $R$, where $\redd^n$ is a reduction that initially synchronizes with $x.\some_{\lfv{N'}}$ when $n \geq 1$, $n + m = k \geq 1$. Type preservation in \spi ensures reducing $\piencodf{ M}_v \redd^m$ does not consume possible synchronizations with $x.\some$ if they occur. Let us consider the the possible sizes of both $m$ and $n$.
            
            \begin{myEnumerate}
                \item For $m = 0$ and $n \geq 1$.
                
                    In this case  $R = \piencodf{M}_u$  and  there are two possibilities of having an unguarded $x.\overline{\some}$ or $x.\overline{\none}$ without internal reductions.  
                    
                    \begin{myEnumerate}
                        \item $M = \fail^{x, \widetilde{y}}$

                            \[
                            \begin{aligned}
                               \piencodf{\expr{N}}_u & =  (\nu x) ( \piencodf{ M }_u \mid   x.\some_{\lfv{N'}};\piencodf{ N' }_x  ) \\
                               & = (\nu x) ( \piencodf{ \fail^{x, \widetilde{y}} }_u \mid   x.\some_{\lfv{N'}};\piencodf{ N' }_x  ) \\
                               & = (\nu x) ( u.\overline{\none} \mid x.\overline{\none} \mid  \widetilde{y}.\overline{\none} \mid x.\some_{\lfv{N'}};\piencodf{ N' }_x  ) \\
                               & \redd u.\overline{\none} \mid \widetilde{y}.\overline{\none} \mid \lfv{N'}.\overline{\none} \\
                            \end{aligned}
                            \]
                            
                            Notice that no further reductions can be performed.
                            
                          Thus, 
                          \[ \piencodf{\expr{N}}_u \redd u.\overline{\none} \mid \widetilde{y}.\overline{\none} \mid \lfv{N'}.\overline{\none}  = Q'.\]
                            We also have that  \[\expr{N} \redd \fail^{ \widetilde{y} \cup \lfv{N'} } = \expr{N}' \text{ and } \piencodf{ \fail^{\widetilde{y} \cup \lfv{N'}} }_u = Q',\]
                            and the result follows.
                            
                        \item $\headf{M} = x $
                    
                            By the diamond property (Proposition~\ref{ch2prop:conf1_lamrsharfail}) we will be reducing each non-deterministic choice of a process simultaneously.
                            Then by Proposition~\ref{ch2prop:NEEDTONAME} we have the following:
                            \[
                            \begin{aligned}
                             \piencodf{\expr{N}}_u     &\redd^* (\nu x) ( \bigoplus_{i \in I}(\nu \widetilde{y})(\piencodf{ x }_{j} \mid P_i) \mid   x.\some_{\lfv{N'}};\piencodf{ N' }_x  ) \\
                                 &= (\nu x) ( \bigoplus_{i \in I}(\nu \widetilde{y})(x.\overline{\some} ;[x \leftrightarrow j ] \mid P_i)  \mid   x.\some_{\lfv{N'}};\piencodf{ N' }_x  ) \\
                                 &\redd (\nu x) ( \bigoplus_{i \in I}(\nu \widetilde{y})([x \leftrightarrow j ] \mid P_i) \mid   \piencodf{ N' }_x  ) \qquad  = Q_1 \\
                                 &\redd  \bigoplus_{i \in I}(\nu \widetilde{y})( \piencodf{ N' }_j \mid P_i ) \qquad = Q_2 \\
                            \end{aligned}
                            \]
                            We also have that 
                            \[
                            \begin{aligned}
                                \expr{N} &=M \linexsub {N' /x} 
                                 \redd  M \headlin{ N' /x }
                                 = \expr{M}'.
                            \end{aligned}
                            \]
                            where by Proposition \ref{ch2prop:NEEDTONAME} we obtain
                            \[
                            \piencodf{M \headlin{ N'/x }}_u \redd^* \bigoplus_{i \in I}(\nu \widetilde{y})( \piencodf{ N' }_j \mid P_i ) = Q_2.
                            \]
                            and finally from Proposition \ref{ch2prop:NEEDTONAME} there exists an $\expr{M}$ with $\expr{M} \pequiv \expr{M}'$ such that:
                            \[
                            \begin{aligned}
                               \piencodf{\expr{M}}_u &= \bigoplus_{i \in I}(\nu \widetilde{y})( \piencodf{ N' }_j \mid P_i ) & = Q_2. 
                            \end{aligned}
                            \]
                            for simplicity we assume 
                            that $\piencodf{\expr{N}}_u \redd Q_1$
                           \begin{myEnumerate}
                                \item  $n = 1$:
                                Then $Q = Q_1$ and  $ \piencodf{\expr{N}}_u \redd^1 Q_1$. Since,
                                $Q_1 \redd^1 Q_2 = Q'$, $\expr{N} \redd^1 M \headlin{ N'/x} \pequiv  \expr{M} = \expr{N}'$ and $\piencodf{\expr{M}}_u = Q_2$, the result follows.
                                 \item  $n \geq 2$:
                                 Then,  $ \piencodf{\expr{N}}_u \redd^2 Q_2 \redd^l Q$, for $l \geq 0$. Also,  $\expr{N} \rightarrow \expr{M}$, $Q_2 = \piencodf{\expr{M}}_u$. By the IH there exist $ Q'$ and $\expr{N}'$ such that $ Q \redd^i Q'$, $\expr{M} \redd_{\pequiv}^j \expr{N}'$ and $\piencodf{\expr{N}'}_u = Q'$ . Finally, $\piencodf{\expr{N}}_u \redd^2 Q_2 \redd^l Q \redd^i Q'$ and $\expr{N} \rightarrow \expr{M}  \redd_{\pequiv}^j \expr{N}' $, and the result follows.
                            \end{myEnumerate}

                    \end{myEnumerate}
 \item For $m \geq 1$ and $ n \geq 0$.
                    
            \begin{myEnumerate}
            \item $n = 0$:
            
               Then,
               \[(\nu x) ( R \mid   x.\some_{\lfv{N'}};\piencodf{ N' }_x  )  = Q ~\text{and }~ \piencodf{M}_u \redd^m R.\] 
               
             By the IH there exist $R'$  and $\expr{M}' $ such that $R \redd^i R'$, $M \redd_{\pequiv}^j \expr{M}'$ and $\piencodf{\expr{M}'}_u = R'$. 
              Hence, 
              \[ 
               \begin{aligned}
                   \piencodf{\expr{N}}_u & = (\nu x) ( \piencodf{M}_u  \mid   x.\some_{\lfv{N'}};\piencodf{ N' }_x  )\\ &\redd^m  (\nu x) ( R \mid   x.\some_{\lfv{N'}};\piencodf{ N' }_x  )  = Q.
                \end{aligned}
                 \]
                Also, 
                \[ 
                \begin{aligned}
                   Q & \redd^i   (\nu x) ( R' \mid   x.\some_{\lfv{N'}};\piencodf{ N' }_x  ) = Q'\\
                \end{aligned}
                \]
                and the term can reduce as follows:
                
                $\expr{N} = M \linexsub {N' /x} \redd_{\pequiv}^j \sum_{M_i' \in \expr{M}'} M_i' \linexsub {N' /x} = \expr{N}'$ and  $\piencodf{\expr{N}'}_u = Q'$.

            \item When $n \geq 1$
            
                Then $R$ has an occurrence of an unguarded $x.\overline{\some}$ or $x.\overline{\none}$, and the case follows by IH.
                        
                    \end{myEnumerate}
            \end{myEnumerate}
    \end{myEnumerate}
\end{proof}

\subsection{Success Sensitiveness}
\label{ch2app:succtwo}

\pressucctwo*

\begin{proof}
In both cases, by induction on the structure of $M$. 
\begin{enumerate}
\item We only need to consider terms of the following form:

    \begin{itemize}

        \item $ M = \checkmark $.
        
        This  case is immediate.
        
        \item $M = N\ B$.
        
       By definition, $\headf{N \ B} = \headf{N}$. Hence we consider that $\headf{N} = \checkmark$. Then,
       
       $$ \piencodf{N \ B}_u = \bigoplus_{B_i \in \perm{B}} (\nu v)(\piencodf{N}_v \mid v.\some_{u, \lfv{B}} ; \outact{v}{x} . ([v \leftrightarrow u] \mid \piencodf{B_i}_x ) ) $$
       and by the IH  $\checkmark$ is unguarded in $\piencodf{N}_u$ after a sequence of reductions.

        \item  $M = (N[\widetilde{x} \leftarrow x])\esubst{ B }{ x }$.
        
        By definition, $\headf{(N[\widetilde{x} \leftarrow x])\esubst{ B }{ x }} = \headf{N[\widetilde{x} \leftarrow x]} = \headf{N} = \checkmark$ where $ \widetilde{x} = x_1 , \cdots , x_k $ and $\#(x,M) = \size{B}$.
        \[
            \begin{aligned}
                \piencodf{M}_u &=  \bigoplus_{B_i \in \perm{B}} (\nu x)( \piencodf{ N[\widetilde{x} \leftarrow x]}_u \mid \piencodf{ B_i}_x ) \\
                &\redd^*  \bigoplus_{B_i \in \perm{B}} (\nu \widetilde{x})( \piencodf{ N}_u \mid x_1.\some_{\lfv{B_i(1)}};\\
                & \qquad \qquad  \qquad \piencodf{ B_i(1) }_{x_1} \mid \cdots \mid x_k.\some_{\lfv{B_i(k)}};\piencodf{ B_i(k) }_{x_k} ) \\
            \end{aligned} 
        \]

        and by the IH  $\checkmark$ is unguarded in $\piencodf{N}_u$ after a sequence of reductions.

        \item $M = M' \linexsub {N /x}$.
        
       By definition,  $\headf{M' \linexsub{N /x}}  = \headf{M'} \checkmark$. Then, $$\piencodf{M' \linexsub {N /x}}_u = (\nu x) ( \piencodf{ M' }_u \mid   x.\some_{\lfv{N}};\piencodf{ N }_x  )$$ and by the IH $\checkmark$ is unguarded in $\piencodf{N}_u$.

    \end{itemize}

   \item  We only need to consider terms of the following form:    
         
    \begin{itemize}
        \item  $M = \checkmark$.
        
        This case is trivial.
        
        \item $M = N\ B$.
        
        Then,
        \[\piencodf{N \ B}_u = \bigoplus_{B_i \in \perm{B}} (\nu v)(\piencodf{N}_v \mid v.\some_{u,\lfv{B}} ; \outact{v}{x} . ([v \leftrightarrow u] \mid \piencodf{B_i}_x ) ).\] 
        The only occurrence of an unguarded $\checkmark$ is within $\piencodf{N}_v$. By the IH we have that $\headf{N} = \checkmark$ and finally $\headf{N \ B} = \headf{N}$.
        
        

        \item $M = (N[\widetilde{x} \leftarrow x])\esubst{ B }{ x }$.
        
        Then,
        $$
            \begin{aligned}
                \piencodf{(N[\widetilde{x} \leftarrow x])\esubst{ B }{ x }}_u &=  \bigoplus_{B_i \in \perm{B}} (\nu x)( \piencodf{ N[\widetilde{x} \leftarrow x]}_u \mid \piencodf{ B_i}_x ) \\
            \end{aligned} 
        $$ However in both $\piencodf{ N[\widetilde{x} \leftarrow x]}_u$ and $\piencodf{ B_i}_x$ we have that both are guarded and hence $\checkmark$ cannot occur without synchronizations.
        
        \item $M = M' \linexsub {N /x}$.
        
        Then,
        $$\piencodf{M' \linexsub {N /x}}_u = (\nu x) ( \piencodf{ M' }_u \mid   x.\some_{\lfv{N}};\piencodf{ N }_x  ),$$ an unguarded occurrence of $\checkmark$ can only occur within $\piencodf{ M' }_u $. By the IH we have $\headf{M'} = \checkmark$ and hence $\headf{M' \linexsub{N /x}}  = \headf{M'}$. 
           \end{itemize}    
\end{enumerate}
\end{proof}

\successsenscetwo*

\begin{proof}
We proceed with the proof in two parts.

\begin{myEnumerate}
    
    \item Suppose that  $\expr{M} \Downarrow_{\checkmark} $. We will prove that $\piencodf{\expr{M}} \Downarrow_{\checkmark}$.

    By \defref{ch2def:app_Suc3}, there exists $ \expr{M}' = M_1 + \cdots + M_k$ such that $\expr{M} \redd^* \expr{M}'$ and with
    $\headf{M_j} = \checkmark$, for some  $j \in \{1, \ldots, k\}$. By completeness there exists $Q$ such that $\piencodf{\expr{M}}_u  \redd^* Q = \piencodf{\expr{M}'}_u$.
    
    We wish to show that there exists $ Q'$ such that $Q \redd^* Q'$ and $Q'$ has an unguarded occurrence of $\checkmark$.
    
    Since $Q = \piencodf{\expr{M}'}_u$ and due to compositionality and the homormorphic preservation of non-determinism, we have that
    \[
        \begin{aligned}
            Q &= \piencodf{M_1}_u \oplus \cdots \oplus \piencodf{M_k}_u\\
        \end{aligned}
    \]
    
    By Proposition \ref{ch2Prop:checkprespi} (1) we have that $$\headf{M_j} = \checkmark \implies \piencodf{M_j}_u \redd^*  (P \mid \checkmark) \oplus Q''$$ 
    for some $Q''$. 
    Hence, $Q \redd^*  (P \mid \checkmark) \oplus Q'' = Q'$, as wanted.
    
    

    \item Suppose that $\piencodf{\expr{M}}_u \Downarrow_{\checkmark}$. We will prove that $ \expr{M} \Downarrow_{\checkmark}$.

    By operational soundness (Theorem~\ref{ch2l:app_soundnesstwo}): if $ \piencodf{\expr{N}}_u \redd^* Q$
    then there exist $Q'$  and $\expr{N}' $ such that 
    $Q \redd^* Q'$, $\expr{N}  \redd^*_{\pequiv} \expr{N}'$ 
    and 
    $\piencodf{\expr{N}'}_u = Q'$.
   Since $\piencodf{\expr{M}}_u \redd^* P_1 \oplus \ldots \oplus P_k$, and $P_j= P_j'' \mid \checkmark$, for some $j$. 
   
   Notice that if $\piencodf{\expr{M}}_u$ is itself a term with unguarded $\checkmark$, say $\piencodf{\expr{M}}_u=P \mid \checkmark$, then $\expr{M}$ is itself headed with $\checkmark$, from Proposition \ref{ch2Prop:checkpres} (2).
   
   In the case $\piencodf{\expr{M}}_u= P_1 \oplus \ldots \oplus P_k$, $k\geq 2$, and $\checkmark$ occurs unguarded in an $P_j$, The encoding acts homomorphically over sums and the reasoning is similar. We have that $P_j = P_j' \mid \checkmark$ we apply Proposition \ref{ch2Prop:checkpres} (2).
\end{myEnumerate}
\end{proof}

\newpage

\chapter{Appendix of Chapter 3}\label{ch3appendix_types}

\section{Appendix to \texorpdfstring{\secref{ch3s:lambda}}{}}\label{ch3appA}

\subsection{Diamond Property for \texorpdfstring{$\lamrfailunres$}{}}

\begin{proposition}[Diamond Property for \lamrfailunres]
\label{ch3app:lambda}
     For all $\expr{N}$, $\expr{N}_1$, $\expr{N}_2$ in $\lamrfailunres$ s.t. $\expr{N} \redd \expr{N}_1$, $\expr{N} \redd \expr{N}_2$ { with } $\expr{N}_1 \neq \expr{N}_2$ { then } $\exists \expr{M}$ s.t. $\expr{N}_1 \redd \expr{M}$, $\expr{N}_2 \redd \expr{M}$.
\end{proposition}

\begin{proof}
    We give a short argument to convince the reader of this. Notice that an expression can only perform a choice of reduction steps when it is a nondeterministic sum of terms in which multiple terms can perform independent reductions. For simplicity sake we will only consider an expression $\expr{N}$ that consist of two terms where $ \expr{N} = N + M $. We also have that $N \redd N'$ and $M \redd M'$. Then we let $\expr{N}_1 = N' + M$ and $\expr{N}_2 = N + M'$ by the $\redlab{R:ECont}$ rules. Finally we prove that $\expr{M}$ exists by letting $\expr{M} = N' + M'$.
\end{proof}

\section{Appendix to \texorpdfstring{\secref{ch3sec:lam_types}}{}}\label{ch3appB}

Here we prove subject reduction (SR) for \lamrfailunres (Theorem~\ref{ch3t:lamrfailsr_unres}). It follows from two substitution lemmas: one for substituting a linear variable (Lemma~\ref{ch3lem:subt_lem_failunres_lin}) and another for an unrestricted variable (Lemma~\ref{ch3lem:subt_lem_failunres_un}). Proofs of both lemmas are standard, by structural induction; we give a complete proof of SR in  Theorem~\ref{ch3t:app_lamrfailsr}.

\begin{lemma}[Linear Substitution Lemma for \lamrfailunres]
\label{ch3lem:subt_lem_failunres_lin}
If $\Theta ; \Gamma ,  {x}:\sigma \wfdash M: \tau$, $\headf{M} =  {x}$, and $\Theta ; \Delta \wfdash N : \sigma$ 
then 
$\Theta ; \Gamma , \Delta \wfdash M \headlin{ N /  {x} }$.
\end{lemma}

\begin{proof}
By structural induction on $M$ with $\headf{M}= {x}$. There are three cases to be analyzed: 

\begin{enumerate}
\item $M= {x}$.

In this case, $\Theta ;  {x}:\sigma \wfdash  {x}:\sigma$ and $\Gamma=\emptyset$.  Observe that $ {x}\headlin{N/ {x}}=N$, since $\Theta ; \Delta\wfdash N:\sigma$, by hypothesis, the result follows.



    \item $M = M'\ B$.
    
    In this case, $\headf{M'\ B} = \headf{M'} =  {x}$, and one has the following derivation:
    
    \begin{prooftree}
         \AxiomC{\( \Theta ;\Gamma_1 ,   {x}:\sigma\wfdash M : (\delta^{j} , \eta ) \rightarrow \tau \quad \Theta ; \Gamma_2 \wfdash B : (\delta^{k} , \epsilon ) \)}
         \AxiomC{\( \eta \relunbag \epsilon \)}
            \LeftLabel{\redlab{F{:}app}}
        \BinaryInfC{\( \Theta ;\Gamma_1 , \Gamma_2 ,  {x}:\sigma \wfdash M\ B : \tau\)}
    \end{prooftree}

 where $\Gamma=\Gamma_1,\Gamma_2$, $\delta$ is a strict type and $j,k$ are non-negative  integers, possibly different.
 
 By IH, we get $\Theta ;\Gamma_1,\Delta\wfdash M'\headlin{N/ {x}}:(\delta^{j} , \eta ) \rightarrow \tau $, which gives the following derivation: 
    \begin{prooftree}
        \AxiomC{$\Theta ;\Gamma_1,\Delta\wfdash M'\headlin{N/ {x}}:(\delta^{j} , \eta ) \rightarrow \tau $}\
        \AxiomC{$\Theta ; \Gamma_2 \wfdash B : (\delta^{k} , \epsilon ) $}
        \AxiomC{\( \eta \relunbag \epsilon \)}
    	\LeftLabel{\redlab{F{:}app}}
        \TrinaryInfC{$\Theta ;\Gamma_1 , \Gamma_2 , \Delta \wfdash ( M'\headlin{ N /  {x} } ) B:\tau $}    
    \end{prooftree}
    Therefore, from \defref{ch3def:linsubfail}, one has $\Theta ;\Gamma_1 , \Gamma_2 , \Delta \wfdash ( M'\headlin{ N /  {x} } ) B:\tau  $, and the result follows.
\item $M = M'\esubst{B }{ y}$.

In this case,  $\headf{M'\esubst{B }{ y}} = \headf{M'} =  {x}$, with $ {x} \not = y$, and one has the following derivation:

    \begin{prooftree}
            \AxiomC{\( \Theta , \banged{y} : \eta ; \Gamma_1 , \hat{y}: \delta^{k}, x:\sigma \wfdash M : \tau \quad \Theta ; \Gamma_2 \wfdash B : (\delta^{j} , \epsilon ) \)}
            \AxiomC{\( \eta \relunbag \epsilon \)}
        \LeftLabel{\redlab{F{:}ex \dash sub}}  
        \BinaryInfC{\( \Theta ; \Gamma_1, \Gamma_2,  x:\sigma \wfdash M' \esubst{B }{ y} : \tau \)}
    \end{prooftree}

 where $\Gamma=\Gamma_1,\Gamma_2$, $\delta$ is a strict type and $j,k$ are positive integers.
By IH, we get $\Theta , \banged{y} : \eta ; \Gamma_1 , \hat{y}: \delta^{k} , \Delta \wfdash M'\headlin{N/ {x}}:\tau$ and

\begin{prooftree}
        \AxiomC{\( \Theta , \banged{y} : \eta ; \Gamma_1 , \hat{y}: \delta^{k} , \Delta \wfdash M'\headlin{N/ {x}}:\tau \quad \Theta ; \Gamma_2 \wfdash B : (\delta^{j} , \epsilon ) \)}
        \AxiomC{\( \eta \relunbag \epsilon \)}
    \LeftLabel{\redlab{F{:}ex \dash sub}}  
    \BinaryInfC{\( \Theta ; \Gamma_1, \Gamma_2,  \Delta \wfdash M' \headlin{ N / {x} } \esubst{ B }{ y} : \tau  \)}
\end{prooftree}

\end{enumerate}

From \defref{ch3def:linsubfail}, $M' \esubst{ B }{ y} \headlin{ N / {x} } = M' \headlin{ N /  {x}} \esubst{ B }{ y}$, therefore, $\Theta ; \Gamma, \Delta \wfdash (M'\esubst{ B }{ y})\headlin{ N / {x} }:\tau$ and  the result follows.
\end{proof}

\begin{lemma}[Unrestricted Substitution Lemma for \lamrfailunres]
\label{ch3lem:subt_lem_failunres_un}
If $\Theta, \banged{x}: \eta ; \Gamma \wfdash M: \tau$, $\headf{M} = {x}[i]$, $\eta_i = \sigma $, and $\Theta ; \cdot \wfdash N : \sigma$
then 
$\Theta, \banged{x}: \eta ; \Gamma  \wfdash M \headlin{ N /  {x}[i] }$.
\end{lemma}

\begin{proof}
By structural induction on $M$ with $\headf{M}= {x}[i]$. There are three cases to be analyzed: 

\begin{enumerate}
\item $M= {x}[i]$.

In this case, 

    \begin{prooftree}
        \AxiomC{}
        \LeftLabel{ \redlab{F{:}var^{ \ell}}}
        \UnaryInfC{\( \Theta , \banged{x}: \eta;  {x}: \eta_i  \wfdash  {x} : \sigma\)}
        \LeftLabel{\redlab{F{:}var^!}}
        \UnaryInfC{\( \Theta, \banged{x}: \eta ; \cdot \wfdash  {x}[i] : \sigma\)}
    \end{prooftree}

 and $\Gamma=\emptyset$.  Observe that $ {x}[i]\headlin{N/ {x}[i]}=N$, since $\Theta, \banged{x}: \eta  ; \Gamma  \wfdash M \headlin{ N /  {x}[i] }$, by hypothesis, the result follows.

    \item $M = M'\ B$.
    
    In this case, $\headf{M'\ B} = \headf{M'} =  {x}[i]$, and one has the following derivation:
    
    \begin{prooftree}
         \AxiomC{\( \Theta, \banged{x}: \eta ;\Gamma_1 \wfdash M : (\delta^{j} , \epsilon ) \rightarrow \tau \quad \Theta, \banged{x}: \sigma ; \Gamma_2 \wfdash B : (\delta^{k} , \epsilon' )  \)}
         \AxiomC{\( \epsilon \relunbag \epsilon' \)}
            \LeftLabel{\redlab{F{:}app}}
        \BinaryInfC{\( \Theta, \banged{x}: \eta ;\Gamma_1 , \Gamma_2 \wfdash M\ B : \tau\)}
    \end{prooftree}
    
 where $\Gamma=\Gamma_1,\Gamma_2$, $\delta$ is a strict type and $j,k$ are non-negative  integers, possibly different.
 
 By IH, we get $\Theta, \banged{x}: \eta ;\Gamma_1 \wfdash M'\headlin{N/ {x}[i]}:(\delta^{j} , \epsilon ) \rightarrow \tau $, which gives the following derivation: 
    {\small
    \begin{prooftree}
        \AxiomC{$\Theta , \banged{x}: \eta;\Gamma_1\wfdash M'\headlin{N/ {x}[i]}:(\delta^{j} , \epsilon ) \rightarrow \tau $}\
        \AxiomC{$\Theta , \banged{x}: \eta; \Gamma_2 \wfdash B : (\delta^{k} , \epsilon' ) $}
        \AxiomC{\( \epsilon \relunbag \epsilon' \)}
    	\LeftLabel{\redlab{F{:}app}}
        \TrinaryInfC{$\Theta , \banged{x}: \eta;\Gamma_1 , \Gamma_2  \wfdash ( M'\headlin{ N /  {x}[i] } ) B:\tau $}    
    \end{prooftree}}
   From \defref{ch3def:linsubfail}, one has $\Theta, \banged{x}: \eta ;\Gamma_1 , \Gamma_2  \wfdash ( M'\headlin{ N /  {x}[i] } ) B:\tau  $, and the result follows.
\item $M = M'\esubst{B }{ y}$.

In this case,  $\headf{M'\esubst{B }{ y}} = \headf{M'} =  {x}[i]$, with $x \not = y$, and one has the following derivation:

    \begin{prooftree}
            \AxiomC{\( \Theta , \banged{y} : \epsilon, x:\eta ; \Gamma_1 , \hat{y}: \delta^{k} \wfdash M : \tau \quad \Theta ,  x:\eta ; \Gamma_2 \wfdash B : (\delta^{j} , \epsilon' ) \)}
            \AxiomC{\( \epsilon \relunbag \epsilon' \)}
        \LeftLabel{\redlab{F{:}ex \dash sub}}  
        \BinaryInfC{\( \Theta,  x:\eta ; \Gamma_1, \Gamma_2 \wfdash M' \esubst{B }{ y} : \tau \)}
    \end{prooftree}
 where $\Gamma=\Gamma_1,\Gamma_2$, $\delta$ is a strict type and $j,k$ are positive integers.
By IH, we get $\Theta , \banged{y} : \epsilon ,  x:\eta ; \Gamma_1 , \hat{y}: \delta^{k} \wfdash M'\headlin{N/ {x}[i]}:\tau$ and 
\begin{prooftree}
    \small
        \AxiomC{\( \Theta , \banged{y} : \epsilon ,  x:\eta ; \Gamma_1 , \hat{y}: \delta^{k} , \wfdash M'\headlin{N/ {x}[i]}:\tau \quad \Theta ,  x:\sigma ; \Gamma_2 \wfdash B : (\delta^{j} , \epsilon' ) \)}
        \AxiomC{\( \epsilon \relunbag \epsilon' \)}
    \LeftLabel{\redlab{F{:}ex \dash sub}}  
    \BinaryInfC{\( \Theta ,  x:\eta ; \Gamma_1, \Gamma_2\wfdash M' \headlin{ N / {x}[i] } \esubst{ B }{ y} : \tau  \)}
\end{prooftree}
Then, $M' \esubst{ B }{ y} \headlin{ N / {x}[i] } = M' \headlin{ N /  {x}[i]} \esubst{ B }{ y}$, and  the result follows.
\end{enumerate}
\end{proof}

\begin{theorem}[SR in \lamrfailunres]
\label{ch3t:app_lamrfailsr}
If $\Theta ; \Gamma \wfdash \expr{M}:\tau$ and $\expr{M} \redd \expr{M}'$ then $\Theta ;\Gamma \wfdash \expr{M}' :\tau$.
\end{theorem}

\begin{proof} By structural induction on the reduction rules. We proceed by analysing the rule applied in $\expr{M}$. There are seven cases:

\begin{enumerate}

	\item {\bf Rule $\redlab{R:Beta}$.}
	
	Then $\expr{M} = (\lambda x . M)B \redd M\ \esubst{B}{x}=\expr{M}'$.
 	Since $\Theta ; \Gamma\wfdash \expr{M}:\tau$, one has the  derivation:
	\begin{prooftree}
			\AxiomC{$\Theta , \banged{x} : \eta ; \Gamma' , \hat{ {x}}: \sigma^{j} \wfdash M : \tau $}
			\LeftLabel{\redlab{F{:}abs}}
            \UnaryInfC{$\Theta ; \Gamma' \wfdash \lambda x. M: (\sigma^{j} , \eta ) \rightarrow \tau $}
            \AxiomC{$\Theta ;\Delta \wfdash B: (\sigma^{k} , \epsilon ) $}
            \AxiomC{\( \eta \relunbag \epsilon \)}
			\LeftLabel{\redlab{F{:}app}}
		\TrinaryInfC{$\Theta ; \Gamma' , \Delta \wfdash (\lambda x. M) B:\tau $}
	\end{prooftree}
	for $\Gamma = \Gamma' , \Delta $. Notice that

    \begin{prooftree}
            \AxiomC{\( \Theta , \banged{x} : \eta ; \Gamma' , \hat{ {x}}: \sigma^{j} \wfdash M : \tau \quad \Theta ;\Delta \wfdash B: (\sigma^{k} , \epsilon ) \)}
            \AxiomC{\( \eta \relunbag \epsilon \)}
        \LeftLabel{\redlab{F{:}ex \dash sub}}  
        \BinaryInfC{\( \Theta ; \Gamma, \Delta \wfdash M \esubst{ B }{ x } : \tau \)}
    \end{prooftree}
    
    Therefore, $ \Theta ; \Gamma \wfdash \expr{M}':\tau$ and the result follows.

    \item {\bf  Rule $\redlab{R:Fetch^{\ell}}$.}
    
    Then $ \expr{M} = M\ \esubst{ C \bagsep U  }{x }$, where $C = {\bag{N_1}}\cdot \dots \cdot {\bag{N_k}}$ , $k\geq 1$, $ \#( {x},M) = k $ and $\headf{M} =  {x}$. The reduction applying the $\redlab{R:Fetch^{\ell}}$ rule is as: 
    {\small{
    \begin{prooftree}
    \AxiomC{$\headf{M} =  {x}$}
    \AxiomC{$C = {\bag{N_1}}\cdot \dots \cdot {\bag{N_k}} \ , \ k\geq 1 $}
    \AxiomC{$ \#( {x},M) = k $}
    \TrinaryInfC{\(
    M\ \esubst{ C \bagsep U  }{x } \redd M \headlin{ N_{1}/ {x} } \esubst{ (C \setminus N_1)\bagsep U}{ x }  + \cdots + M \headlin{ N_{k}/x } \esubst{ (C \setminus N_k)\bagsep U}{x}
    \)}
\end{prooftree}}}

    To simplify the proof we take $k=2$, as the case $k>2$ is similar. Therefore, $c = {\bag{N_1}}\cdot {\bag{N_2}}$ and applying rule $\redlab{F{:}ex \dash sub}$ we obtain:

    \begin{prooftree}
            \AxiomC{\( \Theta , \banged{x} : \eta ; \Gamma' , \hat{x}: \sigma^{2} \wfdash M : \tau \)}
               \AxiomC{\(  \Theta ; \cdot\wfdash  U : \epsilon\)}
                \AxiomC{\(\Pi\)}
                \noLine
                \UnaryInfC{\( \Theta ; \Delta\wfdash {\bag{N_1}}\cdot {\bag{N_2}} : \sigma^2\)}
            \LeftLabel{\redlab{F{:}bag}}
            \BinaryInfC{\( \Theta ; \Delta \wfdash C \bagsep U : (\sigma^{2} , \epsilon ) \)}
            \AxiomC{\(  \eta \relunbag \epsilon  \)}
        \TrinaryInfC{\( \Theta ; \Gamma', \Delta \wfdash M \esubst{ C \bagsep U }{ x } : \tau \)}
    \end{prooftree}
      with  $\Pi$ the derivation
    \begin{prooftree}
            \AxiomC{\( \Theta ; \Delta_1 \wfdash N_1 : \sigma\)}
                \AxiomC{\( \Theta ; \Delta_2 \wfdash N_2 : \sigma\)}
                \AxiomC{\(  \)}
                \LeftLabel{\redlab{F{:}\oneb^{\ell}}}
                \UnaryInfC{\( \Theta ; \dash \wfdash {\oneb} : \omega \)}
            \LeftLabel{\redlab{F{:}bag^{ \ell}}}
            \BinaryInfC{\( \Theta ; \Delta_2 \wfdash {\bag{N_2}} : \sigma\)}
        \LeftLabel{\redlab{F{:}bag^{ \ell}}}
        \BinaryInfC{\( \Theta ; \Delta \wfdash {\bag{N_1}}\cdot {\bag{N_2}} : \sigma^2\)}
    \end{prooftree}
     where $\Delta= \Delta_1,\Delta_2$ and $\Gamma = \Gamma' , \Delta $. By Lemma~\ref{ch3lem:subt_lem_failunres_lin}, there exist derivations $\Pi_1$ of  $
    \Theta , \banged{x} : \eta ; \Gamma' , x: \sigma,  \Delta_1 \wfdash   M \headlin{ N_{1}/  {x} } : \tau $ and  $\Pi_2$ of $\Theta , \banged{x} : \eta ; \Gamma' , x: \sigma,  \Delta_2 \wfdash   M \headlin{ N_{2}/  {x} } : \tau $. Therefore, one has the following derivation where we omit the second case of the sum:

    \begin{prooftree}
    				\AxiomC{\( \Pi_1 \) }
                            \AxiomC{\(  \Theta ; \cdot\wfdash  U : \epsilon\)}
                        \AxiomC{\( \Theta ; \Delta\wfdash  {\bag{N_2}} : \sigma\)}
                    \LeftLabel{\redlab{F{:}bag}}
                    \BinaryInfC{\( \Theta ; \Delta \wfdash \bag{N_2} \bagsep U : (\sigma , \epsilon ) \)}
                \LeftLabel{\redlab{F{:}ex \dash sub}}
    			\BinaryInfC{\( \Theta ; \Gamma' ,  \Delta_1   \wfdash M \headlin{ N_{1}/x } \esubst{ {\bag{N_2}} \bagsep U}{ x }  : \tau\ \) }
    			\AxiomC{\( \vdots \) }
    	\LeftLabel{\redlab{F{:}sum}}
        \BinaryInfC{\(\Theta ; \Gamma' , \Delta  \wfdash     M \headlin{ N_{1}/x } \esubst{ {\bag{N_2}} \bagsep U }{ x } +  M \headlin{ N_{2}/x } \esubst{ {\bag{N_1}} \bagsep U }{x } : \tau\)}
    \end{prooftree}
    
    Assuming  $ \expr{M}'  =    M \headlin{ N_{1}/x } \esubst{ {\bag{N_2}} \bagsep \banged{B} }{ x } +  M \headlin{ N_{2}/x } \esubst{ {\bag{N_1}} \bagsep \banged{B} }{x }$, the result follows.

    \item {\bf Rule $\redlab{R:Fetch^!}$.}
    
    Then $ \expr{M} = M\ \esubst{ C \bagsep U }{x }$, where $U = \banged{\bag{N_1}} \concat \cdots \concat \banged{\bag{N_l}} $ and $\headf{M} =  {x}[i]$. The reduction is as:

    \begin{prooftree}
        \AxiomC{$\headf{M} =  {x}[i]$}
        \AxiomC{$ U_i = \banged{\bag{N_i}} $}
        \LeftLabel{\redlab{R:Fetch^!}}
        \BinaryInfC{\(
        M\ \esubst{ C \bagsep U  }{x } \redd M \headlin{ N_i/ {x}[i] } \esubst{ C \bagsep U}{ x } 
        \)}
    \end{prooftree}
    
     By hypothesis, one has the derivation:
     \begin{prooftree}
            \AxiomC{\( \Theta , \banged{x} : \eta ; \Gamma' , \hat{x}: \sigma^{j} \wfdash M : \tau \)}
               \AxiomC{$\Pi$}
               \noLine
               \UnaryInfC{\(  \Theta ; \cdot\wfdash U : \epsilon\)}
                \AxiomC{\( \Theta ; \Delta \wfdash  {C} : \sigma^k\)}
            \LeftLabel{\redlab{F{:}bag}}
            \BinaryInfC{\( \Theta ; \Delta \wfdash C \bagsep U : (\sigma^{k} , \epsilon ) \)}
            \AxiomC{\(  \eta \relunbag \epsilon  \)}
        \LeftLabel{\redlab{F{:}ex \dash sub}}  
        \TrinaryInfC{\( \Theta ; \Gamma', \Delta \wfdash M \esubst{ C \bagsep U }{ x } : \tau \)}
    \end{prooftree}
    Where $\Pi$ has the form
    \begin{prooftree}
            \AxiomC{\( \Theta ; \cdot \wfdash N_1 : \epsilon_1\)}
            \LeftLabel{\redlab{F{:}bag^!}}
            \UnaryInfC{\( \Theta ; \cdot \wfdash \banged{\bag{N_1}}  : \epsilon_1\)}
            \AxiomC{\( \cdots \)}
            \AxiomC{\( \Theta ; \cdot \wfdash N_l : \epsilon_l \)}
            \LeftLabel{\redlab{F{:}bag^!}}
            \UnaryInfC{\( \Theta ; \cdot \wfdash \banged{\bag{N_l}} : \epsilon_l \)}
        \LeftLabel{\redlab{F{:}\concat bag^!}}
        \TrinaryInfC{\( \Theta ; \cdot  \wfdash \banged{\bag{N_1}} \concat \cdots \concat \banged{\bag{N_l}}  :\epsilon \)}
    \end{prooftree}
    where $\Gamma = \Gamma' , \Delta $. Notice that if $\epsilon_i = \delta$ and $\eta \relunbag \epsilon$ then $\eta_i = \delta$ By Lemma~\ref{ch3lem:subt_lem_failunres_un}, there exists a derivation $\Pi_1$ of  $
    \Theta , \banged{x} : \eta ; \Gamma' , \hat{x}: \sigma^{j} \wfdash   M \headlin{ N_{i}/  {x}[i] } : \tau $. Therefore, one has the following derivation applying rule $\redlab{F{:}ex \dash sub}$:

         \begin{prooftree}
            \AxiomC{\( \Theta , \banged{x} : \eta ; \Gamma' , \hat{x}: \sigma^{j} \wfdash M \headlin{ N_{1}/  {x}[i] } : \tau \)}
               \AxiomC{\(  \Theta ; \cdot\wfdash U : \epsilon\)}
                \AxiomC{\( \Theta ; \Delta \wfdash  {C} : \sigma^k\)}
            \LeftLabel{\redlab{F{:}bag}}
            \BinaryInfC{\( \Theta ; \Delta \wfdash C \bagsep U : (\sigma^{k} , \epsilon ) \)}
            \AxiomC{\(  \eta \relunbag \epsilon  \)}
        \TrinaryInfC{\( \Theta ; \Gamma', \Delta \wfdash M \headlin{ N_{i}/ {x[i]} } \esubst{  C \bagsep U }{ x } : \tau \)}
    \end{prooftree}

	\item  {\bf Rule $\redlab{R:Fail^{ \ell }}$.}
	
	Then $\expr{M} =  M\ \esubst{C \bagsep U}{x } $ where $\#( {x},M) \neq \size{C}$ and we can perform the reduction:
	\begin{prooftree}
        \AxiomC{$\#( {x},M) \neq \size{C}$} 
        \AxiomC{\( \widetilde{y} = (\mlfv{M} \setminus x) \uplus \mlfv{C} \)}
        \LeftLabel{\redlab{R:Fail^{ \ell }}}
        \BinaryInfC{\(  M\ \esubst{C \bagsep U}{x } \redd {}  \sum_{\perm{C}} \fail^{\widetilde{y}} \)}
    \end{prooftree}
	
	with $\expr{M}'=\sum_{\perm{B}} \fail^{\widetilde{y}}$. By hypothesis, one has the derivation:
    
    \begin{prooftree}
            \AxiomC{\( \Theta , \banged{x} : \eta ; \Gamma' , \hat{x}: \sigma^{2} \wfdash M : \tau \)}
            \AxiomC{\( \Theta ; \Delta \wfdash C \bagsep U : (\sigma^{2} , \epsilon ) \)}
            \AxiomC{\(  \eta \relunbag \epsilon  \)}
        \LeftLabel{\redlab{F{:}ex \dash sub}}  
        \TrinaryInfC{\( \Theta ; \Gamma', \Delta \wfdash M \esubst{ C \bagsep U }{ x } : \tau \)}
    \end{prooftree}
    From $\#(x,M) \neq \size{B}$  we have that $j\neq k$. Hence $\Gamma = \Gamma' , \Delta $ and we type the following:
    
    \begin{prooftree}
        \AxiomC{\( \dom{\core{\Gamma}} = \widetilde{y} \)}
        \LeftLabel{\redlab{F{:}fail}}
        \UnaryInfC{$\Theta ; \Gamma \wfdash \fail^{\widetilde{y}} : \tau$}
        \AxiomC{\( \cdots \)}
        \AxiomC{\( \dom{\core{\Gamma}} = \widetilde{y} \)}
        \LeftLabel{\redlab{F{:}fail}}
        \UnaryInfC{$\Theta ;\Gamma \wfdash \fail^{\widetilde{y}} : \tau$}
        \LeftLabel{\redlab{F{:}sum}}
        \TrinaryInfC{$\Theta ; \Gamma \wfdash \sum_{\perm{B}} \fail^{\widetilde{y}}: \tau$}
    \end{prooftree}

\item {\bf Rule $ \redlab{R:Fail^!}$.}

    Then $\expr{M} =  M\ \esubst{C \bagsep U}{x } $ where $\headf{M} =  {x}[i]$, $U_i = \banged{\oneb}$ and we can perform the reduction:
    \begin{prooftree}
        \AxiomC{$\headf{M} =  {x}[i]$}
        \AxiomC{$ U_i = \banged{\oneb} $}
        \AxiomC{\( \)}
        \LeftLabel{\redlab{R:Fail^!}}
        \TrinaryInfC{\(  M\ \esubst{C \bagsep U}{x } \redd  M \headlin{ \fail^{\emptyset} / {x}[i] } \esubst{ C \bagsep U}{ x }\)}
    \end{prooftree}
        with $\expr{M}'=M \headlin{ \fail^{\emptyset} / {x}[i] } \esubst{ C \bagsep U}{ x }$. By hypothesis, one has the derivation:
    
    \begin{prooftree}
            \AxiomC{\( \Theta , \banged{x} : \eta ; \Gamma' , \hat{x}: \sigma^{j} \wfdash M : \tau \)}
            \AxiomC{\( \Theta ; \Delta \wfdash C \bagsep U : (\sigma^{k} , \epsilon ) \)}
            \AxiomC{\(  \eta \relunbag \epsilon  \)}
        \LeftLabel{\redlab{F{:}ex \dash sub}}  
        \TrinaryInfC{\( \Theta ; \Gamma', \Delta \wfdash M \esubst{ C \bagsep U }{ x } : \tau \)}
    \end{prooftree}
    where $\Gamma = \Gamma' , \Delta $. By Lemma~\ref{ch3lem:subt_lem_failunres_un}, there is a derivation $\Pi_1$ of  $
    \Theta , \banged{x} : \eta ; \Gamma' , \hat{x}: \sigma^{j} \wfdash   M \headlin{ \fail^{\emptyset}/ {x[i]} } : \tau $. Therefore, one has the derivation: (the last rule applied is $\redlab{R:ex\dash sub}$)
    \begin{prooftree}
        \small
            \AxiomC{\( \Theta , \banged{x} : \eta ; \Gamma' , \hat{x}: \sigma^{j} \wfdash M \headlin{ \fail^{\emptyset}/ {x}[i] } : \tau \)}
               \AxiomC{\(  \Theta ; \cdot\wfdash U : \epsilon\)}
                \AxiomC{\( \Theta ; \Delta \wfdash  {B} : \sigma^k\)}
            \LeftLabel{\redlab{F{:}bag}}
            \BinaryInfC{\( \Theta ; \Delta \wfdash C \bagsep U : (\sigma^{k} , \epsilon ) \)}
            \AxiomC{\(  \eta \relunbag \epsilon  \)}
        \TrinaryInfC{\( \Theta ; \Gamma', \Delta \wfdash M \headlin{ \fail^{\emptyset} / {x}[i] } \esubst{  C \bagsep U }{ x } : \tau \)}
    \end{prooftree}

\item {\bf Rule $\redlab{R:Cons_1}$.}

Then $\expr{M} =   \fail^{\widetilde{x}} \ B $ where $B = C \bagsep U $ , $ C = \bag{N_1}\cdot \dots \cdot \bag{N_k} $ , $k \geq 0$ and we can perform the following reduction:
\begin{prooftree}
    \AxiomC{$\size{C} = k$} 
    \AxiomC{\( \widetilde{y} = \mlfv{C} \)}
    \LeftLabel{$\redlab{R:Cons_1}$}
    \BinaryInfC{\(\fail^{\widetilde{x}}\ C \bagsep U \redd \sum_{\perm{C}} \fail^{\widetilde{x} \uplus \widetilde{y}} \)}
\end{prooftree}
where $\expr{M}'=\sum_{\perm{C}} \fail^{\widetilde{x} \uplus \widetilde{y}}$. By hypothesis, one has
    \begin{prooftree}
         \AxiomC{\( \dom{\core{\Gamma'}} = \widetilde{x} \)}
        \LeftLabel{\redlab{F{:}fail}}
        \UnaryInfC{\( \Theta ;\Gamma' \wfdash \fail^{\widetilde{x}}: (\sigma^{j} , \eta ) \rightarrow \tau \)}
         \AxiomC{\(  \Theta ;\Delta \wfdash B : (\sigma^{k} , \epsilon )  \)}
         \AxiomC{\( \eta \relunbag \epsilon \)}
            \LeftLabel{\redlab{F{:}app}}
        \TrinaryInfC{\( \Theta ; \Gamma', \Delta \wfdash \fail^{\widetilde{x}}\ B : \tau\)}
    \end{prooftree}

Hence $\Gamma = \Gamma' , \Delta $ and we may type the following:

    \begin{prooftree}
        \AxiomC{\( \dom{\core{\Gamma}} = \widetilde{x} \uplus \widetilde{y} \)}
        \LeftLabel{\redlab{F{:}fail}}
        \UnaryInfC{$ \Theta ;\Gamma \wfdash \fail^{\widetilde{x} \uplus \widetilde{y}} : \tau$}
        \AxiomC{\( \cdots \)}
        \AxiomC{\( \dom{\core{\Gamma}} = \widetilde{x} \uplus \widetilde{y} \)}
        \LeftLabel{\redlab{F{:}fail}}
        \UnaryInfC{$\Theta ;\Gamma \wfdash \fail^{\widetilde{x} \uplus \widetilde{y}} : \tau$}
        \LeftLabel{\redlab{F{:}sum}}
        \TrinaryInfC{$\Theta ; \Gamma \wfdash \sum_{\perm{C}} \fail^{\widetilde{x} \uplus \widetilde{y}}: \tau$}
    \end{prooftree}

\item {\bf Rule $\redlab{R:Cons_2}$.}

Then $\expr{M} =   \fail^{\widetilde{z}}\ \esubst{B}{x} $ where $B = \bag{N_1}\cdot \dots \cdot \bag{N_k} $ , $k \geq 1$ and 
one has the  reduction:

\begin{prooftree}
    \AxiomC{$ \#(z , \widetilde{x}) =  \size{C}\quad \widetilde{y} = \mlfv{C} $} 
    \AxiomC{\( \widetilde{y} = \mlfv{C} \)}
    \LeftLabel{$\redlab{R:Cons_2}$}
    \BinaryInfC{\( \fail^{\widetilde{x}}\ \esubst{C \bagsep U}{z}  \redd \sum_{\perm{C}} \fail^{(\widetilde{x} \setminus z) \uplus\widetilde{y}}  \)}
\end{prooftree}

where $\expr{M}'=\sum_{\perm{B}} \fail^{(\widetilde{z} \setminus x) \uplus \widetilde{y}}$. By hypothesis, there exists a derivation:

     \begin{prooftree}
            \AxiomC{\( \dom{\core{(\Gamma' , \hat{x}:\sigma^{j})}}=\widetilde{z} \)}
            \LeftLabel{\redlab{F{:}fail}}
            \UnaryInfC{\( \Theta , \banged{x} : \eta ; \Gamma' , \hat{x}: \sigma^{j} \wfdash M : \tau  \)}
            \AxiomC{\( \Theta ; \Delta \wfdash B : (\sigma^{k} , \epsilon ) \)}
            \AxiomC{\( \eta \relunbag \epsilon \)}
        \LeftLabel{\redlab{F{:}ex \dash sub}}  
        \TrinaryInfC{\( \Theta ; \Gamma', \Delta \wfdash \fail^{\widetilde{z}} \esubst{ B }{ x } : \tau \)}
    \end{prooftree}

Hence $\Gamma = \Gamma' , \Delta $ and we may type the following:

    \begin{prooftree}
        \AxiomC{\( \dom{\core{\Gamma}} = (\widetilde{z} \setminus x) \uplus\widetilde{y}\)}
        \LeftLabel{\redlab{F{:}fail}}
        \UnaryInfC{$\Theta ; \Gamma \wfdash \fail^{(\widetilde{z} \setminus x) \uplus\widetilde{y}} : \tau$}
        \AxiomC{$ \cdots $}
         \AxiomC{\( \dom{\core{\Gamma}} = (\widetilde{z} \setminus x) \uplus\widetilde{y} \)}
        \LeftLabel{\redlab{F{:}fail}}
        \UnaryInfC{$\Theta ; \Gamma \wfdash \fail^{(\widetilde{z} \setminus x) \uplus\widetilde{y}} : \tau$}
        \LeftLabel{\redlab{F{:}sum}}
        \TrinaryInfC{$\Theta ; \Gamma \wfdash \sum_{\perm{B}} \fail^{(\widetilde{z} \setminus x) \uplus\widetilde{y}} : \tau$}
    \end{prooftree}
    
    \item {\bf Rule $\redlab{R:TCont}$.}

Then $\expr{M} = C[M]$ and the reduction is as follows:

\begin{prooftree}
        \AxiomC{$   M \redd  M'_{1} + \cdots +  M'_{l} $}
        \LeftLabel{\redlab{R:TCont}}
        \UnaryInfC{$ C[M] \redd  C[M'_{1}] + \cdots +  C[M'_{l}] $}
\end{prooftree}

where $\expr{M}'= C[M'_{1}] + \cdots +  C[M'_{l}]$. 
The proof proceeds by analysing the context $C$:

    \begin{enumerate}
    \item $C=[\cdot]\ B$.
    
    In this case $\expr{M}=M \ B$, for some $B$, and the following derivation holds:

    \begin{prooftree}
         \AxiomC{\( \Theta ;\Gamma' \wfdash M: (\sigma^{j} , \eta ) \rightarrow \tau \)}
         \AxiomC{\(  \Theta ;\Delta \wfdash B : (\sigma^{k} , \epsilon )  \)}
         \AxiomC{\( \eta \relunbag \epsilon \)}
            \LeftLabel{\redlab{F{:}app}}
        \TrinaryInfC{\( \Theta ; \Gamma', \Delta \wfdash M\ B : \tau\)}
    \end{prooftree}

where $\Gamma = \Gamma' , \Delta $.
Since $ \Theta ;\Gamma' \wfdash M: (\sigma^{j} , \eta ) \rightarrow \tau $ and $M\redd M_1'+\ldots + M_l'$, it follows by IH that $\Gamma'\wfdash M_1'+\ldots + M_l':(\sigma^{j} , \eta ) \rightarrow \tau $. By applying \redlab{F{:}sum}, one has $\Theta ;\Gamma' \wfdash M_i': (\sigma^{j} , \eta ) \rightarrow \tau $, for $i=1,\ldots, l$.  Therefore, we may type the following:
\begin{prooftree}
    \small
\AxiomC{\(  \forall i \in {1 , \ldots , l} \)}
         \AxiomC{\( \Theta ;\Gamma' \wfdash M_i': (\sigma^{j} , \eta ) \rightarrow \tau \)}
         \AxiomC{\(  \Theta ;\Delta \wfdash B : (\sigma^{k} , \epsilon )  \)}
         \AxiomC{\( \eta \relunbag \epsilon \)}
            \LeftLabel{\redlab{F{:}app}}
        \TrinaryInfC{\( \Theta ; \Gamma', \Delta \wfdash M_i'\ B : \tau\)}
 \LeftLabel{\redlab{F{:}sum}}
    \BinaryInfC{\( \Theta ; \Gamma', \Delta \wfdash (M'_{1}\ B) + \cdots +  (M'_{l} \ B) : \tau\)}
\end{prooftree}

Thus, $\Gamma\wfdash \expr{M'}: \tau$, and the result follows.

    \item $C=([\cdot])\esubst{B}{x}$.
    
    This case is similar to the previous.
\end{enumerate}
	\item {\bf Rule $ \redlab{R:ECont}.$}
	
Then $\expr{M} = D[\expr{M}'']$ where $\expr{M}'' \rightarrow \expr{M}'''$ then we can perform the following reduction:

\begin{prooftree}
        \AxiomC{$ \expr{M}''  \redd \expr{M}'''  $}
        \LeftLabel{$\redlab{R:ECont}$}
        \UnaryInfC{$D[\expr{M}'']  \redd D[\expr{M}''']  $}
\end{prooftree}

Hence $\expr{M}' =  D[\expr{M}'''] $.
The proof proceeds by analysing the context $D$ ($D= [\cdot] + \expr{N}$ or $D= \expr{N} + [\cdot]$), and follows easily by induction hypothesis.



\end{enumerate}
\end{proof}

\subsection{Examples}\label{ch3wf:examples}

This section contains examples illustrating the constructions and results given in Section~\ref{ch3sec:lam_types}.

\begin{example}{}\label{ch3ex:bag_delta4}
The following is a wf-derivation $\Pi_2$ for the bag concatenation $\bag{x}\bagsep \oneb^!:$
{\small 
        \begin{prooftree} 
                \AxiomC{}
                        \LeftLabel{\redlab{F{:}var^{ \ell}}}
                        \UnaryInfC{\(  \Theta' ; x : \sigma \wfdash x : \sigma\)}
                        
                        \AxiomC{\(  \)}
                        \LeftLabel{\redlab{F{:}\oneb^{\ell}}}
                        \UnaryInfC{\( \Theta' ; \dash \wfdash \oneb : \omega \)}
                    
                    \LeftLabel{\redlab{F{:}bag^{ \ell}}}
                    \BinaryInfC{\(  \Theta' ; x : \sigma \wfdash \bag{x}\cdot \oneb :\sigma\)}
                            
                    \AxiomC{\(  \)}
                    \LeftLabel{\redlab{F{:}\oneb^!}}
                    \UnaryInfC{\(   \Theta' ;\dash \wfdash  \banged{\oneb} : \sigma' \)}
                
                \LeftLabel{\redlab{F{:}bag}}
                \BinaryInfC{\( \Theta' ;  x : \sigma \wfdash (\bag{x} \bagsep \banged{\oneb} ) : (\sigma , \sigma' ) \)}
                
        \end{prooftree}
        }
\end{example}

\begin{example}{Cont.\ref{ch3ex:bag_delta4}} The following is a well-formedness derivation (labels of the rules being applied are omitted) for term $\Delta_4=\lambda x. x[1] (\bag{x} \bagsep \banged{\oneb} )$:
     {\small    
    \begin{prooftree}
    \AxiomC{}
    \UnaryInfC{\( \Theta , \banged{x}: (\sigma^{j} , \eta ) \rightarrow \tau ; {x}: (\sigma^{j} , \eta ) \rightarrow \tau  \wfdash  {x} : (\sigma^{j} , \eta ) \rightarrow \tau \)}
    \UnaryInfC{\( \Theta , \banged{x}: (\sigma^{j} , \eta ) \rightarrow \tau ; \dash \wfdash {x}[1] : (\sigma^{j} , \eta ) \rightarrow \tau \)}
\AxiomC{\( \Pi_2 \)}
     \AxiomC{\( \eta \relunbag \sigma' \)}
      \TrinaryInfC{\( \Theta , \banged{x} : (\sigma^{j} , \eta ) \rightarrow \tau ;  x : \sigma \wfdash x[1] (\bag{x} \bagsep \banged{\oneb} ) : \tau \)}
    \UnaryInfC{\( \Theta ; \dash \wfdash \lambda x. (x[1] (\bag{x} \bagsep \banged{\oneb} )) : ( \sigma , (\sigma^{j} , \eta ) \rightarrow \tau )   \rightarrow \tau \)}
        \end{prooftree}
        }
\end{example}

\begin{example}{}\label{ch3ex:bag_a}Below  we show the wf-derivation for the bag
$A=(\bag{x[1]} \cdot  \bag{x}  )  \bagsep \banged{\bag{x[2]}}$.

First, let $\Pi$ be the following derivation:

\begin{prooftree}
   \small 
 \AxiomC{}
 \LeftLabel{\redlab{F{:}var^{ \ell}}}
  \UnaryInfC{\( \Theta , \banged{x}: \eta;   {x}: \sigma_3  \wfdash  {x} : \sigma_3\)}
 \LeftLabel{\redlab{F{:}var^!}}
\UnaryInfC{\( \Theta , \banged{x}: \eta; \dash \wfdash {x}[1] : \sigma_3\)}
\AxiomC{}
\LeftLabel{\redlab{F{:}var^{ \ell}}}
 \UnaryInfC{\( \Theta , \banged{x}:\eta ;  x: \sigma_3  \wfdash x : \sigma_3\)}
 \AxiomC{\(  \)}
 \LeftLabel{\redlab{F{:}\oneb^{ \ell}}}
\UnaryInfC{\( \Theta , \banged{x}:\eta ; \dash \wfdash \oneb : \omega  \)}
 \LeftLabel{\redlab{F{:}bag^{\ell}}}
\BinaryInfC{\( \Theta , \banged{x}:\eta ;  x: \sigma_3  \wfdash \bag{x} \cdot \oneb : \sigma_3 \)}
\LeftLabel{\redlab{F{:}bag^{\ell}}}
\BinaryInfC{\( \Theta , \banged{x}:\eta ;  x: \sigma_3 \wfdash \bag{x[1]} \cdot  \bag{x}  : \sigma_3^2  \)}
\end{prooftree}


    
From $\Pi$ we can obtain the well-formedness derivation $\Pi_A$ for $A$:

   \begin{prooftree}
  \AxiomC{\(\Pi\)}
  \noLine
  \UnaryInfC{\( \Theta , \banged{x}:\eta ;  x: \sigma_3 \wfdash \bag{x[1]} \cdot  \bag{x}   : \sigma_3^2\)}
 \AxiomC{}
 \LeftLabel{\redlab{F{:}var^{\ell}}}
 \UnaryInfC{\( \Theta , \banged{x}:\eta ; x:\sigma_2  \wfdash  x : \sigma_2 \)}
  \LeftLabel{\redlab{F{:}var^!}}
            \UnaryInfC{\( \Theta , \banged{x}:\eta;\dash \wfdash  x[2] : \sigma_2 \)}
            \LeftLabel{\redlab{F{:}bag^!}}
            \UnaryInfC{\( \Theta , \banged{x}:\eta  ;\dash \wfdash  \banged{\bag{x[2]}} : \sigma_2 \)}
        \LeftLabel{\redlab{F{:}bag}}
        \BinaryInfC{\( \Theta , \banged{x}:\eta ;  x: \sigma_3 \wfdash \underbrace{(\bag{x[1]} \cdot  \bag{x}  )  \bagsep \banged{\bag{x[2]}}}_{A}  : (\sigma_3^{2} , \sigma_2 ) \)}
    \end{prooftree}
  where $\eta = \sigma_3 \concat \sigma_2$.

    \end{example}

\begin{example}{}\label{ch3ex:bag_b}
    Below we present the wf-derivation $\Pi_B$ of the bag $B=\bag{x} \bagsep \banged{\oneb}$:

    \begin{prooftree}
 \AxiomC{}
\LeftLabel{\redlab{F{:}var^{\ell}}}
\UnaryInfC{\( \Theta , \banged{x}:\eta ;  x: \sigma_3  \wfdash x : \sigma_3\)}
  \AxiomC{\(  \)}
\LeftLabel{\redlab{F{:}\oneb^{\ell}}}
  \UnaryInfC{\( \Theta , \banged{x}:\eta ; \dash \wfdash \oneb : \omega  \)}
   \LeftLabel{\redlab{F{:}bag^{\ell}}}
  \BinaryInfC{\( \Theta , \banged{x}:\eta ;  x: \sigma_3  \wfdash \bag{x} \cdot \oneb : \sigma_3^1\)}
\AxiomC{\(  \)}
 \LeftLabel{\redlab{F{:}\oneb^!}}
 \UnaryInfC{\(  \Theta , \banged{x}:\eta  ;\dash \wfdash  \banged{\oneb} : \sigma' \)}
 \LeftLabel{\redlab{F{:}bag}}
 \BinaryInfC{\( \Theta , \banged{x}:\eta ; x: \sigma_3 \wfdash (\bag{x} \bagsep \banged{\oneb} ) : (\sigma_3 , \sigma' ) \)}
 \end{prooftree}
    
\end{example}

\begin{example}{}
    To illustrate our well-formed rules, let $M$ be the following $\lamrfailunres$-term:  $$ M= \lambda x. (y ( \underbrace{(\bag{x[1]} \cdot  \bag{x}  )  \bagsep \banged{\bag{x[2]}})}_{A} \underbrace{(\bag{x} \bagsep \banged{\oneb} )}_{B}).$$
    To ease the notation $M$ is an abstraction $\lambda x. ((y A)\ B)$, where $A=(\bag{x[1]} \cdot  \bag{x}  )  \bagsep \banged{\bag{x[2]}}$  and $B=\bag{x} \bagsep \banged{\oneb}$.
      From the derivation $\Pi_A$ (Example~\ref{ch3ex:bag_a}) we obtain the wf-derivation $\Pi_A'$ for the  application $y A$:
           \begin{prooftree}
           \small 
         \AxiomC{}
        \LeftLabel{\redlab{F{:}var^{\ell}}}
        \UnaryInfC{\(  \Theta , \banged{x}:\eta ; \Delta \wfdash y : (\sigma_3^{k} , \eta'' ) \rightarrow ((\sigma_3^{j} , \eta' ) \rightarrow \tau) \)}
         \AxiomC{$\Pi_A$}
         \noLine
         \UnaryInfC{\(  \Theta , \banged{x}:\eta ;  x: \sigma_3 \wfdash A  : (\sigma_3^{2} , \sigma_2 )  \)}
         \AxiomC{\( \eta'' \relunbag \sigma_2 \)}
            \LeftLabel{\redlab{F{:}app}}
        \TrinaryInfC{\( \Theta , \banged{x}:\eta ; x: \sigma_3 , \Delta \wfdash y A: (\sigma_3^{j} , \eta' ) \rightarrow \tau  \)}
    \end{prooftree}
   where  $\eta=\sigma_3\bagsep \sigma_2$, for some list type $\eta'$ and integers $k,j$. From the premise $\eta''\relunbag \sigma_2$ it follows that $\eta'' =  \sigma_2 \concat \eta'''$ for an arbitrary $\eta'''$. From the derivation $\Pi_B$ (Example~\ref{ch3ex:bag_b}) we obtain the well-formed derivation for term $M$:
        \begin{prooftree}
        \AxiomC{$\Pi_A'$}
        \noLine
        \UnaryInfC{\(\Theta , \banged{x}:\eta ;  x: \sigma_3, \Delta \wfdash y \ A: (\sigma^j_3,\eta') \to \tau \)}
        \AxiomC{$\Pi_B$}
        \noLine
        \UnaryInfC{\(\Theta , \banged{x}:\eta ; x:\sigma_3\wfdash B:(\sigma_3,\sigma')\)}
        \AxiomC{\( \eta' \relunbag \sigma'\)}
        \LeftLabel{$\redlab{F:app}$}
        \TrinaryInfC{\(\Theta , \banged{x}:\eta ;  x: \sigma_3, x: \sigma_3,\Delta\wfdash (y A  ) B  : \tau  \qquad x\notin \dom{\Delta} \)}
        \LeftLabel{\redlab{F{:}abs}}
        \UnaryInfC{\( \Theta ;  \Delta  \wfdash  \lambda x. ((y  A)B)   : (\sigma_3^{2} , \eta )   \rightarrow \tau \)}
    \end{prooftree}
    where $\Delta = y : (\sigma_3^{k} , \eta'' ) \rightarrow ((\sigma_3^{j} , \eta' ) \rightarrow \tau)$. From the premise $\eta' \relunbag \sigma'$  we obtain that $\eta' = \sigma' \concat \eta'''' $, where $\sigma'$ is an arbitrary strict type and $\eta''''$ is an arbitrary list type.
    \end{example}

\section{Appendix to \texorpdfstring{\secref{ch3s:pi}}{}}\label{ch3appC}

 \begin{definition}{Structural Congruence}
 Structural congruence 
is defined as the least congruence relation on processes such that:
\[
\begin{array}{c}
\begin{array}{c@{\hspace{1.5cm}}c@{\hspace{1.5cm}}c}
    P \equiv_\alpha Q \Rightarrow P  \equiv Q
& 
P \para \zero  \equiv  P
&
P \para Q \equiv Q \para P 
\\
(\nu x)\zero  \equiv \zero
&
(P \para Q) \para R  \equiv P \para (Q \para R)
&
[x \leftrightarrow y]
 \equiv
[y \leftrightarrow x]
\end{array}
\\
\begin{array}{c@{\hspace{1.5cm}}c}
 x \not \in \fn{P} \Rightarrow ((\nu x )P) \para Q  \equiv (\nu x)(P \para Q) 
 &
 (\nu x)(\nu y)P  \equiv (\nu y)(\nu x)P
 \\
P \oplus (Q \oplus R) \equiv (P \oplus Q) \oplus R
&
P \oplus Q \equiv Q \oplus P
\\
(\nu x)(P \para (Q \oplus R))  \equiv (\nu x)(P \para Q) \oplus (\nu x)(P \para R)
&
\zero \oplus \zero  \equiv  \zero
\end{array}
\end{array}
\]
\end{definition}






\section{Appendix to \texorpdfstring{\secref{ch3ssec:lamshar}}{}} 
\label{ch3app:ssec:lamshar}
We need a few auxiliary notions to formalize reduction for \lamrsharfailunres.

\begin{definition}{Head}
We amend Definition \ref{ch3def:headfailure} for the case of terms in \lamrsharfailunres:
\[
\begin{array}{l}
\begin{array}{l@{\hspace{2.5cm}}l}
\headf{ {x}}  =  {x}   & \headf{{x}[i]}  = {x}[i] 
\\
\headf{M\ B}  = \headf{M} & \headf{\lambda x . (M[  {\widetilde{x}} \leftarrow  {x} ])}  = \lambda x . (M[  {\widetilde{x}} \leftarrow  {x} ])\\
\headf{M \linexsub{N / {x}}}  = \headf{M} & \headf{M \unexsub{U / \unvar{x}}}  = \headf{M}
\end{array}\\
\headf{(M[ {\widetilde{x}} \leftarrow  {x}])\esubst{ B }{ x }} = (M[ {\widetilde{x}} \leftarrow  {x}])\esubst{ B }{ x } \hspace{1cm} \headf{\fail^{\widetilde{x}}}  = \fail^{\widetilde{x}}
\\
\headf{M[ {\widetilde{x}} \leftarrow  {x}]} = 
\begin{cases}
     {x} & \text{If $\headf{M} = y \text{ and } y \in  {\widetilde{x}}$}\\
    \headf{M} & \text{Otherwise}
\end{cases}
\\
\end{array}
\]

\end{definition}
\begin{definition}{Linear Head Substitution}\label{ch3def:headlinfail}
Given an $M$ with $\headf{M} = x$, the linear substitution of a term $N$ for the head variable $x$ of the term $M$, written $M\headlin{ N / x}$ is inductively defined as:
{\small{
\begin{align*}
x \headlin{ N / x}   &= N & 
(M\ B)\headlin{ N/x }  = (M \headlin{ N/x })\ B\\
(M \unexsub{U / \unvar{y}} ) \headlin{ N/x } &= (M\headlin{ N/x })\ \unexsub{U / \unvar{y}}  & x \not = y\\
(M \linexsub{L / {y}} ) \headlin{ N/x } &= (M\headlin{ N/x })\ \linexsub{L / {y}}  & x \not = y\\
((M[ {\widetilde{y}} \leftarrow  {y}])\esubst{ B }{  { {y}} })\headlin{ N/x } &= (M[\widetilde{y} \leftarrow  {y}]\headlin{ N/x })\ \esubst{ B }{ y }  
& x \not = y \\
(M[ {\widetilde{y}} \leftarrow  {y}]) \headlin{ N/x } &=  (M\headlin{ N/x }) [ {\widetilde{y}} \leftarrow  {y}] & x \not = y
\end{align*}
}}
\end{definition}

Following \defref{ch3def:context_lamrfail}, we define contexts  for terms and expressions.
While expression contexts are as in \defref{ch3def:context_lamrfail}; the term contexts for \lamrsharfailunres involve  explicit linear and unrestricted substitutions, rather than an explicit substitution: this is due to the reduction strategy we have chosen to adopt, as we always wish to evaluate explicit substitutions first. 
We assume that the terms that fill in the holes respect the conditions on explicit linear substitutions (i.e., variables appear in a term only once, shared variables must occur in the context), similarly for explicit unrestricted substitutions.

\begin{definition}{Evaluation Contexts}
Contexts for terms and expressions are defined by the following grammar:
{\small
\[
\begin{array}{rl}
     C[\cdot] ,  C'[\cdot] &::=([\cdot])B \mid ([\cdot])\linexsub{N /  {x}}  \mid ([\cdot])\unexsub{U / \unvar{x}} \mid ([\cdot])[ {\widetilde{x}} \leftarrow  {x}]\mid ([\cdot])[ \leftarrow  {x}]\esubst{\oneb}{ x} \\
 D[\cdot] , D'[\cdot] & ::= M + [\cdot] \mid [\cdot] + M
\end{array}
\]}
The result of replacing a hole with a \lamrsharfailunres-term $M$ in a context $C[\cdot]$, denoted with $C[M]$, has to be a term in $\lamrsharfailunres$. 
\end{definition}

This way, e.g., the hole in context $C[\cdot ]= ([\cdot])\linexsub{N/ {x}}$ cannot be filled with $y$, since  $C[y]= (y)\linexsub{N/ {x}}$ is not a well-defined term. 
Indeed, $M\linexsub{N/ {x}}$ requires that $x$ occurs exactly once within $M$.
Similarly, we cannot fill the hole with $\fail^{z}$ with $z\neq x$, since $C[\fail^{z}]= (\fail^{z})\linexsub{N/ {x}}$ is also not a well-defined term, for the same reason.

\begin{figure}[!t]
\small
   \centering
   \[
   \begin{array}{ll}
   \begin{array}{l}
   \llfv{ {x}}  = \{  {x} \} \\
 \llfv{{x}[i]}  = \emptyset \\
 \llfv{ {\oneb}}  = \emptyset \\
      \llfv{ {\bag{M}}}  = \llfv{M}  \\
 \llfv{\banged{\bag{M}}}  = \llfv{M} \\
 \llfv{C \bagsep U}  = \llfv{C} \\
 \llfv{ {\bag{M}}  \cdot C} = \llfv{ {M}} \cup \llfv{ C}  
   \end{array}
        &  
    \begin{array}{l}
    \llfv{M\ B}  =  \llfv{M} \cup \llfv{B} \\
\llfv{\lambda x . M[ {\widetilde{x}} \leftarrow  {x}]}  = \llfv{M[ {\widetilde{x}} \leftarrow  {x}]}\!\setminus\! \{  {x} \}\\
    \llfv{M[ {\widetilde{x}} \leftarrow  {x}] \esubst{B}{x}}  = (\llfv{M[ {\widetilde{x}} \leftarrow  {x}]}\setminus \{  {x} \}) \uplus \llfv{B}  \\
    \llfv{M \linexsub{N /  {x}}} = \llfv{ {M}} \cup \llfv{ N} \\
    \llfv{M \unexsub{U / \unvar{x}}} = \llfv{ {M}} \\
    \llfv{\expr{M}+\expr{N}}  = \llfv{\expr{M}} \cup \llfv{\expr{N}} \\
    \llfv{\fail^{ {x}_1, \cdots ,  {x}_n}}  = \{  {x}_1, \ldots ,  {x}_n \}
        \end{array}
   \end{array}
   \]
    \caption{Free Variables for \lamrsharfailunres.}
    \label{ch3fig:fvarsfail}
\end{figure}

\subsection*{Operational Semantics}
As in \lamrfailunres, the reduction relation $\redd$ on \lamrsharfailunres operates lazily on expressions; it is defined by the rules in \figref{ch3fig:share-reductfailureunres}, and relies on a notion of linear free variables given in \figref{ch3fig:fvarsfail}.

\begin{figure*}[!t]
\centering
  \begin{prooftree}
    \AxiomC{$\raisebox{17.0pt}{}$}
    \LeftLabel{\redlab{RS{:}Beta}}
    \UnaryInfC{\(  (\lambda x . (M[ {\widetilde{x}} \leftarrow  {x}])) B  \redd (M[ {\widetilde{x}} \leftarrow  {x}])\esubst{ B }{ x } \)}
 \end{prooftree}

 \begin{prooftree}
     \AxiomC{$ \headf{M} =  {x}$}
     \LeftLabel{\redlab{RS{:}Fetch^{\ell}}}
     \UnaryInfC{\(  M \linexsub{N /  {x}} \redd  M \headlin{ N/ {x} }  \)}
\DisplayProof\hfill
     \AxiomC{$ \headf{M} = {x}[i]$}
     \AxiomC{$ U_i = \banged{\bag{N}}$}
     \LeftLabel{\redlab{RS{:} Fetch^!}}
     \BinaryInfC{\(  M \unexsub{U / \unvar{x}} \redd  M \headlin{ N /{x}[i] }\unexsub{U / \unvar{x}} \)}
 \end{prooftree}
\vspace{-0.5cm}
 \begin{mathpar}
    \inferrule[\redlab{RS{:}Ex \dash Sub}]
        { C = \bag{M_1} \cdots  \bag{M_k} 
        \\
         M \not= \fail^{\widetilde{y}} }
        {\!M[x_1 , \cdots , x_k \leftarrow  {x}]\esubst{ C \bagsep U }{ x } \redd \displaystyle \sum_{C_i \in \perm{C}}M\linexsub{C_i(1)/ {x_1}} \cdots \linexsub{C_i(k)/ {x_k}} \unexsub{U / \unvar{x} }}
  \end{mathpar}

\begin{prooftree}
     \AxiomC{$ k \neq \size{C} $} 
     \AxiomC{$  \widetilde{y} = (\llfv{M} \setminus \{  \widetilde{x}\} ) \cup \llfv{C} $}
    \LeftLabel{\redlab{RS{:}Fail^{\ell}}}
    \BinaryInfC{\(  M[x_1 , \cdots , x_k \leftarrow  {x}]\ \esubst{C \bagsep U}{ x }  \redd \displaystyle \sum_{C_i \in \perm{C}}  \fail^{\widetilde{y}} \)}
    \DisplayProof
\hfill
         \AxiomC{$\headf{M} = {x}[i]$}
    \AxiomC{$ U_i = \banged{\oneb} $}
    \noLine
    \UnaryInfC{\( \widetilde{y} = \llfv{M} \)}
    \LeftLabel{\redlab{RS{:}Fail^!}}
    \BinaryInfC{\(  M \unexsub{U / \unvar{x} } \redd 
    M \headlin{ \fail^{\emptyset} /{x}[i] } \unexsub{U / \unvar{x} }  \)}
\end{prooftree}

\begin{prooftree}
    \AxiomC{\( \widetilde{y} = \llfv{C} \)}
    \LeftLabel{$\redlab{RS{:}Cons_1}$}
    \UnaryInfC{\(\fail^{\widetilde{x}}\ C \bagsep U \redd \displaystyle \sum_{\perm{C}} \fail^{\widetilde{x} \uplus \widetilde{y}}  \)}
 \DisplayProof
\hfill
    \AxiomC{\(  \size{C} =   |  {\widetilde{x}} |   \)} 
    \AxiomC{\(  \widetilde{z} = \llfv{C} \)}
    \LeftLabel{$\redlab{RS{:}Cons_2}$}
    \BinaryInfC{\(  (\fail^{ {\widetilde{x}} \uplus \widetilde{y}} [ {\widetilde{x}} \leftarrow  {x}])\esubst{ C \bagsep U }{ x }  \redd \displaystyle \sum_{\perm{C}} \fail^{\widetilde{y} \uplus \widetilde{z}}  \)}
\end{prooftree}

\begin{prooftree}
    \AxiomC{\( \widetilde{z} = \llfv{N} \)}
    \LeftLabel{$\redlab{RS{:}Cons_3}$}
    \UnaryInfC{\( \fail^{\widetilde{y}\cup x} \linexsub{N/  {x}} \redd  \fail^{\widetilde{y} \cup \widetilde{z}}  \)}
\DisplayProof
\hfill
    \AxiomC{\(  \)}
    \LeftLabel{$\redlab{RS{:}Cons_4}$}
    \UnaryInfC{\( \fail^{\widetilde{y}} \unexsub{U / \unvar{x}}  \redd  \fail^{\widetilde{y}}  \)}
\end{prooftree}

    \caption{Reduction Rules for \lamrsharfailunres (contextual rules omitted)}
    \label{ch3fig:share-reductfailureunres}
\end{figure*}

 As expected, rule \redlab{RS:Beta} results into an explicit substitution $M[\widetilde{x}\leftarrow x]\esubst{ B }{ x }$, where  $B=C\bagsep U$ is a bag with a linear part $C$ and an unrestricted part $U$.  
 
 In the case $|\widetilde{x}|=k=\size{C}$ and  $M\neq \fail^{\widetilde{y}}$, this explicit substitution  expands into a sum of terms involving explicit linear and unrestricted substitutions $\linexsub {N /x}$ and $\unexsub{U/\unvar{x}}$, which are the ones to reduce into a  head substitution, via rule \redlab{RS{:}Ex\dash Sub}. Intuitively, rule~\redlab{RS{:}Ex \dash Sub} ``distributes'' an  explicit substitution into a sum of terms involving explicit linear substitutions; it considers all possible permutations of the elements in the bag among all shared variables. Explicit linear/unrestricted substitutions evolve either into a head substitution $\headlin{ N / x}$ (with $N \in B$), via rule $\redlab{RS:Fetch^\ell}$, or $\headlin{N/x}\unexsub{U/\unvar{x}}$ (with $U \in B$) via rule $\redlab{RS:Fetch^!}$, depending on whether the head of the term is a linear or an unrestricted variable. 
 
 In the case $|\widetilde{x}|=k\neq \size{C}$ or  $M=\fail^{\widetilde{y}}$, the term $M[\widetilde{x}\leftarrow x]\esubst{ B }{ x }$ will be a redex of either rule $\redlab{RS:Fail^\ell}$  or $\redlab{RS:Cons_2}$. The latter has a side condition $|\widetilde{x}|=\size{C}$, because we want to give priority for application of $\redlab{RS:Fail^\ell}$ when there is a mismatch of linear variables and the number of linear resources.
 Rule $\redlab{RS:Fail^!}$  applies to an unrestricted substitution $M\unexsub{U/\unvar{x}}$  when the head of $M$ is an unrestricted variable, say $x[i]$, that aims to consume the $i$-th component of the bag $U$ which is empty, i.e., $U_i=1^!$; then the term reduces to a term where all the head of $M$ is substituted by  $\fail^\emptyset$, the explicit unrestricted substitution is not consumed and continues in the resulting term. Consuming rules $\redlab{RS:Cons_1}, \redlab{RS:Cons_3}$ and $\redlab{RS:Cons_4}$ the term $\fail$ consume either a bag, or an explicit linear substitution, or an explicit unrestricted substitution, respectively.
 
Notice that the left-hand sides of the reduction rules in $\lamrsharfailunres$  do not interfere with each other. 
 Similarly to $\lamrfailunres$, reduction in \lamrsharfailunres satisfies a \emph{diamond property}. 


\begin{example}{}
\label{ch3ex:id_sem_intermed}

We continue to illustrate the different behaviors of the terms below w.r.t. the reduction rules for $\lamrsharfailunres$ (\figref{ch3fig:share-reductfailureunres}):
\begin{enumerate}
    \item The case with a linear variable $x$ in which the linear bag has size one, is close to the standard {\it meaning} of applying an identity function to a term:
    \[
    \begin{aligned}
        (\lambda x. x_1 [x_1 \leftarrow x] ) \bag{N'}\bagsep U' &\redd_{\redlab{R:Beta}}  x_1 [x_1 \leftarrow x] \esubst{\bag{N'}\bagsep U'}{x} \\ 
        &\redd_{\redlab{RS{:}Ex \dash Sub}}  x_1 \linexsub{ N' / {x_1}} \unexsub{ U' / \unvar{x} } \\
        & \redd_{\redlab{R:Fetch^{\ell}}} x_1 \headlin{ N' / x_1 } \unexsub{ U' / \unvar{x} }\\
        & =N' \unexsub{U' / \unvar{x} }
    \end{aligned}
    \]

\item The case of an abstraction of one unrestricted variable that aims to consume the first element of the unrestricted bag, which fails to contain a resource in the first component. 
    \[
    \begin{aligned}
    (\lambda x. x[1][ \leftarrow x] ) \oneb \bagsep \oneb^! \concat U' &\redd_{\redlab{R:Beta}} x[1][ \leftarrow x] \esubst{\oneb \bagsep \oneb^! \concat U' }{x}\\
    &\redd_{\redlab{RS{:}Ex \dash Sub}}x[1]\unexsub{ \oneb^! \concat U' / \unvar{x} }  \\
    &\redd_{\redlab{RS{:}Fail^!}}x[1]  \headlin{ \fail^{\emptyset} /{x}[1] } \unexsub{ \oneb^! \concat U' / \unvar{x} }\\
    & =\fail^{\emptyset} \unexsub{ \oneb^! \concat U' / \unvar{x} }
    \end{aligned}
    \]

  \item The case of an abstraction of one unrestricted variable that aims to consume the $i$th component of the unrestricted bag $U'$. In the case  $C' = \oneb$ and $U'_i\neq 1^!$ the reduction is:
    \[
    \begin{aligned}
    (\lambda x. x[i][ \leftarrow x] ) C'\concat U' &\redd_{\redlab{R:Beta}}x[i][ \leftarrow x]\esubst{C'\concat U'}{x}\\
    &\redd_{\redlab{RS{:}Ex \dash Sub}}x[i]\unexsub{ C'\concat U' / \unvar{x} }  \\
    &\redd_{\redlab{RS{:} Fetch^!}}x[i] \headlin{ N' /{x}[i] } \unexsub{ C'\concat U' / \unvar{x} }=N\unexsub{ C'\concat U' / \unvar{x} }
    \end{aligned}
    \]
    where $ U'_i=\bag{N'}^!$.
    Otherwise, $U'_i=1^!$ and the reduction relies again on the size of the linear bag $C$: if  $\#(x,x[i])=\size{C'}$ the reduction ends with an application of $\redlab{R:\fail^!}$; otherwise, it ends with an application $\redlab{R:\fail^{\ell }}$.
\end{enumerate}
    
\end{example}

\subsection{Well-formedness rules for  \texorpdfstring{$\lamrsharfailunres$}{}}

Similarly to $\lamrfailunres$ we present a set  ``well-formedness'' rules for $\lamrsharfailunres$-terms, -bags and -expressions,  based on  an intersection type system for \lamrsharfailunres, defined upon strict, multiset, list, tuple types, as introduced for \lamrfailunres and presented in \figref{ch3app_fig:wfsh_rulesunres}. Linear contexts $\Gamma,\Delta$ and unrestricted contexts $\Theta, \Upsilon$ are the same as in $\lamrfailunres$, as well as well-formedness judgements $\Theta; \Gamma \vdash \mathbb{M}:\sigma$.

\begin{definition}{Well-formedness in $\lamrsharfailunres$}
An expression $ \expr{M}$ is well formed if there exists a  $\Theta, \Gamma$ and a   $\tau$ such that $\Theta ; \Gamma \wfdash  \expr{M} : \tau $ is entailed via the rules in \figref{ch3app_fig:wfsh_rulesunres}.
\end{definition}

\begin{figure*}[!h]
    \centering
    
\begin{prooftree}
\AxiomC{}
\LeftLabel{\redlab{FS{:}var^{\ell}}}
\UnaryInfC{\( \Theta ;  {x}: \sigma \wfdash  {x} : \sigma\)}
\DisplayProof
\hfill
\AxiomC{\( \Theta , {x}: \eta;  {x}: \eta_i , \Delta \wfdash  {x} : \sigma\)}
\LeftLabel{\redlab{FS{:}var^!}}
\UnaryInfC{\( \Theta , {x}: \eta; \Delta \wfdash {x}[i] : \sigma\)}
  \DisplayProof
\hfill
  \AxiomC{\(  \)}
\LeftLabel{\redlab{FS{:}\oneb^{\ell}}}
\UnaryInfC{\( \Theta ; \dash \wfdash \oneb : \omega \)}
\end{prooftree}
    
\begin{prooftree}
\AxiomC{\(  \)}
\LeftLabel{\redlab{FS{:}\oneb^!}}
\UnaryInfC{\( \Theta ;  \dash  \wfdash \banged{\oneb} : \sigma \)}
\DisplayProof
\hfill
\AxiomC{\( \Theta ; \Gamma  \wfdash M : \tau\)}
\LeftLabel{ \redlab{FS\!:\!weak}}
\UnaryInfC{\( \Theta ; \Gamma ,  {x}: \omega \wfdash M[\leftarrow  {x}]: \tau \)}
\end{prooftree}

\begin{prooftree}
 \AxiomC{$ \Theta ; \Gamma \wfdash \expr{M} : \sigma$}
 \noLine
 \UnaryInfC{$\Theta ; \Gamma \wfdash \expr{N} : \sigma$}
   \LeftLabel{\redlab{FS{:}sum}}
   \UnaryInfC{$ \Theta ; \Gamma \wfdash \expr{M}+\expr{N}: \sigma$}
     \DisplayProof
  \hfill
    \AxiomC{\( \Theta , {x}:\eta ; \Gamma ,  {x}: \sigma^k \wfdash M[ {\widetilde{x}} \leftarrow  {x}] : \tau \quad  {x} \notin \dom{\Gamma} \)}
        \LeftLabel{\redlab{FS{:}abs\dash sh}}
        \UnaryInfC{\(  \Theta ; \Gamma \wfdash \lambda x . (M[ {\widetilde{x}} \leftarrow  {x}])  : (\sigma^k, \eta )  \rightarrow \tau \)}
\end{prooftree}

\begin{prooftree}
         \AxiomC{\( \Theta ;\Gamma \wfdash M : (\sigma^{j} , \eta ) \rightarrow \tau \)}
         \AxiomC{\( \eta \relunbag \epsilon \)}
         \noLine
         \UnaryInfC{\(  \Theta ;\Delta \wfdash B : (\sigma^{k} , \epsilon )  \)}
        \LeftLabel{\redlab{FS{:}app}}
        \BinaryInfC{\( \Theta ; \Gamma, \Delta \wfdash M\ B : \tau\)}
   \end{prooftree}

    \begin{prooftree}
            \AxiomC{\( \Theta ; \Gamma\wfdash C : \sigma^k\)}
            \AxiomC{\(  \Theta ;\dash \wfdash  U : \eta \)}
        \LeftLabel{\redlab{FS{:}bag}}
        \BinaryInfC{\( \Theta ; \Gamma \wfdash C \bagsep U : (\sigma^{k} , \eta  ) \)}
        \DisplayProof
        \hfill 
            \AxiomC{\( \Theta ; \dash \wfdash U : \epsilon\)}
        \AxiomC{\( \Theta ; \dash \wfdash V : \eta\)}
        \LeftLabel{\redlab{FS{:}\concat-bag^{!}}}
        \BinaryInfC{\( \Theta ; \dash  \wfdash U \concat V :\epsilon \concat \eta \)}
          \end{prooftree}

    \begin{prooftree}
\AxiomC{\( \Theta ; \dash \wfdash M : \sigma\)}
        \LeftLabel{\redlab{FS{:}bag^!}}
        \UnaryInfC{\( \Theta ; \dash  \wfdash \banged{\bag{M}}:\sigma \)}
        \DisplayProof
\hfill
  \AxiomC{\( \Theta ; \Gamma \wfdash M : \sigma\)}
        \AxiomC{\( \Theta ; \Delta \wfdash C : \sigma^k\)}
        \LeftLabel{\redlab{FS{:}bag^{\ell}}}
        \BinaryInfC{\( \Theta ; \Gamma, \Delta \wfdash \bag{M}\cdot C:\sigma^{k+1}\)}
    \end{prooftree}

    \begin{prooftree}
          \AxiomC{\({x}\notin \dom{\Gamma} \quad k \not = 0\)}
        \noLine
        \UnaryInfC{\( \Theta ;  \Gamma ,  {x}_1: \sigma, \cdots,  {x}_k: \sigma \wfdash M : \tau \)}
        \LeftLabel{ \redlab{FS{:}share}}
        \UnaryInfC{\( \Theta ;  \Gamma ,  {x}: \sigma^{k} \wfdash M [  {x}_1 , \cdots ,  {x}_k \leftarrow  {x} ]  : \tau \)}  
        \DisplayProof \hfill
        \AxiomC{\( \Theta ; \Gamma  ,  {x}:\sigma \wfdash M : \tau \)}
        \noLine
        \UnaryInfC{\( \Theta ; \Delta \wfdash N : \sigma \)}
            \LeftLabel{\redlab{FS{:}Esub^{\ell}}}
        \UnaryInfC{\( \Theta ; \Gamma, \Delta \wfdash M \linexsub{N /  {x}} : \tau \)}
         \end{prooftree}
         
\begin{prooftree}
        \AxiomC{\( \Theta , {x} : \eta; \Gamma  \wfdash M : \tau \)}
        \AxiomC{\( \Theta ; \dash \wfdash U : \epsilon \)}
        \AxiomC{\( \eta \relunbag \epsilon \)}
            \LeftLabel{\redlab{FS{:}Esub^!}}
        \TrinaryInfC{\( \Theta ; \Gamma \wfdash M \unexsub{U / \unvar{x}}  : \tau \)}
    \end{prooftree}

    \begin{prooftree}
    \AxiomC{\( \Theta ; \Delta \wfdash B : (\sigma^{k} , \epsilon ) \)}
            \noLine
            \UnaryInfC{\( \Theta , {x} : \eta ; \Gamma ,  {x}: \sigma^{j} \wfdash M[ {\widetilde{x}} \leftarrow  {x}] : \tau  \)}
            \AxiomC{\( \eta \relunbag \epsilon \)}
        \LeftLabel{\redlab{FS{:}Esub}}    
        \BinaryInfC{\( \Theta ; \Gamma, \Delta \wfdash (M[ {\widetilde{x}} \leftarrow  {x}])\esubst{ B }{ x }  : \tau \)}
   \DisplayProof
\hfill
        \AxiomC{\( \dom{\Gamma} = \widetilde{x}\)}
  \LeftLabel{\redlab{FS{:}fail}}
  \UnaryInfC{\( \Theta ; \Gamma \wfdash  \fail^{\widetilde{x}} : \tau \)}
   \end{prooftree}

    \caption{Well-Formedness Rules for \lamrsharfailunres}\label{ch3app_fig:wfsh_rulesunres}
    \vspace{-3mm}
\end{figure*}

Well-formed rules for \lamrsharfailunres are essentially the same as the ones for $\lamrfailunres$.  Rules $\redlab{FS:abs\dash sh}$ and $\redlab{FS{:}Esub}$ are modified to  take into account  the sharing construct $[\widetilde{x}\leftarrow x]$. Rule  $\redlab{FS{:}share}$ is exclusive for $\lamrsharfailunres$ and requires, for each $i=1,\ldots, k$, the variable assignment  $x_i:\sigma$, to derive the well-formedness of $M[x_1,\ldots, x_n \leftarrow x]:\tau$ (in addition to variable assignments in $\Theta$ and $\Gamma$).

\begin{lemma}[Linear Substitution Lemma for \lamrsharfailunres]
\label{ch3l:lamrsharfailsubsunres}
If $\Theta ; \Gamma ,  {x}:\sigma \wfdash M: \tau$, $\headf{M} =  {x}$, and $\Theta ; \Delta \wfdash N : \sigma$ 
then 
$\Gamma , \Delta \wfdash M \headlin{ N /  {x} }:\tau$.
\end{lemma}

\begin{proof}
By structural induction on $M$ with $\headf{M}=  {x}$.
There are six cases to be analyzed:
\begin{enumerate}
\item $M= {x}$

In this case, $\Theta ;  {x}:\sigma \wfdash  {x}:\sigma$ and $\Gamma=\emptyset$.  Observe that $ {x}\headlin{N/ {x}}=N$, since $\Delta\wfdash N:\sigma$, by hypothesis, the result follows.

    \item $M = M'\ B$.
    
    Then $\headf{M'\ B} = \headf{M'} =  {x}$, and the derivation is the following:
    \begin{prooftree}
        \AxiomC{$\Theta ; \Gamma_1 ,  {x}:\sigma \wfdash M': (\delta^{j} , \eta )  \rightarrow \tau$}\
        \AxiomC{$\Theta ; \Gamma_2 \wfdash B : (\delta^{k} , \epsilon ) $}
        \AxiomC{\( \eta \relunbag \epsilon \)}
    	\LeftLabel{\redlab{FS{:}app}}
        \TrinaryInfC{$\Theta ; \Gamma_1 , \Gamma_2 ,  {x}:\sigma \wfdash M'B:\tau $}    
    \end{prooftree}

    where $\Gamma=\Gamma_1,\Gamma_2$, and  $j,k$ are non-negative integers, possibly different.  Since $\Delta \vdash N : \sigma$, by IH, the result holds for $M'$, that is,
    $$\Gamma_1 , \Delta \wfdash M'\headlin{ N /  {x} }: (\delta^{j} , \eta )  \rightarrow \tau$$
    which gives the  derivation:


    \begin{prooftree}
        \AxiomC{$\Theta ; \Gamma_1 , \Delta \wfdash M'\headlin{ N /  {x} }: (\delta^{j} , \eta )  \rightarrow \tau$}\
        \AxiomC{$\Theta ; \Gamma_2 \wfdash B : (\delta^{k} , \epsilon ) $}
        \AxiomC{\( \eta \relunbag \epsilon \)}
    	\LeftLabel{\redlab{FS{:}app}}
        \TrinaryInfC{$\Theta ; \Gamma_1 , \Gamma_2 , \Delta \wfdash ( M'\headlin{ N /  {x} } ) B:\tau $}    
    \end{prooftree}
   
      From \defref{ch3def:headlinfail},   $(M'B) \headlin{ N /  {x} } = ( M'\headlin{ N /  {x} } ) B$, 
      and the result follows.
    
    \item $M = M'[ {\widetilde{y}} \leftarrow  {y}] $.
    
    Then $ \headf{M'[ {\widetilde{y}} \leftarrow  {y}]} = \headf{M'}= {x}$, for  $y\neq x$. Therefore, 
    \begin{prooftree}
        \AxiomC{\(\Theta ; \Gamma_1 ,  {y}_1: \delta, \cdots,  {y}_k: \delta ,  {x}: \sigma \wfdash M' : \tau \quad  {y}\notin \Gamma_1 \quad k \not = 0\)}
        \LeftLabel{ \redlab{FS{:}share}}
        \UnaryInfC{\( \Theta ; \Gamma_1 ,  {y}: \delta^k,  {x}: \sigma \wfdash M'[ {y}_1 , \cdots ,  {y}_k \leftarrow x] : \tau \)}
    \end{prooftree}
    where $\Gamma=\Gamma_1 ,  {y}: \delta^k$. 
    By IH, the result follows for $M'$, that is, 
    $$\Theta ; \Gamma_1 ,  {y}_1: \delta, \cdots,  {y}_k: \delta ,\Delta \wfdash M'\headlin{N/ {x}} : \tau $$
    
    and we have the derivation:
    
    \begin{prooftree}
        \AxiomC{\( \Theta ; \Gamma_1 ,  {y}_1: \delta, \cdots,  {y}_k: \delta , \Delta \wfdash  M' \headlin{ N /  {x}} : \tau \quad  {y}\notin \Gamma_1 \quad k \not = 0\)}
        \LeftLabel{ \redlab{FS{:}share} }
        \UnaryInfC{\( \Theta ; \Gamma_1 ,  {y}: \delta^k, \Delta \wfdash M' \headlin{ N /  {x}} [ {\widetilde{y}} \leftarrow  {y}] : \tau \)}
    \end{prooftree}
    From \defref{ch3def:headlinfail}  $M'[ {\widetilde{y}} \leftarrow  {y}] \headlin{ N /  {x} } = M' \headlin{ N /  {x}} [ {\widetilde{y}} \leftarrow  {y}]$, 
    and the result follows.

    \item $M = M'[ \leftarrow  {y}] $.
    
    Then $ \headf{M'[ \leftarrow  {y}]} = \headf{M'}= {x}$ with  $x \not  = y $, 
    \begin{prooftree}
        \AxiomC{\( \Theta ; \Gamma  ,  {x}: \sigma  \wfdash M : \tau\)}
        \LeftLabel{ \redlab{FS{:}weak} }
        \UnaryInfC{\( \Theta ;  \Gamma  ,  {y}: \omega,  {x}: \sigma  \wfdash M[\leftarrow  {y}]: \tau \)}
    \end{prooftree}
     and $M'[ \leftarrow  {y}] \headlin{ N /  {x} } = M' \headlin{ N /  {x}} [ \leftarrow  {y}]$. Then by the induction hypothesis:
    \begin{prooftree}
        \AxiomC{\( \Theta ;  \Gamma , \Delta  \wfdash M \headlin{ N /  {x}}: \tau\)}
        \LeftLabel{ \redlab{FS{:}weak}}
        \UnaryInfC{\( \Theta ;  \Gamma  ,  {y}: \omega, \Delta \wfdash M\headlin{ N /  {x}}[\leftarrow  {y}]: \tau \)}
    \end{prooftree}

    \item If $M =  M' \linexsub {M'' / {y}} $ then $\headf{M' \linexsub {M'' / {y}}} = \headf{M'} = x \not = y$, 
    
    \begin{prooftree}
        \AxiomC{\( \Theta ;\Gamma  ,  {y}:\delta , x: \sigma \wfdash M : \tau \)}
        \AxiomC{\( \Theta ;\Delta \wfdash M'' : \delta \)}
        \LeftLabel{ \redlab{FS{:}ex \dash sub^{\ell}} }
        \BinaryInfC{\( \Theta ;\Gamma_1, \Gamma_2 , x: \sigma \wfdash M' \linexsub {M'' / {y}} : \tau \)}
    \end{prooftree}
     and $M' \linexsub {M'' / {y}}  \headlin{ N / x } = M'  \headlin{N / x } \linexsub {M'' / {y}}$. Then by the induction hypothesis:
    
    \begin{prooftree}
        \AxiomC{\( \Theta ;\Gamma  ,  {y}:\delta , \Delta  \wfdash M'  \headlin{N / x } : \tau \)}
        \AxiomC{\( \Theta ;\Delta \wfdash M'' : \delta \)}
        \LeftLabel{ \redlab{FS{:}ex \dash sub^{\ell}} }
        \BinaryInfC{\( \Theta ;\Gamma_1, \Gamma_2 , \Delta  \wfdash M'  \headlin{N / x } \linexsub {M'' / {y}} : \tau \)}
    \end{prooftree}

    \item If $M =  M' \unexsub{U / \unvar{y}} $ then $\headf{M' \unexsub{U / \unvar{y}}} = \headf{M'} = x $, and the proofs is similar to the case above.
    
    
    
    \end{enumerate}
\end{proof}

\begin{lemma}[Unrestricted Substitution Lemma for \lamrsharfailunres]
\label{ch3lem:subt_lem_sharefailunres_un}
If $\Theta, \banged{x}: \eta ; \Gamma \wfdash M: \tau$, $\headf{M} = {x}[i]$, $\eta_i = \sigma $, and $\Theta ; \cdot \wfdash N : \sigma$
then 
$\Theta, \banged{x}: \eta ; \Gamma  \wfdash M \headlin{ N / {x}[i] }$.
\end{lemma}

\begin{proof}
By structural induction on $M$ with $\headf{M}= {x}[i]$. There are three cases to be analyzed: 

\begin{enumerate}
\item $M= {x}[i]$.

In this case, 
    \begin{prooftree}
        \AxiomC{}
        \LeftLabel{ \redlab{F{:}var^{ \ell}}}
        \UnaryInfC{\( \Theta , \banged{x}: \eta;  {x}: \eta_i  \wfdash  {x} : \sigma\)}
        \LeftLabel{\redlab{F{:}var^!}}
        \UnaryInfC{\( \Theta, \banged{x}: \eta ; \cdot \wfdash {x}[i] : \sigma\)}
    \end{prooftree}
 and $\Gamma=\emptyset$.  Observe that ${x}[i]\headlin{N/{x}[i]}=N$, since $\Theta, \banged{x}: \eta  ; \Gamma  \wfdash M \headlin{ N / {x}[i] }$, by hypothesis, the result follows.

    \item $M = M'\ B$.
    
    In this case, $\headf{M'\ B} = \headf{M'} =  {x}[i]$, and one has the following derivation:
    
    \begin{prooftree}
         \AxiomC{\( \Theta, \banged{x}: \eta ;\Gamma_1 \wfdash M : (\delta^{j} , \epsilon ) \rightarrow \tau \quad \Theta, \banged{x}: \sigma ; \Gamma_2 \wfdash B : (\delta^{k} , \epsilon' )  \)}
         \AxiomC{\( \epsilon \relunbag \epsilon' \)}
            \LeftLabel{\redlab{F{:}app}}
        \BinaryInfC{\( \Theta, \banged{x}: \eta ;\Gamma_1 , \Gamma_2 \wfdash M\ B : \tau\)}
    \end{prooftree}
    
 where $\Gamma=\Gamma_1,\Gamma_2$, $\delta$ is a strict type and $j,k$ are non-negative  integers, possibly different.
 
 By IH, we get $\Theta, \banged{x}: \eta ;\Gamma_1 \wfdash M'\headlin{N/ {x[i]}}:(\delta^{j} , \epsilon ) \rightarrow \tau $, which gives the derivation:
    \begin{prooftree}
        \small 
        \AxiomC{$\Theta , \banged{x}: \eta;\Gamma_1\wfdash M'\headlin{N/ {x}[i]}:(\delta^{j} , \epsilon ) \rightarrow \tau $}\
        \AxiomC{$\Theta , \banged{x}: \eta; \Gamma_2 \wfdash B : (\delta^{k} , \epsilon' ) $}
        \AxiomC{\( \epsilon \relunbag \epsilon' \)}
    	\LeftLabel{\redlab{F{:}app}}
        \TrinaryInfC{$\Theta , \banged{x}: \eta;\Gamma_1 , \Gamma_2  \wfdash ( M'\headlin{ N / {x}[i] } ) B:\tau $}    
    \end{prooftree}
From \defref{ch3def:linsubfail}, $M' \esubst{ B }{ y} \headlin{ N / {x}[i] } = M' \headlin{ N / {x}[i] } \esubst{ B }{ y}$,
and  the result follows.

    \item $M = M'[ {\widetilde{y}} \leftarrow  {y}] $.
    
    Then $ \headf{M'[ {\widetilde{y}} \leftarrow  {y}]} = \headf{M'}= {x}[i]$, for  $y\neq x$. Therefore, 
    \begin{prooftree}
        \AxiomC{\(\Theta , \banged{x}: \eta; \Gamma_1 ,  {y}_1: \delta, \cdots,  {y}_k: \delta  \wfdash M' : \tau \quad  {y}\notin \Gamma_1 \quad k \not = 0\)}
        \LeftLabel{ \redlab{FS{:}share}}
        \UnaryInfC{\( \Theta , \banged{x}: \eta; \Gamma_1 ,  {y}: \delta^k \wfdash M'[ {y}_1 , \cdots ,  {y}_k \leftarrow y] : \tau \)}
    \end{prooftree}
    where $\Gamma=\Gamma_1 ,  {y}: \delta^k$. 
    By IH, the result follows for $M'$, that is, 
    $$\Theta, \banged{x}: \eta ; \Gamma_1 ,  {y}_1: \delta, \cdots,  {y}_k: \delta  \wfdash M'\headlin{N/ {x}[i] } : \tau $$
    
    and we have the derivation:
    
    \begin{prooftree}
        \AxiomC{\( \Theta , \banged{x}: \eta; \Gamma_1 ,  {y}_1: \delta, \cdots,  {y}_k: \delta  \wfdash  M' \headlin{ N / {x}[i] } : \tau \quad  {y}\notin \Gamma_1 \quad k \not = 0\)}
        \LeftLabel{ \redlab{FS{:}share} }
        \UnaryInfC{\( \Theta , \banged{x}: \eta; \Gamma_1 ,  {y}: \delta^k \wfdash M' \headlin{ N / {x}[i]} [ {\widetilde{y}} \leftarrow  {y}] : \tau \)}
    \end{prooftree}
    From \defref{ch3def:headlinfail}  $M'[ {\widetilde{y}} \leftarrow  {y}] \headlin{ N / {x}[i] } = M' \headlin{ N / {x}[i]} [ {\widetilde{y}} \leftarrow  {y}]$, 
    and the result follows.

\item $M = M'[ \leftarrow  {y}] $.
    
    Then $ \headf{M'[ \leftarrow  {y}]} = \headf{M'}= {x}[i]$ with  $x \not  = y $, 
    \begin{prooftree}
        \AxiomC{\( \Theta , \banged{x}: \eta; \Gamma   \wfdash M : \tau\)}
        \LeftLabel{ \redlab{FS{:}weak} }
        \UnaryInfC{\( \Theta , \banged{x}: \eta;  \Gamma  ,  {y}: \omega \wfdash M[\leftarrow  {y}]: \tau \)}
    \end{prooftree}
     and $M'[ \leftarrow  {y}] \headlin{ N / {x}[i] } = M' \headlin{ N / {x}[i] } [ \leftarrow  {y}]$. Then by the induction hypothesis:
    \begin{prooftree}
        \AxiomC{\( \Theta , \banged{x}: \eta;  \Gamma   \wfdash M \headlin{ N / {x}[i]  }: \tau\)}
        \LeftLabel{ \redlab{FS{:}weak}}
        \UnaryInfC{\( \Theta , \banged{x}: \eta;  \Gamma  ,  {y}: \omega \wfdash M\headlin{ N / {x}[i] }[\leftarrow  {y}]: \tau \)}
    \end{prooftree}

    \item $M =  M' \linexsub {M'' / {y}} $. 
    
    Then $\headf{M' \linexsub {M'' / {y}}} = \headf{M'} = {x}[i]$ with $x \not = y$, 
    
    \begin{prooftree}
        \AxiomC{\( \Theta , \banged{x}: \eta;\Gamma  ,  {y}:\delta \wfdash M : \tau \)}
        \AxiomC{\( \Theta, \banged{x}: \eta ;\Delta \wfdash M'' : \delta \)}
        \LeftLabel{ \redlab{FS{:}ex \dash sub^{\ell}} }
        \BinaryInfC{\( \Theta, \banged{x}: \eta ;\Gamma, \Delta  \wfdash M' \linexsub {M'' / {y}} : \tau \)}
    \end{prooftree}
     and $M' \linexsub {M'' / {y}}  \headlin{ N / {x}[i]  } = M'  \headlin{N / {x}[i]  } \linexsub {M'' / {y}}$. Then by the induction hypothesis:
    
    \begin{prooftree}
        \AxiomC{\( \Theta , \banged{x}: \eta;\Gamma  ,  {y}:\delta \wfdash M'  \headlin{N /  {x}[i] } : \tau \)}
        \AxiomC{\( \Theta , \banged{x}: \eta;\Delta \wfdash M'' : \delta \)}
        \LeftLabel{ \redlab{FS{:}ex \dash sub^{\ell}} }
        \BinaryInfC{\( \Theta ;\Gamma , \Delta  \wfdash M'  \headlin{N /  {x}[i] } \linexsub {M'' / {y}} : \tau \)}
    \end{prooftree}

    \item $M =  M' \unexsub{U / \unvar{y}}$.  
    
    Then $\headf{M' \unexsub {U / \unvar{y}}} = \headf{M'} = {x}[i] $, 
    
    \begin{prooftree}
        \AxiomC{\( \Theta , \banged{x}: \eta, \banged{y} : \epsilon; \Gamma  \wfdash M : \tau \quad  \Theta , \banged{x}: \eta; \dash \wfdash U : \epsilon \)}
            \LeftLabel{\redlab{FS{:}ex \dash sub^!}}
        \UnaryInfC{\( \Theta , \banged{x}: \eta; \Gamma  \wfdash M \unexsub{U / \unvar{y}}  : \tau \)}
    \end{prooftree}
    
     and $M' \unexsub{U /\unvar{y}}  \headlin{ N /  {x}[i] } = M'  \headlin{N /  {x}[i] } \unexsub{U /\unvar{y}}$. Then by the induction hypothesis:
    
    \begin{prooftree}
        \AxiomC{\( \Theta , \banged{x}: \eta, \banged{y} : \epsilon; \Gamma  \wfdash M'  \headlin{N / {x}[i] } : \tau \quad  \Theta , \banged{x}: \eta; \dash \wfdash U : \eta \)}
            \LeftLabel{\redlab{FS{:}ex \dash sub^!}}
        \UnaryInfC{\( \Theta , \banged{x}: \eta ; \Gamma  \wfdash M'  \headlin{N /  {x}[i] } \unexsub {U / \unvar{y}}  : \tau \)}
    \end{prooftree}
\end{enumerate}

\end{proof}

\begin{theorem}[SR in \lamrsharfailunres]
\label{ch3t:app_lamrsharfailsrunres}
If $\Theta ; \Gamma \wfdash \expr{M}:\tau$ and $\expr{M} \redd \expr{M}'$ then $\Theta ; \Gamma \wfdash \expr{M}' :\tau$.
\end{theorem}

\begin{proof} By structural induction on the reduction rule from \figref{ch3fig:share-reductfailureunres} applied in $\expr{M}\redd \expr{N}$.

\begin{enumerate}

	\item Rule $\redlab{RS{:}Beta}$.
	
	Then $\expr{M} = (\lambda x. M[ {\widetilde{x}} \leftarrow  {x}]) B $  and the reduction is:
	  \begin{prooftree}
        \AxiomC{}
        \LeftLabel{\redlab{RS{:}Beta}}
        \UnaryInfC{\((\lambda x. M[ {\widetilde{x}} \leftarrow  {x}]) B \redd M[ {\widetilde{x}} \leftarrow  {x}]\ \esubst{ B }{ x }\)}
     \end{prooftree}

 	where $ \expr{M}'  =  M[ {\widetilde{x}} \leftarrow  {x}]\ \esubst{ B }{ x }$. Since $\Theta ; \Gamma\wfdash \expr{M}:\tau$ we get the following derivation:
	\begin{prooftree}
			\AxiomC{$\Theta , \banged{x} : \eta; \Gamma' ,  {x}_1:\sigma , \cdots ,  {x}_j:\sigma  \wfdash  M: \tau $}
			\LeftLabel{ \redlab{FS{:}share} }
			\UnaryInfC{$\Theta , \banged{x} : \eta;  \Gamma' ,   {x}:\sigma^{j}  \wfdash  M[ {\widetilde{x}} \leftarrow  {x}]: \tau $}
			\LeftLabel{ \redlab{FS{:}abs \dash sh} }
            \UnaryInfC{$\Theta ; \Gamma' \wfdash \lambda x. M[ {\widetilde{x}} \leftarrow  {x}]: (\sigma^{j} , \eta ) \rightarrow \tau $}
            
            \AxiomC{$\Theta ;\Delta \wfdash B: (\sigma^{k} , \epsilon ) $}
            
			\AxiomC{\( \eta \relunbag \epsilon \)}
			\LeftLabel{ \redlab{FS{:}app} }
		\TrinaryInfC{$ \Theta ;\Gamma' , \Delta \wfdash (\lambda x. M[ {\widetilde{x}} \leftarrow  {x}]) B:\tau $}
	\end{prooftree}

	for $\Gamma = \Gamma' , \Delta $ and $x\notin \dom{\Gamma'}$. 
	Notice that: 
    \begin{prooftree}
            \AxiomC{$\Theta , \banged{x} : \eta; \Gamma' ,  {x}_1:\sigma , \cdots ,  {x}_j:\sigma  \wfdash  M: \tau $}
			\LeftLabel{ \redlab{FS{:}share} }
			\UnaryInfC{$\Theta , \banged{x} : \eta;  \Gamma' ,   {x}:\sigma^{j}  \wfdash  M[ {\widetilde{x}} \leftarrow  {x}]: \tau $}
			
            \AxiomC{$\Theta ;\Delta \wfdash B:(\sigma^{k} , \epsilon )  $}
            
            \AxiomC{\( \eta \relunbag \epsilon \)}
            \LeftLabel{ \redlab{FS{:}ex \dash sub} }
        \TrinaryInfC{$ \Theta ;\Gamma' , \Delta \wfdash M[ {\widetilde{x}} \leftarrow  {x}]\ \esubst{ B }{ x }:\tau $}
    \end{prooftree}

    Therefore $ \Theta ; \Gamma',\Delta\wfdash\expr{M}' :\tau$ and the result follows.

    \item Rule $ \redlab{RS{:}Ex \dash Sub}.$
    
    Then $ \expr{M} =  M[ {x}_1, \cdots ,  {x}_k \leftarrow  {x}]\ \esubst{ C \bagsep U }{ x }$ where $C=  \bag{N_1}\cdot \dots \cdot \bag{N_k} $. The reduction applying rule $\redlab{RS{:}Ex \dash Sub}$ is:
    
     \begin{prooftree}
        \AxiomC{$ C = \bag{M_1}
        \cdots  \bag{M_k} $}
        \AxiomC{$ M \not= \fail^{\widetilde{y}} $}
        \BinaryInfC{\( \!M[ {x}_1, \!\cdots\! ,  {x}_k \leftarrow  {x}]\esubst{ C \bagsep U }{ x } \redd \displaystyle \sum_{C_i \in \perm{C}}M\linexsub{C_i(1)/ {x_1}} \cdots \linexsub{C_i(k)/ {x_k}} \unexsub{U / \unvar{x} }   \)}
    \end{prooftree}
    
    and $\expr{M'}= \sum_{C_i \in \perm{C}}M\linexsub{C_i(1)/ {x_1}} \cdots \linexsub{C_i(k)/ {x_k}} \unexsub{U / \unvar{x} }$
    To simplify the proof we take $k=2$, as the case $k>2$ is similar. Therefore,
    \begin{itemize}
        \item $C=\bag{N_1}\cdot \bag{N_2}$; and
        \item $\perm{C}=\{\bag{N_1}\cdot \bag{N_2}, \bag{N_2}\cdot \bag{N_1}\}$.
    \end{itemize} 
    Since $\Theta ; \Gamma\wfdash \expr{M}:\tau$ we get a derivation: (we omit the labels \redlab{FS:ex\dash sub} and \redlab{FS{:}share})
    \begin{prooftree}
            \AxiomC{\( \Theta, \banged{x} : \eta  ;  \Gamma' ,  {x}_1: \sigma,  {x}_2: \sigma \wfdash M : \tau \quad  {x}\notin \dom{\Gamma} \quad k \not = 0\)}
            \UnaryInfC{\(  \Theta , \banged{x} : \eta ; \Gamma' ,  {x}: \sigma^{2} \wfdash M[ {\widetilde{x}} \leftarrow  {x}] : \tau  \)} 
            
            \AxiomC{\( \Theta ; \Delta \wfdash B : (\sigma^{k} , \epsilon ) \)}
            \AxiomC{\( \eta \relunbag \epsilon \)}
        \TrinaryInfC{\( \Theta ; \Gamma', \Delta \wfdash (M[ {\widetilde{x}} \leftarrow  {x}])\esubst{ B }{ x }  : \tau \)}
    \end{prooftree}
    where $\Gamma = \Gamma' , \Delta $. Consider the wf derivation for $\Pi_{1,2}$: (we omit the labels \redlab{FS:ex\dash sub^!} and \redlab{FS:ex\dash sub^{\ell}})
        {\small
        \begin{prooftree}
                    \AxiomC{\( \Theta, \banged{x} : \eta  ;  \Gamma' ,  {x}_1: \sigma,  {x}_2: \sigma \wfdash M : \tau\)}
                    \AxiomC{\( \Theta ; \Delta_1 \wfdash N_1 : \sigma \)}
                \BinaryInfC{\( \Theta, \banged{x} : \eta  ;  \Gamma' ,  {x}_2: \sigma , \Delta_1 \wfdash M\linexsub{N_1/ {x_1}} : \tau \)}
                \AxiomC{\( \Theta ; \Delta_2 \wfdash N_2 : \sigma \)}
            \BinaryInfC{\( \Theta, \banged{x} : \eta  ;  \Gamma' , \Delta \wfdash M\linexsub{N_1/ {x_1}} \linexsub{N_2/ {x_2}} : \tau \)}
            \AxiomC{\( \Theta ; \dash \wfdash U : \epsilon \)}
            \AxiomC{\( \eta \relunbag \epsilon \)}
            \TrinaryInfC{\( \Theta ; \Gamma' , \Delta \wfdash M\linexsub{N_1/ {x_1}} \linexsub{N_2/ {x_2}} \unexsub{U / \unvar{x} }  : \tau \)}
        \end{prooftree}
        }

    Similarly, we can obtain a derivation $\Pi_{2,1}$ of $ \Theta ; \Gamma' , \Delta \wfdash M\linexsub{N_2/ {x_1}} \linexsub{N_1/ {x_2}} \unexsub{U / \unvar{x} }  : \tau$.   Finally, applying \redlab{FS{:}sum}:
    
     \begin{prooftree}
        \small
        \AxiomC{\( \Pi_{1,2} \)}
        \AxiomC{\( \Pi_{2,1} \)}
        \LeftLabel{ \redlab{FS{:}sum} }
        \BinaryInfC{$ \Theta ; \Gamma' , \Delta  \wfdash M\linexsub{N_1/ {x_1}} \linexsub{N_2/ {x_2}} \unexsub{U / \unvar{x} } + M\linexsub{N_2/ {x_1}} \linexsub{N_1/ {x_2}} \unexsub{U / \unvar{x} }   : \tau $}
    \end{prooftree}
    
    and the result follows.

     \item Rule $ \redlab{RS{:}Fetch^{\ell}}$.
    
    Then $ \expr{M} =  M \linexsub{N /  {x}}  $ where  $\headf{M} =  {x}$. The reduction is:
    
        \begin{prooftree}
             \AxiomC{$ \headf{M} =  {x}$}
             \LeftLabel{\redlab{RS{:}Fetch^{\ell}}}
             \UnaryInfC{\(  M \linexsub{N /  {x}} \redd  M \headlin{ N/ {x} }  \)}
         \end{prooftree}
          and $\expr{M'}= M \linexsub{N /  {x}} $.     Since $\Theta ; \Gamma\wfdash \expr{M}:\tau$ we get the following derivation:
        \begin{prooftree}
        \AxiomC{\( \Theta ; \Gamma' ,  {x}:\sigma \wfdash M : \tau \quad  \Theta ; \Delta \wfdash N : \sigma \)}
            \LeftLabel{\redlab{FS{:}ex \dash sub^{\ell}}}
        \UnaryInfC{\( \Theta ; \Gamma', \Delta \wfdash M \linexsub{N /  {x}} : \tau \)}
    \end{prooftree}
      where $\Gamma = \Gamma' , \Delta $. By Lemma~\ref{ch3l:lamrsharfailsubsunres}, we obtain the derivation $ \Theta ; \Gamma' , \Delta \wfdash   M \headlin{ N/ {x} } : \tau $. 

    \item Rule $ \redlab{RS{:} Fetch^!}$.
    
    Then $ \expr{M} =  M \unexsub{U /  \unvar{x}}  $ where  $\headf{M} = {x}[i]$. The reduction is:
    
         \begin{prooftree}
             \AxiomC{$ \headf{M} = {x}[i]$}
             \AxiomC{$ U_i = \banged{\bag{N}}$}
             \LeftLabel{}
             \BinaryInfC{\(  M \unexsub{U /  \unvar{x}} \redd  M \headlin{ N / {x}[i] }\unexsub{U /  \unvar{x}} \)}
         \end{prooftree}
    
    and $\expr{M'}= M \unexsub{U /  \unvar{x}} $.     Since $\Theta ; \Gamma \wfdash \expr{M}:\tau$ we get the following derivation:

    \begin{prooftree}
        \AxiomC{\( \Theta , \banged{x} : \eta; \Gamma  \wfdash M : \tau \)}
        \AxiomC{\( \Theta ; \dash \wfdash U : \epsilon \)}
        \AxiomC{\( \eta \relunbag \epsilon \)}
            \LeftLabel{\redlab{FS{:}ex \dash sub^!}}
        \TrinaryInfC{\( \Theta ; \Gamma \wfdash M \unexsub{U /  \unvar{x}}  : \tau \)}
    \end{prooftree}
    
    By Lemma~\ref{ch3lem:subt_lem_sharefailunres_un}, we obtain the derivation $ \Theta ; \Gamma \wfdash   M \headlin{ N /{x}[i] }\unexsub{U /  \unvar{x}} : \tau $. 

\item Rule $\redlab{RS{:}TCont}$.

Then $\expr{M} = C[M]$ and the reduction is as follows:

\begin{prooftree}
        \AxiomC{$   M \redd M'_{1} + \cdots +  M'_{k} $}
        \LeftLabel{\redlab{RS{:}TCont}}
        \UnaryInfC{$ C[M] \redd  C[M'_{1}] + \cdots +  C[M'_{k}] $}
\end{prooftree}
with $\expr{M'}= C[M] \redd  C[M'_{1}] + \cdots +  C[M'_{k}] $. 
The proof proceeds by analysing the context $C$.
There are four cases:

\begin{enumerate}
    \item $C=[\cdot]\ B$.
    
    In this case $\expr{M}=M \ B$, for some $B$. Since $\Gamma\vdash \expr{M}:\tau$ one has a derivation:
\begin{prooftree}
        \AxiomC{\( \Theta ;\Gamma' \wfdash M : (\sigma^{j} , \eta ) \rightarrow \tau \)}
         \AxiomC{\(  \Theta ;\Delta \wfdash B : (\sigma^{k} , \epsilon )  \)}
         \AxiomC{\( \eta \relunbag \epsilon \)}
        \LeftLabel{\redlab{FS{:}app}}
        \TrinaryInfC{\( \Theta ; \Gamma', \Delta \wfdash M\ B : \tau\)}
\end{prooftree}

where $\Gamma = \Gamma' , \Delta $. From  $\Gamma'\wfdash M:\sigma^j\rightarrow\tau$ and the reduction $M \redd M'_{1} + \cdots +  M'_{k} $, one has by IH that  $\Gamma'\wfdash M_1'+\ldots, M_k':\sigma^j\rightarrow\tau$, which entails $\Gamma'\wfdash M_i':\sigma^j\rightarrow\tau$, for $i=1,\ldots, k$, via rule \redlab{FS{:}sum}. Finally, we may type the following applying the $\redlab{FS{:}sum}$ rule:

\begin{prooftree}
\small
            \AxiomC{\(  \forall i \in {1 , \cdots , l} \)}

			 \AxiomC{\( \Theta ;\Gamma' \wfdash M'_{i} : (\sigma^{j} , \eta ) \rightarrow \tau \)}
            \AxiomC{\(  \Theta ;\Delta \wfdash B : (\sigma^{k} , \epsilon )  \)}
            \AxiomC{\( \eta \relunbag \epsilon \)}
            \LeftLabel{\redlab{FS{:}app}}
            \TrinaryInfC{\( \Theta ; \Gamma', \Delta \wfdash M'_{i}\ B : \tau\)}
			
    \BinaryInfC{\( \Gamma', \Delta \wfdash (M'_{1}\ B) + \cdots +  (M'_{l} \ B) : \tau\)}
\end{prooftree}

Since $ \expr{M}'  =   (C[M'_{1}]) + \cdots +  (C[M'_{l}]) = M_1'B+\ldots+M_k'B$, the result follows.
\item  Cases $C=[\cdot]\linexsub{N/x} $ and $C=[\cdot][\widetilde{x} \leftarrow x]$ are similar to the previous.
\item Other cases proceed similarly.

\end{enumerate}

	\item Rule $ \redlab{RS{:}ECont}$. 
	
	This case is analogous to the previous.
	




\item Rule $ \redlab{RS{:}Fail^{\ell}}.$

Then $\expr{M} =   M[\widetilde{x} \leftarrow  {x}]\ \esubst{C \bagsep U}{ x } $ where $C = \bag{N_1}\cdot \dots \cdot \bag{N_l}  $ and  the reduction is:
\begin{prooftree}
     \AxiomC{$ k \neq \size{C} $} 
     \AxiomC{$  \widetilde{y} = (\llfv{M} \setminus \{  \widetilde{x}\} ) \cup \llfv{C} $}
    \LeftLabel{\redlab{RS{:}Fail^{\ell}}}
    \BinaryInfC{\(  M[x_1 , \cdots , x_k \leftarrow  {x}]\ \esubst{C \bagsep U}{ x }  \redd \displaystyle \sum_{C_i \in \perm{C}}  \fail^{\widetilde{y}} \)}
\end{prooftree}

where $\expr{M'}=\sum_{C_i \in \perm{C}}  \fail^{\widetilde{y}}$. Since $\Theta, x: \eta ; \Gamma' , x_1:\sigma,\ldots, x_k:\sigma \wfdash \expr{M}$, one has a derivation:
\begin{prooftree}
    \small
            \AxiomC{\( \Theta, x: \eta ; \Gamma' , x_1:\sigma,\ldots, x_k:\sigma \wfdash M: \tau \)}
            \LeftLabel{ \redlab{FS{:}ex \dash sub} }    
            \UnaryInfC{\( \Theta, x: \eta ;\Gamma' , x:\sigma^{k} \wfdash M[x_1, \cdots , x_k \leftarrow x] : \tau \)}
            \AxiomC{\(\Theta ; \Delta \wfdash C \bagsep U : (\sigma^{j} , \epsilon ) \)}
            \AxiomC{\( \eta \relunbag \epsilon \)}
        \LeftLabel{ \redlab{FS{:}ex \dash sub} }    
        \TrinaryInfC{\(\Theta ; \Gamma', \Delta \wfdash M[x_1 , \cdots , x_k] \leftarrow  {x}]\ \esubst{C \bagsep U}{ x }  : \tau \)}
    \end{prooftree}

where $\Gamma = \Gamma' , \Delta $. We may type the following:
    \begin{prooftree}
        \AxiomC{\( \)}
        \LeftLabel{ \redlab{FS{:}fail}}
        \UnaryInfC{\(\Theta ; \Gamma' , \Delta \wfdash  \fail^{\widetilde{y}} : \tau  \)}
    \end{prooftree}
since $\Gamma',\Delta$ contain assignments on the free variables in $M$ and $B$.
Therefore, $\Theta ;\Gamma\wfdash \fail^{\widetilde{y}}:\tau$, by applying \redlab{FS{:}sum}, it follows that $\Theta ;\Gamma\wfdash \sum_{B_i\in \perm{B}}\fail^{\widetilde{y}}:\tau$,as required.

\item Rule $ \redlab{RS{:}Fail^!}$.

Then $ M \unexsub{U / \unvar{x} } $ where $\headf{M} = {x}[i]$ and $B = U_i = \banged{\oneb} $ and  the reduction is:

\begin{prooftree}
     \AxiomC{$\headf{M} = {x}[i]$}
    \AxiomC{$ U_i = \banged{\oneb} $}
    \AxiomC{\( \widetilde{y} = \llfv{M} \)}
    \LeftLabel{\redlab{RS{:}Fail^!}}
    \TrinaryInfC{\(  M \unexsub{U / \unvar{x} } \redd 
    M \headlin{ \fail^{\emptyset} /{x}[i] } \unexsub{U / \unvar{x} }  \)}
\end{prooftree}

with $\expr{M}'=M \headlin{ \fail^{\emptyset} /{x}[i] } \unexsub{U / \unvar{x} }$. By hypothesis, one has the derivation:
    
\begin{prooftree}
        \AxiomC{\( \Theta , {x} : \eta; \Gamma  \wfdash M : \tau \)}
        \AxiomC{\( \Theta ; \dash \wfdash U : \epsilon \)}
        \AxiomC{\( \eta \relunbag \epsilon \)}
            \LeftLabel{\redlab{FS{:}Esub^!}}
        \TrinaryInfC{\( \Theta ; \Gamma \wfdash M \unexsub{U / \unvar{x}}  : \tau \)}
\end{prooftree}
By Lemma~\ref{ch3lem:subt_lem_sharefailunres_un}, there exists a derivation $\Pi_1$ of  $
    \Theta , \banged{x} : \eta ; \Gamma'  \wfdash   M \headlin{ \fail^{\emptyset} /{x}[i] }  : \tau $. Thus,
    
\begin{prooftree}
        \AxiomC{\( \Theta , \banged{x} : \eta ; \Gamma \wfdash   M \headlin{ \fail^{\emptyset} /{x}[i] }  : \tau  \)}
        \AxiomC{\( \Theta ; \dash \wfdash U : \epsilon \)}
        \AxiomC{\( \eta \relunbag \epsilon \)}
            \LeftLabel{\redlab{FS{:}Esub^!}}
        \TrinaryInfC{\( \Theta ; \Gamma \wfdash  M \headlin{ \fail^{\emptyset} /{x}[i] } \unexsub{U / \unvar{x}}  : \tau \)}
\end{prooftree}

\item Rule $\redlab{RS{:}Cons_1}$.

Then $\expr{M} =   \fail^{\widetilde{x}}\ B $ where $B = \bag{N_1}\cdot \dots \cdot \bag{N_k} $  and  the  reduction is:

\begin{prooftree}
    \AxiomC{\( \widetilde{y} = \llfv{C} \)}
    \LeftLabel{$\redlab{RS{:}Cons_1}$}
    \UnaryInfC{\(\fail^{\widetilde{x}}\ C \bagsep U \redd \displaystyle \sum_{\perm{C}} \fail^{\widetilde{x} \uplus \widetilde{y}}  \)}
\end{prooftree}
and $ \expr{M}'  =  \sum_{\perm{B}} \fail^{\widetilde{x} \cup \widetilde{y}} $.
Since $\Gamma\wfdash \expr{M}:\tau$, one has the derivation: 

    \begin{prooftree}
         \AxiomC{\( \dom{\core{\Gamma}} = \widetilde{x} \)}
        \LeftLabel{\redlab{F{:}fail}}
        \UnaryInfC{\( \Theta ;\Gamma' \wfdash \fail^{\widetilde{x}}: (\sigma^{j} , \eta ) \rightarrow \tau \)}
         \AxiomC{\(  \Theta ;\Delta \wfdash B : (\sigma^{k} , \epsilon )  \)}
         \AxiomC{\( \eta \relunbag \epsilon \)}
            \LeftLabel{\redlab{F{:}app}}
        \TrinaryInfC{\( \Theta ; \Gamma', \Delta \wfdash \fail^{\widetilde{x}}\ B : \tau\)}
    \end{prooftree}

Hence $\Gamma = \Gamma' , \Delta $ and we may type the following:

    \begin{prooftree}
        \AxiomC{\(\dom{\core{\Gamma}} = \widetilde{x} \)}
        \LeftLabel{\redlab{F{:}fail}}
        \UnaryInfC{$ \Theta ;\Gamma \wfdash \fail^{\widetilde{x} \uplus \widetilde{y}} : \tau$}
        \AxiomC{\( \cdots \)}
        \AxiomC{\(\dom{\core{\Gamma}} = \widetilde{x} \)}
        \LeftLabel{\redlab{F{:}fail}}
        \UnaryInfC{$\Theta ;\Gamma \wfdash \fail^{\widetilde{x} \uplus \widetilde{y}} : \tau$}
        \LeftLabel{\redlab{F{:}sum}}
        \TrinaryInfC{$\Theta ; \Gamma \wfdash \sum_{\perm{C}} \fail^{\widetilde{x} \uplus \widetilde{y}}: \tau$}
    \end{prooftree}
    The proof for the cases of $\redlab{RS{:}Cons_2}$, $\redlab{RS{:}Cons_3}$ and $\redlab{RS{:}Cons_4}$ proceed similarly
    \end{enumerate}
\end{proof}

\section{Appendix to \texorpdfstring{\secref{ch3ssec:first_enc}}{}}\label{ch3app:encodingone}

\subsection{Encodability Criteria}
\label{ch3ss:criteria}
We follow the criteria in~\cite{DBLP:journals/iandc/Gorla10}, a widely studied abstract framework for establishing the \emph{quality} of encodings.
A \emph{language} $\mathcal{L}$ is a pair: a set of terms and a reduction semantics $\redd$ on terms (with reflexive, transitive closure denoted $\tred$).
A correct encoding translates terms of a source language $\mathcal{L}_1= (\mathcal{M}, \redd_1)$ into terms of a target language  $\mathcal{L}_2(\mathcal{P}, \redd_2)$ by respecting certain criteria. 
The criteria in~\cite{DBLP:journals/iandc/Gorla10} concern \emph{untyped} languages; because we treat \emph{typed} languages,   we follow~\cite{DBLP:journals/iandc/KouzapasPY19} in requiring that encodings satisfy the following criteria:
\begin{enumerate}
\item {\bf Type preservation:} For every well-typed $M$, it holds that $\encod{M}{}$ is well-typed.

    \item {\bf Operational Completeness:} For every ${M}, {M}'$ such that ${M} \tred_1 {M}'$, it holds that $\encod{{M}}{} \tred_2 \approx_2 \encod{{M}'}{}$.
    
    \item {\bf Operational Soundness:} For every $M$ and $P$ such that $\encod{M}{} \tred_2 P$, there exists an $M'$ such that $M \redd^*_1 M'$ and $P \tred_2 \approx_2 \encod{M'}{}  $.
    
    \item {\bf Success Sensitiveness:} For every ${M}$, it holds that $M \checkmark_1$ if and only if $\encod{M}{} \checkmark_2$, where $\checkmark_1$ and $\checkmark_2$ denote a success predicate in $\mathcal{M}$ and $\mathcal{P}$, respectively. 
    
\end{enumerate}

In addition to these semantic criteria, we shall also consider \emph{compositionality}:  a composite source term is encoded as the combination of the encodings of its sub-terms. 
Success sensitiveness complements completeness and soundness, giving information about observable behaviors. 
The so-called success predicates $\checkmark_1$ and $\checkmark_2$ serve as a minimal notion of \emph{observables}; the criterion then says that observability of success of a source term implies observability of success in the corresponding target term, and vice-versa.

\subsection{Correctness of  \texorpdfstring{$\recencodopenf{\cdot }$}{}}
The correctness of the encoding from $\recencodopenf{\cdot}$ from $\lamrfailunres$ to $\lamrsharfailunres$ relies on an encoding on contexts (\defref{ch3d:enclamcontfailunres}),  auxiliary propositions (Propositions~\ref{ch3prop:linhed_encfail} and  \ref{ch3prop:wf_linsubunres}) for well-formedness preservation (Theorema~\ref{ch3prop:preservencintolamrfailunres}), operational soundness (Theorem~\ref{ch3l:app_completenessone}) and completeness (Theorem~\ref{ch3l:soundnessoneunres}), and success sensitivity (Theorem~\ref{ch3proof:app_successsensce}).

\begin{definition}{Encoding on Contexts} We define an encoding $\recencod{\cdot }$ on contexts:
\label{ch3d:enclamcontfailunres}
\begin{align*}
\recencod{\Theta} &= \Theta
&
\recencod{\emptyset} = \emptyset
\\
\recencod{ x: \tau , \Gamma }  &=   x:\tau  , \recencod{ \Gamma } &  (x \not \in \dom{\Gamma}) \\
\recencod{ x: \tau , \cdots , x: \tau, \Gamma } &= x : \tau \wedge \cdots \wedge \tau , \recencod{ \Gamma } &  (x \not \in \dom{\Gamma})
    \end{align*}
\end{definition}

\begin{proposition}\label{ch3prop:linhed_encfail}
Let $M, N$ be terms. We have:
 \begin{enumerate}
 \item $ \recencodf{M\headlin{N/x}}=\recencodf{M}\headlin{\recencodf{N}/x}$.
 \item $ \recencodf{M\linsub{\widetilde{x}}{x}}=\recencodf{M}\linsub{\widetilde{x}}{x}$, where $\widetilde{x}=x_1,\ldots, x_k$ is sequence of pairwise distinct fresh variables.
 \end{enumerate}
\end{proposition}

\begin{proof}
By induction of the structure of $M$.
\end{proof}

\begin{proposition}[Well-formedness Preservation under Linear Substitutions] 
\label{ch3prop:wf_linsubunres}
Let ${M} \in \lamrfailunres$.
If $\Theta ; \Gamma, x:\sigma \wfdash {M} : \tau$
and $\Theta ; \Gamma \wfdash x_i : \sigma$ then 
 $\Theta ; \Gamma, x_i:\sigma \wfdash {M}\linsub{x_i}{x} : \tau$.
 \end{proposition}

\begin{proof}
 Standard, by induction on the well-formedness derivation rules in \figref{ch3fig:wf_rules_unres}.
 \end{proof}

\begin{proposition}[Well-formedness preservation for $\recencodf{-}$]
\label{ch3prop:preservencintolamrfailunres}
Let $B$ and  $\expr{M}$  be a bag and a expression in $\lamrfailunres$, respectively. 
\begin{enumerate}
\item
    If $\Theta ; \Gamma \wfdash B:(\sigma^{k} , \eta  )$ 
then $ \recencod{\Theta} ; \recencod{\core{\Gamma}}\wfdash \recencodf{B}:(\sigma^{k} , \eta  )$ and $\forall \  x:\pi \in \Gamma, \ \pi = \tau $ for some $\tau$.

    \item 
    If $\Theta ; \Gamma \wfdash \expr{M}:\sigma$ 
then $ \recencod{\Theta} ; \recencod{\core{\Gamma}}\wfdash \recencodf{\expr{M}}:\sigma$ and $\forall \  x:\pi \in \Gamma, \ \pi = \tau $ for some $\tau$.

\end{enumerate}
\end{proposition}

\begin{theorem}[Well-formedness Preservation for $\recencodopenf{-}$]
\label{ch3thm:preservencintolamrfail}
Let $B$ and  $\expr{M}$  be a bag and an expression in $\lamrfailunres$, respectively. 
\begin{enumerate}
\item
    If $\Theta ; \Gamma \wfdash B:(\sigma^{k} , \eta  )$ 
then $\recencod{ \Theta} ;\recencod{\core{\Gamma}}\wfdash \recencodopenf{B}:(\sigma^{k} , \eta  )$.

    \item 
    If $\Theta ;\Gamma \wfdash \expr{M}:\sigma$ 
then $\recencod{\Theta} ;\recencod{\core{\Gamma}}\wfdash \recencodopenf{\expr{M}}:\sigma$.

\end{enumerate}
\end{theorem}

\begin{proof}
By mutual induction on the typing derivations $\Theta; \Gamma\wfdash B:(\sigma^{k} , \eta  )$ and $\Theta; \Gamma\wfdash \expr{M}:\sigma$, exploiting Proposition~\ref{ch3prop:preservencintolamrfailunres}. The analysis for bags Part 1. follows directly from the IHs and will be omitted. 
As for Part 2. there are two main cases to consider:
\begin{enumerate}
    \item $\expr{M} = M$. 
    
Without loss of generality, assume $\lfv{M} = \{x,y\}$. Then, 
\(\Theta; \hat{x}:\sigma_1^j, \hat{y}:\sigma_2^k \wfdash M : \tau \)
where  $\#(x,M)=j$ and 
$\#(y,M)=k$, for some positive integers $j$ and $k$.

After $j+k$ applications of Proposition~\ref{ch3prop:wf_linsubunres}
we obtain:
\begin{equation*}\label{ch3eq:thmpres2fail}
\Theta; x_1:\sigma_1, \cdots, x_j:\sigma_1, y_1:\sigma_2, \cdots, y_k:\sigma_2 \wfdash M\linsub{\widetilde{x}}{x}\linsub{\widetilde{y}}{y} : \tau    
\end{equation*}
 where  $\widetilde{x}=x_{1},\cdots, x_{j}$
   and $\widetilde{y}=y_{1},\cdots, y_{k}$. 
From Proposition~\ref{ch3prop:preservencintolamrfailunres}
one has
\begin{equation*}\label{ch3eq:thmpres3fail}
\recencod{\Theta}; \recencod{x_1:\sigma_1, \cdots, x_j:\sigma_1, y_1:\sigma_2, \cdots, y_k:\sigma_2} \wfdash \recencodf{M\linsub{\widetilde{x}}{x}\linsub{\widetilde{y}}{y}} : \tau  
\end{equation*}

Since $\recencod{x_1:\sigma_1, \cdots, x_j:\sigma_1, y_1:\sigma_2, \cdots, y_k:\sigma_2}= x_1:\sigma_1, \cdots, x_j:\sigma_1, y_1:\sigma_2, \cdots, y_k:\sigma_2$ and $\recencod{\Theta} = \Theta $, 
we have the following derivation:
\begin{prooftree}
    \AxiomC{$\Theta; {x_1:\sigma_1, \cdots, x_j:\sigma_1, y_1:\sigma_2, \cdots, y_k:\sigma_2} \wfdash \recencodf{M\linsub{\widetilde{x}}{x}\linsub{\widetilde{y}}{y}} : \tau$}
    \LeftLabel{$\redlab{FS:share}$}
    \UnaryInfC{$\Theta; x:\sigma_1^j, y_1:\sigma_2, \cdots, y_k:\sigma_2 \wfdash \recencodf{M\linsub{\widetilde{x}}{x}\linsub{\widetilde{y}}{y}}[\widetilde{x}\leftarrow x] : \tau$}
\LeftLabel{$\redlab{FS:share}$}
    \UnaryInfC{$\Theta; x:\sigma_1^j, y:\sigma_2^k \wfdash \recencodf{M\linsub{\widetilde{x}}{x}\linsub{\widetilde{y}}{y}}[\widetilde{x}\leftarrow x][\widetilde{y}\leftarrow y] : \tau$}
    \end{prooftree}
    
    By expanding \defref{ch3def:enctolamrsharfailunres}, we have 
$$
 \recencodopenf{M} = 
\recencodf{M\linsub{\widetilde{x}}{x}\linsub{\widetilde{y}}{y}}[\widetilde{x}\leftarrow x][\widetilde{y}\leftarrow y] 
$$

    which completes the proof for this case. 
    
    \item $\expr{M} = M_1 + \cdots + M_n$:
    
    This case proceeds easily by IH, using Rule~$\redlab{FS:sum}$.
    \end{enumerate}
\end{proof}

\begin{theorem}[Operational Completeness]
\label{ch3l:app_completenessone}
Let $\expr{M}, \expr{N}$ be well-formed $\lamrfailunres$ expressions. 
Suppose $\expr{N}\redd_{\redlab{R}} \expr{M}$.
\begin{enumerate}
\item If $\redlab{R} =  \redlab{R:Beta}$  then $ \recencodopenf{\expr{N}}  \redd^{\leq 2}\recencodopenf{\expr{M}}$;

\item If $\redlab{R} = \redlab{R:Fetch}$   then   $ \recencodopenf{\expr{N}}  \redd^+ \recencodopenf{\expr{M}'}$, for some $ \expr{M}$. 
\item If $\redlab{R} \neq \redlab{R:Beta}$    and $\redlab{R} \neq \redlab{R:Fetch}$ then   $ \recencodopenf{\expr{N}}  \redd \recencodopenf{\expr{M}}$.
\end{enumerate}
\end{theorem}

\begin{proof}
We proceed by induction on the the rule from \figref{ch3fig:reductions_lamrfailunres} applied to infer $\expr{N}\redd \expr{M}$, distinguishing the three cases: (below $\widetilde{[x_1\leftarrow x_n]}$ abbreviates  
            $[\widetilde{x_1}\leftarrow x_1]\cdots [\widetilde{x_n}\leftarrow x_n]$).

\begin{enumerate}

    \item The rule applied is $\redlab{R}=\redlab{R:Beta}$. 
    
    In this case, 
        $\expr{N}= (\lambda x. M') B$, where  $B = C \bagsep U$, the reduction is 
    \begin{prooftree}
        \AxiomC{}
        \LeftLabel{\redlab{R:Beta}}
        \UnaryInfC{\((\lambda x. M) B \redd M\ \esubst{B}{x}\)}
    \end{prooftree}

          and $\expr{M}= M'\esubst{B}{x}$. Below we assume $\llfv{\expr{N}}=\{x_1,\ldots, x_k\}$ and $\widetilde{x_i}=x_{i_1},\ldots, x_{i_{j_i}}$, where $j_i= \#(x_i, N)$, for $1\leq i\leq k$.
        On the one hand, we have:
        \begin{equation}\label{ch3eq:beta1fail}
            \begin{aligned}
            \recencodopenf{\expr{N}}&= \recencodopenf{(\lambda x. M')B}
            =  \recencodf{((\lambda x. M')B) \langle{ {\widetilde{x_1}}/ {x_1}}\rangle\cdots \langle{ {\widetilde{x_k}}/ {x_k}}\rangle}\widetilde{[x_1\leftarrow x_k]}
            \\
            &=  \recencodf{(\lambda x. M^{''})B'}\widetilde{[x_1\leftarrow x_k]}
            =  (\recencodf{\lambda x. M^{''}}\recencodf{B'})\widetilde{[x_1\leftarrow x_k]} \\
            &=  ((\lambda x.\recencodf{ M^{''}\langle{ {\widetilde{y}}/ {x}}\rangle}[ {\widetilde{y}}\leftarrow  {x}])\recencodf{B'})\widetilde{[x_1\leftarrow x_k]} \\
            &\redd_{\redlab{RS:Beta}} (\recencodf{ M^{''} \langle{ {\widetilde{y}}/ {x}} \rangle} [ {\widetilde{y}} \leftarrow  {x}] \esubst{\recencodf{B'}}{x}) \widetilde{[x_1\leftarrow x_k]}=\expr{L}
            \end{aligned}
        \end{equation}
      
        On the other hand, we have:
        \begin{equation}\label{ch3eq:beta2fail}
            \begin{aligned}
               \recencodopenf{\expr{M}}&=\recencodopenf{M'\esubst{B}{x}}=\recencodf{M'\esubst{B}{x}\langle{ {\widetilde{x_1}}/ {x_1}}\rangle\cdots \langle{ {\widetilde{x_k}}/ {x_k}}\rangle} \widetilde{[x_1\leftarrow x_n]}\\
               &=\recencodf{M^{''}\esubst{B'}{x}} \widetilde{[x_1\leftarrow x_k]}
            \end{aligned}
        \end{equation}

        We need to analyze two sub-cases: either $\#( {x},M) = \size{C} $ or $\#( {x},M) = k \geq 0$ and our first sub-case is not met.
        \begin{enumerate}
            \item If $\#( {x},M) = \size{C}  $ then we can reduce $\expr{L}$ as: (via {\redlab{RS:Ex-sub}})
            \begin{equation*}
                \begin{aligned}
            \expr{L}\redd & \sum_{C_i\in \perm{\recencodf{C}}}\recencodf{M^{''}\linsub{ {\widetilde{y}}}{ {x}}}\linexsub{C_i(1)/ {y_1}}\cdots \linexsub{C_i(n)/ {y_n}} \unexsub{U/ \unvar{x}} \widetilde{[x_1\leftarrow x_k]}  \\
                   =&\recencodopenf{\expr{M}}
                \end{aligned}
            \end{equation*}
            
            From \eqref{ch3eq:beta1fail} and \eqref{ch3eq:beta2fail} and $ {\widetilde{y}}= {y_1}\ldots  {y_n}$, one has the result.

            \item Otherwise,  $\#( {x},M) = n \geq 0$.
            


            Expanding  the encoding in \eqref{ch3eq:beta2fail} :
            \begin{align*}
                \recencodopenf{M}&= \recencodf{M^{''}\esubst{B'}{x}} \widetilde{[x_1\leftarrow x_k]} 
                = (\recencodf{ M^{''} \langle{ {\widetilde{y}}/ {x}} \rangle} [ {\widetilde{y}} \leftarrow  {x}] \esubst{\recencodf{B'}}{x}) \widetilde{[x_1\leftarrow x_k]}
            \end{align*}
           Therefore  $\recencodopenf{M} =\expr{L}$ and $\recencodopenf{\expr{N}}\redd \recencodopenf{\expr{M}}$.
        
        \end{enumerate}

    \item The rule applied is $\redlab{R}=\redlab{R:Fetch^{\ell}}$. 
    
    Then $\expr{N}=M\esubst{ C \bagsep U  }{x } $ and  the reduction applying the $\redlab{R:Fetch^{\ell}}$ rule is: 

    \begin{prooftree}
        \AxiomC{$\headf{M} =  {x}$}
        \AxiomC{$C = {\bag{N_1}}\cdot \dots \cdot {\bag{N_k}} \ , \ k\geq 1 $}
        \AxiomC{$ \#( {x},M) = k $}
        \TrinaryInfC{\(
        M\esubst{ C \bagsep U  }{x } \redd M \headlin{ N_{1}/ {x} } \esubst{ (C \setminus N_1)\bagsep U}{ x }  + \cdots + M \headlin{ N_{k}/ {x} } \esubst{ (C \setminus N_k)\bagsep U}{x}
        \)}
    \end{prooftree}

    with $\expr{M}= M \headlin{ N_{1}/ {x} } \esubst{ (C \setminus N_1)\bagsep U}{ x }  + \cdots + M \headlin{ N_{k}/ {x} } \esubst{ (C \setminus N_k)\bagsep U}{x}$.

    Below we assume $\lfv{\expr{N}}=\{x_1,\ldots, x_k\}$ and $\widetilde{x_i}=x_{i_1},\ldots, x_{i_{j_i}}$, where $j_i= \#(x_i, N)$, for $1\leq i\leq k$.
    On the one hand, we have: (last rule is {\redlab{RS{:}Fetch^{\ell}}})
        \begin{equation*}
            \begin{aligned}
            \recencodopenf{\expr{N}}&= \recencodopenf{M\esubst{ C \bagsep U  }{x}}=  \recencodf{M\esubst{ C \bagsep U  }{x } \langle{ {\widetilde{x_1}}/ {x_1}}\rangle\cdots \langle{ {\widetilde{x_k}}/ {x_k}}\rangle}\widetilde{[x_1\leftarrow x_k]}
            \\
            &=  \recencodf{ M'\esubst{ C' \bagsep U  }{x } }\widetilde{[x_1\leftarrow x_k]}\\
            &=   \sum_{C_i \in \perm{\recencodf{ C'  } }}  (\recencodf{ M' \langle  {\widetilde{y}}/  {x}  \rangle } \linexsub{C_i(1)/ {y}_1} \cdots \linexsub{C_i(k)/ {y}_k}\unexsub{U/\unvar{x}} )\widetilde{[x_1\leftarrow x_k]} \\
            &=  \sum_{C_i \in \perm{\recencodf{ C'  } }}  (\recencodf{ M'' } \linexsub{C_i(1)/ {y}_1} \cdots \linexsub{C_i(k)/ {y}_k}\unexsub{U/\unvar{x}} )\widetilde{[x_1\leftarrow x_k]} \\
            &\redd \sum_{C_i \in \perm{\recencodf{ C'  } }}  (\recencodf{ M''\headlin{ C_i(1) /  {y}_1 } } \linexsub{C_i(2)/ {y}_2} \\
            & \qquad \qquad \qquad \qquad \cdots \linexsub{C_i(k)/ {y}_k}\unexsub{U/\unvar{x}} )\widetilde{[x_1\leftarrow x_k]} \\
             &=\expr{L}
            \end{aligned}
        \end{equation*}
        We assume for simplicity that $  \headf{M''} =  {y}_1$
        On the other hand, we have:
        \begin{equation*}
            \begin{aligned}
               \recencodopenf{\expr{M}}&=\recencodopenf{M \headlin{ N_{1}/ {x} } \esubst{ (C \setminus N_1)\bagsep U}{ x }  + \cdots + M \headlin{ N_{k}/ {x} } \esubst{ (C \setminus N_k)\bagsep U}{x}}\\
               &=\sum_{C_i \in \perm{\recencodf{ C'  } }}  (\recencodf{ M''\headlin{ C_i(1) /  {y}_1 } } \linexsub{C_i(2)/ {y}_2} \\
               &  \qquad \qquad \qquad \qquad \cdots \linexsub{C_i(k)/ {y}_k}\unexsub{U/\unvar{x}} )\widetilde{[x_1\leftarrow x_k]} \\
               &=\expr{L}
            \end{aligned}
        \end{equation*}
        From these developments from $\recencodopenf{\mathbb{N}}$ and $\recencodopenf{\mathbb{M}}$ ,  and $ {\widetilde{y}}= {y_1}\ldots  {y_n}$, one has the result.

    \item The rule applied is $\redlab{R}=\redlab{R:Fetch^!}$. 
    
    Then $\expr{N}=M\esubst{ C \bagsep U  }{x } $ and  the reduction is 

    \begin{prooftree}
        \AxiomC{$\headf{M} = {x}[i] \quad U_i = \banged{\bag{N}} $}
        \LeftLabel{\redlab{R:Fetch^!}}
        \UnaryInfC{\(
        M\ \esubst{ C \bagsep U  }{x } \redd M \headlin{ N/ {x}[i] } \esubst{ C \bagsep U}{ x } 
        \)}
    \end{prooftree}   

    with $\expr{M}= M \headlin{ N/ {x}[i] } \esubst{ C \bagsep U}{ x }$.
    Below we assume $\lfv{\expr{N}}=\{x_1,\ldots, x_k\}$ and $\widetilde{x_i}=x_{i_1},\ldots, x_{i_{j_i}}$, where $j_i= \#(x_i, N)$, for $1\leq i\leq k$.

    On the one hand, we have: (the last rule is {\redlab{RS:Fetch^!}})
        {\small
        \begin{equation}\label{ch3eq:unfetch1fail}
            \begin{aligned}
            \recencodopenf{\expr{N}}&= \recencodopenf{M\esubst{ C \bagsep U  }{x}}=  \recencodf{M\esubst{ C \bagsep U  }{x } \langle{ {\widetilde{x_1}}/ {x_1}}\rangle\cdots \langle{ {\widetilde{x_k}}/ {x_k}}\rangle}\widetilde{[x_1\leftarrow x_k]}
            \\
            &=  \recencodf{ M'\esubst{ C' \bagsep U  }{x } }\widetilde{[x_1\leftarrow x_k]}\\
            &=   \sum_{C_i \in \perm{\recencodf{ C'  } }}  (\recencodf{ M' \langle  {\widetilde{y}}/  {x}  \rangle } \linexsub{C_i(1)/ {y}_1} \cdots \linexsub{C_i(k)/ {y}_k}\unexsub{U/ \unvar{x}} )\widetilde{[x_1\leftarrow x_k]} \\
            &=  \sum_{C_i \in \perm{\recencodf{ C'  } }}  (\recencodf{ M'' } \linexsub{C_i(1)/ {y}_1} \cdots \linexsub{C_i(k)/ {y}_k}\unexsub{U/ \unvar{x}} )\widetilde{[x_1\leftarrow x_k]}
            \\ &\redd \sum_{C_i \in \perm{\recencodf{ C'  } }}  (\recencodf{ M''\headlin{ N / {x}[i]  } } \linexsub{C_i(2)/ {y}_2} \cdots \linexsub{C_i(k)/ {y}_k}\unexsub{U/ \unvar{x}} )\widetilde{[x_1\leftarrow x_k]}\\
              &= \expr{L}
            \end{aligned}
        \end{equation}}
        On the other hand, assuming for simplicity that $  \headf{M''} = {x}[i]$ and $U_i = N$, we have 
        {\small
        \begin{equation}\label{ch3eq:unfetch2fail}
            \begin{aligned}
            \recencodopenf{\expr{M}}&=\recencodopenf{M \headlin{ N/ {x}[i] } \esubst{ C \bagsep U}{ x } }\\
               &=\sum_{C_i \in \perm{\recencodf{ C'  } }}  (\recencodf{ M''\headlin{ N / {x}[i]  } } \linexsub{C_i(2)/ {y}_2} \cdots \linexsub{C_i(k)/ {y}_k}\unexsub{U/\unvar{x}} )\widetilde{[x_1\leftarrow x_k]}=\expr{L}
            \end{aligned}
        \end{equation}}
        From \eqref{ch3eq:unfetch1fail} and \eqref{ch3eq:unfetch2fail}, one has the result.

        \item The rule applied is $ \redlab{R}\neq \redlab{R:Beta}$ and $ \redlab{R}\neq \redlab{R:Fetch}$. There are two possible cases:
            \begin{enumerate}
                 
                \item  $\redlab{R}=\redlab{R:Fail^{\ell}}$
                
                Then $\expr{N}=M\esubst{C \bagsep U}{x }$ and the reduction is 
                
                \begin{prooftree}    
                        \AxiomC{$\#( {x},M) \neq \size{C} $} 
                         \AxiomC{$\widetilde{z} = (\mlfv{M} \!\setminus x) \uplus \mlfv{C} $}
                        \LeftLabel{\redlab{R:Fail^{\ell}}}
                        \BinaryInfC{\(  M\esubst{C \bagsep U}{x } \redd \sum_{\perm{C}} \fail^{\widetilde{z}} \)}
                    \end{prooftree}
                where $\expr{M}=  \sum_{\perm{C}} \fail^{\widetilde{y}}$. Below assume $\lfv{\expr{N}}=\{x_1,\ldots, x_n\}$.
                
                    On the one hand, we have:
     \begin{equation*}\label{ch3eq:fail1fail}
                    \begin{aligned}
                    \recencodopenf{\expr{N}} &= \recencodopenf{M\esubst{C \bagsep U}{x }}= \recencodf{M\esubst{C \bagsep U}{x }\langle  {\widetilde{x_1}}/ {x_1}\rangle\cdots \langle  {\widetilde{x_n}}/ {x_n}\rangle } \widetilde{[x_1\leftarrow x_n]} \\
                     &= \recencodf{M'\esubst{C' \bagsep U}{x } } \widetilde{[x_1\leftarrow x_n]} \\
                    &=  \recencodf{M'\langle y_1, \cdots ,  {y_k} /  {x}  \rangle} [ {y_1}, \cdots ,  {y_k} \leftarrow  {x}] \esubst{C' \bagsep U}{x }\widetilde{[x_1\leftarrow x_n]}\\
                        & \redd_{\redlab{RS:Fail^{\ell}}} \sum_{\perm{C}} \fail^{\widetilde{y},  {\widetilde{x_1}}, \cdots ,  {\widetilde{x_n}}}\widetilde{[x_1\leftarrow x_n]}    =\expr{L}\\
                    \end{aligned}
                \end{equation*}
                
                On the other hand, we have:
            \begin{equation*}\label{ch3eq:fail2fail}
                    \begin{aligned}
                    \recencodopenf{\expr{M}}  &= \sum_{\perm{C}} \recencodopenf{\fail^{\widetilde{z}}}
                    = \sum_{\perm{C}} \fail^{\widetilde{y},  {\widetilde{x_1}}, \cdots ,  {\widetilde{x_n}}}\widetilde{[x_1\leftarrow x_n]}    =\expr{L}
                    \end{aligned}
                \end{equation*}
                
                Therefore, $\recencodopenf{\expr{N}} \redd \recencodopenf{\expr{M}} $ and the result follows.

                \item  $\redlab{R}=\redlab{R:Fail^!}$
                
                Then $\expr{N}=M\esubst{C \bagsep U}{x }$ and the reduction is 
                
                \begin{prooftree} 
                    \AxiomC{$\#( {x},M) = \size{C} $}
                    \AxiomC{$U_i = \banged{\oneb}$}
                    \AxiomC{$\headf{M} = {x}[i]  $}
                    \LeftLabel{\redlab{R:Fail^!}}
                    \TrinaryInfC{\(  M \esubst{C \bagsep U}{x } \redd  M \headlin{ \fail^{\emptyset} / {x}[i] } \esubst{ C \bagsep U}{ x }\)}
                \end{prooftree}
                
                where $\expr{M}=  M \headlin{ \fail^{\emptyset} / {x}[i] } \esubst{ C \bagsep U}{ x } $. 
                

                Below we assume $\lfv{\expr{N}}=\{x_1,\ldots, x_k\}$ and $\widetilde{x_i}=x_{i_1},\ldots, x_{i_{j_i}}$, where $j_i= \#(x_i, N)$, for $1\leq i\leq k$.
            
                On the one hand, we have: (the last rule applied was {\redlab{RS:Fail^!}} )
                   {\small  \begin{equation*}\label{ch3eq:unfetch1failp2}
                        \begin{aligned}
                        \recencodopenf{\expr{N}}&= \recencodopenf{M\esubst{ C \bagsep U  }{x}}=  \recencodf{M\esubst{ C \bagsep U  }{x } \langle{ {\widetilde{x_1}}/ {x_1}}\rangle\cdots \langle{ {\widetilde{x_k}}/ {x_k}}\rangle}\widetilde{[x_1\leftarrow x_k]}
                        \\
                        &=  \recencodf{ M'\esubst{ C' \bagsep U  }{x } }\widetilde{[x_1\leftarrow x_k]}\\
                        &=   \sum_{C_i \in \perm{\recencodf{ C'  } }}  (\recencodf{ M' \langle  {\widetilde{y}}/  {x}  \rangle } \linexsub{C_i(1)/ {y}_1} \cdots \linexsub{C_i(k)/ {y}_k}\unexsub{U/\unvar{x} } )\widetilde{[x_1\leftarrow x_k]}\\
                        &=  \sum_{C_i \in \perm{\recencodf{ C'  } }}  (\recencodf{ M'' } \linexsub{C_i(1)/ {y}_1} \cdots \linexsub{C_i(k)/ {y}_k}\unexsub{U/ \unvar{x} } )\widetilde{[x_1\leftarrow x_k]} \\
                        &\redd \sum_{C_i \in \perm{\recencodf{ C'  } }}  (\recencodf{ M''\headlin{  \fail^{\emptyset} / {x}[i]  } } \linexsub{C_i(2)/ {y}_2} \\
                        & \qquad \qquad \qquad \qquad \cdots \linexsub{C_i(k)/ {y}_k}\unexsub{U/\unvar{x}} )\widetilde{[x_1\leftarrow x_k]}\\
                        &=\expr{L}
                        \end{aligned}
                    \end{equation*}
                    }
                    We assume for simplicity that $  \headf{M''} = {x}[i]$.
                    On the other hand, we have:
                   {\small  \begin{equation*}\label{ch3eq:unfetch2failp2}
                        \begin{aligned}
                           \recencodopenf{\expr{M}}&=\recencodopenf{M \headlin{ \fail^{\emptyset}/ {x}[i] } \esubst{ C \bagsep U}{ x } }\\
                           &=\sum_{C_i \in \perm{\recencodf{ C'  } }}  (\recencodf{ M''\headlin{ \fail^{\emptyset} / {x}[i]  } } \linexsub{C_i(2)/ {y}_2}\\
                           & \qquad \qquad \qquad \qquad \cdots \linexsub{C_i(k)/ {y}_k}\unexsub{U/\unvar{x}} )\widetilde{[x_1\leftarrow x_k]}  \\
                           &=\expr{L}
                        \end{aligned}
                    \end{equation*}
            }
                 From the $\recencodopenf{M}$ and  $\recencodopenf{N}$ above one has the result.

            \item  $\redlab{R}= \redlab{R:Cons_1}$.
            
            Then $\expr{N}=(\fail^{\widetilde{z}})\ C \bagsep U $ and the reduction is

            \begin{prooftree}
                    \AxiomC{$\widetilde{z} = \mlfv{C} $}
                \LeftLabel{$\redlab{R:Cons_1}$}
                \UnaryInfC{\(  (\fail^{\widetilde{y}})\ C \bagsep U \redd {}  \sum_{\perm{C}} \fail^{\widetilde{y} \uplus \widetilde{z}} \)}
            \end{prooftree}
            

           and $\expr{M}'= \sum_{\perm{B}} \fail^{\widetilde{y} \uplus \widetilde{z}}$. Below we assume $\lfv{\expr{N}}=\{x_1,\ldots, x_n\}$.
           
                On the one hand, we have:  
            
            \begin{equation*}\label{ch3eq:consume1fail}
                \begin{aligned}
                \recencodopenf{N} &= \recencodopenf{\fail^{\widetilde{y}}\ B}= \recencodf{ \fail^{\widetilde{y}}\  C \bagsep U \langle \widetilde{x_1}/x_1\rangle\cdots \langle \widetilde{x_n}/x_n\rangle } \widetilde{[x_1\leftarrow x_n]}\\
                 &= \recencodf{ \fail^{\widetilde{y'}}\ \ C' \bagsep U  } \widetilde{[x_1\leftarrow x_n]}
                = \recencodf{ \fail^{\widetilde{y'}}} \ \recencodf{ C' \bagsep U } \widetilde{[x_1\leftarrow x_n]}\\
                &= \fail^{\widetilde{y'}} \ \recencodf{ C' \bagsep U } \widetilde{[x_1\leftarrow x_n]}
                 \redd_{\redlab{RS:Cons_1}} \sum_{\perm{B}} \fail^{\widetilde{y'} \cup \widetilde{z'}}  \widetilde{[x_1\leftarrow x_n]}\\
                 &=\expr{L}\\
                \end{aligned}
            \end{equation*}
            Where $\widetilde{y'} \cup \widetilde{z'} = \widetilde{x_1} , \cdots , \widetilde{x_n}$. On the other hand, we have:
            
            \begin{equation*}\label{ch3eq:consume2fail}
                \begin{aligned}
                \recencodopenf{M}  
                = \sum_{\perm{B}} \recencodf{\fail^{\widetilde{y'} \uplus \widetilde{z'}}}\widetilde{[x_1\leftarrow x_n]}
                = \sum_{\perm{B}} \fail^{\widetilde{y'} \cup \widetilde{z'}}\widetilde{[x_1\leftarrow x_n]}=\expr{L}
                \end{aligned}
            \end{equation*}
            
            Therefore, $\recencodopenf{\expr{N}}\redd \expr{L}= \recencodopenf{ \expr{M}}$, and the result follows.

            \item $\redlab{R}= \redlab{R:Cons_2}$
            
            Then $\expr{N}= \fail^{\widetilde{y}}\ \esubst{C \bagsep U}{z} $ and the reduction is 
            
            \begin{prooftree}
                \AxiomC{$ \#(z , \widetilde{y}) =  \size{C}\quad \widetilde{z} = \mlfv{C} $}
                \LeftLabel{$\redlab{R:Cons_2}$}
                \UnaryInfC{$\fail^{\widetilde{y}}\ \esubst{C \bagsep U}{z}  \redd {}\displaystyle\sum_{\perm{C}} \fail^{(\widetilde{x} \setminus z) \uplus\widetilde{z}}$}
            \end{prooftree}
            
      and $\expr{M}=\sum_{\perm{C}} \fail^{(\widetilde{y} \setminus x) \uplus\widetilde{z}}$. 
            Below we assume $\lfv{\expr{N}}=\{x_1,\ldots, x_n\}$.
            
            
            On the one hand, we have:
            
            \begin{equation}\label{ch3eq:consume3fail}
                \hspace{-10mm}
                \begin{aligned}
                \recencodopenf{\expr{N}} &= \recencodopenf{\fail^{\widetilde{y}}\ \esubst{C \bagsep U}{z}}
                = \recencodf{ \fail^{\widetilde{y}}\ \esubst{C \bagsep U}{z} \langle \widetilde{x_1}/x_1\rangle\cdots \langle \widetilde{x_n}/x_n\rangle } \widetilde{[x_1\leftarrow x_n]}\\
                &=  \sum_{C_i \in \perm{\recencodf{ C'  } }}\recencodf{ \fail^{\widetilde{y'}} \langle  {\widetilde{y}}/  {x}  \rangle } \linexsub{C_i(1)/ {y}_1} \cdots \linexsub{C_i(k)/ {y}_k}\unexsub{U/\unvar{x}} \widetilde{[x_1\leftarrow x_n]}\\
                & \redd^*_{\redlab{RS{:}Cons_3}} \sum_{C_i \in \perm{\recencodf{ C'  } }}\recencodf{ \fail^{(\widetilde{y'} \setminus  {\widetilde{y}} ) \uplus\widetilde{z}}} \unexsub{U/\unvar{x}} \widetilde{[x_1\leftarrow x_n]}\\
                & \redd^*_{\redlab{RS{:}Cons_4}} \sum_{C_i \in \perm{\recencodf{ C'  } }}\recencodf{ \fail^{(\widetilde{y'} \setminus  {\widetilde{y}} ) \uplus\widetilde{z}}} \widetilde{[x_1\leftarrow x_n]}\\
                \end{aligned}
            \end{equation}
            
            As $\widetilde{y}$ consists of free variables, we have that in $\fail^{\widetilde{y}}\ \esubst{C \bagsep U}{x} \langle \widetilde{x_1}/x_1\rangle\cdots \langle \widetilde{x_n}/x_n\rangle$ the substitutions also occur on $ \widetilde{y}$ resulting in a new $\widetilde{y'}$ where all $x_i$'s are replaced with their fresh components in $\widetilde{x_i}$. Similarly $\widetilde{y''}$ is $\widetilde{y'}$ with each $x$ replaced with a fresh $y_i$. On the other hand, we have:
            
            \begin{equation}\label{ch3eq:consume4fail}
                \begin{aligned}
                \recencodopenf{M} = \recencodopenf{\sum_{\perm{C}} \fail^{(\widetilde{y} \setminus x) \uplus\widetilde{z}}}
                &= \sum_{C_i \in \perm{\recencodf{ C'  } }}\recencodf{ \fail^{(\widetilde{y'} \setminus  {\widetilde{y}} ) \uplus\widetilde{z}}} \widetilde{[x_1\leftarrow x_n]}
                \end{aligned}
            \end{equation}
            
            The reductions in \eqref{ch3eq:consume3fail} and \eqref{ch3eq:consume4fail} lead to identical expressions.
        \end{enumerate}
    \end{enumerate}
    
     As before, the reduction via rule $ \redlab{R} $ could occur inside a context (cf. Rules $\redlab{R:TCont}$ and $\redlab{R:ECont}$). We consider only the case when the contextual rule used is $\redlab{R:TCont}$. We have $\expr{N} = C[N]$. When we have $C[N] \redd_{\redlab{R}} C[M] $ such that $N \redd_{\redlab{R}} M$ we need to show that $\recencodopenf{ C[N]} \redd^j \recencodopenf{ C[M] }$for some $j$ dependent on ${\redlab{R}}$. Firstly let us assume ${\redlab{R}} = \redlab{R:Cons_2}$  then we take $j = 1$. Let us take $C[\cdot]$ to be $[\cdot]B$ and $\lfv{NB} = \{ x_1, \cdots , x_k  \}$ then 
     \[
        \begin{aligned}
            \recencodopenf{ N B}  & = \recencodf{NB\linsub{\widetilde{x_{1}}}{x_1}\cdots \linsub{\widetilde{x_k}}{x_k}}\widetilde{[x_1\leftarrow x_k]} = \recencodf{N' B'}\widetilde{[x_1\leftarrow x_k]}  = \recencodf{N'}\recencodf{ B'}\widetilde{[x_1\leftarrow x_k]} \\
        \end{aligned}
     \]
    We take $N'B'= NB\linsub{\widetilde{x_{1}}}{x_1}\cdots \linsub{\widetilde{x_k}}{x_k}$, we have by the IH that $ \recencodf{N}\redd \recencodf{M}$ and hence we can deduce that $\recencodf{N'}\redd \recencodf{M'}$ where $M'B'= MB\linsub{\widetilde{x_{1}}}{x_1}\cdots \linsub{\widetilde{x_k}}{x_k}$. Finally we have
    \(            \recencodf{N'}\recencodf{ B'}\widetilde{[x_1\leftarrow x_k]} \redd \recencodf{M'}\recencodf{ B'}\widetilde{[x_1\leftarrow x_k]}
        \)
     and hence $ \recencodopenf{C[N]} \redd \recencodopenf{C[M]} $.
\end{proof}

\begin{theorem}[Operational Soundness]
\label{ch3l:soundnessoneunres}
Let $\expr{N}$ be a well-formed $\lamrfailunres$ expression. 
Suppose $ \recencodopenf{\expr{N}}  \redd \expr{L}$. Then, there exists $ \expr{N}' $ such that $ \expr{N}  \redd_{\redlab{R}} \expr{N}'$ and 

\begin{enumerate}
 \item If $\redlab{R} = \redlab{R:Beta}$ then $\expr{ L } \redd^{\leq 1} \recencodopenf{\expr{N}'}$;

    \item If $\redlab{R} \neq \redlab{R:Beta}$ then $\expr{ L } \redd^*  \recencodopenf{\expr{N}''}$, for $ \expr{N}''$ such that  $\expr{N}' \pequiv \expr{N}''$.
\end{enumerate}
\end{theorem}

\begin{proof}
By induction on the structure of $\expr{N}$:

\begin{enumerate}

    \item Cases $\expr{N} =  {x}$, $\expr{N} =  x[i]$, $\fail^{\widetilde{y}}$ and $\expr{N} =  \lambda x. N$, are trivial, since no reductions can be performed.
    
    
    



    \item $\expr{N} = N B$:
    
    Suppose  $\llfv{NB} = \{  {x}_1, \cdots ,  {x}_n\}$. Then,

    \begin{equation}\label{ch3eq:app_npfail}
    \begin{aligned}
        \recencodopenf{\expr{N}}=\recencodopenf{NB}  = \recencodf{NB\langle  {\widetilde{x_1}} /  {x}_1 \rangle \cdots \langle  {\widetilde{x_n}} /  {x}_n \rangle} \widetilde{[x_1\leftarrow x_n]}
         &= \recencodf{N' B'} \widetilde{[x_1\leftarrow x_n]}\\
        & = \recencodf{N'} \recencodf{B'} \widetilde{[x_1\leftarrow x_n]}
    \end{aligned}
    \end{equation}
    where $ {\widetilde{x_i}}= {x}_{i1},\ldots,  {x}_{ij_i}$, for $1\leq i \leq n$.
    By the reduction rules in \figref{ch3fig:share-reductfailureunres} there are three possible reductions starting in $\expr{N}$:
    \begin{enumerate}
        \item $\recencodf{N'}\recencodf{B'}\widetilde{[x_1\leftarrow x_n]}$ reduces via a $\redlab{RS:Beta}$.
        
        In this case  $N=\lambda x. N_1$, and the encoding in (\ref{ch3eq:app_npfail}) gives $N'= N\langle  {\widetilde{x_1}} /  {x}_1 \rangle \cdots \langle  {\widetilde{x_n}} /  {x}_n \rangle$, which  implies $N' =\lambda x. N_1^{'}$ and the following holds:
        \begin{equation*}
        \begin{aligned}
            \recencodf{N'}=\recencodf{(\lambda x. N'_1)} &= (\lambda x. \recencodf{N'_1 \langle  {\widetilde{y}} /  {x} \rangle}  {[\widetilde{y}} \leftarrow  {x}])
             = (\lambda x. \recencodf{N^{''}} [ {\widetilde{y}} \leftarrow  {x}])
        \end{aligned}
        \end{equation*}
        Thus, we have the following $\redlab{RS:Beta}$ reduction from \eqref{ch3eq:app_npfail}:
        \begin{equation}\label{ch3eq:sound.appfail}
            \begin{aligned}
                \recencodopenf{\expr{N}} &= \recencodf{N'} \recencodf{B'}\widetilde{[x_1\leftarrow x_n]}
                =(\lambda x. \recencodf{N''} [ {\widetilde{y}} \leftarrow  {x}] \recencodf{B'}) \widetilde{[x_1\leftarrow x_n]}\\
                &\redd_{\redlab{RS:Beta}}  \recencodf{N^{''}} [ {\widetilde{y}} \leftarrow  {x}] \esubst{\recencodf{B'}}{x}  \widetilde{[x_1\leftarrow x_n]} =\expr{L}
            \end{aligned}
        \end{equation}

        Notice that the expression $\expr{N}$ can perform the following $\redlab{R:Beta}$ reduction:
        \[\expr{N}=(\lambda x. N_1) B\redd_{\redlab{R:Beta}} N_1 \esubst{B}{x} \]

        Assuming $\expr{N'}=N_1 \esubst{B}{x}$ and we take $B = C \bagsep U$, there are two cases:
        
        \begin{enumerate}
            
            \item$\#( {x},M) = \size{C} = k$.

            On the one hand, 
            
            \begin{equation*}\label{ch3eq:sound_appn1fail}
            \begin{aligned}
                \recencodopenf{\expr{N'}}&=\recencodopenf{N_1 \esubst{B}{x}}
                = \recencodf{N_1 \esubst{B}{x}\langle  {\widetilde{x_1}} /  {x}_1 \rangle \cdots \langle  {\widetilde{x_n}} /  {x}_n \rangle} \widetilde{[x_1\leftarrow x_n]}\\
                & = \recencodf{N_1' \esubst{B'}{x}}\widetilde{[x_1\leftarrow x_n]}\\
                & = \sum_{C_i \in \perm{\recencodf{ C' }}}\recencodf{ N_1' \langle  {y}_1 , \cdots ,  {y}_k / x  \rangle } \linexsub{C_i(1)/ {y}_1}\\
                & \qquad \qquad \qquad \qquad \cdots \linexsub{C_i(k)/ {y}_k}\unexsub{U /\unvar{x}} \widetilde{[x_1\leftarrow x_n]}\\
                & = \sum_{C_i \in \perm{\recencodf{ C' }}}\recencodf{ N_1''} \linexsub{C_i(1)/ {y}_1}\\
                & \qquad \qquad \qquad \qquad \cdots \linexsub{C_i(k)/ {y}_k}\unexsub{U /\unvar{x}} \widetilde{[x_1\leftarrow x_n]}\\
            \end{aligned}
            \end{equation*}
            On the other hand, via application of rule \redlab{RS:Ex\dash Sub}
            \begin{equation*}\label{ch3eq:sound_appn2fail}
            \begin{aligned}
                \expr{L} =& \recencodf{N''} [ {\widetilde{y}} \leftarrow  {x}] \esubst{\recencodf{B'}}{x}  \widetilde{[x_1\leftarrow x_n]} \\
                \redd& \sum_{C_i \in \perm{\recencodf{ C }}}\recencodf{ N_1''} \linexsub{C_i(1)/ {y}_1} \cdots \linexsub{C_i(k)/ {y}_k}\unexsub{U /\unvar{x}} \widetilde{[x_1\leftarrow x_n]} \\
                =& \recencodopenf{\expr{N}'}
            \end{aligned}
            \end{equation*}
            
         and the result follows.
            
            \item Otherwise $\#( {x},N_1)\neq \size{C}$.

            In this case, 
            \begin{equation*}\label{ch3eq:sound_appn3fail}
            \begin{aligned}
                \recencodopenf{\expr{N'}}&=\recencodopenf{N_1 \esubst{B}{x}}= \recencodf{N_1 \esubst{B}{x}\langle  {\widetilde{x_1}} /  {x}_1 \rangle \cdots \langle  {\widetilde{x_n}} /  {x}_n \rangle}\widetilde{[x_1\leftarrow x_n]}\\
                & = \recencodf{N_1' \esubst{B'}{x}}\widetilde{[x_1\leftarrow x_n]}
                 =  \recencodf{N^{''}} [\widetilde{y} \leftarrow x] \esubst{\recencodf{B'}}{x} \widetilde{[x_1\leftarrow x_n]}\\
                 & = \expr{L} 
            \end{aligned}
            \end{equation*}
            From (\ref{ch3eq:sound.appfail}):  $\recencodopenf{\expr{N}}\redd \expr{L}=\recencodopenf{\expr{N'}}$ and the result follows.

        \end{enumerate}
        
        \item $\recencodf{N'}\recencodf{B'}\widetilde{[x_1\leftarrow x_n]}$ reduces via a $\redlab{RS: Cons_1}$.

        In this case,  $N=\fail^{\widetilde{y}}$, and the encoding in (\ref{ch3eq:app_npfail}) gives $N'= N\linsub{ {\widetilde{x_1}}}{ {x}_1}\ldots \linsub{ {\widetilde{x_n}}}{ {x}_n}$, which implies $N'
        =\fail^{\widetilde{y'}} $, we let $B = C \bagsep U$, $\widetilde{z} = \llfv{C'}$ and the following:
        \begin{equation*}\label{ch3eq:sound.consumfail}
            \begin{aligned}
                \recencodopenf{\expr{N}} &= \recencodf{N'} \recencodf{B'}\widetilde{[x_1\leftarrow x_n]}= \recencodf{\fail^{\widetilde{y'}}} \recencodf{B'}\widetilde{[x_1\leftarrow x_n]}\\
                &= \fail^{\widetilde{y'}} \recencodf{B'}\widetilde{[x_1\leftarrow x_n]} \redd \sum_{\perm{C}} \fail^{\widetilde{y'} \uplus \widetilde{z}}\widetilde{[x_1\leftarrow x_n]}.
            \end{aligned}
        \end{equation*}
         The expression $\expr{N}$ can perform the  reduction:
        
        \begin{equation*}\label{ch3eq:sound.2consumfail}
            \expr{N}=\fail^{\widetilde{y}} \  B\redd_{\redlab{R: Cons1}} \sum_{\perm{C}} \fail^{\widetilde{y}\uplus \widetilde{z}}, \text{  where } \widetilde{z} = \mlfv{C}
        \end{equation*}
        
        Thus, $\expr{L}=\recencodopenf{\expr{N'}}$ and so the result follows.
        
        \item Suppose that $\recencodf{N'} \redd \recencodf{N''}$.  
        This case follows from the induction hypothesis.
    \end{enumerate}

    \item  $\expr{N} = N \esubst{B}{x}$:
    
    Suppose  $\llfv{N \esubst{B}{x}} = \{  {x}_1, \cdots ,  {x}_k\}$. Then,
    
    \begin{equation}\label{ch3eq:sound_expsubfail}
    \begin{aligned}
    \recencodopenf{\expr{N}}= \recencodopenf{N \esubst{B}{x}}
         &= \recencodf{N \esubst{B}{x}\langle  {\widetilde{x_1}} /  {x}_1 \rangle \cdots \langle  {\widetilde{x_k}} /  {x}_k \rangle} \widetilde{[x_1\leftarrow x_k]}\\
         &= \recencodf{N' \esubst{B'}{x}} \widetilde{[x_1\leftarrow x_k]}
        \end{aligned}
    \end{equation}
    
    Let us consider the two possibilities of the encoding where we take $B = C \bagsep U$:
    
    \begin{enumerate}
        
        \item Where $ \#( {x},M) = \size{B} = k $
        
        Then we continue equation (\ref{ch3eq:sound_expsubfail}) as follows
        
        \begin{equation}\label{ch3eq:sound_expsub2fail}
            \begin{aligned}
                \recencodopenf{\expr{N}} &= \recencodf{N' \esubst{B'}{x}} \widetilde{[x_1\leftarrow x_k]}\\
                &=  \sum_{C_i \in \perm{\recencodf{ C' }}}\recencodf{ N' \langle  {y}_1 , \cdots ,  {y}_n /  {x}  \rangle } \linexsub{C_i(1)/ {y}_1} \\
                & \qquad \qquad \qquad \qquad \cdots \linexsub{C_i(n)/ {y}_n} \unexsub{U/\unvar{x}}\widetilde{[x_1\leftarrow x_k]} \\
                &=  \sum_{C_i \in \perm{\recencodf{ C' }}}\recencodf{ N'' } \linexsub{C_i(1)/ {y}_1} \cdots \linexsub{C_i(n)/ {y}_n} \unexsub{U/\unvar{x}} \widetilde{[x_1\leftarrow x_k]} \\
            \end{aligned}
        \end{equation}
        
        There are five possible reductions that can take place, these being $\redlab{RS{:}Fetch^{\ell}}$, $\redlab{RS{:} Fetch^!}$, $\redlab{RS{:}Fail^!}$ , $\redlab{RS:Cons_3}$ and when we apply the \redlab{RS:Cont} rules
        
        \begin{enumerate}
            
            \item Suppose that $\headf{N''} =  {y}_1$ and for simplicity we assume $C'$ has only one element $N_1$ then from (\ref{ch3eq:sound_expsub2fail}) and buy letting $C' = \bag{N'_1}$ we have
                    
                    \begin{equation*}
                        \begin{aligned}
                            \recencodopenf{\expr{N}} 
                            & = \recencodf{ N'' } \linexsub{\recencod{N_1'}/ {y}_1} \unexsub{U/\unvar{x}} \widetilde{[x_1\leftarrow x_k]}\\
                            &\redd \recencodf{N^{''}}\headlin{\recencod{N_1'}/ {y}_1}\unexsub{U/\unvar{x}}\widetilde{[x_1\leftarrow x_k]} = \expr{L}
                        \end{aligned}
                    \end{equation*}
Also,
\(\expr{N} 
= N\esubst{\bag{N_1} \bagsep U}{x}                            \redd N\headlin{N_1/x}\esubst{ \oneb \bagsep U}{x} = \expr{N}'.\)
Then  $\expr{L}'=\recencodopenf{\expr{N'}}$ and the result follows.

        \item Suppose that $\headf{N''} = {x}[i]$ and then from (\ref{ch3eq:sound_expsub2fail}) we have
                    
\begin{equation*}\label{ch3eq:sound_expsub3failp2}
\begin{aligned}
\recencodopenf{\expr{N}} 
 =& \recencodf{ N'' } \linexsub{C_i(1)/ {y}_1} \cdots \linexsub{C_i(n)/ {y}_n} \unexsub{U/\unvar{x}}\widetilde{[x_1\leftarrow x_k]}\\
\redd& \recencodf{N^{''}}\headlin{ U_{i} / {x}[i] } \linexsub{C_i(1)/ {y}_1} \cdots \linexsub{C_i(n)/ {y}_n}\unexsub{U/\unvar{x}}\widetilde{[x_1\leftarrow x_k]}\\
=& \expr{L}
  \end{aligned}
  \end{equation*}
We also have that
                            \(\expr{N} 
                             = N\esubst{C \bagsep U}{x}
                            \redd N\headlin{U_{ind}/\banged{x}}\esubst{ C \bagsep U}{x} = \expr{N}'.\)
                            
                    
                    Then,  $\expr{L}'=\recencodopenf{\expr{N'}}$ and so the result follows.

            \item Suppose that $N'' = \fail^{\widetilde{z'}}$ proceed similarly then from (\ref{ch3eq:sound_expsub2fail})
            
            {\small
            \begin{equation*}\label{ch3eq:sound_expsub99fail}
                \begin{aligned}
                    \recencodopenf{\expr{N}} 
                    & = \sum_{C_i \in \perm{\recencodf{ C' }}}\fail^{\widetilde{z'}} \linexsub{C_i(1)/ {y}_1} \cdots \linexsub{C_i(n)/ {y}_n}\unexsub{U/\unvar{x}}\widetilde{[x_1\leftarrow x_k]}\\
                    &\redd^* \sum_{C_i \in \perm{\recencodf{ C' }}}\fail^{(\widetilde{z'} \setminus  {y}_1, \cdots ,  {y}_n) \uplus\widetilde{y}}\unexsub{U/\unvar{x}} \widetilde{[x_1\leftarrow x_k]} \\
                    &\redd^* \sum_{C_i \in \perm{\recencodf{ C' }}}\fail^{(\widetilde{z'} \setminus  {y}_1, \cdots ,  {y}_n) \uplus\widetilde{y}} \widetilde{[x_1\leftarrow x_k]} 
                    = \expr{L}'
                \end{aligned}
            \end{equation*}}
                    
         where $\widetilde{y} = \llfv{C_i(1)} \uplus \cdots \uplus \llfv{C_i(n)}$.       We also have that
                    
                    \begin{equation*}\label{ch3eq:sound_expsub11fail}
                        \begin{aligned}
                            \expr{N} 
                            & = \fail^{\widetilde{z}} \esubst{B}{x} 
                             \redd \fail^{(\widetilde{z} \setminus x) \uplus\widetilde{y}} 
                            = \expr{N}',  \text{ where }  \widetilde{y} = \mfv{B}. 
                        \end{aligned}
                    \end{equation*}

                Then, $\expr{L}'=\recencodopenf{\expr{N'}}$ and so the result follows.
            
            \item  Suppose that $N'' \redd N'''$.
                This case follows by the induction hypothesis
            
        \end{enumerate}

        \item Otherwise, we continue from equation (\ref{ch3eq:sound_expsubfail}), where $\#(x,M) \not= k$, as follows
        
            \begin{equation*}
            \begin{aligned}
                \recencodopenf{\expr{N}} &= \recencodf{N' \esubst{B'}{x}}\widetilde{[x_1\leftarrow x_k]} \\
                &=  \recencodf{N'\langle  {y}_1. \cdots ,  {y}_k / x  \rangle} [ {y}_1. \cdots ,  {y}_k \leftarrow  {x}] \esubst{ \recencodf{B'} }{ x }\widetilde{[x_1\leftarrow x_k]} \\
                &=  \recencodf{N''} [ {y}_1. \cdots ,  {y}_k \leftarrow  {x}] \esubst{ \recencodf{B'} }{ x }\widetilde{[x_1\leftarrow x_k]}\\
            \end{aligned}
            \end{equation*}
            
            We can perform the reduction
                
            \begin{equation*}\label{ch3eq:sound_expsubotherwise2}
            \begin{aligned}
                \recencodopenf{\expr{N}} &= \recencodf{N''} [ {y}_1. \cdots ,  {y}_k \leftarrow  {x}] \esubst{ \recencodf{B'} }{ x } \widetilde{[x_1\leftarrow x_k]} \\
                &\redd \sum_{C_i \in \perm{C}}  \fail^{\widetilde{z'}}\widetilde{[x_1\leftarrow x_k]}, \text{ where } \widetilde{z'} = \lfv{N''} \uplus \lfv{C'}
                = \expr{L}'
            \end{aligned}
            \end{equation*}
        We also have that
            \begin{equation*}\label{ch3eq:sound_expsubotherwise3}
                \begin{aligned}
                    \expr{N} 
                    & = N \esubst{C}{x}  \redd \sum_{\perm{C}} \fail^{\widetilde{z}}  = \expr{N}' , \text{ where } \widetilde{z} = \mlfv{M} \uplus \mlfv{C}.
                \end{aligned}
            \end{equation*}
        Then, $\expr{L}'=\recencodopenf{\expr{N'}}$ and so the result follows.

    \end{enumerate}

        

    \item $\expr{N} = \expr{N}_1 + \expr{N}_2$: \\ Then this case holds by the induction hypothesis.
\end{enumerate}
\end{proof}

\subsection{Success Sensitiveness of  \texorpdfstring{$\recencodopenf{\cdot}$}{}}

We now consider success sensitiveness, a property that complements (and relies on) operational completeness and soundness. For the purposes of the proof, we consider the extension of $\lamrfailunres$ and $\lamrsharfailunres$ with dedicated constructs and predicates that specify success. 

\begin{definition}{}
We extend the syntax of terms for $\lamrfailunres$ and $\lamrsharfailunres$ with the same $\checkmark$ construct. 
In both cases, we assume $\checkmark$ is well formed. 
Also, we also define $\headf{\checkmark} = \checkmark$ and $\recencodf{\checkmark} = \checkmark$
\end{definition}

An expression $\expr{M}$ has success, denoted
\succp{\expr{M}}{\checkmark}, when there is a sequence of reductions from \expr{M} that leads to an expression that includes a summand that contains an occurrence of $\checkmark$ in head position.

\begin{definition}{Success in \lamrfailunres and \lamrsharfailunres}
\label{ch3def:app_Suc3unres}
In $\lamrfailunres$ and $\lamrsharfailunres $, we define 
\succp{\expr{M}}{\checkmark} if and only if
there exist  $M_1 , \cdots , M_k$ such that 
$\expr{M} \redd^*  M_1 + \cdots + M_k$ and
$\headf{M_j'} = \checkmark$, for some  $j \in \{1, \ldots, k\}$ and term $M_j'$ such that $M_j\pequiv  M_j'$.
\end{definition}

\begin{notation}

    We use the notation $\headsum{\expr{M}}$ to be that $\forall M_i, M_j \in \expr{M}$ we have that $\head{M_i} = \head{M_j}$ hence we say that $\headsum{\expr{M}} = \headf{M_i}$ for some $M_i \in \expr{M}$

\end{notation}

\begin{proposition}[Preservation of Head term]
\label{ch3Prop:checkpres}
The head of a term is preserved when applying the encoding $\recencodf{\dash}$. That is to say:
\[ \forall M \in \lamrfailunres \quad \headf{M} = \checkmark \iff \headsum{\recencodopenf{M}} = \checkmark \]
\end{proposition}

\begin{proof}

By induction on the structure of $M$. We only need to consider terms of the following form.

\begin{enumerate}

    \item When $ M = \checkmark $ the case is immediate.
    
    \item When $ M = NB $ with $\lfv{NB} = \{x_1,\cdots,x_k\}$ and  $\#(x_i,M)=j_i$ we have that: 

        \[ 
            \begin{aligned}
                \headsum{\recencodopenf{NB}} &= \headsum{\recencodf{NB\linsub{\widetilde{x_{1}}}{x_1}\cdots \linsub{\widetilde{x_k}}{x_k}}[\widetilde{x_1}\leftarrow x_1]\cdots [\widetilde{x_k}\leftarrow x_k]}\\
                &= \headsum{\recencodf{NB}}
                 = \headsum{\recencodf{N}} 
            \end{aligned}
        \]
         and $\head{NB}= \head{N} $, by the IH we have $\head{N} = \checkmark \iff \headsum{\recencodf{N}} = \checkmark$.
         
    \item When $M = N \esubst{C \bagsep U}{x}$, we must have that $\#(x,M) = \size{C }$ for the head of this term to be $\checkmark$. Let $\lfv{N \esubst{C \bagsep U}{x}} = \{x_1,\cdots,x_k\}$ and  $\#(x_i,M)=j_i$. We have that: 
        \[\small
            \hspace{-1cm}
            \begin{aligned}
                \headsum{\recencodopenf{N \esubst{C \bagsep U}{x}}} &= \headsum{\recencodf{N \esubst{C \bagsep U}{x}\linsub{\widetilde{x_{1}}}{x_1}\cdots \linsub{\widetilde{x_k}}{x_k}}\widetilde{[x_1\leftarrow x_k]}}\\
                &= \headsum{\recencodf{N \esubst{C \bagsep U}{x}}}\\
                &= \headsum{\sum_{C_i \in \perm{\recencodf{ C }}}\recencodf{ N \langle \widetilde{x} / x  \rangle } \linexsub{C_i(1)/x_1} \cdots \linexsub{C_i(k)/x_k}\unexsub{ U/ \unvar{x}}}\\
                &= \headsum{\recencodf{ N \langle \widetilde{x} / x  \rangle } \linexsub{C_i(1)/x_1} \cdots \linexsub{C_i(k)/x_k}\unexsub{ U/ \unvar{x}}}\\
                &= \headsum{\recencodf{ N \langle \widetilde{x} / x  \rangle } } 
            \end{aligned}
        \]
        
        and $\head{N \esubst{B}{x}} = \head{N}$, by the IH  $\head{N} = \checkmark \iff \headsum{\recencodf{N}} = \checkmark$.
\end{enumerate}
\end{proof}

\begin{theorem}[Success Sensitivity]
\label{ch3proof:app_successsensce}
Let  \expr{M} be a well-formed expression.
We have
$\expr{M} \Downarrow_{\checkmark}$ if and only if $\recencodopenf{\expr{M}} \Downarrow_{\checkmark}$.
\end{theorem}

\begin{proof}
By induction on the structure of expressions $\lamrfailunres$ and $\lamrsharfailunres$.

\begin{enumerate}
    
    \item Suppose that  $\expr{M} \Downarrow_{\checkmark} $. We will prove that $\recencodopenf{\expr{M}} \Downarrow_{\checkmark}$.

    By operational completeness (\thmref{ch3l:app_completenessone}) we have that if $\expr{M}\redd_{\redlab{R}} \expr{M'}$ then

    \begin{enumerate}
        \item If $\redlab{R} =  \redlab{R:Beta}$  then $ \recencodopenf{\expr{M}}  \redd^{\leq 2}\recencodopenf{\expr{M}'}$;

        \item If $\redlab{R} =\redlab{R:Fetch}$   then   $ \recencodopenf{\expr{M}}  \redd^+ \recencodopenf{\expr{M}''}$, for some $ \expr{M}''$ such that  $\expr{M}' \pequiv \expr{M}''$. 
        \item If $\redlab{R} \neq  \redlab{R:Beta}$ and $\redlab{R}\neq \redlab{R:Fetch}$  then $ \recencodopenf{\expr{M}}  \redd\recencodopenf{\expr{M}'}$;
    \end{enumerate}
    Notice that  neither our  reduction rules  (in \defref{ch3fig:share-reductfailureunres}), or our congruence $\pequiv$ (in \figref{ch3def:rsPrecongruencefailure}),  or  our encoding ($\recencodopenf{\checkmark }=\checkmark$)  create or destroy a $\checkmark$ occurring in the head of term. By Proposition \ref{ch3Prop:checkpres} the encoding preserves the head of a term being $\checkmark$. The encoding acts homomorphically over sums, therefore, if a $\checkmark$ appears as the head of a term in a sum, it will stay in the encoded sum. We can iterate the operational completeness lemma and obtain the result.

    \item Suppose that $\recencodopenf{\expr{M}} \Downarrow_{\checkmark}$. We will prove that $ \expr{M} \Downarrow_{\checkmark}$. 

    By operational soundness (\thmref{ch3l:soundnessoneunres}) we have that if $ \recencodopenf{\expr{M}}  \redd \expr{L}$ then there exist $ \expr{M}' $ such that $ \expr{M}  \redd_{\redlab{R}} \expr{M}'$ and 

    \begin{enumerate}
        \item If $\redlab{R} = \redlab{R:Beta}$ then $\expr{ L } \redd^{\leq 1} \recencodopenf{\expr{M}'}$;

    \item If $\redlab{R} \neq \redlab{R:Beta}$ then $\expr{ L } \redd^*  \recencodopenf{\expr{M}''}$, for $ \expr{M}''$ such that  $\expr{M}' \pequiv \expr{M}''$.
    \end{enumerate}
    
   Since $\recencodopenf{\expr{M}}\redd^* M_1+\ldots+M_k$, and $\headf{M_j'}=\checkmark$, for some $j$ and $M_j'$, s.t. $M_j\pequiv M_j'$. 
   
   Notice that if $\recencodopenf{\expr{M}}$ is itself a term headed with $\checkmark$, say $\headf{\recencodopenf{\expr{M}}}=\checkmark$, then $\expr{M}$ is itself headed with $\checkmark$, from Proposition \ref{ch3Prop:checkpres}.
   In the case $\recencodopenf{\expr{M}}= M_1+\ldots+M_k$, $k\geq 2$, and $\checkmark$ occurs in the head of an $M_j$, the reasoning is similar.  $\expr{M}$ has one of the forms:
   \begin{enumerate}
       \item   $\expr{M}= N_1$, then $N_1$ must contain the subterm $ M\esubst{C \bagsep U}{x}$ and $\size{C}=\#(x,M)$.
       
       The encoding of $\expr{M}$ is:
       
       {\small
       $\recencodopenf{M\esubst{C \bagsep U}{x}}=\sum_{C_i \in \perm{\recencodf{ C }}}\recencodf{M\linsub{\widetilde{x}}{x}}\linexsub{C_i(1)/x_i}\ldots \linexsub{C_i(k)/x_i}\unexsub{U / \unvar{x} }$}.

       We can apply Proposition~\ref{ch3Prop:checkpres} and the result follows.
       
       \item $\expr{M}=N_1+\ldots+N_l$ for $l \geq 2$.
       
       This reasoning is similar and uses the fact  that the encoding distributes homomorphically over sums.
   \end{enumerate}
   
   In the case where $\recencodopenf{\expr{M}}\redd^+ M_1+\ldots+M_k$, and $\headf{M_j'}=\checkmark$, for some $j$ and $M_j'$, such that $M_j\pequiv M_j'$, the reasoning is similar to the previous, since our reduction rules do not introduce/eliminate $\checkmark$ occurring in the head of terms. 
\end{enumerate}
\end{proof}


\section{Appendix to \texorpdfstring{\secref{ch3ssec:second_enc}}{}}\label{ch3app:encodingtwo}

\subsection{Type Preservation}

\begin{lemma}\label{ch3prop:app_auxunres}
 $ \piencodf{\sigma^{j}}_{(\tau_1, m)} = \piencodf{\sigma^{k}}_{(\tau_2, n)}$ and $ \piencodf{(\sigma^{j} , \eta)}_{(\tau_1, m)} = \piencodf{(\sigma^{k}, \eta)}_{(\tau_2, n)}$ hold, provided that  $\tau_1,\tau_2,n$ and $m$ are as follows:

        \begin{enumerate}
        \item If $j > k$ then take $\tau_1 $ to be an arbitrary type, $m = 0$,  take $\tau_2 $ to be $\sigma$ and $n = j-k$.
        
        \item If $j < k$ then take $\tau_1 $ to be $\sigma$, $m = k-j$,  take $\tau_2 $ to be an arbitrary type and $n = 0$. 
        
        \item Otherwise, if $j = k$ then take $m = n = 0$. In this case, $\tau_1. \tau_2 $ are unimportant.
    \end{enumerate}
    
\end{lemma}

\begin{proof}
We shall prove the case of $(1)$ for the first equality, and the case for the second equality and of $(2)$ are analogous. The case of $(3)$ follows  by the encoding on types in \defref{ch3def:enc_sestypfailunres}. 

Hence take $j,k,\tau_1,\tau_2, m,n$ satisfying the conditions in (1): $j > k$, $\tau_1 $ to be an arbitrary type,  $m = 0$,  $\tau_2 =\sigma$ and $n = j-k$. 
    We want to show that $ \piencodf{\sigma^{j}}_{(\tau_1, 0)} = \piencodf{\sigma^{k}}_{(\sigma, n)} $. In fact, 
    \[
        \begin{aligned}
            \piencodf{\sigma^{k}}_{(\sigma, n)} &= \oplus(( \with \onef) \ampy ( \oplus  \with (( \oplus \piencodf{\sigma} ) \otimes (\piencodf{\sigma^{k-1}}_{(\sigma, n)}))))\\
            \piencodf{\sigma^{k-1}}_{(\sigma, n)} &= \oplus(( \with \onef) \ampy ( \oplus  \with (( \oplus \piencodf{\sigma} ) \otimes (\piencodf{\sigma^{k-2}}_{(\sigma, n)}))))\\
            \vdots\\
            \piencodf{\sigma^{1}}_{(\sigma, n)} &= \oplus(( \with \onef) \ampy ( \oplus  \with (( \oplus \piencodf{\sigma} ) \otimes (\piencodf{\omega}_{(\sigma, n)}))))
        \end{aligned}
    \]
    and
    \[
        \begin{aligned}
            \piencodf{\sigma^{j}}_{(\tau_1, 0)} &= \oplus(( \with \onef) \ampy ( \oplus  \with (( \oplus \piencodf{\sigma} ) \otimes (\piencodf{\sigma^{j-1}}_{(\tau_1, 0)}))))\\
            \piencodf{\sigma^{j-1}}_{(\tau_1, 0)} &= \oplus(( \with \onef) \ampy ( \oplus  \with (( \oplus \piencodf{\sigma} ) \otimes (\piencodf{\sigma^{j-2}}_{(\tau_1, 0)}))))\\
            \vdots\\
            \piencodf{\sigma^{j-k + 1}}_{(\tau_1, 0)} &= \oplus(( \with \onef) \ampy ( \oplus  \with (( \oplus \piencodf{\sigma} ) \otimes (\piencodf{\sigma^{j-k}}_{(\tau_1, 0)}))))
        \end{aligned}
    \]
    Notice that $n = j-k$, hence we wish to show that $ \piencodf{\sigma^{n}}_{(\tau_1, 0)} = \piencodf{\omega}_{(\sigma, n)} $.  Finally,
    
    \[
        \begin{aligned}
            \piencodf{\omega}_{(\sigma, n)} & = \oplus(( \with \onef) \ampy ( \oplus  \with (( \oplus \piencodf{\sigma} ) \otimes (\piencodf{\omega}_{(\sigma, n-1)})))) \\
            \piencodf{\omega}_{(\sigma, n-1)} & = \oplus(( \with \onef) \ampy ( \oplus  \with (( \oplus \piencodf{\sigma} ) \otimes (\piencodf{\omega}_{(\sigma, n-2)})))) \\
            \vdots\\
            \piencodf{\omega}_{(\sigma, 1)} & = \oplus(( \with \onef) \ampy ( \oplus  \with (( \oplus \piencodf{\sigma} ) \otimes (\piencodf{\omega}_{(\sigma, 0)})))) \\
            \piencodf{\omega}_{(\sigma, 0)} &= \oplus(( \with \onef) \ampy ( \oplus  \with \onef ) \\
        \end{aligned}
    \]
    and 
    \[
        \begin{aligned}
            \piencodf{\sigma^{n}}_{(\tau_1, 0)} &= \oplus(( \with \onef) \ampy ( \oplus  \with (( \oplus \piencodf{\sigma} ) \otimes (\piencodf{\sigma^{n-1}}_{(\tau_1, 0)}))))\\
            \piencodf{\sigma^{n-1}}_{(\tau_1, 0)} &= \oplus(( \with \onef) \ampy ( \oplus  \with (( \oplus \piencodf{\sigma} ) \otimes (\piencodf{\sigma^{n-2}}_{(\tau_1, 0)}))))\\
            \vdots\\
            \piencodf{\sigma^{1}}_{(\tau_1, 0)} &= \oplus(( \with \onef) \ampy ( \oplus  \with (( \oplus \piencodf{\sigma} ) \otimes (\piencodf{\omega}_{(\tau_1, 0)})))) \\
            \piencodf{\omega}_{(\tau_1, 0)} &= \oplus(( \with \onef) \ampy ( \oplus  \with \onef ) \\
        \end{aligned}
    \]
\end{proof}


        
        
    

\begin{lemma}\label{ch3lem:relunbag-typeunres}

If 
$ \eta \relunbag \epsilon $ Then
\begin{enumerate}
    \item If $\piencodf{M}_u\vdash \piencodf{\Gamma} ; \piencodf{\Theta} , \banged{x} : \piencodf{\eta}$
    then 
    $\piencodf{M}_u\vdash \piencodf{\Gamma} ; \piencodf{\Theta}, \banged{x} : \piencodf{\epsilon}$.
    
    \item If $\piencodf{M}_u\vdash \piencodf{\Gamma}, u:\piencodf{(\sigma^{j} , \eta ) \rightarrow \tau} ; \piencodf{\Theta}$
    then 
    $\piencodf{M}_u\vdash \piencodf{\Gamma}, u:\piencodf{(\sigma^{j} , \epsilon ) \rightarrow \tau} ; \piencodf{\Theta}$.

\end{enumerate}

\end{lemma}

\begin{proof}

\begin{enumerate}
    
    \item We consider the first case where if $\piencodf{M}_u\vdash \piencodf{\Gamma} ; \piencodf{\Theta} , \banged{x} : \piencodf{\eta}$
    then 
    $\piencodf{M}_u\vdash \piencodf{\Gamma} ; \piencodf{\Theta}, \banged{x} : \piencodf{\epsilon}$ and by \defref{ch3def:enc_sestypfailunres},  $\piencodf{ \eta } = \&_{\eta_i \in \eta} \{ \mathtt{l}_i ; \piencodf{\eta_i} \}$. We now proceed by induction on the structure of $M$:
    \begin{enumerate}
        
        \item $M =   {x}$.
        \label{ch3proof:relunbag-no}
       
        By \figref{ch3fig:encfailunres},  $\piencodf{ {x}}_u =  {x}.\overline{\some} ; [ {x} \leftrightarrow u] $. We have the following derivation:
                
        \begin{prooftree}
            \AxiomC{}
            \LeftLabel{$\redlab{ (Tid)}$}
            \UnaryInfC{$ [ {x} \leftrightarrow u] \vdash  {x}:  \overline{A}  , u :  A; \piencodf{\Theta} , \banged{x} : \piencodf{\eta}$}
            \LeftLabel{$\redlab{T\with^{x}_{d})}$}
            \UnaryInfC{$  {x}.\overline{\some} ; [ {x} \leftrightarrow u] \vdash  {x}: \with  \overline{A} , u :  A ; \piencodf{\Theta} , \banged{x} : \piencodf{\eta} $}
        \end{prooftree}
        
        For some type A. Notice the derivation is independent of $\banged{x} : \piencodf{\eta}$ , hence holds when $ M =   {x}$. Note that we do not consider $M =  {y}$ where $y \not = x$, this is due to the case being trivial due to the typing of $y$ being independent on $x$.

        \item $ M =  {x}[ind]$.
        \label{ch3proof:relunbag-yes}

            By \figref{ch3fig:encfailunres},  $\piencodf{ {x}[ind]}_u = \outsev{\banged{x}}{{x_i}}. {x}_i.l_{ind}; [{x_i} \leftrightarrow u] $. We have the following derivation applying $\redlab{Tcopy}$:
            
            \begin{prooftree}
                \small
                    \AxiomC{}
                    \LeftLabel{$\redlab{ (Tid)}$}
                    \UnaryInfC{$ [{x_i} \leftrightarrow u]  \vdash  u :  \piencodf{ \tau },  x_i:  \overline{\piencodf{\eta_{ind} }}  ; \banged{x}:\&_{\eta_i \in \eta} \{ \mathtt{l}_i ; \piencodf{\eta_i} \} , \piencodf{\Theta} $}
                \LeftLabel{\redlab{T\oplus_i}}
                \UnaryInfC{$  {x}_i.l_{ind}; [{x_i} \leftrightarrow u] \vdash  u :  \piencodf{ \tau }, {x}_i :  \oplus_{\eta_i \in \eta} \{ \mathtt{l}_i ; \dual{\piencodf{\eta_i}}  \} ; \banged{x}:\oplus_{\eta_i \in \eta} \{ \mathtt{l}_i ; \dual{\piencodf{\eta_i}} \} , \piencodf{\Theta}  $}
                \UnaryInfC{$ \outsev{\banged{x}}{{x_i}}. {x}_i.l_{ind}; [{x_i} \leftrightarrow u] \vdash  u :  \piencodf{ \tau }; \banged{x}:\oplus_{\eta_i \in \eta} \{ \mathtt{l}_i ; \dual{\piencodf{\eta_i}} \} , \piencodf{\Theta} $}
            \end{prooftree}
            
            On the other hand we have derivation:
            
            \begin{prooftree}
                \small
                    \AxiomC{}
                    \LeftLabel{$\redlab{ (Tid)}$}
                    \UnaryInfC{$ [{x_i} \leftrightarrow u]  \vdash  u :  \piencodf{ \tau },  x_i:  \overline{\piencodf{\epsilon_{ind} }}  ; \banged{x}:\&_{\epsilon_i \in \epsilon} \{ \mathtt{l}_i ; \piencodf{\epsilon_i} \} , \piencodf{\Theta} $}
                \LeftLabel{\redlab{T\oplus_i}}
                \UnaryInfC{$  {x}_i.l_{ind}; [{x_i} \leftrightarrow u] \vdash  u :  \piencodf{ \tau }, {x}_i :  \oplus_{\epsilon_i \in \epsilon} \{ \mathtt{l}_i ; \dual{\piencodf{\epsilon_i}}  \} ; \banged{x}:\oplus_{\epsilon_i \in \epsilon} \{ \mathtt{l}_i ; \dual{\piencodf{\epsilon_i}} \} , \piencodf{\Theta}  $}
                \LeftLabel{\redlab{Tcopy}}
                \UnaryInfC{$ \outsev{\banged{x}}{{x_i}}. {x}_i.l_{ind}; [{x_i} \leftrightarrow u] \vdash  u :  \piencodf{ \tau }; \banged{x}:\oplus_{\epsilon_i \in \epsilon} \{ \mathtt{l}_i ; \dual{\piencodf{\epsilon_i}} \} , \piencodf{\Theta} $}
            \end{prooftree}
            
            By $ \eta \relunbag \epsilon $ we have that $ \epsilon_{ind} = \eta_{ind} $. Similarly for the case of $ M =  {y}[ind]$ with $y \not =  x$ we use the argument that the typing of $y$ is independent on $x$.
        
        \item  $M = M' [  {\widetilde{y}} \leftarrow  {y} ] $.
        
        If $y = x$ the case proceeds similarly to (\ref{ch3proof:relunbag-no}) otherwise we proceed by induction on $M'$.
            
        \item $ M =  \lambda x . (M'[ {\widetilde{x}} \leftarrow  {x}])$.
 
            From 
            \defref{ch3def:enc_lamrsharpifailunres} it follows that
            
            $ \piencodf{\lambda x.M'[ {\widetilde{x}} \leftarrow x]}_u = u.\overline{\some}; u(x). x.\overline{\some}; x(\linvar{x}). x(\banged{x}). x. \close ; \piencodf{M'[ {\widetilde{x}}\leftarrow  {x}]}_u$.
            
            We give the final derivation in parts. The first part we name $\Pi_1$ derived by:

            \begin{prooftree}
                \small
                \hspace{-2.4cm}
                \AxiomC{$\piencodf{M'[ {\widetilde{x}} \leftarrow  {x}]}_u \vdash  u:\piencodf{\tau} , \piencodf{\Gamma'} , \linvar{x}: \overline{\piencodf{\sigma^k}_{(\sigma, i)}}  ; \piencodf{\Theta} , \banged{x}: \overline{\piencodf{\eta}}$}
                \LeftLabel{\redlab{T\bot}}
                \UnaryInfC{$ x. \close ; \piencodf{M'[ {\widetilde{x}} \leftarrow  {x}]}_u \vdash x{:}\bot, u:\piencodf{\tau} , \piencodf{\Gamma'} , \linvar{x}: \overline{\piencodf{\sigma^k}_{(\sigma, i)}}  ; \piencodf{\Theta} , \banged{x}: \overline{\piencodf{\eta}}$}
                \LeftLabel{\redlab{T?}}
                \UnaryInfC{$x. \close ; \piencodf{M'[ {\widetilde{x}} \leftarrow  {x}]}_u \vdash x{:}\bot, u:\piencodf{\tau} , \piencodf{\Gamma'} , \linvar{x}: \overline{\piencodf{\sigma^k}_{(\sigma, i)}} , \banged{x}: ? \overline{\piencodf{\eta}} ; \piencodf{\Theta} $}
                \LeftLabel{\redlab{T\ampy}}
                \UnaryInfC{$x(\banged{x}). x. \close ; \piencodf{M'[ {\widetilde{x}} \leftarrow  {x}]}_u \vdash x: (? \overline{\piencodf{\eta}}) \ampy (\bot) , u:\piencodf{\tau} , \piencodf{\Gamma'} , \linvar{x}: \overline{\piencodf{\sigma^k}_{(\sigma, i)}} ; \piencodf{\Theta} $}
                \LeftLabel{\redlab{T\ampy}}
                \UnaryInfC{$  x(\linvar{x}). x(\banged{x}). x. \close ; \piencodf{M'[ {\widetilde{x}} \leftarrow  {x}]}_u \vdash  x: \overline{\piencodf{\sigma^k}_{(\sigma, i)}} \ampy ((? \overline{\piencodf{\eta}}) \ampy (\bot)) , u:\piencodf{\tau} , \piencodf{\Gamma'} ; \piencodf{\Theta} $}
            \end{prooftree}
            
            We take $ P = x(\linvar{x}). x(\banged{x}). x. \close ; \piencodf{M'[ {\widetilde{x}} \leftarrow  {x}]}_u $ and continue the derivation:

            \begin{prooftree}
                \small
                \hspace{-2.4cm}
                \AxiomC{$ \Pi_1 $}
                \noLine
                \UnaryInfC{$ \vdots $}
                \noLine
                \UnaryInfC{$  P \vdash  x: \overline{\piencodf{\sigma^k}_{(\sigma, i)}} \ampy ((? \overline{\piencodf{\eta}}) \ampy (\bot)) , u:\piencodf{\tau} , \piencodf{\Gamma'} ; \piencodf{\Theta} $}
                \LeftLabel{\redlab{T\with_d^x}}
                \UnaryInfC{$ x.\overline{\some}; P \vdash x :\with(  \overline{\piencodf{\sigma^k}_{(\sigma, i)}} \ampy ((? \overline{\piencodf{\eta}}) \ampy (\bot))) , u:\piencodf{\tau} , \piencodf{\Gamma'} ; \piencodf{\Theta} $}
                \LeftLabel{\redlab{T\ampy}}
                \UnaryInfC{$u(x). x.\overline{\some}; P \vdash u: \with(  \overline{\piencodf{\sigma^k}_{(\sigma, i)}} \ampy ((? \overline{\piencodf{\eta}}) \ampy (\bot))) \ampy \piencodf{\tau} , \piencodf{\Gamma'} ; \piencodf{\Theta}  $}
                \LeftLabel{\redlab{T\with_d^x}}
                \UnaryInfC{$u.\overline{\some}; u(x). x.\overline{\some}; P \vdash u :\with (\with(  \overline{\piencodf{\sigma^k}_{(\sigma, i)}} \ampy ((? \overline{\piencodf{\eta}}) \ampy (\bot))) \ampy \piencodf{\tau}) , \piencodf{\Gamma'} ; \piencodf{\Theta} $}
            \end{prooftree}
            
            By Definition \ref{ch3def:enc_sestypfailunres} we have that $ \piencodf{(\sigma^{k} , \eta )   \rightarrow \tau} = \with (\with(  \overline{\piencodf{\sigma^k}_{(\sigma, i)}} \ampy ((? \overline{\piencodf{\eta}}) \ampy (\bot))) \ampy \piencodf{\tau}) $. In this case we must have that the variable names for $ x $ from our hypothesis and $x$ from $ M $ must be distinct.

        \item $ M =  (M'\ B) $, or $ M =   (M[ {\widetilde{x}} \leftarrow  {x}])\esubst{ B }{ x }$, or  $ M =  M' \unexsub{U / \unvar{x}} $.
        
            The proof  follows similarly to that of (\ref{ch3proof:relunbag-yes}).
       
        \item $ M =  M' \linexsub{N /  {x}} $
        
            Case follows by that of (\ref{ch3proof:relunbag-no}) and applying induction hypothesis on $\piencodf{M'}_u$.

        
        
            

        \item When $ M =  \fail^{\widetilde{x}}$
            Case follows by that of (\ref{ch3proof:relunbag-no}).

    \end{enumerate}

    \item If $\piencodf{M}_u\vdash \piencodf{\Gamma}, u:\piencodf{(\sigma^{j} , \eta ) \rightarrow \tau} ; \piencodf{\Theta}$
    then 
    $\piencodf{M}_u\vdash \piencodf{\Gamma}, u:\piencodf{(\sigma^{j} , \epsilon ) \rightarrow \tau} ; \piencodf{\Theta}$ follows from previous case along a similar argument.
\end{enumerate}

\end{proof}

\begin{theorem}[Type Preservation for $\piencodf{\cdot}_u$]
\label{ch3t:preservationtwounres}
Let $B$ and $\expr{M}$ be a bag and an expression in $\lamrsharfailunres$, respectively.
\begin{enumerate}
\item If $\Theta ; \Gamma \wfdash B : (\sigma^{k} , \eta )$
then 
$\piencodf{B}_u \wfdash  \piencodf{\Gamma}, u : \piencodf{(\sigma^{k} , \eta )}_{(\sigma, i)} ; \piencodf{\Theta}$.

\item If $\Theta ; \Gamma \wfdash \expr{M} : \tau$
then 
$\piencodf{\expr{M}}_u \wfdash  \piencodf{\Gamma}, u :\piencodf{\tau} ; \piencodf{\Theta}$.
\end{enumerate}
\end{theorem}

\begin{proof}
The proof is by mutual induction on the typing derivation of $B$ and $\expr{M}$, with an analysis for the last rule applied.
Recall that the encoding of types ($\piencodf{-}$) has been given in 
\defref{ch3def:enc_sestypfailunres}.
    
    \begin{enumerate}
        \item  We have the following derivation where we take $ B =  C \bagsep U$:
        
            \begin{prooftree}
                \AxiomC{\( \Theta ; \Gamma\wfdash C : \sigma^k\)}
                \AxiomC{\(  \Theta ;\cdot \wfdash  U : \eta \)}
            \LeftLabel{\redlab{FS{:}bag}}
            \BinaryInfC{\( \Theta ; \Gamma \wfdash C \bagsep U : (\sigma^{k} , \eta  ) \)}
            \end{prooftree}
        
        Our encoding gives: 

        $ \piencodf{C \bagsep U}_u = x.\some_{\llfv{C}} ; \outact{x}{\linvar{x}} .( \piencodf{ C }_{\linvar{x}} \para \outact{x}{\banged{x}} .( !\banged{x} (x_i). \piencodf{ U }_{x_i} \para x.\overline{\close} ) )$.

        In addition, the encoding of {\small{$(\sigma^{k} , \eta  )$ is:
        $$\piencodf{ (\sigma^{k} , \eta  )  }_{(\sigma, i)} = \oplus( (\piencodf{\sigma^{k} }_{(\sigma, i)}) \otimes ((!\piencodf{\eta}) \otimes (\onef))  )\  \text{(for some  $i \geq 0$ and  strict type $\sigma$)}$$}}

        And one can build the following type derivation in parts (rules from \figref{ch3fig:trulespi}). The first part we name $\Pi_1$ derived by:
        \begin{prooftree}
            \AxiomC{$ \piencodf{ U }_{x_i} \vdash x_i: \piencodf{\eta}; \piencodf{\Theta}$}
            \LeftLabel{\redlab{T!}}
            \UnaryInfC{$!\banged{x} (x_i). \piencodf{ U }_{x_i} \vdash \banged{x}: !\piencodf{\eta} ; \piencodf{\Theta}$}
            
            \AxiomC{\mbox{\ }}
            \LeftLabel{\redlab{T\onef}}
            \UnaryInfC{$x.\dual{\close} \vdash x: \onef ; \piencodf{\Theta}$}
        \LeftLabel{\redlab{T\otimes}}
        \BinaryInfC{$ \outact{x}{\banged{x}} .( !\banged{x}. (x_i). \piencodf{ U }_{x_i} \para x.\overline{\close})  \vdash x: (!\piencodf{\eta}) \otimes (\onef) ; \piencodf{\Theta}$}
        \end{prooftree}

        We take $ P =  \outact{x}{\banged{x}} .( !\banged{x}. (x_i). \piencodf{ U }_{x_i} \para x.\overline{\close}) $ and continue the derivation:
        
        \begin{prooftree}
                \AxiomC{$ \piencodf{ C }_{\linvar{x}} \vdash \piencodf{\Gamma}, \linvar{x}:\piencodf{\sigma^{k} }_{(\sigma, i)} ; \piencodf{\Theta}$}
                
                    \AxiomC{$ \Pi_1$}
            \LeftLabel{\redlab{T\otimes}}
            \BinaryInfC{$\outact{x}{\linvar{x}} .( \piencodf{ C }_{\linvar{x}} \para P ) \vdash  \piencodf{\Gamma}, x:(\piencodf{\sigma^{k} }_{(\sigma, i)}) \otimes ((!\piencodf{\eta}) \otimes (\onef)) ; \piencodf{\Theta} $}
            \LeftLabel{\redlab{T\oplus^x_{\widetilde{w}}}}
            \UnaryInfC{$x.\some_{\llfv{C}} ; \outact{x}{\linvar{x}} .( \piencodf{ C }_{\linvar{x}} \para P ) \vdash \piencodf{\Gamma}, x{:}\piencodf{ (\sigma^{k} , \eta  )  }_{(\sigma, i)}$}
        \end{prooftree}
        
        Hence true provided both $ \piencodf{ C }_{\linvar{x}} \vdash \piencodf{\Gamma}, \linvar{x}:\piencodf{\sigma^{k} }_{(\sigma, i)} ; \piencodf{\Theta}$ and $ \piencodf{ U }_{x_i} \vdash x_i: \piencodf{\eta}; \piencodf{\Theta}$ hold.
        
        Let us consider the two cases:
        
        \begin{enumerate}
        
            \item For $ \piencodf{ C }_{\linvar{x}} \vdash \piencodf{\Gamma}, \linvar{x}:\piencodf{\sigma^{k} }_{(\sigma, i)} ; \piencodf{\Theta}$ to hold we must consider two cases on the shape of $C$:
            
            \begin{enumerate}
                \item When $C = \oneb$ we may type bags with the $\redlab{FS{:}\oneb^{\ell}}$ rule.
                
                That is, 
                
                    \begin{prooftree}
                        \AxiomC{\(  \)}
                        \LeftLabel{\redlab{FS{:}\oneb^{\ell}}}
                        \UnaryInfC{\( \Theta ; \dash \wfdash \oneb : \omega \)}
                    \end{prooftree}
                    
                Our encoding gives:         
                
                $ \piencodf{\oneb}_{\linvar{x}} = \linvar{x}.\some_{\emptyset} ; \linvar{x}(y_n). ( y_n.\overline{\some};y_n . \overline{\close} \para \linvar{x}.\some_{\emptyset} ; \linvar{x}. \overline{\none}). $
                
                and  the encoding of $\omega$ can be either:
                \begin{enumerate}
                \item  $\piencodf{\omega}_{(\sigma,0)} =  \overline{\with(( \oplus \bot )\otimes ( \with \oplus \bot ))}$; or
                \item $\piencodf{\omega}_{(\sigma, i)} =  \overline{   \with(( \oplus \bot) \otimes ( \with  \oplus (( \with  \overline{\piencodf{ \sigma }} )  \ampy (\overline{\piencodf{\omega}_{(\sigma, i - 1)}})))) }$
                \end{enumerate}

                And one can build the following type derivation in parts (rules from \figref{ch3fig:trulespi}). The first part we name $\Pi_1$ derived by:
                
                \begin{prooftree}
                    \hspace{-3.0cm}
                        \AxiomC{\mbox{\ }}
                        \LeftLabel{\redlab{T\onef}}
                        \UnaryInfC{$y_n . \overline{\close} \vdash y_n: \onef; \piencodf{\Theta}$}
                        \LeftLabel{\redlab{T\with_d^x}}
                        \UnaryInfC{$y_n.\overline{\some};y_n . \overline{\close} \vdash  y_n :\with \onef; \piencodf{\Theta}$}
                        \AxiomC{}
                        \LeftLabel{\redlab{T\with^x}}
                        \UnaryInfC{$\linvar{x}.\dual{\none} \vdash \linvar{x} :\with A; \piencodf{\Theta}$}
                        \LeftLabel{\redlab{T\oplus^x_{\widetilde{w}}}}
                        \UnaryInfC{$\linvar{x}.\some_{\emptyset} ; \linvar{x}. \overline{\none} \vdash  \linvar{x}{:}\oplus \with A; \piencodf{\Theta}$}
                    \LeftLabel{\redlab{T\mid}}
                    \BinaryInfC{$ y_n.\overline{\some};y_n . \overline{\close} \para \linvar{x}.\some_{\emptyset} ; \linvar{x}. \overline{\none} \vdash y_n :\with \onef, \linvar{x}{:}\oplus \with A; \piencodf{\Theta}$}
                \end{prooftree}

                We take $ P =   y_n.\overline{\some};y_n . \overline{\close} \para \linvar{x}.\some_{\emptyset} ; \linvar{x}. \overline{\none}$ and continue the derivation:

                \begin{prooftree}
                    \small
                    \AxiomC{$\Pi_1$}
                    \LeftLabel{\redlab{T\ampy}}
                    \UnaryInfC{$\linvar{x}(y_n). P \vdash  \linvar{x}: (\with \onef) \ampy (\oplus \with A) ; \piencodf{\Theta}$}
                    \LeftLabel{\redlab{T\oplus^x_{\widetilde{w}}}}
                    \UnaryInfC{$\linvar{x}.\some_{\emptyset} ; \linvar{x}(y_n). P \vdash  \linvar{x}{:}\oplus ((\with \onef) \ampy (\oplus \with A)); \piencodf{\Theta}$}
                \end{prooftree}

                Since $A$ is arbitrary,  we can take $A=\oneb$ for $\piencodf{\omega}_{(\sigma,0)} $ and  $A= \overline{(( \with  \overline{\piencodf{ \sigma }} )  \ampy (\overline{\piencodf{\omega}_{(\sigma, i - 1)}}))}$  for $\piencodf{\omega}_{(\sigma,i)} $, in both cases, the result follows.

                \item When $C = \bag{M}  \cdot C'$ we may type bags with the $\redlab{FS{:}bag^{\ell}}$ rule. 
                
                \begin{prooftree}
                    \AxiomC{\( \Theta ; \Gamma' \wfdash M : \sigma\)}
                    \AxiomC{\( \Theta ; \Delta \wfdash C' : \sigma^k\)}
                    \LeftLabel{\redlab{FS{:}bag^{\ell}}}
                    \BinaryInfC{\( \Theta ; \Gamma', \Delta \wfdash \bag{M}  \cdot C':\sigma^{k}\)}
                \end{prooftree}
                
                Where $ \Gamma = \Gamma' , \Delta$.
                To simplify the proof, we will consider $k=3$.

                By IH we have
                {\small
                \begin{align*}
                    \piencodf{M}_{x_i} & \vdash \piencodf{\Gamma'}, x_i: \piencodf{\sigma}; \piencodf{\Theta}
                    \qquad 
                    \piencodf{C'}_{\linvar{x}} & \vdash \piencodf{\Delta}, \linvar{x}: \piencodf{\sigma\wedge \sigma}_{(\tau, j)}; \piencodf{\Theta}
                \end{align*}}

                By \defref{ch3def:enc_lamrsharpifailunres},
                \begin{equation} \small
                \begin{aligned}
                    \piencodf{\bag{M} \cdot C'}_{\linvar{x}} =&  \linvar{x}.\some_{\llfv{\bag{M} \cdot C} } ; \linvar{x}(y_i). \linvar{x}.\some_{y_i, \llfv{\bag{M} \cdot C}};\linvar{x}.\overline{\some} ; \outact{\linvar{x}}{x_i}.
                   \\
                   & (x_i.\some_{\llfv{M}} ; \piencodf{M}_{x_i} \para \piencodf{C'}_{\linvar{x}} \para y_i. \overline{\none})
                    \end{aligned}
                \end{equation}

                Let $\Pi_1$ be the derivation:

                \begin{prooftree}
                    \small
                    \hspace{-2cm}
                        \AxiomC{$\piencodf{M}_{x_i} \;{ \vdash} \piencodf{\Gamma'}, x_i: \piencodf{\sigma}; \piencodf{\Theta} $}
                        \LeftLabel{\redlab{T\oplus^x_{\widetilde{w}}}}
                        \UnaryInfC{$x_i.\some_{\llfv{M}} ; \piencodf{M}_{x_i} \vdash \piencodf{\Gamma'} ,x_i: \oplus \piencodf{\sigma} ; \piencodf{\Theta}$}
                        \AxiomC{}
                        \LeftLabel{\redlab{T\with^x}}
                        \UnaryInfC{$ y_i. \overline{\none} \vdash y_i :\with \onef; \piencodf{\Theta}$}
                    \LeftLabel{\redlab{T\para}}
                    \BinaryInfC{$\underbrace{x_i.\some_{\llfv{M}} ; \piencodf{M}_{x_i} \para y_i. \overline{\none}}_{P_1} \vdash \piencodf{\Gamma'} ,x_i: \oplus \piencodf{\sigma}, y_i :\with \onef ; \piencodf{\Theta}$}
                \end{prooftree}
                
                Let $ P_1 = (x_i.\some_{\llfv{M}} ; \piencodf{M}_{x_i} \para y_i. \overline{\none})$, in the the derivation $\Pi_2$ below:

                \begin{prooftree}
                    \small
                    \hspace{-3.1cm}
                        \AxiomC{$ \Pi_1$} 
                        \AxiomC{$ \piencodf{C'}_{\linvar{x}}  \vdash  \piencodf{\Delta}, \linvar{x}: \piencodf{\sigma\wedge \sigma}_{(\tau, j)}; \piencodf{\Theta} $}
                        \LeftLabel{\redlab{T\otimes}}
                    \BinaryInfC{$ \outact{\linvar{x}}{x_i}. (P_1 \para \piencodf{C'}_{\linvar{x}}) \vdash  \piencodf{\Gamma'}  ,  \piencodf{\Delta}, y_i :\with \onef, \linvar{x}: (\oplus \piencodf{\sigma})  \otimes (\piencodf{\sigma\wedge \sigma}_{(\tau, j)}) ; \piencodf{\Theta}$}
                    \LeftLabel{\redlab{T\with_d^x}}
                    \UnaryInfC{$\underbrace{\linvar{x}.\overline{\some} ; \outact{\linvar{x}}{x_i}. (P_1 \para \piencodf{C'}_{\linvar{x}}  )}_{P_2} \vdash \piencodf{\Gamma'}  ,  \piencodf{\Delta}, y_i :\with \onef, \linvar{x}: \with (( \oplus \piencodf{\sigma} ) \otimes (\piencodf{\sigma\wedge \sigma}_{(\tau, j)})) ; \piencodf{\Theta} $}
                \end{prooftree}
                
                Let $P_2 = (\linvar{x}.\overline{\some} ; \outact{\linvar{x}}{x_i}. (P_1 \para \piencodf{A}_{\linvar{x}} ))$ in the derivation below:

                \begin{prooftree}
                    \small
                    \hspace{-3cm}
                    \AxiomC{$ \Pi_2$} 
                    \noLine
                    \UnaryInfC{$\vdots$}
                    \noLine
                    \UnaryInfC{$P_2\vdash \piencodf{\Gamma}, y_i :\with \onef, \linvar{x}: \with (( \oplus \piencodf{\sigma} ) \otimes (\piencodf{\sigma\wedge \sigma}_{(\tau, j)})) ; \piencodf{\Theta} $}
                    \LeftLabel{\redlab{T\oplus^x_{\widetilde{w}}}}
                    \UnaryInfC{$\linvar{x}.\some_{y_i, \llfv{\bag{M} \cdot C'}};P_2  \vdash \piencodf{\Gamma}, y_i :\with \onef, \linvar{x}:\oplus  \with (( \oplus \piencodf{\sigma} ) \otimes (\piencodf{\sigma\wedge \sigma}_{(\tau, j)})); \piencodf{\Theta}$}
                    \LeftLabel{\redlab{T\ampy}}
                    \UnaryInfC{$\linvar{x}(y_i). \linvar{x}.\some_{y_i, \llfv{\bag{M} \cdot C'}};P_2  \vdash \piencodf{\Gamma}, \linvar{x}: ( \with \onef) \ampy ( \oplus  \with (( \oplus \piencodf{\sigma} ) \otimes (\piencodf{\sigma\wedge \sigma}_{(\tau, j)}))); \piencodf{\Theta} $}
                    \LeftLabel{\redlab{T\oplus^x_{\widetilde{w}}}}
                    \UnaryInfC{$\piencodf{\bag{M}\cdot C'}_{\linvar{x}} \vdash   \piencodf{\Gamma}, \linvar{x}: \oplus(( \with \onef) \ampy ( \oplus  \with (( \oplus \piencodf{\sigma} ) \otimes (\piencodf{\sigma\wedge \sigma}_{(\tau, j)})))); \piencodf{\Theta} $}
                \end{prooftree}
                
                From Definitions~\ref{ch3def:duality} (duality) and \ref{ch3def:enc_sestypfailunres}, we infer:
                \begin{equation*}
                    \begin{aligned}
                         \oplus(( \with \onef) \ampy ( \oplus  \with (( \oplus \piencodf{\sigma} ) \otimes (\piencodf{\sigma\wedge \sigma}_{(\tau, j)})))) &=\piencodf{\sigma\wedge \sigma \wedge \sigma}_{(\tau, j)}
                    \end{aligned}
                \end{equation*}
                Therefore, $\piencodf{\bag{M}\cdot C'}_{\linvar{x}} \vdash \piencodf{\Gamma}, \linvar{x}: \piencodf{\sigma\wedge \sigma \wedge \sigma}_{(\tau, j)} $ and the result follows.

            \end{enumerate}
            
             \item For $ \piencodf{ U }_{x_i} \vdash x_i: \piencodf{\eta}; \piencodf{\Theta}$ we consider $U$ to be a binary concatenation of 2 components, one being an empty unrestricted bag and the other being $\banged{\bag{M}}$. Hence we take $U = \banged{\oneb} \concat \banged{\bag{M}}$ with $\eta =  \sigma_1 \concat \sigma_2 $, $\piencodf{\eta_i} = \& \{ \mathtt{l}_1 ; \piencodf{ \sigma_1} , \mathtt{l}_2 ; \piencodf{\sigma_2} \}$ by \defref{ch3def:enc_sestypfailunres} and finally by \defref{ch3def:enc_lamrsharpifailunres} we have $\piencodf{ U }_{x_i} =  {x_i.\mathtt{case} \{ \mathtt{l}_{1} : \piencodf{\banged{\oneb}}_{x_i} , \mathtt{l}_{2} : \piencodf{\banged{\bag{M}}}_{x_i} \}}$, $\piencodf{\banged{\oneb}}_{x_i} =  x_i.\overline{\none} $ and $\piencodf{\banged{\bag{M}}}_{x_i} =  \piencodf{M}_{x_i} $, we can conclude $ \piencodf{ U }_{x_i} =  {x_i.\mathtt{case} \{ \mathtt{l}_{1} : x_i.\overline{\none} , \mathtt{l}_{2} : \piencodf{M}_{x_i} \}} $.

             Hence we have:
             \begin{prooftree}
                    \AxiomC{\(  \)}
                    \LeftLabel{\redlab{FS{:}bag^!}}
                    \UnaryInfC{\( \Theta ;  \dash  \wfdash \banged{\oneb} : \sigma_1 \)}
        
                    \AxiomC{\( \Theta ; \cdot \wfdash M : \sigma_2\)}
                    \LeftLabel{\redlab{FS{:}bag^!}}
                    \UnaryInfC{\( \Theta ; \cdot  \wfdash \banged{\bag{M}}:\sigma_2 \)}
                \LeftLabel{\redlab{FS{:}\concat-bag^{!}}}
                \BinaryInfC{\( \Theta ; \cdot  \wfdash \banged{\oneb} \concat \banged{\bag{M}} : \sigma_1 \concat \sigma_2 \)}
            \end{prooftree}
            
            By the induction hypothesis we have that $ \Theta ; \cdot \wfdash M : \sigma $ implies $ \piencodf{M}_{x_i} \wfdash x_i : \piencodf{\sigma} ;  \piencodf{\Theta}$

             \begin{prooftree}
                    \AxiomC{}
                    \LeftLabel{\redlab{T\with^x}}
                    \UnaryInfC{$ x_i.\dual{\none} \vdash x_i:  \piencodf{ \sigma_1}  ; \piencodf{\Theta} $}
                
                    \AxiomC{$ \piencodf{M}_{x_i} \vdash  x_i:  \piencodf{\sigma_2} ; \piencodf{\Theta} $}
                \LeftLabel{\redlab{T\with}}
                \BinaryInfC{$ {x_i.\mathtt{case} \{ \mathtt{l}_{1} : x_i.\none , \mathtt{l}_{2} : \piencodf{M}_{x_i} \}} \vdash  x_i: \& \{ \mathtt{l}_1 ; \piencodf{ \sigma_1} , \mathtt{l}_2 ; \piencodf{\sigma_2} \} ; \piencodf{\Theta} $}
            \end{prooftree}
            
            Therefore, ${x_i.\mathtt{case} \{ \mathtt{l}_{1} : x_i.\none , \mathtt{l}_{2} : \piencodf{M}_{x_i} \}} \vdash  x_i: \& \{ \mathtt{l}_1 ; \piencodf{ \sigma_1} , \mathtt{l}_2 ; \piencodf{\sigma_2} \} ; \piencodf{\Theta}  $ and the result follows.

        \end{enumerate}

        \item  The proof of type preservation for expressions, relies on the analysis of twelve cases:

        \begin{enumerate}
            
            \item {\bf Rule \redlab{FS{:}var^{\ell}}:}
            Then we have the following derivation:
            
            \begin{prooftree}
                \AxiomC{}
                \LeftLabel{\redlab{FS{:}var^{\ell}}}
                \UnaryInfC{\( \Theta ;  {x}: \tau \wfdash  {x} : \tau\)}
            \end{prooftree}
            
            By \defref{ch3def:enc_sestypfailunres},  $\piencodf{ {x}:\tau}=  {x}:\with \overline{\piencodf{\tau }}$, and by \figref{ch3fig:encfailunres},  $\piencodf{ {x}}_u =  {x}.\overline{\some} ; [ {x} \leftrightarrow u] $. The thesis holds thanks to the following derivation:
            
                \begin{prooftree}
                    \AxiomC{}
                    \LeftLabel{$\redlab{ (Tid)}$}
                    \UnaryInfC{$ [ {x} \leftrightarrow u] \vdash  {x}:  \overline{\piencodf{\tau }}  , u :  \piencodf{ \tau } ; \piencodf{\Theta} $}
                    \LeftLabel{$\redlab{T\with^{x}_{d})}$}
                    \UnaryInfC{$  {x}.\overline{\some} ; [ {x} \leftrightarrow u] \vdash  {x}: \with  \overline{\piencodf{ \tau }} , u :  \piencodf{ \tau } ; \piencodf{\Theta} $}
                \end{prooftree}

            \item {\bf Rule \redlab{FS{:}var^!}:}
            Then we have the following derivation provided $\eta_{ind} = \tau $:
            
            \begin{prooftree}
                \AxiomC{}
                \LeftLabel{\redlab{FS{:}var^{\ell}}}
                \UnaryInfC{\( \Theta , \banged{x}: \eta;  {x}: \eta_{ind}  \wfdash  {x} : \tau\)}
                \LeftLabel{\redlab{FS{:}var^!}}
                \UnaryInfC{\( \Theta , \banged{x}: \eta; \dash \wfdash  {x}[ind] : \tau\)}
            \end{prooftree}
            
            By \defref{ch3def:enc_sestypfailunres},  $\piencodf{\Theta , \banged{x}: \eta}= \piencodf{\Theta} , \banged{x}: \dual{ \&_{\eta_i \in \eta} \{ \mathtt{l}_i ; \piencodf{\eta_i} \} }$, and by \figref{ch3fig:encfailunres},  $\piencodf{ {x}[ind]}_u = \outsev{\banged{x}}{{x_i}}. {x}_i.l_{ind}; [{x_i} \leftrightarrow u] $. The thesis holds thanks to the following derivation:
            
            \begin{prooftree}
                \small
                \hspace{-1cm}
                    \AxiomC{}
                    \LeftLabel{$\redlab{ (Tid)}$}
                    \UnaryInfC{$ [{x_i} \leftrightarrow u]  \vdash  u :  \piencodf{ \tau },  x_i:  \overline{\piencodf{\eta_{ind} }}  ; \banged{x}:\&_{\eta_i \in \eta} \{ \mathtt{l}_i ; \piencodf{\eta_i} \} , \piencodf{\Theta} $}
                \LeftLabel{\redlab{T\oplus_i}}
                \UnaryInfC{$  {x}_i.l_{ind}; [{x_i} \leftrightarrow u] \vdash  u :  \piencodf{ \tau }, {x}_i :  \oplus_{\eta_i \in \eta} \{ \mathtt{l}_i ; \dual{\piencodf{\eta_i}}  \} ; \banged{x}:\oplus_{\eta_i \in \eta} \{ \mathtt{l}_i ; \dual{\piencodf{\eta_i}} \} , \piencodf{\Theta}  $}
                \LeftLabel{\redlab{Tcopy}}
                \UnaryInfC{$ \outsev{\banged{x}}{{x_i}}. {x}_i.l_{ind}; [{x_i} \leftrightarrow u] \vdash  u :  \piencodf{ \tau }; \banged{x}:\oplus_{\eta_i \in \eta} \{ \mathtt{l}_i ; \dual{\piencodf{\eta_i}} \} , \piencodf{\Theta} $}
            \end{prooftree}

            \item {\bf Rule \redlab{FS\!:\!weak}:}
                        Then we have the following derivation:
            
            \begin{prooftree}
                \AxiomC{\( \Theta ; \Gamma  \wfdash M : \tau\)}
                \LeftLabel{ \redlab{FS\!:\!weak}}
                \UnaryInfC{\( \Theta ; \Gamma ,  {x}: \omega \wfdash M[\leftarrow  {x}]: \tau \)}
            \end{prooftree}
            
            By \defref{ch3def:enc_sestypfailunres},  $\piencodf{\Gamma ,  {x}: \omega}= \piencodf{\Gamma}, \linvar{x}: \overline{\piencodf{\omega }_{(\sigma, i_1)}}$, and by \figref{ch3fig:encfailunres}, 
            
            $\piencodf{M[  \leftarrow  {x}]}_u = \linvar{x}. \overline{\some}. \outact{\linvar{x}}{y_i} . ( y_i . \some_{u,\llfv{M}} ;y_{i}.\close; \piencodf{M}_u \para \linvar{x}. \overline{\none}) $. 
            
             By IH, we have $\piencodf{M}_u\vdash  \piencodf{\Gamma }, u:\piencodf{\tau} ; \piencodf{\Theta }$. 
            The thesis holds thanks to the following derivation which we give in parts. The first part we name $\Pi_1$ derived by:

            \begin{prooftree}
                \AxiomC{$\piencodf{M}_u\vdash  \piencodf{\Gamma }, u:\piencodf{\tau} ; \piencodf{\Theta} $}
                \LeftLabel{\redlab{T\bot}}
                \UnaryInfC{$y_{i}.\close; \piencodf{M}_u \vdash y_i{:}\bot, \piencodf{\Gamma }, u:\piencodf{\tau} ; \piencodf{\Theta}$}
                \LeftLabel{\redlab{T\oplus^x_{\widetilde{w}}}}
                \UnaryInfC{$y_i . \some_{u,\llfv{M}} ;y_{i}.\close; \piencodf{M}_u \vdash  y_i{:}\oplus \bot , \piencodf{\Gamma }, u:\piencodf{\tau} ; \piencodf{\Theta}$}
            \end{prooftree}

            We take $ P =  y_i . \some_{u,\llfv{M}} ;y_{i}.\close; \piencodf{M}_u$ and continue the derivation:

            \begin{prooftree}\hspace{-2.5cm}
                    \small
                    \AxiomC{$\Pi_1$}
                    \AxiomC{}
                    \LeftLabel{\redlab{T\with^x}}
                    \UnaryInfC{$\linvar{x}. \overline{\none} \vdash \linvar{x} :\with A$}
                \LeftLabel{\redlab{T\otimes}}
                \BinaryInfC{$\outact{\linvar{x}}{y_i} . ( P \para \linvar{x}. \overline{\none}) \vdash  \linvar{x}: (\oplus \bot) \otimes (\with A) , \piencodf{\Gamma }, u:\piencodf{\tau} ; \piencodf{\Theta}$}
                \LeftLabel{\redlab{T\with_d^x}}
                \UnaryInfC{$\piencodf{M[  \leftarrow  {x}]}_u \vdash  \linvar{x} :\with ((\oplus \bot) \otimes (\with A))  , \piencodf{\Gamma }, u:\piencodf{\tau} ; \piencodf{\Theta} $}
             \end{prooftree}
            
             Since $A$ is arbitrary,  we can take $A=\oneb$ for $\piencodf{\omega}_{(\sigma,0)} $ and  $A= \overline{(( \with  \overline{\piencodf{ \sigma }} )  \ampy (\overline{\piencodf{\omega}_{(\sigma, i - 1)}}))}$  for $\piencodf{\omega}_{(\sigma,i)} $ where $i > 0$, in both cases, the result follows.

            \item {\bf Rule $\redlab{FS:abs \dash sh}$:}
        
            Then $\expr{M} = \lambda x . (M[ {\widetilde{x}} \leftarrow  {x}])$, and the derivation is:
            
            \begin{prooftree}
                \AxiomC{\( \Theta , \banged{x}:\eta ; \Gamma ,  {x}: \sigma^k \wfdash M[ {\widetilde{x}} \leftarrow  {x}] : \tau \quad  {x} \notin \dom{\Gamma} \)}
                \LeftLabel{\redlab{FS{:}abs\dash sh}}
                \UnaryInfC{\( \Theta ; \Gamma \wfdash \lambda x . (M[ {\widetilde{x}} \leftarrow  {x}])  : (\sigma^k, \eta )  \rightarrow \tau \)}
            \end{prooftree}
    
            By IH, we have $\piencodf{M[ {\widetilde{x}} \leftarrow  {x}]}_u \vdash  u:\piencodf{\tau} , \piencodf{\Gamma} , \linvar{x}: \overline{\piencodf{\sigma^k}_{(\sigma, i)}}  ; \piencodf{\Theta} , \banged{x}: \overline{\piencodf{\eta}} $, 
            From 
            \defref{ch3def:enc_lamrsharpifailunres}, it follows 
            $ \piencodf{\lambda x.M[ {\widetilde{x}} \leftarrow x]}_u = u.\overline{\some}; u(x). x.\overline{\some}; x(\linvar{x}). x(\banged{x}). x. \close ; \piencodf{M[ {\widetilde{x}} \leftarrow  {x}]}_u $
            
            We give the final derivation in parts. The first part we name $\Pi_1$ derived by:
            
            \begin{prooftree}
                \small
                \hspace{-2cm}
                \AxiomC{$\piencodf{M[ {\widetilde{x}} \leftarrow  {x}]}_u \vdash  u:\piencodf{\tau} , \piencodf{\Gamma} , \linvar{x}: \overline{\piencodf{\sigma^k}_{(\sigma, i)}}  ; \piencodf{\Theta} , \banged{x}: \overline{\piencodf{\eta}}$}
                \LeftLabel{\redlab{T\bot}}
                \UnaryInfC{$ x. \close ; \piencodf{M[ {\widetilde{x}} \leftarrow  {x}]}_u \vdash x{:}\bot, u:\piencodf{\tau} , \piencodf{\Gamma} , \linvar{x}: \overline{\piencodf{\sigma^k}_{(\sigma, i)}}  ; \piencodf{\Theta} , \banged{x}: \overline{\piencodf{\eta}}$}
                \LeftLabel{\redlab{T?}}
                \UnaryInfC{$x. \close ; \piencodf{M[ {\widetilde{x}} \leftarrow  {x}]}_u \vdash x{:}\bot, u:\piencodf{\tau} , \piencodf{\Gamma} , \linvar{x}: \overline{\piencodf{\sigma^k}_{(\sigma, i)}} , \banged{x}: ? \overline{\piencodf{\eta}} ; \piencodf{\Theta} $}
                \LeftLabel{\redlab{T\ampy}}
                \UnaryInfC{$x(\banged{x}). x. \close ; \piencodf{M[ {\widetilde{x}} \leftarrow  {x}]}_u \vdash x: (? \overline{\piencodf{\eta}}) \ampy (\bot) , u:\piencodf{\tau} , \piencodf{\Gamma} , \linvar{x}: \overline{\piencodf{\sigma^k}_{(\sigma, i)}} ; \piencodf{\Theta} $}
                \LeftLabel{\redlab{T\ampy}}
                \UnaryInfC{$  x(\linvar{x}). x(\banged{x}). x. \close ; \piencodf{M[ {\widetilde{x}} \leftarrow  {x}]}_u \vdash  x: \overline{\piencodf{\sigma^k}_{(\sigma, i)}} \ampy ((? \overline{\piencodf{\eta}}) \ampy (\bot)) , u:\piencodf{\tau} , \piencodf{\Gamma} ; \piencodf{\Theta} $}
            \end{prooftree}
            We take $ P = x(\linvar{x}). x(\banged{x}). x. \close ; \piencodf{M[ {\widetilde{x}} \leftarrow  {x}]}_u $ and continue the derivation:

            \begin{prooftree}
                \small\hspace{-2cm}
                \AxiomC{$ \Pi_1 $}
                \noLine
                \UnaryInfC{$ \vdots $}
                \noLine
                \UnaryInfC{$  P \vdash  x: \overline{\piencodf{\sigma^k}_{(\sigma, i)}} \ampy ((? \overline{\piencodf{\eta}}) \ampy (\bot)) , u:\piencodf{\tau} , \piencodf{\Gamma} ; \piencodf{\Theta} $}
                \LeftLabel{\redlab{T\with_d^x}}
                \UnaryInfC{$ x.\overline{\some}; P \vdash x :\with(  \overline{\piencodf{\sigma^k}_{(\sigma, i)}} \ampy ((? \overline{\piencodf{\eta}}) \ampy (\bot))) , u:\piencodf{\tau} , \piencodf{\Gamma} ; \piencodf{\Theta} $}
                \LeftLabel{\redlab{T\ampy}}
                \UnaryInfC{$u(x). x.\overline{\some}; P \vdash u: \with(  \overline{\piencodf{\sigma^k}_{(\sigma, i)}} \ampy ((? \overline{\piencodf{\eta}}) \ampy (\bot))) \ampy \piencodf{\tau} , \piencodf{\Gamma} ; \piencodf{\Theta}  $}
                \LeftLabel{\redlab{T\with_d^x}}
                \UnaryInfC{$u.\overline{\some}; u(x). x.\overline{\some}; P \vdash u :\with (\with(  \overline{\piencodf{\sigma^k}_{(\sigma, i)}} \ampy ((? \overline{\piencodf{\eta}}) \ampy (\bot))) \ampy \piencodf{\tau}) , \piencodf{\Gamma} ; \piencodf{\Theta} $}
            \end{prooftree}
            
            By Definition \ref{ch3def:enc_sestypfailunres} we have that $ \piencodf{(\sigma^{k} , \eta )   \rightarrow \tau} = \with (\with(  \overline{\piencodf{\sigma^k}_{(\sigma, i)}} \ampy ((? \overline{\piencodf{\eta}}) \ampy (\bot))) \ampy \piencodf{\tau}) $. Hence the case holds by $ \piencodf{\lambda x.M[\linvar{\widetilde{x}} \leftarrow x]}_u \vdash u:\piencodf{(\sigma^{k} , \eta )   \rightarrow \tau} , \piencodf{\Gamma} ; \piencodf{\Theta} $.

            \item {\bf Rule $\redlab{FS:app}$:}
            Then $\expr{M} = M\ B$, where $ B = C \bagsep U $ and the derivation is:
            
            \begin{prooftree}
                \AxiomC{\( \Theta ;\Gamma \wfdash M : (\sigma^{j} , \eta ) \rightarrow \tau \)}
                 \AxiomC{\(  \Theta ;\Delta \wfdash B : (\sigma^{k} , \epsilon )  \)}
                 \AxiomC{\( \eta \relunbag \epsilon \)}
                \LeftLabel{\redlab{FS{:}app}}
                \TrinaryInfC{\( \Theta ; \Gamma, \Delta \wfdash M\ B : \tau\)}
            \end{prooftree}

        
            By IH, we have both 
            
            \begin{itemize}
                \item  $\piencodf{M}_u\vdash \piencodf{\Gamma}, u:\piencodf{(\sigma^{j} , \eta ) \rightarrow \tau} ; \piencodf{\Theta} $
                \item $\piencodf{M}_u\vdash \piencodf{\Gamma}, u:\piencodf{(\sigma^{j} , \epsilon ) \rightarrow \tau} ; \piencodf{\Theta}$,   by Lemma \ref{ch3lem:relunbag-typeunres}
                \item $\piencodf{B}_u\vdash \piencodf{\Delta}, u:\overline{\piencodf{(\sigma^{k} , \epsilon ) }_{(\tau_2, n)}}  ; \piencodf{\Theta} $, for some $\tau_2$ and some $n$.
            \end{itemize}
            
            Therefore, from the fact that $\expr{M}$ is well-formed and Definitions~\ref{ch3def:enc_lamrsharpifailunres} and \ref{ch3def:enc_sestypfailunres}, we  have:
            
            \begin{itemize}
                \item {\small$\displaystyle{\piencodf{M (C \bagsep U)}_u = \hspace{-4mm} \bigoplus_{C_i \in \perm{C}} (\nu v)(\piencodf{M}_v \para v.\some_{u , \llfv{C}} ; \outact{v}{x} . ([v \leftrightarrow u] \para \piencodf{C_i \bagsep U}_x ) )} $};
                \item $\piencodf{(\sigma^{j} , \eta ) \rightarrow \tau}= \oplus( (\piencodf{\sigma^{k} }_{(\tau_1, m)}) \otimes ((!\piencodf{\eta})  $, for some $\tau_1$ and some $m$.
            \end{itemize}
            
            Also, since $\piencodf{B}_u\vdash \piencodf{\Delta}, u:\piencodf{(\sigma^{k} , \epsilon )}_{(\tau_2, n)}$, we have the following derivation $\Pi_i$:
         
        \begin{adjustwidth}{-2cm}{}
          \begin{prooftree}
                \small
                    \AxiomC{$\piencodf{C_i \bagsep U}_x\vdash \piencodf{\Delta}, x:\piencodf{(\sigma^{k} , \epsilon )}_{(\tau_2, n)}  ; \piencodf{\Theta} $ }
                    \AxiomC{\(\)}                                   
                    \LeftLabel{$ \redlab{Tid}$}
                    \UnaryInfC{$ [v \leftrightarrow u]                   \vdash v:  \overline{\piencodf{ \tau }} , u: \piencodf{ \tau }$}
                \LeftLabel{$\redlab{T \otimes}$}
                \BinaryInfC{$\outact{v}{x} . ([v \leftrightarrow u] \para \piencodf{C_i \bagsep U}_x ) \vdash \piencodf{ \Delta }, v:\piencodf{(\sigma^{k} , \epsilon )}_{(\tau_2, n)} \otimes \overline{ \piencodf{ \tau }} , u:\piencodf{ \tau }  ; \piencodf{\Theta} $}
                \LeftLabel{$\redlab{T\oplus^{v}_{w}}$}
                \UnaryInfC{$  v.\some_{u , \llfv{C}} ; \outact{v}{x} . ([v \leftrightarrow u] \para \piencodf{C_i \bagsep U}_x )  \vdash\piencodf{ \Delta }, v:\oplus (\piencodf{(\sigma^{k} , \epsilon )}_{(\tau_2, n)} \otimes  \overline{\piencodf{ \tau }}), u:\piencodf{ \tau } ; \piencodf{\Theta} $} 
            \end{prooftree}
        \end{adjustwidth}
        
            Notice that 
          \(
              \oplus (\piencodf{(\sigma^{k} , \epsilon )}_{(\tau_2, n)} \otimes  \overline{\piencodf{ \tau }}) =  \overline{\piencodf{(\sigma^{k} , \epsilon ) \rightarrow \tau}}.
            \)
            Therefore, by one application of $\redlab{Tcut}$ we obtain the derivations $\nabla_i$, for each $C_i \in \perm{C}$:
    
            \begin{adjustwidth}{-1cm}{}
            \begin{prooftree}
            \small
            \AxiomC{\( \piencodf{M}_v\vdash \piencodf{\Gamma}, v:\piencodf{(\sigma^{j} , \epsilon ) \rightarrow \tau} ; \piencodf{\Theta} \)}
            \AxiomC{$\Pi_i$}
            \LeftLabel{\( \redlab{Tcut} \)}    
            \BinaryInfC{$ (\nu v)( \piencodf{ M}_v \para v.\some_{u , \llfv{C}} ; \outact{v}{x} . ([v \leftrightarrow u] \para \piencodf{B_i}_x ) ) \vdash \piencodf{ \Gamma } ,\piencodf{ \Delta } , u: \piencodf{ \tau }  ; \piencodf{\Theta} $}
            \end{prooftree}
        \end{adjustwidth}
            In order to apply \redlab{Tcut}, we must have that $\piencodf{\sigma^{j}}_{(\tau_1, m)} = \piencodf{\sigma^{k}}_{(\tau_2, n)}$, therefore, the choice of $\tau_1,\tau_2,n$ and $m$, will consider the different possibilities for $j$ and $k$, as in Proposition~\ref{ch3prop:app_auxunres}.
            We can then conclude that $\piencodf{M B}_u \vdash \piencodf{ \Gamma}, \piencodf{ \Delta }, u:\piencodf{ \tau } ; \piencodf{\Theta} $:
            
            \begin{adjustwidth}{-2cm}{}
            \begin{prooftree}
            \small
            \AxiomC{For each $C_i \in \perm{C} \qquad  \nabla_i$}
            \LeftLabel{$\redlab{T\with}$}
            \UnaryInfC{$\displaystyle{\bigoplus_{C_i \in \perm{C}} (\nu v)( \piencodf{ M}_v \para v.\some_{u , \lfv{B}} ; \outact{v}{x} . ([v \leftrightarrow u] \para \piencodf{B_i}_x ) ) \vdash \piencodf{ \Gamma } ,\piencodf{ \Delta } , u: \piencodf{ \tau }} ; \piencodf{\Theta} $}
            \end{prooftree}
            \end{adjustwidth}
            
            and the result follows.

            \item {\bf Rule $\redlab{FS:share}$:}
            Then $\expr{M} = M [  {x}_1, \dots  {x}_k \leftarrow x ]$ and the derivation is:
            \begin{prooftree}
                \AxiomC{\( \Theta ; \Delta ,  {x}_1: \sigma, \cdots,  {x}_k: \sigma \wfdash M : \tau \quad  {x} \notin \Delta \quad k \not = 0\)}
                \LeftLabel{ \redlab{FS:share}}
                \UnaryInfC{\( \Theta ; \Delta ,  {x}: \sigma_{k} \wfdash M[ {x}_1 , \cdots ,  {x}_k \leftarrow  {x}] : \tau \)}
            \end{prooftree}
    
            To simplify the proof we will consider $k=1$ (the case in which $k>1$ follows similarly). 
    
            By IH, we have $\piencodf{M}_u\vdash  \piencodf{\Delta ,  {x}_1:\sigma }, u:\piencodf{\tau} ; \piencodf{\Theta}$. 
            From 
            Definitions~\ref{ch3def:enc_lamrsharpifailunres} and \ref{ch3def:enc_sestypfailunres}, it follows 
            
            \begin{itemize}
             \item $\piencodf{ \Delta ,  {x}_1: \sigma} = \piencodf{\Delta}, \linvar{x}_1:\with\overline{\piencodf{\sigma}} $.
            \item 
            $
            \piencodf{M[ {x}_1, \leftarrow  {x}]}_u =
               \begin{array}[t]{l}
                  \linvar{x}.\overline{\some}. \outact{\linvar{x}}{y_1}. (y_1 . \some_{\emptyset} ;y_{1}.\close;0 
                  \\
                  \para \linvar{x}.\overline{\some};\linvar{x}.\some_{u, (\llfv{M} \setminus  {x}_1 )};
                  \\
                  \quad \linvar{x}( {x}_1) .  \linvar{x}. \overline{\some}. \outact{\linvar{x}}{y_2} . ( y_2 . \some_{u,\llfv{M}} ;
                  \\
                  \quad y_{2}.\close; \piencodf{M}_u \para \linvar{x}. \overline{\none}) )
               \end{array}
                $
           
            \end{itemize}

            We shall split the expression into two parts:
            \[
            \begin{aligned}
                N_1 &= \linvar{x}. \overline{\some}. \outact{\linvar{x}}{y_2} . ( y_2 . \some_{u,\llfv{M}} ;y_{2}.\close; \piencodf{M}_u \para \linvar{x}. \overline{\none}) \\
                N_2 &= \linvar{x}.\overline{\some}. \outact{\linvar{x}}{y_1}. (y_1 . \some_{\emptyset} ;y_{1}.\close;0 \para \\
                & \quad \linvar{x}.\overline{\some};\linvar{x}.\some_{u, (\llfv{M} \setminus  {x}_1 )};\linvar{x}( {x}_1) .N_1)
            \end{aligned}
            \]

            and we obtain the  derivation for term $N_1$ as follows where we omit $; \piencodf{\Theta}$. We give the derivation in two parts, the first we being $\Pi_1$:
            \begin{prooftree}
                \AxiomC{$\piencodf{M}_u \vdash  \piencodf{\Delta ,  {x}_1:\sigma }, u:\piencodf{\tau} $}
                \LeftLabel{\redlab{T\bot}}
                \UnaryInfC{$y_{2}.\close; \piencodf{M}_u \vdash \piencodf{\Delta ,  {x}_1:\sigma }, u:\piencodf{\tau}, y_{2}{:}\bot $}
                \LeftLabel{ \redlab{ T\oplus^x_{ \widetilde{w}}}}
                \UnaryInfC{$ y_2 . \some_{u,\llfv{M}} ;y_{2}.\close; \piencodf{M}_u \vdash \piencodf{\Delta ,  {x}_1:\sigma }, u:\piencodf{\tau}, y_{2}{:}\oplus \bot $}
            \end{prooftree}

            We take $ P = y_2 . \some_{u,\llfv{M}} ;y_{2}.\close; \piencodf{M}_u$ and continue the derivation:

            \begin{prooftree}
                \small
                    \AxiomC{$\Pi_1$}
                    \AxiomC{}
                    \LeftLabel{\redlab{T\with^x}}
                    \UnaryInfC{$\linvar{x}.\dual{\none} \vdash \linvar{x} :\with A $}
                \LeftLabel{\redlab{T\otimes}}
                \BinaryInfC{$  \outact{\linvar{x}}{y_2} . ( P \para \linvar{x}. \overline{\none}) \vdash \piencodf{\Delta ,  {x}_1:\sigma }, u:\piencodf{\tau} , \linvar{x}: ( \oplus \bot )\otimes ( \with A ) $}
                \LeftLabel{\redlab{T\with_d^x}}
                \UnaryInfC{$\underbrace{ \linvar{x}. \overline{\some}. \outact{\linvar{x}}{y_2} . ( P \para \linvar{x}. \overline{\none}) }_{N_1} \vdash \piencodf{\Delta ,  {x}_1:\sigma } , u:\piencodf{\tau} , \linvar{x}: \overline{\piencodf{\omega}_{(\sigma, i)}} $}
            \end{prooftree}
            
            Notice that the last rule applied \redlab{T\with_d^x} assigns $x: \with ((\oplus \bot) \otimes (\with A))$. Again, since $A$ is arbitrary, we can take $A= \oplus (( \with  \overline{\piencodf{ \sigma }} )  \ampy (\overline{\piencodf{\omega}_{(\sigma, i - 1)}}))$, obtaining $x:\overline{\piencodf{\omega}_{(\sigma,i)}}$.
            
            In order to obtain a type derivation for $N_2$, consider the derivation $\Pi_2$:
            \begin{adjustwidth}{-2cm}{}
            \begin{prooftree}
                \small
                    \AxiomC{$N_1 \vdash \piencodf{\Delta},  {x}_1:\with\overline{\piencodf{\sigma}} , u:\piencodf{\tau}, \linvar{x}: \overline{\piencodf{\omega}_{(\sigma, i)}} $}
                    \LeftLabel{\redlab{T\ampy}}
                    \UnaryInfC{$\linvar{x}( {x}_1) .N_1   \vdash \piencodf{\Delta} ,  u:\piencodf{\tau}, \linvar{x}: ( \with\overline{\piencodf{\sigma}} ) \ampy (\overline{\piencodf{\omega}_{(\sigma, i)}}) $}
                    \LeftLabel{\redlab {T\oplus^x_{ \widetilde{w} }}}
                    \UnaryInfC{$ \linvar{x}.\some_{u, (\llfv{M} \setminus  {x}_1 )}; \linvar{x}( {x}_1) .N_1  \vdash \piencodf{\Delta} ,  u:\piencodf{\tau}, \linvar{x} {:}\oplus (( \with \overline{\piencodf{\sigma}} ) \ampy (\overline{\piencodf{\omega}_{(\sigma, i)}}))$}
                    \LeftLabel{\redlab{T\with_d^x}}
                    \UnaryInfC{$ \linvar{x}.\overline{\some};\linvar{x}.\some_{u, (\llfv{M} \setminus  {x}_1 )};\linvar{x}( {x}_1) .N_1  \vdash \piencodf{\Delta},  u:\piencodf{\tau} , \linvar{x} :\with \oplus (( \with\overline{\piencodf{\sigma}} ) \ampy ( \overline{\piencodf{\omega}_{(\sigma, i)}} ))$}
            \end{prooftree}
        \end{adjustwidth}
            We take $ P_1 = \linvar{x}.\overline{\some};\linvar{x}.\some_{u, (\llfv{M} \setminus  {x}_1 )};\linvar{x}( {x}_1) .N_1 $ and $\Gamma_1 =   \piencodf{\Delta},  u:\piencodf{\tau} $ and continue the derivation of $ N_2 $
            
            \begin{adjustwidth}{-2cm}{}
            \begin{prooftree}
                \small
                    \AxiomC{}
                    \LeftLabel{\redlab{T\cdot}}
                    \UnaryInfC{$0\vdash \dash ; \piencodf{\Theta} $}
                    \LeftLabel{\redlab{T\bot}}
                    \UnaryInfC{$ y_{1}.\close;0 \vdash y_{1} : \bot ; \piencodf{\Theta} $}
                    \LeftLabel{\redlab{T\oplus^x_{\widetilde{w}}}}
                    \UnaryInfC{$ y_1 . \some_{\emptyset} ;y_{1}.\close;0 \vdash  y_1{:}\oplus \bot ; \piencodf{\Theta} $}
                    
                    \AxiomC{$\Pi_2 $}
                    \noLine
                    \UnaryInfC{$\vdots$}
                    \noLine
                    \UnaryInfC{$P_1\vdash \Gamma_1, \linvar{x} :\with \oplus (( \with\overline{ \piencodf{\sigma}} ) \ampy ( \overline{\piencodf{\omega}_{(\sigma, i)}} )) ; \piencodf{\Theta} $}
                \LeftLabel{\redlab{T\otimes}}
                \BinaryInfC{$\outact{\linvar{x}}{y_1}. (y_1 . \some_{\emptyset} ;y_{1}.\close;0 \para P_1) \vdash \Gamma_1 ,  \linvar{x} : (\oplus \bot)\otimes (\with \oplus (( \with\overline{\piencodf{\sigma}} ) \ampy ( \overline{\piencodf{\omega}_{(\sigma, i)}} )) ) ; \piencodf{\Theta} $}
                \LeftLabel{\redlab{T\with_d^x}}
                \UnaryInfC{$\underbrace{\linvar{x}.\overline{\some}. \outact{\linvar{x}}{y_1}. (y_1 . \some_{\emptyset} ;y_{1}.\close;0 \para P_1)}_{N_2} \vdash \Gamma_1 , \linvar{x} : \overline{ \piencodf{\sigma \wedge \omega}_{(\sigma, i)}}  ; \piencodf{\Theta}$}
            \end{prooftree}
        \end{adjustwidth}
            Hence the theorem holds for this case.

            \item {\bf Rule $\redlab{FS:ex \dash sub}$:}
            Then $\expr{M} = (M[ {\widetilde{x}} \leftarrow  {x}])\esubst{ B }{ x }$ and
            
            \begin{prooftree}
                    \AxiomC{\( \Theta , \banged{x} : \eta ; \Gamma ,  {x}: \sigma^{j} \wfdash M[ {\widetilde{x}} \leftarrow  {x}] : \tau  \)}
                    \AxiomC{\( \Theta ; \Delta \wfdash B : (\sigma^{k} , \epsilon ) \)}
                    \AxiomC{\( \eta \relunbag \epsilon \)}
                \LeftLabel{\redlab{FS{:}ex \dash sub}}    
                \TrinaryInfC{\( \Theta ; \Gamma, \Delta \wfdash (M[ {\widetilde{x}} \leftarrow  {x}])\esubst{ B }{ x }  : \tau \)}
            \end{prooftree}
        
            By Proposition~\ref{ch3prop:app_auxunres} and IH we have:
            $$
            \begin{array}{rl} 
            \piencodf{ M[x_1, \cdots , x_k \leftarrow x]}_u&\vdash \piencodf{\Gamma}, \linvar{x}: \overline{ \piencodf{ \sigma^j }_{(\tau, n)}} , u:\piencodf{\tau} ; \piencodf{\Theta} , \banged{x} : \dual{\piencodf{\eta}} \\
            \piencodf{ M[x_1, \cdots , x_k \leftarrow x]}_u&\vdash \piencodf{\Gamma}, \linvar{x}: \overline{ \piencodf{ \sigma^j }_{(\tau, n)}} , u:\piencodf{\tau} ; \piencodf{\Theta} , \banged{x} : \dual{\piencodf{\epsilon}} \quad (*)\\
            \piencodf{B}_x&\vdash \piencodf{\Delta}, x:\piencodf{ (\sigma^{k} , \epsilon ) }_{(\tau, m)}  ; \piencodf{\Theta}
            \end{array}
            $$
        
            Where (*) by is derived from Lemma \ref{ch3lem:relunbag-typeunres}. From \defref{ch3def:enc_lamrsharpifailunres}, we have 
            \begin{equation*}
                \hspace{-2cm}
            \small
               \piencodf{ M[ {\widetilde{x}} \leftarrow  {x}]\ \esubst{ B }{ x }}_u = \hspace{-0.5cm} \bigoplus_{C_i \in \perm{C}} \hspace{-0.4cm} (\nu x)( x.\overline{\some}; x(\linvar{x}). x(\banged{x}). x. \close ;\piencodf{ M[ {\widetilde{x}} \leftarrow  {x}]}_u \para \piencodf{ C_i \bagsep U}_x )  
            \end{equation*}

            Therefore, for each $B_i \in \perm{B} $, we obtain the following derivation $\Pi_i$:
            \begin{adjustwidth}{-2cm}{}
            \begin{prooftree}
                \small
                \AxiomC{$ \piencodf{ M[ {\widetilde{x}} \leftarrow  {x}]}_u \vdash \piencodf{\Gamma}, \linvar{x}: \overline{ \piencodf{ \sigma^j }_{(\tau, n)}} , u:\piencodf{\tau} ; \piencodf{\Theta} , \banged{x} : \dual{\piencodf{\epsilon}} $}
                \LeftLabel{\redlab{T\bot}}
                \UnaryInfC{$ x.\close ;\piencodf{ M[ {\widetilde{x}} \leftarrow  {x}]}_u \vdash x{:}\bot, \piencodf{\Gamma}, \linvar{x}: \overline{ \piencodf{ \sigma^j }_{(\tau, n)}} , u:\piencodf{\tau} ; \piencodf{\Theta} , \banged{x} : \dual{\piencodf{\epsilon}}$}
                \LeftLabel{\redlab{T?}}
                \UnaryInfC{$ x.\close ;\piencodf{ M[ {\widetilde{x}} \leftarrow  {x}]}_u \vdash x{:}\bot, \piencodf{\Gamma}, \linvar{x}: \overline{ \piencodf{ \sigma^j }_{(\tau, n)}} , u:\piencodf{\tau} , \banged{x} :! \dual{\piencodf{\epsilon}}; \piencodf{\Theta} $}
                \LeftLabel{\redlab{T\ampy}}
                \UnaryInfC{$x(\banged{x}). x.\close ;\piencodf{ M[ {\widetilde{x}} \leftarrow  {x}]}_u \vdash x:  ( ! \dual{\piencodf{\epsilon}} ) \ampy \bot , \piencodf{\Gamma}, \linvar{x}: \overline{ \piencodf{ \sigma^j }_{(\tau, n)}} , u:\piencodf{\tau} ; \piencodf{\Theta} $}
                \LeftLabel{\redlab{T\ampy}}
                \UnaryInfC{$x(\linvar{x}). x(\banged{x}). x. \close ;\piencodf{ M[ {\widetilde{x}} \leftarrow  {x}]}_u \vdash x: \overline{ \piencodf{ \sigma^j }_{(\tau, n)}} \ampy ( ( ! \dual{\piencodf{\epsilon}} ) \ampy \bot) , \piencodf{\Gamma}, u:\piencodf{\tau} ; \piencodf{\Theta} $}
                \LeftLabel{\redlab{T\with_d^x}}
                \UnaryInfC{$x.\overline{\some}; x(\linvar{x}). x(\banged{x}). x. \close ;\piencodf{ M[ {\widetilde{x}} \leftarrow  {x}]}_u \vdash x : \overline{\piencodf{ (\sigma^{j} , \epsilon  )  }_{(\tau, n)}}  , \piencodf{\Gamma}, u:\piencodf{\tau} ; \piencodf{\Theta}  $}
             \end{prooftree}
            \end{adjustwidth}
            
            We take $ P_1 = x.\overline{\some}; x(\linvar{x}). x(\banged{x}). x. \close ;\piencodf{ M[ {\widetilde{x}} \leftarrow  {x}]}_u$ and continue the derivation of $ \Pi_i $
            \begin{adjustwidth}{-2cm}{}
            \begin{prooftree}
                \small
                \AxiomC{$ P_1 \vdash  x : \overline{\piencodf{ (\sigma^{j} , \epsilon  )  }_{(\tau, n)}}  , \piencodf{\Gamma}, u:\piencodf{\tau} ; \piencodf{\Theta}  $}
                \AxiomC{$  \piencodf{ C_i \bagsep U}_x \vdash \piencodf{\Delta}, x:\piencodf{ (\sigma^{k} , \epsilon ) }_{(\tau, m)}  ; \piencodf{\Theta} $}
            \LeftLabel{$\redlab{Tcut}$}
            \BinaryInfC{$ (\nu x)( P_1 \para \piencodf{ C_i \bagsep U}_x )  \vdash \piencodf{ \Gamma} , \piencodf{ \Delta } , u: \piencodf{ \tau }  ; \piencodf{\Theta}  $}               
            \end{prooftree}
        \end{adjustwidth}
            
            We must have that $\piencodf{\sigma^{j}}_{(\tau, m)} = \piencodf{\sigma^{k}}_{(\tau, n)}$ which by our restrictions allows.
            Therefore, from $\Pi_i$ and multiple applications of $\redlab{T\with}$ it follows that
            
            \begin{prooftree}
                        \AxiomC{$\forall \bigoplus_{C_i \in \perm{C}} \hspace{1cm} \Pi_i$}
                        \LeftLabel{$\redlab{T\with}$}
            \UnaryInfC{$ \bigoplus_{C_i \in \perm{C}}  (\nu x)( P_1 \para \piencodf{ C_i \bagsep U}_x )  \vdash \piencodf{ \Gamma} , \piencodf{ \Delta } , u: \piencodf{ \tau }  ; \piencodf{\Theta} $}
            \end{prooftree}
            that is, $\piencodf{M[  {x}_1 \leftarrow  {x}]\ \esubst{ B }{ x }}\vdash \piencodf{\Gamma, \Delta}, u:\piencodf{\tau}  ; \piencodf{\Theta}$ and the result follows.

            \item {\bf Rule $\redlab{FS{:}ex \dash sub^{\ell}}$: } 
            Then $\expr{M} =  M \linexsub{N /  {x}}$ and
            \begin{prooftree}
                \AxiomC{\( \Theta ; \Gamma  ,  {x}:\sigma \wfdash M : \tau \quad  \Theta ; \Delta \wfdash N : \sigma \)}
                    \LeftLabel{\redlab{FS{:}ex \dash sub^{\ell}}}
                \UnaryInfC{\( \Theta ; \Gamma, \Delta \wfdash M \linexsub{N /  {x}} : \tau \)}
            \end{prooftree}
        
            By IH we have both 
            $$\piencodf{N}_{ {x}}\vdash \piencodf{\Delta},  {x}: \piencodf{\sigma} ; \piencodf{\Theta}$$ 
            $$ \piencodf{M}_u\vdash \piencodf{\Gamma},  {x}: \with\overline{\piencodf{\sigma}}, u:\piencodf{\tau} ; \piencodf{\Theta}$$
        
            From Definition \ref{ch3def:enc_lamrsharpifailunres}, $\piencodf{M \linexsub{N /  {x}} }_u= (\nu  {x}) ( \piencodf{ M }_u \para    {x}.\some_{\llfv{N}};\piencodf{ N }_{ {x}} ) $ and 
            \begin{adjustwidth}{-2cm}{}
            \begin{prooftree}
                \small
                    \AxiomC{\( \piencodf{ M }_u \vdash   \piencodf{\Gamma},  {x}: \with\overline{\piencodf{\sigma}}, u:\piencodf{\tau} ; \piencodf{\Theta}\)}
                    
                    \AxiomC{\( \piencodf{ N }_x  \vdash  \piencodf{\Delta},  {x}: \piencodf{\sigma} ; \piencodf{\Theta} \)}
                    \LeftLabel{$\redlab{T\oplus^x}$}
                    \UnaryInfC{\(  {x}.\some_{\llfv{N}};\piencodf{ N }_x \vdash  \piencodf{ \Delta } ,  {x}: : \oplus \piencodf{\sigma}\)}
                \LeftLabel{$\redlab{TCut}$}
                \BinaryInfC{\(  (\nu  {x}) ( \piencodf{ M }_u \para    {x}.\some_{\llfv{N}};\piencodf{ N }_{ {x}} ) \vdash  \piencodf{ \Gamma} , \piencodf{ \Delta } , u : \piencodf{ \tau }  \)}
            \end{prooftree}
        \end{adjustwidth}
    
            Observe that for the application of rule $\redlab{TCut}$ we used the fact that $\overline{\oplus\piencodf{\sigma}}=\with \overline{\piencodf{\sigma}}$. Therefore, $\piencodf{M \linexsub{N /  {x}} }_u\vdash \piencodf{ \Gamma} , \piencodf{ \Delta } , u : \piencodf{ \tau } $ and the result follows.

            \item {\bf Rule $\redlab{FS{:}ex \dash sub^!}$: }
            Then $\expr{M} =  M \unexsub{U / \unvar{x}}$ and
            \begin{prooftree}
                \AxiomC{\( \Theta , \banged{x} : \eta; \Gamma \wfdash M : \tau \quad  \Theta ; \dash \wfdash U : \eta \)}
                    \LeftLabel{\redlab{FS{:}ex \dash sub^!}}
                \UnaryInfC{\( \Theta ; \Gamma \wfdash M \unexsub{U / \unvar{x}}  : \tau \)}
            \end{prooftree}
        
            By IH we have both 
            $$
            \begin{array}{rl}
            \piencodf{U}_{x_i}&\vdash x_i : \piencodf{\eta}  ; \piencodf{\Theta}\\ 
             \piencodf{M}_u&\vdash \piencodf{\Gamma} , u:\piencodf{\tau} ; \banged{x}: \overline{\piencodf{\eta}} , \piencodf{\Theta}
             \end{array}
             $$
        
            From Definition \ref{ch3def:enc_lamrsharpifailunres}, $ \piencodf{ M \unexsub{U / \unvar{x}}  }_u   =   (\nu \banged{x}) ( \piencodf{ M }_u \para   !\banged{x}. (x_i).\piencodf{ U }_{x_i} ) $ and 
            
            \begin{prooftree}
                \small
                    \AxiomC{$\piencodf{M}_u\vdash \piencodf{\Gamma} , u:\piencodf{\tau}; \banged{x}: \overline{\piencodf{\eta}} , \piencodf{\Theta}$}
                    \LeftLabel{\redlab{T?}}
                    \UnaryInfC{$\piencodf{M}_u \vdash \piencodf{\Gamma} , u:\piencodf{\tau}, \banged{x}: ? \overline{\piencodf{\eta}} ; \piencodf{\Theta}$}
                    
                    \AxiomC{$ \piencodf{ U }_{x_i} \vdash x_i : \piencodf{\eta} ; \piencodf{\Theta} $}
                    \LeftLabel{\redlab{T!}}
                    \UnaryInfC{$!\banged{x}. (x_i).\piencodf{ U }_{x_i} \vdash \banged{x}: !\piencodf{\eta} ; \piencodf{\Theta}  $}
                \LeftLabel{$\redlab{TCut}$}
                \BinaryInfC{\(   (\nu \banged{x}) ( \piencodf{ M }_u \para   !\banged{x}. (x_i).\piencodf{ U }_{x_i} ) \vdash \piencodf{\Gamma} , u:\piencodf{\tau} ; \piencodf{\Theta} \)}
            \end{prooftree}
    
            Observe that for the application of rule $\redlab{TCut}$ we used the fact that $\overline{ !\piencodf{\eta} }= ? \overline{\piencodf{\eta}} $. Therefore, $\piencodf{M \unexsub{U / \unvar{x}} }_u \vdash \piencodf{\Gamma} , u:\piencodf{\tau} ; \piencodf{\Theta} $ and the result follows.

            \item {\bf Rule $\redlab{FS:fail}$:}
            Then $\expr{M} = \fail^{\widetilde{x}}$ where $ \widetilde{x} = x_1, \cdots , x_n$ and
            
                \begin{prooftree}
                    \AxiomC{\( \dom{\Gamma} = \widetilde{x}\)}
                    \LeftLabel{\redlab{FS{:}fail}}
                    \UnaryInfC{\( \Theta ; \Gamma \wfdash  \fail^{\widetilde{x}} : \tau \)}
                \end{prooftree}
            
            From Definition \ref{ch3def:enc_lamrsharpifailunres}, $\piencodf{\fail^{x_1, \cdots , x_n} }_u= u.\overline{\none} \para x_1.\overline{\none} \para \cdots \para x_k.\overline{\none} $ and 
            
            \begin{adjustwidth}{-2cm}{}
            \begin{prooftree}
                \smaller
                    \AxiomC{}
                    \LeftLabel{\redlab{T\with^u}}
                    \UnaryInfC{$u.\overline{\none} \vdash u : \piencodf{ \tau } ; \piencodf{\Theta} $}

                        \AxiomC{}
                        \LeftLabel{\redlab{T\with^{x_1}}}
                        \UnaryInfC{$x_1.\overline{\none} \vdash_1 : \with \overline{\piencodf{\sigma_1}} ; \piencodf{\Theta} $}
                        
                        \AxiomC{}
                        \LeftLabel{\redlab{T\with^{x_n}}}
                        \UnaryInfC{$x_n.\overline{\none} \vdash x_n : \with \overline{\piencodf{\sigma_n}} ; \piencodf{\Theta} $}
                        \UnaryInfC{$\vdots$}
                    \BinaryInfC{$x_1.\overline{\none} \para \cdots \para x_k.\overline{\none} \vdash  x_1 : \with \overline{\piencodf{\sigma_1}}, \cdots  ,x_n : \with \overline{\piencodf{\sigma_n}} ; \piencodf{\Theta} $}
                \LeftLabel{\redlab{T\para}}
                \BinaryInfC{$u.\overline{\none} \para x_1.\overline{\none} \para \cdots \para x_k.\overline{\none} \vdash x_1 : \with \overline{\piencodf{\sigma_1}}, \cdots  ,x_n : \with \overline{\piencodf{\sigma_n}}, u : \piencodf{ \tau } ; \piencodf{\Theta} $}
            \end{prooftree}
        \end{adjustwidth}
            Thus, $\piencodf{\fail^{x_1, \cdots , x_n} }_u\vdash  x_1 : \with \overline{\piencodf{\sigma_1}}, \cdots  ,x_n : \with \overline{\piencodf{\sigma_n}}, u : \piencodf{ \tau } ; \piencodf{\Theta} $ and the result follows.

            \item Rule $\redlab{FS:sum}$: 
            This case follows easily by IH.
        \end{enumerate}
    \end{enumerate}
\end{proof}

\subsection{Operational Correspondence: Completeness and Soundness}

\begin{proposition}
\label{ch3prop:correctformfailunres}
Let  $N$ be a well-formed linearly closed $\lamrsharfailunres$-term with $\headf{N} = x$ ($x$ denoting either linear or unrestricted occurrence of $x$) such that $\llfv{N} = \emptyset$ and $N$ does not fail, that is, there is no $Q\in \lamrsharfailunres$ for which there is a  reduction $N  \redd_{\redlab{RS:Fail}} Q$.
Then,
$$\piencodf{ N }_{u} \redd^* \bigoplus_{i \in I}(\nu \widetilde{y})(\piencodf{ x }_{n} \para P_i) $$ for some index set $I$, names $\widetilde{y}$ and $n$, and processes $P_i$.
\end{proposition}

\begin{proof}
By induction on the structure of $N$.
\begin{enumerate}

    \item $N =  {x}$ or $N=x[j]$:
    
    These cases are trivial, and follow taking 
    $ I = \emptyset$ and $ \widetilde{y} = \emptyset$.
    
    

    \item $N = (M\ B)$:
    
    Then $\headf{M\ B} = \headf{M} = x$ then 
    \[ \piencodf{N}_u = \piencodf{M\ B}_u  = \bigoplus_{B_i \in \perm{B}} (\nu v)(\piencodf{M}_v \para v.\some_{u, \llfv{B}} ; \outact{v}{x} . ([v \leftrightarrow u] \para \piencodf{B^x_i} ) ) \]
   and the proof follows by induction on $\piencodf{M}_u$.
    
    \item  $N = (M[\widetilde{y} \leftarrow y])\esubst{ C \bagsep U }{ y }$:
    
    Then $\headf{(M[\widetilde{y} \leftarrow y])\esubst{ C \bagsep U }{ y }} = \headf{(M[\widetilde{y} \leftarrow y])} = x$. As $N  \redd_{\redlab{R}} $ where $\redlab{R} \not = \redlab{RS:Fail} $ we must have that $\size{\widetilde{y}} = \size{C}$. Thus,

    {\small
    \[
    \begin{aligned}
       \piencodf{N}_u &= \piencodf{(M[\widetilde{y} \leftarrow y])\esubst{ C \bagsep U }{ y }}_u  \\
       & =  \bigoplus_{C_i \in \perm{C}} (\nu y)( y.\overline{\some}; y(\linvar{y}). y(\banged{y}). y. \close ;\piencodf{ M[\widetilde{y} \leftarrow  {y}]}_u \para \piencodf{ C_i \bagsep U}_y )
       \\
       & =  \bigoplus_{C_i \in \perm{C}} (\nu y)( y.\overline{\some}; y(\linvar{y}). y(\banged{y}). y. \close ;\piencodf{ M[\widetilde{y} \leftarrow  {y}]}_u \para \\
       & \hspace{2cm} y.\some_{\llfv{C}} ; \outact{y}{\linvar{y}} .( \piencodf{ C_i }_{\linvar{y}} \para \outact{y}{\banged{y}} .( !\banged{y}. (y_i). \piencodf{ U }_{y_i} \para y.\overline{\close} ) ) )
       \\[4pt]
       & \redd^*  \bigoplus_{C_i \in \perm{C}} (\nu  \linvar{y}, \banged{y})(  \piencodf{ M[\widetilde{y} \leftarrow \linvar{y}]}_u \para  \piencodf{ C_i }_{\linvar{y}} \para  !\banged{y}. (y_i). \piencodf{ U }_{y_i}  )
       \\[4pt]
       &=\bigoplus_{C_i \in \perm{C}} (\nu \linvar{y}, \banged{y})( \linvar{y}.\overline{\some}. \outact{\linvar{y}}{z_1}. (z_1 . \some_{\emptyset} ;z_{1}.\close;0 \para \linvar{y}.\overline{\some};\\
         &\hspace{1cm}\linvar{y}. \some_{u, (\llfv{M} \setminus  {y}_1 , \cdots ,  {y}_n )};\linvar{y}( {y}_1) . \cdots\linvar{y}.\overline{\some}. \outact{\linvar{y}}{z_n} . (z_n . \some_{\emptyset} ; z_{n}.\close;0 \\
        &\hspace{1cm}\para \linvar{y}.\overline{\some};\linvar{y}.\some_{u,(\llfv{M} \setminus  {y}_n )}; \linvar{y}( {y}_n).\linvar{y}.\overline{\some}; \outact{\linvar{y}}{z_{n+1}}. ( z_{n+1} . \some_{u, \llfv{M}} ; \\%
        &\hspace{1cm}z_{n+1}.\close; \piencodf{M}_u \para \linvar{y}.\overline{\none} ) ) \cdots ) \para \linvar{y}.\some_{\llfv{C}} ; \linvar{y}(z_1). \linvar{y}.\some_{z_1,\llfv{C}}; \\ 
              &\hspace{1cm}  \linvar{y}.\overline{\some} ;\outact{\linvar{y}}{ {y}_1}. ( {y}_1.\some_{\llfv{C_i(1)}} ;\piencodf{C_i(1)}_{ {y}_1} \para  \cdots \linvar{y}.\some_{\llfv{C_i(n)}} ; \linvar{y}(z_n).
               \\
              &\hspace{1cm}  \linvar{y}.\some_{z_n,\llfv{C_i(n)}};\linvar{y}.\overline{\some} ;\outact{\linvar{y}}{ {y}_n}. ( {y}_n.\some_{\llfv{C_i(n)}} ; \piencodf{C_i(n)}_{ {y}_n} \para\linvar{y}.\some_{\emptyset} ;  \\
              &\hspace{1cm} \linvar{y}(z_{n+1}). ( z_{n+1}.\overline{\some};z_{n+1} . \overline{\close} \para \linvar{y}.\some_{\emptyset} ; \linvar{y}. \overline{\none})  \para z_1. \overline{\none}) \para \\
              &\hspace{1cm} \cdots\para z_n. \overline{\none}) \para  !\banged{y}. (y_i). \piencodf{ U }_{y_i} )
              \\
       &\redd^* \bigoplus_{C_i \in \perm{C}} (\nu  \widetilde{y}, \banged{y})(  \piencodf{M}_u \para  {y}_1.\some_{\llfv{C_i(1)}} ;\piencodf{C_i(1)}_{ {y}_1} \para   \cdots \para   {y}_n.\some_{\llfv{C_i(n)}} ; \\
       & \hspace{1cm}\piencodf{C_i(n)}_{ {y}_n} \para  !\banged{y}. (y_i). \piencodf{ U }_{y_i}  )\\
    \end{aligned}
    \]}
    and the result follows by induction on $\piencodf{ M }_u $.
    
    \item $N = M \linexsub {N' / {y}}$ and $N = M \unexsub {u /\unvar{y}}$:
    

    
    These cases  follow easily by induction  on $\piencodf{M}_u$.
    
\end{enumerate}
\end{proof}

\subsubsection{Completeness}

Here again, because of the diamond property (Proposition \ref{ch3app:lambda}), it suffices to consider a completeness result based on a single reduction step in $\lamrsharfailunres$:

\begin{notation}
    We use the notation $\llfv{M}.\overline{\none}$ and $\widetilde{x}.\overline{\none}$ where $\llfv{M}$ or $\widetilde{x}$ are equal to $ x_1 , \cdots , x_k$ to describe a process of the form $x_1.\overline{\none} \para \cdots \para x_k.\overline{\none} $
\end{notation}

\begin{theorem}[Well Formed Operational Completeness]
\label{ch3l:app_completenesstwounres}
Let $\expr{N} $ and $ \expr{M} $ be well-formed, linearly closed $\lamrsharfailunres $ expressions. If $ \expr{N}\redd \expr{M}$ then there exists $Q$ such that $\piencodf{\expr{N}}_u  \redd^* Q \equiv \piencodf{\expr{M}}_u$.
\end{theorem}

\begin{proof}

By induction on the reduction rule applied to infer $\expr{N}\redd \expr{M}$.  
We have ten cases.

    \begin{enumerate}
        \item  {\bf Case $\redlab{RS:Beta}$: }
              
        Then  $ \expr{N}= (\lambda x . (M[ {\widetilde{x}} \leftarrow  {x}])) B  \redd (M[ {\widetilde{x}} \leftarrow  {x}])\esubst{ B }{ x }  = \expr{M}$ , where $B = C \bagsep U$. Notice that
        \begin{equation*}\label{ch3eq:compl_lsbeta1failunres}\small
        \begin{aligned}
        \piencodf{\expr{N}}_u =&  \bigoplus_{C_i \in \perm{C}} (\nu v)(\piencodf{\lambda x . (M[ {\widetilde{x}} \leftarrow  {x}])}_v \para v.\some_{u , \llfv{C}} ; \outact{v}{x} . ([v \leftrightarrow u] \para \piencodf{C_i \bagsep U}_x ) )\\
        =&  
            \bigoplus_{C_i \in \perm{C}} (\nu v) 
                 (v.\overline{\some}; v(x). x.\overline{\some}; x(\linvar{x}). x(\banged{x}). x. \close ; \piencodf{M[\widetilde{x} \leftarrow x]}_v \\
                 &\para v.\some_{u,\llfv{C}} ; \outact{v}{x} . ( \piencodf{C_i \bagsep U}_x \para [v \leftrightarrow u] ) )\\
        \redd &  \bigoplus_{C_i \in \perm{C}} (\nu v)( v(x). x.\overline{\some}; x(\linvar{x}). x(\banged{x}). x. \close ; \piencodf{M[\widetilde{x} \leftarrow x]}_v \para \outact{v}{x} . ( \piencodf{ C_i \bagsep U}_x \\
        & \hspace{1cm}\para [v \leftrightarrow u] ) )\\
    \redd &  \bigoplus_{C_i \in \perm{C}} (\nu v, x)( x.\overline{\some}; x(\linvar{x}). x(\banged{x}). x. \close ; \piencodf{M[\widetilde{x} \leftarrow x]}_v \para  \piencodf{ C_i \bagsep U}_x \para [v \leftrightarrow u] )\\ 
    \redd&  \bigoplus_{C_i \in \perm{C}} (\nu x)( x.\overline{\some}; x(\linvar{x}). x(\banged{x}). x. \close ; \piencodf{M[\widetilde{x} \leftarrow x]}_v \para  \piencodf{ C_i \bagsep U}_x )=\piencodf{\expr{M}}_u \\
        \end{aligned}
        \end{equation*}
and the result follows.

        
        \item {\bf Case $ \redlab{RS:Ex \dash Sub}$:}
        
        Then $ \expr{N} =M[ {x}_1, \!\cdots\! ,  {x}_k \leftarrow  {x}]\esubst{ C \bagsep U }{ x }$, with $C = \bag{M_1}
            \cdots  \bag{M_k}$, $k\geq 0$ and $M \not= \fail^{\widetilde{y}}$.

        The reduction is 
        $$
        \begin{aligned}
            \expr{N} =& M[ {x}_1, \!\cdots\! ,  {x}_k \leftarrow  {x}]\esubst{ C \bagsep U }{ x } \\
            & \redd \sum_{C_i \in \perm{C}}M\linexsub{C_i(1)/ {x_1}} \cdots \linexsub{C_i(k)/ {x_k}} \unexsub{U /\unvar{x} } = \expr{M}
        \end{aligned}
        $$
        
        We detail the encodings of $\piencodf{\expr{N}}_u$ and $\piencodf{\expr{M}}_u$. To simplify the proof, we will consider $k=1$ (the case in which $k> 1$ is follows analogously, similarly the case of $k=0$ is contained within the proof of $k=1$). 
        
        On the one hand, we have:
        \begin{equation*}\label{ch3eq:compl_lsbeta3failunres} \small
        \begin{aligned}
        \piencodf{\expr{N}}_u &= \piencodf{M[ {x}_1 \leftarrow  {x}]\esubst{ C \bagsep U }{ x }}_u\\
        &= \bigoplus_{C_i \in \perm{C}} (\nu x)( x.\overline{\some}; x(\linvar{x}). x(\banged{x}). x. \close ;\piencodf{ M[ {x}_1 \leftarrow  {x}]}_u \para \piencodf{ C_i \bagsep U}_x ) \\
        &= \bigoplus_{C_i \in \perm{C}} (\nu x)( x.\overline{\some}; x(\linvar{x}). x(\banged{x}). x. \close ; \piencodf{ M[ {x}_1 \leftarrow  {x}]}_u \para  x.\some_{\llfv{C}} ;\outact{x}{\linvar{x}} .\\
        &\qquad ( \piencodf{ C_i }_{\linvar{x}} \para \outact{x}{\banged{x}} .( !\banged{x}. (x_i). \piencodf{ U }_{x_i} \para x.\overline{\close}   ) ) ) \qquad (:= P_{\mathbb{N}}) \\[4pt]
        \end{aligned}
        \end{equation*}
        
      Note that
        \begin{equation*}\small
        \begin{aligned}
        P_{\mathbb{N}} \redd^*& \bigoplus_{C_i \in \perm{C}} (\nu \linvar{x} , \banged{x} )( \piencodf{ M[ {x}_1 \leftarrow  {x}]}_u \para  \piencodf{ C_i }_{\linvar{x}} \para  !\banged{x}. (x_i). \piencodf{ U }_{x_i}    ) \\
        =&  \bigoplus_{C_i \in \perm{C}} (\nu \linvar{x} , \banged{x} )( \linvar{x}.\overline{\some}. \outact{\linvar{x}}{y_1}. (y_1 . \some_{\emptyset} ;y_{1}.\close;0 \para \linvar{x}.\overline{\some};\\
        &\linvar{x}.\some_{u, (\llfv{M} \setminus  {x}_1 )};\linvar{x}( {x}_1) . \linvar{x}.\overline{\some}; \outact{\linvar{x}}{y_{2}}.( y_{2} .   \some_{u,\llfv{M}} ;y_{2}.\close; \piencodf{M}_u \para \\
         & \linvar{x}.\overline{\none} ) ) \para\linvar{x}.\some_{\llfv{B_i(1)}} ; \linvar{x}(y_1). \linvar{x}.\some_{y_1,\llfv{C_i(1)}};\linvar{x}.\overline{\some} ;   \outact{\linvar{x}}{ {x}_1}.\\
        & ( {x}_1.\some_{\llfv{C_i(1)}} ; \piencodf{C_i(1)}_{ {x}_1} \para y_1. \overline{\none} \para \linvar{x}.\some_{\emptyset} ;\linvar{x}(y_2). ( y_2.\overline{\some};y_2 . \overline{\close} \para \\
        &\linvar{x}.\some_{\emptyset} ; \linvar{x}.\overline{\none}) ) \para !\banged{x}. (x_i). \piencodf{ U }_{x_i})
        \\
         \redd^*& \bigoplus_{C_i \in \perm{C}} (\nu  {x}_1 , \banged{x} )(\piencodf{M}_u \para  {x}_1.\some_{\llfv{C_i(1)}} ; \piencodf{C_i(1)}_{ {x}_1} \para !\banged{x}. (x_i). \piencodf{ U }_{x_i} )= \piencodf{\expr{M}}_u 
        \\
        \end{aligned}
        \end{equation*}
        
and the result follows.
        
        \item {\bf Case $\redlab{RS{:}Fetch^{\ell}}$:}
        
        Then we have
        $\expr{N} = M \linexsub{N /  {x}}$ with $\headf{M} =  {x}$ and $\expr{N} \redd   M \headlin{ N/ {x} } = \expr{M}$.
   Note that
            \begin{equation*}\label{ch3eq:compl_lsbeta5failunres}
            \begin{aligned}
            \piencodf{\expr{N}}_u &= \piencodf{M \linexsub{N /  {x}}}_u\\
            &= (\nu  {x}) ( \piencodf{ M }_u \para   {x}.\some_{\llfv{N}};\piencodf{ N }_{ {x}}  ) \\
            &\redd^* (\nu  {x}) ( \bigoplus_{i \in I}(\nu \widetilde{y})(\piencodf{  {x} }_{j} \para P_i) \para    {x}.\some_{\llfv{N}};\piencodf{ N }_{ {x}}  ) \qquad (*)   
            \\
            &= (\nu  {x}) ( \bigoplus_{i \in I}(\nu \widetilde{y})(\piencodf{  {x} }_{j} \para P_i)  \para    {x}.\some;\piencodf{ N }_{ {x}}  )  \\
            & \redd (\nu  {x}) ( \bigoplus_{i \in I}(\nu \widetilde{y})([ {x} \leftrightarrow j ] \para P_i) \para   \piencodf{ N }_{ {x}}  )\\ &\redd \bigoplus_{i \in I}(\nu \widetilde{y})(P_i \para   \piencodf{ N }_j  )    = \piencodf{\expr{M}}_u
            \end{aligned}
            \end{equation*}
       where the reductions denoted by $(*)$ are inferred via Proposition~\ref{ch3prop:correctformfailunres}, and the result follows.
        

        \item {\bf Case $ \redlab{RS{:} Fetch^!}$: }
        
        Then,
        $\expr{N} = M \unexsub{U / \unvar{x}}$ with $\headf{M} = \banged{x}[k]$, $U_i = \banged{\bag{N}}$ and $\expr{N} \redd  M \headlin{ N /\banged{x} }\unexsub{U / \unvar{x}} = \expr{M}$.
        Note that 
            \begin{equation*}\label{ch3eq:compl_lsbeta5failunresunres}
            \begin{aligned}
            \hspace{-0.5cm}
            \piencodf{\expr{N}}_u &= \piencodf{M \unexsub{U / \unvar{x}}}_u
            = (\nu \banged{x}) ( \piencodf{ M }_u \para   !\banged{x}. (x_k).\piencodf{ U }_{x_k}  ) \\
            & \redd^*(\nu \banged{x}) ( \bigoplus_{i \in I}(\nu \widetilde{y})(\piencodf{ \banged{x}[k] }_{j} \para P_i) \para   !\banged{x}. (x_k).\piencodf{ U }_{x_k}  ) \qquad (*)
            \\
            &  = (\nu \banged{x}) ( \bigoplus_{i \in I}(\nu \widetilde{y})(\outsev{\banged{x}}{{x_k}}. {x}_k.l_{i}; [{x_k} \leftrightarrow j] \para P_i) \para   !\banged{x}. (x_k).\piencodf{ U }_{x_k}  ) \qquad (*)
            \\
            & \redd (\nu \banged{x}) ( \bigoplus_{i \in I}(\nu \widetilde{y})(  (\nu x_k)( x_k.l_{i}; [x_k \leftrightarrow j] \para \piencodf{ U }_{x_k}) \para P_i) \para   !\banged{x}. (x_k).\piencodf{ U }_{x_k}  )
            \\
            & = (\nu \banged{x}) ( \bigoplus_{i \in I}(\nu \widetilde{y})(  (\nu x_k)( x_k.l_{i}; [x_k \leftrightarrow j] \para x_k. case( i.\piencodf{U_i}_{x} )) \para P_i) \para   !\banged{x}. (x_k).\piencodf{ U }_{x_k}  )
            \\
            & \redd (\nu \banged{x}) ( \bigoplus_{i \in I}(\nu \widetilde{y})(  \piencodf{\banged{\bag{N}}}_{j} ) \para P_i) \para   !\banged{x}. (x_k).\piencodf{ U }_{x_k}  )\\
            &= (\nu \banged{x}) ( \bigoplus_{i \in I}(\nu \widetilde{y})(  \piencodf{N}_{j} ) \para P_i) \para   !\banged{x}. (x_k).\piencodf{ U }_{x_k}  ) = \piencodf{\expr{M}}_u
            \end{aligned}
            \end{equation*}
       where the reductions denoted by $(*)$ are inferred via Proposition~\ref{ch3prop:correctformfailunres}.
        

        \item {\bf Cases $\redlab{RS:TCont}$ and $\redlab{RS:ECont}$:}
         
         These cases follow by IH.
         

        \item {\bf Case $\redlab{RS{:}Fail^{\ell}}$:}
        
        Then, 
        $\expr{N} = M[ {x}_1, \!\cdots\! ,  {x}_k \leftarrow  {x}]\ \esubst{C \bagsep U}{ x } $ with $k \neq \size{C}$ and
        
        $\expr{N} \redd  \sum_{C_i \in \perm{C}}  \fail^{\widetilde{y}} = \expr{M}$, where $\widetilde{y} = (\llfv{M} \setminus \{   {x}_1, \cdots ,  {x}_k \} ) \cup \llfv{C}$. 
        
        Let $ \size{C} = l$ and we assume that $k > l$ (proceed similarly for $k > l$). Hence $k = l + m$ for some $m \geq 1$, and 
        {\small
        \begin{equation*}\label{ch3eq:compl_fail1-failunres}
            \hspace{-1cm}
        \begin{aligned}
            \piencodf{N}_u =& \piencodf{M[ {x}_1, \!\cdots\! ,  {x}_k \leftarrow  {x}]\ \esubst{C \bagsep U}{ x } }_u\\
             =&  \bigoplus_{C_i \in \perm{C}} (\nu x)( x.\overline{\some}; x(\linvar{x}). x(\banged{x}). x. \close ;\piencodf{ M[\widetilde{x} \leftarrow  {x}]}_u \para \\
            &  x.\some_{\llfv{C}} ; \outact{x}{\linvar{x}} .( \piencodf{ C }_{\linvar{x}} \para \outact{x}{\banged{x}} .( !\banged{x}. (x_i). \piencodf{ U }_{x_i} \para x.\overline{\close} ) ) )
            \\
             \redd^* &  \bigoplus_{C_i \in \perm{C}} (\nu \linvar{x} , \banged{x})(  \piencodf{ M[\widetilde{x} \leftarrow  {x}]}_u \para \piencodf{ C }_{\linvar{x}} \para !\banged{x}. (x_i). \piencodf{ U }_{x_i}  )\\
              =& \bigoplus_{C_i \in \perm{C}} (\nu \linvar{x} , \banged{x})( \linvar{x}.\overline{\some}. \outact{\linvar{x}}{y_1}. (y_1 . \some_{\emptyset} ;y_{1}.\close;0 \para\linvar{x}.\overline{\some}; \\
              & \linvar{x}.\some_{u,(\llfv{M} \setminus  \widetilde{x} )}; \linvar{x}( {x}_1) . \cdots  \linvar{x}.\overline{\some}. \outact{\linvar{x}}{y_k} . (y_k . \some_{\emptyset} ; y_{k}.\close; \0 \para \\
               & \linvar{x}.\overline{\some};\linvar{x}.\some_{u,(\llfv{M} \setminus   {x}_k )};\linvar{x}( {x}_k) . \linvar{x}.\overline{\some}; \outact{\linvar{x}}{y_{k+1}}. ( y_{k+1} . \some_{u,\llfv{M} } ; \\
               & y_{k+1}.\close; \piencodf{M}_u \para \linvar{x}.\overline{\none} )) \cdots ) \para\linvar{x}.\some_{\llfv{C}} ; \linvar{x}(y_1). \linvar{x}.\some_{y_1,\llfv{C}}; \\
               & \linvar{x}.\overline{\some} ; \outact{\linvar{x}}{ {x}_1}. ( {x}_1.\some_{\llfv{C_i(1)}} ;\piencodf{C_i(1)}_{ {x}_1} \para y_1. \overline{\none} \para \cdots \linvar{x}.\some_{\llfv{C_i(l)}} ;   \\
               & \linvar{x}(y_l). \linvar{x}.\some_{y_l ,\llfv{C_i(l)}};\linvar{x}.\overline{\some} ; \outact{\linvar{x}}{ {x}_l}.( {x}_l.\some_{\llfv{C_i(l)}} ;\piencodf{C_i(l)}_{ {x}_l} \para \\
               & y_l. \overline{\none} \para \linvar{x}.\some_{\emptyset} ; \linvar{x}(y_{l+1}). ( y_{l+1}.\overline{\some};y_{l+1} . \overline{\close}\para \linvar{x}.\some_{\emptyset} ;\linvar{x}. \overline{\none}) ) )\para
               \\
               & !\banged{x}. (x_i). \piencodf{ U }_{x_i} ) \qquad 
            (:= P_\mathbb{N})\\
 \redd^* & \bigoplus_{C_i \in \perm{C}} (\nu \linvar{x} , \banged{x}, y_1,  {x}_1, \cdots  y_l,  {x}_l)( y_1 . \some_{\emptyset} ;y_{1}.\close;0 \para \cdots \para y_l . \some_{\emptyset} ; \\
    & y_{l}.\close;0 \para {x}.\overline{\some}. \outact{\linvar{x}}{y_{l+1}} . (y_{l+1} . \some_{\emptyset} ; y_{l+1}.\close;0 \para \linvar{x}.\overline{\some};\\
                 & \linvar{x}.\some_{u,(\llfv{M} \setminus  {x}_{l+1} , \cdots ,  {x}_k )};\linvar{x}( {x}_{l+1}) . \cdots \linvar{x}.\overline{\some}. \outact{\linvar{x}}{y_k} . (y_k . \some_{\emptyset} ; y_{k}.\close;0 \para  \\
                & \linvar{x}.\overline{\some};\linvar{x}.\some_{u,(\llfv{M} \setminus   {x}_k )};\linvar{x}( {x}_k) .\linvar{x}.\overline{\some}; \outact{\linvar{x}}{y_{k+1}}. ( y_{k+1} . \some_{u,\llfv{M} } ;\\
             &  y_{k+1}.\close; \piencodf{M}_u \para \linvar{x}.\overline{\none} )) \cdots ) \para {x}_1.\some_{\llfv{C_i(1)}} ; \piencodf{C_i(1)}_{ {x}_1} \para \cdots \para  \\
              &    {x}_l.\some_{\llfv{C_i(l)}} ; \piencodf{C_i(l)}_{ {x}_l}\para y_1. \overline{\none} \para \cdots \para y_l. \overline{\none}\linvar{x}.\some_{\emptyset} ; \linvar{x}(y_{l+1}). \\
              &( y_{l+1}.\overline{\some};y_{l+1} . \overline{\close} \para \linvar{x}.\some_{\emptyset} ; \linvar{x}. \overline{\none}) \para !\banged{x}. (x_i). \piencodf{ U }_{x_i}) 
                    \\
 \redd & \bigoplus_{C_i \in \perm{C}} (\nu  \banged{x},   {x}_1, \cdots  {x}_l)(  u . \overline{\none} \para  {x}_1 . \overline{\none} \para  \cdots \para  {x}_{l} . \overline{\none} \para (\llfv{M} \setminus  {x}_{1} , \cdots ,  {x}_k ).  \\
  & \overline{\none}  \para  {x}_1.\some_{\llfv{C_i(1)}} ; \piencodf{C_i(1)}_{ {x}_1} \para \cdots \para   {x}_l.\some_{\llfv{C_i(l)}} ; \piencodf{C_i(l)}_{ {x}_l} \para !\banged{x}. (x_i). \piencodf{ U }_{x_i} ) 
            \\
 \redd^* & \bigoplus_{C_i \in \perm{C}} (\nu  \banged{x}) (  u . \overline{\none} \para (\llfv{M} \setminus  {x}_{1} , \cdots ,  {x}_k ) . \overline{\none}  \para !\banged{x}. (x_i). \piencodf{ U }_{x_i} ) \\
 &\equiv \bigoplus_{C_i \in \perm{C}}   u . \overline{\none} \para (\llfv{M} \setminus  {x}_{1} , \cdots ,  {x}_k ) . \overline{\none} =  \piencodf{\expr{M}}_u
        \end{aligned}
 \end{equation*}}
        
    and  the result follows.

       \item {\bf Case $\redlab{RS{:}Fail^!}$:}
       
        Then,
        $\expr{N} = M \unexsub{U /\unvar{x}}  $ with $\headf{M} =  {x}[i]$, $U_i = \banged{\oneb} $ and
        $\expr{N} \redd   M \headlin{ \fail^{\emptyset} /\banged{x} } \unexsub{U /\unvar{x} } $, where $\widetilde{y} = \llfv{M}  $. 
        Notice that 
        {\small
        \begin{equation*}\label{ch3eq:compl_unfail-failunres}
            \hspace{-0.5cm}
        \begin{aligned}
            \piencodf{N}_u &= \piencodf{ M \unexsub{U /\unvar{x}} }_u  =  (\nu \banged{x}) ( \piencodf{ M }_u \para   !\banged{x}. (x_i).\piencodf{ U }_{x_i}  )  \\
            & \redd^*(\nu \banged{x}) ( \bigoplus_{i \in I}(\nu \widetilde{y})(\piencodf{  {x}[i] }_{j} \para P_i) \para   !\banged{x}. (x_k).\piencodf{ U }_{x_k}  ) \qquad (*)
            \\
            &  = (\nu \banged{x}) ( \bigoplus_{i \in I}(\nu \widetilde{y})(\outsev{\banged{x}}{{x_k}}. {x}_k.l_{i}; [{x_k} \leftrightarrow j] \para P_i) \para   !\banged{x}. (x_k).\piencodf{ U }_{x_k}  ) \qquad (*)
            \\
            & \redd (\nu \banged{x}) ( \bigoplus_{i \in I}(\nu \widetilde{y})(  (\nu x_k)( x_k.l_{i}; [x_k \leftrightarrow j] \para \piencodf{ U }_{x_k}) \para P_i) \para   !\banged{x}. (x_k).\piencodf{ U }_{x_k}  )
            \\
            & = (\nu \banged{x}) ( \bigoplus_{i \in I}(\nu \widetilde{y})(  (\nu x_k)( x_k.l_{i}; [x_k \leftrightarrow j] \para \choice{x_k}{U_i}{U}{i}{\piencodf{U_i}_{x}}) \para P_i) \para   !\banged{x}. (x_k).\piencodf{ U }_{x_k}  )
            \\
            & \redd (\nu \banged{x}) ( \bigoplus_{i \in I}(\nu \widetilde{y})(  \piencodf{\banged{\oneb}}_{j}  \para P_i) \para   !\banged{x}. (x_k).\piencodf{ U }_{x_k}  )\\
          & = (\nu \banged{x}) ( \bigoplus_{i \in I}(\nu \widetilde{y}) ( j.\none \para P_i) \para   !\banged{x}. (x_k).\piencodf{ U }_{x_k}  ) = \piencodf{\expr{M}}_u \\
        \end{aligned}
        \end{equation*}}
        and the result follows.
        
        

        \item {\bf Case $\redlab{RS:Cons_1}$:}
        
        Then,
        $\expr{N} = \fail^{\widetilde{x}}\ C \bagsep U$ and $\expr{N} \redd \sum_{\perm{C}} \fail^{\widetilde{x} \uplus \widetilde{y}}  = \expr{M}$ where $ \widetilde{y} = \llfv{C}$. 
        Notice that 
        {\small
        \begin{equation*}\label{ch3eq:compl_cons1-failunres}
            \hspace{-0.5cm}
        \begin{aligned}
            \piencodf{N}_u &= \piencodf{ \fail^{\widetilde{x}}\ C \bagsep U }_u\\
            &= \bigoplus_{C_i \in \perm{C}} (\nu v)(\piencodf{\fail^{\widetilde{x}}}_v \para v.\some_{u , \llfv{C}} ; \outact{v}{x} . ([v \leftrightarrow u] \para \piencodf{C_i \bagsep U}_x ) ) \\
            &= \bigoplus_{C_i \in \perm{C}} (\nu v)( v . \overline{\none} \para \widetilde{x}. \overline{\none} \para v.\some_{u , \llfv{C}} ; \outact{v}{x} . ([v \leftrightarrow u] \para \piencodf{C_i \bagsep U}_x ) ) \\
            & \redd \bigoplus_{C_i \in \perm{C}} u . \overline{\none} \para \widetilde{x}. \overline{\none} \para \widetilde{y}. \overline{\none}= \bigoplus_{C_i \in \perm{C}} u . \overline{\none} \para \widetilde{x}. \overline{\none} \para \widetilde{y}. \overline{\none} = \piencodf{\expr{M}}_u 
        \end{aligned}
        \end{equation*}}
        and the result follows.
        

        \item {\bf Cases $\redlab{RS:Cons_2}$, $\redlab{RS:Cons_3}$ and \redlab{RS:Cons_4}:}
        
        These cases follow by IH similarly to the previous.

        
        
        
        

        
    \end{enumerate}
\end{proof}

\subsubsection{Soundness}
\begin{theorem}[Well Formed Weak Operational Soundness]
\label{ch3l:app_soundnesstwounres}
Let $\expr{N}$ be a 
well-formed, linearly closed  $ \lamrsharfailunres$ expression. 
If $ \piencodf{\expr{N}}_u \redd^* Q$
then there exist $Q'$  and $\expr{N}' $ such that 
$Q \redd^* Q'$, $\expr{N}  \redd_{\pequiv}^* \expr{N}'$ 
and 
$\piencodf{\expr{N}'}_u \equiv Q'$.
\end{theorem}

\begin{proof}
By induction on the structure of $\expr{N} $ and then induction on the number of reductions of $\piencodf{\expr{N}} \redd^* Q$.

\begin{enumerate}
    \item {\bf Base case:} $\expr{N} =  {x}$, $\expr{N} =  {x}[j]$, $\expr{N} = \fail^{\emptyset}$ and $\expr{N} = \lambda x . (M[ {\widetilde{x}} \leftarrow  {x}])$.
.
    
    No reductions can take place, and the result follows trivially.
    $Q =  \piencodf{\expr{N}}_u \redd^0 \piencodf{\expr{N}}_u = Q'$ and $ {x} \redd^0  {x} = \expr{N}'$.
    
    
    
    
    

    
    \item $\expr{N} =  M (C \bagsep U) $.

        Then, 
        $ \piencodf{M (C \bagsep U)}_u = \bigoplus_{C_i \in \perm{C}} (\nu v)(\piencodf{M}_v \para v.\some_{u , \llfv{C}} ; \outact{v}{x} . ([v \leftrightarrow u] \para \piencodf{C_i \bagsep U}_x ) )$, and we are able to perform the  reductions from $\piencodf{M (C \bagsep U)}_u$. 

        We now proceed by induction on $k$, with  $\piencodf{\expr{N}}_u \redd^k Q$. There are two main cases:
        \begin{enumerate}
            \item When $k = 0$ the thesis follows easily:
            
            We have 
    $Q =  \piencodf{M (C \bagsep U)}_u \redd^0 \piencodf{M (C \bagsep U)}_u = Q'$ and $M (C \bagsep U) \redd^0 M (C \bagsep U) = \expr{N}'$.
    
            \item The interesting case is when $k \geq 1$.
            
            Then, for some process $R$ and $n, m$ such that $k = n+m$, we have the following:
            \[
            \begin{aligned}
               \piencodf{\expr{N}}_u & =  \bigoplus_{C_i \in \perm{C}} (\nu v)(\piencodf{M}_v \para v.\some_{u , \llfv{C}} ; \outact{v}{x} . ([v \leftrightarrow u] \para \piencodf{C_i \bagsep U}_x ) )\\
               & \redd^m  \bigoplus_{C_i \in \perm{C}} (\nu v)(R \para v.\some_{u , \llfv{C}} ; \outact{v}{x} . ([v \leftrightarrow u] \para \piencodf{C_i \bagsep U}_x ) ) \\
               &\redd^n  Q\\
            \end{aligned}
            \]
            Thus, the first $m \geq 0$ reduction steps are  internal to $\piencodf{ M}_v$; type preservation in \spi ensures that, if they occur,  these reductions  do not discard the possibility of synchronizing with $v.\some$. Then, the first of the $n \geq 0$ reduction steps towards $Q$ is a synchronization between $R$ and $v.\some_{u, \llfv{C}}$.
            
            We consider two sub-cases, depending on the values of  $m$ and $n$:
            \begin{enumerate}
                \item $m = 0$ and $n \geq 1$:
                
Then $R = \piencodf{\expr{M}}_v$ as $\piencodf{\expr{M}}_v \redd^0 \piencodf{\expr{M}}_v$. 
 Notice that there are two possibilities of having an unguarded:
                    
 

\begin{enumerate}
\item $M =  (\lambda x . (M'[ {\widetilde{x}} \leftarrow  {x}])) \linexsub{N_1 / y_1} \cdots \linexsub{N_p / y_p} \unexsub{U_1 / \unvar{z_1}} \cdots \unexsub{U_q / \unvar{z_q}}  $ with$ (p, q \geq 0)$

   {\small
   \[
    \hspace{-1cm}
   \begin{aligned}
   \piencodf{M}_v &= \piencodf{ (\lambda x . (M'[ {\widetilde{x}} \leftarrow  {x}])) \linexsub{N_1 / y_1} \cdots \linexsub{N_p / y_p} \unexsub{U_1 / \unvar{z_1}} \cdots \unexsub{U_q / \unvar{z_q}} }_v \\
      &= (\nu y_1, \cdots , y_p , \banged{z}_1, \cdots ,\banged{z}_q) ( \piencodf{\lambda x . (M'[ {\widetilde{x}} \leftarrow  {x}])}_v \para y_1.\some_{\llfv{N_1}};\piencodf{ N_1 }_{y_1} \para \cdots \\
    & \hspace{.5cm}  \para y_p.\some_{\llfv{N_p}};\piencodf{ N_p }_{y_p} \para   !\banged{z}_1. (z_1).\piencodf{ U }_{z_1} \para \cdots  \para   !\banged{z}_q. (z_q).\piencodf{ U }_{z_q}          )\\
  &= (\nu \widetilde{y} ,\widetilde{z} ) ( \piencodf{\lambda x . (M'[ {\widetilde{x}} \leftarrow  {x}])}_v \para Q'' )\\
  &= (\nu \widetilde{y},\widetilde{z}) ( v.\overline{\some}; v(x). x.\overline{\some}; x(\linvar{x}). x(\banged{x}). x. \close ; \piencodf{M'[ {\widetilde{x}} \leftarrow  {x}]}_v \para Q'' )
    \end{aligned}
    \]}
  \noindent where $\widetilde{y} = y_1 , \cdots , y_p$. $\widetilde{z} = \banged{z}_1, \cdots ,\banged{z}_q$ and
  
  $$
  \begin{aligned}
      Q'' =& y_1.\some_{\llfv{N_1}};\piencodf{ N_1 }_{y_1} \para \cdots \para y_p.\some_{\llfv{N_p}};\piencodf{ N_p }_{y_p} \para    \\
      & \quad \para !\banged{z}_1. (z_1).\piencodf{ U }_{z_1} \para \cdots  !\banged{z}_q. (z_q).\piencodf{ U }_{z_q}.
      \end{aligned}$$
With this shape for $M$, we then have the following:
{\small 
\[
    \hspace{-2cm}
 \begin{aligned}
 \piencodf{\expr{N}}_u & = \piencodf{(M\ B)}_u\\
 &= \bigoplus_{C_i \in \perm{C}} (\nu v)(\piencodf{M}_v \para v.\some_{u , \llfv{C}} ; \outact{v}{x} . ([v \leftrightarrow u] \para \piencodf{C_i \bagsep U}_x ) )\\
 & \redd \bigoplus_{C_i \in \perm{C}} (\nu v , \widetilde{y},\widetilde{z})(   v(x). x.\overline{\some}; x(\linvar{x}). x(\banged{x}). x. \close ; \piencodf{M'[ {\widetilde{x}} \leftarrow  {x}]}_v \\
 & \hspace{.5cm} \para Q''  \para \outact{v}{x} . ([v \leftrightarrow u] \para \piencodf{C_i \bagsep U}_x ) ) & = Q_1 \\
& \redd \bigoplus_{C_i \in \perm{C}} (\nu v , \widetilde{y},\widetilde{z}, x)( x.\overline{\some}; x(\linvar{x}). x(\banged{x}). x. \close ; \piencodf{M'[ {\widetilde{x}} \leftarrow  {x}]}_v  \\
 & \hspace{.5cm}\para Q''  \para  [v \leftrightarrow u] \para \piencodf{C_i \bagsep U}_x ) & = Q_2 \\
 & \redd \bigoplus_{C_i \in \perm{C}} (\nu  \widetilde{y},\widetilde{z}, x)( x.\overline{\some}; x(\linvar{x}). x(\banged{x}). x. \close ; \piencodf{M'[ {\widetilde{x}} \leftarrow  {x}]}_u \para Q''\\
 & \hspace{.5cm}\para  \piencodf{C_i \bagsep U}_x ) & = Q_3 \\
\end{aligned}
     \]}
We also have that 
{\small
                    \[
                        \hspace{-1cm}
                    \begin{aligned}
                        \expr{N} &=(\lambda x . (M'[ {\widetilde{x}} \leftarrow  {x}])) \linexsub{N_1 / y_1} \cdots \linexsub{N_p / y_p} \unexsub{U_1 / \unvar{z_1}} \cdots \unexsub{U_q / \unvar{z_q}} (C \bagsep U) \\
                        &\pequiv (\lambda x . (M'[ {\widetilde{x}} \leftarrow  {x}]) (C \bagsep U)) \linexsub{N_1 / y_1} \cdots \linexsub{N_p / y_p} \unexsub{U_1 / \unvar{z_1}} \cdots \unexsub{U_q / \unvar{z_q}} \\
                      & \redd   M'[ {\widetilde{x}} \leftarrow  {x}] \esubst{(C \bagsep U)}{x} \linexsub{N_1 / y_1} \cdots \linexsub{N_p / y_p} \unexsub{U_1 / \unvar{z_1}} \cdots \unexsub{U_q / \unvar{z_q}} = \expr{M}
                    \end{aligned}
                    \]}
  Furthermore, we have:
  {\small
  \[
    \hspace{-1.2cm}
                     \begin{aligned}
                          &\piencodf{\expr{M}}_u = \piencodf{M'[ {\widetilde{x}} \leftarrow  {x}] \esubst{(C \bagsep U)}{x} \linexsub{N_1 / y_1} \cdots \linexsub{N_p / y_p} \unexsub{U_1 / \unvar{z_1}} \cdots \unexsub{U_q / \unvar{z_q}}}_u \\
                          & = \bigoplus_{C_i \in \perm{C}} (\nu  \widetilde{y},\widetilde{z}, x)( x.\overline{\some}; x(\linvar{x}). x(\banged{x}). x. \close ; \piencodf{M'[ {\widetilde{x}} \leftarrow  {x}]}_u   \para  \piencodf{C_i \bagsep U}_x \para Q'' ) 
                    \end{aligned}
                    \]}
                     
        We consider different possibilities for $n \geq 1$; in all  the cases, the result follows.                 
                         \smallskip

 \noindent  {\bf When $n = 1$:}
      We have $Q = Q_1$, $ \piencodf{\expr{N}}_u \redd^1 Q_1$.
          We also have that 
          \begin{itemize}
          \item  $Q_1 \redd^2 Q_3 = Q'$ , 
        \item $\expr{N} \redd^1 M'[\widetilde{x} \leftarrow x]) \esubst{B}{x} = \expr{N}'$
        \item and $\piencodf{M'[\widetilde{x} \leftarrow x]) \esubst{B}{x}}_u = Q_3$.
        \end{itemize}
        
                                \smallskip

 \noindent  {\bf When $n = 2$:} the analysis is similar.


\noindent {\bf When $n \geq 3$:}
 We have $ \piencodf{\expr{N}}_u \redd^3 Q_3 \redd^l Q$, for $l \geq 0$. We also know that $\expr{N} \redd \expr{M}$, $Q_3 = \piencodf{\expr{M}}_u$. By the IH, there exist $ Q' , \expr{N}'$ such that $Q \redd^i Q'$, $\expr{M} \redd_{\pequiv}^j \expr{N}'$ and $\piencodf{\expr{N}'}_u = Q'$ . Finally, $\piencodf{\expr{N}}_u \redd^3 Q_3 \redd^l Q \redd^i Q'$ and $\expr{N} \rightarrow \expr{M}  \redd_{\pequiv}^j \expr{N}'$.
                        
\item $M = \fail^{\widetilde{z}}$. 

Then,                     \(
                        \begin{aligned}
                            \piencodf{M}_v &= \piencodf{\fail^{\widetilde{z}}}_v = v.\overline{\none} \para \widetilde{z}.\overline{\none}.
                        \end{aligned}
                    \)
                    With this shape for $M$, we have:
                    {\small
                    \[\hspace{-1.5cm}
                    \begin{aligned}
                        \piencodf{\expr{N}}_u & = \piencodf{(M\ (C \bagsep U))}_u\\ & = \bigoplus_{C_i \in \perm{C}} (\nu v)(\piencodf{M}_v \para v.\some_{u , \llfv{C}} ; \outact{v}{x} . ([v \leftrightarrow u] \para \piencodf{C_i \bagsep U}_x ) )\\
                        & = \bigoplus_{C_i \in \perm{C}} (\nu v)(v.\overline{\none} \para \widetilde{z}.\overline{\none} \para v.\some_{u , \llfv{C}} ; \outact{v}{x} . ([v \leftrightarrow u] \para \piencodf{C_i \bagsep U}_x ) )\\
                        & \redd \bigoplus_{B_i \in \perm{B}}   u.\overline{\none} \para \widetilde{z}.\overline{\none}  \para \llfv{C_i}.\overline{\none} \\
                        \end{aligned}
                    \]}

                    \end{enumerate}
                    
                    We also have that 
                    \(  \expr{N} = \fail^{\widetilde{x}}\ C \bagsep U \redd  \sum_{\perm{C}} \fail^{\widetilde{x} \uplus \llfv{C}}  = \expr{M}.  \)
                    Furthermore,
                    \[
                     \begin{aligned}
                          \piencodf{\expr{M}}_u &= \piencodf{\sum_{\perm{C}} \fail^{\widetilde{z} \uplus \llfv{C}  } }_u 
                          = \bigoplus_{\perm{C}}\piencodf{ \fail^{\widetilde{z} \uplus \llfv{C} }}_u\\
                          &= \bigoplus_{\perm{C}}    u.\overline{\none} \para \widetilde{z}.\overline{\none}  \para  \llfv{C}.\overline{\none}.
                    \end{aligned}
                    \]

  \item When $m \geq 1$ and $ n \geq 0$, we distinguish two cases:
                   
 \begin{enumerate}
\item When $n = 0$:
                            
Then, $ \bigoplus_{C_i \in \perm{C}} (\nu v)(R \para v.\some_{u , \llfv{C}} ; \outact{v}{x} . ([v \leftrightarrow u] \para \piencodf{C_i \bagsep U}_x ) ) =  Q $ and $\piencodf{M}_u \redd^m R$ where $m \geq 1$. Then by the IH there exist $R'$  and $\expr{M}' $ such that $R \redd^i R'$, $M \redd_{\pequiv}^j \expr{M}'$, and $\piencodf{\expr{M}'}_u = R'$.  Hence we have that 
    {\small
    \[ 
    \begin{aligned}
  \piencodf{\expr{N}}_u & =  \bigoplus_{C_i \in \perm{C}} (\nu v)(\piencodf{M}_v \para v.\some_{u , \llfv{C}} ; \outact{v}{x} . ([v \leftrightarrow u] \para \piencodf{C_i \bagsep U}_x ) )\\
                            & \redd^m  \bigoplus_{C_i \in \perm{C}} (\nu v)(R \para v.\some_{u , \llfv{C}} ; \outact{v}{x} . ([v \leftrightarrow u] \para \piencodf{C_i \bagsep U}_x ) ) \\
                            & = Q
                            \end{aligned}
                             \]}
                            We also know that
                            \[ 
                            \begin{aligned}
                              Q  \redd^i & \bigoplus_{C_i \in \perm{C}} (\nu v)(R' \para v.\some_{u , \llfv{C}} ; \outact{v}{x} . ([v \leftrightarrow u] \para \piencodf{C_i \bagsep U}_x ) )\\
                               = & Q'\\
                            \end{aligned}
                            \]
                            and so the \lamrsharfailunres term can reduce as follows: $\expr{N} = (M\ ( C \bagsep U )) \redd_{\pequiv}^j M'\ ( C \bagsep U ) = \expr{N}'$ and  $\piencodf{\expr{N}'}_u = Q'$.

                        \item When $n \geq 1$:
                        
                            Then  $R$ has an occurrence of an unguarded $v.\overline{\some}$ or $v.\overline{\none}$, hence it is of the form 
                            $ \piencodf{(\lambda x . (M'[ {\widetilde{x}} \leftarrow  {x}])) \linexsub{N_1 / y_1} \cdots \linexsub{N_p / y_p} \unexsub{U_1 / \unvar{z_1}} \cdots \unexsub{U_q / \unvar{z_q}} }_v $ or $ \piencodf{\fail^{\widetilde{x}}}_v. $ 
              This case follows by IH.
                    \end{enumerate}

            \end{enumerate}

        \end{enumerate}

        This concludes the analysis for the case $\expr{N} = (M \, ( C \bagsep U ))$.
        
        \item $\expr{N} = M[ {\widetilde{x}} \leftarrow  {x}]$.

    The sharing variable $ {x}$ is not free and the result follows by vacuity.
        
        \item $\expr{N} = M[ {\widetilde{x}} \leftarrow  {x}] \esubst{ C \bagsep U }{ x}$. Then we have
            
            \[
                \begin{aligned}
                    \piencodf{\expr{N}}_u &=\piencodf{ M[ {\widetilde{x}} \leftarrow  {x}] \esubst{ C \bagsep U }{ x} }_u\\
                    &= \bigoplus_{C_i \in \perm{C}} (\nu x)( x.\overline{\some}; x(\linvar{x}). x(\banged{x}). x. \close ;\piencodf{ M[ {\widetilde{x}} \leftarrow  {x}]}_u \para \piencodf{ C_i \bagsep U}_x )
                \end{aligned}
            \]

            Let us consider three cases.

            \begin{enumerate}
                \item When $\size{ {\widetilde{x}}} = \size{C}$.
                    Then let us consider the shape of the bag $ C$.
                    
  \begin{enumerate}
  \item When $C = \oneb$.
  
  We have the following
    {\small
 \[
    \hspace{-2.2cm}
 \begin{aligned}
 \piencodf{\expr{N}}_u  &=  (\nu x)( x.\overline{\some}; x(\linvar{x}). x(\banged{x}). x. \close ;\piencodf{ M[ \leftarrow  {x}]}_u \para \piencodf{ \oneb \bagsep U}_x )\\
 &=  (\nu x)( x.\overline{\some}; x(\linvar{x}). x(\banged{x}). x. \close ;\piencodf{ M[ \leftarrow  {x}]}_u \para x.\some_{\llfv{C}} ; \outact{x}{\linvar{x}} . \\
 & \hspace{1cm}( \piencodf{ \oneb }_{\linvar{x}} \para\outact{x}{\banged{x}} .( !\banged{x}. (x_i). \piencodf{ U }_{x_i} \para x.\overline{\close} ) ) )\\
&\redd (\nu x)(  x(\linvar{x}). x(\banged{x}). x. \close ;\piencodf{ M[ \leftarrow  {x}]}_u \para
 \outact{x}{\linvar{x}} .( \piencodf{ \oneb }_{\linvar{x}} \para \outact{x}{\banged{x}} . \\
 &\hspace{1cm} ( !\banged{x}. (x_i). \piencodf{ U }_{x_i}\para x.\overline{\close} ) ) ) & = Q_1  
                              \\
  &\redd (\nu x,\linvar{x})(  x(\banged{x}). x. \close ;\piencodf{ M[ \leftarrow  {x}]}_u \para \piencodf{ \oneb }_{\linvar{x}} \para\outact{x}{\banged{x}} .\\
  & \hspace{1cm}( !\banged{x}. (x_i). \piencodf{ U }_{x_i} \para x.\overline{\close} ) ) & = Q_2  
                              \\
  &\redd (\nu x,\linvar{x}, \banged{x})(  x. \close ;\piencodf{ M[ \leftarrow  {x}]}_u \para \piencodf{ \oneb }_{\linvar{x}} \para !\banged{x}. (x_i). \piencodf{ U }_{x_i} \para x.\overline{\close} ) & = Q_3   \\
 &\redd (\nu \linvar{x}, \banged{x})(  \piencodf{ M[ \leftarrow  {x}]}_u \para \piencodf{ \oneb }_{\linvar{x}} \para !\banged{x}. (x_i). \piencodf{ U }_{x_i} ) & = Q_4\\
    & = (\nu \linvar{x}, \banged{x})( \linvar{x}. \overline{\some}. \outact{\linvar{x}}{y_i} . ( y_i . \some_{u,\llfv{M}} ;y_{i}.\close; \piencodf{M}_u \para \linvar{x}. \overline{\none}) \para \\
    & \qquad \linvar{x}.\some_{\emptyset} ; \linvar{x}(y_n). ( y_n.\overline{\some};y_n . \overline{\close} \para \linvar{x}.\some_{\emptyset} ; \linvar{x}. \overline{\none})  \para !\banged{x}. (x_i). \piencodf{ U }_{x_i} )
                            \\
      & \redd (\nu \linvar{x}, \banged{x})(  \outact{\linvar{x}}{y_i} . ( y_i . \some_{u,\llfv{M}} ;y_{i}.\close; \piencodf{M}_u \para \linvar{x}. \overline{\none}) \para \\
                              & \qquad \linvar{x}(y_n). ( y_n.\overline{\some};y_n . \overline{\close} \para \linvar{x}.\some_{\emptyset} ; \linvar{x}. \overline{\none})  \para !\banged{x}. (x_i). \piencodf{ U }_{x_i} )  & = Q_5
                            \\
                            & \redd (\nu \linvar{x}, \banged{x} , y_i)(   y_i . \some_{u,\llfv{M}} ;y_{i}.\close; \piencodf{M}_u \para \linvar{x}. \overline{\none} \para  y_i.\overline{\some};y_i . \overline{\close}  \\
 & \hspace{1cm}\para \linvar{x}.\some_{\emptyset} ;\linvar{x}. \overline{\none}  \para !\banged{x}. (x_i). \piencodf{ U }_{x_i} )  & = Q_6
                            \\
                            & \redd (\nu \linvar{x}, \banged{x} , y_i)(  y_{i}.\close; \piencodf{M}_u \para \linvar{x}. \overline{\none} \para  y_i . \overline{\close} \para \linvar{x}.\some_{\emptyset} ; \linvar{x}. \overline{\none}\\
                            & \hspace{
                            1cm }\para !\banged{x}. (x_i). \piencodf{ U }_{x_i} )  & = Q_7
                            \\
                            & \redd (\nu \linvar{x}, \banged{x} )(  \piencodf{M}_u \para \linvar{x}. \overline{\none} \para  \linvar{x}.\some_{\emptyset} ; \linvar{x}. \overline{\none}  \para !\banged{x}. (x_i). \piencodf{ U }_{x_i} )  & = Q_8
                            \\
                            & \redd (\nu \banged{x})(  \piencodf{M}_u \para !\banged{x}. (x_i). \piencodf{ U }_{x_i} )  
                            =  \piencodf{M \unexsub{U / \unvar{x}}}_u
                            & = Q_9
\end{aligned}
\]}
                            Notice how $Q_8$ has a choice however the $\linvar{x}$ name can be closed at any time so for simplicity we only perform communication across this name once all other names have completed their reductions.
                            
                        Now we proceed by induction on the number of reductions $\piencodf{\expr{N}}_u \redd^k Q$.
                            
                            \begin{enumerate}
                                
                                \item When $k = 0$, the result follows trivially. Just take $\mathbb{N}=\mathbb{N}'$ and $\piencodf{\expr{N}}_u=Q=Q'$.
                                

                                \item When $k = 1$.
                                
                                    We have $Q = Q_1$, $ \piencodf{\expr{N}}_u \redd^1 Q_1$
                                    We also have that $Q_1 \redd^8 Q_9 = Q'$ , $\expr{N} \redd M \unexsub{U / \unvar{x}} = M$ and $\piencodf{ M }_u = Q_9$
                                    
                                \item When $2 \leq  k \leq 8$.
                                
                                    Proceeds similarly to the previous case
                                
                                \item When $k \geq 9$.
                                
      We have $ \piencodf{\expr{N}}_u \redd^9 Q_9 \redd^l Q$, for $l \geq 0$. Since $Q_9 = \piencodf{ M }_u$ we apply the induction hypothesis we have that  there exist $ Q' , \expr{N}' \ s.t. \ Q \redd^i Q' ,  M \redd_{\pequiv}^j \expr{N}'$ and $\piencodf{\expr{N}'}_u = Q'$.                                    Then,  $ \piencodf{\expr{N}}_u \redd^5 Q_5 \redd^l Q \redd^i Q'$ and by the contextual reduction rule it follows that $\expr{N} = (M[ \leftarrow x])\esubst{ 1 }{ x } \redd_{\pequiv}^j  \expr{N}' $ and the case holds.

\end{enumerate}     
                        
\item When $C = \bag{N_1} \cdot \cdots \cdot \bag{N_l}$, for $l \geq 1$.
Then the reduction is shown in \Cref{ch3fig:eroes},

\begin{figure}
    {\small                  
 \[
   \begin{aligned}
   \piencodf{\expr{N}}_u &=\piencodf{ M[ {\widetilde{x}} \leftarrow  {x}] \esubst{ C \bagsep U }{x} }_u\\
   &= \bigoplus_{C_i \in \perm{C}} (\nu x)( x.\overline{\some}; x(\linvar{x}). x(\banged{x}). x. \close ;\piencodf{ M[ {\widetilde{x}} \leftarrow  {x}]}_u \para \piencodf{ C_i \bagsep U}_x ) \\
  &\redd ^{4} \bigoplus_{C_i \in \perm{C}}(\nu \linvar{x}, \banged{x})(  \piencodf{ M[ {\widetilde{x}} \leftarrow  {x}]}_u \para \piencodf{ C_i }_{\linvar{x}} \para !\banged{x}. (x_i). \piencodf{ U }_{x_i} )\\
 &= 
  \bigoplus_{C_i \in \perm{C}} (\nu \linvar{x}, \banged{x})( \linvar{x}.\overline{\some}. \outact{\linvar{x}}{y_1}. (y_1 . \some_{\emptyset} ;y_{1}.\close;0 \para \linvar{x}.\overline{\some};\\
  &\qquad \linvar{x}.\some_{u, (\llfv{M} \setminus  {x}_1 , \cdots ,  {x}_l )}; 
     \linvar{x}( {x}_1) . \cdots \linvar{x}.\overline{\some}. \outact{\linvar{x}}{y_l} . (y_l . \some_{\emptyset} ; y_{l}.\close;0 \\
     &\qquad\para \linvar{x}.\overline{\some};\linvar{x}.\some_{u,(\llfv{M} \setminus  {x}_l )};\linvar{x}( {x}_l) . \linvar{x}.\overline{\some}; \outact{\linvar{x}}{y_{l+1}}. ( y_{l+1} . \some_{u,\llfv{M}} ; \\
    &\qquad y_{l+1}.\close;  \piencodf{M}_u \para \linvar{x}.\overline{\none} )) \cdots ) \para  \linvar{x}.\some_{\llfv{C}} ; \linvar{x}(y_1). \linvar{x}.\some_{y_1, \llfv{C} };\\
   & \qquad \linvar{x}.\overline{\some} ; \outact{\linvar{x}}{ {x}_1}. ( {x}_1.\some_{\llfv{C_i(1)}} ; \piencodf{C_{i}(1)}_{ {x}_1}  \para y_1. \overline{\none}\para \cdots  \linvar{x}.\some_{\llfv{C_{i}(l)}} ; \\
   &\qquad \linvar{x}(y_l). \linvar{x}. \some_{y_l, \llfv{C_{i}(l)}} ;\linvar{x}.\overline{\some} ; \outact{\linvar{x}}{ {x}_l}. ( {x}_l.\some_{\llfv{C_{i}(l)}} ; \piencodf{C_{i}(l)}_{ {x}_l}  \\
    & \qquad  \para y_l. \overline{\none}\para\linvar{x}.\some_{\emptyset} ; \linvar{x}(y_{l+1}). ( y_{l+1}.\overline{\some};y_{l+1} . \overline{\close} \para \linvar{x}.\some_{\emptyset} ; \linvar{x}. \overline{\none})
                                  )
                                  )
 \\
 &\qquad \para !\banged{x}. (x_i). \piencodf{ U }_{x_i} )\\ 
 & \redd ^{5l}
  \bigoplus_{C_i \in \perm{C}} (\nu \linvar{x}, \banged{x} ,  {x}_1,y_1, \cdots ,  {x}_l,y_1)( y_1 . \some_{\emptyset} ;y_{1}.\close;0 \para  \cdots  y_l . \some_{\emptyset} ;\\
 &\hspace{1cm}  y_{l}.\close;0 \para  \linvar{x}.\overline{\some}; \outact{\linvar{x}}{y_{l+1}}. ( y_{l+1} . \some_{u,\llfv{M}} ;y_{l+1}.\close; \piencodf{M}_u \para \linvar{x}.\overline{\none} )
 \para \\
    & \hspace{1cm}  {x}_1.\some_{\llfv{C_{i}(1)}} ; \piencodf{C_{i}(1)}_{ {x}_1}  \para y_1. \overline{\none} \para \cdots    {x}_l.\some_{\llfv{C_{i}(l)}} ; \piencodf{C_{i}(l)}_{ {x}_l}  \para y_l. \overline{\none}\para\\
     & \hspace{1cm}\linvar{x}.\some_{\emptyset} ; \linvar{x}(y_{l+1}). ( y_{l+1}.\overline{\some};y_{l+1} . \overline{\close} \para \linvar{x}.\some_{\emptyset} ; \linvar{x}. \overline{\none})
  \para !\banged{x}. (x_i). \piencodf{ U }_{x_i} )
                               \\
     & \redd ^{5}
  \bigoplus_{C_i \in \perm{C}} (\nu \banged{x} ,  {x}_1,y_1, \cdots ,  {x}_l,y_1)(  y_1 . \some_{\emptyset} ;y_{1}.\close;0 \para  \cdots  y_l . \some_{\emptyset} ; y_{l}.\close;0  \\
   & \qquad   \para \piencodf{M}_u \para {x}_1.\some_{\llfv{C_{i}(1)}} ; \piencodf{C_{i}(1)}_{ {x}_1}  \para y_1. \overline{\none} \para \cdots    {x}_l.\some_{\llfv{C_{i}(l)}} ; \piencodf{C_{i}(l)}_{ {x}_l} \\
   & \qquad \para y_l. \overline{\none} \para !\banged{x}. (x_i). \piencodf{ U }_{x_i})
                           \\
  & \redd ^{l}  \bigoplus_{C_i \in \perm{C}} (\nu \banged{x} ,  {x}_1 \cdots ,  {x}_l)( \piencodf{M}_u \para  {x}_1.\some_{\llfv{C_{i}(1)}} ; \piencodf{C_{i}(1)}_{ {x}_1}  \para  \cdots  \\
  & \hspace{1cm}\para   {x}_l.\some_{\llfv{C_{i}(l)}} ;\piencodf{C_{i}(l)}_{ {x}_l} \para  !\banged{x}. (x_i). \piencodf{ U }_{x_i} )
                                 \\
  & = \piencodf{\sum_{C_i \in \perm{C}}M\linexsub{C_i(1)/ {x_1}} \cdots \linexsub{C_i(l)/ {x_l}} \unexsub{U /\unvar{x} }}_{u}= Q_{6l + 9}\\
  \end{aligned}
           \]}
           \caption{Encoded Reduction of Explicit Substatution (Success)}
           \label{ch3fig:eroes}
\end{figure}

                            The proof follows by induction on the number of reductions $\piencodf{\expr{N}}_u \redd^k Q$.
                            
\begin{enumerate}
\item When $k = 0$, the result follows trivially. Just take $\mathbb{N}=\mathbb{N}'$ and $\piencodf{\expr{N}}_u=Q=Q'$. 
                                
 \item When $1 \leq k \leq 6l + 9$.

 Let $Q_k$ such that $ \piencodf{\expr{N}}_u \redd^k Q_k$.
                                    We also have that $Q_k \redd^{6l + 9 - k} Q_{6l + 9} = Q'$ ,
                                    
                                    $\expr{N} \redd^1 \sum_{C_i \in \perm{C}}M\linexsub{C_i(1)/ {x_1}} \cdots \linexsub{C_i(l)/ {x_l}} \unexsub{U /\unvar{x} } = \expr{N}'$ and 
                                    
                                    $\piencodf{\sum_{C_i \in \perm{C}}M\linexsub{C_i(1)/ {x_1}} \cdots \linexsub{C_i(l)/ {x_l}} \unexsub{U /\unvar{x} }}_u = Q_{6l + 9}$.
                                
\item When $k > 6l + 9$.

Then,  $ \piencodf{\expr{N}}_u \redd^{6l + 9} Q_{6l + 9} \redd^n Q$ for $n \geq 1$. Also, 

\(
\begin{aligned} 
&\expr{N} \redd^1 \sum_{C_i \in \perm{C}}M\linexsub{C_i(1)/ {x_1}} \cdots \linexsub{C_i(l)/ {x_l}} \unexsub{U /\unvar{x} } \text { and } \\
&    Q_{6l + 9} = \piencodf{\sum_{C_i \in \perm{C}}M\linexsub{C_i(1)/ {x_1}} \cdots \linexsub{C_i(l)/ {x_l}} \unexsub{U /\unvar{x} }}_u.
    \end{aligned}
\)

By the induction hypothesis, there exist $ Q'$ and $\expr{N}'$ such that  
$\ Q \redd^i Q'$, 

$\sum_{C_i \in \perm{C}}M\linexsub{C_i(1)/ {x_1}} \cdots \linexsub{C_i(l)/ {x_l}} \unexsub{U /\unvar{x} } \redd_{\pequiv}^j \expr{N}'$ and $\piencodf{\expr{N}'}_u = Q'$. 

Finally, $\piencodf{\expr{N}}_u \redd^{6l + 9} Q_{6l + 9} \redd^n Q \redd^i Q'$ and $$ \expr{N} \rightarrow \sum_{C_i \in \perm{C}}M\linexsub{C_i(1)/ {x_1}} \cdots \linexsub{C_i(l)/ {x_l}} \unexsub{U /\unvar{x} }  \redd_{\pequiv}^j \expr{N}'. $$

                            \end{enumerate}

                    \end{enumerate}

                \item When $\size{\widetilde{x}} > \size{C}$.
                
                    Then we have 
                    $\expr{N} = M[ {x}_1, \cdots ,  {x}_k \leftarrow  {x}]\ \esubst{ C \bagsep U }{x}$ with $C = \bag{N_1}  \cdots  \bag{N_l} \quad k > l$. $\expr{N} \redd  \sum_{C_i \in \perm{C}}  \fail^{\widetilde{z}} = \expr{M}$ and $ \widetilde{z} =  (\llfv{M} \setminus \{   {x}_1, \cdots ,  {x}_k \} ) \cup \llfv{C} $. On the one hand, we have the reduction of \Cref{ch3fig:eroesf}:
                    Hence $k = l + m$ for some $m \geq 1$
                \begin{figure}
                    {\small
                    \[
                    \begin{aligned}
                        \piencodf{N}_u &= \piencodf{M[ {x}_1, \cdots ,  {x}_k \leftarrow  {x}]\ \esubst{ C \bagsep U }{x}}_u \\
                        & = \bigoplus_{C_i \in \perm{C}} (\nu x)( x.\overline{\some}; x(\linvar{x}). x(\banged{x}). x. \close ;\piencodf{ M[ {x}_1, \cdots ,  {x}_k \leftarrow  {x}]}_u \para \piencodf{ C_i \bagsep U}_x ) \\
                              &\redd ^{4} \bigoplus_{C_i \in \perm{C}}(\nu \linvar{x}, \banged{x})(  \piencodf{M[ {x}_1, \cdots ,  {x}_k \leftarrow  {x}]}_u \para \piencodf{ C_i }_{\linvar{x}} \para !\banged{x}. (x_i). \piencodf{ U }_{x_i} )\\
                              &=  \bigoplus_{C_i \in \perm{C}} (\nu \linvar{x}, \banged{x})(  \linvar{x}.\overline{\some}. \outact{\linvar{x}}{y_1}. (y_1 . \some_{\emptyset} ;y_{1}.\close;0 \para \linvar{x}.\overline{\some}; \\
  &\linvar{x}.\some_{u, (\llfv{M} \setminus  {x}_1 , \cdots ,  {x}_k )};\linvar{x}( {x}_1) . \cdots\linvar{x}.\overline{\some}. \outact{\linvar{x}}{y_k} . (y_k . \some_{\emptyset} ; y_{k}.\close;0 \para \linvar{x}.\overline{\some}; \\
 &\linvar{x}.\some_{u,(\llfv{M} \setminus  {x}_k )};\linvar{x}( {x}_k) .\linvar{x}.\overline{\some}; \outact{\linvar{x}}{y_{k+1}}. ( y_{k+1} . \some_{u,\llfv{M}} ;y_{k+1}.\close; \piencodf{M}_u \\
 & \para \linvar{x}.\overline{\none} )) \cdots ) \para \linvar{x}.\some_{\llfv{C}} ; \linvar{x}(y_1). \linvar{x}.\some_{y_1, \llfv{C}} ;\linvar{x}.\overline{\some} ; \outact{\linvar{x}}{ {x}_1}. ( {x}_1.\some_{\llfv{C_i(1)}};  \\
 & \piencodf{C_i(1)}_{ {x}_1} \para  y_1. \overline{\none} \para \cdots\linvar{x}.\some_{\llfv{C_i(l)}} ; \linvar{x}(y_l). \linvar{x}.\some_{y_l, \llfv{C_i(l)} };\linvar{x}.\overline{\some} ; \outact{\linvar{x}}{ {x}_l}.  \\
  & ( {x}_l.\some_{\llfv{C_i(l)}} ; \piencodf{C_i(l)}_{ {x}_l} \para y_l. \overline{\none} \para\linvar{x}.\some_{\emptyset} ; \linvar{x}(y_{l+1}). ( y_{l+1}.\overline{\some};y_{l+1} . \overline{\close} \para \\
  &\linvar{x}.\some_{\emptyset} ; \linvar{x}. \overline{\none}) ) 
                               ) \para !\banged{x}. (x_i). \piencodf{ U }_{x_i})  \\
&\redd^{5l} \bigoplus_{C_i \in \perm{C}} (\nu \linvar{x}, \banged{x} , y_1,  {x}_1, \cdots  y_l,  {x}_l)(   y_1 . \some_{\emptyset} ;y_{1}.\close;0 \para \cdots \para y_l . \some_{\emptyset} ;y_{l}.\close;0 \\
 &\linvar{x}.\overline{\some}. \outact{\linvar{x}}{y_{l+1}} . (y_{l+1} . \some_{\emptyset} ; y_{l+1}.\close;0 \para \linvar{x}.\overline{\some};\linvar{x}.\some_{u,(\llfv{M} \setminus  {x}_{l+1} , \cdots ,  {x}_k )}; \\
&\linvar{x}( {x}_{l+1}) . \cdots\linvar{x}.\overline{\some}. \outact{\linvar{x}}{y_k} . (y_k . \some_{\emptyset} ; y_{k}.\close;0 \para \linvar{x}.\overline{\some};\linvar{x}.\some_{u,(\llfv{M} \setminus  {x}_k )};\linvar{x}( {x}_k) . \\
 &\linvar{x}.\overline{\some}; \outact{\linvar{x}}{y_{k+1}}. ( y_{k+1} . \some_{u,\llfv{M}} ;y_{k+1}.\close; \piencodf{M}_u \para \linvar{x}.\overline{\none} )) \cdots ) \para \\
 &    {x}_1.\some_{\llfv{C_i(1)}} ; \piencodf{C_i(1)}_{ {x}_1} \para \cdots \para   {x}_l.\some_{\llfv{C_i(l)}} ; \piencodf{C_i(l)}_{ {x}_l} \para  y_1. \overline{\none} \para \cdots \para y_l. \overline{\none}\\
& \linvar{x}.\some_{\emptyset} ; \linvar{x}(y_{l+1}). ( y_{l+1}.\overline{\some};y_{l+1} . \overline{\close} \para \linvar{x}.\some_{\emptyset} ; \linvar{x}. \overline{\none}) \para !\banged{x}. (x_i). \piencodf{ U }_{x_i}) \\        
 & \redd^{l+ 5} 
 \bigoplus_{C_i \in \perm{C}} (\nu \linvar{x}, \banged{x} ,  {x}_1, \cdots,  {x}_l) ( \linvar{x}.\some_{u,(\llfv{M} \setminus  {x}_{l+1} , \cdots ,  {x}_k )};\linvar{x}( {x}_{l+1}) . \cdots \\ 
 &\linvar{x}.\overline{\some}. \outact{\linvar{x}}{y_k} . (y_k . \some_{\emptyset} ; y_{k}.\close;0 \para \linvar{x}.\overline{\some};\linvar{x}.\some_{u,(\llfv{M} \setminus  {x}_k )};\linvar{x}( {x}_k) . \\
 &\linvar{x}.\overline{\some}; \outact{\linvar{x}}{y_{k+1}}. ( y_{k+1} . \some_{u,\llfv{M}} ;y_{k+1}.\close; \piencodf{M}_u \para \linvar{x}.\overline{\none} ) )  \para \\
 &    {x}_1.\some_{\llfv{C_i(1)}} ; \piencodf{C_i(1)}_{ {x}_1} \para \cdots \para   {x}_l.\some_{\llfv{C_i(l)}} ; \piencodf{C_i(l)}_{ {x}_l} \para   \linvar{x}. \overline{\none} \para !\banged{x}. (x_i). \piencodf{ U }_{x_i} ) \\
  & \redd
  \bigoplus_{C_i \in \perm{C}} (\nu \banged{x} ,  {x}_1, \cdots,  {x}_l)(  u . \overline{\none} \para  {x}_1 . \overline{\none} \para  \cdots \para  {x}_{l} . \overline{\none} \para   \\
  & (\llfv{M} \setminus \{   {x}_1, \cdots ,  {x}_k \} ). \overline{\none} \para {x}_1.\some_{\llfv{C_i(1)}} ; \piencodf{C_i(1)}_{ {x}_1} \para \cdots \para   {x}_l.\some_{\llfv{C_i(l)}} ; \piencodf{C_i(l)}_{ {x}_l}  ) \\ 
  & \redd^{l}   \bigoplus_{C_i \in \perm{C}}  u . \overline{\none} \para (\llfv{M} \setminus \{  x_1, \cdots , x_k \} ). \overline{\none} \para \llfv{C}. \overline{\none} \\
 &= \piencodf{\sum_{C_i \in \perm{C}}  \fail^{\widetilde{z}}}_u = Q_{7l + 10}  \\
                    \end{aligned}
                    \]}
                    \caption{Encoded Reduction of Explicit Substatution (Failure)}
                    \label{ch3fig:eroesf}
                \end{figure}

The rest of the proof is by induction on the number of reductions $\piencodf{\expr{N}}_u \redd^j Q$.
                            
                            \begin{enumerate}
                                \item When $j = 0$, the result follows trivially. Just take $\mathbb{N}=\mathbb{N}'$ and $\piencodf{\expr{N}}_u=Q=Q'$.
  \item When $1 \leq j \leq 7l + 10$.
                                    
Let $Q_j$ be such that $ \piencodf{\expr{N}}_u \redd^j Q_j$.
By the steps above one has 

\(\begin{aligned}
  &Q_j \redd^{7l + 10 - j} Q_{7l + 6} = Q',\\ &\expr{N} \redd^1 \sum_{C_i \in \perm{C}}  \fail^{\widetilde{z}} = \expr{N}';\text{ and} \piencodf{\sum_{C_i \in \perm{C}}  \fail^{\widetilde{z}}}_u = Q_{7l + 10}.
\end{aligned}
\)
\item When $j > 7l + 10$.

In this case, we have 
$$ \piencodf{\expr{N}}_u \redd^{7l + 10} Q_{7l + 10} \redd^n Q,$$ for $n \geq 1$. 
We also know that 
$\expr{N} \redd^1 \sum_{C_i \in \perm{C}}  \fail^{\widetilde{z}}$. However no further reductions can be performed.

                            \end{enumerate}

                \item When $\size{\widetilde{x}} < \size{C}$, the proof  proceeds similarly to the previous case.
                    
            \end{enumerate}

        \item  $\expr{N} =  M \linexsub{N' /  {x}}$. 
    
           In this case, 
            \(                \piencodf{M \linexsub{N' /  {x}}}_u =  (\nu  {x}) ( \piencodf{ M }_u \para  {x}.\some_{\llfv{N'}};\piencodf{ N' }_{ {x}} ).
            \)
            Therefore,
            \[
            \begin{aligned}
               \piencodf{\expr{N}}_u  = & (\nu  {x}) ( \piencodf{ M }_u \para  {x}.\some_{\llfv{N'}};\piencodf{ N' }_{ {x}} ) 
               \\
               \redd^m & (\nu  {x}) ( R \para  {x}.\some_{\llfv{N'}};\piencodf{ N' }_{ {x}}  ) \\
               \redd^n &  Q,\\
            \end{aligned}
            \]
            
            for some process $R$. Where $\redd^n$ is a reduction that  initially synchronizes with $ {x}.\some_{\llfv{N'}}$ when $n \geq 1$, $n + m = k \geq 1$. Type preservation in \spi ensures reducing $\piencodf{ M}_v \redd^m$ does not consume possible synchronizations with $ {x}.\some$, if they occur. Let us consider the the possible sizes of both $m$ and $n$.
            
            \begin{enumerate}
                \item For $m = 0$ and $n \geq 1$.
                
                    We have that $R = \piencodf{M}_u$ as $\piencodf{M}_u \redd^0 \piencodf{M}_u$. 
                    
                    Notice that there are two possibilities of having an unguarded $x.\overline{\some}$ or $x.\overline{\none}$ without internal reductions: 
                    
                    \begin{enumerate}
                        \item $M = \fail^{ {x}, \widetilde{y}}$.
  \[
  \begin{aligned}
  \piencodf{\expr{N}}_u & =   (\nu  {x}) ( \piencodf{ M }_u \para  {x}.\some_{\llfv{N'}};\piencodf{ N' }_{ {x}} )\\
  &=   (\nu  {x}) ( \piencodf{ \fail^{ {x}, \widetilde{y}}}_u \para  {x}.\some_{\llfv{N'}};\piencodf{ N' }_{ {x}} ) \\
    & =   (\nu  {x}) ( u.\overline{\none} \para  {x}.\overline{\none} \para  \widetilde{y}.\overline{\none} \para  {x}.\some_{\llfv{N'}};\piencodf{ N' }_{ {x}} )\\
    &\redd u.\overline{\none} \para \widetilde{y}.\overline{\none} \para \llfv{N'}.\overline{\none} \\
  \end{aligned}
    \]
  Notice that no further reductions can be performed.
  Thus,                          
 $$ \piencodf{\expr{N}}_u \redd u.\overline{\none} \para \widetilde{y}.\overline{\none} \para \lfv{N'}.\overline{\none}  = Q'.$$
We also have that  $\expr{N} \redd \fail^{ \widetilde{y} \cup \llfv{N'} } = \expr{N}'$ and $\piencodf{ \fail^{\widetilde{y} \cup \llfv{N'}} }_u = Q'$.

                        \item $\headf{M} =  {x}$.
                    
                            By the diamond property we will be reducing each non-deterministic choice of a process simultaneously.
                            Then we have the following
{\small
 \[
 \begin{aligned}
 \piencodf{\expr{N}}_u  & =  (\nu  {x}) ( \bigoplus_{i \in I}(\nu \widetilde{y})(\piencodf{  {x} }_{j} \para P_i) \para    {x}.\some_{\llfv{N'}};\piencodf{ N' }_{ {x}}  ) \\
 & =  (\nu  {x}) ( \bigoplus_{i \in I}(\nu \widetilde{y})(  {x}.\overline{\some} ; [ {x} \leftrightarrow j]  \para P_i) \para    {x}.\some_{\llfv{N'}};\piencodf{ N' }_{ {x}}  ) \\
 & \redd   (\nu  {x}) ( \bigoplus_{i \in I}(\nu \widetilde{y})(  [ {x} \leftrightarrow j]  \para P_i) \para   \piencodf{ N' }_{ {x}}  )& = Q_1 \\
 & \redd \bigoplus_{i \in I}(\nu \widetilde{y})(  \piencodf{ N' }_{j} \para P_i ) & = Q_2\\
 \end{aligned}
 \]}
                            
In addition,
\(
\expr{N} =M \linexsub {N' / {x}}\redd  M \headlin{ N' / {x} } = \expr{M}\).
Finally,
\[
\begin{aligned}                               \piencodf{\expr{M}}_u= \piencodf{M \headlin{ N'/  {x} }}_u &= \bigoplus_{i \in I}(\nu \widetilde{y})(  \piencodf{ N' }_{j} \para P_i ) & = Q_2. 
\end{aligned}
\]
                            
\begin{enumerate}
\item When $n = 1$:

Then, $Q = Q_1$ and  $ \piencodf{\expr{N}}_u \redd^1 Q_1$. Also,

$Q_1 \redd^1 Q_2 = Q'$, $\expr{N} \redd^1 M \headlin{ N'/ {x}} = \expr{N}'$ and $\piencodf{M \headlin{ N'/ {x}}}_u = Q_2$.
\item When $n \geq 2$:

Then  $ \piencodf{\expr{N}}_u \redd^2 Q_2 \redd^l Q$, for $l \geq 0$.  Also, 
$\expr{N} \rightarrow \expr{M}$, $Q_2 = \piencodf{\expr{M}}_u$. By the induction hypothesis, there exist $ Q'$ and $\expr{N}'$ such that $ Q \redd^i Q'$, $\expr{M} \redd_{\pequiv}^j \expr{N}'$ and $\piencodf{\expr{N}'}_u = Q'$. Finally, $\piencodf{\expr{N}}_u \redd^2 Q_2 \redd^l Q \redd^i Q'$ and $\expr{N} \rightarrow \expr{M}  \redd_{\pequiv}^j \expr{N}'$.
                                
                            \end{enumerate}

                    \end{enumerate}
 \item  For $m \geq 1$ and $ n \geq 0$.
                    
            \begin{enumerate}
            \item When $n = 0$.
            
               Then $  (\nu  {x}) ( R \para  {x}.\some_{\llfv{N'}};\piencodf{ N' }_{ {x}} )  = Q$ and $\piencodf{M}_u \redd^m R$ where $m \geq 1$. By the IH there exist $R'$  and $\expr{M}' $ such that $R \redd^i R'$, $M \redd_{\pequiv}^j \expr{M}'$ and $\piencodf{\expr{M}'}_u = R'$. Thus,
              \[ 
               \begin{aligned}
                   \piencodf{\expr{N}}_u =& (\nu  {x}) ( \piencodf{M}_u  \para  {x}.\some_{\llfv{N'}};\piencodf{ N' }_{ {x}}  ) \\
                   \redd^m & (\nu  {x}) ( R \para  {x}.\some_{\llfv{N'}};\piencodf{ N' }_{ {x}} )  = Q
                \end{aligned}
                 \]
                Also,
                \(
               Q  \redd^i  (\nu  {x}) ( R' \para  {x}.\some_{\llfv{N'}};\piencodf{ N' }_{ {x}} )  = Q',
                \)
                and the term can reduce as follows: $\expr{N} = M \linexsub {N' / {x}} \redd_{\pequiv}^j \sum_{M_i' \in \expr{M}'} M_i' \linexsub {N' / {x}} = \expr{N}'$ and  $\piencodf{\expr{N}'}_u = Q'$

            \item When $n \geq 1$.
                Then  $R$ has an occurrence of an unguarded $x.\overline{\some}$ or $x.\overline{\none}$, this case follows by IH.
                        
                    \end{enumerate}
            \end{enumerate}

            \item  $\expr{N} =  M \unexsub{U / \unvar{x}}$. 
    
            In this case, 
            \(
            \begin{aligned}
                \piencodf{M \unexsub{U / \unvar{x}}}_u &=   (\nu \banged{x}) ( \piencodf{ M }_u \para   !\banged{x}. (x_i).\piencodf{ U }_{x_i} ).
            \end{aligned}
            \)
            Then, 
            \[
            \begin{aligned}
               \piencodf{\expr{N}}_u & =   (\nu \banged{x}) ( \piencodf{ M }_u \para   !\banged{x}. (x_i).\piencodf{ U }_{x_i} )  \redd^m  (\nu \banged{x}) ( R \para   !\banged{x}. (x_i).\piencodf{ U }_{x_i} ) \redd^n  Q.
            \end{aligned}
            \]
            for some process $R$. Where $\redd^n$ is a reduction initially synchronises with $!\banged{x}. (x_i)$ when $n \geq 1$, $n + m = k \geq 1$. Type preservation in \spi ensures reducing $\piencodf{ M}_v \redd^m$ doesn't consume possible synchronisations with $!\banged{x}. (x_i)$ if they occur. Let us consider the the possible sizes of both $m$ and $n$.
            
            \begin{enumerate}
                \item For $m = 0$ and $n \geq 1$.
                
                   In this case,  $R = \piencodf{M}_u$ as $\piencodf{M}_u \redd^0 \piencodf{M}_u$. 
                    
                    Notice that the only possibility of having an unguarded $ \outsev{\banged{x}}{{x_i}}$ without internal reductions is when   $\headf{M} =  {x}[ind].$
                           By the diamond property  we will be reducing each non-deterministic choice of a process simultaneously.
                            Then we have the following:
                            
                            \[
                                \hspace{-1.5cm}
                            \begin{aligned}
                            \piencodf{\expr{N}}_u  & =  (\nu \banged{x}) ( \bigoplus_{i \in I}(\nu \widetilde{y})(\piencodf{  {x}[ind] }_{j} \para P_i) \para   !\banged{x}. (x_i).\piencodf{ U }_{x_i} ) \\
                            & =  (\nu \banged{x}) ( \bigoplus_{i \in I}(\nu \widetilde{y})( \outsev{\banged{x}}{{x_i}}. {x}_i.l_{ind}; [{x_i} \leftrightarrow j]  \para P_i) \para   !\banged{x}. (x_i).\piencodf{ U }_{x_i}  ) \\
                            & \redd  (\nu \banged{x}) ( \bigoplus_{i \in I}(\nu \widetilde{y})( (\nu x_i) ({x}_i.l_{ind}; [{x_i} \leftrightarrow j] \para \piencodf{ U }_{x_i} ) \para P_i) \para   !\banged{x}. (x_i).\piencodf{ U }_{x_i}  ) &= Q_1 \\
                            & =  (\nu \banged{x}) ( \bigoplus_{i \in I}(\nu \widetilde{y})( (\nu x_i) ({x}_i.l_{ind}; [{x_i} \leftrightarrow j] \para x_i. case( ind.\piencodf{U_{ind}}_{x_i} ) ) \para P_i) \\
                            &\para   !\banged{x}. (x_i).\piencodf{ U }_{x_i}  ) \\
                            & \redd  (\nu \banged{x}) ( \bigoplus_{i \in I}(\nu \widetilde{y})( (\nu x_i) ( [{x_i} \leftrightarrow j] \para \piencodf{U_{ind}}_{x_i} ) \para P_i) \para   !\banged{x}. (x_i).\piencodf{ U }_{x_i}  ) &= Q_2\\
                            & \redd  (\nu \banged{x}) ( \bigoplus_{i \in I}(\nu \widetilde{y})( \piencodf{U_{ind}}_{j}  \para P_i) \para   !\banged{x}. (x_i).\piencodf{ U }_{x_i}  ) &= Q_3 \\
                            \end{aligned}
                            \]

                We consider the two cases of the form of $U_{ind}$ and show that the choice of $U_{ind}$ is inconsequential
            
                \begin{itemize}
                \item When $ U_i = \banged{\bag{N}}$:
                
                In this case, 
                \(
                \begin{aligned}
                \expr{N} &=M \unexsub{U / \unvar{x}}\redd M \headlin{ N /\banged{x} }\unexsub{U / \unvar{x}} = \expr{M}.
                \end{aligned}
                \)
                 and 
                \[
                 \begin{aligned}
                 \piencodf{\expr{M}}_u=& \piencodf{M \headlin{ N /\banged{x} }\unexsub{U / \unvar{x}}}_u \\
                 =&  (\nu \banged{x}) ( \bigoplus_{i \in I}(\nu \widetilde{y})( \piencodf{\bag{N}}_{j}  \para P_i) \para   !\banged{x}. (x_i).\piencodf{ U }_{x_i}  ) & = Q_3 
                                            \end{aligned}
                 \]
                
                \item When $ U_i = \banged{\oneb} $:
                  
                  In this case,
                        \(
                        \begin{aligned}
                            \expr{N} &=M \unexsub{U / \unvar{x}} \redd M \headlin{ \fail^{\emptyset} /\banged{x} } \unexsub{U /\unvar{x} } = \expr{M}.
                        \end{aligned}
                        \)
                        
                        Notice that $\piencodf{\banged{\oneb}}_{j} =  j.\none$ and that $\piencodf{\fail^{\emptyset}}_j = j.\overline{\none}$. In addition,
 
                            \[
                            \begin{aligned}
                               \piencodf{\expr{M}}_u&= \piencodf{M \headlin{ \fail^{\emptyset} /\banged{x} } \unexsub{U /\unvar{x} }}_u\\
                               &=  (\nu \banged{x}) ( \bigoplus_{i \in I}(\nu \widetilde{y})( \piencodf{\fail^{\emptyset}}_{j}  \para P_i) \para   !\banged{x}. (x_i).\piencodf{ U }_{x_i}  ) \\
                               & = (\nu \banged{x}) ( \bigoplus_{i \in I}(\nu \widetilde{y})( \piencodf{\banged{\oneb}}_{j}  \para P_i) \para   !\banged{x}. (x_i).\piencodf{ U }_{x_i}  ) & = Q_3 
                            \end{aligned}
                            \]
                \end{itemize}
                
                Both choices give an $\expr{M}$ that are equivalent to $Q_3$.

    \begin{enumerate}
    \item When $n \leq 2$.
    
   In this case, $Q = Q_n$ and  $ \piencodf{\expr{N}}_u \redd^n Q_n$.
                                 
Also, $Q_n \redd^{3-n} Q_3 = Q'$, $\expr{N} \redd^1 \expr{M} = \expr{N}'$ and $\piencodf{\expr{M} }_u = Q_2$.
                                 
     \item When $n \geq 3$.
     
     We have $ \piencodf{\expr{N}}_u \redd^3 Q_3 \redd^l Q$ for $l \geq 0$. We also know that $\expr{N} \rightarrow \expr{M}$, $Q_3 = \piencodf{\expr{M}}_u$. By the IH, there exist $ Q$ and $\expr{N}'$ such that $Q \redd^i Q'$, $\expr{M} \redd_{\pequiv}^j \expr{N}'$ and $\piencodf{\expr{N}'}_u = Q'$. Finally, $\piencodf{\expr{N}}_u \redd^2 Q_3 \redd^l Q \redd^i Q'$ and $\expr{N} \rightarrow \expr{M}  \redd_{\pequiv}^j \expr{N}' $.

                    \end{enumerate}
 \item For $m \geq 1$ and $ n \geq 0$.
                    
            \begin{enumerate}
            \item When $n = 0$.
            
               Then $   (\nu \banged{x}) ( R \para   !\banged{x}. (x_i).\piencodf{ U }_{x_i} )  = Q$ and $\piencodf{M}_u \redd^m R$ where $m \geq 1$. By the IH there exist $R'$  and $\expr{M}' $ such that $R \redd^i R'$, $M \redd_{\pequiv}^j \expr{M}'$ and $\piencodf{\expr{M}'}_u = R'$. 
              Hence,
              \[
               \begin{aligned}
                   \piencodf{\expr{N}}_u  =&  (\nu \banged{x}) ( \piencodf{ M }_u \para   !\banged{x}. (x_i).\piencodf{ U }_{x_i} ) \\
                   \redd^m &  (\nu \banged{x}) ( R \para   !\banged{x}. (x_i).\piencodf{ U }_{x_i} )  = Q.
                \end{aligned}
                 \]
                In addition,
                \(
                   Q  \redd^i   (\nu \banged{x}) ( R' \para   !\banged{x}. (x_i).\piencodf{ U }_{x_i} ) = Q\), and the term can reduce as follows: $ \expr{N} = M \unexsub{U / \unvar{x}} \redd_{\pequiv}^j \sum_{M_i' \in \expr{M}'} M_i' \unexsub{U / \unvar{x}} = \expr{N}'$ and  $\piencodf{\expr{N}'}_u = Q'$.

            \item When $n \geq 1$.
            
            Then $R$ has an occurrence of an unguarded $\outsev{\banged{x}}{{x_i}}$, and the case follows by IH.
\end{enumerate}
            \end{enumerate}
               \end{enumerate}
\end{proof}

\subsection{Success Sensitiveness of $\piencodf{\cdot}_u$}


We say that a process occurs \emph{guarded} when it occurs behind a prefix (input, output, closing of channels, servers, server request, choice an selection and non-deterministic session behaviour). Formally,

\begin{definition}{}
 A process $P\in \spi$ is {\em guarded} if  $\alpha.P$ , $ \alpha;P$ or $ \choice{x}{i}{I}{i}{P} $, where $ \alpha = \overline{x}(y), x(y), x.\overline{\close}, x.\close,$ $ x.\overline{\some}, x.\some_{(w_1, \cdots, w_n)} , \case{x}{i} , !x(y) ,  \outsev{x}{y}$. 
We say it occurs \emph{unguarded} if it is not guarded for any prefix.
\end{definition}

\begin{proposition}[Preservation of Success]
\label{ch3Prop:checkprespiunres}
For all $M\in \lamrsharfailunres$, the following hold:
\begin{enumerate}
    \item $ \headf{M} = \checkmark \implies \piencodf{M} = P \para \checkmark \oplus Q $
    \item $ \piencodf{M}_u =  P \para \checkmark \oplus Q \implies \headf{M} = \checkmark$
\end{enumerate}

\end{proposition}

\begin{proof}

Proof of both cases by induction on the structure of $M$. 

\begin{enumerate} 
\item We only need to consider terms of the following form:

    \begin{enumerate}

        \item  $ M = \checkmark $:
        
        This case is immediate.
        
        \item $M = N\ (C \bagsep U)$:
        
        Then, $\headf{N \ (C \bagsep U)} = \headf{N}$. If $\headf{N} = \checkmark$, then  {\small$$ \piencodf{M (C \bagsep U)}_u = \bigoplus_{C_i \in \perm{C}} (\nu v)(\piencodf{M}_v \para v.\some_{u , \llfv{C}} ; \outact{v}{x} . ([v \leftrightarrow u] \para \piencodf{C_i \bagsep U}_x ) ).$$}
        By the IH,  $\checkmark$ is unguarded in $\piencodf{N}_u$.

        \item $M = M' \linexsub {N /x}$
        
        Then we have that $\headf{M' \linexsub{N /  {x}}}  = \headf{M'} = \checkmark$. Then $\piencodf{ M' \linexsub{N /  {x}}  }_u   =   (\nu  {x}) ( \piencodf{ M' }_u \para    {x}.\some_{\llfv{N}};\piencodf{ N }_{ {x}}  )$ and by the IH $\checkmark$ is unguarded in $\piencodf{M'}_u$.
        
        \item $M = M' \unexsub{U / \unvar{x}}$
        
        Then we have that $\headf{M' \unexsub{U / \unvar{x}}}  = \headf{M'} = \checkmark$. Then $\piencodf{ M' \unexsub{U / \unvar{x}}  }_u   =   (\nu \banged{x}) ( \piencodf{ M' }_u \para   !\banged{x}. (x_i).\piencodf{ U }_{x_i} ) $ and by the IH  $\checkmark$ is unguarded in $\piencodf{M'}_u$.

    \end{enumerate}

   \item We only need to consider terms of the following form:    
         
    \begin{enumerate}
    
        \item {\bf Case $M = \checkmark$:}
        
        Then, 
        $\piencodf{\checkmark}_u = \checkmark$ 
        which is an unguarded occurrence of $\checkmark$ and that $\headf{\checkmark} = \checkmark$.
        
        \item {\bf Case $M = N (C \bagsep U)$:}
        
        Then, $\piencodf{N (C \bagsep U)}_u = \bigoplus_{C_i \in \perm{C}} (\nu v)(\piencodf{N}_v \para v.\some_{u , \llfv{C}} ; \outact{v}{x} . ([v \leftrightarrow u] \para \piencodf{C_i \bagsep U}_x ) )$. The only occurrence of an unguarded $\checkmark$ can occur is within $\piencodf{N}_v$. By the IH, $\headf{N} = \checkmark$ and finally $\headf{N \ B} = \headf{N}$.

        \item {\bf Case $M = M' \linexsub{N /  {x}}$:} 
        
        Then,  $\piencodf{ M' \linexsub{N /  {x}}  }_u   =   (\nu  {x}) ( \piencodf{ M' }_u \para    {x}.\some_{\llfv{N}};\piencodf{ N }_{ {x}}  ) $, an unguarded occurrence of $\checkmark$ can only occur within $\piencodf{ M' }_u $. By the IH,  $\headf{M'} = \checkmark$ and hence $\headf{ M' \linexsub{N /  {x}}}  = \headf{M'}$.
        
        \item {\bf Case $M = M' \unexsub{U / \unvar{x}}$:}
        This case is analogous to the previous.
    \end{enumerate}  
\end{enumerate}
\end{proof}

\begin{theorem}[Success Sensitivity]
\label{ch3proof:successsenscetwounres}
The encoding $\piencodf{-}_u:\lamrsharfailunres \rightarrow \spi$ is success sensitive on well formed linearly closed expression if
for any expression we have $\succp{\expr{M}}{\checkmark}$ iff $\succp{\piencodf{\expr{M}}_u}{\checkmark}$.
\end{theorem}

\begin{proof}
We proceed with the proof in two parts.

\begin{enumerate}
    
    \item Suppose that  $\expr{M} \Downarrow_{\checkmark} $. We will prove that $\piencodf{\expr{M}} \Downarrow_{\checkmark}$.

    By \defref{ch3def:app_Suc3unres}, there exists  $ \expr{M}' = M_1 , \cdots , M_k$ such that $\expr{M} \redd^* \expr{M}'$ and
    $\headf{M_j'} = \checkmark$, for some  $j \in \{1, \ldots, k\}$ and term $M_j'$ such that $M_j\pequiv  M_j'$.
    By completeness, there exists $ Q$ such that $\piencodf{\expr{M}}_u  \redd^* Q = \piencodf{\expr{M'}}_u$.
    
    We wish to show that there exists $Q'$such that $Q \redd^* Q'$ and $Q'$ has an unguarded occurrence of $\checkmark$.
    
    From $Q = \piencodf{\expr{M}'}_u$ and due to compositionality and the homomorphic preservation of non-determinism we have that
    \(
        \begin{aligned}
            Q &= \piencodf{M_1}_u \oplus \cdots \oplus \piencodf{M_k}_u\\
        \end{aligned}
    \).
    
    By Proposition \ref{ch3Prop:checkprespiunres} (1) we have that $\headf{M_j} = \checkmark \implies \piencodf{M_j}_u =  P \para \checkmark \oplus Q'$. Hence $Q$ reduces to a process that has an unguarded occurence of $\checkmark$.

    \item Suppose that $\piencodf{\expr{M}}_u \Downarrow_{\checkmark}$. We will prove that $ \expr{M} \Downarrow_{\checkmark}$.

    By operational soundness (Lemma~\ref{ch3l:app_soundnesstwounres}) we have that if $ \piencodf{\expr{N}}_u \redd^* Q$
    then there exist $Q'$  and $\expr{N}' $ such that 
    $Q \redd^* Q'$, $\expr{N}  \redd_{\pequiv}^* \expr{N}'$ 
    and 
    $\piencodf{\expr{N}'}_u = Q'$.
    
   Since $\piencodf{\expr{M}}_u \redd^* P_1 \oplus \ldots \oplus P_k$, and $P_j'= P_j'' \para \checkmark$, for some $j$ and $P_j'$, such that $P_j \equiv P_j'$. 
   
   Notice that if $\piencodf{\expr{M}}_u$ is itself a term with unguarded $\checkmark$, say $\piencodf{\expr{M}}_u=P \para \checkmark$, then $\expr{M}$ is itself headed with $\checkmark$, from Proposition \ref{ch3Prop:checkprespiunres} (2).
   
   In the case $\piencodf{\expr{M}}_u= P_1 \oplus \ldots \oplus P_k$, $k\geq 2$, and $\checkmark$ occurs unguarded in an $P_j$, The encoding acts homomorphically over sums and the reasoning is similar. We have that $P_j = P_j' \para \checkmark$ we apply Proposition \ref{ch3Prop:checkprespiunres} (2).

\end{enumerate}
\end{proof}

\newpage
\chapter{Appendix of Chapter 4}\label{ch4appendix_aplas}

\section{Full \texorpdfstring{\clpi}{Pi}: Replicated Servers and Clients}
\label{ch4as:fullPi}

Full \clpi includes unrestricted session behaviors (replicated servers and clients), not presented in \Cref{ch4s:pi}.
Here we discuss how to add these omitted unrestricted sessions to the system described in \Cref{ch4s:pi}.
The proofs of Type Preservation (\Cref{ch4t:srPi}) and Deadlock-freedom (\Cref{ch4t:dfPi}) in \Cref{ch4s:piProofs} concern Full \clpi.
\begin{itemize}
    \item
        To the syntax of processes in \Cref{ch4f:pilang} (top) we add two prefixes, $\puname{x}{y};P$ and $\guname{x}{y};P$, for client requests and server definitions, respectively.
        Both prefixes bind $y$ in $P$.

    \item
        Now $\fn{P}$ denotes the set of free names of $P$, including those used for unrestricted sessions.
        We write $\fln{P}$ to denote the set of free linear names of $P$, and $\fpn{P}$ for the set of free non-linear names.
        Note that $\fpn{P} = \fn{P} \setminus \fln{P}$.

    \item
        To the structural congruence in \Cref{ch4f:pilang} (bottom) we add a rule that cleans up unused servers:
        \[
            \res{x}(\guname{x}{y};P \| Q) \equiv Q \quad \text{(if $x \notin \fn{Q}$)}.
        \]

    \item
        To the lazy semantics in \Cref{ch4f:redtwo} we add the following rule that initiates a session between a client and a copy of a server:
        \begin{align*}
            \mathsmaller{\rredtwo{{?}{!}}}\quad
            & \res{x} \Big( \bignd_{i \in I} \pctx{C_i}[\puname{x}{y_i};P_i] \| \bignd_{j \in J} \pctx{D_j}[\guname{x}{z};Q_j] \Big)
            \\
            & \redtwo_x
            \bignd_{j \in J} \pctx[\Big]{D_j}[ \res{x} \Big( \res{w} \Big( \bignd_{i \in I} \pctx{C_i}[P_i\{w/z\}] \| Q_j\{w/z\} \Big) \| \guname{x}{z};Q_j \Big) ]
        \end{align*}


    \item
        Writing ${?}\Gamma$ to denote that $\forall x \in \Gamma.~ \exists A.~ \Gamma(x) = {?}A$, to the typing rules in \Cref{ch4fig:trulespi} we add:
        \begin{mathpar}
            \ttype[\scriptsize]{$?$}~
            \inferrule{
                P \vdash \Gamma, y:A
            }{
                \puname{x}{y};P \vdash \Gamma, x:{?}A
            }
            \and
            \ttype[\scriptsize]{$!$}~
            \inferrule{
                P \vdash {?}\Gamma, y:A
            }{
                \guname{x}{y};P \vdash {?}\Gamma, x:{!}A
            }
            \and
            \ttype[\scriptsize]{weaken}~
            \inferrule{
                P \vdash \Gamma
            }{
                P \vdash \Gamma, x:{?}A
            }
            \and
            \ttype[\scriptsize]{contract}~
            \inferrule{
                P \vdash \Gamma, x:{?}A, x':{?}A
            }{
                P\{x/x'\} \vdash \Gamma, x:{?}A
            }
        \end{mathpar}
        Moreover, we replace Rule~$\ttype{${\oplus}\some$}$ with the following:
        \[
            \ttype{${\oplus}\some$}~
            \inferrule{
                P \vdash {\with}\Gamma, {?}\Delta, x:A
            }{
                \gsome{x}{\dom{\Gamma}};P \vdash {\with}\Gamma, {?}\Delta, x:{\oplus}A
            }
        \]
\end{itemize}

\section{An Alternative Eager Semantics for \texorpdfstring{\clpi}{spi+}}\label{ch4s:piEager}
Let us consider a variant of \clpi in which syntax, typing, and structural congruence are as in \secref{ch4s:pi}, but with an \emph{eagerly committing} semantics.
The idea is simple: we fully commit to a non-deterministic choice once a prefix synchronizes.

\begin{figure}[t] \mysmall
        \begin{mathpar}
            \mathsmaller{\rredone{\scc{Id}}}~~
            \inferrule{}{
                \res{x}(\pctx[\big]{N}[\pfwd{x}{y}] \| Q) \redone \pctx*{N}[Q\{y/x\}]
            }
            \and
            \mathsmaller{\rredone{\1 \bot}}~~
            \inferrule{}{
                \res{x}(\pctx{N}[\pclose{x}] \| \pctx{N'}[\gclose{x};Q]) \redone \pctx*{N}[\0] \| \pctx*{N'}[Q]
            }
            \and
            \mathsmaller{\rredone{\tensor \parr}}~~
            \inferrule{}{
                \res{x}(\pctx{N}[\pname{x}{y};(P \| Q)] \| \pctx{N'}[\gname{x}{z};R]) \redone \pctx*[\big]{N}[\res{x}(Q \| \res{y}(P \| \pctx*{N'}[R\{y/z\}]))]
            }
            \and
            \mathsmaller{\rredone{\oplus \with}}~~
            \inferrule{}{
                \forall k' \in K.~ \res{x}(\pctx{N}[\psel{x}{k'};P] \| \pctx{N'}[\gsel{x}\{k:Q^k\}_{k \in K}]) \redone \res{x}(\pctx*{N}[P] \| \pctx*{N'}[Q^{k'}])
            }
            \and
            \mathsmaller{\rredone{{?}{!}}}~~
            \inferrule{}{
                \res{x}(\pctx{N}[ \puname{x}{y};P ] \| \pctx{N'}[ \guname{x}{z};Q ])
                \redone \pctx*[\big]{N'}[ \res{x}\big( \res{y}( \pctx*{N}[P] \| Q\{y/z\} ) \| \guname{x}{z};Q \big) ]
            }
            \and
            \mathsmaller{\rredone{\some}}~~
            \inferrule{}{
                \res{x}(\pctx{N}[\psome{x};P] \| \pctx{N'}[\gsome{x}{w_1, \ldots, w_n};Q]) \redone \res{x}(\pctx*{N}[P] \| \pctx*{N'}[Q])
            }
            \and
            \mathsmaller{\rredone{\none}}~~
            \inferrule{}{
                \res{x}(\pctx{N}[\pnone{x}] \| \pctx{N'}[\gsome{x}{w_1, \ldots, w_n};Q]) \redone \pctx*{N}[\0] \| \pctx*{N'}[\pnone{w_1} \| \ldots \| \pnone{w_n}]
            }
            \and
            \mathsmaller{\rredone{\equiv}}~~
            \inferrule{
                P \equiv P'
                \\
                P' \redone Q'
                \\
                Q' \equiv Q
            }{
                P \redone Q
            }
            \and
            \mathsmaller{\rredone{\nu}}~~
            \inferrule{
                P \redone P'
            }{
                \res{x}(P \| Q) \redone \res{x}(P' \| Q)
            }
            \and
            \mathsmaller{\rredone{\|}}~~
            \inferrule{
                P \redone P'
            }{
                P \| Q \redone P' \| Q
            }
            \and
            \mathsmaller{\rredone{\nd}}~~
            \inferrule{
                P \redone P'
            }{
                P \nd Q \redone P' \nd Q
            }
        \end{mathpar}
    \caption{
        Eager reduction semantics for \texorpdfstring{\clpi}{spi+}.
    }
    \label{ch4f:eagerReductions}
\end{figure}

The eager reduction semantics, denoted $\redone$, is given in \Cref{ch4f:eagerReductions}.
This semantics implements the full commitment of non-deterministic choices by committing ND-contexts to D-contexts as follows:

\begin{definition}{}\label{ch4d:ncoll}
    The \emph{commitment} of an ND-context $\pctx{N}$, denoted $\D{\pctx{N}}$, is defined as follows:
    \begin{align*}
        \D{\hole} &:= \hole
        & \D{\pctx{N} \|  P} &:= \D{\pctx{N}} \| P
        & \D{\res{x}(\pctx{N} \| P)} &:= \res{x}(\D{\pctx{N}} \| P)
        & \D{\pctx{N} \nd P} &:= \D{\pctx{N}}
    \end{align*}
\end{definition}

\begin{proposition}\label{ch4p:ncoll}
    For any ND-context $\pctx{N}$, the context $\D{\pctx{N}}$ is a D-context.
\end{proposition}

Just as \clpi  with lazy semantics,  \clpi with $\redone$ satisfies type preservation and deadlock-freedom.
See \Cref{ch4ss:proofsEager} for details.

\begin{theorem}[Type Preservation: Eager Semantics]\label{ch4t:typePresEager}
    If $P \vdash \Gamma$, then both $P \equiv Q$ and $P \redone Q$ (for any $S$) imply $Q \vdash \Gamma$.
\end{theorem}

\begin{proof}[Proof (Sketch)]
    If $P \equiv Q$, the thesis follows directly from \Cref{ch4t:srPi}.
    If $P \redone Q$, we apply induction on the derivation of the reduction.
    In each case, we show that the commitment of ND-contexts (\Cref{ch4d:ncoll}) preserves typing.
\end{proof}

\begin{restatable}[Deadlock-freedom: Eager Semantics]{theorem}{thmDlfreeOne}\label{ch4t:dlfreeOne}
    If $P \vdash \emptyset$ and $P \not\equiv \0$, then there is $R$ such that $P \redone R$.
\end{restatable}

\begin{proof}[Proof (Sketch)]
    First, we write $P$ in such a way that we can access all its unblocked prefixes.
    Then we inductively show that there must be at least one pair of such prefixes that are connected by a restriction.
    Hence, these prefixes are duals and thus the process can reduce.
\end{proof}

\begin{figure*}[!t] \mysmall
        \begin{align*}
            \mathbin{\rotatebox[origin=r]{30}{$\red$}} &~ (\fail^{\emptyset} \bag{x_2 \bag{x_3\ \oneb}  }) \linexsub{  \bag{y, I} /  x_2,x_3  }
            = N_1
            \\[-5pt]
            M
            =
            (x_1 \!\bag{\!x_2 \!\bag{\!x_3\ \oneb\!}  \!}) \linexsub{\! \bag{\!\fail^{\emptyset} , y , I\!} \!/x_1,x_2,x_3 }~
            \red &~ (y \bag{x_2 \bag{x_3\ \oneb}  }) \linexsub{ \bag{\fail^{\emptyset} , I} /  x_2,x_3  }
            = N_2
            \\[-3pt]
            \mathbin{\rotatebox[origin=r]{-30}{$\red$}} &~ (I \bag{x_2 \bag{x_3\ \oneb}}) \linexsub{ \bag{\fail^{\emptyset} , y}  /  x_2,x_3  }
            = N_3
        \end{align*}
        \vspace{-4ex}
        \begin{align*}
            \piencodfaplas{M}_u
            =
            &~
            \res{z_1}(
            \gsome{z_1}{\emptyset}; \piencodfaplas{ \fail^{\emptyset} }_{z_1}
            \| \res{z_2}(
            \gsome{z_2}{y}; \piencodfaplas{ y }_{z_2}
            \| \res{z_3}(
            \gsome{z_3}{\emptyset}; \piencodfaplas{ I }_{z_3}
            \\
            &~ \quad {}
            \| \bignd_{(x_i,x_j,x_k) \in \pi(\{z_1,z_2,z_3\})}
            \res{v}(
            \psome{x_i}; \pfwd{x_i}{v} \| \gsome{v}{v,x_j,x_k}; \pname{v}{z}; \\
            &~ \qquad {} (
            \piencodfaplas{ \bag{ x_j \bag{ x_k \ \oneb } } }_z
            \| \pfwd{v}{u}
            ) ) ) ) )
        \end{align*}

        \newlength{\sibldist}\setlength{\sibldist}{1.3cm}
            \mbox{}\hfill
            \begin{tikzpicture}
            [
                level 1/.style = {sibling distance = \sibldist},
                level 2/.style = {sibling distance = 0.5cm},
                level distance = 2.2cm,
                squigly/.style = {decorate,decoration={snake,amplitude=1pt,segment length=6pt,pre length=0pt,post length=2pt}}
            ]

                \node {$\piencodfaplas{M}_u$}
                    child { node {$\piencodfaplas{N_1}_u$}
                        edge from parent [->, squigly, cblGreen] node [above, xshift=-.5ex, pos=1.0, text=black] {\scriptsize $\ast$}}
                    child { node {$\piencodfaplas{N_2}_u$}
                        edge from parent [->, squigly, cblGreen] node [left, yshift=1.0ex, xshift=.1ex, pos=1.0, text=black] {\scriptsize $\ast$}}
                    child { node {$\piencodfaplas{N_3}_u$}
                        edge from parent [->, squigly, cblGreen] node [above, xshift=.5ex, pos=1.0, text=black] {\scriptsize $\ast$}};
            \end{tikzpicture}
            \hfill
            \textcolor{gray}{\vrule}
            \hfill
            \begin{minipage}[t]{\textwidth}
                \centering
                \begin{tikzpicture}
                [
                    level 1/.style = {sibling distance = 2.0cm},
                    level 2/.style = {sibling distance = 0.5cm},
                    level distance = 1.1cm
                ]

                    \node {$\piencodfaplas{ M }_u$ }
                        child { node {$P_1(z_2,z_3)$}
                            child { node [text=black] {$\piencodfaplas{N_1}_u$}
                                edge from parent [draw=none] node [draw=none] {$\premattwo$}}
                            edge from parent [->, cblRed] node [above, pos=1.0, text=black] {\scriptsize $\ast$}}
                        child {node {$P_1(z_3,z_2)$}
                            child { node [text=black] {$\piencodfaplas{N_1}_u$}
                                edge from parent [draw=none] node [draw=none] {$\premattwo$}}
                            edge from parent [->, cblRed] node [above, pos=1.0, text=black] {\scriptsize $\ast$}}
                        child {node {$P_2(z_1,z_3)$}
                            child { node [text=black] {$\piencodfaplas{N_2}_u$}
                                edge from parent [draw=none] node [draw=none] {$\premattwo$}}
                            edge from parent [->, cblRed] node [above, xshift=-.5ex, pos=1.0, text=black] {\scriptsize $\ast$}}
                        child {node {$P_2(z_3,z_1)$}
                            child { node [text=black] {$\piencodfaplas{N_2}_u$}
                                edge from parent [draw=none] node [draw=none] {$\premattwo$}}
                            edge from parent [->, cblRed] node [above, xshift=.5ex, pos=1.0, text=black] {\scriptsize $\ast$}}
                        child {node {$P_3(z_1,z_2)$}
                            child { node [text=black] {$\piencodfaplas{N_3}_u$}
                                edge from parent [draw=none] node [draw=none] {$\premattwo$}}
                            edge from parent [->, cblRed] node [above, pos=1.0, text=black] {\scriptsize $\ast$}}
                        child {node {$P_3(z_2,z_1)$}
                            child { node [text=black] {$\piencodfaplas{N_3}_u$}
                                edge from parent [draw=none] node [draw=none] {$\premattwo$}}
                            edge from parent [->, cblRed] node [above, pos=1.0, text=black] {\scriptsize $\ast$}};
                \end{tikzpicture}

                The processes above are as follows:
                \begin{align*}
                    Q(a,b)
                    &=
                    \gsome{v}{u,a,b}; \pname{v}{z}; ( \piencodfaplas{ \bag{a \ \bag{ b \ \oneb}} }_z \| \pfwd{v}{u} )
                    \\[-1mm]
                    P_1(a,b)
                    &=
                    \res{z_2}( \gsome{z_2}{y}; \piencodfaplas{ y }_{z_2} \| \res{z_3}( \gsome{x_3}{\emptyset}; \piencodfaplas{ I }_{z_3} \| \res{v}( \piencodfaplas{ \fail^{\emptyset} }_{v} \| Q(a,b) ) ) )
                    \\[-1mm]
                    P_2(a,b)
                    &=
                    \res{z_1}( \gsome{z_1}{\emptyset}; \piencodfaplas{ \fail^{\emptyset} }_{z_1} \| \res{z_3}( \gsome{x_3}{\emptyset}; \piencodfaplas{ I }_{z_3} \| \res{v}( \piencodfaplas{ y }_{v} \| Q(a,b) ) ) )
                    \\[-1mm]
                    P_3(a,b)
                    &=
                    \res{z_1}( \gsome{z_1}{\emptyset}; \piencodfaplas{ \fail^{\emptyset} }_{z_1} \| \res{z_2}( \gsome{z_2}{y}; \piencodfaplas{ y }_{z_2} \| \res{v}( \piencodfaplas{ I }_{v} \| Q(a,b) ) ) )
                \end{align*}
            \end{minipage}
            \hfill\mbox{}
    \caption{\Cref{ch4ex:looseTight}: Reductions of $M$ and of $\piencodfaplas{M}_u$ under lazy and eager semantics. In $\piencodfaplas{M}_u$, we write `$\pi(X)$' for the permutations of finite set $X$.}
    \label{ch4fig:eager_m_red}\label{ch4f:eager_v_lazy}
\end{figure*}

It is insightful to formally contrast our lazy and eager semantics.
We discuss two different ways.

\paragraph*{Lazy vs Eager, Part I: Translating \lamcoldetshlin.}
One way of comparing $\redtwo$ and $\redone$ is to establish the correctness of the translation
$\piencodf{\cdot}_{-}$ (\Cref{ch4fig:encodinglin}) but now in the eager case.
It turns out that the eager semantics leads to a \emph{strictly weaker} form of correctness, whereby completeness and weak soundness (cf.\ \Cref{ch4d:encCriteria}) hold up to a precongruence $\premat$  instead of an equivalence (as in \Cref{ch4t:correncLazy}).

The  precongruence $\premat$ is defined as follows:
\begin{mathpar}
    \inferrule{ }{
        P \premat P
    }
    \and
    \inferrule{
        P_i \premat P'_i
        \\
        {\scriptstyle i \in \{1,2\}}
    }{
        P_1 \nd P_2 \premat P'_i
    }
    \and
    \inferrule{
        P \premat P'
        \\
        Q \premat Q'
    }{
        P   \| Q \premat P'   \| Q'
    }
    \and
    \inferrule{
        P \premat P'
    }{
        \res{ x }   P \premat  \res{ x }  P'
    }
\end{mathpar}
Intuitively, $P \premat Q$ says that $P$ has at least as many branches as $Q$.
Translation correctness up to $\premat$ thus means that $\redone$ is ``{too eager}'', as it \emph{prematurely commits} to branches.

\appref{ch4a:tight} proves that the translation under the eager semantics satisfies such   criteria.
Before discussing the corresponding completeness and soundness results, we present an example.

\begin{example}{}
\label{ch4ex:looseTight}
    To contrast commitment in eager and lazy semantics (and their effect on the translation's correctness), recall from \Cref{ch4ex:lambdaRed} the term $M$~\eqref{ch4eq:lin_cons_sub2}:
    \[
        M = \big(x_1\ \bag{x_2\ \bag{x_3\ \oneb}}\big) \linexsub{ \bag{\fail^{\emptyset} , y , I} / x_1,x_2,x_3 }
    \]
    \Cref{ch4f:eager_v_lazy} recalls the three branching reductions from $M$ to $N_1$, $N_2$ and $N_3$.
    It also depicts a side-by-side comparison of the reductions of   $\piencodfaplas{M}_u$ under the lazy ($\redtwo$) and eager ($\redone$) semantics.
    In the figure, $\redtwo^\ast$ and $\redone^\ast$ denote the reflexive, transitive closures of $\redtwo$ and $\redone$, respectively.

    Under \redtwo there are three different reduction paths, each resulting directly in the translation of one of $N_1,N_2,N_3$: after the first choice, the following choices are preserved.
    In contrast, under \redone  there are six different reduction paths, each resulting in a process that relates to the translation of one of $N_1,N_2,N_3$ through $\premat$: after the first choice for an item from the bag is made, the semantics commits to choices for the other items.
\end{example}

The correctness properties induced by $\redone$ are \emph{loose} (rather than \emph{tight}, as in \secref{ch4s:trans}; cf.\ \Cref{ch4d:encCriteria}):

\begin{restatable}[Loose Completeness (Under $\redone$)]{theorem}{thmEncLWCompl}\label{ch4thm:opcompletenessweak}
    If $ {N}\red {M}$ for a well-formed closed $\lamcoldetsh$-term $N$, then there exists $Q$ such that $\piencodfaplas{{N}}_u \redone^* Q$ and $\piencodfaplas{{M}}_u \premat Q $.
\end{restatable}

\begin{proof}[Proof (Sketch)]
    By induction on reductions.
    See \secref{ch4a:looscompleteness} for details.
\end{proof}

\begin{restatable}[Loose Weak Soundness (Under $\redone$)]{theorem}{thmEncLWSound}\label{ch4thm:opsoundweak}
    If  $\piencodfaplas{{N}}_u \redone^* Q$ for a well-formed closed $\lamcoldetsh$-term $N$, then there exist ${N}'$ and $Q'$ such that
    (i)~${N} \redone^* {N}'$
    and
    (ii)~$Q \redone^* Q' $ with
    $ \piencodfaplas{{N}'}_u \premat Q'$.
\end{restatable}

While \redtwo reduces multiple branches of a choice in lockstep, \redone  reduces only one branch and discards the rest.
Accordingly, weak soundness under \redtwo (\Cref{ch4t:soundnesstwounres}) relates a sequence of lazy reductions to a sequence of   reductions in \lamcoldetshlin.
In contrast, \Cref{ch4thm:opsoundweak} is weaker: it relates a sequence of eager reductions to a subset of branches, as some branches may have been eagerly discarded.
Hence, the proof of \Cref{ch4thm:opsoundweak} has the added complexity of showing that every branch that is eagerly discarded must also be precongruent to a source reduction. This makes it difficult to apply induction directly, as we do not know which branches have been discarded in $\clpi$.

More in details, to prove \Cref{ch4thm:opsoundweak}, we first show that $\premat$ is stable under reductions.
We need a way of denoting all possible reductions from a process.
We define $P \redone \{P_i\}_{i \in I}$, for a fixed (maximal) finite set $I = \{ i \mid P \redone P_i \}$.
Similarly, we define $P \redone^* \{P_i\}_{i \in I}$ inductively: if $P \redone^* \{P_i\}_{i \in I} $ and $P_i \redone \{P_j\}_{j \in J_i}$ for each $i \in I$, then $P \redone^* \{P_j\}_{j \in J} $ with $J = \bigcup_{ i \in I} J_i  $.
We then have the following:
\begin{restatable}{proposition}{propSoundextra}\label{ch4prop:soundextra}
    If $P \premat Q$ and $P \redone^* \{P_i\}_{i \in I}$,
    then there exist $J$ and $ \{Q_j \}_{j \in J}$ such that $Q \redone^* \{Q_j \}_{j \in J} $, $ J \subseteq I $, and for each $j \in J$, $P_j \premat Q_j$.
\end{restatable}

\begin{proof}[Proof (Sketch)]
   {By induction on the derivation rules of the precongruence $\premat$.
    See \secref{ch4a:loossoundness} for details.}
\end{proof}

Then, using \Cref{ch4prop:soundextra} we prove the following:
\begin{restatable}{lemma}{thmOpsoundone}\label{ch4thm:opsoundone}
    Let ${N}$ be a well-formed closed term.
    If  $\piencodfaplas{{N}}_u \redone^* Q$, then there exist ${N}'$ and $\{Q_i\}_{i \in I}$ such that
    (i)~${N} \red^* {N}'$
    and
    (ii)~$Q \redone^* \{Q_i\}_{i \in I}$ where for each $j \in I$, $ \piencodfaplas{{N}'}_u \premat Q_j$.
\end{restatable}

 \Cref{ch4thm:opsoundone} ensures that the translation does not add behaviors not present in the source term. \Cref{ch4thm:opsoundweak} then follows directly from \Cref{ch4thm:opsoundone} by taking an arbitrary $Q_i$.

\paragraph*{Lazy vs Eager, Part II: Behavioral Equivalence.}
One may ask if the differences between \redtwo and \redone are confined to the ability to correctly translate \lamcoldetshlin, or, relatedly, whether \lamcoldetshlin's formulation is responsible for these differences.

We now compare \redtwo and \redone  \emph{independently from \lamcoldetshlin} by resorting to \emph{behavioral equivalences}.
We define a simple behavioral notion of equivalence on \clpi processes, parametric in \redtwo or \redone; then, we prove that there are classes of processes that are equal with respect to \redtwo, but incomparable with respect to \redone (\Cref{ch4t:piBisim}).
A key ingredient is the following notion of observable on processes:

\begin{definition}{}\label{ch4d:readyPrefix}
    A process $P$ has a \emph{ready-prefix} $\alpha$, denoted $P \readyPrefix{\alpha}$, if and only if there exist $\pctx{N},P'$ such that $P \equiv \pctx{N}[\alpha; P']$.
\end{definition}

We may now define:

\begin{definition}{Ready-Prefix Bisimilarity} \label{ch4d:readyPrefixBisim}
    A relation $\mathbb{B}$ on \clpi processes is a \emph{(strong) ready-prefix bisimulation with respect to \redtwo} if and only if, for every $(P,Q) \in \mathbb{B}$,
    \begin{enumerate}
        \item
            For every $P'$ such that $P \redtwo P'$, there exists $Q'$ such that $Q \redtwo Q'$ and $(P',Q') \in \mathbb{B}$;

        \item
            For every $Q'$ such that $Q \redtwo Q'$, there exists $P'$ such that $P \redtwo P'$ and $(P',Q') \in \mathbb{B}$;

        \item
            For every $\alpha \relalpha \beta$, $P \readyPrefix{\alpha}$ if and only if $Q \readyPrefix{\beta}$.
    \end{enumerate}
    $P$ and $Q$ are \emph{ready-prefix bisimilar with respect to \redtwo}, denoted $P \readyPrefixBisim{L} Q$, if there exists a relation $\mathbb{B}$ that is a
    ready-prefix bisimulation with respect to \redtwo such that $(P,Q) \in \mathbb{B}$.

    A \emph{(strong) ready-prefix bisimulation with respect to \redone} is defined by replacing every occurrence of `$\redtwo$' by `$\redone$' in the definition above.
    We write $P \readyPrefixBisim{E} Q$ if $P$ and $Q$ are \emph{ready-prefix bisimilar with respect to \redone}.
\end{definition}

Ready-prefix bisimulation can highlight a subtle but significant difference between the behavior induced by our lazy and eager semantics.
To demonstrate this, we consider session-typed implementations of a vending machine.
\newcommand{\myMachine}{VM}
\newcommand{\myInterface}{IF}

\begin{example}{Two Vending Machines}\label{ch4ex:prefixReadyBisim}
    Consider vending machines $\sff{\myMachine}_1$ and $\sff{\myMachine}_2$ consisting of three parts:
    (1)~an interface, which interacts with the user to send money and choose between coffee ($\sff{c}$) and tea ($\sff{t}$);
    (2)~a brewer, which produces either beverage;
    (3)~a system, which collects the money and forwards the user's choice to the brewer.
    A \clpi specification follows (below $\textup{€}$ and $\textup{€}2$ stand for names):
    \begin{align*}
        \sff{\myMachine}_1 &:= \res{x} \big( \sff{\myInterface}_1 \| \res{y} ( \sff{Brewer} \| \sff{System} ) \big)
        \\ \displaybreak[1]
        \sff{\myMachine}_2 &:= \res{x} \big( \sff{\myInterface}_2 \| \res{y} ( \sff{Brewer} \| \sff{System} ) \big)
            \\ \displaybreak[1]
        \sff{\myInterface}_1 &:= \pname{x}{\textup{€}2}; \big( \pclose{\textup{€}2} \| ( \psel{x}{\sff{c}}; \pclose{x} \nd \psel{x}{\sff{t}}; \pclose{x} ) \big)
        \\ \displaybreak[1]
        \sff{\myInterface}_2 &:= \pname{x}{\textup{€}2}; ( \pclose{\textup{€}2} \| \psel{x}{\sff{c}}; \pclose{x} ) \nd \pname{x}{\textup{€}2}; ( \pclose{\textup{€}2} \| \psel{x}{\sff{t}}; \pclose{x} )
                    \\ \displaybreak[1]
        \sff{System} &:= \gname{x}{\!\textup{€}}; \gsel{x} \!\left\{\!
            \begin{array}{@{}l@{}}
                \sff{c} : \psel{y}{\sff{c}}; \gclose{x}; \gclose{\textup{€}}; \pclose{y},
                \\
                \sff{t} : \psel{y}{\sff{t}}; \gclose{x}; \gclose{\textup{€}}; \pclose{y}
            \end{array}
        \!\right\}
        \\ \displaybreak[1]
        \sff{Brewer} &:= \gsel{y} \{ \sff{c} : \gclose{y}; \sff{Brew}_{\sff{c}}, \sff{t} : \gclose{y}; \sff{Brew}_{\sff{t}} \}
    \end{align*}
    where $\sff{Brew}_{\sff{c}} \vdash \emptyset$, $\sff{Brew}_{\sff{t}} \vdash \emptyset$, such that $\sff{\myMachine}_1 \vdash \emptyset$, $\sff{\myMachine}_2 \vdash \emptyset$.

    We give two implementations of the interface:
    $\sff{\myInterface}_1$ sends the money and then chooses coffee or tea;
    $\sff{\myInterface}_2$ chooses sending the money and then requesting coffee, or sending the money and then requesting tea.
    Then, $\sff{\myInterface}_1$ and $\sff{\myInterface}_2$ result in two different vending machines, $\sff{\myMachine}_1$ and $\sff{\myMachine}_2$.

    We have $\sff{\myMachine}_1 \nreadyPrefixBisim{E} \sff{\myMachine}_2$: the eager semantics distinguishes between the implementations; e.g., $\sff{\myInterface}_1$ has a single money slot, a button for coffee, and another button for tea, whereas $\sff{\myInterface}_2$ has two money slots, one for coffee, and another for tea.
    In contrast, under the lazy semantics, these machines are indistinguishable:
    $\sff{\myMachine}_1 \readyPrefixBisim{L} \sff{\myMachine}_2$.
\end{example}

\Cref{ch4ex:prefixReadyBisim} highlights a difference in behavior between \redtwo and \redone when a moment of choice is subtly altered.
The following theorem captures this distinction (see \secref{ch4a:piBisim}):

\begin{restatable}{theorem}{thmPiBisim}\label{ch4t:piBisim}
    Take $R \equiv \pctx{N}[\alpha_1; (P \nd Q)] \vdash \emptyset$ and $S \equiv \pctx{N}[\alpha_2; P \nd \alpha_3; Q] \vdash \emptyset$, where $\alpha_1 \relalpha \alpha_2 \relalpha \alpha_3$ and $\alpha_1,\alpha_2,\alpha_3$ require a continuation.
    Suppose that $P \nreadyPrefixBisim{L} Q$ and $P \nreadyPrefixBisim{E} Q$.
    Then (i)~$R \readyPrefixBisim{L} S$ but (ii)~$R \nreadyPrefixBisim{E} S$.
\end{restatable}

Processes~\eqref{ch4ex:sound1} and~\eqref{ch4ex:sound2} from \secref{ch4s:trans} (Page~\pageref{ch4ex:sound1}) provide another example of a change in the moment of choice, different from the one discussed above.
In~\eqref{ch4ex:sound1}, the choice depends on $\alpha_1$ and $\alpha_2$.
In~\eqref{ch4ex:sound2}, the choice does not depend on $\alpha_1$ or $\alpha_2$, but  on choices made in the context in which the process resides.
Though the choice is subtly changed between~\eqref{ch4ex:sound1} and~\eqref{ch4ex:sound2}, the impact is significant: these processes are not ready-prefix bisimilar, with respect to neither semantics.
This is because, under both lazy and eager semantics, in~\eqref{ch4ex:sound1} the two branches evolve in lockstep, whereas in~\eqref{ch4ex:sound2} they evolve independently.


\section{Beyond Linear Resources}
Our results extend to  the language  $\lamcoldetsh$, an extension of $\lamcoldetshlin$ with a more general bag which includes unrestricted resources: resources that may be used zero or more times.
\Cref{ch4f:lambda} gives the syntax of \lamcoldetsh-terms, bags and contexts.

A key difference with \lamcoldetshlin is that variables $x,y,z,\ldots $ have linear and unrestricted occurrences.
Notation $x[\mtt{l}]$ denotes a \emph{linear} occurrence of $x$; we often omit the annotation `$[\mtt{l}]$', and a sequence $\widetilde{x}$ always involves linear occurrences.
Notation~$x[i]$ denotes an \emph{unrestricted} occurrence of $x$, explicitly referencing the $i$-th element of an unrestricted (ordered) bag.
The structure of a bag is now split into a linear and an unrestricted component: as in $\lamcoldetshlin$, linear resources in bags cannot be duplicated, but unrestricted resources are always duplicated when consumed.
The empty unrestricted bag is denoted $\unvar{\oneb}$.
Notation $U_i$ denotes the singleton bag at the $i$-th position in $U$; if there is no $i$-th position in $U$, then $U_i$ defaults to $\unvar{\oneb}$.
We use `$\bagsep$' to combine a linear and an unrestricted bag, and unrestricted bags are joined via the non-commutative `$\concat$'.

To account for the explicit distinction between linear and unrestricted occurrences of variables, we now have two forms of explicit substitution:
\begin{itemize}
    \item $M \linexsub{C /  x_1 , \ldots , x_k}$, a linear substitution as in $\lamcoldetshlin$;
    \item $M \unexsub{U / \unvar{x}}$,  an unrestricted substitution   of an unrestricted  bag $U$ for an unrestricted variable $x^!$ in  $M$, with the assumption that $x^!$ does not appear in another unrestricted substitution in $M$.
\end{itemize}

\begin{figure}[t] \mysmall
        \begin{align*}
            M,N,L ::=~
            & x[*] && \text{variable}
            && \quad
            {}\sepr M[\widetilde{x} \leftarrow x] && \text{sharing}
            \\
            \sepr~
            & \lambda x.M && \text{abstraction}
            && \quad
            {}\sepr \fail^{\tilde{x}} && \text{failure}
            \\
            \sepr~
            & (M\ B) && \text{application}
            && \quad
            {}\sepr M \linexsub{C/x_1,\ldots,x_k} && \text{linear substitution}
            \\
            \sepr~
            & M \esubst{B}{x} && \text{intermediate substitution}
            \span\span\span\span
            \\
            \sepr~
            & M \unexsub{U/\lunvar{x}} && \text{unrestricted substitution}
            \span\span\span\span 
            \\
            [*] ::=~
            & [\mtt{l}]
            \sepr
            [i] ~~ i \in \mathbb{N} && \text{annotations}
            && \quad\hphantom{{}\sepr{}}
            A,B ::=~
            C \bagsep U && \text{bag}
            \\
            U,V ::=~
            & \unvar{\oneb} \sepr {\unvar{\bag{M}}} \sepr {U \concat V} && \text{unrestricted bag}
            && \quad\hphantom{{}\sepr{}}
            C,D ::= \oneb \sepr \bag{M} \cdot\, C && \text{linear bag}
            \\
            \lctx{C} ::=~
            & \hole \sepr (\lctx{C}\ B) \sepr \lctx{C} \linexsub{C/\widetilde{x}} \sepr \lctx{C} \unexsub{U/\lunvar{x}} \sepr \lctx{C}[\widetilde{x} \leftarrow x]
            \span\span\span\span\span
            & \text{context}
        \end{align*}
    \caption{Syntax of $\lamcoldetsh$.}
    \label{ch4f:lambda}
\end{figure}


The reduction rules for \lamcoldetsh, extend the rules given in \Cref{ch4f:lambda_redlin} with some modifications to accomodate the two-component format of bags and dedicated rules for the new constructs. Here, we describe the most interesting new rules: Rule~$\redlab{RS{:}Ex \dash Sub}$ for explicit substitution, and also Rules~$\redlab{R:Fetch^!}$ and $\redlab{R:Fail^!}$ for unrestricted substitution.

\begin{mathpar}
      \inferrule[$\redlab{RS{:}Ex \dash Sub}$]{
                \size{C} = |\widetilde{x}|
                \\
                M \not= \fail^{\tilde{y}}
            }{
                (M\sharing{\widetilde{x}}{x}) \esubst{ C \bagsep U }{ x } \red  M \linexsub{C  /  \widetilde{x}} \unexsub{U / \lunvar{x} }
            }
\and
  \inferrule[$\redlab{RS{:} Fetch^!}$]{
                \headf{M} = {x}[i]
                \\
                U_i = \unvar{\bag{N}}
            }{
                M \unexsub{U / \lunvar{x}} \red  M \headlin{ N /{x}[i] }\unexsub{U / \lunvar{x}}
            }
            \and
            \inferrule[$\redlab{RS{:}Fail^!}$]{
                \headf{M} = {x}[i]
                \\
                U_i = \unvar{\oneb}
            }{
                M \unexsub{U / \lunvar{x} } \red M \headlin{ \fail^{\emptyset} /{x}[i] } \unexsub{U / \lunvar{x} }
            }
\end{mathpar}
An explicit substitution $(M\sharing{\widetilde{x}}{x}) \esubst{ C \bagsep U }{ x }$ reduces to a term in which the linear and unrestricted parts of the bag are separated into their own explicit substitutions $M\linexsub{C  /  \widetilde{x}} \unexsub{U / \lunvar{x} }$, if successful. The fetching of linear/unrestricted resources from their corresponding bags is done by the appropriated fetch rules.
The reduction of an unrestricted substitution $M \unexsub{U/\lunvar{x}}$, where the head variable of $M$ is $x[i]$, depends on $U_i$:
\begin{itemize}
    \item
        If $U_i = \unvar{\bag{N}}$, then the term reduces via Rule~$\redlab{R:Fetch^!}$ by substituting the head occurrence of $x[i]$ in $M$ with $N$, denoted $M\headlin{N/x[i]}$; note that $U_i$ remains available after this reduction.

    \item
        If $U_i = \unvar{\oneb}$, the head variable is instead substituted with failure via Rule~$\redlab{R:Fail^!}$.
\end{itemize}
The definition of $\head{M}$ is as in \Cref{ch4f:lambda_redlin} (bottom), extended with $\head{x[i]} = x[i]$ and $\head{M \unexsub{U/x}} = \head{M}$. The complete set of rules is in\appref{ch4ss:lambdaUnresBags}.


\paragraph{Well-typedness and well-formedness.}

Types for $\lamcoldetsh$ extend the types for \lamcoldetshlin in \Cref{ch4sec:lamTypes} with:
\begin{align*}
    \sigma, \tau, \delta ::=~ &
    \unit \sepr \arrt{ (\pi , \eta) }{\sigma}
\end{align*}
\begin{align*}
    \eta, \epsilon  &::=
    \sigma  \sepr \epsilon \concat \eta
    \qquad \text{list}
    &
    ( \pi , \eta)
    \qquad \text{tuple}
\end{align*}

The list type $\epsilon\concat \eta$ types the concatenation of unrestricted bags.
It can be recursively unfolded into a finite composition of strict types $\sigma_1\concat \ldots\concat \sigma_n$, for some $n\geq 1$, with length $n$ and $\sigma_i$ its $i$-th strict type ($1\leq i\leq n$).
We  write $\unvar{x}:\eta $ to denote for $x[1]:\eta_1 , \ldots , x[k]:\eta_k$ where $\eta$ has length $k$.
The tuple type $(\pi,\eta)$ types concatenation of a linear bag of type $\pi$ with an unrestricted bag of type $\eta$. Finally strict types are amended to allow for unrestricted functional types which go from tuple types to strict types \arrt{ (\pi , \eta) }{\sigma} rather then multiset types to strict types.

We separate contexts into two parts: linear ($\Gamma, \Delta,\ldots$) and unrestricted ($\Theta,\Upsilon,\ldots$):
\begin{align*}
    \Gamma,\Delta &::= \dash \sep \Gamma , x:\pi  \sepr \Gamma, x:\sigma
    &
    \Theta,\Upsilon &::= \dash \sepr \Theta, x^!:\eta
\end{align*}
Both linear and unrestricted occurrences of variables may occur at most once in a context.
Judgments have the form $\Theta;\Gamma\wfdash M:\tau$.
We write $\wfdash M:\tau$ to denote $\dash;\dash\wfdash M:\tau$.

Well-formedness rules for $\lamcoldetsh$ extend the rules given in \Cref{ch4fig:wfsh_ruleslin} with specific rules to handle unrestricted resources, among those, we select  Rules \redlab{FS:var^!}, \redlab{FS{:}abs\dash sh}, \redlab{FS{:}Esub^!} and
\redlab{FS{:} bag^{!}}, for typing unrestricted occurrences of a variable, abstraction of a sharing term, explicit substitution and unrestricted bags, respectively.

\begin{mathpar}
   \inferrule[$\redlab{FS{:}var^!}$]{
                \Theta , x^!: \eta;  {x}: \eta_i , \Delta \wfdash  {x} : \sigma
            }{
                \Theta ,  x^!: \eta; \Delta \wfdash {x}[i] : \sigma
            }
            \and
             \inferrule[$\redlab{FS{:}abs\dash sh}$]{
                \Theta , x^!:\eta ; \Gamma ,  {x}: \sigma^k \wfdash M\sharing{\widetilde{x}}{x} : \tau
                \\
                {x} \notin \dom{\Gamma}
            }{
                \Theta ; \Gamma \wfdash \lambda x . (M[ {\widetilde{x}} \leftarrow  {x}])  : (\sigma^k, \eta )  \rightarrow \tau
            }
               \and
            \inferrule[$\redlab{FS{:}Esub^!}$]{
                \Theta , x^! {:} \eta; \Gamma  \wfdash M : \tau
                \\
                \Theta ; \dash \wfdash U : \epsilon
                \\
                \eta \relunbag \epsilon
            }{
                \Theta ; \Gamma \wfdash M \unexsub{U / \unvar{x}}  : \tau
            }
              \and
            \inferrule[$\redlab{FS{:} bag^{!}}$]{
                \Theta ; \dash \wfdash U : \epsilon
                \\
                \Theta ; \dash \wfdash V : \eta
            }{
                \Theta ; \dash  \wfdash U \concat V :\epsilon \concat \eta
            }
\end{mathpar}

In Rule \redlab{FS{:}Esub!},  $\eta \relunbag \epsilon$ denotes the fact that  $\epsilon=\epsilon_1\concat \ldots \concat \epsilon_k\concat \ldots \concat \epsilon_n$ and $\eta=\eta_1\concat \ldots \concat \eta_k$ are two list types, with $n\geq k$, such that the following hold:   for all $i$, $1\leq i \leq k$, $\epsilon_i = \eta_i $. The complete set of well-formedness rules for $\lamcoldetsh$ is in\appref{ch4ss:lambdaUnresBags}.


\paragraph{Extended translation.}
To extend the translation
in \figref{ch4fig:encodinglin} to \lamcoldet:
The access and use of unrestricted resources in \lamcoldet is codified in   \clpi by combining labeled choices and clients/servers (\appref{ch4ss:transUnres}).

Note: for the sake of generality the proofs in the appendices concern \lamcoldet.


\section{Extensions in Detail: Eager Semantics and Unrestricted Resources}
\label{ch4s:fullAccounts}

\subsection{\texorpdfstring{$\lamcoldetsh$}{Unrestricted Lambda}:  An Extension of \texorpdfstring{\lamcoldetshlin}{Lambda}  with Unrestricted Bags}
\label{ch4ss:lambdaUnresBags}

\Cref{ch4f:lambda_llfv} defines the free linear variables of a term, denoted $\llfv{M}$.
\Cref{ch4f:lambda_red} gives the reduction rules for \lamcoldetsh, extending the rules given in \Cref{ch4f:lambda_redlin} with Rules~$\redlab{R:Fetch^!}$ and $\redlab{R:Fail^!}$ for unrestricted substitution.
%

\begin{figure}[t] 
        \begin{align*}
            \llfv{x}&=\{x\}
            & \llfv{(M\ B)} &= \llfv{M}\cup \llfv{B}\\
            \llfv{x[i]}&=\emptyset
           & \llfv{M\sharing{\widetilde{x}}{x}}&=(\llfv{M}\setminus\{\widetilde{x}\})\cup \{x\} \\
            \llfv{\oneb}&=\emptyset
            & \llfv{\lambda x. (M\sharing{\widetilde{x}}{x})}&=\llfv{M\sharing{\widetilde{x}}{x}}\setminus \{x\}\\
            \llfv{\bag{M}}&=\llfv{M}
            & \llfv{M\linexsub{C/x_1,\ldots, x_k}}&=(\llfv{M}\setminus \{x_1,\ldots, x_k\})\cup \llfv{C}\\
            \llfv{C\bagsep  U}&=\llfv{C}
            & \llfv{ M\esubst{ B }{ x }}&= (\llfv{M}\setminus \{x\})\cup\llfv{B}\\
            \llfv{\fail^{\widetilde{x}}}&=\{\widetilde{x}\}
            & \llfv{\bag{M}\cdot C}&=\llfv{M}\cup \llfv{C}
        \end{align*}
    In the case $\llfv{M}=\emptyset$, the term $M$ is called \emph{linearly closed}.
    \newline
    \caption{Free Linear Variables}
    \label{ch4f:lambda_llfv}
\end{figure}

\begin{figure}[t] \mysmall
        \begin{mathpar}
            \inferrule[$\redlab{RS{:}Beta}$]{ }{
                (\lambda x . M)\ B  \red M \esubst{ B }{ x }
            }
            \and
            \inferrule[$\redlab{RS{:}Ex \dash Sub}$]{
                \size{C} = |\widetilde{x}|
                \\
                M \not= \fail^{\tilde{y}}
            }{
                (M\sharing{\widetilde{x}}{x}) \esubst{ C \bagsep U }{ x } \red  M \linexsub{C  /  \widetilde{x}} \unexsub{U / \lunvar{x} }
            }
            \and
            \inferrule[$\redlab{RS{:}Fetch^{\ell}}$]{
                \headf{M} =  {x}_j
                \\
                0 < i \leq \size{C}
            }{
                M \linexsub{C /  \widetilde{x}, x_j} \red  (M \headlin{ C_i / x_j })  \linexsub{(C \setminus C_i ) /  \widetilde{x}  }
            }
            \and
            \inferrule[$\redlab{RS{:}Fail^{\ell}}$]{
                \size{C} \neq |\widetilde{x}|
                \\
                \widetilde{y} = (\llfv{M} \setminus \{  \widetilde{x}\} ) \cup \llfv{C}
            }{
                (M\sharing{\widetilde{x}}{x}) \esubst{C \bagsep U}{ x }  \red  \fail^{\widetilde{y}}
            }
            \and
            \inferrule[$\redlab{RS{:} Fetch^!}$]{
                \headf{M} = {x}[i]
                \\
                U_i = \unvar{\bag{N}}
            }{
                M \unexsub{U / \lunvar{x}} \red  M \headlin{ N /{x}[i] }\unexsub{U / \lunvar{x}}
            }
            \and
            \inferrule[$\redlab{RS{:}Fail^!}$]{
                \headf{M} = {x}[i]
                \\
                U_i = \unvar{\oneb}
            }{
                M \unexsub{U / \lunvar{x} } \red M \headlin{ \fail^{\emptyset} /{x}[i] } \unexsub{U / \lunvar{x} }
            }
            \and
            \inferrule[$\redlab{RS{:}Cons_1}$]{
                \widetilde{y} = \llfv{C}
            }{
                \fail^{\widetilde{x}}\ (C \bagsep U) \red  \fail^{\widetilde{x} \ \widetilde{y}}
            }
            \and
            \inferrule[$\redlab{RS{:}Cons_2}$]{
                \size{C} =   |  {\widetilde{x}} |
                \\
                \widetilde{z} = \llfv{C}
            }{
                (\fail^{ {\widetilde{x}} \cup \widetilde{y}} \sharing{\widetilde{x}}{x}) \esubst{ C \bagsep U }{ x }  \red  \fail^{\widetilde{y} \cup \widetilde{z}}
            }
            \and
            \inferrule[$\redlab{RS{:}Cons_3}$]{
                \widetilde{z} = \llfv{C}
            }{
                \fail^{\widetilde{y}\cup \widetilde{x}} \linexsub{C /  \widetilde{x}} \red  \fail^{\widetilde{y} \cup \widetilde{z}}
            }
            \and
            \inferrule[$\redlab{RS{:}Cons_4}$]{ }{
                \fail^{\widetilde{y}} \unexsub{U / \lunvar{x}}  \red  \fail^{\widetilde{y}}
            }
            \and
            \inferrule[$\redlab{RS:TCont}$]{
                M \red    N
            }{
                \lctx{C}[M] \red   \lctx{C}[N]
            }
        \end{mathpar}
    \caption{Reduction rules for \texorpdfstring{\lamcoldetsh}{lambda}.}
    \label{ch4fig:reduc_interm}\label{ch4f:lambda_red}
\end{figure}

\paragraph{Well-typedness and well-formedness.}

Types for $\lamcoldetsh$ extend the types for \lamcoldetshlin in \Cref{ch4sec:lamTypes} with:
\begin{align*}
    \sigma, \tau, \delta ::=~ &
    \unit \sepr \arrt{ (\pi , \eta) }{\sigma}
\end{align*}
\begin{align*}
    \eta, \epsilon  &::=
    \sigma  \sepr \epsilon \concat \eta
    \qquad \text{list}
    &
    ( \pi , \eta)
    \qquad \text{tuple}
\end{align*}

The list type $\epsilon\concat \eta$ types the concatenation of unrestricted bags.
It can be recursively unfolded into a finite composition of strict types $\sigma_1\concat \ldots\concat \sigma_n$, for some $n\geq 1$, with length $n$ and $\sigma_i$ its $i$-th strict type ($1\leq i\leq n$).
We  write $\unvar{x}:\eta $ to denote for $x[1]:\eta_1 , \ldots , x[k]:\eta_k$ where $\eta$ has length $k$.
The tuple type $(\pi,\eta)$ types concatenation of a linear bag of type $\pi$ with an unrestricted bag of type $\eta$. Finally strict types are amended to allow for unrestricted functional types which go from tuple types to strict types \arrt{ (\pi , \eta) }{\sigma} rather then multiset types to strict types.

\begin{definition}{$\eta \relunbag \epsilon$}\label{ch4not:ltypes}
    Let $\epsilon$ and $\eta$ be two list types, with the length of $\epsilon$ greater or equal to that of $\eta$.
    We say that $\epsilon$ \emph{embraces} $\eta$, denoted $\eta \relunbag \epsilon$, whenever there exist $  \epsilon'$ and $ \epsilon''$   such that: i) $ \epsilon = \epsilon' \concat \epsilon''$; ii)  the size of $\epsilon' $ is that of $\eta$; iii)  for all $i$, $\epsilon'_i = \eta_i $.

\end{definition}

We separate contexts into two parts: linear ($\Gamma, \Delta,\ldots$) and unrestricted ($\Theta,\Upsilon,\ldots$):
\begin{align*}
    \Gamma,\Delta &::= \dash \sep \Gamma , x:\pi  \sepr \Gamma, x:\sigma
    &
    \Theta,\Upsilon &::= \dash \sepr \Theta, x^!:\eta
\end{align*}
Both linear and unrestricted occurrences of variables may occur at most once in a context.

Judgments have the form $\Theta;\Gamma\wfdash M:\tau$.
We write $\wfdash M:\tau$ to denote $\dash;\dash\wfdash M:\tau$.

\begin{definition}{Well-formedness in $\lamcoldetsh$}
    A $\lamcoldetsh$-term $M$ is \emph{well-formed} if there exists a context $\Theta$ and $\Gamma$ and a  type  $\tau$ such that the rules in \Cref{ch4fig:wfsh_rulesunres} entail $\Theta ; \Gamma \wfdash  M : \tau $.
\end{definition}

\begin{figure}[ht]\mysmall
        \begin{mathpar}
            \inferrule[$\redlab{FS{:}var^{\ell}}$]{ }{
                \Theta ;  {x}: \sigma \wfdash  {x} : \sigma
            }
            \and
            \inferrule[$\redlab{FS{:}var^!}$]{
                \Theta , x^!: \eta;  {x}: \eta_i , \Delta \wfdash  {x} : \sigma
            }{
                \Theta ,  x^!: \eta; \Delta \wfdash {x}[i] : \sigma
            }
            \and
            \inferrule[$\redlab{FS{:}\oneb^{\ell}}$]{ }{
                \Theta ; \dash \wfdash \oneb : \omega
            }
            \and
            \inferrule[$\redlab{FS{:}bag^{\ell}}$]{
                \Theta ; \Gamma \wfdash M : \sigma
                \\
                \Theta ; \Delta \wfdash C : \sigma^k
            }{
                \Theta ; \Gamma , \Delta \wfdash \bag{M}\cdot C:\sigma^{k+1}
            }
            \and
            \inferrule[$\redlab{FS{:}\oneb^!}$]{ }{
                \Theta ;  \dash  \wfdash \unvar{\oneb} : \sigma
            }
            \and
            \inferrule[$\redlab{FS{:} bag^{!}}$]{
                \Theta ; \dash \wfdash U : \epsilon
                \\
                \Theta ; \dash \wfdash V : \eta
            }{
                \Theta ; \dash  \wfdash U \concat V :\epsilon \concat \eta
            }
            \and
            \inferrule[$\redlab{FS{:}bag}$]{
                \Theta ; \Gamma\wfdash C : \sigma^k
                \\
                \Theta ;\dash \wfdash  U : \eta
            }{
                \Theta ; \Gamma \wfdash C \bagsep U : (\sigma^{k} , \eta  )
            }
            \and
            \inferrule[$\redlab{FS{:}fail}$]{
                \dom{\core{\Gamma}} = \widetilde{x}
            }{
                \Theta ; \core{\Gamma} \wfdash  \fail^{\widetilde{x}} : \tau
            }
            \and
            \inferrule[$\redlab{FS{:}weak}$]{
                \Theta ; \Gamma  \wfdash M : \tau
            }{
                \Theta ; \Gamma ,  {x}: \omega \wfdash M\sharing{}{x}: \tau
            }
            \and
            \inferrule[$\redlab{FS{:}shar}$]{
                \Theta ;  \Gamma ,  {x}_1: \sigma, \cdots,  {x}_k: \sigma \wfdash M : \tau
                \\
                {x}\notin \dom{\Gamma}
                \\
                k \not = 0
            }{
                \Theta ;  \Gamma ,  {x}: \sigma^{k} \wfdash M\sharing{{x}_1 , \ldots ,  {x}_k}{x}  : \tau
            }
            \and
            \inferrule[$\redlab{FS{:}abs\dash sh}$]{
                \Theta , x^!:\eta ; \Gamma ,  {x}: \sigma^k \wfdash M\sharing{\widetilde{x}}{x} : \tau
                \\
                {x} \notin \dom{\Gamma}
            }{
                \Theta ; \Gamma \wfdash \lambda x . (M[ {\widetilde{x}} \leftarrow  {x}])  : (\sigma^k, \eta )  \rightarrow \tau
            }
            \and
            \inferrule[$\redlab{FS{:}app}$]{
                \Theta ;\Gamma \wfdash M : (\sigma^{j} , \eta ) \rightarrow \tau
                \\
                \Theta ;\Delta \wfdash B : (\sigma^{k} , \epsilon )
                \\
                \eta \relunbag \epsilon
            }{
                \Theta ; \Gamma , \Delta \wfdash (M\ B) : \tau
            }
            \and
            \inferrule[$\redlab{FS{:}Esub}$]{
                \Theta , x^! : \eta ; \Gamma ,  {x}: \sigma^{j} \wfdash M[ {\widetilde{x}} \leftarrow  {x}] : \tau
                \\
                \Theta ; \Delta \wfdash B : (\sigma^{k} , \epsilon )
                \\
                \eta \relunbag \epsilon
            }{
                \Theta ; \Gamma , \Delta \wfdash (M[ {\widetilde{x}} \leftarrow  {x}])\esubst{ B }{ x }  : \tau
            }
            \and
            \inferrule[$\redlab{FS{:}Esub^{\ell}}$]{
                \Theta ; \Gamma  ,  x_1:\sigma, \cdots , x_k:\sigma \wfdash M : \tau
                \\
                \Theta ; \Delta \wfdash C : \sigma^k
            }{
                \Theta ; \Gamma , \Delta \wfdash M \linexsub{C /  x_1, \cdots , x_k} : \tau
            }
            \and
            \inferrule[$\redlab{FS{:}Esub^!}$]{
                \Theta , x^! {:} \eta; \Gamma  \wfdash M : \tau
                \\
                \Theta ; \dash \wfdash U : \epsilon
                \\
                \eta \relunbag \epsilon
            }{
                \Theta ; \Gamma \wfdash M \unexsub{U / \unvar{x}}  : \tau
            }
        \end{mathpar}
    \caption{Well-Formedness Rules for $\lamcoldetsh$.}
    \label{ch4fig:wfsh_rulesunres}
\end{figure}

\paragraph{A congruence.}
Some terms, though syntactically different, display the same behavior.
For example, assuming $x \notin \lfv{M}$, this holds for $M \unexsub{U/x^!}$ and $M$: the former describes a substitution that ``does nothing'' and would result in $M$ itself.
This notion is formalized through a \emph{congruence} ($\equivlam$) closed under syntax, given in \Cref{ch4fig:rsPrecongruencefailure}.

\begin{figure}[t]
\small
\begin{mathpar}
    \inferrule{
        \widetilde{x} \disj \lfv{B}
    }{
        (M\ B) \linexsub{C/\widetilde{x}}  \equivlam (M \linexsub{C/\widetilde{x}})\ B
    }
    \and
    \inferrule{
        x \not \in \lfv{M}
    }{
        M \unexsub{U/x^!} \equivlam M
    }
    \and
    \inferrule{
        x \not \in \lfv{B}
    }{
        (M\ B)  \unexsub{U/x^!}  \equivlam (M \unexsub{U/x^!})\ B
    }
    \and
    \inferrule{
        \widetilde{x} \disj \lfv{A}
    }{
        \big((M\ A)\sharing{\widetilde{x}}{x}\big) \esubst{B}{x} \equivlam \big((M\sharing{\widetilde{x}}{x}) \esubst{B}{x}\big)\ A
    }
    \and
    \inferrule{
        \widetilde{x} \disj \lfv{A}
        \\
        \widetilde{y} \disj \lfv{B}
    }{
        \Big(\big((M\sharing{\widetilde{y}}{y}) \esubst{A}{y}\big)\sharing{\widetilde{x}}{x}\Big) \esubst{B}{x} \equivlam \Big(\big(M\sharing{\widetilde{y}}{y}) \esubst{B}{y}\big)\sharing{\widetilde{x}}{x}\Big) \esubst{A}{x}
    }
    \and
    \inferrule{
        \widetilde{x} \disj \lfv{B}
        \\
        \widetilde{y} \disj \lfv{C}
    }{
        \big((M\sharing{\widetilde{y}}{y}) \esubst{B}{y}\big) \linexsub{C/\widetilde{x}}  \equivlam \big((M \linexsub{C/\widetilde{x}})\sharing{\widetilde{y}}{y}\big) \esubst{B}{y}
    }
    \and
    \inferrule{
        x \not \in \lfv{B}
        \\
        \widetilde{y} \disj \lfv{U}
    }{
        \big((M\sharing{\widetilde{y}}{y}) \esubst{B}{y}\big)  \unexsub{U/x^!}  \equivlam \big((M \unexsub{U/x^!})\sharing{\widetilde{y}}{y}\big) \esubst{B}{y}
    }
    \and
    \inferrule{
        \widetilde{x} \disj \lfv{C_2}
        \\
        \widetilde{y} \disj \lfv{C_1}
    }{
        (M \linexsub{C_2/\widetilde{y}}) \linexsub{C_1/\widetilde{x}} \equivlam (M \linexsub{C_1/ \widetilde{x} }) \linexsub{C_2/\widetilde{y}}
    }
    \and
    \inferrule{
        \widetilde{x} \disj \lfv{U}
        \\
        y \not \in \lfv{C}
    }{
        (M\linexsub{C/\widetilde{x}}) \unexsub{U/y^!} \equivlam (M  \unexsub{U/y^!})   \linexsub{C/\widetilde{x}}
    }
    \and
    \inferrule{
        x \not \in \lfv{U}
        \\
        y \not \in \lfv{V}
    }{
        (M\unexsub{V/\widetilde{x}}) \unexsub{U/y^!} \equivlam (M  \unexsub{U/y^!})   \unexsub{V/{x}^!}
    }
\end{mathpar}
\caption{Congruence in \lamcoldetsh.}\label{ch4fig:rsPrecongruencefailure}
\end{figure}

\subsection{Translating \texorpdfstring{\lamcoldetsh}{Unrestricted Lambda} into (Full) \texorpdfstring{\clpi}{Pi}}
\label{ch4ss:transUnres}

\Cref{ch4fig:encodinglin} gives the translation of \lamcoldetshlin into \clpi.
Here we extend this translation to consider the extended calculus \lamcoldetsh.
The key differences are in the translation of  unrestricted variable occurrences, intermediate substitution, abstraction, and the new structure of bags.
\Cref{ch4fig:encoding} gives the translation that maps terms in \lamcoldetsh into processes in full \clpi, denoted $\piencodf{\cdot}_u$.

The translation of an unrestricted variable $x[j]$ first connects to a server along channel $x$ via a request $ \puname{\unvar{x}}{{x_i}}$ followed by a selection on $ \psel{x_i}{j}$.

Process $\piencodfaplas{\lambda x. (M\sharing{\widetilde{x}}{x})}_u$ first confirms its behavior  along $u$, followed by the reception of a channel $x$.
The channel $x$ provides a linear channel $\linvar{x}$ and an unrestricted channel $\unvar{x}$ for dedicated substitutions of the linear and unrestricted bag components.
This separation is also present in the translation of $ \piencodfaplas{ M\esubst{B}{x}}_u $, for the same reason.

Process $\piencodfaplas{M\, (C \bagsep U)_u}$ consists of synchronizations between the translation of $\piencodf{M}_v$ and
$\piencodfaplas{C\bagsep U}_x$:  the translation
of  $C \bagsep U$ evolves when $M$ is an abstraction, say
${\lambda x . (M'\sharing{\widetilde{x}}{x})}$.
The channel $ \linvar{x}$ provides the linear behavior of the bag $C$ while $\unvar{x}$ provides the behavior of $U$; this is done by guarding the translation of $U$ with a server connection, such that every time a channel synchronizes with it a fresh copy of $U$ is spawned.

Process $\piencodfaplas{ M \unexsub{U / \unvar{x}}}_u $ consists of  the composition of the translation of $M$ and a server guarding the translation of $U$: in order for $\piencodfaplas{M}_u$ to gain access to $\piencodfaplas{U}_{x_i}$ it must first synchronize with the server channel $\unvar{x}$ to spawn a fresh copy of the translation of $U$.

\begin{figure}[t] \mysmall
        \begin{align*}
            \piencodfaplas{ {x}}_u & = \psome{x}; \pfwd{x}{u} \hspace{1cm}
            \piencodfaplas{{x}[j]}_u  =   \puname{\unvar{x}}{{x_i}}; \psel{x_i}{j}; \pfwd{x_i}{u}
            \\[1mm]
            \piencodfaplas{\lambda x.M}_u & = \psome{u};\gname{u}{x};  \psome{x};\gname{x}{\linvar{x}}; \gname{x}{\unvar{x}};  \gclose{x} ; \piencodfaplas{M}_u
            \\[1mm]
            \piencodfaplas{ M \esubst{ C \bagsep U }{ x} }_u & =  \res{x}( \psome{x}; \gname{x}{\linvar{x}}; \gname{x}{\unvar{x}};  \gclose{x} ;\piencodfaplas{ M}_u \| \piencodfaplas{ C \bagsep U}_x )
            \\[1mm]
            \piencodfaplas{M (C \bagsep U)}_u & = \res{v} (\piencodfaplas{M}_v \| \gsome{v}{u , \llfv{C}};\pname{v}{x}; ( \piencodfaplas{C \bagsep U}_x   \| \pfwd{v}{u}  ) )
            \\[1mm]
            \piencodfaplas{ C \bagsep U }_x & = \gsome{x}{\llfv{C}};  \pname{x}{\linvar{x}}; \big( \piencodfaplas{ C }_{\linvar{x}} \|  \pname{x}{\unvar{x}}; ( \guname{\unvar{x}}{x_i}; \piencodfaplas{ U }_{x_i}  \| \pclose{x} ) \big)
            \\[1mm]
            \piencodfaplas{\bag{M_j} \cdot~ C}_{\linvar{x}}  &=
            \begin{array}{@{}l@{}}
                \gsome{\linvar{x}}{\llfv{C} }; \gname{x}{y_i}; \gsome{\linvar{x}}{y_i, \llfv{C}}; \psome{\linvar{x}}; \\
                \quad \pname{\linvar{x}}{z_i}; ( \gsome{z_i}{\llfv{M_j}};  \piencodfaplas{M_j}_{z_i} \| (\piencodfaplas{(C \setminus M_j)}_{\linvar{x}} \| \pnone{y_i} ))
            \end{array}
            \\[1mm]
            \piencodfaplas{{\oneb}}_{\linvar{x}} & = \gsome{\linvar{x}}{\emptyset};\gname{x}{y_n};  ( \psome{ y_n}; \pclose{y_n}  \| \gsome{\linvar{x}}{\emptyset}; \pnone{\linvar{x}} )
            \\[1mm]
            \span
            \piencodfaplas{\unvar{\oneb}}_{x}  = \pnone{x}
            \qquad\qquad
            \piencodfaplas{\unvar{\bag{N}}}_{x}  =  \piencodfaplas{N}_{x}
            \qquad\qquad
            \piencodfaplas{ U }_{x}  = \gsel{x}\{i:\piencodfaplas{ U_i }_{x} \}_{U_i \in U}
            \\[1mm]
            \piencodfscale{
            M \ltalltriangle \bag{M_1} \cdot \bag{M_2}
                    {} /  x_1, x_2 \rtalltriangle
            }_u    &= \begin{array}{@{}l@{}}
                \res{z_1}( \gsome{z_1}{\llfv{M_{1}}};\piencodfaplas{ M_{1} }_{ {z_1}} \| \res{z_2} ( \gsome{z_2}{\llfv{M_{2}}};\piencodfaplas{ M_{2} }_{ {z_2}} \\
                \qquad \quad {} \| {} {} \bignd_{x_{i_1} \in \{ x_1 , x_2  \}} \bignd_{x_{i_2} \in \{ x_1, x_2 \setminus x_{i_1}  \}} \piencodfaplas{ M }_u \{ z_1 / x_{i_1} \} \{ z_2 / x_{i_2} \} ) \ldots )
            \end{array}
            \\[1mm]
            \piencodfaplas{ M \unexsub{U / \lunvar{x}}  }_u   & =  \res{\unvar{x}} ( \piencodfaplas{ M }_u \|   \guname{\unvar{x}}{x_i}; \piencodfaplas{ U }_{x_i} )
            \\[1mm]
            \piencodfaplas{M[  \leftarrow  {x}]}_u & = \psome{\linvar{x}}; \pname{\linvar{x}}{y_i}; ( \gsome{y_i}{ u , \llfv{M} }; \gclose{ y_{i} } ;\piencodfaplas{M}_u \| \pnone{ \linvar{x} } )
            \\[1mm]
            \piencodfaplas{M[ \widetilde{x} \leftarrow  {x}]}_u & = \begin{array}{@{}l@{}}
                \psome{\linvar{x}}; \pname{\linvar{x}}{y_i}; \big( \gsome{y_i}{ \emptyset }; \gclose{ y_{i} } ; \0 \\
                \quad {} \| \psome{\linvar{x}}; \gsome{\linvar{x}}{u, \llfv{M} \setminus  \widetilde{x} }; \bignd_{x_i \in \widetilde{x}} \gname{\linvar{x}}{{x}_i};\piencodfaplas{M[ (\widetilde{x} \setminus x_i ) \leftarrow  {x}]}_u \big)
            \end{array}
            \\[1mm]
            \piencodfaplas{\fail^{x_1, \ldots, x_k}}_u & = \pnone{ u}  \| \pnone{ x_1} \| \ldots \| \pnone{ x_k}
        \end{align*}
    \caption{Translation of \texorpdfstring{\lamcoldetsh}{lambda} into \texorpdfstring{\clpi}{spi+}.}\label{ch4fig:encoding}
\end{figure}

To complete this section, \Cref{ch4fig:enc_typesunres} gives the translation of intersection types for \lamcoldetsh to session types for full \clpi.

\begin{definition}{}\label{ch4def:enc_sestypfailunres}
    The translation  $\piencodfaplas{\cdot}_{\_}$  in \Cref{ch4fig:enc_typesunres} extends to a linear context
    \[
        \Gamma =  {{x}_1: \sigma_1} , \ldots, {{x}_m : \sigma_m} , {{v}_1: \pi_1} , \ldots ,  {v}_n: \pi_n
    \]
    and an unrestricted context $\Theta =\unvar{x}[1] : \eta_1 , \ldots , \unvar{x}[k] : \eta_k$  as follows:
    \begin{align*}
        \piencodfaplas{\Gamma} &= {x}_1 : \with \overline{\piencodfaplas{\sigma_1}} , \ldots ,   {x}_m : \with \overline{\piencodfaplas{\sigma_m}} ,
        {v}_1:  \overline{\piencodfaplas{\pi_1}_{(\sigma, i_1)}}, \ldots ,  {v}_n: \overline{\piencodfaplas{\pi_n}_{(\sigma, i_n)}}\\
        \piencodfaplas{\Theta}&=\unvar{x}[1] : \dual{\piencodfaplas{\eta_1}} , \ldots , \unvar{x}[k] : \dual{\piencodfaplas{\eta_k}}
    \end{align*}
\end{definition}

\begin{figure}[t] \mysmall
        \begin{align*}
            \piencodfaplas{\unit} &= \with \onef
            &
            \piencodfaplas{ \eta } &= {!} \with_{\eta_i \in \eta} \{ i : \piencodfaplas{\eta_i} \}
            \\
            \piencodfaplas{(\sigma^{k} , \eta )   \rightarrow \tau} &= \with( \dual{\piencodfaplas{ (\sigma^{k} , \eta  )  }_{(\sigma, i)}} \ampy \piencodfaplas{\tau})
            &
            \piencodfaplas{ (\sigma^{k} , \eta  )  }_{(\sigma, i)} &= \oplus( (\piencodfaplas{\sigma^{k} }_{(\sigma, i)}) \otimes (( \piencodfaplas{\eta}) \otimes (\onef))  )
            \\[1mm]
            \piencodfaplas{ \sigma \wedge \pi }_{(\sigma, i)} & = \oplus(( \with \onef) \ampy ( \oplus  \with (( \oplus \piencodfaplas{\sigma} ) \otimes (\piencodfaplas{\pi}_{(\sigma, i)}))))
            \span \span
            \\
            \piencodfaplas{\omega}_{(\sigma, i)} & = \begin{cases}
                \oplus ((\with \1) \parr (\oplus \with \1))
                &  \text{if $i = 0$}
                \\
                \oplus ((\with \1) \parr (\oplus\, \with ((\oplus \piencodfaplas{\sigma}) \tensor (\piencodfaplas{\omega}_{(\sigma, i-1)}))))
                & \text{if $i > 0$}
            \end{cases}
            \span \span
        \end{align*}
    \caption{Translation of intersection types into session types  (cf.\ \defref{ch4def:enc_sestypfailunres}).}
    \label{ch4fig:enc_typesunres}
\end{figure}

\section{Proofs of Type Preservation and Deadlock-freedom for (Full) \texorpdfstring{\clpi}{Pi}}
\label{ch4s:piProofs}

Here we prove \Cref{ch4t:srPi,ch4t:dfPi} (type preservation and deadlock-freedom for the lazy semantics, respectively), as well as the analogue results for the eager semantics.
In fact, deadlock-freedom for the lazy semantics follows from deadlock-freedom for the eager semantics, so we present the proofs for the eager semantics first.

\subsection{Eager Semantics}
\label{ch4ss:proofsEager}

\subsubsection{Subject Congruence}

\begin{theorem}\label{ch4t:subcong}
    If $P \vdash \Gamma$ and $P \equiv Q$, then $Q \vdash \Gamma$.
\end{theorem}

\begin{proof}
    By induction on the derivation of the structural congruence.
    We first detail the base cases:
    \begin{itemize}
        \item
            $P \equiv_\alpha P' \implies P \equiv P'$.
            Since alpha-renaming only affects bound names, it does not affect the names in $\Gamma$, so clearly $P' \vdash \Gamma$.

        \item
            $\pfwd{x}{y} \equiv \pfwd{y}{x}$.
            \begin{mathpar}
                \inferrule{ }{
                    \pfwd{x}{y} \vdash x{:}A, y{:}\ol{A}
                }
                \equiv
                \inferrule{ }{
                    \pfwd{y}{x} \vdash x{:}A, y{:}\ol{A}
                }
            \end{mathpar}

        \item
            $P \| Q \equiv Q \| P$.
            \begin{mathpar}
                \inferrule{
                    P \vdash \Gamma
                    \\
                    Q \vdash \Delta
                }{
                    P \| Q \vdash \Gamma, \Delta
                }
                \equiv
                \inferrule{
                    Q \vdash \Delta
                    \\
                    P \vdash \Gamma
                }{
                    Q \| P \vdash \Gamma, \Delta
                }
            \end{mathpar}

        \item
            $(P \| Q) \| R \equiv P \| (Q \| R)$.
            \begin{mathpar}
                \inferrule{
                    \inferrule*{
                        P \vdash \Gamma
                        \\
                        Q \vdash \Delta
                    }{
                        P \| Q \vdash \Gamma, \Delta
                    }
                    \\
                    R \vdash \Lambda
                }{
                    (P \| Q) \| R \vdash \Gamma, \Delta, \Lambda
                }
                \equiv
                \inferrule{
                    P \vdash \Gamma
                    \\
                    \inferrule*{
                        Q \vdash \Delta
                        \\
                        R \vdash \Lambda
                    }{
                        Q \| R \vdash \Delta, \Lambda
                    }
                }{
                    P \| (Q \| R) \vdash \Gamma, \Delta, \Lambda
                }
            \end{mathpar}

        \item
            $P \| \0 \equiv P$
            \begin{mathpar}
                \inferrule{
                    P \vdash \Gamma
                    \\
                    \inferrule*{ }{
                        \0 \vdash \emptyset
                    }
                }{
                    P \| \0 \vdash \Gamma
                }
                \equiv
                P \vdash \Gamma
            \end{mathpar}

        \item
            $\res{x}(P \| Q) \equiv \res{x}(Q \| P)$.
            \begin{mathpar}
                \inferrule{
                    P \vdash \Gamma, x{:}A
                    \\
                    Q \vdash \Delta, x{:}\ol{A}
                }{
                    \res{x}(P \| Q) \vdash \Gamma, \Delta
                }
                \equiv
                \inferrule{
                    Q \vdash \Delta, x{:}\ol{A}
                    \\
                    P \vdash \Gamma, x{:}A
                }{
                    \res{x}(Q \| P) \vdash \Gamma, \Delta
                }
            \end{mathpar}

        \item
            $x \notin \fn{Q} \implies \res{x}(\res{y}(P \| Q) \| R) \equiv \res{y}(\res{x}(P \| R) \| Q)$.
            \begin{mathpar}
                \inferrule{
                    \inferrule*{
                        P \vdash \Gamma, y{:}A, x{:}B
                        \\
                        Q \vdash \Delta, y{:}\ol{A}
                    }{
                        \res{y}(P \| Q) \vdash \Gamma, \Delta, x{:}B
                    }
                    \\
                    R \vdash \Lambda, x{:}\ol{B}
                }{
                    \res{x}(\res{y}(P \| Q) \| R) \vdash \Gamma, \Delta, \Lambda
                }
                \equiv
                \inferrule{
                    \inferrule*{
                        P \vdash \Gamma, y{:}A, x{:}B
                        \\
                        R \vdash \Lambda, x{:}\ol{B}
                    }{
                        \res{x}(P \| R) \vdash \Gamma, \Lambda, y{:}A
                    }
                    \\
                    Q \vdash \Delta, y{:}\ol{A}
                }{
                    \res{y}(\res{x}(P \| R) \| Q) \vdash \Gamma, \Delta, \Lambda
                }
            \end{mathpar}

        \item
            $x \notin \fn{Q} \implies \res{x}((P \| Q) \| R) \equiv \res{x}(P \| R) \| Q$.
            \begin{mathpar}
                \inferrule{
                    \inferrule*{
                        P \vdash \Gamma, x{:}A
                        \\
                        Q \vdash \Delta
                    }{
                        P \| Q \vdash \Gamma, \Delta, x{:}A
                    }
                    \\
                    R \vdash \Lambda, x{:}\ol{A}
                }{
                    \res{x}((P \| Q) \| R) \vdash \Gamma, \Delta, \Lambda
                }
                \equiv
                \inferrule{
                    \inferrule*{
                        P \vdash \Gamma, x{:}A
                        \\
                        R \vdash \Lambda, x{:}\ol{A}
                    }{
                        \res{x}(P \| R) \vdash \Gamma, \Lambda
                    }
                    \\
                    Q \vdash \Delta
                }{
                    \res{x}(P \| R) \| Q \vdash \Gamma, \Delta, \Lambda
                }
            \end{mathpar}

        \item
            $x \notin \fn{Q} \implies \res{x}(\guname{x}{y};P \| Q) \equiv Q$.
            \begin{mathpar}
                \inferrule{
                    \inferrule*{
                        P \vdash {?}\Gamma, y{:}A
                    }{
                        \guname{x}{y};P \vdash {?}\Gamma, x{:}{!}A
                    }
                    \\
                    \inferrule*{
                        Q \vdash \Delta
                    }{
                        Q \vdash \Delta, x{:}{?}\ol{A}
                    }
                }{
                    \res{x}(\guname{x}{y};P \| Q) \vdash {?}\Gamma, \Delta
                }
                \equiv
                \inferruleDbl[vcenter]{
                    Q \vdash \Delta
                }{
                    Q \vdash {?}\Gamma, \Delta
                }
            \end{mathpar}

        \item
            $P \nd Q \equiv Q \nd P$.
            \begin{mathpar}
                \inferrule{
                    P \vdash \Gamma
                    \\
                    Q \vdash \Gamma
                }{
                    P \nd Q \vdash \Gamma
                }
                \equiv
                \inferrule{
                    Q \vdash \Gamma
                    \\
                    P \vdash \Gamma
                }{
                    Q \nd P \vdash \Gamma
                }
            \end{mathpar}

        \item
            $(P \nd Q) \nd R \equiv P \nd (Q \nd R)$
            \begin{mathpar}
                \inferrule{
                    \inferrule*{
                        P \vdash \Gamma
                        \\
                        Q \vdash \Gamma
                    }{
                        P \nd Q \vdash \Gamma
                    }
                    \\
                    R \vdash \Gamma
                }{
                    (P \nd Q) \nd R \vdash \Gamma
                }
                \equiv
                \inferrule{
                    P \vdash \Gamma
                    \\
                    \inferrule*{
                        Q \vdash \Gamma
                        \\
                        R \vdash \Gamma
                    }{
                        Q \nd R \vdash \Gamma
                    }
                }{
                    P \nd (Q \nd R) \vdash \Gamma
                }
            \end{mathpar}

        \item
            $P \nd P \equiv P$.
            \begin{mathpar}
                \inferrule{
                    P \vdash \Gamma
                    \\
                    P \vdash \Gamma
                }{
                    P \nd P \vdash \Gamma
                }
                \equiv
                P \vdash \Gamma
            \end{mathpar}
    \end{itemize}
    The inductive cases follow from the IH straightforwardly.
    Note that the rules for parallel composition do not apply directly behind the output prefix and restriction.
\end{proof}

\subsubsection{Subject Reduction}

\begin{lemma}\label{ch4l:ctxType}
    Suppose $P \vdash \Gamma, x{:}A$.
    \begin{enumerate}
        \item\label{ch4i:ctxTypePclose}
            If $P = \pctx{N}[\pclose{x}]$, then $A = \1$.

        \item\label{ch4i:ctxTypeGclose}
            If $P = \pctx{N}[\gclose{x};P']$, then $A = \bot$.

        \item\label{ch4i:ctxTypePname}
            If $P = \pctx{N}[\pname{x}{y};(P' \| P'')]$, then $A = B \tensor C$.

        \item\label{ch4i:ctxTypeGname}
            If $P = \pctx{N}[\gname{x}{y};P']$, then $A = B \parr C$.

        \item\label{ch4i:ctxTypePsel}
            If $P = \pctx{N}[\psel{x}{j};P']$, then $A = {\oplus}\{i:B_i\}_{i \in I}$ where $j \in I$.

        \item\label{ch4i:ctxTypeGsel}
            If $P = \pctx{N}[\gsel{x}\{i:P'_i\}_{i \in I}]$, then $A = {\with}\{i:B_i\}_{i \in I}$.

        \item\label{ch4i:ctxTypePsome}
            If $P = \pctx{N}[\psome{x};P']$, then $A = {\with}B$.

        \item\label{ch4i:ctxTypePnone}
            If $P = \pctx{N}[\pnone{x}]$, then $A = {\with}B$.

        \item\label{ch4i:ctxTypeGsome}
            If $P = \pctx{N}[\gsome{x}{w_1,\ldots,w_n}]$, then $A = {\oplus}B$.

        \item\label{ch4i:ctxTypePuname}
            If $P = \pctx{N}[\puname{x}{y};P']$, then $A = {?}B$.

        \item\label{ch4i:ctxTypeGuname}
            If $P = \pctx{N}[\guname{x}{y};P']$, then $A = {!}B$.
    \end{enumerate}
\end{lemma}

\begin{proof}
    Each item follows by induction on the structure of the ND-context.
    The base case follows by inversion of typing, and the inductive cases follow from the IH straightforwardly.
\end{proof}

\begin{lemma}\label{ch4l:ctxRedOne}
    For each of the following items, assume $\Gamma \disj \Delta$.
    \begin{enumerate}
        \item\label{ch4i:ctxRedOneFwd}
            If $\pctx[\big]{N}[\pfwd{x}{y}] \vdash \Gamma, x{:}A$ and $Q \vdash \Delta, x{:}\ol{A}$, then $\pctx*{N}[Q \{y/x\}] \vdash \Gamma, \Delta$.

        \item\label{ch4i:ctxRedOnePclose}
            If $\pctx{N}[\pclose{x}] \vdash \Gamma, x{:}\1$, then $\pctx*{N}[\0] \vdash \Gamma$.

        \item\label{ch4i:ctxRedOneGclose}
            If $\pctx{N}[\gclose{x};Q] \vdash \Gamma, x{:}\bot$, then $\pctx*{N}[Q] \vdash \Gamma$.

        \item\label{ch4i:ctxRedOneName}
            If $\bn{\pctx{N}} \disj \fn{\pctx{N}'}$ and $\pctx{N}[\pname{x}{y};(P \| Q)] \vdash \Gamma, x{:}A \tensor B$ and $\pctx{N'}[\gname{x}{z};R] \vdash \Delta, x{:}\ol{A} \parr \ol{B}$, then $\pctx*{N}[\res{x}(Q \| \res{y}(P \| \pctx*{N'}[R\{y/z\}]))] \vdash \Gamma, \Delta$.

        \item\label{ch4i:ctxRedOnePsel}
            If $\pctx{N}[\psel{x}{j}; P] \vdash \Gamma, x{:}{\oplus}\{i: A_i\}_{i \in I}$ and $j \in I$, then $\pctx*{N}[P] \vdash \Gamma, x{:}A_j$.

        \item\label{ch4i:ctxRedOneGsel}
            If $\pctx{N}[\gsel{x}\{i: P_i\}_{i \in I}] \vdash \Gamma, x{:}\with\{i: A_i\}_{i \in I}$, then $\pctx*{N}[P_i] \vdash \Gamma, x{:}A_i$ for every $i \in I$.

        \item\label{ch4i:ctxRedOnePsome}
            If $\pctx{N}[\psome{x}; P] \vdash \Gamma, x{:}{\with}A$, then $\pctx*{N}[P] \vdash \Gamma, x{:}A$.

        \item\label{ch4i:ctxRedOnePnone}
            If $\pctx{N}[\pnone{x}] \vdash \Gamma, x{:}{\with}A$, then $\pctx*{N}[\0] \vdash \Gamma$.

        \item\label{ch4i:ctxRedOneGsome}
            If $\pctx{N}[\gsome{x}{w_1,\ldots,w_n}; P] \vdash \Gamma, x{:}{\oplus} A$, then $\pctx*{N}[P] \vdash \Gamma, x{:}A$ and $\pctx*{N}[\pnone{w_1} \| \ldots \| \pnone{w_n}] \vdash \Gamma$.

        \item\label{ch4i:ctxRedOneUname}
            If $\bn{\pctx{N'}} \disj \fn{\pctx{N}}$ and $\pctx{N}[\puname{x}{y}; P] \vdash \Gamma, x{:}{?}A$ and $\pctx{N}'[\guname{x}{y}; Q] \vdash \Delta, x{:}{!}\ol{A}$, then $\pctx*{N'}\big[\res{x}(\res{y}(\pctx*{N}[P] \| Q\{y/z\}) \| \guname{x}{z};Q)\big]$.
    \end{enumerate}
\end{lemma}

\begin{proof}
    For each item, we apply induction on the structure of the ND-contexts and detail the base cases.
    For simplicity, we assume no names in $\Gamma$ and $\Delta$ were derived with \ttype{weaken}.
    \begin{enumerate}
        \item
            \begin{mathpar}
                \inferrule*{ }{
                    \pfwd{x}{y} \vdash x{:}A, y{:}\ol{A}
                }
                \and
                \inferrule{}{
                    Q \vdash \Delta, x{:}\ol{A}
                }
                \and
                \implies
                \and
                \inferrule{}{
                    Q\{y/x\} \vdash \Delta, y{:}\ol{A}
                }
            \end{mathpar}

        \item
            \begin{mathpar}
                \inferrule*{ }{
                    \pclose{x} \vdash x{:}\1
                }
                \and\implies\and
                \inferrule*{ }{
                    \0 \vdash \emptyset
                }
            \end{mathpar}

        \item
            \begin{mathpar}
                \inferrule{
                    Q \vdash \Gamma
                }{
                    \gclose{x};Q \vdash \Gamma, x{:}\bot
                }
                \and\implies\and
                \inferrule{}{
                    Q \vdash \Gamma
                }
            \end{mathpar}

        \item
            \begin{mathpar}
                \inferrule{
                    P \vdash \Gamma, y{:}A
                    \\
                    Q \vdash \Gamma', x{:}B
                }{
                    \pname{x}{y};(P \| Q) \vdash \Gamma, \Gamma', x{:}A \tensor B
                }
                \and
                \inferrule{
                    R \vdash \Delta, z{:}\ol{A}, x{:}\ol{B}
                }{
                    \gname{x}{z};R \vdash \Delta, x{:}\ol{A} \parr \ol{B}
                }
                \and\implies\and
                \inferrule{
                    Q \vdash \Gamma', x{:}B
                    \\
                    \inferrule*{
                        P \vdash \Gamma, y{:}A
                        \\
                        R\{y/z\} \vdash \Delta, y{:}\ol{A}, x{:}\ol{B}
                    }{
                        \res{y}(P \| R\{y/z\}) \vdash \Gamma, \Delta, x{:}\ol{B}
                    }
                }{
                    \res{x}(Q \| \res{y}(P \| R\{y/z\})) \vdash \Gamma, \Gamma', \Delta
                }
            \end{mathpar}

        \item
            \begin{mathpar}
                \inferrule{
                    P \vdash \Gamma, x{:}A_j
                    \\
                    j \in I
                }{
                    \psel{x}{j};P \vdash \Gamma, x{:}{\oplus}\{i:A_i\}_{i \in I}
                }
                \and\implies\and
                \inferrule{}{
                    P \vdash \Gamma, x{:}A_j
                }
            \end{mathpar}

        \item
            \begin{mathpar}
                \inferrule{
                    \forall i \in I.~ P_i \vdash \Gamma, x{:}A_i
                }{
                    \gsel{x}\{i:P_i\}_{i \in I} \vdash \Gamma, x{:}{\with}\{i:A_i\}
                }
                \and\implies\and
                \forall i \in I.~ \inferrule*{}{
                    P_i \vdash \Gamma, x{:}A_i
                }
            \end{mathpar}

        \item
            \begin{mathpar}
                \inferrule{
                    P \vdash \Gamma, x{:}A
                }{
                    \psome{x};P \vdash \Gamma, x{:}{\with}A
                }
                \and\implies\and
                \inferrule{}{
                    P \vdash \Gamma, x{:}A
                }
            \end{mathpar}

        \item
            \begin{mathpar}
                \inferrule*{ }{
                    \pnone{x} \vdash x{:}{\with}A
                }
                \and\implies\and
                \inferrule*{ }{
                    \0 \vdash \emptyset
                }
            \end{mathpar}

        \item
            \begin{mathpar}
                \inferrule{
                    P \vdash w_1{:}{\with}B_1, \ldots, w_n{:}{\with}B_n, x{:}A
                }{
                    \gsome{x}{w_1,\ldots,w_n};P \vdash w_1{:}{\with}B_1, \ldots, w_n{:}{\with}B_n, x{:}{\oplus}A
                }
                \and\implies\and
                \inferrule{}{
                    P \vdash w_1{:}{\with}B_1, \ldots, w_n{:}{\with}B_n, x{:}A
                }
                \and
                \inferruleDbl{
                    \inferrule*[fraction={---}]{ }{
                        \pnone{w_1} \vdash w_1{:}{\with}B_1
                    }
                    \\
                    \ldots
                    \\
                    \inferrule*[fraction={---}]{ }{
                        \pnone{w_n} \vdash w_n{:}{\with}B_n
                    }
                }{
                    \pnone{w_1} \| \ldots \| \pnone{w_n} \vdash w_1{:}{\with}B_1, \ldots, w_n{:}{\with}B_n
                }
            \end{mathpar}

        \item
            This item depends on whether $x \in \fn{P}$.
            \begin{itemize}
                \item
                    $x \in \fn{P}$:
                    \begin{mathpar}
                        \and
                        \inferrule*{
                            \inferrule*{
                                P\{x'/x\} \vdash \Gamma, y{:}A, x'{:}{?}A
                            }{
                                \puname{x}{y};(P\{x'/x\}) \vdash \Gamma, x{:}{?}A, x'{:}{?}A
                            }
                        }{
                            \puname{x}{y};P \vdash \Gamma, x{:}{?}A
                        }
                        \and
                        \inferrule*{
                            Q \vdash {?}\Delta, z{:}\ol{A}
                        }{
                            \guname{x}{z};Q \vdash {?}\Delta, x{:}{!}\ol{A}
                        }
                        \and\implies\and
                        \inferruleDbl{
                            \inferrule*[fraction={---}]{
                                \inferrule*{
                                    P \vdash \Gamma, y{:}A, x{:}{?}A
                                    \\
                                    Q\{y/z\}\{w'/w\}_{w \in {?}\Delta} \vdash {?}\Delta', y{:}\ol{A}
                                }{
                                    \res{y}(P \| Q\{y/z\}\{w'/w\}_{w \in {?}\Delta}) \vdash \Gamma, {?}\Delta', x{:}{?}A
                                }
                                \\
                                \guname{x}{z};Q \vdash {?}\Delta, x{:}{!}\ol{A}
                            }{
                                \res{x}(\res{y}(P \| Q\{y/z\}\{w'/w\}_{w \in {?}\Delta}) \| \guname{x}{z};Q) \vdash \Gamma, {?}\Delta, {?}\Delta'
                            }
                        }{
                            \res{x}(\res{y}(P \| Q\{y/z\}) \| \guname{x}{z};Q) \vdash \Gamma, {?}\Delta
                        }
                    \end{mathpar}

                \item
                    $x \notin \fn{P}$:
                    \begin{mathpar}
                        \inferrule*{
                            P \vdash \Gamma, y{:}A
                        }{
                            \puname{x}{y};P \vdash \Gamma, x{:}{?}A
                        }
                        \and
                        \inferrule*{
                            Q \vdash {?}\Delta, z{:}\ol{A}
                        }{
                            \guname{x}{z};Q \vdash {?}\Delta, x{:}{!}\ol{A}
                        }
                        \and\implies\and
                        \inferruleDbl{
                            \inferrule*[fraction={---}]{
                                \inferrule*{
                                    \inferrule*{
                                        P \vdash \Gamma, y{:}A
                                        \\
                                        Q\{y/z\}\{w'/w\}_{w \in {?}\Delta} \vdash {?}\Delta', y{:}\ol{A}
                                    }{
                                        \res{y}(P \| Q\{y/z\}\{w'/w\}_{w \in {?}\Delta}) \vdash \Gamma, {?}\Delta'
                                    }
                                }{
                                    \res{y}(P \| Q\{y/z\}\{w'/w\}_{w \in {?}\Delta}) \vdash \Gamma, {?}\Delta', x{:}{?}A
                                }
                                \\
                                \guname{x}{z};Q \vdash {?}\Delta, x{:}{!}\ol{A}
                            }{
                                \res{x}(\res{y}(P \| Q\{y/z\}\{w'/w\}_{w \in {?}\Delta}) \| \guname{x}{z};Q) \vdash \Gamma, {?}\Delta, {?}\Delta'
                            }
                        }{
                            \res{x}(\res{y}(P \| Q\{y/z\}) \| \guname{x}{z};Q) \vdash \Gamma, {?}\Delta
                        }
                    \end{mathpar}
            \end{itemize}
    \end{enumerate}
    The inductive cases follow straightforwardly.
    Notice that the conditions on the bound and free names of the ND-contexts in \cref{ch4i:ctxRedOneName,ch4i:ctxRedOneUname} make sure that no names are captured when embedding one context in the other.
\end{proof}

\begin{theorem}[SR for the Eager Semantics]\label{ch4t:srOne}
    If $P \vdash \Gamma$ and $P \redone Q$, then $Q \vdash \Gamma$.
\end{theorem}

\begin{proof}
    By induction on the derivation of the reduction.
    \begin{itemize}
        \item
            Rule $\rredone{\scc{Id}}$.
            \begin{mathpar}
                \inferrule{
                    \pctx[\big]{N}[\pfwd{x}{y}] \vdash \Gamma, x{:}A
                    \\
                    Q \vdash \Delta, x{:}\ol{A}
                }{
                    \res{x}(\pctx[\big]{N}[\pfwd{x}{y}] \| Q) \vdash \Gamma, \Delta
                }
                \and\implies\and
                \inferrule{}{
                    \pctx*{N}[Q\{y/x\}] \vdash \Gamma, \Delta
                    ~~\text{(\refitem{l}{ctxRedOne}{Fwd})}
                }
            \end{mathpar}

        \item
            Rule $\rredone{\1\bot}$.
            \begin{mathpar}
                \inferrule{
                    \pctx{N}[\pclose{x}] \vdash \Gamma, x{:}\1
                    ~~\text{(\refitem{l}{ctxType}{Pclose})}
                    \\
                    \pctx{N'}[\gclose{x};Q] \vdash \Delta, x{:}\bot
                    ~~\text{(\refitem{l}{ctxType}{Gclose})}
                }{
                    \nu{x}(\pctx{N}[\pclose{x}] \| \pctx{N'}[\gclose{x};Q]) \vdash \Gamma, \Delta
                }
                \and\implies\and
                \inferrule{
                    \pctx*{N}[\0] \vdash \Gamma
                    ~~\text{(\refitem{l}{ctxRedOne}{Pclose})}
                    \\
                    \pctx*{N'}[Q] \vdash \Delta
                    ~~\text{(\refitem{l}{ctxRedOne}{Gclose})}
                }{
                    \pctx*{N}[\0] \| \pctx*{N'}[Q] \vdash \Gamma, \Delta
                }
            \end{mathpar}

        \item
            Rule $\rredone{\tensor\parr}$.
            \begin{mathpar}
                \inferrule{
                    \pctx{N}[\pname{x}{y};(P \| Q)] \vdash \Gamma, x{:}A \tensor B
                    ~~\text{(\refitem{l}{ctxType}{Pname})}
                    \\
                    \pctx{N'}[\gname{x}{z};R] \vdash \Delta, x{:}\ol{A} \parr \ol{B}
                    ~~\text{(\refitem{l}{ctxType}{Gname})}
                }{
                    \res{x}(\pctx{N}[\pname{x}{y};(P \| Q)] \| \pctx{N'}[\gname{x}{z};R]) \vdash \Gamma, \Delta
                }
                \and\implies\and
                \inferrule{}{
                    \pctx*{N}[\res{x}(Q \| \nu{y}(P \| \pctx*{N'}[R\{y/z\}]))] \vdash \Gamma, \Delta
                    ~~\text{(\refitem{l}{ctxRedOne}{Name})}
                }
            \end{mathpar}

        \item
            Rule $\rredone{{\oplus}{\with}}$.
            \begin{mathpar}
                \inferrule{
                    \pctx{N}[\psel{x}{j};P] \vdash \Gamma, x{:}{\oplus}\{i:A_i\}_{i \in I}
                    ~~ j \in I
                    ~~\text{(\refitem{l}{ctxType}{Psel})}
                    \\
                    \pctx{N'}[\gsel{x}\{i:Q_i\}_{i \in I}] \vdash \Delta, x{:}{\with}\{i:\ol{A_i}\}_{i \in I}
                    ~~\text{(\refitem{l}{ctxType}{Gsel})}
                }{
                    \res{x}(\pctx{N}[\psel{x}{j};P] \| \pctx{N'}[\gsel{x}\{i:Q_i\}_{i \in I}]) \vdash \Gamma, \Delta
                }
                \and\implies\and
                \inferrule{
                    \pctx*{N}[P] \vdash \Gamma, x{:}A_j
                    ~~\text{(\refitem{l}{ctxRedOne}{Psel})}
                    \\
                    \pctx*{N'}[Q_j] \vdash \Delta, x{:}\ol{A_j}
                    ~~\text{(\refitem{l}{ctxRedOne}{Gsel})}
                }{
                    \res{x}(\pctx*{N}[P] \| \pctx*{N'}[Q_j]) \vdash \Gamma, \Delta
                }
            \end{mathpar}

        \item
            Rule $\rredone{\some}$.
            \begin{mathpar}
                \inferrule{
                    \pctx{N}[\psome{x};P] \vdash \Gamma, x{:}{\with}A
                    ~~\text{(\refitem{l}{ctxType}{Psome})}
                    \\
                    \pctx{N'}[\gclose{x}{(w_1,\ldots,w_n)};Q] \vdash \Delta, x{:}{\oplus}\ol{A}
                    ~~\text{(\refitem{l}{ctxType}{Gsome})}
                }{
                    \res{x}(\pctx{N}[\psome{x};P] \| \pctx{N'}[\gclose{x}{(w_1,\ldots,w_n)};Q]) \vdash \Gamma, \Delta
                }
                \and\implies\and
                \inferrule{
                    \pctx*{N}[P] \vdash \Gamma, x{:}A
                    ~~\text{(\refitem{l}{ctxRedOne}{Psome})}
                    \\
                    \pctx*{N'}[Q] \vdash \Delta, x{:}\ol{A}
                    ~~\text{(\refitem{l}{ctxRedOne}{Gsome})}
                }{
                    \res{x}(\pctx*{N}[P] \| \pctx*{N'}[Q]) \vdash \Gamma, \Delta
                }
            \end{mathpar}

        \item
            Rule $\rredone{\none}$.
            \begin{mathpar}
                \inferrule{
                    \pctx{N}[\pnone{x}] \vdash \Gamma, x{:}{\with}A
                    ~~\text{(\refitem{l}{ctxType}{Pnone})}
                    \\
                    \pctx{N'}[\gclose{x}{(w_1,\ldots,w_n)};Q] \vdash \Delta, x{:}{\oplus}\ol{A}
                    ~~\text{(\refitem{l}{ctxType}{Gsome})}
                }{
                    \res{x}(\pctx{N}[\pnone{x}] \| \pctx{N'}[\gclose{x}{(w_1,\ldots,w_n)};Q]) \vdash \Gamma, \Delta
                }
                \and\implies\and
                \inferrule{
                    \pctx*{N}[\0] \vdash \Gamma
                    ~~\text{(\refitem{l}{ctxRedOne}{Pnone})}
                    \\
                    \pctx*{N'}[\pnone{w_1} \| \ldots \| \pnone{w_n}] \vdash \Delta
                    ~~\text{(\refitem{l}{ctxRedOne}{Gsome})}
                }{
                    \pctx*{N}[\0] \| \pctx*{N'}[\pnone{w_1} \| \ldots \| \pnone{w_n}]) \vdash \Gamma, \Delta
                }
            \end{mathpar}

        \item
            Rule $\rredone{{?}{!}}$.
            \begin{mathpar}
                \inferrule{
                    \pctx{N}[\puname{x}{y};P] \vdash \Gamma, x{:}{?}A
                    ~~\text{(\refitem{l}{ctxType}{Puname})}
                    \\
                    \pctx{N'}[\guname{x}{z};Q] \vdash \Delta, x{:}{!}\ol{A}
                    ~~\text{(\refitem{l}{ctxType}{Guname})}
                }{
                    \res{x}(\pctx{N}[\puname{x}{y};P] \| \pctx{N'}[\guname{x}{y};Q]) \vdash \Gamma, \Delta
                }
                \and\implies\and
                \inferrule{}{
                    \pctx*{N'}\big[\res{x}(\res{y}(\pctx*{N}[P] \| Q\{y/z\}) \| \guname{x}{z};Q)\big] \vdash \Gamma, \Delta
                    ~~\text{(\refitem{l}{ctxRedOne}{Uname})}
                }
            \end{mathpar}

        \item
            Rule $\rredone{\equiv}$.
            Assume $P \equiv P'$ and $P' \redone Q'$ and $Q' \equiv Q$.
            By \Cref{ch4t:subcong}, $P' \vdash \Gamma$.
            By the IH, $Q' \vdash \Gamma$.
            By \Cref{ch4t:subcong}, $Q \vdash \Gamma$.

        \item
            Rule $\rredone{\nu}$.
            Assume $P \redone P'$.
            \begin{mathpar}
                \inferrule{
                    P \vdash \Gamma, x{:}A
                    \\
                    Q \vdash \Delta, x{:}\ol{A}
                }{
                    \res{x}(P \| Q) \vdash \Gamma, \Delta
                }
                \and\implies\and
                \inferrule{
                    P' \vdash \Gamma, x{:}A
                    ~~\text{(IH)}
                    \\
                    Q \vdash \Delta, x{:}\ol{A}
                }{
                    \res{x}(P' \| Q) \vdash \Gamma, \Delta
                }
            \end{mathpar}

        \item
            Rule $\rredone{\|}$.
            Assume $P \redone P'$.
            \begin{mathpar}
                \inferrule{
                    P \vdash \Gamma
                    \\
                    Q \vdash \Delta
                }{
                    P \| Q \vdash \Gamma, \Delta
                }
                \and\implies\and
                \inferrule{
                    P' \vdash \Gamma
                    ~~\text{(IH)}
                    \\
                    Q \vdash \Delta
                }{
                    P' \| Q \vdash \Gamma, \Delta
                }
            \end{mathpar}

        \item
            Rule $\rredone{\nd}$.
            Assume $P \redone P'$.
            \begin{mathpar}
                \inferrule{
                    P \vdash \Gamma
                    \\
                    Q \vdash \Gamma
                }{
                    P \nd Q \vdash \Gamma
                }
                \and\implies\and
                \inferrule{
                    P' \vdash \Gamma
                    ~~\text{(IH)}
                    \\
                    Q \vdash \Gamma
                }{
                    P' \nd Q \vdash \Gamma
                }
            \end{mathpar}
    \end{itemize}
\end{proof}

\subsubsection{Deadlock-freedom}

The proof uses several definitions and lemmas, which we summarize:
\begin{itemize}
    \item
        \Cref{ch4d:sctx} defines single-choice multi-hole contexts, where holes may only appear on one side of non-deterministic choices.
        \Cref{ch4d:scoll} yields deterministic multi-hole contexts from single-choice multi-hole contexts by committing non-deterministic choices to the sides of holes.
        \Cref{ch4l:scoll} ensures typing remains consistent when committing a single-choice multi-hole context.

    \item
        \Cref{ch4l:sctxform} states that any typable process not equivalent to $\0$ can be written as an S-context with each hole replaced by a prefixed process.
        Let us refer to this as the \emph{S-context form}.

    \item
        \Cref{ch4l:sformfwd} states that if a process in S-context form is typable under empty context and has a forwarder as one of its prefixes, that process contains a cut on one of the forwarder's subjects.

    \item
        \Cref{ch4l:sctxcuts} states that the number of prefixed processes of a process in S-context form is at least the number of cuts in the S-context.
        This lemma is key to the proof of Deadlock Freedom, as it is necessary to show the next lemma.

    \item
        \Cref{ch4l:sctxsubjs} states that if a process in S-context form is typable under empty context, then there must be two of its prefixed processes that share a subject.
\end{itemize}

\begin{definition}{Single-choice Multi-hole Contexts}\label{ch4d:sctx}
    We define \emph{single-choice multi-hole contexts} (S-contexts, for short) as follows:
    \[
        \pctx{S} ::= {\hole}_i \sepr \res{x}(\pctx{S} \| \pctx{S}) \sepr \pctx{S} \| \pctx{S} \sepr \pctx{S} \nd P
    \]
    An S-context is $n$-ary if it has $n$ holes ${\hole}_1, \ldots, {\hole}_n$.
    We write $\pctx{S}[P_1, \ldots, P_n]$ to denote the process obtained from the $n$-ary multi-hole context $\pctx{S}$ by replacing each $i$-th hole in $\pctx{S}$ with $P_i$.
    Given an S-context $\pctx{S}$ with hole indices $I$ and a sequence of processes ${(P_i)}_{i \in I}$, we write $\pctx{S}[P_i]_{i \in I}$ to denote the process obtained from $\pctx{S}$ by replacing each hole with index $i$ in $\pctx{S}$ with $P_i$.
    We say an S-context is a \emph{deterministic multi-hole context} (DM-context, for short) if its holes do not appear inside any non-deterministic choices.
\end{definition}

\begin{definition}{Commitment of Single-choice Multi-hole Contexts}\label{ch4d:scoll}
    We define the \emph{commitment} of S-context $\pctx{S}$, by abuse of notation denoted $\D{\pctx{S}}$ (cf.\ \Cref{ch4s:disc}), as follows, yielding a deterministic multi-hole context:
    \begin{align*}
        \D{{\hole}_i} &:= {\hole}_i
        & \D{\pctx{S} \| \pctx{S}'} &:= \D{\pctx{S}} \| \D{\pctx{S}'}
        & \D{\res{x}(\pctx{S} \| \pctx{S}')} &:= \res{x}(\D{\pctx{S}} \| \D{\pctx{S}'})
        & \D{\pctx{S} \nd P} &:= \D{\pctx{S}}
    \end{align*}
\end{definition}

\begin{lemma}\label{ch4l:scoll}
    If $\pctx{S}[P_i]_{i \in I} \vdash \Gamma$, then $\D{\pctx{S}}[P_i]_{i \in I} \vdash \Gamma$.
\end{lemma}

\begin{proof}
    Straightforward, by induction on the structure of $\pctx{S}$.
\end{proof}

\begin{lemma}\label{ch4l:sctxform}
    If $P \vdash \Gamma$ and $P \not\equiv \0$, then there exist S-context $\pctx{S}$ with indices $I$ and sequence of prefixed processes ${(\alpha_i;P_i)}_{i \in I}$ such that $P \equiv \pctx{S}[\alpha_i;P_i]_{i \in I}$.
\end{lemma}

\begin{proof}
    Using structural congruence, we first remove all cuts with unused servers and parallel compositions with $\0$, obtaining $P' \equiv P$.
    Since $P \not\equiv \0$, $P' \not\equiv \0$.
    Then, we construct $\pctx{S}$ by induction on the typing derivation of $P'$.
    Rules \ttype{empty} and \ttype{weaken} do not occur, because of how we obtained $P'$ from $P$.
    The structural rules \ttype{mix}, \ttype{cut}, and \ttype{weaken} are simply copied.
    In case of rule \ttype{$\nd$}, we arbitrarily pick a branch to continue the construction of $\pctx{S}$ with, while copying the entire other branch.
    The other rules, which type prefixes, add a hole to $\pctx{S}$; we mark the hole with index $i$ and refer to the prefixed process typed by the rule as $\alpha_i;P_i$.
    Clearly, $P \equiv P' = \pctx{S}[\alpha_i;P_i]_{i \in I}$.
\end{proof}

\begin{lemma}\label{ch4l:sformfwd}
    If $P = \pctx{S}[\alpha_i;P_i]_{i \in I} \vdash \emptyset$ and there is $j \in I$ s.t.\ $\alpha_j = \pfwd{x}{y}$, then there are $\pctx{N},\pctx{N'},Q$ such that $P = \pctx[\Big]{N}[\res{x}(\pctx[\big]{N'}[\pfwd{x}{y}] \| Q)]$.
\end{lemma}

\begin{proof}
    Note that there must be a restriction on $x$ in $P$, because $x$ appears free in $\pfwd{x}{y}$ but $P \vdash \emptyset$.
    First, we obtain $\pctx{N}$ from $P$ by replacing the restriction on $x$ in $P$ with a hole, referring the parallel component in which $\pfwd{x}{y}$ appears as $P'$ and the other parallel component as $Q$.
    Then, we obtain $\pctx{N'}$ from $P'$ by replacing $\pfwd{x}{y}$ with a hole.
    Clearly, $P = \pctx[\Big]{N}[\res{x}(\pctx[\big]{N'}[\pfwd{x}{y}] \| Q)]$.
\end{proof}

\begin{lemma}\label{ch4l:sctxcuts}
    If the derivation of $P = \pctx{S}[\alpha_i;P_i]_{i \in I} \vdash \Gamma$ and $\pctx{S}$ is deterministic and contains $n$ cuts, then $|I| \geq n+1$.
\end{lemma}

\begin{proof}
    We apply strong induction on the number $n$ of cuts in $\pctx{S}$:
    \begin{itemize}
        \item
            Case $n = 0$.
            Any S-context must have at least one hole, so $\pctx{S}$ has at least one hole.
            Hence, $|I| \geq 1 = n + 1$.

        \item
            Case $n = n' + 1$.
            By abuse of notation, $P = P_1 \| \ldots \| P_k$, where for each $1 \leq k' \leq k$, $P_{k'}$ is not a parallel composition.
            By assumption, $m \geq 1$ of the $P_1,\ldots,P_k$ are cuts.
            W.l.o.g., assume $P_1,\ldots,P_m$ are cuts.

            For each $1 \leq j \leq m$, by inversion of rule \ttype{mix}, $P_j \vdash \Gamma_j$, and by construction, there are $\pctx{S_j},I_j$ s.t.\ $P_j = \pctx{S_j}[\alpha_i;P_i]_{i \in I_j}$ where $\pctx{S_j}$ is deterministic.
            We have for each $1 \leq j \leq m$ and $1 \leq j' \leq m$ where $j \neq j'$ that $I_j \disj I_{j'}$, and $\bigcup_{1 \leq j \leq m} I_j \subseteq I$.
            Then, for each $1 \leq j \leq m$, let $1 \leq n_j \leq n$ be the number of cuts in $\pctx{S_j}$.
            Since $P_{m+1},\ldots,P_k$ are not cuts, we have $\sum_{1 \leq j \leq m} n_j = n$.

            Take any $1 \leq j \leq m$.
            We have $P_j = \res{x}(P'_j \| P''_j)$, and by inversion of rule \ttype{cut}, $P'_j \vdash \Gamma'_j, x{:}A$ and $P''_j \vdash \Gamma''_j, x{:}\ol{A}$ where $\Gamma_j = \Gamma'_j, \Gamma''_j$.
            By construction, there are $\pctx{S'_j},\pctx{S''_j},I'_j,I''_j$ s.t.\ $P'_j = \pctx{S'_j}[\alpha_i;P_i]_{i \in I'_j}$ and $P''_j = \pctx{S''_j}[\alpha_i;P_i]_{i \in I''_j}$ and $\pctx{S'_j}$ and $\pctx{S''_j}$ are deterministic.
            We have $I'_j \disj I''_j$ and $I'_j \cup I''_j = I_j$.

            Let $n'_j$ and $n''_j$ be the number of cuts in $\pctx{S'_j}$ and $\pctx{S''_j}$, respectively.
            We have that $n'_j + n''_j + 1 = n_j$.
            Since $n_j \leq n = n' + 1$, then $n'_j,n''_j \leq n'$.
            Then, by the IH, $|I'_j| \geq n'_j + 1$ and $|I''_j| \geq n''_j + 1$.
            Therefore, $|I_j| = |I'_j \cup I''_j| = |I'_j| + |I''_j| \geq n'_j + n''_j + 1 + 1 = n_j + 1$.

            In conclusion,
            \begin{align*}
                |I| \geq & |\bigcup_{1 \leq j \leq m} I_j|
                 = \sum_{1 \leq j \leq m} |I_j| \geq \sum_{1 \leq j \leq m} (n_j + 1) \\
                 = &
                 \sum_{1 \leq j \leq m} n_j + m 
                 = n + m \geq n + 1.
            \end{align*}
    \end{itemize}
\end{proof}

\begin{lemma}\label{ch4l:sctxsubjs}
    If $P = \pctx{S}[\alpha_i;P_i]_{i \in I} \vdash \emptyset$ where for each $i \in I$, $\alpha_i \neq \pfwd{x}{y}$ for any $x$ and $y$, then there are $j,k \in I$ where $j \neq k$ and $x = \subj(\alpha_j) = \subj(\alpha_k)$, and there are $\pctx{N},\pctx{N_j},\pctx{N_k}$ such that $P = \pctx[\big]{N}[\res{x}(\pctx{N_j}[\alpha_j] \| \pctx{N_k}[\alpha_k])]$.
\end{lemma}

\begin{proof}
    Let $Q = \pctx*{S}[{(\alpha_i)}_{i \in I}]$.
    Then $Q$ is deterministic and, by \Cref{ch4l:scoll}, $Q \vdash \emptyset$.
    Let $n$ be the number of cuts in $\pctx{S}$.
    By \Cref{ch4l:sctxcuts}, $|I| \geq n + 1$.

    Suppose, for contradiction, that for every $j,k \in I$ where $j \neq k$, we have $\subj(\alpha_j) \neq \subj(\alpha_k)$.
    Since $Q \vdash \emptyset$, for each $j \in I$, $\subj(\alpha_j)$ must be bound by a cut, so $\pctx{S}$ must contain $|I|$ cuts.
    This means $|I| = n$, contradicting the fact that $|I| \geq n + 1$.
    Therefore, there must be $j,k \in I$ where $j \neq k$ such that $\subj(\alpha_j) = \subj(\alpha_k)$.

    Hence, we can take $x = \subj(\alpha_j) = \subj(\alpha_k)$.
    Since $P \vdash \emptyset$ but $x$ appears free in $\alpha_j;P_j$ and $\alpha_k;P_k$, there must be a restriction on $x$ in $\pctx{S}$ containing the holes ${\hole}_j$ and ${\hole}_k$.
    We now obtain $\pctx{N}$ from $P$ by replacing the restriction on $x$ in $P$ with a hole, referring to the parallel component in which $\alpha_j;P_j$ appears as $P_j$ and the component in which $\alpha_k;P_k$ appears as $P_k$.
    Then, we obtain $\pctx{N_j}$ and $\pctx{N_k}$ from $P_j$ and $P_k$, respectively, by replacing $\alpha_j;P_j$ and $\alpha_k;P_k$ with a hole.
    Clearly, $P = \pctx[\big]{N}[\res{x}(\pctx{N_j}[\alpha_j;P_j] \| \pctx{N_k}[\alpha_k;P_k])]$.
\end{proof}

\thmDlfreeOne*

\begin{proof}
    By \Cref{ch4l:sctxform}, there are S-context $\pctx{S}$ with hole indices $I$ and sequence of prefixed processes ${(\alpha_i;P_i)}_{i \in I}$ such that $P \equiv \pctx{S}[\alpha_i;P_i]_{i \in I}$.
    The next step depends on whether there is a forwarder process among the $\alpha_i$.
    \begin{itemize}
        \item
            If there exists $j \in I$ s.t.\ $\alpha_j = \pfwd{x}{y}$ for some $x$ and $y$, then by \Cref{ch4l:sformfwd} there are $\pctx{N},\pctx{N'},Q$ s.t.\ $\pctx{S}[\alpha_i;P_i] = \pctx[\Big]{N}[\res{x}(\pctx[\big]{N'}[\pfwd{x}{y}] \| Q)]$.
            \begin{align*}
                &\res{x}(\pctx[\big]{N'}[\pfwd{x}{y}] \| Q)\redone \pctx*{N'}[Q\{y/x\}] = R'
                &&\text{(by rule $\rredone{\scc{Id}}$)}
                \\
                &\pctx{S}[\alpha_i;P_i]_{i \in I} = \pctx[\Big]{N}[\res{x}(\pctx[\big]{N'}[\pfwd{x}{y}] \| Q)] \redone \pctx{N}[R'] = R
                &&\text{(by rules $\rredone{\nu},\rredone{\|},\rredone{\nd}$)}
                \\
                &P \redone R
                &&\text{(by rule $\rredone{\equiv}$)}
            \end{align*}

        \item
            If for each $i \in I$, $\alpha_i \neq x \fwd y$ for any $x$ and $y$, then by \Cref{ch4l:sctxsubjs} there are $j,k \in I$ where $j \neq k$ and $x = \subj(\alpha_j) = \subj(\alpha_k)$ for some $x$, and $\pctx{N},\pctx{N_j},\pctx{N_k}$ such that $\pctx{S}[\alpha_i;P_i]_{i \in I} = \pctx[\big]{N}[\res{x}(\pctx{N_j}[\alpha_j;P_j] \| \pctx{N_k}[\alpha_k;P_k])]$.

            We now show by cases on $\alpha_j$ that there is $R'$ such that $\res{x}(\pctx{N_j}[\alpha_j;P_j] \| \pctx{N_k}[\alpha_k;P_k]) \redone R'$.
            First, note that by typability, if the type for $x$ in $\pctx{N_j}[\alpha_j;P_j]$ is $A$, then the type for $x$ in $\pctx{N_k}[\alpha_k;P_k]$ is $\ol{A}$.
            In the following cases, we determine more precisely the form of $A$ by typing inversion on $\pctx{N_j}[\alpha_j;P_j]$, and then determine the form of $\alpha_k$ by typing inversion using the form of $\ol{A}$.
            Note that we can exclude any cases where $\alpha_j$ or $\alpha_k$ are forwarder processes, as we assume they are not.
            \begin{itemize}
                \item
                    If $\alpha_j;P_j = \pclose{x}$, then $A = \1$ and $\ol{A} = \bot$.
                    Hence, $\alpha_k;P_k = \gclose{x};P_k$.
                    By rule $\rredone{\1\bot}$, there is $R'$ such that
                    \begin{align*}
                        \res{x}(\pctx{N_j}[\pclose{x}] \| \pctx{N_k}[\gclose{x};P_k]) \redone R'.
                    \end{align*}
                \item
                    If $\alpha_j;P_j = \pname{x}{y};(P'_j \| P''_j)$ for some $y$, then $A = B \tensor C$ and $\ol{A} = \ol{B} \parr \ol{C}$ for some $B$ and $C$.
                    Hence, $\alpha_k;P_k = \gname{x}{z}; P_k$ for some $z$.
                    By rule $\rredone{\tensor\parr}$, there is $R'$ such that
                    \begin{align*}
                        \res{x}(\pctx{N_j}[\pname{x}{y};(P'_j \| P''_j)] \| \pctx{N_k}[\gname{x}{z};P_k]) \redone R'.
                    \end{align*}

                \item
                    If $\alpha_j;P_j = \psel{x}{l'}; P_j$, then $A = {\oplus}\{l:B_l\}_{l \in L}$ and $\ol{A} = {\with}\{l:\ol{B_l}\}_{l \in L}$ for some ${(B_l)}_{l \in L}$ where $l' \in L$.
                    Hence, $\alpha_k;P_k = \gsel{x}\{l:P_k^l\}_{l \in L}$.
                    By rule $\rredone{\oplus\with}$, there is $R'$ such that
                    \begin{align*}
                        \res{x}(\pctx{N_j}[\psel{x}{l'}; P_j] \| \pctx{N_k}[\gsel{x}\{l:P_k^l\}_{l \in L}]) \redone R'.
                    \end{align*}

                \item
                    If $\alpha_j;P_j = \psome{x}; P_j$, then $A = {\oplus}B$ and $\ol{A} = {\with}\ol{B}$ for some $B$.
                    Hence, $\alpha_k = \gsome{x}{w_1,\ldots,w_n}; P_k$ for some $w_1,\ldots,w_n$.
                    By rule $\rredone{\some}$, there is $R'$ such that
                    \begin{align*}
                        \res{x}(\pctx{N_j}[\psome{x}; P_j] \| \pctx{N_k}[\gsome{x}{w_1,\ldots,w_n}; P_k]) \redone R'.
                    \end{align*}

                \item
                    If $\alpha_j;P_j = \pnone{x}$, then $A = {\oplus}B$ and $\ol{A} = {\with}\ol{B}$ for some $B$.
                    Hence, $\alpha_k = \gsome{x}{w_1,\ldots,w_n}; P_k$ for some $w_1,\ldots,w_n$.
                    By rule $\rredone{\none}$, there is $R'$ such that
                    \begin{align*}
                        \res{x}(\pctx{N_j}[\pnone{x}] \| \pctx{N_k}[\gsome{x}{w_1,\ldots,w_n}; P_k]) \redone R'.
                    \end{align*}

                \item
                    If $\alpha_j;P_j = \puname{x}{y}; P_j$ for some $y$, then $A = {?}B$ and $\ol{A} = {!}\ol{B}$ for some $B$.
                    Hence, $\alpha_k;P_k = \guname{x}{z}; P_k$ for some $z$.
                    By rule $\rredone{{?}{!}}$, there is $R'$ such that
                    \begin{align*}
                        \res{x}(\pctx{N_j}[\puname{x}{y}; P_j] \| \pctx{N_k}[\guname{x}{z}; P_k]) \redone R'.
                    \end{align*}

                \item
                    Otherwise, $\alpha_j$ is a receiving prefix and $\alpha_k$ is thus a sending prefix.
                    By cases on $\alpha_k$, the proof is analogous to above.
            \end{itemize}
            In conclusion,
            \begin{align*}
                \pctx{S}[\alpha_i;P_i]_{i \in I} = \pctx[\big]{N}[\res{x}(\pctx{N_j}[\alpha_j;P_j] \| \pctx{N_k}[\alpha_k;P_k])] \span\\
                \pctx[\big]{N}[\res{x}(\pctx{N_j}[\alpha_j;P_j] \| \pctx{N_k}[\alpha_k;P_k])] &\redone \pctx{N}[R'] = R
                &&\text{(by rules $\rredone{\nu},\rredone{\|},\rredone{\nd}$)}
                \\
                P &\redone R.
                &&\text{(by rule $\rredone{\equiv}$)}
            \end{align*}
    \end{itemize}
\end{proof}

\subsection{Lazy Semantics}
\label{ch4ss:proofsLazy}

\subsubsection{Subject Reduction}
\label{ch4ss:TPLazy}

\begin{lemma}\label{ch4l:ctxRedTwo}
    For both of the following items, assume $\Gamma \disj \Delta$.
    \begin{enumerate}
        \item\label{ch4i:ctxRedTwoName}
            If $\forall i \in I.~ \forall j \in J.~ \bn{\pctx{C_i}} \disj \fn{\pctx{D_j}}$ and $\forall i \in I.~ \pctx{C_i}[\pname{x}{y_i};(P_i \| Q_i)] \vdash \Gamma, x{:}A \tensor B$ and $\forall j \in J.~ \pctx{D_j}[\gname{x}{z};R_j] \vdash \Delta, x{:}\ol{A} \parr \ol{B}$, then $\bignd_{i \in I} \pctx[\Big]{C_i}[\res{x}\Big(Q_i \| \res{w}\Big(P_i\{w/y_i\} \| \bignd_{j \in J} \pctx{D_j}[R_j\{w/z\}]\Big)\Big)] \vdash \Gamma, \Delta$.

        \item\label{ch4i:ctxRedTwoUname}
            If $\forall i \in I.~ \forall j \in J.~ \bn{\pctx{D_j}} \disj \fn{\pctx{C_i}}$ and $\forall i \in I.~ \pctx{C_i}[\puname{x}{y_i};P_i] \vdash \Gamma, x{:}{?}A$ and $\forall j \in J.~ \pctx{D_j}[\guname{x}{z};Q_j] \vdash \Delta, x{:}{!}\ol{A}$, then $\bignd_{j \in J} \pctx[\Big]{D_j}[\res{x}\Big(\res{w}\Big(\bignd_{i \in I} \pctx{C_i}[P_i\{w/y_i\}] \| Q_j\{w/z\}\Big) \| \guname{x}{z};Q_j\Big)] \vdash \Gamma, \Delta$.
    \end{enumerate}
\end{lemma}

\begin{proof}
    Both items follow by induction on the structures of the D-contexts.
    For each item, we detail the base case, where $\forall i \in I.~ \pctx{C_i} = \hole$ and $\forall j \in J.~ \pctx{D_j} = \hole$.
    The inductive cases follow from the IH straightforwardly.
    \begin{enumerate}
        \item
        {\small
        \begin{adjustwidth}{-1cm}{}
            \begin{mathpar}
                \forall i \in I.~
                \inferrule{
                    P_i \vdash \Gamma, y_i{:}A
                    \\
                    Q_i \vdash \Delta, x{:}B
                }{
                    \pname{x}{y_i};(P_i \| Q_i) \vdash \Gamma, \Delta, x{:}A \tensor B
                }
                \and
                \forall j \in J.~
                \inferrule{
                    R_j \vdash \Lambda, z{:}\ol{A}, x{:}\ol{B}
                }{
                    \gname{x}{z};R_j \vdash \Lambda, x{:}\ol{A} \parr \ol{B}
                }
                \and\implies\and
                \inferruleDbl{
                    \inferrule*[fraction={---}]{
                        \forall i \in I.~
                        Q_i \vdash \Delta, x{:}B
                        \\
                        \inferrule*{
                            \forall i \in I.~
                            P_i\{w/y_i\} \vdash \Gamma, w{:}A
                            \\
                            \inferruleDbl{
                                \forall j \in J.~ R_j\{w/z\} \vdash \Lambda, w{:}\ol{A}, x{:}\ol{B}
                            }{
                                \bignd_{j \in J} R_j\{w/z\} \vdash \Lambda, w{:}\ol{A}, x{:}\ol{B}
                            }
                        }{
                            \forall i \in I.~
                            \res{w}\Big(P_i\{w/y_i\} \| \bignd_{j \in J} R_j\{w/z\}\Big) \vdash \Gamma, \Lambda, x{:}\ol{B}
                        }
                    }{
                        \forall i \in I.~
                        \res{x}\Big(Q_i \| \res{w}\Big(P_i\{w/y_i\} \| \bignd_{j \in J} R_j\{w/z\}\Big)\Big) \vdash \Gamma, \Delta, \Lambda
                    }
                }{
                    \bignd_{i \in I} \res{x}\Big(Q_i \| \res{w}\Big(P_i\{w/y_i\} \| \bignd_{j \in J} R_j\{w/z\}\Big)\Big) \vdash \Gamma, \Delta, \Lambda
                }
            \end{mathpar}
        \end{adjustwidth}
        }
        \item
            This item depends on whether $x \in \fn{P_i}$ or not, for each $i \in I$.
            For simplicity, we only consider the cases where either $\forall i \in I.~ x \in \fn{P_i}$ or $\forall i \in I.~ x \notin \fn{P_i}$.
            \begin{itemize}
                \item
                    $\forall i \in I.~ x \in \fn{P_i}$.
                    {\small
                    \begin{adjustwidth}{-2.0cm}{}
                    \begin{mathpar}
                        \forall i \in I.~
                        \inferrule{
                            \inferrule{
                                P_i\{x'/x\} \vdash \Gamma, y_i{:}A, x'{:}{?}A
                            }{
                                \puname{x}{y_i};(P_i\{x'/x\}) \vdash \Gamma, x{:}{?}A, x'{:}{?}A
                            }
                        }{
                            \puname{x}{y_i};P_i \vdash \Gamma, x{:}{?}A
                        }
                        \and
                        \forall j \in J.~
                        \inferrule{
                            Q_j \vdash {?}\Delta, x{:}\ol{A}
                        }{
                            \guname{x}{z};Q_j \vdash {?}\Delta, x{:}{!}\ol{A}
                        }
                        \and\implies\and
                        \inferruleDbl{
                            \inferruleDbl{
                                \inferrule*[fraction={---}]{
                                    \inferrule*{
                                        \inferruleDbl{
                                            {\begin{tarr}[b]{l}
                                                    \forall i \in I.
                                                    \\
                                                    P_i\{w/y_i\} \vdash \Gamma, w{:}A, x{:}{?}A
                                            \end{tarr}}
                                        }{
                                            \bignd_{i \in I} P_i\{w/y_i\} \vdash \Gamma, w{:}A, x{:}{?}A
                                        }
                                        \\
                                        {\begin{tarr}[b]{l}
                                                \forall j \in J.
                                                \\
                                                Q_j\{w/z\}\{v'/v\}_{v \in {?}\Delta} \vdash {?}\Delta', w{:}\ol{A}
                                        \end{tarr}}
                                    }{
                                        {\begin{tarr}[t]{l}
                                                \forall j \in J.
                                                \\
                                                \res{w}\Big(\bignd_{i \in I} P_i\{w/y_i\} \| Q_j\{w/z\}\{v'/v\}_{v \in {?}\Delta}\Big) \vdash \Gamma, {?}\Delta', x{:}{?}A
                                        \end{tarr}}
                                    }
                                    \\
                                    {\begin{tarr}[b]{l}
                                            \forall j \in J.
                                            \\
                                            \guname{x}{z};Q_j \vdash {?}\Delta, x{:}{!}\ol{A}
                                    \end{tarr}}
                                }{
                                    \forall j \in J.~
                                    \res{x}\Big(\res{w}\Big(\bignd_{i \in I} P_i\{w/y_i\} \| Q_j\{w/z\}\{v'/v\}_{v \in {?}\Delta}\Big) \| \guname{x}{z};Q_j\Big) \vdash \Gamma, {?}\Delta, {?}\Delta'
                                }
                            }{
                                \bignd_{j \in J} \res{x}\Big(\res{w}\Big(\bignd_{i \in I} P_i\{w/y_i\} \| Q_j\{w/z\}\{v'/v\}_{v \in {?}\Delta}\Big) \| \guname{x}{z};Q_j\Big) \vdash \Gamma, {?}\Delta, {?}\Delta'
                            }
                        }{
                            \bignd_{j \in J} \res{x}\Big(\res{w}\Big(\bignd_{i \in I} P_i\{w/y_i\} \| Q_j\{w/z\}\Big) \| \guname{x}{z};Q_j\Big) \vdash \Gamma, {?}\Delta
                        }
                    \end{mathpar}
                \end{adjustwidth}}

                \item
                    $\forall i \in I.~ x \notin \fn{P_i}$.
                    {\small
                    \begin{adjustwidth}{-1.5cm}{}
                    \begin{mathpar}
                        \forall i \in I.~
                        \inferrule{
                            P_i \vdash \Gamma, y_i{:}A
                        }{
                            \puname{x}{y_i};P_i \vdash \Gamma, x{:}{?}A
                        }
                        \and
                        \forall j \in J.~
                        \inferrule{
                            Q_j \vdash {?}\Delta, x{:}\ol{A}
                        }{
                            \guname{x}{z};Q_j \vdash {?}\Delta, x{:}{!}\ol{A}
                        }
                        \and\implies\and
                        \inferruleDbl{
                            \inferruleDbl{
                                \inferrule*[fraction={---}]{
                                    \inferrule*{
                                        \inferrule*{
                                            \inferruleDbl{
                                                {\begin{tarr}[b]{l}
                                                        \forall i \in I.
                                                        \\
                                                        P_i\{w/y_i\} \vdash \Gamma, w{:}A
                                                \end{tarr}}
                                            }{
                                                \bignd_{i \in I} P_i\{w/y_i\} \vdash \Gamma, w{:}A
                                            }
                                            \\
                                            {\begin{tarr}[b]{l}
                                                    \forall j \in J.
                                                    \\
                                                    Q_j\{w/z\}\{v'/v\}_{v \in {?}\Delta} \vdash {?}\Delta', w{:}\ol{A}
                                            \end{tarr}}
                                        }{
                                            {\begin{tarr}[t]{l}
                                                    \forall j \in J.
                                                    \\
                                                    \res{w}\Big(\bignd_{i \in I} P_i\{w/y_i\} \| Q_j\{w/z\}\{v'/v\}_{v \in {?}\Delta}\Big) \vdash \Gamma, {?}\Delta'
                                            \end{tarr}}
                                        }
                                    }{
                                        {\begin{tarr}[t]{l}
                                                \forall j \in J.
                                                \\
                                                \res{w}\Big(\bignd_{i \in I} P_i\{w/y_i\} \| Q_j\{w/z\}\{v'/v\}_{v \in {?}\Delta}\Big) \vdash \Gamma, {?}\Delta', x{:}{?}A
                                        \end{tarr}}
                                    }
                                    \\
                                    {\begin{tarr}[b]{l}
                                            \forall j \in J.
                                            \\
                                            \guname{x}{z};Q_j \vdash {?}\Delta, x{:}{!}\ol{A}
                                    \end{tarr}}
                                }{
                                    \forall j \in J.~
                                    \res{x}\Big(\res{w}\Big(\bignd_{i \in I} P_i\{w/y_i\} \| Q_j\{w/z\}\{v'/v\}_{v \in {?}\Delta}\Big) \| \guname{x}{z};Q_j\Big) \vdash \Gamma, {?}\Delta, {?}\Delta'
                                }
                            }{
                                \bignd_{j \in J} \res{x}\Big(\res{w}\Big(\bignd_{i \in I} P_i\{w/y_i\} \| Q_j\{w/z\}\{v'/v\}_{v \in {?}\Delta}\Big) \| \guname{x}{z};Q_j\Big) \vdash \Gamma, {?}\Delta, {?}\Delta'
                            }
                        }{
                            \bignd_{j \in J} \res{x}\Big(\res{w}\Big(\bignd_{i \in I} P_i\{w/y_i\} \| Q_j\{w/z\}\Big) \| \guname{x}{z};Q_j\Big) \vdash \Gamma, {?}\Delta
                        }
                    \end{mathpar}
                \end{adjustwidth}}
            \end{itemize}
    \end{enumerate}
\end{proof}

\begin{theorem}[SR for the Lazy Semantics]\label{ch4t:srTwo}
    If $P \vdash \Gamma$ and $P \redtwo_S Q$, then $Q \vdash \Gamma$.
\end{theorem}

\begin{proof}
    By induction on the derivation of the reduction.
    \begin{itemize}
        \item
            Rule $\rredtwo{\scc{Id}}$.
            \begin{mathpar}
                \inferrule{
                    \inferruleDbl{
                        \forall i \in I.~
                        \pctx[\big]{C_i}[\pfwd{x}{y}] \vdash \Gamma, x{:}A
                    }{
                        \bignd_{i \in I} \pctx[\big]{C_i}[\pfwd{x}{y}] \vdash \Gamma, x{:}A
                    }
                    \\
                    Q \vdash \Delta, x{:}\ol{A}
                }{
                    \res{x}\Big(\bignd_{i \in I} \pctx[\big]{C_i}[\pfwd{x}{y}] \| Q\Big) \vdash \Gamma, \Delta
                }
                \and\implies\and
                \inferruleDbl{
                    \forall i \in I.~
                    \pctx{C_i}[Q\{y/x\}] \vdash \Gamma, \Delta
                    ~~\text{(\refitem{l}{ctxRedOne}{Fwd})}
                }{
                    \bignd_{i \in I} \pctx{C_i}[Q\{y/x\}] \vdash \Gamma, \Delta
                }
            \end{mathpar}

        \item
            Rule $\rredtwo{\1\bot}$.
            \begin{mathpar}
                \inferrule{
                    \inferruleDbl{
                        \forall i \in I.~
                        \pctx{C_i}[\pclose{x}] \vdash \Gamma, x{:}\1
                        ~~\text{(\refitem{l}{ctxType}{Pclose})}
                    }{
                        \bignd_{i \in I} \pctx{C_i}[\pclose{x}] \vdash \Gamma, x{:}\1
                    }
                    \\
                    \inferruleDbl{
                        \forall j \in J.~
                        \pctx{D_j}[\gclose{x};Q_j] \vdash \Delta, x{:}\bot
                        ~~\text{(\refitem{l}{ctxType}{Gclose})}
                    }{
                        \bignd_{j \in J} \pctx{D_j}[\gclose{x};Q_j] \vdash \Delta, x{:}\bot
                    }
                }{
                    \res{x}\Big(\bignd_{i \in I} \pctx{C_i}[\pclose{x}] \| \bignd_{j \in J} \pctx{D_j}[\gclose{x};Q_j]\Big) \vdash \Gamma, \Delta
                }
                \and\implies\and
                \inferrule{
                    \inferruleDbl{
                        \forall i \in I.~
                        \pctx{C_i}[\0] \vdash \Gamma
                        ~~\text{(\refitem{l}{ctxRedOne}{Pclose})}
                    }{
                        \bignd_{i \in I} \pctx{C_i}[\0] \vdash \Gamma
                    }
                    \\
                    \inferruleDbl{
                        \forall j \in J.~
                        \pctx{D_j}[Q_j] \vdash \Delta
                        ~~\text{(\refitem{l}{ctxRedOne}{Gclose})}
                    }{
                        \bignd_{j \in J} \pctx{D_j}[Q_j] \vdash \Delta
                    }
                }{
                    \bignd_{i \in I} \pctx{C_i}[\0] \| \bignd_{j \in J} \pctx{D_j}[Q_j] \vdash \Gamma, \Delta
                }
            \end{mathpar}

        \item
            Rule $\rredtwo{\tensor\parr}$.
            {\small
            \begin{mathpar}
                \mprset{sep=5em}
                \inferrule{
                    \inferruleDbl{
                        \forall i \in I.~
                        \pctx{C_i}[\pname{x}{y_i};(P_i \| Q_i)] \vdash \Gamma, x{:}A \tensor B
                        ~~\text{(\refitem{l}{ctxType}{Pname})}
                    }{
                        \bignd_{i \in I} \pctx{C_i}[\pname{x}{y_i};(P_i \| Q_i)] \vdash \Gamma, x{:}A \tensor B
                    }
                    \\
                    \inferruleDbl{
                        \forall j \in J.~
                        \pctx{D_j}[\gname{x}{z};R_j] \vdash \Delta, x{:}\ol{A} \parr \ol{B}
                        ~~\text{(\refitem{l}{ctxType}{Gname})}
                    }{
                        \bignd_{j \in J} \pctx{D_j}[\gname{x}{z};R_j] \vdash \Delta, x{:}\ol{A} \parr \ol{B}
                    }
                }{
                    \res{x}\Big(\bignd_{i \in I} \pctx{C_i}[\pname{x}{y_i};(P_i \| Q_i)] \| \bignd_{j \in J} \pctx{D_j}[\gname{x}{z};R_j]\Big) \vdash \Gamma, \Delta
                }
                \and\implies\and
                \inferrule{}{
                    \bignd_{i \in I} \pctx[\Big]{C_i}[\res{x}\Big(Q_i \| \res{w}\Big(P_i\{w/y_i\} \| \bignd_{j \in J} \pctx{D_j}[R_j\{w/z\}]\Big)\Big)] \vdash \Gamma, \Delta
                    ~~\text{(\refitem{l}{ctxRedTwo}{Name})}
                }
            \end{mathpar}}

        \item
            Rule $\rredtwo{{\oplus}{\with}}$.
            Take any $k' \in K$.
            \begin{mathpar}
                \inferrule{
                    \inferruleDbl{
                        {\begin{tarr}[b]{l}
                                \forall i \in I.
                                \\
                                \pctx{C_i}[\psel{x}{k'};P_i] \vdash \Gamma, x{:}{\oplus}\{k:A_k\}_{k \in K}
                                \\
                                \text{(\refitem{l}{ctxType}{Psel})}
                        \end{tarr}}
                    }{
                        \bignd_{i \in I} \pctx{C_i}[\psel{x}{k'};P_i] \vdash \Gamma, x{:}{\oplus}\{k:A_k\}_{k \in K}
                    }
                    \\
                    \inferruleDbl{
                        {\begin{tarr}[b]{l}
                                \forall j \in J.
                                \\
                                \pctx{D_j}[\gsel{x}\{k:Q_j^k\}_{k \in K}] \vdash \Delta, x{:}{\with}\{k:\ol{A_k}\}_{k \in K}
                                \\
                                \text{(\refitem{l}{ctxType}{Gsel})}
                        \end{tarr}}
                    }{
                        \bignd_{j \in J} \pctx{D_j}[\gsel{x}\{k:Q_j^k\}_{k \in K}] \vdash \Delta, x{:}{\with}\{k:\ol{A_k}\}_{k \in K}
                    }
                }{
                    \res{x}\Big(\bignd_{i \in I} \pctx{C_i}[\psel{x}{k'};P_i] \| \bignd_{j \in J} \pctx{D_j}[\gsel{x}\{k:Q_j^k\}_{k \in K}]\Big) \vdash \Gamma, \Delta
                }
                \and\implies\and
                \inferrule{
                    \inferruleDbl{
                        \forall i \in I.~
                        \pctx{C_i}[P_i] \vdash \Gamma, x{:}A
                        ~~\text{(\refitem{l}{ctxRedOne}{Psel})}
                    }{
                        \bignd_{i \in I} \pctx{C_i}[P_i] \vdash \Gamma, x{:}A
                    }
                    \\
                    \inferruleDbl{
                        \forall j \in J.~
                        \pctx{D_j}[Q_j^k] \vdash \Delta, x{:}\ol{A}
                        ~~\text{(\refitem{l}{ctxRedOne}{Gsel})}
                    }{
                        \bignd_{j \in J} \pctx{D_j}[Q_j^k] \vdash \Delta, x{:}\ol{A}
                    }
                }{
                    \res{x}\Big(\bignd_{i \in I} \pctx{C_i}[P_i] \| \bignd_{j \in J} \pctx{D_j}[Q_j^k]\Big) \vdash \Gamma, \Delta
                }
            \end{mathpar}

        \item
            Rule $\rredtwo{\some}$.
            \begin{mathpar}
                \inferrule{
                    \inferruleDbl{
                        {\begin{tarr}[b]{l}
                                \forall i \in I.
                                \\
                                \pctx{C_i}[\psome{x};P_i] \vdash \Gamma, x{:}{\with}A
                                \\
                                \text{(\refitem{l}{ctxType}{Psome})}
                        \end{tarr}}
                    }{
                        \bignd_{i \in I} \pctx{C_i}[\psome{x};P_i] \vdash \Gamma, x{:}{\with}A
                    }
                    \\
                    \inferruleDbl{
                        {\begin{tarr}[b]{l}
                                \forall j \in J.
                                \\
                                \pctx{D_j}[\gsome{x}{w_1,\ldots,w_n};Q_j] \vdash \Delta, x{:}{\oplus}\ol{A}
                                \\
                                \text{(\refitem{l}{ctxType}{Gsome})}
                        \end{tarr}}
                    }{
                        \bignd_{j \in J} \pctx{D_j}[\gsome{x}{w_1,\ldots,w_n};Q_j] \vdash \Delta, x{:}{\oplus}\ol{A}
                    }
                }{
                    \res{x}\Big( \bignd_{i \in I} \pctx{C_i}[\psome{x};P_i] \| \bignd_{j \in J} \pctx{D_j}[\gsome{x}{w_1,\ldots,w_n};Q_j]\Big) \vdash \Gamma, \Delta
                }
                \and\implies\and
                \inferrule{
                    \inferruleDbl{
                        \forall i \in I.~
                        \pctx{C_i}[P_i] \vdash \Gamma, x{:}A
                        ~~\text{(\refitem{l}{ctxRedOne}{Psome})}
                    }{
                        \bignd_{i \in I} \pctx{C_i}[P_i] \vdash \Gamma, x{:}A
                    }
                    \\
                    \inferruleDbl{
                        \forall j \in J.~
                        \pctx{D_j}[Q_j] \vdash \Delta, x{:}\ol{A}
                        ~~\text{(\refitem{l}{ctxRedOne}{Gsome})}
                    }{
                        \bignd_{j \in J} \pctx{D_j}[Q_j] \vdash \Delta, x{:}\ol{A}
                    }
                }{
                    \res{x}\Big(\bignd_{i \in I} \pctx{C_i}[P_i] \| \bignd_{j \in J} \pctx{D_j}[Q_j]\Big) \vdash \Gamma, \Delta
                }
            \end{mathpar}

        \item
            Rule $\rredtwo{\none}$.
            \begin{mathpar}
                \inferrule{
                    \inferruleDbl{
                        {\begin{tarr}[b]{l}
                                \forall i \in I.
                                \\
                                \pctx{C_i}[\pnone{x}] \vdash \Gamma, x{:}{\with}A
                                \\
                                \text{(\refitem{l}{ctxType}{Pnone})}
                        \end{tarr}}
                    }{
                        \bignd_{i \in I} \pctx{C_i}[\pnone{x}] \vdash \Gamma, x{:}{\with}A
                    }
                    \\
                    \inferruleDbl{
                        {\begin{tarr}[b]{l}
                                \forall j \in J.
                                \\
                                \pctx{D_j}[\gsome{x}{w_1,\ldots,w_n};Q_j] \vdash \Delta, x{:}{\oplus}\ol{A}
                                \\
                                \text{(\refitem{l}{ctxType}{Gsome})}
                        \end{tarr}}
                    }{
                        \bignd_{j \in J} \pctx{D_j}[\gsome{x}{w_1,\ldots,w_n};Q_j] \vdash \Delta, x{:}{\oplus}\ol{A}
                    }
                }{
                    \res{x}\Big( \bignd_{i \in I} \pctx{C_i}[\pnone{x}] \| \bignd_{j \in J} \pctx{D_j}[\gsome{x}{w_1,\ldots,w_n};Q_j]\Big) \vdash \Gamma, \Delta
                }
                \and\implies\and
                \inferrule{
                    \inferruleDbl{
                        \forall i \in I.~
                        \pctx{C_i}[\0] \vdash \Gamma
                        ~~\text{(\refitem{l}{ctxRedOne}{Pnone})}
                    }{
                        \bignd_{i \in I} \pctx{C_i}[P_i] \vdash \Gamma
                    }
                    \\
                    \inferruleDbl{
                        \forall j \in J.~
                        \pctx{D_j}[\pnone{w_1} \| \ldots \| \pnone{w_n}] \vdash \Delta
                        ~~\text{(\refitem{l}{ctxRedOne}{Gsome})}
                    }{
                        \bignd_{j \in J} \pctx{D_j}[\pnone{w_1} \| \ldots \| \pnone{w_n}] \vdash \Delta
                    }
                }{
                    \bignd_{i \in I} \pctx{C_i}[\0] \| \bignd_{j \in J} \pctx{D_j}[\pnone{w_1} \| \ldots \| \pnone{w_n}] \vdash \Gamma, \Delta
                }
            \end{mathpar}

        \item
            Rule $\rredtwo{{?}{!}}$.
            {\small\begin{mathpar}
                \inferrule{
                    \inferruleDbl{
                        \forall i \in I.~
                        \pctx{C_i}[\puname{x}{y_i};P_i] \vdash \Gamma, x{:}{?}A
                        ~~\text{(\refitem{l}{ctxType}{Puname})}
                    }{
                        \bignd_{i \in I} \pctx{C_i}[\puname{x}{y_i};P_i] \vdash \Gamma, x{:}{?}A
                    }
                    \\
                    \inferruleDbl{
                        \forall j \in J.~
                        \pctx{D_j}[\guname{x}{z};Q_j] \vdash \Delta, x{:}{!}\ol{A}
                        ~~\text{(\refitem{l}{ctxType}{Guname})}
                    }{
                        \bignd_{j \in J} \pctx{D_j}[\guname{x}{z};Q_j] \vdash \Delta, x{:}{!}\ol{A}
                    }
                }{
                    \res{x}\Big(\bignd_{i \in I} \pctx{C_i}[\puname{x}{y_i};P_i] \| \bignd_{j \in J} \pctx{D_j}[\guname{x}{z};Q_j]\Big) \vdash \Gamma, \Delta
                }
                \and\implies\and
                \inferrule{}{
                    \bignd_{j \in J} \pctx[\Big]{D_j}[\res{x}\Big(\res{w}\Big(\bignd_{i \in I} \pctx{C_i}[P_i\{w/z\}] \| Q_j\{w/z\}\Big) \| \guname{x}{z};Q_j\Big)] \vdash \Gamma, \Delta
                    ~~\text{(\refitem{l}{ctxRedTwo}{Uname})}
                }
            \end{mathpar}}
    \end{itemize}
\end{proof}

\subsubsection{Deadlock Freedom}
\label{ch4ss:DFLazy}

\begin{definition}{Multi-hole Non-deterministic Reduction Contexts}
    \begin{align*}
        \pctx{M} ::= \hole \sepr \res{x}(P \| \pctx{M}) \sepr P \| \pctx{M} \sepr \pctx{M} \nd \pctx{M}
    \end{align*}
\end{definition}

\begin{lemma}\label{ch4l:ndCtxMultihole}
    If $\pctx{N}[\alpha;P] \vdash \Gamma, x{:}A$ and $x = \subj(\alpha)$, then there are $\pctx{M}$ and ${(\alpha_i;P_i)}_{i \in I}$ such that $\pctx{N}[\alpha;P] = \pctx{M}[\alpha_i;P_i]_{i \in I}$ where $x \notin \fn{\pctx{M}}$ and $x \in \bigcap_{i \in I} \fn{\alpha_i;P_i}$ and there is $i' \in I$ such that $\alpha_{i'};P_{i'} = \alpha;P$.
\end{lemma}

\begin{lemma}\label{ch4l:multiNdDet}
    For every multi-hole ND-context $\pctx{M}$ with indices $I$:
    \begin{itemize}
        \item
            If $\pctx{M}$ has two or more holes, there are $\pctx{C}$, $\pctx{M_1}$ with indices $I_1$, $\pctx{M_2}$ with indices $I_2$ such that $\pctx{M} = \pctx{C}[\pctx{M_1} \nd \pctx{M_2}]$ where $I_1 \disj I_2$ and $I = I_1 \cup I_2$.

        \item
            If $\pctx{M}$ has only one hole, there is $\pctx{C}$ such that $\pctx{M} = \pctx{C}$.
    \end{itemize}
\end{lemma}

\begin{definition}{}\label{ch4d:ndFlat}
    \begin{align*}
        \flat{\pctx{C}[\pctx{M} \nd \pctx{M'}]}
        &:= \flat{\pctx{C}[\pctx{M}]} \nd \flat{\pctx{C}[\pctx{M'}]}
        &
        \flat{\pctx{C}}
        &:= \pctx{C}
    \end{align*}
\end{definition}

\begin{lemma}\label{ch4l:multiShape}
    If $\pctx{M}[P_i]_{i \in I} \vdash \Gamma$ where $x \notin \fn{\pctx{M}}$ and $\forall i \in I.~ x \in \fn{P_i}$, then there are ${(\pctx{C_i})}_{i \in I}$ such that $\flat{\pctx{M}}[P_i]_{i \in I} = \bignd_{i \in I} \pctx{C_i}[P_i]$ where $\forall i \in I.~ x \notin \fn{\pctx{C_i}}$.
\end{lemma}

\begin{lemma}\label{ch4l:flatRed}
    If
    \begin{align*}
        \res{x}(\pctx[\big]{N}[\flat{\pctx{M}}[\alpha_i;P_i]_{i \in I}] \| \pctx[\big]{N'}[\flat{\pctx{M'}}[\beta_j;Q_j]_{j \in J}]) \redtwo_S R,
    \end{align*}
    then
    \begin{align*}
    \res{x}(\pctx[\big]{N}[\pctx{M}[\alpha_i;P_i]_{i \in I}] \| \pctx[\big]{N'}[\pctx{M'}[\beta_j;Q_j]_{j \in J}]) \redtwo_S R.
    \end{align*}
\end{lemma}

\begin{proof}
    By induction on the structures of $\pctx{M}$ and $\pctx{M'}$.
    By \Cref{ch4l:multiNdDet}, we only have to consider two cases for $\pctx{M}$ ($\pctx{M} = \pctx{C}[\pctx{M_1} \nd \pctx{M_2}]$ and $\pctx{M} = \pctx{C}$), and similarly for $\pctx{M'}$.
    We only detail the base case ($\pctx{M} = \pctx{C}$ and $\pctx{M'} = \pctx{C'}$) and a representative inductive case ($\pctx{M} = \pctx{C}[\pctx{M_1} \nd \pctx{M_2}]$ and $\pctx{M'} = \pctx{C'}$).
    \begin{itemize}
        \item
            $\pctx{M} = \pctx{C}$ and $\pctx{M'} = \pctx{C'}$.
            Note that $\pctx{M}$ and $\pctx{M'}$ have only one hole; w.l.o.g., assume $I = J = \{1\}$.
            \begin{align*}
                \flat{\pctx{M}}[\alpha_1;P_1] &= \flat{\pctx{C}}[\alpha_1;P_1] = \pctx{C}[\alpha_1;P_1] = \pctx{M}[\alpha_1;P_1]
                \\
                \flat{\pctx{M'}}[\beta_1;Q_1] &= \flat{\pctx{C'}}[\beta_1;Q_1] = \pctx{C'}[\beta_1;Q_1] = \pctx{M'}[\beta_1;Q_1]
            \end{align*}
            The thesis follows by assumption and equality.

        \item
            $\pctx{M} = \pctx{C}[\pctx{M_1} \nd \pctx{M_2}]$ and $\pctx{M'} = \pctx{C'}$.
            Note that $\pctx{M'}$ has only one hole; w.l.o.g., assume $J = \{1\}$.
            \begin{align}
                & \res{x}(\pctx[\big]{N}[\flat{\pctx{M}}[\alpha_i;P_i]_{i \in I}] \| \pctx[\big]{N'}[\flat{\pctx{M'}}[\beta_1;Q_1]])
                \nonumber
                \\
                &= \res{x}(\pctx[\big]{N}[\flat{\pctx{C}[\pctx{M_1} \nd \pctx{M_2}]}[\alpha_i;P_i]_{i \in I}] \| \pctx[\big]{N'}[\flat{\pctx{C'}}[\beta_1;Q_1]])
                \nonumber
                \\
                &= \res{x}(\pctx[\big]{N}[(\flat{\pctx{C}[\pctx{M_1}]} \nd \flat{\pctx{C}[\pctx{M_2}]})[\alpha_i;P_i]_{i \in I}] \| \pctx[\big]{N'}[\pctx{C'}[\beta_1;Q_1]])
                \label{ch4eq:oneImplTwoBeforeSplit}
            \end{align}
            There are $I_1$ and $I_2$ such that $I_1 \disj I_2$ and $I = I_1 \cup I_2$ and
            \begin{align}
                & \res{x}(\pctx[\big]{N}[\flat{\pctx{M}}[\alpha_i;P_i]_{i \in I}] \| \pctx[\big]{N'}[\flat{\pctx{M'}}[\beta_1;Q_1]])
                \nonumber
                \\
                &= \res{x}(\pctx[\big]{N}[\flat{\pctx{C}[\pctx{M_1}]}[\alpha_i;P_i]_{i \in I_1} \nd \flat{\pctx{C}[\pctx{M_2}]}[\alpha_i;P_i]_{i \in I_2}] \| \pctx[\big]{N'}[\pctx{C'}[\beta_1;Q_1]]).
                &&\text{(by~\eqref{ch4eq:oneImplTwoBeforeSplit})}
                \label{ch4eq:oneImplTwoAfterSplit}
            \end{align}
            Let $\pctx{N_1} = \pctx[\big]{N}[\hole \nd \flat{\pctx{C}[\pctx{M_2}]}[\alpha_i;P_i]_{i \in I_2}]$.
            \begin{align}
                & \res{x}(\pctx[\big]{N}[\flat{\pctx{M}}[\alpha_i;P_i]_{i \in I}] \| \pctx[\big]{N'}[\flat{\pctx{M'}}[\beta_1;Q_1]])
                \nonumber
                \\
                &= \res{x}(\pctx[\big]{N_1}[\flat{\pctx{C}[\pctx{M_1}]}[\alpha_i;P_i]_{i \in I_1}] \| \pctx[\big]{N'}[\pctx{C'}[\beta_1;Q_1]])
                &&\text{(by~\eqref{ch4eq:oneImplTwoAfterSplit})}
                \nonumber
                \\
                &\redtwo_S R
                &&\text{(by assumption)}
                \label{ch4eq:oneImplTwoLeftRed}
                \\
                & \res{x}(\pctx[\big]{N_1}[\pctx{C}[\pctx{M_1}][\alpha_i;P_i]_{i \in I_1}] \| \pctx[\big]{N'}[\pctx{C'}[\beta_1;Q_1]])
                \nonumber
                \\
                &= \res{x}(\pctx[\big]{N}[\pctx{C}[\pctx{M_1}][\alpha_i;P_i]_{i \in I_1} \nd \flat{\pctx{C}[\pctx{M_2}]}[\alpha_i;P_i]_{i \in I_2}] \| \pctx[\big]{N'}[\pctx{C'}[\beta_1;Q_1]])
                \nonumber
                \\
                &\redtwo_S R
                &&\text{(by IH on~\eqref{ch4eq:oneImplTwoLeftRed})}
                \label{ch4eq:oneImplTwoBeforeRightRed}
            \end{align}
        Let $\pctx{N_2} = \pctx[\big]{N}[\pctx{C}[\pctx{M_1}][\alpha_i;P_i]_{i \in I_1} \nd \hole]$.
            \begin{align}
                & \res{x}(\pctx[\big]{N}[\pctx{C}[\pctx{M_1}][\alpha_i;P_i]_{i \in I_1} \nd \flat{\pctx{C}[\pctx{M_2}]}[\alpha_i;P_i]_{i \in I_2}] \| \pctx[\big]{N'}[\pctx{C'}[\beta_1;Q_1]])
                \nonumber
                \\
                &= \res{x}(\pctx[\big]{N_2}[\flat{\pctx{C}[\pctx{M_2}]}[\alpha_i;P_i]_{i \in I_2}] \| \pctx[\big]{N'}[\pctx{C'}[\beta_1;Q_1]])
                \nonumber
                \\
                &\redtwo_S R
                &&\text{(by~\eqref{ch4eq:oneImplTwoBeforeRightRed})}
                \label{ch4eq:oneImplTwoRightRed}
                \\
                & \res{x}(\pctx[\big]{N_2}[\pctx{C}[\pctx{M_2}][\alpha_i;P_i]_{i \in I_2}] \| \pctx[\big]{N'}[\pctx{C'}[\beta_1;Q_1]])
                \nonumber
                \\
                &= \res{x}(\pctx[\big]{N}[\pctx{C}[\pctx{M_1}][\alpha_i;P_i]_{i \in I_1} \nd \pctx{C}[\pctx{M_2}][\alpha_i;P_i]_{i \in I_2}] \| \pctx[\big]{N'}[\pctx{C'}[\beta_1;Q_1]])
                \nonumber
                \\
                &= \res{x}(\pctx[\Big]{N}[\pctx[\big]{C}[\pctx{M_1}[\alpha_i;P_i]_{i \in I_1}] \nd \pctx[\big]{C}[\pctx{M_2}[\alpha_i;P_i]_{i \in I_2}]] \| \pctx[\big]{N'}[\pctx{C'}[\beta_1;Q_1]])
                \nonumber
                \\
                &\redtwo_S R
                &&\text{(by IH on~\eqref{ch4eq:oneImplTwoRightRed})}
                \label{ch4eq:oneImplTwoIHRed}
                \\
                & \res{x}(\pctx[\Big]{N}[\pctx[\big]{C}[\pctx{M_1}[\alpha_i;P_i]_{i \in I_1} \nd \pctx{M_2}[\alpha_i;P_i]_{i \in I_2}]] \| \pctx[\big]{N'}[\pctx{C'}[\beta_1;Q_1]])
                \nonumber
                \\
                &= \res{x}(\pctx[\Big]{N}[\pctx[\big]{C}[(\pctx{M_1} \nd \pctx{M_2})[\alpha_i;P_i]_{i \in I}]] \| \pctx[\big]{N'}[\pctx{C'}[\beta_1;Q_1]])
                \nonumber
                \\
                &= \res{x}(\pctx[\big]{N}[\pctx{C}[\pctx{M_1} \nd \pctx{M_2}][\alpha_i;P_i]_{i \in I}] \| \pctx[\big]{N'}[\pctx{C'}[\beta_1;Q_1]])
                \nonumber
                \\
                &= \res{x}(\pctx[\big]{N}[\pctx{M}[\alpha_i;P_i]_{i \in I}] \| \pctx[\big]{N'}[\pctx{M'}[\beta_1;Q_1]])
                \nonumber
                \\
                &\redtwo_S R
                && \hspace{-0.5cm}\text{(by rule $\rredtwo{\nu\nd}$ on~\eqref{ch4eq:oneImplTwoIHRed})}
                \nonumber
            \end{align}
    \end{itemize}
\end{proof}

\begin{theorem}\label{ch4t:oneImpliesTwo}
    If $P \vdash \Gamma$ and $P \redone R$, then $P \redtwo_S R$.
\end{theorem}

\begin{proof}
    By induction on the derivation of the reduction.
    The inductive cases of rules $\rredone{\equiv}$, $\rredone{\nu}$, $\rredone{\|}$, and $\rredone{\nd}$ follow from the IH straightforwardly, using the corresponding closure rule for $\redtwo$.
    As representative base case, we consider rule $\rredone{\1\bot}$: $P = \res{x}(\pctx{N}[\pclose{x}] \| \pctx{N'}[\gclose{x};Q]) \redone R$.

    By inversion of typing, $\pctx{N}[\pclose{x}] \vdash \Gamma, x{:}A$, so by \Cref{ch4l:ndCtxMultihole}, there are $\pctx{M}$ and ${(\alpha_i;P_i)}_{i \in I}$ such that $\pctx{N}[\pclose{x}] = \pctx{M}[\alpha_i;P_i]_{i \in I}$ where $x \notin \fn{\pctx{M}}$ and $x \in \bigcap_{i \in I} \fn{\alpha_i;P_i}$ and there is $i' \in I$ such that $\alpha_{i'};P_{i'} = \pclose{x}$.
    Similarly, there are $\pctx{M'}$ and ${(\beta_j;Q_j)}_{j \in J}$ such that $\pctx{N'}[\gclose{x};Q] = \pctx{M'}[\beta_j;Q_j]_{j \in J}$ where $x \notin \fn{\pctx{M'}}$ and $x \in \bigcap_{j \in J} \fn{\beta_j;Q_j}$ and there is $j' \in J$ such that $\beta_{j'};Q_{j'} = \gclose{x};Q$.

    By \Cref{ch4l:multiShape}, $\flat{\pctx{M}}[\alpha_i;P_i]_{i \in I} = \bignd_{i \in I} \pctx{C_i}[\alpha_i;P_i]$ and $\flat{\pctx{M'}}[\beta_j;Q_j]_{j \in J} = \bignd_{j \in J} \pctx{C'_j}[\beta_j;Q_j]$.
    By typability, there is $I' \subseteq I$ such that $\forall i \in I'.~ \alpha_i \relalpha \pclose{x}$ and $\forall i \in I \setminus I'.~ \alpha_i \not\relalpha \pclose{x}$; hence, $i' \in I'$.
    Similarly, there is $J' \subseteq J$ such that $\forall j \in J'.~ \beta_j \relalpha \gclose{x}$ and $\forall j \in J \setminus J'.~ \beta_j \not\relalpha \gclose{x}$; hence, $j' \in J'$.
    Then, by \Cref{ch4d:rpreone}, $\flat{\pctx{M}}[\alpha_i;P_i]_{i \in I} \piprecong{x} \bignd_{i \in I'} \pctx{C_i}[\pclose{x}]$ and $\flat{\pctx{M'}}[\beta_j;Q_j]_{j \in J} \piprecong{x} \bignd_{j \in J'} \pctx{C_j}[\gclose{x};Q_j]$.
    By rule $\rredtwo{\1\bot}$,
    \begin{align*}
        \res{x}(\bignd_{i \in I'} \pctx{C_i}[\pclose{x}] \| \bignd_{j \in J'} \pctx{C_j}[\gclose{x};Q_j]) &\redtwo_x R,
    \end{align*}
    so by rule $\rredtwo{\piprecong{x}}$,
    \begin{align*}
        \res{x}(\flat{\pctx{M}}[\alpha_i;P_i]_{i \in I} \| \flat{\pctx{M'}}[\beta_j;Q_j]_{j \in J}) \redtwo_x R.
    \end{align*}
    Then, by \Cref{ch4l:flatRed},
    \begin{align*}
        \res{x}(\pctx{M}[\alpha_i;P_i]_{i \in I} \| \pctx{M'}[\beta_j;Q_j]_{j \in J}) \redtwo_x R.
    \end{align*}

    As second base case, we consider rule $\rredone{\scc{Id}}$: $P = \res{x}(\pctx[\big]{N}[\pfwd{x}{y}] \| Q) \redone R$.

    By inversion of typing, $ \pctx[\big]{N}[\pfwd{x}{y}]  \vdash \Gamma, x{:}A, y{:}\dual{A}$, so by \Cref{ch4l:ndCtxMultihole}, there are $\pctx{M}$ and ${(\alpha_i;P_i)}_{i \in I}$ such that $\pctx[\big]{N}[\pfwd{x}{y}]  = \pctx{M}[\alpha_i;P_i]_{i \in I}$ where $x \notin \fn{\pctx{M}}$ and $x \in \bigcap_{i \in I} \fn{\alpha_i;P_i}$ and there is $i' \in I$ such that $\alpha_{i'};P_{i'} = \pfwd{x}{y} $.

    By \Cref{ch4l:multiShape}, $\flat{\pctx{M}}[\alpha_i;P_i]_{i \in I} = \bignd_{i \in I} \pctx{C_i}[\alpha_i;P_i]$.
    By typability, there is $I' \subseteq I$ such that $\forall i \in I'.~ \alpha_i = \pfwd{x}{y}$ and $\forall i \in I \setminus I'.~ \alpha_i \not= \pfwd{x}{y}$; hence, $i' \in I'$.

    Then, by \Cref{ch4d:rpreone}, $\flat{\pctx{M}}[\alpha_i;P_i]_{i \in I} \piprecong{x,y} \bignd_{i \in I'} \pctx{C_i}[\pfwd{x}{y}]$.
    By rule $\rredtwo{\scc{Id}}$,
    \begin{align*}
        \res{x} \Big( \bignd_{i \in I}\pctx[\big]{C_i}[\pfwd{x}{y}] \| Q\Big)
                \redtwo_{x,y}
                 R,
    \end{align*}
    so by rule $\rredtwo{\piprecong{x,y}}$,
    \begin{align*}
        \res{x}(\flat{\pctx{M}}[\alpha_i;P_i]_{i \in I} \| Q ) \redtwo_{x,y} R.
    \end{align*}
    Then, by \Cref{ch4l:flatRed},
    \begin{align*}
        \res{x}(\pctx{M}[\alpha_i;P_i]_{i \in I} \| Q ) \redtwo_{x,y} R.
    \end{align*}
\end{proof}

\begin{theorem}[DF: Lazy Semantics]\label{ch4t:dlfreeTwo}
    If $P \vdash \emptyset$ and $P \not\equiv \0$, then $P \redtwo_S R$ for some $S$ and $R$.
\end{theorem}

\begin{proof}
    As a corollary of \Cref{ch4t:dlfreeOne,ch4t:oneImpliesTwo}.
\end{proof}

\section{Proofs of Subject Reduction and Subject Expansion for \texorpdfstring{\lamcoldetsh}{Lambda}}

Here we prove \Cref{ch4t:lamSRShort,ch4t:lamSEShort} (subject reduction and subject expansion, respectively) for \lamcoldetsh.

\subsection{Subject Reduction}\label{ch4a:lamTypes}

\begin{lemma}[Substitution Lemma for $\lamcoldetsh$]\label{ch4l:lamrsharfailsubsunres}
    \leavevmode
    \begin{enumerate}
        \item (Linear) If $\Theta ; \Gamma ,  {x}:\sigma \wfdash M: \tau$, $\headf{M} =  {x}$, and $\Theta ; \Delta \wfdash N : \sigma$
            then
            $\Theta ; \Gamma  , \Delta \wfdash M \headlin{ N /  {x} }:\tau$.
        \item (Unrestricted) If $\Theta, \unvar{x}: \eta ; \Gamma \wfdash M: \tau$, $\headf{M} = {x}[i]$, $\eta_i = \sigma $, and $\Theta ; \cdot \wfdash N : \sigma$
            then
            $\Theta, \unvar{x}: \eta ; \Gamma  \wfdash M \headlin{ N / {x}[i] }$.
    \end{enumerate}
\end{lemma}

\begin{proof}
    \leavevmode
    \begin{enumerate}
        \item By structural induction on $M$ with $\headf{M}=  {x}$.
            There are six cases to be analyzed:
            \begin{enumerate}
                \item $M= {x}$

                    In this case, $\Theta ;  {x}:\sigma \wfdash  {x}:\sigma$ and $\Gamma=\emptyset$.  Observe that $ {x}\headlin{N/ {x}}=N$, since $\Delta\wfdash N:\sigma$, by hypothesis, the result follows.

                \item $M = M'\ B$.

                    Then $\headf{M'\ B} = \headf{M'} =  {x}$, and the derivation is the following:
                    \begin{prooftree}
                        \AxiomC{$\Theta ; \Gamma_1 ,  {x}:\sigma \wfdash M': (\delta^{j} , \eta )  \rightarrow \tau$}\
                        \AxiomC{\quad $\Theta ; \Gamma_2 \wfdash B : (\delta^{k} , \epsilon ) $}
                        \AxiomC{\quad $ \eta \relunbag \epsilon $}
                        \LeftLabel{\redlab{FS{:}app}}
                        \TrinaryInfC{$\Theta ; \Gamma_1 , \Gamma_2 ,  {x}:\sigma \wfdash M'B:\tau $}
                    \end{prooftree}
                    where $\Gamma=\Gamma_1 , \Gamma_2$, and  $j,k$ are non-negative integers, possibly different.  Since $\Delta \vdash N : \sigma$, by IH, the result holds for $M'$, that is,
                    \[\Gamma_1 , \Delta \wfdash M'\headlin{ N /  {x} }: (\delta^{j} , \eta )  \rightarrow \tau\]
                    which gives the  derivation:
                    \begin{prooftree}
                        \AxiomC{$\Theta ; \Gamma_1 , \Delta \wfdash M'\headlin{ N /  {x} }: (\delta^{j} , \eta )  \rightarrow \tau$}\
                        \AxiomC{\quad $\Theta ; \Gamma_2 \wfdash B : (\delta^{k} , \epsilon ) $}
                        \AxiomC{\quad$ \eta \relunbag \epsilon $}
                        \LeftLabel{\redlab{FS{:}app}}
                        \TrinaryInfC{$\Theta ; \Gamma_1 , \Gamma_2 , \Delta \wfdash ( M'\headlin{ N /  {x} } ) B:\tau $}
                    \end{prooftree}
                    From \Cref{ch4f:lambda_red},   $(M'B) \headlin{ N /  {x} } = ( M'\headlin{ N /  {x} } ) B$, and the result follows.
                \item $M = M'[ {\widetilde{y}} \leftarrow  {y}] $.

                    Then $ \headf{M'[ {\widetilde{y}} \leftarrow  {y}]} = \headf{M'}= {x}$, for  $y\neq x$. Therefore,
                    \begin{prooftree}
                        \AxiomC{$\Theta ; \Gamma_1 ,  {y}_1: \delta, \dots,  {y}_k: \delta ,  {x}: \sigma \wfdash M' : \tau \quad  {y}\notin \Gamma_1 \quad k \not = 0$}
                        \LeftLabel{ \redlab{FS{:}share}}
                        \UnaryInfC{$ \Theta ; \Gamma_1 ,  {y}: \delta^k,  {x}: \sigma \wfdash M'[ {y}_1 , \dots ,  {y}_k \leftarrow x] : \tau $}
                    \end{prooftree}
                    where $\Gamma=\Gamma_1 ,  {y}: \delta^k$.
                    By IH, the result follows for $M'$, that is,
                    \[\Theta ; \Gamma_1 ,  {y}_1: \delta, \dots,  {y}_k: \delta ,\Delta \wfdash M'\headlin{N/ {x}} : \tau \]
                    and we have the derivation:
                    \begin{prooftree}
                        \AxiomC{$ \Theta ; \Gamma_1 ,  {y}_1: \delta, \dots,  {y}_k: \delta , \Delta \wfdash  M' \headlin{ N /  {x}} : \tau \quad  {y}\notin \Gamma_1 \quad k \not = 0$}
                        \LeftLabel{ \redlab{FS{:}shar} }
                        \UnaryInfC{$ \Theta ; \Gamma_1 ,  {y}: \delta^k, \Delta \wfdash M' \headlin{ N /  {x}} [ {\widetilde{y}} \leftarrow  {y}] : \tau $}
                    \end{prooftree}
                    From \Cref{ch4f:lambda_red},  $M'[ {\widetilde{y}} \leftarrow  {y}] \headlin{ N /  {x} } = M' \headlin{ N /  {x}} [ {\widetilde{y}} \leftarrow  {y}]$, 
                    and the result follows.

                \item $M = M'[ \leftarrow  {y}] $.

                    Then $ \headf{M'[ \leftarrow  {y}]} = \headf{M'}= {x}$ with  $x \not  = y $,
                    \begin{prooftree}
                        \AxiomC{$ \Theta ; \Gamma  ,  {x}: \sigma  \wfdash M : \tau$}
                        \LeftLabel{ \redlab{FS{:}weak} }
                        \UnaryInfC{$ \Theta ;  \Gamma  ,  {y}: \omega,  {x}: \sigma  \wfdash M[\leftarrow  {y}]: \tau $}
                    \end{prooftree}
                    and $M'[ \leftarrow  {y}] \headlin{ N /  {x} } = M' \headlin{ N /  {x}} [ \leftarrow  {y}]$. Then by the induction hypothesis:
                    \begin{prooftree}
                        \AxiomC{$ \Theta ;  \Gamma , \Delta  \wfdash M \headlin{ N /  {x}}: \tau$}
                        \LeftLabel{ \redlab{FS{:}weak}}
                        \UnaryInfC{$ \Theta ;  \Gamma  ,  {y}: \omega, \Delta \wfdash M\headlin{ N /  {x}}[\leftarrow  {y}]: \tau $}
                    \end{prooftree}

                \item If $M =  M' \linexsub{C /  y_1 , \dots , y_k} $.

                    Then $\headf{M' \linexsub{C /  y_1 , \dots , y_k}} = \headf{M'} = x \not = y_1 , \dots , y_k$,
                    \begin{prooftree}
                            \AxiomC{$ \Theta ; \Gamma_1  ,  y_1:\delta, \dots , y_k:\delta , x: \sigma  \wfdash M' : \tau $}
                        \AxiomC{$\quad \Theta ; \Gamma_2 \wfdash C : \delta^k $}
                            \LeftLabel{\redlab{FS{:}Esub^{\ell}}}
                        \BinaryInfC{$ \Theta ; \Gamma_1 , \Gamma_2, x: \sigma  \wfdash M' \linexsub{C /  y_1 , \dots , y_k} : \tau $}
                    \end{prooftree}
                     and $M' \linexsub{C /  y_1 , \dots , y_k}  \headlin{ N / x } = M' \headlin{ N / x } \linexsub{C /  y_1 , \dots , y_k}  $. Then by the induction hypothesis:
                    \begin{prooftree}
                        \AxiomC{$ \Theta ; \Gamma_1  , \Delta,  y_1:\delta, \dots , y_k:\delta  \wfdash M' \headlin{ N / x } : \tau $}
                        \AxiomC{$\quad \Theta ; \Gamma_2 \wfdash C : \delta^k $}
                        \LeftLabel{\redlab{FS{:}Esub^{\ell}}}
                        \BinaryInfC{$ \Theta ; \Gamma_1 , \Gamma_2 , \Delta  \wfdash M' \headlin{ N / x } \linexsub{C /  y_1 , \dots , y_k} : \tau $}
                    \end{prooftree}

                \item If $M =  M' \unexsub {U / {y}} $ then $\headf{M' \unexsub {U / {y}}} = \headf{M'} = x $, and the proofs is similar to the case above.
            \end{enumerate}

        \item  By structural induction on $M$ with $\headf{M}= {x}[i]$.
            There are three cases to be analyzed:
            \begin{enumerate}
                \item $M= {x}[i]$.

                    In this case,
                    \begin{prooftree}
                        \AxiomC{}
                        \LeftLabel{ \redlab{FS{:}var^{ \ell}}}
                        \UnaryInfC{$ \Theta , \unvar{x}: \eta;  {x}: \eta_i  \wfdash  {x} : \sigma$}
                        \LeftLabel{\redlab{FS{:}var^!}}
                        \UnaryInfC{$ \Theta, \unvar{x}: \eta ; \cdot \wfdash {x}[i] : \sigma$}
                    \end{prooftree}
                    and $\Gamma=\emptyset$.  Observe that ${x}[i]\headlin{N/{x}[i]}=N$, since $\Theta, \unvar{x}: \eta  ; \Gamma  \wfdash M \headlin{ N / {x}[i] }$, by hypothesis, the result follows.

                \item $M = M'\ B$.

                    In this case, $\headf{M'\ B} = \headf{M'} =  {x}[i]$, and one has the following derivation:
                    \begin{prooftree}
                        \AxiomC{$ \Theta, \unvar{x}: \eta ;\Gamma_1 \wfdash M : (\delta^{j} , \epsilon ) \rightarrow \tau \quad \Theta, \unvar{x}: \sigma ; \Gamma_2 \wfdash B : (\delta^{k} , \epsilon' )  $}
                        \AxiomC{$ \epsilon \relunbag \epsilon' $}
                        \LeftLabel{\redlab{FS{:}app}}
                        \BinaryInfC{$ \Theta, \unvar{x}: \eta ;\Gamma_1 , \Gamma_2 \wfdash M\ B : \tau$}
                    \end{prooftree}
                    where $\Gamma=\Gamma_1 , \Gamma_2$, $\delta$ is a strict type and $j,k$ are non-negative  integers, possibly different.

                    By the induction hypothesis, we get $\Theta, \unvar{x}: \eta ;\Gamma_1 \wfdash M'\headlin{N/ {x[i]}}:(\delta^{j} , \epsilon ) \rightarrow \tau $, which gives the derivation:
                    \begin{prooftree}
                        \small
                        \AxiomC{$\Theta , \unvar{x}: \eta;\Gamma_1\wfdash M'\headlin{N/ {x}[i]}:(\delta^{j} , \epsilon ) \rightarrow \tau $}\
                        \AxiomC{$\Theta , \unvar{x}: \eta; \Gamma_2 \wfdash B : (\delta^{k} , \epsilon' ) $}
                        \AxiomC{$ \epsilon \relunbag \epsilon' $}
                        \LeftLabel{\redlab{FS{:}app}}
                        \TrinaryInfC{$\Theta , \unvar{x}: \eta;\Gamma_1 , \Gamma_2  \wfdash ( M'\headlin{ N / {x}[i] } ) B:\tau $}
                    \end{prooftree}
                    From \Cref{ch4f:lambda_red}, $M' \esubst{ B }{ y} \headlin{ N / {x}[i] } = M' \headlin{ N / {x}[i] } \esubst{ B }{ y}$, and  the result follows.

                \item $M = M'[ {\widetilde{y}} \leftarrow  {y}] $.

                    Then $ \headf{M'[ {\widetilde{y}} \leftarrow  {y}]} = \headf{M'}= {x}[i]$, for  $y\neq x$. Therefore,
                    \begin{prooftree}
                        \AxiomC{$\Theta , \unvar{x}: \eta; \Gamma_1 ,  {y}_1: \delta, \dots,  {y}_k: \delta  \wfdash M' : \tau \quad  {y}\notin \Gamma_1 \quad k \not = 0$}
                        \LeftLabel{ \redlab{FS{:}shar}}
                        \UnaryInfC{$ \Theta , \unvar{x}: \eta; \Gamma_1 ,  {y}: \delta^k \wfdash M'[ {y}_1 , \dots ,  {y}_k \leftarrow y] : \tau $}
                    \end{prooftree}
                    where $\Gamma=\Gamma_1 ,  {y}: \delta^k$.
                    By the induction hypothesis, the result follows for $M'$, that is,
                    \[\Theta, \unvar{x}: \eta ; \Gamma_1 ,  {y}_1: \delta, \dots,  {y}_k: \delta  \wfdash M'\headlin{N/ {x}[i] } : \tau \] and we have the derivation:
                    \begin{prooftree}
                        \AxiomC{$ \Theta , \unvar{x}: \eta; \Gamma_1 ,  {y}_1: \delta, \dots,  {y}_k: \delta  \wfdash  M' \headlin{ N / {x}[i] } : \tau \quad  {y}\notin \Gamma_1 \quad k \not = 0$}
                        \LeftLabel{ \redlab{FS{:}shar} }
                        \UnaryInfC{$ \Theta , \unvar{x}: \eta; \Gamma_1 ,  {y}: \delta^k \wfdash M' \headlin{ N / {x}[i]} [ {\widetilde{y}} \leftarrow  {y}] : \tau $}
                    \end{prooftree}
                    From \Cref{ch4f:lambda_red}  $M'[ {\widetilde{y}} \leftarrow  {y}] \headlin{ N / {x}[i] } = M' \headlin{ N / {x}[i]} [ {\widetilde{y}} \leftarrow  {y}]$, 
                    and the result follows.

                \item $M = M'[ \leftarrow  {y}] $.

                    Then $ \headf{M'[ \leftarrow  {y}]} = \headf{M'}= {x}[i]$ with  $x \not  = y $,
                    \begin{prooftree}
                        \AxiomC{$ \Theta , \unvar{x}: \eta; \Gamma   \wfdash M : \tau$}
                        \LeftLabel{ \redlab{FS{:}weak} }
                        \UnaryInfC{$ \Theta , \unvar{x}: \eta;  \Gamma  ,  {y}: \omega \wfdash M[\leftarrow  {y}]: \tau $}
                    \end{prooftree}
                    and $M'[ \leftarrow  {y}] \headlin{ N / {x}[i] } = M' \headlin{ N / {x}[i] } [ \leftarrow  {y}]$.
                    By the induction hypothesis:
                    \begin{prooftree}
                        \AxiomC{$ \Theta , \unvar{x}: \eta;  \Gamma   \wfdash M \headlin{ N / {x}[i]  }: \tau$}
                        \LeftLabel{ \redlab{FS{:}weak}}
                        \UnaryInfC{$ \Theta , \unvar{x}: \eta;  \Gamma  ,  {y}: \omega \wfdash M\headlin{ N / {x}[i] }[\leftarrow  {y}]: \tau $}
                    \end{prooftree}

                \item $M =   M' \linexsub{C /  y_1 , \dots , y_k} $.

                    Then $\headf{ M' \linexsub{C /  y_1 , \dots , y_k} } = \headf{M'} = {x}[i]$ with $x \not = y$,
                    \begin{prooftree}
                        \AxiomC{$ \Theta, \unvar{x}: \eta ; \Gamma  ,  y_1:\delta, \dots , y_k:\delta \wfdash M : \tau $}
                        \AxiomC{$ \Theta, \unvar{x}: \eta ; \Delta \wfdash C : \delta^k $}
                        \LeftLabel{\redlab{FS{:}Esub^{\ell}}}
                        \BinaryInfC{$ \Theta, \unvar{x}: \eta ; \Gamma , \Delta \wfdash M \linexsub{C /  y_1, \dots , y_k} : \tau $}
                     \end{prooftree}
                    and $M' \linexsub{C /  y_1 , \dots , y_k} \headlin{ N / {x}[i]  } = M' \headlin{ N / {x}[i]  }  \linexsub{C /  y_1 , \dots , y_k}  $.
                    By the induction hypothesis:
                    \begin{prooftree}
                        \small
                        \AxiomC{$ \Theta, \unvar{x}: \eta ; \Gamma  ,  y_1:\delta, \dots , y_k:\delta \wfdash M' \headlin{ N / {x}[i]  }  : \tau $}
                        \AxiomC{$ \Theta, \unvar{x}: \eta ; \Delta \wfdash C : \delta^k $}
                        \LeftLabel{\redlab{FS{:}Esub^{\ell}}}
                        \BinaryInfC{$ \Theta, \unvar{x}: \eta ; \Gamma , \Delta \wfdash M' \headlin{ N / {x}[i]  }  \linexsub{C /  y_1 , \dots , y_k} : \tau $}
                    \end{prooftree}

                \item $M =  M' \unexsub {U /\unvar{y}}$.

                    Then $\headf{M' \unexsub {U /\unvar{y}}} = \headf{M'} = {x}[i] $,
                    \begin{prooftree}
                        \AxiomC{$ \Theta , \unvar{x}: \eta , {y} : \epsilon_1; \Gamma  \wfdash M' : \tau $}
                        \AxiomC{$ \Theta , \unvar{x}: \eta; \dash \wfdash U : \epsilon_2 $}
                        \AxiomC{$ \epsilon_1 \relunbag \epsilon_2 $}
                            \LeftLabel{\redlab{FS{:}Esub^!}}
                        \TrinaryInfC{$ \Theta ; \Gamma \wfdash M' \unexsub{U / \unvar{y}}  : \tau $}
                    \end{prooftree}
                    and $M' \unexsub {U /\unvar{y}} \headlin{ N /  {x}[i] } = M'  \headlin{N /  {x}[i] } \unexsub {U /\unvar{y}}$.
                    Then by the induction hypothesis:
                    \begin{prooftree}
                        \small
                        \AxiomC{$ \Theta , \unvar{x}: \eta , {y} : \epsilon_1; \Gamma  \wfdash M'  \headlin{N /  {x}[i] } : \tau $}
                        \AxiomC{$ \Theta , \unvar{x}: \eta; \dash \wfdash U : \epsilon_2 $}
                        \AxiomC{$ \epsilon_1 \relunbag \epsilon_2 $}
                        \LeftLabel{\redlab{FS{:}Esub^!}}
                        \TrinaryInfC{$ \Theta ; \Gamma \wfdash M'  \headlin{N /  {x}[i] } \unexsub{U / \unvar{y}}  : \tau $}
                    \end{prooftree}
            \end{enumerate}
    \end{enumerate}
\end{proof}

\begin{theorem}[SR in $\lamcoldetsh$]\label{ch4t:lamSR}
    If $\Theta ; \Gamma \wfdash M:\tau$ and $M \red M'$ then $\Theta ; \Gamma \wfdash M' :\tau$.
\end{theorem}

\begin{proof}
    By structural induction on the reduction rule from \figref{ch4fig:reduc_interm} applied in $M \red M'$.
    \begin{enumerate}
        \item \textbf{ Rule $\redlab{RS{:}Beta}$.}

            Then $M = (\lambda x. N[ {\widetilde{x}} \leftarrow  {x}]) B $  and the reduction is:

            \begin{prooftree}
                \AxiomC{}
                \LeftLabel{\redlab{RS{:}Beta}}
                \UnaryInfC{$(\lambda x. N[ {\widetilde{x}} \leftarrow  {x}]) B \red N[ {\widetilde{x}} \leftarrow  {x}]\ \esubst{ B }{ x }$}
            \end{prooftree}
            where $ M'  =  N[ {\widetilde{x}} \leftarrow  {x}]\ \esubst{ B }{ x }$. Since $\Theta ; \Gamma\wfdash M:\tau$ we get the following derivation:
            \begin{prooftree}
                \AxiomC{$\Theta , \unvar{x} : \eta; \Gamma' ,  {x}_1:\sigma , \dots ,  {x}_j:\sigma  \wfdash  N: \tau $}
                \LeftLabel{ \redlab{FS{:}share} }
                \UnaryInfC{$\Theta , \unvar{x} : \eta;  \Gamma' ,   {x}:\sigma^{j}  \wfdash  N[ {\widetilde{x}} \leftarrow  {x}]: \tau $}
                \LeftLabel{ \redlab{FS{:}abs \dash sh} }
                \UnaryInfC{$\Theta ; \Gamma' \wfdash \lambda x. N[ {\widetilde{x}} \leftarrow  {x}]: (\sigma^{j} , \eta ) \rightarrow \tau $}
                \AxiomC{$\Theta ;\Delta \wfdash B: (\sigma^{k} , \epsilon ) $}
                \AxiomC{$ \eta \relunbag \epsilon $}
                \LeftLabel{ \redlab{FS{:}app} }
                \TrinaryInfC{$ \Theta ;\Gamma' , \Delta \wfdash (\lambda x. N[ {\widetilde{x}} \leftarrow  {x}]) B:\tau $}
            \end{prooftree}
            for $\Gamma = \Gamma' , \Delta $ and $x\notin \dom{\Gamma'}$.
            Notice that:
            \begin{prooftree}
                \AxiomC{$\Theta , \unvar{x} : \eta; \Gamma' ,  {x}_1:\sigma , \dots ,  {x}_j:\sigma  \wfdash  N: \tau $}
                \LeftLabel{ \redlab{FS{:}share} }
                \UnaryInfC{$\Theta , \unvar{x} : \eta;  \Gamma' ,   {x}:\sigma^{j}  \wfdash  N[ {\widetilde{x}} \leftarrow  {x}]: \tau $}
                \AxiomC{$\Theta ;\Delta \wfdash B:(\sigma^{k} , \epsilon )  $}
                \AxiomC{$ \eta \relunbag \epsilon $}
                \LeftLabel{ \redlab{FS{:}Esub} }
                \TrinaryInfC{$ \Theta ;\Gamma' , \Delta \wfdash N[ {\widetilde{x}} \leftarrow  {x}]\ \esubst{ B }{ x }:\tau $}
            \end{prooftree}
            Therefore $ \Theta ; \Gamma' ,\Delta\wfdash M' :\tau$ and the result follows.

        \item \textbf{ Rule $ \redlab{RS{:}Ex \dash Sub}$}.

            Then $ M =  N[ {x}_1, \dots ,  {x}_k \leftarrow  {x}]\ \esubst{ C \bagsep U }{ x }$ where $C=  \bag{N_1}\cdot \cdots \cdot \bag{N_k} $.
            The reduction is:
            \begin{prooftree}
                \AxiomC{$ C = \bag{N_1}
                \dots  \bag{N_k} \qquad  N \not= \fail^{\widetilde{y}} $}
                \LeftLabel{\redlab{RS{:}Ex \dash Sub}}
                \UnaryInfC{$ N[ {x}_1, \!\dots\! ,  {x}_k \leftarrow  {x}]\esubst{ C \bagsep U }{ x } \red N\linexsub{C  /  x_1 , \dots , x_k} \unexsub{U / \unvar{x} }$}
            \end{prooftree}
            and $M' = N\linexsub{C  /  x_1 , \dots , x_k} \unexsub{U / \unvar{x} }$.
            To simplify the proof we take $k=2$, as the case $k>2$ is similar.
            Therefore $C=\bag{N_1}\cdot \bag{N_2}$.

            Since $\Theta ; \Gamma\wfdash M:\tau$ we get a derivation: (we omit the labels \redlab{FS:Esub} and \redlab{FS{:}share})
            {\small
                \begin{prooftree}
                    \AxiomC{$ \Theta, \unvar{x} : \eta  ;  \Gamma' ,  {x}_1: \sigma,  {x}_2: \sigma \wfdash N : \tau \quad  {x}\notin \dom{\Gamma} \quad k \not = 0$}
                    \UnaryInfC{$  \Theta , \unvar{x} : \eta ; \Gamma' ,  {x}: \sigma^{2} \wfdash N[ {x}_1,  {x}_2 \leftarrow  {x}] : \tau  $}

                    \AxiomC{$ \Theta ; \Delta \wfdash C \bagsep U : (\sigma^{2} , \epsilon ) $}
                    \AxiomC{$ \eta \relunbag \epsilon $}
                    \TrinaryInfC{$ \Theta ; \Gamma' , \Delta \wfdash N[ {x}_1,  {x}_2 \leftarrow  {x}]\esubst{ C \bagsep U }{ x }  : \tau $}
                \end{prooftree}
            }
            where $\Gamma = \Gamma' , \Delta $. Consider the wf derivation for $\Pi_{1,2}$: (we omit the labels \redlab{FS:Esub^!} and \redlab{FS:Esub^{\ell}})
            {\small
                \begin{prooftree}
                    \AxiomC{$ \Theta, \unvar{x} : \eta  ;  \Gamma' ,  {x}_1: \sigma,  {x}_2: \sigma \wfdash N : \tau$}
                    \AxiomC{$ \Theta ; \Delta \wfdash C : \sigma^k $}
                    \BinaryInfC{$ \Theta, \unvar{x} : \eta  ;  \Gamma' , \Delta \wfdash N \linexsub{C  / {x}_1,  {x}_2} : \tau $}
                    \AxiomC{$ \Theta ; \dash \wfdash U : \epsilon $}
                    \AxiomC{$ \eta \relunbag \epsilon $}
                    \TrinaryInfC{$ \Theta ; \Gamma' , \Delta \wfdash N\linexsub{C  / {x}_1,  {x}_2} \unexsub{U / \unvar{x} }  : \tau $}
                \end{prooftree}
            }
            and the result follows.

         \item Rule $ \redlab{RS{:}Fetch^{\ell}}$.

            Then $ M =  N \linexsub{C /  \widetilde{x}, x_j}  $ where  $\headf{N} =  {x}_j$.
            The reduction is:
            \begin{prooftree}
                \AxiomC{$ \headf{N} =  {x}_j$}
                 \LeftLabel{\redlab{RS{:}Fetch^{\ell}_{{i}}}}
                 \UnaryInfC{$  N \linexsub{C /  \widetilde{x}, x_j} \red  N \headlin{ C_i / x_j }  \linexsub{(C \setminus C_i ) /  \widetilde{x}  } $}
            \end{prooftree}
            and $M' =  N \headlin{ C_i / x_j }  \linexsub{(C \setminus C_i ) /  \widetilde{x}  } $.
            Since $\Theta ; \Gamma\wfdash M:\tau$ we get the following derivation:
            \begin{prooftree}
                \AxiomC{$ \Theta ; \Gamma'  ,  \widetilde{x}:\sigma^{k-1},  x_j:\sigma \wfdash N : \tau $}
                \AxiomC{$ \Theta ; \Delta \wfdash C : \sigma^k $}
                \LeftLabel{\redlab{FS{:}Esub^{\ell}}}
                \BinaryInfC{$ \Theta ; \Gamma' , \Delta \wfdash N \linexsub{C /  \widetilde{x}, x_j}  : \tau $}
            \end{prooftree}
            where $\Gamma = \Gamma' , \Delta $ and $\Delta = \Delta_i , \Delta_i'$ with $ \Theta ;  \Delta_i \wfdash  C_i : \sigma $.
            By \Cref{ch4l:lamrsharfailsubsunres}, we obtain the derivation $ \Theta ; \Gamma' , \Delta \wfdash   N \headlin{ C_i / x_j }  \linexsub{(C \setminus C_i ) /  \widetilde{x}  } : \tau $ via:
            \begin{prooftree}
                \AxiomC{$ \Theta ; \Gamma' , \Delta_i ,  \widetilde{x}:\sigma^{k-1} \wfdash N \headlin{ C_i / x_j }: \tau $}
                \AxiomC{$ \Theta ; \Delta'_i \wfdash C \setminus C_i : \sigma^{k-1} $}
                \LeftLabel{\redlab{FS{:}Esub^{\ell}}}
                \BinaryInfC{$ \Theta ; \Gamma', \Delta \wfdash N \headlin{ C_i / x_j }  \linexsub{(C \setminus C_i ) /  \widetilde{x}  }  : \tau $}
            \end{prooftree}

        \item Rule $ \redlab{RS{:} Fetch^!}$.

            Then $ M =  N \unexsub{U /  \unvar{x}}  $ where  $\headf{M} = {x}[i]$. The reduction is:
            \begin{prooftree}
                \AxiomC{$ \headf{N} = {x}[i]$}
                \AxiomC{$ U_i = \unvar{\bag{N_i}}$}
                \LeftLabel{\redlab{RS{:} Fetch^!}}
                \BinaryInfC{$  N \unexsub{U / \unvar{x}} \red  N \headlin{ N_i /{x}[i] }\unexsub{U / \unvar{x}} $}
            \end{prooftree}
            and $M' =  N \headlin{ N_i /{x}[i] }\unexsub{U / \unvar{x}}  $.     Since $\Theta ; \Gamma \wfdash M:\tau$ we get the following derivation:
            \begin{prooftree}
                \AxiomC{$ \Theta , {x} : \eta; \Gamma  \wfdash N : \tau $}
                \AxiomC{$ \Theta ; \dash \wfdash U : \epsilon $}
                \AxiomC{$ \eta \relunbag \epsilon $}
                \LeftLabel{\redlab{FS{:}Esub^!}}
                \TrinaryInfC{$ \Theta ; \Gamma \wfdash N \unexsub{U / \unvar{x}}  : \tau $}
            \end{prooftree}
            By \Cref{ch4l:lamrsharfailsubsunres}, we obtain the derivation $  \Theta , \unvar{x} : \eta; \Gamma  \wfdash  N \headlin{ N_i /{x}[i] } $, and the result follows from:
            \begin{prooftree}
                \AxiomC{$ \Theta , \unvar{x} : \eta; \Gamma  \wfdash N \headlin{ N_i /{x}[i] } : \tau $}
                \AxiomC{$ \Theta ; \dash \wfdash U : \epsilon $}
                \AxiomC{$ \eta \relunbag \epsilon $}
                \LeftLabel{\redlab{FS{:}Esub^!}}
                \TrinaryInfC{$ \Theta ; \Gamma \wfdash N \headlin{ N_i /{x}[i] }\unexsub{U / \unvar{x}}  : \tau $}
            \end{prooftree}

        \item Rule $\redlab{RS{:}TCont}$.

            Then $M = C[N]$ and the reduction is as follows:
            \begin{prooftree}
                    \AxiomC{$   N \red N' $}
                    \LeftLabel{\redlab{RS{:}TCont}}
                    \UnaryInfC{$ C[N] \red  C[N'] $}
            \end{prooftree}
            with $M'=  C[N'] $.
            The proof proceeds by analysing the context $C$.
            There are three cases:
            \begin{enumerate}
                \item $C=[\cdot]\ B$.

                    In this case $ M = N \ B$, for some $B$.
                    Since $\Gamma\vdash M:\tau$ one has a derivation:
                    \begin{prooftree}
                        \AxiomC{$ \Theta ;\Gamma' \wfdash N : (\sigma^{j} , \eta ) \rightarrow \tau $}
                        \AxiomC{$  \Theta ;\Delta \wfdash B : (\sigma^{k} , \epsilon )  $}
                        \AxiomC{$ \eta \relunbag \epsilon $}
                        \LeftLabel{\redlab{FS{:}app}}
                        \TrinaryInfC{$ \Theta ; \Gamma' , \Delta \wfdash N\ B : \tau$}
                    \end{prooftree}
                    where $\Gamma = \Gamma' , \Delta $.
                    From  $\Gamma'\wfdash N:\sigma^j\rightarrow\tau$ and the reduction $N \red N' $, one has by IH that  $\Gamma'\wfdash N':\sigma^j\rightarrow\tau$.
                    Finally, we may type the following:
                    \begin{prooftree}
                        \AxiomC{$ \Theta ;\Gamma' \wfdash N' : (\sigma^{j} , \eta ) \rightarrow \tau $}
                        \AxiomC{$  \Theta ;\Delta \wfdash B : (\sigma^{k} , \epsilon )  $}
                        \AxiomC{$ \eta \relunbag \epsilon $}
                        \LeftLabel{\redlab{FS{:}app}}
                        \TrinaryInfC{$ \Theta ; \Gamma' , \Delta \wfdash N' \ B : \tau$}
                    \end{prooftree}
                    Since $ M'  =   (C[N']) = N'B $, the result follows.

                \item  Cases $C=[\cdot]\linexsub{N/x} $ and $C=[\cdot][\widetilde{x} \leftarrow x]$ are similar to the previous.
            \end{enumerate}

        \item Rule $ \redlab{RS{:}Fail^{\ell}}$.

            Then $M =    N[x_1 , \dots , x_k \leftarrow  {x}]\ \esubst{C \bagsep U}{ x } $ where $ k \neq \size{C} $, $  \widetilde{y} = (\llfv{N} \setminus \{  \widetilde{x}\} ) \cup \llfv{C} $ and  the reduction is:
            \begin{prooftree}
                \AxiomC{$ k \neq \size{C} $}
                \AxiomC{$  \widetilde{y} = (\llfv{N} \setminus \{  \widetilde{x}\} ) \cup \llfv{C} $}
                \LeftLabel{\redlab{RS{:}Fail^{\ell}}}
                \BinaryInfC{$  N[x_1 , \dots , x_k \leftarrow  {x}]\ \esubst{C \bagsep U}{ x }  \red  \fail^{\widetilde{y}} $}
            \end{prooftree}
            where $M' =   \fail^{\widetilde{y}}$.
            Since $\Theta, x: \eta ; \Gamma' , x_1:\sigma,\ldots, x_k:\sigma \wfdash M$, one has a derivation:
            \begin{prooftree}
                \AxiomC{$ \Theta, x: \eta ; \Gamma' , x_1:\sigma,\ldots, x_k:\sigma \wfdash N: \tau $}
                \LeftLabel{ \redlab{FS{:}Esub} }
                \UnaryInfC{$ \Theta, x: \eta ;\Gamma' , x:\sigma^{k} \wfdash N[x_1, \dots , x_k \leftarrow x] : \tau $}
                \AxiomC{$\Theta ; \Delta \wfdash C \bagsep U : (\sigma^{j} , \epsilon ) $}
                \AxiomC{$ \eta \relunbag \epsilon $}
                \LeftLabel{ \redlab{FS{:}Esub} }
                \TrinaryInfC{$\Theta ; \Gamma' , \Delta \wfdash N[x_1 , \dots , x_k \leftarrow  {x}]\ \esubst{C \bagsep U}{ x }  : \tau $}
            \end{prooftree}
            where $\Gamma = \Gamma' , \Delta $.
            We may type the following:
            \begin{prooftree}
                \AxiomC{$ \dom{\Gamma' , \Delta} = \widetilde{x} $}
                \LeftLabel{ \redlab{FS{:}fail}}
                \UnaryInfC{$\Theta ; \Gamma' , \Delta \wfdash  \fail^{\widetilde{y}} : \tau  $}
            \end{prooftree}
            since $\Gamma' , \Delta$ contain assignments on the free variables in $M$ and $B$.
            Therefore, $\Theta ;\Gamma\wfdash \fail^{\widetilde{y}}:\tau$, by applying \redlab{FS{:}sum} as required.

        \item Rule $ \redlab{RS{:}Fail^!}$.

            Then $M = N \unexsub{U / \unvar{x} } $ where $\headf{M} = {x}[i]$ and $U_i = \unvar{\oneb} $ and  the reduction is:
            \begin{prooftree}
                \AxiomC{$\headf{N} = {x}[i]$}
                \AxiomC{$ U_i = \unvar{\oneb} $}
                \AxiomC{$  $}
                \LeftLabel{\redlab{RS{:}Fail^!}}
                \TrinaryInfC{$  N \unexsub{U / \unvar{x} } \red N \headlin{ \fail^{\emptyset} /{x}[i] } \unexsub{U / \unvar{x} }  $}
            \end{prooftree}
            with $M'=N \headlin{ \fail^{\emptyset} /{x}[i] } \unexsub{U / \unvar{x} }$.
            By the induction hypothesis, one has the derivation:
            \begin{prooftree}
                \AxiomC{$ \Theta , {x} : \eta; \Gamma  \wfdash N : \tau $}
                \AxiomC{$ \Theta ; \dash \wfdash U : \epsilon $}
                \AxiomC{$ \eta \relunbag \epsilon $}
                \LeftLabel{\redlab{FS{:}Esub^!}}
                \TrinaryInfC{$ \Theta ; \Gamma \wfdash N \unexsub{U / \unvar{x}}  : \tau $}
            \end{prooftree}
            By \Cref{ch4l:lamrsharfailsubsunres}, there exists a derivation $\Pi_1$ of  $\Theta , \unvar{x} : \eta ; \Gamma'  \wfdash   N \headlin{ \fail^{\emptyset} /{x}[i] }  : \tau $.
            Thus,
            \begin{prooftree}
                \AxiomC{$ \Theta , \unvar{x} : \eta ; \Gamma \wfdash   N \headlin{ \fail^{\emptyset} /{x}[i] }  : \tau  $}
                \AxiomC{$ \Theta ; \dash \wfdash U : \epsilon $}
                \AxiomC{$ \eta \relunbag \epsilon $}
                \LeftLabel{\redlab{FS{:}Esub^!}}
                \TrinaryInfC{$ \Theta ; \Gamma \wfdash  N \headlin{ \fail^{\emptyset} /{x}[i] } \unexsub{U / \unvar{x}}  : \tau $}
            \end{prooftree}

        \item Rule $\redlab{RS{:}Cons_1}$.

            Then $M =   \fail^{\widetilde{x}}\ B $  and  the  reduction is:
            \begin{prooftree}
                \AxiomC{$ \widetilde{y} = \llfv{C} $}
                \LeftLabel{$\redlab{RS{:}Cons_1}$}
                \UnaryInfC{$\fail^{\widetilde{x}}\ C \bagsep U \red \displaystyle \fail^{\widetilde{x} \cup \widetilde{y}}  $}
            \end{prooftree}
            and $ M'  =   \fail^{\widetilde{x} \cup \widetilde{y}} $.
            Since $\Theta ; \Gamma\wfdash M:\tau$, one has the derivation:

            \begin{prooftree}
                \AxiomC{$ $}
                \LeftLabel{\redlab{F{:}fail}}
                \UnaryInfC{$ \Theta ;\Gamma' \wfdash \fail^{\widetilde{x}}: (\sigma^{j} , \eta ) \rightarrow \tau $}
                \AxiomC{$\Theta ; \Delta \wfdash C : \sigma^k $}
                \AxiomC{$\Theta ;\dash \wfdash  U : \epsilon $}
                \LeftLabel{$\redlab{FS{:}bag}$}
                \BinaryInfC{$ \Theta ;\Delta \wfdash C \bagsep U : (\sigma^{k} , \epsilon ) \quad \eta \relunbag \epsilon $}
                \LeftLabel{\redlab{FS{:}app}}
                \BinaryInfC{$\Theta ; \Gamma' , \Delta \wfdash \fail^{\widetilde{x}}:\ C \bagsep U : \tau$}
            \end{prooftree}

            Hence $\Gamma = \Gamma' , \Delta $ and we may type the following:
            \begin{prooftree}
                \AxiomC{$ $}
                \LeftLabel{\redlab{F{:}fail}}
                \UnaryInfC{$ \Theta ;\Gamma \wfdash \fail^{\widetilde{x} \cup \widetilde{y}} : \tau$}
            \end{prooftree}

        \item The proof for the cases of $\redlab{RS{:}Cons_2}$, $\redlab{RS{:}Cons_3}$ and $\redlab{RS{:}Cons_4}$ proceed similarly.
    \end{enumerate}
\end{proof}

\subsection{Subject Expansion}\label{ch4a:subexpan}

The full well-typed rules can be seen in \Cref{ch4fig:wtsh_rulesunres}.

\begin{figure}[t]
\small
\begin{mathpar}
    \inferrule[$\redlab{TS{:}var^{\ell}}$]{ }{
        \Theta;  {x}: \sigma \wtdash  {x} : \sigma
    }
    \and
    \inferrule[$\redlab{TS{:}var^!}$]{
        \Theta , x^!: \eta;  {x}: \eta_i , \Delta \wtdash  {x} : \sigma
    }{
        \Theta ,  x^!: \eta; \Delta \wtdash {x}[i] : \sigma
    }
    \and
    \inferrule[$\redlab{TS{:}\oneb^{\ell}}$]{ }{
        \Theta ; \dash \wtdash \oneb : \omega
    }
    \and
    \inferrule[$\redlab{TS\!:\!weak}$]{
        \Theta ; \Gamma  \wtdash M : \tau
    }{
        \Theta ; \Gamma ,  {x}: \omega \wtdash M[\leftarrow  {x}]: \tau
    }
    \and
    \inferrule[$\redlab{TS{:}abs\dash sh}$]{
        \Theta , x^!:\eta ; \Gamma ,  {x}: \sigma^k \wtdash M[ {\widetilde{x}} \leftarrow  {x}] : \tau
        \\
        {x} \notin \dom{\Gamma}
    }{
        \Theta ; \Gamma \wtdash \lambda x . (M[ {\widetilde{x}} \leftarrow  {x}])  : (\sigma^k, \eta )  \rightarrow \tau
    }
    \and
    \inferrule[$\redlab{TS{:}bag^{\ell}}$]{
        \Theta ; \Gamma \wtdash M : \sigma
        \\
        \Theta ; \Delta \wtdash C : \sigma^k
    }{
        \Theta ; \Gamma , \Delta \wtdash \bag{M}\cdot C:\sigma^{k+1}
    }
    \and
    \inferrule[$\redlab{TS{:}app}$]{
        \Theta ;\Gamma \wtdash M : (\sigma^{j} , \eta ) \rightarrow \tau
        \\
        \Theta ;\Delta \wtdash B : (\sigma^{j} , \eta )
    }{
        \Theta ; \Gamma , \Delta \wtdash M\ B : \tau
    }
    \and
    \inferrule[$\redlab{TS{:} bag^{!}}$]{
        \Theta ; \dash \wtdash U : \epsilon
        \\
        \Theta ; \dash \wtdash V : \eta
    }{
        \Theta ; \dash  \wtdash U \concat V :\epsilon \concat \eta
    }
    \and
    \inferrule[$\redlab{TS{:}shar}$]{
        \Theta ;  \Gamma ,  {x}_1: \sigma, \dots,  {x}_k: \sigma \wtdash M : \tau
        \\
        {x}\notin \dom{\Gamma}
        \\
        k \not = 0
    }{
        \Theta ;  \Gamma ,  {x}: \sigma^{k} \wtdash M [  {x}_1 , \dots ,  {x}_k \leftarrow  {x} ]  : \tau
    }
    \and
    \inferrule[$\redlab{TS{:}bag}$]{
        \Theta ; \Gamma\wtdash C : \sigma^k
        \\
        \Theta ;\dash \wtdash  U : \eta
    }{
        \Theta ; \Gamma \wtdash C \bagsep U : (\sigma^{k} , \eta  )
    }
    \and
    \inferrule[$\redlab{TS{:}Esub^{\ell}}$]{
        \Theta ; \Gamma  ,  x_1:\sigma, \dots , x_k:\sigma \wtdash M : \tau ~~  \Theta ; \Delta \wtdash C : \sigma^k
    }{
        \Theta ; \Gamma , \Delta \wtdash M \linexsub{C /  x_1, \dots , x_k} : \tau
    }
    \and
    \inferrule[$\redlab{TS{:}Esub^!}$]{
        \Theta , x^! {:} \eta; \Gamma  \wtdash M : \tau
        \\
        \Theta ; \dash \wtdash U : \eta
    }{
        \Theta ; \Gamma \wtdash M \unexsub{U / \unvar{x}}  : \tau
    }
    \and
    \inferrule[$\redlab{TS{:}Esub}$]{
        \Theta , x^! : \eta ; \Gamma ,  {x}: \sigma^{j} \wtdash M[ {\widetilde{x}} \leftarrow  {x}] : \tau
        \\
        \Theta ; \Delta \wtdash B : (\sigma^{k} , \eta )
    }{
        \Theta ; \Gamma , \Delta \wtdash (M[ {\widetilde{x}} \leftarrow  {x}])\esubst{ B }{ x }  : \tau
    }
\end{mathpar}
\caption{Well-Typed Rules for $\lamcoldetsh$.}
\label{ch4fig:wtsh_rulesunres}
\end{figure}

\begin{lemma}[Anti-Substitution Lemma for $\lamcoldetsh$]\label{ch4l:lamrsharfailantisub}
    \leavevmode
    \begin{enumerate}
        \item (Linear)
        If $\Theta ; \Gamma \wtdash M \headlin{ N /  {x} }:\tau$ then there exists $\Gamma' , \Delta$ and $ \sigma$ such that $\Theta ; \Gamma' ,  {x}:\sigma \wtdash M: \tau$ and $\Theta ; \Delta \wtdash N : \sigma$ with $\Gamma = \Gamma' , \Delta $.
        \item (Unrestricted)
        If $\Theta, \unvar{x}: \eta ; \Gamma  \wtdash M \headlin{ N / {x}[i] }$ with $\eta_i = \sigma $ then $\Theta, \unvar{x}: \eta ; \Gamma \wtdash M: \tau$ and $\Theta ; \cdot \wtdash N : \sigma$.
    \end{enumerate}
\end{lemma}

\begin{proof}
\leavevmode
\begin{enumerate}
\item By structural induction on $M$ with $\headf{M}=  {x}$.
There are six cases to be analyzed:
\begin{enumerate}
\item $M= {x}$.
\item
In this case, ${x}\headlin{N/ {x}}=N$ and so $\Gamma \wtdash N:\sigma$. Take $\Gamma' = \dash$ and $\Delta = \Gamma$ then $\Theta ;  {x}:\sigma \wtdash  {x}:\sigma$ and $\Delta \wtdash N:\sigma$ by hypothesis, the result follows.

\item $M = M'\ B$.

From \Cref{ch4f:lambda_red}, $(M'B) \headlin{ N /  {x} } = ( M'\headlin{ N /  {x} } ) B$. Let $\Gamma = \Gamma_1, \Gamma_2$ for some $\Gamma_1,\Gamma_2$ and the derivation is the following:
\begin{prooftree}
\AxiomC{$\Theta ; \Gamma_1 \wtdash M'\headlin{ N /  {x} }: (\delta^{j} , \eta )  \rightarrow \tau$}\
\AxiomC{\quad $\Theta ; \Gamma_2 \wtdash B : (\delta^{j} , \eta ) $}
\LeftLabel{\redlab{TS{:}app}}
\BinaryInfC{$\Theta ; \Gamma_1 , \Gamma_2 \wtdash ( M'\headlin{ N /  {x} } ) B:\tau $}
\end{prooftree}
where $j \geq 0$. By IH there exists $\Gamma_1', \Delta, \sigma $ such that $\Gamma_1 = \Gamma_1' , \Delta$ with $\Theta ; \Gamma_1' ,  {x}:\sigma \wtdash M': (\delta^{j} , \eta )  \rightarrow \tau$ and $\Theta ; \Delta  \wtdash N: \sigma$. Which gives the derivation:
\begin{prooftree}
\AxiomC{$\Theta ; \Gamma_1' ,  {x}:\sigma \wtdash M': (\delta^{j} , \eta )  \rightarrow \tau$}\
\AxiomC{\quad $\Theta ; \Gamma_2 \wtdash B : (\delta^{j} , \eta ) $}
\LeftLabel{\redlab{TS{:}app}}
\BinaryInfC{$\Theta ; \Gamma_1' , \Gamma_2,  {x}:\sigma \wtdash M'B:\tau $}
\end{prooftree}
By taking $\Gamma' = \Gamma_1' , \Gamma_2$ the result follows.

\item $M = M'[ {\widetilde{y}} \leftarrow  {y}] $ with $y\neq x$.

Then from \Cref{ch4f:lambda_red},  $M'[ {\widetilde{y}} \leftarrow  {y}] \headlin{ N /  {x} } = M' \headlin{ N /  {x}} [ {\widetilde{y}} \leftarrow  {y}]$. Let $\Gamma =  \Gamma_1 ,  {y}: \delta^k$ with $k \not = 0$. Therefore,
\begin{prooftree}
\AxiomC{$ \Theta ; \Gamma_1 ,  {y}_1: \delta, \dots,  {y}_k: \delta  \wtdash  M' \headlin{ N /  {x}} : \tau \quad  {y}\notin \Gamma_1 \quad k \not = 0$}
\LeftLabel{ \redlab{TS{:}shar} }
\UnaryInfC{$ \Theta ; \Gamma_1 ,  {y}: \delta^k \wtdash M' \headlin{ N /  {x}} [ {\widetilde{y}} \leftarrow  {y}] : \tau $}
\end{prooftree}
By IH there exists $ \Gamma_1 ' , \Delta, \sigma$ such that $\Gamma_1 , {y}_1: \delta, \dots,  {y}_k: \delta = \Gamma_1' , {y}_1: \delta, \dots,  {y}_k: \delta , \Delta$ with $\Theta ; \Gamma_1' , {y}_1: \delta, \dots,  {y}_k: \delta ,  {x}:\sigma \wtdash M': \tau$ and $\Theta ; \Delta  \wtdash N: \sigma$. Which gives the derivation:
\begin{prooftree}
\AxiomC{$\Theta ; \Gamma_1' ,  {y}_1: \delta, \dots,  {y}_k: \delta ,  {x}: \sigma \wtdash M' : \tau \quad  {y}\notin \Gamma_1 \quad k \not = 0$}
\LeftLabel{ \redlab{TS{:}share}}
\UnaryInfC{$ \Theta ; \Gamma_1' ,  {y}: \delta^k,  {x}: \sigma \wtdash M'[ {y}_1 , \dots ,  {y}_k \leftarrow x] : \tau $}
\end{prooftree}
By taking $\Gamma' = \Gamma_1' , {y}: \delta^k$ the result follows.

\item $M = M'[ \leftarrow  {y}] $ with  $x \not  = y $.

Then $M'[ \leftarrow  {y}] \headlin{ N /  {x} } = M' \headlin{ N /  {x}} [ \leftarrow  {y}]$. Therefore,
\begin{prooftree}
\AxiomC{$ \Theta ;  \Gamma  \wtdash M \headlin{ N /  {x}}: \tau$}
\LeftLabel{ \redlab{TS{:}weak}}
\UnaryInfC{$ \Theta ;  \Gamma  ,  {y}: \omega \wtdash M\headlin{ N /  {x}}[\leftarrow  {y}]: \tau $}
\end{prooftree}
By IH there exists $\Gamma' , \Delta, \sigma$ such that $\Gamma = \Gamma' , \Delta$ with $\Theta ; \Gamma' ,  {x}:\sigma \wtdash M': \tau$ and $\Theta ; \Delta  \wtdash N: \sigma$. Which gives the derivation:
\begin{prooftree}
\AxiomC{$ \Theta ; \Gamma'  ,  {x}: \sigma  \wtdash M : \tau$}
\LeftLabel{ \redlab{TS{:}weak} }
\UnaryInfC{$ \Theta ;  \Gamma'  ,  {y}: \omega,  {x}: \sigma  \wtdash M[\leftarrow  {y}]: \tau $}
\end{prooftree}
By taking $\Gamma' = \Gamma'  ,  {y}: \omega$ the result follows.

\item If $M =  M' \linexsub{C /  y_1 , \dots , y_k} $  with $ x \not = y_1 , \dots , y_k$.

Then $M' \linexsub{C /  y_1 , \dots , y_k}  \headlin{ N / x } = M' \headlin{ N / x } \linexsub{C /  y_1 , \dots , y_k}  $ and
\begin{prooftree}
\AxiomC{$ \Theta ; \Gamma_1 ,  y_1:\delta, \dots , y_k:\delta  \wtdash M' \headlin{ N / x } : \tau $}
\AxiomC{$\quad \Theta ; \Gamma_2 \wtdash C : \delta^k $}
\LeftLabel{\redlab{TS{:}Esub^{\ell}}}
\BinaryInfC{$ \Theta ; \Gamma_1 , \Gamma_2  \wtdash M' \headlin{ N / x } \linexsub{C /  y_1 , \dots , y_k} : \tau $}
\end{prooftree}
By IH there exists $ \Gamma_1 ' , \Delta, \sigma$ such that $\Gamma_1 , {y}_1: \delta, \dots,  {y}_k: \delta = \Gamma_1' , {y}_1: \delta, \dots,  {y}_k: \delta , \Delta$ with $\Theta ; \Gamma_1' , {y}_1: \delta, \dots,  {y}_k: \delta ,  {x}:\sigma \wtdash M': \tau$ and $\Theta ; \Delta  \wtdash N: \sigma$. Which gives the derivation:
\begin{prooftree}
    \AxiomC{$ \Theta ; \Gamma_1'  ,  y_1:\delta, \dots , y_k:\delta , x: \sigma  \wtdash M' : \tau $}
\AxiomC{$\quad \Theta ; \Gamma_2 \wtdash C : \delta^k $}
    \LeftLabel{\redlab{TS{:}Esub^{\ell}}}
\BinaryInfC{$ \Theta ; \Gamma_1' , \Gamma_2, x: \sigma  \wtdash M' \linexsub{C /  y_1 , \dots , y_k} : \tau $}
\end{prooftree}
By taking $\Gamma' = \Gamma_1' , \Gamma_2 $ the result follows.

\item If $M =  M' \unexsub {U / {y}} $ then $\headf{M' \unexsub {U / {y}}} = \headf{M'} = x $, and the proofs is similar to the case above.
\end{enumerate}

\item  By structural induction on $M$ with $\headf{M}= {x}[i]$.
There are three cases to be analyzed:
\begin{enumerate}
\item $M= {x}[i]$.

Observe that ${x}[i]\headlin{N/{x}[i]}=N$ and let $\Theta, \unvar{x}: \eta  ; \Gamma  \wtdash {x}[i]\headlin{N/{x}[i]}:\sigma$ with $\eta_i = \sigma$
In this case $\Gamma = \dash$, and we have both
\begin{prooftree}
\AxiomC{}
\LeftLabel{ \redlab{TS{:}var^{ \ell}}}
\UnaryInfC{$ \Theta , \unvar{x}: \eta;  {x}: \eta_i  \wtdash  {x} : \sigma$}
\LeftLabel{\redlab{TS{:}var^!}}
\UnaryInfC{$ \Theta, \unvar{x}: \eta ; \cdot \wtdash {x}[i] : \sigma$}
\end{prooftree}
and $\Theta, \unvar{x}: \eta  ; \Gamma  \wtdash N:\sigma$ and the result follows.

\item $M = M'\ B$.

From \Cref{ch4f:lambda_red}, $M' \esubst{ B }{ y} \headlin{ N / {x}[i] } = M' \headlin{ N / {x}[i] } \esubst{ B }{ y}$, let $\Theta , \unvar{x}: \eta;\Gamma_1 , \Gamma_2  \wtdash ( M'\headlin{ N / {x}[i] } ) B:\tau$ where $\Gamma = \Gamma_1 , \Gamma_2 $ and $\eta_i = \sigma$ we derrive:
\begin{prooftree}
\AxiomC{$\Theta , \unvar{x}: \eta;\Gamma_1\wtdash M'\headlin{N/ {x}[i]}:(\delta^{j} , \epsilon ) \rightarrow \tau $}\
\AxiomC{$\Theta , \unvar{x}: \eta; \Gamma_2 \wtdash B : (\delta^{j} , \epsilon ) $}
\AxiomC{}
\LeftLabel{\redlab{TS{:}app}}
\TrinaryInfC{$\Theta , \unvar{x}: \eta;\Gamma_1 , \Gamma_2  \wtdash ( M'\headlin{ N / {x}[i] } ) B:\tau $}
\end{prooftree}
for some strict type $\delta$. By IH $\Theta, \unvar{x}: \eta ; \Gamma_1 \wtdash M': (\delta^{j} , \epsilon ) \rightarrow \tau $ and $\Theta ; \cdot \wtdash N : \sigma$ and we derrive:
\begin{prooftree}
\AxiomC{$ \Theta, \unvar{x}: \eta ;\Gamma_1 \wtdash M' : (\delta^{j} , \epsilon ) \rightarrow \tau \quad \Theta, \unvar{x}: \sigma ; \Gamma_2 \wtdash B : (\delta^{j} , \epsilon' )  $}
\AxiomC{}
\LeftLabel{\redlab{TS{:}app}}
\BinaryInfC{$ \Theta, \unvar{x}: \eta ;\Gamma_1 , \Gamma_2 \wtdash M'\ B : \tau$}
\end{prooftree}
and the result follows.

\item $M = M'[ {\widetilde{y}} \leftarrow  {y}] $.

From \Cref{ch4f:lambda_red}  $M'[ {\widetilde{y}} \leftarrow  {y}] \headlin{ N / {x}[i] } = M' \headlin{ N / {x}[i]} [ {\widetilde{y}} \leftarrow  {y}]$, let $\Theta , \unvar{x}: \eta; \Gamma_1 ,  {y}: \delta^k \wtdash M' \headlin{ N / {x}[i]} [ {\widetilde{y}} \leftarrow  {y}] : \tau$ with $\Gamma = \Gamma_1 ,  {y}: \delta^k$ and $\eta_i = \sigma$, then we derive:
\begin{prooftree}
\small
\AxiomC{$ \Theta , \unvar{x}: \eta; \Gamma_1 ,  {y}_1: \delta, \dots,  {y}_k: \delta  \wtdash  M' \headlin{ N / {x}[i] } : \tau \quad  {y}\notin \Gamma_1 \quad k \not = 0$}
\LeftLabel{ \redlab{TS{:}shar} }
\UnaryInfC{$ \Theta , \unvar{x}: \eta; \Gamma_1 ,  {y}: \delta^k \wtdash M' \headlin{ N / {x}[i]} [ {\widetilde{y}} \leftarrow  {y}] : \tau $}
\end{prooftree}

By IH $\Theta, \unvar{x}: \eta ; \Gamma_1 ,  {y}_1: \delta, \dots,  {y}_k: \delta \wtdash M': \tau $ and $\Theta ; \cdot \wtdash N : \sigma$ and we derrive:

\begin{prooftree}
\AxiomC{$\Theta , \unvar{x}: \eta; \Gamma_1 ,  {y}_1: \delta, \dots,  {y}_k: \delta  \wtdash M' : \tau \quad  {y}\notin \Gamma_1 \quad k \not = 0$}
\LeftLabel{ \redlab{TS{:}shar}}
\UnaryInfC{$ \Theta , \unvar{x}: \eta; \Gamma_1 ,  {y}: \delta^k \wtdash M'[ {y}_1 , \dots ,  {y}_k \leftarrow y] : \tau $}
\end{prooftree}
and the result follows.

\item $M = M'[ \leftarrow  {y}] $.

From \Cref{ch4f:lambda_red}  $M'[ \leftarrow  {y}] \headlin{ N / {x}[i] } = M' \headlin{ N / {x}[i] } [ \leftarrow  {y}]$, let $\Theta , \unvar{x}: \eta;  \Gamma  ,  {y}: \omega \wtdash M\headlin{ N / {x}[i] }[\leftarrow  {y}]: \tau$ with $\Gamma = \Gamma  ,  {y}: \omega $ and $\eta_i = \sigma$, then we derive:
\begin{prooftree}
\AxiomC{$ \Theta , \unvar{x}: \eta;  \Gamma   \wtdash M \headlin{ N / {x}[i]  }: \tau$}
\LeftLabel{ \redlab{TS{:}weak}}
\UnaryInfC{$ \Theta , \unvar{x}: \eta;  \Gamma  ,  {y}: \omega \wtdash M\headlin{ N / {x}[i] }[\leftarrow  {y}]: \tau $}
\end{prooftree}

By IH $\Theta , \unvar{x}: \eta;  \Gamma \wtdash M': \tau $ and $\Theta ; \cdot \wtdash N : \sigma$ and we derrive:
\begin{prooftree}
\AxiomC{$ \Theta , \unvar{x}: \eta; \Gamma   \wtdash M : \tau$}
\LeftLabel{ \redlab{TS{:}weak} }
\UnaryInfC{$ \Theta , \unvar{x}: \eta;  \Gamma  ,  {y}: \omega \wtdash M[\leftarrow  {y}]: \tau $}
\end{prooftree}
and the result follows.

\item $M =   M' \linexsub{C /  y_1 , \dots , y_k} $.

From \Cref{ch4f:lambda_red}  $M' \linexsub{C /  y_1 , \dots , y_k} \headlin{ N / {x}[i]  } = M' \headlin{ N / {x}[i]  }  \linexsub{C /  y_1 , \dots , y_k}  $, let $ \Theta, \unvar{x}: \eta ; \Gamma' , \Delta \wtdash M' \headlin{ N / {x}[i]  }  \linexsub{C /  y_1 , \dots , y_k} : \tau $ with $\Gamma =  \Gamma' , \Delta $ and $\eta_i = \sigma$, then we derive:
\begin{prooftree}
    \small
\AxiomC{$ \Theta, \unvar{x}: \eta ; \Gamma'  ,  y_1:\delta, \dots , y_k:\delta \wtdash M' \headlin{ N / {x}[i]  }  : \tau $}
\AxiomC{$ \Theta, \unvar{x}: \eta ; \Delta \wtdash C : \delta^k $}
\LeftLabel{\redlab{TS{:}Esub^{\ell}}}
\BinaryInfC{$ \Theta, \unvar{x}: \eta ; \Gamma' , \Delta \wtdash M' \headlin{ N / {x}[i]  }  \linexsub{C /  y_1 , \dots , y_k} : \tau $}
\end{prooftree}
By IH $\Theta , \unvar{x}: \eta;  \Gamma \wtdash M': \tau $ and $\Theta ; \cdot \wtdash N : \sigma$ and we derrive:
\begin{prooftree}
\AxiomC{$ \Theta, \unvar{x}: \eta ; \Gamma'  ,  y_1:\delta, \dots , y_k:\delta \wtdash M : \tau $}
\AxiomC{$ \Theta, \unvar{x}: \eta ; \Delta \wtdash C : \delta^k $}
\LeftLabel{\redlab{TS{:}Esub^{\ell}}}
\BinaryInfC{$ \Theta, \unvar{x}: \eta ; \Gamma' , \Delta \wtdash M \linexsub{C /  y_1, \dots , y_k} : \tau $}
\end{prooftree}
and the result follows.

\item $M =  M' \unexsub {U /\unvar{y}}$.

From \Cref{ch4f:lambda_red}  $ M' \unexsub {U /\unvar{y}} \headlin{ N /  {x}[i] } = M'  \headlin{N /  {x}[i] } \unexsub {U /\unvar{y}} $, let $  \Theta ; \Gamma \wtdash M'  \headlin{N /  {x}[i] } \unexsub{U / \unvar{y}}  : \tau  $ and $\eta_i = \sigma$, then we derive:
\begin{prooftree}
\AxiomC{$ \Theta , \unvar{x}: \eta , {y} : \epsilon; \Gamma  \wtdash M'  \headlin{N /  {x}[i] } : \tau $}
\AxiomC{$ \Theta , \unvar{x}: \eta; \dash \wtdash U : \epsilon $}
\AxiomC{$ $}
\LeftLabel{\redlab{TS{:}Esub^!}}
\TrinaryInfC{$ \Theta ; \Gamma \wtdash M'  \headlin{N /  {x}[i] } \unexsub{U / \unvar{y}}  : \tau $}
\end{prooftree}
By IH $ \Theta , \unvar{x}: \eta , {y} : \epsilon; \Gamma  \wtdash M' : \tau  $ and $\Theta ; \cdot \wtdash N : \sigma$ and we derive:
\begin{prooftree}
\AxiomC{$ \Theta , \unvar{x}: \eta , {y} : \epsilon; \Gamma  \wtdash M' : \tau $}
\AxiomC{$ \Theta , \unvar{x}: \eta; \dash \wtdash U : \epsilon $}
\AxiomC{$  $}
    \LeftLabel{\redlab{TS{:}Esub^!}}
\TrinaryInfC{$ \Theta ; \Gamma \wtdash M' \unexsub{U / \unvar{y}}  : \tau $}
\end{prooftree}
and the result follows.
\end{enumerate}
\end{enumerate}
\end{proof}

\begin{theorem}[Subject Expansion in $\lamcoldetsh$]\label{ch4t:lamSE}
If $\Theta ; \Gamma \wtdash M':\tau$ and $M \red M'$ then $\Theta ; \Gamma \wtdash M :\tau$.
\end{theorem}

\begin{proof}
By structural induction on the reduction rule from \figref{ch4fig:reduc_interm} applied in $M \red M'$.
\begin{enumerate}
\item \textbf{ Rule $\redlab{RS{:}Beta}$.}

Then $M' = N[ {\widetilde{x}} \leftarrow  {x}]\ \esubst{ B }{ x } $  and the reduction is:

\begin{prooftree}
\AxiomC{}
\LeftLabel{\redlab{RS{:}Beta}}
\UnaryInfC{$(\lambda x. N[ {\widetilde{x}} \leftarrow  {x}]) B \red N[ {\widetilde{x}} \leftarrow  {x}]\ \esubst{ B }{ x }$}
\end{prooftree}

where $ M  =  (\lambda x. N[ {\widetilde{x}} \leftarrow  {x}]) B$. Since $\Theta ; \Gamma\wtdash M':\tau$ we get the following derivation:

\begin{prooftree}
\AxiomC{$\Theta , \unvar{x} : \eta; \Gamma' ,  {x}_1:\sigma , \dots ,  {x}_j:\sigma  \wtdash  N: \tau $}
\LeftLabel{ \redlab{TS{:}share} }
\UnaryInfC{$\Theta , \unvar{x} : \eta;  \Gamma' ,   {x}:\sigma^{j}  \wtdash  N[ {\widetilde{x}} \leftarrow  {x}]: \tau $}
\AxiomC{$\Theta ;\Delta \wtdash B:(\sigma^{j} , \eta )  $}
\AxiomC{$  $}
\LeftLabel{ \redlab{TS{:}Esub} }
\TrinaryInfC{$ \Theta ;\Gamma' , \Delta \wtdash N[ {\widetilde{x}} \leftarrow  {x}]\ \esubst{ B }{ x }:\tau $}
\end{prooftree}
for $\Gamma = \Gamma' , \Delta $. Notice that:
\begin{prooftree}
\AxiomC{$\Theta , \unvar{x} : \eta; \Gamma' ,  {x}_1:\sigma , \dots ,  {x}_j:\sigma  \wtdash  N: \tau $}
\LeftLabel{ \redlab{TS{:}share} }
\UnaryInfC{$\Theta , \unvar{x} : \eta;  \Gamma' ,   {x}:\sigma^{j}  \wtdash  N[ {\widetilde{x}} \leftarrow  {x}]: \tau $}
\LeftLabel{ \redlab{TS{:}abs \dash sh} }
\UnaryInfC{$\Theta ; \Gamma' \wtdash \lambda x. N[ {\widetilde{x}} \leftarrow  {x}]: (\sigma^{j} , \eta ) \rightarrow \tau $}
\AxiomC{$\Theta ;\Delta \wtdash B: (\sigma^{j} , \eta ) $}
\AxiomC{$  $}
\LeftLabel{ \redlab{TS{:}app} }
\TrinaryInfC{$ \Theta ;\Gamma' , \Delta \wtdash (\lambda x. N[ {\widetilde{x}} \leftarrow  {x}]) B:\tau $}
\end{prooftree}
Therefore $\Theta ;\Gamma' , \Delta \wtdash (\lambda x. N[ {\widetilde{x}} \leftarrow  {x}]) B:\tau$ and the result follows.

Then $M = (\lambda x. N[ {\widetilde{x}} \leftarrow  {x}]) B $  and the reduction is:

\item \textbf{ Rule $ \redlab{RS{:}Ex \dash Sub}.$}

Then $ M' =  N\linexsub{C  /  x_1 , \dots , x_k} \unexsub{U / \unvar{x} }$ where $C=  \bag{N_1}\cdot \dots \cdot \bag{N_k} $.
The reduction is:
\begin{prooftree}
\AxiomC{$ C = \bag{N_1}
\dots  \bag{N_k} \qquad  N \not= \fail^{\widetilde{y}} $}
\LeftLabel{\redlab{RS{:}Ex \dash Sub}}
\UnaryInfC{$ N[ {x}_1, \!\dots\! ,  {x}_k \leftarrow  {x}]\esubst{ C \bagsep U }{ x } \red N\linexsub{C  /  x_1 , \dots , x_k} \unexsub{U / \unvar{x} }$}
\end{prooftree}
and $M = N[ {x}_1, \!\dots\! ,  {x}_k \leftarrow  {x}]\esubst{ C \bagsep U }{ x }$.
To simplify the proof we take $k=2$, as the case $k>2$ is similar.
Therefore $C=\bag{N_1}\cdot \bag{N_2}$.

Since $\Theta ; \Gamma\wtdash M':\tau$ we get a derivation(we omit the labels \redlab{TS:Esub^!} and \redlab{TS:Esub^{\ell}}):
{\small
\begin{prooftree}
\AxiomC{$ \Theta, \unvar{x} : \eta  ;  \Gamma' ,  {x}_1: \sigma,  {x}_2: \sigma \wtdash N : \tau$}
\AxiomC{$ \Theta ; \Delta \wtdash C : \sigma^2 $}
\BinaryInfC{$ \Theta, \unvar{x} : \eta  ;  \Gamma' , \Delta \wtdash N \linexsub{C  / {x}_1,  {x}_2} : \tau $}
\AxiomC{$ \Theta ; \dash \wtdash U : \eta $}
\AxiomC{$  $}
\TrinaryInfC{$ \Theta ; \Gamma' , \Delta \wtdash N\linexsub{C  / {x}_1,  {x}_2} \unexsub{U / \unvar{x} }  : \tau $}
\end{prooftree}
}
where $\Gamma = \Gamma' , \Delta $. Consider the typing derivation:(we omit the labels \redlab{TS:Esub} and \redlab{TS{:}share})
{\small
\begin{prooftree}
\AxiomC{$ \Theta, \unvar{x} : \eta  ;  \Gamma' ,  {x}_1: \sigma,  {x}_2: \sigma \wtdash N : \tau $}
\UnaryInfC{$  \Theta , \unvar{x} : \eta ; \Gamma' ,  {x}: \sigma^{2} \wtdash N[ {x}_1,  {x}_2 \leftarrow  {x}] : \tau  $}

\AxiomC{$ \Theta ; \Delta \wtdash C \bagsep U : (\sigma^{2} , \eta ) $}
\AxiomC{$ $}
\TrinaryInfC{$ \Theta ; \Gamma' , \Delta \wtdash N[ {x}_1,  {x}_2 \leftarrow  {x}]\esubst{ C \bagsep U }{ x }  : \tau $}
\end{prooftree}
}
and the result follows.

\item Rule $ \redlab{RS{:}Fetch^{\ell}}$.

Then $ M' =   N \headlin{ C_i / x_j }  \linexsub{(C \setminus C_i ) /  \widetilde{x}  }  $ where  $\headf{N} =  {x}_j$.
The reduction is:
\begin{prooftree}
\AxiomC{$ \headf{N} =  {x}_j$}
\LeftLabel{\redlab{RS{:}Fetch^{\ell}_{{i}}}}
\UnaryInfC{$  N \linexsub{C /  \widetilde{x}, x_j} \red  N \headlin{ C_i / x_j }  \linexsub{(C \setminus C_i ) /  \widetilde{x}  } $}
\end{prooftree}
and $M =  N \linexsub{C /  \widetilde{x}, x_j} $.
Since $\Theta ; \Gamma\wtdash M':\tau$ we get the following derivation:
\begin{prooftree}
\AxiomC{$ \Theta ; \Gamma' , \Delta_i ,  \widetilde{x}:\sigma^{k-1} \wtdash N \headlin{ C_i / x_j }: \tau $}
\AxiomC{$ \Theta ; \Delta'_i \wtdash C \setminus C_i : \sigma^{k-1} $}
\LeftLabel{\redlab{TS{:}Esub^{\ell}}}
\BinaryInfC{$ \Theta ; \Gamma', \Delta \wtdash N \headlin{ C_i / x_j }  \linexsub{(C \setminus C_i ) /  \widetilde{x}  }  : \tau $}
\end{prooftree}
where $\Gamma = \Gamma' , \Delta $ and $\Delta = \Delta_i , \Delta_i'$.
By \Cref{ch4l:lamrsharfailantisub}, we obtain the derivation $ \Theta ; \Gamma'  ,  \widetilde{x}:\sigma^{k-1},  x_j:\sigma \wtdash N : \tau $ and $ \Theta ;  \Delta_i \wtdash  C_i : \sigma $ via:
\begin{prooftree}
\AxiomC{$ \Theta ; \Gamma'  ,  \widetilde{x}:\sigma^{k-1},  x_j:\sigma \wtdash N : \tau $}
\AxiomC{$ \Theta ; \Delta \wtdash C : \sigma^k $}
\LeftLabel{\redlab{TS{:}Esub^{\ell}}}
\BinaryInfC{$ \Theta ; \Gamma' , \Delta \wtdash N \linexsub{C /  \widetilde{x}, x_j}  : \tau $}
\end{prooftree}

\item Rule $ \redlab{RS{:} Fetch^!}$.

Then $ M' = N \headlin{ N_i /{x}[i] }\unexsub{U / \unvar{x}}  $ where  $\headf{M} = {x}[i]$. The reduction is:
\begin{prooftree}
\AxiomC{$ \headf{N} = {x}[i]$}
\AxiomC{$ U_i = \unvar{\bag{N_i}}$}
\LeftLabel{\redlab{RS{:} Fetch^!}}
\BinaryInfC{$  N \unexsub{U / \unvar{x}} \red  N \headlin{ N_i /{x}[i] }\unexsub{U / \unvar{x}} $}
\end{prooftree}
and $M =  N \unexsub{U / \unvar{x}}  $.     Since $\Theta ; \Gamma \wtdash M':\tau$ we get the following derivation:
\begin{prooftree}
\AxiomC{$ \Theta , \unvar{x} : \eta; \Gamma  \wtdash N \headlin{ N_i /{x}[i] } : \tau $}
\AxiomC{$ \Theta ; \dash \wtdash U : \eta $}
\AxiomC{$  $}
\LeftLabel{\redlab{TS{:}Esub^!}}
\TrinaryInfC{$ \Theta ; \Gamma \wtdash N \headlin{ N_i /{x}[i] }\unexsub{U / \unvar{x}}  : \tau $}
\end{prooftree}
By \Cref{ch4l:lamrsharfailantisub}, we obtain the derivation $ \Theta , {x} : \eta; \Gamma  \wtdash N : \tau  $, and the result follows from:
\begin{prooftree}
\AxiomC{$ \Theta , {x} : \eta; \Gamma  \wtdash N : \tau $}
\AxiomC{$ \Theta ; \dash \wtdash U : \eta $}
\AxiomC{$ $}
\LeftLabel{\redlab{TS{:}Esub^!}}
\TrinaryInfC{$ \Theta ; \Gamma \wtdash N \unexsub{U / \unvar{x}}  : \tau $}
\end{prooftree}

\item Rule $\redlab{RS{:}TCont}$.

Then $M' = C[N']$ and the reduction is as follows:
\begin{prooftree}
\AxiomC{$   N \red N' $}
\LeftLabel{\redlab{RS{:}TCont}}
\UnaryInfC{$ C[N] \red  C[N'] $}
\end{prooftree}
with $M=  C[N] $.
The proof proceeds by analysing the context $C$.
There are three cases:
\begin{enumerate}
\item $C=[\cdot]\ B$.

In this case $ M' = N' \ B$, for some $B$.
Since $\Gamma\vdash M':\tau$ one has a derivation:
\begin{prooftree}
\AxiomC{$ \Theta ;\Gamma' \wtdash N' : (\sigma^{j} , \eta ) \rightarrow \tau $}
\AxiomC{$  \Theta ;\Delta \wtdash B : (\sigma^{j} , \eta )  $}
\AxiomC{$ $}
\LeftLabel{\redlab{TS{:}app}}
\TrinaryInfC{$ \Theta ; \Gamma' , \Delta \wtdash N' \ B : \tau$}
\end{prooftree}
where $\Gamma = \Gamma' , \Delta $.
From  $\Gamma'\wtdash N':(\sigma^{j} , \eta ) \rightarrow \tau$ and the reduction $N \red N' $, one has by IH that  $\Gamma'\wtdash N:(\sigma^{j} , \eta ) \rightarrow \tau$.
Finally, we may type the following:
\begin{prooftree}
\AxiomC{$ \Theta ;\Gamma' \wtdash N : (\sigma^{j} , \eta ) \rightarrow \tau $}
\AxiomC{$  \Theta ;\Delta \wtdash B : (\sigma^{j} , \eta )  $}
\AxiomC{$  $}
\LeftLabel{\redlab{TS{:}app}}
\TrinaryInfC{$ \Theta ; \Gamma' , \Delta \wtdash N\ B : \tau$}
\end{prooftree}
Since $ M  =   (C[N]) = NB $, the result follows.

\item  Cases $C=[\cdot]\linexsub{N/x} $ and $C=[\cdot][\widetilde{x} \leftarrow x]$ are similar to the previous.
\end{enumerate}

\item Rules $ \redlab{RS{:}Fail^{\ell}}$, $ \redlab{RS{:}Fail^!}$ and $\redlab{RS{:}Cons_1}$.
These cases are trivial, since $\oneb$, $\oneb^!$ and $\fail^{\widetilde{x} \cup \widetilde{y}}$ are not well-typed.

\end{enumerate}
\end{proof}

\section{Proof of Tight Correctness of the Translation under the Lazy Semantics}\label{ch4a:tight}

Here we prove \Cref{ch4t:correncLazy}.

\begin{definition}{Success}\label{ch4def:app_Suc3unres}
    We define define $\headf{\sucs{\lambda}} = \sucs{\lambda}$ and $\piencodfaplas{\sucs{\lambda}}_u = \sucs{\pi}$.
    We define {\em success} for terms  $M\in  \lamcoldet$ and $P\in \clpi$.
    \begin{itemize}
        \item \succp{{M}}{\sucs{\lambda}} if and only if, there exist  $M'_1 , \cdots , M_k'\in \lamcoldetsh$ such that ${M} \red^*  M'$ and $\headf{M'} = \sucs{\lambda}$.

        \item  ${P}\succone{\sucs{\pi}}$ if and only if, there exist  $Q_1 , Q_2\in \clpi$ such that $P \redone^*   (Q_1   \| \sucs{\pi}) \nd Q_2 $.

        \item  ${P}\succtwo{\sucs{\pi}}$ if and only if, there exist  $S$ and $ Q_1 , Q_2\in \clpi$ such that $P \redtwo_S^*   (Q_1   \| \sucs{\pi}) \nd Q_2 $.
    \end{itemize}
\end{definition}

\subsection{Type Preservation}\label{ch4a:typepres}

\begin{lemma}\label{ch4prop:app_auxunres}
 $ \piencodfaplas{\sigma^{j}}_{(\tau_1, m)} = \piencodfaplas{\sigma^{k}}_{(\tau_2, n)}$ and $ \piencodfaplas{(\sigma^{j} , \eta)}_{(\tau_1, m)} = \piencodfaplas{(\sigma^{k}, \eta)}_{(\tau_2, n)}$ hold, provided that  $\tau_1,\tau_2,n$ and $m$ are as follows:

        \begin{enumerate}
        \item If $j > k$ then take $\tau_1 $ to be an arbitrary type, $m = 0$,  take $\tau_2 $ to be $\sigma$ and $n = j-k$.

        \item If $j < k$ then take $\tau_1 $ to be $\sigma$, $m = k-j$,  take $\tau_2 $ to be an arbitrary type and $n = 0$.

        \item Otherwise, if $j = k$ then take $m = n = 0$. In this case, $\tau_1. \tau_2 $ are unimportant.
    \end{enumerate}

\end{lemma}

\begin{proof}
This proof  proceeds by analyzing the conditions on types, following \Cref{ch3}.
\end{proof}

\begin{lemma}\label{ch4lem:relunbag-typeunres}
If
$ \eta \relunbag \epsilon $ then the following hold:
\begin{enumerate}
    \item If $\piencodfaplas{M}_u\vdash \piencodfaplas{\Gamma} , \piencodfaplas{\Theta} , \unvar{x} : \dual{\piencodfaplas{\eta}}$
    then
    $\piencodfaplas{M}_u\vdash \piencodfaplas{\Gamma} , \piencodfaplas{\Theta}, \unvar{x} : \dual{\piencodfaplas{\epsilon}}$.

    \item If $\piencodfaplas{M}_u\vdash \piencodfaplas{\Gamma}, u:\piencodfaplas{(\sigma^{j} , \eta ) \rightarrow \tau} , \piencodfaplas{\Theta}$
    then
    $\piencodfaplas{M}_u\vdash \piencodfaplas{\Gamma}, u:\piencodfaplas{(\sigma^{j} , \epsilon ) \rightarrow \tau} , \piencodfaplas{\Theta}$.

\end{enumerate}

\end{lemma}

\begin{proof} The proof is  by mutual induction on the the derivations of If $\piencodfaplas{M}_u\vdash \piencodfaplas{\Gamma} , \piencodfaplas{\Theta} , \unvar{x} : \dual{\piencodfaplas{\eta}}$ and  $\piencodfaplas{M}_u\vdash \piencodfaplas{\Gamma}, u:\piencodfaplas{(\sigma^{j} , \eta ) \rightarrow \tau} , \piencodfaplas{\Theta}$ and on the structure of $M$.

We use \defref{ch4def:enc_sestypfailunres} that establishes   \(  \piencodfaplas{ \eta } = ! \with_{\eta_i \in \eta} \{ i : \piencodfaplas{\eta_i} \}\) and by duality $\dual{\piencodfaplas{ \eta }} = ? \oplus_{\eta_i \in \eta} \{ i : \dual{\piencodfaplas{\eta_i}} \}$.
\begin{enumerate}

\item $M =   {x}$.
\label{ch4proof:relunbag-no}

By the translation in \figref{ch4fig:encoding}:  $\piencodfaplas{ {x}}_u =  \psome{x} ; \pfwd{x}{u} $. We have the following derivation,  for some type $A$:

\begin{prooftree}
    \AxiomC{}
    \LeftLabel{\ttype{id}}
    \UnaryInfC{$ \pfwd{x}{u} \vdash  {x}:  \overline{A}  , u :  A  $}
    \LeftLabel{\ttype{weaken}}
    \UnaryInfC{$ \pfwd{x}{u} \vdash  {x}:  \overline{A}  , u :  A, \unvar{x} : \dual{\piencodfaplas{\eta}}$}
    \LeftLabel{\ttype{$\with$}}
    \UnaryInfC{$  \psome{x}; \pfwd{x}{u} \vdash  {x}: \with  \overline{A} , u :  A , \unvar{x} : \dual{\piencodfaplas{\eta}} $}
\end{prooftree}

The derivation is independent of $\unvar{x} : \dual{\piencodfaplas{\eta}}$, hence the result  trivially holds for $ \piencodfaplas{\Gamma} , \piencodfaplas{\Theta}={x}: \with  \overline{A} , u :  A$.
\item $ M =  {x}[k]$.
\label{ch4proof:relunbag-yes}

By the translation in \figref{ch4fig:encoding}:  $\piencodfaplas{ {x}[k]}_u = \puname{ \unvar{x} }{ {x_i} }; \psel{ {x}_i }{k };  \pfwd{x_i}{u} $. We have the following derivation:

\begin{prooftree}
\AxiomC{}
\LeftLabel{\ttype{id}}
\UnaryInfC{$ \pfwd{x_i}{u}  \vdash  u :  \piencodfaplas{ \tau },  x_i:  \overline{\piencodfaplas{\eta_{k} }}  $}`'
\LeftLabel{ \ttype{$\oplus$}}
\UnaryInfC{$  \psel{ {x}_i }{ k }; \pfwd{x_i}{u} \vdash  u :  \piencodfaplas{ \tau }, {x}_i :  \oplus_{\eta_i \in \eta} \{ i : \dual{\piencodfaplas{\eta_i}}  \} $}
\LeftLabel{\ttype{$?$}}
\UnaryInfC{$ \puname{ \unvar{x} }{ x_i }; \psel{ {x}_i }{ k }; \pfwd{x_i}{u} \vdash  u :  \piencodfaplas{ \tau }, \unvar{x}:? \oplus_{\eta_i \in \eta} \{ i : \dual{\piencodfaplas{\eta_i}} \}  $}
\end{prooftree}

  Since $ \eta \relunbag \epsilon $ we have that $ \epsilon_{k} = \eta_{k} $ for each $k=1,\ldots |\eta|$. Thus, the  same derivation above, replacing $\eta_i$'s for $\epsilon_i$'s entails $ \puname{ \unvar{x} }{ x_i }; \psel{ {x}_i }{ k}; \pfwd{x_i}{u} \vdash  u :  \piencodfaplas{ \tau }; \unvar{x}:\dual{\piencodf{\epsilon}}$, and the result follows. For the case of $ M =  {y}[k]$ with $y \not =  x$ we use the argument that the typing of $y$ is independent on $x$.

\item $ M =  \lambda y . (M'[ {\widetilde{y}} \leftarrow  {y}])$.

From the translation in
\figref{ch4fig:encoding}:

$ \piencodfaplas{\lambda y.M'[ {\widetilde{y}} \leftarrow y]}_u = \psome{u}; \gname{u}{y}; \psome{y};\underbrace{ \gname{ y }{ \linvar{y} }; \gname{ y }{ \unvar{y} }; \gclose{ y } ;  \piencodfaplas{M'[ {\widetilde{y}}\leftarrow  {y}]}_u}_{P}$.

    \begin{prooftree}
        \small
    \AxiomC{$ \Pi_1 $}
        \noLine
        \UnaryInfC{$ \vdots $}
        \noLine
        \UnaryInfC{$\piencodfaplas{M'[ {\widetilde{y}} \leftarrow  {y}]}_u \vdash  u:\piencodfaplas{\tau} , \piencodfaplas{\Gamma'} , \linvar{y}: \overline{\piencodfaplas{\sigma^k}_{(\sigma, i)}}  , \piencodfaplas{\Theta} , \unvar{y}: \overline{\piencodfaplas{\eta}}$}
        \LeftLabel{\ttype{$\bot$}}
        \UnaryInfC{$ \gclose{ y } ; \piencodfaplas{M'[ {\widetilde{y}} \leftarrow  {y}]}_u \vdash y{:}\bot, u:\piencodfaplas{\tau} , \piencodfaplas{\Gamma'} , \linvar{y}: \overline{\piencodfaplas{\sigma^k}_{(\sigma, i)}}  , \piencodfaplas{\Theta} , \unvar{y}: \overline{\piencodfaplas{\eta}}$}
        \LeftLabel{\ttype{$\parr$}}
        \UnaryInfC{$\gname{ y }{ \unvar{y} }; \gclose{ y } ; \piencodfaplas{M'[ {\widetilde{y}} \leftarrow  {y}]}_u \vdash y: ( \overline{\piencodfaplas{\eta}}) \ampy (\bot) , u:\piencodfaplas{\tau} , \piencodfaplas{\Gamma'} , \linvar{y}: \overline{\piencodfaplas{\sigma^k}_{(\sigma, i)}} , \piencodfaplas{\Theta} $}
        \LeftLabel{\ttype{$\parr$}}
        \UnaryInfC{$ \gname{ y }{ \linvar{y} }; \gname{ y }{ \unvar{y} }; \gclose{ y } ; \piencodfaplas{M'[ {\widetilde{y}} \leftarrow  {y}]}_u \vdash  y: \overline{\piencodfaplas{\sigma^k}_{(\sigma, i)}} \ampy (( \overline{\piencodfaplas{\eta}}) \ampy (\bot)) , u:\piencodfaplas{\tau} , \piencodfaplas{\Gamma'} , \piencodfaplas{\Theta} $}
        \LeftLabel{\ttype{${\with}\some$}}
        \UnaryInfC{$ \psome{y}; P \vdash y :\with(  \overline{\piencodfaplas{\sigma^k}_{(\sigma, i)}} \ampy (( \overline{\piencodfaplas{\eta}}) \ampy (\bot))) , u:\piencodfaplas{\tau} , \piencodfaplas{\Gamma'} , \piencodfaplas{\Theta} $}
        \LeftLabel{\ttype{$\parr$}}
        \UnaryInfC{$\gname{ u }{ y }; \psome{y}; P \vdash u: \with(  \overline{\piencodfaplas{\sigma^k}_{(\sigma, i)}} \ampy (( \overline{\piencodfaplas{\eta}}) \ampy (\bot))) \ampy \piencodfaplas{\tau} , \piencodfaplas{\Gamma'} , \piencodfaplas{\Theta}  $}
        \LeftLabel{\ttype{${\with}\some$}}
        \UnaryInfC{$\psome{u}; \gname{ u }{ y }; \psome{y}; P \vdash u :\underbrace{\with (\with(  \overline{\piencodfaplas{\sigma^k}_{(\sigma, i)}} \ampy (( \overline{\piencodfaplas{\eta}}) \ampy (\bot))) \ampy \piencodfaplas{\tau})}_{\piencodfaplas{(\sigma^{k} , \eta )   \rightarrow \tau} \qquad \defref{ch4def:enc_sestypfailunres}} , \piencodfaplas{\Gamma'} , \piencodfaplas{\Theta} $}
    \end{prooftree}

    Let us consider the following two cases:
    \begin{itemize}
        \item $y = x$

        By the IH  there exists a derivation $\Pi_1'$ of \(\piencodfaplas{M'[ {\widetilde{y}} \leftarrow  {y}]}_u \vdash  u:\piencodfaplas{\tau} , \piencodfaplas{\Gamma'} , \linvar{y}: \overline{\piencodfaplas{\sigma^k}_{(\sigma, i)}}  , \piencodfaplas{\Theta} , \unvar{y}: \overline{\piencodfaplas{\epsilon}}\). Following the steps above we obtain \(\psome{u}; \gname{ u }{ y }; \psome{y}; P \vdash u:\piencodf{(\sigma^{k} , \epsilon )   \rightarrow \tau}, \piencodfaplas{\Gamma'} , \piencodfaplas{\Theta}\).

        \item $ y \not = x$

        Then we have that $\piencodfaplas{\Theta} = \piencodfaplas{\Theta'}, \unvar{x} : \dual{\piencodfaplas{\epsilon}}$ By the IH  there exists a derivation $\Pi_1'$ of \(\piencodfaplas{M'[ {\widetilde{y}} \leftarrow  {y}]}_u \vdash  u:\piencodfaplas{\tau} , \piencodfaplas{\Gamma'} , \linvar{y}: \overline{\piencodfaplas{\sigma^k}_{(\sigma, i)}}  , \piencodfaplas{\Theta'} , \unvar{y}: \overline{\piencodfaplas{\eta}}, \unvar{x} : \dual{\piencodfaplas{\epsilon}} \) and then redo the steps above and obtain \(\psome{u}; \gname{ u }{ y }; \psome{y}; P \vdash u:\piencodf{(\sigma^{k} , \eta )   \rightarrow \tau}, \piencodfaplas{\Gamma'} , \piencodfaplas{\Theta'}, \unvar{x} : \dual{\piencodfaplas{\epsilon}} \).

    \end{itemize}

    \item The analysis of the other cases proceeds similarly. 

    \end{enumerate}

\end{proof}

\begin{restatable}[Type Preservation]{theorem}{thmEncTypePres}\label{ch4t:preservationencode}
    Let $B$ and ${M}$ be a bag and an term in \lamcoldetsh, respectively.
    \begin{enumerate}
        \item If $\Theta ; \Gamma \wfdash B : (\sigma^{k} , \eta )$
        then
        $\piencodfaplas{B}_u \wfdash  \piencodfaplas{\Gamma}, u : \piencodfaplas{(\sigma^{k} , \eta )}_{(\sigma, i)} , \piencodfaplas{\Theta}$.

        \item If $\Theta ; \Gamma \wfdash M : \tau$
        then
        $\piencodfaplas{{M}}_u \wfdash  \piencodfaplas{\Gamma}, u :\piencodfaplas{\tau} , \piencodfaplas{\Theta}$.
    \end{enumerate}
\end{restatable}

\begin{proof}

The proof is by mutual induction on the typing derivation of $B$ and ${M'}$, with an analysis for the last rule applied.
Recall that the translation of types ($\piencodfaplas{-}$) has been given in
\defref{ch4def:enc_sestypfailunres}. We will be silently performing \ttype{contract} to split the translation of unrestricted context in derivation trees of \clpi processes as well as combining multiple \ttype{weaken} when convenient.

\begin{enumerate}
\item For $ B =  C \bagsep U$:

\begin{prooftree}
    \AxiomC{\( \Theta ; \Gamma\wfdash C : \sigma^k\)}
    \AxiomC{\(  \Theta ;\cdot \wfdash  U : \eta \)}
\LeftLabel{\redlab{FS{:}bag}}
\BinaryInfC{\( \Theta ; \Gamma \wfdash C \bagsep U : (\sigma^{k} , \eta  ) \)}
\end{prooftree}

        Our translation gives:
$ \piencodfaplas{C \bagsep U}_u = \gsome{ x }{ \llfv{C} }; \pname{x}{\linvar{x}} .( \piencodfaplas{ C }_{\linvar{x}}   \| \pname{x}{\unvar{x}} .(  \guname{ \unvar{x} }{ x_i };  \piencodfaplas{ U }_{x_i}   \|  \pclose{ x } ) )$.
        In addition, the translation of $(\sigma^{k} , \eta  )$ is:
        $$\piencodfaplas{ (\sigma^{k} , \eta  )  }_{(\sigma, i)} = \oplus( (\piencodfaplas{\sigma^{k} }_{(\sigma, i)}) \otimes ((\piencodfaplas{\eta}) \otimes (\onef))  )\quad \text{(for some  $i \geq 0$ and  strict type $\sigma$)}$$

        And one can build the following type derivation (rules from \figref{ch4fig:trulespi}):
{
\small
\begin{adjustwidth}{-1cm}{}
\begin{prooftree}
\AxiomC{$ \piencodfaplas{ C }_{\linvar{x}} \vdash \piencodfaplas{\Gamma}, \linvar{x}:\piencodfaplas{\sigma^{k} }_{(\sigma, i)} , \piencodfaplas{\Theta}$}

\AxiomC{$ \piencodfaplas{ U }_{x_i} \vdash x_i: \with_{\eta_i \in \eta} \{ i ; \piencodfaplas{\eta_i} \}, \piencodfaplas{\Theta}$}
\LeftLabel{\ttype{$!$}}
\UnaryInfC{$ \guname{ \unvar{x} }{ x_i } ;  \piencodfaplas{ U }_{x_i} \vdash \unvar{x}: \piencodfaplas{\eta} , \piencodfaplas{\Theta}$}

\AxiomC{\mbox{\ }}
\LeftLabel{\ttype{$\1$}}
\UnaryInfC{$ \pclose{ x } \vdash x: \onef $}
\UnaryInfC{$ \pclose{ x } \vdash x: \onef , \piencodfaplas{\Theta}$}
\LeftLabel{\ttype{$\tensor$}}
\BinaryInfC{$ \pname{x}{\unvar{x}} .(  \guname{ \unvar{x} }{ x_i } ;  \piencodfaplas{ U }_{x_i}   \|  \pclose{ x } ) \vdash x: (\piencodfaplas{\eta}) \otimes (\onef) , \piencodfaplas{\Theta}$}
\LeftLabel{\ttype{$\tensor$}}
\BinaryInfC{$\pname{x}{\linvar{x}} .( \piencodfaplas{ C }_{\linvar{x}}   \| \pname{x}{\unvar{x}} .(  \guname{ \unvar{x} }{ x_i } ;  \piencodfaplas{ U }_{x_i}   \|  \pclose{ x } ) ) \vdash  \piencodfaplas{\Gamma}, x:(\piencodfaplas{\sigma^{k} }_{(\sigma, i)}) \otimes ((\piencodfaplas{\eta}) \otimes (\onef)) , \piencodfaplas{\Theta} $}
\LeftLabel{\ttype{${\oplus}\some$}}
\UnaryInfC{$\gsome{ x }{ \llfv{C} };  \pname{x}{\linvar{x}} .( \piencodfaplas{ C }_{\linvar{x}}   \| \pname{x}{\unvar{x}} .(  \guname{ \unvar{x} }{ x_i } ;  \piencodfaplas{ U }_{x_i}   \|  \pclose{ x } ) ) \vdash \piencodfaplas{\Gamma}, x{:}\piencodfaplas{ (\sigma^{k} , \eta  )  }_{(\sigma, i)}, \piencodfaplas{\Theta}$}
\end{prooftree}
\end{adjustwidth}
}

The result follows  provided both $ \piencodfaplas{ C }_{\linvar{x}} \vdash \piencodfaplas{\Gamma}, \linvar{x}:\piencodfaplas{\sigma^{k} }_{(\sigma, i)} , \piencodfaplas{\Theta}$ and $ \piencodfaplas{ U }_{x_i} \vdash  x_i: \with_{\eta_i \in \eta} \{ i ; \piencodfaplas{\eta_i} \}, \piencodfaplas{\Theta}$ hold.

\begin{enumerate}

\item For $ \piencodfaplas{ C }_{\linvar{x}} \vdash \piencodfaplas{\Gamma}, \linvar{x}:\piencodfaplas{\sigma^{k} }_{(\sigma, i)} , \piencodfaplas{\Theta}$ to hold analyze the shape of $C$:

\noindent{\bf For $C = \oneb$:}
\begin{prooftree}
    \AxiomC{\(  \)}
    \LeftLabel{\redlab{FS{:}\oneb^{\ell}}}
    \UnaryInfC{\( \Theta ; \dash \wfdash \oneb : \omega \)}
\end{prooftree}

Our translation gives:         $ \piencodfaplas{\oneb}_{\linvar{x}} = \gsome{ \linvar{x} }{ \emptyset }; \gname{ \linvar{x} }{ y_n }; ( \psome{y_n}; \pclose{ y_n }    \| \gsome{ \linvar{x} }{ \emptyset };   \pnone{ \linvar{x} }  ). $

and  the translation of $\omega$ can be either:
\begin{enumerate}
\item  $\piencodfaplas{\omega}_{(\sigma,0)} =  \overline{\with(( \oplus \bot )\otimes ( \with \oplus \bot ))}$; or
\item $\piencodfaplas{\omega}_{(\sigma, i)} =  \overline{   \with(( \oplus \bot) \otimes ( \with  \oplus (( \with  \overline{\piencodfaplas{ \sigma }} )  \ampy (\overline{\piencodfaplas{\omega}_{(\sigma, i - 1)}})))) }$
\end{enumerate}

And one can build the following type derivation (rules from \figref{ch4fig:trulespi}):

\begin{prooftree}
\small
\AxiomC{\mbox{\ }}
\UnaryInfC{$ \pclose{ y_n } \vdash y_n: \onef$}
\UnaryInfC{$ \pclose{ y_n } \vdash y_n: \onef, \piencodfaplas{\Theta}$}
\UnaryInfC{$\psome{y_n}; \pclose{ y_n }  \vdash  y_n :\with \onef , \piencodfaplas{\Theta}$}
\AxiomC{}
\UnaryInfC{$  \pnone{ \linvar{x} } \vdash \linvar{x} :\with A $}
\UnaryInfC{$  \pnone{ \linvar{x} } \vdash \linvar{x} :\with A , \piencodfaplas{\Theta}$}
\UnaryInfC{$\gsome{ \linvar{x} }{ \emptyset };   \pnone{ \linvar{x} }\vdash  \linvar{x}{:}\oplus \with A, \piencodfaplas{\Theta}$}
\BinaryInfC{$(\psome{y_n}; \pclose{ y_n }   \| \gsome{ \linvar{x} }{ \emptyset };   \pnone{ \linvar{x} } ) \vdash y_n :\with \onef, \linvar{x}{:}\oplus \with A , \piencodfaplas{\Theta}$}
\UnaryInfC{$\gname{\linvar{x}}{y_n}; ( \psome{y_n}; \pclose{ y_n }    \| \gsome{ \linvar{x} }{ \emptyset };   \pnone{ \linvar{x} } ) \vdash  \linvar{x}: (\with \onef) \ampy (\oplus \with A) , \piencodfaplas{\Theta}$}
\UnaryInfC{$\gsome{ \linvar{x} }{ \emptyset };  \gname{ \linvar{x} }{y_n  };  ( \psome{y_n} ; \pclose{ y_n }    \| \gsome{ \linvar{x} }{ \emptyset };    \pnone{ \linvar{x} }) \vdash  \linvar{x}{:}\oplus ((\with \onef) \ampy (\oplus \with A)), \piencodfaplas{\Theta}$}
\end{prooftree}

Since $A$ is arbitrary,  we can take $A=\oneb$ for $\piencodfaplas{\omega}_{(\sigma,0)} $ and  $A= \overline{(( \with  \overline{\piencodfaplas{ \sigma }} )  \ampy (\overline{\piencodfaplas{\omega}_{(\sigma, i - 1)}}))}$  for $\piencodfaplas{\omega}_{(\sigma,i)} $, in both cases, the result follows.

\item
\noindent{\bf For  $C = \bag{M'}  \cdot C'$:}

\begin{prooftree}
\AxiomC{\( \Theta ; \Gamma' \wfdash M' : \sigma\)}
\AxiomC{\( \Theta ; \Delta \wfdash C' : \sigma^k\)}
\LeftLabel{\redlab{FS{:}bag^{\ell}}}
\BinaryInfC{\( \Theta ; \Gamma' \contexcat \Delta \wfdash \bag{M'}  \cdot C':\sigma^{k}\)}
\end{prooftree}

Where $ \Gamma = \Gamma' \contexcat \Delta$.
To simplify the proof, we will consider $k=3$.

By IH we have \(
\piencodfaplas{M'}_{x_i}  \vdash \piencodfaplas{\Gamma'}, x_i: \piencodfaplas{\sigma}; \piencodfaplas{\Theta} \) and
\( \piencodfaplas{C'}_{\linvar{x}}  \vdash \piencodfaplas{\Delta}, \linvar{x}: \piencodfaplas{\sigma\wedge \sigma}_{(\tau, j)}; \piencodfaplas{\Theta}\).

Our translation  \figref{ch4fig:encoding} gives:
\begin{equation}
\begin{aligned}
\piencodfaplas{C}_{\linvar{x}} & =
        \gsome{ \linvar{x} }{ \llfv{C} }; \gname{\linvar{x}}{y_i}; \gsome{ \linvar{x} }{ y_i, \llfv{C} }; \psome{\linvar{x}} ;  \pname{\linvar{x}}{x_i}. \\
        & \qquad  (\gsome{ x_i }{ \llfv{M'} };  \piencodfaplas{M'}_{x_i}   \| \piencodfaplas{(C' )}_{\linvar{x}}   \|   \pnone{ y_i })
\end{aligned}
\end{equation}

Let $\Pi_{1}$ be the derivation:

\begin{prooftree}
\AxiomC{$\piencodfaplas{M'}_{x_i} \;{ \vdash} \piencodfaplas{\Gamma'}, x_i: \piencodfaplas{\sigma}, \piencodfaplas{\Theta} $}
\UnaryInfC{$\gsome{ x_i }{ \llfv{M'} };  \piencodfaplas{M'}_{x_i} \vdash \piencodfaplas{\Gamma'} ,x_i: \oplus \piencodfaplas{\sigma} , \piencodfaplas{\Theta}$}
\AxiomC{}
\UnaryInfC{$   \pnone{ y_i }  \vdash y_i :\with \onef $}
\UnaryInfC{$   \pnone{ y_i }  \vdash y_i :\with \onef, \piencodfaplas{\Theta}$}
\BinaryInfC{$\gsome{  x_i }{ \llfv{M'} }; \piencodfaplas{M'}_{x_i}   \|   \pnone{ y_i } \vdash \piencodfaplas{\Gamma'} ,x_i: \oplus \piencodfaplas{\sigma}, y_i :\with \onef , \piencodfaplas{\Theta}$}
\end{prooftree}

Let $ P =   \piencodfaplas{M'}_{x_i}   \| \piencodfaplas{C'}_{\linvar{x}}   \|   \pnone{ y_i }$, in the derivation $\Pi_{2}$ below:

\begin{adjustwidth}{-1cm}{}
\begin{prooftree}
    \small
\AxiomC{$ \Pi_{1}$}

\AxiomC{$ \piencodfaplas{C' }_{\linvar{x}}  \vdash  \piencodfaplas{\Delta}, \linvar{x}: \piencodfaplas{\sigma\wedge \sigma}_{(\tau, j)}, \piencodfaplas{\Theta} $}

\LeftLabel{\ttype{$\tensor$}}
\BinaryInfC{$ \pname{\linvar{x}}{x_i}. P \vdash  \piencodfaplas{\Gamma'}  ,  \piencodfaplas{\Delta}, y_i :\with \onef, \linvar{x}: (\oplus \piencodfaplas{\sigma})  \otimes (\piencodfaplas{\sigma\wedge \sigma}_{(\tau, j)}) , \piencodfaplas{\Theta}$}
\LeftLabel{\ttype{${\with}\some$}}
\UnaryInfC{$\underbrace{ \psome{\linvar{x}} ;   \pname{\linvar{x}}{x_i}. P }_{P_{2}} \vdash \piencodfaplas{\Gamma'}  ,  \piencodfaplas{\Delta}, y_i :\with \onef, \linvar{x}: \with (( \oplus \piencodfaplas{\sigma} ) \otimes (\piencodfaplas{\sigma\wedge \sigma}_{(\tau, j)})) , \piencodfaplas{\Theta} $}
\end{prooftree}
\end{adjustwidth}

Let $P_2 =   \psome{\linvar{x}} ;  \pname{\linvar{x}}{x_i}. P$ in the derivation below:

\begin{adjustwidth}{-1.5cm}{}
\begin{prooftree}
    \small
\AxiomC{$ \Pi_{2}$}
\noLine
\UnaryInfC{$\vdots$}
\noLine
\UnaryInfC{$P_2\vdash \piencodfaplas{\Gamma}, y_i :\with \onef, \linvar{x}: \with (( \oplus \piencodfaplas{\sigma} ) \otimes (\piencodfaplas{\sigma\wedge \sigma}_{(\tau, j)})) , \piencodfaplas{\Theta} $}
\LeftLabel{\ttype{${\oplus}\some$}}
\UnaryInfC{$\gsome{ \linvar{x} }{ y_i, \llfv{C} };P_2  \vdash \piencodfaplas{\Gamma}, y_i :\with \onef, \linvar{x}:\oplus  \with (( \oplus \piencodfaplas{\sigma} ) \otimes (\piencodfaplas{\sigma\wedge \sigma}_{(\tau, j)})), \piencodfaplas{\Theta}$}
\LeftLabel{\ttype{$\parr$}}
\UnaryInfC{$
\gname{\linvar{x}}{y_i}; \gsome{ \linvar{x} }{ y_i, \llfv{C} }; P_2
\vdash \piencodfaplas{\Gamma}, \linvar{x}: ( \with \onef) \ampy ( \oplus  \with (( \oplus \piencodfaplas{\sigma} ) \otimes (\piencodfaplas{\sigma\wedge \sigma}_{(\tau, j)}))), \piencodfaplas{\Theta} $}
\LeftLabel{\ttype{${\oplus}\some$}}
\UnaryInfC{$\piencodfaplas{C}_{\linvar{x}} \vdash   \piencodfaplas{\Gamma}, \linvar{x}: \underbrace{\oplus(( \with \onef) \ampy ( \oplus  \with (( \oplus \piencodfaplas{\sigma} ) \otimes (\piencodfaplas{\sigma\wedge \sigma}_{(\tau, j)}))))}_{\piencodfaplas{\sigma\wedge \sigma \wedge \sigma}_{(\tau, j)} \qquad \defref{ch4def:enc_sestypfailunres}}, \piencodfaplas{\Theta} $}
\end{prooftree}
\end{adjustwidth}

Therefore, $ \piencodfaplas{C}_{\linvar{x}} \vdash \piencodfaplas{\Gamma}, \linvar{x}: \piencodfaplas{\sigma\wedge \sigma \wedge \sigma}_{(\tau, j)} , \piencodfaplas{\Theta} $ and the result follows.

             \item For $ \piencodfaplas{ U }_{x_i} \vdash x_i: \with_{\eta_i \in \eta} \{ i : \piencodfaplas{\eta_i} \}, \piencodfaplas{\Theta}$.

To simplify the proof, we  consider $U= \unvar{\oneb} \concat \unvar{\bag{M'}}$ with $\eta =  \sigma_1 \concat \sigma_2 $ and $\with_{\eta_i \in \eta} = \with \{ 1 : \piencodfaplas{ \sigma_1} , 2 : \piencodfaplas{\sigma_2} \}$.

Our translation gives $ \piencodfaplas{ U }_{x_i} = \gsel{ x_i }\{ {1} :   \pnone{ x_i }  , {2} : \piencodfaplas{M'}_{x_i}  \}_{  }  $. Hence, we have:

\begin{prooftree}
\AxiomC{\(  \)}
\LeftLabel{\redlab{FS{:}bag^!}}
\UnaryInfC{\( \Theta ;  \dash  \wfdash \unvar{\oneb} : \sigma_1 \)}

\AxiomC{\( \Theta ; \cdot \wfdash M' : \sigma_2\)}
\LeftLabel{\redlab{FS{:}bag^!}}
\UnaryInfC{\( \Theta ; \cdot  \wfdash \unvar{\bag{M'}}:\sigma_2 \)}
\LeftLabel{\redlab{FS{:}\concat-bag^{!}}}
\BinaryInfC{\( \Theta ; \cdot  \wfdash \unvar{\oneb} \concat \unvar{\bag{M'}} : \sigma_1 \concat \sigma_2 \)}
\end{prooftree}

By the induction hypothesis we have that $ \Theta ; \cdot \wfdash M' : \sigma $ implies $ \piencodfaplas{M'}_{x_i} \wfdash x_i : \piencodfaplas{\sigma} ,  \piencodfaplas{\Theta}$. Thus,

\begin{prooftree}
\AxiomC{}
\LeftLabel{\ttype{${\with}\none$}}
\UnaryInfC{$   \pnone{ x_i } \vdash x_i:  \piencodfaplas{ \sigma_1}  $}
\LeftLabel{\ttype{weaken}}
\UnaryInfC{$   \pnone{ x_i } \vdash x_i:  \piencodfaplas{ \sigma_1}  , \piencodfaplas{\Theta} $}

\AxiomC{$ \piencodfaplas{M'}_{x_i} \vdash  x_i:  \piencodfaplas{\sigma_2} , \piencodfaplas{\Theta} $}
\LeftLabel{\ttype{$\with$}}
\BinaryInfC{$ \gsel{ x_i }\{ {1} :   \pnone{ x_i }  , {2} : \piencodfaplas{M'}_{x_i} \}_{  } \vdash  x_i: \with \{ 1 : \piencodfaplas{ \sigma_1} , 2 : \piencodfaplas{\sigma_2} \} , \piencodfaplas{\Theta} $}
\end{prooftree}

Therefore, $\gsel{ x_i }\{ {1} :   \pnone{  x_i } , {2} : \piencodfaplas{M'}_{x_i}  \}_{  } \vdash  x_i: \with \{ 1 : \piencodfaplas{ \sigma_1} , 2 : \piencodfaplas{\sigma_2} \} , \piencodfaplas{\Theta}  $ and the result follows.

\end{enumerate}

\item  The proof of type preservation for terms, relies on the analysis of ten cases:

\begin{enumerate}

\item {\bf Rule \redlab{FS{:}var^{\ell}}:}
Then we have the following derivation:

\begin{prooftree}
\AxiomC{}
\LeftLabel{\redlab{FS{:}var^{\ell}}}
\UnaryInfC{\( \Theta;  {x}: \tau \wfdash  {x} : \tau\)}
\end{prooftree}

By \defref{ch4def:enc_sestypfailunres},  $\piencodfaplas{ {x}:\tau}=  {x}:\with \overline{\piencodfaplas{\tau }}$, and by \figref{ch4fig:encoding},  $\piencodfaplas{ {x}}_u = \psome{x} ;   \pfwd{x}{u} $. The result follows from the derivation:

\begin{prooftree}
    \AxiomC{}
    \LeftLabel{\ttype{id}}
    \UnaryInfC{$ \pfwd{x}{u} \vdash  {x}:  \overline{\piencodfaplas{\tau }}  , u :  \piencodfaplas{ \tau } , \piencodfaplas{\Theta} $}
    \LeftLabel{\ttype{${\with}\some$}}
    \UnaryInfC{$  \psome{x} ; \pfwd{x}{u} \vdash  {x}: \with  \overline{\piencodfaplas{ \tau }} , u :  \piencodfaplas{ \tau } , \piencodfaplas{\Theta} $}
\end{prooftree}

\item {\bf Rule \redlab{FS{:}var^!}:}
Then we have the following derivation provided $\eta_{k} = \tau $:

\begin{prooftree}
\AxiomC{}
\LeftLabel{\redlab{FS{:}var^{\ell}}}
\UnaryInfC{\( \Theta , \unvar{x}: \eta;  {x}: \eta_{k}  \wfdash  {x} : \tau\)}
\LeftLabel{\redlab{FS{:}var^!}}
\UnaryInfC{\( \Theta , \unvar{x}: \eta; \dash \wfdash  {x}[k] : \tau\)}
\end{prooftree}

By \defref{ch4def:enc_sestypfailunres},  $\piencodfaplas{\Theta , \unvar{x}: \eta}= \piencodfaplas{\Theta} , \unvar{x}: \dual{!  \with_{\eta_i \in \eta} \{ i ; \piencodfaplas{\eta_i} \} }$, and by our translation (in \figref{ch4fig:encoding}),  $\piencodfaplas{ {x}[k]}_u = \puname{ \unvar{x} }{ x_i };\psel{ {x}_i }{ k }; \pfwd{x_i}{u} $. The result follows from the derivation:

\begin{prooftree}
    \AxiomC{}
    \LeftLabel{\ttype{id}}
    \UnaryInfC{$ \pfwd{x_i}{u}  \vdash  u :  \piencodfaplas{ \tau },  x_i:  \overline{\piencodfaplas{\eta_{k} }}   , \piencodfaplas{\Theta} $}
\LeftLabel{ \ttype{$\oplus$}}
\UnaryInfC{$  \psel{ {x}_i }{ k }; \pfwd{x_i}{u} \vdash  u :  \piencodfaplas{ \tau }, {x}_i :  \oplus_{\eta_i \in \eta} \{  i ; \dual{\piencodfaplas{\eta_i}}  \} ,  \piencodfaplas{\Theta}  $}
\LeftLabel{\ttype{$?$}}
\UnaryInfC{$ \puname{ \unvar{x} }{ x_i }; \psel{ {x}_i }{ k }; \pfwd{x_i}{u} \vdash  u :  \piencodfaplas{ \tau } , \unvar{x}: ? \oplus_{\eta_i \in \eta} \{  i ; \dual{\piencodfaplas{\eta_i}} \} , \piencodfaplas{\Theta} $}
\end{prooftree}

\item {\bf Rule \redlab{FS\!:\!weak}:}
        Then we have the following derivation:

\begin{prooftree}
\AxiomC{\( \Theta ; \Gamma  \wfdash M' : \tau\)}
\LeftLabel{ \redlab{FS\!:\!weak}}
\UnaryInfC{\( \Theta ; \Gamma ,  {x}: \omega \wfdash M'[\leftarrow  {x}]: \tau \)}
\end{prooftree}

By \defref{ch4def:enc_sestypfailunres},  $\piencodfaplas{\Gamma ,  {x}: \omega}= \piencodfaplas{\Gamma}, \linvar{x}: \overline{\piencodfaplas{\omega }_{(\sigma, i_1)}}$, and by our translation \figref{ch4fig:encoding},

$\piencodfaplas{M'[  \leftarrow  {x}]}_u = \psome{\linvar{x}} ;  \pname{\linvar{x}}{y_i} . (\gsome{  y_i }{ u,\llfv{M'} }; \gclose{ y_{i} } ; \piencodfaplas{M'}_u   \|   \pnone{ \linvar{x} } ) $.

By IH, we have $\piencodfaplas{M'}_u\vdash  \piencodfaplas{\Gamma }, u:\piencodfaplas{\tau} , \piencodfaplas{\Theta }$.
The result follows from the derivation, omitting labels:

    \begin{prooftree}
            \small
    \AxiomC{$\piencodfaplas{M'}_u\vdash  \piencodfaplas{\Gamma }, u:\piencodfaplas{\tau} , \piencodfaplas{\Theta} $}
    \UnaryInfC{$\gclose{ y_{i} } ; \piencodfaplas{M'}_u \vdash y_i{:}\bot, \piencodfaplas{\Gamma }, u:\piencodfaplas{\tau} , \piencodfaplas{\Theta}$}
    \UnaryInfC{$\gsome{ y_i }{ u,\llfv{M'} };  \gclose{ y_{i} } ; \piencodfaplas{M'}_u \vdash  y_i{:}\oplus \bot , \piencodfaplas{\Gamma }, u:\piencodfaplas{\tau} , \piencodfaplas{\Theta}$}

    \AxiomC{}
    \UnaryInfC{$  \pnone{ \linvar{x} }  \vdash \linvar{x} :\with A $}
    \UnaryInfC{$  \pnone{ \linvar{x} }  \vdash \linvar{x} :\with A , \piencodfaplas{\Theta}$}
\BinaryInfC{$\pname{\linvar{x}}{y_i} . ( \gsome{ y_i }{ u,\llfv{M'} }; \gclose{ y_{i} } ; \piencodfaplas{M'}_u   \|   \pnone{ \linvar{x} } ) \vdash  \linvar{x}: (\oplus \bot) \otimes (\with A) , \piencodfaplas{\Gamma }, u:\piencodfaplas{\tau} , \piencodfaplas{\Theta}$}
\UnaryInfC{$\piencodfaplas{M'[  \leftarrow  {x}]}_u \vdash  \linvar{x} :\with ((\oplus \bot) \otimes (\with A))  , \piencodfaplas{\Gamma }, u:\piencodfaplas{\tau} , \piencodfaplas{\Theta} $}
\end{prooftree}

Since $A$ is arbitrary,  we can take $A=\oneb$ for $\piencodfaplas{\omega}_{(\sigma,0)} $ and  $A= \overline{(( \with  \overline{\piencodfaplas{ \sigma }} )  \ampy (\overline{\piencodfaplas{\omega}_{(\sigma, i - 1)}}))}$  for $\piencodfaplas{\omega}_{(\sigma,i)} $ where $i > 0$, in both cases, the result follows.

\item {\bf Rule $\redlab{FS:abs \dash sh}$:}

Then $M= \lambda x . (M'[ {\widetilde{x}} \leftarrow  {x}])$, and the derivation is:

\begin{prooftree}
\AxiomC{\( \Theta , \unvar{x}:\eta ; \Gamma ,  {x}: \sigma^k \wfdash M'[ {\widetilde{x}} \leftarrow  {x}] : \tau \quad  {x} \notin \dom{\Gamma} \)}
\LeftLabel{\redlab{FS{:}abs\dash sh}}
\UnaryInfC{\( \Theta ; \Gamma \wfdash \lambda x . (M'[ {\widetilde{x}} \leftarrow  {x}])  : (\sigma^k, \eta )  \rightarrow \tau \)}
\end{prooftree}

By IH, we have $\piencodfaplas{M'[ {\widetilde{x}} \leftarrow  {x}]}_u \vdash  u:\piencodfaplas{\tau} , \piencodfaplas{\Gamma} , \linvar{x}: \overline{\piencodfaplas{\sigma^k}_{(\sigma, i)}}  , \piencodfaplas{\Theta} , \unvar{x}: \overline{\piencodfaplas{\eta}} $ and our translation (in
\figref{ch4fig:encoding}) gives
$ \piencodfaplas{\lambda x.M'[ {\widetilde{x}} \leftarrow x]}_u = \psome{u}; \gname{u}{x}; \psome{x}; \gname{x}{\linvar{x}}; \gname{x}{\unvar{x}}; \gclose{ x } ; \piencodfaplas{M'[ {\widetilde{x}} \leftarrow  {x}]}_u $, whose type derivation $\piencodfaplas{M'[ {\widetilde{x}} \leftarrow  {x}]}_u \vdash u:\piencodfaplas{(\sigma^{k} , \eta )   \rightarrow \tau}  , \piencodfaplas{\Gamma} , \piencodfaplas{\Theta} $,  was given in the proof of Lemma~\ref{ch4lem:relunbag-typeunres}, item 3.

\item {\bf Rule $\redlab{FS:app}$:}
Then $M = M'\ B$, where $ B = C \bagsep U $ and the derivation is:

\begin{prooftree}
\AxiomC{\( \Theta ;\Gamma \wfdash M' : (\sigma^{j} , \eta ) \rightarrow \tau \)}
\AxiomC{\(  \Theta ;\Delta \wfdash B : (\sigma^{k} , \epsilon )  \)}
\AxiomC{\( \eta \relunbag \epsilon \)}
\LeftLabel{\redlab{FS{:}app}}
\TrinaryInfC{\( \Theta ; \Gamma, \Delta \wfdash M'\ B : \tau\)}
\end{prooftree}

By IH, we have both

\begin{itemize}
\item  $\piencodfaplas{M'}_u\vdash \piencodfaplas{\Gamma}, u:\piencodfaplas{(\sigma^{j} , \eta ) \rightarrow \tau} , \piencodfaplas{\Theta}$;
\item $\piencodfaplas{M'}_u\vdash \piencodfaplas{\Gamma}, u:\piencodfaplas{(\sigma^{j} , \epsilon ) \rightarrow \tau} , \piencodfaplas{\Theta}$,   by Lemma \ref{ch4lem:relunbag-typeunres};
\item $\piencodfaplas{B}_u\vdash \piencodfaplas{\Delta}, u:\overline{\piencodfaplas{(\sigma^{k} , \epsilon ) }_{(\tau_2, n)}}  , \piencodfaplas{\Theta} $, for some $\tau_2$ and some $n$.
\end{itemize}

Therefore, from the fact that $M$ is well-formed and \figref{ch4fig:encoding} and \ref{ch4def:enc_sestypfailunres}, we  have:

\begin{itemize}
\item $\displaystyle{\piencodfaplas{M' (C \bagsep U)}_u =   \res{ v } (\piencodfaplas{M'}_v   \| \gsome{ v }{ u , \llfv{C} };  \pname{v}{x} . (\pfwd{v}{u}   \| \piencodfaplas{C \bagsep U}_x ) )} $;
\item $\piencodfaplas{(\sigma^{j} , \eta ) \rightarrow \tau}= \oplus( (\piencodfaplas{\sigma^{k} }_{(\tau_1, m)}) \otimes ((!\piencodfaplas{\eta})  $, for some $\tau_1$ and some $m$.
\end{itemize}

Also, since $\piencodfaplas{B}_u\vdash \piencodfaplas{\Delta}, u:\piencodfaplas{(\sigma^{k} , \epsilon )}_{(\tau_2, n)}, \piencodfaplas{\Theta}$, we have the following derivation $\Pi$:
\begin{adjustwidth}{-1cm}{}
\begin{prooftree}
\small
\AxiomC{$\piencodfaplas{C \bagsep U}_x\vdash \piencodfaplas{\Delta}, x:\piencodfaplas{(\sigma^{k} , \epsilon )}_{(\tau_2, n)}  , \piencodfaplas{\Theta} $ }
\AxiomC{\(\)}
\UnaryInfC{$ \pfwd{v}{u}  \vdash v:  \overline{\piencodfaplas{ \tau }} , u: \piencodfaplas{ \tau }  $}
\UnaryInfC{$ \pfwd{v}{u}  \vdash v:  \overline{\piencodfaplas{ \tau }} , u: \piencodfaplas{ \tau }  , \piencodfaplas{\Theta}$}
\BinaryInfC{$\pname{v}{x} . (\pfwd{v}{u}   \| \piencodfaplas{C \bagsep U}_x ) \vdash \piencodfaplas{ \Delta }, v:\piencodfaplas{(\sigma^{k} , \epsilon )}_{(\tau_2, n)} \otimes \overline{ \piencodfaplas{ \tau }} , u:\piencodfaplas{ \tau }  , \piencodfaplas{\Theta} $}
\UnaryInfC{$ \gsome{ v }{ u , \llfv{C} };   \pname{v}{x} . (\pfwd{v}{u}   \| \piencodfaplas{C \bagsep U}_x )  \vdash\piencodfaplas{ \Delta }, v:\underbrace{\oplus (\piencodfaplas{(\sigma^{k} , \epsilon )}_{(\tau_2, n)} \otimes  \overline{\piencodfaplas{ \tau }})}_{\overline{\piencodfaplas{(\sigma^{k} , \epsilon ) \rightarrow \tau}}}, u:\piencodfaplas{ \tau } , \piencodfaplas{\Theta} $}
\end{prooftree}
\end{adjustwidth}
  In order to apply \ttype{cut}, we must have that $\piencodfaplas{\sigma^{j}}_{(\tau_1, m)} = \piencodfaplas{\sigma^{k}}_{(\tau_2, n)}$, therefore, the choice of $\tau_1,\tau_2,n$ and $m$, will consider the different possibilities for $j$ and $k$, as in Proposition~\ref{ch4prop:app_auxunres}.
\begin{prooftree}
\small
\AxiomC{\( \piencodfaplas{M'}_v\vdash \piencodfaplas{\Gamma}, v:\piencodfaplas{(\sigma^{j} , \epsilon ) \rightarrow \tau} ; \piencodfaplas{\Theta} \)}
\AxiomC{$\Pi$}
\LeftLabel{\ttype{cut}}
\BinaryInfC{$  \res{ v }  ( \piencodfaplas{ M'}_v   \| \gsome{ v }{ u , \llfv{C} };  \pname{v}{x} . (\pfwd{v}{u}   \| \piencodfaplas{B}_x ) ) \vdash \piencodfaplas{ \Gamma } ,\piencodfaplas{ \Delta } , u: \piencodfaplas{ \tau }  ; \piencodfaplas{\Theta} $}
\end{prooftree}

We can then conclude that $\piencodfaplas{M' B}_u \vdash \piencodfaplas{ \Gamma}, \piencodfaplas{ \Delta }, u:\piencodfaplas{ \tau } , \piencodfaplas{\Theta} $ and the result follows.

\item {\bf Rule $\redlab{FS:share}$:}
Then $M = M' [  {x}_1, \dots  {x}_k \leftarrow x ]$ and the derivation is:
\begin{prooftree}
    \AxiomC{\( \Theta ; \Delta ,  {x}_1: \sigma, \cdots,  {x}_k: \sigma \wfdash M' : \tau \quad  {x} \notin \Delta \quad k \not = 0\)}
    \LeftLabel{ \redlab{FS:share}}
    \UnaryInfC{\( \Theta ; \Delta ,  {x}: \sigma_{k} \wfdash M'[ {x}_1 , \cdots ,  {x}_k \leftarrow  {x}] : \tau \)}
\end{prooftree}

To simplify the proof we will consider $k=1$ (the case in which $k>1$ follows similarly).

By IH, we have $\piencodfaplas{M'}_u\vdash  \piencodfaplas{\Delta ,  {x}_1:\sigma }, u:\piencodfaplas{\tau} ; \piencodfaplas{\Theta}$.
From
\figref{ch4fig:encoding} and \ref{ch4def:enc_sestypfailunres}, it follows

\begin{itemize}
    \item $\piencodfaplas{ \Delta ,  {x}_1: \sigma} = \piencodfaplas{\Delta}, \linvar{x}_1:\with\overline{\piencodfaplas{\sigma}} $.
\item
$
\piencodfaplas{M'[ {x}_1, \leftarrow  {x}]}_u =
    \begin{array}[t]{l}
        \psome{\linvar{x}}; \pname{\linvar{x}}{y_1}. (\gsome{ y_1 }{ \emptyset };   \gclose{ y_{1} } ;\0
            \| \psome{\linvar{x}};
        \\
        \gsome{ \linvar{x} }{ u, \llfv{M'} \setminus  {x}_1  };\bignd_{x_i \in x_1}\gname{x}{{x}_i}; \psome{\linvar{x}}; \pname{\linvar{x}}{y_2} .\\
         (\gsome{ y_2 }{ u,\llfv{M'} };  \gclose{ y_{2} } ; \piencodfaplas{M'}_u   \|   \pnone{ \linvar{x} } ) )
    \end{array}
    $

\end{itemize}

We shall split the expression into two parts:
\[
\begin{aligned}
N_1 &= \psome{\linvar{x}}; \pname{\linvar{x}}{y_2} . ( \gsome{ y_2 }{ u,\llfv{M'} }; \gclose{ y_{2} } ; \piencodfaplas{M'}_u   \|   \pnone{ \linvar{x} } ) \\
N_2 &= \psome{\linvar{x}}; \pname{\linvar{x}}{y_1}. (\gsome{ y_1 }{ \emptyset }; \gclose{ y_{1} } ;\0   \| \psome{\linvar{x}}; \gsome{ \linvar{x} }{ u, \llfv{M'} \setminus  {x}_1  }; \gname{x}{{x}_1};N_1)
\end{aligned}
\]

and we obtain the  derivation for term $N_1$ as follows where we omit $, \piencodfaplas{\Theta}$ and derivation labels:
\begin{adjustwidth}{-1cm}{}
\begin{prooftree}
    \small
\AxiomC{$\piencodfaplas{M'}_u \vdash  \piencodfaplas{\Delta ,  {x}_1:\sigma }, u:\piencodfaplas{\tau}  $}
\UnaryInfC{$\gclose{ y_{2} } ; \piencodfaplas{M'}_u \vdash \piencodfaplas{\Delta ,  {x}_1:\sigma }, u:\piencodfaplas{\tau}, y_{2}{:}\bot   $}
\UnaryInfC{$\gsome{ y_2 }{ u,\llfv{M'} };   \gclose{ y_{2} } ; \piencodfaplas{M'}_u \vdash \piencodfaplas{\Delta ,  {x}_1:\sigma }, u:\piencodfaplas{\tau}, y_{2}{:}\oplus \bot    $}

\AxiomC{}
\UnaryInfC{$  \pnone{ \linvar{x} }  \vdash \linvar{x} :\with A  $}
\BinaryInfC{$  \pname{\linvar{x}}{y_2} . (  \gsome{ y_2 }{ u,\llfv{M'} }; \gclose{ y_{2} } ; \piencodfaplas{M'}_u   \|   \pnone{ \linvar{x} } ) \vdash \piencodfaplas{\Delta ,  {x}_1:\sigma }, u:\piencodfaplas{\tau} , \linvar{x}: ( \oplus \bot )\otimes ( \with A )    $}
\UnaryInfC{$\underbrace{ \psome{\linvar{x}}; \pname{\linvar{x}}{y_2} . (  \gsome{ y_2 }{ u,\llfv{M'} };  \gclose{ y_{2} } ; \piencodfaplas{M'}_u   \|   \pnone{ \linvar{x} } ) }_{N_1} \vdash \piencodfaplas{\Delta ,  {x}_1:\sigma } , u:\piencodfaplas{\tau} , \linvar{x}: \overline{\piencodfaplas{\omega}_{(\sigma, i)}}   $}
\end{prooftree}
\end{adjustwidth}
Notice that the last rule applied \ttype{${\with}\some$} assigns $x: \with ((\oplus \bot) \otimes (\with A))$. Again, since $A$ is arbitrary, we can take $A= \oplus (( \with  \overline{\piencodfaplas{ \sigma }} )  \ampy (\overline{\piencodfaplas{\omega}_{(\sigma, i - 1)}}))$, obtaining $x:\overline{\piencodfaplas{\omega}_{(\sigma,i)}}$.

In order to obtain a type derivation for $N_2$, consider the derivation $\Pi_1$:
{
\begin{adjustwidth}{-2cm}{}
\begin{prooftree}
    \small
\AxiomC{$N_1 \vdash \piencodfaplas{\Delta},  {x}_1:\with\overline{\piencodfaplas{\sigma}} , u:\piencodfaplas{\tau}, \linvar{x}: \overline{\piencodfaplas{\omega}_{(\sigma, i)}}  , \piencodfaplas{\Theta} $}
\LeftLabel{\ttype{$\parr$}}
\UnaryInfC{$\gname{x}{{x}_i};N_1   \vdash \piencodfaplas{\Delta} ,  u:\piencodfaplas{\tau}, \linvar{x}: ( \with\overline{\piencodfaplas{\sigma}} ) \ampy (\overline{\piencodfaplas{\omega}_{(\sigma, i)}})  , \piencodfaplas{\Theta} $}
\LeftLabel{\ttype{$\nd$}}
\UnaryInfC{$\bignd_{x_i \in x_1} \gname{x}{{x}_i};N_1   \vdash \piencodfaplas{\Delta} ,  u:\piencodfaplas{\tau}, \linvar{x}: ( \with\overline{\piencodfaplas{\sigma}} ) \ampy (\overline{\piencodfaplas{\omega}_{(\sigma, i)}})  , \piencodfaplas{\Theta} $}
\UnaryInfC{$  \gsome{ \linvar{x} }{ u, \llfv{M'} \setminus  {x}_1  };  \bignd_{x_i \in x_1} \gname{x}{{x}_i};N_1  \vdash \piencodfaplas{\Delta} ,  u:\piencodfaplas{\tau}, \linvar{x} {:}\oplus (( \with \overline{\piencodfaplas{\sigma}} ) \ampy (\overline{\piencodfaplas{\omega}_{(\sigma, i)}})) , \piencodfaplas{\Theta} $}
\UnaryInfC{$ \psome{\linvar{x}}; \gsome{ \linvar{x} }{ u, \llfv{M'} \setminus  {x}_1  };  \bignd_{x_i \in x_1} \gname{x}{{x}_i};N_1  \vdash \piencodfaplas{\Delta},  u:\piencodfaplas{\tau} , \linvar{x} :\with \oplus (( \with\overline{\piencodfaplas{\sigma}} ) \ampy ( \overline{\piencodfaplas{\omega}_{(\sigma, i)}} ))  , \piencodfaplas{\Theta} $}
\end{prooftree}
\end{adjustwidth}
}

We take $ P_1 = \psome{\linvar{x}};  \gsome{ \linvar{x} }{ u, \llfv{M'} \setminus  {x}_1  }; \bignd_{x_i \in x_1} \gname{x}{{x}_i};N_1 $ and $\Gamma_1 =   \piencodfaplas{\Delta},  u:\piencodfaplas{\tau} $ and continue the derivation of $ N_2 $

\begin{adjustwidth}{-1cm}{}
\begin{prooftree}
    \small
\AxiomC{}
\LeftLabel{\redlab{T\cdot}}
\UnaryInfC{$\0\vdash \dash , \piencodfaplas{\Theta} $}
\LeftLabel{\ttype{$\bot$}}
\UnaryInfC{$ \gclose{ y_{1} } ;\0 \vdash y_{1} : \bot , \piencodfaplas{\Theta} $}
\UnaryInfC{$  \gsome{ y_1 }{ \emptyset };  \gclose{ y_{1} };\0 \vdash  y_1{:}\oplus \bot , \piencodfaplas{\Theta} $}

\AxiomC{$\Pi_1 $}
\noLine
\UnaryInfC{$\vdots$}
\noLine
\UnaryInfC{$P_1\vdash \Gamma_1, \linvar{x} :\with \oplus (( \with\overline{ \piencodfaplas{\sigma}} ) \ampy ( \overline{\piencodfaplas{\omega}_{(\sigma, i)}} )) , \piencodfaplas{\Theta} $}
\LeftLabel{\ttype{$\tensor$}}
\BinaryInfC{$\pname{\linvar{x}}{y_1}. ( \gsome{ y_1 }{ \emptyset };  \gclose{ y_{1} } ;\0   \| P_1) \vdash \Gamma_1 ,  \linvar{x} : (\oplus \bot)\otimes (\with \oplus (( \with\overline{\piencodfaplas{\sigma}} ) \ampy ( \overline{\piencodfaplas{\omega}_{(\sigma, i)}} )) ) , \piencodfaplas{\Theta} $}
\UnaryInfC{$\underbrace{ \psome{\linvar{x}}; \pname{\linvar{x}}{y_1}. ( \gsome{ y_1 }{ \emptyset }; \gclose{ y_{1} } ;\0   \| P_1)}_{N_2} \vdash \Gamma_1 , \linvar{x} : \overline{ \piencodfaplas{\sigma \wedge \omega}_{(\sigma, i)}}  , \piencodfaplas{\Theta}$}
\end{prooftree}
\end{adjustwidth}
Hence the theorem holds for this case.

\item {\bf Rule $\redlab{FS:Esub}$:}
Then $M = (M'[ {\widetilde{x}} \leftarrow  {x}])\esubst{ B }{ x }$ and

\begin{prooftree}
\AxiomC{\( \Theta , \unvar{x} : \eta ; \Gamma ,  {x}: \sigma^{j} \wfdash M'[ {\widetilde{x}} \leftarrow  {x}] : \tau  \)}
\AxiomC{\( \Theta ; \Delta \wfdash B : (\sigma^{k} , \epsilon ) \)}
\AxiomC{\( \eta \relunbag \epsilon \)}
\LeftLabel{\redlab{FS{:}Esub}}
\TrinaryInfC{\( \Theta ; \Gamma, \Delta \wfdash (M'[ {\widetilde{x}} \leftarrow  {x}])\esubst{ B }{ x }  : \tau \)}
\end{prooftree}

By Proposition~\ref{ch4prop:app_auxunres} and IH we have:
$$
\hspace{-1.5cm}
\begin{array}{rl}
\piencodfaplas{ M'[x_1, \cdots , x_k \leftarrow x]}_u&\vdash \piencodfaplas{\Gamma}, \linvar{x}: \overline{ \piencodfaplas{ \sigma^j }_{(\tau, n)}} , u:\piencodfaplas{\tau} , \piencodfaplas{\Theta} , \unvar{x} : \dual{\piencodfaplas{\eta}} \\
\piencodfaplas{ M'[x_1, \cdots , x_k \leftarrow x]}_u&\vdash \piencodfaplas{\Gamma}, \linvar{x}: \overline{ \piencodfaplas{ \sigma^j }_{(\tau, n)}} , u:\piencodfaplas{\tau} , \piencodfaplas{\Theta} , \unvar{x} : \dual{\piencodfaplas{\epsilon}},\text{ by Lemma \ref{ch4lem:relunbag-typeunres} }\\
\piencodfaplas{B}_x&\vdash \piencodfaplas{\Delta}, x:\piencodfaplas{ (\sigma^{k} , \epsilon ) }_{(\tau, m)}  , \piencodfaplas{\Theta}
\end{array}
$$

From \figref{ch4fig:encoding}, we have
\begin{equation*}
\piencodfaplas{ M'[ {\widetilde{x}} \leftarrow  {x}]\ \esubst{ B }{ x }}_u =  \res{ x } (\psome{x}; \gname{x}{\linvar{x}}; \gname{x}{\unvar{x}}; \gclose{ x } ;\piencodfaplas{ M'[ {\widetilde{x}} \leftarrow  {x}]}_u   \| \piencodfaplas{ C \bagsep U}_x )
\end{equation*}

Therefore we obtain the following derivation $\Pi$:
\begin{adjustwidth}{-1cm}{}
\begin{prooftree}
    \small
\AxiomC{$ \piencodfaplas{ M'[ {\widetilde{x}} \leftarrow  {x}]}_u \vdash \piencodfaplas{\Gamma}, \linvar{x}: \overline{ \piencodfaplas{ \sigma^j }_{(\tau, n)}} , u:\piencodfaplas{\tau} , \piencodfaplas{\Theta} , \unvar{x} : \dual{\piencodfaplas{\epsilon}} $}
\LeftLabel{\ttype{$\bot$}}
\UnaryInfC{$ \gclose{ x } ;\piencodfaplas{ M'[ {\widetilde{x}} \leftarrow  {x}]}_u \vdash x{:}\bot, \piencodfaplas{\Gamma}, \linvar{x}: \overline{ \piencodfaplas{ \sigma^j }_{(\tau, n)}} , u:\piencodfaplas{\tau} , \piencodfaplas{\Theta} , \unvar{x} : \dual{\piencodfaplas{\epsilon}}$}
\LeftLabel{\ttype{$\parr$}}
\UnaryInfC{$\gname{x}{\unvar{x}}; \gclose{ x } ;\piencodfaplas{ M'[ {\widetilde{x}} \leftarrow  {x}]}_u \vdash x:  (  \dual{\piencodfaplas{\epsilon}} ) \ampy \bot , \piencodfaplas{\Gamma}, \linvar{x}: \overline{ \piencodfaplas{ \sigma^j }_{(\tau, n)}} , u:\piencodfaplas{\tau} , \piencodfaplas{\Theta} $}
\LeftLabel{\ttype{$\parr$}}
\UnaryInfC{$\gname{x}{\linvar{x}}; \gname{x}{\unvar{x}}; \gclose{ x } ;\piencodfaplas{ M'[ {\widetilde{x}} \leftarrow  {x}]}_u \vdash x: \overline{ \piencodfaplas{ \sigma^j }_{(\tau, n)}} \ampy ( (  \dual{\piencodfaplas{\epsilon}} ) \ampy \bot) , \piencodfaplas{\Gamma}, u:\piencodfaplas{\tau} , \piencodfaplas{\Theta} $}
\LeftLabel{\ttype{${\with}\some$}}
\UnaryInfC{$\psome{x}; \gname{x}{\linvar{x}}; \gname{x}{\unvar{x}};\gclose{ x } ; \piencodfaplas{ M'[ {\widetilde{x}} \leftarrow  {x}]}_u \vdash x : \overline{\piencodfaplas{ (\sigma^{j} , \epsilon  )  }_{(\tau, n)}}  , \piencodfaplas{\Gamma}, u:\piencodfaplas{\tau} , \piencodfaplas{\Theta}  $}
\end{prooftree}
\end{adjustwidth}

We take $ P_1 = \psome{x}; \gname{x}{\linvar{x}}; \gname{x}{\unvar{x}}; \gclose{ x } ;\piencodfaplas{ M'[ {\widetilde{x}} \leftarrow  {x}]}_u$ and continue the derivation of $ \Pi $
\begin{adjustwidth}{-1cm}{}
\begin{prooftree}
    \small
\AxiomC{$ P_1 \vdash  x : \overline{\piencodfaplas{ (\sigma^{j} , \epsilon  )  }_{(\tau, n)}}  , \piencodfaplas{\Gamma}, u:\piencodfaplas{\tau} ; \piencodfaplas{\Theta}  $}
\AxiomC{$  \piencodfaplas{ C \bagsep U}_x \vdash \piencodfaplas{\Delta}, x:\piencodfaplas{ (\sigma^{k} , \epsilon ) }_{(\tau, m)}  , \piencodfaplas{\Theta} $}
\LeftLabel{\ttype{cut}}
\BinaryInfC{$  \res{ x }  ( P_1   \| \piencodfaplas{ C \bagsep U}_x )  \vdash \piencodfaplas{ \Gamma} , \piencodfaplas{ \Delta } , u: \piencodfaplas{ \tau }  , \piencodfaplas{\Theta}  $}
\end{prooftree}
\end{adjustwidth}
We must have that $\piencodfaplas{\sigma^{j}}_{(\tau, m)} = \piencodfaplas{\sigma^{k}}_{(\tau, n)}$ which by our restrictions allows. It follows that $\piencodfaplas{M'[  {x}_1 \leftarrow  {x}]\ \esubst{ B }{ x }} $
$ \vdash \piencodfaplas{\Gamma, \Delta}, u:\piencodfaplas{\tau}  , \piencodfaplas{\Theta}$ and the result follows.

\item {\bf Rule $\redlab{FS{:}Esub^{\ell}}$: }
Then $M =  M' \linexsub{C /  {x}_1 , \cdots , x_k}$, with $C = \bag{N_1} \cdot \cdots \cdot \bag{N_k}$ and

\begin{prooftree}
\AxiomC{\( \Theta ; \Delta ,  x_1:\sigma, \cdots , x_k:\sigma \wfdash M' : \tau \)}
\AxiomC{\( \Theta ; \Gamma,  \wfdash C:\sigma^{k}\)}
\LeftLabel{\redlab{FS{:}Esub^{\ell}}}
\BinaryInfC{\( \Theta ; \Gamma, \Delta \wfdash M' \linexsub{C /  x_1, \cdots , x_k} : \tau \)}
\end{prooftree}

with $ \Gamma = \Gamma_1 , \cdots ,\Gamma_k $ and:

\begin{prooftree}
\AxiomC{\( \Theta ; \Gamma_1 \wfdash N_1 : \sigma\)}
\AxiomC{\( \Theta ; \Gamma_k \wfdash N_k : \sigma\)}
\AxiomC{\(  \)}
\LeftLabel{\redlab{FS{:}\oneb^!}}
\UnaryInfC{\( \Theta ;  \dash  \wfdash \unvar{\oneb} : \sigma \)}
\LeftLabel{\redlab{FS{:}bag^{\ell}}}
\BinaryInfC{\( \vdots \)}
\LeftLabel{\redlab{FS{:}bag^{\ell}}}
\UnaryInfC{\( \Theta ; \Gamma_2 , \cdots ,\Gamma_k \wfdash C:\sigma^{k-1}\)}
\LeftLabel{\redlab{FS{:}bag^{\ell}}}
\BinaryInfC{\( \Theta ; \Gamma_1 , \cdots ,\Gamma_k  \wfdash \bag{M'}\cdot C:\sigma^{k}\)}
\end{prooftree}

By IH we have both
\[\piencodfaplas{ N_1 }_{{x}_1} \vdash \piencodfaplas{\Gamma_1}, {x}_1:\piencodfaplas{\sigma } , \piencodfaplas{\Theta} \qquad \cdots \qquad
\piencodfaplas{ N_k }_{{x}_k} \vdash \piencodfaplas{\Gamma_k}, {x}_k:\piencodfaplas{\sigma } , \piencodfaplas{\Theta}\]
\[\hspace{-1cm}\small\piencodfaplas{ C }_{{x}} \vdash \piencodfaplas{\Gamma}, {x}:\piencodfaplas{\sigma^{k} }_{(\sigma, i)} , \piencodfaplas{\Theta}
\ \text{ and } \  \piencodfaplas{M'}_u\vdash \piencodfaplas{\Gamma},  {x}_1: \with\overline{\piencodfaplas{\sigma}} , \cdots , {x}_k: \with\overline{\piencodfaplas{\sigma}}, u:\piencodfaplas{\tau} , \piencodfaplas{\Theta}\]

From \figref{ch4fig:encoding},

\[ 
    \hspace{-1cm}
    \small
\begin{aligned}
\piencodfaplas{ M' \linexsub{\bag{N_1} \cdot \cdots \cdot \bag{N_k} /  x_1 , \cdots , x_k}  }_u    & =
\res{z_1}( \gsome{z_1}{\llfv{N_{1}}};\piencodfaplas{ N_{1} }_{ {z_1}}  \|  \cdots\\
& \qquad \res{z_k} ( \gsome{z_k}{\llfv{N_{k}}};\piencodfaplas{ N_{k} }_{ {z_k}} \\
& \qquad \qquad  \| \bignd_{x_{i_1} \in \{ x_1 ,\cdots , x_k  \}} \cdots \bignd_{x_{i_k} \in \{ x_1 ,\cdots , x_k \setminus x_{i_1} , \cdots , x_{i_{k-1}}  \}}\\
& \qquad \qquad \qquad  \piencodfaplas{ M' }_u \{ z_1 / x_{i_1} \} \cdots \{ z_k / x_{i_k} \} ) \cdots )
\\
\end{aligned}
\]
Let us take $k = 1$ and for $k > 1$ cases follow similarly omitting labels:
{
\small
\begin{adjustwidth}{-2cm}{}
\begin{prooftree}
\AxiomC{\( \piencodfaplas{ M' }_u \{ z_1 / x_{i} \}  \vdash   \piencodfaplas{\Delta},  {z_1}: \with\overline{\piencodfaplas{\sigma}}, u:\piencodfaplas{\tau} , \piencodfaplas{\Theta}\)}
\UnaryInfC{\( \bignd_{x_{i} \in \{ x_1 \}}\piencodfaplas{ M' }_u \{ z_1 / x_{i} \}  \vdash   \piencodfaplas{\Delta},  {z_1}: \with\overline{\piencodfaplas{\sigma}}, u:\piencodfaplas{\tau} , \piencodfaplas{\Theta}\)}

\AxiomC{\( \piencodfaplas{ N_1 }_x  \vdash  \piencodfaplas{\Gamma},  {x}: \piencodfaplas{\sigma} , \piencodfaplas{\Theta} \)}
\UnaryInfC{\(   \gsome{ {z_1} }{ \llfv{N_1} }; \piencodfaplas{ N_1 }_{z_1} \vdash  \piencodfaplas{ \Gamma } ,  {z_1}: : \oplus \piencodfaplas{\sigma}, \piencodfaplas{\Theta}\)}
\BinaryInfC{\(   \res{ {z_1} }  ( \bignd_{x_{i} \in \{ x_1 \}}\piencodfaplas{ M' }_u \{ z_1 / x_{i} \}  \|   \gsome{ {z_1} }{ \llfv{N_1} };  \piencodfaplas{ N_1 }_{ {z_1}} ) \vdash  \piencodfaplas{ \Gamma} , \piencodfaplas{ \Delta } , u : \piencodfaplas{ \tau } , \piencodfaplas{\Theta} \)}
\end{prooftree}
\end{adjustwidth}
}

Therefore, $\piencodfaplas{M' \linexsub{N_1 /  {x_1}} }_u\vdash \piencodfaplas{ \Gamma} , \piencodfaplas{ \Delta } , u : \piencodfaplas{ \tau }, \piencodfaplas{\Theta} $ and the result follows.

\item {\bf Rule $\redlab{FS{:}Esub^!}$: }
Then $M =  M' \unexsub{U / \unvar{x}}$ and
\begin{prooftree}
\AxiomC{\( \Theta , \unvar{x} : \eta; \Gamma \wfdash M' : \tau \quad  \Theta ; \dash \wfdash U : \eta \)}
\LeftLabel{\redlab{FS{:}Esub^!}}
\UnaryInfC{\( \Theta ; \Gamma \wfdash M' \unexsub{U / \unvar{x}}  : \tau \)}
\end{prooftree}

By IH we have both
$$
\begin{array}{c}
\piencodfaplas{U}_{x_i}\vdash x_i : \with_{\eta_i \in \eta} \{ i ; \piencodfaplas{\eta_i} \}  , \piencodfaplas{\Theta}\quad \text{ and } \quad
\piencodfaplas{M'}_u\vdash \piencodfaplas{\Gamma} , u:\piencodfaplas{\tau} , \unvar{x}: \overline{\piencodfaplas{\eta}} , \piencodfaplas{\Theta}
\end{array}
$$

From Definition \figref{ch4fig:encoding}, $ \piencodfaplas{ M' \unexsub{U / \unvar{x}}  }_u   =    \res{ \unvar{x} }  ( \piencodfaplas{ M' }_u   \|   \guname{ \unvar{x} }{ x_i } ; \piencodfaplas{ U }_{x_i} ) $ and

\begin{prooftree}
    \small
\AxiomC{$\piencodfaplas{M'}_u\vdash \piencodfaplas{\Gamma} , u:\piencodfaplas{\tau}, \unvar{x}: \overline{\piencodfaplas{\eta}} , \piencodfaplas{\Theta}$}

\AxiomC{$ \piencodfaplas{ U }_{x_i} \vdash x_i : \with_{\eta_i \in \eta} \{ i ; \piencodfaplas{\eta_i} \} , \piencodfaplas{\Theta} $}
\LeftLabel{\ttype{$!$}}
\UnaryInfC{$ \guname{ \unvar{x} }{ x_i } ; \piencodfaplas{ U }_{x_i} \vdash \unvar{x}: \piencodfaplas{\eta} , \piencodfaplas{\Theta}  $}
\LeftLabel{\ttype{cut}}
\BinaryInfC{\(    \res{ \unvar{x} } ( \piencodfaplas{ M' }_u   \|    \guname{ \unvar{x} }{ x_i } ; \piencodfaplas{ U }_{x_i} ) \vdash \piencodfaplas{\Gamma} , u:\piencodfaplas{\tau} , \piencodfaplas{\Theta} \)}
\end{prooftree}

Therefore, $\piencodfaplas{M' \unexsub{U / \unvar{x}} }_u \vdash \piencodfaplas{\Gamma} , u:\piencodfaplas{\tau} , \piencodfaplas{\Theta} $ and the result follows.

\item {\bf Rule $\redlab{FS:fail}$:}
Then $M= \fail^{\widetilde{x}}$ where $ \widetilde{x} = x_1, \cdots , x_n$ and

\begin{prooftree}
\AxiomC{\( \dom{\Gamma} = \widetilde{x}\)}
\LeftLabel{\redlab{FS{:}fail}}
\UnaryInfC{\( \Theta ; \Gamma \wfdash  \fail^{\widetilde{x}} : \tau \)}
\end{prooftree}

From \figref{ch4fig:encoding}, $\piencodfaplas{\fail^{x_1, \cdots , x_n} }_u=   \pnone{ u }    \|   \pnone{ x_1 }   \| \cdots   \|   \pnone{ x_k }  $ and

\begin{adjustwidth}{-1.5cm}{}
\begin{prooftree}
    \small
\AxiomC{}
\UnaryInfC{$  \pnone{ u }  \vdash u : \piencodfaplas{ \tau }  $}
\UnaryInfC{$  \pnone{ u }  \vdash u : \piencodfaplas{ \tau } , \piencodfaplas{\Theta} $}

\AxiomC{}
\UnaryInfC{$  \pnone{ x_1 }  \vdash_1 : \with \overline{\piencodfaplas{\sigma_1}} $}
\UnaryInfC{$  \pnone{ x_1 }  \vdash_1 : \with \overline{\piencodfaplas{\sigma_1}} , \piencodfaplas{\Theta} $}

\AxiomC{}
\UnaryInfC{$  \pnone{ x_n }  \vdash x_n : \with \overline{\piencodfaplas{\sigma_n}}  $}
\UnaryInfC{$  \pnone{ x_n }  \vdash x_n : \with \overline{\piencodfaplas{\sigma_n}} , \piencodfaplas{\Theta} $}
\UnaryInfC{$\vdots$}
\BinaryInfC{$  \pnone{ x_1 }    \| \cdots   \|   \pnone{ x_k }  \vdash  x_1 : \with \overline{\piencodfaplas{\sigma_1}}, \cdots  ,x_n : \with \overline{\piencodfaplas{\sigma_n}} , \piencodfaplas{\Theta} $}
\LeftLabel{\ttype{mix}}
\BinaryInfC{$  \pnone{ u }    \|   \pnone{ x_1 }    \| \cdots   \|   \pnone{ x_k }  \vdash x_1 : \with \overline{\piencodfaplas{\sigma_1}}, \cdots  ,x_n : \with \overline{\piencodfaplas{\sigma_n}}, u : \piencodfaplas{ \tau } , \piencodfaplas{\Theta} $}
\end{prooftree}
\end{adjustwidth}

Thus, $\piencodfaplas{\fail^{x_1, \cdots , x_n} }_u\vdash  x_1 : \with \overline{\piencodfaplas{\sigma_1}}, \cdots  ,x_n : \with \overline{\piencodfaplas{\sigma_n}}, u : \piencodfaplas{ \tau } , \piencodfaplas{\Theta} $ and the result follows.

\end{enumerate}
\end{enumerate}
\end{proof}

\subsection{Completeness}\label{ch4a:tcompletness}

\begin{definition}{Linearly Partially Open Terms}
    We say that a $\lamcoldetsh$-term $M$ is \emph{linearly partially open} if $\forall x \in \llfv{M}$ implies that $x$ is not a sharing variable.
\end{definition}

\begin{proposition}
    \label{ch4prop:correctformfailunres}
    Suppose $N$ is a well-formed linearly partially open $\lamcoldetsh$-term with $\headf{N} = x$ ($x$ denoting either linear or unrestricted occurrence of $x$) .
    Then,
    \[
    \piencodfaplas{ N }_{u} = \res{ y_1 }( \cdots \res{y_m}( \piencodfaplas{ x }_{n} \| P_m ) \cdots \| P_1  )
    \]
    which we shall denote as:
    \(
    \piencodfaplas{ N }_{u} =  \res{ \widetilde{y} } (\piencodfaplas{ x }_{n}   \| P)
    \),
    for some names $\widetilde{y} =$ and $n$, and processes $P$.
    \end{proposition}

    \begin{proof}
    The proof is by induction on the structure of $N$.

    \begin{enumerate}

        \item $N =  {x}$:  then $\piencodfaplas{ {x}}_u  =  \psome{x}; \pfwd{x}{u} $. Hence $ P = \zero$ and $ \widetilde{y} = \emptyset$.

        \item $N =  {x}[j]$:  then $\piencodfaplas{ {x}[j]}_u = \puname{ \unvar{x} }{ x_i };\psel{ {x}_i }{j }; \pfwd{x_i}{u}$. Hence $ P = \zero$ and $ \widetilde{y} = \emptyset$.

        \item  $N = M\ (C \bagsep U)$: then $\headf{M\ (C \bagsep U)} = \headf{M} = x$ and
        \[ \piencodfaplas{N}_u = \piencodfaplas{M\ (C \bagsep U)}_u  =   \res{ v } (\piencodfaplas{M}_v   \|  \gsome{ v }{ u, \llfv{C} };  \pname{v}{x} . (\pfwd{v}{u}   \| \piencodfaplas{(C \bagsep U)}_x ) ) \]
        The result follows by  induction hypothesis applied on $\piencodfaplas{M}_u$.

        \item $N = (M[\widetilde{y} \leftarrow y])\esubst{ C \bagsep U }{ y }$:  this case does not  apply since $\head{N}\neq x$.

        \item $N = M \linexsub{C /  x_1 , \cdots , x_k}$:  then $\headf{M \linexsub{C /  x_1 , \cdots , x_k}} = \headf{M} = x$. Let $C = \bag{M_1} \cdot \cdots \cdot \bag{M_k} $
        \[
        \begin{aligned}
            \piencodfaplas{N}_u &= \res{z_1}( \gsome{z_1}{\llfv{M_{1}}};\piencodfaplas{ M_{1} }_{ {z_1}}  \|      \cdots \res{z_k} ( \gsome{z_k}{\llfv{M_{k}}};\piencodfaplas{ M_{k} }_{ {z_k}} \\
            & \quad  \| \bignd_{x_{i_1} \in \{ x_1 ,\cdots , x_k  \}} \cdots \bignd_{x_{i_k} \in \{ x_1 ,\cdots , x_k \setminus x_{i_1} , \cdots , x_{i_{k-1}}  \}} \piencodfaplas{ M }_u \{ z_1 / x_{i_1} \} \cdots \{ z_k / x_{i_k} \} ) \cdots )
        \end{aligned}
           \]
        The result follows by  induction hypothesis applied on $\piencodfaplas{M}_u$.

        \item  $N = M \unexsub{U / \unvar{x}} $: then $\headf{M \unexsub{U / \unvar{x}} } = \headf{M} = x$ and
        \[ \piencodfaplas{N}_u = \piencodfaplas{M \unexsub{U / \unvar{x}} }_u  =   \res{ \unvar{x} } ( \piencodfaplas{ M }_u   \|   ~ \guname{ \unvar{x} }{ x_i } ; \piencodfaplas{ U }_{x_i} ) \]
        The result follows by  induction hypothesis applied on $\piencodfaplas{M}_u$.


    \end{enumerate}

    \end{proof}

\begin{restatable}[Completeness (Under $\redtwo$)]{theorem}{thmEncTWCompl}\label{ch4l:app_completenesstwo}
    If $ {N}\red {M}$ for a well-formed closed $\lamcoldetsh$-term $N$, then $\piencodfaplas{{N}}_u \redtwo^\ast \piencodfaplas{{M}}_u$.
\end{restatable}

\begin{proof}
    By induction on the reduction rule applied to infer ${N}\red {M}$.
    We have five cases.

 \begin{enumerate}
        \item  Case $\redlab{RS:Beta}$:

        Then  $ N= (\lambda x . (M'[ {\widetilde{x}} \leftarrow  {x}])) B  \red (M' [ {\widetilde{x}} \leftarrow  {x}])\esubst{ B }{ x }  = M$ , where $B = C \bagsep U$.
    The result follows from

        \begin{equation*}\label{ch4eq:comp_sh_beta}
        \begin{aligned}
        \piencodfaplas{N}_u & = \res{v} (\piencodfaplas{\lambda x . (M'[ {\widetilde{x}} \leftarrow  {x}])}_v \| \gsome{v}{u , \llfv{C}};\pname{v}{x}; ( \piencodfaplas{C \bagsep U}_x   \| \pfwd{v}{u}  ) )\\
        & = \res{v} (\psome{v};\gname{v}{x};  \psome{x};\gname{x}{\linvar{x}}; \gname{x}{\unvar{x}};  \gclose{x} ; \piencodfaplas{M'[ {\widetilde{x}} \leftarrow  {x}]}_v \| \\
        & \qquad \gsome{v}{u , \llfv{C}};\pname{v}{x}; ( \piencodfaplas{C \bagsep U}_x   \| \pfwd{v}{u}  ) )\\
        & \redtwo \res{v} (\gname{v}{x};  \psome{x};\gname{x}{\linvar{x}}; \gname{x}{\unvar{x}};  \gclose{x} ; \piencodfaplas{M'[ {\widetilde{x}} \leftarrow  {x}]}_v \| \pname{v}{x}; ( \piencodfaplas{C \bagsep U}_x   \| \pfwd{v}{u}  ) )\\
        & \redtwo \res{v} ( \res{x}( \psome{x};\gname{x}{\linvar{x}}; \gname{x}{\unvar{x}};  \gclose{x} ; \piencodfaplas{M'[ {\widetilde{x}} \leftarrow  {x}]}_v \|   \piencodfaplas{C \bagsep U}_x )  \| \pfwd{v}{u}   )\\
        & \redtwo \res{x}( \psome{x};\gname{x}{\linvar{x}}; \gname{x}{\unvar{x}};  \gclose{x} ; \piencodfaplas{M'[ {\widetilde{x}} \leftarrow  {x}]}_u \|   \piencodfaplas{C \bagsep U}_x ) =   \piencodfaplas{M}_u    \\
        \end{aligned}
        \end{equation*}

        \item Case $ \redlab{RS:Ex \dash Sub}$: Then $ N =M'[ {x}_1, \!\cdots\! ,  {x}_k \leftarrow  {x}]\esubst{ C \bagsep U }{ x }$, with $C = \bag{M_1}
            \cdots  \bag{M_k}$, $k\geq 0$ and $M' \not= \fail^{\widetilde{y}}$.

        The reduction is $N = M'[ {x}_1, \!\cdots\! ,  {x}_k \leftarrow  {x}]\esubst{ C \bagsep U }{ x } \red  M'\linexsub{C  /  x_1 , \cdots , x_k} \unexsub{U / \unvar{x} } = M.$

        We detail the translations of $\piencodfaplas{N}_u$ and $\piencodfaplas{M}_u$. To simplify the proof, we will consider $k=2$ (the case in which $k> 2$ is follows analogously. Similarly the case of $k < 2$ it contained within $k = 2$). The result follows from:

        \begin{equation*}\label{ch4eq:comp_sh_esub}
        \small\begin{aligned}
        \piencodfaplas{N}_u
        &= \piencodfaplas{M'[ {x}_1 \leftarrow  {x}]\esubst{ C \bagsep U }{ x }}_u
        = \res{x}( \psome{x}; \gname{x}{\linvar{x}}; \gname{x}{\unvar{x}};  \gclose{x} ;\piencodfaplas{ M'[ {\widetilde{x}} \leftarrow  {x}]}_u \| \piencodfaplas{ C \bagsep U}_x ) \\
        &= \res{x}( \psome{x}; \gname{x}{\linvar{x}}; \gname{x}{\unvar{x}};  \gclose{x} ;\piencodfaplas{ M'[ {\widetilde{x}} \leftarrow  {x}]}_u \| \gsome{x}{\llfv{C}};  \pname{x}{\linvar{x}}; \big( \piencodfaplas{ C }_{\linvar{x}} \|  \pname{x}{\unvar{x}};\\
        & \qquad \qquad \qquad ( \guname{\unvar{x}}{x_i}; \piencodfaplas{ U }_{x_i} \| \pclose{x} ) \big) ) (:= P_{\mathbb{N}})\\
            & \redtwo^* \res{\unvar{x}} ( \res{\linvar{x}}( \piencodfaplas{ M'[ x_1, x_2 \leftarrow  {x}]}_u \|   \piencodfaplas{ \bag{M_1} \cdot \bag{M_2} }_{\linvar{x}} ) \|   \guname{\unvar{x}}{x_i}; \piencodfaplas{ U }_{x_i}   )  \\
            & = \res{\unvar{x}} ( \res{\linvar{x}}( \psome{\linvar{x}}; \pname{\linvar{x}}{y_1}; \big( \gsome{y_1}{ \emptyset }; \gclose{ y_{1} } ; \0   \\
            & \qquad \qquad \|\psome{\linvar{x}}; \gsome{\linvar{x}}{u, \llfv{M'} \setminus  \widetilde{x} }; \bignd_{x_{i_1} \in \widetilde{x}} \gname{x}{{x}_{i_1}};\piencodfaplas{M'[ (\widetilde{x} \setminus x_{i_1} ) \leftarrow  {x}]}_u \big) \| \\
            & \qquad \qquad \gsome{\linvar{x}}{\llfv{C} }; \gname{x}{y_1}; \gsome{\linvar{x}}{y_i, \llfv{C}}; \psome{\linvar{x}}; \pname{\linvar{x}}{z_1}; \\
            & \qquad   ( \gsome{z_1}{\llfv{M_1}};  \piencodfaplas{M_1}_{z_1} \| \piencodfaplas{\bag{M_2}}_{\linvar{x}} \| \pnone{y_1} ) ) \|   \guname{\unvar{x}}{x_i}; \piencodfaplas{ U }_{x_i}   )  \\
            & \redtwo^* \res{\unvar{x}} ( \res{\linvar{x}}( \piencodfaplas{\bag{M_2}}_{\linvar{x}}  \|  \res{z_1}  \big( \bignd_{x_{i_1} \in \widetilde{x}} \piencodfaplas{M'[ (\widetilde{x} \setminus x_{i_1} ) \leftarrow  {x}]}_u \{ z_1 / x_{i_1} \}  \| \\
            & \qquad \qquad    \gsome{z_1}{\llfv{M_1}};  \piencodfaplas{M_1}_{z_1}  \big) ) \|   \guname{\unvar{x}}{x_i}; \piencodfaplas{ U }_{x_i}   )  \\
            & = \res{\unvar{x}} ( \res{\linvar{x}}( \gsome{\linvar{x}}{\llfv{C} }; \gname{x}{y_2}; \gsome{\linvar{x}}{y_2, \llfv{C}}; \psome{\linvar{x}}; \pname{\linvar{x}}{z_2}; \\
            & \qquad   ( \gsome{z_2}{\llfv{M_2}};  \piencodfaplas{M_2}_{z_2} \| \piencodfaplas{\oneb}_{\linvar{x}} \| \pnone{y_2} )  \|\\
            &   \res{z_1}  \big( \bignd_{x_{i_1} \in \widetilde{x}} \psome{\linvar{x}}; \pname{\linvar{x}}{y_2}; \big( \gsome{y_2}{ \emptyset }; \gclose{ y_{2} } ; \0   \\
            & \qquad \qquad \|\psome{\linvar{x}}; \gsome{\linvar{x}}{u, \llfv{M'} \setminus  (\widetilde{x} \setminus x_{i_1}  )}; \bignd_{x_{i_2} \in (\widetilde{x} \setminus x_{i_1} )} \gname{x}{{x}_i};\piencodfaplas{M'[  \leftarrow  {x}]}_u\{ z_1 / x_{i_1} \} \big)   \| \\
            & \qquad \qquad    \gsome{z_1}{\llfv{M_1}};  \piencodfaplas{M_1}_{z_1}  \big) ) \|   \guname{\unvar{x}}{x_i}; \piencodfaplas{ U }_{x_i}   )  \\
            & \red^* \res{\unvar{x}} ( \res{\linvar{x}}(  \piencodfaplas{\oneb}_{\linvar{x}} \| \\
            & \qquad \res{z_2}  ( \res{z_1}  \big( \bignd_{x_{i_1} \in \widetilde{x}}  \bignd_{x_{i_2} \in (\widetilde{x} \setminus x_{i_1} )} \gname{x}{{x}_i};\piencodfaplas{M'[  \leftarrow  {x}]}_u\{ z_1 / x_{i_1} \}   \| \\
            & \qquad \qquad    \gsome{z_1}{\llfv{M_1}};  \piencodfaplas{M_1}_{z_1}  \big) \|
            \gsome{z_2}{\llfv{M_2}};  \piencodfaplas{M_2}_{z_2} )      ) \|   \guname{\unvar{x}}{x_i}; \piencodfaplas{ U }_{x_i}   )  \\
            & \red^* \res{\unvar{x}} (  \res{z_2}  ( \res{z_1}  \big( \bignd_{x_{i_1} \in \widetilde{x}}  \bignd_{x_{i_2} \in (\widetilde{x} \setminus x_{i_1} )} \gname{x}{{x}_i};\piencodfaplas{M'}_u\{ z_1 / x_{i_1} \}   \| \\
            & \qquad \qquad    \gsome{z_1}{\llfv{M_1}};  \piencodfaplas{M_1}_{z_1}  \big) \|
            \gsome{z_2}{\llfv{M_2}};  \piencodfaplas{M_2}_{z_2} ) \|   \guname{\unvar{x}}{x_i}; \piencodfaplas{ U }_{x_i}   ) =    \piencodfaplas{M}_u \\
        \end{aligned}
        \end{equation*}

        \item Case $\redlab{RS{:}Fetch^{\ell}}$:

        Then we have
        $N = M' \linexsub{C /  x_1 , \cdots , x_k} $ with $\headf{M'} =  {x}_j$, $C = M_1 , \cdots , M_k$ and $N \red  M' \headlin{ M_i / x_j }  \linexsub{(C \setminus M_i ) /  x_1 , \cdots , x_k \setminus x_j }  = M$, for some $M_i \in C$.

        On the one hand, we have:
            \begin{equation}\label{ch4eq:comp_sh_linfet1}
            \begin{aligned}
            \piencodfaplas{N}_u &= \piencodfaplas{M' \linexsub{C /  x_1 , \cdots , x_k, x_j}}_u \\
            &= \res{z_1}( \gsome{z_1}{\llfv{M_{1}}};\piencodfaplas{ M_{1} }_{ {z_1}}  \|  \cdots \res{z_k} ( \gsome{z_k}{\llfv{M_{k}}};\piencodfaplas{ M_{k} }_{ {z_k}} \\
            &  \quad  \| \bignd_{x_{i_1} \in \{ x_1 ,\cdots , x_k  \}} \cdots \\
            &  \quad \qquad  \bignd_{x_{i_k} \in \{ x_1 ,\cdots , x_k \setminus x_{i_1} , \cdots , x_{i_{k-1}}  \}} \piencodfaplas{ M' }_u \{ z_1 / x_{i_1} \} \cdots \{ z_k / x_{i_k} \} ) \cdots )\\
            &= \res{z_1}( \gsome{z_1}{\llfv{M_{1}}};\piencodfaplas{ M_{1} }_{ {z_1}}  \|  \cdots \res{z_k} ( \gsome{z_k}{\llfv{M_{k}}};\piencodfaplas{ M_{k} }_{ {z_k}} \\
            &  \quad  \| \bignd_{x_{i_1} \in \{ x_1 ,\cdots , x_k  \}} \cdots \\
            &  \quad \qquad \bignd_{x_{i_k} \in \{ x_1 ,\cdots , x_k \setminus x_{i_1} , \cdots , x_{i_{k-1}}  \}} \res{y}(\piencodfaplas{  {x}_{j} }_y \| P) \{ z_1 / x_{i_1} \} \cdots \{ z_k / x_{i_k} \} )\\
            &  \quad \cdots ) \qquad (*)\\
            &= \res{z_1}( \gsome{z_1}{\llfv{M_{1}}};\piencodfaplas{ M_{1} }_{ {z_1}}  \| \cdots \res{z_k} ( \gsome{z_k}{\llfv{M_{k}}};\piencodfaplas{ M_{k} }_{ {z_k}} \\
            &  \quad  \| \bignd_{x_{i_1} \in \{ x_1 ,\cdots , x_k  \}} \cdots \\
            &  \quad \qquad \bignd_{x_{i_k} \in \{ x_1 ,\cdots , x_k \setminus x_{i_1} , \cdots , x_{i_{k-1}}  \}}\\
            &  \quad \qquad \qquad \res{y}(  \psome{x_j}; \pfwd{x_j}{y}  \| P) \{ z_1 / x_{i_1} \} \cdots \{ z_k / x_{i_k} \} )\cdots )\\
            \end{aligned}
            \end{equation}
        where $(*)$ is inferred via Proposition~\ref{ch4prop:correctformfailunres}.

        Let us consider the case when $j = k$ and $ M_i = M_1$ the other cases proceed similarly. Then we have the following reduction:

        \begin{equation}\label{ch4eq:comp_sh_linfet2}
            \begin{aligned}
                &= \res{z_1}( \gsome{z_1}{\llfv{M_{1}}};\piencodfaplas{ M_{1} }_{ {z_1}}  \| \cdots \res{z_k} ( \gsome{z_k}{\llfv{M_{k}}};\piencodfaplas{ M_{k} }_{ {z_k}} \\
                & \qquad \qquad  \| \bignd_{x_{i_1} \in \{ x_1 ,\cdots , x_k  \}} \cdots \\
                & \qquad \qquad \qquad \bignd_{x_{i_k} \in \{ x_1 ,\cdots , x_k \setminus x_{i_1} , \cdots , x_{i_{k-1}}  \}}\\
                & \qquad \qquad \qquad \qquad \res{y}(\psome{x_k}; \pfwd{x_k}{y} \| P) \{ z_1 / x_{i_1} \} \cdots \{ z_k / x_{i_k} \} ) \cdots )\\
                & \redtwo \res{z_1}( \piencodfaplas{ M_{1} }_{ {z_1}}  \|  \cdots \res{z_k} ( \gsome{z_k}{\llfv{M_{k}}};\piencodfaplas{ M_{k} }_{ {z_k}} \\
                & \qquad \qquad  \| \bignd_{x_{i_2} \in \{ x_1 ,\cdots , x_{k-1}  \}} \cdots \\
                & \qquad \qquad \qquad \bignd_{x_{i_k} \in \{ x_1 ,\cdots , x_{k-1} \setminus x_{i_1} , \cdots , x_{i_{k-1}}  \}}\\
                & \qquad \qquad \qquad \qquad \res{y}(  \pfwd{z_1}{y} \| P) \{ z_2 / x_{i_2} \} \cdots \{ z_k / x_{i_k} \} )\cdots )\\
                & \redtwo \res{z_2}( \gsome{z_2}{\llfv{M_{2}}};\piencodfaplas{ M_{2} }_{ {z_2}}  \|  \cdots \res{z_k} ( \gsome{z_k}{\llfv{M_{k}}};\piencodfaplas{ M_{k} }_{ {z_k}} \\
                & \qquad \qquad  \| \bignd_{x_{i_2} \in \{ x_1 ,\cdots , x_{k-1}  \}} \cdots \\
                & \qquad \qquad \qquad \bignd_{x_{i_k} \in \{ x_1 ,\cdots , x_{k-1} \setminus x_{i_1} , \cdots , x_{i_{k-1}}  \}}\\
                & \qquad \qquad \qquad \qquad \res{y}( \piencodfaplas{ M_{1} }_{y}  \| P) \{ z_2 / x_{i_2} \} \cdots \{ z_k / x_{i_k} \} )\cdots )\\
            \end{aligned}
            \end{equation}

         On the other hand, we have:
        \begin{equation}\label{ch4eq:comp_sh_linfet3}
        \begin{aligned}
        \piencodfaplas{M}_u &= \piencodfaplas{M' \headlin{ M_1 / x_k }  \linexsub{(C \setminus M_1 ) /  x_1 , \cdots , x_{k-1} } }_u \\
            & =   \res{z_1}( \gsome{z_1}{\llfv{M_{2}}};\piencodfaplas{ M_{2} }_{ {z_1}}  \|  \cdots \res{z_{k-1}} ( \gsome{z_{k-1}}{\llfv{M_{k}}};\piencodfaplas{ M_{k} }_{ {z_{k-1}}} \\
            & \quad  \| \bignd_{x_{i_1} \in \{ x_1 ,\cdots , x_{k-1}  \}}
            \cdots \bignd_{x_{i_{k-1}} \in \{ x_1 ,\cdots , x_{k-1} \setminus x_{i_1} , \cdots , x_{i_{k-2}}  \}}
            \piencodfaplas{ M' }_u \{ z_1 / x_{i_1} \} \cdots \\
            & \quad \{ z_{k-1} / x_{i_{k-1}} \} ) \cdots ) \\
        \end{aligned}
        \end{equation}
                Therefore, by \eqref{ch4eq:comp_sh_linfet2}, \eqref{ch4eq:comp_sh_linfet3} and taking $M_{j_k} = M_i$ the result follows.

        \item Case $ \redlab{RS{:} Fetch^!}$:

        Then,
        $N = M' \unexsub{U / \unvar{x}}$ with $\headf{M' } = \unvar{x}[k]$, $U_k = \unvar{\bag{N}}$ and $N \red  M' \headlin{ N /\unvar{x} }\unexsub{U / \unvar{x}} = M$. The result follows from

            \begin{equation}\label{ch4eq:compl_sh_fetchun1}
            \hspace{-0.5cm}
            \begin{aligned}
                \small
            \piencodfaplas{N}_u &= \piencodfaplas{M' \unexsub{U / \unvar{x}}}_u = \res{\unvar{x}} ( \piencodfaplas{ M' }_u \|   ~ \guname{\unvar{x}}{x_i}; \piencodfaplas{ U }_{x_i} ) \\
            & = \res{\unvar{x}} ( \res{y}(\piencodfaplas{ \unvar{x}[k] }_{j} \| P) \|   ~ \guname{\unvar{x}}{x_i}; \piencodfaplas{ U }_{x_i} ) \qquad (*)\\
            & = \res{\unvar{x}} ( \res{y}( \puname{\unvar{x}}{{x_i}}; \psel{x_i}{k}; \pfwd{x_i}{j} \| P) \|   ~ \guname{\unvar{x}}{x_i}; \piencodfaplas{ U }_{x_i} )
            \\
            & \redtwo \res{\unvar{x}} ( \res{y}( \puname{\unvar{x}}{{x_i}}; \psel{x_i}{k}; \pfwd{x_i}{j} \| P) \|   ~ \guname{\unvar{x}}{x_i}; \piencodfaplas{ U }_{x_i} )
            \\
             & \redtwo \res{\unvar{x}} ( \guname{\unvar{x}}{x_i}; \piencodfaplas{ U }_{x_i} \| \res{x_i} ( \res{y}( \psel{x_i}{k}; \pfwd{x_i}{j} \| P) \|  \piencodfaplas{ U }_{x_i}   )  ~  )
            \\
             & = \res{\unvar{x}} ( \guname{\unvar{x}}{x_i}; \piencodfaplas{ U }_{x_i} \| \res{x_i} ( \res{y}( \psel{x_i}{k}; \pfwd{x_i}{j} \| P) \|  \gsel{x_i}\{i:\piencodfaplas{U_i}_{x_i}\}_{U_i \in U}   )  ~  )
            \\
            & \redtwo \res{\unvar{x}} ( \guname{\unvar{x}}{x_i}; \piencodfaplas{ U }_{x_i} \| \res{x_i} ( \res{y}( \pfwd{x_i}{j} \| P) \| \piencodfaplas{U_i}_{x_i}   )  ~  )
            \\
             & \redtwo \res{\unvar{x}} ( \guname{\unvar{x}}{x_i}; \piencodfaplas{ U }_{x_i} \| \res{y}( \piencodfaplas{U_i}_{j} \| P)  ~  ) \\
             & \redtwo \res{\unvar{x}} ( \guname{\unvar{x}}{x_i}; \piencodfaplas{ U }_{x_i} \| \res{y}( \piencodfaplas{N}_{j} \| P)  ~  ) =\piencodfaplas{M}_u
            \end{aligned}
            \end{equation}
       where the reductions denoted by $(*)$ are inferred via Proposition~\ref{ch4prop:correctformfailunres}.

        \item Case $\redlab{RS:TCont}$:
            This case follows by IH.

        \item Case $\redlab{RS{:}Fail^{\ell}}$:

        Then we have
        $N = M' [ {x}_1, \!\cdots\! ,  {x}_k \leftarrow  {x}]\ \esubst{C \bagsep U}{ x } $ with $k \neq \size{C}$ and
        $N \red  \fail^{\widetilde{y}} = M$, where $\widetilde{y} = (\llfv{M' } \setminus \{   {x}_1, \cdots ,  {x}_k \} ) \cup \llfv{C}$. Let $ C = \bag{M_1} \cdot \cdots \cdot \bag{M_l}$ and we assume that $k > l$ and we proceed similarly for $k > l$. Hence $k = l + m$ for some $m \geq 1$
        \begin{equation*}\label{ch4eq:compl_sh_faillin1}
        \small
        \begin{aligned}
            \piencodfaplas{N}_u
            &= \piencodfaplas{M' [ {x}_1, \!\cdots\! ,  {x}_k \leftarrow  {x}]\ \esubst{C \bagsep U}{ x } }_u\\
            &= \res{x}( \psome{x}; \gname{x}{\linvar{x}}; \gname{x}{\unvar{x}};  \gclose{x} ;\piencodfaplas{ M' [ {\widetilde{x}} \leftarrow  {x}]}_u \| \piencodfaplas{ C \bagsep U}_x )\\
            &= \res{x}( \psome{x}; \gname{x}{\linvar{x}}; \gname{x}{\unvar{x}};  \gclose{x} ;\piencodfaplas{ M' [ {\widetilde{x}} \leftarrow  {x}]}_u \| \\
            & \qquad \qquad \gsome{x}{\llfv{C}};  \pname{x}{\linvar{x}}; \big( \piencodfaplas{ C }_{\linvar{x}} \|  \pname{x}{\unvar{x}}; ( \guname{\unvar{x}}{x_i}; \piencodfaplas{ U }_{x_i} \| \pclose{x} ) \big) )\\
            & \redtwo^*  \res{\unvar{x}} ( \res{\linvar{x} } \big( \piencodfaplas{ M' [ {\widetilde{x}} \leftarrow  {x}]}_u \|  \piencodfaplas{ C }_{\linvar{x}}   \big) \|  \guname{\unvar{x}}{x_i}; \piencodfaplas{ U }_{x_i}  )  \\
            & = \res{\unvar{x}} ( \res{\linvar{x} } \big( \psome{\linvar{x}}; \pname{\linvar{x}}{y_1}; \big( \gsome{y_1}{ \emptyset }; \gclose{ y_{1} } ; \0   \\
            & \qquad \qquad \|\psome{\linvar{x}}; \gsome{\linvar{x}}{u, \llfv{M' } \setminus  \widetilde{x} }; \bignd_{x_{i_1} \in \widetilde{x}} \gname{x}{{x}_{i_1}}; \cdots \\
            & \qquad \qquad \psome{\linvar{x}}; \pname{\linvar{x}}{y_k}; \big( \gsome{y_k}{ \emptyset }; \gclose{ y_{k} } ; \0   \\
            & \qquad \qquad \|\psome{\linvar{x}}; \gsome{\linvar{x}}{u, \llfv{M' } \setminus  (\widetilde{x} \setminus x_{i_1} , \cdots , x_{i_{k-1}}   )}; \bignd_{x_{i_k} \in (\widetilde{x} \setminus x_{i_1} , \cdots , x_{i_{k-1}}  )} \gname{x}{{x}_{i_k}};\\
            & \qquad \qquad  \piencodfaplas{M' [  \leftarrow  {x}]}_u \big)
            \big) \| \gsome{\linvar{x}}{\llfv{C} }; \gname{x}{y_1}; \gsome{\linvar{x}}{y_1, \llfv{C}}; \psome{\linvar{x}}; \pname{\linvar{x}}{z_1}; \\
            & \qquad \qquad  ( \gsome{z_1}{\llfv{M_1}};  \piencodfaplas{M_1}_{z_1} \| \pnone{y_1}  \| \cdots  \gsome{\linvar{x}}{\llfv{C} }; \gname{x}{y_l}; \gsome{\linvar{x}}{y_l, \llfv{M_l}}; \\
            & \qquad \qquad   \psome{\linvar{x}}; \pname{\linvar{x}}{z_l}; ( \gsome{z_l}{\llfv{M_l}};  \piencodfaplas{M_l}_{z_l} \| \piencodfaplas{ \oneb }_{\linvar{x}} \| \pnone{y_l} ) \cdots   )   \big) \|  \guname{\unvar{x}}{x_i}; \piencodfaplas{ U }_{x_i}  ) \\
            &
            (:= P_\mathbb{N}) \\
            \end{aligned}
            \end{equation*}

            we reduce $P_\mathbb{N}$ arbitrarily synchronising along channels $\linvar{x} , y_1, \cdots y_l$.

            \begin{equation*}
                \small
            \begin{aligned}
                P_\mathbb{N}
                & \redtwo^* \res{\unvar{x}} ( \res{\linvar{x} } \big( \piencodfaplas{ \oneb }_{\linvar{x}} \|
                \res{z_1} (\gsome{z_1}{\llfv{M_1}};  \piencodfaplas{M_1}_{z_1}   \| \cdots
                \res{z_l} (\gsome{z_l}{\llfv{M_l}};  \piencodfaplas{M_l}_{z_l} \| \\
                &  \qquad \bignd_{x_{i_1} \in \widetilde{x}} \cdots \bignd_{x_{i_l} \in (\widetilde{x} \setminus x_{i_1} , \cdots , x_{i_{l-1}} )}  \psome{\linvar{x}}; \pname{\linvar{x}}{y_{l+1}}; \big( \gsome{y_{l+1}}{ \emptyset }; \gclose{ y_{{l+1}} } ; \0   \\
                & \qquad \|\psome{\linvar{x}}; \gsome{\linvar{x}}{u, \llfv{M' } \setminus  (\widetilde{x} \setminus x_{i_1} , \cdots , x_{i_{l}}  )}; \bignd_{x_{i_{l+1}} \in (\widetilde{x} \setminus x_{i_1} , \cdots , x_{i_{l}} )} \gname{x}{{x}_{i_{l+1}}}; \cdots   \\
                & \qquad  \psome{\linvar{x}}; \pname{\linvar{x}}{y_k}; \big( \gsome{y_k}{ \emptyset }; \gclose{ y_{k} } ; \0   \\
                & \qquad \|\psome{\linvar{x}}; \gsome{\linvar{x}}{u, \llfv{M' } \setminus  (\widetilde{x} \setminus x_{i_1} , \cdots , x_{i_{k-1}}   )}; \bignd_{x_{i_k} \in (\widetilde{x} \setminus x_{i_1} , \cdots , x_{i_{k-1}}  )} \gname{x}{{x}_{i_k}};\\
                & \qquad \piencodfaplas{M' [  \leftarrow  {x}]}_u \{ z_1 / x_{i_1} \} \cdots \{ z_l / {x}_{i_l} \}
                \big) \cdots \big)  ) \cdots ) \big) \|\guname{\unvar{x}}{x_i}; \piencodfaplas{ U }_{x_i}  )\\
                & = \res{\unvar{x}} ( \res{\linvar{x} } \big(
               \gsome{\linvar{x}}{\emptyset}; \gname{x}{y_{l+1}};  ( \psome{ y_{l+1}}; \pclose{y_{l+1}}  \| \gsome{\linvar{x}}{\emptyset}; \pnone{\linvar{x}} ) \\
                & \qquad  \|
                \res{z_1} (\gsome{z_1}{\llfv{M_1}};  \piencodfaplas{M_1}_{z_1}   \| \cdots
                \res{z_l} (\gsome{z_l}{\llfv{M_l}};  \piencodfaplas{M_l}_{z_l} \| \\
                & \qquad \bignd_{x_{i_1} \in \widetilde{x}} \cdots \bignd_{x_{i_l} \in (\widetilde{x} \setminus x_{i_1} , \cdots , x_{i_{l-1}} )}\psome{\linvar{x}}; \pname{\linvar{x}}{y_{l+1}}; \big( \gsome{y_{l+1}}{ \emptyset }; \gclose{ y_{{l+1}} } ; \0   \\
                & \qquad  \|\psome{\linvar{x}}; \gsome{\linvar{x}}{u, \llfv{M' } \setminus  (\widetilde{x} \setminus x_{i_1} , \cdots , x_{i_{l}}  )}; \bignd_{x_{i_{l+1}} \in (\widetilde{x} \setminus x_{i_1} , \cdots , x_{i_{l}} )} \gname{x}{{x}_{i_{l+1}}}; \cdots   \\
                & \qquad \psome{\linvar{x}}; \pname{\linvar{x}}{y_k}; \big( \gsome{y_k}{ \emptyset }; \gclose{ y_{k} } ; \0   \\
                & \qquad \|\psome{\linvar{x}}; \gsome{\linvar{x}}{u, \llfv{M' } \setminus  (\widetilde{x} \setminus x_{i_1} , \cdots , x_{i_{k-1}}   )}; \bignd_{x_{i_k} \in (\widetilde{x} \setminus x_{i_1} , \cdots , x_{i_{k-1}}  )} \gname{x}{{x}_{i_k}};\\
                & \qquad  \piencodfaplas{M' [  \leftarrow  {x}]}_u \{ z_1 / x_{i_1} \} \cdots \{ z_l / {x}_{i_l} \}
                \big) \cdots \big)  ) \cdots ) \big) \| \guname{\unvar{x}}{x_i}; \piencodfaplas{ U }_{x_i}  )\\
                & \redtwo^* \res{\unvar{x}} ( \res{\linvar{x} } \big( \pnone{\linvar{x}}  \|
                \res{z_1} (\gsome{z_1}{\llfv{M_1}};  \piencodfaplas{M_1}_{z_1}   \| \cdots
                \res{z_l} (\gsome{z_l}{\llfv{M_l}};  \piencodfaplas{M_l}_{z_l} \| \\
                &  \qquad  \bignd_{x_{i_1} \in \widetilde{x}} \cdots \bignd_{x_{i_l} \in (\widetilde{x} \setminus x_{i_1} , \cdots , x_{i_{l-1}} )} \gsome{\linvar{x}}{u, \llfv{M' } \setminus  (\widetilde{x} \setminus x_{i_1} , \cdots , x_{i_{l}}  )};   \\
                & \qquad  \bignd_{x_{i_{l+1}} \in (\widetilde{x} \setminus x_{i_1} , \cdots , x_{i_{l}} )} \gname{x}{{x}_{i_{l+1}}}; \cdots \psome{\linvar{x}}; \pname{\linvar{x}}{y_k}; \big( \gsome{y_k}{ \emptyset }; \gclose{ y_{k} } ; \0   \\
                & \qquad  \|\psome{\linvar{x}}; \gsome{\linvar{x}}{u, \llfv{M' } \setminus  (\widetilde{x} \setminus x_{i_1} , \cdots , x_{i_{k-1}}   )}; \bignd_{x_{i_k} \in (\widetilde{x} \setminus x_{i_1} , \cdots , x_{i_{k-1}}  )} \gname{x}{{x}_{i_k}};\\
                & \qquad  \piencodfaplas{M' [  \leftarrow  {x}]}_u \{ z_1 / x_{i_1} \} \cdots \{ z_l / {x}_{i_l} \}
                \big) \cdots \big)  ) \cdots ) \big) \|  \guname{\unvar{x}}{x_i}; \piencodfaplas{ U }_{x_i}  )\\
                & \redtwo \res{\unvar{x}} (
                \res{z_1} (\gsome{z_1}{\llfv{M_1}};  \piencodfaplas{M_1}_{z_1}   \| \cdots
                \res{z_l} (\gsome{z_l}{\llfv{M_l}};  \piencodfaplas{M_l}_{z_l} \| \\
                &  \qquad  \bignd_{x_{i_1} \in \widetilde{x}} \cdots \bignd_{x_{i_l} \in (\widetilde{x} \setminus x_{i_1} , \cdots , x_{i_{l-1}} )} \pnone{u } \| \pnone{(\llfv{M' } \setminus  \widetilde{x}) } \|  \pnone{(z_1 , \cdots ,z_l ) }
                ) \cdots ) \| \\
                & \qquad \guname{\unvar{x}}{x_i}; \piencodfaplas{ U }_{x_i}  )\\
                & \redtwo^* \res{\unvar{x}} (
                 \bignd_{x_{i_1} \in \widetilde{x}} \cdots \bignd_{x_{i_l} \in (\widetilde{x} \setminus x_{i_1} , \cdots , x_{i_{l-1}} )} \pnone{u } \| \pnone{(\llfv{M' } \setminus  \widetilde{x}) }  \|\guname{\unvar{x}}{x_i}; \piencodfaplas{ U }_{x_i}  )\\
                & \equiv \pnone{u } \| \pnone{(\llfv{M' } \setminus  \widetilde{x}) }   =\piencodfaplas{M}_u\\
        \end{aligned}
        \end{equation*}

       \item Case $\redlab{RS{:}Fail^!}$:

        Then,
        $N = M' \unexsub{U /\unvar{x}}  $ with $\headf{M' } =  {x}[i]$, $U_i = \unvar{\oneb} $ and
        $N \red   M' \headlin{ \fail^{\emptyset} /\unvar{x} } \unexsub{U /\unvar{x} } $. The result follows from

        \begin{equation*}\label{ch4eq:compl_sh_failun1}
        \begin{aligned}
            \piencodfaplas{N}_u &= \piencodfaplas{ M' \unexsub{U /\unvar{x}} }_u  =   \res{\unvar{x}} ( \piencodfaplas{ M' }_u \|   ~ \guname{\unvar{x}}{x_i}; \piencodfaplas{ U }_{x_i} )  \\
            & = \res{\unvar{x}} (  \res{\widetilde{y}}(\piencodfaplas{  {x}[i] }_{j} \| P)   \|   ~ \guname{\unvar{x}}{x_i}; \piencodfaplas{ U }_{x_i} ) \qquad (*)
            \\
            & = \res{\unvar{x}} (  \res{\widetilde{y}}(  \puname{\unvar{x}}{{x_i}}; \psel{x_i}{i}; \pfwd{x_i}{j} \| P)   \|   ~ \guname{\unvar{x}}{x_i}; \piencodfaplas{ U }_{x_i} )
            \\
            & \redtwo \res{\unvar{x}} ( \res{x_i} ( \res{\widetilde{y}}(  \psel{x_i}{i}; \pfwd{x_i}{j} \| P) \|  \piencodfaplas{ U }_{x_i} )   \|   ~ \guname{\unvar{x}}{x_i}; \piencodfaplas{ U }_{x_i} )
            \\
            & = \res{\unvar{x}} ( \res{x_i} ( \res{\widetilde{y}}(  \psel{x_i}{i}; \pfwd{x_i}{j} \| P) \|  \gsel{x_i}\{i:\piencodfaplas{ U_i }_{x_i} \}_{U_i \in U} )   \|   ~ \guname{\unvar{x}}{x_i}; \piencodfaplas{ U }_{x_i} )
            \\
            & \redtwo* \res{\unvar{x}} (  \res{\widetilde{y}}( \piencodfaplas{ U_i }_{j}  \| P)     \|   ~ \guname{\unvar{x}}{x_i}; \piencodfaplas{ U }_{x_i} )
            \\
            & = \res{\unvar{x}} (  \res{\widetilde{y}}( \piencodfaplas{ \unvar{\oneb} }_{j}  \| P)    \|   ~ \guname{\unvar{x}}{x_i}; \piencodfaplas{ U }_{x_i} )
            \\
            & = \res{\unvar{x}} (  \res{\widetilde{y}}( \pnone{j}   \| P)    \|   ~ \guname{\unvar{x}}{x_i}; \piencodfaplas{ U }_{x_i} ) =   \piencodfaplas{M}_u
            \\
        \end{aligned}
        \end{equation*}

        \item Case $\redlab{RS:Cons_1}$:
        Then we have
        $N = \fail^{\widetilde{x}}\ C \bagsep U$ and $N \red \fail^{\widetilde{x} \cup \widetilde{y}}  = M$ where $ \widetilde{y} = \llfv{C}$. Also,

        \begin{equation*}\label{ch4eq:compl_sh_con1}
        \begin{aligned}
            \piencodfaplas{N}_u &= \piencodfaplas{ \fail^{\widetilde{x}}\ C \bagsep U }_u = \res{v} (\piencodfaplas{\fail^{\widetilde{x}}}_v \| \gsome{v}{u , \llfv{C}};\pname{v}{x}; ( \piencodfaplas{C \bagsep U}_x  \| \pfwd{v}{u}  ) ) \\
            & = \res{v} ( \pnone{v } \| \pnone{\widetilde{x}}  \| \gsome{v}{u , \llfv{C}};\pname{v}{x}; ( \piencodfaplas{C \bagsep U}_x  \| \pfwd{v}{u}  ) ) \\
            & \redtwo  \pnone{\widetilde{x}} \| \pnone{u} \| \pnone{\widetilde{y}} =  \piencodfaplas{M}_u
        \end{aligned}
        \end{equation*}

        \item Cases $\redlab{RS:Cons_2}$ and $\redlab{RS:Cons_3}$: These cases follow by IH similarly to Case 7.

        \item Case $\redlab{RS{:}Cons_4}$:
        Then we have
        $N =  \fail^{\widetilde{y}} \unexsub{U / \unvar{x}} $ and $N \red \fail^{\widetilde{y}}  = M$, and

        \begin{align}\label{ch4eq:compl_sh_con4_1}
            \piencodfaplas{N}_u &= \piencodfaplas{ \fail^{\widetilde{y}} \unexsub{U / \unvar{x}}}_u = \res{\unvar{x}} ( \piencodfaplas{ \fail^{\widetilde{y}} }_u \|   ~ \guname{\unvar{x}}{x_i}; \piencodfaplas{ U }_{x_i} ) \\
            &= \res{\unvar{x}} ( \pnone{u }  \| \pnone{\widetilde{x}} \|   ~ \guname{\unvar{x}}{x_i}; \piencodfaplas{ U }_{x_i} )
            \equiv   \pnone{u }  \| \pnone{\widetilde{x}}=  \piencodfaplas{M}_u
        \end{align}

    \end{enumerate}

\end{proof}

\subsection{Soundness}\label{ch4a:tsoundness}

\begin{restatable}[Weak Soundness (Under $\redtwo$)]{theorem}{thmEncTWSound}\label{ch4t:soundnesstwounres}
    If $ \piencodfaplas{N}_u \redtwo^* Q$ for a well-formed closed $\lamcoldetsh$-term $N$, then there exist $Q'$  and $N' $ such that $Q \redtwo^* Q'$, $N  \red^* N'$ and $\piencodfaplas{N'}_u \equiv Q'$.
\end{restatable}

\begin{proof}
By induction on the structure of $N $ and then induction on the number of reductions of $\piencodfaplas{N} \redtwo^* Q$.

\begin{enumerate}
    \item {\bf Base case:} $N =  {x}$, $N =  {x}[j]$, $N = \fail^{\emptyset}$ and $N = \lambda x . (M[ {\widetilde{x}} \leftarrow  {x}])$.
.

    No reductions can take place, and the result follows trivially.
    $Q =  \piencodfaplas{N}_u \redtwo^0 \piencodfaplas{N}_u = Q'$ and $ N \red^0  N = N'$.

    \item $N =  M (C \bagsep U) $.

        Then,
        $ \piencodfaplas{M (C \bagsep U)}_u = \res{v} (\piencodfaplas{M}_v \| \gsome{v}{u , \llfv{C}};\pname{v}{x}; ( \piencodfaplas{C \bagsep U}_x  \| \pfwd{v}{u}  ) )$, and we are able to perform the  reductions from $\piencodfaplas{M (C \bagsep U)}_u$.

        We now proceed by induction on $k$, with  $\piencodfaplas{N}_u \redtwo^k Q$. There are two main cases:
        \begin{enumerate}
            \item When $k = 0$ the thesis follows easily:

            We have
    $Q =  \piencodfaplas{M (C \bagsep U)}_u \redtwo^0 \piencodfaplas{M (C \bagsep U)}_u = Q'$ and $M (C \bagsep U) \red^0 M (C \bagsep U) = N'$.

            \item The interesting case is when $k \geq 1$.

            Then, for some process $R$ and $n, m$ such that $k = n+m$, we have the following:
            \[
            \begin{aligned}
               \piencodfaplas{{N}}_u & =  \res{v} (\piencodfaplas{M}_v \| \gsome{v}{u , \llfv{C}};\pname{v}{x}; ( \piencodfaplas{C \bagsep U}_x  \| \pfwd{v}{u}  ) )\\
               & \redtwo^m  \res{v} ( R \| \gsome{v}{u , \llfv{C}};\pname{v}{x}; ( \piencodfaplas{C \bagsep U}_x  \| \pfwd{v}{u}  ) ))
               \redtwo^n  Q\\
            \end{aligned}
            \]
            Thus, the first $m \geq 0$ reduction steps are  internal to $\piencodfaplas{ M}_v$; type preservation in \clpi ensures that, if they occur,  these reductions  do not discard the possibility of synchronizing with $\psome{v}$. Then, the first of the $n \geq 0$ reduction steps towards $Q$ is a synchronization between $R$ and $\gsome{v}{u , \llfv{C}}$.

            We consider two sub-cases, depending on the values of  $m$ and $n$:
            \begin{enumerate}
                \item $m = 0$ and $n \geq 1$:

            Then $R = \piencodfaplas{M}_v$ as $\piencodfaplas{M}_v \redtwo^0 \piencodfaplas{M}_v$.
            Notice that there are two possibilities of having an unguarded:

            \begin{enumerate}
            \item $M =  (\lambda x . (M'[ {\widetilde{x}} \leftarrow  {x}])) \linexsub{C_1 / \widetilde{ y}_1} \cdots \linexsub{C_p / \widetilde{ y}_p} \unexsub{U_1 / \unvar{z}_1} \cdots \unexsub{U_q / \unvar{z}_q}  $ $ (p, q \geq 0)$
            \[
                \small
            \begin{aligned}
            \piencodfaplas{M}_v &= \piencodfaplas{ (\lambda x . (M'[ {\widetilde{x}} \leftarrow  {x}])) \linexsub{C_1 / \widetilde{ y}_1} \cdots \linexsub{C_p / \widetilde{ y}_p} \unexsub{U_1 / \unvar{z}_1} \cdots \unexsub{U_q / \unvar{z}_q}  }_v \\
            &=  \res{\unvar{z}_q} ( \cdots \res{\unvar{z}_1} ( \piencodfaplas{(\lambda x . (M'[ {\widetilde{x}} \leftarrow  {x}])) \linexsub{C_1 / \widetilde{ y}_1} \cdots \linexsub{C_p / \widetilde{ y}_p}}_v  \| \\
            & \qquad \qquad   ~ \guname{\unvar{z}_1}{z_1}; \piencodfaplas{ U_1 }_{z_1}  ) \cdots   \|   ~ \guname{\unvar{z}_q}{z_q}; \piencodfaplas{ U_q }_{z_q} ) \quad = Q
            \\
            \end{aligned}
            \]
            Which we shall write as:
            \[
            \begin{aligned}
            Q =& \res{\unvar{z}_q,  \cdots , \unvar{z}_1} ( \piencodfaplas{(\lambda x . (M'[ {\widetilde{x}} \leftarrow  {x}])) \linexsub{C_1 / \widetilde{ y}_1} \cdots \linexsub{C_p / \widetilde{ y}_p}}_v  \| \\
            &  ~ \guname{\unvar{z}_1}{z_1}; \piencodfaplas{ U_1 }_{z_1}  \cdots   \|   ~ \guname{\unvar{z}_q}{z_q}; \piencodfaplas{ U_q }_{z_q} )
            \end{aligned}
                \]
                for simplicity to represent the process. We also use this to simplify the translation of linear explicit substitutions of bags from:
                \[
                \hspace{-2.5cm}
                \small
                \begin{aligned}
                \piencodfaplas{ M \linexsub{\bag{M_1} \cdot \cdots \cdot \bag{M_k} /  x_1 , \cdots , x_k}  }_v    & =
                    \res{z_1}( \gsome{z_1}{\llfv{M_{1}}};\piencodfaplas{ M_{1} }_{ {z_1}}  \| \cdots \\
                    & \qquad \res{z_k} ( \gsome{z_k}{\llfv{M_{k}}};\piencodfaplas{ M_{k} }_{ {z_k}} \\
                    & \qquad \| \bignd_{x_{i_1} \in \{ x_1 ,\cdots , x_k  \}} \cdots \bignd_{x_{i_k} \in \{ x_1 ,\cdots , x_k \setminus x_{i_1} , \cdots , x_{i_{k-1}}  \}} \\
                    & \qquad \quad  \piencodfaplas{ M }_v \{ z_1 / x_{i_1} \} \cdots \{ z_k / x_{i_k} \} ) \cdots )
                \end{aligned}
                \]
                to be represented as:
                \[
                \begin{aligned}
                \piencodfaplas{ M \linexsub{C /  \widetilde{x}}  }_v    & =
                    \res{\widetilde{z}}( \gsome{\widetilde{z}}{\llfv{C}};\piencodfaplas{ C }_{\widetilde{z} }  \| \bignd_{\widetilde{x}_{i} \in \perm{\widetilde{x}}} \piencodfaplas{ M }_v \{ \widetilde{z} / \widetilde{x}_{i} \}).
                \end{aligned}
                \]
                These representations are purely for simplicity and are not an alternative to the actual translation. We continue expanding the sub-process\\
                $\piencodfaplas{(\lambda x . (M'[ {\widetilde{x}} \leftarrow  {x}])) \linexsub{C_1 / \widetilde{ y}_1} \cdots \linexsub{C_p / \widetilde{ y}_p}}_v$ which we shall denote $P$ using the above shortened and simplified notation:
                \[
            \begin{aligned}
            P & =
                    \res{\widetilde{w}_p}( \gsome{\widetilde{w}_p}{\llfv{C_p}};\piencodfaplas{ C_p }_{\widetilde{w}_p }  \| \\
                    & \qquad  \bignd_{\widetilde{y}_{p_i} \in \perm{\widetilde{y}_p}} \cdots   \res{\widetilde{w}_1}( \gsome{\widetilde{w}_1}{\llfv{C_1}};\piencodfaplas{ C_1 }_{\widetilde{w}_1 }  \| \\
                    & \qquad  \bignd_{\widetilde{y}_{1_i} \in \perm{\widetilde{y}_1}}  \piencodfaplas{ \lambda x . (M'[ {\widetilde{x}} \leftarrow  {x}]) }_v \{ \widetilde{w}_1 / \widetilde{y}_{{1_i}} \}  ) \cdots   \{ \widetilde{w}_p / \widetilde{y}_{{p_i}} \}  )
            \\
            \end{aligned}
            \]
            Hence we represent $\piencodfaplas{M}_v$ as:

            \[
                \small
            \begin{aligned}
            \piencodfaplas{M}_v &= \res{\unvar{z}_q,  \cdots , \unvar{z}_1} ( \res{\widetilde{w}_p}( \gsome{\widetilde{w}_p}{\llfv{C_p}};\piencodfaplas{ C_p }_{\widetilde{w}_p }  \| \\
                    & \qquad \qquad \bignd_{\widetilde{y}_{p_i} \in \perm{\widetilde{y}_p}} \cdots   \res{\widetilde{w}_1}( \gsome{\widetilde{w}_1}{\llfv{C_1}};\piencodfaplas{ C_1 }_{\widetilde{w}_1 }  \| \\
                    & \qquad \qquad \bignd_{\widetilde{y}_{1_i} \in \perm{\widetilde{y}_1}}  \piencodfaplas{ \lambda x . (M'[ {\widetilde{x}} \leftarrow  {x}]) }_v \{ \widetilde{w}_1 / \widetilde{y}_{{1_i}} \}  ) \cdots   \{ \widetilde{w}_p / \widetilde{y}_{{p_i}} \}  ) \\
                    & \qquad \qquad \|   ~ \guname{\unvar{z}_1}{z_1}; \piencodfaplas{ U_1 }_{z_1}  \cdots   \|   ~ \guname{\unvar{z}_q}{z_q}; \piencodfaplas{ U_q }_{z_q} )
            \\
            \end{aligned}
            \]
            Finally we shall simplify the process to become:
            \[
                \piencodfaplas{M}_v  = \res{\widetilde{z} , \widetilde{w} } ( \bignd_{i \in I}  \piencodfaplas{ \lambda x . (M'[ {\widetilde{x}} \leftarrow  {x}]) }_v \{ \widetilde{w} / \widetilde{y}_{i} \}  \|  Q'' )
            \\
            \]
            With this shape for $M$, we then have the following:
            {
            \small
            \[
                \hspace{-3cm}
            \begin{aligned}
            \piencodfaplas{N}_u & = \piencodfaplas{M(C \bagsep U)}_u= \res{v} (\piencodfaplas{M}_v \| \gsome{v}{u , \llfv{C}};\pname{v}{x}; ( \piencodfaplas{C \bagsep U}_x  \| \pfwd{v}{u}  ) )\\
            &= \res{v} (\res{\widetilde{z} , \widetilde{w} } ( \bignd_{i \in I}  \piencodfaplas{ \lambda x . (M'[ {\widetilde{x}} \leftarrow  {x}]) }_v \{ \widetilde{w} / \widetilde{y}_{i} \}  \|  Q'' )   \\
            &\qquad  \|  \gsome{v}{u , \llfv{C}};\pname{v}{x}; ( \piencodfaplas{C \bagsep U}_x  \| \pfwd{v}{u}  ) )\\
            &= \res{v} (\res{\widetilde{z} , \widetilde{w} } ( \bignd_{i \in I}  \psome{v};\gname{v}{x};  \psome{x};\gname{x}{\linvar{x}}; \gname{x}{\unvar{x}};  \gclose{x} ;  \\
            & \qquad \piencodfaplas{M[ {\widetilde{x}} \leftarrow  {x}]}_v \{ \widetilde{w} / \widetilde{y}_{i} \}  \|  Q'' ) \| \gsome{v}{u , \llfv{C}};\pname{v}{x}; ( \piencodfaplas{C \bagsep U}_x  \| \pfwd{v}{u}  ) )\\
            & \redtwo \res{v} (\res{\widetilde{z} , \widetilde{w} } ( \bignd_{i \in I}  \gname{v}{x};  \psome{x};\gname{x}{\linvar{x}}; \gname{x}{\unvar{x}};  \gclose{x} ; \piencodfaplas{M[ {\widetilde{x}} \leftarrow  {x}]}_v \{ \widetilde{w} / \widetilde{y}_{i} \}  \|  Q'' ) & = Q_1 \\
            & \qquad  \| \pname{v}{x}; ( \piencodfaplas{C \bagsep U}_x  \| \pfwd{v}{u}  ) )\\
            & \redtwo \res{v} (   \pfwd{v}{u} \| \res{x}( \res{\widetilde{z} , \widetilde{w} } ( \bignd_{i \in I}  \psome{x};\gname{x}{\linvar{x}}; \gname{x}{\unvar{x}};  \gclose{x} ;  & = Q_2\\
            & \qquad \piencodfaplas{M[ {\widetilde{x}} \leftarrow  {x}]}_v \{ \widetilde{w} / \widetilde{y}_{i} \}  \|  Q'' ) \|  \piencodfaplas{C \bagsep U}_x     ) )\\
            & \redtwo  \res{x}( \res{\widetilde{z} , \widetilde{w} } ( \bignd_{i \in I}  \psome{x};\gname{x}{\linvar{x}}; \gname{x}{\unvar{x}};  \gclose{x} ; & = Q_3\\
            & \qquad \piencodfaplas{M[ {\widetilde{x}} \leftarrow  {x}]}_u \{ \widetilde{w} / \widetilde{y}_{i} \}  \|  Q'' )  \|  \piencodfaplas{C \bagsep U}_x     ) \\
            \end{aligned}
            \]
            }
            We also have that
                \[
                    \small
                \begin{aligned}
                    N &=(\lambda x . (M'[ {\widetilde{x}} \leftarrow  {x}]))  \linexsub{C_1 / \widetilde{ y}_1} \cdots \linexsub{C_p / \widetilde{ y}_p} \unexsub{U_1 / \unvar{z}_1} \cdots \unexsub{U_q / \unvar{z}_q}  (C \bagsep U) \\
                    & \equivlam  (\lambda x . (M'[ {\widetilde{x}} \leftarrow  {x}]) (C \bagsep U))  \linexsub{C_1 / \widetilde{ y}_1} \cdots \linexsub{C_p / \widetilde{ y}_p} \unexsub{U_1 / \unvar{z}_1} \cdots \unexsub{U_q / \unvar{z}_q}  \\
                & \red   M'[ {\widetilde{x}} \leftarrow  {x}] \esubst{(C \bagsep U)}{x}  \linexsub{C_1 / \widetilde{ y}_1} \cdots \linexsub{C_p / \widetilde{ y}_p} \unexsub{U_1 / \unvar{z}_1} \cdots \unexsub{U_q / \unvar{z}_q}  \\
                &= {M}
                \end{aligned}
                \]
            Furthermore, we have:
            \[
                \hspace{-0.5cm}
                \small
            \begin{aligned}
                \piencodfaplas{M}_u &= \piencodfaplas{M'[ {\widetilde{x}} \leftarrow  {x}] \esubst{(C \bagsep U)}{x}  \linexsub{C_1 / \widetilde{ y}_1} \cdots \linexsub{C_p / \widetilde{ y}_p} \unexsub{U_1 / \unvar{z}_1} \cdots \unexsub{U_q / \unvar{z}_q} }_u \\
                & = \res{x}( \res{\widetilde{z} , \widetilde{w} } ( \bignd_{i \in I}  \psome{x};\gname{x}{\linvar{x}}; \gname{x}{\unvar{x}};  \gclose{x} ; \piencodfaplas{M[ {\widetilde{x}} \leftarrow  {x}]}_u \{ \widetilde{w} / \widetilde{y}_{i} \}  \|  Q'' ) \\
                & \qquad \qquad \|  \piencodfaplas{C \bagsep U}_x
            \end{aligned}
            \]

                    We consider different possibilities for $n \geq 1$; in all  the cases, the result follows.
                                    \smallskip

            \noindent  {\bf When $n = 1$:}
                We have $Q = Q_1$, $ \piencodfaplas{N}_u \redtwo^1 Q_1$.
                    We also have that
                    \begin{itemize}
                    \item  $Q_1 \redtwo^2 Q_3 = Q'$ ,
                    \item {\small${N} \red^1 M'[ {\widetilde{x}} \leftarrow  {x}] \esubst{(C \bagsep U)}{x}  \linexsub{C_1 / \widetilde{ y}_1} \cdots \linexsub{C_p / \widetilde{ y}_p} \unexsub{U_1 / \unvar{z}_1} \cdots \unexsub{U_q / \unvar{z}_q}$  $ =~{N}'$}
                    \item and {\small$\piencodfaplas{M'[ {\widetilde{x}} \leftarrow  {x}] \esubst{(C \bagsep U)}{x}  \linexsub{C_1 / \widetilde{ y}_1} \cdots \linexsub{C_p / \widetilde{ y}_p} \unexsub{U_1 / \unvar{z}_1} \cdots \unexsub{U_q / \unvar{z}_q}}_u$  $ =~Q_3$}.
                    \end{itemize}

                                            \smallskip

            \noindent  {\bf When $n = 2$:} the analysis is similar.

            \noindent {\bf When $n \geq 3$:}
            We have $ \piencodfaplas{{N}}_u \redtwo^3 Q_3 \redtwo^l Q$, for $l \geq 0$. We also know that ${N} \red {M}$, $Q_3 = \piencodfaplas{{M}}_u$. By the IH, there exist $ Q' , {N}'$ such that $Q \redtwo^i Q'$, ${M} \red^j {N}'$ and $\piencodfaplas{{N}'}_u = Q'$ . Finally, $\piencodfaplas{{N}}_u \redtwo^3 Q_3 \redtwo^l Q \redtwo^i Q'$ and ${N} \rightarrow {M}  \red^j {N}'$.

            \item $M = \fail^{\widetilde{z}}$.

            Then,                     \(
                        \begin{aligned}
                            \piencodfaplas{M}_v &= \piencodfaplas{\fail^{\widetilde{z}}}_v = \pnone{ v}  \| \pnone{ \widetilde{z}} .
                        \end{aligned}
                    \)
                    With this shape for $M$, we have:
                    \[
                        \hspace{-1cm}
                        \small
                    \begin{aligned}
                        \piencodfaplas{{N}}_u & = \piencodfaplas{(M\ (C \bagsep U))}_u =\res{v} (\piencodfaplas{M}_v \| \gsome{v}{u , \llfv{C}};\pname{v}{x}; ( \piencodfaplas{C \bagsep U}_x  \| \pfwd{v}{u}  ) )
                        \\
                        & =\res{v} (\pnone{ v}  \| \pnone{ \widetilde{z}} \| \gsome{v}{u , \llfv{C}};\pname{v}{x}; ( \piencodfaplas{C \bagsep U}_x  \| \pfwd{v}{u}  ) )
                        \\
                        & \redtwo  \pnone{ \widetilde{z}} \| \pnone{u} \| \pnone{ \llfv{C} }
                        \\
                        \end{aligned}
                    \]

                    \end{enumerate}

                    We also have that
                    \(  {N} = \fail^{\widetilde{z}}\ C \bagsep U \red   \fail^{\widetilde{z} \cup \llfv{C}}  = {M}.  \)
                    Furthermore,
                    \[
                     \begin{aligned}
                          \piencodfaplas{{M}}_u &= \piencodfaplas{ \fail^{\widetilde{z} \cup \llfv{C}  } }_u
                          = \pnone{ \widetilde{z}} \| \pnone{u} \| \pnone{ \llfv{C} }\\
                    \end{aligned}
                    \]

  \item When $m \geq 1$ and $ n \geq 0$, we distinguish two cases:

 \begin{enumerate}
\item When $n = 0$:

Then, $ \res{v} ( R \| \gsome{v}{u , \llfv{C}};\pname{v}{x}; ( \piencodfaplas{C \bagsep U}_x  \| \pfwd{v}{u}  ) )) =  Q $ and $\piencodfaplas{M}_u \redtwo^m R$ where $m \geq 1$. Then by the IH there exist $R'$  and ${M}' $ such that $R \redtwo^i R'$, $M \red^j {M}'$, and $\piencodfaplas{{M}'}_u = R'$.  Hence we have that

            \[
                \begin{aligned}
                   \piencodfaplas{{N}}_u
                   & = \res{v} (\piencodfaplas{M}_v \| \gsome{v}{u , \llfv{C}};\pname{v}{x}; ( \piencodfaplas{C \bagsep U}_x  \| \pfwd{v}{u}  ) )\\
                   & \redtwo^m  \res{v} ( R \| \gsome{v}{u , \llfv{C}};\pname{v}{x}; ( \piencodfaplas{C \bagsep U}_x  \| \pfwd{v}{u}  ) )  = Q
                \end{aligned}
             \]
            We also know that
            \[
            \begin{aligned}
              Q & \redtwo^i  \res{v} ( R \| \gsome{v}{u , \llfv{C}};\pname{v}{x}; ( \piencodfaplas{C \bagsep U}_x  \| \pfwd{v}{u}  ) ) = Q'\\
            \end{aligned}
            \]
            and so the \lamcoldetsh term can reduce as follows: ${N} = (M\ ( C \bagsep U )) \red^j M'\ ( C \bagsep U ) = {N}'$ and  $\piencodfaplas{ {N}'}_u = Q'$.

                        \item When $n \geq 1$:

                            Then  $R$ has an occurrence of an unguarded $ \psome{v} $ or $\pnone{v}$, hence it is of the form

                            $ \piencodfaplas{(\lambda x . (M'[ {\widetilde{x}} \leftarrow  {x}])) \linexsub{C_1 / \widetilde{ y}_1} \cdots \linexsub{C_p / \widetilde{ y}_p} \unexsub{U_1 / \unvar{z}_1} \cdots \unexsub{U_q / \unvar{z}_q}  }_v $ or $ \piencodfaplas{\fail^{\widetilde{x}}}_v. $
              This case follows by IH.
                    \end{enumerate}

            \end{enumerate}

        \end{enumerate}

        This concludes the analysis for the case ${N} = (M \, ( C \bagsep U ))$.

        \item ${N} = M[ {\widetilde{x}} \leftarrow  {x}]$.

    The sharing variable $ {x}$ is not free and the result follows by vacuity.

        \item ${N} = M[ {\widetilde{x}} \leftarrow  {x}] \esubst{ C \bagsep U }{ x}$. Then we have
            \[
                \small
                \begin{aligned}
                    \piencodfaplas{{N}}_u &=\piencodfaplas{ M[ {\widetilde{x}} \leftarrow  {x}] \esubst{ C \bagsep U }{ x} }_u= \res{x}( \psome{x}; \gname{x}{\linvar{x}}; \gname{x}{\unvar{x}};  \gclose{x} ;\piencodfaplas{ M[ {\widetilde{x}} \leftarrow  {x}]}_u \| \piencodfaplas{ C \bagsep U}_x )
                \end{aligned}
            \]
            Let us consider three cases.

            \begin{enumerate}
                \item When $\size{ {\widetilde{x}}} = \size{C}$.
                    Then let us consider the shape of the bag $ C$.

  \begin{enumerate}
  \item When $C = \oneb$.

                              We have the following
                             \[
                             \hspace{-2.0cm}
                             \small
                             \begin{aligned}
                             \piencodfaplas{{N}}_u
                             &= \res{x}( \psome{x}; \gname{x}{\linvar{x}}; \gname{x}{\unvar{x}};  \gclose{x} ;\piencodfaplas{ M[  \leftarrow  {x}]}_u \| \piencodfaplas{ \oneb \bagsep U}_x )\\
                             &= \res{x}( \psome{x}; \gname{x}{\linvar{x}}; \gname{x}{\unvar{x}};  \gclose{x} ;\piencodfaplas{ M[  \leftarrow  {x}]}_u \| \\
                             & \qquad \gsome{x}{\llfv{C}};  \pname{x}{\linvar{x}}; \big( \piencodfaplas{ \oneb }_{\linvar{x}} \|  \pname{x}{\unvar{x}}; ( \guname{\unvar{x}}{x_i}; \piencodfaplas{ U }_{x_i} \| \pclose{x} ) \big) )\\
                             & \redtwo \res{x}(  \gname{x}{\linvar{x}}; \gname{x}{\unvar{x}};  \gclose{x} ;\piencodfaplas{ M[  \leftarrow  {x}]}_u \| \pname{x}{\linvar{x}}; \big( \piencodfaplas{ \oneb }_{\linvar{x}} \|  \pname{x}{\unvar{x}}; ( \guname{\unvar{x}}{x_i}; \piencodfaplas{ U }_{x_i} \| \pclose{x} ) \big) )  & = Q_1  \\
                             & \redtwo \res{x}(  \pname{x}{\unvar{x}}; ( \guname{\unvar{x}}{x_i}; \piencodfaplas{ U }_{x_i} \| \pclose{x} ) \|  \res{\linvar{x}} (  \gname{x}{\unvar{x}};  \gclose{x} ;\piencodfaplas{ M[  \leftarrow  {x}]}_u \| \piencodfaplas{ \oneb }_{\linvar{x}}   ))  & = Q_2  \\
                             & \redtwo \res{x}( \pclose{x} \|  \res{\unvar{x}} ( \guname{\unvar{x}}{x_i}; \piencodfaplas{ U }_{x_i}  \|  \res{\linvar{x}} (   \gclose{x} ;\piencodfaplas{ M[  \leftarrow  {x}]}_u \| \piencodfaplas{ \oneb }_{\linvar{x}}   )))  & = Q_3  \\
                             & \redtwo  \res{\unvar{x}} ( \guname{\unvar{x}}{x_i}; \piencodfaplas{ U }_{x_i}  \|  \res{\linvar{x}} (  \piencodfaplas{ M[  \leftarrow  {x}]}_u \| \piencodfaplas{ \oneb }_{\linvar{x}}   )))  & = Q_4  \\
                             & =  \res{\unvar{x}} ( \guname{\unvar{x}}{x_i}; \piencodfaplas{ U }_{x_i}  \|  \res{\linvar{x}} (  \psome{\linvar{x}}; \pname{\linvar{x}}{y_i}; ( \gsome{y_i}{ u,\llfv{M} }; \gclose{ y_{i} } ;\piencodfaplas{M}_u \| \| \\
                             & \qquad  \pnone{ \linvar{x} } )  \gsome{\linvar{x}}{\emptyset};\gname{x}{y_i};  ( \psome{ y_i}; \pclose{y_i}  \| \gsome{\linvar{x}}{\emptyset}; \pnone{\linvar{x}} )   )))  \\
                             & \redtwo  \res{\unvar{x}} ( \guname{\unvar{x}}{x_i}; \piencodfaplas{ U }_{x_i}  \|
                             \res{\linvar{x}} ( \pname{\linvar{x}}{y_i}; ( \gsome{y_i}{ u,\llfv{M} }; \gclose{ y_{i} } ;\piencodfaplas{M}_u \| \pnone{ \linvar{x} } ) \| \\
                             & \qquad   \gname{x}{y_i};  ( \psome{ y_i}; \pclose{y_i}  \| \gsome{\linvar{x}}{\emptyset}; \pnone{\linvar{x}} )   )))   & = Q_5 \\
                             & \redtwo  \res{\unvar{x}} ( \guname{\unvar{x}}{x_i}; \piencodfaplas{ U }_{x_i}  \|
                             \res{\linvar{x}} ( \pnone{ \linvar{x} } \| \res{y_i} ( \gsome{y_i}{ u,\llfv{M} }; \gclose{ y_{i} } ;\piencodfaplas{M}_u   \| \\
                             & \qquad  \psome{ y_i}; \pclose{y_i}  \| \gsome{\linvar{x}}{\emptyset}; \pnone{\linvar{x}} )   ))   & = Q_6 \\
                             & \redtwo  \res{\unvar{x}} ( \guname{\unvar{x}}{x_i}; \piencodfaplas{ U }_{x_i}  \|
                              \res{y_i} ( \gsome{y_i}{ u,\llfv{M} }; \gclose{ y_{i} } ;\piencodfaplas{M}_u   \|   \psome{ y_i}; \pclose{y_i}  ) )   & = Q_7 \\
                              & \redtwo  \res{\unvar{x}} ( \guname{\unvar{x}}{x_i}; \piencodfaplas{ U }_{x_i}  \|
                              \res{y_i} ( \ \gclose{ y_{i} } ;\piencodfaplas{M}_u   \|    \pclose{y_i}  ) )   & = Q_8 \\
                              & \redtwo  \res{\unvar{x}} ( \guname{\unvar{x}}{x_i}; \piencodfaplas{ U }_{x_i}  \|
                              \piencodfaplas{M}_u   ) =  \piencodfaplas{M \unexsub{U / \unvar{x}}}_u   & = Q_9 \\
                            \end{aligned}
                            \]
                            Notice how $Q_6$ has a choice however the $\linvar{x}$ name can be closed at any time so for simplicity we perform communication across this name first followed by all other comunications that can take place.

                        Now we proceed by induction on the number of reductions $\piencodfaplas{{N}}_u \redtwo^k Q$.

                            \begin{enumerate}

                                \item When $k = 0$, the result follows trivially. Just take ${N}={N}'$ and $\piencodfaplas{{N}}_u=Q=Q'$.

                                \item When $k = 1$.

                                    We have $Q = Q_1$, $ \piencodfaplas{{N}}_u \redtwo^1 Q_1$
                                    We also have that $Q_1 \redtwo^8 Q_9 = Q'$ , ${N} \red M \unexsub{U / \unvar{x}} = N'$ and $\piencodfaplas{ N' }_u = Q_9$

                                \item When $2 \leq  k \leq 8$.

                                    Proceeds similarly to the previous case

                                \item When $k \geq 9$.

                                We have $ \piencodfaplas{{N}}_u \redtwo^9 Q_9 \redtwo^l Q$, for $l \geq 0$. Since $Q_9 = \piencodfaplas{ M \unexsub{U / \unvar{x}} }_u$ we apply the induction hypothesis we have that  there exist $ Q' , {N}' \ s.t. \ Q \redtwo^i Q' ,  M \unexsub{U / \unvar{x}}  \red^j {N}'$ and $\piencodfaplas{{N}'}_u = Q'$.                                    Then,  $ \piencodfaplas{{N}}_u \redtwo^5 Q_5 \redtwo^l Q \redtwo^i Q'$ and by the contextual reduction rule it follows that ${N} = (M[ \leftarrow x])\esubst{ 1 }{ x } \red  M \unexsub{U / \unvar{x}} \red^j  {N}' $ and the case holds.

\end{enumerate}

\item When $C = \bag{N_1} \cdot \cdots \cdot \bag{N_l}$, for $l \geq 1$.
                    Then,

 \[
    \hspace{-1.6cm}
    \small
   \begin{aligned}
   \piencodfaplas{{N}}_u &=\piencodfaplas{ M[ {\widetilde{x}} \leftarrow  {x}] \esubst{ C \bagsep U }{x} }_u\\
   & =  \res{x}( \psome{x}; \gname{x}{\linvar{x}}; \gname{x}{\unvar{x}};  \gclose{x} ;\piencodfaplas{ M[ {\widetilde{x}} \leftarrow  {x}]}_u \| \piencodfaplas{ C \bagsep U}_x )
    \\
   & \redtwo^4  \res{\unvar{x}} ( \guname{\unvar{x}}{x_i}; \piencodfaplas{ U }_{x_i}  \|  \res{\linvar{x}} (  \piencodfaplas{ M[ {\widetilde{x}} \leftarrow  {x}]}_u \| \piencodfaplas{ C }_{\linvar{x}}   )))   \\
   & =  \res{\unvar{x}} ( \guname{\unvar{x}}{x_i}; \piencodfaplas{ U }_{x_i}  \|  \res{\linvar{x}} (
    \psome{\linvar{x}}; \pname{\linvar{x}}{y_1}; \big( \gsome{y_1}{ \emptyset }; \gclose{ y_{1} } ; \0   \\
    & \qquad  \|\psome{\linvar{x}}; \gsome{\linvar{x}}{u, \llfv{M} \setminus  \widetilde{x} }; \bignd_{x_{i_1} \in \widetilde{x}} \gname{x}{{x}_{i_1}};
    \cdots
    \psome{\linvar{x}}; \pname{\linvar{x}}{y_l}; \big( \gsome{y_l}{ \emptyset }; \gclose{ y_{l} } ; \0   \\
    & \qquad \|\psome{\linvar{x}}; \gsome{\linvar{x}}{u, \llfv{M} \setminus  (\widetilde{x} \setminus x_1, \cdots , x_{i_{l-1}} )}; \bignd_{x_{i_l} \in (\widetilde{x} \setminus x_1, \cdots , x_{i_{l-1}})} \gname{x}{{x}_{i_l}};\piencodfaplas{M[  \leftarrow  {x}]}_u \big)
    \cdots
    \big)
    \\
    & \qquad  \| \gsome{\linvar{x}}{\llfv{C} }; \gname{x}{y_1}; \gsome{\linvar{x}}{y_1, \llfv{C}}; \psome{\linvar{x}}; \pname{\linvar{x}}{z_1}; \\
    & \qquad  ( \gsome{z_1}{\llfv{M_1}};  \piencodfaplas{M_1}_{z_1} \| \pnone{y_1} \|
    \cdots \\
    & \qquad  \gsome{\linvar{x}}{\llfv{M_l} }; \gname{x}{y_l}; \gsome{\linvar{x}}{y_l, \llfv{M_l}}; \psome{\linvar{x}}; \pname{\linvar{x}}{z_l}; \\
    & \qquad  ( \gsome{z_l}{\llfv{M_l}};  \piencodfaplas{M_l}_{z_l} \| \pnone{y_l} \| \piencodfaplas{\oneb}_{\linvar{x}}  ) \cdots  )     ))   \\
\end{aligned}
\]
We shall now perform multiple non-committing reductions at once. Notice that non-determinism guards the same prefixes denying the use of the reduction $\rredtwo{\piprecong{x}}$ hence denying the commitment of non-determinism.

\[
\hspace{-2cm}
    \small
   \begin{aligned}
    & \redtwo^{6l}  \res{\unvar{x}} ( \guname{\unvar{x}}{x_i}; \piencodfaplas{ U }_{x_i}  \|  \res{\linvar{x}} ( \piencodfaplas{\oneb}_{\linvar{x}} \|
    \\
    & \qquad \qquad \res{z_1}  (  \gsome{z_1}{\llfv{M_1}};  \piencodfaplas{M_1}_{z_1} \| \cdots
    \res{z_l} ( \gsome{z_l}{\llfv{M_l}};  \piencodfaplas{M_l}_{z_l} \| \\
    & \qquad \qquad  \bignd_{x_{i_1} \in \widetilde{x}} \bignd_{x_{i_l} \in (\widetilde{x} \setminus x_1, \cdots , x_{i_{l-1}})} \piencodfaplas{M[  \leftarrow  {x}]}_u \{ z_1 / x_{i_1} \} \cdots \{ z_l / x_{i_l} \} )
    \cdots    )     ))   \\
    & =  \res{\unvar{x}} ( \guname{\unvar{x}}{x_i}; \piencodfaplas{ U }_{x_i}  \|  \res{\linvar{x}} ( \gsome{\linvar{x}}{\emptyset};\gname{x}{y_{l+1}};  ( \psome{ y_{l+1}}; \pclose{y_{l+1}}  \| \gsome{\linvar{x}}{\emptyset}; \pnone{\linvar{x}} ) \|
    \\
    & \qquad \qquad \res{z_1}  (  \gsome{z_1}{\llfv{M_1}};  \piencodfaplas{M_1}_{z_1} \| \cdots
    \res{z_l} ( \gsome{z_l}{\llfv{M_l}};  \piencodfaplas{M_l}_{z_l} \| \\
    & \qquad \qquad  \bignd_{x_{i_1} \in \widetilde{x}} \bignd_{x_{i_l} \in (\widetilde{x} \setminus x_1, \cdots , x_{i_{l-1}})} \psome{\linvar{x}}; \\
    & \qquad \qquad \pname{\linvar{x}}{y_{l+1}}; ( \gsome{y_{l+1}}{ u,\llfv{M} }; \gclose{ y_{{l+1}} } ;\piencodfaplas{M}_u \{ z_1 / x_{i_1} \} \cdots \{ z_l / x_{i_l} \} \| \pnone{ \linvar{x} } )  )
    \cdots    )     ))   \\
    & \redtwo^{5}  \res{\unvar{x}} ( \guname{\unvar{x}}{x_i}; \piencodfaplas{ U }_{x_i}  \|
    \\
    & \qquad \qquad \res{z_1}  (  \gsome{z_1}{\llfv{M_1}};  \piencodfaplas{M_1}_{z_1} \| \cdots
    \res{z_l} ( \gsome{z_l}{\llfv{M_l}};  \piencodfaplas{M_l}_{z_l} \| \\
    & \qquad \qquad  \bignd_{x_{i_1} \in \widetilde{x}} \bignd_{x_{i_l} \in (\widetilde{x} \setminus x_1, \cdots , x_{i_{l-1}})} \piencodfaplas{M}_u  \{ z_1 / x_{i_1} \} \cdots \{ z_l / x_{i_l} \} )  \cdots    )     )   \\
   & = \piencodfaplas{ M\linexsub{\bag{M_1} \cdot \cdots \cdots \bag{M_l}/ {x_1}, \cdots , x_1}  \unexsub{U /\unvar{x} }}_{u}= Q_{6l + 9}\\
  \end{aligned}
 \]

    The proof follows by induction on the number of reductions $\piencodfaplas{{N}}_u \redtwo^k Q$.

\begin{enumerate}
\item When $k = 0$, the result follows trivially. Just take ${N}={N}'$ and $\piencodfaplas{{N}}_u=Q=Q'$.

 \item When $1 \leq k \leq 6l + 9$.

 Let $Q_k$ such that $ \piencodfaplas{{N}}_u \redtwo^k Q_k$.
    We also have that $Q_k \redtwo^{6l + 9 - k} Q_{6l + 9} = Q'$ ,

    ${N} \red^1 M\linexsub{\bag{M_1} \cdot \cdots \cdots \bag{M_l}/ {x_1}, \cdots , x_1}  \unexsub{U /\unvar{x} } = {N}'$ and

    $\piencodfaplas{M\linexsub{\bag{M_1} \cdot \cdots \cdots \bag{M_l}/ {x_1}, \cdots , x_1}  \unexsub{U /\unvar{x} }}_u = Q_{6l + 9}$.

\item When $k > 6l + 9$.

Then,  $ \piencodfaplas{{N}}_u \redtwo^{6l + 9} Q_{6l + 9} \redtwo^n Q$ for $n \geq 1$. Also,

\(
\begin{aligned}
&  {N} \red^1 M\linexsub{\bag{M_1} \cdot \cdots \cdots \bag{M_l}/ {x_1}, \cdots , x_1}  \unexsub{U /\unvar{x} } \text { and } \\
&    Q_{6l + 9} = \piencodfaplas{M\linexsub{\bag{M_1} \cdot \cdots \cdots \bag{M_l}/ {x_1}, \cdots , x_1}  \unexsub{U /\unvar{x} }}_u.
    \end{aligned}
\)

By the induction hypothesis, there exist $ Q'$ and ${N}'$ such that
$\ Q \redtwo^i Q'$,

$M\linexsub{\bag{M_1} \cdot \cdots \cdots \bag{M_l}/ {x_1}, \cdots , x_1}  \unexsub{U /\unvar{x} } \red^j {N}'$ and $\piencodfaplas{{N}'}_u = Q'$.

Finally, $\piencodfaplas{{N}}_u \redtwo^{6l + 9} Q_{6l + 9} \redtwo^n Q \redtwo^i Q'$ and $$ {N} \red M\linexsub{\bag{M_1} \cdot \cdots \cdots \bag{M_l}/ {x_1}, \cdots , x_1}  \unexsub{U /\unvar{x} }  \red^j {N}'. $$

                            \end{enumerate}

                    \end{enumerate}

                \item When $\size{\widetilde{x}} > \size{C}$.

                    Then we have
                    ${N} = M[ {x}_1, \cdots ,  {x}_k \leftarrow  {x}]\ \esubst{ C \bagsep U }{x}$ with $C = \bag{N_1}  \cdots  \bag{N_l} \quad k > l$. ${N} \red   \fail^{\widetilde{z}} = {M}'$ and $ \widetilde{z} =  (\llfv{M} \setminus \{   {x}_1, \cdots ,  {x}_k \} ) \cup \llfv{C} $. On the one hand, we have:
                    Hence $k = l + m$ for some $m \geq 1$

        \begin{equation}\label{ch4eq:compl_sh_faillin2}
            \hspace{-1cm}
            \small
        \begin{aligned}
            \piencodfaplas{N}_u
            &= \piencodfaplas{M[ {x}_1, \!\cdots\! ,  {x}_k \leftarrow  {x}]\ \esubst{C \bagsep U}{ x } }_u\\
            &= \res{x}( \psome{x}; \gname{x}{\linvar{x}}; \gname{x}{\unvar{x}};  \gclose{x} ;\piencodfaplas{ M[ {\widetilde{x}} \leftarrow  {x}]}_u \| \piencodfaplas{ C \bagsep U}_x )\\
            &= \res{x}( \psome{x}; \gname{x}{\linvar{x}}; \gname{x}{\unvar{x}};  \gclose{x} ;\piencodfaplas{ M[ {\widetilde{x}} \leftarrow  {x}]}_u \| \\
            & \qquad \qquad \gsome{x}{\llfv{C}};  \pname{x}{\linvar{x}}; \big( \piencodfaplas{ C }_{\linvar{x}} \|  \pname{x}{\unvar{x}}; ( \guname{\unvar{x}}{x_i}; \piencodfaplas{ U }_{x_i} \| \pclose{x} ) \big) )\\
            & \redtwo^4  \res{\unvar{x}} ( \res{\linvar{x} } \big( \piencodfaplas{ M[ {\widetilde{x}} \leftarrow  {x}]}_u \|  \piencodfaplas{ C }_{\linvar{x}}   \big) \|  \guname{\unvar{x}}{x_i}; \piencodfaplas{ U }_{x_i}  )  \\
            & = \res{\unvar{x}} ( \res{\linvar{x} } \big( \psome{\linvar{x}}; \pname{\linvar{x}}{y_1}; \big( \gsome{y_1}{ \emptyset }; \gclose{ y_{1} } ; \0   \\
            & \qquad \qquad \|\psome{\linvar{x}}; \gsome{\linvar{x}}{u, \llfv{M} \setminus  \widetilde{x} }; \bignd_{x_{i_1} \in \widetilde{x}} \gname{x}{{x}_{i_1}}; \cdots \\
            & \qquad \qquad \psome{\linvar{x}}; \pname{\linvar{x}}{y_k}; \big( \gsome{y_k}{ \emptyset }; \gclose{ y_{k} } ; \0 \|\psome{\linvar{x}};  \\
            & \qquad \qquad  \gsome{\linvar{x}}{u, \llfv{M} \setminus  (\widetilde{x} \setminus x_{i_1} , \cdots , x_{i_{k-1}}   )}; \bignd_{x_{i_k} \in (\widetilde{x} \setminus x_{i_1} , \cdots , x_{i_{k-1}}  )} \gname{x}{{x}_{i_k}};\piencodfaplas{M[  \leftarrow  {x}]}_u \big)
            \big) \| \\
            & \qquad \qquad  \gsome{\linvar{x}}{\llfv{C} }; \gname{x}{y_1}; \gsome{\linvar{x}}{y_1, \llfv{C}}; \psome{\linvar{x}}; \pname{\linvar{x}}{z_1}; \\
            & \qquad \qquad  ( \gsome{z_1}{\llfv{M_1}};  \piencodfaplas{M_1}_{z_1} \| \pnone{y_1}  \| \cdots  \gsome{\linvar{x}}{\llfv{C} }; \gname{x}{y_l}; \gsome{\linvar{x}}{y_l, \llfv{M_l}};  \\
            & \qquad \qquad  \psome{\linvar{x}}; \pname{\linvar{x}}{z_l}; ( \gsome{z_l}{\llfv{M_l}};  \piencodfaplas{M_l}_{z_l} \| \piencodfaplas{ \oneb }_{\linvar{x}} \| \pnone{y_l} ) \cdots   )   \big) \|  \guname{\unvar{x}}{x_i}; \piencodfaplas{ U }_{x_i}  ) \\
            & (:= Q_4) \\
            \end{aligned}
            \end{equation}

            we reduce $Q_4$ arbitrarily synchronising along channels $\linvar{x} , y_1, \cdots y_l$.

            \begin{equation*}
                \small
            \begin{aligned}
                Q_4
                & \redtwo^{6l} \res{\unvar{x}} ( \res{\linvar{x} } \big( \piencodfaplas{ \oneb }_{\linvar{x}} \|
                \res{z_1} (\gsome{z_1}{\llfv{M_1}};  \piencodfaplas{M_1}_{z_1}   \| \cdots
                \res{z_l} (\gsome{z_l}{\llfv{M_l}};   \\
                &  \qquad  \piencodfaplas{M_l}_{z_l} \|\bignd_{x_{i_1} \in \widetilde{x}} \cdots \bignd_{x_{i_l} \in (\widetilde{x} \setminus x_{i_1} , \cdots , x_{i_{l-1}} )}  \psome{\linvar{x}}; \pname{\linvar{x}}{y_{l+1}}; \big( \gsome{y_{l+1}}{ \emptyset }; \gclose{ y_{{l+1}} } ; \0   \\
                & \qquad  \|\psome{\linvar{x}}; \gsome{\linvar{x}}{u, \llfv{M} \setminus  (\widetilde{x} \setminus x_{i_1} , \cdots , x_{i_{l}}  )}; \bignd_{x_{i_{l+1}} \in (\widetilde{x} \setminus x_{i_1} , \cdots , x_{i_{l}} )} \gname{x}{{x}_{i_{l+1}}}; \cdots   \\
                & \qquad  \psome{\linvar{x}}; \pname{\linvar{x}}{y_k}; \big( \gsome{y_k}{ \emptyset }; \gclose{ y_{k} } ; \0   \\
                & \qquad  \|\psome{\linvar{x}}; \gsome{\linvar{x}}{u, \llfv{M} \setminus  (\widetilde{x} \setminus x_{i_1} , \cdots , x_{i_{k-1}}   )}; \bignd_{x_{i_k} \in (\widetilde{x} \setminus x_{i_1} , \cdots , x_{i_{k-1}}  )} \gname{x}{{x}_{i_k}};\\
                & \qquad  \piencodfaplas{M[  \leftarrow  {x}]}_u \{ z_1 / x_{i_1} \} \cdots \{ z_l / {x}_{i_l} \}
                \big) \cdots \big)  ) \cdots ) \big) \| \guname{\unvar{x}}{x_i}; \piencodfaplas{ U }_{x_i}  )\\
                & = \res{\unvar{x}} ( \res{\linvar{x} } \big(
                 \gsome{\linvar{x}}{\emptyset}; \gname{x}{y_{l+1}};  ( \psome{ y_{l+1}}; \pclose{y_{l+1}}  \| \gsome{\linvar{x}}{\emptyset}; \pnone{\linvar{x}} ) \\
                & \qquad  \|
                \res{z_1} (\gsome{z_1}{\llfv{M_1}};  \piencodfaplas{M_1}_{z_1}   \| \cdots
                \res{z_l} (\gsome{z_l}{\llfv{M_l}};  \piencodfaplas{M_l}_{z_l} \| \\
                &  \qquad  \bignd_{x_{i_1} \in \widetilde{x}} \cdots \bignd_{x_{i_l} \in (\widetilde{x} \setminus x_{i_1} , \cdots , x_{i_{l-1}} )}\psome{\linvar{x}}; \pname{\linvar{x}}{y_{l+1}}; \big( \gsome{y_{l+1}}{ \emptyset }; \gclose{ y_{{l+1}} } ; \0   \\
                & \qquad  \|\psome{\linvar{x}}; \gsome{\linvar{x}}{u, \llfv{M} \setminus  (\widetilde{x} \setminus x_{i_1} , \cdots , x_{i_{l}}  )}; \bignd_{x_{i_{l+1}} \in (\widetilde{x} \setminus x_{i_1} , \cdots , x_{i_{l}} )} \gname{x}{{x}_{i_{l+1}}}; \cdots   \\
                & \qquad \psome{\linvar{x}}; \pname{\linvar{x}}{y_k}; \big( \gsome{y_k}{ \emptyset }; \gclose{ y_{k} } ; \0   \\
                & \qquad  \|\psome{\linvar{x}}; \gsome{\linvar{x}}{u, \llfv{M} \setminus  (\widetilde{x} \setminus x_{i_1} , \cdots , x_{i_{k-1}}   )}; \bignd_{x_{i_k} \in (\widetilde{x} \setminus x_{i_1} , \cdots , x_{i_{k-1}}  )} \gname{x}{{x}_{i_k}};\\
                & \qquad \piencodfaplas{M[  \leftarrow  {x}]}_u \{ z_1 / x_{i_1} \} \cdots \{ z_l / {x}_{i_l} \}
                \big) \cdots \big)  ) \cdots ) \big) \| \guname{\unvar{x}}{x_i}; \piencodfaplas{ U }_{x_i}  )\\
                & \redtwo^5 \res{\unvar{x}} ( \res{\linvar{x} } \big( \pnone{\linvar{x}}  \|
                \res{z_1} (\gsome{z_1}{\llfv{M_1}};  \piencodfaplas{M_1}_{z_1}   \| \cdots
                \res{z_l} (\gsome{z_l}{\llfv{M_l}};   \\
                &  \qquad \piencodfaplas{M_l}_{z_l} \|\bignd_{x_{i_1} \in \widetilde{x}} \cdots \bignd_{x_{i_l} \in (\widetilde{x} \setminus x_{i_1} , \cdots , x_{i_{l-1}} )} \gsome{\linvar{x}}{u, \llfv{M} \setminus  (\widetilde{x} \setminus x_{i_1} , \cdots , x_{i_{l}}  )};   \\
                & \qquad \bignd_{x_{i_{l+1}} \in (\widetilde{x} \setminus x_{i_1} , \cdots , x_{i_{l}} )}  \gname{x}{{x}_{i_{l+1}}}; \cdots \psome{\linvar{x}}; \pname{\linvar{x}}{y_k}; \big( \gsome{y_k}{ \emptyset }; \gclose{ y_{k} } ; \0   \\
                & \qquad \|\psome{\linvar{x}}; \gsome{\linvar{x}}{u, \llfv{M} \setminus  (\widetilde{x} \setminus x_{i_1} , \cdots , x_{i_{k-1}}   )}; \bignd_{x_{i_k} \in (\widetilde{x} \setminus x_{i_1} , \cdots , x_{i_{k-1}}  )} \gname{x}{{x}_{i_k}};\\
                & \qquad \piencodfaplas{M[  \leftarrow  {x}]}_u \{ z_1 / x_{i_1} \} \cdots \{ z_l / {x}_{i_l} \}
                \big) \cdots \big)  ) \cdots ) \big) \| \guname{\unvar{x}}{x_i}; \piencodfaplas{ U }_{x_i}  )\\
                & \redtwo \res{\unvar{x}} (
                \res{z_1} (\gsome{z_1}{\llfv{M_1}};  \piencodfaplas{M_1}_{z_1}   \| \cdots
                \res{z_l} (\gsome{z_l}{\llfv{M_l}};  \piencodfaplas{M_l}_{z_l} \| \\
                &   \qquad \bignd_{x_{i_1} \in \widetilde{x}} \cdots \bignd_{x_{i_l} \in (\widetilde{x} \setminus x_{i_1} , \cdots , x_{i_{l-1}} )} \pnone{u } \| \pnone{(\llfv{M} \setminus  \widetilde{x}) } \| \\
                &  \qquad  \pnone{(z_1 , \cdots ,z_l ) }
                ) \cdots ) \|  \guname{\unvar{x}}{x_i}; \piencodfaplas{ U }_{x_i}  )\\
                & \redtwo^l \res{\unvar{x}} (
                 \bignd_{x_{i_1} \in \widetilde{x}} \cdots \bignd_{x_{i_l} \in (\widetilde{x} \setminus x_{i_1} , \cdots , x_{i_{l-1}} )} \pnone{u } \| \pnone{(\llfv{M} \setminus  \widetilde{x}) }  \| \pnone{ \llfv{C}  } \|\\
                & \qquad  \guname{\unvar{x}}{x_i}; \piencodfaplas{ U }_{x_i}  )\\
                & \equiv \pnone{u } \| \pnone{(\llfv{M} \setminus  \widetilde{x}) } \| \pnone{ \llfv{C}  }
                = \piencodfaplas{ \fail^{\widetilde{z}}}_u = Q_{7l + 10}  \\
        \end{aligned}
        \end{equation*}

The rest of the proof is by induction on the number of reductions $\piencodfaplas{{N}}_u \redtwo^j Q$.

                            \begin{enumerate}
                                \item When $j = 0$, the result follows trivially. Just take ${N}={N}'$ and $\piencodfaplas{{N}}_u=Q=Q'$.
  \item When $1 \leq j \leq 7l + 10$.

Let $Q_j$ be such that $ \piencodfaplas{{N}}_u \redtwo^j Q_j$.
By the steps above one has

\(\begin{aligned}
  &Q_j \redtwo^{7l + 10 - j} Q_{7l + 6} = Q',\\ &{N} \red^1   \fail^{\widetilde{z}} = {N}';\text{ and} \piencodfaplas{ \fail^{\widetilde{z}}}_u = Q_{7l + 10}.
\end{aligned}
\)
\item When $j > 7l + 10$.

In this case, we have
$ \piencodfaplas{{N}}_u \redtwo^{7l + 10} Q_{7l + 10} \redtwo^n Q,$ for $n \geq 1$.
We also know that
${N} \redtwo^1  \fail^{\widetilde{z}}$. However no further reductions can be performed.

                            \end{enumerate}

                \item When $\size{\widetilde{x}} < \size{C}$, the proof  proceeds similarly to the previous case.

            \end{enumerate}

        \item  ${N} =  M \linexsub{C /  \widetilde{x}}$.

           In this case let us consider $ C = \bag{M_1} \cdot \cdots \cdot \bag{M_k} $,
            \[
                \small
            \begin{aligned}
                \piencodfaplas{ M \linexsub{C /  \widetilde{x}}}_u &=  \res{z_1}( \gsome{z_1}{\llfv{M_{1}}};\piencodfaplas{ M_{1} }_{ {z_1}}  \|  \cdots \res{z_k} ( \gsome{z_k}{\llfv{M_{k}}};\piencodfaplas{ M_{k} }_{ {z_k}} \\
                & \qquad  \| \bignd_{x_{i_1} \in \{ x_1 ,\cdots , x_k  \}} \cdots \bignd_{x_{i_k} \in \{ x_1 ,\cdots , x_k \setminus x_{i_1} , \cdots , x_{i_{k-1}}  \}} \piencodfaplas{ M }_u \{ z_1 / x_{i_1} \} \cdots \{ z_k / x_{i_k} \} ) \cdots )
            \end{aligned}
            \]
            Therefore,
            \[
                \small
            \begin{aligned}
               \piencodfaplas{{N}}_u
                &=  \res{z_1}( \gsome{z_1}{\llfv{M_{1}}};\piencodfaplas{ M_{1} }_{ {z_1}}  \|  \cdots \res{z_k} ( \gsome{z_k}{\llfv{M_{k}}};\piencodfaplas{ M_{k} }_{ {z_k}} \\
                & \qquad  \| \bignd_{x_{i_1} \in \{ x_1 ,\cdots , x_k  \}} \cdots \bignd_{x_{i_k} \in \{ x_1 ,\cdots , x_k \setminus x_{i_1} , \cdots , x_{i_{k-1}}  \}} \piencodfaplas{ M }_u \{ z_1 / x_{i_1} \} \cdots \{ z_k / x_{i_k} \} ) \cdots )\\
                & \redtwo^m  \res{z_1}( \gsome{z_1}{\llfv{M_{1}}};\piencodfaplas{ M_{1} }_{ {z_1}}  \|  \cdots \res{z_k} ( \gsome{z_k}{\llfv{M_{k}}};\piencodfaplas{ M_{k} }_{ {z_k}} \\
                & \qquad   \| \bignd_{x_{i_1} \in \{ x_1 ,\cdots , x_k  \}} \cdots \bignd_{x_{i_k} \in \{ x_1 ,\cdots , x_k \setminus x_{i_1} , \cdots , x_{i_{k-1}}  \}} R \{ z_1 / x_{i_1} \} \cdots \{ z_k / x_{i_k} \} ) \cdots )\\
                &  \redtwo^n  Q,\\
            \end{aligned}
            \]

            for some process $R$. Where $\redtwo^n$ is a reduction that  initially synchronizes with $ \gsome{z_i}{\llfv{M_{i}}}$ for some $i \leq k$ when $n \geq 1$, $n + m = k \geq 1$. Type preservation in \clpi ensures reducing $\piencodfaplas{ M}_v \redtwo^m$ does not consume possible synchronizations with $\psome{z_i} $, if they occur. Let us consider the the possible sizes of both $m$ and $n$.

            \begin{enumerate}
                \item For $m = 0$ and $n \geq 1$.

                    We have that $R = \piencodfaplas{M}_u$ as $\piencodfaplas{M}_u \redtwo^0 \piencodfaplas{M}_u$.

                    Notice that there are two possibilities of having an unguarded $\psome{z_i}$ or $\pnone{z_i}$ without internal reductions:

                    \begin{enumerate}
                        \item $M = \fail^{ \widetilde{x}, \widetilde{y}}$.
  \[
    \small
    \hspace{-2cm}
  \begin{aligned}
  \piencodfaplas{ {N}}_u
    & = \res{z_1}( \gsome{z_1}{\llfv{M_{1}}};\piencodfaplas{ M_{1} }_{ {z_1}}  \|  \cdots \res{z_k} ( \gsome{z_k}{\llfv{M_{k}}};\piencodfaplas{ M_{k} }_{ {z_k}} \\
        &  \qquad  \| \bignd_{x_{i_1} \in \{ x_1 ,\cdots , x_k  \}} \cdots \bignd_{x_{i_k} \in \{ x_1 ,\cdots , x_k \setminus x_{i_1} , \cdots , x_{i_{k-1}}  \}} \piencodfaplas{ M }_u \{ z_1 / x_{i_1} \} \cdots \{ z_k / x_{i_k} \} ) \cdots )\\
  & = \res{z_1}( \gsome{z_1}{\llfv{M_{1}}};\piencodfaplas{ M_{1} }_{ {z_1}}  \|  \cdots \res{z_k} ( \gsome{z_k}{\llfv{M_{k}}};\piencodfaplas{ M_{k} }_{ {z_k}} \\
        &  \qquad  \| \bignd_{x_{i_1} \in \{ x_1 ,\cdots , x_k  \}} \cdots \bignd_{x_{i_k} \in \{ x_1 ,\cdots , x_k \setminus x_{i_1} , \cdots , x_{i_{k-1}}  \}} \piencodfaplas{\fail^{ \widetilde{x}, \widetilde{y}} }_u \{ z_1 / x_{i_1} \} \cdots \{ z_k / x_{i_k} \} ) \cdots )\\
  & = \res{z_1}( \gsome{z_1}{\llfv{M_{1}}};\piencodfaplas{ M_{1} }_{ {z_1}}  \|  \cdots \res{z_k} ( \gsome{z_k}{\llfv{M_{k}}};\piencodfaplas{ M_{k} }_{ {z_k}} \\
        &  \qquad  \| \bignd_{x_{i_1} \in \{ x_1 ,\cdots , x_k  \}} \cdots \bignd_{x_{i_k} \in \{ x_1 ,\cdots , x_k \setminus x_{i_1} , \cdots , x_{i_{k-1}}  \}}  \pnone{\widetilde{x}} \| \pnone{\widetilde{y}} \| \pnone{u}  \{ z_1 / x_{i_1} \} \cdots \\
        &  \qquad \{ z_k / x_{i_k} \}  ) \cdots )\\
  & \equiv  \res{z_1}( \gsome{z_1}{\llfv{M_{1}}};\piencodfaplas{ M_{1} }_{ {z_1}}  \| \\
        &  \qquad \cdots \res{z_k} ( \gsome{z_k}{\llfv{M_{k}}};\piencodfaplas{ M_{k} }_{ {z_k}}  \|   \pnone{z_1} \| \cdots \| \pnone{z_k} \| \pnone{\widetilde{y}} \| \pnone{u}  ) \cdots )\\
  & \redtwo^k \pnone{\llfv{C}} \| \pnone{\widetilde{y}} \| \pnone{u} =Q',
  \end{aligned}
    \]
and  no further reductions can be performed.
We also have that  ${N} \red \fail^{ \widetilde{y} \cup \llfv{C} } = {N}'$ and $\piencodfaplas{ \fail^{\widetilde{y} \cup \llfv{C}} }_u = Q'$.

                        \item $\headf{M} =  {x}_i$.

                            Then we have the following
 \[
    \small\hspace{-2cm}
 \begin{aligned}
 \piencodfaplas{{N}}_u
  &=  \res{z_1}( \gsome{z_1}{\llfv{M_{1}}};\piencodfaplas{ M_{1} }_{ {z_1}}  \|  \cdots \res{z_k} ( \gsome{z_k}{\llfv{M_{k}}};\piencodfaplas{ M_{k} }_{ {z_k}} \\
                & \quad  \| \bignd_{x_{i_1} \in \{ x_1 ,\cdots , x_k  \}} \cdots \bignd_{x_{i_k} \in \{ x_1 ,\cdots , x_k \setminus x_{i_1} , \cdots , x_{i_{k-1}}  \}} \res{\widetilde{y}} (\piencodfaplas{  {x}_i }_{j} \| P) \{ z_1 / x_{i_1} \} \cdots \{ z_k / x_{i_k} \} ) \cdots )\\
 &=  \res{z_1}( \gsome{z_1}{\llfv{M_{1}}};\piencodfaplas{ M_{1} }_{ {z_1}}  \|  \cdots \res{z_k} ( \gsome{z_k}{\llfv{M_{k}}};\piencodfaplas{ M_{k} }_{ {z_k}} \\
                & \quad  \| \bignd_{x_{i_1} \in \{ x_1 ,\cdots , x_k  \}} \cdots \bignd_{x_{i_k} \in \{ x_1 ,\cdots , x_k \setminus x_{i_1} , \cdots , x_{i_{k-1}}  \}} \\
                & \qquad \qquad \qquad \res{\widetilde{y}} ( \psome{x_i}; \pfwd{x_i}{j} \| P) \{ z_1 / x_{i_1} \} \cdots \{ z_k / x_{i_k} \} ) \cdots )\\
 \end{aligned}
 \]

 Notice that multiple reductions can take place along any of the channels $z_l$. Let us consider for simplicity that $i = l = k$.
  \[
    \small
    \hspace{-2.5cm}
 \begin{aligned}
 \piencodfaplas{{N}}_u
 &=  \res{z_1}( \gsome{z_1}{\llfv{M_{1}}};\piencodfaplas{ M_{1} }_{ {z_1}}  \|  \cdots \res{z_k} ( \gsome{z_k}{\llfv{M_{k}}};\piencodfaplas{ M_{k} }_{ {z_k}} \\
                & \quad  \| \bignd_{x_{i_1} \in \{ x_1 ,\cdots , x_k  \}} \cdots \bignd_{x_{i_k} \in \{ x_1 ,\cdots , x_k \setminus x_{i_1} , \cdots , x_{i_{k-1}}  \}}  \res{\widetilde{y}} ( \psome{x_k}; \pfwd{x_k}{j} \| P) \{ z_1 / x_{i_1} \} \cdots\\
                & \quad \{ z_k / x_{i_k} \} ) \cdots )\\
\end{aligned}
\]
We show the full process of the reduction in this case for correctness. Applying multiple reductions of the form $\rredtwo{\nu}$ we obtain:

{
\smaller
\begin{prooftree}
    \hspace*{\dimexpr-\leftmargini-\leftmarginii-\leftmarginiii}
        \AxiomC{$\begin{aligned}
                &\res{z_k} ( \gsome{z_k}{\llfv{M_{k}}};\piencodfaplas{ M_{k} }_{ {z_k}} \\
                & \qquad   \| \bignd_{x_{i_1} \in \{ x_1 ,\cdots , x_k  \}} \cdots \bignd_{x_{i_k} \in \{ x_1 ,\cdots , x_k \setminus x_{i_1} , \cdots , x_{i_{k-1}}  \}} \\
                & \qquad  \res{\widetilde{y}} ( \psome{x_k}; \pfwd{x_k}{j} \| P) \{ z_1 / x_{i_1} \} \cdots \{ z_k / x_{i_k} \} )
                \end{aligned} \redtwo_{z_k} R_{k}
                $}
        \LeftLabel{$ \rredtwo{\nu} $}
        \UnaryInfC{$ \vdots $}
        \noLine
        \UnaryInfC{$
        \begin{aligned}
            &\res{z_2}( \gsome{z_2}{\llfv{M_{2}}};\piencodfaplas{ M_{2} }_{ {z_2}}  \| \\
                & \qquad  \cdots \res{z_k} ( \gsome{z_k}{\llfv{M_{k}}};\piencodfaplas{ M_{k} }_{ {z_k}} \\
                & \qquad   \| \bignd_{x_{i_1} \in \{ x_1 ,\cdots , x_k  \}} \cdots \bignd_{x_{i_k} \in \{ x_1 ,\cdots , x_k \setminus x_{i_1} , \cdots , x_{i_{k-1}}  \}} \\
                & \qquad  \res{\widetilde{y}} ( \psome{x_k}; \pfwd{x_k}{j} \| P) \{ z_1 / x_{i_1} \} \cdots \{ z_k / x_{i_k} \} ) \cdots )\\
        \end{aligned} \redtwo_{z_k} R_1$}
        \LeftLabel{$ \rredtwo{\nu} $}
        \UnaryInfC{$
        \begin{aligned}
            &\res{z_1}( \gsome{z_1}{\llfv{M_{1}}};\piencodfaplas{ M_{1} }_{ {z_1}}  \| \\
                & \quad \cdots \res{z_k} ( \gsome{z_k}{\llfv{M_{k}}};\piencodfaplas{ M_{k} }_{ {z_k}} \\
                & \quad   \| \bignd_{x_{i_1} \in \{ x_1 ,\cdots , x_k  \}} \cdots \bignd_{x_{i_k} \in \{ x_1 ,\cdots , x_k \setminus x_{i_1} , \cdots , x_{i_{k-1}}  \}} \\
                & \quad  \res{\widetilde{y}} ( \psome{x_k}; \pfwd{x_k}{j} \| P) \{ z_1 / x_{i_1} \} \cdots \{ z_k / x_{i_k} \} ) \cdots )\\
        \end{aligned} \redtwo_{z_k} \res{z_1}( \gsome{z_1}{\llfv{M_{1}}};\piencodfaplas{ M_{1} }_{ {z_1}}  \| R_1 )$}
\end{prooftree}
}

where $R_i = \res{z_i}( \gsome{z_i}{\llfv{M_{i}}};\piencodfaplas{ M_{i} }_{ {z_i}}  \| R_{i+1} ) $ for $ i < k$

Hence we wish to show the following reduction:

\[
\begin{aligned}
                &\res{z_k} ( \gsome{z_k}{\llfv{M_{k}}};\piencodfaplas{ M_{k} }_{ {z_k}} \\
                & \qquad   \| \bignd_{x_{i_1} \in \{ x_1 ,\cdots , x_k  \}} \cdots \bignd_{x_{i_k} \in \{ x_1 ,\cdots , x_k \setminus x_{i_1} , \cdots , x_{i_{k-1}}  \}} \\
                & \qquad  \res{\widetilde{y}} ( \psome{x_k}; \pfwd{x_k}{j} \| P) \{ z_1 / x_{i_1} \} \cdots \{ z_k / x_{i_k} \} )
\end{aligned} \redtwo_{z_k} R_{k}
\]

We apply the reduction rule

\begin{prooftree}
        \AxiomC{$ P \piprecong{z_k} P' $}
        \AxiomC{$ Q \piprecong{z_k} Q' $}
        \AxiomC{$ \res{z_k}(P' \| Q') \redtwo_{z_k} R_k $}
        \TrinaryInfC{$ \res{z_k}(P \| Q) \redtwo_{z_k} R_k $}
\end{prooftree}

Where we take $P = \gsome{z_k}{\llfv{M_{k}}};\piencodfaplas{ M_{k} }_{ {z_k}} \piprecong{z_k } \gsome{z_k}{\llfv{M_{k}}};\piencodfaplas{ M_{k} }_{ {z_k}} = P'$
and
Notice that when $ x_{i_k} = x_k$ we have that the substitution $\{ z_k / x_{i_k} \} $ takes place
\[
    \small
    \hspace{-2.5cm}
\begin{aligned}
    Q & = \bignd_{x_{i_1} \in \{ x_1 ,\cdots , x_k  \}} \cdots \bignd_{x_{i_k} \in \{ x_1 ,\cdots , x_k \setminus x_{i_1} , \cdots , x_{i_{k-1}}  \}}  \res{\widetilde{y}} ( \psome{x_k}; \pfwd{x_k}{j} \| P) \{ z_1 / x_{i_1} \} \cdots \{ z_k / x_{i_k} \} \\
    & = \bignd_{x_{i_1} \in \{ x_1 ,\cdots , x_{k-1}  \}} \cdots \bignd_{x_{i_{k-1}} \in \{ x_1 ,\cdots , x_{k-1} \setminus x_{i_1} , \cdots , x_{i_{k-2}}  \}}  \res{\widetilde{y}} ( \psome{z_k}; \pfwd{z_k}{j} \| P) \{ z_1 / x_{i_1} \} \cdots  \\
    & \quad \{ z_{k-1} / x_{i_{k-1}} \}\nd \bignd_{x_{i_k} \in \{ x_1 ,\cdots , x_{k-1} \} }\bignd_{x_{i_1} \in \{ x_1 ,\cdots , x_k \setminus x_{i_k} \}} \cdots \bignd_{x_{i_{k-1}} \in \{ x_1 ,\cdots , x_{k-1} \setminus x_{i_1} , \cdots , x_{i_{k-2}} , x_{i_k} \}}  \\
    & \qquad \res{\widetilde{y}} ( \psome{x_k}; \pfwd{x_k}{j} \| P) \{ z_1 / x_{i_1} \} \cdots \{ z_k / x_{i_k} \} \\
    & \piprecong{z_k}  \bignd_{x_{i_1} \in \{ x_1 ,\cdots , x_{k-1}  \}} \cdots \bignd_{x_{i_{k-1}} \in \{ x_1 ,\cdots , x_{k-1} \setminus x_{i_1} , \cdots , x_{i_{k-2}}  \}}  \res{\widetilde{y}} ( \psome{z_k}; \pfwd{z_k}{j} \| P) \{ z_1 / x_{i_1} \} \cdots \\
    & \quad \{ z_{k-1} / x_{i_{k-1}} \} \\
    & = Q'
\end{aligned}
\]

Hence we wish to show the following reduction
\(
\res{z_k}(\gsome{z_k}{\llfv{M_{k}}};\piencodfaplas{ M_{k} }_{ {z_k}} \| Q' ) \redtwo_{z_k} R_k
\)
We do this via the rule $\rredtwo{\some}$ to obtain

$
R_k = \res{z_k}\Big(\piencodfaplas{ M_{k} }_{ {z_k}} \|
\begin{aligned}
    &\bignd_{x_{i_1} \in \{ x_1 ,\cdots , x_{k-1}  \}} \cdots \bignd_{x_{i_{k-1}} \in \{ x_1 ,\cdots , x_{k-1} \setminus x_{i_1} , \cdots , x_{i_{k-2}}  \}} \\
    & \qquad \res{\widetilde{y}} (  \pfwd{z_k}{j} \| P) \{ z_1 / x_{i_1} \} \cdots \{ z_{k-1} / x_{i_{k-1}} \}
\end{aligned}
\Big)
$

Hence we continue with the following reductions:
\[
 \begin{aligned}
 & \redtwo  \res{z_1}( \gsome{z_1}{\llfv{M_{1}}};\piencodfaplas{ M_{1} }_{ {z_1}}  \| \cdots \res{z_k} ( \piencodfaplas{ M_{k} }_{ {z_k}} \\
                & \qquad \qquad  \| \bignd_{x_{i_1} \in \{ x_1 ,\cdots , x_{k-1}  \}} \cdots \bignd_{x_{i_{k-1}} \in \{ x_1 ,\cdots , x_{k-1} \setminus x_{i_1} , \cdots , x_{i_{k-2}}  \}} \\
                & \qquad \qquad \qquad \res{\widetilde{y}} (  \pfwd{x_k}{j} \| P) \{ z_1 / x_{i_1} \} \cdots \{ z_{k-1} / x_{i_{k-1}} \} ) \cdots ) = Q_1\\
 & \redtwo \res{z_1}( \gsome{z_1}{\llfv{M_{1}}};\piencodfaplas{ M_{1} }_{ {z_1}}  \| \cdots \res{z_{k-1}} ( \piencodfaplas{ M_{k-1} }_{ {z_{k-1}}} \\
                & \qquad \qquad  \| \bignd_{x_{i_1} \in \{ x_1 ,\cdots , x_{k-1}  \}} \cdots \bignd_{x_{i_{k-1}} \in \{ x_1 ,\cdots , x_{k-1} \setminus x_{i_1} , \cdots , x_{i_{k-2}}  \}} \\
                & \qquad \qquad \qquad \res{\widetilde{y}} (  \piencodfaplas{ M_{k} }_{ {j}} \| P) \{ z_1 / x_{i_1} \} \cdots \{ z_{k-1} / x_{i_{k-1}} \} ) \cdots )  = Q_2\\
 \end{aligned}
 \]

In addition,
\(
 {N} = M \linexsub{C /  \widetilde{x}} \red  M \headlin{ M_k / x_k }\linexsub{C \setminus M_k / \widetilde{x} \setminus x_k}  = {M}'\).
Finally,
\[
\begin{aligned}
\piencodfaplas{{M}'}_u &= \piencodfaplas{M \headlin{ M_k / x_k }\linexsub{C \setminus M_k / \widetilde{x} \setminus x_k}}_u\\
&=  \res{z_1}( \gsome{z_1}{\llfv{M_{1}}};\piencodfaplas{ M_{1} }_{ {z_1}}  \|  \cdots \res{z_{k-1}} ( \piencodfaplas{ M_{k-1} }_{ {z_{k-1}}} \\
                & \qquad  \| \bignd_{x_{i_1} \in \{ x_1 ,\cdots , x_{k-1}  \}} \cdots \bignd_{x_{i_{k-1}} \in \{ x_1 ,\cdots , x_{k-1} \setminus x_{i_1} , \cdots , x_{i_{k-2}}  \}} \\
                & \qquad \qquad \res{\widetilde{y}} (  \piencodfaplas{ M_{k} }_{ {j}} \| P) \{ z_1 / x_{i_1} \} \cdots \{ z_{k-1} / x_{i_{k-1}} \} ) \cdots )
     = Q_2
\end{aligned}
\]

\begin{enumerate}
\item When $n = 1$:

Then, $Q = Q_1$ and  $ \piencodfaplas{{N}}_u \redtwo^1 Q_1$. Also,

$Q_1 \redtwo^1 Q_2 = Q'$, ${N} \red^1 M \headlin{ M_k / x_k }\linexsub{C \setminus M_k / \widetilde{x} \setminus x_k} = {N}'$ and\\
 $\piencodfaplas{M \headlin{ M_k / x_k }\linexsub{C \setminus M_k / \widetilde{x} \setminus x_k}}_u = Q_2$.
\item When $n \geq 2$:

Then  $ \piencodfaplas{{N}}_u \redtwo^2 Q_2 \redtwo^l Q$, for $l \geq 0$.  Also,
${N} \red {M}'$, $Q_2 = \piencodfaplas{{M}'}_u$. By the induction hypothesis, there exist $ Q'$ and ${N}'$ such that $ Q \redtwo^i Q'$, ${M}' \red^j {N}'$ and $\piencodfaplas{{N}'}_u = Q'$. Finally, $\piencodfaplas{{N}}_u \redtwo^2 Q_2 \redtwo^l Q \redtwo^i Q'$ and ${N} \rightarrow {M}'  \red^j {N}'$.

                            \end{enumerate}

                    \end{enumerate}
 \item  For $m \geq 1$ and $ n \geq 0$.

            \begin{enumerate}
            \item When $n = 0$.

               Then
               \[
                \small
                \begin{aligned}
                   Q& =  \res{z_1}( \gsome{z_1}{\llfv{M_{1}}};\piencodfaplas{ M_{1} }_{ {z_1}}  \| \cdots \res{z_k} ( \gsome{z_k}{\llfv{M_{k}}};\piencodfaplas{ M_{k} }_{ {z_k}} \\
                & \quad  \| \bignd_{x_{i_1} \in \{ x_1 ,\cdots , x_k  \}} \cdots \bignd_{x_{i_k} \in \{ x_1 ,\cdots , x_k \setminus x_{i_1} , \cdots , x_{i_{k-1}}  \}} R \{ z_1 / x_{i_1} \} \cdots \{ z_k / x_{i_k} \} ) \cdots )\\
                \end{aligned}
               \]
                and $\piencodfaplas{M}_u \redtwo^m R$ where $m \geq 1$. By the IH there exist $R'$  and ${M}' $ such that $R \redtwo^i R'$, $M \red^j {M}'$ and $\piencodfaplas{{M}'}_u = R'$. Thus,
              \[
                \small
                \hspace{-1cm}
               \begin{aligned}
                   \piencodfaplas{{N}}_u & \redtwo^m \res{z_1}( \gsome{z_1}{\llfv{M_{1}}};\piencodfaplas{ M_{1} }_{ {z_1}}  \| \cdots \res{z_k} ( \gsome{z_k}{\llfv{M_{k}}};\piencodfaplas{ M_{k} }_{ {z_k}} \\
                    & \quad  \| \bignd_{x_{i_1} \in \{ x_1 ,\cdots , x_k  \}} \cdots \bignd_{x_{i_k} \in \{ x_1 ,\cdots , x_k \setminus x_{i_1} , \cdots , x_{i_{k-1}}  \}} R \{ z_1 / x_{i_1} \} \cdots \{ z_k / x_{i_k} \} ) \cdots ) & = Q\\
                    & \redtwo^i \res{z_1}( \gsome{z_1}{\llfv{M_{1}}};\piencodfaplas{ M_{1} }_{ {z_1}}  \| \cdots \res{z_k} ( \gsome{z_k}{\llfv{M_{k}}};\piencodfaplas{ M_{k} }_{ {z_k}} \\
                    & \quad  \| \bignd_{x_{i_1} \in \{ x_1 ,\cdots , x_k  \}} \cdots \bignd_{x_{i_k} \in \{ x_1 ,\cdots , x_k \setminus x_{i_1} , \cdots , x_{i_{k-1}}  \}} R' \{ z_1 / x_{i_1} \} \cdots \{ z_k / x_{i_k} \} ) \cdots ) & = Q' \\
                \end{aligned}
                 \]
                Also, ${N} = M \linexsub{C /  \widetilde{x}} \red^j  M' \linexsub{C /  \widetilde{x}}  = {N}'$ and  $\piencodfaplas{{N}'}_u = Q'$

            \item When $n \geq 1$.
                Then  $R$ has an occurrence of an unguarded $\psome{x_i}$ or $\pnone{x_i}$, this case follows by IH.

                    \end{enumerate}
            \end{enumerate}

            \item  ${N} =  M \unexsub{U / \unvar{x}}$.

            In this case,
            \(
            \begin{aligned}
                \piencodfaplas{M \unexsub{U / \unvar{x}}}_u &=   \res{\unvar{x}} ( \piencodfaplas{ M }_u \|   ~ \guname{\unvar{x}}{x_i}; \piencodfaplas{ U }_{x_i} ).
            \end{aligned}
            \)
            Then,
            \[
            \begin{aligned}
               \piencodfaplas{{N}}_u & =   \res{\unvar{x}} ( \piencodfaplas{ M }_u \|   ~ \guname{\unvar{x}}{x_i}; \piencodfaplas{ U }_{x_i} )  \redtwo^m  \res{\unvar{x}} ( R \|   ~ \guname{\unvar{x}}{x_i}; \piencodfaplas{ U }_{x_i} ) \redtwo^n  Q.
            \end{aligned}
            \]
            for some process $R$. Where $\redtwo^n$ is a reduction initially synchronises with $\guname{\unvar{x}}{x_i}$ when $n \geq 1$, $n + m = k \geq 1$. Type preservation in \clpi ensures reducing $\piencodfaplas{ M}_v \redtwo^m$ doesn't consume possible synchronisations with $ \guname{ \unvar{x} }{ x_i }  $ if they occur. Let us consider the the possible sizes of both $m$ and $n$.

            \begin{enumerate}
                \item For $m = 0$ and $n \geq 1$.

                   In this case,  $R = \piencodfaplas{M}_u$ as $\piencodfaplas{M}_u \red^0 \piencodfaplas{M}_u$.

                    Notice that the only possibility of having an unguarded $ \puname{ \unvar{x} }{ x_i }$ without internal reductions is when   $\headf{M} =  {x}[j]$ for some index $j$.
                            Then we have the following:

                            \[
                                \small
                                \hspace{-0.5cm}
                            \begin{aligned}
                            \piencodfaplas{{N}}_u
                            & =\res{\unvar{x}} ( \res{\widetilde{y}} ( \piencodfaplas{ x[j] }_k \| P ) \|   ~ \guname{\unvar{x}}{x_i}; \piencodfaplas{ U }_{x_i} )\\
                            & =\res{\unvar{x}} ( \res{\widetilde{y}} ( \puname{\unvar{x}}{{x_i}}; \psel{x_i}{j}; \pfwd{x_i}{k}  \| P ) \|   ~ \guname{\unvar{x}}{x_i}; \piencodfaplas{ U }_{x_i} )\\
                            & \redtwo \res{\unvar{x}} (  \res{x_i}(  \res{\widetilde{y}} ( \psel{x_i}{j}; \pfwd{x_i}{k}  \| P ) \| \piencodfaplas{ U }_{x_i}  ) \|   ~ \guname{\unvar{x}}{x_i}; \piencodfaplas{ U }_{x_i} ) & = Q_1\\
                            & = \res{\unvar{x}} (  \res{x_i}(  \res{\widetilde{y}} ( \psel{x_i}{j}; \pfwd{x_i}{k}  \| P ) \| \gsel{x_i}\{i:\piencodfaplas{ U_i }_{x_i} \}_{U_i \in U}  ) \| \\
                            & \quad \guname{\unvar{x}}{x_i}; \piencodfaplas{ U }_{x_i} )\\
                            & \redtwo \res{\unvar{x}} (  \res{x_i}(  \res{\widetilde{y}} ( \ \pfwd{x_i}{k}  \| P ) \| \piencodfaplas{ U_j }_{x_i}  ) \|   ~ \guname{\unvar{x}}{x_i}; \piencodfaplas{ U }_{x_i} ) & = Q_2\\
                            & \redtwo \res{\unvar{x}} (    \res{\widetilde{y}} ( \piencodfaplas{ U_j }_{k}   \| P )  \|   ~ \guname{\unvar{x}}{x_i}; \piencodfaplas{ U }_{x_i} ) & = Q_3\\
                            \end{aligned}
                            \]

                We consider the two cases of the form of $U_{j}$ and show that the choice of $U_{j}$ is inconsequential

                \begin{itemize}
                \item When $ U_j = \unvar{\bag{N}}$:

                In this case,
                \(
                \begin{aligned}
                {N} &=M \unexsub{U / \unvar{x}}\red M \headlin{ N /\unvar{x} }\unexsub{U / \unvar{x}} = {M}'.
                \end{aligned}
                \)
                 and
                \[
                 \begin{aligned}
                 \piencodfaplas{{M}'}_u
                 &= \piencodfaplas{M \headlin{ N /\unvar{x} }\unexsub{U / \unvar{x}}}_u\\
                 &=  \res{\unvar{x}} (    \res{\widetilde{y}} ( \piencodfaplas{ N }_{k}   \| P )  \|   ~ \guname{\unvar{x}}{x_i}; \piencodfaplas{ U }_{x_i} )  & = Q_3
                                            \end{aligned}
                 \]

                \item When $ U_i = \unvar{\oneb} $:

                  In this case,
                        \(
                        \begin{aligned}
                            {N} &=M \unexsub{U / \unvar{x}} \red M \headlin{ \fail^{\emptyset} /\unvar{x} } \unexsub{U /\unvar{x} } = {M}'.
                        \end{aligned}
                        \)

                        Notice that $\piencodfaplas{\unvar{\oneb}}_{k} =  \pnone{k}$ and that $\piencodfaplas{\fail^{\emptyset}}_k = \pnone{k} $. In addition,

                            \[
                            \begin{aligned}
                               \piencodfaplas{{M}'}_u &= \piencodfaplas{M \headlin{ \fail^{\emptyset} /\unvar{x} } \unexsub{U /\unvar{x} }}_u\\
                               &=  \res{\unvar{x}} (    \res{\widetilde{y}} ( \piencodfaplas{\fail^{\emptyset}}_{k}   \| P )  \|   ~ \guname{\unvar{x}}{x_i}; \piencodfaplas{ U }_{x_i} )\\
                               &=  \res{\unvar{x}} (    \res{\widetilde{y}} ( \piencodfaplas{\unvar{\oneb}}_{k}   \| P )  \|   ~ \guname{\unvar{x}}{x_i}; \piencodfaplas{ U }_{x_i} ) & = Q_3
                            \end{aligned}
                            \]
                \end{itemize}

                Both choices give an ${M}$ that are equivalent to $Q_3$.

    \begin{enumerate}
    \item When $n \leq 2$.

   In this case, $Q = Q_n$ and  $ \piencodfaplas{{N}}_u \redtwo^n Q_n$.

Also, $Q_n \redtwo^{3-n} Q_3 = Q'$, ${N} \red^1 {M}' = {N}'$ and $\piencodfaplas{{M}' }_u = Q_2$.

     \item When $n \geq 3$.

     We have $ \piencodfaplas{{N}}_u \redtwo^3 Q_3 \redtwo^l Q$ for $l \geq 0$. We also know that ${N} \rightarrow {M}'$, $Q_3 = \piencodfaplas{{M}'}_u$. By the IH, there exist $ Q$ and ${N}'$ such that $Q \redtwo^i Q'$, ${M}' \red^j {N}'$ and $\piencodfaplas{{N}'}_u = Q'$. Finally, $\piencodfaplas{{N}}_u \redtwo^3 Q_3 \redtwo^l Q \redtwo^i Q'$ and ${N} \rightarrow {M}'  \red^j {N}' $.

                    \end{enumerate}
 \item For $m \geq 1$ and $ n \geq 0$.

            \begin{enumerate}
            \item When $n = 0$.

               Then $ \res{\unvar{x}} ( R \|   ~ \guname{\unvar{x}}{x_i}; \piencodfaplas{ U }_{x_i} )  = Q$ and $\piencodfaplas{M}_u \redtwo^m R$ where $m \geq 1$. By the IH there exist $R'$  and ${M}' $ such that $R \redtwo^i R'$, $M \red^j {M}'$ and $\piencodfaplas{{M}'}_u = R'$.
              Hence,
              \[
               \begin{aligned}
                   \piencodfaplas{{N}}_u & =   \res{\unvar{x}} ( \piencodfaplas{M}_u \|   ~ \guname{\unvar{x}}{x_i}; \piencodfaplas{ U }_{x_i} )\redtwo^m   \res{\unvar{x}} ( R \|   ~ \guname{\unvar{x}}{x_i}; \piencodfaplas{ U }_{x_i} )
                   & = Q.
                \end{aligned}
                 \]
                In addition,
                \(
                   Q  \redtwo^i  \res{\unvar{x}} ( R' \|   ~ \guname{\unvar{x}}{x_i}; \piencodfaplas{ U }_{x_i} ) = Q \), and the term can reduce as follows: $ {N} = M \unexsub{U / \unvar{x}} \red^j  M' \unexsub{U / \unvar{x}} = {N}'$ and  $\piencodfaplas{{N}'}_u = Q'$.

            \item When $n \geq 1$.

            Then $R$ has an occurrence of an unguarded $\guname{\unvar{x}}{x_i}$, and the case follows by IH.
\end{enumerate}
            \end{enumerate}
               \end{enumerate}
\end{proof}

\subsection{Success Sensitivity}\label{ch4a:tsuccess}

    \begin{proposition}[Preservation of Success]
        \label{ch4Prop:checkprespiunres}
        For all closed $M\in \lamcoldetsh$, the following hold:
        \begin{enumerate}
            \item $ \headf{M} = \checkmark \implies \piencodfaplas{M} =  \res{ \widetilde{x} }  (P   \| \checkmark) $
            \item $ \piencodfaplas{M}_u =  P   \| \checkmark \nd Q \implies \headf{M} = \checkmark$
        \end{enumerate}

        \end{proposition}

        \begin{proof}

        Proof of both cases by induction on the structure of $M$.

        \begin{enumerate}
        \item We only need to consider terms of the following form:

            \begin{enumerate}

                \item  $ M = \checkmark $:

                This case is immediate.

                \item $M = N\ (C \bagsep U)$:

                Then, $\headf{N \ (C \bagsep U)} = \headf{N}$. If $\headf{N} = \checkmark$, then  $$ \piencodfaplas{M (C \bagsep U)}_u =   \res{ v }  (\piencodfaplas{M}_v   \|  \gsome{ v }{ u , \llfv{C} }; \pname{v}{x} . (\pfwd{v}{u}   \| \piencodfaplas{C \bagsep U}_x ) ).$$
                By the IH,  $\checkmark$ is unguarded in $\piencodfaplas{N}_u$.

                \item $M = M' \linexsub{C /  x_1 , \cdots , x_k}$

                Then we have that $\headf{M' \linexsub{C /  x_1 , \cdots , x_k}}  = \headf{M'} = \checkmark$. Then

                \[
                    \small
                \begin{aligned}
                    \piencodfaplas{ M  }_u    & =
                    \res{z_1}( \gsome{z_1}{\llfv{M_{1}}};\piencodfaplas{ M_{1} }_{ {z_1}}  \| \cdots \res{z_k} ( \gsome{z_k}{\llfv{M_{k}}};\piencodfaplas{ M_{k} }_{ {z_k}} \\
                    & \qquad  \| \bignd_{x_{i_1} \in \{ x_1 ,\cdots , x_k  \}} \cdots \bignd_{x_{i_k} \in \{ x_1 ,\cdots , x_k \setminus x_{i_1} , \cdots , x_{i_{k-1}}  \}} \piencodfaplas{ M' }_u \{ z_1 / x_{i_1} \} \cdots \{ z_k / x_{i_k} \} ) \cdots )
                \end{aligned}
                \]

                and by the IH $\checkmark$ is unguarded in $\piencodfaplas{M'}_u$ hence ungaurded in every summand in $\piencodfaplas{ M \linexsub{C /  x_1 , \cdots , x_k}  }_u$.

                \item $M = M' \unexsub{U / \unvar{x}}$

                Then we have that $\headf{M' \unexsub{U / \unvar{x}}}  = \headf{M'} = \checkmark$. Then $\piencodfaplas{ M' \unexsub{U / \unvar{x}}  }_u   =    \res{ \unvar{x} }  ( \piencodfaplas{ M' }_u   \|    \guname{ \unvar{x} }{ x_i } ; \piencodfaplas{ U }_{x_i} ) $ and by the IH  $\checkmark$ is unguarded in $\piencodfaplas{M'}_u$.

            \end{enumerate}

           \item We only need to consider terms of the following form:

            \begin{enumerate}

                \item {\bf Case $M = \checkmark$:}

                Then,
                $\piencodfaplas{\checkmark}_u = \checkmark$
                which is an unguarded occurrence of $\checkmark$ and that $\headf{\checkmark} = \checkmark$.

                \item {\bf Case $M = N (C \bagsep U)$:}

                Then, $\piencodfaplas{N (C \bagsep U)}_u =  \res{ v }  (\piencodfaplas{N}_v   \| \gsome{ v }{ u , \llfv{C} };   \pname{v}{x} . (\pfwd{v}{u}   \| \piencodfaplas{C \bagsep U}_x ) )$. The only occurrence of an unguarded $\checkmark$ can occur is within $\piencodfaplas{N}_v$. By the IH, $\headf{N} = \checkmark$ and finally $\headf{N \ (C \bagsep U)} = \headf{N}$.

                \item {\bf Case $M = M' \linexsub{C /  x_1 , \cdots , x_k}$:}

                Then:
                \[
                    \small
                \begin{aligned}
                    \piencodfaplas{ M  }_u    & =
                    \res{z_1}( \gsome{z_1}{\llfv{M_{1}}};\piencodfaplas{ M_{1} }_{ {z_1}}  \|  \cdots \res{z_k} ( \gsome{z_k}{\llfv{M_{k}}};\piencodfaplas{ M_{k} }_{ {z_k}} \\
                & \quad \| \bignd_{x_{i_1} \in \{ x_1 ,\cdots , x_k  \}} \cdots \bignd_{x_{i_k} \in \{ x_1 ,\cdots , x_k \setminus x_{i_1} , \cdots , x_{i_{k-1}}  \}} \piencodfaplas{ M' }_u \{ z_1 / x_{i_1} \} \cdots \{ z_k / x_{i_k} \} ) \cdots )
                \end{aligned}
                \]
                 An unguarded occurrence of $\checkmark$ can only occur within $\piencodfaplas{ M' }_u $. By the IH,  $\headf{M'} = \checkmark$ and hence $\headf{ M' \linexsub{C /  x_1 , \cdots , x_k}}  = \headf{M'}$.

                \item {\bf Case $M = M' \unexsub{U / \unvar{x}}$:}
                This case is analogous to the previous.
            \end{enumerate}
        \end{enumerate}
        \end{proof}

\begin{restatable}[Success Sensitivity (Under $\redtwo$)]{theorem}{thmEncLazySucc}\label{ch4proof:successsenscetwo}
    $\succp{{M}}{\checkmark}$ if and only if ${\piencodfaplas{{M}}_u}\succtwo_{\checkmark}$ for well-formed closed terms $M$.
\end{restatable}

\begin{proof}
We proceed with the proof in two parts.

\begin{enumerate}

    \item Suppose that  ${M} \Downarrow_{\checkmark} $. We will prove that $\piencodfaplas{{M}} \succtwo_{\checkmark}$.

    By \defref{ch4def:app_Suc3unres}, there exists  $ {M}'$ such that $ M \red^* {M}'$ and
    $\headf{M'} = \checkmark$.
    By completeness, if $ M \red M'$ then there exist $Q, Q'$ such that $\piencodfaplas{M}_u \equiv Q \redtwo^* Q'$ and $\piencodfaplas{M'}_u \equiv Q' $.

    We wish to show that there exists $Q''$such that $Q' \redtwo^* Q''$ and $Q''$ has an unguarded occurrence of $\checkmark$.

    By Proposition \ref{ch4Prop:checkprespiunres} (1) we have that $\headf{M'} = \checkmark \implies \piencodfaplas{M'}_u =   \res{ \widetilde{x} } (P   \| \checkmark)$. Finally $\piencodfaplas{M'}_u  =   \res{ \widetilde{x} } (P   \| \checkmark) \equiv Q'$. Hence $Q$ reduces to a process that has an unguarded occurrence of $\checkmark$.

    \item Suppose that $\piencodfaplas{{M}}_u \succtwo_{\checkmark}$. We will prove that $ {M} \Downarrow_{\checkmark}$.

    By operational soundness we have that if  $\piencodfaplas{{M}}_u \redtwo^* Q$ then there exist ${M}'$ and $Q'$ such that:
    (i)~${M} \red^* {M}'$
    and
    (ii)~$Q \redtwo^* Q' $ with
    $ \piencodfaplas{{M}'}_u \equiv Q'$.

   Since $\piencodfaplas{{M}}_u \redtwo^*  Q$, and $Q= Q'' \| \checkmark$ and $Q \redtwo^* Q'$ we must have that $Q'' \redtwo^* Q''' $ with $ Q' = Q''' \| \checkmark $. As $ \piencodfaplas{{M}'}_u \equiv Q'$ we have that $ \piencodfaplas{{M}'}_u  = $. Finally applying Proposition \ref{ch4Prop:checkprespiunres} (2) we have that $\piencodfaplas{{M}'}_u  = Q''' \| \checkmark$ is itself a term with unguarded $\checkmark$, then ${M}$ is itself headed with $\checkmark$.

\end{enumerate}
\end{proof}

\section{Proof of Loose Correctness of the Translation under the Eager Semantics}\label{ch4a:secloose}

\subsection{Completeness}\label{ch4a:looscompleteness}

\thmEncLWCompl*

\begin{proof}
By induction on the reduction rule applied to infer $N\red M$.
We have five cases.

 \begin{enumerate}
        \item  Case $\redlab{RS:Beta}$:
        Then  $ N= (\lambda x . (M'[ {\widetilde{x}} \leftarrow  {x}])) B  \red (M'[ {\widetilde{x}} \leftarrow  {x}])\esubst{ B }{ x }  = M$ , where $B = C \bagsep U$. The result folows easily, since
\begin{equation}\label{ch4eq:compl_lsbeta1failunres}
    \small
\begin{aligned}
\piencodfaplas{N}_u &=    \res{ v }  (\piencodfaplas{\lambda x . (M'[ {\widetilde{x}} \leftarrow  {x}])}_v   \|  \gsome{ v }{ u , \llfv{C} };  \pname{v}{x} . (\pfwd{v}{u}   \| \piencodfaplas{C \bagsep U}_x ) )  \\
&=   \res{ v }  (\psome{v}; \gname{v}{x}; \psome{x}; \gname{x}{\linvar{x}}; \gname{x}{\unvar{x}};\gclose{ x } ; \piencodfaplas{M'[ {\widetilde{x}} \leftarrow  {x}]}_v   \|  \\
& \qquad \gsome{ v }{ u , \llfv{C} };  \pname{v}{x} . (\pfwd{v}{u}   \| \piencodfaplas{C \bagsep U}_x ) )  \\
& \redone   \res{ v }  ( \gname{v}{x}; \psome{x}; \gname{x}{\linvar{x}}; \gname{x}{\unvar{x}};\gclose{ x } ; \piencodfaplas{M'[ {\widetilde{x}} \leftarrow  {x}]}_v   \|  \pname{v}{x} . (\pfwd{v}{u}   \| \piencodfaplas{C \bagsep U}_x ) )  \\
& \redone   \res{ x }  ( \res{v}( \psome{x}; \gname{x}{\linvar{x}}; \gname{x}{\unvar{x}}; \gclose{ x } ; \piencodfaplas{M'[ {\widetilde{x}} \leftarrow  {x}]}_v   \|  \pfwd{v}{u} )  \| \piencodfaplas{C \bagsep U}_x  )  \\
& \redone   \res{ x }  (  \psome{x}; \gname{x}{\linvar{x}}; \gname{x}{\unvar{x}};  \gclose{ x } ; \piencodfaplas{M'[ {\widetilde{x}} \leftarrow  {x}]}_u   \|  \piencodfaplas{C \bagsep U}_x  )  =   \piencodfaplas{M}_u\\
\end{aligned}
\end{equation}

\item  Case $ \redlab{RS:Ex \dash Sub}$: Then $ N =M'[ {x}_1, \!\cdots\! ,  {x}_k \leftarrow  {x}]\esubst{ C \bagsep U }{ x }$, with $C = \bag{M_1}
\cdots  \bag{M_k}$, $k\geq 0$ and $M'\not= \fail^{\widetilde{y}}$.

\begin{equation}\label{ch4eq:compl_lsbeta1failunres2}
    \small
    \begin{aligned}
    \piencodfaplas{N}_u &=    \res{ v }  (\piencodfaplas{\lambda x . (M'[ {\widetilde{x}} \leftarrow  {x}])}_v   \|  \gsome{ v }{ u , \llfv{C} };  \pname{v}{x} . (\pfwd{v}{u}   \| \piencodfaplas{C \bagsep U}_x ) )  \\
    &=   \res{ v }  (\psome{v}; \gname{v}{x}; \psome{x}; \gname{x}{\linvar{x}}; \gname{x}{\unvar{x}};\gclose{ x } ; \piencodfaplas{M'[ {\widetilde{x}} \leftarrow  {x}]}_v   \|  \\
    & \qquad \gsome{ v }{ u , \llfv{C} };  \pname{v}{x} . (\pfwd{v}{u}   \| \piencodfaplas{C \bagsep U}_x ) )  \\
    & \redone   \res{ v }  ( \gname{v}{x}; \psome{x}; \gname{x}{\linvar{x}}; \gname{x}{\unvar{x}};\gclose{ x } ; \piencodfaplas{M'[ {\widetilde{x}} \leftarrow  {x}]}_v   \|  \pname{v}{x} . (\pfwd{v}{u}   \| \piencodfaplas{C \bagsep U}_x ) )  \\
    & \redone   \res{ x }  ( \res{v}( \psome{x}; \gname{x}{\linvar{x}}; \gname{x}{\unvar{x}}; \gclose{ x } ; \piencodfaplas{M'[ {\widetilde{x}} \leftarrow  {x}]}_v   \|  \pfwd{v}{u} )  \| \piencodfaplas{C \bagsep U}_x  )  \\
    & \redone   \res{ x }  (  \psome{x}; \gname{x}{\linvar{x}}; \gname{x}{\unvar{x}};  \gclose{ x } ; \piencodfaplas{M'[ {\widetilde{x}} \leftarrow  {x}]}_u   \|  \piencodfaplas{C \bagsep U}_x  )  =   \piencodfaplas{M}_u\\
    \end{aligned}
    \end{equation}

\item  Case $ \redlab{RS:Ex \dash Sub}$: Then $ N =M'[ {x}_1, \!\cdots\! ,  {x}_k \leftarrow  {x}]\esubst{ C \bagsep U }{ x }$, with $C = \bag{M_1}
    \cdots  \bag{M_k}$, $k\geq 0$ and $M'\not= \fail^{\widetilde{y}}$.

    The reduction is $N = M'[ {x}_1, \!\cdots\! ,  {x}_k \leftarrow  {x}]\esubst{ C \bagsep U }{ x } \red  M' \linexsub{C  /  x_1 , \cdots , x_k} \unexsub{U / \unvar{x} } = M.$

    We detail the translations of $\piencodfaplas{N}_u$ and $\piencodfaplas{M}_u$. To simplify the proof, we  consider $k=2$ (the case in which $k> 2$ is follows analogously. Similarly the case of $k =0,1$ it contained within $k = 2$). On the one hand, we have:

    \begin{equation}\label{ch4eq:compl_lsbeta3failunres0}
    \begin{aligned}
    \piencodfaplas{N}_u &= \piencodfaplas{M'[ {x}_1 \leftarrow  {x}]\esubst{ C \bagsep U }{ x }}_u\\
    &= \res{ x }  ( \psome{x}; \gname{x}{\linvar{x}}; \gname{x}{\unvar{x}}; \gclose{ x } ; \piencodfaplas{ M'[ x_1, x_2 \leftarrow  {x}]}_u   \| \piencodfaplas{ C \bagsep U}_x )   \\
    &= \res{ x }  ( \psome{x}; \gname{x}{\linvar{x}}; \gname{x}{\unvar{x}};  \gclose{ x } ;\piencodfaplas{ M'[ x_1, x_2 \leftarrow  {x}]}_u   \|  \gsome{ x }{ \llfv{C} };  \pname{x}{\linvar{x}}.\big( \piencodfaplas{ C }_{\linvar{x}}\\
    &  \qquad \qquad   \| \pname{x}{\unvar{x}} .(  \guname{ \unvar{x} }{ x_i } ;  \piencodfaplas{ U }_{x_i}   \|  \pclose{ x } ) \big) )   \qquad (:= P_{\mathbb{N}}) \\[4pt]
    \end{aligned}
    \end{equation}

The reduction is $N = M'[ {x}_1, \!\cdots\! ,  {x}_k \leftarrow  {x}]\esubst{ C \bagsep U }{ x } \red  M' \linexsub{C  /  x_1 , \cdots , x_k} \unexsub{U / \unvar{x} } = M.$

We detail the translations of $\piencodfaplas{N}_u$ and $\piencodfaplas{M}_u$. To simplify the proof, we  consider $k=2$ (the case in which $k> 2$ is follows analogously. Similarly the case of $k =0,1$ it contained within $k = 2$). On the one hand, we have:

\begin{equation}\label{ch4eq:compl_lsbeta3failunres}
\begin{aligned}
\piencodfaplas{N}_u &= \piencodfaplas{M'[ {x}_1 \leftarrow  {x}]\esubst{ C \bagsep U }{ x }}_u\\
&= \res{ x }  ( \psome{x}; \gname{x}{\linvar{x}}; \gname{x}{\unvar{x}}; \gclose{ x } ; \piencodfaplas{ M'[ x_1, x_2 \leftarrow  {x}]}_u   \| \piencodfaplas{ C \bagsep U}_x )   \\
&= \res{ x }  ( \psome{x}; \gname{x}{\linvar{x}}; \gname{x}{\unvar{x}};  \gclose{ x } ;\piencodfaplas{ M'[ x_1, x_2 \leftarrow  {x}]}_u   \|  \gsome{ x }{ \llfv{C} };  \pname{x}{\linvar{x}}.\big( \piencodfaplas{ C }_{\linvar{x}}\\
&  \qquad \qquad   \| \pname{x}{\unvar{x}} .(  \guname{ \unvar{x} }{ x_i } ;  \piencodfaplas{ U }_{x_i}   \|  \pclose{ x } ) \big) )   \qquad (:= P_{\mathbb{N}}) \\[4pt]
\end{aligned}
\end{equation}

        \begin{equation*}
            \small
        \begin{aligned}
            P_{\mathbb{N}}& \redone^*  \res{  \unvar{x} } ( \res{\linvar{x}} (  \piencodfaplas{ M'[ x_1, x_2 \leftarrow  {x}]}_u   \|  \piencodfaplas{\bag{M_1} \cdot \bag{M_2} }_{\linvar{x}}  ) \|   \guname{ \unvar{x} }{ x_i } ;  \piencodfaplas{ U }_{x_i}   ) \\
            & =  \res{  \unvar{x} } (
            \res{\linvar{x}} (
            \psome{\linvar{x}}; \pname{\linvar{x}}{y_1}; \big( \gsome{y_1}{ \emptyset }; \gclose{ y_{1} } ; \0   \\
            &  \qquad \| \psome{\linvar{x}}; \gsome{\linvar{x}}{u, \llfv{M'} \setminus  \widetilde{x} }; \bignd_{x_{i_1} \in x_1, x_2} \gname{x}{{x}_{i_1}};
            \psome{\linvar{x}}; \pname{\linvar{x}}{y_2};   \\
            & \qquad \big( \gsome{y_2}{ \emptyset }; \gclose{ y_{2} } ; \0 \| \psome{\linvar{x}}; \gsome{\linvar{x}}{u, \llfv{M'} \setminus  \widetilde{x} }; \bignd_{x_{i_2} \in (x_1, x_2 \setminus x_{i_1} )} \gname{x}{{x}_{i_2}};\\
            &  \qquad \psome{\linvar{x}}; \pname{\linvar{x}}{y_3}; ( \gsome{y_3}{ u, \llfv{M'} }; \gclose{ y_{3} } ;\piencodfaplas{M'}_u \| \pnone{ \linvar{x} } )
            \big)
            \big)
            \\
            &  \qquad\| 
            \gsome{\linvar{x}}{\llfv{C} }; \gname{x}{y_1}; \gsome{\linvar{x}}{y_1, \llfv{C}}; \psome{\linvar{x}}; \pname{\linvar{x}}{z_1}; \\
            &  \qquad   ( \gsome{z_1}{\llfv{M_j}};  \piencodfaplas{M_j}_{z_1} \|  \pnone{y_1} \| \\
            & \qquad \gsome{\linvar{x}}{\llfv{C} }; \gname{x}{y_2}; \gsome{\linvar{x}}{y_2, \llfv{C}}; \psome{\linvar{x}}; \pname{\linvar{x}}{z_2}; \\
            & \qquad   ( \gsome{z_2}{\llfv{M_j}};  \piencodfaplas{M_j}_{z_2} \| \pnone{y_2} \| \\
            & \qquad \gsome{\linvar{x}}{\emptyset};\gname{x}{y_3};  ( \psome{ y_3}; \pclose{y_3}  \| \gsome{\linvar{x}}{\emptyset}; \pnone{\linvar{x}} )
            )
            )
            )\|   \guname{ \unvar{x} }{ x_i } ;  \piencodfaplas{ U }_{x_i}   ) \\
            & \redone^*  \res{  \unvar{x} }  (
            \res{z_1}( \gsome{z_1}{\llfv{M_{1}}};\piencodfaplas{ M_{1} }_{ {z_1}}  \|  \res{z_2} ( \gsome{z_2}{\llfv{M_{2}}};\piencodfaplas{ M_{2} }_{ {z_2}} \| \\
            & \qquad \qquad  \piencodfaplas{ M'}_u \{ z_1 / x_{i_1} \} \{ z_2 / x_{i_2} \} )  )
            \|  \guname{ \unvar{x} }{ x_i } ;  \piencodfaplas{ U }_{x_i} )  \\
        \end{aligned}
        \end{equation*}

        Where $x_{i_1}, x_{i_2}$ is an arbitrary permutation of $x_1,x_2$. On the other hand, we have:

        \begin{equation}\label{ch4eq:compl_lsbeta4failunres}
        \begin{aligned}
            \piencodfaplas{M}_u &= \piencodfaplas{ M' \linexsub{M_1/ {x_1}} \unexsub{U /\unvar{x} } }_u \\
            &=   \res{  \unvar{x} }  (
            \res{z_1}( \gsome{z_1}{\llfv{M_{1}}};\piencodfaplas{ M_{1} }_{ {z_1}}  \| \res{z_2} ( \gsome{z_2}{\llfv{M_{2}}};\piencodfaplas{ M_{2} }_{ {z_2}} \\
        & \quad  \| \bignd_{x_{i_1} \in \{ x_1 , x_2  \}}  \bignd_{x_{i_2} \in \{ x_1 , x_2 \} \setminus \{ x_{i_1}  \}} \piencodfaplas{ M' }_u \{ z_1 / x_{i_1} \} \{ z_2 / x_{i_2} \} )  )
            \|  \guname{ \unvar{x} }{ x_i } ;  \piencodfaplas{ U }_{x_i} )  \\
            & \premat   \res{  \unvar{x} }  (
            \res{z_1}( \gsome{z_1}{\llfv{M_{1}}};\piencodfaplas{ M_{1} }_{ {z_1}}  \| \\
            &\quad   \res{z_2} ( \gsome{z_2}{\llfv{M_{2}}};\piencodfaplas{ M_{2} }_{ {z_2}}           \|  \piencodfaplas{ M' }_u \{ z_1 / x_{i_1} \} \{ z_2 / x_{i_2} \} )  )
            \|  \guname{ \unvar{x} }{ x_i } ;  \piencodfaplas{ U }_{x_i} )
        \end{aligned}
        \end{equation}
        Therefore, by \eqref{ch4eq:compl_lsbeta3failunres}
        and  \eqref{ch4eq:compl_lsbeta4failunres} the result follows.

        \item Case $\redlab{RS{:}Fetch^{\ell}}$:
        Then, we have
        $N = M' \linexsub{C /  x_1 , \cdots , x_k} $ with $\headf{M'} =  {x}_j$, $C = M_1 , \cdots , M_k$ and $N \red  M' \headlin{ M_i / x_j }  \linexsub{(C \setminus M_i ) /  x_1 , \cdots , x_k \setminus x_j }  = M$, for some $M_i \in C$.
        On the one hand, we have:
            {
            \small
            \begin{equation}\label{ch4eq:compl_lsbeta5failunres0}
                \hspace{-1cm}
            \begin{aligned}
            \piencodfaplas{N}_u &= \res{z_1}( \gsome{z_1}{\llfv{M_{1}}};\piencodfaplas{ M_{1} }_{ {z_1}}  \|
              \cdots \res{z_k} ( \gsome{z_k}{\llfv{M_{k}}};\piencodfaplas{ M_{k} }_{ {z_k}} \\
                & \quad \| \bignd_{x_{i_1} \in \{ x_1 ,\cdots , x_k  \}} \cdots \bignd_{x_{i_k} \in \{ x_1 ,\cdots , x_k \setminus x_{i_1} , \cdots , x_{i_{k-1}}  \}} \piencodfaplas{ M' }_u \{ z_1 / x_{i_1} \} \cdots \{ z_k / x_{i_k} \} ) \cdots )
               \\
            &= \res{z_1}( \gsome{z_1}{\llfv{M_{1}}};\piencodfaplas{ M_{1} }_{ {z_1}}  \|
              \cdots \res{z_k} ( \gsome{z_k}{\llfv{M_{k}}};\piencodfaplas{ M_{k} }_{ {z_k}} \\
                & \quad  \| \bignd_{x_{i_1} \in \{ x_1 ,\cdots , x_k  \}} \cdots \bignd_{x_{i_k} \in \{ x_1 ,\cdots , x_k \setminus x_{i_1} , \cdots , x_{i_{k-1}}  \}} \res{ \widetilde{y} }  (\piencodfaplas{  {x}_{j} }_v   \| P) \{ z_1 / x_{i_1} \} \cdots \{ z_k / x_{i_k} \} ) \cdots ) \quad (*)
               \\
            &= \res{z_1}( \gsome{z_1}{\llfv{M_{1}}};\piencodfaplas{ M_{1} }_{ {z_1}}  \|
             \cdots \res{z_k} ( \gsome{z_k}{\llfv{M_{k}}};\piencodfaplas{ M_{k} }_{ {z_k}} \\
                & \quad   \| \bignd_{x_{i_1} \in \{ x_1 ,\cdots , x_k  \}} \cdots \bignd_{x_{i_k} \in \{ x_1 ,\cdots , x_k \setminus x_{i_1} , \cdots , x_{i_{k-1}}  \}} \\
                & \quad  \res{ \widetilde{y} } (\psome{x_j};  \pfwd{x_j}{v}   \| P)  \{ z_1 / x_{i_1} \} \cdots \{ z_k / x_{i_k} \} ) \cdots )
               \\
            \end{aligned}
            \end{equation}
            }
        where $(*)$ is inferred via Proposition~\ref{ch4prop:correctformfailunres}.
        Let us consider the case when $x_{i_k} = x_j$ the other cases proceed similarly. Then, we have  reduction:

        \begin{equation}\label{ch4eq:compl_lsbeta5failunres}
            \begin{aligned}
            &\redone \res{z_1}( \gsome{z_1}{\llfv{M_{1}}};\piencodfaplas{ M_{1} }_{ {z_1}}  \| \\
            & \qquad \cdots \res{z_k} ( \piencodfaplas{ M_{k} }_{ {z_k}}   \|  \res{ \widetilde{y} } (  \pfwd{x_k}{v}   \| P)  \{ z_1 / x_{i_1} \} \cdots \{ z_{k-1} / x_{i_{k-1}} \} ) \cdots )
               \\
            &\redone \res{z_1}( \gsome{z_1}{\llfv{M_{1}}};\piencodfaplas{ M_{1} }_{ {z_1}}  \| \cdots \res{z_{k-1}}( \gsome{z_{k-1}}{\llfv{M_{{k-1}}}};\piencodfaplas{ M_{{k-1}} }_{ {z_{k-1}}}  \| \\
            & \qquad \res{ \widetilde{y} } (  \piencodfaplas{ M_{k} }_{ {v}}   \| P)  \{ z_1 / x_{i_1} \} \cdots \{ z_{k-1} / x_{i_{k-1}} \} ) \cdots )
               \\
            \end{aligned}
            \end{equation}

        for some permutation $ x_{i_1}, \cdots , x_{i_{k-1}} $ of the bag $x_1, \cdots , x_{k-1}$.
         On the other hand, we have:
        \begin{equation}\label{ch4eq:compl_lsbeta6failunres}
        \begin{aligned}
        \piencodfaplas{M}_u &= \piencodfaplas{M' \headlin{ M_k / x_k }  \linexsub{(C \setminus M_k ) /  x_1 , \cdots , x_{k-1} } }_u \\
        & = \res{z_1}( \gsome{z_1}{\llfv{M_{1}}};\piencodfaplas{ M_{1} }_{ {z_1}}  \| \cdots \res{z_{k-1}} ( \gsome{z_{k-1}}{\llfv{M_{k-1}}};\piencodfaplas{ M_{k-1} }_{ {z_{k-1}}} \\
                & \qquad \qquad  \| \bignd_{x_{i_1} \in \{ x_1 ,\cdots , x_{k-1}  \}} \cdots \bignd_{x_{i_{k-1}} \in \{ x_1 ,\cdots , x_{k-1} \setminus x_{i_1} , \cdots , x_{i_{k-2}}  \}} \\
                & \qquad \qquad \res{ \widetilde{y} } (  \piencodfaplas{ M_{k} }_{ {v}}   \| P)  \{ z_1 / x_{i_1} \} \cdots \{ z_{k-1} / x_{i_{k-1}} \} ) \cdots )
               \\
        & \premat  \res{z_1}( \gsome{z_1}{\llfv{M_{1}}};\piencodfaplas{ M_{1} }_{ {z_1}}  \|  \cdots \res{z_{k-1}} ( \gsome{z_{k-1}}{\llfv{M_{k-1}}};\piencodfaplas{ M_{k-1} }_{ {z_{k-1}}} \\
                & \qquad \qquad  \| \res{ \widetilde{y} } (  \piencodfaplas{ M_{k} }_{ {v}}   \| P)  \{ z_1 / x_{i_1} \} \cdots \{ z_{k-1} / x_{i_{k-1}} \} ) \cdots )
               \\
        \end{aligned}
        \end{equation}
                Therefore, by \eqref{ch4eq:compl_lsbeta5failunres}
        and  \eqref{ch4eq:compl_lsbeta6failunres} the result follows.

        \item Case $ \redlab{RS{:} Fetch^!}$:
        Then we have
        $N = M' \unexsub{U / \unvar{x}}$ with $\headf{M'} = \unvar{x}[k]$, $U_k = \unvar{\bag{N}}$ and $N \red  M' \headlin{ N /\unvar{x} }\unexsub{U / \unvar{x}} = M$.
    The result follows easily from
            \begin{equation}\label{ch4eq:compl_lsbeta5failunresunres}
                \small
            \begin{aligned}
            \piencodfaplas{N}_u &= \piencodfaplas{M' \unexsub{U / \unvar{x}}}_u
            =  \res{ \unvar{x} } ( \piencodfaplas{ M' }_u   \|    \guname{ \unvar{x} }{ x_k } ;\piencodfaplas{ U }_{x_k}  ) \\
            & \redone^* \res{ \unvar{x} } (  \res{ \widetilde{y} }  (\piencodfaplas{ \unvar{x}[k] }_{j}   \| P)   \|   \guname{ \unvar{x} }{ x_k } ;  \piencodfaplas{ U }_{x_k}  ) \qquad (*)
            \\
            &  =  \res{ \unvar{x} }   (  \res{ \widetilde{y} }  (\puname{ \unvar{x} }{ x_k }; \psel{ {x}_k }{ k }; \pfwd{x_k}{j}   \| P)   \|   \guname{ \unvar{x} }{ x_k } ;  \piencodfaplas{ U }_{x_k}  )
            \\
            & \redone  \res{ \unvar{x} }  (  \res{ \widetilde{y} }  (   \res{ x_k } ( \psel{ {x}_k }{ k }; \pfwd{x_k}{j}   \| \piencodfaplas{ U }_{x_k})   \| P)   \|   \guname{ \unvar{x} }{ x_k } ;  \piencodfaplas{ U }_{x_k}  )
            \\
            & =  \res{ \unvar{x} }  (  \res{ \widetilde{y} } (   \res{ x_k } ( \psel{ {x}_k }{ k }; \pfwd{x_k}{j}   \| \gsel{x_k}\{i:\piencodfaplas{ U_i }_{x} \}_{U_i \in U} )   \| P)   \|   \guname{ \unvar{x} }{ x_k } ;  \piencodfaplas{ U }_{x_k}  )
            \\
            & \redone  \res{ \unvar{x} }   (  \res{ \widetilde{y} }  (  \piencodfaplas{\unvar{\bag{N}}}_{j} )   \| P)   \|   \guname{ \unvar{x} }{ x_k } ;  \piencodfaplas{ U }_{x_k}  )\\
            &=  \res{ \unvar{x} }   (  \res{ \widetilde{y} } (  \piencodfaplas{N}_{j} )   \| P)   \|   \guname{ \unvar{x} }{ x_k } ;  \piencodfaplas{ U }_{x_k}  ) = \piencodfaplas{M}_u
            \end{aligned}
            \end{equation}
       where the reductions denoted by $(*)$ are inferred via Proposition~\ref{ch4prop:correctformfailunres}.

        \item Cases $\redlab{RS:TCont}$ :
        This case follows by IH.

        \item Case $\redlab{RS{:}Fail^{\ell}}$:
        Then we have
        $N = M'[ {x}_1, \!\cdots\! ,  {x}_k \leftarrow  {x}]\ \esubst{C \bagsep U}{ x } $ with $k \neq \size{C}$ and
        $N \red  \fail^{\widetilde{y}} = M$, where $\widetilde{y} = (\llfv{M'} \setminus \{   {x}_1, \cdots ,  {x}_k \} ) \cup \llfv{C}$. Let $ \size{C} = l$ and we assume that $k > l$ and we proceed similarly for $k > l$. Hence $k = l + m$ for some $m \geq 1$
        {
        \small
        \begin{equation}\label{ch4eq:compl_fail1-failunres}
        \begin{aligned}
            \piencodfaplas{N}_u &= \piencodfaplas{M'[ {x}_1, \!\cdots\! ,  {x}_k \leftarrow  {x}]\ \esubst{C \bagsep U}{ x } }_u
            =    \res{ x }  ( \psome{x}; \gname{x}{\linvar{x}}; \gname{x}{\unvar{x}};  \gclose{ x } ;\piencodfaplas{ M'[\widetilde{x} \leftarrow  {x}]}_u   \| \\
            & \quad \piencodfaplas{ C \bagsep U}_x )
            \\
            & =   \res{  x }  ( \psome{x}; \gname{x}{\linvar{x}}; \gname{x}{\unvar{x}};  \gclose{ x } ;\piencodfaplas{ M'[\widetilde{x} \leftarrow  {x}]}_u   \| \\
            & \quad  \gsome{ x }{ \llfv{C} };  \pname{x}{\linvar{x}} .( \piencodfaplas{ C }_{\linvar{x}}   \| \pname{x}{\unvar{x}} .(  \guname{ \unvar{x} }{ x_i } ;  \piencodfaplas{ U }_{x_i}   \| \pclose{ x } ) ) )
            \\
            & \redone^*    \res{ \unvar{x} } ( \res{\linvar{x}} ( \piencodfaplas{ M'[\widetilde{x} \leftarrow  {x}]}_u   \| \piencodfaplas{ C }_{\linvar{x}} )  \|  \guname{ \unvar{x} }{ x_i } ;  \piencodfaplas{ U }_{x_i}  )\\
            & =    \res{ \unvar{x} } (
            \res{\linvar{x}} (
            \psome{\linvar{x}}; \pname{\linvar{x}}{y_1}; \big( \gsome{y_1}{ \emptyset }; \gclose{ y_{1} } ; \0   \\
            &  \qquad \| \psome{\linvar{x}}; \gsome{\linvar{x}}{u, \llfv{M'} \setminus  \widetilde{x} }; \bignd_{x_{i_1} \in \widetilde{x}} \gname{x}{{x}_{i_1}}; \cdots   \psome{\linvar{x}}; \pname{\linvar{x}}{y_k};   \\
            &  \qquad \big( \gsome{y_k}{ \emptyset }; \gclose{ y_{k} } ; \0 \| \psome{\linvar{x}}; \gsome{\linvar{x}}{u, \llfv{M'} \setminus  \widetilde{x} }; \bignd_{x_{i_k} \in (\widetilde{x} \setminus x_{i_1} , \cdots , x_{i_{k-1}}    )} \gname{x}{{x}_{i_k}}; \\
            &  \qquad \psome{\linvar{x}}; \pname{\linvar{x}}{y_{k+1}}; ( \gsome{y_{k+1}}{ u, \llfv{M'} }; \gclose{ y_{{k+1}} } ;\piencodfaplas{M'}_u \| \pnone{ \linvar{x} } )
            \big) \cdots
            \big)
            \\
            &  \qquad \| \gsome{\linvar{x}}{\llfv{C} }; \gname{x}{y_1}; \gsome{\linvar{x}}{y_1, \llfv{C}}; \psome{\linvar{x}}; \pname{\linvar{x}}{z_1}; \\
            &  \qquad   ( \gsome{z_1}{\llfv{M_1}};  \piencodfaplas{M_1}_{z_1} \| \pnone{y_1}  \| \\
            &  \qquad \cdots \gsome{\linvar{x}}{\llfv{M_l} }; \gname{x}{y_l}; \gsome{\linvar{x}}{y_l, \llfv{C}}; \psome{\linvar{x}}; \pname{\linvar{x}}{z_l}; \cdots\\
            &  \qquad  ( \gsome{z_l}{\llfv{M_l}};  \piencodfaplas{M_l}_{z_l} \| \pnone{y_l}  \| \\
            &  \qquad  \gsome{\linvar{x}}{\emptyset};\gname{x}{y_{l+1}};  ( \psome{ y_{l+1}}; \pclose{y_{l+1}}  \| \gsome{\linvar{x}}{\emptyset}; \pnone{\linvar{x}} )
            ) \cdots
            )
            )
            \|
            \guname{ \unvar{x} }{ x_i } ;  \piencodfaplas{ U }_{x_i}
            ) \\
            &
            (:= P_\mathbb{N})
               \\
            \end{aligned}
        \end{equation}
        }

            we reduce $P_\mathbb{N}$ arbitrarily discarding non-deterministic sums.

            \begin{equation*}
                \small
                \hspace{-1cm}
            \begin{aligned}
            P_\mathbb{N}
            & \redone^* \res{ \unvar{x} } (
            \res{\linvar{x}} ( \res{z_1} ( \gsome{z_1}{\llfv{M_1}};  \piencodfaplas{M_1}_{z_1} \| \cdots  \res{z_k} ( \gsome{z_k}{\llfv{M_k}};  \piencodfaplas{M_k}_{z_k} \| \\
            &  \qquad \psome{\linvar{x}}; \pname{\linvar{x}}{y_{k+1}}; ( \gsome{y_{k+1}}{ u, \llfv{M'} }; \gclose{ y_{{k+1}} } ;\piencodfaplas{M'}_u\{ z_1 / x_{i_1} \} \cdots \{ z_k / x_{i_k} \} \| 
            \\
            &  \qquad \pnone{ \linvar{x} } )) \cdots   ) \| \gsome{\linvar{x}}{\llfv{C} \setminus (M_1 , \cdots , M_k) }; \gname{x}{y_{k+1}}; \gsome{\linvar{x}}{y_{k+1}, \llfv{C}\setminus (M_1 , \cdots , M_k)}; \psome{\linvar{x}};  \\
            & \qquad  \pname{\linvar{x}}{z_{k+1}}; ( \gsome{z_{k+1}}{\llfv{M_{k+1}}};  \piencodfaplas{M_{k+1}}_{z_{k+1}} \| \pnone{y_1}  \| \\
            & \qquad \cdots \gsome{\linvar{x}}{\llfv{M_l} }; \gname{x}{y_l}; \gsome{\linvar{x}}{y_l, \llfv{M_l}}; \psome{\linvar{x}}; \pname{\linvar{x}}{z_l}; \cdots\\
            & \qquad  ( \gsome{z_l}{\llfv{M_l}};  \piencodfaplas{M_l}_{z_l} \| \pnone{y_l}  \| \\
            & \qquad  \gsome{\linvar{x}}{\emptyset};\gname{x}{y_{l+1}};  ( \psome{ y_{l+1}}; \pclose{y_{l+1}}  \| \gsome{\linvar{x}}{\emptyset}; \pnone{\linvar{x}} )
            ) \cdots
            )
            )
            \|
            \guname{ \unvar{x} }{ x_i } ;  \piencodfaplas{ U }_{x_i}
            )
            \\
            & \redone^* \res{ \unvar{x} } (
              \res{z_1} ( \gsome{z_1}{\llfv{M_1}};  \piencodfaplas{M_1}_{z_1} \| \cdots  \res{z_k} ( \gsome{z_k}{\llfv{M_k}};  \piencodfaplas{M_k}_{z_k} \| \\
            & \qquad   \pnone{u} \| \pnone{(\llfv{M'} \setminus \widetilde{x})} \|  \pnone{z_1} \| \cdots \| \pnone{z_k} \| \pnone{( \llfv{C} \setminus (M_1 , \cdots , M_k) )}  ) \cdots   )
            \\
            & \qquad \|
            \guname{ \unvar{x} }{ x_i } ;  \piencodfaplas{ U }_{x_i}
            )
            \\
            & \redone^* \res{ \unvar{x} } ( ( \pnone{u} \| \pnone{(\llfv{M'} \setminus \widetilde{x})} \|   \| \pnone{ \llfv{C} }  )     \|
            \guname{ \unvar{x} }{ x_i } ;  \piencodfaplas{ U }_{x_i}
            )
        \\
        & \equiv  \pnone{u} \| \pnone{(\llfv{M'} \setminus \widetilde{x})} \|   \| \pnone{ \llfv{C} }=   \piencodfaplas{M}_u   \\
        \end{aligned}
        \end{equation*}

       \item Case $\redlab{RS{:}Fail^!}$:         Then we have
        $N = M' \unexsub{U /\unvar{x}}  $ with $\headf{M'} =  {x}[i]$, $U_i = \unvar{\oneb} $ and
        $N \red   M' \headlin{ \fail^{\emptyset} /\unvar{x} } \unexsub{U /\unvar{x} } $.
The result follows easily from

        \begin{equation}\label{ch4eq:compl_unfail-failunres}
            \small
        \begin{aligned}
            \piencodfaplas{N}_u &= \piencodfaplas{ M' \unexsub{U /\unvar{x}} }_u  =   \res{ \unvar{x} }  ( \piencodfaplas{ M' }_u   \|   \guname{ \unvar{x} }{ x_i } ;  \piencodfaplas{ U }_{x_i}  )  \\
            & \redone^* \res{ \unvar{x} }  (  \res{ \widetilde{y} } (\piencodfaplas{  {x}[i] }_{j}   \| P)   \|   \guname{ \unvar{x} }{ x_k } ;  \piencodfaplas{ U }_{x_k}  ) \qquad (*)
            \\
            &  =  \res{ \unvar{x} }   (  \res{ \widetilde{y} } (\puname{ \unvar{x} }{ x_k };\psel{ {x}_k }{ i }; \pfwd{x_k}{j}   \| P)   \|   \guname{ \unvar{x} }{ x_k } ;  \piencodfaplas{ U }_{x_k}  ) \qquad (*)
            \\
            & \redone  \res{ \unvar{x} }  (  \res{ \widetilde{y} }(   \res{ x_k } ( x_k.l_{i}; \pfwd{x_k}{j}   \| \piencodfaplas{ U }_{x_k})   \| P)   \|   \guname{ \unvar{x} }{ x_k } ;  \piencodfaplas{ U }_{x_k}  )
            \\
            & =  \res{ \unvar{x} } (  \res{ \widetilde{y} } (   \res{  x_k }  ( x_k.l_{i}; \pfwd{x_k}{j}   \| x_k. case( i.\piencodfaplas{U_i}_{x} ))   \| P)   \|   \guname{ \unvar{x} }{ x_k } ;  \piencodfaplas{ U }_{x_k}  )
            \\
            & \redone  \res{ \unvar{x} }  (  \res{ \widetilde{y} }  (  \piencodfaplas{\unvar{\oneb}}_{j}    \| P)   \|   \guname{ \unvar{x} }{ x_k } ; \piencodfaplas{ U }_{x_k}  )
           =  \res{ \unvar{x} }  (  \res{ \widetilde{y} }  (   \pnone{ j }    \| P)   \|   \guname{ \unvar{x} }{ x_k } ;  \piencodfaplas{ U }_{x_k}  ) \\
           & = \piencodfaplas{M}_u
        \end{aligned}
        \end{equation}

        \item Case $\redlab{RS:Cons_1}$:
        Then we have
        $N = \fail^{\widetilde{x}}\ C \bagsep U$ and $N \red \fail^{\widetilde{x} \cup \widetilde{y}}  = M$ where $ \widetilde{y} = \llfv{C}$. The result follows easily from

        \begin{equation}\label{ch4eq:compl_cons1-failunres}
        \begin{aligned}
            \piencodfaplas{N}_u &= \piencodfaplas{ \fail^{\widetilde{x}}\ C \bagsep U }_u=   \res{ v }  (\piencodfaplas{\fail^{\widetilde{x}}}_v   \|  \gsome{ v }{ u , \llfv{C} };  \pname{v}{x} . (\pfwd{v}{u}   \| \piencodfaplas{C \bagsep U}_x ) ) \\
            &=  \res{ v } (   \pnone{ v }    \|   \pnone{ \widetilde{x} }    \|  \gsome{ v }{ u , \llfv{C} };  \pname{v}{x} . (\pfwd{v}{u}   \| \piencodfaplas{C \bagsep U}_x ) ) \\
            & \redone    \pnone{ u }   \|   \pnone{ \widetilde{x} }    \|  \pnone{ \widetilde{y} }= \piencodfaplas{M}_u   \\
        \end{aligned}
        \end{equation}

        \item Cases $\redlab{RS:Cons_2}$ and $\redlab{RS:Cons_3}$: These cases follow by IH similarly to Case 7.

        \item Case $\redlab{RS{:}Cons_4}$:
        Then we have
        $N =  \fail^{\widetilde{y}} \unexsub{U / \unvar{x}} $ and $N \red \fail^{\widetilde{y}}  = M$. The result follows easily from

        \begin{align}\label{ch4eq:compl_cons4-failunres}
            \piencodfaplas{N}_u &= \piencodfaplas{ \fail^{\widetilde{y}} \unexsub{U / \unvar{x}}}_u
            =  \res{ \unvar{x} }  ( \piencodfaplas{ \fail^{\widetilde{y}} }_u   \|   \guname{ \unvar{x} }{ x_i } ;  \piencodfaplas{ U }_{x_i} )  \\
            &=  \res{ \unvar{x} } (   \pnone{ u }    \|   \pnone{ \widetilde{x} }    \|    \guname{ \unvar{x} }{ x_i } ; \piencodfaplas{ U }_{x_i} )
             \equiv    \pnone{ u }   \|   \pnone{ \widetilde{x} }=   \piencodfaplas{M}_u
         \end{align}

    \end{enumerate}

\end{proof}

\subsection{Soundness} \label{ch4a:loossoundness}

We define $P \red \{P_i\}_{i \in I}$, for a fixed finite set $I$ where $ I = \{ i \ s.t. \  P \red P_i \}$. Similarly we define $P \red^* \{P_i\}_{i \in I}$ to be defined inductively by $P \red^* \{P_i\}_{i \in I} $ and $P_i \red \{P_j\}_{j \in J_i}$ for each $i \in I$ then $P \red^* \{P_j\}_{j \in J} $ with $J = \cup_{ i \in I} J_i  $

\propSoundextra*

\begin{proof}

Proof by induction on the precongruence rules.

\begin{itemize}
    \item When $P =P \premat P = Q$ then all reductions in $P$ are matched in $Q$.

    \item When $ P = P_1 \nd P_2 \premat Q$ with $P_i \premat Q \quad i \in \{ 1 , 2 \} $. Let us take $i = 1$ Then by the rule:

    \begin{prooftree}
            \infAss{
                $P_1\redone P_1' $
            }
            \infUn{
                $P_1 \nd P_2 \redone P_1' \nd P_2 $
            }{$\rredone{\nd}$}
    \end{prooftree}

    we have $P_1 \nd P_2 \redone \{P'_{1_i}\}_{i \in I_1}  \nd P_2 $  and $P_1 \nd P_2\redone P_1 \nd \{ P'_{2_i} \}_{i \in I_2} $
    for some $I_1, I_2$ Hence we have that $ P_1 \nd P_2\redone \{ P'_i \}_{i \in I_1 \cup I_2}$ where $P_i' = P'_{1_i} \nd P_2$ if $i \in I_1$ and $P_i' = P_{1} \nd P_{2_i}'$ if $i \in I_2$ . By the induction hypothesis $P_1 \premat Q$ and $P_1\redone \{ P_{1_i} \}_{i \in I_1}' $ imply $\exists J', \{ Q_j \}_{j \in J'} \ s.t. \  Q \redone^* \{Q_j\}_{j \in J'} $ , $ J' \subseteq I_1 $ and $P_j \premat Q_j \  , \ \forall j \in J'$. We take $J = J'$ and hence we can deduce $ P \redone \{P'_{i}\}_{i \in I_1 \cup I_2}$ , $Q \redone^* \{Q_j\}_{j \in J}$. Finally we have that $\forall j \in J$.

    \begin{prooftree}
        \AxiomC{$ P_{1_j}' \premat Q_j  $}
        \UnaryInfC{$ P_{1_j}' \nd P_2 \premat Q$}
    \end{prooftree}

    \item When $P = P_1   \| P_2  $ with $ P_1   \| P_2 \premat Q $. Then by the rule:

        \begin{prooftree}
            \infAss{
                $P_1\redone P_1'$
            }
            \infUn{
                $P_1   \| P_2\redone P_1'   \| P_2 $
            }{$\rredone{  \| }$}
        \end{prooftree}

    we have $P_1   \| P_2\redone \{P'_{1_{i}}\}_{i  \in I_1}   \| P_2 $  and $P_1   \| P_2\redone P_1   \| \{P'_{2_{i}}\}_{i \in I_1} $ for some $I_1, I_2$ Hence we have that $ P_1   \| P_2\redone \{P'_i \}_{i \in I_1 \cup I_2}$ where $P_i' = P_{1_i}'   \| P_2$ if $i \in I_1$ and $P_i' = P_{1}   \| P'_{2_i}$. By the induction hypothesis $P_1   \| P_2 \premat Q$
    and $P_1\redone \{P'_{1_i} \}_{i \in I_1} $ imply $\exists J_1', \{Q_j\}_{j \in J'_1} \ s.t. \  Q \redone^* \{Q_j\}_{j \in J_1'} $ ,
    $ J_1' \subseteq I_1 $ and $P'_{1_j} \premat Q_j \  , \ \forall j \in J_1'$. Similarly we have that $P_2\redone \{P'_2\}_{i \in I_2} $ imply $\exists J_2', \{Q_j\}_{j \in J'_2} \ s.t. \  Q \redone^* \{Q_j\}_{j \in J_2'} $ , $ J_2' \subseteq I_2 $ and $P_{2_j} \premat Q_j \  , \ \forall j \in J_2'$.
    We take $Q_{j \in J} = P_{1}   \| \{P'_{2_{i}}\}_{i \in I_2} \cup \{P'_{1_{i}}\}{i \in I_1}   \| P_{2}$ and hence we can deduce $ P\redone \{P'_{i}\}_{i \in I_1 \cup I_2}$ , $Q \redone^* \{Q_j\}_{j \in J}$. Finally we have that $\forall j \in J$ $ P_j \premat Q_j $

    \item When $ P =  \res{  x }  P_1 $ with $   \res{  x }   P_1 \premat Q $. Then by the rule:

        \begin{prooftree}
            \infAss{
                $P_1\redone P'$
            }
            \infUn{
                $ \res{ x }  P_1\redone  \res{ x }  P'$
            }{$\rredone{ \nu }$}
        \end{prooftree}

    we have $ \res{ x }  P_1 \redone  \res{ x }   \{P'_i\}_{{i \in I}} $ for some $I$. By the induction hypothesis $ \res{ x }   P_1 \ \premat Q$
    and $ \res{ x }  P_1\redone  \res{ x }  \{P'_i\}_{{i \in I}} $ imply $\exists J', \{Q_j\}_{j \in J'} \ s.t. \  Q \redone^* \{Q_j\}_{j \in J'} $ ,
    $ J' \subseteq I $ and $P'_{j} \premat Q_j \  , \ \forall j \in J'$.
     We take $J = J'$ and hence we can deduce $ P\redone \{P'_i\}_{i \in I}$ , $Q \redone^* Q_{j \in J}$. Finally we have that $\forall j \in J$

    \begin{prooftree}
            \AxiomC{$ P_j' \premat Q_j $}
            \UnaryInfC{$  \res{ x }   P_j' \premat  \res{ x }   Q_j  $}
    \end{prooftree}

\end{itemize}

\end{proof}

\thmOpsoundone*

\begin{proof}
By induction on the structure of $N $ and then induction on the number of reductions of $\piencodfaplas{N} \redone^* Q$.

\begin{enumerate}
    \item {\bf Base case:} $N =  {x}$, $N =  {x}[j]$, $N = \fail^{\emptyset}$ and $N = \lambda x . (M'[ {\widetilde{x}} \leftarrow  {x}])$.
.

    No reductions can take place, and the result follows trivially. Take $I = \{ a \}$,

    $Q =  \piencodfaplas{N}_u \redone^0 \piencodfaplas{N}_u = Q_a$ and $ {x} \red^0  {x} = N'$.

    \item $N =  M'(C \bagsep U) $.

        Then,
        $ \piencodfaplas{M'(C \bagsep U)}_u =   \res{ v }  (\piencodfaplas{M'}_v   \|  \gsome{ v }{ u , \llfv{C} };  \pname{v}{x} . (\pfwd{v}{u}   \| \piencodfaplas{C \bagsep U}_x ) )$, and we are able to perform the  reductions from $\piencodfaplas{M'(C \bagsep U)}_u$.

        We now proceed by induction on $k$, with  $\piencodfaplas{N}_u \redone^k Q$. There are two main cases:
        \begin{enumerate}
        \item When $k = 0$ the thesis follows easily:

            We have $i = \{ a \} $,
    $Q =  \piencodfaplas{M'(C \bagsep U)}_u \redone^0 \piencodfaplas{M'(C \bagsep U)}_u = Q_a$ and $M'(C \bagsep U) \red^0 M'(C \bagsep U) = N'$.

            \item The interesting case is when $k \geq 1$.

            Then, for some process $R$ and $n, m$ such that $k = n+m$, we have the following:
            \[
            \begin{aligned}
               \piencodfaplas{N}_u & =    \res{ v } (\piencodfaplas{M'}_v   \|  \gsome{ v }{ u , \llfv{C} }; \pname{v}{x} . (\pfwd{v}{u}   \| \piencodfaplas{C \bagsep U}_x ) )\\
               & \redone^m   \res{ v }  (R   \|  \gsome{ v }{ u , \llfv{C} };  \pname{v}{x} . (\pfwd{v}{u}   \| \piencodfaplas{C \bagsep U}_x ) ) \redone^n  Q\\
            \end{aligned}
            \]
            Thus, the first $m \geq 0$ reduction steps are  internal to $\piencodfaplas{ M'}_v$; type preservation in \clpi ensures that, if they occur,  these reductions  do not discard the possibility of synchronizing with $\psome{v}$. Then, the first of the $n \geq 0$ reduction steps towards $Q$ is a synchronization between $R$ and $ \gsome{ v }{ u, \llfv{C} }$.

            We consider two sub-cases, depending on the values of  $m$ and $n$:

\noindent{\bf (b.1) Case $m = 0$ and $n \geq 1$:}

Then $R = \piencodfaplas{M}_v$ as $\piencodfaplas{M}_v \redone^0 \piencodfaplas{M}_v$.
 Notice that there are two possibilities of having an unguarded:

\begin{enumerate}
\item $M'=  (\lambda x . (M''[ {\widetilde{x}} \leftarrow  {x}])) \linexsub{C_1 / \widetilde{y_1}} \cdots \linexsub{C_p / \widetilde{y_p}} \unexsub{U_1 / \unvar{z}_1} \cdots \unexsub{U_q / \unvar{z}_q} $  $ (p, q \geq 0)$
   \[
   \begin{aligned}
   \piencodfaplas{M'}_v &= \piencodfaplas{ (\lambda x . (M''[ {\widetilde{x}} \leftarrow  {x}])) \linexsub{C_1 / \widetilde{y_1}} \cdots \linexsub{C_p / \widetilde{y_p}} \unexsub{U_1 / \unvar{z}_1} \cdots \unexsub{U_q / \unvar{z}_q} }_v \\
          \end{aligned}
    \]

    For simplicity we shall denote
    \[\piencodfaplas{M' }_v = \res{ \widetilde{y} ,\widetilde{z} } ( \piencodfaplas{\lambda x . (M''[ {\widetilde{x}} \leftarrow  {x}])}_v   \| Q'' )\]
     where $\widetilde{y} = \widetilde{y_1} , \cdots , \widetilde{y_p}$. $\widetilde{z} = \unvar{z}_1, \cdots ,\unvar{z}_q$ and we continue the evaluation as:

      \[
      \begin{aligned}
      &=  \res{ \widetilde{y} ,\widetilde{z} } ( \piencodfaplas{\lambda x . (M''[ {\widetilde{x}} \leftarrow  {x}])}_v   \| Q'' )\\
      &=  \res{ \widetilde{y},\widetilde{z} }  ( \psome{v}; \gname{v}{x}; \psome{x}; \gname{x}{\linvar{x}}; \gname{x}{\unvar{x}}; \gclose{ x } ;  \piencodfaplas{M''[ {\widetilde{x}} \leftarrow  {x}]}_v   \| Q'' )
        \end{aligned}
        \]
  \noindent

With this shape for $M$, we then have the following:
 \[
    \small
 \begin{aligned}
 \piencodfaplas{N}_u & = \piencodfaplas{(M'\ B)}_u\\
 &=   \res{ v }  (\piencodfaplas{M'}_v   \|  \gsome{ v }{ u , \llfv{C} }; \pname{v}{x} . (\pfwd{v}{u}   \| \piencodfaplas{C \bagsep U}_x ) )\\
 & \redone  \res{ v  }( \res{ \widetilde{y},\widetilde{z}} (  \gname{v}{x}; \psome{x}; \gname{x}{\linvar{x}}; \gname{x}{\unvar{x}}; \gclose{ x } ;  \piencodfaplas{M''[ {\widetilde{x}} \leftarrow  {x}]}_v \\
 & \hspace{.5cm}   \| Q'' )   \| \pname{v}{x} . (\pfwd{v}{u}   \| \piencodfaplas{C \bagsep U}_x ) ) & = Q_1 \\
& \redone  \res{ x}  ( \res{ v }(  \res{ \widetilde{y},\widetilde{z}}( \psome{x}; \gname{x}{\linvar{x}}; \gname{x}{\unvar{x}}; \gclose{ x } ;  \piencodfaplas{M''[ {\widetilde{x}} \leftarrow  {x}]}_v  \\
 & \hspace{.5cm}  \| Q'' )   \|  \pfwd{v}{u}  ) \| \piencodfaplas{C \bagsep U}_x ) & = Q_2 \\
 & \redone   \res{ x } ( \res{ \widetilde{y},\widetilde{z}}(\psome{x}; \gname{x}{\linvar{x}}; \gname{x}{\unvar{x}};  \gclose{ x } ; \piencodfaplas{M''[ {\widetilde{x}} \leftarrow  {x}]}_u   \| Q'')\\
 & \hspace{.5cm}  \|  \piencodfaplas{C \bagsep U}_x ) & = Q_3 \\
\end{aligned}
     \]
We also have that
\[
    \small
\begin{aligned}
    N &=(\lambda x . (M''[ {\widetilde{x}} \leftarrow  {x}])) \linexsub{C_1 /  \widetilde{y_1}} \cdots \linexsub{C_p / \widetilde{y_p}} \unexsub{U_1 / \unvar{z}_1} \cdots \unexsub{U_q / \unvar{z}_q} (C \bagsep U) \\
    & \equivlam (\lambda x . (M''[ {\widetilde{x}} \leftarrow  {x}]) (C \bagsep U)) \linexsub{C_1 / \widetilde{y_1}} \cdots \linexsub{C_p / \widetilde{y_p}} \unexsub{U_1 / \unvar{z}_1} \cdots \unexsub{U_q / \unvar{z}_q} \\
    & \red   M''[ {\widetilde{x}} \leftarrow  {x}] \esubst{(C \bagsep U)}{x} \linexsub{C_1 / \widetilde{y_1}} \cdots \linexsub{C_p / \widetilde{y_p}} \unexsub{U_1 / \unvar{z}_1} \cdots \unexsub{U_q / \unvar{z}_q} = M
\end{aligned}
\]
Furthermore, we have:
\[\hspace{-1cm}
\begin{aligned}
\small
&\piencodfaplas{M}_u = \piencodfaplas{M''[ {\widetilde{x}} \leftarrow  {x}] \esubst{(C \bagsep U)}{x} \linexsub{C_1 / \widetilde{y_1}} \cdots \linexsub{C_p / \widetilde{y_p}} \unexsub{U_1 / \unvar{z}_1} \cdots \unexsub{U_q / \unvar{z}_q}}_u \\
& =  \res{  x } ( \res{ \widetilde{y},\widetilde{z} } ( \psome{x}; \gname{x}{\linvar{x}}; \gname{x}{\unvar{x}}; \gclose{ x } ;  \piencodfaplas{M''[ {\widetilde{x}} \leftarrow  {x}]}_u  \| Q'' )  \|  \piencodfaplas{C \bagsep U}_x    )
\end{aligned}
\]

We consider different possibilities for $n \geq 1$; in all  the cases, the result follows.
        \smallskip

 \noindent  {\bf When $n = 1$:}

We have $I = \{a \}$, $Q = Q_1$, $ \piencodfaplas{N}_u \redone^1 Q_1$.
We also have that
\begin{itemize}
\item  $Q_1 \redone^2 Q_3 = Q_a$ ,
\item {\small$N \red^1 M''[ {\widetilde{x}} \leftarrow  {x}] \esubst{(C \bagsep U)}{x} \linexsub{C_1 / \widetilde{y_1}} \cdots \linexsub{C_p / \widetilde{y_p}} \unexsub{U_1 / \unvar{z}_1} \cdots \unexsub{U_q / \unvar{z}_q} $  $= N'$}
\item and {\small$\piencodfaplas{M''[ {\widetilde{x}} \leftarrow  {x}] \esubst{(C \bagsep U)}{x} \linexsub{C_1 / \widetilde{y_1}} \cdots \linexsub{C_p / \widetilde{y_p}} \unexsub{U_1 / \unvar{z}_1} \cdots \unexsub{U_q / \unvar{z}_q}}_u $  $= Q_3$.}
\end{itemize}

                    \smallskip

 \noindent  {\bf When $n = 2,3$:} the analysis is similar.

\noindent {\bf When $n \geq 4$:}

 We have $ \piencodfaplas{N}_u \redone^3 Q_3 \redone^*_{} Q$. We also know that $N \red M$, $Q_3 = \piencodfaplas{M}_u$. By the IH, there exist $\{Q_i\}_{i \in I}$ , $ N'$ such that $Q \redone^* \{ Q_i \}_{i \in I}$, $M \red^* N'$ and $\piencodfaplas{N'}_u \premat Q_i \quad \forall i \in I $ . Finally, $\piencodfaplas{N}_u \redone^3 Q_3 \redone^l Q \redone_I Q_i$ and $N \red M  \red^* N'$.

\item $M'= \fail^{\widetilde{z}}$.

Then,                     \(
\begin{aligned}
\piencodfaplas{M'}_v &= \piencodfaplas{\fail^{\widetilde{z}}}_v =   \pnone{ v }    \|   \pnone{ \widetilde{z} } .
\end{aligned}
\)
With this shape for $M$, we have:

\[
\begin{aligned}
\piencodfaplas{N}_u & = \piencodfaplas{(M'\ (C \bagsep U))}_u\\
&=   \res{ v } (\piencodfaplas{M'}_v   \|  \gsome{ v }{ u , \llfv{C} };  \pname{v}{x} . (\pfwd{v}{u}   \| \piencodfaplas{C \bagsep U}_x ) )\\
& =  \res{  v }  (  \pnone{ v }    \|   \pnone{ \widetilde{z} }    \|  \gsome{ v }{ u , \llfv{C} };  \pname{v}{x} . (\pfwd{v}{u}   \| \piencodfaplas{C \bagsep U}_x ) )\\
& \redone   \pnone{ u }     \|  \pnone{ \widetilde{z} }      \|   \pnone{ \llfv{C} }  \\
\end{aligned}
\]

\end{enumerate}

We also have that
\(  N = \fail^{\widetilde{x}}\ C \bagsep U \red  \fail^{\widetilde{x} \cup \llfv{C}}  = N'.  \)
Furthermore,
\[
\begin{aligned}
\piencodfaplas{N'}_u &= \piencodfaplas{ \fail^{\widetilde{z} \cup \llfv{C}  } }_u
= \piencodfaplas{ \fail^{\widetilde{z} \cup \llfv{C} }}_u
=    \pnone{ u }   \|   \pnone{ \widetilde{z} }    \|    \pnone{ \llfv{C} } .
\end{aligned}
\]

\noindent{ \bf (b.2) Case  $m \geq 1$ and $ n \geq 0$:}

We distinguish two cases:

 \begin{enumerate}
\item When $n = 0$:

Then, $  \res{ v }  (R   \|  \gsome{ v }{ u , \llfv{C} };  \pname{v}{x} . (\pfwd{v}{u}   \| \piencodfaplas{C \bagsep U}_x ) ) =  Q $ and $\piencodfaplas{M'}_u \redone^m R$ where $m \geq 1$. Then by the IH there exist  $\{R_i'\}_{i \in I}$  and $M'' $ such that $R \redone^* \{ R_i' \}_{i \in I}$, $M'\red^* M''$, and $\piencodfaplas{M''}_u \premat R_i \quad \forall i \in I $.  Hence we have that

            \[
            \begin{aligned}
                \piencodfaplas{N}_u & =   \res{ v } (\piencodfaplas{M'}_v   \|  \gsome{ v }{ u , \llfv{C} };  \pname{v}{x} . (\pfwd{v}{u}   \| \piencodfaplas{C \bagsep U}_x ) )\\
                    & \redone^m   \res{ v }  (R   \|  \gsome{ v }{ u , \llfv{C} };  \pname{v}{x} . (\pfwd{v}{u}   \| \piencodfaplas{C \bagsep U}_x ) )  = Q
            \end{aligned}
             \]
            We also know that
            \[
            \begin{aligned}
              Q & \redone^* \{  \res{ v } (R'_i   \|  \gsome{ v }{ u , \llfv{C} };  \pname{v}{x} . (\pfwd{v}{u}   \| \piencodfaplas{C \bagsep U}_x ) ) \}_{i \in I} = \{ Q_i \}_{i \in I}\\
            \end{aligned}
            \]

            and so the \lamcoldetsh term can reduce as follows: $N = (M'\ ( C \bagsep U )) \red^* M''\ ( C \bagsep U ) = N'$ and  $\piencodfaplas{N'}_u \premat Q_i \quad \forall i \in I$ via the $\premat$ rules.

                \item When $n \geq 1$:

                    Then $R$ has an occurrence of an unguarded $\psome{v}$ or $  \pnone{ v } $, hence it is of the form

                    $ \piencodfaplas{(\lambda x . (M''[ {\widetilde{x}} \leftarrow  {x}]))  \linexsub{N_1 / y_1} \cdots \linexsub{N_p / y_p} \unexsub{U_1 / \unvar{z}_1} \cdots \unexsub{U_q / \unvar{z}_q} }_v $ or $ \piencodfaplas{\fail^{\widetilde{x}}}_v. $

                    \end{enumerate}

        \end{enumerate}

        This concludes the analysis for the case $N = (M'\, ( C \bagsep U ))$.

        \item $N = M'[ {\widetilde{x}} \leftarrow  {x}]$.
    The sharing variable $ {x}$ is not free and the result follows by vacuity.

        \item $N = M'[ {\widetilde{x}} \leftarrow  {x}] \esubst{ C \bagsep U }{ x}$. Then we have
            \[
                \begin{aligned}
                    \piencodfaplas{N}_u &=\piencodfaplas{ M'[ {\widetilde{x}} \leftarrow  {x}] \esubst{ C \bagsep U }{ x} }_u \\
                    &=  \res{ x } ( \psome{x}; \gname{x}{\linvar{x}}; \gname{x}{\unvar{x}};  \gclose{ x } ;\piencodfaplas{ M'[ {\widetilde{x}} \leftarrow  {x}]}_u   \| \piencodfaplas{ C \bagsep U}_x )
                \end{aligned}
            \]
            Let us consider three cases.

            \begin{enumerate}
                \item When $\size{ {\widetilde{x}}} = \size{C}$.
                    Then let us consider the shape of the bag $ C$.

  \begin{enumerate}
  \item When $C = \oneb$.

  We have the following
 \[
    \small\hspace{-2cm}
 \begin{aligned}
 \piencodfaplas{N}_u  &=   \res{ x } ( \psome{x}; \gname{x}{\linvar{x}}; \gname{x}{\unvar{x}};  \gclose{ x } ;\piencodfaplas{ M'[ \leftarrow  {x}]}_u   \| \piencodfaplas{ \oneb \bagsep U}_x )\\
 &=   \res{ x }  ( \psome{x}; \gname{x}{\linvar{x}}; \gname{x}{\unvar{x}}; \gclose{ x } ; \piencodfaplas{ M'[ \leftarrow  {x}]}_u   \|  \gsome{ x }{ \llfv{C} };  \pname{x}{\linvar{x}} . \\
 & \hspace{1cm}( \piencodfaplas{ \oneb }_{\linvar{x}}   \|\pname{x}{\unvar{x}} .(  \guname{ \unvar{x} }{ x_i } ;  \piencodfaplas{ U }_{x_i}   \|  \pclose{ x } ) ) )\\
&\redone  \res{ x }  (  \gname{x}{\linvar{x}}; \gname{x}{\unvar{x}}; \gclose{ x } ; \piencodfaplas{ M'[ \leftarrow  {x}]}_u   \|
 \pname{x}{\linvar{x}} .( \piencodfaplas{ \oneb }_{\linvar{x}}   \| \pname{x}{\unvar{x}} .  (  \guname{ \unvar{x} }{ x_i } ;  \piencodfaplas{ U }_{x_i}  \|  \pclose{ x } ) ) ) & = Q_1
                              \\
  &\redone  \res{ x,\linvar{x} } (  \gname{x}{\unvar{x}}; \gclose{ x } ; \piencodfaplas{ M'[ \leftarrow  {x}]}_u   \| \piencodfaplas{ \oneb }_{\linvar{x}}   \|\pname{x}{\unvar{x}} .(  \guname{ \unvar{x} }{ x_i } ;  \piencodfaplas{ U }_{x_i}   \|  \pclose{ x } ) ) & = Q_2
                              \\
  &\redone  \res{ x,\linvar{x}, \unvar{x} } (  \gclose{ x } ; \piencodfaplas{ M'[ \leftarrow  {x}]}_u   \| \piencodfaplas{ \oneb }_{\linvar{x}}   \|  \guname{ \unvar{x} }{ x_i } ;  \piencodfaplas{ U }_{x_i}   \|  \pclose{ x } ) & = Q_3   \\
 &\redone  \res{ \linvar{x}, \unvar{x} } (  \piencodfaplas{ M'[ \leftarrow  {x}]}_u   \| \piencodfaplas{ \oneb }_{\linvar{x}}   \|  \guname{ \unvar{x} }{ x_i } ;  \piencodfaplas{ U }_{x_i} ) & = Q_4\\
    & =  \res{ \linvar{x}, \unvar{x} } ( \psome{\linvar{x}}; \pname{\linvar{x}}{y_i} . (  \gsome{ y_i }{ u,\llfv{M'} };  \gclose{ y_{i} } ; \piencodfaplas{M'}_u   \|   \pnone{ \linvar{x} } )   \| \\
    & \qquad  \gsome{ \linvar{x} }{ \emptyset };  \gname{\linvar{x}}{y_n}; ( \psome{y_n}; \pclose{ y_n }    \|  \gsome{ \linvar{x} }{ \emptyset };   \pnone{ \linvar{x} } )    \|  \guname{ \unvar{x} }{ x_i } ;  \piencodfaplas{ U }_{x_i} )
                            \\
      & \redone  \res{ \linvar{x}, \unvar{x} }  (  \pname{\linvar{x}}{y_i} . ( \gsome{  y_i }{ u,\llfv{M'} };  \gclose{ y_{i} } ; \piencodfaplas{M'}_u   \|   \pnone{ \linvar{x} } )   \| \\
                              & \qquad \gname{\linvar{x}}{y_n}; ( \psome{y_n}; \pclose{ y_n }    \|  \gsome{ \linvar{x} }{ \emptyset };   \pnone{  \linvar{x} } )    \|  \guname{ \unvar{x} }{ x_i } ;  \piencodfaplas{ U }_{x_i} )  & = Q_5
                            \\
                            & \redone  \res{ \linvar{x}, \unvar{x} , y_i } (   \gsome{ y_i }{ u,\llfv{M'} }; \gclose{ y_{i} } ; \piencodfaplas{M'}_u   \|   \pnone{ \linvar{x} }    \|  \psome{y_i}; \pclose{ y_i }   \\
 & \hspace{1cm}  \|  \gsome{ \linvar{x} }{ \emptyset };   \pnone{ \linvar{x} }     \|  \guname{ \unvar{x} }{ x_i } ;  \piencodfaplas{ U }_{x_i} )  & = Q_6
                            \\
                            & \redone  \res{ \linvar{x}, \unvar{x} , y_i }  (  \gclose{ y_{i} } ;  \piencodfaplas{M'}_u   \|   \pnone{  \linvar{x} }   \|   \pclose{ y_i }    \|  \gsome{ \linvar{x} }{ \emptyset };   \pnone{ \linvar{x} }    \|  \guname{ \unvar{x} }{ x_i } ;  \piencodfaplas{ U }_{x_i} )  & = Q_7
                            \\
                            & \redone  \res{ \linvar{x}, \unvar{x} } (  \piencodfaplas{M'}_u   \|   \pnone{ \linvar{x} }    \|   \gsome{ \linvar{x} }{ \emptyset };    \pnone{ \linvar{x} }     \|  \guname{ \unvar{x} }{ x_i } ;  \piencodfaplas{ U }_{x_i} )  & = Q_8
                            \\
                            & \redone  \res{ \unvar{x} }  (  \piencodfaplas{M'}_u   \|  \guname{ \unvar{x} }{ x_i } ;  \piencodfaplas{ U }_{x_i} )
                            =  \piencodfaplas{M'\unexsub{U / \unvar{x}}}_u
                            & = Q_9
                            \end{aligned}
                            \]
                            Notice how $Q_8$ has a choice however the $\linvar{x}$ name can be closed at any time so for simplicity we only perform communication across this name once all other names have completed their reductions.

                        Now we proceed by induction on the number of reductions $\piencodfaplas{N}_u \red^k Q$.

                            \begin{enumerate}

                                \item When $k = 0$, the result follows trivially. Just take $I = \{a\}$, $\mathbb{N}=\mathbb{N}'$ and $\piencodfaplas{N}_u=Q=Q_a$.

                                \item When $k = 1$.

                                    We have $Q = Q_1$, $ \piencodfaplas{N}_u \redone^1 Q_1$. Let us take $I = a$, we also have that $Q_1 \redone^8 Q_9 = Q_a$ , $N \red M'\unexsub{U / \unvar{x}} = M$ and $\piencodfaplas{ M'}_u \premat Q_9$

                                \item When $2 \leq  k \leq 8$.

                                    Proceeds similarly to the previous case

                                \item When $k \geq 9$.

      We have $ \piencodfaplas{N}_u \redone^9 Q_9 \redone^l Q$, for $l \geq 1$. Since $Q_9 = \piencodfaplas{ M'}_u$ we apply the induction hypothesis we have that  there $\exists \{Q_i\}_{ i \in I } $, $N' \ s.t. \ Q \redone^* \{ Q_i\}_{ i \in I} ,  M'\red^* N'$ and $\piencodfaplas{N'}_u \premat Q_i \quad \forall i \in I$.                                    Then,  $ \piencodfaplas{N}_u \redone^9 Q_9 \redone^l Q \redone^* \{ Q_i \}_{i \in I}$ and by the contextual reduction rule it follows that $N = (M'[ \leftarrow x])\esubst{ 1 }{ x } \red^*  N' $ and the case holds.

\end{enumerate}

\item When $C = \bag{N_1} \cdot \cdots \cdot \bag{N_l}$, for $l \geq 1$.
                    Then,

 \[
    \small
    \hspace{-2.2cm}
   \begin{aligned}
   \piencodfaplas{N}_u &=\piencodfaplas{ M'[ {\widetilde{x}} \leftarrow  {x}] \esubst{ C \bagsep U }{x} }_u\\
   &=   \res{ x }  ( \psome{x}; \gname{x}{\linvar{x}}; \gname{x}{\unvar{x}}; \gclose{ x } ; \piencodfaplas{ M'[ {\widetilde{x}} \leftarrow  {x}]}_u   \| \piencodfaplas{ C \bagsep U}_x ) \\
  &\redone ^{4}  \res{ \unvar{x} }  ( \res{\linvar{x}} ( \piencodfaplas{ M'[ {\widetilde{x}} \leftarrow  {x}]}_u   \| \piencodfaplas{ C }_{\linvar{x}} )   \|  \guname{ \unvar{x} }{ x_i } ;  \piencodfaplas{ U }_{x_i} )\\
  &=
  \res{ \unvar{x} }  (
  \res{\linvar{x}} (
  \psome{\linvar{x}}; \pname{\linvar{x}}{y_1}; \big( \gsome{y_1}{ \emptyset }; \gclose{ y_{1} } ; \0   \\
    &  \qquad \| \psome{\linvar{x}}; \gsome{\linvar{x}}{u, \llfv{M'} \setminus  \widetilde{x} }; \bignd_{x_{i_1} \in \widetilde{x}} \gname{x}{{x}_{i_1}};
    \cdots \\
    &  \qquad \psome{\linvar{x}}; \pname{\linvar{x}}{y_l}; \big( \gsome{y_l}{ \emptyset }; \gclose{ y_{l} } ; \0   \\
    &  \qquad \| \psome{\linvar{x}}; \gsome{\linvar{x}}{u, \llfv{M'} \setminus  (\widetilde{x} \setminus (x_{i_1} , \cdots , x_{i_{l-1}}) )}; \bignd_{x_{i_l} \in \widetilde{x}} \gname{x}{{x}_{i_l}};\\
    &  \qquad \psome{\linvar{x}}; \pname{\linvar{x}}{y_{l+1}}; ( \gsome{y_{l+1}}{ u , \llfv{M'} }; \gclose{ y_{l+1} } ;\piencodfaplas{M'}_u \| \pnone{ \linvar{x} } )\big) \cdots
    \big)
  \\
  &  \qquad \|  \gsome{\linvar{x}}{\llfv{C} }; \gname{x}{y_1}; \gsome{\linvar{x}}{y_1, \llfv{C}}; \psome{\linvar{x}}; \pname{\linvar{x}}{z_1}; \\
       &  \qquad  ( \gsome{z_1}{\llfv{M_1}};  \piencodfaplas{M_1}_{z_1} \| \pnone{y_1}  \| \\
  &  \qquad \cdots \gsome{\linvar{x}}{\llfv{M_l} }; \gname{x}{y_l}; \gsome{\linvar{x}}{y_l, \llfv{M_l}}; \psome{\linvar{x}}; \pname{\linvar{x}}{z_l}; \\
       &  \qquad  ( \gsome{z_l}{\llfv{M_l}};  \piencodfaplas{M_l}_{z_l} \| \pnone{y_l}  \| \\
  &  \qquad \gsome{\linvar{x}}{\emptyset};\gname{x}{y_{l+1}};  ( \psome{ y_{l+1}}; \pclose{y_{l+1}}  \| \gsome{\linvar{x}}{\emptyset}; \pnone{\linvar{x}} )) \cdots )
  )\|
  \guname{ \unvar{x} }{ x_i } ;  \piencodfaplas{ U }_{x_i} )\\
 & \redone^{6l}
    \res{ \unvar{x} }  ( \res{\linvar{x}} ( \res{ z_1 }( \cdots \res{ z_l } ( \psome{\linvar{x}}; \pname{\linvar{x}}{y_{l+1}}. (  \gsome{ y_{l+1} }{ u,\llfv{M'} };   \\
    &\qquad  \gclose{ y_{l+1} } ;  \piencodfaplas{M'}_u\{ z_1 / x_{i_1} \} \cdots \{ z_l / x_{i_l} \}   \|   \pnone{ \linvar{x} }  )\\
    & \qquad   \|  \gsome{ z_l }{ \llfv{M_{j_l}} };  \piencodfaplas{M_{j_l}}_{z_l} ) \cdots \|  \gsome{ z_1 }{ \llfv{M_{j_1}} }; \piencodfaplas{M_{j_1}}_{z_1} )       \|
    \\
    &\qquad  \gsome{ \linvar{x} }{ \emptyset };  \linvar{x}(y_{l+1}). ( \psome{y_{l+1}}; \pclose{ y_{l+1} }    \|  \gsome{ \linvar{x} }{ \emptyset };    \pnone{ \linvar{x} } ) )
   \|  \guname{ \unvar{x} }{ x_i } ; \piencodfaplas{ U }_{x_i} )\\
& \redone^{5}
    \res{ \unvar{x} }  (  \res{ z_1 }(\gsome{ x_1 }{ \llfv{M_{j_1}} };  \piencodfaplas{M_{j_1}}_{x_1} \| \cdots \\
    & \qquad \qquad \res{ z_l }  ( \gsome{ x_l }{ \llfv{M_{j_l}} };  \piencodfaplas{M_{j_l}}_{x_l} \| \piencodfaplas{M'}_u \{ z_1 / x_{i_1} \} \cdots \{ z_l / x_{i_l} \}  )  \cdots      )
    \|  \guname{ \unvar{x} }{ x_i } ;  \piencodfaplas{ U }_{x_i} )\\
  \end{aligned}
 \]

                            The proof follows by induction on the number of reductions $\piencodfaplas{N}_u \redone^k Q$.

\begin{enumerate}
\item When $k = 0$, the result follows trivially. Just take $I = \{ a\} $, $\mathbb{N}=\mathbb{N}'$ and $\piencodfaplas{N}_u=Q=Q_a$.

 \item When $1 \leq k \leq 6l + 9$.

 Let $I = \{ a \}$, $Q_k$ be such that $ \piencodfaplas{N}_u \redone^k Q_k$.
            We also have that $Q_k \redone^{6l + 9 - k} Q_{6l + 9} = Q_a$ ,

            $N \red  M'\linexsub{C  /  x_1 , \cdots , x_l} \unexsub{U / \unvar{x} }= N'$ and

            $
                \hspace{-1.0cm}
                \small
            \begin{aligned}
                \piencodfaplas{  M'\linexsub{C  /  x_1 , \cdots , x_l} \unexsub{U / \unvar{x} } }_u & =  \res{  \unvar{x} }  \Big( \res{z_1}( \gsome{z_1}{\llfv{M_{1}}};\piencodfaplas{ M_{1} }_{ {z_1}}  \| \\
                & \quad  \cdots \res{z_l} ( \gsome{z_l}{\llfv{M_{l}}};\piencodfaplas{ M_{l} }_{ {z_l}} \\
                & \quad   \| \bignd_{x_{i_1} \in \{ x_1 ,\cdots , x_l  \}} \cdots \bignd_{x_{i_l} \in \{ x_1 ,\cdots , x_l \setminus x_{i_1} , \cdots , x_{i_{l-1}}  \}}\\
                & \quad  \piencodfaplas{ M'}_u \{ z_1 / x_{i_1} \} \cdots \{ z_l / x_{i_l} \} ) \cdots )
                     \|  \guname{ \unvar{x} }{ x_i } ;  \piencodfaplas{ U }_{x_i} \Big )\\
            & \premat  \res{  \unvar{x} }  \Big( \res{z_1}( \gsome{z_1}{\llfv{M_{1}}};\piencodfaplas{ M_{1} }_{ {z_1}}  \| \\
                & \quad  \cdots \res{z_l} ( \gsome{z_l}{\llfv{M_{l}}};\piencodfaplas{ M_{l} }_{ {z_l}} \\
                & \quad   \| \piencodfaplas{ M'}_u \{ z_1 / x_{i_1} \} \cdots \{ z_l / x_{i_l} \} ) \cdots )
                    \\
                   &  \quad   \|  \guname{ \unvar{x} }{ x_i } ;  \piencodfaplas{ U }_{x_i} \Big )
                    = Q_{6l + 9}
            \end{aligned}
            $.

\item When $k > 6l + 9$.

Then,  $ \piencodfaplas{N}_u \redone^{6l + 9} Q_{6l + 9} \redone^n Q$ for $n \geq 1$. Also,

\(
\begin{aligned}
&N \red^1   M'\linexsub{C  /  x_1 , \cdots , x_l} \unexsub{U / \unvar{x} } \text { and } \\
&   \piencodfaplas{ M'\linexsub{C  /  x_1 , \cdots , x_l} \unexsub{U / \unvar{x} } }_u \premat Q_{6l + 9} .
    \end{aligned}
\)

$\exists  \{ P_j \}_{ j \in J}, N', \ s.t. \ M'\linexsub{C  /  x_1 , \cdots , x_l} \unexsub{U / \unvar{x} } \red^* N',$ $ \piencodfaplas{M'\linexsub{C  /  x_1 , \cdots , x_l} \unexsub{U / \unvar{x} }}$ $ \redone^n P \redone^* \{ P_j \}_{j \in J}  $ and
$ \piencodfaplas{N'}_u \premat P_j \quad \forall j \in J $. We also have by Prop.\ref{ch4prop:soundextra} that as $\piencodfaplas{M'\linexsub{C  /  x_1 , \cdots , x_l} \unexsub{U / \unvar{x} }} \premat Q_{6l + 9} $ and $\piencodfaplas{M'\linexsub{C  /  x_1 , \cdots , x_l} \unexsub{U / \unvar{x} }} \redone^* \{ Q_j \}_{j \in J} $ implies $\exists  \{Q_i\}_{i \in I} \ s.t. \  Q_{6l + 9} \redone^* Q_{i \in I} $ , $ I \subset J $ and $P_i \premat Q_i \  , \ \forall i \in I$

                        \end{enumerate}

                    \end{enumerate}

                \item When $\size{\widetilde{x}} > \size{C}$.

                    Then we have
                    $N = M'[ {x}_1, \cdots ,  {x}_k \leftarrow  {x}]\ \esubst{ C \bagsep U }{x}$ with $C = \bag{M_1}  \cdots  \bag{M_l} \quad k > l$, $N \red  \fail^{\widetilde{z}} = M$ and $ \widetilde{z} =  (\llfv{M'} \setminus \{   {x}_1, \cdots ,  {x}_k \} ) \cup \llfv{C} $. On the one hand, we have:
                    Hence $k = l + m$ for some $m \geq 1$

                    \[
                        \small
                        \hspace{-2cm}
                    \begin{aligned}
                        \piencodfaplas{N}_u &= \piencodfaplas{M'[ {x}_1, \cdots ,  {x}_k \leftarrow  {x}]\ \esubst{ C \bagsep U }{x}}_u \\
                        & =   \res{ x } ( \psome{x}; \gname{x}{\linvar{x}}; \gname{x}{\unvar{x}}; \gclose{ x } ; \piencodfaplas{ M'[ {x}_1, \cdots ,  {x}_k \leftarrow  {x}]}_u   \| \piencodfaplas{ C \bagsep U}_x ) \\
                              &\redone ^{4}  \res{ \unvar{x} } ( \res{\linvar{x}} ( \piencodfaplas{M'[ {x}_1, \cdots ,  {x}_k \leftarrow  {x}]}_u   \| \piencodfaplas{ C }_{\linvar{x}}  ) \|  \guname{ \unvar{x} }{ x_i } ;  \piencodfaplas{ U }_{x_i} )\\
            & = \res{ \unvar{x} } (
            \res{\linvar{x}} (
            \psome{\linvar{x}}; \pname{\linvar{x}}{y_1}; \big( \gsome{y_1}{ \emptyset }; \gclose{ y_{1} } ; \0   \\
            & \qquad \qquad \|\psome{\linvar{x}}; \gsome{\linvar{x}}{u, \llfv{M'} \setminus  \widetilde{x} }; \bignd_{x_{i_1} \in \widetilde{x}} \gname{x}{{x}_{i_1}}; \cdots  \\
            & \qquad \qquad \psome{\linvar{x}}; \pname{\linvar{x}}{y_k}; \big( \gsome{y_k}{ \emptyset }; \gclose{ y_{k} } ; \0   \\
            & \qquad \qquad \| \psome{\linvar{x}}; \gsome{\linvar{x}}{u, \llfv{M'} \setminus  \widetilde{x} }; \bignd_{x_{i_k} \in (\widetilde{x} \setminus x_{i_1} , \cdots , x_{i_{k-1}}    )} \gname{x}{{x}_{i_k}}; \\
            & \qquad \qquad \psome{\linvar{x}}; \pname{\linvar{x}}{y_{k+1}}; ( \gsome{y_{k+1}}{ u, \llfv{M'} }; \gclose{ y_{{k+1}} } ;\piencodfaplas{M'}_u \| \pnone{ \linvar{x} } )
            \big) \cdots
            \big)
            \\
            & \qquad \qquad \| \gsome{\linvar{x}}{\llfv{C} }; \gname{x}{y_1}; \gsome{\linvar{x}}{y_1, \llfv{C}}; \psome{\linvar{x}}; \pname{\linvar{x}}{z_1}; \\
            & \qquad \qquad   ( \gsome{z_1}{\llfv{M_1}};  \piencodfaplas{M_1}_{z_1} \| \pnone{y_1}  \| \\
            & \qquad \qquad \cdots \gsome{\linvar{x}}{\llfv{M_l} }; \gname{x}{y_l}; \gsome{\linvar{x}}{y_l, \llfv{C}}; \psome{\linvar{x}}; \pname{\linvar{x}}{z_l}; \cdots\\
            & \qquad \qquad  ( \gsome{z_l}{\llfv{M_l}};  \piencodfaplas{M_l}_{z_l} \| \pnone{y_l}  \| \\
            & \qquad \qquad  \gsome{\linvar{x}}{\emptyset};\gname{x}{y_{l+1}};  ( \psome{ y_{l+1}}; \pclose{y_{l+1}}  \| \gsome{\linvar{x}}{\emptyset}; \pnone{\linvar{x}} )
            ) \cdots
            )
            )
            \|
            \guname{ \unvar{x} }{ x_i } ;  \piencodfaplas{ U }_{x_i}
            ) \\
            & \redone^{6k}
             \res{ \unvar{x} } (
            \res{\linvar{x}} ( \res{z_1} ( \gsome{z_1}{\llfv{M_1}};  \piencodfaplas{M_1}_{z_1} \| \cdots  \res{z_k} ( \gsome{z_k}{\llfv{M_k}};  \piencodfaplas{M_k}_{z_k} \| \\
            & \qquad \qquad \psome{\linvar{x}}; \pname{\linvar{x}}{y_{k+1}}; ( \gsome{y_{k+1}}{ u, \llfv{M'} }; \gclose{ y_{{k+1}} } ;\piencodfaplas{M'}_u\{ z_1 / x_{i_1} \} \cdots \{ z_k / x_{i_k} \} \| 
            \\
            & \qquad \qquad \pnone{ \linvar{x} } )) \cdots   ) \| \gsome{\linvar{x}}{\llfv{C} \setminus (M_1 , \cdots , M_k) }; \gname{x}{y_{k+1}}; \gsome{\linvar{x}}{y_{k+1}, \llfv{C}\setminus (M_1 , \cdots , M_k)};   \\
            & \qquad \qquad  \psome{\linvar{x}};\pname{\linvar{x}}{z_{k+1}}; ( \gsome{z_{k+1}}{\llfv{M_{k+1}}};  \piencodfaplas{M_{k+1}}_{z_{k+1}} \| \pnone{y_1}  \| \\
            & \qquad \qquad \cdots \gsome{\linvar{x}}{\llfv{M_l} }; \gname{x}{y_l}; \gsome{\linvar{x}}{y_l, \llfv{M_l}}; \psome{\linvar{x}}; \pname{\linvar{x}}{z_l}; \cdots\\
            & \qquad \qquad  ( \gsome{z_l}{\llfv{M_l}};  \piencodfaplas{M_l}_{z_l} \| \pnone{y_l}  \| \\
            & \qquad \qquad  \gsome{\linvar{x}}{\emptyset};\gname{x}{y_{l+1}};  ( \psome{ y_{l+1}}; \pclose{y_{l+1}}  \| \gsome{\linvar{x}}{\emptyset}; \pnone{\linvar{x}} )
            ) \cdots
            )
            )
            \|
            \guname{ \unvar{x} }{ x_i } ;  \piencodfaplas{ U }_{x_i}
            )
\\
 & \redone^{6}  \res{ \unvar{x} } (
              \res{z_1} ( \gsome{z_1}{\llfv{M_1}};  \piencodfaplas{M_1}_{z_1} \| \cdots  \res{z_k} ( \gsome{z_k}{\llfv{M_k}};  \piencodfaplas{M_k}_{z_k} \| \\
            & \qquad \qquad   \pnone{u} \| \pnone{(\llfv{M'} \setminus \widetilde{x})} \|  \pnone{z_1} \| \cdots \| \pnone{z_k} \| 
            \\
            & \qquad \qquad \pnone{( \llfv{C} \setminus (M_1 , \cdots , M_k) )}  ) \cdots   ) \|
            \guname{ \unvar{x} }{ x_i } ;  \piencodfaplas{ U }_{x_i}
            ) \\
  & \redone^{k}   \pnone{ u }     \|   \pnone{ (\llfv{M'} \setminus \{  x_1, \cdots , x_k \} ) }    \|   \pnone{ \llfv{C} }  \\
 &= \piencodfaplas{  \fail^{\widetilde{z}}}_u = Q_{7l + 10}  \\
 \end{aligned}
 \]
The rest of the proof is by induction on the number of reductions $\piencodfaplas{N}_u \redone^j Q$.

                            \begin{enumerate}
                                \item When $j = 0$, the result follows trivially. Just take $I = \{a\}$ $\mathbb{N}=\mathbb{N}'$ and $\piencodfaplas{N}_u=Q=Q_a$.
  \item When $1 \leq j \leq 7k + 10$.

Let $I = \{ a\}$ and  $Q_j$ be such that $ \piencodfaplas{N}_u \redone^j Q_j$.
By the steps above one has

\(\begin{aligned}
  &Q_j \redone^{7k + 10 - j} Q_{7k + 10} = Q_a,\\ &N \red^1 \fail^{\widetilde{z}} = N';\text{ and} \piencodfaplas{ \fail^{\widetilde{z}}}_u = Q_{7k + 10}.
\end{aligned}
\)
\item When $j > 7k + 10$.

In this case, we have
$ \piencodfaplas{N}_u \redone^{7k + 10} Q_{7k + 10} \redone^n Q,$ for $n \geq 1$.
We also know that

$N \red^1 \fail^{\widetilde{z}}$. However no further reductions can be performed.

                            \end{enumerate}

                \item When $\size{\widetilde{x}} < \size{C}$, the proof  proceeds similarly to the previous case.

            \end{enumerate}

        \item  $N =   M'\linexsub{C /  x_1 , \cdots , x_k} $.

           In this case we let $C = \bag{M_1} \cdot \cdots \cdot \bag{M_k}$,
            \[
            \begin{aligned}
               \piencodfaplas{ M'\linexsub{C /  x_1 , \cdots , x_k} }_u =&  \res{z_1}( \gsome{z_1}{\llfv{M_{1}}};\piencodfaplas{ M_{1} }_{ {z_1}}  \|  \cdots \res{z_k} (  \\
                & \quad \gsome{z_k}{\llfv{M_{k}}};\piencodfaplas{ M_{k} }_{ {z_k}} \| \bignd_{x_{i_1} \in \{ x_1 ,\cdots , x_k  \}} \cdots \\
                & \quad \bignd_{x_{i_k} \in \{ x_1 ,\cdots , x_k \setminus x_{i_1} , \cdots , x_{i_{k-1}}  \}} \piencodfaplas{ M'}_u \{ z_1 / x_{i_1} \} \cdots \{ z_k / x_{i_k} \} ) \cdots ) \\
            \end{aligned}
            \]
            Therefore,
            \[
            \begin{aligned}
               \piencodfaplas{N}_u & =   \res{z_1}( \gsome{z_1}{\llfv{M_{1}}};\piencodfaplas{ M_{1} }_{ {z_1}}  \|
              \cdots \res{z_k} ( \gsome{z_k}{\llfv{M_{k}}};\piencodfaplas{ M_{k} }_{ {z_k}} \\
                & \quad  \| \bignd_{x_{i_1} \in \{ x_1 ,\cdots , x_k  \}} \cdots \bignd_{x_{i_k} \in \{ x_1 ,\cdots , x_k \setminus x_{i_1} , \cdots , x_{i_{k-1}}  \}} \piencodfaplas{ M'}_u \{ z_1 / x_{i_1} \} \cdots \{ z_k / x_{i_k} \} ) \cdots ) \\
               & \redone^m   \res{z_1}( \gsome{z_1}{\llfv{M_{1}}};\piencodfaplas{ M_{1} }_{ {z_1}}  \| \cdots \res{z_k} ( \gsome{z_k}{\llfv{M_{k}}};\piencodfaplas{ M_{k} }_{ {z_k}} \\
                & \quad  \| \bignd_{x_{i_1} \in \{ x_1 ,\cdots , x_k  \}} \cdots \bignd_{x_{i_k} \in \{ x_1 ,\cdots , x_k \setminus x_{i_1} , \cdots , x_{i_{k-1}}  \}} R \{ z_1 / x_{i_1} \} \cdots \{ z_k / x_{i_k} \} ) \cdots ) \\
               & \redone^n  Q\\
            \end{aligned}
            \]

            for some process $R$. Where $\redone^n$ is a reduction that  initially synchronizes with $  \gsome{ {x}_i }{ \llfv{M_{J_i}} } $ for some $ i \in \{ 1 , \cdots , k \}$, when $n \geq 1$, $n + m = k \geq 1$. Type preservation in \clpi ensures reducing $\piencodfaplas{ M'}_v \redone^m$ does not consume possible synchronizations with $ \psome{{x}_i} $, if they occur. Let us consider the the possible sizes of both $m$ and $n$.

            \begin{enumerate}
                \item For $m = 0$ and $n \geq 1$.

                    We have that $R = \piencodfaplas{M'}_u$ as $\piencodfaplas{M'}_u \redone^0 \piencodfaplas{M'}_u$.

                    Notice that there are two possibilities of having an unguarded $\psome{x_i}$ or $\pnone{ x_i }$ without internal reductions:

                    \begin{enumerate}
                        \item $M'= \fail^{ {x}_i, \widetilde{y}}$.
    \[
        \small
        \hspace{-1.5cm}
    \begin{aligned}
  \piencodfaplas{N}_u & =
                \res{z_1}( \gsome{z_1}{\llfv{M_{1}}};\piencodfaplas{ M_{1} }_{ {z_1}}  \|  \cdots \res{z_k} ( \gsome{z_k}{\llfv{M_{k}}};\piencodfaplas{ M_{k} }_{ {z_k}} \\
                & \quad  \| \bignd_{x_{i_1} \in \{ x_1 ,\cdots , x_k  \}} \cdots \bignd_{x_{i_k} \in \{ x_1 ,\cdots , x_k \setminus x_{i_1} , \cdots , x_{i_{k-1}}  \}} \piencodfaplas{ M'}_u \{ z_1 / x_{i_1} \} \cdots \{ z_k / x_{i_k} \} ) \cdots ) \\
                & = \res{z_1}( \gsome{z_1}{\llfv{M_{1}}};\piencodfaplas{ M_{1} }_{ {z_1}}  \|  \cdots \res{z_k} ( \gsome{z_k}{\llfv{M_{k}}};\piencodfaplas{ M_{k} }_{ {z_k}} \\
                & \quad  \| \bignd_{x_{i_1} \in \{ x_1 ,\cdots , x_k  \}} \cdots \bignd_{x_{i_k} \in \{ x_1 ,\cdots , x_k \setminus x_{i_1} , \cdots , x_{i_{k-1}}  \}} \piencodfaplas{ \fail^{ {x}_i, \widetilde{y}}}_u \{ z_1 / x_{i_1} \} \cdots \{ z_k / x_{i_k} \} ) \cdots ) \\
                & = \res{z_1}( \gsome{z_1}{\llfv{M_{1}}};\piencodfaplas{ M_{1} }_{ {z_1}}  \| \cdots \res{z_k} ( \gsome{z_k}{\llfv{M_{k}}};\piencodfaplas{ M_{k} }_{ {z_k}} \\
                & \quad  \| \bignd_{x_{i_1} \in \{ x_1 ,\cdots , x_k  \}} \cdots \bignd_{x_{i_k} \in \{ x_1 ,\cdots , x_k \setminus x_{i_1} , \cdots , x_{i_{k-1}}  \}} \\
                & \qquad
                \pnone{ u }    \|   \pnone{ {x}_i }     \|    \pnone{ \widetilde{y} }
                \{ z_1 / x_{i_1} \} \cdots \{ z_k / x_{i_k} \} ) \cdots ) \\
    \end{aligned}
    \]

    by type preservation we have that $\widetilde{y} = \{x_1 , \cdots , x_k \} \setminus x_i , \widetilde{y}' $ for some $\widetilde{y}'$

    \[
    \begin{aligned}
    & =\res{z_1}( \gsome{z_1}{\llfv{M_{1}}};\piencodfaplas{ M_{1} }_{ {z_1}}  \|  \cdots \res{z_k} ( \gsome{z_k}{\llfv{M_{k}}};\piencodfaplas{ M_{k} }_{ {z_k}} \\
                & \qquad \qquad  \| \bignd_{x_{i_1} \in \{ x_1 ,\cdots , x_k  \}} \cdots \bignd_{x_{i_k} \in \{ x_1 ,\cdots , x_k \setminus x_{i_1} , \cdots , x_{i_{k-1}}  \}}
                \pnone{ u }   \|    \pnone{ \widetilde{y}' } \\
                & \qquad \qquad \|   \pnone{ \{ x_1 , \cdots , x_k  \} }
                \{ z_1 / x_{i_1} \} \cdots \{ z_k / x_{i_k} \} ) \cdots ) \\
    & =\res{z_1}( \gsome{z_1}{\llfv{M_{1}}};\piencodfaplas{ M_{1} }_{ {z_1}}  \| \cdots \res{z_k} ( \gsome{z_k}{\llfv{M_{k}}};\piencodfaplas{ M_{k} }_{ {z_k}} \\
                & \qquad \qquad  \| \bignd_{x_{i_1} \in \{ x_1 ,\cdots , x_k  \}} \cdots \bignd_{x_{i_k} \in \{ x_1 ,\cdots , x_k \setminus x_{i_1} , \cdots , x_{i_{k-1}}  \}}
                \pnone{ u }   \|    \pnone{ \widetilde{y}' } \\
                & \qquad \qquad \|   \pnone{ \{ z_1 , \cdots , z_k  \} } ) \cdots ) \\
    &\redone^k   \pnone{ u }    \|   \pnone{ \widetilde{y}' }    \|   \pnone{ \llfv{C} }  \\
  \end{aligned}
    \]
  Notice that no further reductions can be performed.
  Thus we take $I = \{ a \}$ and,
 $$ \piencodfaplas{N}_u \redone   \pnone{ u }   \|   \pnone{ \widetilde{y} }    \|   \pnone{ \lfv{C} }  = Q_a.$$

We also have that  $N \red \fail^{ \widetilde{y} \cup \llfv{C} } = N'$ and $\piencodfaplas{ \fail^{\widetilde{y} \cup \llfv{C}} }_u = Q_a$.

                        \item $\headf{M'} =  {x}_i$ with $i \in \{ 1 , \cdots , k \}$.

 Then we have the following

 \[
    \small
    \hspace{-1.7cm}
    \begin{aligned}
  \piencodfaplas{N}_u
  & =
                \res{z_1}( \gsome{z_1}{\llfv{M_{1}}};\piencodfaplas{ M_{1} }_{ {z_1}}  \| \cdots \res{z_k} ( \gsome{z_k}{\llfv{M_{k}}};\piencodfaplas{ M_{k} }_{ {z_k}} \\
                & \quad  \| \bignd_{x_{i_1} \in \{ x_1 ,\cdots , x_k  \}} \cdots \bignd_{x_{i_k} \in \{ x_1 ,\cdots , x_k \setminus x_{i_1} , \cdots , x_{i_{k-1}}  \}} \piencodfaplas{ M'}_u \{ z_1 / x_{i_1} \} \cdots \{ z_k / x_{i_k} \} ) \cdots ) \\
  & =
                \res{z_1}( \gsome{z_1}{\llfv{M_{1}}};\piencodfaplas{ M_{1} }_{ {z_1}}  \|  \cdots \res{z_k} ( \gsome{z_k}{\llfv{M_{k}}};\piencodfaplas{ M_{k} }_{ {z_k}} \\
                & \quad  \| \bignd_{x_{i_1} \in \{ x_1 ,\cdots , x_k  \}} \cdots \bignd_{x_{i_k} \in \{ x_1 ,\cdots , x_k \setminus x_{i_1} , \cdots , x_{i_{k-1}}  \}}
                \res{ \widetilde{y} } (\piencodfaplas{  {x_i} }_{j}   \| P)
                \{ z_1 / x_{i_1} \} \cdots \{ z_k / x_{i_k} \} ) \cdots ) \\
    & =
                \res{z_1}( \gsome{z_1}{\llfv{M_{1}}};\piencodfaplas{ M_{1} }_{ {z_1}}  \| \cdots \res{z_k} ( \gsome{z_k}{\llfv{M_{k}}};\piencodfaplas{ M_{k} }_{ {z_k}} \\
                & \quad  \| \bignd_{x_{i_1} \in \{ x_1 ,\cdots , x_k  \}} \cdots \bignd_{x_{i_k} \in \{ x_1 ,\cdots , x_k \setminus x_{i_1} , \cdots , x_{i_{k-1}}  \}}\\
                & \quad
                \res{ \widetilde{y} } (  \psome{{x_i}}; \pfwd{x_i}{j}    \| P)
                \{ z_1 / x_{i_1} \} \cdots \{ z_k / x_{i_k} \} ) \cdots ) \\
    \end{aligned}
    \]
 Let us consider a arbitrary sum where $x_{i_k} = x_i$ , other cases follow similarly.

 \[
    \hspace{-1cm}
    \small
 \begin{aligned}
    & =
                \res{z_1}( \gsome{z_1}{\llfv{M_{1}}};\piencodfaplas{ M_{1} }_{ {z_1}}  \| \cdots \res{z_k} ( \gsome{z_k}{\llfv{M_{k}}};\piencodfaplas{ M_{k} }_{ {z_k}} \\
                & \qquad \qquad  \| \bignd_{x_{i_1} \in \{ x_1 ,\cdots , x_k  \}} \cdots \bignd_{x_{i_k} \in \{ x_1 ,\cdots , x_k \setminus x_{i_1} , \cdots , x_{i_{k-1}}  \}} \\
                & \qquad \qquad
                \res{ \widetilde{y} } (  \psome{{x_i}}; \pfwd{x_i}{j}    \| P)
                \{ z_1 / x_{i_1} \} \cdots \{ z_k / x_{i_k} \} ) \cdots ) \\
     & \redone
     \res{z_1}( \gsome{z_1}{\llfv{M_{1}}};\piencodfaplas{ M_{1} }_{ {z_1}}  \| \cdots \res{z_k} ( \piencodfaplas{ M_{k} }_{ {z_k}} \\
                & \qquad \qquad  \|
                \res{ \widetilde{y} } (  \ \pfwd{x_k}{j}    \| P)
                \{ z_1 / x_{i_1} \} \cdots \{ z_{k-1} / x_{i_{k-1}} \} ) \cdots ) & = Q_1\\
    & \redone
     \res{z_1}( \gsome{z_1}{\llfv{M_{1}}};\piencodfaplas{ M_{1} }_{ {z_1}}  \| \cdots \res{z_{k-1}}( \gsome{z_{k-1}}{\llfv{M_{{k-1}}}};\piencodfaplas{ M_{{k-1}} }_{ {z_{k-1}}}    \\
                & \qquad \qquad \| \res{ \widetilde{y} } (  \piencodfaplas{ M_{k} }_{j}\    \| P)
                \{ z_1 / x_{i_1} \} \cdots \{ z_{k-1} / x_{i_{k-1}} \} ) \cdots ) & = Q_2\\
 \end{aligned}
 \]

In addition,
\(
N = M'\linexsub{C /  x_1 , \cdots , x_k} \redone  M'\headlin{ C_i / x_k }  \linexsub{(C \setminus C_i ) /  x_1 , \cdots , x_{k-1}  } = M\).
Finally,
\[
    \small
\begin{aligned}
    \hspace{-1cm}
    \piencodfaplas{M}_u &= \piencodfaplas{ M'\headlin{ C_i / x_k }  \linexsub{(C \setminus C_i ) /  x_1 , \cdots , x_{k-1}  } }_u \\
    &=  \res{z_1}( \gsome{z_1}{\llfv{M_{1}}};\piencodfaplas{ M_{1} }_{ {z_1}}  \|  \cdots \res{z_{k-1}}( \gsome{z_{k-1}}{\llfv{M_{{k-1}}}};\piencodfaplas{ M_{{k-1}} }_{ {z_{k-1}}}  \|  \\
                & \qquad \qquad \bignd_{x_{i_1} \in \{ x_1 ,\cdots , x_{k-1}  \}} \cdots \bignd_{x_{i_{k-1}} \in \{ x_1 ,\cdots , x_{k-1} \setminus x_{i_1} , \cdots , x_{i_{k-2}}  \}} \\
                & \qquad \qquad \res{ \widetilde{y} } (  \piencodfaplas{ M_{k} }_{j}\    \| P)
                \{ z_1 / x_{i_1} \} \cdots \{ z_{k-1} / x_{i_{k-1}} \} ) \cdots )\\
    & \premat Q_2.
\end{aligned}
\]

\begin{enumerate}
\item When $n = 1$:

Then, $I = \{ a \} $, $Q = Q_1$ and  $ \piencodfaplas{N}_u \redone^1 Q_1$. Also,

$Q_1 \redone^1 Q_2 = Q_a$, $N \red^1  M'\headlin{ C_i / x_k }  \linexsub{(C \setminus C_i ) /  x_1 , \cdots , x_{k-1}  } = N'$ and

$\piencodfaplas{M'\headlin{ C_i / x_k }  \linexsub{(C \setminus C_i ) /  x_1 , \cdots , x_{k-1}  }}_u \premat Q_a$.

\item When $n = 2$:

Then, $I = \{ a \} $, $Q = Q_2$ and  $ \piencodfaplas{N}_u \redone^2 Q_2$. Also,

$Q_2 \redone^0 Q_2 = Q_a$, $N \red^1  M'\headlin{ C_i / x_k }  \linexsub{(C \setminus C_i ) /  x_1 , \cdots , x_{k-1}  } = N'$ and

$\piencodfaplas{M'\headlin{ C_i / x_k }  \linexsub{(C \setminus C_i ) /  x_1 , \cdots , x_{k-1}  }}_u \premat Q_a$.

\item When $n > 2$:

Then  $ \piencodfaplas{N}_u \redone^2 Q_2 \redone^l Q$, for $l \geq 1$.  Also,
$N \rightarrow^1 M$, $ \piencodfaplas{M}_u \premat Q_2 $.
$\exists  \{P_j\}_{ j \in J}, N', \ s.t. M \red^* N', $ $ \ \piencodfaplas{M} \redone^l P \redone^* \{P_j\}_{j \in J}  $ and
$ \piencodfaplas{N'}_u \premat P_j \quad \forall j \in J $. We also have by Prop.\ref{ch4prop:soundextra} that as $\piencodfaplas{M} \premat Q_{2} $ and $\piencodfaplas{M} \redone^* \{ Q_j \}_{j \in J}  $ implies $\exists \{ Q_i \}_{i \in I} \ s.t. \  Q_{2} \redone^* \{Q_i\}_{i \in I} $ , $ I \\subseteq J $ and $P_i \premat Q_i \  , \ \forall i \in I$

                            \end{enumerate}

                    \end{enumerate}
 \item  For $m \geq 1$ and $ n \geq 0$.

            \begin{enumerate}
            \item When $n = 0$.

               Then $   \res{ {x} }   ( R   \|   \gsome{ {x} }{ \llfv{N'} }; \piencodfaplas{ N' }_{ {x}} )  = Q$ and $\piencodfaplas{M'}_u \redone^m R$ where $m \geq 1$. By the IH $\exists  \{R_i\}_{ i \in I}, $ $ M'', \ s.t.\ M'\red^* M'', \  R \redone^* \{R_i\}_{i \in I}  $ and

               $ \piencodfaplas{M''}_u \premat R_i \quad \forall i \in I $. Thus,
              \[
                \small
                \hspace{-0.5cm}
               \begin{aligned}
                   \piencodfaplas{N}_u
                   & =   \res{z_1}( \gsome{z_1}{\llfv{M_{1}}};\piencodfaplas{ M_{1} }_{ {z_1}}  \|  \cdots \res{z_k} ( \gsome{z_k}{\llfv{M_{k}}};\piencodfaplas{ M_{k} }_{ {z_k}} \\
                & \quad  \| \bignd_{x_{i_1} \in \{ x_1 ,\cdots , x_k  \}} \cdots \bignd_{x_{i_k} \in \{ x_1 ,\cdots , x_k \setminus x_{i_1} , \cdots , x_{i_{k-1}}  \}} \piencodfaplas{ M'}_u \{ z_1 / x_{i_1} \} \cdots \{ z_k / x_{i_k} \} ) \cdots ) \\
               & \redone^m   \res{z_1}( \gsome{z_1}{\llfv{M_{1}}};\piencodfaplas{ M_{1} }_{ {z_1}}  \|  \cdots \res{z_k} ( \gsome{z_k}{\llfv{M_{k}}};\piencodfaplas{ M_{k} }_{ {z_k}} \\
                & \quad  \| \bignd_{x_{i_1} \in \{ x_1 ,\cdots , x_k  \}} \cdots \bignd_{x_{i_k} \in \{ x_1 ,\cdots , x_k \setminus x_{i_1} , \cdots , x_{i_{k-1}}  \}} R \{ z_1 / x_{i_1} \} \cdots \{ z_k / x_{i_k} \} ) \cdots ) \\
               & \redone^n  Q\\
                \end{aligned}
                 \]
                Also,
                \(
               Q  \redone^*  \{ \res{  {x} }   ( R_i   \|  \gsome{ {x} }{ \llfv{N'} };  \piencodfaplas{ N' }_{ {x}} ) \}_{i \in I} = Q_i,
                \)
                and the term can reduce as follows:

                $N = M'\linexsub{C /  x_1 , \cdots , x_k} \red^* M'' \linexsub{C /  x_1 , \cdots , x_k} = N'$ and  $\piencodfaplas{N'}_u \premat Q_i$ $\forall i \in I$

            \item When $n \geq 1$.
                Then  $R$ has an occurrence of an unguarded $\psome{x}$ or $  \pnone{ x } $, this case follows by IH and applying Proposition \ref{ch4prop:soundextra}.

                    \end{enumerate}
            \end{enumerate}

            \item  $N =  M'\unexsub{U / \unvar{x}}$.

            In this case,
            \(
            \begin{aligned}
                \piencodfaplas{M'\unexsub{U / \unvar{x}}}_u &=   \res{ \unvar{x} }  ( \piencodfaplas{ M'}_u   \|    \guname{ \unvar{x} }{ x_i } ; \piencodfaplas{ U }_{x_i} ).
            \end{aligned}
            \)
            Then,
            \[
            \begin{aligned}
               \piencodfaplas{N}_u & =   \res{ \unvar{x} }  ( \piencodfaplas{ M'}_u   \|    \guname{ \unvar{x} }{ x_i } ; \piencodfaplas{ U }_{x_i} )  \redone^m   \res{ \unvar{x} }  ( R   \|   \guname{ \unvar{x} }{ x_i } ;  \piencodfaplas{ U }_{x_i} ) \redone^n  Q.
            \end{aligned}
            \]
            for some process $R$. Where $\redone^n$ is a reduction initially synchronises with $ \guname{ \unvar{x} }{ x_i }$ when $n \geq 1$, $n + m = k \geq 1$. Type preservation in \clpi ensures reducing $\piencodfaplas{ M'}_v \redone^m$ doesn't consume possible synchronisations with $ \guname{ \unvar{x} }{ x_i } $ if they occur. Let us consider the the possible sizes of both $m$ and $n$.

            \begin{enumerate}
                \item For $m = 0$ and $n \geq 1$.

                   In this case,  $R = \piencodfaplas{M'}_u$ as $\piencodfaplas{M'}_u \redone^0 \piencodfaplas{M'}_u$.

                    Notice that the only possibility of having an unguarded $ \puname{ \unvar{x} }{ x_i }$ without internal reductions is when   $\headf{M'} =  {x}[ind].$
                           By the diamond property  we will be reducing each non-deterministic choice of a process simultaneously.
                            Then we have the following:

                            \[
                            \begin{aligned}
                            \piencodfaplas{N}_u  & =   \res{ \unvar{x} }  (  \res{ \widetilde{y} }  (\piencodfaplas{  {x}[ind] }_{j}   \| P)   \|    \guname{ \unvar{x} }{ x_i } ; \piencodfaplas{ U }_{x_i} ) \\
                            & =   \res{ \unvar{x} }   (  \res{ \widetilde{y} }  ( \puname{ \unvar{x} }{ x_i }; \psel{ {x}_i }{ ind }; \pfwd{x_i}{j}    \| P)   \|    \guname{ \unvar{x} }{ x_i } ; \piencodfaplas{ U }_{x_i}  ) \\
                            & \redone   \res{ \unvar{x} }  (  \res{ \widetilde{y} }  (  \res{ x_i }  (\psel{ {x}_i }{ ind }; \pfwd{x_i}{j}   \| \piencodfaplas{ U }_{x_i} )   \| P)   \|   \guname{ \unvar{x} }{ x_i } ;  \piencodfaplas{ U }_{x_i}  ) &= Q_1 \\
                            & =   \res{ \unvar{x} }   (  \res{ \widetilde{y} }  (  \res{ x_i }   (\psel{ {x}_i }{ ind };\pfwd{x_i}{j}    \| x_i. case( ind.\piencodfaplas{U_{ind}}_{x_i} ) )   \| P) \\
                            &  \|   \guname{ \unvar{x} }{ x_i } ; \piencodfaplas{ U }_{x_i}  ) \\
                            & \redone   \res{ \unvar{x} }   (  \res{ \widetilde{y} } (  \res{ x_i } (\pfwd{x_i}{j}   \| \piencodfaplas{U_{ind}}_{x_i} )   \| P)   \|    \guname{ \unvar{x} }{ x_i } ; \piencodfaplas{ U }_{x_i}  ) &= Q_2\\
                            & \redone   \res{ \unvar{x} }   (  \res{ \widetilde{y} }  ( \piencodfaplas{U_{ind}}_{j}    \| P)   \|  \guname{ \unvar{x} }{ x_i } ;   \piencodfaplas{ U }_{x_i}  ) &= Q_3 \\
                            \end{aligned}
                            \]

                We consider the two cases of the form of $U_{ind}$ and show that the choice of $U_{ind}$ is inconsequential

                \begin{itemize}
                \item When $ U_{ind} = \unvar{\bag{N}}$:

                In this case,
                \(
                \begin{aligned}
                N &=M'\unexsub{U / \unvar{x}} \red M'\headlin{ N /\unvar{x} }\unexsub{U / \unvar{x}} = M.
                \end{aligned}
                \)
                 and
                \[
                 \begin{aligned}
                 \piencodfaplas{M}_u &= \piencodfaplas{M'\headlin{ N /\unvar{x} }\unexsub{U / \unvar{x}}}_u \\
                 &=   \res{ \unvar{x} }  ( \res{ \widetilde{y} }  ( \piencodfaplas{\bag{N}}_{j}    \| P)   \|   \guname{ \unvar{x} }{ x_i } ; \piencodfaplas{ U }_{x_i}  ) & = Q_3
                                            \end{aligned}
                 \]

                \item When $ U_i = \unvar{\oneb} $:

                  In this case,
                        \(
                        \begin{aligned}
                            N &=M'\unexsub{U / \unvar{x}} \red M'\headlin{ \fail^{\emptyset} /\unvar{x} } \unexsub{U /\unvar{x} } = M.
                        \end{aligned}
                        \)

                        Notice that $\piencodfaplas{\unvar{\oneb}}_{j} =    \pnone{ j } $ and that $\piencodfaplas{\fail^{\emptyset}}_j =   \pnone{ j } $. In addition,

                            \[
                            \begin{aligned}
                               \piencodfaplas{M}_u&= \piencodfaplas{M'\headlin{ \fail^{\emptyset} /\unvar{x} } \unexsub{U /\unvar{x} }}_u\\
                               &=   \res{ \unvar{x} }   (  \res{ \widetilde{y} } ( \piencodfaplas{\fail^{\emptyset}}_{j}    \| P)   \|   \guname{ \unvar{x} }{ x_i } ;  \piencodfaplas{ U }_{x_i}  ) \\
                               & =  \res{ \unvar{x} }   (  \res{ \widetilde{y} } ( \piencodfaplas{\unvar{\oneb}}_{j}    \| P)   \|   \guname{ \unvar{x} }{ x_i } ;  \piencodfaplas{ U }_{x_i}  ) & = Q_3
                            \end{aligned}
                            \]
                \end{itemize}

                Both choices give an $M$ that are equivalent to $Q_3$.

    \begin{enumerate}
    \item When $n \leq 2$.

   In this case, $Q = Q_n$ and  $ \piencodfaplas{N}_u \redone^n Q_n$.

Also, $Q_n \redone^{3-n} Q_3 = Q'$, $N \red^1 M = N'$ and $\piencodfaplas{M }_u = Q_2$.

\item When $n \geq 3$.

We have $ \piencodfaplas{N}_u \redone^3 Q_3 \redone^l Q$ for $l \geq 0$. We also know that $N \rightarrow M$, $Q_3 = \piencodfaplas{M}_u$.  By the IH, there exist $\{ Q_i \}_{i \in I}$ , $ N'$ such that $Q \redone^* \{Q_i\}_{i \in I}$, $M \red^* N'$ and $\piencodfaplas{N'}_u \premat Q_i \quad \forall i \in I $ . Finally, $\piencodfaplas{N}_u \redone^3 Q_3 \redone^l Q \redone \{Q_i\}_{i \in I}$ and $N \red M  \red^* N'$.

\end{enumerate}
 \item For $m \geq 1$ and $ n \geq 0$.

{\bf (b.1) Case $n = 0$.}

Then $    \res{ \unvar{x} }   ( R   \|   \guname{ \unvar{x} }{ x_i } ; \piencodfaplas{ U }_{x_i} )  = Q$ and $\piencodfaplas{M'}_u \redone^m R$ where $m \geq 1$. Then by the IH there exist $\{R_i'\}_{i \in I}$  and $M'' $ such that $R \redone^* \{R_i'\}_{i \in I}$, $M'\red^* M''$, and $\piencodfaplas{M''}_u \premat R_i \quad \forall i \in I $.  Hence we have that

\[
\begin{aligned}
\piencodfaplas{N}_u & =   \res{ \unvar{x} }  ( \piencodfaplas{M'}_u   \|   \guname{ \unvar{x} }{ x_i } ; \piencodfaplas{ U }_{x_i} ) \redone^m   \res{ \unvar{x} }  ( R   \|   \guname{ \unvar{x} }{ x_i } ; \piencodfaplas{ U }_{x_i} )  = Q
\end{aligned}
\]
We also know that
\[
\begin{aligned}
Q & \redone^*  \{ \res{ \unvar{x} } ( R'_i   \|    \guname{ \unvar{x} }{ x_i } ; \piencodfaplas{ U }_{x_i} )\}_{i \in I} = \{ Q_i\}_{i \in I} \\
\end{aligned}
\]
and so the \lamcoldetsh term can reduce as follows: $N =  M'\unexsub{U / \unvar{x}} \red^*  M'' \unexsub{U / \unvar{x}} = N'$ and  $\piencodfaplas{N'}_u \premat Q_i \quad \forall i \in I$ via the $\premat$ rules

{\bf (b.2) Case  $n \geq 1$.}

Then $R$ has an occurrence of an unguarded $\puname{ \unvar{x} }{ x_i }$, and the case follows by IH.
\end{enumerate}
\end{enumerate}
\end{proof}

\thmEncLWSound*

\begin{proof}
Immediate from \Cref{ch4thm:opsoundone}.
\end{proof}

\subsection{Success Sensitivity}\label{ch4a:loosuccess}

\begin{restatable}[Success Sensitivity (Under $\redone$)]{theorem}{thmEncEagerSucc}\label{ch4proof:successsenscetwounres}
    $\succp{{M}}{\sucs{\lambda}}$ iff ${\piencodfaplas{{M}}_u}\succone{\sucs{\pi}}$ for well-formed closed terms $M$.
\end{restatable}

\begin{proof}
We proceed with the proof in two parts.

\begin{enumerate}

    \item Suppose that  ${M} \Downarrow_{\checkmark} $. We will prove that $\piencodfaplas{{M}} \succone_{\checkmark}$.

    By \defref{ch4def:app_Suc3unres}, there exists  $ {M}' $ such that $ M \red^* {M}'$ and
    $\headf{M'} = \checkmark$.
    By completeness, if $ M\red M'$ then there exist $Q, Q'$ such that $\piencodfaplas{M}_u \equiv Q \redone^* Q'$ and $\piencodfaplas{M'}_u \premat Q' $.

    We wish to show that there exists $Q''$such that $Q' \redone^* Q''$ and $Q''$ has an unguarded occurrence of $\checkmark$.

    By Proposition \ref{ch4Prop:checkprespiunres} (1) we have that $\headf{M'} = \checkmark \implies \piencodfaplas{M'}_u =   \res{ \widetilde{x} } (P   \| \checkmark)$. Finally $\piencodfaplas{M'}_u  =   \res{ \widetilde{x} }  (P   \| \checkmark) \premat Q'$ hence $Q$  must be of the form $ \res{ \widetilde{x} } (P'   \| \checkmark) $ where $P \premat P' $ . Hence $Q$ reduces to a process that has an unguarded occurence of $\checkmark$.

    \item Suppose that $\piencodfaplas{{M}}_u \succone_{\checkmark}$. We will prove that $ {M} \Downarrow_{\checkmark}$.

    By operational soundness we have that if  $\piencodfaplas{{M}}_u \redone^* Q$ then there exist ${M}'$ and $\{Q_i\}_{i \in I}$ such that:
    (i)~${M} \red^* {M}'$
    and
    (ii)~$Q \redone^* \{Q_i\}_{i \in I}$ with
    $ \piencodfaplas{{M}'}_u \premat Q_j$,  for all $j \in I$.

   Since $\piencodfaplas{{M}}_u \redone^* P_1 \nd \ldots \nd P_k (:= Q)$, and $P_j= P_j'   \| \checkmark$ and $Q \redone^* \{Q_i\}_{i \in I}$ we must have that each $Q_i$ is of the form $P'_{i_1} \nd \ldots \nd P'_{i_k}$ with $P'_{i_j}= P''_{i_j}   \| \checkmark$. As $ \piencodfaplas{{M}'}_u \premat Q_j$,  for all $j \in I$ we have that $ \piencodfaplas{{M}'}_u  = (\bignd_{i \in I} Q_i  ) \nd R$ for some $R$. Finally applying Proposition \ref{ch4Prop:checkprespiunres} (2) we have that $\piencodfaplas{{M}'}_u$ is itself a term with unguarded $\checkmark$, then ${M}$ is itself headed with $\checkmark$.
\end{enumerate}
\end{proof}

\section{Proof of Separation of Lazy and Eager Semantics}
\label{ch4a:piBisim}

\begin{definition}{Dual Prefix}\label{ch4d:readyPrefixDual}
    Given prefixes $\alpha$ and $\beta$ (\Cref{ch4d:prefix}), we say $\alpha$ and $\beta$ are duals, denoted $\alpha \dualPrefix \beta$, if and only if
    $\sub{\alpha} \cap \sub{\beta} \neq \emptyset$.
\end{definition}

\begin{lemma}\label{ch4l:readyPrefixDual}
    Given $P \vdash \emptyset$, if $P \readyPrefix{\alpha}$, then there exist $P',\beta$ such that $P \redone^\ast P' \readyPrefix{\beta}$ and $\alpha \dualPrefix \beta$.
\end{lemma}

\begin{proof}
    By well-typedness, there appears $\beta$ in $P$ with $\alpha \dualPrefix \beta$.
    However, we may have $P \nreadyPrefix{\beta}$, because $\beta$ is blocked by other prefixes.
    Hence, we need to find reductions from $P$ such that we unblock $\beta$.
    However, the prefixes blocking $\beta$ are connected to dual prefixes, that may be blocked themselves.
    The crux of this proof is thus to show that we can reduce $P$, such that we eventually unblock $\beta$.

    The proof is by induction on the number of names that may block $\beta$ (\ih{1}).
    Initially, this number corresponds to the total number of names appearing in $P$.
    Suppose $\beta$ is blocked by $n$ prefixes $\gamma_i$, where $\gamma_n$ blocks $\beta$, and $\gamma_1$ is not blocked.
    We apply another layer of induction on $n$ (\ih{2}).

    In the inductive case, $n \geq 1$.
    The goal is to perform a reduction that synchronizes $\gamma_1$ with its dual, say $\ol{\gamma_1}$.
    The prefix $\ol{\gamma_1}$ may be blocked by a number of prefixes itself.
    However, the type system of \clpi is based on \scc{Cut}, so $\ol{\gamma_1}$ appears in parallel with the duals of $\gamma_2,\ldots,\gamma_n$ and $\alpha$.
    We then may apply \ih{1} to find $P \redone^\ast P_0 \readyPrefix{\ol{\gamma_1}}$.
    We can then reduce $P_0$ by synchronizing between $\gamma_1$ and $\ol{\gamma_1}$: $P \redone^\ast P_0 \redone P_1$.
    In $P_1$, $\beta$ is blocked by one less prefix.
    Hence, by \ih{2}, $P \redone^\ast P_0 \redone P_1 \redone^\ast P' \readyPrefix{\beta}$, proving the thesis.

    In the base case, $\beta$ is not blocked: $P \readyPrefix{\beta}$.
    Let $P' := P$; trivially, $P \redone^\ast P' \readyPrefix{\beta}$, proving the thesis.
\end{proof}

\thmPiBisim*

\begin{proof}
    For~(i), we construct a relation $\mathbb{B}$ as follows:
    \begin{align*}
        \sff{Id}^\equiv
        &:=
        \{ (T,U) \mid T \equiv U \}
        \\
        \mathbb{B}'
        &:=
        \{
            (T,U) \mid \begin{array}[t]{@{}l@{}}
                T \equiv \pctx{M}[\beta_1; (V \nd W)] \vdash \emptyset
                \text{ and}
                \\
                U \equiv \pctx{M}[\beta_2; V \nd \beta_3; W] \vdash \emptyset
                \text{ and}
                \\
                V \nreadyPrefixBisim{L} W
                \text{ and}
                \\
                \beta_1 \relalpha \beta_2 \relalpha \beta_3
                \text{ and}
                \\
                \text{$\beta_1,\beta_2,\beta_3$ require a continuation}
            \}
        \end{array}
        \\
        \mathbb{B}
        &:=
        \sff{Id}^\equiv \cup \mathbb{B}'
    \end{align*}

    We prove that $\mathbb{B}$ is a strong ready-prefix bisimulation w.r.t.\ the lazy semantics by proving the three conditions of \Cref{ch4d:readyPrefixBisim} for each $(T,U) \in \mathbb{B}$.
    We distinguish cases depending on whether $(T,U) \in \sff{Id}^\equiv$ or $(T,U) \in \mathbb{B}'$.
    \begin{itemize}
        \item
            $(T,U) \in \sff{Id}^\equiv$.
            The three conditions hold trivially.

        \item
            $(T,U) \in \mathbb{B}'$.
            Then $T \equiv \pctx{M}[\beta_1; (V \nd W)]$, $U \equiv \pctx{M}[\beta_2; V \nd \beta_3; W]$, $V \nreadyPrefixBisim{L} W$, and $\beta_1 \relalpha \beta_2 \relalpha \beta_3$.
            We prove each condition separately.
            \begin{enumerate}
                \item
                    Suppose $T \redtwo T'$.
                    Note that the hole in $\pctx{M}$ may appear inside a non-deterministic choice.
                    We distinguish three cases: (a)~the reduction is inside $\pctx{M}$ and maintains the branch with the hole, (b)~the reduction is inside $\pctx{M}$ and discards the branch with the hole, or (c)~the reduction synchronizes on $\beta_1$.
                    \begin{enumerate}
                        \item
                            The reduction is inside $\pctx{M}$ and maintains the branch with the hole.
                            Then $T' \equiv \pctx{M'}[\beta_1; (V \nd W)]$ and $U \redtwo U' \equiv \pctx{M'}[\beta_2; V \nd \beta_3; W]$.
                            Clearly, $(T',U') \in \mathbb{B}'$, so $(T',U') \in \mathbb{B}$.

                        \item
                            The reduction is inside $\pctx{M}$ and discards the branch with the hole.
                            Then there exists $U'$ such that $U \redtwo U' \equiv T'$, so $(T',U') \in \sff{Id}^\equiv$, and thus $(T',U') \in \mathbb{B}$.

                        \item
                            The reduction synchronizes on $\beta_1$.
                            Then $T' \equiv \pctx{M'}[V \nd W]$ and, since $\beta_1 \relalpha \beta_2 \relalpha \beta_3$, $U \redtwo U' \equiv \pctx{M'}[V \nd W]$.
                            Then $T' \equiv U'$, so $(T',U') \in \sff{Id}^\equiv$ and thus $(T',U') \in \mathbb{B}$.
                    \end{enumerate}

                \item
                    Suppose $U \redtwo U'$.
                    By reasoning analogous to above, $T \redtwo T'$ and $(T',U') \in \mathbb{B}$.

                \item
                    Suppose $T \readyPrefix{\gamma}$.
                    If the prefix $\gamma$ appears in $\pctx{M}$, then clearly also $U \readyPrefix{\gamma}$.
                    Otherwise, $\gamma = \beta_1$.
                    We have, e.g., $\gamma \relalpha \beta_2$ and clearly $U \readyPrefix{\beta_2}$.
                    The other direction is analogous.
            \end{enumerate}
    \end{itemize}

    It remains to show that $(R,S) \in \mathbb{B}$ which trivially holds.

    \medskip
    For~(ii), toward a contradiction, assume there exists a strong ready-prefix bisimulation w.r.t.\ \redone $\mathbb{B}$ where $(R,S) \in \mathbb{B}$.

    By \Cref{ch4l:readyPrefixDual}, there exist $R',\beta_1$ such that $R \redone^\ast R' \readyPrefix{\beta_1}$, and $\alpha_1 \dualPrefix \beta_1$.
    By the well-typedness of $R$ and $S$, $\beta_1$ must appear in $\pctx{N}$, and the reduction $R \redone^\ast R'$ takes place in $\pctx{N}$.
    Take $x \in \sub{\alpha_1} \cap \sub{\beta_1}$ (which is non-empty by \Cref{ch4l:readyPrefixDual}).
    Then $R' \equiv \pctx[\big]{N'_1}[\res{x}(\pctx{N'_2}[\beta_1; R'_2] \| \pctx{N'_3}[\alpha_1; (P \nd Q)])]$.
    Moreover, clearly $S \redone^\ast S'$ following the same reductions, resulting in $S' \equiv \pctx[\big]{N'_1}[\res{x}(\pctx{N'_2}[\beta_1; R'_2] \| \pctx{N'_3}[\alpha_2; P \nd \alpha_3; Q])]$; note that, since $\alpha_1 \relalpha \alpha_2 \relalpha \alpha_3$, also $\alpha_2 \dualPrefix \beta_1$ and $\alpha_3 \dualPrefix \beta_1$.
    At this point, we must have $(R',S') \in \mathbb{B}$.

    The synchronization between $\beta_1$ and $\alpha_1$ gives $R' \redone R'' \equiv \pctx{N''}[P \nd Q]$.
    Then by the bisimulation, there exists $S''$ such that $S' \redone S''$ with $(R'',S'') \in \mathbb{B}$.
    By clause~3 of the bisimulation, $R''$ and $S''$ must have the same ready-prefixes, so clearly the reduction $S' \redone S''$ results from a synchronization between $\beta_1$ and either of $\alpha_2$ and $\alpha_3$.
    W.l.o.g., let us assume this was $\alpha_3$.
    Then $S'' \equiv \pctx{N''}[Q]$.
    By assumption, $P \nreadyPrefixBisim{E} Q$ and thus $P \nd Q \nreadyPrefixBisim{E} Q$, so clearly $R'' \nreadyPrefixBisim{E} S''$.
    Hence, $\mathbb{B}$ cannot be a strong ready-prefix bisimulation w.r.t.\ \redone.
    In other words, $R \nreadyPrefixBisim{E} S$.
\end{proof}

\newpage

\newpage

\chapter{Appendix of Chapter 5}\label{ch5appendices_ch5}

\section{Proofs of Section~\ref{ch5s:weight}}

This section contains the proofs of the main results in \Cref{ch5s:weight}.
\begin{proposition}
	\label{ch5prop:lvlweightreduce}
	Suppose $ \Gamma \lt P $ and $ P \lts{\tau} P' $ then $ \weight{P'} \wtless \weight{P}$.
\end{proposition}

\begin{proof}
	Follows similarly to that of \cite{DS06}. As $ P \lts{\tau} P' $ then by rule $\lvltype{Tau}$ we have:
	\begin{mathpar}
		\inferrule* [ left = \lvltype{Tau} ]{ Q_1  \lts{ \res {\widetilde{b}}\ov x\out { \tilde{v} } } Q_1' \\ Q_2 \lts{ x\inp {\tilde{v}} }  Q_2' \\ \widetilde{b} \cap \fn{Q} = \emptyset }
		{ Q_1 \pp Q_2 \lts{ \tau } \res{\widetilde{b}} (Q_1'\pp Q_2')}
	\end{mathpar}
	Where $P = Q_1 \pp Q_2$ and $P' =  \res{\widetilde{b}} (Q_1'\pp Q_2')$. By \Cref{ch5thm:lvltypepres} $ \Gamma \lt P' $ and hence $\weight{P'}$ is defined.	
	We have that $ \weight{ Q_1 \pp Q_2} = \weight{ Q_1 } +  \weight{ Q_2} $ similarly that $ \weight{\res{\widetilde{b}} Q'_1 \pp Q'_2} = \weight{ Q'_1 } +  \weight{ Q'_2}$.

	We wish to show the following:
	\[
	\begin{aligned}
		\weight{ Q'_1 } +  \weight{ Q'_2}
								& \wtless \weight{ Q_1 } +  \weight{ Q_2}\\
								& = \weight{ Q_1 } +  \weight{ Q_2} + \zerovec_{\levelof{x}} -\zerovec_{\levelof{x}} \\
	\end{aligned}
	\]
Hence it is sufficient to show that $ \weight{ Q'_1 } \wtleq \weight{ Q_1 } - \zerovec_{\levelof{x}} $ and $ \weight{ Q'_2 } \wtless \weight{ Q_2 } + \zerovec_{\levelof{x}} $ which can be reduced to the following two statements:
	\begin{enumerate}

		\item $\Gamma \lt P $ and $ P \lts{ x\inp {\tilde{v}} } P $'then $ \weight{P'} \wtless  \weight{P}  + \zerovec_{\levelof{x}}  $

		\item $\Gamma \lt P $ and $ P \lts{\ov x\out { \tilde{v} }} P $'then $ \weight{P'} \wtleq  \weight{P}  - \zerovec_{\levelof{x}}  $
	\end{enumerate}

	We shall show only (1),   as (2) follows similarly.
	As $\Gamma \lt P $ and $ P \lts{ x\inp {\tilde{v}} } P' $ then there are two cases to consider. Firstly by the rule 
	\begin{mathpar}
		\inferrule{  }
		{ x\inp {\tilde{y}}.Q \lts{x\inp {\tilde{y}}} Q }
	\end{mathpar}
	with $\weight{x\inp {\tilde{y}}.Q} = \weight{Q} \wtless \weight{Q} + \zerovec_{\levelof{x}}$.

	Secondly by the rule:
	\begin{mathpar}
		\inferrule{  }
		{ \bang x \inp {\tilde{y}}.Q \lts{x\inp {\tilde{y}}} \bang x \inp {\tilde{y}}.Q  \pp Q  }
	\end{mathpar}
	with $\weight{\bang x \inp {\tilde{y}}.Q } = \zerovec$ and $\weight{Q} \wtless \zerovec_{\levelof{x}}$
	as by typing $\forall b \in \outsubj{P}, \  \levelof{b} < \levelof{x}$.
	Finally we have that $\weight{\bang x \inp {\tilde{y}}.Q \pp Q} = \zerovec + \weight{Q} \wtless \zerovec + \zerovec_{\levelof{x}} $.
\end{proof}

\begin{theorem}[Termination]
\label{ch5t:lvlsn}
	If $\Gamma \lt P$ then $P$ terminates.
\end{theorem}

\begin{proof}
	Let us consider the following two cases:
	\begin{enumerate}
		\item When $ \weight{ P } = \zerovec $ then $P$ has no active outputs. Therefore, $P$ has no enabled synchronizations, and so no reductions can take place. Thus, $P$ terminates.
		\item When $ \zerovec \wtless \weight{ P } $ then the set of active inputs is non-empty. If none of the channels in the set of active inputs enables a synchronization then the process must terminate. If syncronizations are enabled then, by \Cref{ch5prop:lvlweightreduce}, any synchronization strictly reduces the weight of the reduced process. As processes are of finite length and hence weights are finite we may inductively reduce until we reach a process of weight $\zerovec$ or such that the set of active outputs do not allow for synchronizations.
	\end{enumerate}
\end{proof}

\section{Proofs of Section~\ref{ch5s:lvllang}}

In this section we present the proofs of the main results in \Cref{ch5s:lvllang}. Namely,  operational completeness~(\Cref{ch5thm:op_completeness}) and operational soundness~(\Cref{ch5thm:op_soundness})  of the translation $\sttoltj{\cdot }{l}{}$, which require the substitution lemma (\Cref{ch5lem:substitution}) below. We also present the detailed proof of the proper inclusion $\lvllang\subset \vaslang$ (\Cref{ch5t:wsubsets}).

\begin{lemma}[Substitution Lemma]\label{ch5lem:substitution}
Suppose $\Gamma , u: T , v : S \lt P$, with $T = \dual{S}$	and $ \contun{T}$ for a level function $l$. Then  $\Gamma , z: \dualjoin{T}{S} \lt \lvljoinsub{P}{z}{u,v}$ for $l$.
\end{lemma}

\begin{proof}
	By induction on the structure of $P$.
	\begin{enumerate}
		\item Case $P = \nil$.  
		
		Then $\Gamma , u: T , v : S \lt \nil$ with  $T = \dual{S}$ and $ \contun{T}$ implies  that the following derivation is possible:
			\begin{mathpar}
				\inferrule* [ left = \lvltype{Nil} ]{ \contun{\Gamma , u: T , v : S \lt \nil} }
				{ \Gamma , u: T , v : S \lt \nil\lt \nil }
			\end{mathpar}
			The result follows trivially, since one has the derivation:
			\begin{mathpar}
				\inferrule* [ left = \lvltype{Nil} ]{ \contun{\Gamma , z: \dualjoin{T}{S}} }
				{ \Gamma , z: \dualjoin{T}{S} \lt \nil }
			\end{mathpar}
		\item Case $P = \lvlinpoly{x}{y}{z}{P'}$. 
		
		Then there are  three cases to consider, based on the rule applied in the derivation of $\Gamma , u: T , v : S \lt \lvlinpoly{x}{y}{z}{P'}$,  which can be \lvltype{Lin-In_1},  or \lvltype{Lin-In_2} or  \lvltype{Lin-In_3}, depending on  whether $\Gamma=\Gamma_1, x: \lvlintpoly{n}{V}{V_2} \contcomp \Gamma_2 $, or $\Gamma=\Gamma_1, x: \lvlunst{n}{V} \contcomp \Gamma_2, x: \lvlunst{n}{V} $, or $\Gamma=\Gamma_1, \lvlsplit{x}{\lvlunst{n}{V}} \contcomp \Gamma_2,\lvlsplit{x}{\lvlunst{n}{V}}$, respectively.
			\begin{enumerate}
				\item\label{ch5proof:jointypes1} In the case $\Gamma=\Gamma_1,  x: \lvlintpoly{n}{V}{V_2} \contcomp \Gamma_2$, for some contexts $\Gamma_1,\Gamma_2$, the rule applied is  \lvltype{Lin-In_1}. Thus, by hypothesis, $\Gamma_1 , x: \lvlintpoly{n}{V}{V_2}, u: T , v : S \contcomp \Gamma_2 , u: T , v : S  \lt \lvlinpoly{x}{y}{z}{P'}$ with $T=\dual{S}$ and $\contun{T}$. It follows from the definition of $\contun{T}$  that it is not the case that $x= u$ or $x=v$. The derivation is as  follows:			 
					\begin{mathpar}
						\mprset{ flushleft }
						\inferrule* [ left = \lvltype{Lin-In_1} ]{ 
							\Gamma_1 , x: \lvlintpoly{n}{V}{V_2} , u: T , v : S  \lt x: \lvlintpoly{n}{V}{V_2} \\
							\Gamma_2 ,   \tilde{y}:\tilde{V}, u: T , v : S \lt P' \\\tilde{y} = y_1,y_2\\ \levelof{x} = \levelof{y_2} }
						{(\Gamma_1 , x: \lvlintpoly{n}{V}{V_2} , u: T , v : S) \contcomp (\Gamma_2 , u: T , v : S) \lt \lvlinpoly{x}{y}{z}{P'}}
					\end{mathpar} 
					By the induction hypothesis 
					and $\lvljoinsub{\lvlinpoly{x}{y}{z}{P'}}{z}{u,v} = \lvlinpoly{x}{y}{z}{(\lvljoinsub{P'}{z}{u,v})}$, one has the derivation:
					\begin{mathpar}
						\mprset{ flushleft }
						\inferrule* [ left = \lvltype{Lin-In_1} ]{ 
							\Gamma_1 , x: \lvlintpoly{n}{V}{V_2} ,  z: \dualjoin{T}{S} \lt x: \lvlintpoly{n}{V}{V_2}
						\\	
						\Gamma_2 ,   \tilde{y}:\tilde{V},   z: \dualjoin{T}{S} \lt \lvljoinsub{P'}{z}{u,v} \\\tilde{y} = y_1,y_2\\ \levelof{x} = \levelof{y_2} }
						{(\Gamma_1 , x: \lvlintpoly{n}{V}{V_2} ,  z: \dualjoin{T}{S}) \contcomp (\Gamma_2,  z: \dualjoin{T}{S}) \lt \lvlinpoly{x}{y}{z}{\lvljoinsub{P'}{z}{u,v}} }
					\end{mathpar} 
					and the result follows.
				\item  In the case $\Gamma=\Gamma_1, \lvlsplit{x}{\lvlunst{n}{V}} \contcomp \Gamma_2,\lvlsplit{x}{\lvlunst{n}{V}}$, for some $\Gamma_1,\Gamma_2$, the rule is \lvltype{Lin-In_3}.
				  The result follows analogously to case (\ref{ch5proof:jointypes1}).
				
			\item In the case  $\Gamma= \Gamma_1, x: \lvlunst{n}{V} \contcomp \Gamma_2, x: \lvlunst{n}{V} $, either  \lvltype{Lin-In_2}  or \lvltype{Lin-In_3} could be applied:
			
			Since $\Gamma_1 , x: \lvlunst{n}{V} , u: T , v : S \contcomp \Gamma_2, x: \lvlunst{n}{V}, u: T , v : S  \lt \lvlinpoly{x}{y}{z}{P'}$ with  $T = \dual{S}$ and $ \contun{T}$, there are two possibilities: 				
				\begin{enumerate}
					\item Case $u = x$.
					
					 Then, the context boils down to $\Gamma_1 , x: \lvlunst{n}{V} , v : \dual{\lvlunst{n}{V}} \contcomp \Gamma_2,x: \lvlunst{n}{V} , v : \dual{\lvlunst{n}{V}}  $, and the derivation is as follows:
						\begin{mathpar}
							\hspace{-0.8cm}
							\small
							\mprset{ flushleft }
							\inferrule* [ left = \lvltype{Lin-In_2} ]{  
							\Gamma_1 , x: \lvlunst{n}{V} , v : \dual{\lvlunst{n}{V}} \lt x: \lvlunst{n}{V} \\
							\Gamma_2 , x: \lvlunst{n}{V} ,   y_1:V  , y_2: \lvlnilT , v : \dual{\lvlunst{n}{V}} \lt P' \\ \tilde{y} = y_1,y_2 
							}
							{ (\Gamma_1 , x: \lvlunst{n}{V} , v : \dual{\lvlunst{n}{V}}) \contcomp (\Gamma_2,x: \lvlunst{n}{V} , v : \dual{\lvlunst{n}{V}}) \lt \lvlinpoly{x}{y}{z}{P'}  }
						\end{mathpar}
						By the induction hypothesis, we obtain:
						\begin{mathpar}
							\mprset{ flushleft }
							\inferrule* [ left = \lvltype{Lin-In_3} ]{
							\Gamma_1 , \lvlsplit{z}{\lvlunst{n}{V}} \lt  z: \lvlunst{n}{V}\\
							\Gamma_2 , \lvlsplit{z}{ \lvlunst{n}{V}} ,   y_1:V  , y_2: \lvlnilT \lt \lvljoinsub{P'}{z}{x,v} \\ \tilde{y} = y_1,y_2 }
							{ (\Gamma_1 , \lvlsplit{z}{\lvlunst{n}{V}}) \contcomp (\Gamma_2 , \lvlsplit{z}{\lvlunst{n}{V}} ) \lt \lvlinpoly{z}{y}{z}{\lvljoinsub{P'}{z}{x,v}}  }
						\end{mathpar}
					\item Case $u \not = x$.
					Then the proof  follows analogously as in case (\ref{ch5proof:jointypes1}).
				\end{enumerate}
			\end{enumerate}

		\item Case $P = \lvlout{x}{y_1}{y_2}{P'}$.
		
		 Then there are  three cases to consider, based on the rule applied to the derivation $ \Gamma , u: T , v : S \lt \lvlout{x}{y_1}{y_2}{P'}$, which can be \lvltype{Lin-Out}, or  \lvltype{Lin-Out_1}, or  \lvltype{Lin-Out_2}, depending on the context $\Gamma$.
		
			\begin{enumerate}
				
				\item\label{ch5proof:jointypesout1} The rule is \lvltype{Lin-Out}. 

				Since $T=\dual{S}$ and  $ \contun{T}$, it follows that  we need not consider the case of $x = u$, but we still need to consider  the following cases:
				
				\begin{enumerate}
					\item $u \not = y_1$.
					
					Then, there exist $\Gamma_1, \Gamma_2, \Gamma_3$ such that $\contun{\Gamma_1,\Gamma_2}$ and  $ \Gamma , u: T , v : S \lt \lvlout{x}{y_1}{y_2}{P'}$ can be written as  $ (\Gamma_1, u: T , v : S, x: \lvloutt{n}{V_1}{V_2}) \contcomp (\Gamma_2, u: T , v : S , y_1:V_1) \contcomp (\Gamma_3, u: T , v : S , y_2: V_2) \lt \lvlout{x}{y_1}{y_2}{P'}$. Thus, the following derivation holds:
\begin{mathpar}
	\mprset{ flushleft }
    \inferrule* [ left = \lvltype{Lin-Out} ]{ 
		\Gamma_1, u: T , v : S, x: \lvloutt{n}{V_1}{V_2} \lt x: \lvloutt{n}{V_1}{V_2} \
		\\
		\Gamma_2, u: T , v : S , y_1:V_1 \lt  y_1:V_1
		\\
		\Gamma_3, u: T , v : S , y_2:V_2  \lt P' \\ \levelof{x} = \levelof{y_2} }
	{  \Gamma_1' \contcomp \Gamma_2' \contcomp 
	 \Gamma_3'  \lt \lvlout{x}{y_1}{y_2}{P'} }
\end{mathpar}
where 
\begin{itemize}
\item $\Gamma_1'=\Gamma_1, u: T , v : S, x: \lvloutt{n}{V_1}{V_2}$, 
\item $\Gamma_2'=\Gamma_2, u: T , v : S , y_1:V_1$
\item $\Gamma_3'=\Gamma_3, u: T , v : S , y_2: V_2$
\end{itemize}

%
%
%
						By applying the induction hypothesis in $\Gamma_3, u: T , v : S , y_2: V_2  \lt P'$, the judgment 
						$ \Gamma_3, z: \dualjoin{T}{S} , y_2:V_2  \lt  \lvljoinsub{P'}{z}{u,v}$ holds. Then  there exists a derivation
						\begin{mathpar}
							\mprset{ flushleft }
							\inferrule* [ left = \lvltype{Lin-Out} ]{
								\Gamma_1, z: \dualjoin{T}{S} , x: \lvloutt{n}{V_1}{V_2} \lt x: \lvloutt{n}{V_1}{V_2}
								\\
								\Gamma_2, z: \dualjoin{T}{S} , y_1:V_1 \lt y_1:V_1
								\\
								\Gamma_3, z: \dualjoin{T}{S} , y_2: V_2 \lt  \lvljoinsub{P'}{z}{u,v}\\ \levelof{x} = \levelof{y_2} }
							{ (\Gamma_1, z: \dualjoin{T}{S} , x: \lvloutt{n}{V_1}{V_2}) \contcomp (\Gamma_2, z: \dualjoin{T}{S} , y_1:V_1) \contcomp \\\\  (\Gamma_3, z: \dualjoin{T}{S} , y_2: V_2)  \lt \lvlout{x}{y_1}{y_2}{\lvljoinsub{P'}{z}{u,v}} }
						\end{mathpar}
and the result follows.

\item $u = y_1$

Then there exist $\Gamma_1, \Gamma_2, \Gamma_3$  such that $\contun{\Gamma_1,\Gamma_2}$ and  $ \Gamma , u: T , v : S \lt \lvlout{x}{y_1}{y_2}{P'}$ can be written as  $ (\Gamma_1, y_1:V_1 , v : \dual{V_1}, x: \lvloutt{n}{V_1}{V_2}) \contcomp (\Gamma_2, y_1:V_1 , v : \dual{V_1}) \contcomp  (\Gamma_3, y_1:V_1 , v : \dual{V_1} , y_2: V_2)  \lt \lvlout{x}{y_1}{y_2}{P'}$. Thus, the following derivation holds:
\begin{mathpar}
	\mprset{ flushleft }
      \inferrule* [ left = \lvltype{Lin-Out} ]{
		\Gamma_1, y_1:V_1 , v : \dual{V_1}, x: \lvloutt{n}{V_1}{V_2} \lt x: \lvloutt{n}{V_1}{V_2}
		\\
		\Gamma_2, y_1:V_1 , v : \dual{V_1} \lt y_1:V_1
		\\
		\Gamma_3, y_1:V_1 , v : \dual{V_1} , y_2: V_2 \lt P' \\ \levelof{x} = \levelof{y_2} }
	{ (\Gamma_1, y_1:V_1 , v : \dual{V_1}, x: \lvloutt{n}{V_1}{V_2}) \contcomp (\Gamma_2, y_1:V_1 , v : \dual{V_1}) \contcomp 
	\\  \quad (\Gamma_3, y_1:V_1 , v : \dual{V_1} , y_2: V_2)   \lt \lvlout{x}{y_1}{y_2}{P'} }
\end{mathpar}

%

						By applying the induction hypothesis in $\Gamma_3, y_1:V_1 , v : \dual{V_1} , y_2: V_2 \lt P'$ we obtain $\Gamma_3, z: \dualjoin{V_1}{\dual{V_1}} , y_2:V_2  \lt \lvljoinsub{P'}{z}{y_1,v} $.  Since $\contun{\Gamma''} $ and $l(x)=l(y_2)$, we have:
						\begin{mathpar}
							\mprset{ flushleft }
							\inferrule* [ left = \lvltype{Lin-Out} ]{
								\Gamma_1, z: \dualjoin{V_1}{\dual{V_1}} , x: \lvloutt{n}{V_1}{V_2} \lt x: \lvloutt{n}{V_1}{V_2}
								\\
								\Gamma_2, z: \dualjoin{V_1}{\dual{V_1}}  \lt z: V_1
								\\
								\Gamma_3, z: \dualjoin{V_1}{\dual{V_1}} , y_2:V_2  \lt P' \\ \levelof{x} = \levelof{y_2} }
							{ 
								(\Gamma_1, z: \dualjoin{V_1}{\dual{V_1}} , x: \lvloutt{n}{V_1}{V_2}) \contcomp (\Gamma_2, z: \dualjoin{V_1}{\dual{V_1}} ) \contcomp 
								\\  \quad (\Gamma_3, z: \dualjoin{V_1}{\dual{V_1}} , y_2: V_2)  \lt \lvlout{x}{z}{y_2}{\lvljoinsub{P'}{z}{y_1,v}} }
						\end{mathpar}

				\end{enumerate}

				\item The rule is  \lvltype{Un-Out_1}. 
				
				There are two cases: 
				
\begin{enumerate}
\item  $u  = x$

Then there exist $\Gamma_1, \Gamma_2,\Gamma_3$  such that $\contun{\Gamma_1,\Gamma_2}$ and  $ \Gamma , u: T , v : S \lt \lvlout{x}{y_1}{y_2}{P'}$ can be written as  $ (\Gamma_1, x: \lvlunct{n}{V}  , v:\dual{\lvlunct{n}{V} }) \contcomp  (\Gamma_2 , x: \lvlunct{n}{V} , v:\dual{\lvlunct{n}{V} } , y_1:V_1 ) \contcomp (\Gamma_3 , x: \lvlunct{n}{V}  , v:\dual{\lvlunct{n}{V} } ) \lt \lvlout{x}{y_1}{y_2}{P'}$. By the hypothesis $T=\dual{S}$, we conclude that  $ , v:\dual{\lvlunct{n}{V} } $ is in both $\Gamma'$ and $\Gamma''$. Thus, the following derivation holds:

\begin{mathpar}\hspace{-1.0cm}
	\small
      \inferrule* [ left = \lvltype{Un-Out_1} ]{ 
		\Gamma_1, x: \lvlunct{n}{V} , v:\dual{\lvlunct{n}{V} } \lt x: \lvlunct{n}{V}  \\
		\Gamma_2 , x: \lvlunct{n}{V} , v:\dual{\lvlunct{n}{V} } , y_1:V_1 \lt y_1:V_1 \\
		\Gamma_3, v : \dual{\lvlunct{n}{V}}, x: \lvlunct{n}{V},  y_2:\lvlnilT  \lt P'  }
	{ (\Gamma_1, x: \lvlunct{n}{V}  , v:\dual{\lvlunct{n}{V} } )\contcomp  (\Gamma_2 , x: \lvlunct{n}{V} , v:\dual{\lvlunct{n}{V} } , y_1:V_1  )\contcomp \\
	\Gamma_3 , x: \lvlunct{n}{V}  , v:\dual{\lvlunct{n}{V} } \lt \lvlout{x}{y_1}{y_2}{P'} }
\end{mathpar}

%

					By applying the induction hypothesis on $\Gamma_3, v : \dual{\lvlunct{n}{V}}, x: \lvlunct{n}{V},  y_2:\lvlnilT \lt P' $ we obtain $\Gamma_3, z: \dualjoin{\lvlunct{n}{V}}{\dual{\lvlunct{n}{V}}} , y_2: \lvlnilT   \lt \lvljoinsub{P'}{z}{u,v}  $:
					\begin{mathpar}
						\hspace{-0.5cm}
						\small
						\inferrule* [ left = \lvltype{Un-Out_2} ]{ 
							\Gamma_1, z: \dualjoin{\lvlunct{n}{V}}{\dual{\lvlunct{n}{V}}} \lt z: \lvlunct{n}{V} \\
							\Gamma_2 , z: \dualjoin{\lvlunct{n}{V}}{\dual{\lvlunct{n}{V}}} , y_1:V_1 \lt y_1:V_1 \\
							\Gamma_3, z: \dualjoin{\lvlunct{n}{V}}{\dual{\lvlunct{n}{V}}} , y_2: \lvlnilT   \lt P'\lvljoinsub{P'}{z}{u,v}  
						}
						{  (\Gamma_1, z: \dualjoin{\lvlunct{n}{V}}{\dual{\lvlunct{n}{V}}} ) \contcomp  (\Gamma_2 , z: \dualjoin{\lvlunct{n}{V}}{\dual{\lvlunct{n}{V}}} , y_1:V_1  )\contcomp \\
						\Gamma_3 , z: \dualjoin{\lvlunct{n}{V}}{\dual{\lvlunct{n}{V}}} \lt \lvlout{z}{y_1}{y_2}{\lvljoinsub{P'}{z}{u,v}}
						}
					\end{mathpar}
and the result follows. 
\item $u \not = x$ 

This case is analogous to  (\ref{ch5proof:jointypesout1}). 
\end{enumerate}
				\item The rule is \lvltype{Un-Out_2}. 

				This  case is  analogous to (\ref{ch5proof:jointypesout1}).

			\end{enumerate}

		\item Case  $P = P' \pp Q$. 
		
		Then there exist $\Gamma_1, \Gamma_2$ such that   $\Gamma, u:T, v:S\lt P'\pp Q$ has the form  $(\Gamma_1 , u: T , v : S)  \contcomp (\Gamma_2 , u: T , v : S)  \lt P' \pp Q$, since    $T = \dual{S}$ and $ \contun{T}$. Thus, the following derivation holds  
			\begin{mathpar}
				\inferrule* [ left = \lvltype{Par} ]{\Gamma_1, u: T , v : S  \lt P'  \\ \Gamma_2, u: T , v : S \lt Q }
				{ (\Gamma_1 , u: T , v : S)  \contcomp (\Gamma_2 , u: T , v : S)  \lt P' \pp Q }
			\end{mathpar} 
			By  applying he induction hypothesis on $\Gamma_1, u: T , v : S  \lt P'$ and $\Gamma_2, u: T , v : S \lt Q$, we obtain 
			\begin{mathpar}
				\inferrule* [ left = \lvltype{Par} ]{\Gamma_1,z: \dualjoin{T}{S}   \lt \lvljoinsub{P'}{z}{u,v}  \\ \Gamma_2, z: \dualjoin{T}{S} \lt \lvljoinsub{Q}{z}{u,v} }
				{ (\Gamma_1 , z: \dualjoin{T}{S} ) \contcomp (\Gamma_2 , z: \dualjoin{T}{S}) \lt \lvljoinsub{P'}{z}{u,v} \pp \lvljoinsub{Q}{z}{u,v} }
			\end{mathpar}
			and the result follows.
		\item Case $P = \res {x}P' $. 
		
		 Then $\Gamma  , u: T , v : S  \lt \res {x}P'$ with  $T = \dual{S}$ and $ \contun{T}$ which implies that the following derivation holds:
			\begin{mathpar}
				\inferrule* [ left = \lvltype{Res} ]{ \Gamma , x: \dualjoin{V}{\dual{V}}  , u: T , v : S  \lt P'  }
				{ \Gamma  , u: T , v : S  \lt \res {x}P' }
			\end{mathpar}
			By  applying the induction hypothesis on $\Gamma , x: \dualjoin{V}{\dual{V}}  , u: T , v : S  \lt P'$ we obtain that $\Gamma, x: \dualjoin{V}{\dual{V}} , z: \dualjoin{T}{S}  \lt \lvljoinsub{P'}{z}{u,v} $. Thus, 
			\begin{mathpar}
				\inferrule* [ left = \lvltype{Res} ]{ \Gamma, x: \dualjoin{V}{\dual{V}} , z: \dualjoin{T}{S}  \lt \lvljoinsub{P'}{z}{u,v}  }
				{ \Gamma , z: \dualjoin{T}{S}  \lt \res {x}\lvljoinsub{P'}{z}{u,v} }
			\end{mathpar}
			
			and the result follows.

		\item Case $P =  \lvlservpoly{x}{y}{z}{P'}$.
		
		There are two  cases to consider:

		\begin{enumerate}
			\item $u = x$. 
			
			Then $\Gamma  , x: \lvlunst{n}{V} , v : \dual{\lvlunst{n}{V}}  \lt \lvlservpoly{x}{y}{z}{P'}$ then
				\begin{mathpar}
					\mprset{flushleft}
					\inferrule* [ left = \lvltype{Un-In_1} ]{ 
					\Gamma  , x: \lvlunst{n}{V} \lt x: \lvlunst{n}{V} \\
					\Gamma, x: \lvlunst{n}{V} , v : \dual{\lvlunst{n}{V}} , y_1:V , y_2: \lvlnilT \lt P' \\  \forall b \in \outsubj{P'}, \ \levelof{b} < n }
					{ \Gamma  , x: \lvlunst{n}{V} , v : \dual{\lvlunst{n}{V}}  \lt \lvlserv{x}{y_1}{y_2}{P'}}
				\end{mathpar}
	
				By the induction hypothesis on $\Gamma, x: \lvlunst{n}{V} , v : \dual{\lvlunst{n}{V}}, y_1:V , y_2: \lvlnilT \lt P'$ we have $\Gamma, \lvlsplit{z}{ \lvlunst{n}{V} }, y_1:V , y_2: \lvlnilT \lt \lvljoinsub{P'}{z}{u,v}$ hence:
				\begin{mathpar}
					\mprset{flushleft}
					\inferrule* [ left = \lvltype{Un-In_2} ]{
					\Gamma, \lvlsplit{z}{ \lvlunst{n}{V} } \lt z:{ \lvlunst{n}{V} }
					\\	
					\Gamma, \lvlsplit{z}{ \lvlunst{n}{V} }, y_1:V , y_2: \lvlnilT \lt \lvljoinsub{P'}{z}{u,v} \\  \forall b \in \outsubj{P}, \ \levelof{b} < n }
					{ \Gamma, \lvlsplit{z}{ \lvlunst{n}{V} } \lt  \lvlserv{x}{y_1}{y_2}{\lvljoinsub{P'}{z}{u,v}} }
				\end{mathpar}

				and the result follows. 
				\item $u \not = x$.
				
				 This case follows anogously to (\ref{ch5proof:jointypes1}).

		\end{enumerate}

	\end{enumerate}

\end{proof}

\begin{theorem}[Operational Completeness]
	Let $ P \in \lvllang$ such that  $\sttoltt{\Gamma}{l}{} \lt \sttoltp{P}{}$,  for some level function $l$. Then there exists $R \in \lvllang$ such that $P \vasred Q \implies\sttoltp{P}{}{} \lts{ \tau } \sttoltp{R}{}{}$  and $R \equiv Q$.
\end{theorem}

\begin{proof}
By induction on the reduction rules $\vasred$.
	\begin{enumerate}
		\item When $P = \vasres{xy}{( \vaspara{\vasout{x}{v}{P'}} {\vaspara{\vasin{\lin}{y}{w}{Q'}}{R}}  )} $ then $P \vasred \vasres{xy}{(\vaspara{P'}{\vaspara{Q'\substj{v}{w}}{R}} )} = Q$.

		Let us consider the following cases:
		
		\begin{enumerate}
			\item\label{ch5proof:encod1comp1} When $x: \lin \oc T.S$, typing ensures $y \not \in \fn{P'}$, $x \not \in \fn{Q'}$ and $x,y \not \in \fn{R'}$.

			 Hence
			
			\[
			\begin{aligned}
				\sttoltp{ P }{} &=   \res {z} ( \res{u} \lvlout{z}{v}{u}{\sttoltp{(P'\substj{u}{x})\substj{z}{x}}{}} 
				\pp \lvlin{z}{w}{u}{\sttoltp{(Q'\substj{u}{y})\substj{z}{y}}{}} \pp \sttoltp{R'}{} )
				\\
				& \lts{ \tau } \res {z}  \res{u} ( \sttoltp{(P'\substj{u}{x})\substj{z}{x}}{} 
				\pp \sttoltp{(Q'\substj{u}{y})\substj{z}{y}\substj{v}{w}}{} \pp \sttoltp{R'}{} ) 
			\end{aligned}
			\]

			and
			\[
				Q \equiv \res{rs} Q = R'
			\]
			\[
				\sttoltp{ \res{rs} Q }{} = \res{r} \res {z} ( \sttoltp{(P')\substj{z}{x}}{} 
				\pp \sttoltp{(Q')\substj{z}{y}\substj{v}{w}}{} \pp \sttoltp{R'}{} )
			\]

			\item When $ x: \recur{\oc T} $. Then
			
			$\sttoltp{ P }{} =   \res {z} ( \lvlout{z}{v}{u}{\sttoltp{P'\substj{z}{x}\substj{z}{y}}{}} 
			\pp \lvlin{x}{w}{u}{\sttoltp{Q'\substj{z}{x}\substj{z}{y}}{}}\pp R\substj{z}{x}\substj{z}{y} )
		   $ and the result follow analogously to \ref{ch5proof:encod1comp1}.
		\end{enumerate}

		\item When $P = \vasres{xy}{(\vaspara{\vasout{x}{v}{P'}}	{\vaspara{\vasin{\un}{y}{w}{Q'}}	{R'}})} $ then $ P \vasred \vasres{xy}{(	\vaspara{P'}	{\vaspara{Q'\substj{v}{w}}	{\vaspara{\vasin{\un}{y}{w}{Q'}}	{R'}}}	)} = Q = R$ and typing enforces $x: \recur{\oc T}$.
		
		\[
			\begin{aligned}
				\sttoltp{ P }{} &=   \res {z} ( \lvlout{z}{v}{u}{\sttoltp{P'\substj{z}{x}\substj{z}{y}}{}} 
				\pp \lvlserv{z}{w}{u}{\sttoltp{Q'\substj{z}{x}\substj{z}{y}}{}} \pp \sttoltp{R'}{}\substj{z}{x}\substj{z}{y} )
				\\
				& \lts{ \tau } \res {z}  ( \sttoltp{(P')\substj{z}{x}\substj{z}{y}}{} 
				\pp \sttoltp{(Q')\substj{z}{x}\substj{z}{y}\substj{v}{w}}{} \pp \lvlserv{z}{w}{u}{\sttoltp{Q'\substj{z}{x}\substj{z}{y}}{}} \pp \\
				& \qquad \sttoltp{R'}{}\substj{z}{x}\substj{z}{y} ) \\
				& = \sttoltp{R}{}
			\end{aligned}
		\]

		\item When $P = \vaspara{P'}{R'} $ then $ P \vasred \vaspara{Q'}{R'} = Q$ when $P' \vasred Q'$. 
		
		By the induction hypothesis $P' \vasred Q'$ implies there exist a $R''$ such that $\sttoltp{P'}{} \lts{\tau} \sttoltp{R''}{} $ with $ Q' \equiv R''$. As  $ Q' \pp R' \equiv R'' \pp R'$ by \lvltype{Par} $ \sttoltp{P'}{} = \sttoltp{P'}{} \pp \sttoltp{R'}{} \lts{\tau} \sttoltp{R''}{} \pp \sttoltp{R'}{} = R$

		\item When $P = \vasres{xy}{P'} $ then $ P  \vasred \vasres{xy}{Q'}= Q$ when $P' \vasred Q'$. 
		
		By the induction hypothesis $P' \vasred Q'$ implies there exist a $R'$ such that $\sttoltp{P'}{} \lts{\tau} \sttoltp{R'}{} $ with $ Q' \equiv R'$. As  $ \vasres{xy}Q' \equiv \vasres{xy}R'$ by \lvltype{Res} $ \sttoltp{P'}{} = \sttoltp{ \vasres{xy}P'}{} \lts{\tau} \sttoltp{ \vasres{xy}R'}{}  = R$ 

		\item When $P  \vasred Q $ given $P \equiv P' $, $ P'  \vasred Q'$and $Q' \equiv Q$ then consider the shape of $P$:

			\begin{enumerate}
				\item When  either $P = \vaspara{R}{S} \equiv \vaspara{S}{R}$, $ P = \vaspara{(\vaspara{R}{S})}{T} \equiv \vaspara{R}{(\vaspara{S}{T})}  $ or $ P =\vaspara{P}{\nil}  \equiv P $ then by \lvltype{Par} the result follow.
				
				\item When $P = \vasres{xy}{\nil}  \equiv \nil $ then no reduction is avaliable.
				
				\item When $ P =\vasres{xy}\vasres{zw}{R}  \equiv \vasres{zw}\vasres{xy}{R} $ then by \lvltype{Res} the result follow.
				
				\item When $P = \vaspara{(\vasres{xy}R)}{S}  \equiv \vasres{xy}(\vaspara{R}{S}) \quad \text{if } x,y \not \in \fv{S} $ then by applications of \lvltype{Par} and \lvltype{Res} the result follow.
				
				\item When $P = \vasres{xy}{R}  \equiv \vasres{yx}{R}  = P'$ then $\sttoltp{ P  }{} 	=  \res {z} \sttoltp{R\substj{z}{x}\substj{z}{y}}{} = \sttoltp{ P'  }{} $.
				
			\end{enumerate}

	\end{enumerate}

\end {proof}

\begin{theorem}[Operational Soundness]
	Let $P \in \lvllang$ with  $\sttoltt{\Gamma}{l}{} \lt \sttoltp{P}{}$, for some level function $l$. If  $\sttoltp{P}{}{} \lts{ \tau } U$ Then there exists $R,Q \in \lvllang$ such that $P \vasred Q \land R \equiv Q \land U = \sttoltp{R}{}{} $. 
\end{theorem}

\begin{proof}
	Proof by induction on the reduction rules. We will omit the symmetry of \lvltype{Par} and \lvltype{Tau}. Consider the following cases:

	\begin{enumerate}
		
		\item When $\sttoltp{P}{}{}$ reduces via $\lvltype{Tau}$.
		
		 Then 
		 \begin{itemize}
			\item  ${P = \vasres{xy}{( \vaspara{{P'}}{{Q'}}  )}}$
			\item ${\sttoltp{P}{}{} = \res {z} ( {\sttoltp{P'\substj{z}{x}\substj{z}{y}}{}} \pp {\sttoltp{Q'\substj{z}{x}\substj{z}{y}}{}} )  \lts{ \tau } \res{\widetilde{b}} (\sttoltp{P''}{}{}\pp \sttoltp{Q''}{}{})}$ 
		 \end{itemize}
		 with  $\sttoltp{P'}{}{}  \lts{ \res {\widetilde{b}}\ov z\out {v,u } } \sttoltp{P''}{}{} $, $ \sttoltp{Q'}{}{} \lts{ z \inp {w,u} } \sttoltp{Q''}{}{} $ and $ \widetilde{b} \cap \fn{\sttoltp{Q'}{}{}} = \emptyset$. We must consider the following cases in the reductions on $\sttoltp{P'}{}{}$ and $\sttoltp{Q'}{}{}$ respectively.

		\begin{enumerate}

			\item When $\sttoltp{P'}{}{}$ reduces via \lvltype{Out} and $\sttoltp{Q'}{}{}$ reduces via \lvltype{In}. 
			
			Then  $P' = \vasout{x}{v}{R'}$ and $Q' = \vasin{\lin}{x}{w}{S}$. Let us consider the following cases for the type of $x$, this being linear or unrestricted:

			\begin{enumerate}
				\item\label{ch5proof:encod1sound1} When $x: \lin \oc T.S'$, typing ensures $y \not \in \fn{P'}$ and $x \not \in \fn{Q'}$. Hence
				
				\[
				\begin{aligned}
					\sttoltp{ P }{} &=   \res {z} ( \res{u} \lvlout{z}{v}{u}{\sttoltp{(R\substj{u}{x})\substj{z}{x}}{}} 
					\pp \lvlin{z}{w}{u}{\sttoltp{(S\substj{u}{y})\substj{z}{y}}{}}  )
					\\
					& \lts{ \tau } \res {z}  \res{u} ( \sttoltp{(R\substj{u}{x})\substj{z}{x}}{} 
					\pp \sttoltp{(S\substj{u}{y})\substj{z}{y}\substj{v}{w}}{} \pp  ) 
				\end{aligned}
				\]
	
				and
				\[
					R = \res{rs} \vasres{xy}{(\vaspara{R'}{S\substj{v}{w}} )} \equiv \vasres{xy}{(\vaspara{R'}{S\substj{v}{w}} )} = Q
				\]
				\[
					\sttoltp{ R }{} = \res{r} \res {z} ( \sttoltp{(R')\substj{z}{x}}{} 
					\pp \sttoltp{(S)\substj{z}{y}\substj{v}{w}}{}  )
				\]

				Finally we have that $P \vasred \vasres{xy}{(\vaspara{R'}{S\substj{v}{w}}{} )} = Q$ via \Did{R-LinCom}.
	
				\item When $ x: \recur{\oc T} $. Then
				
				$\sttoltp{ P }{} =   \res {z} ( \lvlout{z}{v}{u}{\sttoltp{R'\substj{z}{x}\substj{z}{y}}{}} 
				\pp \lvlin{x}{w}{u}{\sttoltp{S\substj{z}{x}\substj{z}{y}}{}} \substj{z}{x}\substj{z}{y} )
			    $ and follow analogously to \ref{ch5proof:encod1sound1}.
 
			\end{enumerate}

			\item\label{ch5proof:encod1sound2} When $\sttoltp{P'}{}{}$ reduces via \lvltype{Out} and $\sttoltp{Q'}{}{}$ reduces via \lvltype{Rep}. 
			
			Then  $P' = \vasout{x}{v}{R'}$, $Q' = \vasin{\un}{x}{w}{S}$ and typing enforces $x: \recur{\oc T}$ with:

			\[
			\begin{aligned}
				\sttoltp{ P }{} &=   \res {z} ( \lvlout{z}{v}{u}{\sttoltp{R'\substj{z}{x}\substj{z}{y}}{}} 
				\pp \lvlserv{z}{w}{u}{\sttoltp{S\substj{z}{x}\substj{z}{y}}{}}  )
				\\
				& \lts{ \tau } \res {z}  ( \sttoltp{(R')\substj{z}{x}\substj{z}{y}}{} 
				\pp \sttoltp{(S)\substj{z}{x}\substj{z}{y}\substj{v}{w}}{} \pp \lvlserv{z}{w}{u}{\sttoltp{S\substj{z}{x}\substj{z}{y}}{}}  ) \\
				& = \sttoltp{\vasres{xy}{(	\vaspara{R'}	{\vaspara{S\substj{v}{w}}	{\vasin{\un}{y}{w}{S}}}	)} }{}\\
				& = \sttoltp{ R }{}
			\end{aligned}
			\]

			Finally we have that $ P \vasred \vasres{xy}{(	\vaspara{R'}	{\vaspara{S\substj{v}{w}}	{\vasin{\un}{y}{w}{S}}}	)} = Q = R$ via \Did{R-UnCom}.

			\item\label{ch5proof:encod1} When either $\sttoltp{P'}{}{}$ or $\sttoltp{Q'}{}{}$ reduces via \lvltype{Par}. 
			
			Consider the case of $\sttoltp{P'}{}{}$ and $\sttoltp{Q'}{}{}$ proceeds similarly. Then $\sttoltp{P'}{}{} = \sttoltp{R'}{}{} \pp \sttoltp{S'}{}{}$ and by \lvltype{Par} $\sttoltp{R'}{}{} \lts{ \res {\widetilde{b}}\ov z\out {v,u } }  \sttoltp{R''}{}{}$. Hence $\sttoltp{R'}{}{} \lts{tau}  \res{\widetilde{b}} (( \sttoltp{R''}{}{} \pp \sttoltp{S'}{}{} )\pp \sttoltp{Q''}{}{}) $. By the induction hypothesis, $\equiv$ being associative and commutative and applications of \Did{R-Str} the result follows.

			\item When $\sttoltp{P'}{}{}$ reduces via \lvltype{Out} and $\sttoltp{Q'}{}{}$ reduces via \lvltype{Res}. Then case proceeds analogously to \ref{ch5proof:encod1sound2}.

		\end{enumerate}

		\item\label{ch5proof:encod1sound3} When $\sttoltp{P}{}{}$ reduces via $\lvltype{Par}$  then $P =  P' \pp Q' $, ${\sttoltp{P}{}{} = {\sttoltp{P'}{}} 
		\pp {\sttoltp{Q'}{}}  \lts{ \tau } \sttoltp{P''}{}{} \pp \sttoltp{Q'}{}{}}$ with  $\sttoltp{P'}{}{}  \lts{ \tau }  \sttoltp{P''}{}{} $. By the induction hypothesis and application of \Did{R-Par} the result follows.

		\item When $\sttoltp{P}{}{}$ reduces via $ \lvltype{Res} $  then $P =  \res {xy}P'  $. By the induction hypothesis and application of \Did{R-Res} the result follows analogously to \ref{ch5proof:encod1sound3}.
		
	\end{enumerate}

\end{proof}

\begin{theorem}$\lvllang \subset \vaslang$.
\end{theorem}
\begin{proof}
The inclusion $\lvllang \subseteq \vaslang$ is immediate by definition. 
To prove that the inclusion is strict, it is sufficient to consider a counterexample, i.e., a process $P$ typable in $\vaslang$ but not typable in $\pilvl$.

	Consider the process $P =  \vasres{xy}{( \vaspara{\vasout{y}{w}{\nil}}{\vasin{\un}{x}{z}{\vasout{y}{w}{\nil}}} )}$  which invokes itself ad infinitum. This process is typed typed  in $\vaslang$ by $w: \nilT \st P$ with derivation as
	
		\begin{mathpar}
			\inferrule*[left=\vastype{Res}]{
				\inferrule*[left=\vastype{Par}]{ 
					\Pi \\
					\inferrule*[left=\vastype{Un-In}]{ 
						\inferrule{
							\contun{\Gamma}
						}{
							\Gamma  \st x: \recur{\wn \nilT}  
						}
						\\
						\Pi
					}
					{  \Gamma \st   \vasin{\un}{x}{z}{\vasout{y}{w}{\nil}}  }
				}
				{ \Gamma \st \vasout{y}{w}{\nil} \pp \vasin{\un}{x}{z}{\vasout{y}{w}{\nil}} }
			}
			{ 
				w: \nilT \st  \vasres{xy}{( \vaspara{\vasout{y}{w}{\nil}}{\vasin{\un}{x}{z}{\vasout{y}{w}{\nil}}} )}
			}
		\end{mathpar}		
where 	 $  \Gamma = x:\recur{\wn \nilT} , y: \recur{\oc \nilT}, w: \nilT $ and $\Pi$ is the following derivation:		
	
		\begin{mathpar}
			\inferrule*[left=\vastype{Un-Out}]{ 
				\inferrule{
					\contun{\Gamma' }
				}
				{
					\Gamma' \st y: \recur{\oc \nilT }  
				}
				\\
				\inferrule{
					\contun{\Gamma' }
				}{
					\Gamma' \st w: \nilT  
				}\\
				\inferrule{ 
					\contun{\Gamma' } 
				}
				{ 
					\Gamma' \st \nil  
				}
			}
			{  \Gamma' \st \vasout{y}{w}{\nil} } 
		\end{mathpar}
	with $ \Gamma' = x:\recur{\wn \nilT} , y: \recur{\oc \nilT}, w: \nilT, z: \nilT $.

\noindent{\bf Claim.}	 $P \notin \lvllang$.

 In fact, if  $P \in \lvllang$ then there would exist of a function $l$ such that $\sttoltj{w: \nilT \st  P}{l}{}$. Thus, we shall show that this $l$ does not exist. 
 
 Unfolding the encoding of the derivation of $w: \nilT \st  P $ we have that 
 $$\sttoltj{\inferrule{\vdots}{w: \nilT \st  P}}{l}{}=\inferrule{\vdots}{\sttoltt{w: \nilT}{l}{} \lt \sttoltp{\vasres{xy}{( \vaspara{\vasout{y}{w}{\nil}}{\vasin{\un}{x}{z}{\vasout{y}{w}{\nil}}} )}}{}}$$
And the derivation in $\pilvl$  would be as follows:
	\begin{mathpar}
		\inferrule*[left=\lvltype{Res}]{  
			\inferrule*[left=\lvltype{Par}]{
				\Pi_1
				\\
				\Pi_3  
			}
			{ 
				w: \lvlnilT , \lvlsplit{a}{\lvlunct{\levelof{x}}{\sttoltt{\lvlnilT}{l}{\gamma}}} \lt  \lvlout{a}{w}{b}{\nil}   \pp \bang\lvlin{a}{z}{c}{\lvlout{a}{w}{d}{\nil}}{}  
			}
		}
		{ 
			w: \lvlnilT \lt \res {a} (\lvlout{a}{w}{b}{\nil} \pp \bang\lvlin{a}{z}{c}{\lvlout{a}{w}{d}{\nil}}{})
		}
	\end{mathpar}
where $\Pi_1$ and $\Pi_3$ are subderivations given below.

\begin{itemize}	
\item $\Pi_1$ is the derivation of $\Gamma_{w}  \st \lvlout{a}{w}{z}{\nil}$:
	\begin{mathpar}
		\inferrule*[left=\lvltype{Un-Out_1}]{
			\inferrule{
			}{
				\Gamma_{w} \lt {a}:{\lvlunct{\levelof{x}}{\sttoltt{\lvlnilT}{l}{\gamma}}}
			}
			\\
			\inferrule{
			}{
				\Gamma_{w} \lt w: \lvlnilT
			}
			\\
			\inferrule{
			}{
				\Gamma_{w}, b:\lvlnilT \lt \nil
			}
		}{
			\Gamma_{w} \lt \lvlout{a}{w}{b}{\nil}  
		}
	\end{mathpar}
where $\Gamma_{w}=  \lvlsplit{a}{\lvlunct{\levelof{x}}{\sttoltt{\lvlnilT}{l}{\gamma}}} , w : \lvlnilT $.

\item $\Pi_3 $ is the subderivation of $\Gamma_{w}  \st \bang \lvlin{a}{z}{c}{\lvlout{a}{w}{d}{\nil}}{}$:
	\begin{mathpar}
			\inferrule*[left=\lvltype{Un-In_1}]{ 
				\inferrule{
					}{
						\Gamma_{w} \lt {a}:{\lvlunst{\levelof{x}}{\sttoltt{\lvlnilT}{l}{\gamma}}}
					}
				\\
				\Pi_2
				\\
				\forall e \in \outsubj{\lvlout{a}{w}{d}{\nil}}, \ \levelof{e} < \levelof{x} }
			{ \Gamma_{w} \lt \bang \lvlin{a}{z}{c}{\lvlout{a}{w}{d}{\nil}}{}}
	\end{mathpar}
 where $\Gamma_w=  \lvlsplit{a}{\lvlunct{\levelof{x}}{\sttoltt{\lvlnilT}{l}{\gamma}}} , w : \lvlnilT $ and  $\Pi_2$ is  the subderivation of $\Gamma_{w,z,c} \lt \lvlout{a}{w}{d}{\nil} $:
	\begin{mathpar}
				\inferrule*[left=\lvltype{Un-Out_1}]{
					\inferrule{
					}{
						\Gamma_{w,z,c} \lt {a}:{\lvlunct{\levelof{x}}{\sttoltt{\lvlnilT}{l}{\gamma}}}
					}
					\\
					\inferrule{
					}{
						\Gamma_{w,z,c} \lt w: \lvlnilT
					}
					\\
					\inferrule{
					}{
						\Gamma_{w,z,c,d} \lt \nil
					}
				}{
					\Gamma_{w,z,c} \lt \lvlout{a}{w}{d}{\nil}  
				}\\
	\end{mathpar}
	where $\Gamma_{w,z,c}=\Gamma_w, z: \lvlnilT, c: \lvlnilT$ and $\Gamma_{w,z,c,d}=\Gamma_{w,z,c}, d:\lvlnilT$.

	\end{itemize}
	However, notice that $\outsubj{\lvlout{a}{w}{d}{\nil}} = \{  a \}$ and $\Pi_2$ imposes  the condition $\levelof{x} < \levelof{x} $ on $l$. Hence, there is no $l$ that can satisfy this condition.
\end{proof}

\section{Proofs of Section~\ref{ch5s:dilllang}}\label{ch5app:appC}
In this section we present the detailed proofs of the results in \Cref{ch5s:dilllang}. Namely, type preservation (\Cref{ch5thm:second_type_pres}) and the proper inclusion $\dilllang \subset \lvllang$(\Cref{ch5thm:main_result}). The latter requires the construction on strict partial orders that will induce levels, described in \Cref{ch5ssecapp:levels}, and the auxiliary results  in \Cref{ch5prop:join2} - \Cref{ch5lem:subfunction}. The section concludes with the proof that $  \lvllang \not\subset \dilllang$ (\Cref{ch5thm:wnotinl}).

\begin{theorem}[Type preservation (cf. \Cref{ch5thm:second_type_pres})]\label{ch5thm:dilltypepres}
If $\Gamma \vaslinunsplit \Delta \st P$ with then there exist $\Gamma'$, $\Delta'$, and $A$ such that $ \dillnotun{\sttodillt{\Gamma'}{}{}} ;  \sttodillt{\Delta'}{}{} \dill \sttodillp{P}{} :: u: A$ and one of the following holds:
\begin{enumerate}
	\item $A= \sttodillt{\dual{T}}{}{}$ when $\{ u: T \} \subset \Gamma , \Delta$ with $ (\Gamma \dillcontrel{u}{T} \Gamma') \land (\Delta \dillcontrel{u}{T} \Delta')$
	\item $A= \dillnilT $ when $ u \not\in \dom{\Gamma , \Delta}$ with $ (\Gamma = \Gamma') \land (\Delta = \Delta')$.
\end{enumerate}
\end{theorem}

\begin{proof}
	By induction on the entries of \Cref{ch5d:secondtablep1}. Thus, we consider 11 possibilities. 

\begin{enumerate}

	\item {\bf Inaction:}
	\begin{mathpar}
			\sttodillj{\inferrule{ 
			 \contun{\Gamma} }
			{ \Gamma \st \nil  }}_u = 
			\inferrule{  }
			{ \dillnotun{\sttodillt{\Gamma}} ;  \cdot \dill \nil :: u : \dillnilT }
	\end{mathpar}
	\item\label{ch5encod:2} {\bf Independent parallel composition:} If $w \not \in \dom{\Gamma, \Delta_1, \Delta}\land
			(
				\abconditionO{u}{\Gamma}{\Delta}{\Gamma'}{\Delta'}{A}
				\lor 
				\abconditionTstar{u}{R}{\Gamma}{\Delta}{\Gamma'}{\Delta'}{A}{P}{u}  )  $
	
	\[
	\begin{array}{rcl}
		\hspace{-1cm}
		\small
		\sttodillj{\inferrule
		{ \Gamma, \Delta_1 \st P \\ \Gamma , \Delta  \st Q }
		{ (\Gamma, \Delta_1) \contcomp (\Gamma , \Delta) \st P \pp Q  }}_u
		&=&
		\inferrule{ 
				\sttodillj{ \Gamma', \Delta_1 \st P}_w  
				\\
				\inferrule{
					\sttodillj{\Gamma , \Delta  \st Q }_u
				}{
					\dillnotun{\sttodillt{\Gamma'}} ;  \sttodillt{\Delta'} ,  w : \dillnilT \dill  \sttodillp{Q}     :: A
				}
			 }
			{ \dillnotun{\sttodillt{\Gamma'}} ; \sttodillt{\Delta_1} , \sttodillt{\Delta'} \dill \res{w} ( \sttodillp{P} \pp \sttodillp{Q}   )  ::  u: A }
	\end{array}
	\]

	The condition $u \not \in \fn{P}$ of $\abconditionTstar{u}{R}{\Gamma}{\Delta}{\Gamma'}{\Delta'}{A}{P}{u}$ follows from the fact that  $\Gamma, \Delta_1 \st P $ implies $ \Gamma', \Delta_1 \st P$, if $u \not \in \fn{P}$.

	\item[(3a)] {\bf Linear composition ($\neg \contun{V}$):} if $\neg \contun{V}\land
			(
				\abconditionO{u}{\Gamma}{\Delta}{\Gamma'}{\Delta'}{A}
				\lor 
				\abconditionTstar{u}{R}{\Gamma}{\Delta}{\Gamma'}{\Delta'}{A}{P}{u}  )  $.
This involves typing rules $\vastype{Res}$ and \vastype{Par}.
	\[
	\begin{array}{l}
		\hspace{-1.5cm}
		\small
		\sttodillj{\inferrule{ 
			\inferrule
			{ \Gamma, \Delta_1, z:V \st P   \\  \Gamma , \Delta ,v: \dual{V} \st  Q   }
			{\Gamma, \Delta_1, z:V \contcomp \Gamma , \Delta ,v: \dual{V} \st P \pp Q}
		}
		{ 
			\Gamma, \Delta_1 \contcomp \Gamma,  \Delta \st  \res {zv} (P \pp Q) 
		}}_u \\=
		\inferrule{ \sttodillj{ \Gamma', \Delta_1, z:V \st P }_z  \\
			\sttodillj{ \Gamma , \Delta ,z: \dual{V} \st  Q\substj{z}{v} }_u }
			{ \dillnotun{\sttodillt{\Gamma'}} ; \sttodillt{\Delta_1} , \sttodillt{\Delta'}  \dill \res{z}( \sttodillp{P} \pp \sttodillp{Q}\substj{z}{v} ) ::  u: A }
	
	\end{array}
		\]

	\item[(3b)]\label{ch5encod:5} {\bf Unrestricted composition ($\contun{V}$):} If  $\contun{V} \land v \not \in \fn{P} \land z \not \in \fn{Q} \land
			(
				\abconditionO{u}{\Gamma}{\Delta}{\Gamma'}{\Delta'}{A}
				\lor 
				\abconditionTstar{u}{R}{\Gamma}{\Delta}{\Gamma'}{\Delta'}{A}{P}{u}  )  $

	\[
	\begin{array}{l}
		\small
		\sttodillj{\inferrule{ 
			\inferrule{ 
			 \Gamma,  \Delta_1, z:V, v:\dual{V} \st P   \\  \Gamma,  \Delta ,z:V,v: \dual{V} \st  Q   }
			{\Gamma,  \Delta_1, z:V, v:\dual{V} \contcomp \Gamma,  \Delta ,z:V, v:\dual{V} \st P \pp Q }}
		{ \Gamma,  \Delta_1 \contcomp \Gamma,  \Delta \st  \res {zv} (P \pp Q)}}_u \\=
		\inferrule{ \sttodillj{ \Gamma', \Delta_1, z:V \st P }_z  \\
			\sttodillj{ \Gamma , \Delta ,z: \dual{V} \st  Q\substj{z}{v} }_u }
			{ \dillnotun{\sttodillt{\Gamma'}} ; \sttodillt{\Delta_1} , \sttodillt{\Delta'}  \dill \res{z}( \sttodillp{P} \pp \sttodillp{Q}\substj{z}{v} ) ::  u: A } 
	\end{array}
	\]

Conditions $v \not \in \fn{P}$ and $z \not \in \fn{Q}$ follow with a similar argument as in case \ref{ch5encod:2} above. Premise $\sttodillj{ \Gamma', \Delta_1, z:V \st P }_z$ can only hold if $v \not \in \fn{P}$ (and similarly for $Q$).

	\item[(4)] {\bf Unrestricted input (server):}
We consider three sub-cases. 
	\begin{enumerate}
	\item  If $u = x$, $x \not \in \fn{P}$, $\notserver{T}\land \notclient{T} $

	\[
	\begin{array}{l}
	\small
		\sttodillj{
				\inferrule{
								\Gamma, x: \recur{\wn T} , y : T \st P \\\\
					\Gamma , x: \recur{\wn T} \st x: \recur{\wn T} 
				}
				{ \Gamma , x: \recur{\wn T} \st   \vasin{\un}{x}{y}{P}  }}_u
			\\= 
	\inferrule{ 
					\inferrule{
						\inferrule{
							\sttodillj{ \Gamma , y:T \st   P }_w
						}{
							\dillnotun{\sttodillt{\Gamma}} ; \cdot  \dill   	\dillin{w}{y}{ \sttodillp{P} } :: w: \dillint{\sttodillt{T}}{\sttodillt{\dillnilT}}
						}
						\\
						\sttodillj{ \Gamma , y:T \st   P }_y
					}
					{
						\dillnotun{\sttodillt{\Gamma}} ; \cdot  \dill   \dillchoice{w}{	\dillin{w}{y}{ \sttodillp{P} }	}{	\sttodillp{P\substj{w}{y}} } :: w: \dillselet{\dillint{\sttodillt{T}}{\sttodillt{\dillnilT}}}{\sttodillt{\dual{T}}}
					}
				}
				{ \dillnotun{\sttodillt{\Gamma}} ; \cdot  \dill   \dillserv{x}{w}{  
					\dillchoice{w}{	\dillin{w}{y}{ \sttodillp{P} }	}{	\sttodillp{P\substj{w}{y}} }} ::  u: \dillunt{(\dillselet{\dillint{\sttodillt{T}}{\sttodillt{\dillnilT}}}{\sttodillt{\dual{T}}})}  
				}
	\end{array}
	\]

	From an application of $\vastype{Var}$ one obtains that $\Gamma , x: \recur{\wn T} \st x: \recur{\wn T} $ holds,  when  \( \contun{\Gamma}\).
	\item  If $u = x$, $x \not \in \fn{P}$, $\server{T} \land \notclient{T} $.

	\[
	\begin{array}{lcl}
	\mprset{flushleft}
		\sttodillj{
				\inferrule{
								\Gamma, x: \recur{\wn T} , y : T \st P \\\\
					\Gamma , x: \recur{\wn T} \st x: \recur{\wn T} 
				}
				{ \Gamma , x: \recur{\wn T} \st   \vasin{\un}{x}{y}{P}  }}_u
			&=& 
	\inferrule{
					\sttodillj{ \Gamma , y:T \st   P }_y
				}
				{ \dillnotun{\sttodillt{\Gamma}} ; \cdot  \dill   \dillserv{x}{w}{  
					{	\sttodillp{P\substj{w}{y}} }} ::  u: \dillunt{\sttodillt{\dual{T}}} 
				}
	\end{array}
	\]
	From an application of $\vastype{Var}$ one obtains that $\Gamma , x: \recur{\wn T} \st x: \recur{\wn T} $ holds,  when  \( \contun{\Gamma}\).

	\item  If $u = x$, $x \not \in \fn{P}$, $\notserver{T} \land \client{T}$

	\[
	\begin{array}{lcl}
		\hspace{-0.5cm}
		\small
		\sttodillj{
				\inferrule{
								\Gamma, x: \recur{\wn T} , y : T \st P \\\\
					{ \Gamma , x: \recur{\wn T} \st x: \recur{\wn T} } 
				}
				{ \Gamma , x: \recur{\wn T} \st   \vasin{\un}{x}{y}{P}  }}_u
			&=& 
			\inferrule{ 
					\inferrule{
						\sttodillj{ \Gamma , y:T \st   P }_w
					}{
						\dillnotun{\sttodillt{\Gamma}} ; \cdot  \dill   	\dillin{w}{y}{ \sttodillp{P} } :: w: \dillint{\sttodillt{T}}{\sttodillt{\dillnilT}}
					}
				}
				{ \dillnotun{\sttodillt{\Gamma}} ; \cdot  \dill   \dillserv{x}{w}{  
					\dillin{w}{y}{ \sttodillp{P} }	} ::  u: \dillunt{\dillint{\sttodillt{T}}{\sttodillt{\dillnilT}}}  
				}
	\end{array}
	\]
	\end{enumerate}

Some intuitions follow:
\begin{itemize}
	\item 
	The first sub-case requires $\notserver{T}$ and $\notclient{T} $ as the server is unable to distinguish if it will connect to a bound output or a free output; hence, we need to account for both behaviors: the encoding will be $\sttodillj{ \Gamma , y:T \st   P }_w$ and $\sttodillj{ \Gamma , y:T \st   P }_y$ (i.e., on $w$ and $y$, respectively), as both can only be true with this condition. 
	\item The second sub-case requires $T$ to produce a server behavior, which is only possible by means of a bound output. When this is the case notice that the left branch of the choice, (i.e., $\sttodillj{ \Gamma , y:T \st   P }_w$) would not be well typed. 
	\item The third sub-case is when $T$ is a client: then, the only possibility is a synchronization with a free output. As a consequence, the right-hand side $\sttodillj{ \Gamma , y:T \st   P }_y$  is not well typed. Also, notice that $u$ can only correspond to $x$: this is because server behaviours must appear on the right-hand side of a judgment in DILL. 
\end{itemize}
Finally, the case $\notserver{T} = \ffalse$ and $\notserver{T} = \ffalse$ is not typable: this is because neither $\sttodillj{ \Gamma , y:T \st   P }_w$ or $\sttodillj{ \Gamma , y:T \st   P }_y$ can hold.

	\item[(5)] {\bf  Linear input:}
	We consider two sub-cases.
	\begin{enumerate}
	\item If $\abconditionO{u}{\Gamma}{\Delta}{\Gamma'}{\Delta'}{A} \lor \abconditionT{u}{R}{\Gamma}{\Delta}{\Gamma'}{\Delta'}{A}$:
	\[
	\begin{array}{lc}
	\sttodillj{\inferrule{ \Gamma , x : \lin \vasreci{T}{S} \st x: \lin \vasreci{T}{S}  \\\\
		\Gamma , \Delta, x: S , y : T \st P}
		{ \Gamma , x: \lin \vasreci{T}{S} \contcomp \Gamma , \Delta\st  \vasin{\lin}{x}{y}{P}   }
		}_u 
		\\=	
		\inferrule{ 
				\sttodillj{ \Gamma , \Delta , x: S , y : T \st P }_u }
			{ \dillnotun{\sttodillt{\Gamma'}} ; \sttodillt{\Delta'} , x:\dilloutt{\sttodillt{T}}{\sttodillt{S}}  \dill \dillin{x}{y}{\sttodillp{P}} :: u: A }	
	\end{array}
	\]
         \item If $u = x$:
         
         \[
         	\begin{array}{lcl}
				\hspace{-1cm}
				\small
	\sttodillj{\inferrule{ \Gamma , x : \lin \vasreci{T}{S} \st x: \lin \vasreci{T}{S}  \\\\
		\Gamma , \Delta, x: S , y : T \st P}
		{ \Gamma , x: \lin \vasreci{T}{S} \contcomp \Gamma , \Delta\st  \vasin{\lin}{x}{y}{P}   }
		}_u 
		&=&
	\inferrule{ 
				\sttodillj{ \Gamma , \Delta , x: S , y : T \st P }_x  }
			{ \dillnotun{\sttodillt{\Gamma}} ; \sttodillt{\Delta} ,  \dill \dillin{x}{y}{\sttodillp{P}} :: x:\dillint{\sttodillt{T}}{\sttodillt{\dual{S}}}   }
	\end{array}
	\]

	\end{enumerate}
	Notice that there is no counterpart to judgement by applying the rule \vastype{Lin-In_2} within $\pidill$. That is:
	\begin{mathpar}
		\sttodillj{\inferrule{  \Gamma_1 , x: \recur{\wn T} \st x: \recur{\wn T}  \\
		\Gamma_2 , x: \recur{\wn T} , y : T \st P
		}
		{  (\Gamma_1 , x: \recur{\wn T}) \contcomp (\Gamma_2 , x: \recur{\wn T}) \st  \vasin{\lin}{x}{y}{P} }
		}_u 
		= \text{Undefined}	
	\end{mathpar}
	This is because the linearity of the input is not respected by  its typing.

	\item[(6a)]\label{ch5encod:7} {\bf Server request, when the type of the request ($y$) is linear ($\neg \contun{T}$):}
	We consider two possibilities:
	\begin{enumerate}
         \item 	If $\notserver{T} \land \notclient{T} \land \neg \contun{T} \land 
			(\abconditionO{u}{\Gamma}{\Delta}{\Gamma'}{\Delta'}{A} \lor \abconditionT{u}{R}{\Gamma}{\Delta}{\Gamma'}{\Delta'}{A})
			$. This involves typing rule $\vastype{Res}$:
         \[
         \begin{array}{l}
         \small
         \sttodillj{
         \mprset{flushleft}
			\inferrule{ 
				\inferrule{
					\Gamma_1 \st z: \recur{\oc T}  \\\\ 
					\Gamma_1,  y: T  \st y: T  \\
					\Gamma_1, x: \dual{T}  , \Delta \st P
				}{
					\Gamma_1 \contcomp 
					(\Gamma_1 , y: T)  \contcomp 
					(\Gamma_1  , x:  \dual{T} , \Delta) 
					\st \vasout{z}{y}{P}
				}
			}
			{  \Gamma_1 \contcomp \Gamma_1  \contcomp (\Gamma_1 , \Delta) \st  \res{xy}\vasout{z}{y}{P} }
		}_u\\
		=
				\inferrule{ 
				\inferrule{
					\sttodillj{ \Gamma, z: \recur{\oc T} , x: \dual{T} , \Delta \st P }_u
				}
				{	
					\dillnotun{\sttodillt{\Gamma'_1}} ; \sttodillt{\Delta'} , x : \dillselet{\dillint{\sttodillt{T}}{\sttodillt{\dillnilT}}}{\sttodillt{\dual{T}}} \dill 
						\dillseler{x}{\sttodillp{P}} :: u: A
				}
			}
			{ 
				\dillnotun{\sttodillt{\Gamma'_1}}  ; \sttodillt{\Delta'}  \dill \dillbout{z}{x}{
				\dillseler{x}{\sttodillp{P}}} ::  u: A 
			}
         \end{array}
         \]
         
      where    $\Gamma_1=\Gamma,  z: \recur{\oc T} $ and $\dillnotun{\sttodillt{\Gamma'_1}} =\dillnotun{\sttodillt{\Gamma'}} , z:\dillselet{\dillint{\sttodillt{T}}{\sttodillt{\dillnilT}}}{\sttodillt{\dual{T}}} $.
      
        \item If $\server{T} \land \notclient{T} \land \neg \contun{T} \land 
			(\abconditionO{u}{\Gamma}{\Delta}{\Gamma'}{\Delta'}{A} \lor \abconditionT{u}{R}{\Gamma}{\Delta}{\Gamma'}{\Delta'}{A} )
			$:
   \[
         \begin{array}{l}
			\hspace{-1cm}
		\small
         \sttodillj{
         \mprset{flushleft}
			\inferrule{ 
				\inferrule{
					\Gamma_1 \st z: \recur{\oc T}  \\\\ 
					\Gamma_1,  y: T  \st y: T  \\
					\Gamma_1, x: \dual{T}  , \Delta \st P
				}{
					\Gamma_1 \contcomp 
					(\Gamma_1 , y: T)  \contcomp 
					(\Gamma_1  , x:  \dual{T} , \Delta) 
					\st \vasout{z}{y}{P}
				}
			}
			{  \Gamma_1 \contcomp \Gamma_1  \contcomp (\Gamma_1 , \Delta) \st  \res{xy}\vasout{z}{y}{P} }
		}_u = 
		\inferrule{ 
					\sttodillj{ \Gamma, z: \recur{\oc T} , x: \dual{T}, \Delta\st P }_u
			}
			{
				\dillnotun{\sttodillt{\Gamma'}} , z:\sttodillt{\dual{T}}  ; \sttodillt{\Delta'}  \dill  \dillbout{z}{x}{
				\sttodillp{P}} :: u: A
			}
         \end{array}
         \]
         
	\end{enumerate}

	Note that the case $u = z$ is not allowed; this is the dual situation to why only $u=z$ is allowed in \ref{ch5encod:5} above.

	\item[(6b)] {\bf Server request, when the type of the request ($y$) is unrestricted ($\contun{T}$):}
We consider two possibilities:
\begin{enumerate} 
\item If \( \notserver{T} \land \notclient{T} , y \not \in \fn{P}, \contun{T} \land 
		(\abconditionO{u}{\Gamma}{\Delta}{\Gamma'}{\Delta'}{A} \lor \abconditionT{u}{R}{\Gamma}{\Delta}{\Gamma'}{\Delta'}{A})\).
		This involves typing rule \vastype{Res}:

\[\small
\begin{array}{l}
\mprset{flushleft}
		\sttodillj{
			\inferrule
			{  
				\inferrule{
					\Gamma_1, y: T \st z: \recur{\oc T}  \\\\  
					\Gamma_1,  y: T  \st y: T  \\
					\Gamma_1 , y: T, x: \dual{T}  , \Delta \st P
				}{
					(\Gamma_1 , y: T)\contcomp 
					(\Gamma_1 , y: T ) \contcomp 
					(\Gamma_1, y: T, x:  \dual{T} , \Delta )
					\st \vasout{z}{y}{P}
				}
			}
			{  \Gamma _1 \contcomp \Gamma_1  \contcomp (\Gamma , z: \recur{\oc T} , \Delta) \st  \res{xy}\vasout{z}{y}{P} }
		}_u
\\=
\inferrule{ 
				\inferrule{
					\inferrule{
					\sttodillj{ \Gamma, z: \recur{\oc T} , x: \dual{T} , \Delta \st P }_u
					}
					{	
						\dillnotun{\sttodillt{\Gamma'_!}}  ; \sttodillt{\Delta'} , x : \sttodillt{\dual{T}} \dill 
							\sttodillp{P} :: u: A
					}
				}
				{	
					\dillnotun{\sttodillt{\Gamma'_1}}  ; \sttodillt{\Delta'} , x : \dillselet{\dillint{\sttodillt{T}}{\sttodillt{\dillnilT}}}{\sttodillt{\dual{T}}} \dill 
						\dillseler{x}{\sttodillp{P}} :: u: A
				}
			}
			{ 
				\dillnotun{\sttodillt{\Gamma'_1}}  ; \sttodillt{\Delta'}  \dill \dillbout{z}{x}{
				\dillseler{x}{\sttodillp{P}}} ::  u: A 
			}
\end{array}
\]
where $\Gamma_1=\Gamma,z: \recur{\oc T} $ and $\dillnotun{\sttodillt{\Gamma'_1}}=\dillnotun{\sttodillt{\Gamma'}} , z:\dillselet{\dillint{\sttodillt{T}}.{\sttodillt{\dillnilT}}}{\sttodillt{\dual{T}}}$.

\item If $\server{T} \land \notclient{T} $, $y \not \in \fn{P}$, $\contun{T} \land 
			(\abconditionO{u}{\Gamma}{\Delta}{\Gamma'}{\Delta'}{A} \lor \abconditionT{u}{R}{\Gamma}{\Delta}{\Gamma'}{\Delta'}{A})$:

\[\small
\begin{array}{lcl}
\mprset{flushleft}
		\sttodillj{
			\inferrule
			{  
				\inferrule{
					\Gamma_1, y: T \st z: \recur{\oc T}  \\\\  
					\Gamma_1,  y: T  \st y: T  \\
					\Gamma_1 , y: T, x: \dual{T}  , \Delta \st P
				}{
					(\Gamma_1 , y: T)\contcomp 
					(\Gamma_1 , y: T ) \contcomp 
					(\Gamma_1, y: T, x:  \dual{T} , \Delta )
					\st \vasout{z}{y}{P}
				}
			}
			{  \Gamma _1 \contcomp \Gamma_1  \contcomp (\Gamma , z: \recur{\oc T} , \Delta) \st  \res{xy}\vasout{z}{y}{P} }
		}_u
\\=
\inferrule{ 
				\inferrule{ 
					\sttodillj{ \Gamma, z: \recur{\oc T} , x: \dual{T}, \Delta\st P }_u
				}
				{
					\dillnotun{\sttodillt{\Gamma'}} , z:\sttodillt{\dual{T}}  ; \sttodillt{\Delta'}  , x:\sttodillt{\dual{T}}  \dill \sttodillp{P} :: u: A
				}
			}
			{
				\dillnotun{\sttodillt{\Gamma'}} , z:\sttodillt{\dual{T}}  ; \sttodillt{\Delta'}  \dill  \dillbout{z}{x}{
				\sttodillp{P}} :: u: A
			}
\end{array}
\]
where $\Gamma_1=\Gamma,z: \recur{\oc T}$.

	Notice here that also only $u = z$. Also the condition $y \not \in \fn{P}$ is required (yet not in \ref{ch5encod:7}).
\end{enumerate}

	\item[(7a)] {\bf Linear bound output, sending a linear channel $T$.}
We have two possibilities:
\begin{enumerate}
\item  If $z \not = u \land z \not \in \fn{P} \land y \not \in \fn{Q} \land
			(
				\abconditionO{u}{\Gamma}{\Delta}{\Gamma'}{\Delta'}{A}
				\lor 
				\abconditionTstar{u}{R}{\Gamma}{\Delta}{\Gamma'}{\Delta'}{A}{P}{u}  ) 
			$.
\[
\begin{array}{l}
\mprset{flushleft}
\small
\sttodillj{
			\inferrule{
				\inferrule{
					\Gamma_1 \st z: \lin \vassend{T}{S}  \\  
					\Gamma, x: T \st x: T  \\
					\Upsilon_1				}{
					\Gamma_1\contcomp  (\Gamma, x: T)  \contcomp  (\Gamma, \Delta_1' , \Delta) \st \vasout{z}{x}{(P \pp Q)}
				}
			}
			{  \Gamma_1 \contcomp \Gamma   \contcomp (\Gamma, \Delta_1, \Delta)  \st  \res{xy}\vasout{z}{x}{(P \pp Q)} }
		}_u 
		\\=	
\inferrule
			{
				\sttodillj{ \Gamma', \Delta_1, y:  \dual{T} \st P }_y\\ 
				\sttodillj{ \Gamma, \Delta  , z:S \st Q }_u
			}
			{ \dillnotun{\sttodillt{\Gamma'}} ; \sttodillt{\Delta_1} , \sttodillt{\Delta'}  , z : \dillint{\sttodillt{T}}{\sttodillt{S}} \dill \dillbout{z}{y}{(\sttodillp{P} \pp \sttodillp{Q})} :: u: A }
\end{array}			  
\]

\item If $z = u$, $z \not \in \fn{P}$ and $y \not \in \fn{Q}$.

\[
\begin{array}{l}
\small
\mprset{flushleft}
\sttodillj{
			\inferrule{
				\inferrule{
					\Gamma_1 \st z: \lin \vassend{T}{S}  \\  
					\Gamma, x: T \st x: T  \\
					\Upsilon_1
				}{
					\Gamma_1\contcomp  (\Gamma, x: T)  \contcomp  (\Gamma, \Delta_1' , \Delta) \st \vasout{z}{x}{(P \pp Q)}
				}
			}
			{  \Gamma_1 \contcomp \Gamma   \contcomp (\Gamma, \Delta_1, \Delta)  \st  \res{xy}\vasout{z}{x}{(P \pp Q)} }
		}_u 
		\\=	
\inferrule
			{
				\sttodillj{ \Gamma, \Delta_1, y:  \dual{T} \st P }_y\\ 
				\sttodillj{ \Gamma, \Delta  , z:S \st Q }_z
			}
			{ \dillnotun{\sttodillt{\Gamma}} ; \sttodillt{\Delta_1} , \sttodillt{\Delta}  \dill \dillbout{z}{y}{(\sttodillp{P} \pp \sttodillp{Q})} ::   z : \dilloutt{\sttodillt{T}}{\sttodillt{\dual{S}}} }
\end{array}			  
\]

where $\Gamma_1= \Gamma , z: \lin \vassend{T}{S}$ and $\Delta_1'=\Delta_1, y:  \dual{T}$. 
Moreover, $\Upsilon_1$ is the following derivation:
\begin{mathpar}
\inferrule{
						\Gamma'' \dillcontrel{z}{S} \Gamma
						\\
						\Gamma'', \Delta_1' \st P
						\\
						\Gamma, \Delta  , z:S \st Q
					}
					{\Gamma'' , \Delta_1'  \contcomp \Gamma, \Delta , z:S \st P \pp Q}
\end{mathpar}

	We have that $\Gamma \dillcontrel{z}{S} \Gamma'$ in our typing condition as it collapses the case of $S$ being linear or unrestricted into a single case, rather then having two cases to handle this when applying parallel composition on $P \pp Q$.

\end{enumerate}

	\item[(7b)] {\bf Linear bound output, sending unrestricted channel $T$.}
	We consider two possibilities:
	\setcounter{enumi}{7}

\begin{enumerate}
\item If $z \not = u$, $z \not \in \fn{P}$, $y \not \in \fn{Q}$ , $x \not \in \fn{P} \cup \fn{Q} \land
			(
				\abconditionO{u}{\Gamma}{\Delta}{\Gamma'}{\Delta'}{A}
				\lor 
				\abconditionTstar{u}{R}{\Gamma}{\Delta}{\Gamma'}{\Delta'}{A}{P}{u}  )  
			$:

\[
\begin{array}{l}
\small
\mprset{flushleft}
\sttodillj{
			\inferrule{ 
				\inferrule{
					\Upsilon_1
					\\
					\Gamma_1, z: \lin \vassend{T}{S} \st z: \lin \vassend{T}{S}  \\  
					\Gamma_1\st x: T 
				}{
					(\Gamma_1, z: \lin \vassend{T}{S}) \contcomp  \Gamma_1 \contcomp  (\Gamma_1, \Delta_1, \Delta) \st \vasout{z}{x}{(P \pp Q)}
				}
			}
			{ ( \Gamma , z: \lin \vassend{T}{S}) \contcomp \Gamma   \contcomp (\Gamma, \Delta_1, \Delta)  \st  \res{xy}\vasout{z}{x}{(P \pp Q)} }
		}_u 
		\\=	
\inferrule
			{  
				\sttodillj{ \Gamma, \Delta_1, y:  \dual{T} \st P }_y\\
				\sttodillj{ \Gamma, \Delta  , z:S \st Q }_u
			}
			{ \dillnotun{\sttodillt{\Gamma'}} ; \sttodillt{\Delta_1} , \sttodillt{\Delta'}  , z : \dillint{\sttodillt{T}}{\sttodillt{S}} \dill  \dillbout{z}{y}{(\sttodillp{P} \pp \sttodillp{Q})} :: u: A }

\end{array}
\]

\item  If $z = u$, $z \not \in \fn{P}$, $y \not \in \fn{Q}$ and $x \not \in \fn{P} \cup \fn{Q}$:

\[
\begin{array}{l}
\small
\mprset{flushleft}
\sttodillj{
			\inferrule{ 
				\inferrule{
					\Upsilon_1
					\\
					\Gamma_1, z: \lin \vassend{T}{S} \st z: \lin \vassend{T}{S}  \\  
					\Gamma_1\st x: T 
				}{
					(\Gamma_1, z: \lin \vassend{T}{S}) \contcomp  \Gamma_1 \contcomp  (\Gamma_1, \Delta_1, \Delta) \st \vasout{z}{x}{(P \pp Q)}
				}
			}
			{ ( \Gamma , z: \lin \vassend{T}{S}) \contcomp \Gamma   \contcomp (\Gamma, \Delta_1, \Delta)  \st  \res{xy}\vasout{z}{x}{(P \pp Q)} }
		}_u
		\\=
\inferrule
			{ 
				\sttodillj{ \Gamma, \Delta_1, y:  \dual{T} \st P }_y\\
				\sttodillj{ \Gamma, \Delta  , z:S \st Q }_z
			}
			{ \dillnotun{\sttodillt{\Gamma}} ; \sttodillt{\Delta_1} , \sttodillt{\Delta}  \dill \dillbout{z}{y}{(\sttodillp{P} \pp \sttodillp{Q})} ::   z : \dilloutt{\sttodillt{T}}{\sttodillt{S}} }
\end{array}
\]

where 
\begin{itemize}
\item $\Gamma_1=\Gamma, x: T, y:  \dual{T} $
\item $\Gamma_1''=\Gamma'', x: T, y:  \dual{T} $
\item $\Upsilon_1$ is the following derivation:
\begin{mathpar}
\inferrule{
						\Gamma'' \dillcontrel{z}{S} \Gamma
						\\
						\Gamma'', \Delta_1 \st P
						\\
						\Gamma_1, \Delta  , z:S \st Q
					}
					{(\Gamma''_1, \Delta_1, \Delta) \contcomp \Gamma, z:S \st P \pp Q}
\end{mathpar}
\end{itemize}
\end{enumerate}

	\item {\bf Linear free output ($\neg \contun{T}$).}
We consider two possibilities:
\begin{enumerate}
\item If  $\neg \contun{T} \land \notserver{T} \land 
			(\abconditionO{u}{\Gamma}{\Delta}{\Gamma'}{\Delta'}{A} \lor \abconditionT{u}{R}{\Gamma}{\Delta}{\Gamma'}{\Delta'}{A})$.
\[			
\begin{array}{l}
\mprset{flushleft}
\sttodillj{
			\inferrule{ 
				\Gamma_1 \st x: \lin \oc (T).S 	\\\\
				\Gamma, v:T \st v:T  \\
				\Gamma, \Delta + x:S \st P  }
			{ \Gamma_1 \contcomp
			 (\Gamma, v:T) \contcomp
			  \Gamma , \Delta \st \ov x\out v.P }
		}_u
		\\= 
\inferrule{  
\dillnotun{\sttodillt{\Gamma'}} ; v: \sttodillt{T} \dill \dillforward{v}{y} :: y : \sttodillt{T} \\
				\sttodillj{\Gamma, \Delta + x:S \st P }_u
			}
			{ \dillnotun{\sttodillt{\Gamma'}} ;  v: \sttodillt{T},  \sttodillt{\Delta'}  , x : \dillint{\sttodillt{T}}{\sttodillt{S}} \dill \dillbout{x}{y}{(\dillforward{v}{y} \pp P)} ::  u: \sttodillt{\dual{R}} }
\end{array}
\]		
where $\Gamma_1=\Gamma , x: \lin \oc (T).S$.

\item If $u = x$, $\neg \contun{T}$ and $\notserver{T}$:

\[			
\begin{array}{l}
\mprset{flushleft}
\sttodillj{
			\inferrule{ 
				\Gamma_1 \st x: \lin \oc (T).S 	\\\\
				\Gamma, v:T \st v:T  \\
				\Gamma, \Delta + x:S \st P  }
			{ \Gamma_1 \contcomp
			 (\Gamma, v:T) \contcomp
			  \Gamma , \Delta \st \ov x\out v.P }
		}_u
		\\= 
	\inferrule{ 	\dillnotun{\sttodillt{\Gamma}} ; v: \sttodillt{T} \dill \dillforward{v}{y} :: y : \sttodillt{T}\\
				\sttodillj{\Gamma, \Delta + x:S \st P }_x
			}
			{ \dillnotun{\sttodillt{\Gamma}} ;  v: \sttodillt{T}, \sttodillt{\Delta}  \dill \dillbout{x}{y}{(\dillforward{v}{y}  \pp  \sttodillp{P})} ::   u : \dilloutt{\sttodillt{T}}{\sttodillt{\dual{S}}}} 
\end{array}
\]		
where $\Gamma_1=\Gamma , x: \lin \oc (T).S$	

	We adopt the condition  $\neg\server{T}$ for the sake of simplicity in proofs. 
	If the condition $\server{T}$ would be allowed, to preserve typability, $v$ would not  make any syncronization. That is, we may send a free server in a free output only if the server is never instantiated. Because this is not very meaningful, we exclude this possibility. 
\end{enumerate}

	\item {\bf Linear free output on $x$ of an unrestricted value $v$ ($\contun{T}$).} We have two possibilities:
	\begin{enumerate}
\item	If  $\contun{T}$, $\notserver{T}\land 
			(\abconditionO{u}{\Gamma}{\Delta}{\Gamma'}{\Delta'}{A} \lor \abconditionT{u}{R}{\Gamma}{\Delta}{\Gamma'}{\Delta'}{A})$:
\[
\begin{array}{l}
\sttodillj{
			\inferrule{ 
				\Gamma_1 \st x: \lin \oc T.S \\\\
				\Gamma_2 \st v:T  \\\\
				\Gamma_2 , \Delta + x:S \st P  }
			{ \Gamma_1 \contcomp
			\Gamma_2\contcomp
			 ( \Gamma_2, \Delta) \st \ov x\out v.P }
		}_u
		\\=
\inferrule{ 
\dillnotun{\sttodillt{\Gamma'}} ,v: \sttodillt{T}; \cdot \dill  \dillfwdbang{v}{y} :: y : \sttodillt{T} \\
				\sttodillj{\Gamma,  v:T , \Delta + x:S \st P }_u
			}
			{ \dillnotun{\sttodillt{\Gamma'}},  v:T  ; \sttodillt{\Delta'}  , x : \dillint{\sttodillt{T}}{\sttodillt{S}} \dill \dillbout{x}{y}{(\dillfwdbang{v}{y} \pp P)} ::  u: A } \\
\end{array}
\]

\item If $u = x$, $\contun{T}$ and $\notserver{T}$"
\[
\begin{array}{l}
\sttodillj{
			\inferrule{ 
				\Gamma_1 \st x: \lin \oc T.S \\\\
				\Gamma_2 \st v:T  \\\\
				\Gamma_2 , \Delta + x:S \st P  }
			{ \Gamma_1 \contcomp
			\Gamma_2\contcomp
			 ( \Gamma_2, \Delta) \st \ov x\out v.P }
		}_u
		\\= 
\inferrule{ 
\inferrule{ \ }{\dillnotun{\sttodillt{\Gamma}} ,v: \sttodillt{T}; \cdot \dill  \dillfwdbang{v}{y} :: y : \sttodillt{T}}\\
				\sttodillj{\Gamma,  v:T , \Delta + x:S \st P }_x
			}
			{ \dillnotun{\sttodillt{\Gamma}} , v: \sttodillt{T} ;  \sttodillt{\Delta}  \dill \dillbout{x}{y}{(\dillfwdbang{v}{y}  \pp  \sttodillp{P})} ::   x : \dilloutt{\sttodillt{T}}{\sttodillt{\dual{S}}} } 
\end{array}
\]
\end{enumerate}
where, for both cases:
	\begin{itemize}
\item $\Gamma_1=\Gamma,  v:T  , x: \lin \oc (T).S$
\item $\Gamma_2=\Gamma,  v:T $
\end{itemize}

	\item {\bf Unrestricted free output of a linear $v$ ($\neg \contun{T}$).}
	We have two possibilities:

	\begin{enumerate}
	\item  If $\notserver{T} \land \notclient{T} $, $\neg \contun{T}$, $z \not \in \fn{P} \land 
			(\abconditionO{u}{\Gamma}{\Delta}{\Gamma'}{\Delta'}{A} \lor \abconditionT{u}{R}{\Gamma}{\Delta}{\Gamma'}{\Delta'}{A})$:
	
	\[
	\begin{array}{l}
		\hspace{-1cm}
	\footnotesize
	\sttodillj{
			\inferrule{ \Gamma_1 \st x: \recur{\oc T}	\\\\ 
			\Gamma_1 , v:T \st v:T  \\
			\Gamma_1 , \Delta \st P  }
			{ \Gamma_1\contcomp  (\Gamma_1 , v:T) \contcomp (\Gamma_1 , \Delta) \st \ov x\out v.P }
		}_u
		\\
		\hspace{-1cm}=
\inferrule{
				\inferrule{ 
					\inferrule{
						\inferrule{
						}{
							\dillnotun{\sttodillt{\Gamma'_1}} ;  v:\sttodillt{T}  \dill \dillforward{v}{w} :: w:\sttodillt{T}
						}\\ 
						\inferrule{
							\sttodillj{ \Gamma , x:\recur{\oc T} , \Delta  \st {{P}} }_u
						}{
							\dillnotun{\sttodillt{\Gamma'_1}} ; \sttodillt{\Delta'} , z:\dillnilT \dill \sttodillp{{P}} :: u: A
						}
					}
					{ 
						\dillnotun{\sttodillt{\Gamma'_1}} ;  v:\sttodillt{T} , \sttodillt{\Delta'} , z:\dillint{\sttodillt{T}}{\dillnilT} \dill \dillbout{z}{w}{
						( \dillforward{v}{w}  \pp  \sttodillp{{P}}   )}:: u: A
					}
				}
				{ 
					\dillnotun{\sttodillt{\Gamma'_1}} ;  v:\sttodillt{T} , \sttodillt{\Delta'} , z:\dillselet{\dillint{\sttodillt{T}}{\dillnilT}}{\sttodillt{\dual{T}}} \dill \dillselel{z}{\dillbout{z}{w}{
						( \dillforward{v}{w}  \pp  \sttodillp{{P}}   )}}:: u: A
				}
			}
			{ \dillnotun{\sttodillt{\Gamma'_1}} ;  v:\sttodillt{T} , \sttodillt{\Delta'}  \dill \dillbout{x}{z}{ \dillselel{z}{
				\dillbout{z}{w}{
					( \dillforward{v}{w}  \pp  \sttodillp{{P}}   )
					}
				}
				} :: u: A
			}		
	\end{array}
	\]
	
	where $\Gamma_1=\Gamma,  x: \recur{\oc T}$ and $\dillnotun{\sttodillt{\Gamma'_1}}= \dillnotun{\sttodillt{\Gamma'}} , x:\dillselet{\dillint{\sttodillt{T}}{\dillnilT}}{\sttodillt{\dual{T}}}$. 
	
	\item  If $\notserver{T} \land \client{T} $, $\neg \contun{T}$, $z \not \in \fn{P}\land 
			(\abconditionO{u}{\Gamma}{\Delta}{\Gamma'}{\Delta'}{A} \lor \abconditionT{u}{R}{\Gamma}{\Delta}{\Gamma'}{\Delta'}{A})$:
	\[
	\begin{array}{l}
		\hspace{-0.5cm}
	\small
	\sttodillj{
			\inferrule{ \Gamma_1 \st x: \recur{\oc T}	\\\\ 
			\Gamma_1 , v:T \st v:T  \\
			\Gamma_1 , \Delta \st P  }
			{ \Gamma_1\contcomp  (\Gamma_1 , v:T) \contcomp (\Gamma_1 , \Delta) \st \ov x\out v.P }
		}_u
		\\
		\hspace{-0.5cm}=	
		\inferrule{
				\inferrule{
					\inferrule{
					}{
						\dillnotun{\sttodillt{\Gamma'_1}} ;  v:\sttodillt{T}  \dill \dillforward{v}{w} :: w:\sttodillt{T}
					}\\ 
					\inferrule{
						\sttodillj{ \Gamma , x:\recur{\oc T} , \Delta  \st {{P}} }_u
					}{
						\dillnotun{\sttodillt{\Gamma'_1}} ; \sttodillt{\Delta'} , z:\dillnilT \dill \sttodillp{{P}} :: u: A
					}
				}
				{ 
					\dillnotun{\sttodillt{\Gamma'_1}};  v:\sttodillt{T} , \sttodillt{\Delta'} , z:\dillint{\sttodillt{T}}{\dillnilT} \dill \dillbout{z}{w}{
					( \dillforward{v}{w}  \pp  \sttodillp{{P}}   )}::  u: A
				}
			}
			{ 
				\dillnotun{\sttodillt{\Gamma'_!}};  v:\sttodillt{T} , \sttodillt{\Delta'}  \dill \dillbout{x}{z}{ 
				\dillbout{z}{w}{
					( \dillforward{v}{w}  \pp  \sttodillp{{P}}   )
					}
				} ::  u: A
			} \\
	\end{array}
	\]
	
	where $\Gamma_1=\Gamma,  x: \recur{\oc T}$ and $\dillnotun{\sttodillt{\Gamma'_1}} = \dillnotun{\sttodillt{\Gamma'}} , x:\dillint{\sttodillt{T}}{\dillnilT}$.

	\end{enumerate}

	\item {\bf Unrestricted free output of an unrestricted $v$ ($\contun{T}$).} 
	In this case, we have
	 $\notserver{T} \land \client{T} $, $\contun{T}$, $z \not \in \fn{P} \land 
			(\abconditionO{u}{\Gamma}{\Delta}{\Gamma'}{\Delta'}{A} \lor \abconditionT{u}{R}{\Gamma}{\Delta}{\Gamma'}{\Delta'}{A})$:
\[
\begin{array}{l}
\small
\mprset{flushleft}
		\sttodillj{
			\inferrule{\Gamma_1 \st x: \recur{\oc T}	\\\\ 
			\Gamma_1 \st v:T  \\ 
			\Gamma_1 , \Delta \st P  }
			{ \Gamma_1 \contcomp \Gamma_1 \contcomp (\Gamma_1, \Delta )\st \ov x\out v.P }
		}_u
		\\= 
\inferrule{
				\inferrule{
					\inferrule{
					}{
						\dillnotun{\sttodillt{\Gamma'_1}} ; \cdot   \dill \dillfwdbang{v}{w} :: w:\sttodillt{T}
					}\\ 
					\inferrule{
						\sttodillj{ \Gamma , x:\recur{\oc T}, v:T , \Delta  \st {{P}} }_u
					}{
						\dillnotun{\sttodillt{\Gamma'_1}}; \sttodillt{\Delta'} , z:\dillnilT \dill \sttodillp{{P}} :: u: A
					}
				}
				{ 
					\dillnotun{\sttodillt{\Gamma'_1}} ; \sttodillt{\Delta'} , z:\dillint{\sttodillt{T}}{\dillnilT} \dill \dillbout{z}{w}{
					( \dillfwdbang{v}{w}  \pp  \sttodillp{{P}}   )}:: u: A
				}
			}
			{ 
				\dillnotun{\sttodillt{\Gamma'_1}} ;   \sttodillt{\Delta'}  \dill \dillbout{x}{z}{ 
				\dillbout{z}{w}{
					( \dillfwdbang{v}{w}  \pp  \sttodillp{{P}}   )
					}
				} :: u: A
			}
\end{array}
\]
where $\Gamma_1=\Gamma , x: \recur{\oc T}, v:T$ and $\dillnotun{\sttodillt{\Gamma'_1}}=\dillnotun{\sttodillt{\Gamma'}} , x:\dillint{\sttodillt{T}}{\dillnilT}, v:\sttodillt{T}$.

\end{enumerate}
This concludes the proof of \Cref{ch5thm:second_type_pres} / \Cref{ch5thm:dilltypepres}.
\end{proof}

\subsection{Proving that linear logic typing induces levels}\label{ch5ssecapp:levels}

In this subsection we present the entire construction for the proof of \Cref{ch5thm:main_result}.

\subsubsection{Defining Strict Partial Orders}
We define a strict partial order with maximum element as follows.

\begin{definition}{}
	Given a finite set $S$, a {\em strict partial order (\spo) with maximum element} is a  binary relation  $R$ on $S$ that is irreflexive, asymmetric, transitive and has a maximum element: 
	\begin{description}
		\item[(Maximum Element)] There exists $x \in S$ such that for all  $y \in S $ we have $(x,y) \in R$.
	\end{description}
\end{definition}

We will adopt a strict partial order $>$  and denote $(x,y)\in >$ as  $x>y$, and read it as  
`$x$ is greater than $y$'.

\begin{definition}{}
Let $S$ be a finite set (with elements $w,u,\ldots$), and let  $>$ be a strict partial order on $S$. 
\begin{itemize}
	\item 
We define  $>\partremove{w}$ as  the following restriction of $>$:  $>\partremove{w} \defeq \{ (x,y) \mid  (x, y) \in > \land x \not = w \land y \not = w \}$.	
\item We define the relation  $\gtrdot^u$  based on $>$ as   $\gtrdot^u \defeq \{(u,y) \mid \exists t.~ y >t \vee t >y\}$. 
\item We use indexes to decorate strict partial orders; this way, e.g., $>_1, >_2, \ldots$ denote different partial orders. \\ We use $\gtrdot^u_i$ to denote the  relation  $\gtrdot^u$  based on the strict partial order $>_i$, for some $i$.
\item $>^+$ denotes the transitive closure of $>$.
\end{itemize}

\end{definition}

Intuitively, $>\partremove{w}$ is $>$ but without the pairs that include $w$.
Also, $\gtrdot^u$ declares $u$ to be greater than any element occurring in $>$.

\begin{notation}
	We say that $\Gamma \st P \in \dilllang(u)$ if $\sttodillj{\Gamma \st P}_{u}{}$ is defined.
\end{notation}

\begin{figure}[!t]
	\small
	\begin{mathpar}

		\inferrule*[left=\orvastype{Nil}]{ > \defeq \{ (u , x) \mid   x \in \Gamma \setminus u \} }
		{ \Gamma \stpre{>}{u} \nil  }

		\and

		\inferrule*[left=\orvastype{LinIn}]{ 
			\Gamma ,  x: S , y : T \stpre{>}{u} P
		}
		{  \Gamma , x: \lin \vasreci{T}{S}  \stpre{>}{u}  \vasin{\lin}{x}{y}{P}   }

		\and

		\inferrule*[left=\orvastype{Par_1}]{ 
			\Gamma \setminus u \vaslinunsplit \Delta_1 \stpre{>_1}{w} P
			\\
			\Gamma \vaslinunsplit \Delta_2 \stpre{>_2}{u} Q
			\\
			>\,\defeq (>_1\partremove{w})\, \cup >_2\cup\, ((\gtrdot^u_1)\partremove{w})
			\\
			w \text{ fresh wrt $P, Q, \Gamma$} 
		}
		{ \Gamma \vaslinunsplit \Delta_1, \Delta_2 \stpre{>}{u} P \pp Q  }

		\and

		\inferrule*[left=\orvastype{Par_2}]{ 
			z:V , (\Gamma \setminus u \vaslinunsplit \Delta_1) \stpre{>_1}{z} P
			\\
			v:\dual{V} , (\Gamma \vaslinunsplit \Delta_2) \stpre{>_2}{u} Q
			\\
			> \defeq >_1  \cup >_2  \cup \gtrdot^u_1
		}
		{ \Gamma \vaslinunsplit \Delta_1, \Delta_2 \stpre{>}{u} \res{z v}( P \pp Q ) }
		\and
		\inferrule*[left=\orvastype{UnIn_1}]{ 
			\Gamma,  y : T \stpre{>_1}{w} P
			\\
			> \defeq  >_1\partremove{w}  \cup (\gtrdot^x_1)\partremove{w}\\
			\notserver{T}
			\\
			w \text{ fresh} 
		}
		{\Gamma , x: \recur{\wn T} \stpre{>}{x}  \vasin{\un}{x}{y}{P}   }
		\and
		\inferrule*[left=\orvastype{UnIn_2}]{ 
			\Gamma,  y : T \stpre{>_1}{y} P
			\\
			>\defeq >_1\cup (\gtrdot^x_1)
			\\
			\notclient{T} \land \server{T}
		}
		{\Gamma , x: \recur{\wn T} \stpre{>}{x}  \vasin{\un}{x}{y}{P}   }
		\and
		\inferrule*[left=\orvastype{LinOut_1}]{ 
			\Gamma, v:T ,  x:S \stpre{>}{u} P
			\quad
			\contun{T}
		}
		{ \Gamma, v:T ,  x: \lin \oc (T).S \stpre{>}{u} \ov x\out v.P  }
				\and
		\inferrule*[left=\orvastype{LinOut_2}]{ 
			\Gamma,  x:S \stpre{>_1}{u} P
			\quad 
			> \defeq >_1  \cup \{(u , v)\}
			\quad
			\neg \contun{T}
		}
		{ \Gamma, v:T ,  x: \lin \oc (T).S \stpre{>}{u} \ov x\out v.P  }
		
		\and
		
		\inferrule*[left=\orvastype{LinOut_3}]{ 
			\Gamma\setminus u \vaslinunsplit \Delta_1, y:  \dual{T} \stpre{>_1}{y} P
			\\ 
			\Gamma \vaslinunsplit \Delta  , z:S \stpre{>_2}{u} Q
			\\
                        > \defeq >_1  \cup >_2  \cup \gtrdot^u_1\cup \{(u,x)\}
			\\
			u \not = z
		}
		{ \Gamma \vaslinunsplit \Delta_1, \Delta_2 , z: \lin \vassend{T}{S}  \stpre{>}{u}  \res{xy}\vasout{z}{x}{(P \pp Q)}  }
		\and
		\inferrule*[left=\orvastype{LinOut_4}]{ 
			\Gamma \vaslinunsplit \Delta_1, y:  \dual{T} \stpre{>_1}{y} P
			\\ 
			\Gamma \vaslinunsplit \Delta_2  , z:S \stpre{>_2}{z} Q
		       \\
		       > \defeq  >_1  \cup >_2  \cup \gtrdot^z_1 \cup \{(z,x)\}
		}
		{ \Gamma \vaslinunsplit \Delta_1, \Delta_2 , z: \lin \vassend{T}{S}  \stpre{>}{z}  \res{xy}\vasout{z}{x}{(P \pp Q)}  }
		\and
		\inferrule*[left=\orvastype{UnOut_1}]{ 
			\Gamma, z: \recur{\oc T} , x: \dual{T} \stpre{>_1}{u} P
			\\
			> \defeq  >_1 \cup \{(u , y)\}
			\\
			u \not = z
		}
		{ \Gamma , z: \recur{\oc T}  \stpre{>}{u}  \res{xy}\vasout{z}{y}{P}  }
		\and
		\inferrule*[left=\orvastype{UnOut_2}]{ 
			\Gamma , x:\recur{\oc T}  , v:T \stpre{>}{u} {{P}} 
			\\
			u \not = x
			\\ 
			\contun{T}
		}
		{ \Gamma ,  x: \recur{\oc T} , v:T  \stpre{>}{u} \ov x\out v.P  }
				\and
		\inferrule*[left=\orvastype{UnOut_3}]{ 
			\Gamma , x:\recur{\oc T}   \stpre{>_1}{u} {{P}} 
			\\
			\neg \contun{T}
			\\
			> \defeq >_1  \cup\,\{(u , v)\}
			\\
			u \not = x
		}
		{ \Gamma ,  x: \recur{\oc T} , v:T  \stpre{>}{u} \ov x\out v.P  }
	\end{mathpar}	
\caption{A type system for $\vasco$ that induces a strict partial order $>$.}\label{ch5fig:orderedtypes}
\end{figure}

We consider a refinement of the type system for \vasco (cf. \Cref{ch5f:vasrules}) which allows us to infer a strict partial order $>$ on names for processes. 
This set of rules is given in Figure~\ref{ch5fig:orderedtypes}: it obeys typing judgments of the form $$\Gamma \stpre{>}{u} P $$ for some name $u$. 

We often shall appeal to the rules in \Cref{ch5fig:orderedtypes} in combination with the assumption that the process under consideration is typable in DILL (via an encoding). 
This way, e.g., in rule $\orvastype{UnIn}$  we do not  consider the case $\server{T}\wedge \client{T}$ because such a possibility is not typable in DILL.

Given $ \Gamma \stpre{>}{u} P $, we shall say that $u$ is the \emph{maximum element} of order $>$.
We always order with strict partial orders, referred simply to as  \spo~or ``orders'' in the following. 

\begin{example}{An order for a process}\label{ch5ex:build_order}
Consider the process $P=\res{xz}(\vasin{\un}{x}{y}{\nil}\pp \vasout{z}{w}{\nil})$, typable as follows:
\begin{mathpar}
\small
\inferrule
{
\inferrule{
 \inferrule{
  w,\nilT, x:\recur{\wn \nilT }\st x:\recur{\wn \nilT }\\\\   w:\nilT, x:\recur{\wn \nilT },y:\nilT\st \nil }
 { w:\nilT, x:\recur{\wn \nilT }\st \vasin{\un}{x}{y}{\nil}} \\ 
\inferrule{
w:\nilT, z:\recur{\oc \nilT}\st z:\recur{\oc \nilT} \\\\ w:\nilT\st w:\nilT
}{w:\nilT, z: \recur{\oc \nilT} \st \vasout{z}{w}{\nil}}}
{w:\nilT, x: \recur{\wn \nilT}, z: \recur{\oc \nilT} \st \vasin{\un}{x}{y}{\nil}\pp \vasout{z}{w}{\nil})}}
{w:\nilT\st \res{xz}(\vasin{\un}{x}{y}{\nil}\pp \vasout{z}{w}{\nil})}
\end{mathpar}
Now, since $w:\nilT \st P \in \dilllang(u)$, we can build the ordering $>$ such that $w:\nilT\stpre{>}{u}P$.
\begin{mathpar}
\small
\inferrule{
\inferrule{
\inferrule{>_{11}\defeq \{(a,w), (a,y)\}\quad
a \text{~fresh }}{w:\nilT, y:\nilT\stpre{>_{11}}{a} \nil} \quad >_1= \{(x,w),(x,y)\}}
{x:\recur{\wn \nilT, w:\nilT}\stpre{>_1}{x} \vasin{\un}{x}{y}{\nil}}\\
\inferrule{
\inferrule{>_2\defeq \{(u,z),(u,w)\}}{z:\recur{\oc \nilT} , w:\nilT\stpre{>_2}{u} \nil}}
{z:\recur{\oc \nilT} , w:\nilT\stpre{>_2}{u}\vasout{z}{w}{\nil}}
\quad
>=>_1\cup>_2\cup \{(u,y), (u,x)\} }
{w:\nilT\stpre{>}{u} \res{xz}(\vasin{\un}{x}{y}{\nil}\pp \vasout{z}{w}{\nil})}
\end{mathpar}
The final relation is $>\defeq \{ (x,w), (x,y), (u,z),(u,w), (u,y),(u,x)\}$, which  is illustrated by the diagram below:
\begin{center}
\begin{tikzpicture}
\node (startu) {$u$};
\node (nextx) [right of=startu] {$x$};
\node (nextw) [right of=nextx] {$w$};
\node (nexty) [below of= nextw] {$y$};
\node (nextz) [below of= nextx] {$z$};
\draw[->] (startu) -- (nextx);
\draw[->] (nextx) -- (nextw);
\draw[->] (nextx) -- (nexty);
\draw[->] (startu) -- (nextz);
\end{tikzpicture}
\end{center}

Intuitively, $u$ is a name that does not occur in the process, and is greater than both $x$ and $z$, the names originating from the parallel threads. The order also captures that $x$ is greater than $y$ and $w$. 
\end{example}

\subsubsection{Properties of Strict Partial Orders}
We prove an invariant property of the maximum element of the order with respect to the free names of its associated process $P$ (i.e., every free name in the context is minimal); the property does not hold for the \emph{bound} names of $P$.

\begin{proposition}\label{ch5prop:ontleaf}
	If $ \Gamma \stpre{>}{u} P $ for some $u$ then for all $x \in \dom{\Gamma}\setminus u$ there is no $y$ such that $x >y$. 
\end{proposition}

\begin{proof}
	By structural induction on $P$. There are 8 cases.

	\begin{enumerate}
		\item When $ P = \nil$.
		
		 Then, by definition $ > \defeq \{ (u , x) \mid  x \in \Gamma \setminus u \}$, and the result  holds trivially.
		
		\item\label{ch5proof:contleaf1} When $ P = P' \pp Q$.
		
		Then, the rule applied is $\orvastype{Par_1}$ and $\Gamma$ can be split into $\Gamma',\Delta_1, \Delta_2$ and the derivation is as follows: 
		 \begin{mathpar}
		 \inferrule*[left=\orvastype{Par_1}]{ 
			\Gamma' \setminus u \vaslinunsplit \Delta_1 \stpre{>_1}{w} P'
			\\
			\Gamma' \vaslinunsplit \Delta_2 \stpre{>_2}{u} Q
			\\
			>\defeq >_1\partremove{w}\cup >_2\cup (\gtrdot^u_1)\partremove{w}
			\\
			w \text{ fresh} 
		}
		{ \Gamma' \vaslinunsplit \Delta_1, \Delta_2 \stpre{>}{u} P' \pp Q  }
		 \end{mathpar}
		 Let $x\in \dom{\Gamma}\setminus u$, i.e., $x\in \dom{\Gamma'\vaslinunsplit \Delta_1,\Delta_2}\setminus u$.
		 
		 From  $x\in \dom{\Gamma'\vaslinunsplit \Delta_1,\Delta_2}\setminus u$ we have that $x\in \dom{\Gamma'}$, or $x\in \dom{\Delta_1}$ or $x\in \dom{\Delta_2}$ (not simultaneously), and $x\neq u$. 
		 \begin{itemize} 
		 \item If $x\in \dom{\Gamma'}$ then we can apply the IH in both $\Gamma' \setminus u \vaslinunsplit \Delta_1 \stpre{>_1}{w} P'$ and $\Gamma'  \vaslinunsplit \Delta_2 \stpre{>_2}{u} Q$ and conclude that there is no $y$ such that $x>_1y$ and $x>_2y$.  Here, we need $x\neq w$ to apply the IH on  the type derivation for $P'$. Therefore, there is no $y$ such that $(x,y) \in >_1\partremove{w}\cup >_2$. Since $\gtrdot^u_1$ only add relations to $u$, it follows easily that   there is no $y$ such that $x>y$.         
		 \item If $x\in \dom{\Delta_1}$ then $x\notin \dom{\Gamma'\vaslinunsplit \Delta_2}$ and we can only apply IH in $\Gamma' \setminus u \vaslinunsplit \Delta_1 \stpre{>_1}{w} P'$, and conclude that there is no $y$ such that $x>_1y$. Similarly, $x\neq w$. Therefore, there is no $y$ such that $(x,y) \in >_1\partremove{w}\cup (\gtrdot^u_1)\partremove{w}$. By construction, $>_2$ is defined  $\dom{\Gamma'\vaslinunsplit \Delta_2}$, since $x$ is not a member, it follows that there is no $y$ such that $x>_2y$, and the result follows.
		 \item If $x\in \dom{\Delta_2}$ then  then $x\notin \dom{\Gamma'\vaslinunsplit \Delta_1}$ we can only apply IH in  $\Gamma'  \vaslinunsplit \Delta_2 \stpre{>_2}{u} Q$, and conclude that there is no $y$ such that $x>_2y$.  By definition, there is no $y$ such that $(x,y) \in >_1\partremove{w} \cup (\gtrdot^u_1)\partremove{w}$.
		 \end{itemize}
	 Therefore,  there is no $y$ such that $x>y$.

		\item When $ P = \res{z v}( P' \pp Q )$.
		
		Then, the rule applied is $\orvastype{Par_2}$ and $\Gamma$ can be split into $\Gamma',\Delta_1,\Delta_2$ and the derivation is as follows:
		\begin{mathpar}
				\inferrule*[left=\orvastype{Par_2}]{ 
			z:V , (\Gamma' \setminus u \vaslinunsplit \Delta_1) \stpre{>_1}{z} P'
			\\
			v:\dual{V} , (\Gamma' \vaslinunsplit \Delta_2) \stpre{>_2}{u} Q
			\\
			> \defeq >_1  \cup >_2  \cup \gtrdot^u_1
		}
		{ \Gamma' \vaslinunsplit \Delta_1, \Delta_2 \stpre{>}{u} \res{z v}( P' \pp Q ) }
		\end{mathpar}
		
		By applying the IH on both premises we get 
		\begin{itemize}
		\item for all  $x \in \dom{\Gamma'\setminus u \vaslinunsplit \Delta_1} $  and $x\neq z$ there is no $y$ such that $x>_1y$. 
		\item for all  $x \in \dom{v:\dual{V},\Gamma' \vaslinunsplit \Delta_2} $  and $x\neq u$ there is no $y$ such that $x>_2y$. 
		\end{itemize}
		
		 Let $x\in \dom{\Gamma}\setminus u$, i.e., $x\in \dom{\Gamma'\vaslinunsplit \Delta_1,\Delta_2}\setminus u$. Then, it follows from both cases above that there is no $y$ such that $(x,y)\in >_1\cup >_2$. Since $\gtrdot^u_1$ only add relations to $u$, it follows that   there is no $y$ such that $x>y$.

		 \item When $ P = \vasin{\un}{x}{y}{P'} $. 
		
		There are two cases to consider, depending on whether $\neg\server{T}$ or $\server{T}$.
		\begin{itemize}
		\item In the first case the derivation is as follows:
	\begin{mathpar}
		\inferrule*[left=\orvastype{UnIn_1}]{ 
			\Gamma',  y : T \stpre{>_1}{w} P'
			\\
			> \defeq  >_1\partremove{w}  \cup (\gtrdot^x_1)\partremove{w}\\
			\notserver{T}
			\\
			w \text{ fresh} 
		}
		{\Gamma' , x: \recur{\wn T} \stpre{>}{x}  \vasin{\un}{x}{y}{P'}   }
	\end{mathpar}
		
Then  $ \Gamma',  y : T \stpre{>_1}{w} P' $ for a fresh $w$, by the IH, for all $x'\in \dom{\Gamma', y:T}\setminus w$, there is no $y'$ such that $x'>_1y'$. Thus, for  $z\in \dom{\Gamma',x:\recur{\wn{T}}}\setminus x$,  since $z\neq w$ ( $w$ is fresh),  it follows that there is no $y'$ such that $(z,y') \in >_1\partremove{w}$, which implies that there is no $y'$ such that $z>y'$. 

\item In the second case, the derivation ends with an application of \orvastype{UnIn_2}, if we have, in addition, that $\notclient{T}$ and the proof is analogous to the previous case.
			\end{itemize}	 
		 \item When $ P =  \vasin{\lin}{x}{y}{P'}$.
		 
		 Then the derivation is as follows:
		 \begin{mathpar}
		 \inferrule*[left=\orvastype{LinIn}]{ 
			\Gamma' ,  x: S , y : T \stpre{>}{u} P'
		}
		{  \Gamma', x: \lin \vasreci{T}{S}  \stpre{>}{u}  \vasin{\lin}{x}{y}{P'}   }
		 \end{mathpar}
		 
		  This case follows easily by  the induction hypothesis.

		\item When $ P = \res{xy}\vasout{z}{y}{P'}  $.
		
		Then the derivation has the form:
		\begin{mathpar}
		\inferrule*[left=\orvastype{UnOut_1}]{ 
			\Gamma', z: \recur{\oc T} , x: \dual{T}  \stpre{>_1}{u} P'
			\\
			> \defeq  >_1 \cup \{(u , y)\}
			\\
			u \not = z
		}
		{ \Gamma' , z: \recur{\oc T}   \stpre{>}{u}  \res{xy}\vasout{z}{y}{P'}  }
		\end{mathpar}
		
		By the IH, for all $z\in \dom{\Gamma', z: \recur{\oc T} , x: \dual{T}  }\setminus u$, there is no $y'$ such that  $x>_1y'$. Since $\dom{\Gamma', z: \recur{\oc T}  }\subseteq \dom{\Gamma', z: \recur{\oc T} , x: \dual{T}  }$ and $>_1\subseteq >$ the result follows trivially.

		\item When $ P = \res{xy}\vasout{z}{x}{(P' \pp Q)}$.
		
		 Then we consider the two cases depending on whether $u=z$ or $u\neq z$:
	
			\begin{enumerate}
				\item\label{ch5proof:contleaf2} When $u \not = z$.
				\begin{mathpar}
		\inferrule*[left=\orvastype{LinOut_3}]{ 
			\Gamma'\setminus u \vaslinunsplit \Delta_1, y:  \dual{T} \stpre{>_1}{y} P'
			\\ 
			\Gamma' \vaslinunsplit \Delta_2  , z:S \stpre{>_2}{u} Q
			\\
                        > \defeq >_1  \cup >_2  \cup \gtrdot^u_1\cup \{(u,x)\}
			\\
			u \not = z
		}
		{ \Gamma' \vaslinunsplit \Delta_1, \Delta_2 , z: \lin \vassend{T}{S}  \stpre{>}{u}  \res{xy}\vasout{z}{x}{(P' \pp Q)}  }

\end{mathpar}

	By the induction hypothesis: 
	\begin{itemize}
	\item for all $x'\in \dom{\Gamma'\setminus u \vaslinunsplit \Delta_1,y:\dual{T}}\setminus y$ there is no $y'$ such that $x'>_1y'$;
	\item for all $x'\in \dom{\Gamma' \vaslinunsplit \Delta_2,z :S}\setminus u$ there is no $y'$ such that $x'>_2y'$;
	\end{itemize} 
	
	Let $x'\in \dom{\Gamma'\vaslinunsplit \Delta_1,\Delta_2, z: \lin \vassend{T}{S} }\setminus u$. Then $x'\neq y$ and $x'\neq u$, and by the IH it follows that there is no $y'$ such that $(x',y')\in >_1\cup >_2$. Notice that $ \gtrdot^u_1\cup \{(u,x)\}$ only adds relations to $u$. Therefore, there is no $y'$ such that $x'>y'$.

				\item When $u=z$  the rule applied is \orvastype{LinOut4} and the result  follows analogously.
			\end{enumerate}

		\item When $ P = x\out v.P $ then we consider the type of $x$.

			\begin{enumerate}
				\item  If $x:\lin \oc (T).S $ then we consider the structure of $T$. If $\contun{T}$ we apply rule \orvastype{LinOut1}; otherwise, we apply rule \orvastype{LinOut2}. In both cases the result follows following the same ideas of the cases above.
				
				\item If $x:\recur{\oc T}$ then we consider the structure of $T$.  If $\contun{T}$ we apply rule \orvastype{UnOut2}; otherwise, we apply rule \orvastype{UnOut3}. In both cases the result follows following the same ideas of the cases above.
			\end{enumerate}
	\end{enumerate}
This concludes the proof of \Cref{ch5prop:ontleaf}.
\end{proof}


The following property says that (typable) processes in $\dilllang$ have an ordering $>$ defined by the rules of 
Figure~\ref{ch5fig:orderedtypes}. Notice that the converse property does not hold in general: the rules for constructing the partial order are less stringent than the typing rules, essentially because linear/unrestricted considerations on the contexts are not needed for constructing the partial order.
They are only minimally used; see for instance, the occurrence of `$\vaslinunsplit$' in rules $\orvastype{Par_1}$ and $\orvastype{Par_2}$.
Later on, Proposition~\ref{ch5p:spo} will ensure that $>$ is a strict partial order.

\begin{proposition}\label{ch5prop:dillmakesorder}
	If $\Gamma\st P \in \dilllang(u)$ then $ \Gamma \stpre{>}{u} P $.
\end{proposition}

\begin{proof}
	
	By induction on the structure of $P$. There are 8 cases. 
	
	\begin{enumerate}
		\item When $P = \nil$.
		
		 Then $\Gamma \st \nil $ and $ \Gamma\st \nil \in \dilllang(u)$ with  $\dillnotun{\sttodillt{\Gamma \setminus u }} ;  \cdot \dill \nil :: u : \dillnilT$. Also,
		\[ \inferrule*[left=\orvastype{Nil}]{ > = \{ (u , x) \mid  x \in \Gamma \setminus u \} }
			{ \Gamma \stpre{>}{u} \nil  }\]
	
		\item\label{ch5proof:constuctpartial1} When $P = P' \pp Q$.

		 Suppose we have $ (\Gamma', \Delta_1) \contcomp (\Gamma' , \Delta) \st P' \pp Q $ and $(\Gamma', \Delta_1) \contcomp (\Gamma' , \Delta) \st P' \pp Q   \in \dilllang(u)$. Then $\sttodillj{(\Gamma', \Delta_1) \contcomp (\Gamma' , \Delta) \st P' \pp Q }_u$ holds and by case (2) of  \Cref{ch5d:secondtablep1}, it follows that $\dillnotun{\sttodillt{\Gamma'}};\sttodillt{\Delta_1},\sttodillt{\Delta'}\dill \res{w}(\sttodillp{P'}\pp \sttodillp{Q})::u:A$ holds, for some $\Gamma', \Delta_1,\Delta'$ satisfying \Cref{ch5d:transjudgdill},  with $u \not \in \fn{P'}$ and $w$ fresh. Thus, there are derivations $\Pi_1$ and $\Pi_2$ such that the following holds (cf. proof of \Cref{ch5thm:dilltypepres})
		 \begin{mathpar}
		 \inferrule*[left=\dilltype{cut}]{
		 \inferrule{\Pi_1}{\dillnotun{\sttodillt{\Gamma'}};\sttodillt{\Delta_1}\dill \sttodillp{P'}::w:\dillnilT}\\
		 \inferrule{\Pi_2}{\dillnotun{\sttodillt{\Gamma'}};\sttodillt{\Delta'},w:\dillnilT\dill \sttodillp{Q}::u:A}
		 }
		 {\dillnotun{\sttodillt{\Gamma'}};\sttodillt{\Delta_1},\sttodillt{\Delta'}\dill \res{w}(\sttodillp{P'}\pp \sttodillp{Q})::u:A}
		 \end{mathpar}
		  From type preservation (\Cref{ch5thm:dilltypepres}) it follows that  (i) $ \Gamma', \Delta_1 \st P'$ and $ \Gamma', \Delta_1 \st P'  \in \dilllang(w) $; (ii) $ \Gamma' , \Delta \st Q$ and $ \Gamma' , \Delta \st Q  \in \dilllang(u) $,  and by \Cref{ch5thm:propertiesvasco}, we have by  that $\Gamma' \setminus u , \Delta_1 \st P'  \in \dilllang(w)$.
		  
	   By the induction hypothesis there exist  $>_1$ and $>_2$ such that:
		\[ \inferrule*[left=\orvastype{Par_1}]{ 
			\Gamma' \setminus u \vaslinunsplit \Delta_1 \stpre{>_1}{w} P'
			\\
			\Gamma' \vaslinunsplit \Delta \stpre{>_2}{u} Q
			\\
			> = >_1 \partremove{w} \cup >_2  \cup \gtrdot^u_1 \partremove{w}
			\\
			w \text{ fresh} 
		}
		{ \Gamma' \vaslinunsplit \Delta_1, \Delta \stpre{>}{u} P' \pp Q  } \]
	and the result follows.
		\item When $P = \res{z v}( P' \pp Q )$.

		 Suppose that  $(\Gamma', \Delta_1) \contcomp (\Gamma',  \Delta) \st  \res {zv} (P' \pp Q) $ and  $(\Gamma', \Delta_1) \contcomp (\Gamma',  \Delta) \st  \res {zv} (P' \pp Q)  \in \dilllang(u)$. Then $\sttodillj{(\Gamma', \Delta_1) \contcomp (\Gamma' , \Delta) \st \res{zv}(P' \pp Q) }_u$ holds and following case (3) of  the proof of \Cref{ch5thm:dilltypepres} one has $v \not \in \fn{P'} \land z \not \in \fn{Q}$. For $z:V$ we distinguish two cases:
		 
		 \begin{itemize}
		 \item $\neg\contun{V}$:
		 
		  Then, we have (i) $ \Gamma', \Delta_1, z:V \st P'$ and $ \Gamma', \Delta_1, z:V \st P'  \in \dilllang(z) $; (ii) $ \Gamma' , \Delta , v: \dual{V} \st  Q$ and $ \Gamma' , \Delta , v: \dual{V} \st  Q \in \dilllang(u) $. 
		 \item $\contun{V}$: 
		 
		  Then, we have (i)$ \Gamma', \Delta_1, z:V, v: \dual{V} \st P'$ and $ \Gamma', \Delta_1, z:V, v: \dual{V} \st P' \in \dilllang(z) $; (ii)  $ \Gamma' , \Delta ,  z:V, v: \dual{V} \st  Q $ and $ \Gamma' , \Delta ,  z:V, v: \dual{V} \st  Q \in \dilllang(u)$. Further, by strengthening (\Cref{ch5thm:propertiesvasco}) it follows that   $\Gamma', \Delta_1, z:V \st P' \in \dilllang(z) $ and  $ \Gamma' , \Delta , v: \dual{V} \st  Q \in \dilllang(u) $.
		 \end{itemize}
		  In any case, by applying the induction hypothesis, there exist  $>_1$ and $>_2$ such that:
		\[
			\hspace{-1cm}
			\inferrule*[left=\orvastype{Par_2}]{ 
				z:V , (\Gamma' \setminus u \vaslinunsplit \Delta_1) \stpre{>_1}{z} P'
				\\
				v:\dual{V} , (\Gamma' \vaslinunsplit \Delta) \stpre{>_2}{u} Q
				\\
				> = >_1  \cup >_2  \cup \gtrdot^u_1
			}
			{ \Gamma' \vaslinunsplit \Delta_1, \Delta \stpre{>}{u} \res{z v}( P' \pp Q ) }
		\]
		and the result follows.

		\item When $P =  \vasin{\un}{x}{y}{P'} $.

		 Suppose that $\Gamma,  x: \recur{\wn T} \st \vasin{\un}{x}{y}{P'} $ and $\Gamma,  x: \recur{\wn T} \st \vasin{\un}{x}{y}{P'}  \in \dilllang(x)$. Then, $\sttodillj{\Gamma',x:\recur{\wn T} \st \vasin{\un}{x}{y}{P'}}_u$ holds and following case (4) of the proof of \Cref{ch5thm:dilltypepres}, it follows that  $u=x$
and $x \not \in \fn{P'}$. Additionally, one has to consider the structure of $T$:
		
		\begin{enumerate}
			\item\label{ch5proof:constuctpartial2} $\notserver{T}$:
			
				Then  $ \Gamma' , y:T \st   P'$ and $ \Gamma' , y:T \st   P' \in \dilllang(w) $ for $w$ fresh. By the induction hypothesis there exists a $>_1$ such that:
				\begin{mathpar}
					\inferrule*[left=\orvastype{UnIn_1}]{ 
						\Gamma',  y : T \stpre{>_1}{w} P'
						\\
						{> = >_1\partremove{w}  \cup (\gtrdot^u_1)\partremove{w}}\\
						\notserver{T}
						\\
						w \text{ fresh} 
					}
					{\Gamma' , x: \recur{\wn T} \stpre{>}{x}  \vasin{\un}{x}{y}{P'}   }
				\end{mathpar}
	
			\item $\server{T} \land \notclient{T} $:
			
			Then  (i) $ \Gamma' , y:T \st   P'$ and $ \Gamma' , y:T \st   P' \in \dilllang(w) $,  where $w$ is fresh; (ii) $ \Gamma' , y:T \st   P'$ and  $ \Gamma' , y:T \st   P' \in \dilllang(y) $. 
			By the induction hypothesis  there exists $>_1$ such that:
			\begin{mathpar}
			\inferrule*[left=\orvastype{UnIn_2}]{ 
				\Gamma',  y : T \stpre{>_1}{y} P'
				\\
				{>=>_1\cup \gtrdot^x_1}
				\\
				\notclient{T} \land \server{T}
				}
				{\Gamma' , x: \recur{\wn T} \stpre{>}{x}  \vasin{\un}{x}{y}{P'}   }
			\end{mathpar}
			
			and the result follows.

			%
	
		\end{enumerate}

		\item When $P = \vasin{\lin}{x}{y}{P'}$.

		 Suppose that  $\Gamma' , x: \lin \vasreci{T}{S}  \st  \vasin{\lin}{x}{y}{P'} $ and $\Gamma' , x: \lin \vasreci{T}{S}  \st  \vasin{\lin}{x}{y}{P'}  \in \dilllang(u)$. Then, $\sttodillj{\Gamma', x: \lin \vasreci{T}{S}  \st \vasin{\lin}{x}{y}{P'}  }_u$ holds, and by case (5) of the proof of \Cref{ch5thm:dilltypepres} one has $\Gamma' ,  x: S , y : T  \st  P' $ and $\Gamma' ,  x: S , y : T  \st  P' \in \dilllang(u) $. By the induction hypothesis, there exists a $>$ such that:
		\begin{mathpar}
	
			\inferrule*[left=\orvastype{LinIn}]{ 
				\Gamma' ,  x: S , y : T \stpre{>}{u} P'
			}
			{  \Gamma' , x: \lin \vasreci{T}{S}  \stpre{>}{u}  \vasin{\lin}{x}{y}{P'}   }
	
		\end{mathpar}

		\item When $P = \res{xy}\vasout{z}{y}{P'}$.

		 Suppose that  $\Gamma' , z: \recur{\oc T}  \st  \res{xy}\vasout{z}{y}{P'}$ and  $\Gamma' , z: \recur{\oc T}  \st  \res{xy}\vasout{z}{y}{P'}  \in \dilllang(u)$. Then $ u \not = z $ and  $\sttodillj{\Gamma', z: \recur{\oc T}   \st \res{xy}\vasout{z}{y}{P'} }_u$ holds. Then, by  case (6) of the proof of \Cref{ch5thm:dilltypepres} there are two cases: 
		 \begin{itemize} 
		 \item $\contun{T}$:  
		 
		 Then, $\Gamma', z: \recur{\oc T} , x: \dual{T} \st P'$ and $\Gamma', z: \recur{\oc T} , x: \dual{T} \st P' \in \dilllang(u) $;
		 \item $\neg \contun{T}$:   
		 
		 Then, $\Gamma', z: \recur{\oc T} , x: \dual{T} , y: \dual{T} \st P' $ and $\Gamma', z: \recur{\oc T} , x: \dual{T} , y: \dual{T} \st P' \in \dilllang(u) $ with $y \not \in \fn{P'} $.  By strengthening (\Cref{ch5thm:propertiesvasco}) it follows that  $\Gamma', z: \recur{\oc T} , x: \dual{T} \st P' \in \dilllang(u) $. 
		 \end{itemize} 
		 
		 In any case, by the induction hypothesis, there exists a $>_1$ such that:
		\begin{mathpar}
			\inferrule*[left=\orvastype{UnOut_1}]{ 
				\Gamma', z: \recur{\oc T} , x: \dual{T}  \stpre{>_1}{u} P'
				\\
				> = >_1 \cup \{(u , y)\}
				\\
				u \not = z
			}
			{ \Gamma' , z: \recur{\oc T}  \stpre{>}{u}  \res{xy}\vasout{z}{y}{P'}  }
		\end{mathpar}
	and the result follows.
		\item When $P = \res{xy}\vasout{z}{x}{(P' \pp Q)}$.
		
		 Suppose that $ \Gamma' \vaslinunsplit \Delta_1, \Delta_2 , z: \lin \vassend{T}{S}  \st  \res{xy}\vasout{z}{x}{(P' \pp Q)}$ and $ \Gamma' \vaslinunsplit \Delta_1, \Delta_2 , z: \lin \vassend{T}{S}  \st  \res{xy}\vasout{z}{x}{(P' \pp Q)} \in \dilllang(u)$. Then, $\sttodillj{\Gamma' \vaslinunsplit \Delta_1, \Delta_2 , z: \lin \vassend{T}{S}  \st  \res{xy}\vasout{z}{x}{(P' \pp Q)} }_u$ holds and, by case (7) of the proof of \Cref{ch5thm:dilltypepres} and we need to consider the following two cases:

		\begin{enumerate}
			\item $u \not = z$:
			
			Then, we have that $u \not \in \fn{P'}$ and we distinguish  two additional cases:
			\begin{enumerate}
				\item $ \neg \contun{T} $:
				
				Then $\Gamma' , \Delta_1, y:\dual{T} \st P' \in \dilllang(y) $ and $\Gamma',  \Delta_2 , z: {S}, x:{t}  \st  P'  \in \dilllang(u)$. By \Cref{ch5thm:propertiesvasco}, it follows that $\Gamma' \setminus u , \Delta_1, y:\dual{T} \st P' \in \dilllang(y) $.
				
				\item $ \contun{T} $:
				
				Then $y \not \in \fn{Q}$, $x \not \in \fn{P'} \cup \fn{Q}$ and we have both $\Gamma' , \Delta_1, y:\dual{T} , x:{t} \st P' \in \dilllang(y) $ and $\Gamma',  \Delta_2 , z: {S}, y:\dual{T}, x:{t}  \st  P'  \in \dilllang(u)$.  By \Cref{ch5thm:propertiesvasco} it follows that $\Gamma' \setminus u , \Delta_1, y:\dual{T} \st P' \in \dilllang(y) $ and $\Gamma',  \Delta_2 , z: {S}, x:{t}  \st  P'  \in \dilllang(u)$.
			\end{enumerate}
			By the induction hypothesis there exist  $>_1$ and $>_2$ such that:
			\begin{mathpar}
	\inferrule*[left=\orvastype{LinOut_3}]{ 
					\Gamma'\setminus u \vaslinunsplit \Delta_1, y:  \dual{T} \stpre{>_1}{y} P'
					\\ 
					\Gamma' \vaslinunsplit \Delta_2  , z:S \stpre{>_2}{u} Q
					\\
					{> = >_1  \cup >_2  \cup \gtrdot^u_1\cup \{(u,x)\}}
					\\
					u \not = z
				}
				{ \Gamma' \vaslinunsplit \Delta_1, \Delta_2 , z: \lin \vassend{T}{S}  \stpre{>}{u}  \res{xy}\vasout{z}{x}{(P' \pp Q)}  }
			\end{mathpar}
and the result follows.
			\item $u = z$:
			
			Then we distinguish two additional cases:
			\begin{enumerate}
				\item  $ \neg \contun{T} $:
				
				 Then both $\Gamma' , \Delta_1, y:\dual{T} \st P' \in \dilllang(y) $ and $\Gamma',  \Delta_2 , z: {S}, x:{t}  \st  P'  \in \dilllang(z)$.
				
				\item $ \contun{T} $:

				 Then $y \not \in \fn{Q}$, $x \not \in \fn{P'} \cup \fn{Q}$ and we have both $\Gamma' , \Delta_1, y:\dual{T} , x:{t} \st P' \in \dilllang(y) $ and $\Gamma',  \Delta_2 , z: {S}, y:\dual{T}, x:{t}  \st  P'  \in \dilllang(z)$. By \Cref{ch5thm:propertiesvasco}, it follows that $\Gamma' , \Delta_1, y:\dual{T} \st P' \in \dilllang(y) $ and $\Gamma',  \Delta_2 , z: {S}, x:{t}  \st  P'  \in \dilllang(z)$.
				By the induction hypothesis there exist  $>_1$ and $>_2$ such that:
				\begin{mathpar}			
					\inferrule*[left=\orvastype{LinOut_4}]{ 
						\Gamma' \vaslinunsplit \Delta_1, y:  \dual{T} \stpre{>_1}{y} P'
						\\ 
						\Gamma' \vaslinunsplit \Delta_2  , z:S \stpre{>_2}{z} Q
						\\
						{> = >_1  \cup >_2 \cup \gtrdot^z_1 \cup \{(z,x)\}
						}
					}
					{ \Gamma' \vaslinunsplit \Delta_1, \Delta_2 , z: \lin \vassend{T}{S}  \stpre{>}{z}  \res{xy}\vasout{z}{x}{(P' \pp Q)}  }
				\end{mathpar}
and the result follows.

			\end{enumerate}

		\end{enumerate}

		\item When $P = \ov x\out v.P' $.

		 Then, we need to consider the typing of $x$. There are four possibilities:
		
		\begin{enumerate}
		\item $x: \lin \oc (T).S $ and $\neg \contun{T}$:
			
			 Then,  by case (8) \Cref{ch5thm:dilltypepres} that $\Gamma',  x:S  \st P' $ and  $\Gamma',  x:S  \st P' \in\dilllang(u)$.  By the induction hypothesis, there exists a $>_1$ such that:
			\begin{mathpar}
						\inferrule*[left=\orvastype{LinOut_2}]{ 
							\Gamma', x:S \stpre{>_1}{u} P'
							\quad 
							> = >_1  \cup \{(u , v)\}
							\quad
							\neg \contun{T}
						}
						{ \Gamma', v:T ,  x: \lin \oc (T).S \stpre{>}{u} \ov x\out v.P'  }
						
			\end{mathpar}

			\item $x: \lin \oc (T).S$ and $\contun{T}$:
			
			 Then, by case (9) \Cref{ch5thm:dilltypepres} that $\Gamma', v:T ,  x:S \st P' $ and $\Gamma', v:T ,  x:S \st P' \in \dilllang(u)$.  By the induction hypothesis, there exists a $>$ such that:
			\begin{mathpar}
						\inferrule*[left=\orvastype{LinOut_1}]{ 
							\Gamma', v:T ,  x:S \stpre{>}{u} P'
							\quad
							\contun{T}
						}
						{ \Gamma', v:T ,  x: \lin \oc (T).S \stpre{>}{u} \ov x\out v.P'  }

			\end{mathpar}

			\item  $ x: \recur{\oc T} $ and $ \neg \contun{T} $:
			
			Then, by case (10) of \Cref{ch5thm:dilltypepres} that $\Gamma' , x:\recur{\oc T}  \st P' $ and $\Gamma' , x:\recur{\oc T}  \st P' \in \dilllang(u) $.  By the induction hypothesis there exists a $>_1$ such that:
			\begin{mathpar}
				
						\inferrule*[left=\orvastype{UnOut_2}]{ 
							\Gamma' , x:\recur{\oc T}  \stpre{>_1}{u} {{P'}} 
							\\
							> = >_1  \cup \{(u , v)\}
							\\
							u \not = x
							\\
							\neg \contun{T}
						}
						{ \Gamma_1 ,  x: \recur{\oc T} , v:T \stpre{>}{u} \ov x\out v.P'  }
				
			\end{mathpar}

			\item $ x: \recur{\oc T} $ and $ \contun{T} $:
			
			 Then, by  case (11) of \Cref{ch5thm:dilltypepres} that $\Gamma' , x:\recur{\oc T} , v:T \st P' $ and $\Gamma' , x:\recur{\oc T} , v:T \st P' \in \dilllang(u)$. By the induction hypothesis there exists a $>$ such that:
			\begin{mathpar}
			
						\inferrule*[left=\orvastype{UnOut_3}]{ 
							\Gamma' , x:\recur{\oc T} , v:T  \stpre{>}{u} {{P'}} 
							\\
							u \not = x
							\\
							\contun{T}
						}
						{ \Gamma_1 ,  x: \recur{\oc T} , v:T  \stpre{>}{u} \ov x\out v.P'  }
			\end{mathpar}	
	and the result follows.	
		\end{enumerate}			
	\end{enumerate}
This concludes the proof of	\Cref{ch5prop:dillmakesorder}.
	\end{proof}


The following property ensures that the rules from \Cref{ch5fig:orderedtypes} induce a strict partial order:
\begin{proposition}\label{ch5prop:itisapartialorder}
\label{ch5p:spo}
	If $ \Gamma \stpre{>}{u} P $ then $>$ is a strict partial order with maximum element $u$.
\end{proposition}

\begin{proof}
	By induction on the last applied rule from \Cref{ch5fig:orderedtypes}; we consider 13 cases:

	\begin{enumerate}
		\item When $\orvastype{Nil}$ then:
		\begin{mathpar}
			\inferrule*[left=\orvastype{Nil}]{ > \defeq \{ (u , x) \mid   x \in \Gamma \setminus u \} }
			{ \Gamma \stpre{>}{u} \nil  }
		\end{mathpar}

		and $>$ is an \spo~with maximum element $u$ by construction.
		
		\item\label{ch5proof:constuctpartialorder1} When $\orvastype{Par_1}$ then:
		\begin{mathpar}
			\inferrule*[left=\orvastype{Par_1}]{ 
				\Gamma \setminus u \vaslinunsplit \Delta_1 \stpre{>_1}{w} P
				\\
				\Gamma \vaslinunsplit \Delta_2 \stpre{>_2}{u} Q
				\\
				>\defeq >_1\partremove{w}\cup >_2\cup (\gtrdot^u_1)\partremove{w}
				\\
				w \text{ fresh} 
			}
			{ \Gamma \vaslinunsplit \Delta_1, \Delta_2 \stpre{>}{u} P \pp Q  }
		\end{mathpar}

		By the induction hypothesis both $>_1$ and $>_2$ are \spo~with maximum element $w$ and $u$ respectively. First we show that $ >_1 \partremove{w} \cup >_2  $ is an \spo:
	
	\begin{itemize}
 	\item (Irreflexive) The union of two irreflexive relations is also irreflexive. 
	\item (Transitive) Suppose that $>_1 \partremove{w} \cup >_2$ is not transitive. Then there exists a pair $(a,b) \in >_1 \partremove{w} \land (b,c) \in >_2$ or $(a,b) \in >_2 \land (b,c) \in >_1 \partremove{w}$ with $(a,c) \not \in >_1 \partremove{w} \cup >_2$. Hence we need to find a common $b$ in both $>_1 \partremove{w}$ and $>_2$, however, typing ensures that these are the names within $\Gamma$. By \Cref{ch5prop:ontleaf} we have that for all names $d \in  \Gamma \setminus u $  there exist no name $e$ that either $(d,e) \in >_1 \partremove{w}$ or $(d,e) \in >_2$. 
	\item (Asymmetric) Similarly, let us assume that $>_1 \partremove{w} \cup >_2$ is not asymmetric. Then there exists a pair $(a,b)$ such that $(a,b) \in >_1 \partremove{w} \land (b,a) \in >_2$, however, following a similar argument to  that of transitivity, we obtain the same contradiction to \Cref{ch5prop:ontleaf}.
	\end{itemize}
		Secondly we show that $> = >_1 \partremove{w} \cup >_2  \cup \gtrless^u_1 \partremove{w}$ is an \spo~with maximum element $u$. 
		\begin{itemize}
		\item (Irreflexive) Notice how $u$ is not in any connections within $>_1 \partremove{w}$.  Then $\gtrless^u_1 \partremove{w} $ is irreflexive which implies that  $>$ is irreflexive. 
		\item  (Asymmetric) Notice that for this to hold we must have for all  $(u , a) \in \gtrless^u_1 \partremove{w}$  that $(a,u) \not \in >_1 \partremove{w} \cup >_2 $. As there are no connections with $u$ in $>_1 \partremove{w}$ and that $u$ is the maximum element of $>_2$ there can be no relation $(b,u)$ for any $b$ in $>_1 \partremove{w} \cup >_2 $. Hence the relation is asymmetric.  
		\item (Transitive)  Notice that $\gtrless^u_1 \partremove{w}$ is transitive: To contradict transitivity we would need to show  that there exist  names $a$ and $b$ such that $(u , a) \in \gtrless^u_1 \partremove{w}$ and $(a,b) \in >_1 \partremove{w} \cup >_2$ such that  $(u,b) \not \in >$. However,  $(u , a) \in \gtrless^u_1 \partremove{w}$ we get that  $(a,b) \in >_1 \partremove{w}$ as $a$ is an element of $>_1$ and $b$ is an element of $>_1$, which imply that  $(u,b) \in \gtrless^u_1 $ contradicting that $(u,b) \not \in >$.
	
		\item (Maximum) The maximum element is $u$ as it is both the maximum element of $>_2$ and $(u , a) \in \gtrless^u_1 \partremove{w}$. Finally, the construction of $(u , a) \in \gtrless^u_1 \partremove{w}$ ensures that $u$ is larger then every element within $>_1$.
\end{itemize}

		\item When $\orvastype{Par_2}$ then:
		\begin{mathpar}
			\inferrule*[left=\orvastype{Par_2}]{ 
			z:V , (\Gamma \setminus u \vaslinunsplit \Delta_1) \stpre{>_1}{z} P
			\\
			v:\dual{V} , (\Gamma \vaslinunsplit \Delta_2) \stpre{>_2}{u} Q
			\\
			> \defeq >_1  \cup >_2  \cup \gtrdot^u_1
		}
		{ \Gamma \vaslinunsplit \Delta_1, \Delta_2 \stpre{>}{u} \res{z v}( P \pp Q ) }
		\end{mathpar}

		By the induction hypothesis both $>_1$ and $>_2$ are \spo~with maximum element $w$ and $u$. We argue that $>$ is an \spo~with maximum element $u$  follows analogously to \ref{ch5proof:constuctpartialorder1} with the only difference being that the maximum element $z$ of $>_1$ is preserved.

		\item When $\orvastype{LinIn}$ then:
		\begin{mathpar}
			\inferrule*[left=\orvastype{LinIn}]{ 
				\Gamma ,  x: S , y : T \stpre{>}{u} P
			}
			{  \Gamma , x: \lin \vasreci{T}{S}  \stpre{>}{u}  \vasin{\lin}{x}{y}{P}   }
		\end{mathpar}

		By the induction hypothesis both $>$ is an \spo~with maximum element $u$.

		\item\label{ch5proof:constuctpartialorder2} When $\orvastype{UnIn_1}$ then:
		\begin{mathpar}
			\inferrule*[left=\orvastype{UnIn_1}]{ 
				\Gamma,  y : T \stpre{>_1}{w} P
				\\
				> \defeq  >_1\partremove{w}  \cup (\gtrdot^x_1)\partremove{w}\\
				\notserver{T}
				\\
				w \text{ fresh} 
			}
			{\Gamma , x: \recur{\wn T} \stpre{>}{x}  \vasin{\un}{x}{y}{P}   }
		\end{mathpar}

		By the induction hypothesis $>_1\partremove{w}$ is a strict partial order with maximum element $w$. We show that $>$ is a strict partial orders with maximum element $x$. As both $>_1 \partremove{w}$ and $\gtrdot^x_1$ are irreflexive then so is their union. Both $>_1 \partremove{w}$ and $\gtrdot^x_1$ are asymmetric, there are no connections in $>_1 \partremove{w}$ regarding $x$ and $\gtrdot^x_1$ only contain connection in $x$ hence $>$ is asymmetric. Assume $>$ is not transitive, then there exists names $a$ and $b$ such that $(x , a) \in \gtrdot^x_1 $ and $(a,b) \in >_1 \partremove{w}  $ with $(x,b) \not \in >$ however $(a,b) \in >_1 \partremove{w} $ implies that $(x,b) \in \gtrdot^x_1$. Finally by construction $x$ is the largest element of $\gtrdot^x_1$ which contains all elements of $>_1 \partremove{w}$.

		\item When $\orvastype{UnIn_2}$ then:
		\begin{mathpar}
			\inferrule*[left=\orvastype{UnIn_2}]{ 
				\Gamma,  y : T \stpre{>_1}{y} P
				\\
				>\defeq >_1\cup (\gtrdot^x_1)
				\\
				\notclient{T} \land \server{T}
			}
			{\Gamma , x: \recur{\wn T} \stpre{>}{x}  \vasin{\un}{x}{y}{P}   }
		\end{mathpar}

		By the induction hypothesis $>_1$ is an \spo~with maximum element $y$. The case follows analogously to case \ref{ch5proof:constuctpartialorder2}.

		\item When $\orvastype{LinOut_1}$ then:
		\begin{mathpar}
	\inferrule*[left=\orvastype{LinOut_1}]{ 
				\Gamma, v:T ,  x:S \stpre{>}{u} P
				\quad
				\contun{T}
			}
			{ \Gamma, v:T ,  x: \lin \oc (T).S \stpre{>}{u} \ov x\out v.P  }
		\end{mathpar}
		By the induction hypothesis $>$ is an \spo~with maximum element $u$.

		\item\label{ch5proof:constuctpartialorder8} When $ \orvastype{LinOut_2} $ then:
		\begin{mathpar}
		\inferrule*[left=\orvastype{LinOut_2}]{ 
				\Gamma,  x:S \stpre{>_1}{u} P
				\quad 
				> \defeq >_1  \cup \{(u , v)\}
				\quad
				\neg \contun{T}
			}
			{ \Gamma, v:T ,  x: \lin \oc (T).S \stpre{>}{u} \ov x\out v.P  }
		\end{mathpar}

		By the induction hypothesis $>_1$ is an \spo~with maximum element $u$. We show that $>$ is an \spo~with maximum element $u$. Clearly $>$ is irreflexive. As $>_1$ is asymmetric we must show that adding the pair $ \{(u , v)\}$ preserves this, but as $v\not \in  \fn{P}$ then there are no pairs of the form $(v,a)$ or $(a,v)$ for some name $a$ in $>_1$. This same argument is made to show transitivity.

		\item When $\orvastype{LinOut_3}$ then:
		\begin{mathpar}
			\inferrule*[left=\orvastype{LinOut_3}]{ 
				\Gamma\setminus u \vaslinunsplit \Delta_1, y:  \dual{T} \stpre{>_1}{y} P
				\\ 
				\Gamma \vaslinunsplit \Delta  , z:S \stpre{>_2}{u} Q
				\\
							> \defeq >_1  \cup >_2  \cup \gtrdot^u_1\cup \{(u,x)\}
				\\
				u \not = z
			}
			{ \Gamma \vaslinunsplit \Delta_1, \Delta_2 , z: \lin \vassend{T}{S}  \stpre{>}{u}  \res{xy}\vasout{z}{x}{(P \pp Q)}  }
		\end{mathpar}

		By the induction hypothesis both $>_1$ and $>_2$ are strict partial orders with maximum element $y$ and $u$ respectively. $ >_1 \partremove{w} \cup >_2  \cup \, \gtrdot^u_1 \partremove{w}$ is a strict partial order with maximum element $u$ which is argued analogously to that of case~\ref{ch5proof:constuctpartialorder1}. Secondly, we show that $>$ is a strict partial order with maximum element $u$ which is then argued analogously to that of case~\ref{ch5proof:constuctpartialorder8}.

		\item When $ \orvastype{LinOut_4} $ then:
		\begin{mathpar}
			\inferrule*[left=\orvastype{LinOut_4}]{ 
				\Gamma \vaslinunsplit \Delta_1, y:  \dual{T} \stpre{>_1}{y} P
				\\ 
				\Gamma \vaslinunsplit \Delta_2  , z:S \stpre{>_2}{z} Q
				\\
				> \defeq  >_1  \cup >_2 
			}
			{ \Gamma \vaslinunsplit \Delta_1, \Delta_2 , z: \lin \vassend{T}{S}  \stpre{>}{z}  \res{xy}\vasout{z}{x}{(P \pp Q)}  }
		\end{mathpar}

		By the induction hypothesis both $>_1$ and $>_2$ are strict partial orders with maximum element $y$ and $z$ respectively. $>$ follows from the first step of case~\ref{ch5proof:constuctpartialorder1}.

		\item When $ \orvastype{UnOut_1} $ then:
		\begin{mathpar}
			\inferrule*[left=\orvastype{UnOut_1}]{ 
				\Gamma, z: \recur{\oc T} , x: \dual{T} \stpre{>_1}{u} P
				\\
				> \defeq  >_1 \cup \{(u , y)\}
				\\
				u \not = z
			}
			{ \Gamma , z: \recur{\oc T}  \stpre{>}{u}  \res{xy}\vasout{z}{y}{P}  }
		\end{mathpar}

		By the induction hypothesis $>_1$ is a strict partial order with maximum element $u$. We argue $>$ is a strict partial order with maximum element $u$ analogously to case~\ref{ch5proof:constuctpartialorder8}.

		\item When $ \orvastype{UnOut_2} $ then:
		\begin{mathpar}
			\inferrule*[left=\orvastype{UnOut_2}]{ 
				\Gamma , x:\recur{\oc T}  , v:T \stpre{>}{u} {{P}} 
				\\
				u \not = x
				\\ 
				\contun{T}
			}
			{ \Gamma_1 ,  x: \recur{\oc T} , v:T  \stpre{>}{u} \ov x\out v.P  }
		\end{mathpar}
		By the induction hypothesis $>$ is a strict partial order with maximum element $u$.

		\item When $ \orvastype{UnOut_3} $ then: 
		\begin{mathpar}
			\inferrule*[left=\orvastype{UnOut_3}]{ 
				\Gamma , x:\recur{\oc T}   \stpre{>_1}{u} {{P}} 
				\\
				\neg \contun{T}
				\\
				> \defeq >_1  \cup \{(u , v)\}
				\\
				u \not = x
			}
			{ \Gamma_1 ,  x: \recur{\oc T} , v:T  \stpre{>}{u} \ov x\out v.P  }
		\end{mathpar}

		By the induction hypothesis $>_1$ is an \spo~with maximum element $u$. We argue $>$ is an \spo~with maximum element $u$ analogously to case~\ref{ch5proof:constuctpartialorder8}.
	\end{enumerate}
	This concludes the proof of	\Cref{ch5prop:itisapartialorder}.
\end{proof}


\subsubsection{Operations on Strict Partial Orders}
Up to here, we have defined the strict partial order induced by a typable process. 
Because we need to reason compositionally about processes, we should define corresponding ways of combining their partial orders.

\begin{notation}
\label{ch5not:prefix}
	We give a shorthand notation on input and output prefixes within $\vaslang$ on a name $x$ to be:
		$$
		\begin{aligned}
			\guardpref{ch5x}{\widetilde{y}}{} :=&  \res{yz}\ov x \out y && \text{when } \widetilde{y} = \{y , z\}
			\\
			& \sep \ov x \out y \sep q \  x\inp y  && \text{when } \widetilde{y} = \{y\}
		\end{aligned}
		$$
	\end{notation}
	
	The following definition serves to extract causality relations between names of a process, both free and bound.

\begin{definition}{Connected Names}\label{ch5def:con_names}
Let $P$ be a process and $x$ be a name (either bound or free in $P$). 
The {\em connected names} of $ P$ with respect to $x$, written $\cc{x}{P}$, is a finite list of pairs $(n_1 , \widetilde{m}_1), \ldots, (n_k , \widetilde{m}_k)$, where each $n_i$ is a  name and $\widetilde{m}_i$ is  a list of names connected to $n_i$ in $P$. This list  of pairs is defined by induction on $P$ as follows:
	\[
	\begin{aligned}
		\cc{x}{\guardpref{ch5x}{\{y\}}{}.{P'} }&=
		( y , \cc{y}{  P'  } ) ::
			\cc{x}{ P'}
			&
		\\
		\cc{x}{ \guardpref{ch5x}{\{y,z\}}{}.{P'} }&=
		(z  , \cc{z}{  P'  }) :: \cc{x}{ P'}
			& 
		\\
					\cc{x}{\guardpref{ch5w}{\widetilde{y}}{}.{P'}}&= \cc{x}{ P'}
			& \text{if }   w \not = x 
		\\
		\cc{x}{P}&= \emptyset 
			& \text{if } x \not \in \names{P} 
		\\
		\cc{x}{(P'\pp Q)}&= \cc{x}{P'} 
			& \text{if }    x \in \names{P'}
		\\
		\cc{x}{(P'\pp Q)}&= \cc{x}{Q} 
			& \text{if }  x \not\in \names{P'}
		\\
		\cc{x}{ \res{yz} P'} &= \cc{x}{P'}
			& 
		\\
	\end{aligned}
	\]

\end{definition}	
	
\begin{example}
Consider the process $P=\vasin{\un}{x}{y}{\vasout{y}{z}{\nil}}$ which,
following \Cref{ch5not:prefix}, 
 can be written as $P=\guardpref{ch5x}{\{y\}}{}.\guardpref{ch5y}{\{z\}}{}.{\nil}$. The connected names of $P$ w.r.t. $x$ are as follows:
\[
\begin{aligned}
\cc{x}{\guardpref{ch5x}{\{y\}}{}.\guardpref{ch5y}{\{z\}}{}.{\nil}} &= (y,\cc{y}{\guardpref{ch5y}{\{z\}}{}.\nil}):: \cc{x}{\guardpref{ch5y}{\{z\}}{}.\nil}\\
&= (y, (z, \cc{z}{\nil})::  \cc{z}{\nil} ):: \cc{x}{\nil}\\
&= (y, (z, \emptyset):: \emptyset ):: \emptyset\\
& = (y, (z, \emptyset))
\end{aligned}
\]
The connected names of $P$ w.r.t. $y$ are as follows:
\[
\begin{aligned}
\cc{y}{\guardpref{ch5x}{\{y\}}{}.\guardpref{ch5y}{\{z\}}{}.{\nil}} &= \cc{y}{\guardpref{ch5y}{\{z\}}{}.\nil}\\
&=  (z, \cc{z}{\nil})::  \cc{z}{\nil} \\
&= (y,  \emptyset ):: \emptyset = (y,  \emptyset ) \\
\end{aligned}
\]
\end{example}

\begin{example}{Cont. \Cref{ch5ex:build_order}}\label{ch5ex:cc_process}
Consider the process $ P = \res{xz}( \vasin{\un}{x}{y}{\nil} \pp   \vasout{z}{w}{\nil} )$ which, 
following \Cref{ch5not:prefix}, 
 can be written as $P=\res{xz}(\guardpref{ch5x}{\{y\}}{}\pp \guardpref{ch5z}{\{w\}}{})$. Then, the connected names of $P$ w.r.t. $x$ and $z$ are as follows: 

\[
\begin{aligned}
\cc{x}{P}&= \cc{x}{\guardpref{ch5x}{\{y\}}{}}&&=(y,\emptyset)::\emptyset=(y,\emptyset)\\
\cc{z}{P}&= \cc{z}{\guardpref{ch5z}{\{w\}}{}}&&=(w,\emptyset)::\emptyset=(w,\emptyset)
\end{aligned}
\]
\end{example}
	
Given a strict partial order $>$, we now define an extension of $>$ that arises by ``joining'' (or merging) two elements in $>$ by combining their respective pairs. 

\begin{definition}{Join of An Order}
Let  $>$ be  an \spo~with maximal element on a set $S$, and two elements $x, y\in S$. 
We define the function $\partjoin{>}{x}{y}$ as follows:
\begin{align*}
		\partjoin{>}{x}{y} =  (> & \cup \{ (x, a) \mid  (y,a) \in >  \} \\
		& \cup \{ (a, x) \mid  (a,y) \in >  \}
		\\ 
		& \cup \{ (y, a) \mid  (x,a) \in >  \}
		\\
		& \cup \{ (a, y) \mid  (a,x) \in >  \})^+	\end{align*}
\end{definition}

In the above definition, the transitive closure ($^+$) is needed: if we intend to join $x$ and $y$, and we have $a>x$ and $y>b$ then we add $a>y$ and $x>b$ but also $a>b$ needs to be added to preserve a strict partial order.

\medskip

We now define functions that enable us to (i)~simplify the pairs contained in the ordering
and (ii)~potentially extend this ordering by inspecting the process structure.
This is done by starting from a basic operation that relates connected names.
\begin{definition}{Flattening of An Order}\label{ch5def:flaten}
Let $A, B, \ldots$ be lists of pairs of names, in the sense of \Cref{ch5def:con_names}, and  $>$ be an \spo. 
First, we define the function $\flatcc{>}{A}{B}$ as follows:
\[
\begin{aligned}
\flatcc{>}{ (x, A) : B }{ (y,C) : D } &= \partjoin{\flatcc{\flatcc{>}{A}{C}}{ B }{ D }}{x}{y}
\\
\flatcc{>}{ \emptyset }{\emptyset} &= 
\flatcc{>}{ A }{\emptyset} = 
\flatcc{>}{ \emptyset }{A} = ~>
\end{aligned}
\]
Second, given a process $P \in \dilllang$,
we define the function $\flaten{>}{P}$ as follows:
\[
\begin{aligned}
\flaten{>}{  (P \pp Q)  } &= \flaten{\flaten{>}{P}}{ Q }\\
\flaten{>}{ \res{xy}\vasout{z}{y}{P} } &= \partjoin{\flaten{>}{P}}{x}{y}
\\
\flaten{>}{  \res{xy}\vasout{z}{x}{(P \pp Q)}  } &= \partjoin{\flaten{>}{P \pp Q}}{x}{y}\\
\flaten{>}{ \res{z v}( P \pp Q ) } & = \partjoin{\flaten{\flatcc{>}{\cc{ z}{P}}{\cc{ v }{Q}}}{  P \pp Q }}{z}{v}, (*)
\\
\flaten{>}{ \vasin{\lin}{x}{y}{P} } &= \flaten{>}{ {P} }\\
\flaten{>}{ \vasin{\un}{x}{y}{P} } &= \flaten{>}{ {P} }\\
\flaten{>}{  \ov x\out v.P } &= \flaten{>}{  P }\\
\flaten{>}{\nil}&= ~>\\
\end{aligned}
\]
\end{definition}

Where $(*)$ requires $ z \in \fn{P} \land v \in \fn{Q}$.
The fact that flattening preserves some relevant structure will be addressed later on, via \Cref{ch5prop:flatpreserves}.

\begin{example}{Cont. \Cref{ch5ex:cc_process}}\label{ch5ex:flat_proc} Consider the flattening of the process $P=\res{xz}(\vasin{\un}{x}{y}{\nil}\pp \vasout{z}{w}{\nil})$ w.r.t. the spo $>\defeq \{ (x,w), (x,y), (u,z),(u,w), (u,y),(u,x)\}$  built in \Cref{ch5ex:build_order}.

To compute $\flaten{>}{ \res{xz}(\vasin{\un}{x}{y}{\nil}\pp \vasout{z}{w}{\nil})}$, we first obtain:
\[
\begin{aligned}
\flatcc{>}{\cc{x}{\vasin{\un}{x}{y}{\nil}}}{\cc{z}{\vasout{z}{w}{\nil}}}&=\flatcc{>}{(y,\emptyset)::\emptyset}{(w,\emptyset)::\emptyset}\\
&= \partjoin{\flatcc{\flatcc{>}{\emptyset}{\emptyset}}{\emptyset}{\emptyset}}{y}{w}\\
&= \partjoin{>}{y}{w}\\
&= ~>
\end{aligned}
\]
Now we have:
\[
\small
\begin{aligned}
\flaten{>}{ P } &= \partjoin{\flaten{\flaten{\flatcc{>}{\cc{x}{\vasin{\un}{x}{y}{\nil}}}{\cc{z}{\vasout{z}{w}{\nil}}}}{\vasin{\un}{x}{y}{\nil}}}{\vasout{z}{w}{\nil}}}{x}{z}\\
&= \partjoin{\flaten{\flaten{>}{\vasin{\un}{x}{y}{\nil}}}{\vasout{z}{w}{\nil}}}{x}{z}\\
&= \partjoin{\flaten{>}{\vasout{z}{w}{\nil}}}{x}{z}\\
&= \partjoin{>}{x}{z}\\
&= (> \cup \{(z,w),(z,y)\})\\
\end{aligned}
\]
Hence, in this case, flattening extends the ordering. This can be illustrated by the diagram:
\begin{center}
	\begin{tikzpicture}
\node (startu) {$u$};
\node (nextx) [right of=startu] {$x$};
\node (nextw) [right of=nextx] {$w$};
\node (nexty) [below of= nextw] {$y$};
\node (nextz) [below of= nextx] {$z$};
\draw[->] (startu) -- (nextx);
\draw[->] (nextx) -- (nextw);
\draw[->] (nextz) -- (nextw);
\draw[->] (nextz) -- (nexty);
\draw[->] (nextx) -- (nexty);
\draw[->] (startu) -- (nextz);
\end{tikzpicture}
\end{center}
which can be simplified to
\begin{center}
\begin{tikzpicture}
\node (startu) {$u$};
\node (nextx) [right of=startu] {$x,z$};
\node[xshift=4pt] (nextw) [right of=nextx] {$w,y$};
\draw[->] (startu) -- (nextx);
\draw[->] (nextx) -- (nextw);
\end{tikzpicture}
\end{center}
\end{example}

\begin{notation}
Consider the following notations:
\begin{itemize}
	\item 
We write $w,v \pairrel x, y$ to denote $(w  = x \land v = y) \lor (v  = x \land w = y)$ (that is, if $\{ w , v\} \cap \{ x , y \} = \{ x , y \}$).  
\item Similarly, we write $w,v \not \pairrel x, y$ to denote $w \not = x \land v \not = x \land w \not = y \land  v \not = y$ (that is, if $\{ w , v\} \cap \{ x , y \} = \emptyset$).
\end{itemize}

\end{notation}

We now define a predicate that indicates whether that two names are \emph{compatible} and may be joined.
\begin{definition}{Compatibility}\label{ch5def:compatible}

The auxiliary predicate $\compatiblecc{A}{B}{x}{y}$ is defined on two lists of connected channels $A$ and $B$ as:
	\[
	\begin{aligned}
		\small
			\compatiblecc{ (w, A) : B }{ (v,C) : D }{x}{y} &=
			\begin{cases}
			\true  & \text{if } w,v \pairrel x, y \\
			\begin{aligned}
				\compatiblecc{ A}{C }{x}{y} \lor \\\compatiblecc{ B}{D }{x}{y}
			\end{aligned}
			 & \text{ otherwise }
			\end{cases}
			\\
		\compatiblecc{\emptyset}{\emptyset}{x}{y} & = 
		\compatiblecc{A}{\emptyset}{x}{y} = 
		\compatiblecc{\emptyset}{A}{x}{y} = \false
				\\
	\end{aligned}
	\]
	
Let $P \in \dilllang$ be a process, and let $x,y$ be two names.
	The predicate $\compatible{P}{x}{y}$ is defined on the structure of $P$ as follows:
	\[
		\small
	\begin{aligned}
		\compatible{\nil}{x}{y} &= \false\\
		\compatible{\vasout{w}{v}{P'}}{x}{y} &= 
			\begin{cases}
				\compatible{{P'}}{x}{y} & \text{if } w,v \not\pairrel x, y \\
				\false  & \text{ otherwise } \\
			\end{cases}\\
		\compatible{\res{w,v}\vasout{z}{v}{P'}}{x}{y} &= 
			\begin{cases}
				\true & \text{if } w,v \pairrel x, y \\
			 \compatible{{P'}}{x}{y} & \text{if }  w,v \not\pairrel x, y \\
				\false  & \text{ otherwise } \\
			\end{cases}\\
		\compatible{\res{wv}\vasout{z}{w}{(P' \pp Q)}}{x}{y} &= 
			\begin{cases}
				\true & \text{if } w,v \pairrel x, y \\
				\compatible{{P'}}{x}{y}  \lor \compatible{ Q }{x}{y} & \text{if } w,v \not\pairrel x, y \\
				\false  & \text{ otherwise } \\
			\end{cases}\\
		\compatible{\vasin{\lin}{w}{v}{P'}}{x}{y} &= 
			\begin{cases}
				\compatible{{P'}}{x}{y} & \text{if } w,v \not\pairrel x, y \\
				\false  & \text{ otherwise } \\
			\end{cases}\\
		\compatible{\vasin{\un}{w}{{v}}{P'}}{x}{y} &= 
			\begin{cases}
				\compatible{{P'}}{x}{y} & \text{if } w,v \not\pairrel x, y \\
				\false  & \text{ otherwise } \\
			\end{cases}\\
		\compatible{ \res{w v}(P' \pp Q) }{x}{y} &= 
			\begin{cases}
				\true & \text{if } w,v \pairrel x, y \\
				\begin{aligned}
					\mathsf{compcc}(\cc{w}{ \res{w v}(P' \pp Q)},\\
					\cc{ v}{\res{w v}(P' \pp Q)},x, y)
				\end{aligned}
				 & \text{ otherwise }\\
				\quad \lor \compatible{ P' }{x}{y} \lor \compatible{ Q }{x}{y} 
				\\
			\end{cases}\\
		\compatible{ P' \pp Q }{x}{y} &= 
			\compatible{ P' }{x}{y} \lor \compatible{ Q }{x}{y} 
				\\
	\end{aligned}
	\]

\end{definition}	
\noindent Note that to verify when  the compatible predicate holds we need to check the pairs of names in a restriction, say $\res{wv}$ above, and also the connected names $\cc{\cdot}{P}$ of the process.

To illustrate this predicate, consider the following example.
\begin{example}{Cont. \Cref{ch5ex:flat_proc}} 
\label{ch5ex:compat}
Recall once again process $P=\res{xz}(\vasin{\un}{x}{y}{\nil}\pp \vasout{z}{w}{\nil})$. 

From \Cref{ch5ex:cc_process}, we have $\cc{x}{P}=(y,\emptyset)::\emptyset$ and $\cc{z}{P}=(w,\emptyset)::\emptyset$. Thus,
\[
\begin{aligned}
\compatible{P}{x}{z}&=\true\\
\compatible{P}{y}{w}&=\compatiblecc{\cc{x}{P}}{\cc{z}{P}}{y}{w} \vee \compatible{\vasin{\un}{x}{y}{\nil}}{y}{w}\\
& \qquad \vee \compatible{\vasout{z}{w}{\nil}}{y}{w}\\
&=\compatiblecc{(y,\emptyset)::\emptyset}{(w,\emptyset)::\emptyset}{y}{w} \vee \compatible{\vasin{\un}{x}{y}{\nil}}{y}{w} \\
& \qquad \vee \compatible{\vasout{z}{w}{\nil}}{y}{w}\\
&= \true \vee \compatible{\vasin{\un}{x}{y}{\nil}}{y}{w} \vee \compatible{\vasout{z}{w}{\nil}}{y}{w}\\
&=\true\\
\compatible{P}{u}{y}&= \compatiblecc{\cc{x}{P}}{\cc{z}{P}}{u}{y} \vee \compatible{\vasin{\un}{x}{y}{\nil}}{u}{y}\\
& \qquad \vee \compatible{\vasout{z}{w}{\nil}}{u}{y}\\
&=  \compatiblecc{\cc{x}{P}}{\cc{z}{P}}{u}{y} \vee \false \lor \false\\
&=\compatiblecc{(y,\emptyset)::\emptyset}{(w,\emptyset)::\emptyset}{u}{y}\\
&= \false
\end{aligned}
\]
Note that the $\compatible{P}{\cdot}{\cdot}$ predicate holds on the names in $P$  that were joined when flattening $>$ in  \Cref{ch5ex:flat_proc}:
\begin{center}
\begin{tikzpicture}
\node (startu) {$u$};
\node[blue] (nextx) [right of=startu] {$x,z$};
\node[xshift=4pt,blue] (nextw) [right of=nextx] {$w,y$};
\draw[->] (startu) -- (nextx);
\draw[->] (nextx) -- (nextw);
\end{tikzpicture}
\end{center}

\end{example}

The next definition allows for collecting the names that are compatible w.r.t. the \Cref{ch5def:compatible}:
\begin{definition}{Set of compatible names}\label{ch5def:compN}
	The set of {\em compatible names in a process $P$}, denoted $\varcompat{P}$,
	 is defined inductively as follows, using 
	 \Cref{ch5def:con_names} (the connected names of $ P$ with respect to $x$, written $\cc{x}{P}$)
	 and \Cref{ch5def:compatible}:
	\[
	\begin{aligned}
	\small
	\varcompat{\nil} &= \emptyset\\
	\varcompat{\vasout{w}{v}{P}} &= \varcompat{P} \\
	\varcompat{\res{wv}\vasout{z}{v}{P}} &=\{ (w,v) , (v,w) \} \cup\varcompat{P} \\
	\varcompat{\res{wv}\vasout{z}{w}{(P \pp Q)}} &=\{ (w,v) , (v,w) \} \cup\varcompat{P}\cup \varcompat{Q}\\
	\varcompat{\vasin{\lin}{w}{v}{P}} &=\varcompat{P} \\
	\varcompat{\vasin{\un}{w}{{v}}{P}} &=\varcompat{P} \\
	\varcompat{\res{w v}(P \pp Q) } &= \{ (w,v) , (v,w) \} \cup \varcompat{P} \cup \varcompat{Q} \cup \{ (x, y) \mid\\
	&  \quad  \compatiblecc{ \cc{w}{ \res{w v}(P \pp Q)}}{\cc{ v}{\res{w v}(P \pp Q)}}{x}{y} \} \\
	\varcompat{P \pp Q} &= \varcompat{P} \cup \varcompat{Q}
	\end{aligned}
	\]

\end{definition}
\noindent For example, the set of compatible names for the process in \Cref{ch5ex:flat_proc}, i.e.,  $\varcompat{P}=\{(x,z), (z,x), (y,w), (w,y)\}$.

The next property establishes the relation between (i)~the names within a process that can be joined and (ii)~the names that are compatible. This result is used in the proof of \Cref{ch5prop:flatpreserves2}.

\begin{proposition}[{\sf flat} vs. {\sf compN}] \label{ch5prop:flat_comp}
Let  $P\in \dilllang$  a process and  $>$ be an \spo.  The set of compatible names on a process $P$, $\varcompat{P}$, is exactly the set of names in which the function $\flaten{>}{P}$ applies the join function on.

\end{proposition}

\begin{proof}
Follows trivially from \Cref{ch5def:flaten,ch5def:compN}.
\end{proof}

\begin{example}{}
	From \Cref{ch5ex:flat_proc} we can see that when evaluating $\flaten{>}{ }$ with $P = \res{xz}(\vasin{\un}{x}{y}{\nil}\pp \vasout{z}{w}{\nil})$ then $\partjoin{>}{-}{-}$ is applied twice, the first time on the  pair of channels are $y$ and $w$ and the second time it is applied on on $x$ and $z$. Secondly as prevesly stated $\varcompat{P}=\{(x,z), (z,x), (y,w), (w,y)\}$ which is exactly the pair of channels that are joined. 
\end{example}

We define a notion of transitive closure up-to compatibility, in the sense defined above. 

\begin{definition}{Closure of $>$ under $P$} \label{ch5def:closure}
Let $>$ be an  \spo~and $P$ be a process.
	We write $a>_P^*b$ 
	if one of the following holds:
	\begin{enumerate}
		\item $a> b$
		\item $\exists c.~a >c \land c >_P^* b  $
		\item $\exists c.~\compatible{ P }{a}{c} \land c >_P^* b$
		\item $\exists c.~a >_P^* c \land \compatible{ P }{c}{b}$
	\end{enumerate}
	When the process $P$ is clear from the context, we shall write `$a >^* b$' rather than `$a >_P^* b$'.
\end{definition}
Intuitively, $a>_P^* b$ identifies `paths' between names $a$ and $b$, induced by the relation $>$ and also by the compatible names in $P$.

\begin{example}{}
Recalling the \spo~$>\defeq \{ (x,w), (x,y), (u,z),(u,w), (u,y),(u,x)\}$  built in \Cref{ch5ex:build_order}. It is easy to see that $u>_P^* y$ since  $u>x \wedge x>y$. Unfortunately, the example does not add any interesting insight with the compatible names,  in fact,  $x>y \wedge \compatible{P}{y}{w}$ implies $x>_P^*w$, since $x>w$ the compatibility does not add any new information.
\end{example}

We are interested in \spo s which are acyclic (i.e., without loops), a notion that we define considering compatibility between two names, and the closure up-to compatibility just given.

\begin{definition}{Acyclic Orders}\label{ch5def:acyclic}
Let $>$ be an F,  $P$ be a process, and  $x$ and $y$ be names in $P$. 
We define the following predicate.
\[
\small
\acyclic{>}{P}{x}{y} = \left\{
\begin{array}{ll}
\true, &
 \begin{array}{l}
 i)~ x\not >_P^* y; and \\
ii)~ (x >_P^* a  \implies a \not  >_P^* y) \land (a >_P^* x \implies y \not >_P^*  a);  and \\
iii)~ (a>_P^*x \land  y>_P^* b) \implies \neg\compatible{P}{a}{b};  and \\
iv)~  a>_P^*x >_P^* b \implies \neg\compatible{P}{a}{b}.
\end{array}
\\ 
\false, & \text{otherwise}
\end{array}
\right.
\]
	
\end{definition}

\begin{example}{Cont. \Cref{ch5ex:flat_proc}} Consider the process $P=\res{xz}(\vasin{\un}{x}{y}{\nil}\pp \vasout{z}{w}{\nil})$ and  the 
\spo $$>\defeq \{ (x,w), (x,y), (u,z),(u,w), (u,y),(u,x), (z,w),(z,y)\}$$ We have checked that $\compatible{P}{x}{z}$ and $\compatible{P}{y}{w}$. Below we check that  $\acyclic{>}{P}{x}{z}$:

\begin{itemize}
\item $x\not > z$;
\item ($x>w$ but $w\not >z$) and ($x>y$ but $y\not > z$). Similarly, $u>x$ but $z\not >u$; 
\item $u>x \land z>y \land \neg \compatible{P}{u}{y}$.
\end{itemize}

Similarly, one can verify that $\acyclic{>}{P}{y}{w}$. In fact, $y\not > w$. Also,
 $ x,u,z>y$ but $\not \exists a. w> a$.
 
 Note that for the relation to hold the red arrows are disallowed (they would induce a cycle):
 \begin{center}
\begin{tikzpicture}
\node (startu) at (0,0) {$u$};
\node[blue] (nextx) at (1,0) {$x$ ,};
\node[blue] (nextz) at (1.3,0) {$z$};
\node[blue] (nextw) at (2.3,0) {$w$,};
\node[blue] (nexty) at (2.6,0) {$y$};
\draw[->,thick] (startu) -- (nextx);
\draw[->,thick] (nextz) -- (nextw);
\draw[->,thick,red] (0.9,-0.12) arc (181 : 360 : 2mm);
\draw[->,thick,red] (2.2,-0.12) arc (181 : 360 : 2mm);
\draw[->,thick,red] (nextz) edge [bend right = 50] (startu);
\draw[->,thick,red] (nextw) edge [bend right = 50] (nextz);
\draw[->,red,thick] (nexty) edge [bend right = 50] (nextz);
\end{tikzpicture}
\end{center}

\end{example}

The following definition lifts our definition of acyclicity from pairs of names to entire processes:

\begin{definition}{}
\label{ch5def:nocycle}
Let $P \in \dilllang$  be a process and $ > $ an \spo.
Recall that $\varcompat{P}$ has been defined in \Cref{ch5def:compN}. 
The predicate $\preserves{>}{P}$ is then defined as:
	\[
		\preserves{>}{P} = \bigwedge\limits_{ (x,y) \in \varcompat{P} } \acyclic{>}{P}{x}{y}
	\]
\end{definition}

\subsubsection{Properties}

We now state a number of useful properties; all items follow immediately from the definitions.

\begin{proposition}\label{ch5prop:auxiliary}
 The following properties hold:
\begin{enumerate}
\item $\compatible{P}{x}{y} \iff  \compatible{P}{y}{x}$.  
\item $\compatible{P}{x}{y} \implies x,y \in \names{P}$.
\item $\acyclic{>}{P}{x}{y} \iff \acyclic{>}{P}{y}{x}$.
\item $\preserves{>}{P} \implies \preserves{>}{P'}$, where $P'$ is a subprocess of $P$. 
\item If $P'$ is a subprocess of $P$ with $a~{>}_{P'}^*~b$ then $a~{>}_P^*~b$
\item\label{ch5prop:longprop6} If $\Gamma \stpre{>}{u} P$ and $x \not \in \dom{\Gamma} \cup \names{P}$ then $\nexists y$ such that  $x> y \lor y > x \lor \compatible{P}{x}{y}$.

\item \label{ch5prop:case_compatible} If $\Gamma \stpre{>}{u} P$ then there is no $x$ such that $ \compatible{P}{u}{x}$.

\item  If $\res{w v}(P \pp Q) \in \dilllang$ with $\Gamma \stpre{>}{u} \res{w v}(P \pp Q)$ then 
\begin{align*}
\cc{w}{ \res{w v}(P \pp Q)} &= 	 \cc{w}{ P }
\\
\cc{ v}{\res{w v}(P \pp Q)} &=  
\cc{ v}{ Q }
\end{align*}

\item If $> \,=\, >_1 \cup >_2$ then $\gtrdot^u \,=\, \gtrdot^u_1 \cup \gtrdot^u_2 $

\item If $\Gamma \stpre{>}{u} P$ with $a \in \dom{\Gamma}$ and $ \compatible{P}{a}{b}$ then $b \in \bn{P}$. 

\item If $>_1 \,\subseteq\, >$ then $\flaten{>_1}{P} \subseteq \flaten{>}{P}$

\item If $\sttoltj{\Gamma  \st P}{l }{}$ and $l$ is a subfunction of $l'$ then  $\sttoltj{\Gamma  \st P}{l'}{}$.

\item Given a strict partial order $>$, we have $> \, \subseteq \flaten{>}{P}$, for any $P$.

\item Suppose $\Gamma \stpre{>}{u} P$ with $a \in \dom{\Gamma}\setminus u$. Assume given a sequence of distinct, channels of compatible channels $b_1, \cdots b_n,b$ such that $ \compatible{P}{a}{b_1},\ \compatible{P}{b_1}{b_2},\ \cdots \compatible{P}{b_n}{b_{n+1}}$.
 Then $b_{n+1} \in \bn{P}$ and $b_{n+1}\not>c$ for any $c$. 

 \item Suppose $\Gamma \stpre{>}{u} P$  with $a \in \dom{\Gamma}\setminus u$, then $a\not>^*_Pc$ for any $c$. 

 \item Suppose $\Gamma \stpre{>}{u} P$ and \preserves{>}{P} then $a\not>^*_P a$ for any name $a$

\end{enumerate}
\end{proposition}

\begin{proof}
Follows easily by the definition. We give a short description for the following cases:
\begin{enumerate}
	\item[(6)] First, notice that the rules for constructing $>$ in $\Gamma \stpre{>}{u} P$ add pairs  exclusively from $\Gamma$ or the names of $P$. Hence, if $x$ occurs within neither of them then $x$ can be neither greater nor less than any channel. Second, compatible channels are only defined on the names of $P$, this being either bound or free which $x$ is neither hence is not compatible to any channel either.
	\item[(8)] Typing ensures that $w \in \fn{P} \land w \not\in \names{P}$; similarly, $v \not\in \names{P} \land v \in \fn{P}$. This way, the proof then follows by definition.
	\item[(12)] The level function defines a hierarchy on channels that avoids cyclic dependencies; adding channels to this function will not affect the dependencies on weights. Intuitively, if a channel does not have any weight dependencies nor is dependent on any weight (such as a channel not occurring in the free or bound names of $P$) then these weights are not of consequence to the typing.
	\item[(14)]  If $a$ is free then, by the definition of compatibility, we must have that $a$ is sent within an output prefix and that the compatible channel $b_1$ is received via an input prefix (say, $\vasin{q}{w}{b_1}{P'}$). 
	Now, $b_1$ is also free within $P'$ and by \Cref{ch5prop:ontleaf} we have that there does not exist a $c_1$ such that $b_1>'c_1$ for some \spo used in the type derivation of $\vasin{q}{w}{b_1}{P'}$ and hence $ b_1>c_1 $.  
	Therefore, for $b_1$ to be compatible to $b_2$ we argue similarly that $b_2$ must also be bound; this argument may be made inductively along the length of the chain of compatible channels showing that $b$ must also be a bound name and that $b \not > c$ for any $c$.
	\item[(15)] Fisrt note that from \cref{ch5prop:ontleaf} $a \not > d$. Suppose by contradiction that  $a >_P^*c$, then there must exist $a'$ such that either  $a>a' \wedge a'>^*_Pc$ or $\compatible{P}{a}{a'}\wedge a'>_P^*c$. From \cref{ch5prop:ontleaf} we now that the first case does not hold, therefore, we check $\compatible{P}{a}{a'}\wedge a'>_P^*c$. From \Cref{ch5prop:auxiliary}(14) we know that $a'\not>d$ for any $d$ hence for $\compatible{P}{a}{a'}\wedge a'>_P^*c$ to hold then it must be the case that $\compatible{P}{a}{a'}\wedge\compatible{P}{a'}{a''}\wedge a''>_P^*c$. By performing induction on the chain of compatible channels one can see that every channel in the chain cannot have an element smaller within the \spo $>$ and hence $a {\not >}_P^*c$.

\end{enumerate}
\end{proof}
\begin{notation}
We will write `$>$ is an $\spom(u)$' to abbreviate that $>$ is a strict partial order with  maximal element $u$.
\end{notation}

The following proposition shows that joining compatible channels in an acyclic partial order does not jeorpardize acyclicity. 

\begin{lemma}\label{ch5prop:join2}
	Let $P\in \dilllang$  and  $>$  be  its $\spom(u)$. If  $\preserves{>}{P}$ and $\compatible{P}{x}{y}$ then $\partjoin{>}{x}{y}$ is an $\spom(u)$ such that $\preserves{\partjoin{>}{x}{y}}{P}$.
\end{lemma}

\begin{proof}
We prove two  points separately: first, that $\partjoin{>}{x}{y}$ a strict partial order; then,  we prove that $\preserves{\partjoin{>}{x}{y}}{P}$ holds.
	\begin{itemize}
\item 	 $\partjoin{>}{x}{y}$ is a strict partial order: 
	\begin{description}
		\item[(Irreflexive)] Suppose, by contradiction, that there exists a $z$ such that 
		$(z,z) \in \partjoin{<}{x}{y}$. As $>$ is irreflexive, $(z,z) \notin >$, then
		it was added by the transitive closure of the sets added in the definition:   
		$\{ (x, a) \mid (y,a) \in >  \}$, $\{ (a, x) \mid (a,y) \in >  \}$, 
		$\{ (y, a) \mid  (x,a) \in >  \}$ and $\{ (a, y) \mid (a,x) \in >  \}$.
		 
		 Then, either $(z,x), (x,z)$ or $(z,y), (y,z)$ are in $\partjoin{>}{x}{y}$. Suppose, w.l.o.g, that the pair is $(z,x), (x,z)$.  Suppose w.l.o.g. $(x,z)\in >$ and $(z,x)\notin >$, i.e., $(z,x)$ was added in $\partjoin{>}{x}{y}$ after closure. Then, $(z,y)\in >$ and, by transitivity, we would have $(x,y)\in >$.
		 However, by hypothesis, we have $\preserves{>}{P}$ and  $\compatible{P}{x}{y}$.
 which imply that  $\acyclic{>}{P}{x}{y}$, and consequently, $x\not > y$. Contradiction.
	 
	    \item[(Asymmetric)] 
		 Suppose, by contradiction, that  $\partjoin{>}{x}{y}$ is symmetric. Then,  either $(a,y) , (y,a) $ or $(a,x) , (x,a)$ are  in $\partjoin{>}{x}{y}$. 
		  Let us assume w.l.o.g. that
		   $(a,y) , (y,a) \in \partjoin{>}{x}{y}$, and let us, once again,
		    assume w.l.o.g. that $(a,y) \in >$.  Then 
			$(y,a) \in \partjoin{>}{x}{y}$ implies that $(x,a) \in >$, 
			however this contradicts case  (ii)
			 $ (a,y) \in > \implies (x,a) \not \in > $ comming from  the definition of $\acyclic{>}{P}{x}{y} $.
			  
		\item[(Transitive)] The relation is transitive by definition as it is the transitive closure. Finally if $b$ is a maximum element of $>$ then by construction $b$ is a maximal element of $>$. 
    \end{description}

\item $\preserves{\partjoin{>}{x}{y}}{P}$ holds: \\
	Let $>' = \partjoin{>}{x}{y}$.
	As $\preserves{>}{P}$ already holds we must show that the added relations within the closure with $ \{ (x, a) \mid  (y,a) \in >  \} \cup \{ (a, x) \mid  (a,y) \in >  \} \cup \{ (y, a) \mid  (x,a) \in >  \} \cup \{ (a, y) \mid  (a,x) \in >  \}$ satisfy $\preserves{>'}{P}$.

	By definition, $\preserves{>'}{P}$  holds if, for all pairs  $(c,d)$ of compatible channels $\compatible{P}{c}{d}$,   we have $ \acyclic{>'}{P}{c}{d}$.

	We consider the following two cases.
	\begin{enumerate}
		\item When $c,d \not \in \{x,y\}$.
		
		Then we must show $\acyclic{>'}{P}{c}{d}$. Following \Cref{ch5def:acyclic} we consider four sub-cases.
		\begin{enumerate}[label=(\roman*)]
		\item  $c \not {>'}^*_P d$. 
		
		First, notice that from $\acyclic{>}{P}{c}{d}$ we have that $c \not >^*_P d$.  
		
		Suppose, by contradiction that $c\,{>'}^*_P\, d$. Then, $(c,d)$ was added via closure of the sets defining $>'$ ($:= \partjoin{>}{x}{y}$). Thus,  $c \,{>'}^*_P\, x$, $x \,{>'}^*_P\, d$, $c\,{>'}^*_P\, y $ and $ y \,{>'}^*_P\, d $. However, this implies that $ c \not >^*_P d $.

		\item  $(c {>'}^*_P e  \implies e \not  {>'}^*_P d) \land (e {>'}^*_P c \implies d \not {>'}^*_P  e)$. 
		
		We shall show $(c {>'}^*_P e  \implies e \not  {>'}^*_P d)$ as the other case follows analogously.
		
		Suppose, by contradiction, that $ c {>'}^*_P e $ and $ e {>'}^*_P d $.  Assume, w.l.o.g., that $c >^*_P e$ (the case $c\not >^*_P e$ is similar) and  note that 
		
		\[
		\begin{aligned}
		(1)~&\preserves{>}{P} \land \compatible{P}{c}{d} \implies \acyclic{>}{P}{c}{d}\\
		(2)~& \acyclic{>}{P}{c}{d} \wedge c>e \implies e\not  >^*_P  d
		\end{aligned}
		\]
		Thus, $e{>'}^*_P d$ must have been introduced via the transitive closure of the defining sets for $>'$. Then both $e{>'}^*_P x\wedge x{>'}^*_P d$ and $e{>'}^*_P y \wedge y{>'}^*_P d$ hold. There are three cases to consider:
		\begin{itemize}
		\item $e\not >^*_Px \wedge x\not >^*_Pd$:
		
		Then, $e{>'}^*_P x$ because $e>^*_Py$ holds, and $x{>'}^*_P d$ because $y>^*_Pd$ holds. By transitivity, we would have $e>^*_Pd$, which is a contradiction w.r.t. (2).
		\item $e\not >^*_Px \wedge x >^*_Pd$: 
		
		Then, $e{>'}^*_P x$ because $e{>}^*_Py$ holds.
		\[
		\begin{aligned}
		(1)~&c>_P^*e \wedge e>_P^*y \implies c >_P^* y ~\text{ (by transitivity and } \\
		& \text{ \Cref{ch5def:closure})}\\
 	(2)~&\compatible{P}{y}{x} \wedge x>^*_Pd \implies y>_P^* d ~\text{ (by \Cref{ch5def:closure})}\\
	       (3)~& \acyclic{>}{P}{y}{x}\wedge \compatible{P}{c}{d} \implies c\not >_P^* y\vee y\not >_P^* d ~\text{ (by } \\
		   &\text{\Cref{ch5def:acyclic})}
		\end{aligned}
		\]
	From (1), (2), and (3) a contradiction follows.
		\item $e >_P^* x \wedge x\not >_P^* d$:

		Then, $x{>'}^*_P d$ because $y{>}^*_Pd$ holds.
			\[
		\begin{aligned}
		(1)~&c{>}^*_Pe \wedge  e{>}^*_P x \implies ~\text{ (by \Cref{ch5def:closure})}\\
		(2)~&\compatible{P}{x}{y} \wedge y>^*_Pd \implies x>_P^* d ~\text{ (by \Cref{ch5def:closure})}\\
		(3)~& \acyclic{>}{P}{x}{y}\wedge \compatible{P}{c}{d} \implies c\not >_P^* x\vee x\not >_P^* d\\
		\end{aligned}
		\]
                From (1), (2) and (3) follow the contradiction.
		\end{itemize}

		\item  $ e{>'}_P^*c\wedge d{>'}_P^* f \implies \neg\compatible{P}{e}{f}$.

		Suppose, by contradiction, that $\compatible{P}{e}{f}$
		Consider the following cases:

		\begin{itemize}

			\item $ e{>}_P^*c\wedge d{>}_P^* f$:
			
			Then $\preserves{>}{P} \land \compatible{P}{c}{d}$ imply  $\acyclic{>}{P}{c}{d}$ giving, by (iii), the contradiction $\neg\compatible{P}{e}{f}$.
			
			\item $ e{>}_P^*c\wedge d\not {>}_P^* f$:
			
			Then $>'$ must introduce $d {>}_P^* f$. Therefore, we can suppose  w.l.o.g  that $d>_P^*x$ with either $x>_P^*f$ or $y>_P^*f$ But then this implies that $d>_P^*f$.

			\item  $( e\not{>}_P^*c\wedge d{>}_P^* f) $ or $(e\not{>}_P^*c\wedge d\not{>}_P^* f)$.
		
			These cases are treated analogously to the previous.

		\end{itemize}

		\item  $ e{>'}_P^* x{ >'}_P^* f \implies \neg\compatible{P}{e}{f}$.
		
		 This can be argued analogously to the case above.
		\end{enumerate}

		\item $c\in \{x,y\} \vee d \in \{x,y\}$. 
		
		The following cases need to be considered, and the are verified similarly to the case 1).
		
		\begin{enumerate}
		\item $c = x $ and $ d \not = y$. 
		
                 Then we must show $\acyclic{>'}{P}{x}{d}$. 

		\item $c \not = x$ and $ d = y$.
		
       Then we must show $\acyclic{>'}{P}{c}{y}$.  

		\item $c= x$ and $d = y$
		
		Then we must show $\acyclic{>'}{P}{x}{y}$, which follows easily from $\acyclic{>}{P}{x}{y}$.
		\end{enumerate}
	\end{enumerate}
	\end{itemize}
	This concludes the proof of \Cref{ch5prop:join2}.
\end{proof}


The following property concerns the preservation of ordering (and maximal element) with respect to flattening, in the important specific case in which a typable process is derived from a cut-rule (and flattening involves the two connected channels).

\begin{lemma}\label{ch5prop:flatpreserves}
Let $>$ be an \spo~with maximum element $u$ and $\res{xy}(P\pp Q) \in \dilllang$ such that  $\preserves{>}{\res{xy}(P\pp Q) }$. Then, the following hold:

\begin{enumerate}
\item $\flatcc{>}{\cc{x}{P}}{\cc{y}{Q}}$  is an $\spom(u)$.  
\item $\preserves{\flatcc{>}{\cc{x}{P}}{\cc{y}{Q}}}{\res{xy}(P\pp Q)}$.
\end{enumerate}

\end{lemma}

\begin{proof}
By induction on the structure of $P$. We have 7 cases.
\begin{enumerate}
   \item When $P = \nil$.
		
   Then $\cc{\nil}{x} = \emptyset$ and, by definition, $\flatcc{>}{\emptyset}{\cc{y}{Q}} = >$. Thus, the result follows by hypothesis. 

   \item When $P = \vasout{w}{v}{P'}$.
		
	If $w \not = x$ then $\cc{x}{P} = \cc{x}{P'} $, which implies  $\flatcc{>}{\cc{x}{P}}{\cc{y}{Q}} = \flatcc{>}{\cc{x}{P'}}{\cc{y}{Q}}$. The result follows by applying the induction hypothesis. If $w=x$ then  $P = \vasout{x}{v}{P'}$ and $\cc{x}{P} = (v  , \cc{v}{  P'  }) : \cc{x}{ P'} $ and we proceed by induction  on the structure of $Q$. We distinguish 8 sub-cases:
	\begin{enumerate}
	\item When $Q = \nil$. 
				 
				 Then $\cc{\nil}{y} = \emptyset$ and $\flatcc{>}{ (w  , \cc{w}{  P'  }) : \cc{z}{ P'}}{\emptyset} = >$ and the result follows by hypothesis.

				\item When $Q = \vasout{w'}{v'}{Q'}$.
				
				If $w' = y$ then $\res{xy}(P\pp Q) \not \in \dilllang$  which contradicts our hypothesis.
				Then, it must be that $w' \not = y$ and by definition $\cc{y}{Q} = \cc{y}{Q'} $.  Hence, $\flatcc{>}{\cc{x}{P}}{\cc{y}{Q}} = \flatcc{>}{\cc{x}{P}}{\cc{y}{Q'}}$, applying the induction hypothesis the case follows.

				\item When $Q = \res{w'v'}\vasout{z'}{v'}{Q'}$. 
				
				If  $z' = y$ then $\res{xy}(P\pp Q) \not \in \dilllang$ which contradicts our hypothesis.
				Then $z' \not = y$  and by definition  $\cc{y}{Q} = \cc{x}{Q'} $.  Hence, $\flatcc{>}{\cc{x}{P}}{\cc{y}{Q}} = \flatcc{>}{\cc{x}{P}}{\cc{y}{Q'}}$, applying the induction hypothesis the case follows.

\item \label{ch5item:out_par} When $Q = \res{w'v'}\vasout{z'}{v'}{(Q' \pp Q'')}$.
				
If  $z' = y$ then $\res{xy}(P\pp Q) \not \in \dilllang$ which contradicts our hypothesis. Then $z' \not = y$ and  we need to consider the following three cases:
			
\begin{itemize}
\item $y \in \fn{Q'}$:
						
Then $\cc{y}{Q} = \cc{y}{Q'} $.  Hence, $\flatcc{>}{\cc{x}{P}}{\cc{y}{Q}} = \flatcc{>}{\cc{x}{P}}{\cc{y}{Q'}}$, applying the induction hypothesis the case follows.
\item $y \not \in \fn{Q'} \land y \in \fn{Q''}$:
						
 Then $\cc{y}{Q} = \cc{y}{Q''} $.  Hence, $\flatcc{>}{\cc{P}{x}}{\cc{Q}{y}} = \flatcc{>}{\cc{x}{P}}{\cc{y}{Q''}}$, applying the induction hypothesis the case follows.
\item $y \not \in \fn{Q'} \land y \not \in \fn{Q''}$. 

 Then $\cc{y}{Q} = \emptyset $, which implies $\flatcc{>}{\cc{x}{P}}{ \emptyset } = >$ and the result follows by hypothesis.
\end{itemize} 
\item\label{ch5proof:bignest} When $Q = \vasin{\lin}{w'}{v'}{Q'}$ (case $Q = \vasin{\un}{w'}{v'}{Q'}$ is analogous).

When $w' \not = y$, then $\cc{y}{Q} = \cc{y}{Q'} $.  Hence, $\flatcc{>}{\cc{x}{P}}{\cc{y}{Q}} = \flatcc{>}{\cc{x}{P}}{\cc{y}{Q'}}$, applying the induction hypothesis the case follows. Otherwise, if  $w' = y$, then  $Q = \vasin{\lin}{y}{v'}{Q'}$ and $\cc{y}{Q} = (v'  , \cc{v'}{  Q'  }) :: \cc{y}{ Q'} $.

We want to prove that:
\begin{itemize}
\item $\flatcc{>}{\cc{x}{P}}{\cc{y}{Q}}$  is an \spo~with maximal element $u$.
\item $\preserves{\flatcc{>}{\cc{x}{P}}{\cc{y}{Q}}}{\res{xy}(P\pp Q)}$.
\end{itemize}

 In fact,
  $$
  \begin{aligned}
  \flatcc{>}{  (v  , \cc{v}{  P'  }) :: \cc{x}{ P'}   }{ (v'  , \cc{v'}{  Q'  }) :: \cc{y}{ Q'} } \hspace{2cm} \\
  = \partjoin{\flatcc{\flatcc{>}{ \cc{v}{  P'  }}{\cc{v'}{  Q'  }}}{ \cc{x}{ P'}  }{ \cc{y}{ Q'} }}{v}{v'}.
  \end{aligned}
  $$

 We want to use the \Cref{ch5prop:join2}. Note the following:
 \begin{enumerate}
\item By definition,  

$$\hspace{-1cm}\small\preserves{>}{\res{xy}(P\pp Q) } = \true \iff \bigwedge\limits_{ (x,y) \in S } \acyclic{>}{\res{xy}(P\pp Q)}{x}{y} $$
 where $ S= \{(x,y) \mid  \compatible{\res{xy}(P\pp Q)}{x}{y} \}$.
 Notice that from the definition of $P$ and $Q$ we have  $\{(x,y) \mid  \compatible{P'}{x}{y} \} \subset S $, $\{(x,y) \mid  \compatible{Q'}{x}{y} \} \subset S $ and that $(v,v') \in S$.

\item  Note that 
  $\preserves{>}{\res{v v'}(P'\pp Q')}$.  In fact,
  $$\hspace{-1cm}\small\preserves{>}{\res{v v'}(P'\pp Q')} = \true \iff \bigwedge\limits_{ R } \acyclic{>}{\res{v v'}(P'\pp Q') }{x}{y} $$
   with $ R= \{(x,y) \mid  \compatible{\res{v v'}(P'\pp Q')}{x}{y} \}= \{(x,y) \mid  \compatible{P'}{x}{y} \} \cup \{(x,y) \mid  \compatible{Q'}{x}{y} \} \cup \{(v,v')\} $.
  
  The hypothesis that $\res{xy}(P\pp Q) \in \dilllang$  implies that {\bf (H1)} $\res{v v'}(P'\pp Q')\in \dilllang $. Also, by the definition of compatible names we have $\varcompat{\res{v v'}(P'\pp Q')}\subseteq \varcompat{\res{xy}(P\pp Q)}$, which implies that   {\bf (H2)} $\preserves{>}{\res{v v'}(P'\pp Q')}$ holds. We can apply the induction hypothesis in  {\bf (H1)} and  {\bf (H2)}  and obtain that 
  
 \begin{itemize}
 \item $>'\triangleq \flatcc{>}{\cc{v}{P'}}{\cc{v'}{Q'}}$  is an \spo~with maximal element $u$.
 \item $\preserves{>'}{\res{vv'}(P'\pp Q')}$.
 \end{itemize}

Next $\preserves{>'}{\res{v v'}(P'\pp Q')} \implies \preserves{>'}{\res{xy}(P'\pp Q')}$. 

 Applying the induction hypothesis again we obtain:
 \begin{itemize}
\item  $>''\triangleq \flatcc{>'}{ \cc{x}{ P'}  }{ \cc{y}{ Q'} }$, and 
\item $\preserves{>''}{\res{x y}(P'\pp Q')}$.
 \end{itemize}
  
 Finally, as  
 $\preserves{>''}{\res{xy}(P \pp Q)}$, and  $\compatible{P}{v}{v'}$
we can apply \Cref{ch5prop:join2} which implies  $>'''\triangleq \partjoin{>''}{v}{v'} $ is an \spo~with maximum element $u$ and 
$\preserves{>'''}{\res{xy}(P \pp Q)}$.
  
\end{enumerate}
	\item \label{ch5item:res_par}When $ Q = \res{wv}(Q' \pp Q'')$.
	
As $\res{xy}(P \pp Q) \in \dilllang$ we have $y \not \in \{ w, v\}$. Hence

\begin{itemize}
\item  $\cc{y}{Q} = \cc{y}{(Q' \pp Q'')} $, and 
\item $\flatcc{>}{\cc{x}{P}}{\cc{y}{Q}} = \flatcc{>}{\cc{x}{P}}{\cc{y}{Q' \pp Q''}}$
\end{itemize}
The case follows by  the induction hypothesis.

\item When $Q =Q' \pp Q''$.

This case is similar to \ref{ch5item:out_par} above.		
\end{enumerate}

\item\label{ch5proof:flatcc2} When $P = \res{wv}\vasout{z}{v}{P'}$.
		
 Then let us consider the following cases:
		
\begin{enumerate}
\item When $z \not = x$. 

Then $\cc{x}{P} = \cc{x}{P'} $. Hence, $\flatcc{>}{\cc{P}{x}}{\cc{Q}{y}} = \flatcc{>}{\cc{x}{P'}}{\cc{y}{Q}}$, applying the induction hypothesis the case follows.

\item When $z = x$.

 Then  $P = \res{wv}\vasout{x}{v}{P'}$ and $\cc{x}{P} = (w  , \cc{w}{  P'  }) : \cc{x}{ P'} $ and we proceed by induction on the structure of $Q$, where there are multiple cases that follow as in case 2.

\end{enumerate}
		
\item When $P = \res{wv}\vasout{z}{v}{(P' \pp P'')}$.

Then let us consider the following cases:
		
\begin{enumerate}

\item When $z \not = x$.

This case is similar to part \ref{ch5item:out_par}, depending whether $x\in \fn{P'}$ or $x\in \fn{P''}$.

\item When $z = x$. 

Then $P = \res{wv}\vasout{x}{v}{(P' \pp P'')}$ and $\cc{x}{P} = (w  , \cc{w}{  P' \pp P''  }) : \cc{z}{ P' \pp P''} $ and we proceed by induction on the structure of $Q$,  where there are multiple cases that follow as in part 2.

		\end{enumerate}
		
		\item When $P = \vasin{q}{w}{v}{P'}$.
		
		 Then let us consider the following cases:
		
		\begin{enumerate}
			\item When $w \not = x$.
			
			 Then $\cc{x}{P} = \cc{x}{P'} $.  Hence, $\flatcc{>}{\cc{x}{P}}{\cc{y}{Q}} = \flatcc{>}{\cc{x}{P'}}{\cc{y}{Q}}$ and the case follows by applying the induction hypothesis.

			\item When $w = x$.
			
			 Then  $P = \vasin{q}{x}{v}{P'}$ and $\cc{x}{P} = (v  , \cc{v}{  P'  }) : \cc{x}{ P'} $ and we proceed by induction on the structure of $Q$, where there are multiple cases that follow analogously to part 2. %

		\end{enumerate}

		\item When $P = \res{wv}(P' \pp P'')$.
		
		This case is analogous to case~\ref{ch5item:res_par}.

		\item When $P =P' \pp P''$.

This case is similar to case \ref{ch5item:out_par}.			
\end{enumerate}
This concludes the proof of \Cref{ch5prop:flatpreserves}.
\end{proof}
The following property also concerns the preservation of ordering (and maximal element) with respect to flattening, but now in a more general setting (beyond two compatible channels).

\begin{lemma}\label{ch5prop:flatpreserves2}
Let $P \in \dilllang$ and  $>$ an  $\spom(u)$ such that $\preserves{>}{P}$.
Also, let $>_\mathsf{f}\, = \flaten{>}{P}$.
Then $>_\mathsf{f}$
is an  $\spom(u)$ and $\preserves{>_\mathsf{f}}{P}$ holds.
\end{lemma}


\begin{proof}
By induction on the structure of $P$. There are seven cases to consider.
\begin{enumerate}
\item When $P=\nil$.

Then $\flaten{>}{ \nil } = >$ and  the result follows trivially.

\item When $P=P'\pp P$.

By definition, $\flaten{>}{  P' \pp Q } = \flaten{\flaten{>}{  P' }}{ Q }$. 
Also $P' \in \dilllang$ and, by \Cref{ch5prop:auxiliary} (4), $\preserves{>}{P'}$.
Denote $ >':=\flaten{>}{  P' }$. 

By the induction hypothesis $>'$ is an $\spom(u)$ such that ({\bf H1}) $\preserves{>'}{P'}$ holds. Note that by definition of $>':=\flaten{>}{P'}$ in combination with \Cref{ch5prop:flat_comp}, only the compatible  names in $P'$ are joined via $\partjoin{P'}{\cdot}{\cdot}$. Thus, we can conclude that ({\bf H2}) $\preserves{>'}{P}$.

Now we will  show that the following holds:
\[\preserves{>'}{Q}=\bigwedge_{(x,y)\in \varcompat{Q}} \acyclic{>'}{Q}{x}{y}\]

Note that
\begin{itemize}
\item By hypothesis that  $\preserves{>}{P}$, it follows that   $\preserves{>}{Q}=\bigwedge_{(x,y)\in \varcompat{Q}} \acyclic{>}{Q}{x}{y}$ holds. 
\item We need to analyze the pairs of names that were added in $>'$:
by \Cref{ch5prop:flat_comp}, as $>'=\flaten{>}{P'}$, only pairs of compatible names of  $P'$ were joined in $\flaten{>}{P'}$.
Thus, by \Cref{ch5prop:join2}, it follows that $>'$ is an  $\spom(u)$ and that ({\bf H3})  $\preserves{>'}{Q}$ holds.
\end{itemize}
We still need to prove that $>_{\sf f}:=\flaten{>'}{Q}$ is an  $\spom(u)$ and that  $\preserves{>_{\sf f}}{P'\pp Q}$ holds. Note that:
\begin{itemize}
\item By applying the induction hypothesis again, with $>_{\sf f}=\flaten{>'}{Q}$, it follows that ({\bf H4}) $>_{\sf f}$ is an  $\spom(u)$ and that ({\bf H5})  $\preserves{>_{\sf f}}{Q}$ holds.
\item By definition of $>_{\sf f}:=\flaten{>'}{Q}$ and  \Cref{ch5prop:flat_comp} only pairs of compatible names were joined via $\partjoin{Q}{\cdot }{\cdot}$, which is included in the set of compatible names $\varcompat{P'\pp Q}$. By applying \Cref{ch5prop:join2} again, we obtain 
\(\preserves{>_{\sf f}}{P'\pp Q}\) and the result follows.
\end{itemize}

\item When $P=\res{zv}(P'\pp Q)$.

By definition, $ \flaten{>}{ \res{z v}( P' \pp Q ) } = \partjoin{\flaten{\flaten{\flatcc{>}{\cc{z}{P' }}{\cc{v}{ Q }}}{  P' }}{ Q }}{z}{v}$. 

Denote $>':=\flatcc{>}{\cc{z}{P'}}{\cc{v}{Q}}$.

			By \Cref{ch5prop:flatpreserves},
			$>'$ is an  $\spom(u)$ such that $\preserves{>'}{P}$ holds.  By \Cref{ch5prop:auxiliary} (2), it follows that  {\bf (H1)}~$\preserves{>'}{P'}$ and {\bf (H2)}~$\preserves{>'}{Q}$ also hold.

\begin{itemize} 
\item By the induction hypothesis in {\bf (H1)} with $>'':=\flaten{>'}{P'}$, it follows that $>''$ is an  $\spom(u)$ and {\bf (H3)}~$\preserves{>''}{P'}$.
\item Note that, by definition,  $>''':=\flaten{>''}{ Q } $  only joins the compatible names of $Q$. Therefore, it follows that  $>'''$ is an $\spom(u)$ and both $\preserves{>''}{P}$ and {\bf (H4)}$~\preserves{>'''}{P}$. 
  \end{itemize}
  Finally, as $\compatible{P}{z}{v}$, we apply \Cref{ch5prop:join2} in {\bf (H4)} and $>_f:=\partjoin{>'''}{z}{v} $ and obtain  that $>_f$ is an \spom(u) with $\preserves{>_f}{P}$, and the result follows.
  
  		\item When $P = \vasin{q}{x}{y}{P'}$ or $P = \ov x\out v.P' $.
		
		Then  $\flaten{>}{ P} = \flaten{>}{ {P'} }$ and by applying the induction hypothesis on $\flaten{>}{ {P'} }$ the result follows easily.

		\item When $P = \res{xy}\vasout{z}{x}{(P' \pp Q)} $. 
		
		Then $\flaten{>}{  \res{xy}\vasout{z}{x}{(P' \pp Q)}  } = \partjoin{\flaten{\flaten{>}{P'}}{ Q }}{x}{y}$.

		From the hypothesis $\preserves{>}{P}$ it follows that  $\preserves{>}{P'}$. 
		\begin{itemize}
		\item By the induction hypothesis it follows that  $\flaten{>}{P'}$ is an $\spom(u)$ and $\preserves{\flaten{>}{P'}}{P'}$ holds. Similarly to the argument of  case (2) above,  we also argue that {\bf (H1)}~$\preserves{\flaten{>}{P'}}{Q}$ holds. 
		\item Again, by the induction hypothesis $ >'':=\flaten{>'\flaten{>}{P'}}{ Q }$ with both $>''$ being an $\spom(u)$ and $\preserves{>''}{Q}$. Again we argue similarly to case (2) to show that {\bf (H2)}~$\preserves{>''}{P'}$.
		\item  Therefore, from {\bf (H1)}  and {\bf (H2)}, it follows that both $\preserves{>''}{P' \pp Q}$ and $\preserves{>''}{P}$ hold.
		\end{itemize}
		As $\compatible{P}{x}{y}$ we apply \Cref{ch5prop:join2} and $ >_f:=\partjoin{>''}{x}{y} $ where  $>_f$ is an $\spom(u)$  with $\preserves{>_f}{P}$.
		
\item When $P =  \res{xy}\vasout{z}{y}{P'}$.

		This case is similar to the previous case.
\end{enumerate}
\end{proof}

We shall be interested in showing that the composition of two acyclic strict partial orders (associated to specific connected channels) leads to another acyclic strict partial order. Before addressing this important property, we give an extended example that further illustrates partial orders and checking acyciclity in them.

\begin{example}{}
\label{ch5e:hardlemmas}
	To further illustrate strict partial orders, let us consider the  process $P = \vasres{xy}{(
				\vaspara{Q_1}{Q_2})}$, where $Q_1, Q_2$ are as follows:
	\[
		\begin{aligned}
			Q_1 &= \vasin{\un}{x}{z}{
				\vasres{ab}{ \vasout{z}{a}{(
					\vaspara{
						\vasin{\un}{b}{c}{P_1} 
					}
					{
						P_2
					}
				)}
				}}\\
			Q_2 &= \vasres{ts}\vasout{u}{t}{(
				\vaspara{
					\vasin{\un}{s}{d}{
						\vasres{wv}{
							\vasout{y}{w}{
								\vasin{\lin}{v}{e}{P_3
								}
							}
						}
					}
				}
				{
					P_4
				}
			)}
			\\
		\end{aligned}
	\]

Notice that processes $P_1, \ldots, P_4$ have been left unspecified.
We first consider the $\spo$   for $Q_1$, denoted $>_5$, which is constructed using the rules from \Cref{ch5fig:orderedtypes} with the following derivations: 	
	\begin{mathpar}
		\Pi_1 = 
		\inferrule*[left=\orvastype{UnIn_1}]{ 
			\Gamma,  c : T \stpre{>_1}{w} P_1
			\\
			>_2 \defeq  >_1\partremove{w}  \cup (\gtrdot^b_1)\partremove{w}\\
			\notserver{T}
			\\
			w \text{ fresh} 
		}
		{\Gamma , b: \recur{\wn T} \stpre{>_2}{b}  \vasin{\un}{b}{c}{P_1} }

		\and

		\Pi_2 = 
		\inferrule*[left=\orvastype{LinOut_4}]{ 
			\Pi_1
			\\ 
			\Gamma, z:S \stpre{>_3}{z} P_2
		       \\
		       >_4 \defeq  >_2  \cup >_3 \cup \gtrdot^z_3 \cup \{(z,a)\}
		}
		{ \Gamma , z: \lin \vassend{ (\recur{\oc {T}} )}{S}  \stpre{>_4}{z}  \vasres{ab}{ \vasout{z}{a}{( \vaspara{ \vasin{\un}{b}{c}{P_1} } { P_2 } )} }  }

		\and

		\inferrule*[left=\orvastype{UnIn_2}]{ 
			\Pi_2
			\\
			>_5 \defeq >_4\cup (\gtrdot^x_4)
		}
		{\Gamma , x: \recur{\wn (\lin \vassend{ (\recur{\oc {T}} )}{S})} \stpre{>_5}{x}  \vasin{\un}{x}{z}{ \vasres{ab}{ \vasout{z}{a}{ \vaspara{ \vasin{\un}{b}{c}{P_1} } { P_2 } } }} }
	\end{mathpar}

The corresponding strict partial orders are depicted in \Cref{ch5f:hardlemmas}, using the diagram format of \Cref{ch5ex:build_order}. 
We also have added dotted arrows: these represent additional dependencies involving the names in the unspecified processes $P_1$ and $P_2$. 
For simplicity, we do not show all these dependencies; for example, within $>_1$ process $P_1$ may have (bound) channels that are greater then $c$. For simplicity, we assume that this is not the case.

\begin{figure}[!t]	
	\[
		\begin{aligned}
			\begin{tikzpicture}
			\node (name)   {$>_1 =$};
			\node (startw) [right of= name] {$w$};
			\node (nextGamma) [right of= startw] {$\Gamma$};
			\node (nextc) [above of= nextGamma] {$c$};
			\node (nextP1) [below of= nextGamma] {$P_1$};
			\draw[->] (startw) -- (nextc);
			\draw[->] (startw) -- (nextGamma);
			\draw[dotted, ->]  (nextP1)-- (nextGamma);
			\draw[->] (startw) -- (nextP1);
			\end{tikzpicture}
			\hspace{1cm}
			\begin{tikzpicture}
			\node (startb) {$b$};
			\node (name) [left of= startb]  {$>_2 =$};
			\node (nextGamma) [right of= startb] {$\Gamma$};
			\node (nextc) [above of= nextGamma] {$c$};
			\node (nextP1) [below of= nextGamma] {$P_1$};
			\draw[->] (startb) -- (nextc);
			\draw[->] (startb) -- (nextGamma);
			\draw[dotted, ->]  (nextP1)-- (nextGamma);
			\draw[->] (startb) -- (nextP1);
			\end{tikzpicture}
			\hspace{1cm} 
			\begin{tikzpicture}
			\node (startz) {$z$};
			\node (name) [left of= startz]  {$>_3 =$};
			\node (nextGamma) [right of= startz] {$\Gamma$};
			\node (nextP2) [below of= nextGamma] {$P_2$};
			\draw[->] (startz) -- (nextGamma);
			\draw[dotted, ->]  (nextP2)-- (nextGamma);
			\draw[->] (startz) -- (nextP2);
			\end{tikzpicture}
			\\
			\begin{tikzpicture}
			\node (startz) {$z$};
			\node (name) [left of= startz]  {$>_4 =$};
			\node (startb) [right of= startz] {$b$};
			\node (nextGamma) [right of= startb] {$\Gamma$};
			\node (nextc) [below of= startb] {$c$};
			\node (nextP1) [above of= nextGamma] {$P_1$};
			\node (nexta) [above of= startb] {$a$};
			\node (nextP2) [below of= nextc] {$P_2$};
			\draw[->] (startz) -- (startb);
			\draw[->] (startz) -- (nexta);
			\draw[->] (startz) -- (nextP2);
			\draw[->] (startb) -- (nextc);
			\draw[->] (startb) -- (nextGamma);
			\draw[->] (startb) -- (nextP1);
			\draw[dotted, ->]  (nextP2)-- (nextGamma);
			\draw[dotted, ->]  (nextP1)-- (nextGamma);
			\end{tikzpicture}
			\hspace{1cm} 
			\begin{tikzpicture}
			\node (startx) {$x$};
			\node (name) [left of= startx]  {$>_5 =$};
			\node (startz) [right of= startx] {$z$};
			\node (startb) [right of= startz] {$b$};
			\node (nextGamma) [right of= startb] {$\Gamma$};
			\node (nextc) [below of= startb] {$c$};
			\node (nextP1) [above of= nextGamma] {$P_1$};
			\node (nexta) [above of= startb] {$a$};
			\node (nextP2) [below of= nextc] {$P_2$};
			\draw[->] (startx) -- (startz);
			\draw[->] (startz) -- (startb);
			\draw[->] (startz) -- (nexta);
			\draw[->] (startz) -- (nextP2);
			\draw[->] (startb) -- (nextc);
			\draw[->] (startb) -- (nextGamma);
			\draw[->] (startb) -- (nextP1);
			\draw[dotted, ->]  (nextP2)-- (nextGamma);
			\draw[dotted, ->]  (nextP1)-- (nextGamma);
			\end{tikzpicture}
			\\
		\end{aligned}
	\]
	\caption{Partial orders for process $Q_1$ from \Cref{ch5e:hardlemmas}.\label{ch5f:hardlemmas}}
\end{figure}

Similarly, the $\spo$   for $Q_2$, denoted $>_{10}$, is constructed from the following derivations: 	

	\begin{mathpar}
		\Pi_3 =
		\inferrule*[left=\orvastype{LinIn}]{ 
			\Gamma,  d : W, y: \recur{\oc(\lin \vassend{ (\recur{\oc {T}} )}{S}) } ,  v:\dual{S} , e : \recur{\oc {T}} \stpre{>_6}{d} P_3
		}
		{ \Gamma,  d : W, y: \recur{\oc(\lin \vassend{ (\recur{\oc {T}} )}{S}) }  , v: \lin \vasreci{ (\recur{\oc {T}} )}{\dual{S}} \stpre{>_6}{d} \vasin{\lin}{v}{e}{P_3 }   }

		\and
		
		\Pi_4 =
		\inferrule*[left=\orvastype{UnOut_1}]{ 
			\Pi_3
			\\
			>_7 \defeq  >_6 \cup \{(d , w)\}
		}
		{ \Gamma,  d : W , y: \recur{\oc(\lin \vassend{ (\recur{\oc {T}} )}{S}) }  \stpre{>_7}{d}  \vasres{wv}{ \vasout{y}{w}{ \vasin{\lin}{v}{e}{P_3 } } }  }

		\and

		\Pi_5 =
		\inferrule*[left=\orvastype{UnIn_2}]{ 
			\Pi_4
			\\
			>_8 \defeq >_7\cup (\gtrdot^s_7)
			\\
			\notclient{W} \land \server{W}
		}
		{\Gamma , s: \recur{\wn W} , y:  \recur{\oc (\lin \vassend{ (\recur{\oc {T}} )}{S} ) }  \stpre{>_8}{s}  \vasin{\un}{s}{d}{ \vasres{wv}{ (\vasout{y}{w}{ \vasin{\lin}{v}{e}{P_3 } }) } }  }

		\and 

		\inferrule*[left=\orvastype{LinOut_4}]{ 
			\Pi_5
			\\ 
			\Gamma , u:V \stpre{>_9}{u}  P_4 
		       \\
		       >_{10} \defeq  >_8  \cup >_9 \cup \gtrdot^u_8 \cup \{(u,t)\}
		}
		{ \Gamma , u: \lin \vassend{ \recur{\oc {W}}}{V} , y:  \recur{\oc (\lin \vassend{ (\recur{\oc {T}} )}{S} ) }  \stpre{>_{10}}{u}  Q_2 }
	\end{mathpar}

\begin{figure}[!t]
			\[
		\begin{aligned} 
			\begin{tikzpicture}
			\node (startd) {$d$};
			\node [left of=startd] {$>_6 =$};
			\node (nextGamma) [right of= startd] {$\Gamma$};
			\node (nextv) [above of= nextGamma] {$v$};
			\node (nexte) [left of= nextv] {$e$};
			\node (nextP3) [below of= nextGamma] {$P_3$};
			\node (nexty) [left of= nextP3] {$y$};
			\draw[->] (startd) -- (nextv);
			\draw[->] (startd) -- (nexty);
			\draw[->] (startd) -- (nexte);
			\draw[->] (startd) -- (nextGamma);
			\draw[dotted, ->]  (nextP3)-- (nextGamma);
			\draw[->] (startd) -- (nextP3);
			\end{tikzpicture}
			\hspace{1cm} 
			\begin{tikzpicture}
				\node (startd) {$d$};
							\node [left of=startd] {$>_7 =$};
				\node (nextGamma) [right of= startd] {$\Gamma$};
				\node (nextv) [above of= nextGamma] {$v$};
				\node (nexte) [left of= nextv] {$e$};
				\node (nextP3) [below of= nextGamma] {$P_3$};
				\node (nexty) [below of= nextP3] {$y$};
				\node (nextw) [left of= nexty] {$w$};
				\draw[->] (startd) -- (nextv);
				\draw[->] (startd) -- (nexty);
				\draw[->] (startd) -- (nextw);
				\draw[->] (startd) -- (nexte);
				\draw[->] (startd) -- (nextGamma);
				\draw[dotted, ->]  (nextP3)-- (nextGamma);
				\draw[->] (startd) -- (nextP3);
			\end{tikzpicture}
			\hspace{1cm}  
			\begin{tikzpicture}
				\node (starts) {$s$};
							\node [left of=starts] {$>_8 =$};
				\node (startd) [right of= starts] {$d$};
				\node (nextGamma) [right of= startd] {$\Gamma$};
				\node (nextv) [above of= nextGamma] {$v$};
				\node (nexte) [left of= nextv] {$e$};
				\node (nextP3) [below of= nextGamma] {$P_3$};
				\node (nexty) [below of= nextP3] {$y$};
				\node (nextw) [left of= nexty] {$w$};
				\draw[->] (starts) -- (startd);
				\draw[->] (startd) -- (nextv);
				\draw[->] (startd) -- (nexty);
				\draw[->] (startd) -- (nextw);
				\draw[->] (startd) -- (nexte);
				\draw[->] (startd) -- (nextGamma);
				\draw[dotted, ->]  (nextP3)-- (nextGamma);
				\draw[->] (startd) -- (nextP3);
			\end{tikzpicture}
			\\ 
			\begin{tikzpicture}
				\node (startu) {$u$};
							\node [left of=startu] {$>_9 =$};
				\node (nextGamma) [right of= startd] {$\Gamma$};
				\node (nextP4) [below of= nextGamma] {$P_4$};
				\draw[->] (startu) -- (nextGamma);
				\draw[dotted, ->]  (nextP4)-- (nextGamma);
				\draw[->] (startu) -- (nextP4);
			\end{tikzpicture}
			\hspace{1cm} 
			\begin{tikzpicture}
				\node (startu) {$u$};
				\node [left of=startu] {$>_{10} =$};
				\node (starts) [right of= startu]  {$s$};
				\node (nextt) [above of= starts] {$t$};
				\node (startd) [right of= starts] {$d$};
				\node (nextP3) [right of= startd] {$P_3$};
				\node (nextGamma) [below of= nextP3] {$\Gamma$};
				\node (nexte) [above of= nextP3] {$e$};
				\node (nextP4) [below of= starts] {$P_4$};
				\node (nextw) [above of= nexte] {$w$};
				\node (nextv) [left of= nextw] {$v$};
				\node (nexty) [left of= nextv] {$y$};
				\draw[->] (startu) -- (starts);
				\draw[->] (starts) -- (startd);
				\draw[->] (startd) -- (nextv);
				\draw[->] (startd) -- (nexty);
				\draw[->] (startd) -- (nextw);
				\draw[->] (startd) -- (nexte);
				\draw[->] (startd) -- (nextGamma);
				\draw[dotted, ->]  (nextP3)-- (nextGamma);
				\draw[dotted, ->]  (nextP4)-- (nextGamma);
				\draw[->] (startd) -- (nextP3);
				\draw[->] (startu) -- (nextP4);
				\draw[->] (startu) -- (nextt);
			\end{tikzpicture}
		\end{aligned}
	\]
	\caption{Partial orders for process $Q_2$ from \Cref{ch5e:hardlemmas}.\label{ch5f:hardlemmass}}
\end{figure}

These partial orders are given in \Cref{ch5f:hardlemmass}.
	Finally, we obtain the partial order $>_{11}$ for $P$ is obtained as follows and shown in \Cref{ch5f:hardlemmasss}.

	\begin{mathpar}
\inferrule*[left=\orvastype{Par_2}]{ 
			\Gamma , x: \recur{\wn (\lin \vassend{ (\recur{\oc {T}} )}{S})}  \stpre{>_5}{x} Q_1
			\\
			\Gamma , u: \lin \vassend{ \recur{\oc {W}}}{V} , y:  \recur{\oc (\lin \vassend{ (\recur{\oc {T}} )}{S} ) }  \stpre{>_{10}}{u} Q_2
			\\
			>_{11} \defeq >_5  \cup >_{10}  \cup \gtrdot^u_5
		}
		{ \Gamma , u: \lin \vassend{ \recur{\oc {W}}}{V} \stpre{>_{11}}{u} 
		P}
	\end{mathpar}

\begin{figure}[!t]
	\[
			\begin{tikzpicture}
			\node (name) [left of = startu]{$>_{11} ~= $};
				\node (startu) {$u$};
				\node (starts) [right of= startu]  {$s$};
				\node (nextt) [above of= starts] {$t$};
				\node (startd) [right of= starts] {$d$};
				\node (nextP3) [right of= startd] {$P_3$};
				\node (dummy) [below of= nextP3] {$ $};
				\node (nextGamma) [right of= dummy] {$\Gamma$};
				\node (nexte) [above of= nextP3] {$e$};
				\node (nextP4) [below of= starts] {$P_4$};
				\node (nextw) [above of= nexte] {$w$};
				\node (nextv) [left of= nextw] {$v$};
				\node (nexty) [left of= nextv] {$y$};
				\node (dummy2) [below of= nextP4] {$ $};
				\node (startx) [below of= dummy2] {$x$};
				\node (startz) [right of= startx] {$z$};
				\node (startb) [right of= startz] {$b$};
				\node (nextc) [below of= startb] {$c$};
				\node (nextP1) [above of= startb] {$P_1$};
				\node (nexta) [below of= startz] {$a$};
				\node (nextP2) [above of= startz] {$P_2$};
				\draw[->] (startu) -- (startx);
				\draw[->] (startx) -- (startz);
				\draw[->] (startz) -- (startb);
				\draw[->] (startz) -- (nexta);
				\draw[->] (startz) -- (nextP2);
				\draw[->] (startb) -- (nextc);
				\draw[->] (startb) -- (nextGamma);
				\draw[->] (startb) -- (nextP1);
				\draw[dotted, ->]  (nextP2)-- (nextGamma);
				\draw[dotted, ->]  (nextP1)-- (nextGamma);
				\draw[->] (startu) -- (starts);
				\draw[->] (starts) -- (startd);
				\draw[->] (startd) -- (nextv);
				\draw[->] (startd) -- (nexty);
				\draw[->] (startd) -- (nextw);
				\draw[->] (startd) -- (nexte);
				\draw[->] (startd) -- (nextGamma);
				\draw[dotted, ->]  (nextP3)-- (nextGamma);
				\draw[dotted, ->]  (nextP4)-- (nextGamma);
				\draw[->] (startd) -- (nextP3);
				\draw[->] (startu) -- (nextP4);
				\draw[->] (startu) -- (nextt);
			\end{tikzpicture}
	\]
	\caption{Partial order for process $P$ from \Cref{ch5e:hardlemmas}.\label{ch5f:hardlemmasss}}
\end{figure}

	This example shows that more complex structures can be formed that those shown in \Cref{ch5ex:build_order}. For example: both $d$ and $b$ have no relation to each other; however, they both must be larger than $\Gamma$ and also smaller then $u$. Furthermore, the dotted lines emerging from $P_1$ to $P_4$ show that this diamond/lattice-like shape can occur multiple times within the structure.

It is easy to see that the partial order in \Cref{ch5f:hardlemmasss} is acyclic. 
However, this is not the ordered structure we should focus on: to show acyclicity of $P$ with respect to $>_{11}$, i.e., $\preserves{>_{11}}{P}$ (\Cref{ch5def:nocycle}), we should actually consider the closure $>^*_{11}$, as given by \Cref{ch5def:closure}.
	Indeed, checking acyclicity ultimately depends on this closure, which can be 
non-trivial: we should not only show that there are no cycles in  $>_{11}$ but also through connected compatible channels. 
By the relation between these connected names  and the flattening of $>_{11}$ with respect to $P$ (\Cref{ch5prop:flat_comp}) the extension of $>_{11}$ with respect to these connected names of is of great importance. 

To obtain $>^*_{11}$, we first determine  the set of connected names of $P$ (\Cref{ch5def:compN}):
	\[
	\begin{aligned}
		\varcompat{P} &  = \varcompat{\vasres{xy}{( \vaspara{ Q_1 }{ Q_2 } )}}
		\\
		&= \{ (x,y) , (y,x) \} \cup \varcompat{Q_1} \cup \varcompat{Q_2} 
		\ \cup
		\\
		& \qquad
		\{ (f, g) \mid \compatiblecc{ \cc{x}{ \vasres{xy}{( \vaspara{ Q_1 }{ Q_2 } )}  }}{\cc{ y}{\vasres{xy}{( \vaspara{ Q_1 }{ Q_2 } )}}}{f}{g} \}
		\\
		&= \{ (x,y) , (y,x) \} 
		\cup 
		\varcompat{
			\vasin{\un}{x}{z}{ \vasres{ab}{ \vasout{z}{a}{( \vaspara{ \vasin{\un}{b}{c}{P_1} }{ P_2 } )} }}
		} 
		\ \cup
		\\
		& \qquad
		\varcompat{
			\vasres{ts}\vasout{u}{t}{( \vaspara{	\vasin{\un}{s}{d}{ \vasres{wv}{ \vasout{y}{w}{ \vasin{\lin}{v}{e}{P_3} } } } }{ P_4 } )}
		} 
		\ \cup
		\\
		& \qquad
		\{ (f, g) \mid \compatiblecc{ \cc{x}{ Q_1 } }{\cc{ y}{ Q_2}}{f}{g} \}
		\\
		&= \{ (x,y) , (y,x) \} 
		\cup 
		\varcompat{
			\vasres{ab}{ \vasout{z}{a}{( \vaspara{ \vasin{\un}{b}{c}{P_1} }{ P_2 } )} }
		} 
		\ \cup 
		\\
		& \qquad
		\{ (t,s) , (s,t) \} \cup \varcompat{\vasin{\un}{s}{d}{ \vasres{wv}{ \vasout{y}{w}{ \vasin{\lin}{v}{e}{P_3} } } }} \cup \varcompat{ P_4 }
		\ 
		\\
		& \qquad
		\cup \{ (f, g) \mid \compatiblecc{ (z, (b, \emptyset):: \emptyset):: \emptyset }{ (v , (e,\emptyset )::\emptyset)::\emptyset  }{f}{g} \}
		\\
		&= \{ (x,y) , (y,x) , (t,s) , (s,t) \} 
		\cup 
		\{ (a,b) , (b,a) \} \cup \varcompat{\vasin{\un}{b}{c}{P_1}} \cup 
		\\
		& \qquad \varcompat{P_2}
		\ \cup 
		\varcompat{\vasin{\un}{s}{d}{ \vasres{wv}{ \vasout{y}{w}{ \vasin{\lin}{v}{e}{P_3} } } }} \cup \varcompat{ P_4 }
		\ \cup
		\\
		& \qquad
		\{ (f, g) \mid \compatiblecc{ (z, (b, \emptyset))}{ (v , (e,\emptyset ))  }{f}{g} \}
		\\
		&= \{ (x,y) , (y,x) , (t,s) , (s,t), (a,b) , (b,a) \} 
		\cup 
		\varcompat{P_1} \cup \varcompat{P_2}
		\ \cup 
		\\
		& \qquad
		\varcompat{ \vasres{wv}{ \vasout{y}{w}{ \vasin{\lin}{v}{e}{P_3} } } } \cup \varcompat{ P_4 }
		\ \cup
		\\
		& \qquad
		\{(z,v), (v,z) , (b,e), (e,b) \} 
		\\
		&= \{ (x,y) , (y,x) , (t,s) , (s,t), (a,b) , (b,a) , (z,v), (v,z) , (b,e), (e,b) \} 
		\cup 
		\\
		& \qquad
		\varcompat{P_1} \cup \varcompat{P_2}
		\ \cup \{ ( w, v) , ( v, w) \} \cup  \varcompat{  \vasin{\lin}{v}{e}{P_3} }
		\\
		& \qquad \cup \varcompat{ P_4 }
		\\
		&= \{ (x,y) , (y,x) , (t,s) , (s,t), (a,b) , (b,a) , ( w, v) , ( v, w) , (z,v), (v,z) , (b,e),  
		\ \cup 
		\\
		& \qquad (e,b)\}
		  \varcompat{P_1} \cup \varcompat{P_2} \cup \varcompat{ P_3 } \cup \varcompat{ P_4 }
		\\
	\end{aligned}
	\]

	Here again, for the sake of readability, we have done some simplifications. For example, we have assumed that $\cc{x}{ \vasres{ab}{ \vasout{z}{a}{( \vaspara{ \vasin{\un}{b}{c}{P_1} }{ P_2 } )} } } = \emptyset$ and made similar assumptions for all  occurrences within $\{ (f, g) \mid \compatiblecc{ (z, (b, \emptyset):: \emptyset):: \emptyset }{ (v , (e,\emptyset )::\emptyset)::\emptyset  }{f}{g} \}$. Strictly speaking, the structure of $P_1, \ldots, P_4$ would determine the shape of the diagram. 
	
	\begin{figure}[!t]
	\[
			\begin{tikzpicture}
			\node (name) [left of= startu] {${>^*_{11}} = $};
				\node (startu) {$u$};
				\node (starts) [right of= startu]  {$s$};
				\node (nextt) [above of= starts] {$t$};
				\node (startd) [right of= starts] {$d$};
				\node (nextP3) [right of= startd] {$P_3$};
				\node (dummy) [below of= nextP3] {$ $};
				\node (nextGamma) [right of= dummy] {$\Gamma$};
				\node (nexte) [above of= nextP3] {$e$};
				\node (nextP4) [below of= starts] {$P_4$};
				\node (nextw) [above of= nexte] {$w$};
				\node (nextv) [left of= nextw] {$v$};
				\node (nexty) [left of= nextv] {$y$};
				\node (dummy2) [below of= nextP4] {$ $};
				\node (startx) [below of= dummy2] {$x$};
				\node (startz) [right of= startx] {$z$};
				\node (startb) [right of= startz] {$b$};
				\node (nextc) [below of= startb] {$c$};
				\node (nextP1) [above of= startb] {$P_1$};
				\node (nexta) [below of= startz] {$a$};
				\node (nextP2) [above of= startz] {$P_2$};
				\draw[->] (startu) -- (startx);
				\draw[->] (startx) -- (startz);
				\draw[->] (startz) -- (startb);
				\draw[->] (startz) -- (nexta);
				\draw[->] (startz) -- (nextP2);
				\draw[->] (startb) -- (nextc);
				\draw[->] (startb) -- (nextGamma);
				\draw[->] (startb) -- (nextP1);
				\draw[dotted, ->]  (nextP2)-- (nextGamma);
				\draw[dotted, ->]  (nextP1)-- (nextGamma);
				\draw[->] (startu) -- (starts);
				\draw[->] (starts) -- (startd);
				\draw[->] (startd) -- (nextv);
				\draw[->] (startd) -- (nexty);
				\draw[->] (startd) -- (nextw);
				\draw[->] (startd) -- (nexte);
				\draw[->] (startd) -- (nextGamma);
				\draw[dotted, ->]  (nextP3)-- (nextGamma);
				\draw[dotted, ->]  (nextP4)-- (nextGamma);
				\draw[->] (startd) -- (nextP3);
				\draw[->] (startu) -- (nextP4);
				\draw[->] (startu) -- (nextt);
				\draw[<->,thick,red] (startx) edge [bend right = 20] (nexty);
				\draw[<->,thick,red] (starts) edge [bend right = 10] (nextt);
				\draw[<->,thick,red] (startb) edge [bend right = -10] (nexta);
				\draw[<->,thick,red] (nextw) edge [bend right = 20] (nextv);
				\draw[<->,thick,red] (startz) edge [bend right = 20] (nextv);
				\draw[<->,thick,red] (startb) edge [bend right = 70] (nexte);
			\end{tikzpicture}
	\]
		\caption{Extended partial order for process $P$ from \Cref{ch5e:hardlemmas}.\label{ch5f:vhardlemma}}
\end{figure}


\Cref{ch5f:vhardlemma} depicts the extension of $>_{11}$ with the pairs on connected names, depicted with red arrows.
	Showing that there are no cycles within $>^*_{11}$ is non-trivial. 
	As we will see, treating the compositionality of parallel composition can be complex. 
	By virtue of connected names we may now jump between unrelated branches of the diagram, and so we need to ensure that there is no path back to where we began. 
	
	To see this, consider the case of $y$: it is not greater than any element, but is it connected to $x$ (i.e., $(x,y) \in \varcompat{P}$). Because we also have $x>_{11} z$ with $(z,v) \in \varcompat{P}$, by \Cref{ch5def:closure} we obtain $y {>^*_{11}} v$. 
	This relation between $y$ and $v$ is not evident purely from $>_{11}$ and arises from the parallel composition of $Q_1$ and $Q_2$. Still, it is crucial to see that even if the added pairs (in red) entail new bidirectional connections between names, no cycle is induced; intuitively, this is because the existing connections (in black) are kept uni-directional and the operation $>_{11}^*$ requires at least one step in $>_{11}$. That is, the extension with compatible names does not break asymmetry of the given strict partial order(s). 
	
	We close this example by observing that every pair of names $(x,y)$ added via compatible names satisfy three useful properties, related to acyclicity (and induced by typing):
	\begin{enumerate}
		\item At least one of them is minimal in the ordering. That is, one or both of them are ``leaves'' in the tree induced by the partial ordering. 
		\item Both $x$ and $y$ originate in different ``branches'' of the tree. That is, there is a successor (such as $u$) is that is greater than both of them. 
		\item There are no ``crisscross connections'' between different branches. In our example, we have connections between $x$ and $y$ and between $v$ and $z$; a crisscross connection would be formed by $x$ and $v$, and by $y$ and $z$. 
	\end{enumerate}
\end{example}

\begin{example}{}
	Another important observation is that for every pair of connected channels, at least one of them is minimal in that there is no element smaller then them in the partial order. This alone does not enforce acyciclity take for example the following two \spo's:
	\[
		\begin{tikzpicture}
			\node (name) [left of= startu] {${>^*_{a}} = $};
				\node (startu) {$u$};
				\node (dummy) [right of= startu]  {};
				\node (nexta) [above of= dummy]  {$a$};
				\node (nextb) [right of= nexta] {$b$};
				\node (nextc) [below of= dummy] {$c$};
				\node (nextd) [right of= nextc] {$d$};
				\draw[->] (startu) -- (nexta);
				\draw[->] (nexta) -- (nextb);
				\draw[->] (startu) -- (nextc);
				\draw[->] (nextc) -- (nextd);
				\draw[<->,thick,red] (nexta) edge [bend right = 50] (nextb);
				\draw[<->,thick,red] (nextc) edge [bend right = -50] (nextd);
		\end{tikzpicture}
		\hspace{2cm}
		\begin{tikzpicture}
		\node (name) [left of= startu] {${>^*_{b}} = $};
			\node (startu) {$u$};
			\node (dummy) [right of= startu]  {};
			\node (nexta) [above of= dummy]  {$a$};
			\node (nextb) [right of= nexta] {$b$};
			\node (nextc) [below of= dummy] {$c$};
			\node (nextd) [right of= nextc] {$d$};
			\draw[->] (startu) -- (nexta);
			\draw[->] (nexta) -- (nextb);
			\draw[->] (startu) -- (nextc);
			\draw[->] (nextc) -- (nextd);
			\draw[<->,thick,red] (nexta) edge [bend right = 10] (nextd);
			\draw[<->,thick,red] (nextc) edge [bend right = 10] (nextb);
		\end{tikzpicture}
\]

The fist example is excluded as we also have that connected channels never appear in the same branch of an \spo, structualy connected channels occur via restriction or via compatible names and the rules of \Cref{ch5fig:orderedtypes} ensure this. However cycles of the second form are much more delicate, a rought intuition is that this structure can only be formed if there is a paralel composition on two channles with a with one of the connected channels being from restriction and the other comming from connected channels, however this would require these two processes to be in paralled sharing two channels rather then one which \dilllang ensures. 
\end{example}

We may now move to ensure that the composition of two acyclic strict partial orders (associated to specific connected channels) leads to another acyclic strict partial order on the pairs of connected channels. 
Recall 
that the predicate $\compatiblecc{A}{B}{x}{y}$ has been given in \Cref{ch5def:compatible}.

\begin{lemma}\label{ch5prop:bigproofforonething}
Assume 	$P_0,P,Q \in \dilllang$ such that 
\begin{itemize}
	\item 
	$z:V , (\Gamma' \setminus u \vaslinunsplit \Delta_1) \stpre{>_1}{z} P$
	\item 	
	$v:\dual{V} , (\Gamma' \vaslinunsplit \Delta) \stpre{>_2}{u} Q$
	\item $\preserves{>_1}{P}$
	\item $\preserves{>_2}{Q}$
	\item $P_0 = \res {z v}( P \pp Q )$
	\item $\Gamma' \vaslinunsplit \Delta_1, \Delta \stpre{>}{u} P_0$, where $> = >_1  \cup >_2  \cup \,\gtrdot^u_1$
\end{itemize}
Furthermore, let $
C = \{ (a, b)\,|\, \compatiblecc{ \cc{z}{ P_0}}{\cc{ v}{P_0}}{a}{b} \}
$.
 Then $\acyclic{>}{P_0}{x}{y}$, for all 	$(x,y) \in  C$.
 \end{lemma}

\begin{proof}
By induction, first on the structure of $P$ and then on $Q$. 
	We have six main cases (numbered (1) to (6)), each with multiple sub-cases.
	Among these, the case of linear output $P = \res{ws}\vasout{t}{s}{(P' \pp P'')}$ (case \ref{ch5proof:cutacyclic6}) is the most technically involved. 
\begin{enumerate}
	\item {\bf Case $P = \nil$}. 
	
	Then $\cc{z}{ \res{z v}(\nil \pp Q)} = \emptyset $ and $ \{ (a, b)\,|\, \compatiblecc{ \emptyset }{\cc{ v}{\res{z v}(\nil \pp Q)}}{a}{b} \} = \emptyset$, and the result follows trivially.
	
	\item {\bf Case $P = \vasout{w}{s}{P'}$ (unrestricted output).}
	
	 Then,  let $(\Gamma'' , w:\recur{\oc T}  , s:T)  = z:V , (\Gamma' \setminus u \vaslinunsplit \Delta_1)$. We consider two possibilities:
		\[
				\inferrule*[left=\orvastype{UnOut_2}]{ 
				\Gamma'' , w:\recur{\oc T}  , s:T \stpre{{>_1}}{u} {{P'}} 
				\\
				u \not = w
				\\ 
				\contun{T}
			}
			{ \Gamma'' ,  w: \recur{\oc T} , s:T  \stpre{{>_1}}{u} \ov w\out s.P'  }
		\]
		\[
			\small
			\inferrule*[left=\orvastype{UnOut_3}]{ 
				\Gamma'' , w:\recur{\oc T}   \stpre{>_{(1,1)}}{u} {{P'}} 
				\\
				\neg \contun{T}
				\\
				{>_1} \defeq >_{(1,1)}  \cup \{(u , s)\}
				\\
				u \not = w
			}
			{ \Gamma'' ,  w: \recur{\oc T} , s:T  \stpre{{>_1}}{u} \ov w\out s.P'  }
		\]

		Based on these two rules, we have:
	
		\begin{enumerate}

			\item The sub-case $z = w$ is not possible, because none of the rules allow us to have $\stpre{>}{w} $.

			\item The sub-case $z \not = w$ is possible, and so we distinguish between which of the two rules was 
			applied:
			
			\begin{enumerate}

			\item\label{ch5proof:cutacyclic1} The rule is $\orvastype{UnOut_2}$.
			
			Notice that $\Gamma' \vaslinunsplit \Delta_1, \Delta \stpre{>}{u} P'_0 $ where $P'_0 = \res{z v}( P' \pp Q )$, $> = >_1  \cup >_2  \cup \gtrdot^u_1$ and $\preserves{>_1}{P'}$. Hence by applying the induction hypothesis, we obtain  $\acyclic{>}{P'_0}{x}{y}$ for all 
				$(x,y) \in  \{ (a, b) \mid  \compatiblecc{ \cc{z}{ P'_0 }}{\cc{ v}{ P'_0 }}{a}{b} \} $.

				Let us consider the following sets constructed by exploiting the fact that $\cc{z}{ \ov w\out s.P' } =  \cc{z}{ P' }$:
				
				\begin{equation}\label{ch5eq:form1}
					\hspace{-2cm}
					\small
					\begin{aligned}
						\{ (a, b) \mid \compatiblecc{ \cc{z}{ P'_0 }}{\cc{ v}{ P'_0 }}{a}{b} \} &= \{ (a, b) \mid  \compatiblecc{ \cc{z}{ P' }}{\cc{ v}{ Q }}{a}{b} \}\\
					\end{aligned}
				\end{equation}

				\begin{equation}\label{ch5eq:form2}
					\hspace{-2cm}
					\small
					\begin{aligned}
						\{ (a, b)\mid  \compatiblecc{ \cc{z}{ P_0 }}{\cc{ v}{ P_0 }}{a}{b} \} 
						&= \{ (a, b) \mid \compatiblecc{ \cc{z}{ P }}{\cc{ v}{ Q }}{a}{b}\}	\\
						&= \{ (a, b)\mid  \compatiblecc{ \cc{z}{ P' }}{\cc{ v}{ Q }}{a}{b}\}	\\
					\end{aligned}
				\end{equation}

				As the sets defined by \eqref{ch5eq:form1} and \eqref{ch5eq:form2} are equal, then
				 $\acyclic{>}{ P_0 }{x}{y}$ for all $(x,y) \in  \{ (a, b)\,|\, \compatiblecc{ \cc{z}{ P_0 }}{\cc{ v}{ P_0 }}{a}{b} \} $, and the result follows.

				\item The rule is $\orvastype{UnOut_3}$. 
				
				This case follows similarly to \ref{ch5proof:cutacyclic1} with two key differences. First, rather than $\Gamma' \vaslinunsplit \Delta_1, \Delta \stpre{>}{u} \res{z v}( P' \pp Q )$ we have $\Gamma' \vaslinunsplit (\Delta_1, \Delta  \setminus s:T) \stpre{>'}{u} \res{z v}( P' \pp Q )$ where $>' = >_{(1,1)}  \cup >_2  \cup \gtrdot^u_{(1,1)}$. Second, because typing ensures that $s$ does not occur within $P'$, then it is easy to show that the added relation $(u,s)$ does not affect $\acyclic{>}{\res{z v}( P' \pp Q )}{x}{y}$.

			\end{enumerate}

		\end{enumerate}
	%
	\item\label{ch5proof:cutacyclic6} {\bf Case $P = \res{ws}\vasout{t}{s}{(P' \pp P'')}$ (linear output). }
	
	 This is the most involved case.

	  Let $ \Gamma'' \vaslinunsplit \Delta_1', \Delta_2' , t: \lin \vassend{T}{S}  = z:V , (\Gamma' \setminus u \vaslinunsplit \Delta_1)$. Then, either one of the rules 
			\[
			\inferrule*[left=\orvastype{LinOut_3}]{ 
				\Gamma'' \setminus u \vaslinunsplit \Delta_1', w:  \dual{T} \stpre{>_{(1,1)}}{w} P'
				\\ 
				\Gamma'' \vaslinunsplit \Delta'  , t:S \stpre{>_{(1,2)}}{u} P''
				\\
							{>_1} \defeq >_{(1,1)}  \cup >_{(1,2)}  \cup \gtrdot^u_1\cup \{(u,s)\}
				\\
				u \not = t
			}
			{ \Gamma'' \vaslinunsplit \Delta_1', \Delta_2' , t: \lin \vassend{T}{S}  \stpre{{>_1}}{u}  \res{ws}\vasout{t}{s}{(P' \pp P'')}  }
		\] 
		or
		\[
			\inferrule*[left=\orvastype{LinOut_4}]{ 
				\Gamma'' \vaslinunsplit \Delta_1', w:  \dual{T} \stpre{>_{(1,1)}}{w} P'
				\\ 
				\Gamma'' \vaslinunsplit \Delta_2'  , t:S \stpre{>_{(1,2)}}{t} P''
				\\
				{>_1} \defeq  >_{(1,1)}  \cup >_{(1,2)}  \cup \gtrdot^t_1\cup \{(u,s)\}
			}
			{ \Gamma'' \vaslinunsplit \Delta_1', \Delta_2' , t: \lin \vassend{T}{S}  \stpre{{>_1}}{t}  \res{ws}\vasout{t}{s}{(P' \pp P'')}  }
		\]
		
		seems to apply. Note that $\orvastype{LinOut_3}$ does not respect $\stpre{{>_1}}{t}$, we can disregard this sub-case. Hence, we only consider the sub-case when  $\orvastype{LinOut_4}$ is applied. There are two possibilities: $z = t$ and $z \neq t$. The case $z\neq t$ is similar to \Cref{ch5proof:cutacyclic1}, therefore, we will only  expand the proof for the case $z=t$:


			
			 Then let us consider the structure of $Q$. We have seven cases (denoted \ref{ch5item:start} -- \ref{ch5item:end}).
			 
			\begin{enumerate}

				\item \label{ch5item:start} {\bf When $Q= \nil$.}
				
				 Then $\res{z v}( P \pp Q ) \not \in \dilllang$ and typing is contradicted and the case does not apply.
		
				\item\label{ch5proof:cutacyclic2} {\bf When $Q = \vasout{w'}{s'}{Q'}$.}

					If $w' \not = v$ we consider both cases for the rule being applied and the proof follows analogously to \Cref{ch5proof:cutacyclic1} by applying the induction hypothesis.  If $w' = v$ then $\res{z v}( P \pp Q )$ would be not typable in $\dilllang$. Hence, this case can be disregarded.

				\item {\bf When $Q = \res{w's'}\vasout{t'}{s'}{Q'}$.} 
				
				 Then the case is analogous to \Cref{ch5proof:cutacyclic2} (above).

				\item {\bf  When $Q = \res{w's'}\vasout{t'}{s'}{(Q' \pp Q'')}$.} 
								
				 Then,  $ \Gamma''' \vaslinunsplit \Delta_1'', \Delta_2'' , t': \lin \vassend{T'}{S'} = v:\dual{V} , (\Gamma' \vaslinunsplit \Delta) $ and either one of the rules applies: 
				\[
					\inferrule*[left=\orvastype{LinOut_3}]{ 
						\Gamma''' \setminus u \vaslinunsplit \Delta_1'', s':  \dual{T'} \stpre{>_{(2,1)}}{s'} Q'
						\\\\
						\Gamma''' \vaslinunsplit \Delta''  , t':S' \stpre{>_{(2,2)}}{u} Q''
						\\
									>_2 \defeq >_{(2,1)}  \cup >_{(2,2)}  \cup \gtrdot^u_{(2,1)}\cup \{(u,w')\}
						\\
						u \not = t'
					}
					{ \Gamma''' \vaslinunsplit \Delta_1'', \Delta_2'' , t': \lin \vassend{T'}{S'}  \stpre{>_2}{u}  \res{w's'}\vasout{t'}{w'}{(Q' \pp Q'')}  }
				\]
				or
				\[
					\inferrule*[left=\orvastype{LinOut_4}]{ 
						\Gamma''' \vaslinunsplit \Delta_1'', s':  \dual{T'} \stpre{>_{(2,1)}}{s'} Q'
						\\\\
						\Gamma''' \vaslinunsplit \Delta_2''  , t':S' \stpre{>_{(2,2)}}{t'} Q''
						\\\\
						>_2 \defeq  >_{(2,1)}  \cup >_{(2,2)}  \cup \gtrdot^{t'}_{(2,1)} \cup \{(u,w')\}
					}
					{ \Gamma''' \vaslinunsplit \Delta_1'', \Delta_2'' , t': \lin \vassend{T'}{S'}  \stpre{>_2}{t'}  \res{w's'}\vasout{t'}{w'}{(Q' \pp Q'')}  }
				\]

				Note that  $t' = v$ then we run into a contradiction, as $\res{z v}( P \pp Q ) \not \in \dilllang$.
				Then, we only need to consider the case  $t' \not = v$.  
				
				 Let us assume w.l.o.g. that $v \in \dom{\Delta_1''}$ and consider the application of rule $\orvastype{LinOut_3}$. The case involving $\orvastype{LinOut_4}$ is similar.

					  When applying $\orvastype{LinOut_3}$ we first notice that we may construct the following derivation: 
						
						\[
							\Gamma'' \vaslinunsplit \Delta_1', \Delta_2' , t: \lin \vassend{T}{S} + \Gamma''' \vaslinunsplit \Delta_1'', s':  \dual{T'}  \stpre{>_3}{u} P'_0
						\]

						where $P'_0 = \res{z v}( P \pp Q' )$,  $>_3 = >_1  \cup >_{(2,1)}\,  \cup~ \gtrdot^u_1$ 
						and $\preserves{>_{(2,1)}}{Q'}$.
						Hence, by applying the induction hypothesis we obtain $\acyclic{>_3}{ P'_0 }{x}{y}$ for all 
						$(x,y) \in  \{ (a, b)\mid  \compatiblecc{ \cc{z}{ P'_0 }}{\cc{ v}{ P'_0 }}{a}{b} \} $.

						Now as we consider the following sets constructed making use that $$ \cc{v}{ \res{w's'}\vasout{t'}{w'}{(Q' \pp Q'')}   } = \cc{v}{ (Q' \pp Q'')   } = \cc{v}{ Q' } $$ as by typing $v \not \in \names{Q''}$:

						\begin{equation}\label{ch5eq:form3}
							\hspace{-2cm}
							\small
							\{ (a, b)\mid  \compatiblecc{ \cc{z}{ P'_0 }}{\cc{ v}{P'_0 }}{a}{b} \} = \{ (a, b)\mid  \compatiblecc{ \cc{z}{ P }}{\cc{ v}{ Q' }}{a}{b} \}
						\end{equation}

						\begin{equation}\label{ch5eq:form4}
							\hspace{-2cm}
							\small
							\begin{aligned}
								\{ (a, b) \mid  \compatiblecc{ \cc{z}{ P_0 }}{\cc{ v}{ P_0 }}{a}{b} \} 
								&= \{ (a, b)\mid  \compatiblecc{ \cc{z}{ P }}{\cc{ v}{ Q }}{a}{b}\}	\\
								&= \{ (a, b)\mid  \compatiblecc{ \cc{z}{ P }}{\cc{ v}{ Q' }}{a}{b}\}	\\
							\end{aligned}
						\end{equation}

						We need to show that  $\acyclic{>}{ P_0 }{x}{y}$ for all $(x,y) \in  \{ (a, b) \mid  \compatiblecc{ \cc{z}{P}}{\cc{ v}{Q'}}{a}{b} \} $, where $> = >_3 \cup >_{(2,2)}$. By expanding the definition, we have four cases:

						\begin{enumerate}

							\item { $x \not \pclosu{P_0} y$.}  
							
							First, because $\acyclic{>_3}{ P'_0 }{x}{y}$ we deduce $x \not >_{3,P'_0}^* y$. This in turn implies that if $x \pclosu{P_0} y$ were to hold it would imply that we must make use of $>_{(2,2)}$ in the seqence. However we can proove by induction on the length of the sequence $x \pclosu{P_0} y$ that this must not be the case and hence $x \not\pclosu{P_0} y$. If only one step is made then  in the \spo then $x >_{(2,2),P'_0}^* y$ however then this implies that both $x$ and $y$ occur in the domain of $\Gamma'''$; this in turn implies that  there is no $c$ (via \Cref{ch5prop:auxiliary}(14) and \Cref{ch5prop:ontleaf}) such that $x >_{(2,2),P'_0}^* c$, which contradicts $ x >_{(2,2)} y$. Applying the induction hypotheses we apply the same reasoning that any the step taken within $>_{(2,2)}$ must start from a channel that apears in the names of both $P$ and $Q$ hence must appear in $\Gamma'''$.
			
							\item  {$(x >_{P_0}^* c  \implies c \not  \pclosu{P_0} y) \land
							 (c \pclosu{P_0} x \implies y \not \pclosu{P_0}  c) $.}
							
							We only show that  $(x \pclosu{P_0} c  \implies c \not  >_{P_0}^* y) $ holds; the proof for the  right conjunct is similar. 
							
							Since 
							$\acyclic{>_3}{P'_0}{x}{y}$ then  $(x \, \pclosu{3,P_0'} \, c  \implies c \pclosu{3,{P'_0}}  y) $. {We will show inductively that if $ x {>^*}_{P_0} c$ then $ x \pclosu{3,P_0'}  c$ with  $c  \pclosu{P_0}  y$, which will lead to a contradiction}. 
							
							Following \Cref{ch5def:closure}, $ x  \pclosu{P_0} c$ can hold in the following 4 situations:
																					\begin{enumerate}
								
								\item\label{ch5proof:whyxnotcompy} {\bf If $x > c$.}
								
								 
								We have that $x \not \in \dom{\Gamma' \vaslinunsplit \Delta_1, \Delta}\setminus u $, {otherwise, by \Cref{ch5prop:ontleaf}, we would have a contradiction  with $x > c $}. Secondly as  $(x,y)\in\{ (a, b)\,|\, \compatiblecc{ \cc{z}{ P_0}}{\cc{ v}{P_0}}{a}{b} \}$ then $x\in \names{P_0}$ and hence $x\in \names{P'_0}$.
								We may then deduce that $x >_3 c$ with $c \in \names{P'_0}$ (notice $x \not >_{(2,2)} c$ by \Cref{ch5prop:auxiliary} (6)) and hence also $x \pclosu{3,P'_0} c$. 
								
								Next, we show that  $c\pclosu{P_0}y$  cannot hold. There are  four cases to be analyzed:
								
								\begin{myEnumerate}
									\item[(A.1)] Suppose $c > y$: Then we must have that $c >_3 y$ as $ c>_{(2,2)} $ implies that $c \in \dom{\Gamma'''}$ contradicting \Cref{ch5prop:ontleaf}. If $c >_3 y$ then $c \pclosu{3,P'_0} y$ contradicing $\acyclic{>_3}{P'_0}{x}{y}$. 
									
									
									\item[(A.2)]\label{ch5proof:casebigger} Suppose there is an $e$ such that $ c > e \land e >_{P_0}^* y  $: Then,  $e \not \in \dom{\Gamma' \vaslinunsplit \Delta_1, \Delta}\setminus u $, {otherwise,  by \Cref{ch5prop:auxiliary} (15), we would have a contradiction with} $e >_{P_0}^* y $. Let us consider the three remaining possibilities for $e$:
									
									\begin{itemize}
										\item Case $e = u$:  we would have  $c > u$, which contradicts the fact that   $u$ is  greatest element w.r.t. $>$, since $\stpre{>}{u}$.
									
										\item Case $e \in \bn{P'_0}$: {then as $c >e$ we also have by definition $c >_{P_0}^* e$}. Hence, as $x >_{P_0}^* c >_{P_0}^* e  \implies x >_{P_0}^* e$, we have $x >_{P_0}^* e  \implies e  >_{P_0}^* y$ but by the induction hypothesis on the length of $c >_{P_0}^* y$ we have $x >_{P_0}^* e  \implies e \not >_{P_0}^* y$.

										\item Case $e \in \bn{Q''} $:  then $c > e$, which  is a contradiction, as firstly $c >_3e$ cannot hold (\Cref{ch5prop:auxiliary} (6)).  Then, we must have that $c>_{(2,2)}e$ holds instead.  This requires that $c \in \dom{\Gamma'''}$ and {hence, by \Cref{ch5prop:ontleaf}, $c$ cannot be greater than $e$.}

									\end{itemize}

									\item[(A.3)]\label{ch5proof:casecomp} Suppose there is an $e$ such that $\compatible{ {P_0} }{c}{e} \land e >_{P_0}^* y$: Then $e \not \in \dom{\Gamma' \vaslinunsplit \Delta_1, \Delta}\setminus u $ as by \Cref{ch5prop:ontleaf} this would contradict $e >_{P_0}^* y $. Let us consider the three remaining possibilities:
									\begin{itemize}
										\item Case $e = u$: this case contradicts $\compatible{ {P_0} }{c}{u} $ as there does not exist a channel $c$ that is compatible with $u$.
									
										\item Case $e \in \bn{P'_0}$: then  $c >_{P_0}^* e$ which implies   $x >_{P_0}^* c \land  \compatible{ {P_0} }{c}{e}  \implies x >_{P_0}^* e$. Thus, we have $x >_{P_0}^* e  \implies e  >_{P_0}^* y$,  but by the induction hypothesis on the length of $c >_{P_0}^* y$ we have $x >_{P_0}^* e  \implies e \not >_{P_0}^* y$.

										\item Case  $e \in \bn{Q''}$:  then $c \in \dom{\Gamma'''}$ and hence contradicting $c >_{P_0}^* y $ as no element is smaller then $c$.

									\end{itemize}

									\item[(A.4)]  Suppose there is $e$ such that $c >_{P_0}^* e \land \compatible{ {P_0} }{e}{y}$. Then, there are three cases to consider:
									\begin{itemize}
										\item There is an $f$ such that  $c > f \land f >_{P_0}^* e \land \compatible{ {P_0} }{e}{y}$. Then we have $x >_{P_0}^* c > f$  and hence $ x >_{P_0}^* f$, we reson inductivly that $ x >_{P_0}^* f \implies f \not>_{P_0}^* e$.
										
										
										\item There is an $f$ such that  $\compatible{ {P_0} }{c}{f} \land f  >_{P_0}^* e \land \compatible{ {P_0} }{e}{y}$. Then we argue simulalyas we have $x >_{P_0}^* c \land \compatible{ {P_0} }{c}{f}$  and hence $ x >_{P_0}^* f$, we reson inductivly that $ x >_{P_0}^* f \implies f \not>_{P_0}^* e$.
										
										
										\item\label{ch5proof:caseaftercomp} There is no $f$ such that  $c > f \land f >_{P_0}^* e \land \compatible{ {P_0} }{e}{y}$ or $\compatible{ {P_0} }{c}{f} \land f  >_{P_0}^* e \land \compatible{ {P_0} }{e}{y}$. Hence $c >_{P_0}^* e \land \compatible{ {P_0} }{e}{y}$ may be considered as $c > e_1 \land \compatible{ {P_0} }{e_1}{e_2} \land  \cdots \land \compatible{ {P_0} }{e_n}{e}  \land \compatible{ {P_0} }{e}{y} $. That is,  $c>e_1$ and then there is a sequence of compatible channels that leads to $y$.
									\end{itemize}
									
									We now proceed to show \ref{ch5proof:caseaftercomp}:

									 First, note  that $e_i \not = u$ for $i \in \{1,\dots ,n\} $ and $e \not = u$ as this would contradict $\compatible{ {P_0} }{e_i}{e_{i+1}}$ and $\compatible{ {P_0} }{e}{y}$, respectively. Second, $c \not \in \dom{\Gamma' \vaslinunsplit \Delta_1, \Delta}\setminus u  \cup \bn{Q''} $ as this would contradict $c >_{P_0}^* e$, hence $c \in \bn{P'_0}$. We proceed by induction on $n$: 
						
						\noindent ({\bf Base case}) When $n= 0$.
						
						 Then, it must be the case that  $c > e \land \compatible{ {P_0} }{e}{y}$ and we consider $e$:

									\begin{itemize}
										\item $e \in \dom{\Gamma' \vaslinunsplit \Delta_1, \Delta}\setminus u  $: we must have that either $y \in \bn{P'_0}$ or $y \in \bn{Q''}$. If $y \in \bn{Q''}$ then this contradicts $(x,y) \in \{ (a, b) \mid  \compatiblecc{ \cc{z}{P}}{\cc{ v}{Q'}}{a}{b} \}$. If $y \in \bn{P'_0}$  then we have $c \pclosu{P'_0} e$ and  $c \pclosu{3,P'_0} e$. Also, we have  that $\compatible{ {P_0} }{e}{y} \implies \compatible{ {P'_0} }{e}{y}$ but by the induction hypothesis we have that {$c\not\pclosu{3,P_0'}y$}.
										
										\item $e \in \bn{P'_0}$: then  $y \in \names{P'_0} \cup \dom{\Gamma'' \vaslinunsplit \Delta_1', \Delta_2' , t: \lin \vassend{T}{S} + \Gamma''' \vaslinunsplit \Delta_1'', s':  \dual{T'}}$. Then,  we have that $c >_{P'_0}^* e$ and  $c \pclosu{3,P'_0} e$. Also,  $\compatible{ {P_0} }{e}{y} \implies \compatible{ {P'_0} }{e}{y}$ but by the induction hypothesis we have that { $c \not >_{3,P'_0}^* y$.}
										
										\item $e \in \bn{Q''}$ then as $c$ must be within the bound names of $P'_0$ but then we cannot have that $c >_{(2,2)} e $ hence we must have $ c {>_3} e $ which then implies that $e \not \in \bn{Q''}$.
										
									\end{itemize}

									\noindent ({\bf Inductive Step}) When $n \geq 1$.
									
									 We have that  $c >_{P_0}^* e_1 \land \compatible{P_0}{e_1}{e_2} \land \cdots \land \compatible{P_0}{e_{n-1}}{e_n} \land \compatible{ {P_0} }{e}{y}$ then we consider $e_1$:

									\begin{itemize}
										\item\label{ch5proof:whenitisifree1} $e_1 \in \dom{\Gamma' \vaslinunsplit \Delta_1, \Delta}\setminus u  $ we must have that either $e_2 \in \bn{P'_0}$ or $e_2 \in \bn{Q''}$.
										
										If $e_2 \in \bn{Q''}$ then as $P_0 = \res {z v}( P \pp \res{w's'}\vasout{t'}{w'}{(Q' \pp Q'')}  )$ and we have:
										
										$$
										 \begin{aligned}
										 	\compatible{ P_0 }{e_1}{e_2} 
										 	& = 
										 	\compatiblecc{ \cc{z}{ P }}{\cc{ v}{ Q' }}{e_1}{e_2} \lor 
											\\
											& \quad \compatible{ P }{e_1}{e_2} \lor \compatible{ Q }{e_1}{e_2} \\
										 	& = 
										 	\false \lor \false \lor \compatible{ Q }{e_1}{e_2}
										 \end{aligned}
										$$
										
										 with 
										 
										$$
										 \begin{aligned}
											\small
										 	\compatible{ Q }{e_1}{e_2} &= 
										 	\{ ( w', s' ) \} \lor \compatible{ Q' }{e_1}{e_2} \\
											& \quad \lor \compatible{ Q'' }{e_1}{e_2} \\
										 	& = \{ ( w', s' ) \}  \lor \false \lor \compatible{ Q'' }{e_1}{e_2} 
										 \end{aligned}
										$$
										
										As $e_1 \not \in \{ w', s' \}$ then we must have that $\compatible{ Q'' }{e_1}{e_2}$, which implies $n = 1$ then $e_2= e$. Thus,  we must have that $e_2 \in \bn{Q''}$ and hence all $e_i$, $e$ and $y$ must be bound names in $Q''$. However,  this contradicts that $(x,y) \in \{ (a, b) \mid \compatiblecc{ \cc{z}{ P }}{\cc{ v}{ Q' }}{a}{b}\}$. When $e_2 \in \bn{Q'}$ the case follows by the induction hypothesis.

										\item $e_1 \in \bn{P'_0}$ then the case follows by the induction hypothesis.

										\item $e_1 \in \bn{Q''}$ then the case follows analagously to that of \ref{ch5proof:whenitisifree1} when $e_2 \in \bn{Q''}$.

									\end{itemize}

								\end{myEnumerate}
								
								\item {\bf  If $ x >e \land e >_{P_0}^* c  $ for some $e$}.
								
								 Then,  both $x$ and $e$ are in the names of ${P'_0}$, which implies that  $x \pclosu{3,P'_0} e$. By the induction hypothesis we have that $e \pclosu{3,P'_0} c$ and hence $ x \pclosu{3,P'_0} c  $. We then apply induction on $ c  >_{P_0}^* y$ showing that $x$ and $y$ cannot be compatible, the argument of which follows from \Cref{ch5proof:whyxnotcompy}

								\item\label{ch5proof:whytheyarecompatible} {\bf  If $\compatible{ P_0 }{x}{e} \land e >_{P_0}^* c$ for some $e$.}
								
								 Then $x$ cannot be in the domain of $\dom{\Gamma' \vaslinunsplit \Delta_1, \Delta}\setminus u$ and similarly $x \not \in \bn{Q''}$ as this would contradict $(x,y) \in \{ (a, b)\mid \compatiblecc{ \cc{z}{ P }}{\cc{ v}{ Q' }}{a}{b}\}$. Therefore,  both $x$ and $e$ are in the bound names of $P'_0$, which implies that  $\compatible{ P'_0 }{x}{e}$. By the induction hypothesis it follows that  $e >_{P_0}^* c$, which implies $ e \pclosu{3,P'_0} c$. We then apply induction on $ c  >_{P_0}^* y$ showing that $x$ and $y$ cannot be compatible, the argument of which follows from~\Cref{ch5proof:whyxnotcompy}.

								\item {\bf If $x >_{P_0}^* e \land \compatible{ P_0 }{e}{c}$ for some $e$.}
								
			 By the induction hypothesis, if $x >_{P_0}^* e$ then $x \pclosu{3,P'_0} e$. Since $\compatible{ P_0 }{e}{c}$ and $c >_{P_0}^* y$ we can argue analogously to \Cref{ch5proof:whytheyarecompatible} that $\compatible{ P'_0 }{e}{c}$ and hence $x {{>_3}}_{P'_0}^* c$. We then apply induction on $ c  >_{P_0}^* y$ showing that $x$ and $y$ cannot be compatible, the argument of which follows from~\Cref{ch5proof:whyxnotcompy}.

							\end{enumerate}

							\item\label{ch5proof:casethatisusedtosimplify} {$ (c>_{P_0}^*x \land  y>_{P_0}^* d) \implies \neg\compatible{P_0}{c}{d}$.}
							
							As $\acyclic{>_3}{ P'_0 }{x}{y}$ then for $(x,y) \in C $ we have that $ (c\pclosu{3,P'_0}x \land  y\pclosu{3,P'_0} d) \implies \neg\compatible{P'_0}{c}{d}$.
							 Let us consider the following cases for $c$:

							\begin{enumerate}

								\item\label{ch5proof:extendedstuff1} When $ c \in \dom{\Gamma'' \vaslinunsplit \Delta_1', \Delta_2' , t: \lin \vassend{T}{S} + \Gamma''' \vaslinunsplit \Delta_1'', \Delta_2'' , t': \lin \vassend{T'}{S'} }$. 
								
								Then, there doesn't any channel smaller then  $c$ hence we don't have {$c>_{P_0}^*x$} and this case is not considered.
Similarly for $ c \in \dom{ \Delta''  , t':S'}$.

								\item When $c \in \bn{\res{z v}( P \pp Q' )} $.
								
								Then, we show inductively that if $ c >_{P_0}^* x$ then $ c \pclosu{3,P'_0} x$.  Following \Cref{ch5def:closure}, $ c >_{P_0}^* x$ can hold in the following 4 situations:
								
								\begin{myEnumerate}
									
									\item[(B.1)]\label{ch5proof:whyxnotcompy2} {\bf If $c > x$.}

									 Then,  $c >_{P'_0}^* x$ since $c \in \bn{P'_0} $ and $x \in \names{P'_0}$. As $c \not \in  \dom{\Gamma' \vaslinunsplit \Delta_1, \Delta}\setminus u \cup \names{Q''} $, then  $c \not >_{(2,2)} x $  which implies  $c >_3 x$ and  also $c \pclosu{3,P'_0} x$. As $y>_{P_0}^* d$ and  $(x,y) \in \{ (a, b) \mid  \compatiblecc{ \cc{z}{ P }}{\cc{ v}{ Q' }}{a}{b}\}$, it follows  that $y \not \in \dom{\Gamma' \vaslinunsplit \Delta_1, \Delta}\setminus u $ and we must have that $y \in \bn{P'_0}$.
									
	Next we show that if $e \in \bn{P'_0}$ and $e >_{P_0}^* d$ then $e \pclosu{3,P'_0} d$;  and if $y = e$ then $\neg\compatible{P_0}{c}{d}$:
									
									\begin{myEnumerate}
										\item Suppose $y > d$: Then as $y \in \bn{P'_0} $ we have $y \not >_{(2,2)} d $ and we may deduce that $y >_3 d$ and hence also $y \pclosu{3,P'_0}d$. By the induction hypothesis this implies that $\neg\compatible{P'_0}{c}{d}$ and as $c \not \in \names{Q''}$ we may deduce $\neg\compatible{Q''}{c}{d}$. Hence we obtain  $\neg\compatible{P_0}{c}{d}$.
										
										\item\label{ch5proof:casebigger2} Suppose there is an $e$ such that $ y > e \land e >_{P_0}^* d  $: Then,  $e \not \in \dom{\Gamma' \vaslinunsplit \Delta_1, \Delta}\setminus u $ as by \Cref{ch5prop:ontleaf} this would contradict $e >_{P_0}^* d $. Let us consider the three remaining possibilities for $e$:
										
										\begin{itemize}
											\item  $e = u$: this case is not possible, as we would have  $y > u$  and there is no channel larger than $u$ from $\stpre{>}{u}$.
											
											\item $e \in \bn{P'_0}$: then $y >_{P_0}^* e$ and $y \pclosu{3,P'_0}e$ and, by the induction hypothesis, $e \pclosu{3,P'_0}^* d$. Hence, $y \pclosu{3,P'_0} d$ and  $\neg\compatible{P'_0}{c}{d}$.
											As $c \not \in \names{Q''}$ we may deduce $\neg\compatible{Q''}{c}{d}$. Hence we obtain  $\neg\compatible{P_0}{c}{d}$.

											\item $e \in \bn{Q''} $: then $y > e$ is a contradiction as  this will require $y \in \dom{\Gamma'''}$ and hence {$y$} cannot be larger than $e$ by \Cref{ch5prop:ontleaf}.

										\end{itemize}

										\item\label{ch5proof:casecomp2} Suppose there is an $e$ such that $\compatible{ {P_0} }{y}{e} \land e >_{P_0}^* d$: Then $y , e \not \in \dom{\Gamma' \vaslinunsplit \Delta_1, \Delta}\setminus u $ as by \Cref{ch5prop:ontleaf} this would contradict $y >_{P_0}^* d $ and $e >_{P_0}^* d $ respectively. Let us consider the three remaining possibilities:
										\begin{itemize}
											\item $e = u$: this case contradicts $\compatible{ {P_0} }{y}{u} $ as there does not exist a channel $y$ that is compatible with $u$. 
										
											\item $e \in \bn{P'_0}$: then $\compatible{ {P'_0} }{y}{e}$ as $y$ is a bound name in $P'_0$, therefore, it  cannot occur in $Q''$. Thus,  there will be no channels compatible with $y$ in $Q''$. By the induction hypothesis, since  $e >_{P_0}^* d$ it follows that  $e \pclosu{3,P'_0} d$. Then, $y \pclosu{3,P'_0} d$ with $\neg\compatible{P'_0}{c}{d}$.
											As $c \not \in \names{Q''}$ we may deduce $\neg\compatible{Q''}{c}{d}$. Hence we obtain  $\neg\compatible{P_0}{c}{d}$.
											
											
											\item $e \in \bn{Q''}$: then  $e \in \dom{\Gamma'''}$ and hence contradicting $e >_{P_0}^* d $ as no element is smaller then $e$.

										\end{itemize}

										\item  $y >_{P_0}^* e \land \compatible{ {P_0} }{e}{d}$ then 
										firstly notice that there are three cases here:
										\begin{itemize}
											\item There is an $f$ such that  $y > f \land f >_{P_0}^* e \land \compatible{ {P_0} }{e}{d}$, in this case we may apply case \ref{ch5proof:casebigger2}. 
											
											\item There is an $f$ such that  $\compatible{ {P_0} }{y}{f} \land f  >_{P_0}^* e \land \compatible{ {P_0} }{e}{d}$, in this case we may apply case \ref{ch5proof:casecomp2}. 
											
											\item\label{ch5proof:caseaftercomp2} There is no $f$ such that  $y > f \land f >_{P_0}^* e \land \compatible{ {P_0} }{e}{d}$ or $\compatible{ {P_0} }{y}{f} \land f  >_{P_0}^* e \land \compatible{ {P_0} }{e}{d}$, hence $y >_{P_0}^* e \land \compatible{ {P_0} }{e}{d}$ may be considered as $y > e_1 \land \compatible{ {P_0} }{e_1}{e_2} \land  \cdots \land \compatible{ {P_0} }{e_n}{e}  \land \compatible{ {P_0} }{e}{d} $. That is that $y>e_1$ and then there is a sequence of compatible channels that leads to $d$.
										\end{itemize}
										
										We now proceed to show \ref{ch5proof:caseaftercomp2}.

										Note that  $e_i \not = u$ for $i \in \{1,\dots ,n\} $ and $e \not = u$ as this would contradict $\compatible{ {P_0} }{e_i}{e_{i+1}}$ and $\compatible{ {P_0} }{e}{d}$, respectively. Also,  $y \not \in \dom{\Gamma' \vaslinunsplit \Delta_1, \Delta}\setminus u  \cup \bn{Q''} $ as this would contradict $y >_{P_0}^* e$. Hence $y \in \bn{P'_0}$. 
										
				We proceed by induction on $n$: 
					
					\noindent {\bf (Base Case)} When $n= 0$.
					
					 We have $y > e \land \compatible{ {P_0} }{e}{d}$ then we consider $e$:

										\begin{itemize}
											\item $e \in \dom{\Gamma' \vaslinunsplit \Delta_1, \Delta}\setminus u  $ we must have that either $d \in \bn{P'_0}$ or $d \in \bn{Q''}$. If $d \in \bn{Q''}$ then $\neg\compatible{ {P_0} }{c}{d}$. If $d \in \bn{P'_0}$, first, notice that  $y >_{P'_0}^* e$ and  $y \pclosu{3,P'_0} d$; second, that $\compatible{ {P_0} }{e}{d} \implies \compatible{ {P'_0} }{e}{d}$.  By the induction hypothesis we have that $\neg\compatible{ {P'_0} }{c}{d}$ and hence $\neg\compatible{ {P_0} }{c}{d}$.


											\item $e \in \bn{P'_0}$ then  $d \in \names{P'_0} \cup \dom{\Gamma'' \vaslinunsplit \Delta_1', \Delta_2' , t: \lin \vassend{T}{S} + \Gamma''' \vaslinunsplit \Delta_1'', s':  \dual{T'}}$, which implies  that $y >_{P'_0}^* e$ and $y \pclosu{3,P'_0}^* e$.  Also, since  $\compatible{ {P_0} }{e}{d} $ and $e \in \bn{P'_0} $ then $d \in \names{P'_0}$. Thus,   $\compatible{ {P'_0} }{e}{d}$ but, by the induction hypothesis $\neg\compatible{ {P'_0} }{c}{d}$, and hence $\neg\compatible{ {P_0} }{c}{d}$.

											\item $e \in \bn{Q''}$ then as $y$ must be within the names of $P'_0$ as  $(x,y) \in \{ (a, b)| \compatiblecc{ \cc{z}{ P }}{\cc{ v}{ Q' }}{a}{b}\}$ but then we cannot have that $y >_{(2,2)} e $ as this would imply that $ y \in \dom{\Gamma' \vaslinunsplit \Delta_1, \Delta}\setminus u  $ and then $y \not >_{(2,2)} e $.  Thus,  we must have $ y >' e $ which then implies that $e \not \in \bn{Q''}$.
											 
										\end{itemize}

										\noindent ({\bf Inductive  Step}) When $n \geq 1$. 
										
										 We have that  $y >_{P_0}^* e_1 \land \compatible{P_0}{e_1}{e_2} \land \cdots \land \compatible{P_0}{e_{n-1}}{e_n} \land \compatible{ {P_0} }{e}{d}$ then we consider $e_1$:

										\begin{itemize}
											\item\label{ch5proof:whenitisifree2} $e_1 \in \dom{\Gamma' \vaslinunsplit \Delta_1, \Delta}\setminus u  $ we must have that either $e_2 \in \bn{P'_0}$ or $e_2 \in \bn{Q''}$.
											
											If $e_2 \in \bn{Q''}$ then as $P_0 = \res {z v}( P \pp \res{w's'}\vasout{t'}{w'}{(Q' \pp Q'')}  )$ and we have:
											
											$
											\begin{aligned}
												\compatible{ P_0 }{e_1}{e_2} 
												& = 
												\compatiblecc{ \cc{z}{ P }}{\cc{ v}{ Q' }}{e_1}{e_2}\\
												& \quad \lor \compatible{ P }{e_1}{e_2} \lor \compatible{ Q }{e_1}{e_2} \\
												& = 
												\false \lor \false \lor \compatible{ Q }{e_1}{e_2}
											\end{aligned}
											$
											
											with 
											$$
											\begin{aligned}
												\compatible{ Q }{e_1}{e_2} &= 
												\{ ( w', s' ) \} \lor \compatible{ Q' }{e_1}{e_2}\\
												& \quad \lor \compatible{ Q'' }{e_1}{e_2} \\
												& = \{ ( w', s' ) \}  \lor \false \lor \compatible{ Q'' }{e_1}{e_2} 
											\end{aligned}
											$$
											
											Since $e_1 \not \in \{ w', s' \}$ then $\compatible{ Q'' }{e_1}{e_2}$, which implies that if $n = 1$ then  $e_2= e$ and we must have that $e_2 \in \bn{Q''}$. Thus,  all $e_i$, $e$ and $d$ must be bound names in $Q''$. As  $(x,y) \in \{ (a, b) \mid  \compatiblecc{ \cc{z}{ P }}{\cc{ v}{ Q' }}{a}{b}\}$ it follows that at least one of these channels is bound within $P$ or $Q'$. Let us assume that we have $\compatible{P_0}{c}{d}$ and show that this cannot be the case. As $d$ is bound in $Q''$ we then must also have $\compatible{Q''}{c}{d}$ which implies that $c$ must also be in the names of $Q''$ which contradicts $c \in \bn{\res{z v}( P \pp Q' )} $.

											\item $e_1 \in \bn{P'_0}$ then the case follows by the induction hypothesis.

											\item $e_1 \in \bn{Q''}$ then the case follows analogously to that of \ref{ch5proof:whenitisifree2} when $e_2 \in \bn{Q''}$.

										\end{itemize}

									\end{myEnumerate}
									
									\item[(B.2)] {\bf If $ c >e \land e >_{P_0}^* x  $ for some $e$.}

									 Then, we have that $e \in \names{P'_0}$ and hence $c \pclosu{3,P'_0} e$. By the induction hypothesis we have that $e \pclosu{3,P'_0} x$ and hence $ c \pclosu{3,P'_0} x  $. We then apply induction on $ y  >_{P_0}^* d$ showing that $c$ and $d$ cannot be compatible, the argument of which follows from \Cref{ch5proof:whyxnotcompy2}.

									\item[(B.3)]{\bf If  $\compatible{ P_0 }{c}{e} \land e >_{P_0}^* x$}.
									
									 Then $c$ cannot be in $\dom{\Gamma' \vaslinunsplit \Delta_1, \Delta}\setminus u$. As $c \in \bn{\res{z v}( P \pp Q' )} $ then $\compatible{ P_0 }{c}{e}$ implies that $e \in \names{P'_0}$ hence $\compatible{ P'_0 }{c}{e}$. By the induction hypothesis  $e \pclosu{3,P'_0} x$ and we may deduce $c \pclosu{3,P'_0}^* x$. We then apply induction on $ y  >_{P_0}^* d$ showing that $c$ and $d$ cannot be compatible, the argument of which follows from \Cref{ch5proof:whyxnotcompy2}.

									\item[(B.4)] {\bf If $c >_{P_0}^* e \land \compatible{ P_0 }{e}{x}$.}
									
									 By the induction hypothesis if $x >_{P_0}^* e$ then $c \pclosu{3,P'_0}^* e$. Therefore, we can deduce that $e \in \names{P'_0}$, as $ ( x , y ) \in \{ (a, b) \mid  \compatiblecc{ \cc{z}{ P }}{\cc{ v}{ Q' }}{a}{b}\}$ we also have that $x \in \names{P'_0}$.  Thus, since  $\compatible{ P_0 }{e}{x}$ it follows that  $\compatible{ P'_0 }{e}{x}$. We then apply induction on $ y  >_{P_0}^* d$ showing that $c$ and $d$ cannot be compatible, the argument of which follows from \Cref{ch5proof:whyxnotcompy2}.

								\end{myEnumerate}

								\item When $c \in \bn{Q''}$.
								
								 Then we show that we cannot have that $c>_{P_0}^*x \land  y>_{P_0}^* d$. We show the case of $c>x$ as all other cases follow similarly in that $x$ will have to be in both $Q''$ and $P'_0$:
								
								Suppose that $c>x$.  Since $c \in \bn{Q''}$, this implies that $c >_{(2,2)} x$, therefore,  $x$ is in the names of both $P'_0$ and $Q''$. Hence, $x \in \dom{\Gamma' \vaslinunsplit \Delta_1, \Delta}\setminus u$. In addition,  as $ ( x , y ) \in \{ (a, b) \mid  \compatiblecc{ \cc{z}{ P }}{\cc{ v}{ Q' }}{a}{b}\}$ and $y>_{P_0}^* d$ then we have $x>_{P_0}^* d$ but as $x \in \dom{\Gamma' \vaslinunsplit \Delta_1, \Delta}\setminus u$ we know that this cannot be the case.

							\end{enumerate}

							\item\label{ch5proof:casethatisusedtoalsosimp} { $ c>_{P_0}^*x >_{P_0}^* d \implies \neg\compatible{{P_0}}{c}{d}$. }
							
							As shown in \ref{ch5proof:casethatisusedtosimplify} we have that $ (c>_{P_0}^*x \land  y>_{P_0}^* d) \implies \neg\compatible{P_0}{c}{d}$ and as $\compatible{P_0}{x}{y}$ then we have so $x>_{P_0}^* d \Leftrightarrow y>_{P_0}^* d$ and the result follows.

						\end{enumerate}

				\item\label{ch5trytostopduplication} {\bf When $Q = \vasin{q}{w'}{s'}{Q'}$.}
				
				 Then let us consider the cases of $q$.
				
				\begin{enumerate}

					\item \label{ch5proof:cutacyclic4} When $q = \lin$.
					
					 Then $ \Gamma''' , w': \lin \vasreci{T'}{S'} = v:\dual{V} , (\Gamma' \vaslinunsplit \Delta) $ and: 
					
					\[
						\inferrule*[left=\orvastype{LinIn}]{ 
							\Gamma''' ,  w': S' , s' : T' \stpre{>_2}{u} Q'
						}
						{  \Gamma''' , w': \lin \vasreci{T'}{S'}  \stpre{>_2}{u}  \vasin{\lin}{w'}{s'}{Q'}   }
					\]

					Let us consider the case of $w'$ and assume $\orvastype{LinOut_3}$ was applied, as the case involving $\orvastype{LinOut_4}$ is similar.

					\begin{enumerate}

						\item When $w' = v$ then $T' = T$ and $S' = \dual{S}$.

						Notice the following two properties which allow us to apply the induction hypothesis:
						
						\begin{itemize}
							\item $\Gamma'' \vaslinunsplit \Delta_2'  , z:S  +  \Gamma''' ,  w': S' , s' : T'   \stpre{>_3}{u} \res{z v}( P'' \pp Q' )$,  where $>_3= >_{(1,2)}  \cup >_2  \cup \, \gtrdot^u_{(1,2)}$,  $\preserves{>_{(1,2)} }{P''}$ and $\preserves{>_2}{Q'}$.  By the induction hypothesis, we have  $\acyclic{>_3}{\res{z v}( P'' \pp Q' )}{x}{y}$, for all 
							$(x,y) \in  \{ (a, b) \mid  \compatiblecc{ \cc{z}{ \res{z v}(P'' \pp Q')}}{\cc{ v}{\res{z v}(P'' \pp Q')}}{a}{b} \} $.

							\item $\Gamma'' \setminus u \vaslinunsplit \Delta_1', w:  \dual{T}  +  \Gamma''' ,  w': S' , s' : T'    \stpre{>_4}{u} \res{w s'}( P' \pp Q' )$, where $>_4 = >_{(1,1)}  \cup >_2  \cup \, \gtrdot^u_{(1,1)}$,  $\preserves{>_{(1,1)} }{P'}$ and $\preserves{>_2}{Q'}$. By the induction hypothesis, we have  $\acyclic{>_4}{ \res{w s'}( P' \pp Q' )}{x}{y}$, for all 
							$(x,y) \in  \{ (a, b) \mid \compatiblecc{ \cc{w}{  \res{w s'}( P' \pp Q' ) }}{\cc{ s'}{  \res{w s'}( P' \pp Q' ) }}{a}{b} \} $. 
						\end{itemize}
						
						First, notice that 
						\[
							\begin{aligned}
								\cc{z}{ \res{z v}(P \pp Q)} 
							&= \cc{z}{ P}\\
							&= \cc{z}{ \res{ws}\vasout{z}{s}{P'} }\\
							&= (w  , \cc{w}{  P'  }) :: \cc{z}{ P'}
							\end{aligned}
						\]
						and 
						\[ 
							\begin{aligned}
								\cc{v}{ \res{z v}(P \pp Q)} 
							&= \cc{v}{ Q}\\
							&= \cc{v}{  \vasin{\lin}{v}{s'}{Q'}  }\\
							&= (s'  , \cc{s'}{  Q'  }) :: \cc{v}{ Q'}
							\end{aligned}
						\]

						Second, notice that 
						\[
							\hspace{-3.5cm}
							\small
							\begin{aligned}
								\{ (a, b)| \compatiblecc{ \cc{z}{ P_0}}{\cc{ v}{P_0}}{a}{b} \}
								= &
								\{ ( w , s'  ) , ( s' , w ) \}
								\cup \\
								& \{ (a, b) \mid  \compatiblecc{  \cc{w}{  P'  } }{ \cc{s'}{  Q'  } }{a}{b}  \}
								\cup \\
								& \{ (a, b) \mid  \compatiblecc{ \cc{z}{ P''} }{  \cc{v}{ Q'} }{a}{b}  \}
							\end{aligned}
						\]
						
						We will  show that  $\acyclic{>}{\underbrace{\res{z v}( P \pp Q )}_{P_0}}{x}{y}$, for all $(x,y) \in  \{ (a, b)| \compatiblecc{ \cc{z}{ P_0}}{\cc{ v}{P_0}}{a}{b} \} $, where we take  $> = >_3 \cup >_4$:

						\begin{myEnumerate}

							\item {\bf $x\not >_{P_0}^* y$ consider the following:}
							
							\begin{itemize}

								\item $(x , y) =  ( w , s'  )$, the case of $(s' , w)$ follows similarly. 
								
								Then, since  $\acyclic{>_4}{ \res{w s'}( P' \pp Q' )}{x}{y}$ and $\compatible{P}{x}{y}$, it follows that  $w \not >_{4,\res{w s'}( P' \pp Q' )}^* s'$. This in turn implies that if $w \pclosu{P_0} s'$ were to hold it would imply that we must make use of $>_{3}$ in the seqence. As $>_3= >_{(1,2)}  \cup >_2  \cup \, \gtrdot^u_{(1,2)}$ and $>_4 = >_{(1,1)}  \cup >_2  \cup \, \gtrdot^u_{(1,1)}$ we may restrict this to $>_{(1,2)}  \cup  \gtrdot^u_{(1,2)}$. As $u$ is the greatest element in $>$ we may also eliminate $\gtrdot^u_{(1,2)}$ by definition hence it leavesus only to show that if $w \pclosu{P_0} s'$ were to hold it would imply that we must make use of $>_{(1,2)}$ in the seqence and then we must contradict this. We prove by induction on the length of the sequence $w \pclosu{P_0} s'$ that this must not be the case and hence $w \not\pclosu{P_0} s'$. If $w \pclosu{P_0} s'$ is made in one step then we would have $w > s'$ and hence we are forced to apply $>_{(1,2)}$. However $w >_{(1,2)} s'$ cannot hold as $w \in \names{\res{w s'}( P' \pp Q' )}$ and hence $w \not\in \names{P''}$ so is contradicted by \Cref{ch5prop:auxiliary} (6). Applying the induction hypotheses we apply the same reasoning that any the step taken within $>_{(1,2)}$ must start from a channel cannot occur within the names of $P''$ and hence \Cref{ch5prop:auxiliary} (6) contradiscts the assumption.

								\item\label{ch5proof:step1forgoodcut} $(x , y)  \in  \{ (a, b) \mid  \compatiblecc{  \cc{w}{  P'  } }{ \cc{s'}{  Q'  } }{a}{b}  \} $.
								
								Then as $\acyclic{>_4}{\underbrace{ \res{w s'}( P' \pp Q' )}_{{P_0'}}}{x}{y}$ and $\compatible{P_0'}{x}{y}$ we have $x \not >_{4,P_0'}^* y$. As in the previouse case we need to show that $w \pclosu{P_0} s'$ were to hold it would imply that we must make use of $>_{3}$ and hence by the same resoning as the previouse case that we  must make use of $>_{(1,2)}$ which we will contradict. We prove by induction on the length of the sequence $w \pclosu{P_0} s'$ that this must not be the case and hence $w \not\pclosu{P_0} s'$. If $w \pclosu{P_0} s'$ is made in one step then we would have $x > y$ and hence we are forced to apply $>_{(1,2)}$ to have the step $x >_(1,2) y $. In addition, we have that either both $x$ and $y$ are bound within $P'$ or $Q'$, or that one of $x$ or $y$ is free (by \Cref{ch5prop:auxiliary} (10)). In the case that both are bound then we know that neither $x$ nor $y$ occur within $P''$ then we know that $x\not >_{(1,2)}y$. If $x$ is free we know that there is no channel that $x$ is greater than. Finally, if $y$ is free then let us consider the following two cases: {\bf (1.a)} When $y$ is free in $Q'$ then $x$ is bound in $P'$ and hence $x$ is not in the names of either $P''$ or $Q'$ and we have that $x \not >_{(1,2)} y$. {\bf (1.b)} When $y$ is free in $P'$ then by typing we have that $y \in \dom{ \Gamma'' \vaslinunsplit \Delta_1', w:  \dual{T} } $. If $y \in \dom{\Delta_1', w:  \dual{T} }$ then by typing $y \not \in \names{P''}$ hence $x \not >' y$.  If $y \in \dom{\Gamma''} $ then we must have that $y$ occurs within an output of $P'$ and $x$ occurs within an unresrticted input of $Q'$ i.e. there is a contect $C_1$ and $C_2$ such that $P' = C_1[\ov a\out y.P''']$ and $Q' = C_2[\vasin{\un}{b}{y}{Q''}]$ for some channel $a$ and $b$ with dual types as only outputs may send bound channels and simularly inorder for $x$ to be greater then any chanel then it must be recived within a unretriced input by \Cref{ch5fig:orderedtypes}, however we would need that $ a: \recur{\wn T} \land\notserver{T}\land \client{T} $ and $ b: \recur{\wn T} \land\server{T}\land \notclient{T} $ hence $a$ and $b$ do not have dual types and contradict $P_0 \in \dilllang$. Applying the induction hypotheses we apply the same reasoning that any the step taken within $>_{(1,2)}$ must contradict that $P_0 \in \dilllang$.						
								
								\item $(x , y) \in \{ (a, b)|  \compatiblecc{ \cc{z}{ P'} }{  \cc{v}{ Q'} }{a}{b}  \} $ then the case is the dual to that of \ref{ch5proof:step1forgoodcut} in the sense that \ref{ch5proof:step1forgoodcut} considers the sub process $\res{w s'}( P' \pp Q' )$ where as here we are considering $\res{z v}( P'' \pp Q' )$ and hence will be ommited.

							\end{itemize}

							\item {\bf $(x >_{P_0}^* c  \implies c \not  >_{P_0}^* y) \land (c >_{P_0}^* x \implies y \not >_{P_0}^*  c) $.}
							
							 We only show that the conjunct $(x >_{P_0}^* c  \implies c \not  >_{P_0}^* y) $ holds.  Consider the following:

							\begin{myEnumerate}

								\item 
								$(x , y) =  ( w , s'  )$, the case of $(s' , w)$ follows similarly. 
								
								Then as $\acyclic{>_4}{ \res{w s'}( P' \pp Q' )}{w}{s'}$ and $\compatible{P'_0}{w}{s'}$ where $P'_0 = \res{w s'}( P' \pp Q' ) $ we have $(w \pclosu{4,P'_0} c  \implies c \not  \pclosu{4,P'_0} s')$. We argue similarly to previous cases via contradiction, showing firstly that when we have $ c  >_{P_0}^* e  >_{P_0}^* s'$ we may consider instead $w >_{P_0}^* e \implies e  >_{P_0}^* s'$ where $e  >_{P_0}^* s' $ as $w >_{P_0}^* c  >_{P_0}^* e$. This argument is applied until we obtain an $e$ such that we have that $e  >_{P_0}^* s' $ may be represented as $e>f_1$ followed by a chain of pairs of compatible channels with respect to $P_0$ that leads to $s'$ (this being $f_1,f_2$ being compatible, $f_2, f_3$ being compatible, ..., $f_n, s'$ being compatible). Next all channels that enable $ w >_{P_0}^* e $ must be bound within $P'$ and we have that $ w \pclosu{4,P'_0} e $ and $ w \not\pclosu{3,P_0} g $ for any $g$. Finally, we show by induction on the length of the chain of compatible channels $f_1, \cdots, f_n$ to $s'$ that we cannot have $e >_{P_0}^* s' $ without a contradiction this either being done by the induction hypothesis from $\acyclic{>_4}{ P_0'}{w}{s'}$ or by contradicting $ \compatiblecc{ \cc{z}{ P_0}}{\cc{ v}{P_0}}{w}{s'}$.

								\item $(x , y)  \in  \{ (a, b) \mid   \compatiblecc{  \cc{w}{  P'  } }{ \cc{s'}{  Q'  } }{a}{b}  \} $.
								
								 Let $ P'_0 =  \res{w s'}( P' \pp Q' ) $, then as $\acyclic{>_4}{ P'_0 }{x}{y}$ and $\compatible{P'_0}{x}{y}$ we have $(x \pclosu{4,P'_0} c  \implies c \not  \pclosu{4,P'_0} y)$.

								We show inductively that if $ x >_{P_0}^* c$ then $ x \pclosu{4,P'_0} c$ where $c >_{P_0}^* y$.  Following \Cref{ch5def:closure}, $ x >_{P_0}^* c$ can hold in the following 4 situations:
								\begin{myEnumerate}
									
									\item\label{ch5proof:whyxnotcompy3} {\bf If $x > c$.}
									
									 Then we also have that $x >_{P'_0}^* c$ and as $x \not \in \dom{\Gamma' \vaslinunsplit \Delta_1, \Delta}\setminus u $ as by \Cref{ch5prop:ontleaf} this would contradict $x > c $ we may deduce that $x \in \bn{P'_0} $ and $x >_4 c$ and hence also $x \pclosu{4,P'_0} c$.
									Next we show that $c >_{P_0}^* y$ cannot hold:

									\begin{myEnumerate}
										\item Suppose $c > y$: Then, this implies  $c  >_{P'_0}^* y$ and similarly that $c  \pclosu{4,P'_0} y$, which contradicts $\acyclic{>_4}{ P'_0 }{x}{y}$ as $c \not  \pclosu{4,P'_0} y$.

										\item\label{ch5proof:casebigger3} Suppose there is an $e$ such that $ c > e \land e >_{P_0}^* y  $: Then,  $e \not \in \dom{\Gamma' \vaslinunsplit \Delta_1, \Delta}\setminus u $ as by \Cref{ch5prop:ontleaf} this would contradict $e >_{P_0}^* y $. Let us consider the three remaining possibilities for $e$:
										
										\begin{itemize}
											\item Case $e = u$: we would have  $c > u$  which contradicts maximality of $u$ w.r.t. $>$, from  typing $\stpre{>}{u}$. 
										
											\item Case $e \in \bn{P'_0}$: then $c >_{P_0}^* e$ hence we have $x >_{P_0}^* e$ and by the induction hypothesis $e \not  >_{P_0}^* y$.
											
											\item Case $e \in \bn{P''} $: then $c > e$ is a contradiction as  this will require $c \in \dom{\Gamma'''}$ and {hence $C$ cannot be larger then $e$} by \Cref{ch5prop:ontleaf}.

										\end{itemize}
										
										\item\label{ch5proof:casecomp3} Suppose there is an $e$ such that $\compatible{ {P_0} }{c}{e} \land e >_{P_0}^* y$: Then $e \not \in \dom{\Gamma' \vaslinunsplit \Delta_1, \Delta}\setminus u $ as by \Cref{ch5prop:ontleaf} this would contradict $e >_{P_0}^* y $. Let us consider the three remaining posibilities:
										\begin{itemize}
											\item $e = u$: this case contradicts $\compatible{ {P_0} }{c}{u} $ as there does not exist a channel $c$ that is compatible with $u$ hence we do not consider the case.
										
											\item $e \in \bn{P'_0}$: then $c >_{P_0}^* e$ hence we have $x >_{P_0}^* e$ and by the induction hypothesis $e \not  >_{P_0}^* y$.
											
											\item $e \in \bn{P''}$: then  $c \in \dom{\Gamma'''}$ and hence contradicting $c >_{P_0}^* y $ as no element is smaller then $c$.

										\end{itemize}
										
										\item  $c >_{P_0}^* e \land \compatible{ {P_0} }{e}{y}$ then firstly notice that there are three cases here:
										\begin{itemize}
											\item There is an $f$ such that  $c > f \land f >_{P_0}^* e \land \compatible{ {P_0} }{e}{y}$, in this case we may apply case \ref{ch5proof:casebigger3}. 
											
											\item There is an $f$ such that  $\compatible{ {P_0} }{c}{f} \land f  >_{P_0}^* e \land \compatible{ {P_0} }{e}{y}$, in this case we may apply case \ref{ch5proof:casecomp3}.
											
											\item\label{ch5proof:caseaftercomp3} There is no $f$ such that  $c > f \land f >_{P_0}^* e \land \compatible{ {P_0} }{e}{y}$ or $\compatible{ {P_0} }{c}{f} \land f  >_{P_0}^* e \land \compatible{ {P_0} }{e}{y}$, hence $c >_{P_0}^* e \land \compatible{ {P_0} }{e}{y}$ may be considered as $c > e_1 \land \compatible{ {P_0} }{e_1}{e_2} \land  \cdots \land \compatible{ {P_0} }{e_n}{e}  \land \compatible{ {P_0} }{e}{y} $. That is that $c>e_1$ and then there is a sequence of compatible channels that leads to $y$.
										\end{itemize}
										
										We now proceed to show \ref{ch5proof:caseaftercomp3}.

										Note that $e_i \not = u$ for $i \in \{1,\dots ,n\} $ and $e \not = u$ as these would contradict $\compatible{ {P_0} }{e_i}{e_{i+1}}$ and $\compatible{ {P_0} }{e}{y}$, respectively. Also, we have that $c \not \in \dom{\Gamma' \vaslinunsplit \Delta_1, \Delta}\setminus u  \cup \bn{P''} $ as this would contradict $c >_{P_0}^* e$. Thus,  $c \in \bn{P'_0}$. We proceed by induction on $n$:
										
										\noindent ({\bf Base Case}) when $n= 0$.
										
										 We have $c > e \land \compatible{ {P_0} }{e}{y}$ then we consider $e$

										\begin{itemize}
											\item $e \in \dom{\Gamma' \vaslinunsplit \Delta_1, \Delta}\setminus u  $:  we must have that either $y \in \bn{P'_0}$ or $y \in \bn{P''}$. If $y \in \bn{P''}$ then this contradicts $(x , y)  \in  \{ (a, b) \mid  \compatiblecc{  \cc{w}{  P'  } }{ \cc{s'}{  Q'  } }{a}{b}  \} $. If $y \in \bn{P'_0}$ then we have $c >_{P'_0}^* e$ and  $c \pclosu{4,P'_0} e$. Also,  $\compatible{ {P_0} }{e}{y} \implies \compatible{ {P'_0} }{e}{y}$, but by the induction hypothesis, $c \not \pclosu{4,P'_0} y$.

											\item $e \in \bn{P'_0}$ then we must have that $y \in \names{P'_0} \cup \dom{\Gamma'' \setminus u \vaslinunsplit \Delta_1', w:  \dual{T}  +  \Gamma''' ,  w': S' , s' : T' }$ and we have that $c >_{P'_0}^* e$ and also $c \pclosu{4,P'_0} e$ secondly that $\compatible{ {P_0} }{e}{y} \implies \compatible{ {P'_0} }{e}{y}$ but by the induction hypothesis we have that $c \not \pclosu{4,P'_0} y$.

											\item $e \in \bn{P''}$: then as $c$ must be within the bound names of $P'_0$ but then we cannot have that $c >_3  e $. Hence we must have $ c >_4 e $ which then implies that $e \not \in \bn{P''}$.

										\end{itemize}

										\noindent ({\bf Inductive Step}) When $n \geq 1$.
										
										 We have that  $c > e_1 \land \compatible{P_0}{e_1}{e_2} \land \cdots \land \compatible{P_0}{e_{n-1}}{e_n} \land \compatible{ {P_0} }{e}{y}$ then we consider $e_1$:
										
										\begin{itemize}

											\item\label{ch5proof:whenitisifree3} $e_1 \in \dom{\Gamma' \vaslinunsplit \Delta_1, \Delta}\setminus u  $ we must have that either $e_2 \in \bn{P'_0}$ or $e_2 \in \bn{P''}$.
											
											If $e_2 \in \bn{P''}$ then as $P_0 = \res {z v}( \res{ws}\vasout{z}{s}{(P' \pp P'')} \pp \vasin{\lin}{v}{s'}{Q'}  )$ and we have:
											
											\[
												\hspace{-5cm}
												\small
											\begin{aligned}
												\compatible{ P_0 }{e_1}{e_2} 
												& = 
												\compatiblecc{ \cc{z}{  P  }}{\cc{ v}{ Q }}{e_1}{e_2}  \lor \compatible{ P }{e_1}{e_2} \lor \compatible{ Q }{e_1}{e_2} \\
												& = 
												\compatiblecc{   (w , \cc{w}{ P' } ) :: \cc{z}{ P'' }}{ ( s' , \cc{s'}{Q'} )  :: \cc{ v}{ Q' }}{e_1}{e_2} \\
												& \quad \lor \compatible{ P }{e_1}{e_2} \lor \compatible{ Q }{e_1}{e_2} \\
												& = 
												\compatiblecc{ \cc{z}{ P'' }}{ \cc{ v}{ Q' }}{e_1}{e_2}  \lor \compatible{ P }{e_1}{e_2} \lor \compatible{ Q }{e_1}{e_2} \\
												& = 
												\compatiblecc{ \cc{z}{ P'' }}{ \cc{ v}{ Q' }}{e_1}{e_2} \lor \compatible{ P }{e_1}{e_2} \lor \false \\
											\end{aligned}
											\]
											with 
											\[
												\small
											\begin{aligned}
												\compatible{ P }{e_1}{e_2} &= 
												\{ ( w, s ) \}\\
												& \quad \lor \compatible{ P' }{e_1}{e_2} \lor \compatible{ P'' }{e_1}{e_2} \\
												& = \{ ( w, s ) \}  \lor \false \lor \compatible{ P'' }{e_1}{e_2} 
											\end{aligned}
											\]
											
											As $e_1 \not \in \{ w, s'\}$ then we must have that either $\compatiblecc{ \cc{z}{ P'' }}{ \cc{ v}{ Q' }}{e_1}{e_2}$, which holds  by the induction hypothesis, or $\compatible{ P'' }{e_1}{e_2}$, and hence when $n = 1$ then $e_2= e$ and we must have that $e_2 \in \bn{P''}$. Therefore,  all $e_i$, $e$ and $y$ must be bound names in $P''$ but then this contradicts that $(x , y)  \in  \{ (a, b) \mid  \compatiblecc{  \cc{w}{  P'  } }{ \cc{s'}{  Q'  } }{a}{b}  \} $.

											When $e_2 \in \bn{Q'}$ the case follows by the induction hypothesis.

											\item $e_1 \in \bn{P'_0}$ then the case follows by the induction hypothesis.

											\item $e_1 \in \bn{Q''}$ then the case follows analagously to that of \ref{ch5proof:whenitisifree3} when $e_2 \in \bn{Q''}$.

										\end{itemize}

									\end{myEnumerate}
									
									\item {\bf If $ x >e \land e >_{P_0}^* c  $ for some $e$.}
									
									 Then we have that $e \in \names{P'_0}$ and hence $x\pclosu{3,P_0'} e$.  By the induction hypothesis, $e\pclosu{3,P_0'} c$ which implies $ x\pclosu{3,P_0'} c  $. We then reason by induction  on $ c  >_{P_0}^* y$ showing that $x$ and $y$ cannot be compatible, the argument of which follows from \ref{ch5proof:whyxnotcompy3}

									\item\label{ch5proof:whytheyarecompatible3} {\bf If $\compatible{ P_0 }{x}{e} \land e >_{P_0}^* c$ for some $e$}.
									
									 Then $x \notin \dom{\Gamma' \vaslinunsplit \Delta_1, \Delta}\setminus u \cup \bn{Q''}$ as this would contradict $(x,y) \in \{ (a, b) \mid \compatiblecc{ \cc{z}{ P }}{\cc{ v}{ Q' }}{a}{b}\}$. Hence we have that both $x$ and $e$ are in the bound names of $P'_0$, which implies that $\compatible{ P'_0 }{x}{e}$.  By the induction hypothesis, we have $e >_{P_0}^* c$, then finally we have $ e\pclosu{3,P_0'} c$. We then reason by induction on $ c  >_{P_0}^* y$ showing that $x$ and $y$ cannot be compatible, the argument of which follows from \ref{ch5proof:whyxnotcompy3}

									\item {\bf $x >_{P_0}^* e \land \compatible{ P_0 }{e}{c}$ for some $e$.}
									
									 Then by the induction hypothesis if $x >_{P_0}^* e$ then $x\pclosu{3,P_0'} e$ as $\compatible{ P_0 }{e}{c}$ and $c >_{P_0}^* y$ we can argue analagously to \ref{ch5proof:whytheyarecompatible3} that $\compatible{ P'_0 }{e}{c}$ and hence $x >_{P_0}^* c$. We then apply induction on $ c  >_{P_0}^* y$ showing that $x$ and $y$ cannot be compatible, the argument of which follows from \ref{ch5proof:whyxnotcompy3}

								\end{myEnumerate}

							\end{myEnumerate}

							\item  $(c>_{P_0}^*x \land  y>_{P_0}^* d) \implies \neg\compatible{P_0}{c}{d}$ and $ c>_{P_0}^*x >_{P_0}^* d \implies \neg\compatible{{P_0}}{c}{d}$ are shown similarly to \Cref{ch5proof:casethatisusedtosimplify} and \Cref{ch5proof:casethatisusedtoalsosimp} respectively.

						\end{myEnumerate}

						\item When $w' \not= v$ the results follow analagously to \ref{ch5proof:cutacyclic1} applying the induction hypothesis.

					\end{enumerate}

					\item When $q = \un$.
					
					  Then $ \Gamma''' , w': \recur{\wn T'} = v:\dual{V} , (\Gamma' \vaslinunsplit \Delta) $ and either: 
					\[
						\inferrule*[left=\orvastype{UnIn_1}]{ 
							\Gamma''',  s' : T' \stpre{>_{(2,1)}}{w} Q'
							\\
							>_2 \defeq  >_{(2,1)}\partremove{w}  \cup (\gtrdot^{w'}_{(2,1)})\partremove{w}\\
							\notserver{T'}
							\\
							w \text{ fresh} 
						}
						{\Gamma''' , w': \recur{\wn T'} \stpre{>_2}{w'}  \vasin{\un}{w'}{s'}{Q'}   }
					\]
					or
					\[\hspace{-2cm}
					\small
						\inferrule*[left=\orvastype{UnIn_2}]{ 
							\Gamma''',  s' : T' \stpre{>_{(2,1)}}{s'} Q'
							\\
							>_2\defeq >_{(2,1)}\cup (\gtrdot^{w'}_{(2,1)})
							\\
							\notclient{T'} \land \server{T'}
						}
						{\Gamma''' , w': \recur{\wn T'} \stpre{>_2}{w'}  \vasin{\un}{w'}{s'}{Q'}   }
					\]

					The case proceeds analogously to that of \Cref{ch5proof:cutacyclic4} with the excetion that in the case of $w' = v$ that $\res{z v}( P \pp Q ) \not \in \dilllang$.

				\end{enumerate}

				\item\label{ch5proof:cutacyclic5} {\bf When $Q = \res{w's'}(Q' \pp Q'')$.} 
				
				Then $ \Gamma''' \vaslinunsplit \Delta''_1, \Delta''_2 = v:\dual{V} , (\Gamma' \vaslinunsplit \Delta)  $ and:

				\[
					\inferrule*[left=\orvastype{Par_2}]{ 
						w':V , (\Gamma''' \setminus u \vaslinunsplit \Delta''_1) \stpre{>_1}{z} Q'
						\\
						s':\dual{V} , (\Gamma''' \vaslinunsplit \Delta''_2) \stpre{>_2}{u} Q''
						\\
						> \defeq >_1  \cup >_2  \cup \gtrdot^u_1
					}
					{ \Gamma''' \vaslinunsplit \Delta''_1, \Delta''_2 \stpre{>}{u}  \res{w's'}(Q' \pp Q'') }
				\]

				Since $z$ and $v$ have a linear type then $v$ must occur either within $Q'$ or $Q''$ but not both. Assume w.l.o.g. that $v \in \fn{Q''}$ then $v \not \in \dom{w':V , (\Gamma''' \setminus u \vaslinunsplit \Delta''_1)} \lor \names{Q'} $. By the induction hypothesis, it follows that 
				 $$\Gamma'' \vaslinunsplit \Delta_1', \Delta_2' , t: \lin \vassend{T}{S}   + s':\dual{V} , (\Gamma''' \vaslinunsplit \Delta''_2)  \stpre{>_3}{u} \res{z v}( P \pp Q'' ),$$
				  where $>_3 = >_1  \cup >_{(2,2)}  \cup \gtrdot^u_{1}$ and both $\preserves{>_{1} }{P}$ and $\preserves{>_{(2,2)} }{Q''}$. Thus, by  the induction hypothesis  we have $\acyclic{>_3}{\res{z v}( P \pp Q'' )}{x}{y}$,  for all 
				$(x,y) \in  \{ (a, b) \mid  \compatiblecc{ \cc{z}{ \res{z v}(P \pp Q'')}}{\cc{ v}{\res{z v}(P \pp Q'')}}{a}{b} \} $. Therefore,  since  $\acyclic{>_3}{\res{z v}( P \pp Q'' )}{x}{y}$ and $\not \exists c $ such that $v >_{(2,1)}  c \lor c >_{(2,1)}  v$ the result holds.

				\item \label{ch5item:end}  {\bf When $Q = Q' \pp Q''$}.
				
				  Then $  \Gamma''' \vaslinunsplit \Delta''_1, \Delta''_2  = v:\dual{V} , (\Gamma' \vaslinunsplit \Delta)  $ and:

				\[
					\hspace{-2cm}
					\small
					\inferrule*[left=\orvastype{Par_1}]{ 
						\Gamma''' \setminus u \vaslinunsplit \Delta''_1 \stpre{>_{(2,1)}}{w} Q'
						\\
						\Gamma''' \vaslinunsplit \Delta''_2 \stpre{>_{(2,2)}}{u} Q''
						\\
						>_2 \defeq >_{(2,1)} \partremove{w}\cup >_{(2,2)} \cup (\gtrdot^u_{(2,1)})\partremove{w}
						\\
						w \text{ fresh} 
					}
					{ \Gamma''' \vaslinunsplit \Delta''_1, \Delta''_2 \stpre{>_2}{u} (Q' \pp Q'')  }
				\]

				Result follows analogously from \Cref{ch5proof:cutacyclic5}.

			\end{enumerate}


                 
	\item {\bf Case $P = \res{ws}\vasout{t}{s}{P'} $.}

	 Then the case follows for the most part analogously to \Cref{ch5proof:cutacyclic6}. 
	\item {\bf Case $P = \vasin{q}{w}{v}{P'}$.}
	
	 Then by duality, the key cases have already been considered, otherwise, induction is applied on the structure of $Q$ and the cases are analogous to that of \ref{ch5proof:cutacyclic2}
		
      \item {\bf Case $P = \res{wv}(P' \pp P'')$ or $P =P' \pp P''$.}
       
       We apply induction on the structure of $Q$ and all cases follow analogously from \ref{ch5proof:cutacyclic5}.

	\end{enumerate}
\end{proof}


We can now prove that typable processes (in $\dilllang$) induce acyclic partial orders:

\begin{lemma}\label{ch5proof:dillispres}
	If $P \in \dilllang(u)$ with $ \Gamma \stpre{>}{u} P $ then $\preserves{>}{P}$.
\end{lemma}

\begin{proof}
The proof is by induction on the structure of $P$.
\begin{enumerate}
\item  {\bf  When $ P=\nil$.}

	Then,
		by \Cref{ch5prop:dillmakesorder}, we have the following derivation:
			\begin{mathpar}
				\inferrule*[left=\orvastype{Nil}]{ > \defeq \{ (u , x) \mid   x \in \Gamma \setminus u \} }
				{ \Gamma \stpre{>}{u} \nil  }

			\end{mathpar}

			Since $\varcompat{\nil}  = \emptyset $, one has  $\preserves{>}{\nil}$ and the result follows.
\item {\bf When $P =  \vasin{\un}{x}{y}{P'} $. }

Then we  need to consider the type of $x$:

\begin{enumerate}
\item \label{ch5proof:dillispres4} $\notclient{T} \land \server{T} $:

Then, by the proof of \Cref{ch5prop:dillmakesorder}, we have the following derivation:
\begin{mathpar}
\inferrule*[left=\orvastype{UnIn_2}]{ 
\Gamma',  y : T \stpre{>_1}{y} P' \\
{>=>_1\cup \gtrdot^x_1}
\\
\notclient{T} \land \server{T}
}
{\Gamma' , x: \recur{\wn T} \stpre{>}{x}  \vasin{\un}{x}{y}{P'}   }
\end{mathpar}
By the induction hypothesis, we have  $\preserves{>_1}{P'}$, and  by \Cref{ch5def:compN}  we have   $\varcompat{P'} = \varcompat{P} $. It remains to show  that $\preserves{>}{P}. $ 

By definition $\preserves{>}{P} =\bigwedge\limits_{ (a,b) \in \varcompat{P}}\acyclic{>}{P}{a}{b}$, and we will verify the conditions in \Cref{ch5def:acyclic}: 
\begin{itemize}
\item $a \not >^*_P b$.

		As $\preserves{>_1}{P'}$ then $a  {{\not >}_1}^*_{P'} b$. 
		
		Let us consider the following two cases w.l.o.g.

		\begin{enumerate}

			\item When $x = a$ then $\Gamma' , x: \recur{\wn T} \stpre{>}{x}  \vasin{\un}{x}{y}{P'} $, by applying  \Cref{ch5prop:auxiliary} $(vii)$ implies $\neg\compatible{P}{a}{b}$ and the condition follows.

			\item When $x \not = a$ then $a \not \gtrdot^x_1 b$ by definition hence we must have $a{>_1}^*_Pb$. Secondly for $\compatible{\vasin{\un}{x}{{y}}{P'}}{a}{b} $ to hold then $ \compatible{{P'}}{a}{b} $. Finally by the induction hypothesis $\preserves{>_1}{P'}$ implies $a{>_1}^*_{P'} b $ if $ \neg\compatible{{P'}}{a}{b} $ and hence $a{>_1}^*_{P} b $.

		\end{enumerate}

		\item $(a >^*_P c  \implies c \not  >^*_P b) \land (c >^*_P a \implies b \not >^*_P  c) $

		We show w.l.o.g. that $(a >^*_P c  \implies c \not  >^*_P b)$.
		
			As $\preserves{>_1}{P'}$ then $(a >_{1,P'}^* c  \implies c \not >^*_{1,P'} b)$ and as already stated $x= a$ or $x = b$ contradicts $\compatible{P}{a}{b}$ and hence $a \not = x$. As $\varcompat{P}=\varcompat{P'}$ and no channel is either greater then or compatible to $x$ it follows that $(a >^*_P c  \implies c \not  >^*_P b)$.
			

		\item $ (c>_P^*a \land  b>_P^* d) \implies \neg\compatible{P}{c}{d}$ 

		Firstly as $\preserves{>_1}{P'}$ then $ (c>_{1,P'}^*a \land  b>_{1,P'}^* d) \implies \neg\compatible{P'}{c}{d}$. Let us consider w.l.o.g. the following two cases:
		
		\begin{enumerate}
			\item $c = x$ then $x {\gtrdot^x_1}_P^*a $ however when $b>_P^* d$ then as $\Gamma' , x: \recur{\wn T} \stpre{>}{x}  \vasin{\un}{x}{y}{P'}$ applying  \Cref{ch5prop:auxiliary} $(vii)$ we have that $x$ is not compatible with any channel hence $\neg \compatible{P}{x}{d}$ .
			
			\item When $c \not = x$ the case holds as shown previously no channel can be greater then or compatible to $x$ meaning applying the induction hypothesis the case follows.
		\end{enumerate}

		\item $ c>_P^*a >_P^* d \implies \neg\compatible{P}{c}{d}$

		As $\preserves{>_1}{P'}$ then $ c>_{1,P'}^*a >_{1,P'}^* d \implies \neg\compatible{P'}{c}{d}$. Let us consider w.l.o.g. the following two cases:
		
		\begin{enumerate}
			
			\item $c = x$ then $x {\gtrdot^x_1}_P^*a $ however when $a>_P^* d$ then $\neg \compatible{P}{x}{d}$ is implied by  \Cref{ch5prop:auxiliary} $(vii)$ .
			
			\item When $c \not = x$ the case holds similarly to previous cases by the induction hypothesis.

		\end{enumerate}

\end{itemize}

\item When $\notserver{T}$ then by \Cref{ch5prop:dillmakesorder}: 
\begin{mathpar}
	\inferrule*[left=\orvastype{UnIn_1}]{ 
		\Gamma',  y : T \stpre{>_1}{w} P'
		\\
		{> = >_1\partremove{w}  \cup (\gtrdot^u_1)\partremove{w}}\\
		\notserver{T}
		\\
		w \text{ fresh} 
	}
	{\Gamma' , x: \recur{\wn T} \stpre{>}{x}  \vasin{\un}{x}{y}{P'}   }
\end{mathpar}

The case holds analogously to \ref{ch5proof:dillispres4}.

\end{enumerate}
\item {\bf  When $P = \vasin{\lin}{x}{y}{P'}$.}

 Then, by \Cref{ch5prop:dillmakesorder}, we have the following derivation:
		\begin{mathpar}
	
			\inferrule*[left=\orvastype{LinIn}]{ 
				\Gamma' ,  x: S , y : T \stpre{>}{u} P'
			}
			{  \Gamma' , x: \lin \vasreci{T}{S}  \stpre{>}{u}  \vasin{\lin}{x}{y}{P'}   }
	
		\end{mathpar}

		By the induction hypothesis, it follows that  $\preserves{>}{P'}$. By \Cref{ch5def:compN}, we have that $\varcompat{P}=\varcompat{P'}$. Thus, 
		
		$\preserves{>}{P}=\bigwedge\limits_{ (a,b) \in \varcompat{P}}\acyclic{>}{P}{a}{b}=\bigwedge\limits_{ (a,b) \in \varcompat{P'}}\acyclic{>}{P}{a}{b}$ holds, since $\preserves{>}{P'}$ holds.
		
%
%
\item {\bf When $P = \ov x\out v.P' $.}

 Then there are four cases:

		\begin{enumerate}
			\item $x: \lin \oc (T).S $ and $\neg \contun{T}$:
			
			Then, the derivation is as
				\begin{mathpar}
					\inferrule*[left=\orvastype{LinOut_2}]{ 
						\Gamma', x:S \stpre{>_1}{u} P'
						\quad 
						> = >_1  \cup \{(u , v)\}
						\quad
						\neg \contun{T}
					}
					{ \Gamma', v:T ,  x: \lin \oc (T).S \stpre{>}{u} \ov x\out v.P'  }		
				\end{mathpar}	

		By the induction hypothesis, it follows that $\preserves{>_1}{P'}$. 		
		We want to prove  that 
		$$\preserves{>}{P}=\bigwedge_{(a,b)\in \varcompat{P}} \acyclic{>}{P}{a}{b}$$
		holds.

                 \begin{enumerate}
                 \item $a \not >_P^*b$.
                 
                 Since $\preserves{>_1}{P'}$, it follows that $a \not >_{1,P'}^*b$. By definition, we have $\varcompat{P}=\varcompat{P'}$ and the condition follows.
                 \item $(a >_P^* c \implies c \not >_P^* b) \wedge (c >_P^* a \implies b \not >_P^* c)$.
                 
                 Note that $u,v \notin \varcompat{P}$. Thus, $(a >_P^* c \implies a >_{1,P'}^* c )$.
                 
                  Since $\preserves{>_1}{P'}$, it follows that  $(a >_{1,P'}^* c \implies c \not >_{1,P'}^* b)$. Finally,  $ c \not >_{1,P'}^* b$ implies   $c \not >_{P}^* b$, as  $(u,v)\in >$ does not relate with any other pairs in $>_1$. The proof is similar for $(c>_P^* a)$.
                 \item $(c>_P^* a \wedge b>_P^* d) \implies \neg \compatible{P}{c}{d}$.
                 
                 Suppose $(c>_P^* a \wedge b>_P^* d) $. Since $u,v \notin \varcompat{P}$, it follows that $(c>_{1,P'}^* a \wedge b>_{1,P'}^* d) $ and $\neg \compatible{P'}{c}{d}$. Hence, $\neg \compatible{P}{c}{d}$, since $>$ does not add new compatible names.
                   \item $(c>_P^* a >_P^* d) \implies \neg \compatible{P}{c}{d}$.
                   
                   This case is similar to the previous one.
               \end{enumerate}
			\item $x: \lin \oc (T).S$ and $\contun{T}$:
				\begin{mathpar}
							\inferrule*[left=\orvastype{LinOut_1}]{ 
								\Gamma', v:T ,  x:S \stpre{>}{u} P'
								\quad
								\contun{T}
							}
							{ \Gamma', v:T ,  x: \lin \oc (T).S \stpre{>}{u} \ov x\out v.P'  }

				\end{mathpar}

			\item  $ x: \recur{\oc T} $ and $ \neg \contun{T} $:
				\begin{mathpar}
						
					\inferrule*[left=\orvastype{UnOut_2}]{ 
						\Gamma' , x:\recur{\oc T}  \stpre{>_1}{u} {{P'}} 
						\\
						> = >_1  \cup \{(u , v)\}
						\\
						u \not = x
						\\
						\neg \contun{T}
					}
					{ \Gamma_1 ,  x: \recur{\oc T} , v:T \stpre{>}{u} \ov x\out v.P'  }
			
				\end{mathpar}

			\item $ x: \recur{\oc T} $ and $ \contun{T} $:
				\begin{mathpar}

						\inferrule*[left=\orvastype{UnOut_3}]{ 
							\Gamma' , x:\recur{\oc T} , v:T  \stpre{>}{u} {{P'}} 
							\\
							u \not = x
							\\
							\contun{T}
						}
						{ \Gamma_1 ,  x: \recur{\oc T} , v:T  \stpre{>}{u} \ov x\out v.P'  }
				\end{mathpar}

		\end{enumerate}
	Cases (b) - (d)  follow by applying the induction hypothesis in $P'$, followed by verifying that the conditions (i) - (iv) for acyclicity hold.

\item {\bf When $P = \res{xy}\vasout{z}{y}{P'}$.}

 Then, by \Cref{ch5prop:dillmakesorder}, we have the following derivation:
		\begin{mathpar}
			\inferrule*[left=\orvastype{UnOut_1}]{ 
				\Gamma', z: \recur{\oc T} , x: \dual{T}  \stpre{>_1}{u} P'
				\\
				> = >_1 \cup \{(u , y)\}
				\\
				u \not = z
			}
			{ \Gamma' , z: \recur{\oc T}  \stpre{>}{u}  \res{xy}\vasout{z}{y}{P'}  }
		\end{mathpar}

		By the induction hypothesis  we have $\preserves{>_1}{P'}$. By definition, 
		$\varcompat{P}= \varcompat{P'} \cup  \{ (x,y) , (y,x)\}.$
		
		We need to show $\preserves{>}{P}$:
                 
                 Let $(a,b) \in \varcompat{P}$. We will verify that $\acyclic{>}{P}{a}{b}$.

                There are two cases to consider:
                 \begin{enumerate}
                 \item $(a,b) \in \varcompat{P'}$.
                     
                      By \Cref{ch5prop:auxiliary}(7), it follows that there is no $x'$ such that $\compatible{P'}{u}{x'}$, thus  $u \notin \{a,b\}$.   Since  $\Gamma\stpre{>_1}{u} P'$, it follows that {\bf (H1)}~$y\notin \names{P'}\cup \dom{\Gamma'}$, thus $y\notin \{a,b\}$. Also, by \Cref{ch5prop:auxiliary}(6) and {\bf (H1)}, it follows that there does not exist $x'$ such that {\bf (H2)} $x'>_1y\vee y>_1x'\vee \compatible{P'}{x'}{y}$.
                     \begin{enumerate}
                      \item  $a \not >_P^* b$. 
                      
                      Suppose, by contradiction, that $a>_P^*b$.
						 Then, there exists a path from $a$ to $b$ via $>_P^*$, and it must be the case, by \Cref{ch5def:closure}, that either $\compatible{P}{x}{y}$, or $\compatible{P}{y}{x}$ or $u>y$ where used in the path, otherwise, we would have  $a>_{1,P'}^*b$, and this contradicts the fact that $a\not>_{1,P'}^*b$, from the hypothesis that $\preserves{>_1}{P'}$ holds. Note that $u$ does not occur in the path due to \Cref{ch5prop:auxiliary}(7). Also notice that $a \not = x$ by \Cref{ch5prop:ontleaf} and \Cref{ch5prop:auxiliary}(15) and similarly $a \not = y$ by {\bf (H2)}. Therefore, the path is, w.l.o.g., either as 
						 $$
						 \begin{aligned}
						 &a>_{1,P'}^*x\wedge \compatible{P}{x}{y}\wedge y>_{1}a_{i+1} >_{1,P'}^*b\\
				   \text{or as}~~       &a>_{1,P'}^*x\wedge \compatible{P}{x}{y}\wedge \compatible{P'}{y}{ a_{i+1}}\wedge  a_{i+1} >_{1,P'}^*b\\
						\end{aligned}
						 $$
						 contradicting {\bf (H2)}.  Hence,  $a \not >_P^* b$.
                 \item  $(a >_P^* c  \implies c \not  >_P^* b) \land (c >_P^* a \implies b \not >_P^*  c)$.
                      
                  Suppose that $a >_P^* c$. If  there were a path $c>^*_P b$ it has to use either $\compatible{P}{x}{y}$, or $\compatible{P}{y}{x}$ or $u>y$, which is a contradiction (similar to the case above). Therefore,   $c\not >_P^* b$ . The proof is similar for the implication on the right.
                       
 \item $(c>_P^*a \land  b>_P^* d) \implies \neg\compatible{P}{c}{d}$.
 
 Firstly, as $\preserves{>_1}{P'}$ then $ (c{>}_{1,P'}^*a \land  b{>}_{1,P'}^* d) \implies \neg\compatible{P'}{c}{d}$. Secondly, suppose  $(c>_P^*a \land  b>_P^* d) $ using the pairs $(x,y)$, or  $(y,x)$ or $(u,y)$. Let us consider w.l.o.g. the following cases:
				
				\begin{enumerate}

					\item When $c = u$ with $ (u>_P^*a \land  b>_P^* d)$ then $\neg\compatible{P}{u}{d}$ holds by \Cref{ch5prop:auxiliary}(7).
					
					\item When $c=x$ with $ (x>_P^*a \land  b>_P^* d)$. Note that $y$ cannot occur in the path $x>_P^*y$, since it contradicts {\bf (H2)}. Therefore, the path is $ x>_{1,P'}^*a$ and $y$ has to occur somewhere in $ b>_P^* d$. By {\bf (H1)} we have that  $y\neq b$; also,  $y$ cannot occur internally in the path $b>_{P}^*d$ without contradicting {\bf (H2)}. It remains to check whether $d=y$ can occur: if that were the case, then we would have the path $b>_{P}^*y$  as $b>_P^*x\wedge \compatible{P}{x}{y}$ (since $u$ is not compatible with any name). Then, we would have $x>_{P}^*a\wedge b>_{P}^*x$ which contradicts case (ii) above. Therefore, this case is only possible in $ (x>_{1,P'}^*a \land  b>_{1,P'}^* d)$, and the result follows by the hypothesis that $\preserves{>_1}{P'}$ holds.

					
					\item When $c = y$ with  $  y>_P^*a \land  b>_P^* d$. 
					
					Then it must be the case that $\compatible{P}{y}{x}\wedge x>^*_{1,P'} a \land  b>_P^* d$. Since $x,y$ cannot occur within $ b>_P^* d$ (otherwise it would contradict case (ii)), then we must have  $\compatible{P}{y}{x}\wedge x>^*_{1,P'} a \land  b>_{1,P'}^* d$. Suppose, by contradiction, that $\compatible{P}{y}{d}$. Since $y\notin \dom{\Gamma'}\cup \names{P'}$, by \Cref{ch5prop:auxiliary} (6), it follows that there is no $ y'$ such that $y>y'\vee  y'>y \vee \compatible{P}{y}{y'}$. Therefore, it must be the case that $d=x$, and we would have $\compatible{P}{y}{x}\wedge x>_{1,P'}^*a \wedge b>^*_{1,P'} x $, which contradicts case (ii), and the result follows.

					\item When $c \not \in \{ u,x,y \} $ with $ (c>_P^*a \land  b>_P^* d)$. 
										
					  Let us assume the contradiction that we have a pair such that $\compatible{P}{c}{d}$. Then $d$ cannot be $u$ as this contradicts that no element is larger than $u$. Similarly $d$ cannot be $y$ as this would imply that $c=x$ as that is the only compatible channel with $y$ contradicting $c\not \in \{ u,x,y \} $ . Hence we must have that $\compatible{P'}{c}{d}$ but the induction hypotheses though we have that if $(c>_{P'}^*a \land  b>_{P'}^* d) \implies \neg\compatible{P'}{c}{d}$ and  $ \neg\compatible{P'}{c}{d} \implies \neg\compatible{P}{c}{d} $ contradicting this.


				\end{enumerate}

\item $ c>_P^*a >_P^* d \implies \neg\compatible{P}{c}{d}$.

Suppose  $(c>_P^*a >_P^* d) $ using the pairs $(x,y)$, or  $(y,x)$ or $(u,y)$.  The cases where $c=u$ or $c=x$ are analogous to the analysis above. 

 It remains to analyze the case $c=y$, for which, we have $\compatible{P}{y}{x}\wedge x>_{1,P'}^*a >_P^* d $.  Since $y\notin \dom{\Gamma'}\cup \names{P'}$, by \Cref{ch5prop:auxiliary} (6), it follows that there is no $ y'$ such that $y>y'\vee  y'>y \vee \compatible{P}{y}{y'}$. Therefore, it must be the case that $d=x$, and we would have $\compatible{P}{y}{x}\wedge x>_{1,P'}^*a >_P^* x $. Since $y$ cannot occur in $a>_P^*x$, it follows that $\compatible{P}{y}{x}\wedge x>_{1,P'}^*a >_{1,P'}^* x $, which contradicts the fact that $\preserves{P'}{>_1}$.

                      \end{enumerate}
                 \item $(a,b)\in \{ (x,y) , (y,x)\}$.
                 
                 We consider $(a,b) = (x,y)$ as the case of $(y,x)$ follows analogously.

			\begin{enumerate}

				\item $x \not >_P^* y$.
				
				As $\Gamma', z: \recur{\oc T} , x: \dual{T}  \stpre{>_1}{u} P'$ then $x \not >_1 y$ and hence $ ( x , y ) \not \in >_1 \cup \{(u , y)\} $.

				\item\label{ch5proof:6(2)p1} $(x >_P^* c  \implies c \not  >_P^* y) \land (c >_P^* x \implies y \not >_P^*  c) $
				
				As $\Gamma', z: \recur{\oc T} , x: \dual{T}  \stpre{>_1}{u} P'$ then $\nexists c$ such that $c >_1 y \lor y >_1 c$. We show w.l.o.g. that $(x > c  \implies c \not  > y)$. The only relation on $y$ is that $u>y$ however we also have that $u >_1 x$ and hence $u>x$. Hence there exists no $c$.

				\item $ (c>_P^*x \land  y>_P^* d) \implies \neg\compatible{P}{c}{d}$ 
				
				Analogous to \ref{ch5proof:6(2)p1}.

				\item $ c>_P^*x >_P^* d \implies \neg\compatible{P}{c}{d}$

				As $\preserves{>_1}{P'}$ then $ c>_{1,P'}^*x {>_1}_P^* d \implies \neg\compatible{P'}{c}{d}$. Let us consider w.l.o.g. the following cases:
				
				\begin{enumerate}

					\item When $c = u$ then $\nexists d$ such that $\compatible{P}{u}{d}$.
					
					\item\label{ch5proof:6(2)B1} When $c = z$ then $z>_P^*a >_P^* d \implies z>_{P'}^*a >_{P'}^* d  $ and $\compatible{P}{z}{d}$ if $\compatible{P'}{z}{d}$ which does not hold, hence $\neg\compatible{P}{z}{d}$.
					
					\item When $c \not \in \{ u,z \}$ the case proceeds analagously to \ref{ch5proof:6(2)B1}.

				\end{enumerate}

			\end{enumerate}

                 \end{enumerate}


				 \item {\bf When $P = \res{xy}\vasout{z}{y}{P'}$.}

				Then, by \Cref{ch5prop:dillmakesorder}, we have the following derivation:
						\begin{mathpar}
							\inferrule*[left=\orvastype{UnOut_1}]{ 
								\Gamma', z: \recur{\oc T} , x: \dual{T}  \stpre{>_1}{u} P'
								\\
								> = >_1 \cup \{(u , y)\}
								\\
								u \not = z
							}
							{ \Gamma' , z: \recur{\oc T}  \stpre{>}{u}  \res{xy}\vasout{z}{y}{P'}  }
						\end{mathpar}

						By the induction hypothesis  we have $\preserves{>_1}{P'}$. By definition, 
						$\varcompat{P}= \varcompat{P'} \cup  \{ (x,y) , (y,x)\}.$
						
						We need to show $\preserves{>}{P}$:
                 
                 Let $(a,b) \in \varcompat{P}$. We will verify that $\acyclic{>}{P}{a}{b}$.

                There are two cases to consider:

				\begin{enumerate}
					\item $(a,b) \in \varcompat{P'}$.
						
						 By \Cref{ch5prop:auxiliary}(7), it follows that there is no $x'$ such that $\compatible{P'}{u}{x'}$, thus  $u \notin \{a,b\}$.   Since  $\Gamma\stpre{>_1}{u} P'$, it follows that {\bf (H1)}~$y\notin \names{P'}\cup \dom{\Gamma'}$, thus $y\notin \{a,b\}$. Also, by \Cref{ch5prop:auxiliary}(6) and {\bf (H1)}, it follows that there does not exist $x'$ such that {\bf (H2)} $x'>_1y\vee y>_1x'\vee \compatible{P'}{x'}{y}$.
						\begin{enumerate}
						 \item  $a \not >_P^* b$. 
						 
						 Suppose, by contradiction, that $a>_P^*b$.
						 Then, there exists a path from $a$ to $b$ via $>_P^*$, and it must be the case, by \Cref{ch5def:closure}, that either $\compatible{P}{x}{y}$, or $\compatible{P}{y}{x}$ or $u>y$ where used in the path, otherwise, we would have  $a>_{1,P'}^*b$, and this contradicts the fact that $a\not>_{1,P'}^*b$, from the hypothesis that $\preserves{>_1}{P'}$ holds. Note that $u$ does not occur in the path due to \Cref{ch5prop:auxiliary}(7). Also notice that $a \not = x$ by \Cref{ch5prop:ontleaf}/\Cref{ch5prop:auxiliary}(15) and similarly $a \not = y$ by {\bf (H2)}. Therefore, the path is, w.l.o.g., either as 
						 $$
						 \begin{aligned}
						 &a>_{1,P'}^*x\wedge \compatible{P}{x}{y}\wedge y>_{1}a_{i+1} >_{1,P'}^*b\\
				   \text{or as}~~       &a>_{1,P'}^*x\wedge \compatible{P}{x}{y}\wedge \compatible{P'}{y}{ a_{i+1}}\wedge  a_{i+1} >_{1,P'}^*b\\
						\end{aligned}
						 $$
						 contradicting {\bf (H2)}.  Hence,  $a \not >_P^* b$.
					\item  $(a >_P^* c  \implies c \not  >_P^* b) \land (c >_P^* a \implies b \not >_P^*  c)$.
						 
						  Suppose that $a >_P^* c$. If  there were a path $c>^*_P b$ it has to use either $\compatible{P}{x}{y}$, or $\compatible{P}{y}{x}$ or $u>y$, which is a contradiction (similar to the case above). Therefore,   $c\not >_P^* b$ . The proof is similar for the implication on the right.
						  
	\item $(c>_P^*a \land  b>_P^* d) \implies \neg\compatible{P}{c}{d}$.
	
	Firstly, as $\preserves{>_1}{P'}$ then $ (c{>}_{1,P'}^*a \land  b{>}_{1,P'}^* d) \implies \neg\compatible{P'}{c}{d}$. Let us consider w.l.o.g. the following cases:
				   
				   \begin{enumerate}
   
					   \item When $c = u$ with $ (u>_P^*a \land  b>_P^* d)$ then $\neg\compatible{P}{u}{d}$ holds by \Cref{ch5prop:auxiliary}(7).
					   
					   \item When $c=x$ with $ (x>_P^*a \land  b>_P^* d)$. Note that $y$ cannot occur in the path $x>_P^*y$, since it contradicts {\bf (H2)}. Therefore, the path is $ x>_{1,P'}^*a$ and $y$ has to occur somewhere in $ b>_P^* d$. By {\bf (H1)} we have that  $y\neq b$; also,  $y$ cannot occur internally in the path $b>_{P}^*d$ without contradicting {\bf (H2)}. It remains to check whether $d=y$ can occur: if that were the case, then we would have the path $b>_{P}^*y$  as $b>_P^*x\wedge \compatible{P}{x}{y}$ (since $u$ is not compatible with any name). Then, we would have $x>_{P}^*a\wedge b>_{P}^*x$ which contradicts case (ii) above. Therefore, this case is only possible in $ (x>_{1,P'}^*a \land  b>_{1,P'}^* d)$, and the result follows by the hypothesis that $\preserves{>_1}{P'}$ holds.

					   
					   \item When $c = y$ with  $  y>_P^*a \land  b>_P^* d$. 
					   
					   Then it must be the case that $\compatible{P}{y}{x}\wedge x>^*_{1,P'} a \land  b>_P^* d$. Since $x,y$ cannot occur within $ b>_P^* d$ (otherwise it would contradict case (ii)), then we must have  $\compatible{P}{y}{x}\wedge x>^*_{1,P'} a \land  b>_{1,P'}^* d$. Suppose, by contradiction, that $\compatible{P}{y}{d}$. Since $y\notin \dom{\Gamma'}\cup \names{P'}$, by \Cref{ch5prop:auxiliary} (6), it follows that there is no $ y'$ such that $y>y'\vee  y'>y \vee \compatible{P}{y}{y'}$. Therefore, it must be the case that $d=x$, and we would have $\compatible{P}{y}{x}\wedge x>_{1,P'}^*a \wedge b>^*_{1,P'} x $, which contradicts case (ii), and the result follows.

					   \item When $c \not \in \{ u,x,y \} $ with $ (c>_P^*a \land  b>_P^* d)$. 
					   
					   Let us assume the contradiction that we have a pair such that $\compatible{P}{c}{d}$. Then $d$ cannot be $u$ as this contradicts that no element is larger than $u$. Similarly $d$ cannot be $y$ as this would imply that $c=x$ as that is the only compatible channel with $y$ contradicting $c\not \in \{ u,x,y \} $ . Hence we must have that $\compatible{P'}{c}{d}$ but the induction hypotheses though we have that if $(c>_{P'}^*a \land  b>_{P'}^* d) \implies \neg\compatible{P'}{c}{d}$ and  $ \neg\compatible{P'}{c}{d} \implies \neg\compatible{P}{c}{d} $ contradicting this.
   
				   \end{enumerate}
   
   \item $ c>_P^*a >_P^* d \implies \neg\compatible{P}{c}{d}$.
   
   Suppose  $(c>_P^*a >_P^* d) $ using either $\compatible{P}{x}{y}$, or $\compatible{P}{y}{x}$ or $u>y$.  The cases where $c=u$ or $c=x$ are analogous to the analysis above. 
   
	It remains to analyze the case $c=y$, for which, we have $\compatible{P}{y}{x}\wedge x>_{1,P'}^*a >_P^* d $.  Since $y\notin \dom{\Gamma'}\cup \names{P'}$, by \Cref{ch5prop:auxiliary} (6), it follows that there is no $ y'$ such that $y>y'\vee  y'>y \vee \compatible{P}{y}{y'}$. Therefore, it must be the case that $d=x$, and we would have $\compatible{P}{y}{x}\wedge x>_{1,P'}^*a >_P^* x $. Since $y$ cannot occur in $a>_P^*x$, it follows that $\compatible{P}{y}{x}\wedge x>_{1,P'}^*a >_{1,P'}^* x $, which contradicts the fact that $\preserves{P'}{>_1}$.

						 \end{enumerate}
					\item $(a,b)\in \{ (x,y) , (y,x)\}$.
					
					We consider $(a,b) = (x,y)$ as the case of $(y,x)$ follows analogously.

			   \begin{enumerate}
   
				   \item $x \not >_P^* y$.
				   
				   As $\Gamma', z: \recur{\oc T} , x: \dual{T}  \stpre{>_1}{u} P'$ then $x \not >_1 y$ and hence $ ( x , y ) \not \in >_1 \cup \{(u , y)\} $.

				   \item\label{ch5proof:6(2)p2} $(x >_P^* c  \implies c \not  >_P^* y) \land (c >_P^* x \implies y \not >_P^*  c) $
				   
				   As $\Gamma', z: \recur{\oc T} , x: \dual{T}  \stpre{>_1}{u} P'$ then $\nexists c$ such that $c >_1 y \lor y >_1 c$. We show w.l.o.g. that $(x > c  \implies c \not  > y)$. The only relation on $y$ is that $u>y$ however we also have that $u >_1 x$ and hence $u>x$. Hence there exists no $c$.

				   \item $ (c>_P^*x \land  y>_P^* d) \implies \neg\compatible{P}{c}{d}$ 
				   
				   Analogous to \ref{ch5proof:6(2)p2}.

				   \item $ c>_P^*x >_P^* d \implies \neg\compatible{P}{c}{d}$
   
				   As $\preserves{>_1}{P'}$ then $ c>_{1,P'}^*x {>_1}_P^* d \implies \neg\compatible{P'}{c}{d}$. Let us consider w.l.o.g. the following cases:
				   
				   \begin{enumerate}
   
					   \item When $c = u$ then $\nexists d$ such that $\compatible{P}{u}{d}$.
					   
					   \item\label{ch5proof:6(2)B2} When $c = z$ then $z>_P^*a >_P^* d \implies z>_{P'}^*a >_{P'}^* d  $ and $\compatible{P}{z}{d}$ if $\compatible{P'}{z}{d}$ which does not hold, hence $\neg\compatible{P}{z}{d}$.
					   
					   \item When $c \not \in \{ u,z \}$ the case proceeds analogously to \ref{ch5proof:6(2)B2}.

				   \end{enumerate}
   
			   \end{enumerate}

					\end{enumerate}

           \item {\bf When $P=P'\pp Q$.}
           
           Then, by \Cref{ch5prop:dillmakesorder}, we have the following derivation:
			\[ \inferrule*[left=\orvastype{Par_1}]{ 
				\Gamma' \setminus u \vaslinunsplit \Delta_1 \stpre{>_1}{w} P'
				\\
				\Gamma' \vaslinunsplit \Delta_2 \stpre{>_2}{u} Q
				\\
				> = >_1 \partremove{w} \cup >_2  \cup \gtrdot^u_1 \partremove{w}
				\\
				w \text{ fresh} 
			}
			{ \Gamma' \vaslinunsplit \Delta_1, \Delta_2 \stpre{>}{u} P' \pp Q  } \]

		By the induction hypothesis both $\preserves{>_1}{P'}$ and $\preserves{>_2}{Q}$ hold. 
		
		We need to show $\preserves{>}{P}$:
                 
                 Let $(a,b) \in \varcompat{P} = \varcompat{P'}  \cup   \varcompat{Q}$. We will verify that $\acyclic{>}{P}{a}{b}$: 			To simplify the notations, in the following, we will adopt the abbreviations $\partremoveabb{1}\triangleq >_1 \partremove{w} $ and $\partremdotabb\triangleq \gtrdot^u_1 \partremove{w}=\{(u,y) \mid \exists t. y >_1 t \vee t >_1 y\wedge y\neq w\}$.


\begin{enumerate}
\item \label{ch5proof:dillpres1} When $(a,b)\in \varcompat{P'}\setminus \varcompat{Q} $:
\begin{enumerate}
\item $a \not >_P^* b$.
				
				From $\preserves{>_1}{P'}$ we have $a \not >_{1,P'}^*  b$.  We need to  consider the following cases:

				\begin{enumerate}

					\item \label{ch5case:a_eq_u_iA} When $a = u$.  From $\Gamma' \setminus u \vaslinunsplit \Delta_1 \stpre{>_1}{w} P'$ and \Cref{ch5prop:auxiliary}~(vi) we have a contradiction with the fact that $\compatible{P'}{u}{b} $.
					
					\item \label{ch5case:a_eq_w_iB}  When $a = w$.  From $\Gamma' \setminus u \vaslinunsplit \Delta_1 \stpre{>_1}{w} P'$ and \Cref{ch5prop:auxiliary}~(vii) we have a contradiction with the fact that $\compatible{P'}{w}{b} $.
					
					\item When $a \in \dom{\Gamma' \setminus u}\cup \dom{\Delta_1}\cup \dom{\Delta_2}$.  By \Cref{ch5prop:ontleaf} there is no $b$ such that $a>b$, and the result follows.
										
					\item When $a \in \bn{P'}$. 
					
					Then firslty as $\preserves{>_1}{P'}$ we have $a \not >_1  \partremove{w} b$. Secondly as $a \not \in (\dom{\Gamma' \vaslinunsplit \Delta_2} \setminus u) \cup \bn{Q}$ we apply \Cref{ch5prop:ontleaf} implying $a \not >_2 b$. Finally $a \not \gtrdot^u_1 \partremove{w} b$ by definition and the case follows as required.  

				\end{enumerate}
\item $(a >_P^* c  \implies c \not  >_P^* b) \land (c >_P^* a \implies b \not >_P^*  c) $

				We show w.l.o.g. that $(a >_P^* c  \implies c \not  >_P^* b)$ as the other conjunct is similar. 
				 We consider the following cases:

			\begin{enumerate}
			\item When $b=u$ or $b=w$. 
			
			Then, it follows similarly to \Cref{ch5case:a_eq_u_iA} and \Cref{ch5case:a_eq_w_iB}, that we have a contradiction with the fact that $\compatible{P'}{a}{b}$.
					
			\item When $b \in \dom{\Gamma' \setminus u}$.
			
			Assume  $a >_{P}^* c$. 
			
			Since  $\preserves{>_1}{P'}$, it follows  that $(a >_{1,P'}^* c  \implies c \not  >_{1,P'}^* b )$, which implies 
			$(a ~\partremovestar{1,P'} c  \implies  c \not  \partremovestar{1,P'}^* b)$.

			We must show that $(c,b)\not \in >_2  \cup ~\partremdotabb$: Firstly,  note that it must be the case that $(c,b)\not \in >_2 $, since  if $(c,b) \in >_2 $ then it must be the case that $c$ occurs in both $P'$ and $Q$, hence $c \in \dom{\Gamma'}$. However, from $\Gamma' \vaslinunsplit \Delta \stpre{>_2}{u} Q$, \Cref{ch5prop:ontleaf} and $c\neq u$, it follows that there is no $b$ such that $c>_2b$, which gives us a contradiction.  
			
			Secondly, $(c,b)\not \in \partremdotabb$ holds by definition. In fact, by definition, 
			
			$\partremdotabb=\gtrdot^u_1 \partremove{w}= \{(u,y) \mid \exists t. y >_1 t \vee t >_1 y\}\partremove{w}=\{(u,y) \mid \exists t. y >_1 t \vee t >_1 y\wedge y\neq w\}$,
			therefore, $(c,b) \in \partremdotabb$ iff $c=u$. However, we cannot have $c=u$ because $u$ is a maximal element (cf. \Cref{ch5prop:itisapartialorder}) and this would contradict the fact that $a >_{P}^* c$.

	\item\label{ch5proof:prespart2(2)} When $b \in \dom{\Delta_1}$.

	 Since $\preserves{>_1}{P'}$, it follows that  $(a >_{1,P'}^* c  \implies c \not  >_{1,P'}^* b)$,  hence also $(a \partremovestar{1,P'} c  \implies c \not  \partremovestar{1,P'} b)$.  Note that, as $b \in \dom{\Delta_1}$ then $b \not \in \dom{\Gamma' \vaslinunsplit \Delta_2 }$, therefore, there does not exist $c$ such that $c >_2 b$. Similarly to the case above, $( c,b)  \notin\partremdotabb $ holds by definition.
\item \label{ch5case:b_in_delta2} When $b \in \dom{\Delta_2}$.

 This case contradicts the hypothesis that $\compatible{P'}{a}{b} $.
\item \label{ch5case:b_in_bnP} When $b \in \bn{P'}$. 

This case is analogous to \Cref{ch5proof:prespart2(2)}.
	        \end{enumerate}
	        
	      \item $ (c>_P^*a \land  b>_P^* d) \implies \neg\compatible{P}{c}{d}$
				
				We consider the following cases:

		\begin{enumerate}
		\item When $b=u$ or $b=w$. 
			
			Then, it follows similarly to \Cref{ch5case:a_eq_u_iA} and \Cref{ch5case:a_eq_w_iB}, that we have a contradiction with the fact that $\compatible{P'}{a}{b}$.
			
%
					
			\item When $b \in \dom{\Gamma' \setminus u} \cup \dom{\Delta_1} \cup \dom{\Delta_2}$.

			The case follows from \Cref{ch5prop:auxiliary}(15) as there is no channel $d$ such that $b >_P^*d$.



						

					
					\item When $b \in \bn{P'}$.
					
					Then the proof follows analogously to case \Cref{ch5case:b_in_bnP} above.
%
\end{enumerate}
\item $ c>_P^*a >_P^* d \implies \neg\compatible{P}{c}{d}$
				
				We consider the following cases:

				\begin{enumerate}

%
					
					\item When $c = u$.
					
					From   $\Gamma' \vaslinunsplit \Delta_1, \Delta_2 \stpre{>}{u} P' \pp Q  $, then from \Cref{ch5prop:auxiliary}(\ref{ch5prop:case_compatible}), it follows that there is no $x$ such that $\compatible{P}{u}{x}$, therefore, $\neg\compatible{P}{u}{d}$ holds.
					
					\item When $c = w$.
					
					Since $\Gamma' \setminus u \vaslinunsplit \Delta_1 \stpre{>_1}{w} P'$, then from \Cref{ch5prop:auxiliary}(\ref{ch5prop:case_compatible}), it follows that there is no $x$ such that $\compatible{P'}{w}{x}$, therefore, $\neg\compatible{P}{w}{d}$ holds. 
					
					\item\label{ch5proof:dillpreserve2} When $c \in \dom{\Gamma' \setminus u} \cup \dom{\Delta_1}\cup \dom{\Delta_2}$.
					
					The case follows from \Cref{ch5prop:auxiliary}(15) as there is no channel $a$ such that $c>_P^*a$.

					
					

					
					\item When $c \in \bn{P'}$.
					
Then firstly we must have that $ c>_{1,P}^*a $ as  $ c \not >_{2,P}^*a $ as  we have just seen above that there is no $ e$ such that $c >_2 e$. Let us consider the following two cases:

If  $a >_{1,P'}^* d$ then by the induction hypothesis $ \neg\compatible{P'}{c}{d}$ holds, which implies   $ \neg\compatible{P}{c}{d}$.

						If  $a \not >_{1,P}^* d$ then $a \not >_P^* d$ must be introduced via $>_2$. Then there is some sequence $a >_{1,P}^* d_1 >_{2,P}^*  .... >_{1,P}^* d_n >_{2,P}^* d$, In this senario we have that $d_1$ is in the relations of both $>_1$ and $>_2$ hence $d_1 \in \dom{\Gamma}$ however we are now contradicting that there exists a name $e$ such that $d_1 >_{2,P}^* e$.

				\end{enumerate}
\end{enumerate}
\item When $\compatible{Q}{a}{b} \land \neg\compatible{P'}{a}{b} $ then the case follows analogously to \Cref{ch5proof:dillpres1}.

			\item When $\compatible{P'}{a}{b} \land \compatible{Q}{a}{b} $ then $P \not \in \dilllang$. This is as typing ensures that $a,b \in \Gamma'$, however then this implies that $a$ and $b$ are not compatible as at most $1$ channel may be free.
\end{enumerate}           
		\item {\bf When $P = \res{z v}( P' \pp Q )$.}
		
          Then, by \Cref{ch5prop:dillmakesorder}, we have the following derivation:

		\[
			\small
			\inferrule*[left=\orvastype{Par_2}]{ 
				z:V , (\Gamma' \setminus u \vaslinunsplit \Delta_1) \stpre{>_1}{z} P'
				\\
				v:\dual{V} , (\Gamma' \vaslinunsplit \Delta) \stpre{>_2}{u} Q
				\\
				> = >_1  \cup >_2  \cup \gtrdot^u_1
			}
			{ \Gamma' \vaslinunsplit \Delta_1, \Delta \stpre{>}{u} \res{z v}( P' \pp Q ) }
		\]

		By the induction hypothesis both $\preserves{>_1}{P'}$ and $\preserves{>_2}{Q}$  hold. 
		
		We need to show $\preserves{>}{P}$:

		By \Cref{ch5def:compN} the set of compatible names is as 
		\[ 
			\begin{aligned}
				\varcompat{P}  = &\{ (z,v) , (v,z) \} \cup \varcompat{P'} \cup \varcompat{Q} \cup \\
				&   \{ (x, y) \mid \compatiblecc{ \cc{z}{ \res{z v}(P \pp Q)}}{\cc{ v}{\res{z v}(P \pp Q)}}{x}{y} \} \\
			\end{aligned}
		 \]
		 We will verify that $\acyclic{>}{P}{a}{b}$: To simplify the notations, in the following, we will adopt the abbreviation $\abbgrdot \triangleq \gtrdot^u_1$

		\begin{enumerate}

			\item When $(a,b) \in \{ (z,v) , (v,z)\}$ . We consider the case $(a,b) = (z,v)$ as the case of $(v,z)$ follows similarly.

			\begin{myEnumerate}

				\item $v \not >_P^* z$.
				
				From $v:\dual{V} , (\Gamma' \vaslinunsplit \Delta) \stpre{>_2}{u} Q$ and \Cref{ch5prop:ontleaf} we know there is no $c$ such that  {\bf (H1)} $ v >_2 c$.  From $z:V , (\Gamma' \setminus u \vaslinunsplit \Delta_1) \stpre{>_1}{z} P'$ and applying \Cref{ch5prop:auxiliary}(6) we have that there is no $c$ such that {\bf (H2)} $ v >_1 c$ (which we may extend to there not existing a $c$ such that {\bf (H3)} $v\gtrdot^u_1c$ by definition) or  {\bf (H4)} $\compatible{P'}{v}{c}$.
				
				 Suppose,  by contradiction, assume that $v >_P^* z$. Then, from {\bf (H1)} - {\bf (H4)}, we must have $\compatible{P}{v}{d} \land d >_P^* z$ for some $d$. We have one of two options: (i)  $d = z$; and (ii)  $(v,d) \in \varcompat{Q}$ as, by definition, we have  $(v,d) \not \in \{ (x, y) \mid \compatiblecc{ \cc{z}{ \res{z v}(P \pp Q)}}{\cc{ v}{\res{z v}(P \pp Q)}}{x}{y} \}$.

				\begin{enumerate}
					\item When $d = z$.
					
					Irreflexivity of $>_2$ implies $z \not>_2 z$ and  \Cref{ch5def:acyclic}(ii) implies $z {\not>}_{2,P}^* z$. By \Cref{ch5prop:auxiliary}(6) $z \not>_1 e \land z\not\gtrdot^u_1e$ for any $e$ and as already shown, $z$ is not compatible to any names in $P'$. Hence we must have $\compatible{Q}{z}{f} \land f \not >_P^* z$ with $(z,f) \in \varcompat{Q}$. Let us consider the next step taken that $f$ can make:
					\begin{enumerate}
						\item If $f \not>_2 z$ as this introduces the contradiction $z \not>_2 z$.
						\item If $f >_1 g$ then this implies that $f\in\dom{\Gamma}$ contradicting \Cref{ch5prop:ontleaf}.
						\item If $\compatible{P'}{f}{g}$ then this implies that $f\in\dom{\Gamma}$ and by \Cref{ch5prop:auxiliary}(15) there doesn't exist a $h$ such that $g >_P^* h$.
						\item If $f >_2 g $ then the case follows by induction on the length of $f \not >_P^* z$ showing that no path exists.
						\item If $\compatible{Q}{f}{g}$ then the case follows by induction on the length of $f \not >_P^* z$ showing that no path exists.
					\end{enumerate}

					\item When $(v,d) \in \varcompat{Q}$.
					 
					 Then we must have that $d \in \names{Q}$. 
					 
					 If $d \not \in \dom{\Gamma}$  then by \Cref{ch5prop:auxiliary}(15) we have that $d\not >_1 e$ for any $e$ and by \Cref{ch5prop:auxiliary}(6) $d\not >_2 e$. It follows that we must have $\compatible{P}{d}{e} \land e \not >_P^* z$ and by induction on the length of the chain of connected channels we can see that every channel cannot perform a step within the partial order and hence we may never reach $z$. In the case $d \in \dom{\Gamma}$ then it follows from \Cref{ch5prop:auxiliary}(15).
				\end{enumerate}



				\item $(z >_{P}^* c  \implies c \not  > _{P}^* v) \land (c >_{P}^* z \implies v \not >_{P}^*  c) $
				
				As $\Gamma', z: \recur{\oc T} , x: \dual{T}  \stpre{>_1}{u} P'$ then as $v \not \in \dom{\Gamma', z: \recur{\oc T} , x: \dual{T} } \cup \names{P'}$ we apply \Cref{ch5prop:auxiliary}(6) and  there doesn't exist a $c$ such that $c >_{1,P'}^* v \lor v >_{1,P'}^* c \lor \compatible{P'}{v}{c}$. We show w.l.o.g. that $z >_{P}^* c  \implies c \not  > _{P}^* v$. The only relations on $v$ in $>_2$ are that $u >_2 v$ (which we may exclude as $u$ is the greatest channel in $>$ contradicting $z >_{P}^* u$) or $e >_2  v$ where $e$ is bound (the second condition comming from \Cref{ch5prop:auxiliary}(6) implying that $e \in \dom{v:\dual{V} , (\Gamma' \vaslinunsplit \Delta)} \cup \names{Q}$ and \Cref{ch5prop:ontleaf} implies that $e \not \in \dom{v:\dual{V} , (\Gamma' \vaslinunsplit \Delta)} \cup \fn{Q}$). Hence any compatible channel to $e$ must also be bound (as any free channel $f$ compatible to $e$ would give $f >_{2,P'}^*  v$ contradicting \Cref{ch5prop:ontleaf}), and hence both $e$ and all compatible channels to $e$ must be in the names of $Q$. Finally any bound name cannot be compatible with the bound names in $P'$ hence $c \not > _{P}^* v$.

				\item\label{ch5proof:case6(2)} $ (c>_P^*z \land  v>_P^* d) \implies \neg\compatible{P}{c}{d}$.

				 Note that the only channel larger then $z$ is $u$ (obtained from $\gtrdot^u_1$) and hence $c = u$.  From $\Gamma' \vaslinunsplit \Delta_1, \Delta \stpre{>}{u} \res{z v}( P' \pp Q )$  and \Cref{ch5prop:auxiliary} we know that no channel is compatible with $u$ hence the case holds.

				\item $ c>_P^* z >_P^* d \implies \neg\compatible{P}{c}{d}$.
				
				 Again, we note that the only channel larger than $z$ is $u$ and this case proceeds analogously to \ref{ch5proof:case6(2)}.

			\end{myEnumerate}

			\item When $(a,b) \in \{ (x, y) \mid \compatiblecc{ \cc{z}{ \res{z v}(P \pp Q)}}{\cc{ v}{\res{z v}(P \pp Q)}}{x}{y} \} $.
			
			 Then case holds by \Cref{ch5prop:bigproofforonething}.
			
			\item\label{ch5proof:case6(3)} When $(a,b) \in  \varcompat{P'} $ then:
			
			\begin{myEnumerate}

				\item $a \not >_P^* b$.

				As $(a,b) \in  \varcompat{P'} $ we must have either i) both $a$ and $b$ are bound names within $P'$ or ii) exactly one of $a$ or $b$ is free within $P'$ (from \Cref{ch5prop:auxiliary}(10)).

				When i) then $a \not >_{P'}^* b$ by the induction hypothesis from $\preserves{>_1}{P'}$. Let us assume the contradiction $a >_{P'}^* b$ then as $\compatible{P}{a}{b}$ we also have $b >_{P'}^* b$ contradiciting \Cref{ch5prop:auxiliary}(16).

				When ii) then the case holds by \Cref{ch5prop:auxiliary}(15).

				By the induction hypothesis both $\preserves{>_1}{P'}$ hence $a \not >_1 b$, we have that $a \not \in \dom{\Gamma' \vaslinunsplit \Delta_1, \Delta}\setminus u$ (if $a=u$ then not $b$ would be compatible) hence $a \in \bn{P'}$. As $a \in \bn{P'}$ then there does not exist an $c$ such that $a >_2 c $ hence $a \not > b$.

				\item $(a >_{P}^* c  \implies c \not  > _{P}^* b) \land (a >_{P}^* z \implies v \not >_{P}^*  b) $

				We show $(a >_{P}^* c  \implies c \not  > _{P}^* b)$ as the other case follows similarly. As $\preserves{>_1}{P'}$ by the induction hypothesis then $(a >_{1,P'}^* c  \implies c \not  >_{1,P'}^* b)$. Let us consider $a$:

				\begin{myEnumerate}
					
					\item $b\in \{u,z,v\} $ then this contradicts $\compatible{ P' }{a}{b}$.
					
%
%
%

					\item\label{ch5proof:case6(3:1)} $b \in \dom{ \Gamma' \vaslinunsplit \Delta_1, \Delta }$ then $(a >_{1,P'}^* c  \implies c \not  >_{1,P'}^* b)$. We show that both $c \not  >_{2,P'}^* b$ and $c \not \abbgrdotstar{1,P'}  b$. As $c \not = u$ by $a >_{P}^* c$ then we have that $c \not  \abbgrdotstar{1,P'} b$. As $c \in \bn{P'}$ then $c \not  >_{2,P'}^* b$.
					
					\item $b \in \bn{P'}$ then we proceed analogously to \ref{ch5proof:case6(3:1)}.
					
					\item $b \in \bn{Q}$ then then this contradicts $\compatible{ P' }{a}{b}$.

				\end{myEnumerate}

				\item $ (c>_P^*a \land  b>_P^* d) \implies \neg\compatible{P}{c}{d}$ then we consider the following cases:

				\begin{myEnumerate}
					
					\item When $b \in \{u,z,v\}$ then this contradicts $\compatible{ P' }{a}{b}$
					
%
					
					\item When $b \in \dom{ \Gamma' \vaslinunsplit \Delta_1, \Delta }$ then this contradicts $ b>_P^* d$.

					\item When $b \in \bn{P'}$ then  by the induction hypothesis we have that $ (c>_{1,P'}^*a \land  b>_{1,P'}^* d) \implies \neg\compatible{P'}{c}{d}$. As $b \in \bn{P'}$ then there does not exist a $d$ such that $b>_{2,P'}^* d$ and similarly by definition $b \not\abbgrdotstar{1,P'} d$.

					\item When $b \in \bn{Q}$ then then this contradicts $\compatible{ P' }{a}{b}$.

				\end{myEnumerate}
				

				\item $ c>_P^* a >_P^* d \implies \neg\compatible{P}{c}{d}$ then as $\compatible{ P' }{a}{b}$ and $ (c>_P^*a \land  b>_P^* d) \implies \neg\compatible{P}{c}{d}$ the case follows.

			\end{myEnumerate}
			
			\item When $(a,b) \in  \varcompat{Q}$ then case proceeds similarly to \ref{ch5proof:case6(3)}.

		\end{enumerate}

		\item {\bf When $P = \res{xy}\vasout{z}{x}{(P' \pp Q)}$.}

		\begin{enumerate}
			\item\label{ch5proof:dillpreserve7} Case $u \not = z$.
			
			 By \Cref{ch5prop:dillmakesorder}, we have the following derivation:
				\begin{mathpar}
					\inferrule*[left=\orvastype{LinOut_3}]{ 
						\Gamma'\setminus u \vaslinunsplit \Delta_1, y:  \dual{T} \stpre{>_1}{y} P'
						\\ 
						\Gamma' \vaslinunsplit \Delta_2  , z:S \stpre{>_2}{u} Q
						\\
						{> = >_1  \cup >_2  \cup \gtrdot^u_1\cup \{(u,x)\}}
						\\
						u \not = z
					}
					{ \Gamma' \vaslinunsplit \Delta_1, \Delta_2 , z: \lin \vassend{T}{S}  \stpre{>}{u}  \res{xy}\vasout{z}{x}{(P' \pp Q)}  }
				\end{mathpar}

				By the induction hypothesis, it follows that $\preserves{>_1}{P'}$ and $\preserves{>_2}{Q}$.

				We  want to show that $\preserves{>}{P}=\bigwedge_{(a,b)\in\varcompat{P}} \acyclic{>}{P}{a}{b}$ where,  by \Cref{ch5def:compN},  we have that $\varcompat{P}=\{ (x,y) , (y,x) \} \cup \varcompat{P'} \cup \varcompat{Q}$.

				\begin{myEnumerate}
					\item When $ (a,b) \in \{ (c, d)| \compatible{{P'}}{c}{d} \} \cup \{ (c, d)| \compatible{ Q }{c}{d} \} $

					\begin{myEnumerate}

						\item\label{ch5proof:dillpreserve5} When $\compatible{P'}{a}{b} \land \neg\compatible{Q}{a}{b} $ then we show $\acyclic{>}{P}{a}{b}$:
						
						\begin{myEnumerate}
			
							\item\label{ch5proof:ispreord2(1)2} $a \not >^*_P b$.
							
							Then as $\preserves{>_1}{P'}$ we have $a {\not >}_{1,P'}^* b$. We prove by contradiction assuming that $a >^*_P b$. Let us consider the following cases w.l.o.g.:
			
							\begin{myEnumerate}
			
								\item When $a = u$ then $\Gamma' \vaslinunsplit \Delta_1, \Delta_2 , z: \lin \vassend{T}{S}  \stpre{>}{u}  \res{xy}\vasout{z}{x}{(P' \pp Q)}$ and by \Cref{ch5prop:auxiliary}(7) we have the contradiction $\compatible{P}{u}{b}$.
								
								\item  When $a = y$ then $\Gamma'\setminus u \vaslinunsplit \Delta_1, y:  \dual{T} \stpre{>_1}{y} P'$ and by \Cref{ch5prop:auxiliary}(7) we have the contradiction $\compatible{P'}{a}{b} $.
								
								
								\item  When $a = z$ then as $\Gamma'\setminus u \vaslinunsplit \Delta_1, y:  \dual{T} \stpre{>_1}{y} P'$ this contradicts $\compatible{P}{z}{b} $ by \Cref{ch5prop:auxiliary}(6).
								
								\item When $a \in \dom{\Gamma' \setminus u}\cup \dom{\Delta_1}\cup \dom{\Delta_2}$ then case holds as no channel is smaller  then $a$ by \Cref{ch5prop:auxiliary}(15), hence $b$ cannot exist such that $a>b$.
								
								
								
								\item When $a \in \bn{P'}$ then firslty  $a \not >_1 b$ secondly as $a \not >_2 b$ as $a$ does not occur in $Q$, thirdly $a \not \gtrdot^u_1 b$ by definition and finally $\{(u,x)\}$ does not introduce $a>b$. Let us assume the contradiction $a >_{P'}^* b$ then as $\compatible{P}{a}{b}$ we also have $b >_{P'}^* b$ contradiciting \Cref{ch5prop:auxiliary}(16).
								
								\item When $a \in \bn{Q}$ then this contradicts $\compatible{P'}{a}{b} $ by \Cref{ch5prop:auxiliary}(6).
			
							\end{myEnumerate}

							\item $(a >_{P}^* e  \implies e \not  >_{P}^* b) \land (e >_{P}^* a \implies b \not >_{P}^*  e) $ 
			
							We show that $(a > e  \implies e \not  > b)$ and the other case follows similarly. As established in \ref{ch5proof:ispreord2(1)2} let $a \in \bn{P'}$ and we consider the following cases:

							\begin{myEnumerate}
			
								\item When $b = u$ then $\Gamma'\setminus u \vaslinunsplit \Delta_1, y:  \dual{T} \stpre{>_1}{y} P'$ contradicts $\compatible{P'}{a}{u} $.
								
								\item When $b = y$ then this contradicts $a > y$ as $y$ is the greatest element in $>_1$.
								
								\item When $b = z$ then this contradicts $\compatible{P'}{a}{z} $.
								
								\item When $b \in \dom{\Gamma' \setminus u}$ then as $\preserves{>_1}{P'}$ we have $(a >_{1,P'}^* e  \implies e \not  >_{1,P'}^* b)$. We show that both $e \not  >_{2,P'}^* b$ and $e \not \abbgrdotstar{1,P'}  b$. As $e \not = u$ by $a >_{P}^* e$ then we have that $e \not  \abbgrdotstar{1,P'} b$. As $e \in \bn{P'}$ then $e \not  >_{2,P'}^* b$.

								\item\label{ch5proof:dillpreserve8} When $b \in \dom{\Delta_1}$ then as $\preserves{>_1}{P'}$ we have $(a >_{1,P'}^* e  \implies e \not  >_{1,P'}^* b)$. As $b \in \dom{\Delta_1}$ then $b \not \in \dom{\Gamma' \vaslinunsplit \Delta }$ hence there doesn't exist $e$ such that $e >_{2,Q}^* b$. Thus, $ e \not\gtrdot^u_1 b$ holds by definition as $e$ must be $u$ resulting in the contradiction $a >_{P}^* u$ and finally $ (c,b)\not \in \{(u,x)\}$ holds trivially.
								
								\item When $b \in \dom{\Delta_2}$ then this contradicts $\compatible{P'}{a}{b} $.
								
								\item When $b \in \bn{P'}$ then analogous to \ref{ch5proof:dillpreserve8}.
			
							\end{myEnumerate}
			
							\item $ (e>_P^*a \land  b>_P^* d) \implies \neg\compatible{P}{e}{d}$ 
							
								We consider the following cases:

								\begin{myEnumerate}

									\item When $b = u$ then $\Gamma' \vaslinunsplit \Delta_1, \Delta_2 , z: \lin \vassend{T}{S}  \stpre{>}{u}  \res{xy}\vasout{z}{x}{(P' \pp Q)} $ contradicts $\compatible{P}{a}{u} $.
									
									\item When $b = y$ then $\Gamma'\setminus u \vaslinunsplit \Delta_1, y:  \dual{T} \stpre{>_1}{y} P'$ contradicts $\compatible{P'}{a}{y} $.

									\item When $b = z$ then this contradicts $\compatible{P'}{a}{z} $.

									\item When $b \in \dom{\Gamma' \setminus u}\cup \dom{\Delta_1}$ then there is no $d$ such that $b>_P^* d$

									
									\item When $b \in \dom{\Delta_2}$ then this contradicts $\compatible{P'}{a}{b} $.
									
									\item When $b \in \bn{P'}$ then as $\preserves{>_1}{P'}$ we have $ (e>_{1,P'}^*a \land  b>_{1,P'}^* d) \implies \neg\compatible{P}{e}{d}$ . As $b \in \bn{P'}$ then there doesnt exist $d$ such that $b >_{2,P'}^* d$ or $ b \gtrdot^u_1 d$ holds by definition and $\{(u,x)\}$ holds trivially.

								\end{myEnumerate}

							\item $ c>_P^*a >_P^* d \implies \neg\compatible{P}{c}{d}$  then as $\compatible{ P' }{a}{b}$ and $ (c>_P^*a \land  b>_P^* d) \implies \neg\compatible{P}{c}{d}$ the case follows.

						\end{myEnumerate}

						\item When $\compatible{Q}{a}{b} \land \neg\compatible{P'}{a}{b} $ then the case follows analogously to \ref{ch5proof:dillpreserve5}.

						\item When $\compatible{P'}{a}{b} \land \compatible{Q}{a}{b} $ then $P \not \in \dilllang$. This is if typing ensures that $a,b \in \Gamma'$, however then this implies that $a$ and $b$ are not compatible as at most $1$ channel may be free.
						
					\end{myEnumerate}

					\item When $(a,b) \in \{ (x,y) , (x,y) \}$.

					 We consider $(a,b) = (x,y)$ as the case of $(y,x)$ follows analogously.
		
					\begin{myEnumerate}
		
						\item $x \not >^*_P y$. 
						
						Then from $\Gamma'\setminus u \vaslinunsplit \Delta_1, y:  \dual{T} \stpre{>_1}{y} P'$ and $\Gamma' \vaslinunsplit \Delta_2  , z:S \stpre{>_2}{u} Q$ and by \Cref{ch5prop:auxiliary}(6) we have that $\not \exists c$ such that $  x >_1 c \lor x >_2 c $. By the definition of 
						$\gtrdot^u_1$ then $ x \not >_1 c$ implies $x \not \gtrdot^u_1 c$ and $x \not = u$ hence $\{(u,x)\}$ does not need to be considered. Let us assume the contradiction $x >^*_P y$, then we must have that $ \compatible{P}{x}{d} \land d >^*_P y$. The only channel compatible with $x$ is $y$ giving $d = y$ and the contradiction $y >^*_P y$.


						\item\label{ch5proof:dillpreserve6} $(x > c  \implies c \not  > y) \land (c > x \implies y \not >  c) $
						
						As $\Gamma'\setminus u \vaslinunsplit \Delta_1, y:  \dual{T} \stpre{>_1}{y} P'$ and $\Gamma' \vaslinunsplit \Delta_2  , z:S \stpre{>_2}{u} Q$ then $\nexists c$ such that $c >_1 x \lor x >_1 c \lor c >_2 x \lor x >_2 c$. We show w.l.o.g. that $(c > x \implies y \not >  c)$. The only relation on $x$ is that $u>x$ however $u$ is the greatest element hence we must have that $y \not >  c$.

						\item $ (c>_P^*x \land  y>_P^* d) \implies \neg\compatible{P}{c}{d}$ 
						
						Analogous to \ref{ch5proof:dillpreserve6}.

						\item $ c>_P^*x >_P^* d \implies \neg\compatible{P}{c}{d}$

						As $\Gamma'\setminus u \vaslinunsplit \Delta_1, y:  \dual{T} \stpre{>_1}{y} P'$ and $\Gamma' \vaslinunsplit \Delta_2  , z:S \stpre{>_2}{u} Q$ then as established the we only have the relation $u > x$ and as there does not exist any name $e$ such that $\compatible{P}{u}{e}$ the case holds.

					\end{myEnumerate}

				\end{myEnumerate}

			\item When  $u  = z$.
			
			 Then, by \Cref{ch5prop:dillmakesorder}, we have the following derivation:
				\begin{mathpar}			
					\inferrule*[left=\orvastype{LinOut_4}]{ 
						\Gamma' \vaslinunsplit \Delta_1, y:  \dual{T} \stpre{>_1}{y} P'
						\\ 
						\Gamma' \vaslinunsplit \Delta_2  , z:S \stpre{>_2}{z} Q
						\\
						{> = >_1  \cup >_2  \cup \gtrdot^z_1 \cup \{(z,x)\}
						}
					}
					{ \Gamma' \vaslinunsplit \Delta_1, \Delta_2 , z: \lin \vassend{T}{S}  \stpre{>}{z}  \res{xy}\vasout{z}{x}{(P' \pp Q)}  }
				\end{mathpar}

				This case follows analagously to case  \ref{ch5proof:dillpreserve7} above.

		\end{enumerate}

\end{enumerate}
\end{proof}


We now define how to obtain a level function based on a (flattened) partial order:

\begin{definition}{}
Let $>$ be a strict partial order.
	The level function constructed from $>$,   denoted $\lvlpartord{>}$, is defined as:
\[
	\lvlpartord{>}({x}) = 
		\begin{cases}
			1 & \text{if }  \nexists \,  y \text{ such that } x > y \\ 
			max{\{ \lvlpartord{>}({y}) \mid  x>y \}} + 1 & \text{otherwise}
		\end{cases}
\]
\end{definition}

\begin{lemma}\label{ch5lem:subfunction}
If $>_1 \,\subset\, >$ then $\lvlpartord{>_1}$ is a subfunction of $\lvlpartord{>}$.
\end{lemma}

\subsubsection{Main Result}

Finally, we can prove \Cref{ch5thm:main_result}, our main result:
\begin{theorem}
Suppose $\sttodillj{\Gamma  \st P}_u$ with $ \Gamma \stpre{>}{u} P $ and 
let $>_\mathsf{f} \,= \flaten{>}{P}$.
Then $\sttoltj{\Gamma  \st P}{ \lvlpartord{>_\mathsf{f}}}{}$.
\end{theorem}


\begin{proof}
	The proof proceeds by induction on $P$. Note the following:
	\begin{description}
		\item[(H1)] the existence of $>$ is guaranteed by \Cref{ch5prop:dillmakesorder}.
		\item[(H2)] $>$ is a strict partial order with maximum element $u$, as ensured by \Cref{ch5prop:itisapartialorder}.
		\item[(H3)] $\preserves{>}{P}$ holds, by  \Cref{ch5proof:dillispres}.
		\item[(H4)] $>_\mathsf{f}$ is a strict partial order with maximal element $u$ and $\preserves{>_\mathsf{f}}{P}$ by \Cref{ch5prop:flatpreserves2}.
	\end{description}
	We consider only two key cases (input and composition) and the base case of $P = \nil$. All other cases follow by the induction hypothesis.

\begin{enumerate}
	\item When $P = \nil$.
		
		 Then we have the following derivations:
		
		\begin{itemize}
		\item By definition of $\sttodillj{\Gamma  \st \nil}_u$ we have 
			\begin{mathpar}
		\inferrule{  }
			{ \dillnotun{\sttodillt{\Gamma'}} ;  \cdot \dill \nil :: u : \dillnilT }
		\end{mathpar}
		\item From the proof of \Cref{ch5prop:dillmakesorder}, we know
		\begin{mathpar}
			\inferrule*[left=\orvastype{Nil}]{ > \defeq \{ (u , x) \mid   x \in \Gamma \setminus u \} }
		{ \Gamma \stpre{>}{u} \nil  }
		\end{mathpar}
		\end{itemize}
		By definition, $>_\mathsf{f} = \flaten{>}{\nil} = >$. Hence we may construct $\sttoltj{\Gamma  \st \nil}{ \lvlpartord{>_\mathsf{f}}}{}$ to be:
		\begin{mathpar}
			\inferrule*[left= \lvltype{Nil}]
			{ \contun{\sttoltt{\Gamma}{\lvlpartord{>}}{}}}
			{ \sttoltt{\Gamma}{\lvlpartord{>}}{} \lt \nil  }
		\end{mathpar}

		which holds regardless of the level function.

		\item When $P =  \vasin{\un}{x}{y}{P'}$.
		
		 Then $\sttodillj{\Gamma , x: \recur{\wn T} \st P}_u$ with $u = x$, hence $ \Gamma , x: \recur{\wn T}\stpre{>}{x} P $ and $x$ is the maximum element within $>$. As	
		\[
			\inferrule*[left=\orvastype{UnIn_2}]{ 
				\Gamma,  y : T \stpre{>_1}{y} P
				\\
				>\defeq >_1\cup (\gtrdot^x_1)
				\\
				\notclient{T} \land \server{T}
			}
			{\Gamma , x: \recur{\wn T} \stpre{>}{x}  \vasin{\un}{x}{y}{P}   }
		\]

		By applying the induction hypothesis on $\sttodillj{\Gamma,  y : T \st P'}_u$ and $\Gamma,  y : T \stpre{>_1}{y} P'$ it follows that $\sttoltj{\Gamma,  y : T  \st P'}{ \lvlpartord{\flaten{>_1}{P'}} }{}$. 
		
		Also, since  $>_1 \subset >$ it follows that  $\flaten{>_1}{P'} \subset \flaten{>}{P'} = \flaten{>}{P} $ and  $\lvlpartord{\flaten{>_1}{P'}}$ is a subfunction of $\lvlpartord{\flaten{>}{P}}$; thus, $\sttoltj{\Gamma,  y : T  \st P'}{ \lvlpartord{\lvlpartord{\flaten{>}{P}}} }{}$. Then we have the following derivation:

		\[ 
			\mprset{flushleft}
			\inferrule 
			{
			\sttoltt{\Gamma'}{\lvlpartord{\flaten{>}{P}}}{} \lt x:\lvlunst{\lvlpartord{\flaten{>}{P}}(x)}{\sttoltt{T}{\lvlpartord{\flaten{>}{P}}}{y}} \\\\
			\sttoltt{\Gamma'}{\lvlpartord{\flaten{>}{P}}}{} , y:\sttoltt{T}{\lvlpartord{\flaten{>}{P}}}{y},  z : \lvlnilT   \lt \sttoltp{P'}{}  \\
			\forall b \in \outsubj{\sttoltp{P'}{}}, \ \lvlpartord{\flaten{>}{P}}({b}) < \lvlpartord{\flaten{>}{P}}({x})
			}
			{
			\sttoltt{\Gamma}{\lvlpartord{\flaten{>}{P}}}{} ,  x:\lvlunst{\lvlpartord{\flaten{>}{P}}(x)}{\sttoltt{T}{\lvlpartord{\flaten{>}{P}}}{y}}  \lt \bang x \inp {y , z}.\sttoltp{P'}{} 
			}
		\]

		as the side condition $\forall b \in \outsubj{\sttoltp{P'}{}}, \ \lvlpartord{\flaten{>}{P}}({b}) < \lvlpartord{\flaten{>}{P}}({x})$ holds due to $x$ being maximum.

		\item When $P = \res{z v}( P' \pp Q )$. 
		
		Then $\sttodillj{\Gamma \vaslinunsplit \Delta_1, \Delta_2 \st P}_u$, hence $\Gamma \vaslinunsplit \Delta_1, \Delta_2 \stpre{>}{u} \res{z v}( P' \pp Q ) $ and $u$ is the maximum element within $>$. As we have both the following:

		\[
			\inferrule{
				\sttodillj{ \Gamma', \Delta_1, z:V \st P' }_z  \\
				\sttodillj{ \Gamma , \Delta_2 ,z: \dual{V} \st  Q\substj{z}{v} }_u 
				}
			{ \dillnotun{\sttodillt{\Gamma'}} ; \sttodillt{\Delta_1} , \sttodillt{\Delta_2'}  \dill \res{z}( \sttodillp{P'} \pp \sttodillp{Q}\substj{z}{v} ) ::  u: A }
		\]

		\[
			\small
			\inferrule*[left=\orvastype{Par_2}]{ 
				z:V , (\Gamma \setminus u \vaslinunsplit \Delta_1) \stpre{>_1}{z} P'
				\\
				v:\dual{V} , (\Gamma \vaslinunsplit \Delta_2) \stpre{>_2}{u} Q
				\\
				> \defeq >_1  \cup >_2  \cup \gtrdot^u_1
			}
			{ \Gamma \vaslinunsplit \Delta_1, \Delta_2 \stpre{>}{u} \res{z v}( P' \pp Q ) }
		\]
  where $\Gamma' = \Gamma \setminus u$ and $\Delta_2' = \Delta_2\setminus u$.

		By applying  the induction hypothesis on $\sttodillj{ \Gamma', \Delta_1, z:V \st P' }_z$ and $z:V , (\Gamma \setminus u \vaslinunsplit \Delta_1) \stpre{>_1}{z} P'$ it follows  that $\sttoltj{ \Gamma', \Delta_1, z:V \st P'}{ \lvlpartord{\flaten{>_1}{P'}} }{}$.

		Since  $>_1 \subset \ > \ \subset \flatcc{>}{\cc{ z}{P'}}{\cc{ v }{Q}}$, it follows that 

		\[
			\begin{aligned}
				\flaten{>_1}{P'}  & \subset \flaten{>}{P'} \\
				& \subset \flaten{\flatcc{>}{\cc{ z}{P'}}{\cc{ v }{Q}}}{P'} \\
				& \subset  \partjoin{\flaten{\flaten{\flatcc{>}{\cc{ z}{P'}}{\cc{ v }{Q}}}{  P' }}{ Q }}{z}{v} \\
				& = \flaten{>}{P} 
			\end{aligned}
		\]

		and $\lvlpartord{\flaten{>_1}{P'}}$ is a subfunction of $\lvlpartord{\flaten{>}{P}}$ so  $\sttoltj{ \Gamma', \Delta_1, z:V \st P'}{ \lvlpartord{\flaten{>}{P}} }{}$.

		Secondly, by the induction hypothesis  $\sttodillj{ \Gamma , \Delta_2 ,v: \dual{V} \st  Q }_u $ and $v:\dual{V} , (\Gamma \vaslinunsplit \Delta_2) \stpre{>_2}{u} Q$ implies that $\sttoltj{ \Gamma , \Delta_2 ,v: \dual{V} \st  Q}{ \lvlpartord{\flaten{>_2}{Q}} }{}$.

		As $>_2 \subset > \subset \flatcc{>}{\cc{ z}{P'}}{\cc{ v }{Q}} \subset \flaten{\flatcc{>}{\cc{ z}{P'}}{\cc{ v }{Q}}}{  P' }$ then:

		\[
			\begin{aligned}
				\flaten{>_2}{Q}  & \subset \flaten{>}{Q} \\
				& \subset \flaten{\flaten{\flatcc{>}{\cc{ z}{P'}}{\cc{ v }{Q}}}{  P' }}{Q} \\
				& \subset  \partjoin{\flaten{\flaten{\flatcc{>}{\cc{ z}{P'}}{\cc{ v }{Q}}}{  P' }}{ Q }}{z}{v} \\
				& = \flaten{>}{P} 
			\end{aligned}
		\]

		and $\lvlpartord{\flaten{>_2}{Q}}$ is a subfunction of $\lvlpartord{\flaten{>}{P}}$ so  $\sttoltj{ \Gamma , \Delta_2 , v: \dual{V} \st  Q }{ \lvlpartord{\flaten{>}{P}} }{}$.

		Hence we construct the following derivation:
		\[
			\hspace{-0.5cm}
			\small
			\inferrule*[left=\lvltype{Res}]{ 
				\inferrule*[left=\lvltype{Par}]
				{
					\sttoltj{ \Gamma', \Delta_1, z:V \st P'}{ \lvlpartord{\flaten{>}{P}} }{} 
				\\
					\sttoltj{ \Gamma , \Delta_2 , v: \dual{V} \st  Q }{ \lvlpartord{\flaten{>}{P}} }{}\substj{z}{v}
				}
				{ \sttoltt{\Gamma}{\lvlpartord{\flaten{>}{P}}}{} \vaslinunsplit \sttoltt{\Delta_1}{\lvlpartord{\flaten{>}{P}}}{} , \sttoltt{\Delta_2}{\lvlpartord{\flaten{>}{P}}}{} , \lvlsplit{z}{\sttoltt{V}{\lvlpartord{\flaten{>}{P}}}{z}} \lt \sttoltp{P'}{} \pp (\sttoltp{Q}{}\substj{z}{v})
				}
			}
			{ \sttoltt{\Gamma}{\lvlpartord{\flaten{>}{P}}}{} \vaslinunsplit \sttoltt{\Delta_1}{\lvlpartord{\flaten{>}{P}}} {}, \sttoltt{\Delta_2}{\lvlpartord{\flaten{>}{P}}}{} \lt \res {z} (\sttoltp{P'}{} \pp (\sttoltp{Q}{}\substj{z}{v}))  }\\
		\]

		provided that $\sttoltt{V}{\lvlpartord{\flaten{>}{P}}}{z} = \sttoltt{V}{\lvlpartord{\flaten{>}{P}}}{v}$. 
		
		We show this by induction on the structure $V$, however, this is evident by the construction on the connected channels and the application of the join function. 
		\begin{itemize}
		\item First, notice that within $\flaten{>}{P} $ the last application performed is $ \partjoin{ - }{z}{v} $ which ensures that the channels $z$ and $v$ share all the same channes that are less then them; hence $\lvlpartord{\flaten{>}{P}}(z) = \lvlpartord{\flaten{>}{P}}(v)$.
		\item  Second, notice that $\flatcc{ > }{\cc{ z}{P'}}{\cc{ v }{Q}}$ is applied first. As the structure of the connected channels is that of the structure of the name constraints on the type of $V$ and that these channels are joined, then the level of these channels also match as required.
		\end{itemize}
\end{enumerate}
\end{proof}

\subsection{Proof for \Cref{ch5thm:wnotinl}}
Below we present a proof  for \Cref{ch5thm:wnotinl}:
\begin{theorem}
	 $\exists P \in \lvllang$  with $\Gamma \st P$ and $\sttoltj{\Gamma \st P}{l}{} $ for some $l$  such that $ \nexists  \ z \ s.t. \ \sttodillj{\Gamma \st P}_z $.
\end{theorem}

\begin{proof}\label{ch5proof:wnotinl}
We provide a counter-example.
	Consider the process
	\[
		P = 	
		\vasres{xy}{(
			\vaspara{ 
				\vasin{\lin}{x}{z}{\vasin{\un}{z}{w}{\nil}} 
			}
			{
				\vasres{st}{
					\vasout{y}{s}{(
						\vaspara{
							\vasres{uv}{(\vasout{t}{u}{\nil})}
						}
						{
							\nil
						}
					)}
				}
			}
		)}
	\]
Clearly, $P$ has the following reduction behaviour:
	\[
		\begin{aligned}
			P &\reddpp 
					\vasres{st}{(
						\vaspara{ 
							\vasin{\un}{s}{w}{\nil}
						}
						{
							\vasres{uv}{(\vasout{t}{u}{\nil})}
						}
					)}
				\\
				&\reddpp 
					\vasres{st}{
						\vasin{\un}{s}{w}{\nil}
					}
		\end{aligned}
	\]
Hence, $P$ terminates. First we type the process in $\vaslang$ then we show that there exists a level function that types the process in $\pilvl$ finally we show that there exists no channel such that the process can be typed in $\pidill$
	\begin{enumerate}
		\item $P \in \vaslang$

		\vasres{xy}{(\vaspara{\vasin{\lin}{x}{z}{\vasin{\un}{z}{w}{\nil}}}{\vasres{st}{\vasout{y}{s}{(\vaspara{\vasres{uv}{(\vasout{t}{u}{\nil})}}{\nil})}}})}
			\begin{mathpar}
				\Pi_1 = 
				\inferrule{ 
					\inferrule{ 
						\inferrule{
						}
						{x: \nilT ,z: \recur{\wn \nilT } \st z: \recur{\wn \nilT }}
						\\
						\inferrule{
						}
						{x: \nilT , z: \recur{\wn \nilT }, w:\nilT  \st \nil }
					}
					{ x: \nilT ,z: \recur{\wn \nilT } \st \vasin{\un}{z}{w}{\nil} }
					\\
					x: \lin \vasreci{(\recur{\wn \nilT })}{\nilT} \st x: \lin \vasreci{(\recur{\wn \nilT })}{\nilT}  
				}
				{ 
					 x: \lin \vasreci{(\recur{\wn \nilT })}{\nilT} \st  \vasin{\lin}{x}{z}{\vasin{\un}{z}{w}{\nil}}  
				}
                               \end{mathpar}

                                \begin{mathpar}
				\Pi_2 = 
				\inferrule{
					\inferrule{
						\inferrule{
							\Gamma \st t:  \recur{\oc \nilT}
							\\
							\Gamma \st u : \nilT
							\\
							\inferrule{ }{s:\recur{\wn \nilT } , t:  \recur{\oc \nilT} , y:\nilT , u : \nilT , v:\nilT \st \nil}
						}
						{
							\Gamma \st \vasout{t}{u}{\nil}
						}
					}{
						s:\recur{\wn \nilT } , t:  \recur{\oc \nilT} , y:\nilT \st \vasres{uv}{(\vasout{t}{u}{\nil})} 
					}
					\\
					s:\recur{\wn \nilT } , t:  \recur{\oc \nilT} , y:\nilT \st {\nil}
				}
				{ 
					s:\recur{\wn \nilT } , t:  \recur{\oc \nilT} , y:\nilT \st \vaspara{\vasres{uv}{(\vasout{t}{u}{\nil})}}{\nil}
				} 
			   \end{mathpar}
where $\Gamma=s:\recur{\wn \nilT } , t:  \recur{\oc \nilT} , y:\nilT , u : \nilT , v:\nilT$.
                                \begin{mathpar}
				\Pi_3 = 
				\inferrule{
					\inferrule{  
						y: \lin \vassend{(\recur{\wn \nilT })}{\nilT},\Gamma_1\st x: \lin \vassend{T}{S}  
						\\
						\Gamma_1 \st s: \recur{\wn \nilT }  
						\\
						\inferrule{
							\Pi_2
						}
						{
							s:\recur{\wn \nilT } , t:  \recur{\oc \nilT} , y:\nilT \st \vaspara{\vasres{uv}{(\vasout{t}{u}{\nil})}}{\nil}
						}
					}
					{ 
						y: \lin \vassend{(\recur{\wn \nilT })}{\nilT}, s:\recur{\wn \nilT } , t:  \recur{\oc \nilT}   \st   \vasout{y}{s}{(\vaspara{\vasres{uv}{(\vasout{t}{u}{\nil})}}{\nil})}  
					}
				}
				{
					y: \lin \vassend{(\recur{\wn \nilT })}{\nilT} \st \vasres{st}{\vasout{y}{s}{(\vaspara{\vasres{uv}{(\vasout{t}{u}{\nil})}}{\nil})}}
				}
			\end{mathpar}
		\begin{mathpar}
				\hspace{-1cm}
				{\small
				\inferrule{
					\inferrule{
						\inferrule{\Pi_1}{
							x:\lin \vasreci{(\recur{\wn \nilT })}{\nilT}  \st \vasin{\lin}{x}{z}{\vasin{\un}{z}{w}{\nil}}
						}
						\\
						\inferrule{\Pi_3}{
							y:\lin \vassend{(\recur{\wn \nilT })}{\nilT} \st \vasres{st}{\vasout{y}{s}{(\vaspara{\vasres{uv}{(\vasout{t}{u}{\nil})}}{\nil})}}}
					}
					{
						x:\lin \vasreci{(\recur{\wn \nilT })}{\nilT} , y: \lin \vassend{(\recur{\wn \nilT })}{\nilT}  \st \vaspara{\vasin{\lin}{x}{z}{\vasin{\un}{z}{w}{\nil}}}{\vasres{st}{\vasout{y}{s}{(\vaspara{\vasres{uv}{(\vasout{t}{u}{\nil})}}{\nil})}}}
					}
				}
				{
					 \cdot \st  \vasres{xy}{(\vaspara{\vasin{\lin}{x}{z}{\vasin{\un}{z}{w}{\nil}}}{\vasres{st}{\vasout{y}{s}{(\vaspara{\vasres{uv}{(\vasout{t}{u}{\nil})}}{\nil})}}})}
				}}		
			\end{mathpar}
	where $\Gamma_1=  s:\recur{\wn \nilT } , t:  \recur{\oc \nilT} $.	
			\item $P \in \lvllang$. 
			
			We show the existence of a function $\levelof{\cdot}$ such that $\sttoltj{ \cdot \st P}{l}{}$:
			\begin{mathpar}
				\hspace{-1.2cm}
				\small
				\sttoltj{
					\inferrule{
						\inferrule{
							\inferrule{\Pi_1}{
								x:\lin \vasreci{(\recur{\wn \nilT })}{\nilT}  \st \vasin{\lin}{x}{z}{\vasin{\un}{z}{w}{\nil}}
							}
							\\
							\inferrule{\Pi_3}{
								y:\lin \vassend{(\recur{\wn \nilT })}{\nilT} \st \vasres{st}{\vasout{y}{s}{(\vaspara{\vasres{uv}{(\vasout{t}{u}{\nil})}}{\nil})}}}
						}
						{
							x:\lin \vasreci{(\recur{\wn \nilT })}{\nilT} , y: \lin \vassend{(\recur{\wn \nilT })}{\nilT}  \st \vaspara{\vasin{\lin}{x}{z}{\vasin{\un}{z}{w}{\nil}}}{\vasres{st}{\vasout{y}{s}{(\vaspara{\vasres{uv}{(\vasout{t}{u}{\nil})}}{\nil})}}}
						}
					}
					{
						\cdot \st  \vasres{xy}{(\vaspara{\vasin{\lin}{x}{z}{\vasin{\un}{z}{w}{\nil}}}{\vasres{st}{\vasout{y}{s}{(\vaspara{\vasres{uv}{(\vasout{t}{u}{\nil})}}{\nil})}}})}
					}
				}{l}{}  \\

				=\inferrule{ 
					\inferrule{
						\inferrule{
							\sttoltj{\Pi_1}{l}{}
						}{
							x:\sttoltt{\lin \vasreci{(\recur{\wn \nilT })}{\nilT}}{l}{x} \lt \lvlin{x}{z}{a}{\lvlserv{z}{w}{b}{\nil}}
						}
						\\
						\inferrule{
							\sttoltj{\Pi_3}{l}{}
						}{
							x:\dual{\sttoltt{\lin \vasreci{(\recur{\wn \nilT })}{\nilT}}{l}{x}} \lt \res{s}{\lvlout{x}{s}{c}{(\res{u}{(\lvlout{s}{u}{d}{\nil})} \pp\nil )}{}}  
						}
					}
					{ 
						\lvlsplit{x}{\sttoltt{\lin \vasreci{(\recur{\wn \nilT })}{\nilT}}{l}{x} } \lt 
						\lvlin{x}{z}{a}{\lvlserv{z}{w}{b}{\nil}}
						\pp {\res{s}{\lvlout{x}{s}{c}{(\res{u}{(\lvlout{s}{u}{d}{\nil})} \pp\nil )}{}}}  
					}
				}
				{  \lt \res{x}{(\lvlin{x}{z}{a}{\lvlserv{z}{w}{b}{\nil}}
				\pp{\res{s}{\lvlout{x}{s}{c}{(\res{u}{(\lvlout{s}{u}{d}{\nil})} \pp\nil )}{}}}
				)} 
				}
			\end{mathpar}
			$\sttoltj{\Pi_1}{l}{}$ is:
			\begin{mathpar}
				\sttoltj{\inferrule{ 
					\inferrule{ 
						\inferrule{
						}
						{x: \nilT ,z: \recur{\wn \nilT } \st z: \recur{\wn \nilT }}
						\\
						\inferrule{
						}
						{x: \nilT , z: \recur{\wn \nilT }, w:\nilT  \st \nil }
					}
					{ x: \nilT ,z: \recur{\wn \nilT } \st \vasin{\un}{z}{w}{\nil} }
					\\
					x: \lin \vasreci{(\recur{\wn \nilT })}{\nilT} \st x: \lin \vasreci{(\recur{\wn \nilT })}{\nilT}  
				}
				{ 
					 x: \lin \vasreci{(\recur{\wn \nilT })}{\nilT} \st  \vasin{\lin}{x}{z}{\vasin{\un}{z}{w}{\nil}}  
				}
				}{l}{}
				=
				\inferrule{
					\sttoltj{\Pi_1}{l}{}
				}{
					x:\sttoltt{\lin \vasreci{(\recur{\wn \nilT })}{\nilT}}{l}{x} \lt \lvlin{x}{z}{a}{\lvlserv{z}{w}{b}{\nil}}
				}

				\inferrule{   z:\sttoltt{(\recur{\wn \nilT })}{l}{z} , a:\sttoltt{\nilT}{l}{a}  \lt \lvlserv{z}{w}{b}{\nil}  \\ l(x) = l(a) }
				{ x: \lvlint{l(x)}{\sttoltt{(\recur{\wn \nilT })}{l}{z}}{\sttoltt{\nilT}{l}{a}} \lt  \lvlin{x}{z}{a}{\lvlserv{z}{w}{b}{\nil}} }
			\end{mathpar}
			\[
				\res{x}{(\lvlin{x}{z}{a}{\lvlserv{z}{w}{b}{\nil}}
					\pp{\res{s}{\lvlout{x}{s}{c}{(\res{u}{(\lvlout{s}{u}{d}{\nil})} \pp\nil )}{}}}
				)}
			\]

			\item $P \not \in \dilllang$.
			
			We show that there is no channel $a$ such that $\sttodillj{\cdot \st  \vasres{xy}{(\vaspara{\vasin{\lin}{x}{z}{\vasin{\un}{z}{w}{\nil}}}{\vasres{st}{(\vasout{y}{s}{\vaspara{\vasres{uv}{(\vasout{t}{u}{\nil})}}{\nil}})}})}}_a$. First notice as this is a closed process we must have that $a$ is a fresh name immediately.
			\begin{mathpar}
			\hspace{-1.2cm}
			\small
				\sttodillj{
					\inferrule{
						\inferrule{
							\inferrule{\Pi_1}{
								x:\lin \vasreci{(\recur{\wn \nilT })}{\nilT}  \st \vasin{\lin}{x}{z}{\vasin{\un}{z}{w}{\nil}}
							}
							\\
							\inferrule{\Pi_3}{
								y:\lin \vassend{(\recur{\wn \nilT })}{\nilT} \st \vasres{st}{\vasout{y}{s}{(\vaspara{\vasres{uv}{(\vasout{t}{u}{\nil})}}{\nil})}}}
						}
						{
							x:\lin \vasreci{(\recur{\wn \nilT })}{\nilT} , y: \lin \vassend{(\recur{\wn \nilT })}{\nilT}  \st \vaspara{\vasin{\lin}{x}{z}{\vasin{\un}{z}{w}{\nil}}}{\vasres{st}{\vasout{y}{s}{(\vaspara{\vasres{uv}{(\vasout{t}{u}{\nil})}}{\nil})}}}
						}
					}
					{
							\cdot \st  \vasres{xy}{(\vaspara{\vasin{\lin}{x}{z}{\vasin{\un}{z}{w}{\nil}}}{\vasres{st}{\vasout{y}{s}{(\vaspara{\vasres{uv}{(\vasout{t}{u}{\nil})}}{\nil})}}})}
					}
				}_a 

				\\
				= \inferrule{ 
					\sttodillj{ x:\lin \vasreci{(\recur{\wn \nilT })}{\nilT} \st \vasin{\lin}{x}{z}{\vasin{\un}{z}{w}{\nil}} }_x  \\
					\sttodillj{ x: \dual{\lin \vasreci{(\recur{\wn \nilT })}{\nilT}} \st \vasres{st}{\vasout{x}{s}{(\vaspara{\vasres{uv}{(\vasout{t}{u}{\nil})}}{\nil})}}  }_a }
				{ \cdot ; \cdot  \dill \res{x}( 
					\sttodillp{\vasin{\lin}{x}{z}{\vasin{\un}{z}{w}{\nil}}} 
					\pp 
					\sttodillp{\vasres{st}{\vasout{x}{s}{(\vaspara{\vasres{uv}{(\vasout{t}{u}{\nil})}}{\nil})}}} 
					) ::  a : \dillnilT }
			\end{mathpar}
			for $\Pi_3$ and part of $\Pi_2$:
			\begin{mathpar}
				\sttodillj{
					\inferrule{
						\inferrule{  
							y: \lin \vassend{(\recur{\wn \nilT })}{\nilT}, s:\recur{\wn \nilT } , t:  \recur{\oc \nilT} \st x: \lin \vassend{T}{S}  
							\\
							s:\recur{\wn \nilT } , t:  \recur{\oc \nilT} \st s: \recur{\wn \nilT }  
							\\
							\inferrule{
								\Pi_2
							}
							{
								s:\recur{\wn \nilT } , t:  \recur{\oc \nilT} , y:\nilT \st \vaspara{\vasres{uv}{(\vasout{t}{u}{\nil})}}{\nil}
							}
						}
						{ 
							y: \lin \vassend{(\recur{\wn \nilT })}{\nilT}, s:\recur{\wn \nilT } , t:  \recur{\oc \nilT}   \st   \vasout{y}{s}{(\vaspara{\vasres{uv}{(\vasout{t}{u}{\nil})}}{\nil})}  
						}
					}
					{
						y: \lin \vassend{(\recur{\wn \nilT })}{\nilT} \st \vasres{st}{\vasout{y}{s}{(\vaspara{\vasres{uv}{(\vasout{t}{u}{\nil})}}{\nil})}}
					}
				}_a = 
				\inferrule{
					\inferrule{
						\text{No rule hence not typeable}
					}{
						\sttodillj{
						s : \recur{\oc \nilT}  \st \vasres{uv}{(\vasout{s}{u}{\nil})} 
						}_s
					}
					\\
					\inferrule{
					}{
						\cdot ; y: \dillnilT \dill \nil  ::  a : \dillnilT 
					}
				}
				{
					\cdot; y: \dillint{\sttodillt{\recur{\oc \nilT} }}{\dillnilT}  \dill \res {s} \dillout{y}{s}{( \sttodillp{\vasres{uv}{(\vasout{s}{u}{\nil})}} \pp \nil )} ::  a : \dillnilT 
				}
				\\
				\\
			\end{mathpar}
			for $\Pi_1$:
			\begin{mathpar}
				\sttodillj{
					\inferrule{ 
						\inferrule{ 
							\inferrule{
							}
							{x: \nilT ,z: \recur{\wn \nilT } \st z: \recur{\wn \nilT }}
							\\
							\inferrule{
							}
							{x: \nilT , z: \recur{\wn \nilT }, w:\nilT  \st \nil }
						}
						{ x: \nilT ,z: \recur{\wn \nilT } \st \vasin{\un}{z}{w}{\nil} }
						\\
						x: \lin \vasreci{(\recur{\wn \nilT })}{\nilT} \st x: \lin \vasreci{(\recur{\wn \nilT })}{\nilT}  
					}
					{ 
						x:\lin \vasreci{(\recur{\wn \nilT })}{\nilT} \st \vasin{\lin}{x}{z}{\vasin{\un}{z}{w}{\nil}}
					}
				}_x =
				\inferrule{ 
					\inferrule{
						\text{No rule here hence not typeable! }
					}{
						\sttodillj{ \Gamma , \Delta , x: \nilT , z : \recur{\wn \nilT } \st \vasin{\un}{z}{w}{\nil} }_x  
					}
				}
				{ 
					\cdot ; \cdot  \dill \dillin{x}{z}{\sttodillp{\vasin{\un}{z}{w}{\nil}}} :: x:\dillint{\sttodillt{\recur{\wn \nilT }}}{\sttodillt{\dual{\nilT}}} 
				}
			\end{mathpar}
	\end{enumerate}

\end{proof}

\begin{remark}
	There are many interesting counter examples of $\lvllang \not \subset \dilllang$:
	\[
	\begin{aligned}
		1~ & \vasin{\un}{x}{y}{\vasin{\un}{x}{z}{P}} &
		2~ & \vasin{\un}{x}{y}{\vasin{\un}{z}{w}{P}} \\
		3~ & (\vaspara{\vasin{\un}{x}{y}{P}}{\vasin{\un}{x}{z}{Q}}) &
		4~ & (\vaspara{\vasin{\un}{x}{y}{P}}{\vasin{\un}{z}{w}{Q}}) \\
		5~ & \vasin{\lin}{x}{y}{\vasin{\un}{y}{z}{P}} &
		6~ & \vasres{yz}{(\vasout{x}{{y}}{(
			\vaspara{\vasres{wv}{\vasout{z}{w}{P}}}{Q_x}
			)})}  \\
		7~ & \vasres{yz}{\vasout{x}{y}{\vasin{\un}{z}{w}{P}}} &
		8~ & \vasin{\un}{x}{y}{ \vasres{wv}{\vasout{y}{w}{P}}} \\
	\end{aligned}
	\]
	The proof of \Cref{ch5thm:wnotinl} can be seen as $5$ and $6$ cut together; they illustrate more interesting examples of the differences between the two type systems.
\end{remark}

\section{Operational Correspondence }
This section contains the proofs of operational correspondence for the encoding $\sttodillp{\cdot }{}{}$ for typed process in $\dilllang$. The main theorems are operational completeness (cf. \Cref{ch5thm:op.compl}) and weak operational soundness (cf. \Cref{ch5thm:op.sound}). 
\begin{definition}{Substitution Lifting}
	For  processes $P , Q \in \pidill$, we say that { \em $P$ lifts the substitutions of $Q$ on names $y,v$}, written $Q \lifts P\substj{v}{y}$, if $Q \equiv \res{y} (\dillfwdbang{v}{y}  \pp P)$.
\end{definition}

\begin{theorem}[Operational Completeness]\label{ch5thm:op.compl}
	Let $ P \in \dilllang$ such that  $\sttodillj{\Gamma \st P}_u$,  for some name $u$. Then there exists $R \in \pidill$ such that $P \vasred Q \implies \sttodillp{P}{}{} \dillred^* R$  and $R \lifts \sttodillp{Q}$.
\end{theorem}

\begin{proof} The proof is standard, by analyzing the rule applied in $P$. We will present the interesting cases below:
\begin{enumerate}
	
\item\label{ch5proof:dillcompleteness1}  {\bf The rule is $\Did{R-LinCom}$.}

Then $P=\vasres{xy}{(  \vaspara{\vasout{x}{v}{P'}}{\vaspara{\vasin{\lin}{y}{z}{Q'}}{R'}}   )}$ and $Q= \vasres{xy}{(
				\vaspara{P'}	{\vaspara{Q'\substj{v}{z}}{R'}})}$, and the reduction is as follows:
	\[
		\Did{R-LinCom} \quad 
		\vasres{xy}{(  
				\vaspara{\vasout{x}{v}{P'}}
					{\vaspara{\vasin{\lin}{y}{z}{Q'}}
						{R'}} 
			   )}  \vasred \vasres{xy}{(
				\vaspara{P'}
					{\vaspara{Q'\substj{v}{z}}
						{R'}}
			)}
	\]
By hypothesis, $\sttodillj{\Gamma \st P}_u$, and by \Cref{ch5d:secondtablep1} we have 
\[\dillnotun{\sttodillt{\Gamma'}} ; \sttodillt{\Delta_1} , \sttodillt{\Delta'}  \dill \res{x}( \sttodillp{\vasout{x}{y}{P'}} \pp \sttodillp{\vasin{\lin}{x}{y}{Q'\substj{x}{y}}\pp R'\substj{x}{y}} ) ::  u: A, \]

 where  \(x:V \land (u \not \in \fn{P'} )	\land ( (\neg \contun{V}) \lor ( \contun{V} \land y \not \in \fn{P'} \land x \not \in \fn{Q'} ))\).  In addition, since  $\sttodillp{\vasin{\lin}{x}{y}{Q'\substj{x}{y}}}$ we must have that $x: \lin \vasreci{T}{S}$, which implies  $ \neg \contun{V}$. Let us consider the following two possibilities:

 \begin{enumerate}
	\item When $ x: \lin \oc (T).S \land \neg \contun{T}\land \notserver{T}$.
	
	 Then $\sttodillp{P} = \res{x}( \dillbout{x}{a}{( \dillforward{v}{a}  \pp  \sttodillp{{P'}}   )}  \pp  \res{w}(  \dillin{x}{z}{\sttodillp{Q'}\substj{x}{y}}  \pp \sttodillp{R'}\substj{x}{y} ) )$ with $w$ fresh and reductions:

	\[
		\begin{aligned}
			\sttodillp{P} 
			& \dillred \res{x}( \res{a}( \dillforward{v}{a}  \pp  \res{w}( \sttodillp{Q'}\substj{x}{y}\substj{a}{z} \pp  \sttodillp{R'}\substj{x}{y} ) )  \pp  \sttodillp{{P''}}   )  \\
			& \dillred \res{x}(  \sttodillp{{P''}}  \pp  \res{w}( \sttodillp{Q'}\substj{x}{y}\substj{a}{z}\substj{v}{a} \pp \sttodillp{R'}\substj{x}{y} )) \qquad = R \\
		\end{aligned}	
	\]
	
	With 
	\[\sttodillp{\res{x y}(P'' \pp Q'\substj{v}{z} \pp R')} = \res{x}(  \sttodillp{{P''}}  \pp \res{w}(  \sttodillp{Q'}\substj{x}{y}\substj{v}{z} \pp \sttodillp{R'}\substj{x}{y} ) )\] 
	as required.

	\item When $ x: \lin \oc (T).S \land \contun{T} \land \notserver{T}$.
	
	 Then $\sttodillp{P} = \res{x}( \dillbout{x}{a}{( \dillfwdbang{v}{a}  \pp  \sttodillp{{P}}   )}  \pp  \res{w}(  \dillin{x}{z}{\sttodillp{Q'}\substj{x}{y}} \pp \sttodillp{R'}\substj{x}{y} ) )$ with $w$ fresh and reductions:
	
	\[
		\begin{aligned}
			\hspace{-1cm}
			\sttodillp{P} 
			& \dillred \res{x}( \res{a}( \dillfwdbang{v}{a}  \pp   \res{w}(  \sttodillp{Q'}\substj{x}{y}\substj{a}{z})  \pp  \sttodillp{R'}\substj{x}{y} ) \pp  \sttodillp{{P''}}   )  = R  \\
		\end{aligned}	
	\]

	Notice that
		\[
			\hspace{-2cm}
			\small
			\begin{aligned}
				\sttodillp{\res{x y}(P'' \pp Q'\substj{v}{z} \pp R')} 
				& = \res{x}(  \sttodillp{{P''}}  \pp \res{w}(  \sttodillp{Q'}\substj{x}{y}\substj{v}{z}  \pp \sttodillp{R'}\substj{x}{y} ) ) \\
				& \lifts \res{x}( \sttodillp{{P''}} \pp \res{z}( \dillfwdbang{v}{z}  \pp   \res{w}( \sttodillp{Q'}\substj{x}{y}  \pp \sttodillp{R'}\substj{x}{y}  ) )      ) \\
				& \equiv \res{x}( \sttodillp{{P''}} \pp \res{a}( \dillfwdbang{v}{a}  \pp \res{w}(  \sttodillp{Q'}\substj{x}{y}\substj{a}{z}  \pp   \sttodillp{R'}\substj{x}{y} ))  )  \\
				&\equiv U_1
			\end{aligned}
		\]
	as required.
	
 \end{enumerate}

	\item {\bf The rule is $\Did{R-UnCom}$.}

	Then $P=\vasres{xy}{(
		\vaspara{\vasout{x}{b}{P'}}
			{\vaspara{\vasin{\un}{y}{d}{Q'}}
				{R'}}
	)}$ and $Q= \vasres{xy}{(
		\vaspara{P'}
			{\vaspara{Q'\substj{b}{d}}
				{\vaspara{\vasin{\un}{y}{d}{Q'}}
					{R'}}
			}
	)}$, and the reduction is as follows:
		\[
			\small
			\Did{R-UnCom} \, 
				\vasres{xy}{(
						\vaspara{\vasout{x}{b}{P'}}
							{\vaspara{\vasin{\un}{y}{d}{Q'}}
								{R'}}
					)}  \vasred
				\vasres{xy}{(
						\vaspara{P'}
							{\vaspara{Q'\substj{b}{d}}
								{\vaspara{\vasin{\un}{y}{d}{Q'}}
									{R'}}
							}
					)}
		\]

	The proof proceeds similarly to  case \Cref{ch5proof:dillcompleteness1}, where typing ensures that \(x:V \land (u \not \in \fn{P'} )	\land ( \contun{V} \land y \not \in \fn{P'} \land x \not \in \fn{Q'} )\). There are four cases to consider: 
	\begin{enumerate}
		\item $ x: \recur{\oc T} \land \neg \contun{T} \land  \notserver{T} \land \client{T}$
		\item $ x: \recur{\oc T} \land \contun{T} \land \notserver{T} \land \client{T}$
		\item $ x: \recur{\oc T} \land \neg \contun{T}\land \notserver{T} \land \notclient{T} $
		\item $ x: \recur{\oc T} \land \contun{T}\land \notserver{T} \land \notclient{T} $
	\end{enumerate}

	We shall only consider the case (a) with $ x: \recur{\oc T} \land \neg \contun{T} \land  \notserver{T} \land \client{T}$.
	
	In this case,  $\sttodillp{P} = \res{x}(  \dillbout{x}{z}{ \dillbout{z}{w}{ ( \dillforward{b}{w}  \pp  \sttodillp{{P}}   ) }}  \pp   \res{e}( \dillserv{x}{z}{ \dillin{z}{d}{ \sttodillp{Q'}\substj{x}{y} }} \pp \sttodillp{R'}\substj{x}{y}  )   )$ with $e$ fresh and reductions:

	\[
					\begin{aligned}
						\sttodillp{P} 
						& \dillred \res{x}(  \res{z}( \dillbout{z}{w}{ ( \dillforward{b}{w}  \pp  \sttodillp{{P}}   ) } \pp \dillin{z}{d}{ \sttodillp{Q'}\substj{x}{y} } )  \pp \\
						& \quad  \res{e}(  \dillserv{x}{z}{ \dillin{z}{d}{ \sttodillp{Q'}\substj{x}{y} }}  \pp \sttodillp{R'}\substj{x}{y}  )  )  && = U_1  \\
						& \dillred \res{x}(  \res{z}(    \sttodillp{{P}}  \pp \res{w}( \dillforward{b}{w}  \pp \sttodillp{Q'}\substj{x}{y}\substj{w}{d} ) )  \pp \\
						&\quad  \res{e}(  \dillserv{x}{z}{ \dillin{z}{d}{ \sttodillp{Q'}\substj{x}{y} }}  \pp \sttodillp{R'}\substj{x}{y}  )   )  && = U_2  \\
						& \dillred \res{x}(  \res{z}(    \sttodillp{{P}}  \pp \sttodillp{Q'}\substj{x}{y}\substj{w}{d}\substj{b}{w} )  \pp  \\
						&\quad \res{e}(  \dillserv{x}{z}{ \dillin{z}{d}{ \sttodillp{Q'}\substj{x}{y} }}  \pp \sttodillp{R'}\substj{x}{y}  )    )  && = U_3  \\
					\end{aligned}	
				\]

	Notice that
				\[
					\begin{aligned}
						\sttodillp{\vasres{xy}{(
							\vaspara{P'}
								{\vaspara{Q'\substj{b}{d}}
									{\vaspara{\vasin{\un}{y}{d}{Q'}}
										{R'}}
								}
						)}} 
						& = \res{x}(  \res{z}(    \sttodillp{{P}}  \pp \sttodillp{Q'}\substj{x}{y}\substj{b}{d} )  \pp \\
						& \quad  \res{e}(  \dillserv{x}{z}{ \dillin{z}{d}{ \sttodillp{Q'}\substj{x}{y} }}  \pp \sttodillp{R'}\substj{x}{y}  )    )   
					\end{aligned}
				\]
	as required.

\end{enumerate}

\end{proof}



The following technical result is auxiliary to the proof of Operational Soundness (c.f. \Cref{ch5thm:op.sound}), and it guarantees the preservation of reductions within restricted parallel compositions in $\dilllang$.
\begin{proposition}\label{ch5prop:indillandreduces}
	If $ \res{x y}(P \pp \res{c d}(Q \pp R)) \in \dilllang$ then the following hold:
	\begin{enumerate}
		\item $ \res{x y}(P \pp Q ) \in \dilllang$ and $\res{x y}(P \pp R ) \in \dilllang $.
		\item If $ \res{x y}(P \pp Q ) \reddpp  \res{x y}(P' \pp Q' )$ then $ \res{x y}(P \pp \res{c d}(Q \pp R)) \reddpp  \res{x y}(P' \pp \res{c d}(Q' \pp R))$;
		\item If $ \res{x y}(P \pp R ) \reddpp  \res{x y}(P' \pp R' )$ then $ \res{x y}(P \pp \res{c d}(Q \pp R)) \reddpp  \res{x y}(P' \pp \res{c d}(Q \pp R'))$.
	\end{enumerate}
	 
\end{proposition}

\begin{proof}

	We show each case as follows:

	\begin{enumerate}
		\item We show the case of $ \res{x y}(P \pp Q ) \in \dilllang$ as $ \res{x y}(P \pp R ) \in \dilllang$ follows analogously. 
		
		As $ \res{x y}(P \pp \res{c d}(Q \pp R)) \in \dilllang$, it follows that,  for some contexts $\Gamma, \Delta_1, \Delta_2, \Delta_3$ and types $V_1$, $V_2$ and $A$:
		
		\[\small
			\inferrule{ \sttodillj{ \Gamma, \Delta_1, x:V_1 \st P }_x  
			\\
				\inferrule{ \sttodillj{ \Gamma', \Delta_2,y: \dual{V_1} , c:V_2 \st Q }_c \substj{x}{y}  \\
				\sttodillj{ \Gamma , \Delta , d: \dual{V_2} \st  R \substj{c}{d} }_u \substj{x}{y}
				}
				{ \dillnotun{\sttodillt{\Gamma}} ; \sttodillt{\Delta_1} , \sttodillt{\Delta'}  \dill  \res{c}(\sttodillp{Q} \pp \sttodillp{R}\substj{c}{d})\substj{x}{y} ::  u: A }
			}
			{ 
				\dillnotun{\sttodillt{\Gamma}} ; \sttodillt{\Delta_1} , \sttodillt{\Delta_2} ,\sttodillt{\Delta_3}  \dill \res{x}( \sttodillp{P} \pp \res{c}(\sttodillp{Q} \pp \sttodillp{R}\substj{c}{d})\substj{x}{y} ) ::  u: A 
			}
		\]

		Here for simplicity, we assume that the types of $x,y,c$ and $d$ are linear and that $y \in \fn{Q}$, the permutations of $x,y $ and $c,d$ being linear and $y \not\in \fn{Q}$ cases follow similarly. Finally, we will assume for this case that $A$ is linear.

		\[\small
			\inferrule{ \sttodillj{ \Gamma, \Delta_1, x:V_1 \st P }_x  
			\\
			\sttodillj{ \Gamma', \Delta_2,y: \dual{V_1} , c:V_2 \st Q }_c \substj{x}{y} 
			}
			{ 
				\dillnotun{\sttodillt{\Gamma}} ; \sttodillt{\Delta_1} , \sttodillt{\Delta_2} ,\sttodillt{\Delta_3}, y:\sttodillt{V_1}   \dill \res{x}( \sttodillp{P} \pp \res{c}(\sttodillp{Q} \pp \sttodillp{R}\substj{c}{d})\substj{x}{y} ) ::  c: \sttodillt{\dual{V_2}}{}{} 
			}
		\]

		and hence $ \res{x y}(P \pp Q ) \in \dilllang$ provided  $\res{x y}(P \pp \res{c d}(Q \pp R)) \in \dilllang$.

		\item Suppose $\res{x y}(P \pp Q ) \reddpp  \res{x y}(P' \pp Q' )$.
		
		Then the only possible reductions that may be performed are those of the form $\Did{R-LinCom}$ or $\Did{R-UnCom}$. Let us assume that $\Did{R-LinCom}$ is performed, $P$ perfoms the output syncronisation and $Q$ the input and that $x \in \fn{P}, x\not \in \fn{Q} , y\in \fn{Q}$ and $y \not \in \fn{P}$, implying that $P$ is of the form $(\vasout{x}{v}{P''} \pp P''')$ and $Q$ is of the form $(\vasin{\lin}{y}{v}{Q''} \pp Q''')$. By the congruence rules in \Cref{ch5f:vascosymantics}:
		\[
			\begin{aligned}
				\res{x y}(P \pp \res{c d}(Q \pp R)) &\equiv \res{c d}\res{x y}(P \pp Q \pp R)\\
				& = \res{c d}\res{x y}(\vasout{x}{v}{P''} \pp P''' \pp \vasin{\lin}{y}{z}{Q''} \pp Q''' \pp R)\\
				& \equiv \res{c d}\res{x y}(\vasout{x}{v}{P''}  \pp \vasin{\lin}{y}{v}{Q''} \pp P''' \pp Q''' \pp R)
			\end{aligned}
		\]
		By rules $\Did{R-Str}, \Did{R-Res}$ and $\Did{R-LinCom}$ of \Cref{ch5f:vascosymantics} we have that:

		\[
			\small
			\begin{aligned}
				\res{c d}\res{x y}(\vasout{x}{v}{P''}  \pp \vasin{\lin}{y}{v}{Q''} \pp P''' \pp Q''' \pp R) &\reddpp  \res{c d}\res{x y}( {P''}  \pp {Q''} \pp P''' \pp Q''' \pp R) \\
				& \equiv \res{x y}( {P''} \pp P'''  \pp  \res{c d} (  {Q''} \pp Q''' \pp R) ) \\
				& = \res{x y}( P'  \pp  \res{c d} (  Q' \pp R) ) \\
			\end{aligned}
		\]

		
      and the result follows.
         \item   This case is similar to the previous one.
	\end{enumerate}

\end{proof}

\begin{theorem}[Weak Operational Soundness]\label{ch5thm:op.sound}
	Let $P \in \dilllang$ with   $\sttodillj{\Gamma \st P}_u$,  for some name $u$. If  $\sttodillp{P}{}{} \dillred^* U$ then there exists $R \in \dilllang$ and $V \in \pidill$ such that $P \vasred^*  R \land U \dillred^* V \land V \lifts \sttodillp{R}{}{} $.
\end{theorem}

\begin{proof}
	By induction on the structure of $P$. Consider the following cases:

	\begin{enumerate}
		\item {\bf  When $P =  \nil$.}
		
		  Then $\sttodillp{\nil}  = \nil$ and no reductions are available.
		
		\item {\bf When $P =  \vasout{x}{y}{P'}$.}
		
		 There are many options for $\sttodillp{\vasout{x}{y}{P'}}$ depending on typing. In the case  $\sttodillp{\vasout{x}{y}{P'}} = \dillbout{x}{z}{( \dillforward{y}{z}  \pp  \sttodillp{{P'}}   )} $ then no reduction can be performed, and the result follows trivially. Similarly for all other cases.
		
		\item {\bf When $P = \res{xy}\vasout{z}{y}{P' }$.} 

		Then, either $\sttodillp{\res{xy}\vasout{z}{y}{P'}  } = \dillbout{z}{x}{\dillseler{x}{\sttodillp{P'}}}$ or $ \sttodillp{\res{xy}\vasout{z}{y}{P'}  }  = \dillbout{z}{x}{\sttodillp{P'}} $, neither with any reductions.
		
		\item {\bf When $P = \res{xy}\vasout{z}{x}{(P' \pp Q)}$}.
		
		 Then $\sttodillp{\res{xy}\vasout{z}{x}{(P' \pp Q)}}  = 
		\dillbout{z}{y}{(\sttodillp{P'} \pp \sttodillp{Q})} $ and no reductions are avaliable.
		
		\item {\bf When $P = \vasin{\lin}{x}{y}{P'}$}. 
		
		 Then $ \sttodillp{\vasin{\lin}{x}{y}{P'}} = 
				\dillin{x}{y}{\sttodillp{P'}} $ and no reductions are avaliable.

		\item {\bf When $P =  \vasin{\un}{x}{{y}}{P'}$.}
		
		 There are many options for  $\sttodillp{\vasin{\un}{x}{{y}}{P'}}$ depending on typing. In the case $\sttodillp{\vasout{x}{y}{P'}} = \dillserv{x}{z}{\sttodillp{P' \substj{z}{y}}}  $, then no reduction can be performed.  Similarly for all other cases.

		\item {\bf When $P =  \res{x y}(P' \pp Q)$.}
		
		 Then $\sttodillp{\res{x y}(P' \pp Q)}  = \res{x}( \sttodillp{P'} \pp \sttodillp{Q}\substj{x}{y})$ with $ y \not \in \fn{P}$ and $ x \not \in \fn{Q}$. 
		 
		 Now we proceed by induction on the structure of $P'$:
		
		\begin{enumerate}
			\item When $P' = \nil$.
			
			 Then $\sttodillp{P} = \res{x}( \nil \pp \sttodillp{Q}\substj{x}{y})$.
			 
			   \begin{itemize}
			  \item By applying rule  $\Did{R$\nu$}$  one has $\res{x}( \nil \pp \sttodillp{Q}\substj{x}{y}) \dillred \res{x}(U)$,  if $\nil \pp \sttodillp{Q}\substj{x}{y}  \dillred Q  $.
			  \item By applying rule  $\Did{R$\pp$}$ one has 
			$\sttodillp{Q}\substj{x}{y} \dillred U' 
			\implies 
			\nil \pp \sttodillp{Q}\substj{x}{y} \dillred \nil \pp  U'$.
			\end{itemize} 
			
			By the induction hypothesis, as $\sttodillp{Q}\substj{x}{y} \dillred^* U' $ then there exists $R' \in \dilllang$ and $V' \in \pidill$ such that
			
			\begin{itemize} 
			\item  $Q \vasred^*  R' $;
			\item $ U' \dillred^* V'$;
			\item $ V' \lifts \sttodillp{R'}{}{} $. 
			\end{itemize}
			Hence $\res{x y}( \nil \pp Q ) \dillred^* \res{x}( \nil \pp R')$ and both $\res{x y}( \nil \pp R' ) \in \dilllang$ and $ \res{x}( \nil \pp V' ) \in \pidill $ hold. Therefore, 
			\begin{itemize}
			\item  $ \res{x y}( \nil \pp Q) \vasred^*   \res{x y}( \nil \pp R')$;
			\item $  \res{x}( \nil \pp U' ) \dillred^* \res{x}( \nil \pp V' )$; 
			\item  $\res{x}( \nil \pp V' ) \lifts \sttodillp{\res{x y}( \nil \pp R') }{}{} $.
			\end{itemize}
                        and the result follows.

			\item\label{ch5proof:soundnessfirstoutputcase} When $P' = \vasout{a}{b}{P''}$.
			
			 Then let us consider the following cases:
			
			\begin{enumerate}
				
				\item\label{ch5proof:sound2bi} When $ a: \lin \oc (T).S \land \neg \contun{T}\land \notserver{T}$.
				
				 Then $\sttodillp{\vasout{a}{b}{P''}} = \dillbout{a}{z}{( \dillforward{b}{z}  \pp  \sttodillp{{P''}}   )} $.
				 
				  We now perform induction on the structure of $Q$:
				
				\begin{myEnumerate}
					\item[(i.1)] When $ Q = \nil$.
					
					 In this case $\sttodillp{P} = \res{x}( \dillbout{a}{z}{( \dillforward{b}{z}  \pp  \sttodillp{{P''}}   )}  \pp  \nil )$,  and the process does not reduce. The result follows trivially.
					
					\item[(i.2)] When $Q =  \vasout{c}{d}{Q'}$. 
					
					Then let us consider the following:
					
					\begin{itemize}
						\item When $x= a \land y = c$ then $x$ and $y$ do not have dual types and $P \notin \dilllang$+
						. 
						\item When $\neg(x= a \land y = c)$.
						
						 In the case $ c: \lin \oc (T').S' \land \contun{T'} \land \notserver{T'}$, we have  
						 
						 $
						  \sttodillp{\vasout{c}{d}{Q''}}  = \dillbout{c}{z'}{( \dillforward{d}{z'}  \pp  \sttodillp{{Q'}}   )} $ and  
						 
						  $\sttodillp{P} = \res{x}( \dillbout{a}{z}{( \dillforward{b}{z}  \pp  \sttodillp{{P''}}   )}  \pp   \dillbout{c}{z'}{( \dillforward{d}{z'}  \pp  \sttodillp{{Q'}}\substj{x}{y}   )}  )$
						  and no reduction can be performed. 
						 
						 Other cases follow analogously not allowing reductions to be performed.

					\end{itemize}

					\item[(i.3)] When $Q =  \res{e d}\vasout{c}{d}{Q'} $.
					
					 Then let us consider the following:
					
					\begin{itemize}
						\item When $x= a \land y = c$ then $y: \recur{\oc T'}$, $x$ and $y$ do not have dual types and  hence not in $\dilllang$.
						\item When $\neg(x= a \land y = c)$ then we consider the following cases:
						
						\begin{myEnumerate}
							\item When $c:\recur{\oc T'} \land \notserver{T'} \land \notclient{T'} $.

							 Then $\sttodillp{\res{ed}\vasout{c}{d}{Q'}  }  = \dillbout{c}{e}{\dillseler{e}{\sttodillp{Q'}}}$ and  $\sttodillp{P} = \res{x}( \dillbout{a}{z}{( \dillforward{b}{z}  \pp  \sttodillp{{P''}}   )}  \pp  \dillbout{c}{e}{\dillseler{e}{\sttodillp{Q'}\substj{x}{y}}}   )$ and no reductions are avaliable.

							\item When $c:\recur{\oc T'} \land \server{T'} \land \notclient{T'} $.
							
							 Then $\sttodillp{\res{ed}\vasout{c}{d}{Q'}  }  = \dillbout{c}{e}{\sttodillp{Q'}}  $ and  $\sttodillp{P} = \res{x}( \dillbout{a}{z}{( \dillforward{b}{z}  \pp  \sttodillp{{P''}}   )}  \pp \dillbout{c}{e}{\sttodillp{Q'}\substj{x}{y}}   )$ and no reductions are avaliable.

						\end{myEnumerate}

					\end{itemize}

					\item[(i.4)] When $Q =  \res{de}\vasout{c}{d}{(Q_1 \pp Q_2)} $.
					
					 Then $ \sttodillp{\res{de}\vasout{c}{d}{(Q_1 \pp Q_2)}}  = 
					\dillbout{c}{e}{(\sttodillp{Q_1} \pp \sttodillp{Q_2})}
					$.  Let us consider the following:
					
					\begin{itemize}
						\item When $x= a \land y = c$ then $x$ and $y$ do not have dual types and hence not in $\dilllang$.
						\item When $\neg(x= a \land y = c)$ then 
							$\sttodillp{P} = \res{x}( \dillbout{a}{z}{( \dillforward{b}{z}  \pp  \sttodillp{{P''}}   )}  \pp \dillbout{c}{e}{(\sttodillp{Q_1} \pp \sttodillp{Q_2})}\substj{x}{y} )$ and no reductions are avaliable.

					\end{itemize}

					\item[(i.5)] When $Q =  \vasin{\lin}{c}{d}{Q'} $.
					
					 Then $ \sttodillp{\vasin{\lin}{c}{d}{Q'}}  = \dillin{c}{d}{\sttodillp{Q'}}
					$ let us consider the following:

					\begin{itemize}
						\item When $x= a \land y = c$.

						 Then $x$ and $y$ have dual types and hence  in $\dilllang$.

						 On the one hand, the encoding is as  $\sttodillp{P} = \res{x}( \dillbout{x}{z}{( \dillforward{b}{z}  \pp  \sttodillp{{P''}}   )}  \pp  \dillin{x}{d}{\sttodillp{Q'}\substj{x}{y}} )$ with reductions:

						\[
							\begin{aligned}
								\sttodillp{P} 
								& \dillred \res{x}( \res{z}( \dillforward{b}{z}  \pp  \sttodillp{Q'}\substj{x}{y}\substj{z}{d})  \pp  \sttodillp{{P''}}   ) && = U_1  \\
								& \dillred \res{x}(  \sttodillp{{P''}}  \pp  \sttodillp{Q'}\substj{x}{y}\substj{z}{d}\substj{b}{z} ) && = U_2  \\
							\end{aligned}	
						\]

						(note that reductions may be performed within $\sttodillp{Q'}$ from $U_1$ however we  leverage the confluence of reductions within $\pidill$ )

						On the other hand,
						\[
							\begin{aligned}
								P = \res{x y}(\vasout{x}{b}{P''} \pp \vasin{\lin}{y}{d}{Q'}) \reddpp \res{x y}(P'' \pp Q'\substj{b}{d})
							\end{aligned}	
						\]

						Notice that $\sttodillp{\res{x y}(P'' \pp Q'\substj{b}{d})} = \res{x}(  \sttodillp{{P''}}  \pp  \sttodillp{Q'}\substj{x}{y}\substj{b}{d} ) \equiv U_2$.

						We now perform induction on the length $n$ of $\sttodillp{P}{}{} \dillred^n U$:
						
							\noindent{\bf (Base Case)}When $n \leq 2$.
							
							 Then $\sttodillp{P}{}{} \dillred^n U_n$. 
							 
							Take $R = \res{x y}(P'' \pp Q'\substj{b}{d})$ and $V = U_2$. Then  $P \vasred^*  R \land U_n \dillred^* U_2 \land U_2 \equiv \sttodillp{R}{}{} $, and the result follows.

							\noindent{\bf (Inductive Step)} When $n > 2$.
							
							 Then $\sttodillp{P}{}{} \dillred^2 U_2 \dillred^m U $ and $P \vasred^*  R'  \land U_2 \equiv \sttodillp{R'}{}{} $.  By the induction hypothesis $\sttodillp{R'}{}{} \dillred^m U $ implies that there exist $R \in \dilllang$ and $V \in \pidill$ such that $R' \vasred^*  R \land U \dillred^* V \land V \lifts \sttodillp{R}{}{} $. Hence $P \vasred^*  R' \vasred^*  R \land U \dillred^* V \land V \lifts \sttodillp{R}{}{}$, and the result follows.

						\item When $\neg(x= a \land y = c)$.
						
						 Then 
							$\sttodillp{P} = \res{x}( \dillbout{a}{z}{( \dillforward{b}{z}  \pp  \sttodillp{{P''}}   )}  \pp \dillin{c}{d}{\sttodillp{Q'}}\substj{x}{y} )$ and no reductions are avaliable.

					\end{itemize}

					\item[(i.6)] When $Q =  \vasin{\un}{c}{{d}}{Q'} $.

					 Let us consider the following:

					\begin{itemize}
						\item When $x= a \land y = c$ then $x$ and $y$ do not have dual types and hence not in $\dilllang$.						

						\item When $\neg(x= a \land y = c)$. 
						
						 In the case $ c: \recur{\wn T'} \land\server{T'}\land \notclient{T'} $, we have 
						 
						  $\sttodillp{ \vasin{\un}{c}{{d}}{Q'} } = \dillserv{c}{z'}{\sttodillp{Q'\substj{z'}{d}}}  $ and $\sttodillp{P} = \res{x}( \dillbout{a}{z}{( \dillforward{b}{z}  \pp  \sttodillp{{P''}}   )}  \pp  \dillserv{c}{z'}{\sttodillp{Q'\substj{z'}{d}}}\substj{x}{y}   )$ and no reduction can be performed. 
						  
						  Other cases follow analogously not allowing reductions to be performed.

					\end{itemize}

					\item[(i.7)]\label{ch5proof:opsound2G1} When $Q =  \res{c d}(Q_1 \pp Q_2) $.
					
					 Then,  
					$\sttodillp{\res{cd}(Q_1 \pp Q_2)} = 
					\res{c}( \sttodillp{Q_1} \pp \sttodillp{Q_2}\substj{c}{d}) $
					 and 
					 
					 $$
					 \begin{aligned}
					 \sttodillp{P}& = \res{x}( \dillbout{a}{z}{( \dillforward{b}{z}  \pp  \sttodillp{{P''}}   )}  \pp  \res{c}( \sttodillp{Q_1} \pp \sttodillp{Q_2}\substj{c}{d})  ) \\
					 &\equiv \res{x}\res{c}( \dillbout{a}{z}{( \dillforward{b}{z}  \pp  \sttodillp{{P''}}   )}  \pp   \sttodillp{Q_1} \pp \sttodillp{Q_2}\substj{c}{d} ). 
					 \end{aligned} $$ 
					 
					 Now we perform induction on the number of reductions $n$ performed in $\sttodillp{P}{}{} \dillred^n U$.

					Let us first consider the first reduction:

					\begin{myEnumerate}

						\item When either 
						
						{\bf I.1)} $\sttodillp{Q_1} \dillred U'$; or
						
						{\bf I.2)} $ \sttodillp{Q_2}\substj{c}{d} \reddpp U'$; or 
						 
						{\bf I.3)} $\res{c}( \sttodillp{Q_1} \pp \sttodillp{Q_2}\substj{c}{d})  \dillred U' $.

						   Then, by the induction hypothesis there exists $R' \in \dilllang$ and $V' \in \pidill$ such that either
						   
						    {\bf I.1)} $Q_1 \vasred^*  R' \land U' \dillred^* V' \land V' \lifts \sttodillp{R'}{}{} $; or 
						    
						     {\bf I.2)} $Q_2 \vasred^*  R' \land U' \dillred^* V' \land V' \lifts \sttodillp{R'}{}{} $; or
						     
						      {\bf I.3)} $\res{c d}(Q_1 \pp Q_2) \vasred^*  R' \land U' \dillred^* V' \land V' \lifts \sttodillp{R'}{}{} $, respectively. 
						      
						      Hence either
						
						\noindent {\bf I.1)}
						\[
							\small
							\begin{aligned}
							\sttodillp{P} &= \res{x}( \dillbout{a}{z}{( \dillforward{b}{z}  \pp  \sttodillp{{P''}}   )}  \pp \res{c}( \sttodillp{Q_1} \pp \sttodillp{Q_2}\substj{c}{d})\substj{x}{y} ) \\
								& \dillred \res{x}( \dillbout{a}{z}{( \dillforward{b}{z}  \pp  \sttodillp{{P''}}   )}  \pp \res{c}( U' \pp \sttodillp{Q_2}\substj{c}{d})\substj{x}{y} ) \\
								&= U''
								\\
								& \dillred^m \res{x}( \dillbout{a}{z}{( \dillforward{b}{z}  \pp  \sttodillp{{P''}}   )}  \pp \res{c}( V' \pp \sttodillp{Q_2}\substj{c}{d})\substj{x}{y} ) \\
								&= V''
							\end{aligned}
						\]
						or
						 {\bf I.2)}
						\[
						\small
							\begin{aligned}
								\sttodillp{P} 
								&\dillred \res{x}( \dillbout{a}{z}{( \dillforward{b}{z}  \pp  \sttodillp{{P''}}   )}  \pp \res{c}( \sttodillp{Q_1} \pp U'\substj{c}{d})\substj{x}{y} ) \\
								& = U''
								\\
								&\dillred^m \res{x}( \dillbout{a}{z}{( \dillforward{b}{z}  \pp  \sttodillp{{P''}}   )}  \pp \res{c}( \sttodillp{Q_1} \pp V'\substj{c}{d})\substj{x}{y} ) \\
								 & = V''
							\end{aligned}	
						\] 
						or {\bf I.3)}
						\[
							\begin{aligned}
								\sttodillp{P}
								&\dillred \res{x}( \dillbout{a}{z}{( \dillforward{b}{z}  \pp  \sttodillp{{P''}}   )}  \pp U'\substj{x}{y} ) = U''
								\\
								&\dillred^m \res{x}( \dillbout{a}{z}{( \dillforward{b}{z}  \pp  \sttodillp{{P''}}   )}  \pp V'
								\substj{x}{y} )  = V''
							\end{aligned}	
						\] 

						with either 
						
						{\bf I.1)}
						\(
						 \qquad
							\begin{aligned}
								 P & =  \res{x y}(\vasout{a}{b}{P''} \pp \res{c d}(Q_1 \pp Q_2)) \\
								 &\reddpp  \res{x y}(\vasout{a}{b}{P''} \pp \res{c d}(R' \pp Q_2)) & = R''
							\end{aligned}
						\); or 
						
					 {\bf I.2)}
						\(
						 \qquad
							\begin{aligned}
								P & \reddpp  \res{x y}(\vasout{a}{b}{P''} \pp \res{c d}(Q_1 \pp R'))  & = R''
							\end{aligned}
						\)
						or;
						
						 {\bf I.3)}
						\(
						 \qquad
							\begin{aligned}
								P & \reddpp  \res{x y}(\vasout{a}{b}{P''} \pp R')  & = R''
							\end{aligned}
						\).
						
						Hence $R'' \in \dilllang$, $V'' \in \pidill$ and $P \vasred^*  R'' \land U'' \dillred^* V'' \land V'' \lifts \sttodillp{R''}{}{} $. 

						We now consider the number of reductions taken by $\sttodillp{P}{}{} \dillred^n U$. If $n \leq m + 1$ then we take $R = R''$ and the case follows. If $n > 1 + m $ then we apply the induction hypothesis on $\sttodillp{R''}{}{} \dillred^{n-m-1} U$ and the case follows.

						\item When $a = x$ and 
						\[
							\small
							\begin{aligned}
								\sttodillp{P} & \dillred \res{x} (   \sttodillp{{P''}}    \pp  \res{z}( \dillforward{b}{z}  \pp  \res{c}( \sttodillp{Q'_1} \pp \sttodillp{Q_2}\substj{c}{d}) \substj{x}{y}  ))\\
								& \dillred \res{x} (   \sttodillp{{P''}}  \pp  \pp  \res{c}( \sttodillp{Q'_1}\substj{b}{z} \pp \sttodillp{Q_2}\substj{c}{d}) \substj{x}{y}  )
							\end{aligned}
						\] or \[ \small
							\begin{aligned}
								\sttodillp{P} & \dillred \res{x} (   \sttodillp{{P''}}    \pp  \res{z}( \dillforward{b}{z}  \pp  \res{c}( \sttodillp{Q_1} \pp \sttodillp{Q'_2}\substj{c}{d}) \substj{x}{y}  ))\\
								& \dillred \res{x} (   \sttodillp{{P''}}  \pp  \pp  \res{c}( \sttodillp{Q_1} \pp \sttodillp{Q'_2}\substj{c}{d}\substj{b}{z}) \substj{x}{y}  )
							\end{aligned}
						\]
						We consider only the first case, then by \Cref{ch5prop:indillandreduces} $\res{x y}(P' \pp Q_1)  \in \dilllang$ and 
						\[
							\begin{aligned}
								\sttodillp{\res{x y}(P' \pp Q_1) } &=   \res{x}( \dillbout{a}{z}{( \dillforward{b}{z}  \pp  \sttodillp{{P''}}   )}  \pp   \sttodillp{Q_1}  ) \\
								& \reddpp  \res{x} (   \sttodillp{{P''}}    \pp  \res{z}( \dillforward{b}{z}  \pp  \sttodillp{Q'_1}\substj{x}{y}  )) \\
								& \reddpp  \res{x} (   \sttodillp{{P''}}    \pp  \sttodillp{Q'_1}\substj{x}{y}\substj{b}{z}  )\\
								& = \sttodillp{\res{x y}(P'' \pp Q'_1) } 
							\end{aligned}
						\]
						By \Cref{ch5prop:indillandreduces} as $\res{x y}(P' \pp Q_1) \reddpp \res{x y}(P'' \pp Q'_1) $ we have that $\res{x y}(P' \pp \res{c d}(Q_1 \pp Q_2)) \reddpp \res{x y}(P'' \pp \res{c d}(Q'_1 \pp Q_2))$ and the rest of the case now follows by induction on the number of reduction steps. 



					\end{myEnumerate}

					\item[(i.8)] When $Q =  Q_1 \pp Q_2 $ then the case follows analogously to \ref{ch5proof:opsound2G1}.

				\end{myEnumerate}

				\item When $ a: \lin \oc (T).S \land \contun{T} \land \notserver{T}$.
				
				 Then $\sttodillp{\vasout{a}{b}{P''}} = \dillbout{a}{z}{( \dillfwdbang{b}{z}  \pp  \sttodillp{{P}}   )} $.
				 
				  We now perform induction on the structure of $Q$ analogously to that of \Cref{ch5proof:sound2bi} with the exception of following case $Q$:
				
				When $Q =  \vasin{\lin}{c}{d}{Q'} $.

				 Then $ \sttodillp{\vasin{\lin}{c}{d}{Q'}}  = \dillin{c}{d}{\sttodillp{Q'}}$.  Let us consider the following:

					\begin{myEnumerate}
						\item[(ii.1)] When $x= a \land y = c$.
						
						 Then $x$ and $y$ have dual types and hence in $\dilllang$.

						 On the one hand, the encoding  is as as 
						$\sttodillp{P} = \res{x}( \dillbout{x}{z}{( \dillfwdbang{b}{z}  \pp  \sttodillp{{P}}   )}  \pp  \dillin{x}{d}{\sttodillp{Q'}\substj{x}{y}} )$  with reduction:

						\[
							\begin{aligned}
								U_0 = \sttodillp{P} 
								& \dillred \res{x}( \res{z}( \dillfwdbang{b}{z}  \pp  \sttodillp{Q'}\substj{x}{y}\substj{z}{d})  \pp  \sttodillp{{P''}}   ) = U_1  \\
							\end{aligned}	
						\]

						On the other hand,

						\[
							\begin{aligned}
								P = \res{x y}(\vasout{x}{b}{P''} \pp \vasin{\lin}{y}{d}{Q'}) \reddpp \res{x y}(P'' \pp Q'\substj{b}{d})
							\end{aligned}	
						\]

						Notice that
						\[
							\small
							\begin{aligned}
								\sttodillp{\res{x y}(P'' \pp Q'\substj{b}{d})} 
								& = \res{x}(  \sttodillp{{P''}}  \pp  \sttodillp{Q'}\substj{x}{y}\substj{b}{d} ) \\
								& \lifts \res{x}( \sttodillp{{P''}} \pp \res{d}( \dillfwdbang{b}{d}  \pp  \sttodillp{Q'}\substj{x}{y})      ) \\
								& \equiv \res{x}( \sttodillp{{P''}} \pp \res{z}( \dillfwdbang{b}{z}  \pp  \sttodillp{Q'}\substj{x}{y}\substj{z}{d})      )\\
								&   \equiv U_1
							\end{aligned}
						\]
						 We now perform induction on the length $n$ of $\sttodillp{P}{}{} \dillred^n U$: 

							\noindent{\bf (Base Case)} When $n \leq 1$ then $\sttodillp{P}{}{} \dillred^n U_n$ by taking $R = \res{x y}(P'' \pp Q'\substj{b}{d})$ and $V = U_1$ we have that $P \vasred^*  R \land U_n \dillred^* U_1 \land U_1 \lifts \sttodillp{R}{}{}$.

							\noindent{\bf (Inductive Step)}  When $n > 1$ then $\sttodillp{P}{}{} \dillred^1 U_1 \dillred^m U $ and $P \vasred^*  R'  \land U_2 \equiv \sttodillp{R'}{}{} $. By the induction hypothesis $\sttodillp{R'}{}{} \dillred^m U $ implies that there exists $R \in \dilllang$ and $V \in \pidill$ such that $R' \vasred^*  R \land U \dillred^* V \land V \lifts \sttodillp{R}{}{} $. Hence $P \vasred^*  R' \vasred^*  R \land U \dillred^* V \land V \lifts \sttodillp{R}{}{}$.

						\item[(ii.2)] When $\neg(x= a \land y = c)$ then 
							$\sttodillp{P} = \res{x}( \dillbout{a}{z}{( \dillfwdbang{b}{z}  \pp  \sttodillp{{P''}}   )}  \pp \dillin{c}{d}{\sttodillp{Q'}}\substj{x}{y} )$ and no reductions are avaliable.

					\end{myEnumerate}

				\item\label{ch5proofsoundnessunresoutput}  When $ a: \recur{\oc T} \land \neg \contun{T} \land  \notserver{T} \land \client{T}$.
				
				 Then $\sttodillp{\vasout{a}{b}{P''}} = \dillbout{a}{z}{ \dillbout{z}{w}{ ( \dillforward{b}{w}  \pp  \sttodillp{{P}}   ) }}  $. 
				 
				 We proceed by  induction on the structure of $Q$ analogously to  \Cref{ch5proof:sound2bi} with the following exceptional cases for $Q$:

				\begin{myEnumerate}
					\item[(iii.1)] When $Q =  \vasin{\lin}{c}{d}{Q'} $.
					
					 Then $ \sttodillp{\vasin{\lin}{c}{d}{Q'}}  = \dillin{c}{d}{\sttodillp{Q'}}
					$. Let us consider the following:

					\begin{itemize}
						\item When $x= a \land y = c$  then $x$ and $y$ do not have dual types and hence not in $\dilllang$.
						\item When $\neg(x= a \land y = c)$ then 
							$\sttodillp{P} = \res{x}( \dillbout{a}{z}{ \dillbout{z}{w}{ ( \dillforward{b}{w}  \pp  \sttodillp{{P}}   ) }}   \pp \dillin{c}{d}{\sttodillp{Q'}}\substj{x}{y} )$ and no reductions are available.

					\end{itemize}

					\item[(iii.2)] When $Q =  \vasin{\un}{c}{{d}}{Q'}$.
					
					Let us consider the following:

					\begin{itemize}
						\item When $x= a \land y = c$ then let us consider the following cases:
						
						\begin{itemize}
							\item When $ x: \recur{\wn T} \land\server{T}\land \notclient{T} $ then $x$ and $y$ do not have dual types and hence not in $\dilllang$.

							\item When $ x: \recur{\wn T} \land\notserver{T}\land \client{T} $.

							Then $x$ and $y$ have dual types and hence in $\dilllang$.

							On the one hand, the  encoding is as
							$\sttodillp{P} = \res{x}(  \dillbout{x}{z}{ \dillbout{z}{w}{ ( \dillforward{b}{w}  \pp  \sttodillp{{P}}   ) }}  \pp   \dillserv{x}{z}{ \dillin{z}{d}{ \sttodillp{Q'}\substj{x}{y} }}    )$   with reductions:

								\[
									\hspace{-4cm}
									\small
									\begin{aligned}
										\sttodillp{P} 
										& \dillred \res{x}(  \res{z}( \dillbout{z}{w}{ ( \dillforward{b}{w}  \pp  \sttodillp{{P}}   ) } \pp \dillin{z}{d}{ \sttodillp{Q'}\substj{x}{y} } )  \pp   \dillserv{x}{z}{ \dillin{z}{d}{ \sttodillp{Q'}\substj{x}{y} }}    )  && = U_1  \\
										& \dillred \res{x}(  \res{z}(    \sttodillp{{P}}  \pp \res{w}( \dillforward{b}{w}  \pp \sttodillp{Q'}\substj{x}{y}\substj{w}{d} ) )  \pp   \dillserv{x}{z}{ \dillin{z}{d}{ \sttodillp{Q'}\substj{x}{y} }}    )  && = U_2  \\
										& \dillred \res{x}(  \res{z}(    \sttodillp{{P}}  \pp \sttodillp{Q'}\substj{x}{y}\substj{w}{d}\substj{b}{w} )  \pp   \dillserv{x}{z}{ \dillin{z}{d}{ \sttodillp{Q'}\substj{x}{y} }}    )  && = U_3  \\
									\end{aligned}	
								\]

 On the other hand,
								\[
									\begin{aligned}
										P = \res{x y}(\vasout{x}{b}{P''} \pp \vasin{\lin}{y}{d}{Q'}) \reddpp \res{x y}(P'' \pp Q'\substj{b}{d})
									\end{aligned}	
								\]

								Notice that
								\[
									\hspace{-2cm}
									\begin{aligned}
										\sttodillp{\res{x y}(P'' \pp Q'\substj{b}{d})} 
										& = \res{x}(  \sttodillp{{P''}}  \pp  \sttodillp{Q'}\substj{x}{y}\substj{b}{d} ) \\
										& \lifts \res{x}( \sttodillp{{P''}} \pp \res{d}( \dillfwdbang{b}{d}  \pp  \sttodillp{Q'}\substj{x}{y})      ) \\
										& \equiv \res{x}( \sttodillp{{P''}} \pp \res{z}( \dillfwdbang{b}{z}  \pp  \sttodillp{Q'}\substj{x}{y}\substj{z}{d})      )  \\
										& \equiv U_1
									\end{aligned}
								\]
								
								We now perform induction on the length $n$ of $\sttodillp{P}{}{} \dillred^n U$:

							\noindent{\bf (Base Case)} When $n \leq 1$ then $\sttodillp{P}{}{} \dillred^n U_n$.  Take $R = \res{x y}(P'' \pp Q'\substj{b}{d})$ and $V = U_1$. Then, we have that $P \vasred^*  R \land U_n \dillred^* U_1 \land U_1 \lifts \sttodillp{R}{}{} $, and the result follows.

									\noindent{\bf (Inductive Step)}  When $n > 1$ then $\sttodillp{P}{}{} \dillred^1 U_1 \dillred^m U $ and $P \vasred^*  R'  \land U_2 \equiv \sttodillp{R'}{}{} $.  By the induction hypothesis $\sttodillp{R'}{}{} \dillred^m U $ implies that there exist $R \in \dilllang$ and $V \in \pidill$ such that $R' \vasred^*  R \land U \dillred^* V \land V \lifts \sttodillp{R}{}{} $. Hence, $P \vasred^*  R' \vasred^*  R \land U \dillred^* V \land V \lifts \sttodillp{R}{}{}$ and the result follows.

							\item If $ x: \recur{\wn T} \land\notserver{T}\land \notclient{T} $ then $x$ and $y$ do not have dual types and hence not in $\dilllang$.

						\end{itemize}

						\item When $\neg(x= a \land y = c)$.
						
						In the case  $ c: \recur{\wn T'} \land\server{T'}\land \notclient{T'} $, it follows that  $\sttodillp{ \vasin{\un}{c}{{d}}{Q'} } = \dillserv{c}{z'}{\sttodillp{Q'\substj{z'}{d}}}  $ and $\sttodillp{P} = \res{x}( \dillbout{a}{z}{( \dillforward{b}{z}  \pp  \sttodillp{{P''}}   )}  \pp  \dillserv{c}{z'}{\sttodillp{Q'\substj{z'}{d}}}\substj{x}{y}   )$ and no reduction can be performed. 
						
						Other cases follow analogously not allowing reductions to be performed.

					\end{itemize}

				\end{myEnumerate}

				\item The cases below follow analogously to \Cref{ch5proofsoundnessunresoutput}.
				
				\begin{itemize}
					\item When $ a: \recur{\oc T} \land \contun{T} \land \notserver{T} \land \client{T}$.
					\item When $ a: \recur{\oc T} \land \neg \contun{T}\land \notserver{T} \land \notclient{T} $.
					\item When $ a: \recur{\oc T} \land \contun{T}\land \notserver{T} \land \notclient{T} $.
				\end{itemize}

				Except for the case of  $x= a \land y = c$ when considering the type of $y$ the encoding of $Q =  \vasin{\un}{c}{{d}}{Q'}$ changes due to $c$ having dual type to $b$ but the syncronisations follow simularly. 

			\end{enumerate}


			\item When $P' = \res{ab}\vasout{c}{b}{P''}$.
			
			 Let us consider the following cases:
			
			\begin{myEnumerate}
				
				\item When $c:\recur{\oc T} \land \notserver{T} \land \notclient{T} $.

				 Then, the encoding gives  $\sttodillp{\res{ab}\vasout{c}{b}{P''}} = \dillbout{c}{a}{\dillseler{a}{\sttodillp{P''}}}  $. We proceed by induction on the structure of $Q$ analogously to \Cref{ch5proof:soundnessfirstoutputcase}.

				 We shall only consider the case of $Q = \vasin{\un}{y}{{d}}{Q'}$ with $c= x$.

				 By duality on typing we have that $\sttodillp{\vasin{\un}{y}{{d}}{Q'}} = \dillserv{y}{e}{  
					\dillchoice{e}{	\dillin{z}{d}{ \sttodillp{Q'} }	}{	\sttodillp{Q'\substj{e}{d}} }
					} 
				$.
				
				On the one hand, 

				\[
					\hspace{-2cm}
					\small
					\begin{aligned}
						\sttodillp{\res{x y}(P' \pp Q)} & = \res{x}( \sttodillp{P'} \pp \sttodillp{Q}\substj{x}{y})\\
					& = \res{x}( \dillbout{x}{a}{\dillseler{a}{\sttodillp{P''}}} \pp   \dillserv{x}{e}{  
						\dillchoice{e}{	\dillin{e}{d}{ \sttodillp{Q'} }	}{	\sttodillp{Q'\substj{e}{d}} }} ) && = U_0
					\\
					& \dillred \res{x}( 
						\res{a}( \dillseler{a}{\sttodillp{P''}} 
						\pp
						\dillchoice{a}{	\dillin{e}{d}{ \sttodillp{Q'} }	}{	\sttodillp{Q'\substj{a}{d}} })
						\pp \\
					& \qquad  \dillserv{x}{e}{  
						\dillchoice{e}{	\dillin{e}{d}{ \sttodillp{Q'} }	}{	\sttodillp{Q'\substj{e}{d}} }} ) && = U_1
					\\
					& \dillred \res{x}( 
						\res{a}( \sttodillp{P''} 
						\pp
						\sttodillp{P\substj{a}{d}} )\\
						&\qquad
						\pp   \dillserv{x}{e}{  
						\dillchoice{e}{	\dillin{e}{d}{ \sttodillp{Q'} }	}{	\sttodillp{Q'\substj{e}{d}} }} ) && = U_2
					\end{aligned}
				\]

				On the other hand,

				\[
					\begin{aligned}
						\res{x y}(P' \pp Q) & = \res{x y}(\res{ab}\vasout{x}{b}{P''} \pp \vasin{\un}{y}{{d}}{Q'})
						\\
						& \reddpp  \res{x y}(\res{ab}( {P''} \pp {Q'}\substj{b}{d} )\pp \vasin{\un}{y}{{d}}{Q'})
					\end{aligned}	
				\]
				
				Notice that

				\[
					\hspace{-2cm}
					\begin{aligned}
						\sttodillp{\res{x y}(\res{ab}( {P''} \pp {Q'}\substj{b}{d} )\pp \vasin{\un}{y}{{d}}{Q'})} &= \res{x}( 
						\res{a}( \sttodillp{P''} 
						\pp
						\sttodillp{Q'\substj{a}{d}} )
						\pp  \\
						& \dillserv{x}{e}{  
						\dillchoice{e}{	\dillin{e}{d}{ \sttodillp{Q'} }	}{	\sttodillp{Q'\substj{e}{d}} }} ) 
					\end{aligned}
				\]

				We now perform induction on the length $n$ of $\sttodillp{P}{}{} \dillred^n U$:

					\noindent{\bf (Base Case)}   When $n \leq 2$ then $\sttodillp{P}{}{} \dillred^n U_n$ by taking $R = \res{x y}(\res{ab}( {P''} \pp {Q'}\substj{b}{d} )\pp \vasin{\un}{y}{{d}}{Q'})$ and $V = U_2$ we have that $P \vasred^*  R \land U_n \dillred^* U_2 \land U_2 \lifts \sttodillp{R}{}{} $, and the result follows.
					
 				\noindent{\bf (Inductive Step)}  When $n > 2$ then $\sttodillp{P}{}{} \dillred^2 U_2 \dillred^m U $ and $P \vasred^*  R'  \land U_2 \equiv \sttodillp{R'}{}{} $.  By the induction hypothesis $\sttodillp{R'}{}{} \dillred^m U $ implies that there exist $R \in \dilllang$ and $V \in \pidill$ such that $R' \vasred^*  R \land U \dillred^* V \land V \lifts \sttodillp{R}{}{} $. Thus,  $P \vasred^*  R' \vasred^*  R \land U \dillred^* V \land V \lifts \sttodillp{R}{}{}$, and the result follows.

				\item When $c:\recur{\oc T} \land \server{T} \land \notclient{T} $.

				 Then, the encoding gives  $\sttodillp{\res{ab}\vasout{c}{b}{P''}} = \dillbout{c}{a}{\sttodillp{P''}} $.
				  We perform induction on the structure of $Q$ analogously to \Cref{ch5proof:soundnessfirstoutputcase}. 
				  
				  We shall only consider the case of $Q = \vasin{\un}{y}{{d}}{Q'}$ with $c= x$.

				  By duality on typing we have that $\sttodillp{\vasin{\un}{y}{{d}}{Q'}} = \dillserv{y}{d}{\sttodillp{Q'}}
				$.
				
				On the one hand, 

				\[
					\small
					\begin{aligned}
						\sttodillp{\res{x y}(P' \pp Q)} & = \res{x}( \sttodillp{P'} \pp \sttodillp{Q}\substj{x}{y})\\
					& = \res{x}( \dillbout{x}{a}{\sttodillp{P''}} \pp  \dillserv{x}{d}{\sttodillp{Q'}}  ) && = U_0
					\\
					& \dillred \res{x}( 
						\res{a}( \sttodillp{P''} 
						\pp
						\sttodillp{P\substj{a}{d}} )
						\pp   \dillserv{x}{d}{\sttodillp{Q'}}  ) && = U_1
					\end{aligned}
				\]

				On the other hand,

				\[
					\begin{aligned}
						\res{x y}(P' \pp Q) & = \res{x y}(\res{ab}\vasout{x}{b}{P''} \pp \vasin{\un}{y}{{d}}{Q'})
						\\
						& \reddpp  \res{x y}(\res{ab}( {P''} \pp {Q'}\substj{b}{d} )\pp \vasin{\un}{y}{{d}}{Q'})
					\end{aligned}	
				\]
				
				Notice that:

				\[
					\hspace{-2cm}
					\small
					\begin{aligned}
						\sttodillp{\res{x y}(\res{ab}( {P''} \pp {Q'}\substj{b}{d} )\pp \vasin{\un}{y}{{d}}{Q'})} &= \res{x}( 
						\res{a}( \sttodillp{P''} 
						\pp
						\sttodillp{Q'\substj{a}{d}} )
						\pp \\
						& \quad  \dillserv{x}{d}{\sttodillp{Q'}}  ) 
					\end{aligned}
				\]

				We now perform induction on the length $n$ of $\sttodillp{P}{}{} \dillred^n U$:
					
		\noindent{\bf (Base Case)}  When $n \leq 1$ then $\sttodillp{P}{}{} \dillred^n U_n$.  Take  $R = \res{x y}(\res{ab}( {P''} \pp {Q'}\substj{b}{d} )\pp \vasin{\un}{y}{{d}}{Q'})$ and $V = U_1$, then  we have that $P \vasred^*  R \land U_n \dillred^* U_1 \land U_1 \lifts \sttodillp{R}{}{} $ and the result follows.

	 		\noindent{\bf (Inductive Step)}  When $n > 1$ then $\sttodillp{P}{}{} \dillred^1 U_1 \dillred^m U $ and $P \vasred^*  R'  \land U_1 \equiv \sttodillp{R'}{}{} $.  By the induction hypothesis $\sttodillp{R'}{}{} \dillred^m U $ implies that there exist $R \in \dilllang$ and $V \in \pidill$ such that $R' \vasred^*  R \land U \dillred^* V \land V \lifts \sttodillp{R}{}{} $. Thus, $P \vasred^*  R' \vasred^*  R \land U \dillred^* V \land V \lifts \sttodillp{R}{}{}$, and the result follows.

			\end{myEnumerate}

			\item When $P' = \res{ab}\vasout{c}{a}{(P_1 \pp P_2)}$.

			 Then we must have that $z \not \in \fn{P_1} \land y \not \in \fn{P_2}$ and $\sttodillp{ \res{ab}\vasout{c}{a}{(P \pp Q)} } = \dillbout{c}{b}{(\sttodillp{P_1} \pp \sttodillp{P_2})} $. 
			 
			 We perform induction on the structure of $Q$ analogously to  \Cref{ch5proof:soundnessfirstoutputcase}. 
			 
			 We shall only consider the case of $Q = \vasin{\lin}{y}{d}{Q'}$ with $c= x$ and we have that $\sttodillp{ \vasin{\lin}{y}{d}{Q'}} = \dillin{y}{d}{\sttodillp{Q'}}
			$.
			
 On the one hand, 			
			\[
				\begin{aligned}
					\sttodillp{\res{x y}(P' \pp Q)} & = \res{x}( \sttodillp{P'} \pp \sttodillp{Q}\substj{x}{y})\\
				& = \res{x}( \dillbout{c}{b}{(\sttodillp{P_1} \pp \sttodillp{P_2})}  \pp  \dillin{x}{d}{\sttodillp{Q'}}  ) && = U_0
				\\
				& \dillred \res{x}( 
					\res{a}( \sttodillp{P_1} 
					\pp
					\sttodillp{Q'\substj{a}{d}} )
					\pp   {\sttodillp{Q_2}}  ) && = U_1
				\end{aligned}
			\]

On the other hand,
			\[
				\begin{aligned}
					\res{x y}(P' \pp Q) & = \res{x y}(
						\res{ab}\vasout{c}{a}{(P_1 \pp P_2)} \pp 
						\vasin{\lin}{y}{d}{Q'} )
					\\
					& \reddpp  \res{x y}(
						\res{ab}(P_1 \pp 
						{Q'}\substj{a}{d}  ) \pp P_2
						 )
				\end{aligned}	
			\]
			
			Notice that:

			\[
				\sttodillp{ \res{x y}(
					\res{ab}(P_1 \pp 
					{Q'}\substj{a}{d}  ) \pp P_2
					 )} = \res{x}( 
						\res{a}( \sttodillp{P_1} 
						\pp
						\sttodillp{Q'\substj{a}{d}} )
						\pp   {\sttodillp{Q_2}}  ) 
			\]

			We now perform induction on the length $n$ of $\sttodillp{P}{}{} \dillred^n U$:

					\noindent{\bf (Base Case)} When $n \leq 1$ then $\sttodillp{P}{}{} \dillred^n U_n$.  Take $R = \res{x y}(\res{ab}(P_1 \pp 
						{Q'}\substj{a}{d}  ) \pp P_2
						 )  $ and $V = U_1$. Then,  we have that $P \vasred^*  R \land U_n \dillred^* U_1 \land U_1 \lifts \sttodillp{R}{}{} $ and the result follows.

					\noindent {\bf (Inductive Step)} When $n > 1$ then $\sttodillp{P}{}{} \dillred^1 U_1 \dillred^m U $ and $P \vasred^*  R'  \land U_1 \equiv \sttodillp{R'}{}{} $.  By the induction hypothesis $\sttodillp{R'}{}{} \dillred^m U $ implies that there exist $R \in \dilllang$ and $V \in \pidill$ such that $R' \vasred^*  R \land U \dillred^* V \land V \lifts \sttodillp{R}{}{} $. Thus, $P \vasred^*  R' \vasred^*  R \land U \dillred^* V \land V \lifts \sttodillp{R}{}{}$ and the result follows.

			\item When $P' = \vasin{\lin}{a}{b}{P''}$ or $P' = \vasin{\un}{a}{{b}}{P''}$ the cases follow by duality on the output cases already shown.
			
			\item When $P' = \res{x y}(P_1 \pp P_2)$ or $P' = P_1 \pp P_2$ then these cases follow by the induction hypothesis.

		\end{enumerate}

	\end{enumerate}
\end{proof}


}
\else 
\newpage
\clearemptydoublepage
\fi










\iffulldoc
\else
%
%
%
%
%
%
%
%

\newcommand*{\promitem}[4]{\noindent \textbf{#1}. \emph{#2}. #3.~\mbox{#4}\medskip}

\clearpage \pagestyle{empty}

\setlength{\columnsep}{2em}
\begin{multicols}{2}
        [\subsection*{Titles in the IPA Dissertation Series since 2021}]

\promitem{D. Frumin}
         {Concurrent Separation Logics for Safety, Refinement, and
Security}
         {Faculty of Science, Mathematics and Computer Science, RU}
		 {2021-01}

\promitem{A. Bentkamp}
         {Superposition for Higher-Order Logic}
         {Faculty of Sciences, Department of Computer Science, VU}
         {2021-02}

\promitem{P. Derakhshanfar}
         {Carving Information Sources to Drive Search-based Crash Reproduction and Test Case Generation}
         {Faculty of Electrical Engineering, Mathematics, and Computer Science, TUD}
         {2021-03}

\promitem{K. Aslam}
         {Deriving Behavioral Specifications of Industrial Software Components}
         {Faculty of Mathematics and Computer Science, TU/e}
         {2021-04}

\promitem{W. Silva Torres}
         {Supporting Multi-Domain Model Management}
         {Faculty of Mathematics and Computer Science, TU/e}
         {2021-05}

\promitem{A. Fedotov}
         {Verification Techniques for xMAS}
         {Faculty of Mathematics and Computer Science, TU/e}
         {2022-01}

\promitem{M.O. Mahmoud}
         {GPU Enabled Automated Reasoning}
         {Faculty of Mathematics and Computer Science, TU/e}
         {2022-02}

\promitem{M. Safari}
         {Correct Optimized GPU Programs}
         {Faculty of Electrical Engineering, Mathematics \& Computer Science, UT}
         {2022-03}

\promitem{M. Verano Merino}
         {Engineering Language-Parametric End-User Programming Environments for DSLs}
         {Faculty of Mathematics and Computer Science, TU/e}
         {2022-04}

\promitem{G.F.C. Dupont}
         {Network Security Monitoring in Environments where Digital and Physical Safety are Critical}
         {Faculty of Mathematics and Computer Science, TU/e}
         {2022-05}
		 
\promitem{T.M. Soethout}
         {Banking on Domain Knowledge for Faster Transactions}
         {Faculty of Mathematics and Computer Science, TU/e}
         {2022-06}
		
\promitem{P. Vukmirovi\'{c}}
         {Implementation of Higher-Order Superposition}
         {Faculty of Sciences, Department of Computer Science, VU}
         {2022-07}

\promitem{J. Wagemaker}
         {Extensions of (Concurrent) Kleene Algebra}
         {Faculty of Science, Mathematics and Computer Science, RU}
		 {2022-08}
		 
\promitem{R. Janssen}
         {Refinement and Partiality for Model-Based Testing}
         {Faculty of Science, Mathematics and Computer Science, RU}
		 {2022-09}

\promitem{M. Laveaux}
         {Accelerated Verification of Concurrent Systems}
         {Faculty of Mathematics and Computer Science, TU/e}
         {2022-10}
		 
\promitem{S. Kochanthara}
         {A Changing Landscape: On Safety \& Open Source in Automated and Connected Driving}
         {Faculty of Mathematics and Computer Science, TU/e}
         {2023-01}
		 
\promitem{L.M. Ochoa Venegas}
         {Break the Code? Breaking Changes and Their Impact on Software Evolution}
         {Faculty of Mathematics and Computer Science, TU/e}
         {2023-02}

\promitem{N. Yang}
         {Logs and models in engineering complex embedded production software systems}
         {Faculty of Mathematics and Computer Science, TU/e}
         {2023-03}
		 
\promitem{J. Cao}
         {An Independent Timing Analysis for Credit-Based Shaping in Ethernet TSN}
         {Faculty of Mathematics and Computer Science, TU/e}
         {2023-04}

\promitem{K. Dokter}
         {Scheduled Protocol Programming}
         {Faculty of Mathematics and Natural Sciences, UL}
         {2023-05}

\promitem{J. Smits}
         {Strategic Language Workbench Improvements}
         {Faculty of Electrical Engineering, Mathematics, and Computer Science, TUD}
         {2023-06}

\promitem{A. Arslanagi\'{c}}
         {Minimal Structures for Program Analysis and Verification}
         {Faculty of Science and Engineering, RUG}
         {2023-07}

\promitem{M.S. Bouwman}
         {Supporting Railway Standardisation with Formal Verification}
         {Faculty of Mathematics and Computer Science, TU/e}
         {2023-08}

\promitem{S.A.M. Lathouwers}
         {Exploring Annotations for Deductive Verification}
         {Faculty of Electrical Engineering, Mathematics \& Computer Science, UT}
         {2023-09}
		 
\promitem{J.H. Stoel}
         {Solving the Bank, Lightweight Specification and Verification Techniques for Enterprise Software}
         {Faculty of Mathematics and Computer Science, TU/e}
         {2023-10}

\promitem{D.M. Groenewegen}
         {WebDSL: Linguistic Abstractions for Web Programming}
         {Faculty of Electrical Engineering, Mathematics, and Computer Science, TUD}
         {2023-11}

\promitem{D.R. do Vale}
         {On Semantical Methods for Higher-Order Complexity Analysis}
         {Faculty of Science, Mathematics and Computer Science, RU}
         {2024-01}

\promitem{M.J.G. Olsthoorn}
         {More Effective Test Case Generation with Multiple Tribes of AI}
         {Faculty of Electrical Engineering, Mathematics, and Computer Science, TUD}
         {2024-02}

\promitem{B. van den Heuvel}
         {Correctly Communicating Software: Distributed, Asynchronous, and Beyond}
         {Faculty of Science and Engineering, RUG}
         {2024-03}
         
\promitem{H.A. Hiep}
         {New Foundations for Separation Logic}
         {Faculty of Mathematics and Natural Sciences, UL}
         {2024-04}
         
\promitem{C.E. Brandt}
         {Test Amplification For and With Developers}
         {Faculty of Electrical Engineering, Mathematics, and Computer Science, TUD}
         {2024-05}
         
\promitem{J.I. Hejderup}
         {Fine-Grained Analysis of Software Supply Chains}
         {Faculty of Electrical Engineering, Mathematics, and Computer Science, TUD}
         {2024-06}

\promitem{J. Jacobs}
         {Guarantees by construction}
         {Faculty of Science, Mathematics and Computer Science, RU}
         {2024-07}

\promitem{O. Bunte}
         {Cracking OIL: A Formal Perspective on an Industrial DSL for Modelling Control Software}
         {Faculty of Mathematics and Computer Science, TU/e}
         {2024-08}
         
\promitem{R.J.A. Erkens}
         {Automaton-based Techniques for Optimized Term Rewriting}
         {Faculty of Mathematics and Computer Science, TU/e}
         {2024-09}
         
\promitem{J.J.M. Martens}
         {The Complexity of Bisimilarity by Partition Refinement}
         {Faculty of Mathematics and Computer Science, TU/e}
         {2024-10}

\promitem{L.J. Edixhoven}
         {Expressive Specification and Verification of Choreographies}
         {Faculty of Science, OU}
         {2024-11}

\promitem{J.W.N. Paulus}
         {On the Expressivity of Typed Concurrent Calculi}
         {Faculty of Science and Engineering, RUG}
         {2024-12}

\end{multicols}

\fi

\end{document}